\documentclass[a4paper,10pt,twoside,onecolumn,final]{article}

\usepackage{amssymb, amsbsy, amsthm, amsmath, bm, bbm, array, enumerate, calc, scalerel, float, graphicx} 

\usepackage[retainorgcmds]{IEEEtrantools}

\newcommand{\ib}{\bfseries\itshape}
\newcommand{\mi}{\mathit} 
\DeclareMathAlphabet{\mib}{T1}{cmr}{bx}{it}
\newcommand{\mc}{\mathcal}

\newcommand{\mf}{\mathfrak}

\newcommand{\mbb}{\mathbb} 
\newcommand{\mbbm}{\mathbbm}
\newcommand{\bs}{\boldsymbol}
\newcommand{\bmm}{\boldmath}

\newcommand{\cref}[1]{\arabic{#1}}
\newcommand{\ctag}[2]{\addtocounter{equation}{1}\setcounter{#1}{\value{equation}}
                      \tag*{#2 (\cref{#1})}}

\newcommand{\geql}{\eqcirc}
\newcommand{\gleq}{\preceq}
\newcommand{\gle}{\prec}   

\newcommand{\feql}{\operatorname{\bs{\eqcirc}}}
\newcommand{\fleq}{\operatorname{\bs{\preceq}}}
\newcommand{\fle}{\operatorname{\bs{\prec}}}

\newcommand{\gz}{\mi{0}}
\newcommand{\gu}{\mi{1}}

\newcommand{\fcom}[1]{\pmb{\overline{\phantom{#1}}}\mspace{-9.75mu}#1}
\newcommand{\fcomd}[1]{\pmb{\overline{\phantom{#1}}}\mspace{-47.50mu}#1}
\newcommand{\fcoml}{\pmb{\overline{\parbox{7pt}{\mbox{}\vspace{6.6pt}\mbox{}}}}}

\newcommand{\fvee}{\operatorname{\bs{\vee}}}
\newcommand{\bigfvee}{\operatornamewithlimits{\bs{\bigvee}}}

\newcommand{\fwedge}{\operatorname{\bs{\wedge}}}
\newcommand{\bigfwedge}{\operatornamewithlimits{\bs{\bigwedge}}}

\newcommand{\frightarrow}{\operatorname{\bs{\Rightarrow}}}

\newcommand{\fdiamond}{\operatorname{\bs{\diamond}}}

\newcommand{\del}{\operatorname{\Delta}}
\newcommand{\fdel}{\operatorname{\bs{\Delta}}}

\newcommand{\eqvl}{\equiv}

\newcommand{\worder}{\operatorname{\trianglelefteq}}

\newcommand{\swedge}{\operatorname{\&}}

\newcommand{\fswedge}{\operatorname{\cdot}}
\newcommand{\bigfswedge}{\prod}

\newcommand{\Cn}{\mi{Cn}}

\newcommand{\mult}{\odot}
\newcommand{\bigmult}{\operatornamewithlimits{\scaleto{\odot}{8pt}}}
\newcommand{\bigO}{\scaleto{O}{10pt}}

\newcommand{\bigsetver}{\scaleto[4pt]{|}{45pt}}

\newcommand{\bsquare}{\bs{\square}}

\newlength{\llll}

\newtheorem{theorem}{Theorem}
\newtheorem{lemma}{Lemma}
\newtheorem{corollary}{Corollary}

\begin{document}

\markboth{\small Du\v{s}an Guller}{\small A {\it DPLL} Procedure with Dichotomous Branching for Propositional Product Logic}


\title{\bf A {\ib DPLL} Procedure with Dichotomous Branching for Propositional Product Logic}

\author{Du\v{s}an Guller \\[2mm]
        Department of Applied Informatics, Comenius University \\
        Mlynsk\'a dolina, Bratislava, 84248, Slovakia \\
        guller@fmph.uniba.sk}

\date{}

\maketitle


\begin{abstract}
The propositional product logic is one of the basic fuzzy logics with continuous $t$-norms, exploiting the multiplication $t$-norm on the unit interval $[0,1]$.
Our aim is to combine well-established automated deduction (theorem proving) with fuzzy inference.
As a first step, we devise a modification of the procedure of Davis, Putnam, Logemann, and Loveland ({\it DPLL}) with dichotomous branching inferring in the product logic.
We prove that the procedure is refutation sound and finitely complete.
As a consequence, solutions to the deduction, satisfiability, and validity problems will be proposed for the finite case.
The presented results are applicable to a design of intelligent systems, exploiting some kind of multi-step fuzzy inference.
\\[2mm]
\end{abstract}


\allowdisplaybreaks

\section{Introduction}
\label{S1}

Intelligent (artificial) systems have a rather long development history, surprisingly for sixty to seventy years.
They are designed to solve complex real-world problems as well as for performing sophisticated decision making and dealing with uncertainty.
Since traditional algorithmic techniques (hard computing) are of limited use because of description and computational complexity, 
intelligent systems rely on methods of soft computing such as machine learning, Bayesian and Markov networks, fuzzy logics, artificial neural networks, evolutionary computation, 
metaheuristics, swarm intelligence-based techniques, inductive reasoning, etc.
A fuzzy logic approach seems to be very promising for two key reasons.
Its strong declarative power is based on the many-valued logics \cite{NOPEMO99,Haj01,ESGO01,NOPEDV16}, 
which provide the mathematically correct and advanced semantics of uncertainty. 
In addition, it can offer computational efficiency exploiting methods of automated deduction (theorem proving) \cite{BIHEMAWA09,SCTO13},
extended to the many-valued case.

The product, G\"{o}del, and {\L}ukasiewicz logics are the three basic fuzzy logics with continuous $t$-norms \cite{MOSH57}.
However, the product logic appears to be the most difficult one \cite{HAGOES96,CITO00}.
It is not finitely approximatisable \cite{Mun87,AGCI00}, opposite to G\"{o}del and {\L}ukasiewicz logics.
The G\"{o}del (minimum) $t$-norm is idempotent, hence not Archimedean \cite{KLME05,CIOTMU13,KLMEPA13}.
The product and {\L}ukasiewicz logics are both Archimedean.
The {\L}ukasiewicz $t$-norm is nilpotent but the multiplication $t$-norm not; it is strict.

Fuzzy rules and one-step fuzzy inference is widely exploited in fuzzy controllers since the seventies.
From a viewpoint of artificial intelligence, such a kind of inference has a reactive behavior.
In contrast to control purposes, one-step fuzzy inference is not sufficient for so-called fuzzy reasoning, 
where some kind of abstract inference is needed to reach a reasonable conclusion, which stipulates multiple inference steps.
The satisfiability ({\it SAT}), validity ({\it VAL}), and deduction ({\it DED}) problems are the primary ones studied in the area of automated deduction.
{\it SAT} solvers are commonly powered by the procedure of Davis, Putnam, Logemann, and Loveland (algorithm) ({\it DPLL}) \cite{DAPU60,DALOLO62} and its refinements
(e.g. conflict-driven clause learning ({\it CDCL}) \cite{SISA96,SISA99}).
In the Boolean (two-valued) propositional case, both the {\it VAL} ($\models \phi$) and {\it DED} ($T\models \phi$)%
\footnote{$\phi$ is a propositional formula and $T$ a propositional theory.}
problems can be reduced to the unsatisfiability problem of $(T\cup \mbox{}) \{\neg \phi\}$.
The underlying rule of the {\it DPLL} procedure is a branching one of the following form:%
\footnote{$l$ is a classical literal; an atom or the negation of an atom; and $S$ is a set of classical clauses; sets of classical literals.}
\begin{alignat*}{1}
\tag*{({\it Branching rule})} \\[1mm]
& \dfrac{S}
        {S\cup \{l\}\ \big|\ S\cup \{\neg l\}}; \\[1mm]
& l\ \text{\it occurs in}\ S.
\end{alignat*}
In \cite{Guller2018a}, we have introduced a {\it DPLL} procedure for the propositional G\"{o}del logic, 
augmented by connectives $\geql$, equality, $\gle$, strict order, and truth constants $\gz$, the false, $\gu$, the true (cf. Section \ref{S2}).
The {\it DED} problem $T\models \phi$ is reducible to the unsatisfiability problem of $T\cup \{\phi\gle \gu\}$, but not of $T\cup \{\neg \phi\}$. 
The classical notion of literal has been replaced by a notion of order literal.
An order literal is a formula of the form $\varepsilon_1\diamond \varepsilon_2$ 
where $\varepsilon_i$ is an atom or a truth constant $\gz$ or $\gu$, and $\diamond$ is a connective $\geql$ or $\gle$.
An order clause is a finite set of order literals as usual.
An order clause $\varepsilon_1\gle \varepsilon_2\vee \varepsilon_1\geql \varepsilon_2\vee \varepsilon_2\gle \varepsilon_1$ is called a trichotomy.
The linearity of the standard strict order $<$ on $[0,1]$ naturally induces a branching rule of a trichotomy form%
\footnote{$l_i$ is an order literal; $C$ an order clause; $S$ a set of order clauses.}
\begin{alignat*}{1}
\tag*{({\it Trichotomy branching rule})} \\[1mm]
& \dfrac{S}
        {(S-\{l_1\vee C\})\cup \{l_1\}\ \big|\ (S-\{l_1\vee C\})\cup \{C\}\cup \{l_2\}\ \big|\ (S-\{l_1\vee C\})\cup \{C\}\cup \{l_3\}}; \\[1mm]
& l_1\vee C\in S, C\neq \square; l_1\vee l_2\vee l_3\ \text{\it is a trichotomy}.
\end{alignat*}
In the propositional product logic, the situation is more complicated.
We again augment the propositional logic by the connectives $\geql$, $\gle$, and the truth constants $\gz$, $\gu$.
For technical purposes, we add a connective $\gleq$, order, as well; 
a formula $\varepsilon_1\gleq \varepsilon_2$ may be considered as an abbreviation for the order clause $\varepsilon_1\gle \varepsilon_2\vee \varepsilon_1\geql \varepsilon_2$. 
The multiplication $t$-norm is not idempotent, in contrast to the G\"{o}del (minimum) $t$-norm.
Hence, the structure of order literal has to allow for conjunctions of powers of atoms.
An order literal is a formula of the form $\varepsilon_1\diamond \varepsilon_2$ 
where $\varepsilon_i$ is a truth constant or a conjunction of powers of atoms of the form $a^n$,%
\footnote{$a$ is an atom and $n\geq 1$ a natural number.}
interpreted using the multiplication $t$-norm, and $\diamond$ is a connective $\geql$ or $\gleq$ or $\gle$.
In the product logic, the standard cancellation law holds:%
\footnote{$\bs{\Pi}$ is the standard product algebra (cf. Section \ref{S2}).}
\begin{alignat*}{1}
\tag*{({\it Cancellation law})} \\[1mm]
\text{for all}\ x, y, z\in \bs{\Pi},\ 0<x\longrightarrow (x\fswedge y=x\fswedge z\longrightarrow y=z).
\end{alignat*}
The cancellation law together with the linearity of the standard strict order $<$ on $[0,1]$ permits us to introduce some dichotomous tautologies and related branching rules:
an order clause $a\geql \gz\vee \gz\gle a$ is called a $\gz$-dichotomy (tautology);
\begin{alignat*}{1}
\tag*{({\it $\gz$-dichotomy branching rule})} \\[1mm]
& \dfrac{S}
        {S\cup \{a\geql \gz\}\ \big|\ S\cup \{\gz\gle a\}}; \\[1mm]
& a\in \mi{atoms}(S).
\end{alignat*}
an order clause $\Cn_1\gleq \Cn_2\vee \Cn_2\gle \Cn_1$, $\Cn_1\neq \Cn_2$, is called a pure dichotomy (tautology);%
\footnote{$\Cn$, $\Cn_i$ are conjunctions of powers of atoms.} 
\begin{alignat*}{1}
\tag*{({\it Pure dichotomy branching rule})} \\[1mm]
& \dfrac{S}
        {(S-\{l_1\vee C\})\cup \{l_1\}\ \big|\ (S-\{l_1\vee C\})\cup \{C\}\cup \{l_2\}}; \\[1mm]
& l_1\vee C\in S, C\neq \square; l_1\vee l_2\ \text{\it is a pure dichotomy}.
\end{alignat*}
\begin{alignat*}{1}
\tag*{({\it Pure $\geql$-dichotomy branching rule})} \\[1mm]
& \dfrac{S}
        {\begin{array}{l}
         (S-\{\Cn_1\geql \Cn_2\vee C\})\cup \{\Cn_1\geql \Cn_2\}\ \big| \\ 
         (S-\{\Cn_1\geql \Cn_2\vee C\})\cup \{C\}\cup \{\Cn_1\gle \Cn_2\vee \Cn_2\gle \Cn_1\}
         \end{array}}; \\[1mm]
& \Cn_1\geql \Cn_2\vee C\in S, \Cn_1\neq \Cn_2, C\neq \square.
\end{alignat*}
To obtain a basic calculus of the {\it DPLL} procedure, we add some auxiliary simplification rules.
So, it is sufficient to detect a contradiction only in the case of a unit order clausal theory not containing the truth constant $\gz$;
we assume that all atoms occurring in the unit order clausal theory will be evaluated with truth values greater than $0$.
If there exists a product of powers of order literals occurring in the theory of the form $\Cn\gle \Cn$,   
%
%
%
%
then the theory is unsatisfiable; $\Cn\gle \Cn$ can be cancelled to a contradiction $\gu\gle \gu$.
To solve this case, we propose a unit contradiction rule (cf. Rule (23), Section \ref{S4}).
Our {\it DPLL} procedure will be proved refutation sound and finitely complete.
Some preliminary results have been presented in \cite{Guller2013,Guller2016a,Guller2016d,Guller2019a}.
Other approaches to the solutions to the {\it SAT}, {\it VAL}, and {\it DED} problems in fuzzy logics with continuous $t$-norms exploit 
mixed integer programming ({\it MIP}) \cite{Hah94a,Hah94b,Hah01} or satisfiability modulo theories ({\it SMT}) \cite{ANBOMAVI12,ANBOMAVI16,BOMAVIVI18}.
Formulae are translated to mixed integer programs or to {\it SMT} instances modulo quantifier-free (non-)linear real arithmetic.

The paper is arranged as follows.
Section \ref{S2} introduces the propositional product logic.
Section \ref{S3} describes translation to clausal form.
Section \ref{S4} proposes a basic calculus of the {\it DPLL} procedure.
%
%
Section \ref{S6} summarises the presented results.

\subsection{Preliminaries}
\label{S1.1}

$\mbb{N}$ ($\mbb{Z}$, $\mbb{Q}$, $\mbb{R}$) designates the set of natural (integer, rational, real) numbers, and
$=$ ($\leq$, $<$) denotes the standard equality (order, strict order) on $\mbb{N}$ ($\mbb{Z}$, $\mbb{Q}$, $\mbb{R}$).
We denote 
$\mbb{Q}_0^+=\{c \,|\, 0\leq c\in \mbb{Q}\}$, $\mbb{Q}^+=\{c \,|\, 0<c\in \mbb{Q}\}$,
$\mbb{R}_0^+=\{c \,|\, 0\leq c\in \mbb{R}\}$, $\mbb{R}^+=\{c \,|\, 0<c\in \mbb{R}\}$,
$[0,1]=\{c \,|\, c\in \mbb{R}, 0\leq c\leq 1\}$; $[0,1]$ is called the unit interval;
$(0,1]=\{c \,|\, c\in \mbb{R}, 0<c\leq 1\}$, $(0,1)=\{c \,|\, c\in \mbb{R}, 0<c<1\}$.
%
%
%
%
%
Let $X$, $Y$, $Z$ be sets and $f : X\longrightarrow Y$ be a mapping.
By $X\subseteq_{\mc F} Y$ we denote the relation that $X$ is a finite subset of $Y$.
We denote
${\mc P}(X)=\{x \,|\, x\subseteq X\}$; ${\mc P}(X)$ is called the power set of $X$;
${\mc P}_{\mc F}(X)=\{x \,|\, x\subseteq_{\mc F} X\}$.
By $\mi{card}(X)$ we denote the set-theoretic cardinality of $X$.
By $X\times Y$ we denote the Cartesian product of $X$ and $Y$.
Let $Z\subseteq X$.
We denote 
$f|_Z=\{(z,f(z)) \,|\, z\in Z\}$; $f|_Z$ is called the restriction of $f$ onto $Z$.
Let $\{x_1,\dots,x_n\}$ be a set.
We assume that for all $1\leq i, i'\leq n$ satisfying $i\neq i'$, $x_i\neq x_{i'}$.         
Let $\gamma\leq \omega$.
A sequence $\delta$ of $X$ is a bijection $\delta : \gamma\longrightarrow X$.
Recall that $X$ is countable if and only if there exists a sequence of $X$.
%
%
Let $c\in \mbb{R}^+$.
By $\log c$ we denote the binary logarithm of $c$.
Let $f, g : \mbb{N}\longrightarrow \mbb{R}_0^+$.
$f$ is of the order of $g$, in symbols $f\in O(g)$, iff there exist $n_0$ and $c^*\in \mbb{R}_0^+$ such that for all $n\geq n_0$, $f(n)\leq c^*\cdot g(n)$.

\section{Propositional product logic}
\label{S2}

Throughout the paper, we shall use the common notions and notation of propositional logic. 
The set of all propositional atoms of the product logic will be designated as $\mi{PropAtom}$.
We assume two truth constants $\gz$ and $\gu$, $\gz, \gu\not\in \mi{PropAtom}$; $\gz$ denotes the false and $\gu$ the true in the product logic.
By $\mi{PropForm}$ we designate the set of all propositional formulae of the product logic built up from $\mi{PropAtom}$ and $\{\gz,\gu\}$ using the connectives: 
$\neg$, negation, $\del$, Delta, 
$\wedge$, conjunction, $\swedge$, strong conjunction, $\vee$, disjunction, $\rightarrow$, implication, $\leftrightarrow$, equivalence, 
$\geql$, equality, $\gleq$, order, $\gle$, strict order.%
\footnote{We assume a decreasing connective precedence: $\neg$, $\del$, $\swedge$, $\geql$, $\gleq$, $\gle$, $\wedge$, $\vee$, $\rightarrow$, $\leftrightarrow$.}
In the paper, we shall assume that $\mi{PropAtom}$ is countably infinite; hence, $\mi{PropForm}$ is also countably infinite. 
Let $\varepsilon_i$, $1\leq i\leq n$, be either an expression or a set of expressions or a set of sets of expressions, in general.
By $\mi{atoms}(\varepsilon_1,\dots,\varepsilon_n)\subseteq \mi{PropAtom}$ we denote the set of all propositional atoms occurring in $\varepsilon_1,\dots,\varepsilon_n$.
%
%
%
%
Let $\phi\in \mi{PropForm}$ and $T\subseteq_{\mc F} \mi{PropForm}$.
We define the size $|\phi|$ of $\phi$ by recursion on the structure of $\phi$ as follows:
\begin{alignat*}{1}
|\phi| &= \left\{\begin{array}{ll}
                 1                   &\ \text{\it if}\ \phi\in \mi{PropAtom}\cup \{\gz,\gu\}, \\[1mm]
                 1+|\phi_1|          &\ \text{\it if}\ \phi=\diamond \phi_1, \phi_1\in \mi{PropForm}, \diamond\in \{\neg,\del\}, \\[1mm] 
                 1+|\phi_1|+|\phi_2| &\ \text{\it if}\ \phi=\phi_1\diamond \phi_2, \phi_i\in \mi{PropForm}, \diamond\in \{\wedge,\swedge,\vee,\rightarrow,\leftrightarrow,\geql,\gleq,\gle\};   
                 \end{array}
          \right. \\[1mm]
       &\geq 1.
\end{alignat*}
The size $|T|$ of $T$ is defined as $|T|=\sum_{\phi\in T} |\phi|$.

The product logic is interpreted by the standard $\bs{\Pi}$-algebra augmented with operators $\feql$, $\fleq$, $\fle$, $\fdel$ for the connectives $\geql$, $\gleq$, $\gle$, $\del$, respectively.
\begin{equation}
\notag
\bs{\Pi}=([0,1],\leq,\fvee,\fwedge,\fswedge,\frightarrow,\fcoml,\feql,\fleq,\fle,\fdel,0,1)
\end{equation}
where $\fvee$ ($\fwedge$) denotes the supremum (infimum) operator on $[0,1]$;
the operators $\frightarrow, \fcoml, \feql, \fleq, \fle, \fdel : [0,1]\times [0,1]\longrightarrow [0,1]$ are defined as follows:
\begin{alignat*}{2}
a\frightarrow b &= \left\{\begin{array}{ll}
                          1           &\ \text{\it if}\ a\leq b, \\[1mm]
                          \frac{b}{a} &\ \text{\it else};
                          \end{array}
                   \right. 
& 
\fcom{a}        &= \left\{\begin{array}{ll}
                          1 &\ \text{\it if}\ a=0, \\[1mm]
                          0 &\ \text{\it else};
                          \end{array}
                   \right. 
\\[1mm]
a\feql b        &= \left\{\begin{array}{ll}
                          1 &\ \text{\it if}\ a=b, \\[1mm]
                          0 &\ \text{\it else};
                          \end{array}
                   \right. 
& \qquad
a\fleq b        &= \left\{\begin{array}{ll}
                          1 &\ \text{\it if}\ a\leq b, \\[1mm]
                          0 &\ \text{\it else};
                          \end{array}
                   \right. 
\\[1mm]
a\fle b         &= \left\{\begin{array}{ll}
                          1 &\ \text{\it if}\ a<b, \\[1mm]
                          0 &\ \text{\it else};
                          \end{array}
                   \right. 
& 
\fdel a         &= \left\{\begin{array}{ll}
                          1 &\ \text{\it if}\ a=1, \\[1mm]
                          0 &\ \text{\it else}.
                          \end{array}
                   \right.
\end{alignat*}
Recall that $\bs{\Pi}$ is a complete linearly ordered lattice algebra;
$\fvee$ ($\fwedge$) is commutative, associative, idempotent, monotone; $0$ ($1$) is its neutral element; 
the multiplication $t$-norm $\fswedge$ is commutative, associative, monotone; $1$ is its neutral element;
the residuum operator $\frightarrow$ of $\fswedge$ satisfies the condition of residuation:
\begin{equation}
\label{eq0a}
\text{for all}\ a, b, c\in \bs{\Pi},\ a\fswedge b\leq c\ \text{if and only if}\ a\leq b\frightarrow c;
\end{equation}
G\"{o}del negation $\fcoml$ satisfies the condition:
\begin{equation}
\label{eq0b}
\text{for all}\ a\in \bs{\Pi},\ \fcom{a}=a\frightarrow 0;
\end{equation}
the order operator $\fleq$ satisfies the conditions:
\begin{alignat}{1}
\label{eq0k}
& \text{for all}\ a, b\in \bs{\Pi},\ a\fleq b=a\fle b\fvee a\feql b, \\
\ctag{ceq0kk}{({\it Antisymmetry})}
& \phantom{\text{for all}\ a, b\in \bs{\Pi},\ \mbox{}}
                                     a\fleq b\fwedge b\fleq a=a\feql b;
\end{alignat}
the projection operator $\fdel$ satisfies the condition:%
\footnote{We assume a decreasing operator precedence: $\fcoml$, $\fdel$, $\fswedge$, $\feql$, $\fleq$, $\fle$, $\fwedge$, $\fvee$, $\frightarrow$.}
\begin{equation}
\label{eq0kkk}   
\text{for all}\ a\in \bs{\Pi},\ \fdel a=a\feql 1.
\end{equation}

A valuation ${\mc V}$ of propositional atoms is a mapping ${\mc V} : \mi{PropAtom}\longrightarrow [0,1]$. 
Let ${\mc V}$ be a valuation.
We define the truth value $\|\phi\|^{\mc V}$ of $\phi$ in ${\mc V}$ by recursion on the structure of $\phi$ as follows:
\begin{alignat*}{1}
\|\phi\|^{\mc V} &= \left\{\begin{array}{ll}
                           {\mc V}(\phi)              &\ \text{\it if}\ \phi\in \mi{PropAtom}, \\[1mm]
                           0                          &\ \text{\it if}\ \phi=\gz, \\[1mm]
                           1                          &\ \text{\it if}\ \phi=\gu, \\[1mm]
                           \fcomd{\|\phi_1\|^{\mc V}} &\ \text{\it if}\ \phi=\neg \phi_1, \phi_1\in \mi{PropForm}, \\[1mm]
                           \fdel \|\phi_1\|^{\mc V}   &\ \text{\it if}\ \phi=\del \phi_1, \phi_1\in \mi{PropForm}, \\[1mm]
                           \|\phi_1\|^{\mc V}\fdiamond \|\phi_2\|^{\mc V} &\ \text{\it if}\ \phi=\phi_1\diamond \phi_2, \phi_i\in \mi{PropForm}, \\
                                                                          &\ \phantom{\text{\it if}\ \mbox{}} 
                                                                                                     \diamond\in \{\wedge,\swedge,\vee,\rightarrow,\geql,\gleq,\gle\}, \\[1mm]
                           (\|\phi_1\|^{\mc V}\frightarrow \|\phi_2\|^{\mc V})\fwedge (\|\phi_2\|^{\mc V}\frightarrow \|\phi_1\|^{\mc V}) &\ \text{\it if}\ \phi=\phi_1\leftrightarrow \phi_2,
                                                                                                                                             \phi_i\in \mi{PropForm};
                           \end{array}
                    \right. \\[1mm]
                 &\in [0,1].
\end{alignat*}
A propositional theory is a subset of $\mi{PropForm}$.
Let $\phi'\in \mi{PropForm}$ and $T\subseteq \mi{PropForm}$.
$\phi$ is true in ${\mc V}$, written as ${\mc V}\models \phi$, iff $\|\phi\|^{\mc V}=1$.
${\mc V}$ is a model of $T$, in symbols ${\mc V}\models T$, iff, for all $\phi\in T$, ${\mc V}\models \phi$.
$\phi$ is a tautology iff, for every valuation ${\mc V}$, ${\mc V}\models \phi$.
$\phi$ is equivalent to $\phi'$, in symbols $\phi\eqvl \phi'$, iff, for every valuation ${\mc V}$, $\|\phi\|^{\mc V}=\|\phi'\|^{\mc V}$.

\section{Translation to clausal form}
\label{S3}

In this section, we propose translation to clausal form.
We develop a fragment of order clauses for the product logic in a gradual way.
We firstly introduce a notion of power of a propositional atom and a notion of conjunction of powers of propositional atoms.
Let $r\geq 1$ and $a\in \mi{PropAtom}$.
An $r$th power of $a$ ($a$ raised to the power of $r$) is a pair $(a,r)$, written as $a^r$.
A power $a^1$ is denoted as $a$; if it does not cause the ambiguity with the denotation of the single atom $a$ in a given context.
The set of all powers of propositional atoms is designated as $\mi{PropPow}$.
Let $a^r\in \mi{PropPow}$ and ${\mc V}$ be a valuation.
The truth value $\|a^r\|^{\mc V}$ of $a^r$ in ${\mc V}$ is defined as follows:
\begin{equation}
\notag
\|a^r\|^{\mc V}=\underbrace{\|a\|^{\mc V}\fswedge\cdots\fswedge \|a\|^{\mc V}}_r\in [0,1].
\end{equation}
We define the size $|a^r|$ of $a^r$ as $|a^r|=1+r\geq 2$.
A conjunction $\Cn$ of powers of propositional atoms is a non-empty finite subset of $\mi{PropPow}$ such that 
for all $a_1^{r_1}, a_2^{r_2}\in \Cn$, $a_i\in \mi{PropAtom}$, $r_i\geq 1$, satisfying $a_1^{r_1}\neq a_2^{r_2}$, $a_1\neq a_2$.
$\{a_0^{r_0},\dots,a_n^{r_n}\}\subseteq \mi{PropPow}$, $a_i\in \mi{PropAtom}$, $r_i\geq 1$, is written in the form $a_0^{r_0}\swedge\cdots\swedge a_n^{r_n}$.
$\{p\}\subseteq \mi{PropPow}$ is called unit and denoted as $p$; if it does not cause the ambiguity with the denotation of the single power $p$ in a given context.
The set of all conjunctions of powers of propositional atoms is designated as $\mi{PropConj}$.
Let $\Cn, \Cn_1, \Cn_2\in \mi{PropConj}$.
The truth value $\|\Cn\|^{\mc V}$ of $\Cn=a_0^{r_0}\swedge\cdots\swedge a_n^{r_n}$, $a_i\in \mi{PropAtom}$, $r_i\geq 1$, in ${\mc V}$ is defined as follows:
\begin{equation}
\notag
\|\Cn\|^{\mc V}=\|a_0^{r_0}\|^{\mc V}\fswedge\cdots\fswedge \|a_n^{r_n}\|^{\mc V}\in [0,1].    
\end{equation}
We define the size $|\Cn|$ of $\Cn$ as $|\Cn|=\sum_{p\in \Cn} |p|\geq 2$.
We denote $a^r\swedge \Cn=\{a^r\}\cup \Cn$ such that $a\not\in \mi{atoms}(\Cn)$.
$\Cn_1$ is a subconjunction of $\Cn_2$, in symbols $\Cn_1\sqsubseteq \Cn_2$, iff, 
for all $b^{r_1}\in \Cn_1$, $b\in \mi{PropAtom}$, $r_1\geq 1$, there exists $b^{r_2}\in \Cn_2$, $r_2\geq 1$, such that $r_1\leq r_2$.
$\Cn_1$ is a proper subconjunction of $\Cn_2$, in symbols $\Cn_1\sqsubset \Cn_2$, iff $\Cn_1\sqsubseteq \Cn_2$ and $\Cn_1\neq \Cn_2$.

We finally introduce order clauses in the product logic.
Let $\varepsilon_1, \varepsilon_2\in \{\gz,\gu\}\cup \mi{PropConj}$, $\diamond\in \{\geql,\gleq,\gle\}$, $\diamond^\#\in \{\gleq,\gle\}$.
A literal is an expression of the form $\varepsilon_1\diamond \varepsilon_2$.
The set of all literals is designated as $\mi{PropLit}$.
Let $l\in \mi{PropLit}$.
$l=\varepsilon_1\diamond \varepsilon_2$ is a pure literal iff, for both $i$, $\varepsilon_i\in \mi{PropConj}$; $l$ does not contain $\gz$ and $\gu$.
The set of all pure literals is designated as $\mi{PurPropLit}\subseteq \mi{PropLit}$.
The truth value $\|l\|^{\mc V}$ of $l=\varepsilon_1\diamond \varepsilon_2$ in ${\mc V}$ is defined by $\|l\|^{\mc V}=\|\varepsilon_1\|^{\mc V}\fdiamond \|\varepsilon_2\|^{\mc V}\in [0,1]$.
An order literal is a pair either $(\geql,\{\varepsilon_1,\varepsilon_2\})$, written as $\varepsilon_1\geql \varepsilon_2$, 
or $(\diamond^\#,(\varepsilon_1,\varepsilon_2))$, written as $\varepsilon_1\diamond^\# \varepsilon_2$.
Note that $\varepsilon_1\geql \varepsilon_2=\varepsilon_2\geql \varepsilon_1=(\geql,\{\varepsilon_1,\varepsilon_2\})$, and if $\varepsilon_1=\varepsilon_2$, 
$\varepsilon_1\geql \varepsilon_2=(\geql,\{\varepsilon_1\})$.
The set of all order literals is designated as $\mi{OrdPropLit}$.
Let $l\in \mi{OrdPropLit}$.
$l=\varepsilon_1\diamond \varepsilon_2$ is a pure order literal iff, for both $i$, $\varepsilon_i\in \mi{PropConj}$; $l$ does not contain $\gz$ and $\gu$.
The set of all pure order literals is designated as $\mi{PurOrdPropLit}\subseteq \mi{OrdPropLit}$.
If $l=\varepsilon_1\geql \varepsilon_2$, then we denote $\langle l\rangle=\varepsilon_1\geql \varepsilon_2\in \mi{PropLit}$ or $\langle l\rangle=\varepsilon_2\geql \varepsilon_1\in \mi{PropLit}$.
If $l=\varepsilon_1\diamond^\# \varepsilon_2$, then we denote $\langle l\rangle=\varepsilon_1\diamond^\# \varepsilon_2\in \mi{PropLit}$.
The truth value $\|l\|^{\mc V}$ of $l=\varepsilon_1\diamond \varepsilon_2$ in ${\mc V}$ is defined by $\|l\|^{\mc V}=\|\varepsilon_1\|^{\mc V}\fdiamond \|\varepsilon_2\|^{\mc V}\in [0,1]$.
Note that $\feql$ is commutative and $\|\varepsilon_1\|^{\mc V}\feql \|\varepsilon_2\|^{\mc V}=\|\varepsilon_2\|^{\mc V}\feql \|\varepsilon_1\|^{\mc V}$.
$l$ is true in ${\mc V}$, written as ${\mc V}\models l$, iff $\|l\|^{\mc V}=1$.
Note that ${\mc V}\models l$ if and only if 
either $l=\varepsilon_1\geql \varepsilon_2$, $\|l\|^{\mc V}=\|\varepsilon_1\geql \varepsilon_2\|^{\mc V}=\|\varepsilon_1\|^{\mc V}\feql \|\varepsilon_2\|^{\mc V}=1$, 
$\|\varepsilon_1\|^{\mc V}=\|\varepsilon_2\|^{\mc V}$, 
or $l=\varepsilon_1\gleq \varepsilon_2$, $\|l\|^{\mc V}=\|\varepsilon_1\gleq \varepsilon_2\|^{\mc V}=\|\varepsilon_1\|^{\mc V}\fleq \|\varepsilon_2\|^{\mc V}=1$, 
$\|\varepsilon_1\|^{\mc V}\leq \|\varepsilon_2\|^{\mc V}$, 
or $l=\varepsilon_1\gle \varepsilon_2$, $\|l\|^{\mc V}=\|\varepsilon_1\gle \varepsilon_2\|^{\mc V}=\|\varepsilon_1\|^{\mc V}\fle \|\varepsilon_2\|^{\mc V}=1$, 
$\|\varepsilon_1\|^{\mc V}<\|\varepsilon_2\|^{\mc V}$.
We define the size $|l|$ of $l=\varepsilon_1\diamond \varepsilon_2$ as $|l|=1+|\varepsilon_1|+|\varepsilon_2|\geq 3$.
An (pure) order clause is a finite subset of $\mi{OrdPropLit}$ ($\mi{PurOrdPropLit}$).
$\{l_0,\dots,l_n\}\subseteq \mi{OrdPropLit}$ is written in the form $l_0\vee\cdots\vee l_n$.
$\emptyset\subseteq \mi{OrdPropLit}$ is called the empty order clause and denoted as $\square$.
$\{l\}\subseteq \mi{OrdPropLit}$ is called unit and denoted as $l$; if it does not cause the ambiguity with the denotation of the single order literal $l$ in a given context.
The set of all (pure) order clauses is designated as $\mi{OrdPropCl}$ ($\mi{PurOrdPropCl}$).
Note that $\mi{PurOrdPropCl}\subseteq \mi{OrdPropCl}$.
Let $l_0,\dots,l_n\in \mi{OrdPropLit}$ and $C, C'\in \mi{OrdPropCl}$.
We define the size $|C|$ of $C$ as $|C|=\sum_{l\in C} |l|$.
We denote $l_0\vee\cdots\vee l_n\vee C=\{l_0,\dots,l_n\}\cup C$ such that for all $i, i'\leq n$ satisfying $i\neq i'$, $l_i\not\in C$ and $l_i\neq l_{i'}$.
We denote $C\vee C'=C\cup C'$.
$C$ is a subclause of $C'$, in symbols $C\sqsubseteq C'$, iff $C\subseteq C'$.
An (pure) order clausal theory is a subset of $\mi{OrdPropCl}$ ($\mi{PurOrdPropCl}$).
A unit order clausal theory is a subset of $\mi{OrdPropCl}$ consisting of unit order clauses.

Let $\phi, \phi'\in \mi{PropForm}$, $T, T'\subseteq \mi{PropForm}$, $S, S'\subseteq \mi{OrdPropCl}$.
$C$ is true in ${\mc V}$, written as ${\mc V}\models C$, iff there exists $l^*\in C$ such that ${\mc V}\models l^*$.
${\mc V}$ is a model of $S$, in symbols ${\mc V}\models S$, iff, for all $C\in S$, ${\mc V}\models C$.
Let $\upsilon\in \{\phi,T,C,S\}$ and $\upsilon'\in \{\phi',T',C',S'\}$. 
$\upsilon'$ is a propositional consequence of $\upsilon$, in symbols $\upsilon\models \upsilon'$, iff, for every valuation ${\mc V}$, if ${\mc V}\models \upsilon$, then ${\mc V}\models \upsilon'$.
$\upsilon$ is satisfiable iff there exists a valuation ${\mc V}^*$ such that ${\mc V}^*\models \upsilon$.
$\upsilon$ is unsatisfiable iff $\upsilon$ is not satisfiable.
$\upsilon$ is equisatisfiable to $\upsilon'$ iff $\upsilon$ is satisfiable if and only if $\upsilon'$ is satisfiable.
Let $S\subseteq_{\mc F} \mi{OrdPropCl}$.
We define the size $|S|$ of $S$ as $|S|=\sum_{C\in S} |C|$.

We denote $\mbb{I}=\mbb{N}\times \mbb{N}$; $\mbb{I}$ will serve as a countably infinite index set.
Since $\mi{PropAtom}$ is countably infinite, there exist $\mbb{O}, \tilde{\mbb{A}}\subseteq \mi{PropAtom}$ such that $\mbb{O}\cup \tilde{\mbb{A}}=\mi{PropAtom}$, $\mbb{O}\cap \tilde{\mbb{A}}=\emptyset$,
both are countably infinite, $\tilde{\mbb{A}}=\{\tilde{a}_\mbbm{i} \,|\, \mbbm{i}\in \mbb{I}\}$.
Note that $\gz, \gu\not\in \mbb{O}$ and $\gz, \gu\not\in \tilde{\mbb{A}}$.
The purpose of $\mbb{O}$ and $\tilde{\mbb{A}}$ is as follows.
During translation to clausal form, we need to introduce fresh atoms, not occurring in the input and in previous intermediate results.
Therefore, we confine every input to contain atoms only from $\mbb{O}$ and choose fresh atoms from $\tilde{\mbb{A}}$.
Let $A\subseteq \tilde{\mbb{A}}$.
We denote 
\begin{alignat*}{1}
\mi{PropForm}_A   &= \{\phi \,|\, \phi\in \mi{PropForm}, \mi{atoms}(\phi)\subseteq \mbb{O}\cup A\}, \\
\mi{OrdPropLit}_A &= \{l \,|\, l\in \mi{OrdPropLit}, \mi{atoms}(l)\subseteq \mbb{O}\cup A\}, \\
\mi{OrdPropCl}_A  &= \{C \,|\, C\in \mi{OrdPropCl}, \mi{atoms}(C)\subseteq \mbb{O}\cup A\}.
\end{alignat*}

\subsection{A technical treatment}
\label{S3.2}

Informally speaking, there are two ways how to translate a formula or theory to clausal form.
One way is to propose a suitable conjunctive normal form ({\it CNF}) and then to translate a formula $\phi$ to an equivalent $\mi{CNF}_\phi$.
However, to preserve the equivalence $\phi\eqvl \mi{CNF}_\phi$ often stipulates unrestricted applications of the distributivity law of $\vee$ over $\wedge$, 
which produces the output $\mi{CNF}_\phi$ of exponential size with respect to the size of $\phi$.
To avoid this disadvantage, another way is to propose translation to clausal form via interpolation using new atoms,
which produces an output order clausal theory $S^\phi$ of linear size respect to the size of $\phi$ at the cost of being only equisatisfiable to $\phi$.
The translation is of linearithmic time complexity.
Such equisatisfiability preserving translation is though sufficient for our purposes.
A similar renaming subformulae technique has been presented in \cite{Tse70,PLGR86,Boy92,Hah94b,NOROWE98,She04}.
Translation of a formula or theory to clausal form is based on the following lemmata:

\begin{lemma}
\label{le111}
Let $n_\theta\in \mbb{N}$ and $\theta\in \mi{PropForm}_\emptyset$.
There exists $\xi\in \mi{PropForm}_\emptyset$ such that 
\begin{enumerate}[\rm (a)]
\item
$\xi\eqvl \theta$; 
\item 
$|\xi|\leq 2\cdot |\theta|$,
$\xi$ can be built up from $\theta$ via a postorder traversal of $\theta$ with $\#{\mc O}(\theta)\in O(|\theta|)$ and the time complexity in $O(|\theta|\cdot (\log (1+n_\theta)+\log |\theta|))$;
\item
$\xi$ does not contain $\neg$ and $\del$; 
\item
either $\xi\in \{\gz,\gu\}$, 
or for every subformula of $\xi$ of the form $\xi_1\diamond \xi_2$, $\xi_i\in \mi{PropForm}_\emptyset$, $\diamond\in \{\wedge,\swedge,\vee,\leftrightarrow,\gleq\}$, 
for both $i$, $\xi_i\neq \gz, \gu$, and
for every subformula of $\xi$ of the form $\xi_1\rightarrow \xi_2$, $\xi_i\in \mi{PropForm}_\emptyset$, $\xi_1\neq \gz, \gu$, $\xi_2\neq \gu$, and
for every subformula of $\xi$ of the form $\xi_1\geql \xi_2$, $\xi_i\in \mi{PropForm}_\emptyset$, $\xi_1\neq \gz, \gu$, and
for every subformula of $\xi$ of the form $\xi_1\gle \xi_2$, $\xi_i\in \mi{PropForm}_\emptyset$, $\xi_1\neq \gu$, $\xi_2\neq \gz$, $\xi_1\gle \xi_2\neq \gz\gle \gu$.
\end{enumerate}
\end{lemma}

\begin{proof}
The proof is by induction on the structure of $\theta$ using (\ref{eq0b}), (\ref{eq0kkk}), the obvious simplification identities over $\bs{\Pi}$. 
%
%
%
\end{proof}

\begin{lemma}
\label{le11}
Let $n_\theta, j_\mbbm{i}\in \mbb{N}$, $\theta\in \mi{PropForm}_\emptyset-\{\gz,\gu\}$, {\rm (c,d)} of Lemma \ref{le111} hold for $\theta$,
$\mbbm{i}=(n_\theta,j_\mbbm{i})\in \{(n_\theta,j) \,|\, j\in \mbb{N}\}\subseteq \mbb{I}$, $\tilde{a}_\mbbm{i}\in \tilde{\mbb{A}}$.
There exist $n_J\geq j_\mbbm{i}$, $J=\{(n_\theta,j) \,|\, j_\mbbm{i}+1\leq j\leq n_J\}\subseteq \{(n_\theta,j) \,|\, j\in \mbb{N}\}\subseteq \mbb{I}$, $\mbbm{i}\not\in J$, 
$S\subseteq_{\mc F} \mi{OrdPropCl}_{\{\tilde{a}_\mbbm{i}\}\cup \{\tilde{a}_\mbbm{j} \,|\, \mbbm{j}\in J\}}$ such that 
\begin{enumerate}[\rm (a)]
\item
$\mi{card}(J)\leq |\theta|-1$; 
\item
there exists a valuation ${\mf A}$ satisfying ${\mf A}\models \tilde{a}_\mbbm{i}\leftrightarrow \theta\in \mi{PropForm}_{\{\tilde{a}_\mbbm{i}\}}$ if and only if 
there exists a valuation ${\mf A}'$ satisfying ${\mf A}'\models S$; 
${\mf A}|_{\mbb{O}\cup \{\tilde{a}_\mbbm{i}\}}={\mf A}'|_{\mbb{O}\cup \{\tilde{a}_\mbbm{i}\}}$;
\item
$|S|\leq 25\cdot |\theta|$, 
$S$ can be built up from $\theta$ via a preorder traversal of $\theta$ with $\#{\mc O}(\theta)\in O(|\theta|)$; 
\item
for all $C\in S$, $\emptyset\neq \mi{atoms}(C)\cap \tilde{\mbb{A}}\subseteq \{\tilde{a}_\mbbm{i}\}\cup \{\tilde{a}_\mbbm{j} \,|\, \mbbm{j}\in J\}$;
$\tilde{a}_\mbbm{i}\geql \gu, \tilde{a}_\mbbm{i}\gle \gu\not\in S$.
\end{enumerate}
\end{lemma}

\begin{proof}
The proof is by induction on the structure of $\theta$ using the binary and unary interpolation rules in Tables \ref{tab2} and \ref{tab3}.
\begin{table}
\caption{Binary interpolation rules for $\wedge$, $\swedge$, $\vee$, $\rightarrow$, $\leftrightarrow$, $\geql$, $\gleq$, and $\gle$}
\label{tab2}
\begin{minipage}[t]{\linewidth}
\footnotesize
\begin{IEEEeqnarray}{*LL}
\hline \hline \notag \\[1mm]
\notag 
\text{\bf Case} & \\[1mm]
\hline \notag \\[2mm]
\label{eq0rr1+}
\text{\bmm $\theta=\theta_1\wedge \theta_2, \theta_i\in \mi{PropForm}$} & 
\dfrac{\tilde{a}_\mbbm{i}\leftrightarrow (\theta_1\wedge \theta_2)}
      {\left\{\begin{array}{l}
              \tilde{a}_{\mbbm{i}_1}\gleq \tilde{a}_{\mbbm{i}_2}\vee \tilde{a}_\mbbm{i}\geql \tilde{a}_{\mbbm{i}_2}, 
              \tilde{a}_{\mbbm{i}_2}\gle \tilde{a}_{\mbbm{i}_1}\vee \tilde{a}_\mbbm{i}\geql \tilde{a}_{\mbbm{i}_1}, \\               
              \tilde{a}_{\mbbm{i}_1}\leftrightarrow \theta_1, \tilde{a}_{\mbbm{i}_2}\leftrightarrow \theta_2
              \end{array}\right\}} \\[2mm]
\IEEEeqnarraymulticol{2}{l}{
|\text{Consequent}|=12+|\tilde{a}_{\mbbm{i}_1}\leftrightarrow \theta_1|+|\tilde{a}_{\mbbm{i}_2}\leftrightarrow \theta_2|<
                    25+|\tilde{a}_{\mbbm{i}_1}\leftrightarrow \theta_1|+|\tilde{a}_{\mbbm{i}_2}\leftrightarrow \theta_2|} \notag \\[4mm]
\label{eq0rr11+}
\text{\bmm $\theta=\theta_1\swedge \theta_2, \theta_i\in \mi{PropForm}$} & 
\dfrac{\tilde{a}_\mbbm{i}\leftrightarrow (\theta_1\swedge \theta_2)}
      {\left\{\tilde{a}_\mbbm{i}\geql \tilde{a}_{\mbbm{i}_1}\swedge \tilde{a}_{\mbbm{i}_2}, 
              \tilde{a}_{\mbbm{i}_1}\leftrightarrow \theta_1, \tilde{a}_{\mbbm{i}_2}\leftrightarrow \theta_2\right\}} \\[2mm]
\IEEEeqnarraymulticol{2}{l}{
|\text{Consequent}|=5+|\tilde{a}_{\mbbm{i}_1}\leftrightarrow \theta_1|+|\tilde{a}_{\mbbm{i}_2}\leftrightarrow \theta_2|<
                    25+|\tilde{a}_{\mbbm{i}_1}\leftrightarrow \theta_1|+|\tilde{a}_{\mbbm{i}_2}\leftrightarrow \theta_2|} \notag \\[4mm]
\label{eq0rr2+}
\text{\bmm $\theta=\theta_1\vee \theta_2, \theta_i\in \mi{PropForm}$} & 
\dfrac{\tilde{a}_\mbbm{i}\leftrightarrow (\theta_1\vee \theta_2)}
      {\left\{\begin{array}{l}
              \tilde{a}_{\mbbm{i}_1}\gleq \tilde{a}_{\mbbm{i}_2}\vee \tilde{a}_\mbbm{i}\geql \tilde{a}_{\mbbm{i}_1}, \\
              \tilde{a}_{\mbbm{i}_2}\gle \tilde{a}_{\mbbm{i}_1}\vee \tilde{a}_\mbbm{i}\geql \tilde{a}_{\mbbm{i}_2}, \\              
              \tilde{a}_{\mbbm{i}_1}\leftrightarrow \theta_1, \tilde{a}_{\mbbm{i}_2}\leftrightarrow \theta_2
              \end{array}\right\}} \\[2mm]
\IEEEeqnarraymulticol{2}{l}{
|\text{Consequent}|=12+|\tilde{a}_{\mbbm{i}_1}\leftrightarrow \theta_1|+|\tilde{a}_{\mbbm{i}_2}\leftrightarrow \theta_2|<
                    25+|\tilde{a}_{\mbbm{i}_1}\leftrightarrow \theta_1|+|\tilde{a}_{\mbbm{i}_2}\leftrightarrow \theta_2|} \notag \\[4mm]
\label{eq0rr3+}
\text{\bmm $\theta=\theta_1\rightarrow \theta_2, \theta_i\in \mi{PropForm}, \theta_2\neq \gz$} & 
\dfrac{\tilde{a}_\mbbm{i}\leftrightarrow (\theta_1\rightarrow \theta_2)}
      {\left\{\begin{array}{l}
              \tilde{a}_{\mbbm{i}_1}\gleq \tilde{a}_{\mbbm{i}_2}\vee \tilde{a}_{\mbbm{i}_1}\swedge \tilde{a}_\mbbm{i}\geql \tilde{a}_{\mbbm{i}_2}, \\     
              \tilde{a}_{\mbbm{i}_2}\gle \tilde{a}_{\mbbm{i}_1}\vee \tilde{a}_\mbbm{i}\geql \gu,                
              \tilde{a}_{\mbbm{i}_1}\leftrightarrow \theta_1, \tilde{a}_{\mbbm{i}_2}\leftrightarrow \theta_2
              \end{array}\right\}} \\[2mm]
\IEEEeqnarraymulticol{2}{l}{
|\text{Consequent}|=14+|\tilde{a}_{\mbbm{i}_1}\leftrightarrow \theta_1|+|\tilde{a}_{\mbbm{i}_2}\leftrightarrow \theta_2|<
                    25+|\tilde{a}_{\mbbm{i}_1}\leftrightarrow \theta_1|+|\tilde{a}_{\mbbm{i}_2}\leftrightarrow \theta_2|} \notag \\[4mm]
\label{eq0rr33+}
\text{\bmm $\theta=\theta_1\leftrightarrow \theta_2, \theta_i\in \mi{PropForm}$} & 
\dfrac{\tilde{a}_\mbbm{i}\leftrightarrow (\theta_1\leftrightarrow \theta_2)}
      {\left\{\begin{array}{l}
              \tilde{a}_{\mbbm{i}_1}\gleq \tilde{a}_{\mbbm{i}_2}\vee \tilde{a}_{\mbbm{i}_1}\swedge \tilde{a}_\mbbm{i}\geql \tilde{a}_{\mbbm{i}_2}, \\
              \tilde{a}_{\mbbm{i}_2}\gleq \tilde{a}_{\mbbm{i}_1}\vee \tilde{a}_{\mbbm{i}_2}\swedge \tilde{a}_\mbbm{i}\geql \tilde{a}_{\mbbm{i}_1}, \\
              \tilde{a}_{\mbbm{i}_1}\gle \tilde{a}_{\mbbm{i}_2}\vee \tilde{a}_{\mbbm{i}_2}\gle \tilde{a}_{\mbbm{i}_1}\vee \tilde{a}_\mbbm{i}\geql \gu, \\
              \tilde{a}_{\mbbm{i}_1}\leftrightarrow \theta_1, \tilde{a}_{\mbbm{i}_2}\leftrightarrow \theta_2
              \end{array}\right\}} \\[2mm]
\IEEEeqnarraymulticol{2}{l}{
|\text{Consequent}|=25+|\tilde{a}_{\mbbm{i}_1}\leftrightarrow \theta_1|+|\tilde{a}_{\mbbm{i}_2}\leftrightarrow \theta_2|} \notag \\[4mm]
\label{eq0rr7+}
\text{\bmm $\theta=\theta_1\geql \theta_2, \theta_i\in \mi{PropForm}, \theta_2\neq \gz, \gu$} \qquad & 
\dfrac{\tilde{a}_\mbbm{i}\leftrightarrow (\theta_1\geql \theta_2)}
      {\left\{\begin{array}{l}
              \tilde{a}_{\mbbm{i}_1}\geql \tilde{a}_{\mbbm{i}_2}\vee \tilde{a}_\mbbm{i}\geql \gz, \\
              \tilde{a}_{\mbbm{i}_1}\gle \tilde{a}_{\mbbm{i}_2}\vee \tilde{a}_{\mbbm{i}_2}\gle \tilde{a}_{\mbbm{i}_1}\vee \tilde{a}_\mbbm{i}\geql \gu, \\
              \tilde{a}_{\mbbm{i}_1}\leftrightarrow \theta_1, \tilde{a}_{\mbbm{i}_2}\leftrightarrow \theta_2
              \end{array}\right\}} \\[2mm]
\IEEEeqnarraymulticol{2}{l}{
|\text{Consequent}|=15+|\tilde{a}_{\mbbm{i}_1}\leftrightarrow \theta_1|+|\tilde{a}_{\mbbm{i}_2}\leftrightarrow \theta_2|<
                    25+|\tilde{a}_{\mbbm{i}_1}\leftrightarrow \theta_1|+|\tilde{a}_{\mbbm{i}_2}\leftrightarrow \theta_2|} \notag \\[4mm]
\label{eq0rr77+}
\text{\bmm $\theta=\theta_1\gleq \theta_2, \theta_i\in \mi{PropForm}$} & 
\dfrac{\tilde{a}_\mbbm{i}\leftrightarrow (\theta_1\gleq \theta_2)}
      {\left\{\begin{array}{l}
              \tilde{a}_{\mbbm{i}_1}\gleq \tilde{a}_{\mbbm{i}_2}\vee \tilde{a}_\mbbm{i}\geql \gz, 
              \tilde{a}_{\mbbm{i}_2}\gle \tilde{a}_{\mbbm{i}_1}\vee \tilde{a}_\mbbm{i}\geql \gu, \\
              \tilde{a}_{\mbbm{i}_1}\leftrightarrow \theta_1, \tilde{a}_{\mbbm{i}_2}\leftrightarrow \theta_2
              \end{array}\right\}} \\[2mm]
\IEEEeqnarraymulticol{2}{l}{
|\text{Consequent}|=12+|\tilde{a}_{\mbbm{i}_1}\leftrightarrow \theta_1|+|\tilde{a}_{\mbbm{i}_2}\leftrightarrow \theta_2|<
                    25+|\tilde{a}_{\mbbm{i}_1}\leftrightarrow \theta_1|+|\tilde{a}_{\mbbm{i}_2}\leftrightarrow \theta_2|} \notag \\[4mm]
\label{eq0rr8+}
\text{\bmm $\begin{array}{l}
            \theta=\theta_1\gle \theta_2, \theta_i\in \mi{PropForm}, \\ 
            \theta_1\neq \gz, \theta_2\neq \gu
            \end{array}$} & 
\dfrac{\tilde{a}_\mbbm{i}\leftrightarrow (\theta_1\gle \theta_2)}
      {\left\{\begin{array}{l}
              \tilde{a}_{\mbbm{i}_1}\gle \tilde{a}_{\mbbm{i}_2}\vee \tilde{a}_\mbbm{i}\geql \gz, 
              \tilde{a}_{\mbbm{i}_2}\gleq \tilde{a}_{\mbbm{i}_1}\vee \tilde{a}_\mbbm{i}\geql \gu, \\
              \tilde{a}_{\mbbm{i}_1}\leftrightarrow \theta_1, \tilde{a}_{\mbbm{i}_2}\leftrightarrow \theta_2
              \end{array}\right\}} \\[2mm]
\IEEEeqnarraymulticol{2}{l}{
|\text{Consequent}|=12+|\tilde{a}_{\mbbm{i}_1}\leftrightarrow \theta_1|+|\tilde{a}_{\mbbm{i}_2}\leftrightarrow \theta_2|<
                    25+|\tilde{a}_{\mbbm{i}_1}\leftrightarrow \theta_1|+|\tilde{a}_{\mbbm{i}_2}\leftrightarrow \theta_2|} \notag \\[2mm]
\hline \hline \notag
\end{IEEEeqnarray}
\end{minipage}
\end{table}
\begin{table}[t]
\caption{Unary interpolation rules for $\rightarrow$, $\geql$, and $\gle$}
\label{tab3}
\begin{minipage}[t]{\linewidth}
\footnotesize
\begin{IEEEeqnarray}{LL}
\hline \hline \notag \\[1mm]
\notag 
\text{\bf Case} & \\[1mm]
\hline \notag \\[2mm]
\label{eq0rr4+}
\text{\bmm $\theta=\theta_1\rightarrow \gz, \theta_1\in \mi{PropForm}$} \qquad &
\dfrac{\tilde{a}_\mbbm{i}\leftrightarrow (\theta_1\rightarrow \gz)}
      {\left\{\tilde{a}_{\mbbm{i}_1}\geql \gz\vee \tilde{a}_\mbbm{i}\geql \gz,
              \gz\gle \tilde{a}_{\mbbm{i}_1}\vee \tilde{a}_\mbbm{i}\geql \gu, 
              \tilde{a}_{\mbbm{i}_1}\leftrightarrow \theta_1\right\}} \\[2mm]
\IEEEeqnarraymulticol{2}{l}{
|\text{Consequent}|=12+|\tilde{a}_{\mbbm{i}_1}\leftrightarrow \theta_1|<25+|\tilde{a}_{\mbbm{i}_1}\leftrightarrow \theta_1|} \notag \\[4mm]
\label{eq0rr777+}
\text{\bmm $\theta=\theta_1\geql \gz, \theta_1\in \mi{PropForm}$} &
\dfrac{\tilde{a}_\mbbm{i}\leftrightarrow (\theta_1\geql \gz)}
      {\left\{\tilde{a}_{\mbbm{i}_1}\geql \gz\vee \tilde{a}_\mbbm{i}\geql \gz, 
              \gz\gle \tilde{a}_{\mbbm{i}_1}\vee \tilde{a}_\mbbm{i}\geql \gu, 
              \tilde{a}_{\mbbm{i}_1}\leftrightarrow \theta_1\right\}} \\[2mm]
\IEEEeqnarraymulticol{2}{l}{
|\text{Consequent}|=12+|\tilde{a}_{\mbbm{i}_1}\leftrightarrow \theta_1|<25+|\tilde{a}_{\mbbm{i}_1}\leftrightarrow \theta_1|} \notag \\[4mm]
\label{eq0rr7777+}
\text{\bmm $\theta=\theta_1\geql \gu, \theta_1\in \mi{PropForm}$} &
\dfrac{\tilde{a}_\mbbm{i}\leftrightarrow (\theta_1\geql \gu)}
      {\left\{\tilde{a}_{\mbbm{i}_1}\geql \gu\vee \tilde{a}_\mbbm{i}\geql \gz, 
              \tilde{a}_{\mbbm{i}_1}\gle \gu\vee \tilde{a}_\mbbm{i}\geql \gu, 
              \tilde{a}_{\mbbm{i}_1}\leftrightarrow \theta_1\right\}} \\[2mm]
\IEEEeqnarraymulticol{2}{l}{
|\text{Consequent}|=12+|\tilde{a}_{\mbbm{i}_1}\leftrightarrow \theta_1|<25+|\tilde{a}_{\mbbm{i}_1}\leftrightarrow \theta_1|} \notag \\[4mm]
\label{eq0rr88+}
\text{\bmm $\theta=\gz\gle \theta_1, \theta_1\in \mi{PropForm}$} &
\dfrac{\tilde{a}_\mbbm{i}\leftrightarrow (\gz\gle \theta_1)}
      {\left\{\gz\gle \tilde{a}_{\mbbm{i}_1}\vee \tilde{a}_\mbbm{i}\geql \gz, 
              \tilde{a}_{\mbbm{i}_1}\geql \gz\vee \tilde{a}_\mbbm{i}\geql \gu,                              
              \tilde{a}_{\mbbm{i}_1}\leftrightarrow \theta_1\right\}} \\[2mm]
\IEEEeqnarraymulticol{2}{l}{
|\text{Consequent}|=12+|\tilde{a}_{\mbbm{i}_1}\leftrightarrow \theta_1|<25+|\tilde{a}_{\mbbm{i}_1}\leftrightarrow \theta_1|} \notag \\[4mm] 
\label{eq0rr888+}
\text{\bmm $\theta=\theta_1\gle \gu, \theta_1\in \mi{PropForm}$} &
\dfrac{\tilde{a}_\mbbm{i}\leftrightarrow (\theta_1\gle \gu)}
      {\left\{\tilde{a}_{\mbbm{i}_1}\gle \gu\vee \tilde{a}_\mbbm{i}\geql \gz, 
              \tilde{a}_{\mbbm{i}_1}\geql \gu\vee \tilde{a}_\mbbm{i}\geql \gu, 
              \tilde{a}_{\mbbm{i}_1}\leftrightarrow \theta_1\right\}} \\[2mm]
\IEEEeqnarraymulticol{2}{l}{
|\text{Consequent}|=12+|\tilde{a}_{\mbbm{i}_1}\leftrightarrow \theta_1|<25+|\tilde{a}_{\mbbm{i}_1}\leftrightarrow \theta_1|} \notag \\[2mm]
\hline \hline \notag
\end{IEEEeqnarray}
\end{minipage}
\end{table}
%
%
%
%
\end{proof}

\begin{lemma}
\label{le1}
Let $n_\phi\in \mbb{N}$ and $\phi\in \mi{PropForm}_\emptyset$.
There exist either $J_\phi=\emptyset$, or $n_{J_\phi}$, $J_\phi=\{(n_\phi,j) \,|\, j\leq n_{J_\phi}\}$,
$J_\phi\subseteq_{\mc F} \{(n_\phi,j) \,|\, j\in \mbb{N}\}\subseteq \mbb{I}$, 
$S_\phi\subseteq_{\mc F} \mi{OrdPropCl}_{\{\tilde{a}_\mbbm{j} \,|\, \mbbm{j}\in J_\phi\}}$ such that
\begin{enumerate}[\rm (a)]
\item
$\mi{card}(J_\phi)\leq 2\cdot |\phi|$; 
\item
either $J_\phi=\emptyset$, $S_\phi=\{\square\}$, or $J_\phi=S_\phi=\emptyset$, or $J_\phi\neq \emptyset$, $\square\not\in S_\phi\neq \emptyset$; 
\item
there exists a valuation ${\mf A}$ satisfying ${\mf A}\models \phi$ if and only if 
there exists a valuation ${\mf A}'$ satisfying ${\mf A}'\models S_\phi$; 
${\mf A}|_\mbb{O}={\mf A}'|_\mbb{O}$; 
\item
$|S_\phi|\in O(|\phi|)$, 
the number of all elementary operations of the translation of $\phi$ to $S_\phi$ is in $O(|\phi|)$;
the time complexity of the translation of $\phi$ to $S_\phi$ is in $O(|\phi|\cdot (\log (1+n_\phi)+\log |\phi|))$;
\item
if $S_\phi\neq \emptyset, \{\square\}$, then $J_\phi\neq \emptyset$;
for all $C\in S_\phi$, $\emptyset\neq \mi{atoms}(C)\cap \tilde{\mbb{A}}\subseteq \{\tilde{a}_\mbbm{j} \,|\, \mbbm{j}\in J_\phi\}$. 
\end{enumerate}
\end{lemma}

\begin{proof}
A straightforward consequence of Lemmata \ref{le111} and \ref{le11}.
%
%
%
\end{proof}

\begin{corollary}
\label{cor12}
Let $n_0\in \mbb{N}$ and $T\subseteq \mi{PropForm}_\emptyset$.
There exist $J_T\subseteq \{(i,j) \,|\, i\geq n_0, j\in \mbb{N}\}\subseteq \mbb{I}$ and 
$S_T\subseteq \mi{OrdPropCl}_{\{\tilde{a}_\mbbm{j} \,|\, \mbbm{j}\in J_T\}}$ such that
\begin{enumerate}[\rm (a)]
\item
either $J_T=\emptyset$, $S_T=\{\square\}$, or $J_T=S_T=\emptyset$, or $J_T\neq \emptyset$, $\square\not\in S_T\neq \emptyset$; 
\item
there exists a valuation ${\mf A}$ satisfying ${\mf A}\models T$ if and only if 
there exists a valuation ${\mf A}'$ satisfying ${\mf A}'\models S_T$; 
${\mf A}|_\mbb{O}={\mf A}'|_\mbb{O}$; 
\item
if $T\subseteq_{\mc F} \mi{PropForm}_\emptyset$, then $J_T\subseteq_{\mc F} \{(i,j) \,|\, i\geq n_0, j\in \mbb{N}\}\subseteq \mbb{I}$, $\mi{card}(J_T)\leq 2\cdot |T|$, 
$S_T\subseteq_{\mc F} \mi{OrdPropCl}_{\{\tilde{a}_\mbbm{j} \,|\, \mbbm{j}\in J_T\}}$, $|S_T|\in O(|T|)$, 
the number of all elementary operations of the translation of $T$ to $S_T$ is in $O(|T|)$;
the time complexity of the translation of $T$ to $S_T$ is in $O(|T|\cdot \log (1+n_0+|T|))$; 
\item
if $S_T\neq \emptyset, \{\square\}$, then $J_T\neq \emptyset$;
for all $C\in S_T$, $\emptyset\neq \mi{atoms}(C)\cap \tilde{\mbb{A}}\subseteq \{\tilde{a}_\mbbm{j} \,|\, \mbbm{j}\in J_T\}$.
\end{enumerate}
\end{corollary}

\begin{proof}
A straightforward consequence of Lemma \ref{le1}.
%
%
%
\end{proof}

\begin{theorem}
\label{T1}
Let $n_0\in \mbb{N}$, $\phi\in \mi{PropForm}_\emptyset$, $T\subseteq \mi{PropForm}_\emptyset$.
There exist $J_T^\phi\subseteq \{(i,j) \,|\, i\geq n_0, j\in \mbb{N}\}\subseteq \mbb{I}$ and
$S_T^\phi\subseteq \mi{OrdPropCl}_{\{\tilde{a}_\mbbm{j} \,|\, \mbbm{j}\in J_T^\phi\}}$ such that
\begin{enumerate}[\rm (i)]
\item
there exists a valuation ${\mf A}$ satisfying ${\mf A}\models T$, ${\mf A}\not\models \phi$ if and only if 
there exists a valuation ${\mf A}'$ satisfying ${\mf A}'\models S_T^\phi$; 
${\mf A}|_\mbb{O}={\mf A}'|_\mbb{O}$;
\item
$T\models \phi$ if and only if $S_T^\phi$ is unsatisfiable;
\item
if $T\subseteq_{\mc F} \mi{PropForm}_\emptyset$, then $J_T^\phi\subseteq_{\mc F} \{(i,j) \,|\, i\geq n_0, j\in \mbb{N}\}\subseteq \mbb{I}$, $\mi{card}(J_T^\phi)\in O(|\phi|+|T|)$,
$S_T^\phi\subseteq_{\mc F} \mi{OrdPropCl}_{\{\tilde{a}_\mbbm{j} \,|\, \mbbm{j}\in J_T^\phi\}}$, $|S_T^\phi|\in O(|\phi|+|T|)$,
the number of all elementary operations of the translation of $\phi$ and $T$ to $S_T^\phi$ is in $O(|\phi|+|T|)$;
the time complexity of the translation of $\phi$ and $T$ to $S_T^\phi$ is in $O(|\phi|\cdot (\log (1+n_0)+\log |\phi|)+|T|\cdot \log (1+n_0+|T|))$.
\end{enumerate}
\end{theorem}

\begin{proof}
A straightforward consequence of Lemmata \ref{le111}, \ref{le11}, and Corollary \ref{cor12}.
%
%
%
\end{proof}

For illustration purposes, we translate $\phi=\gz\gle a\rightarrow (a\swedge b\gleq a\swedge c\rightarrow b\gleq c)\in \mi{PropForm}_\emptyset$ to clausal form 
using the binary and unary interpolation rules, Tables \ref{tab4} and \ref{tab5}.
\begin{table}
\caption{Translation of the formula $\phi$ to clausal form}
\label{tab4}
\begin{minipage}[t]{\linewidth}
\footnotesize
\begin{IEEEeqnarray*}{LL}
\hline \hline \\[2mm]
\Big\{\tilde{a}_0\geql \gu,
      \tilde{a}_0\leftrightarrow \big(\underbrace{\gz\gle a}_{\tilde{a}_1}\rightarrow (\underbrace{a\swedge b\gleq a\swedge c\rightarrow b\gleq c}_{\tilde{a}_2})\big)\Big\}
& (\ref{eq0rr3+}) \\
\Big\{\tilde{a}_0\geql \gu,
      \tilde{a}_1\gleq \tilde{a}_2\vee \tilde{a}_1\swedge \tilde{a}_0\geql \tilde{a}_2,
      \tilde{a}_2\gle \tilde{a}_1\vee \tilde{a}_0\geql \gu,
& \\
\phantom{\Big\{}
      \tilde{a}_1\leftrightarrow \gz\gle \underbrace{a}_{\tilde{a}_3},
      \tilde{a}_2\leftrightarrow (\underbrace{a\swedge b\gleq a\swedge c}_{\tilde{a}_4}\rightarrow \underbrace{b\gleq c}_{\tilde{a}_5})\Big\} 
& (\ref{eq0rr88+})\ (\ref{eq0rr3+}) \\
\Big\{\tilde{a}_0\geql \gu,
      \tilde{a}_1\gleq \tilde{a}_2\vee \tilde{a}_1\swedge \tilde{a}_0\geql \tilde{a}_2,
      \tilde{a}_2\gle \tilde{a}_1\vee \tilde{a}_0\geql \gu,
& \\
\phantom{\Big\{}
      \gz\gle \tilde{a}_3\vee \tilde{a}_1\geql \gz,
      \tilde{a}_3\geql \gz\vee \tilde{a}_1\geql \gu,
      \tilde{a}_3\geql a,
& \\
\phantom{\Big\{}
      \tilde{a}_4\gleq \tilde{a}_5\vee \tilde{a}_4\swedge \tilde{a}_2\geql \tilde{a}_5,
      \tilde{a}_5\gle \tilde{a}_4\vee \tilde{a}_2\geql \gu,
      \tilde{a}_4\leftrightarrow \underbrace{a\swedge b}_{\tilde{a}_6}\gleq \underbrace{a\swedge c}_{\tilde{a}_7},
      \tilde{a}_5\leftrightarrow \underbrace{b}_{\tilde{a}_8}\gleq \underbrace{c}_{\tilde{a}_9}\Big\} \quad
& (\ref{eq0rr77+}) \\
\Big\{\tilde{a}_0\geql \gu,
      \tilde{a}_1\gleq \tilde{a}_2\vee \tilde{a}_1\swedge \tilde{a}_0\geql \tilde{a}_2,
      \tilde{a}_2\gle \tilde{a}_1\vee \tilde{a}_0\geql \gu,
& \\
\phantom{\Big\{}
      \gz\gle \tilde{a}_3\vee \tilde{a}_1\geql \gz,
      \tilde{a}_3\geql \gz\vee \tilde{a}_1\geql \gu,
      \tilde{a}_3\geql a,
& \\
\phantom{\Big\{}
      \tilde{a}_4\gleq \tilde{a}_5\vee \tilde{a}_4\swedge \tilde{a}_2\geql \tilde{a}_5,
      \tilde{a}_5\gle \tilde{a}_4\vee \tilde{a}_2\geql \gu,
& \\
\phantom{\Big\{}
      \tilde{a}_6\gleq \tilde{a}_7\vee \tilde{a}_4\geql \gz,
      \tilde{a}_7\gle \tilde{a}_6\vee \tilde{a}_4\geql \gu,
      \tilde{a}_6\leftrightarrow \underbrace{a}_{\tilde{a}_{10}}\swedge \underbrace{b}_{\tilde{a}_{11}},
      \tilde{a}_7\leftrightarrow \underbrace{a}_{\tilde{a}_{12}}\swedge \underbrace{c}_{\tilde{a}_{13}},
& \\
\phantom{\Big\{}
      \tilde{a}_8\gleq \tilde{a}_9\vee \tilde{a}_5\geql \gz,
      \tilde{a}_9\gle \tilde{a}_8\vee \tilde{a}_5\geql \gu,
      \tilde{a}_8\geql b, 
      \tilde{a}_9\geql c\Big\}
& (\ref{eq0rr11+}) \\[2mm]
\hline \hline 
\end{IEEEeqnarray*}
\end{minipage}
\end{table}
\begin{table}
\caption{Translation of the formula $\phi$ to clausal form}
\label{tab5}
\begin{minipage}[t]{\linewidth}
\begin{IEEEeqnarray*}{L}
\hline \hline 
\end{IEEEeqnarray*}
\end{minipage}
\begin{minipage}[t]{0.49\linewidth}
\footnotesize
\begin{IEEEeqnarray*}{RLL}
S^\phi=\Bigg\{
& \tilde{a}_0\geql \gu
& [1] \\
& \tilde{a}_1\gleq \tilde{a}_2\vee \tilde{a}_1\swedge \tilde{a}_0\geql \tilde{a}_2 \quad 
& [2] \\
& \tilde{a}_2\gle \tilde{a}_1\vee \tilde{a}_0\geql \gu
& [3] \\
& \gz\gle \tilde{a}_3\vee \tilde{a}_1\geql \gz
& [4] \\
& \tilde{a}_3\geql \gz\vee \tilde{a}_1\geql \gu
& [5] \\
& \tilde{a}_3\geql a
& [6] \\
& \tilde{a}_4\gleq \tilde{a}_5\vee \tilde{a}_4\swedge \tilde{a}_2\geql \tilde{a}_5
& [7] \\
& \tilde{a}_5\gle \tilde{a}_4\vee \tilde{a}_2\geql \gu
& [8] \\
& \tilde{a}_6\gleq \tilde{a}_7\vee \tilde{a}_4\geql \gz
& [9] \\
& \tilde{a}_7\gle \tilde{a}_6\vee \tilde{a}_4\geql \gu
& [10] 
\end{IEEEeqnarray*} 
\end{minipage}
\begin{minipage}[t]{0.49\linewidth}
\footnotesize
\begin{IEEEeqnarray*}{RLL}
& \tilde{a}_6\geql \tilde{a}_{10}\swedge \tilde{a}_{11}
& [11] \\
& \tilde{a}_{10}\geql a
& [12] \\
& \tilde{a}_{11}\geql b
& [13] \\
& \tilde{a}_7\geql \tilde{a}_{12}\swedge \tilde{a}_{13}
& [14] \\
& \tilde{a}_{12}\geql a
& [15] \\
& \tilde{a}_{13}\geql c
& [16] \\
& \tilde{a}_8\gleq \tilde{a}_9\vee \tilde{a}_5\geql \gz \quad 
& [17] \\
& \tilde{a}_9\gle \tilde{a}_8\vee \tilde{a}_5\geql \gu
& [18] \\
& \tilde{a}_8\geql b
& [19] \\
& \tilde{a}_9\geql c\Bigg\}
& [20] 
\end{IEEEeqnarray*} 
\end{minipage}
\begin{minipage}[t]{\linewidth}
\begin{IEEEeqnarray*}{L}
\hline \hline 
\end{IEEEeqnarray*}
\end{minipage}
\end{table}
By Lemma \ref{le1}(c), $\phi$ is equisatisfiable to $S^\phi\subseteq \mi{OrdPropCl}_{\{\tilde{a}_0,\dots,\tilde{a}_{13}\}}$.
During the translation, we have introduced some fresh atoms $\tilde{a}_0,\dots,\tilde{a}_{13}$.

\section{A {\ib DPLL} procedure over order clauses}
\label{S4}

In this section, we devise a variant of the {\it DPLL} procedure operating over finite order clausal theories and prove its refutational soundness and finite completeness.
At first, we introduce some basic notions and notation.
Let $\Cn\in \mi{PropConj}$, $\varepsilon\in \{\gz,\gu\}\cup \mi{PropConj}$, $l\in \mi{OrdPropLit}$, $C\in \mi{OrdPropCl}$.
$l$ is a contradiction iff either $l=\gz\geql \gu$ or $l=\gu\gleq \gz$ or $l=\gu\gle \gz$ or $l=\Cn\gle \gz$ or $l=\gu\gle \Cn$ or $l=\varepsilon\gle \varepsilon$.
$l$ is a tautology iff either $l=\varepsilon\geql \varepsilon$ or $l=\gz\gleq \gu$ or $l=\gz\gleq \Cn$ or $l=\Cn\gleq \gu$ or $l=\varepsilon\gleq \varepsilon$ or $l=\gz\gle \gu$.
We define an auxiliary ternary operator 
$\mi{simplify} : (\{\gz,\gu\}\cup \mi{PropConj}\cup \mi{OrdPropLit}\cup \mi{OrdPropCl})\times \mi{PropAtom}\times \{\gz,\gu\}\longrightarrow 
                 \{\gz,\gu\}\cup \mi{PropConj}\cup \mi{OrdPropLit}\cup \mi{OrdPropCl}$ as follows:
\begin{alignat*}{1} 
\mi{simplify}(\gz,a,\tau) &= \gz; \\
\mi{simplify}(\gu,a,\tau) &= \gu; \\[1mm]
\mi{simplify}(\Cn,a,\gz)  &= \left\{\begin{array}{ll}
                                    \gz &\ \text{\it if}\ a\in \mi{atoms}(\Cn), \\[1mm]
                                    \Cn &\ \text{\it if}\ a\not\in \mi{atoms}(\Cn);
                                    \end{array}
                             \right. \\[1mm]
\mi{simplify}(\Cn,a,\gu)  &= \left\{\begin{array}{ll}
                                    \gu         &\ \text{\it if there exists}\ r^*\geq 1\ \text{\it such that} \\
                                                &\phantom{\ \text{\it if there exists}\ \mbox{}}
                                                                               a^{r^*}\in \Cn, \Cn-a^{r^*}=\emptyset, \\[1mm]
                                    \Cn-a^{r^*} &\ \text{\it if there exists}\ r^*\geq 1\ \text{\it such that} \\ 
                                                &\phantom{\ \text{\it if there exists}\ \mbox{}}
                                                                               a^{r^*}\in \Cn, \Cn-a^{r^*}\neq \emptyset, \\[1mm]
                                    \Cn         &\ \text{\it if}\ a\not\in \mi{atoms}(\Cn);
                                    \end{array}
                             \right. \\[1mm]
\mi{simplify}(l,a,\tau)   &= \mi{simplify}(\varepsilon_1,a,\tau)\diamond \mi{simplify}(\varepsilon_2,a,\tau) \\
                          &\phantom{\mbox{}=\mbox{}} \quad 
                             \text{\it if}\ l=\varepsilon_1\diamond \varepsilon_2, \varepsilon_i\in \{\gz,\gu\}\cup \mi{PropConj}, \diamond\in \{\geql,\gleq,\gle\}; \\[1mm]
\mi{simplify}(C,a,\tau)   &= \{\mi{simplify}(l,a,\tau) \,|\, l\in C\}.
\end{alignat*}                              
For an input expression $\mu$, atom $a$, and a truth constant $\tau$, the operator $\mi{simplify}$ replaces every occurrence of $a$ with $\tau$ in $\mu$, and 
the modified $\mu$ is partially evaluated in $\bs{\Pi}$ as the result.
Let $\Cn_1, \Cn_2\in \mi{PropConj}$ and $l, l_1, l_2\in \mi{PropLit}$.
Another auxiliary binary operator $\mult : (\{\gz,\gu\}\cup \mi{PropConj})\times (\{\gz,\gu\}\cup \mi{PropConj})\longrightarrow \{\gz,\gu\}\cup \mi{PropConj}$ is defined as follows:
\begin{alignat*}{1}
\gz\mult \varepsilon=\varepsilon\mult \gz &= \gz; \\
\gu\mult \varepsilon=\varepsilon\mult \gu &= \varepsilon; \\[1mm]
\Cn_1\mult \Cn_2                          &= \{a^{r+s} \,|\, a^r\in \Cn_1, a^s\in \Cn_2, a\in \mi{PropAtom}, r, s\geq 1\}\cup \\
                                          &\phantom{\mbox{}=\mbox{}}
                                             \{a^r \,|\, a^r\in \Cn_1, a\in \mi{PropAtom}, r\geq 1, a\not\in \mi{atoms}(\Cn_2)\}\cup \\
                                          &\phantom{\mbox{}=\mbox{}}
                                             \{a^s \,|\, a^s\in \Cn_2, a\in \mi{PropAtom}, s\geq 1, a\not\in \mi{atoms}(\Cn_1)\}.
\end{alignat*}                          
The operator $\mult$ returns the product of two input expressions (truth constants or conjunctions of powers).
Note that $\mult$ is commutative and associative.
Let $r\in \mbb{N}$.
We define an $r$th power of $\varepsilon$ as follows:
\begin{alignat*}{1}
\varepsilon^r &= \left\{\begin{array}{ll}
                        \gu                                                    &\ \text{\it if}\ r=0, \\[1mm]
                        \underbrace{\varepsilon\mult\cdots\mult \varepsilon}_r &\ \text{\it if}\ r\geq 1;
                        \end{array}
                 \right. \\[1mm]
              &\in \{\gz,\gu\}\cup \mi{PropConj}.     
\end{alignat*}
Note that $\gz^0=\gu^0=\gu$, and $\gz^r=\gz$, $\gu^r=\gu$, $r\geq 1$.
The operator $\mult$ can be extended to literals component-wisely.
$\mult : (\{\gz,\gu\}\cup \mi{PropLit})\times (\{\gz,\gu\}\cup \mi{PropLit})\longrightarrow \{\gz,\gu\}\cup \mi{PropLit}$ is defined as follows:
\begin{alignat*}{1}
\gz\mult \mu=\mu\mult \gz &= \gz; \\
\gu\mult \mu=\mu\mult \gu &= \mu; \\[1mm]
l_1\mult l_2              &= (\varepsilon_1^1\mult \varepsilon_1^2)\diamond (\varepsilon_2^1\mult \varepsilon_2^2) \\ 
                          &\phantom{\mbox{}=\mbox{}} \quad
                             \text{\it if}\ l_i=\varepsilon_1^i\diamond^i \varepsilon_2^i, \varepsilon_j^i\in \{\gz,\gu\}\cup \mi{PropConj}, \diamond^i\in \{\geql,\gleq,\gle\}, \\[1mm]
                          &\phantom{\mbox{}=\mbox{}}
                             \diamond=\left\{\begin{array}{ll}
                                             \geql &\ \text{\it if}\ \diamond^1=\diamond^2=\geql, \\[1mm]
                                             \gleq &\ \text{\it if either}\ \diamond^1=\geql, \diamond^2=\gleq,\
                                                      \text{\it or}\ \diamond^1=\gleq, \diamond^2=\geql,\
                                                      \text{\it or}\ \diamond^1=\diamond^2=\gleq, \\[1mm]
                                             \gle  &\ \text{\it if}\ \diamond^1=\gle\ \text{\it or}\ \diamond^2=\gle.
                                             \end{array}
                                      \right.
\end{alignat*}
Note that $\mult$ is still commutative and associative.
We define an $r$th power of $l$ as follows:
\begin{alignat*}{1}
l^r &= \left\{\begin{array}{ll}
              \gu                                &\ \text{\it if}\ r=0, \\[1mm]
              \underbrace{l\mult\cdots\mult l}_r &\ \text{\it if}\ r\geq 1;
              \end{array}
       \right. \\[1mm]
    &\in \{\gu\}\cup \mi{PropLit}.     
\end{alignat*}
Let $I\subseteq_{\mc F} \mbb{N}$, $l_i\in \mi{PropLit}$, $\alpha_i\geq 1$, $i\in I$.
We define the product $\bigmult_{i\in I} l_i^{\alpha_i}$ of powers of literals $l_i^{\alpha_i}$, $i\in I$, by recursion on $I$ as follows:
\begin{alignat*}{1}
\bigmult_{i\in I} l_i^{\alpha_i} &= \left\{\begin{array}{ll}
                                           \gu                                                                  &\ \text{\it if}\ I=\emptyset, \\[1mm]   
                                           l_{i^*}^{\alpha_{i^*}}\mult \bigmult_{i\in I-\{i^*\}} l_i^{\alpha_i} &\ \text{\it if there exists}\ i^*\in I;
                                           \end{array}
                                    \right. \\[1mm]
                                 &\in \{\gu\}\cup \mi{PropLit}.
\end{alignat*}
We now introduce a notion of guard, which lays some constraint on the truth value eventually assigned to an atom.
Such constraints will efficiently be used to confine the form of temporary order clausal theories during inference of the {\it DPLL} procedure. 
Let $a\in \mi{PropAtom}$.
$C$ is a guard iff either $C=a\geql \gz$ or $C=a\gleq \gz$ or $C=\gz\gle a$ or $C=a\gle \gu$ or $C=a\geql \gu$ or $C=\gu\gleq a$.
Note that a guard is unit.
Let $S\subseteq \mi{OrdPropCl}$.
We denote
$\mi{guards}(a)=\{a\geql \gz,a\gleq \gz,\gz\gle a,a\gle \gu,a\geql \gu,\gu\gleq a\}\subseteq \mi{OrdPropCl}$,
$\mi{guards}(S)=\{C \,|\, C\in S\ \text{\it is a guard}\}$,
$\mi{guards}(S,a)=S\cap \mi{guards}(a)$.
Note that $\mi{guards}(S,a)\subseteq \mi{guards}(a)$, $\mi{guards}(S,a)\subseteq \mi{guards}(S)\subseteq S$, $(S-\mi{guards}(S))\cup \mi{guards}(S)=S$, $(S-\mi{guards}(S))\cap \mi{guards}(S)=\emptyset$,
$\mi{guards}(S)=\bigcup_{a\in \mi{atoms}(S)} \mi{guards}(S,a)$.
$S$ is a contradictory set of guards for $a$ iff 
$\mi{guards}(S,a)\supseteq \{a\geql \gz,\gz\gle a\}$ or $\mi{guards}(S,a)\supseteq \{a\gleq \gz,\gz\gle a\}$ or
$\mi{guards}(S,a)\supseteq \{a\geql \gz,a\geql \gu\}$ or $\mi{guards}(S,a)\supseteq \{a\geql \gz,\gu\gleq a\}$ or 
$\mi{guards}(S,a)\supseteq \{a\gleq \gz,a\geql \gu\}$ or $\mi{guards}(S,a)\supseteq \{a\gleq \gz,\gu\gleq a\}$ or
$\mi{guards}(S,a)\supseteq \{a\gle \gu,a\geql \gu\}$ or $\mi{guards}(S,a)\supseteq \{a\gle \gu,\gu\gleq a\}$.
$a$ is $\gz$-guarded in $S$ iff either $\mi{guards}(S,a)=\{a\geql \gz\}$ or $\mi{guards}(S,a)=\{\gz\gle a\}$ or $\mi{guards}(S,a)=\{\gz\gle a,a\gle \gu\}$ or $\mi{guards}(S,a)=\{a\geql \gu\}$.
$a$ is semi-positively guarded in $S$ iff either $\mi{guards}(S,a)=\{\gz\gle a\}$ or $\mi{guards}(S,a)=\{\gz\gle a,a\gle \gu\}$ or $\mi{guards}(S,a)=\{a\geql \gu\}$.
$a$ is positively guarded in $S$ iff either $\mi{guards}(S,a)=\{\gz\gle a\}$ or $\mi{guards}(S,a)=\{\gz\gle a,a\gle \gu\}$.
Note that if $a$ is $\gz$-guarded (\mbox{(semi-)}positively guarded) in $S$, then $a\in \mi{atoms}(\mi{guards}(S,a))\subseteq \mi{atoms}(\mi{guards}(S))\subseteq \mi{atoms}(S)$.
Note that if $a$ is positively guarded in $S$, then $a$ is semi-positively guarded in $S$, further $a$ is $\gz$-guarded in $S$.
$S$ is simplified iff $\square\not\in S-\mi{guards}(S)$, and for all $C\in S-\mi{guards}(S)$, $C$ does not contain contradictions and tautologies.
Note that obviously, $\square\not\in \mi{guards}(S)$, and for all $C\in \mi{guards}(S)$, $C$ does not contain contradictions and tautologies.
Hence, $S$ is simplified iff $\square\not\in S$, and for all $C\in S$, $C$ does not contain contradictions and tautologies.
$S$ is $\gz$-guarded ((semi-)positively guarded) iff $S$ is simplified, and for all $a\in \mi{atoms}(S)$, $a$ is $\gz$-guarded ((semi-)positively guarded) in $S$.
Note that if $S$ is $\gz$-guarded ((semi-)positively guarded), then, for all $a\in \mi{atoms}(S)$, $a\in \mi{atoms}(\mi{guards}(S))$ and $S$ is not a contradictory set of guards for $a$; 
$\mi{atoms}(\mi{guards}(S))=\mi{atoms}(S)$.
Note that a positively guarded order clausal theory is a semi-positively guarded order clausal theory, and hence, a $\gz$-guarded order clausal theory.
$a\geql \gz\vee \gz\gle a\in \mi{OrdPropCl}$ is called a $\gz$-dichotomy.
$\Cn_1\gleq \Cn_2\vee \Cn_2\gle \Cn_1\in \mi{PurOrdPropCl}$, $\Cn_1\neq \Cn_2$, is called a pure dichotomy.
Let $A\subseteq \mi{PropAtom}$.
We denote
\begin{alignat*}{1}
\mi{PropPow}_A             &= \{p \,|\, p\in \mi{PropPow}, \mi{atoms}(p)\subseteq A\}, \\
\mi{PropConj}_A            &= \{\Cn \,|\, \Cn\in \mi{PropConj}, \mi{atoms}(\Cn)\subseteq A\}, \\[1mm]
\mi{PropLit}_A^\gu         &= \{\varepsilon_1\diamond \varepsilon_2 \,|\, \varepsilon_1\diamond \varepsilon_2\in \mi{PropLit},
                                                                          \varepsilon_i\in \{\gu\}\cup \mi{PropConj}_A, \\
                           &\phantom{\mbox{}=\{\varepsilon_1\diamond \varepsilon_2 \,|\, \mbox{}}
                                                                          \diamond\in \{\geql,\gleq,\gle\}\}, \\[1mm]
\mi{PropLit}^\gu           &= \mi{PropLit}_\mi{PropAtom}^\gu, \\[1mm]
\mi{OrdPropLit}_A^\gu      &= \{\varepsilon_1\diamond \varepsilon_2 \,|\, \varepsilon_1\diamond \varepsilon_2\in \mi{OrdPropLit}, 
                                                                          \varepsilon_i\in \{\gu\}\cup \mi{PropConj}_A, \\
                           &\phantom{\mbox{}=\{\varepsilon_1\diamond \varepsilon_2 \,|\, \mbox{}}
                                                                          \diamond\in \{\geql,\gleq,\gle\}\}, \\[1mm]
\mi{OrdPropLit}^\gu        &= \mi{OrdPropLit}_\mi{PropAtom}^\gu, \\
\mi{OrdPropCl}_A^\gu       &= \{C \,|\, C\subseteq_{\mc F} \mi{OrdPropLit}_A^\gu\}, \\
\mi{OrdPropCl}^\gu         &= \mi{OrdPropCl}_\mi{PropAtom}^\gu, \\
\mi{OrdPropCl}_A^{\gz,\gu} &= \{C \,|\, C\in \mi{OrdPropCl}, \mi{atoms}(C)\subseteq A\}, \\
\mi{PurPropLit}_A          &= \{l \,|\, l\in \mi{PurPropLit}, \mi{atoms}(l)\subseteq A\}, \\
\mi{PurOrdPropLit}_A       &= \{l \,|\, l\in \mi{PurOrdPropLit}, \mi{atoms}(l)\subseteq A\}, \\
\mi{PurOrdPropCl}_A        &= \{C \,|\, C\in \mi{PurOrdPropCl}, \mi{atoms}(C)\subseteq A\}.
\end{alignat*}
$S$ is positive iff $S$ is positively guarded, and for all $C\in S-\mi{guards}(S)$, 
either $C\in \mi{PurOrdPropCl}$, or $C=a_0\swedge\cdots\swedge a_n\gle \gu\vee C^\natural$, $a_i\in \mi{PropAtom}$, $\mi{guards}(S,a_i)=\{\gz\gle a_i\}$, $C^\natural\in \mi{PurOrdPropCl}$.
Note that $S-\mi{guards}(S)\subseteq \mi{OrdPropCl}^\gu$.

The basic rules of the {\it DPLL} procedure are as follows:                                                                                                                                
\begin{alignat}{1}
\ctag{ceq4hr0}{({\it Unit contradiction rule})} \\[1mm]
\notag
& \dfrac{S}
        {S\cup \{\square\}}; \\[1mm]
\notag
& \begin{array}{l}
  S\ \text{\it is unit}; \\[1mm]
  \text{\it there exist}\ \gz\gle a_0,\dots,\gz\gle a_m\in \mi{guards}(S), a_j\in \mi{PropAtom}, \\
  \phantom{\text{\it there exist}\ \mbox{}}
                          \{l_0,\dots,l_n\}\subseteq (S\cap \mi{OrdPropCl}_{\{a_0,\dots,a_m\}}^\gu)\cup \\
  \phantom{\text{\it there exist}\ \{l_0,\dots,l_n\}\subseteq \mbox{}} \quad 
                                                     \{a\gleq \gu \,|\, a\in \{a_0,\dots,a_m\}, a\gle \gu\not\in \mi{guards}(S)\}, \\
  \phantom{\text{\it there exist}\ \mbox{}}
                          \langle l_0\rangle,\dots,\langle l_n\rangle\in \mi{PropLit}_{\{a_0,\dots,a_m\}}^\gu\ \text{\it such that} \\
  \quad \mi{atoms}(l_0,\dots,l_n,\langle l_0\rangle,\dots,\langle l_n\rangle)=\{a_0,\dots,a_m\}; \\[1mm]
  \text{\it there exist}\ \alpha_0,\dots,\alpha_n, \alpha_i\geq 1,\ \text{\it and}\ \Cn\in \mi{PropConj}_{\{a_0,\dots,a_m\}}\ \text{\it such that} \\ 
  \quad \bigmult_{i=0}^n \langle l_i\rangle^{\alpha_i}=\Cn\gle \Cn.
  \end{array}
\end{alignat}
If there exists a product of powers of the input literals $\langle l_0\rangle,\dots,\langle l_n\rangle$ which is a contradiction of the form $\Cn\gle \Cn$, 
then the input unit order clausal theory $S$ is unsatisfiable; we can derive $\square$.
In practical inference, we shall replace the application subcondition:
{\it $S$ is unit}; 
{\it there exist $\gz\gle a_0,\dots,\gz\gle a_m\in \mi{guards}(S)$, $a_j\in \mi{PropAtom}$,
                 $\{l_0,\dots,l_n\}\subseteq (S\cap \mi{OrdPropCl}_{\{a_0,\dots,a_m\}}^\gu)\cup \{a\gleq \gu \,|\, a\in \{a_0,\dots,a_m\}, a\gle \gu\not\in \mi{guards}(S)\}$} with a weaker one:
{\it there exists $S'\subseteq S$ such that $S'$ is unit}; 
{\it there exist $\gz\gle a_0,\dots,\gz\gle a_m\in \mi{guards}(S')=\mi{guards}(S)$, $a_j\in \mi{PropAtom}$,
                 $\{l_0,\dots,l_n\}\subseteq (S'\cap \mi{OrdPropCl}_{\{a_0,\dots,a_m\}}^\gu)\cup \{a\gleq \gu \,|\, a\in \{a_0,\dots,a_m\}, a\gle \gu\not\in \mi{guards}(S')=\mi{guards}(S)\}$}.
Such a modification of the application condition is admissible with respect to satisfiability; it implies that $S$ is unsatisfiable, as the original application condition does.
\begin{alignat}{1}
\ctag{ceq4hr1}{({\it $\gz$-dichotomy branching rule})} \\[1mm]
\notag
& \dfrac{S}
        {S\cup \{a\geql \gz\}\ \big|\ S\cup \{\gz\gle a\}}; \\[1mm]
\notag
& a\in \mi{atoms}(S).
\end{alignat}
The linearity of the standard strict order $<$ on $[0,1]$ together with the bounds $0$ and $1$ ensures that a $\gz$-dichotomy $a\geql \gz\vee \gz\gle a$ is obviously true in every valuation.
This means that either the guard $a\geql \gz$ or the guard $\gz\gle a$ is true in every valuation.
Rule (\cref{ceq4hr1}) is a branching one.
It splits derivation into two branches corresponding to the two subcases of the $\gz$-dichotomy;
i.e. in the first branch, we suppose that $a\geql \gz$ is true, and in the second one, $\gz\gle a$ is true.                                                                                \pagebreak[4]
\begin{alignat}{1}
\ctag{ceq4hr1x}{({\it $\gu$-dichotomy branching rule})} \\[1mm]
\notag
& \dfrac{S}
        {(S-\{l\vee C\})\cup \{\Cn\geql \gu\}\ \big|\ (S-\{l\vee C\})\cup \{C\}\cup \{\Cn\gle \gu\}}; \\[1mm]
\notag
& \begin{array}{l}
  l\vee C\in S-\mi{guards}(S),\ \text{\it either}\ l=\Cn\geql \gu\ \text{\it or}\ l=\gu\gleq \Cn, \\
  \Cn=a_0\swedge\cdots\swedge a_n, a_i\in \mi{PropAtom}, \square\neq C\in \mi{OrdPropCl}. 
  \end{array}
\end{alignat}
The linearity of the standard strict order $<$ on $[0,1]$ together with the bounds $0$ and $1$ guarantees in every valuation that 
either the input order literal $l=\Cn\geql \gu$ ($l=\gu\gleq \Cn$) or the order literal $\Cn\gle \gu$ is true, depending on whether all the atoms $a_i$ are assigned the truth value $1$, or not. 
In addition, $l$ belongs to the input order clause $l\vee C$.
Hence, either $l$ and also $l\vee C$ are true, or $\Cn\gle \gu$ and the remainder order clause $C$ are true in every model of the input order clausal theory $S$.
Rule (\cref{ceq4hr1x}) is a branching one as well.
It splits derivation into two branches: in the first branch, we suppose that $\Cn\geql \gu$ is true, and in the second one, $C$ and $\Cn\gle \gu$ are true.
\begin{alignat}{1}
\ctag{ceq4hr11}{({\it Pure dichotomy branching rule})} \\[1mm]
\notag
& \dfrac{S}
        {(S-\{l_1\vee C\})\cup \{l_1\}\ \big|\ (S-\{l_1\vee C\})\cup \{C\}\cup \{l_2\}}; \\[1mm]
\notag
& \begin{array}{l}
  l_1\vee C\in S-\mi{guards}(S), l_1, l_2\in \mi{PurOrdPropLit}, \square\neq C\in \mi{OrdPropCl}; \\
  l_1\vee l_2\ \text{\it is a pure dichotomy}.
  \end{array}
\end{alignat}
Analogously to the previous rule, the linearity of the standard strict order $<$ on $[0,1]$ ensures that the pure dichotomy $l_1\vee l_2$ is obviously true in every valuation.
Moreover, the input order literal $l_1$ belongs to the input order clause $l_1\vee C$.
This yields that either $l_1$ and also $l_1\vee C$ are true, or the order literal $l_2$ and the remainder order clause $C$ are true in every model of the input order clausal theory $S$.
Rule (\cref{ceq4hr11}) is a branching one as well, splitting derivation into two branches corresponding to the two subcases of the pure dichotomy;
i.e. in the first branch, we suppose that $l_1$ is true, and in the second one, $C$ and $l_2$ are true.
\begin{alignat}{1}
\ctag{ceq4hr111}{({\it Pure $\geql$-dichotomy branching rule})} \\[1mm]
\notag
& \dfrac{S}
        {\begin{array}{l}
         (S-\{\Cn_1\geql \Cn_2\vee C\})\cup \{\Cn_1\geql \Cn_2\}\ \big| \\ 
         (S-\{\Cn_1\geql \Cn_2\vee C\})\cup \{C\}\cup \{\Cn_1\gle \Cn_2\vee \Cn_2\gle \Cn_1\}
         \end{array}}; \\[1mm]
\notag
& \begin{array}{l}
  \Cn_1\geql \Cn_2\vee C\in S-\mi{guards}(S), \Cn_i\in \mi{PropConj}, \square\neq C\in \mi{OrdPropCl}, \\
  \Cn_1\neq \Cn_2.
  \end{array}
\end{alignat}
Once again, the linearity of the standard strict order $<$ on $[0,1]$ ensures that the pure order clause $\Cn_1\geql \Cn_2\vee \Cn_1\gle \Cn_2\vee \Cn_2\gle \Cn_1$ is obviously true in every valuation.
Moreover, the input order literal $\Cn_1\geql \Cn_2$ belongs to the input order clause $\Cn_1\geql \Cn_2\vee C$.
This yields that either $\Cn_1\geql \Cn_2$ and also $\Cn_1\geql \Cn_2\vee C$ are true, 
or the pure order clause $\Cn_1\gle \Cn_2\vee \Cn_2\gle \Cn_1$ and the remainder order clause $C$ are true in every model of the input order clausal theory $S$.
Rule (\cref{ceq4hr111}) is a branching one as well, splitting derivation into two branches corresponding to these two subcases:
in the first branch, we suppose that $\Cn_1\geql \Cn_2$ is true, and in the second one, $C$ and $\Cn_1\gle \Cn_2\vee \Cn_2\gle \Cn_1$ are true.                                            
\begin{alignat}{1}
\ctag{ceq4hr1111}{({\it Merging rule})} \\[1mm]
\notag
& \dfrac{S}
        {(S-\{(\bigvee_{i=0}^n \Cn_i\gle \gu)\vee C\})\cup \{(\bigmult_{i=0}^n \Cn_i)\gle \gu\vee C\}}; \\[1mm]
\notag
& \begin{array}{l}
  (\bigvee_{i=0}^n \Cn_i\gle \gu)\vee C\in S-\mi{guards}(S), \Cn_i\in \mi{PropConj}, n\geq 1; \\
  C\in \mi{OrdPropCl}\ \text{\it does not contain an order literal of the form}\ \Cn\gle \gu, \\
  \Cn\in \mi{PropConj}.
  \end{array}
\end{alignat}
Note that $(\bigmult_{i=0}^n \Cn_i)\gle \gu\not\in C$.
In every valuation, the input order subclause $\bigvee_{i=0}^n \Cn_i\gle \gu$ is true if and only if the order literal $(\bigmult_{i=0}^n \Cn_i)\gle \gu$ is true.
This yields in every valuation that the input order clause $(\bigvee_{i=0}^n \Cn_i\gle \gu)\vee C$ is true if and only if the order clause $(\bigmult_{i=0}^n \Cn_i)\gle \gu\vee C$ is true.  
So, the input order literals $\Cn_0\gle \gu,\dots,\Cn_n\gle \gu$ can be merged into the one $(\bigmult_{i=0}^n \Cn_i)\gle \gu$, and 
$(\bigvee_{i=0}^n \Cn_i\gle \gu)\vee C$ can be replaced with $(\bigmult_{i=0}^n \Cn_i)\gle \gu\vee C$ in the input order clausal theory $S$.
\begin{alignat}{1}
\ctag{ceq4hr11111}{({\it Simplification rule})} \\[1mm]
\notag
& \dfrac{S}
        {(S-\{l\vee C\})\cup \{C\cup \{a_0\swedge\cdots\swedge a_n\diamond^* \gu\}\}}; \\[1mm]
\notag
& \begin{array}{l}
  l\vee C\in S-\mi{guards}(S),\ \text{\it either}\ l=\Cn\diamond \gu\ \text{\it or}\ l=\gu\gleq \Cn, \Cn=a_0^{\alpha_0}\swedge\cdots\swedge a_n^{\alpha_n}, \\ 
  a_i\in \mi{PropAtom}, \alpha_i\geq 1, \diamond\in \{\geql,\gle\}, C\in \mi{OrdPropCl}; \\[1mm]
  \text{\it there exists}\ i^*\leq n\ \text{\it such that}\ \alpha_{i^*}\geq 2; \\[1mm]
  \diamond^*=\left\{\begin{array}{ll}
                    \geql &\ \text{\it if either}\ l=\Cn\geql \gu\ \text{\it or}\ l=\gu\gleq \Cn, \\[1mm]
                    \gle  &\ \text{\it if}\ l=\Cn\gle \gu.
                    \end{array}
             \right. 
  \end{array}
\end{alignat}
In every valuation, the input order literal $l$ is true if and only if the order literal $a_0\swedge\cdots\swedge a_n\diamond^* \gu$ is true (the exponents $\alpha_0,\dots,\alpha_n$ have been removed).
This yields in every valuation that the input order clause $l\vee C$ is true if and only if the order clause $C\cup \{a_0\swedge\cdots\swedge a_n\diamond^* \gu\}$ is true.  
So, $l\vee C$ can be replaced with $C\cup \{a_0\swedge\cdots\swedge a_n\diamond^* \gu\}$ in the input order clausal theory $S$.
\begin{alignat}{1}
\ctag{ceq4hr111111}{({\it Expansion rule})} \\[1mm]
\notag
& \dfrac{S}
        {(S-\{l\})\cup \{a_0\geql \gu,\dots,a_n\geql \gu\}}; \\[1mm]
\notag
& \begin{array}{l}
  l\in S-\mi{guards}(S),\ \text{\it either}\ l=\Cn\geql \gu\ \text{\it or}\ l=\gu\gleq \Cn, \\
  \Cn=a_0\swedge\cdots\swedge a_n, a_i\in \mi{PropAtom}, n\geq 1.
  \end{array}
\end{alignat}
For every valuation ${\mc V}$, the input unit order clause $l$ is true in ${\mc V}$ if and only if 
${\mc V}$ is a model of the order clausal theory $\{a_0\geql \gu,\dots,a_n\geql \gu\}$, consisting of guards.
So, $l$ can be replaced with the guards $a_0\geql \gu,\dots,a_n\geql \gu$ in the input order clausal theory $S$.                                                                           \pagebreak[4]
\begin{alignat}{1}
\ctag{ceq4hr1111111}{({\it Guard $\gz$-replacement rule})} \\[1mm]
\notag
& \dfrac{S}
        {(S-\{a\gleq \gz\})\cup \{a\geql \gz\}}; \\[1mm]
\notag
& a\gleq \gz\in \mi{guards}(S), a\in \mi{PropAtom}.
\end{alignat}
In every valuation, the input guard $a\gleq \gz$ is true if and only if the guard $a\geql \gz$ is true.
So, $a\gleq \gz$ can be replaced with $a\geql \gz$ in the input order clausal theory $S$.
Analogously, a guard $\gu\gleq a$ can be replaced with the guard $a\geql \gu$ in $S$.
\begin{alignat}{1}
\ctag{ceq4hr11111111}{({\it Guard $\gu$-replacement rule})} \\[1mm]
\notag
& \dfrac{S}
        {(S-\{\gu\gleq a\})\cup \{a\geql \gu\}}; \\[1mm]
\notag
& \gu\gleq a\in \mi{guards}(S), a\in \mi{PropAtom}.
\end{alignat}
\begin{alignat}{1}
\ctag{ceq4hr2}{({\it Contradiction rule})} \\[1mm]
\notag
& \dfrac{S}
        {(S-\{l\vee C\})\cup \{C\}}; \\[1mm]
\notag
& l\vee C\in S-\mi{guards}(S); l\in \mi{OrdPropLit}\ \text{\it is a contradiction};\ C\in \mi{OrdPropCl}.
\end{alignat}
If the input order literal $l$ is a contradiction, then it may be removed from the input order clause $l\vee C$.
\begin{alignat}{1}
\ctag{ceq4hr22}{({\it Tautology rule})} \\[1mm]    
\notag
& \dfrac{S}
        {S-\{l\vee C\}}; \\[1mm]            
\notag
& l\vee C\in S-\mi{guards}(S); l\in \mi{OrdPropLit}\ \text{\it is a tautology};\ C\in \mi{OrdPropCl}.    
\end{alignat}
If the input order literal $l$ is a tautology, then the input order clause $l\vee C$ may be removed from the input order clausal theory $S$ altogether. 
\begin{alignat}{1}
\ctag{ceq4hr3}{({\it $\gz$-simplification rule})} \\[1mm]
\notag
& \dfrac{S}
        {(S-\{C\})\cup \{\mi{simplify}(C,a,\gz)\}}; \\[1mm]
\notag
& a\geql \gz\in \mi{guards}(S), a\in \mi{atoms}(C), C\in S, a\geql \gz\neq C.
\end{alignat}
If the input guard $a\geql \gz$ and the input order clause $C$ contain some atom $a$ in common,
then $C$ can be simplified in the input order clausal theory $S$ using the auxiliary function application $\mi{simplify}(C,a,\gz)$.
Analogously, $C$ can be simplified in $S$ exploiting a guard $a\geql \gu$.
\begin{alignat}{1}
\ctag{ceq4hr4}{({\it $\gu$-simplification rule})} \\[1mm]
\notag
& \dfrac{S}
        {(S-\{C\})\cup \{\mi{simplify}(C,a,\gu)\}}; \\[1mm]
\notag
& a\geql \gu\in \mi{guards}(S), a\in \mi{atoms}(C), C\in S, a\geql \gu\neq C.
\end{alignat}
\begin{alignat}{1}
\ctag{ceq4hr5}{({\it $\gz$-contradiction rule})} \\[1mm]
\notag
& \dfrac{S}
        {(S-\{l\vee C\})\cup \{C\}}; \\[1mm]
\notag
& \begin{array}{l}
  l\vee C\in S-\mi{guards}(S),\ \text{\it either}\ l=\Cn\geql \gz\ \text{\it or}\ l=\Cn\gleq \gz, \Cn=a_0^{\alpha_0}\swedge\cdots\swedge a_n^{\alpha_n}, \\ 
  a_i\in \mi{PropAtom}, \alpha_i\geq 1, \gz\gle a_i\in \mi{guards}(S), C\in \mi{OrdPropCl}.
  \end{array}
\end{alignat}
\begin{alignat}{1}
\ctag{ceq4hr6}{({\it $\gu$-contradiction rule})} \\[1mm] 
\notag
& \dfrac{S}
        {(S-\{l\vee C\})\cup \{C\}}; \\[1mm]
\notag
& \begin{array}{l}
  l\vee C\in S-\mi{guards}(S),\ \text{\it either}\ l=\Cn\geql \gu\ \text{\it or}\ l=\gu\gleq \Cn, \Cn=a_0^{\alpha_0}\swedge\cdots\swedge a_n^{\alpha_n}, \\
  a_i\in \mi{PropAtom}, \alpha_i\geq 1, C\in \mi{OrdPropCl}; \\
  \text{\it there exists}\ i^*\leq n\ \text{\it such that}\ a_{i^*}\gle \gu\in \mi{guards}(S).
  \end{array}
\end{alignat}
These two rules detect an unsatisfiable set of the input guard(s) and the unit order clause $l$ of the form either $\{\gz\gle a_0,\dots,\gz\gle a_n,l\}$ or $\{a_{i^*}\gle \gu,l\}$.
In either case, the input order literal $l$ may be removed from the input order clause $l\vee C$.                                                                                         
\begin{alignat}{1}
\ctag{ceq4hr55}{({\it $\gz$-consequence rule})} \\[1mm]
\notag
& \dfrac{S}
        {S-\{\gz\gle a_0^{\alpha_0}\swedge\cdots\swedge a_n^{\alpha_n}\vee C\}}; \\[1mm]
\notag
& \begin{array}{l}
  \gz\gle a_0^{\alpha_0}\swedge\cdots\swedge a_n^{\alpha_n}\vee C\in S-\mi{guards}(S), a_i\in \mi{PropAtom}, \alpha_i\geq 1, \\
  \gz\gle a_i\in \mi{guards}(S), C\in \mi{OrdPropCl}.
  \end{array}
\end{alignat}
\begin{alignat}{1}
\ctag{ceq4hr66}{({\it $\gu$-consequence rule})} \\[1mm]
\notag
& \dfrac{S}
        {S-\{a_0^{\alpha_0}\swedge\cdots\swedge a_n^{\alpha_n}\gle \gu\vee C\}}; \\[1mm]
\notag
& \begin{array}{l}
  a_0^{\alpha_0}\swedge\cdots\swedge a_n^{\alpha_n}\gle \gu\vee C\in S-\mi{guards}(S), a_i\in \mi{PropAtom}, \alpha_i\geq 1, C\in \mi{OrdPropCl}; \\
  \text{\it there exists}\ i^*\leq n\ \text{\it such that}\ a_{i^*}\gle \gu\in \mi{guards}(S).
  \end{array}
\end{alignat}
These two rules detect that the unit order clause $\gz\gle a_0^{\alpha_0}\swedge\cdots\swedge a_n^{\alpha_n}$ ($a_0^{\alpha_0}\swedge\cdots\swedge a_n^{\alpha_n}\gle \gu$) is
a consequence of the input guard(s):
either $\{\gz\gle a_0,\dots,\gz\gle a_n\}\models \gz\gle a_0^{\alpha_0}\swedge\cdots\swedge a_n^{\alpha_n}$ or $a_{i^*}\gle \gu\models a_0^{\alpha_0}\swedge\cdots\swedge a_n^{\alpha_n}\gle \gu$.
Then the input order clause $\gz\gle a_0^{\alpha_0}\swedge\cdots\swedge a_n^{\alpha_n}\vee C$ ($a_0^{\alpha_0}\swedge\cdots\swedge a_n^{\alpha_n}\gle \gu\vee C$), 
which is a consequence of the input guard(s), may be removed from the input order clausal theory $S$ altogether.
\begin{alignat}{1}
\ctag{ceq4hr7}{({\it $\gz$-annihilation rule})} \\[1mm]
\notag
& \dfrac{S}  
        {S-\{a\geql \gz\}}; \\[1mm]
\notag
& a\geql \gz\in \mi{guards}(S), a\in \mi{PropAtom}, a\not\in \mi{atoms}(S-\{a\geql \gz\}).
\end{alignat}
\begin{alignat}{1}
\ctag{ceq4hr8}{({\it $\gu$-annihilation rule})} \\[1mm]
\notag
& \dfrac{S}
        {S-\{a\geql \gu\}}; \\[1mm]                                 
\notag
& a\geql \gu\in \mi{guards}(S), a\in \mi{PropAtom}, a\not\in \mi{atoms}(S-\{a\geql \gu\}).
\end{alignat}
If the only occurrence of the atom $a$ in the input order clausal theory $S$ is in the input guard $a\geql \gz$ ($a\geql \gu$), 
then $S$ is equisatisfiable to the order clausal theory $S-\{a\geql \gz\}$ ($S-\{a\geql \gu\}$), and $a\geql \gz$ ($a\geql \gu$) may be removed from $S$.

\begin{lemma}
\label{le333}  
Let $n\in \{1,2\}$, $S, S_1,\dots,S_n\subseteq \mi{OrdPropCl}$, 
$\dfrac{S}
       {S_1\ \big|\cdots\big|\ S_n}$ be an application of Rule {\rm (\cref{ceq4hr0})}--{\rm (\cref{ceq4hr66})}; ${\mf A}$ be a valuation.
\begin{enumerate}[\rm (i)]
\item
For all $1\leq i\leq n$, $\mi{atoms}(S)\supseteq \mi{atoms}(S_i)$;
\item
${\mf A}\models S$ if and only if there exists $1\leq i^*\leq n$ satisfying ${\mf A}\models S_{i^*}$. 
\end{enumerate}
\end{lemma}

\begin{proof}
The proof is by case analysis for every Rule (\cref{ceq4hr0})--(\cref{ceq4hr66}).
%
%
%
\end{proof}

\begin{lemma}
\label{le3333}
Let $S\subseteq \mi{OrdPropCl}$, $a\geql \tau\in \mi{guards}(S)$, $a\in \mi{PropAtom}$, $\tau\in \{\gz,\gu\}$, $a\not\in \mi{atoms}(S-\{a\geql \tau\})$, 
$\dfrac{S}
       {S-\{a\geql \tau\}}$ be an application of Rule either {\rm (\cref{ceq4hr7})} or {\rm (\cref{ceq4hr8})}.
\begin{enumerate}[\rm (i)]
\item
$\mi{atoms}(S-\{a\geql \tau\})=\mi{atoms}(S)-\{a\}$;
\item
if there exists a valuation ${\mf A}$ satisfying ${\mf A}\models S$, then ${\mf A}\models S-\{a\geql \tau\}$;
\item
if there exists a valuation ${\mf A}'$ satisfying ${\mf A}'\models S-\{a\geql \tau\}$, then there exists a valuation ${\mf A}$ satisfying ${\mf A}\models S$,
${\mf A}|_{\mi{atoms}(S)}={\mf A}'|_{\mi{atoms}(S)-\{a\}}\cup \{(a,\|\tau\|^{{\mf A}'})\}$.
\end{enumerate}
\end{lemma}

\begin{proof}
Straightforward.
%
%
%
\end{proof}

All Rules (\cref{ceq4hr0})--(\cref{ceq4hr8}) are sound with respect to satisfiability.

\begin{lemma}
\label{le33333}
Let $n\in \{1,2\}$, $S, S_1,\dots,S_n\subseteq \mi{OrdPropCl}$,
$\dfrac{S}
       {S_1\ \big|\cdots\big|\ S_n}$ be an application of Rule {\rm (\cref{ceq4hr0})}--{\rm (\cref{ceq4hr8})}.
$S$ is satisfiable if and only if there exists $1\leq i^*\leq n$ such that $S_{i^*}$ is satisfiable.
\end{lemma}

\begin{proof}
An immediate consequence of Lemmata \ref{le333} and \ref{le3333}.
%
%
%
\end{proof}

Using the basic rules, one can construct a {\it DPLL} tree derivation so as the classical {\it DPLL} procedure does.
A {\it DPLL}-tree $\mi{Tree}$ is a labelled rooted (arborescence) tree.
Vertices are labelled with order clausal theories, called label order clausal theories; the root is labelled with the input order clausal theory.
For every inner vertex $v$ of $\mi{Tree}$ and its successor vertices $v_1,\dots,v_n$, $n\in \{1,2\}$, 
there exists an application $\dfrac{S}
                                   {S_1\ \big|\cdots\big|\ S_n}$, $S, S_1,\dots,S_n\subseteq \mi{OrdPropCl}$, of Rule (\cref{ceq4hr0})--(\cref{ceq4hr8}) such that
$v$ is labelled with $S$, the antecedent order clausal theory of this application, and
$v_1,\dots,v_n$ are labelled with $S_1,\dots,S_n$, respectively, the consequent order clausal theories of this application corresponding to its branches.
Note that a {\it DPLL}-tree is a finitely generated tree; every inner vertex has at most two successor vertices.
A branch of a {\it DPLL}-tree is closed iff it has a leaf $v$ labelled with $\square\in S_v\subseteq \mi{OrdPropCl}$; such a branch is finite.
A branch of a {\it DPLL}-tree is open iff it is not closed.
A {\it DPLL}-tree is closed iff every its branch is closed.
Note that a closed {\it DPLL}-tree is finite by K\"{o}nig's Lemma.
A {\it DPLL}-tree is open iff it is not closed.
A {\it DPLL}-tree is linear iff it consists of only one branch, beginning from its root and ending in its only leaf.
Notice that phrases such as "with the root $v$ labelled with $S$", "for every leaf $v$ labelled with $S$", "for its only leaf $v$ labelled with $S$" will be abbreviated 
as "with the root $S$", "for every leaf $S$", "for its only leaf $S$", respectively.

The following lemma solves the satisfiability problem for the basic case where the order clausal theory in question is positive and unit.
It introduces a criterion for satisfiability: 
if there does not exist an application of Rule (\cref{ceq4hr0}) to the order clausal theory, then it is satisfiable, and
we are able to construct a model of it.

\begin{lemma}
\label{le2}
Let $S\subseteq_{\mc F} \mi{OrdPropCl}$ be positive and unit, and there not exist an application of Rule {\rm (\cref{ceq4hr0})} to $S$.
$S$ is satisfiable.
\end{lemma}

\begin{proof}
Let $\mi{atoms}(S)=\{a_1,\dots,a_m\}$.
We construct a sequence of partial valuations ${\mc V}_\iota$, $\iota\leq m$, so that ${\mc V}_m\models S$. 
The proof is by induction on $\iota\leq m$.
%
%
%
\end{proof}

\begin{lemma}
\label{le33} 
Let $S\subseteq \mi{OrdPropCl}$ and $\mi{Tree}$ be a finite {\it DPLL}-tree with the root $S$ constructed using Rules {\rm (\cref{ceq4hr0})}--{\rm (\cref{ceq4hr8})}.
\begin{enumerate}[\rm (i)]
\item
If there exists a valuation ${\mf A}$ satisfying ${\mf A}\models S$, then there exists a leaf $S^*$ of $\mi{Tree}$ satisfying ${\mf A}\models S^*$;
\item
if there exist a leaf $S^*$ of $\mi{Tree}$ and a valuation ${\mf A}'$ satisfying ${\mf A}'\models S^*$, then there exists a model ${\mf A}$ of $S$ related to $\mi{Tree}$.
\end{enumerate}
\end{lemma}

\begin{proof}
The proof is by induction on the structure of $\mi{Tree}$ using Lemmata \ref{le333} and \ref{le3333}.                                                                                      \linebreak[4]
%
%
%
\end{proof}

\begin{lemma}
\label{le3}  
Let $S\subseteq \mi{OrdPropCl}$.
If there exists a closed {\it DPLL}-tree with the root $S$ constructed using Rules {\rm (\cref{ceq4hr0})}--{\rm (\cref{ceq4hr8})}, then $S$ is unsatisfiable.
\end{lemma}

\begin{proof}
An immediate consequence of Lemma \ref{le33}(i).
%
%
%
\end{proof}

\begin{lemma}
\label{le44}
Let $S\subseteq_{\mc F} \mi{OrdPropCl}$ be positive, $a^*\in \mi{atoms}(S)$, $\mi{guards}(S,a^*)=\{\gz\gle a^*\}$.
There exists a finite linear {\it DPLL}-tree with the root $S\cup \{a^*\gle \gu\}$ constructed using Rule {\rm (\cref{ceq4hr66})} such that
for its only leaf $S'$, $S'\subseteq_{\mc F} \mi{OrdPropCl}$ is positive, 
$S'-\mi{guards}(S')\subseteq S-\mi{guards}(S)$, $\mi{guards}(S')=\mi{guards}(S)\cup \{a^*\gle \gu\}$, $a^*\gle \gu\not\in S$.
\end{lemma}

\begin{proof}
We define a binary measure operator $\mi{undeleted} : \mi{PropAtom}\times {\mc P}(\mi{OrdPropCl}^\gu)\longrightarrow {\mc P}(\mi{OrdPropCl}^\gu)$,
$\mi{undeleted}(a^F,S^F)=\{C \,|\, C=a_0\swedge\cdots\swedge a_n\gle \gu\vee C^\natural\in S^F, a_i\in \mi{PropAtom}, C^\natural\in \mi{OrdPropCl}^\gu, a^F\in \{a_0,\dots,a_n\}\}$.
The proof is by induction on $\mi{undeleted}(a^*,S-\mi{guards}(S))\subseteq_{\mc F} \mi{OrdPropCl}^\gu$.
%
%
%
\end{proof}

The following lemma solves the satisfiability problem for the case where the order clausal theory in question is positive.

\begin{lemma}
\label{le4}
Let $S\subseteq_{\mc F} \mi{OrdPropCl}$ be positive.
There exists a finite {\it DPLL}-tree $\mi{Tree}$ with the root $S$ constructed using Rules {\rm (\cref{ceq4hr0})}, {\rm (\cref{ceq4hr11})}, {\rm (\cref{ceq4hr111})}, {\rm (\cref{ceq4hr66})} 
with the following properties:
\begin{alignat}{1}
\label{eq5a}
& \text{if}\ S\ \text{is unsatisfiable, then}\ \mi{Tree}\ \text{is closed}; \\[1mm]
\label{eq5b}
& \text{if}\ S\ \text{is satisfiable, then}\ \mi{Tree}\ \text{is open, and} \\
\notag
& \phantom{\text{if}\ S\ \text{is satisfiable, then}\ \mbox{}}
                                             \text{there exists a model}\ {\mf A}\ \text{of}\ S\ \text{related to}\ \mi{Tree}.
\end{alignat}
\end{lemma}

\begin{proof}
We exploit the excess literal technique \cite{ANBL70}. 
Let $l^F\in \mi{OrdPropLit}$, $C^F\in \mi{OrdPropCl}$, $\square\not\in S^F\subseteq_{\mc F} \mi{OrdPropCl}$.
We define three measure operators as follows:
\begin{alignat*}{1}
\mi{count}(l^F)     &= \left\{\begin{array}{ll}
                              2 &\ \text{\it if}\ l^F=\varepsilon_1\geql \varepsilon_2, \varepsilon_i\in \{\gz,\gu\}\cup \mi{PropConj}, \\[1mm]
                              1 &\ \text{\it if}\ l^F=\varepsilon_1\diamond \varepsilon_2, \varepsilon_i\in \{\gz,\gu\}\cup \mi{PropConj}, \diamond\in \{\gleq,\gle\};
                              \end{array}
                       \right. \\[1mm]
\mi{count}(C^F)     &= \sum_{l\in C^F} \mi{count}(l); \\
\mi{elmeasure}(S^F) &= \sum_{C\in S^F} \mi{count}(C)-1.
\end{alignat*}
The proof is by induction on $\mi{elmeasure}(S)$.
%
%
%
\end{proof}

\begin{lemma}
\label{le555}
Let $S\subseteq_{\mc F} \mi{OrdPropCl}$ and $C^*\in S$.
There exists a finite linear {\it DPLL}-tree with the root $S$ constructed using Rules {\rm (\cref{ceq4hr2})} and {\rm (\cref{ceq4hr22})} such that
for its only leaf $S'$, either $\square\in S'$, or $\square\not\in S'\subseteq_{\mc F} \mi{OrdPropCl}$ and exactly one of the following points holds.
\begin{enumerate}[\rm (a)]
\item
$S'=S$, $C^*\neq \square$ does not contain contradictions and tautologies;
\item
$S'=S-\{C^*\}$;
\item
there exists $C^{**}\in \mi{OrdPropCl}$ satisfying that $S'=(S-\{C^*\})\cup \{C^{**}\}$, $C^{**}\not\in S$, $\square\neq C^{**}\subset C^*$ does not contain contradictions and tautologies.
\end{enumerate}
\end{lemma}

\begin{proof}
Let $C^F\in \mi{OrdPropCl}$.
We define a measure operator $\mi{unsimplified}(C^F)=\{l \,|\, l\in C^F\ \text{\it is either a contradiction or tautology}\}$. 
The proof is by induction on $\mi{unsimplified}(C^*)\subseteq_{\mc F} \mi{OrdPropLit}$.
%
%
%
\end{proof}

We are in position to propose reduction of a general order clausal theory to a positive order clausal theory.
We reduce step-by-step the general order clausal theory to a simplified theory, Lemma \ref{le55}, a $\gz$-guarded theory, Lemma \ref{le5}, a positively guarded theory, Lemma \ref{le7}, and
finally to a positive one, Lemma \ref{le9}. 
The satisfiability problem for the case where the resulting order clausal theory is positive is then solved, as is shown by Lemma \ref{le4}.

\begin{lemma}
\label{le55}  
Let $S\subseteq_{\mc F} \mi{OrdPropCl}$.
There exists a finite linear {\it DPLL}-tree with the root $S$ constructed using Rules {\rm (\cref{ceq4hr2})} and {\rm (\cref{ceq4hr22})} such that
for its only leaf $S'$, either $\square\in S'$ or $S'\subseteq_{\mc F} \mi{OrdPropCl}$ is simplified.
\end{lemma}

\begin{proof}
Let $S^F\subseteq \mi{OrdPropCl}$.
We define a measure operator $\mi{unsimplified}(S^F)=\{C \,|\, C\in S^F\ \text{\it contains a contradiction or tautology}\}$.
The proof is by induction on $\mi{unsimplified}(S)\subseteq_{\mc F} \mi{OrdPropCl}$ using Lemma \ref{le555}.
%
%
%
\end{proof}

\begin{lemma}
\label{le5}  
Let $S\subseteq_{\mc F} \mi{OrdPropCl}$ be simplified.
There exists a finite {\it DPLL}-tree with the root $S$ constructed using Rules {\rm (\cref{ceq4hr1})}, {\rm (\cref{ceq4hr1111111})}--{\rm (\cref{ceq4hr4})} such that
for every leaf $S'$, either $\square\in S'$ or $S'\subseteq_{\mc F} \mi{OrdPropCl}$ is $\gz$-guarded.
\end{lemma}

\begin{proof}
Let $S^F\subseteq \mi{OrdPropCl}$.
We define a measure operator $\mi{unguarded}(S^F)=\{a \,|\, a\in \mi{atoms}(S^F)\ \text{\it is not $\gz$-guarded in}\ S^F\}$.
The proof is by induction on $\mi{unguarded}(S)\subseteq_{\mc F} \mi{PropAtom}$.
%
%
%
\end{proof}

\begin{lemma}
\label{le6}  
Let $S\subseteq_{\mc F} \mi{OrdPropCl}$ be $\gz$-guarded.
\begin{enumerate}[\rm (i)]
\item
Let $a^*\in \mi{atoms}(S)$ and $\mi{guards}(S,a^*)=\{a^*\geql \gz\}$.
There exists a finite linear {\it DPLL}-tree with the root $S$ constructed using Rules {\rm (\cref{ceq4hr1111111})}--{\rm (\cref{ceq4hr4})}, {\rm (\cref{ceq4hr7})} such that
for its only leaf $S'$, either $\square\in S'$, or $S'\subseteq_{\mc F} \mi{OrdPropCl}$ is $\gz$-guarded, $\mi{atoms}(S')\subseteq \mi{atoms}(S)-\{a^*\}$.
\item
Let $a^*\in \mi{atoms}(S)$ and $\mi{guards}(S,a^*)=\{a^*\geql \gu\}$.
There exists a finite linear {\it DPLL}-tree with the root $S$ constructed using Rules {\rm (\cref{ceq4hr1111111})}--{\rm (\cref{ceq4hr4})}, {\rm (\cref{ceq4hr8})} such that
for its only leaf $S'$, either $\square\in S'$, or $S'\subseteq_{\mc F} \mi{OrdPropCl}$ is $\gz$-guarded, $\mi{atoms}(S')\subseteq \mi{atoms}(S)-\{a^*\}$.
\end{enumerate}
\end{lemma}

\begin{proof}
Let $S^F\subseteq \mi{OrdPropCl}$.
We distinguish two cases for $\mi{guards}(S,a^*)$.

Case 1:
$\mi{guards}(S,a^*)=\{a^*\geql \gz\}$.
We define a measure operator $\mi{restricted}(S^F)=\{C \,|\, C\in S^F, a^*\in \mi{atoms}(C), C\neq a^*\geql \gz\}$.

Case 2:
$\mi{guards}(S,a^*)=\{a^*\geql \gu\}$.
We define a measure operator $\mi{restricted}(S^F)=\{C \,|\, C\in S^F, a^*\in \mi{atoms}(C), C\neq a^*\geql \gu\}$.

In both Cases 1 and 2, the proof is by induction on $\mi{restricted}(S)\subseteq_{\mc F} \mi{OrdPropCl}$.                                                                                  \linebreak[4]
%
%
%
\end{proof}

\begin{lemma}
\label{le66}  
Let $S\subseteq_{\mc F} \mi{OrdPropCl}$ be semi-positively guarded, $a^*\in \mi{atoms}(S)$, $\mi{guards}(S,a^*)=\{a^*\geql \gu\}$.
There exists a finite linear {\it DPLL}-tree with the root $S$ constructed using Rules {\rm (\cref{ceq4hr1111111})}--{\rm (\cref{ceq4hr4})}, {\rm (\cref{ceq4hr8})} such that
for its only leaf $S'$, either $\square\in S'$, or $S'\subseteq_{\mc F} \mi{OrdPropCl}$ is semi-positively guarded, $\mi{atoms}(S')\subseteq \mi{atoms}(S)-\{a^*\}$.
\end{lemma}

\begin{proof}
Let $S^F\subseteq \mi{OrdPropCl}$.
We define a measure operator $\mi{restricted}(S^F)=\{C \,|\, C\in S^F, a^*\in \mi{atoms}(C), C\neq a^*\geql \gu\}$.
The proof is by induction on $\mi{restricted}(S)\subseteq_{\mc F} \mi{OrdPropCl}$.
%
%
%
\end{proof}

\begin{lemma}
\label{le7}  
Let $S\subseteq_{\mc F} \mi{OrdPropCl}$ be $\gz$-guarded.
There exists a finite linear {\it DPLL}-tree with the root $S$ constructed using Rules {\rm (\cref{ceq4hr1111111})}--{\rm (\cref{ceq4hr4})}, {\rm (\cref{ceq4hr7})}, {\rm (\cref{ceq4hr8})} such that
for its only leaf $S'$, either $\square\in S'$ or $S'\subseteq_{\mc F} \mi{OrdPropCl}$ is positively guarded.  
\end{lemma}

\begin{proof}
The proof is by induction on $\mi{atoms}(S)\subseteq_{\mc F} \mi{PropAtom}$ using Lemma \ref{le6}.
%
%
%
\end{proof}

\begin{lemma}
\label{le88}  
Let $S\subseteq_{\mc F} \mi{OrdPropCl}$ be simplified, and $\{a_1,\dots,a_n\}\subseteq \mi{atoms}(S)$, $\mi{guards}(S,a_i)=\{\gz\gle a_i,a_i\geql \gu\}$.
There exists a finite linear {\it DPLL}-tree with the root $S$ constructed using Rules {\rm (\cref{ceq4hr22})} and {\rm (\cref{ceq4hr4})} such that
for its only leaf $S'$, $S'=S-\{\gz\gle a_1,\dots,\gz\gle a_n\}\subseteq_{\mc F} \mi{OrdPropCl}$ is simplified.
\end{lemma}

\begin{proof}
The proof is by induction on $n$.
%
%
%
\end{proof}

We define two auxiliary binary relations as follows:
\begin{IEEEeqnarray*}{RL}
\IEEEeqnarraymulticol{2}{l}{
\mi{valid}\subseteq \mi{OrdPropLit}\times {\mc P}(\mi{OrdPropCl});} \\[1mm]
\mi{valid}(l^F,S^F) &\longleftrightarrow \left\{\begin{array}{l}
                                                l^F\in \mi{OrdPropLit}^\gu; \\[1mm]
                                                \text{\it if}\ \text{\it either}\ l^F=\Cn\diamond \gu\ \text{\it or}\ l^F=\gu\gleq \Cn, \\
                                                \phantom{\text{\it if}\ \mbox{}}
                                                               \Cn\in \mi{PropConj}, \diamond\in \{\geql,\gle\}, \\ 
                                                \quad \text{\it then}\ \Cn=a_0\swedge\cdots\swedge a_n, a_i\in \mi{PropAtom}, \\
                                                \quad \phantom{\text{\it then}\ \mbox{}}
                                                                       \mi{guards}(S^F,a_i)=\{\gz\gle a_i\};
                                                \end{array}
                                         \right. \\[1mm] 
\IEEEeqnarraymulticol{2}{l}{
\mi{valid}\subseteq \mi{OrdPropCl}\times {\mc P}(\mi{OrdPropCl});} \\[1mm]
\mi{valid}(C^F,S^F) &\longleftrightarrow \left\{\begin{array}{l}
                                                C^F\neq \square; \\
                                                \text{\it for all}\ l\in C^F,\ \mi{valid}(l,S^F); \\[1mm]
                                                \text{\it if}\ C^F=\Cn\gle \gu\vee C, \Cn\in \mi{PropConj}, \square\neq C\in \mi{OrdPropCl}, \\
                                                \quad \text{\it then}\ C\ \text{\it does not contain an order literal of the form} \\ 
                                                \quad \phantom{\text{\it then}\ \mbox{}}
                                                                       \Cn'\gle \gu, \Cn'\in \mi{PropConj}.
                                                \end{array}
                                         \right. 
\end{IEEEeqnarray*}

\begin{lemma}
\label{le8}
Let $S\subseteq_{\mc F} \mi{OrdPropCl}$ be positively guarded and $C^*\in S-\mi{guards}(S)$.
There exists a finite linear {\it DPLL}-tree with the root $S$ constructed using Rules {\rm (\cref{ceq4hr1111})}--{\rm (\cref{ceq4hr66})} such that
for its only leaf $S'$, either $\square\in S'$, or $S'\subseteq_{\mc F} \mi{OrdPropCl}$ is semi-positively guarded, and exactly one of the following points holds.
\begin{enumerate}[\rm (a)]
\item
$S'=S$, $\mi{valid}(C^*,S)$;
\item
$S'=S-\{C^*\}$, $\mi{guards}(S')=\mi{guards}(S)$;
\item
there exists $C^{**}\in \mi{OrdPropCl}^\gu$, $\mi{atoms}(C^{**})\subseteq \mi{atoms}(C^*)$, satisfying 
$S'=(S-\{C^*\})\cup \{C^{**}\}$, $\mi{guards}(S')=\mi{guards}(S)$, $C^{**}\not\in S$, $\mi{valid}(C^{**},S')$;
\item
there exists $a^*\gle \gu\in \mi{OrdPropCl}^\gu$, $a^*\in \mi{atoms}(C^*)$, satisfying 
$S'=(S-\{C^*\})\cup \{a^*\gle \gu\}$, $\mi{guards}(S')=\mi{guards}(S)\cup \{a^*\gle \gu\}$, $a^*\gle \gu\not\in S$;
\item
there exists $S^{**}=\{a_0\geql \gu,\dots,a_n\geql \gu\}\subseteq \mi{OrdPropCl}^\gu$, $\{a_0,\dots,a_n\}\subseteq \mi{atoms}(C^*)$, $\{\gz\gle a_0,\dots,\gz\gle a_n\}\subseteq \mi{guards}(S)$, satisfying 
$S'=(S-(\{C^*\}\cup \{\gz\gle a_0,\dots,\gz\gle a_n\}))\cup S^{**}$, $\mi{guards}(S')=(\mi{guards}(S)-\{\gz\gle a_0,\dots,\gz\gle a_n\})\cup S^{**}$, $S^{**}\cap S=\emptyset$. 
\end{enumerate}
\end{lemma}

\begin{proof}
Let $S^F\subseteq \mi{OrdPropCl}$ and $C^F\in S^F-\mi{guards}(S^F)$.
We define a binary measure partial operator $\mi{invalid} : \mi{OrdPropCl}\times {\mc P}(\mi{OrdPropCl})\longrightarrow {\mc P}(\mi{OrdPropLit})$,
$\mi{invalid}(C^F,S^F)=\{l \,|\, l\in C^F,\ \text{\it not}\ \mi{valid}(l,S^F)\}$.
The proof is by induction on $\mi{invalid}(C^*,S)\subseteq_{\mc F} \mi{OrdPropLit}$.
%
%
%
\end{proof}

\begin{lemma}
\label{le9}  
Let $S\subseteq_{\mc F} \mi{OrdPropCl}$ be semi-positively guarded.
There exists a finite {\it DPLL}-tree with the root $S$ constructed using Rules {\rm (\cref{ceq4hr1x})}, {\rm (\cref{ceq4hr1111})}--{\rm (\cref{ceq4hr66})}, {\rm (\cref{ceq4hr8})} such that
for every leaf $S'$, either $\square\in S'$ or $S'\subseteq_{\mc F} \mi{OrdPropCl}$ is positive.  
\end{lemma}

\begin{proof}
Let $l^F\in \mi{OrdPropLit}$, $C^F\in \mi{OrdPropCl}$, $S^F\subseteq_{\mc F} \mi{OrdPropCl}$.
We define six measure operators and two auxiliary binary relations as follows:
\begin{IEEEeqnarray*}{RL}
\mi{count}(l^F)       &= \left\{\begin{array}{ll}
                                1 &\ \text{\it if either}\ l^F=\Cn\geql \gu\ \text{\it or}\ l^F=\gu\gleq \Cn, \\
                                  &\ \phantom{\text{\it if}\ \mbox{}}
                                                  \Cn\in \mi{PropConj}, \\[1mm]
                                0 &\ \text{\it else};
                                \end{array}
                         \right. \\[1mm]
\mi{count}(C^F)       &= \sum_{l\in C^F} \mi{count}(l); \\
\mi{count}(S^F)       &= \sum_{C\in S^F-\mi{guards}(S^F)} \mi{count}(C); \\
\mi{invalid}(S^F)     &= \{C \,|\, C\in S^F-\mi{guards}(S^F),\ \text{\it not}\ \mi{valid}(C,S^F)\}; \\
\mi{unsaturated}(S^F) &= \{a \,|\, a\in \mi{atoms}(S^F), \mi{guards}(S^F,a)=\{\gz\gle a\}\}; \\[1mm]
\IEEEeqnarraymulticol{2}{l}{
\mi{measure} : {\mc P}_{\mc F}(\mi{OrdPropCl})\longrightarrow {\mc P}_{\mc F}(\mi{PropAtom})\times {\mc P}_{\mc F}(\mi{PropAtom})\times} \\
\IEEEeqnarraymulticol{2}{l}{
\phantom{\mi{measure} : {\mc P}_{\mc F}(\mi{OrdPropCl})\longrightarrow \mbox{}} \quad 
                                                              {\mc P}_{\mc F}(\mi{OrdPropCl})\times \mbb{N};} \\
\IEEEeqnarraymulticol{2}{l}{
\mi{measure}(S^F)=(\mi{atoms}(S^F),\mi{unsaturated}(S^F),\mi{invalid}(S^F),\mi{count}(S^F));} \\[1mm]
\IEEEeqnarraymulticol{2}{l}{
\preceq, \prec\ \subseteq ({\mc P}_{\mc F}(\mi{PropAtom})\times {\mc P}_{\mc F}(\mi{PropAtom})\times {\mc P}_{\mc F}(\mi{OrdPropCl})\times \mbb{N})^2;} \\[1mm]
(x_1,x_2,x_3,x_4)\preceq (y_1,y_2,y_3,y_4) &\longleftrightarrow \text{\it either}\ x_1\subset y_1,\ \text{\it or}\ x_1=y_1, x_2\subset y_2,\ \text{\it or} \\ 
                                           &\phantom{\mbox{}\longleftrightarrow \mbox{}} \quad
                                                                x_1=y_1, x_2=y_2, x_3\subset y_3,\ \text{\it or} \\ 
                                           &\phantom{\mbox{}\longleftrightarrow \mbox{}} \quad 
                                                                x_1=y_1, x_2=y_2, x_3=y_3, x_4\leq y_4; \\[1mm]
(x_1,x_2,x_3,x_4)\prec (y_1,y_2,y_3,y_4)   &\longleftrightarrow (x_1,x_2,x_3,x_4)\preceq (y_1,y_2,y_3,y_4), \\
                                           &\phantom{\mbox{}\longleftrightarrow \mbox{}} \quad
                                                                (x_1,x_2,x_3,x_4)\neq (y_1,y_2,y_3,y_4).
\end{IEEEeqnarray*}
Note that $\preceq$ is reflexive, antisymmetric, transitive, a well-founded order, which arranges tuples in the lexicographic manner, and 
$\prec$ is irreflexive, transitive, a strict order.
The proof is by induction on $\mi{measure}(S)$ with respect to $\preceq$ using Lemmata \ref{le66} and \ref{le8}.
%
%
%
\end{proof}

The main theorem of this section solves the satisfiability problem for a general order clausal theory exploiting the described reduction to the case of a positive order clausal theory.

\begin{theorem}
\label{T2}
Let $S\subseteq_{\mc F} \mi{OrdPropCl}$.
There exists a finite {\it DPLL}-tree $\mi{Tree}$ with the root $S$ constructed using Rules {\rm (\cref{ceq4hr0})}--{\rm (\cref{ceq4hr8})} with the following properties:
\begin{alignat}{1}
\label{eq55a}
& \text{if}\ S\ \text{is unsatisfiable, then}\ \mi{Tree}\ \text{is closed}; \\[1mm]
\label{eq55b}
& \text{if}\ S\ \text{is satisfiable, then}\ \mi{Tree}\ \text{is open, and} \\ 
\notag
& \phantom{\text{if}\ S\ \text{is satisfiable, then}\ \mbox{}}
                                             \text{there exists a model}\ {\mf A}\ \text{of}\ S\ \text{related to}\ \mi{Tree}.
\end{alignat}
\end{theorem}

\begin{proof}
We construct a finite {\it DPLL}-tree $\mi{Tree}$ with the root $S$ using Lemmata \ref{le55}, \ref{le5}, \ref{le7}, \ref{le9}, \ref{le4} so that
the input order clausal theory $S$ is subsequently refined to simplified, $\gz$-guarded, positively guarded, and positive order clausal theories during the arborescence construction of $\mi{Tree}$. 
%
%
%
\end{proof}

The deduction problem of a formula from a finite theory can be solved as follows.

\begin{corollary}
\label{cor1}
Let $n_0\in \mbb{N}$, $\phi\in \mi{PropForm}_\emptyset$, $T\subseteq_{\mc F} \mi{PropForm}_\emptyset$.
There exist $J_T^\phi\subseteq_{\mc F} \{(i,j) \,|\, i\geq n_0, j\in \mbb{N}\}\subseteq \mbb{I}$,
$S_T^\phi\subseteq_{\mc F} \mi{OrdPropCl}_{\{\tilde{a}_\mbbm{j} \,|\, \mbbm{j}\in J_T^\phi\}}$,
a finite {\it DPLL}-tree $\mi{Tree}$ with the root $S_T^\phi$ constructed using Rules {\rm (\cref{ceq4hr0})}--{\rm (\cref{ceq4hr8})} such that
\begin{enumerate}[\rm (i)]
\item
$T\models \phi$ if and only if $\mi{Tree}$ is closed;
\item
$T\not\models \phi$ if and only if $\mi{Tree}$ is open, and there exists a model ${\mf A}$ of $S_T^\phi$ related to $\mi{Tree}$ satisfying ${\mf A}\models T$, ${\mf A}\not\models \phi$.
\end{enumerate}
\end{corollary}

\begin{proof}
By Theorem \ref{T1} and \ref{T1}(iii), there exist $J_T^\phi\subseteq_{\mc F} \{(i,j) \,|\, i\geq n_0, j\in \mbb{N}\}\subseteq \mbb{I}$ and
$S_T^\phi\subseteq_{\mc F} \mi{OrdPropCl}_{\{\tilde{a}_\mbbm{j} \,|\, \mbbm{j}\in J_T^\phi\}}$ satisfying that
\begin{enumerate}[\rm \ref{T1}(i)]
\item
there exists a valuation ${\mf A}$ satisfying ${\mf A}\models T$, ${\mf A}\not\models \phi$ if and only if 
there exists a valuation ${\mf A}'$ satisfying ${\mf A}'\models S_T^\phi$; 
${\mf A}|_\mbb{O}={\mf A}'|_\mbb{O}$;
\item
$T\models \phi$ if and only if $S_T^\phi$ is unsatisfiable;
\end{enumerate}
by Theorem \ref{T2} for $S_T^\phi$, there exists a finite {\it DPLL}-tree $\mi{Tree}$ with the root $S_T^\phi$ constructed using Rules (\cref{ceq4hr0})--(\cref{ceq4hr8}) with the following properties:
\begin{alignat}{1}
\label{eq6a}
& \text{if}\ S_T^\phi\ \text{is unsatisfiable, then}\ \mi{Tree}\ \text{is closed}; \\[1mm]
\label{eq6b}
& \text{if}\ S_T^\phi\ \text{is satisfiable, then}\ \mi{Tree}\ \text{is open, and} \\ 
\notag
& \phantom{\text{if}\ S_T^\phi\ \text{is satisfiable, then}\ \mbox{}}
                                                    \text{there exists a model}\ {\mf A}\ \text{of}\ S_T^\phi\ \text{related to}\ \mi{Tree}.
\end{alignat}
We get two cases for $\phi$ and $T$.

Case 1:
$T\models \phi$.
Then, by \ref{T1}(ii), $S_T^\phi$ is unsatisfiable; 
by (\ref{eq6a}), $\mi{Tree}$ is closed;
if $\mi{Tree}$ is closed,
by Lemma \ref{le3} for $S_T^\phi$ and $\mi{Tree}$, $S_T^\phi$ is unsatisfiable;
by \ref{T1}(ii), $T\models \phi$;
(i) holds.

Case 2:
$T\not\models \phi$.
Then, by \ref{T1}(ii), $S_T^\phi$ is satisfiable;
by (\ref{eq6b}), $\mi{Tree}$ is open; 
there exists a model ${\mf A}$ of $S_T^\phi$ related to $\mi{Tree}$;
by \ref{T1}(i) for ${\mf A}$, there exists a valuation ${\mf A}^\#$ satisfying ${\mf A}^\#\models T$, ${\mf A}^\#\not\models \phi$, ${\mf A}^\#|_\mbb{O}={\mf A}|_\mbb{O}$;
$\mi{atoms}(T), \mi{atoms}(\phi)\subseteq \mbb{O}$, ${\mf A}\models T$, ${\mf A}\not\models \phi$;
if $\mi{Tree}$ is open, and there exists a model ${\mf A}$ of $S_T^\phi$ related to $\mi{Tree}$ satisfying ${\mf A}\models T$, ${\mf A}\not\models \phi$,
$T\not\models \phi$;
(ii) holds.
%
%
%
\end{proof}

Continuing the example from the end of Section \ref{S3}, we can show that the formula $\phi$ is a tautology.
Since $\phi=\gz\gle a\rightarrow (a\swedge b\gleq a\swedge c\rightarrow b\gleq c)\neq \gz, \gu$,
by Theorem \ref{T1}, Case 3, we slightly modify the order clausal theory $S^\phi$ so that the order clause $\tilde{a}_0\geql \gu\ [1]$ is replaced with $\tilde{a}_0\gle \gu\ [1]$.
By Theorem \ref{T1}(ii), $\models \phi$ if and only if $S^\phi$ is unsatisfiable.
The corresponding {\it DPLL} derivation at the code level appears in Tables \ref{tab6} and \ref{tab7}, and 
a closed {\it DPLL}-tree for $\phi$ with the root $S^\phi$ is outlined in Figure \ref{fig1}, in Appendix, Section \ref{S7.0}.
By Corollary \ref{cor1}(i), $\models \phi$; 
$\phi$ is a tautology.

As a second example, we can show that $\psi=a\swedge b\gleq a\swedge c\rightarrow b\gleq c\in \mi{PropForm}_\emptyset$ is not a tautology.
Its translation to clausal form is given in Tables \ref{tab8} and \ref{tab9};
the corresponding {\it DPLL} derivation at the code level appears in Table \ref{tab10}, and
an open {\it DPLL}-tree $\mi{Tree}$ for $\psi$ with the root $S^\psi$ is outlined in Figure \ref{fig2}, in Appendix, Section \ref{S7.0}.
$\mi{Tree}$ contains an open branch ending in a leaf labelled with a unit order clausal theory $S_R^\psi$.
Moreover, $S_R^\psi$ is a unit positive order clausal theory, and  
there does not exist an application of Rule (\cref{ceq4hr0}) to $S_R^\psi$.
Then, using Lemma \ref{le2} for $S_R^\psi$, we can construct a model ${\mf A}$ of $S_R^\psi$, and
using Lemma \ref{le33}(ii) for $S^\psi$, $S_R^\psi$, and ${\mf A}$, we obtain a model ${\mf A}^*$ of $S^\psi$ related to $\mi{Tree}$, more in details in Appendix, Section \ref{S7.0}:
\begin{alignat*}{1}
{\mf A}|_{\mi{atoms}(S_R^\psi)}
&= \left\{\left(b,\dfrac{1}{2}\right),\left(c,\dfrac{1}{4}\right), 
          \left(\tilde{a}_5,\dfrac{1}{2}\right),\left(\tilde{a}_6,\dfrac{1}{4}\right), 
          \left(\tilde{a}_8,\dfrac{1}{2}\right),\left(\tilde{a}_{10},\dfrac{1}{4}\right)\right\}, \\[1mm]
{\mf A}^*|_{\mi{atoms}(S^\psi)}
&= {\mf A}|_{\mi{atoms}(S_R^\psi)}\cup \{(a,0),(\tilde{a}_0,0),(\tilde{a}_1,1),(\tilde{a}_2,0), \\
&\phantom{\mbox{}={\mf A}|_{\mi{atoms}(S_R^\psi)}\cup \{}
                                         (\tilde{a}_3,0),(\tilde{a}_4,0),(\tilde{a}_7,0),(\tilde{a}_9,0)\}.
\end{alignat*}
By Theorem \ref{T1}(i) for $\psi$, $S^\psi$, and ${\mf A}^*$, there exists a valuation ${\mf A}^\#$ such that ${\mf A}^\#\not\models \psi$, ${\mf A}^\#|_\mbb{O}={\mf A}^*|_\mbb{O}$;
$\psi$ is not a tautology; 
indeed $\mi{atoms}(\psi)=\{a,b,c\}\subseteq \mbb{O}$, ${\mf A}^*\not\models \psi$,
$\|\psi\|^{{\mf A}^*}=\|a\swedge b\gleq a\swedge c\rightarrow b\gleq c\|^{{\mf A}^*}=
                      {\mf A}^*(a)\fswedge {\mf A}^*(b)\fleq {\mf A}^*(a)\fswedge {\mf A}^*(c)\frightarrow {\mf A}^*(b)\fleq {\mf A}^*(c)=
                      0\fswedge \frac{1}{2}\fleq 0\fswedge \frac{1}{4}\frightarrow \frac{1}{2}\fleq \frac{1}{4}=1\frightarrow 0=0$.

\section{Conclusions}
\label{S6}

In the paper, we have proposed a {\it DPLL}-based calculus in the propositional product logic and translation to clausal form in linearithmic time.
We have formulated a notion of order literal:
an order literal is a formula of the form $\varepsilon_1\diamond \varepsilon_2$ 
where $\varepsilon_i$ is a truth constant or a conjunction of powers of atoms, and $\diamond$ is a connective $\geql$ or $\gleq$ or $\gle$.
Then an order clause is a finite set of order literals as usual.
$\geql$, $\gleq$, and $\gle$ are interpreted by the equality and standard (strict) linear order on $[0,1]$, respectively.
The {\it DPLL}-based calculus has four dichotomous branching rules.
The first rule splits a derivation into two subcases: either an atom is equal to $\gz$ or is greater than $\gz$.
The second one implements a split: either $a_0\swedge\cdots\swedge a_n\geql \gu$ or $a_0\swedge\cdots\swedge a_n\gle \gu$;
$a_i$ are atoms.
The third rule exploits a pure dichotomy $\Cn_1\gleq \Cn_2\vee \Cn_2\gle \Cn_1$, $\Cn_1\neq \Cn_2$, for splitting into two subcases:
either $\Cn_1\gleq \Cn_2$ is true, or $\Cn_2\gle \Cn_1$ is true; 
$\Cn_i$ is a conjunction of powers of atoms.
Finally, the fourth rule splits a derivation into two branches:
either $\Cn_1\geql \Cn_2$ is true, or $\Cn_1\gle \Cn_2\vee \Cn_2\gle \Cn_1$ is true, $\Cn_1\neq \Cn_2$.
The calculus is refutation sound and complete in the finite case.
In addition, the finite deduction problem has been reduced to the finite {\it SAT} problem,
which can be tested using the {\it DPLL} procedure.
The presented calculus provides technical foundations for the design of a {\it DPLL}-based {\it SAT} solver.

\bibliographystyle{plain}
\bibliography{ijuf23}

\clearpage
\newpage

\section{Appendix}
\label{S7}

\subsection{A computational point of view}
\label{S7.00}

From a computational point of view, the worst case time and space complexity will be estimated using the logarithmic cost measurement.
Notice that if the estimated upper bound on the space complexity is equal to the estimated upper bound on the time complexity for some algorithm, then it will not explicitly be stated.
Since our computational framework is only slightly different from that in \cite{Guller2018a}, analogous considerations will be shorten.
Let $n_s\in \mbb{N}$ ($n_s$ can be viewed as an offset in the memory) and $E$ be either a formula or an order clause or a finite theory or a finite order clausal theory.
$E$ can be represented by a tree-like data structure ${\mc D}(E)$ having vertices being data records.
In Table \ref{tab0}, we introduce all possible forms of data record.
\begin{table}
\caption{Forms of data record}
\label{tab0}
\begin{minipage}[t]{\linewidth}
\footnotesize
\begin{IEEEeqnarray*}{LL}
\hline \hline \\[1mm]
\text{\bf Expression} & \text{\bf Data record} \\[1mm]
\hline \\[2mm]
a_\mi{index}\in \mi{PropAtom}, \mi{index}\in \mbb{I} 
& \framebox{$a,\mi{index}$} \\[1mm]
c\in \{\gz,\gu\}
& \framebox{$c$} \\[1mm]
\diamond \phi_1, \phi_1\in \mi{PropForm}, \diamond\in \{\neg,\del\}
& \framebox{$\diamond,\mi{pointer}_{\phi_1}$} \\[1mm]
\phi_1\diamond \phi_2, \phi_i\in \mi{PropForm}, \diamond\in \{\wedge,\swedge,\vee,\rightarrow,\leftrightarrow,\geql,\gleq,\gle\} \qquad
& \framebox{$\diamond,\mi{pointer}_{\phi_1},\mi{pointer}_{\phi_2}$} \\[1mm] 
a^r, a\in \mi{PropAtom}, r\geq 1
& \framebox{$\uparrow,\mi{pointer}_a,r$} \\[1mm]
\{p\}\subseteq \mi{PropPow} 
& \framebox{$\&,\mi{pointer}_p$} \\[1mm]
p\swedge \Cn, p\in \mi{PropPow}, \Cn\in \mi{PropConj}
& \framebox{$\&,\mi{pointer}_p,\mi{pointer}_\Cn$} \\[1mm] 
\varepsilon_1\diamond \varepsilon_2, \varepsilon_i\in \{\gz,\gu\}\cup \mi{PropConj}, \diamond\in \{\geql,\gleq,\gle\}
& \framebox{$\diamond,\mi{pointer}_{\varepsilon_1},\mi{pointer}_{\varepsilon_2}$} \\[1mm]
\square\in \mi{OrdPropCl} 
& \framebox{$\square$} \\[1mm]
l\vee C, l\in \mi{OrdPropLit}, C\in \mi{OrdPropCl} 
& \framebox{$|,\mi{pointer}_l,\mi{pointer}_C$} \\[1mm]
\emptyset\subseteq \mi{PropForm}, \mi{OrdPropCl} 
& \framebox{$\emptyset$} \\[1mm]
\{\phi\}\cup T, \phi\in \mi{PropForm}, T\subseteq_{\mc F} \mi{PropForm}, \phi\not\in T 
& \framebox{$\&,\mi{pointer}_\phi,\mi{pointer}_T$} \\[1mm]
\{C\}\cup S, C\in \mi{OrdPropCl}, S\subseteq_{\mc F} \mi{OrdPropCl}, C\not\in S 
& \framebox{$\&,\mi{pointer}_C,\mi{pointer}_S$} \\[2mm] 
\hline \hline
\end{IEEEeqnarray*}
$\mi{pointer}_E$ denotes a pointer which references ${\mc D}(E)$.
\end{minipage}
\end{table}
Concerning Table \ref{tab0}, a data record consists of a field of length in $O(1)$ and of a constant number (with respect to the size of the input) of indices, pointers, and exponents.
The atoms occurring in $E$ can be indexed by indices of the form $(n_s,j)\in \mbb{I}$.
The number of indices occurring in ${\mc D}(E)$ is in $O(|E|)$ and the length of an index in $O(\log (1+n_s)+\log (1+|E|))$.
We set an upper bound on the length of a pointer to $O(\log (1+n_s)+\log (1+|E|))$; which can be verified as sufficient.
The data record for a power $a^r$, $a\in \mi{PropAtom}$, $r\geq 1$, occurring in $E$ is of the form \framebox{$\uparrow,\mi{pointer}_a,r$} and $|a^r|=1+r\leq |E|$.
Hence, the length of the exponent $r$ is in $O(\log r)=O(\log |a^r|)=O(\log (1+r))\leq O(\log (1+|E|))$.
So, the length of a data record of ${\mc D}(E)$ is in $O(\log (1+n_s)+\log (1+|E|))$.
The time complexity of an elementary operation on ${\mc D}(E)$ is in $O(\log (1+n_s)+\log (1+|E|))$.
The number of data records occurring in ${\mc D}(E)$ is in $O(|E|)$, and the size of ${\mc D}(E)$ is in $O(|E|\cdot (\log (1+n_s)+\log (1+|E|)))$.

Let ${\mc A}$ be an algorithm with inputs $E_0$, $E_1$ which uses only $E_i$, $i=0,\dots,q$, 
where $E_i$ is either a formula or an order clause or a finite theory or a finite order clausal theory;
$q\geq 1$ is a constant (with respect to the size of the input);
there exists a constant $r\geq 1$ satisfying, for all $i\leq q$, $|E_i|\in O(|E_0|^r+|E_1|^r)$.
$\#{\mc O}_{\mc A}(E_0,E_1)\geq 1$ denotes the number of all elementary operations executed by ${\mc A}$;%
\footnote{If the algorithm in question is not explicitly designated, we shall only write $\#{\mc O}(E_0,E_1)$.}
we assume that ${\mc A}$ executes at least one elementary operation.
The length of a data record is in $O(\log (1+n_s)+\log (1+|E_0|+|E_1|))$.
The time complexity of an elementary operation on ${\mc D}(E_i)$, $i=0,\dots,q$, executed by ${\mc A}$ is in $O(\log (1+n_s)+\log (1+|E_0|+|E_1|))$.
The number of data records occurring in ${\mc D}(E_i)$, $i=0,\dots,q$, is in $O(|E_0|^r+|E_1|^r)$.
The size of ${\mc D}(E_i)$, $i=0,\dots,q$, {\it area of data records}, is in $O((|E_0|^r+|E_1|^r)\cdot (\log (1+n_s)+\log (1+|E_0|+|E_1|)))$.

${\mc A}$ also uses several auxiliary data structures: {\it stack}, {\it index generator}, and {\it addressing unit}.
{\it stack} consists of a finite number of frames.
A frame is of the form \framebox{$\mi{field},\mi{pointer}$} where $\mi{field}$ is of length in $O(1)$, and 
$\mi{pointer}$ is a copy of that occurring in ${\mc D}(E_i)$, $i=0,\dots,q$, of length in $O(\log (1+n_s)+\log (1+|E_0|+|E_1|))$.
The length of a frame of {\it stack} is in $O(\log (1+n_s)+\log (1+|E_0|+|E_1|))$.
The time complexity of an elementary operation on {\it stack} executed by ${\mc A}$ is in $O(\log (1+n_s)+\log (1+|E_0|+|E_1|))$.
The size of {\it stack} is in $O(\#{\mc O}_{\mc A}(E_0,E_1)\cdot (\log (1+n_s)+\log (1+|E_0|+|E_1|)))$.

{\it index generator} serves for generating fresh atoms of the form $\tilde{a}_\mbbm{i}\in \tilde{\mbb{A}}$.
It consists of a constant number of indices of length in $O(\log (1+n_s)+\log (1+|E_0|+|E_1|))$. 
The size of {\it index generator} is in $O(\log (1+n_s)+\log (1+|E_0|+|E_1|))$.
The time complexity of an elementary operation on {\it index generator} executed by ${\mc A}$ is in $O(\log (1+n_s)+\log (1+|E_0|+|E_1|))$.

{\it addressing unit} consists of a constant number of address registers. 
The memory can be arranged as follows: 
\begin{equation}
\notag
\Big[\ \framebox{{\it stack\vphantom{g}}} \qquad \framebox{{\it index generator}} \qquad \framebox{{\it addressing unit}} \qquad \framebox{{\it area of data records}}\ \Big].     
\end{equation}
The length of an address register and the size of {\it addressing unit} is in $O(\log (1+n_s)+\log (\#{\mc O}_{\mc A}(E_0,E_1)+|E_0|+|E_1|))$.
The time complexity of an elementary operation on {\it addressing unit} executed by ${\mc A}$ is in $O(\log (1+n_s)+\log (\#{\mc O}_{\mc A}(E_0,E_1)+|E_0|+|E_1|))$.
The size of the memory is the total size of {\it stack}, {\it index generator}, {\it addressing unit}, and {\it area of data records}
in $O((\#{\mc O}_{\mc A}(E_0,E_1)+|E_0|^r+|E_1|^r)\cdot (\log (1+n_s)+\log (1+|E_0|+|E_1|)))$.

We assume that ${\mc A}$ executes only elementary operations on {\it stack}, {\it index generator}, {\it addressing unit}, and {\it area of data records}.
We get that the time complexity of an elementary operation executed by ${\mc A}$ is in $O(\log (1+n_s)+\log (\#{\mc O}_{\mc A}(E_0,E_1)+|E_0|+|E_1|))$.
We conclude that
\begin{alignat}{1}
\label{eq00t}   
& \begin{minipage}[t]{\linewidth-15mm}
  the time complexity of ${\mc A}$ on $E_0$ and $E_1$ is in $O(\#{\mc O}_{\mc A}(E_0,E_1)\cdot (\log (1+n_s)+\log (\#{\mc O}_{\mc A}(E_0,E_1)+|E_0|+|E_1|)))$;                  
  \end{minipage}
\end{alignat}
\begin{alignat}{1}
\label{eq00s}     
& \begin{minipage}[t]{\linewidth-15mm}
  the space complexity of ${\mc A}$ on $E_0$ and $E_1$ is in $O((\#{\mc O}_{\mc A}(E_0,E_1)+|E_0|^r+|E_1|^r)\cdot (\log (1+n_s)+\log (1+|E_0|+|E_1|)))$.
  \end{minipage}
\end{alignat}

\subsection{Examples}
\label{S7.0}

We finally introduce some admissible rules for the {\it DPLL} procedure, which are useful to get smaller and more readable {\it DPLL}-trees, but superfluous for the refutational completeness argument.
\begin{alignat}{1}
\ctag{ceq4hr100}{({\it Guard propagation rule} I)} \\[1mm]
\notag
& \dfrac{S}
        {S\cup \{\gz\gle b\}}; \\[1mm]
\notag
& \gz\gle a\in \mi{guards}(S), a\diamond b\in S-\mi{guards}(S), a, b\in \mi{PropAtom}, \diamond\in \{\geql,\gleq,\gle\}.
\end{alignat}
The transitivity of the standard order $\leq$ and strict order $<$ on $[0,1]$ ensures that 
the guard $\gz\gle b$ is a consequence of the input guard $\gz\gle a$ and the input order literal $a\diamond b$, and it can be derived from the input order clausal theory $S$.            
\begin{alignat}{1}
\ctag{ceq4hr101}{({\it Guard propagation rule} II)} \\[1mm]
\notag
& \dfrac{S}
        {S\cup \{\gz\gle c\}}; \\[1mm]           
\notag
& \begin{array}{l}
  \gz\gle a, \gz\gle b\in \mi{guards}(S), a\swedge b\diamond c\in S-\mi{guards}(S), \\
  a, b, c\in \mi{PropAtom}, \diamond\in \{\geql,\gleq,\gle\}.               
  \end{array}
\end{alignat}
The multiplication $t$-norm satisfies the condition:
\begin{equation}
\notag
\text{for all}\ x, y\in \bs{\Pi},\ 0<x, y\longrightarrow 0<x\fswedge y.
\end{equation}
Then, using the transitivity of the standard order $\leq$ and strict order $<$ on $[0,1]$, we get that 
the guard $\gz\gle c$ is a consequence of the input guards $\gz\gle a$, $\gz\gle b$, and the input order literal $a\swedge b\diamond c$, and it can be derived from the input order clausal theory $S$. 
\begin{alignat}{1}
\ctag{ceq4hr102}{({\it $\gz$-dichotomy simplification rule})} \\[1mm]
\notag
& \dfrac{S}
        {(S-\{l_1\vee C\})\cup \{C\}}; \\[1mm]
\notag
& \begin{array}{l}
  l_1\vee C\in S-\mi{guards}(S), l_2\in S, l_i\in \mi{OrdPropLit}, \square\neq C\in \mi{OrdPropCl}; \\
  l_1\vee l_2\ \text{\it is a $\gz$-dichotomy}.
  \end{array}
\end{alignat}
The linearity of the standard strict order $<$ on $[0,1]$ together with the bounds $0$ and $1$ ensures that the $\gz$-dichotomy $l_1\vee l_2$ is obviously true in every valuation.
This means that either the input order literal $l_1$ or the input order literal $l_2$ is true in every valuation.
In addition, $l_2$ belongs to the input order clausal theory $S$.
Hence, $l_2$ is true in every model of $S$, but $l_1$ not, and it may be removed from the input order clause $l_1\vee C$.
\begin{alignat}{1}
\ctag{ceq4hr103}{({\it Pure dichotomy simplification rule})} \\[1mm]
\notag
& \dfrac{S}
        {(S-\{l_1\vee C\})\cup \{C\}}; \\[1mm]         
\notag
& \begin{array}{l}
  l_1\vee C\in S-\mi{guards}(S), l_2\in S, l_i\in \mi{PurOrdPropLit}, \square\neq C\in \mi{OrdPropCl}; \\
  l_1\vee l_2\ \text{\it is a pure dichotomy}.
  \end{array}
\end{alignat}
Analogously to the previous rule, the linearity of the standard strict order $<$ on $[0,1]$ guarantees that the pure dichotomy $l_1\vee l_2$ is obviously true in every valuation.
This yields that either the input order literal $l_1$ or the input order literal $l_2$ is true in every valuation.
Moreover, $l_2$ belongs to the input order clausal theory $S$.
So, $l_2$ is true in every model of $S$, but $l_1$ not, and it may be removed from the input order clause $l_1\vee C$ as well.
All Rules (\cref{ceq4hr100})--(\cref{ceq4hr103}) are sound with respect to satisfiability; i.e. Lemmata \ref{le333} and \ref{le33333} hold also for them.

\begin{figure}
\centering
\vspace*{-20mm}
\hspace*{-38mm} 
\includegraphics[width=200mm,height=180mm,angle=0]{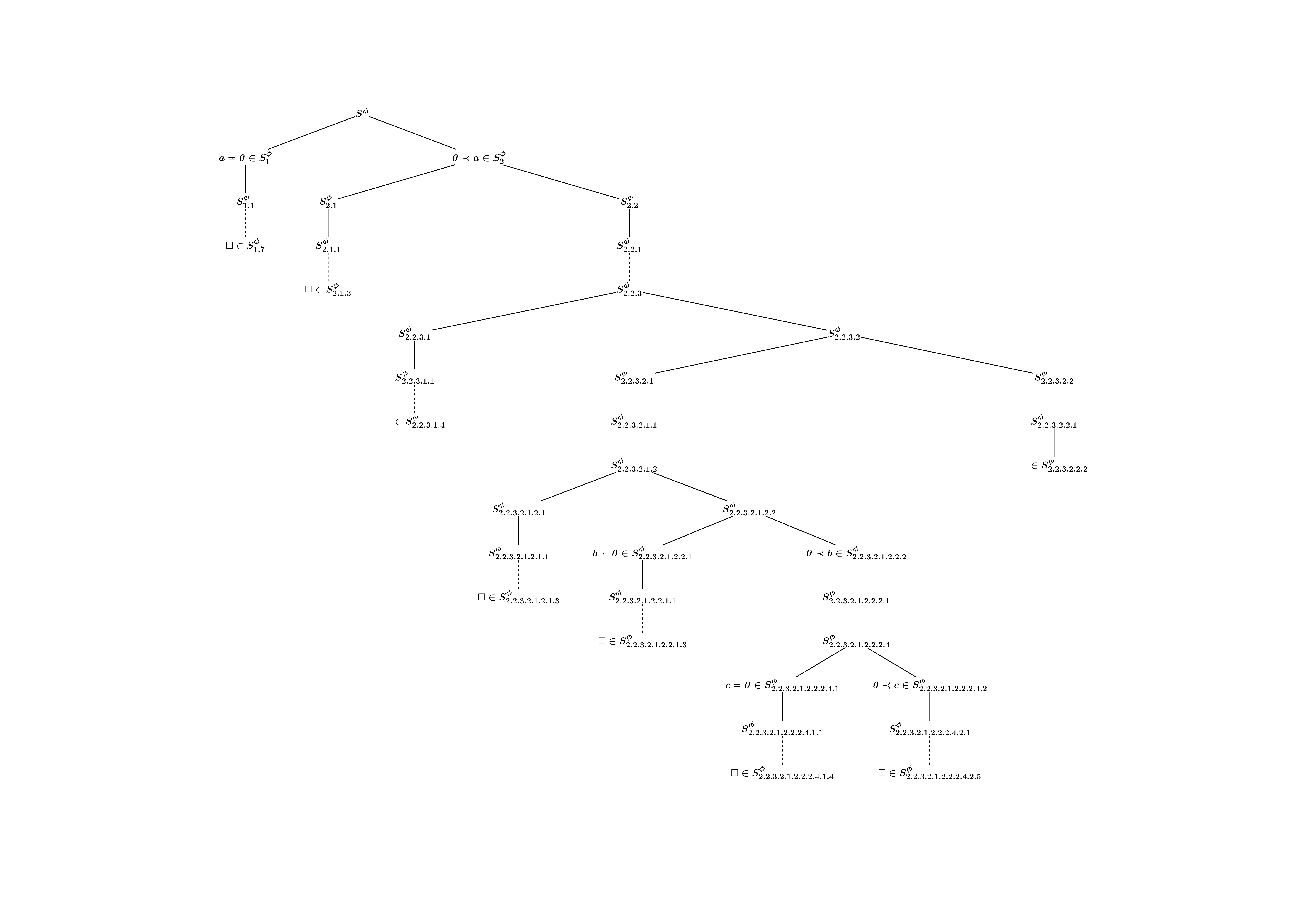}
\vspace{-30mm}
\caption{A closed {\it DPLL}-tree for $\phi$.}
\label{fig1} 
\end{figure}

We illustrate the solution to the deduction problem with two examples.
Continuing the example from the end of Section \ref{S3}, we show that the formula $\phi$ is a tautology.
Since $\phi=\gz\gle a\rightarrow (a\swedge b\gleq a\swedge c\rightarrow b\gleq c)\neq \gz, \gu$,
by Theorem \ref{T1}, Case 3, we slightly modify the order clausal theory $S^\phi$ so that the order clause $\tilde{a}_0\geql \gu\ [1]$ is replaced with $\tilde{a}_0\gle \gu\ [1]$.
By Theorem \ref{T1}(ii), $\models \phi$ if and only if $S^\phi$ is unsatisfiable.
In Figure \ref{fig1}, it is outlined a closed {\it DPLL}-tree for $\phi$ with the root $S^\phi$.
The corresponding {\it DPLL} derivation at the code level appears in Tables \ref{tab6} and \ref{tab7}.
By Corollary \ref{cor1}(i), $\models \phi$; 
$\phi$ is a tautology.
\begin{table}
\caption{{\it DPLL} derivation from the order clausal theory $S^\phi$}
\label{tab6}  
\begin{minipage}[t]{\linewidth}
\begin{IEEEeqnarray*}{L}
\hline \hline 
\end{IEEEeqnarray*}
\end{minipage}
\hspace*{-10mm}
\mbox{
\begin{minipage}[t]{0.52\linewidth}
\tiny
\begin{IEEEeqnarray*}{*LL*}
\IEEEeqnarraymulticol{2}{l}{
\text{\bf Rule (\cref{ceq4hr1})} : a : S_1^\phi : S_2^\phi :} \\[2mm]
S_1^\phi=S^\phi\cup \{a\geql \gz\}
& [21] \\
\IEEEeqnarraymulticol{2}{l}{
\text{\bf Rule (\cref{ceq4hr3})} : [21] [6] :} \\
S_{1.1}^\phi=(S_1^\phi-[6])\cup \{\tilde{a}_3\geql \gz\}
& [22] \\
\IEEEeqnarraymulticol{2}{l}{
\text{\bf Rule (\cref{ceq4hr3})} : [22] [4] :} \\
S_{1.2}^\phi=(S_{1.1}^\phi-[4])\cup \{\gz\gle \gz\vee \tilde{a}_1\geql \gz\}
& [23] \\
\IEEEeqnarraymulticol{2}{l}{
\text{\bf Rule (\cref{ceq4hr2})} : [23] :} \\
S_{1.3}^\phi=(S_{1.2}^\phi-[23])\cup \{\tilde{a}_1\geql \gz\}
& [24] \\
\IEEEeqnarraymulticol{2}{l}{
\text{\bf Rule (\cref{ceq4hr3})} : [24] [3] :} \\
S_{1.4}^\phi=(S_{1.3}^\phi-[3])\cup \{\tilde{a}_2\gle \gz\vee \tilde{a}_0\geql \gu\}
& [25] \\
\IEEEeqnarraymulticol{2}{l}{
\text{\bf Rule (\cref{ceq4hr2})} : [25] :} \\
S_{1.5}^\phi=(S_{1.4}^\phi-[25])\cup \{\tilde{a}_0\geql \gu\}
& [26] \\
\IEEEeqnarraymulticol{2}{l}{
\text{\bf Rule (\cref{ceq4hr4})} : [26] [1] :} \\
S_{1.6}^\phi=(S_{1.5}^\phi-[1])\cup \{\gu\gle \gu\}
& [27] \\
\IEEEeqnarraymulticol{2}{l}{
\text{\bf Rule (\cref{ceq4hr2})} : [27] :} \\
\IEEEeqnarraymulticol{2}{l}{
S_{1.7}^\phi=(S_{1.6}^\phi-[27])\cup \{\bsquare\}}
\\[2mm]
S_2^\phi=S^\phi\cup \{\gz\gle a\} 
& [28] \\
\IEEEeqnarraymulticol{2}{l}{
\text{\bf Rule (\cref{ceq4hr11})} : [2] : S_{2.1}^\phi : S_{2.2}^\phi :} \\[2mm]
S_{2.1}^\phi=(S_2^\phi-[2])\cup \{\tilde{a}_1\gleq \tilde{a}_2\}
& [29] \\
\IEEEeqnarraymulticol{2}{l}{
\text{\bf Rule (\cref{ceq4hr103})} : [29] [3] :} \\
S_{2.1.1}^\phi=(S_{2.1}^\phi-[3])\cup \{\tilde{a}_0\geql \gu\}
& [30] \\
\IEEEeqnarraymulticol{2}{l}{
\text{\bf Rule (\cref{ceq4hr4})} : [30] [1] :} \\
S_{2.1.2}^\phi=(S_{2.1.1}^\phi-[1])\cup \{\gu\gle \gu\}
& [31] \\
\IEEEeqnarraymulticol{2}{l}{
\text{\bf Rule (\cref{ceq4hr2})} : [31] :} \\
\IEEEeqnarraymulticol{2}{l}{
S_{2.1.3}^\phi=(S_{2.1.2}^\phi-[31])\cup \{\bsquare\}}
\\[2mm]
S_{2.2}^\phi=(S_2^\phi-[2])\cup \{\tilde{a}_1\swedge \tilde{a}_0\geql \tilde{a}_2
& [32] \\
\phantom{S_{2.2}^\phi=(S_2^\phi-[2])\cup \{}
                                  \tilde{a}_2\gle \tilde{a}_1\}
& [33] \\
\IEEEeqnarraymulticol{2}{l}{
\text{\bf Rule (\cref{ceq4hr100})} : [28] [6] :} \\
S_{2.2.1}^\phi=S_{2.2}^\phi\cup \{\gz\gle \tilde{a}_3\}
& [34] \\
\IEEEeqnarraymulticol{2}{l}{
\text{\bf Rule (\cref{ceq4hr102})} : [34] [5] :} \\
S_{2.2.2}^\phi=(S_{2.2.1}^\phi-[5])\cup \{\tilde{a}_1\geql \gu\}
& [35] \\
\IEEEeqnarraymulticol{2}{l}{
\text{\bf Rule (\cref{ceq4hr4})} : [35] [32] :} \\
S_{2.2.3}^\phi=(S_{2.2.2}^\phi-[32])\cup \{\tilde{a}_0\geql \tilde{a}_2\}
& [36] \\
\IEEEeqnarraymulticol{2}{l}{
\text{\bf Rule (\cref{ceq4hr11})} : [7] : S_{2.2.3.1}^\phi : S_{2.2.3.2}^\phi :} \\[2mm]
S_{2.2.3.1}^\phi=(S_{2.2.3}^\phi-[7])\cup \{\tilde{a}_4\gleq \tilde{a}_5\}
& [37] \\
\IEEEeqnarraymulticol{2}{l}{
\text{\bf Rule (\cref{ceq4hr103})} : [37] [8] :} \\
S_{2.2.3.1.1}^\phi=(S_{2.2.3.1}^\phi-[8])\cup \{\tilde{a}_2\geql \gu\}
& [38] \\
\IEEEeqnarraymulticol{2}{l}{
\text{\bf Rule (\cref{ceq4hr4})} : [38] [36] :} \\
S_{2.2.3.1.2}^\phi=(S_{2.2.3.1.1}^\phi-[36])\cup \{\tilde{a}_0\geql \gu\} \quad
& [39] \\
\IEEEeqnarraymulticol{2}{l}{
\text{\bf Rule (\cref{ceq4hr4})} : [39] [1] :} \\
S_{2.2.3.1.3}^\phi=(S_{2.2.3.1.2}^\phi-[1])\cup \{\gu\gle \gu\}
& [40] \\
\IEEEeqnarraymulticol{2}{l}{
\text{\bf Rule (\cref{ceq4hr2})} : [40] :} \\
\IEEEeqnarraymulticol{2}{l}{
S_{2.2.3.1.4}^\phi=(S_{2.2.3.1.3}^\phi-[40])\cup \{\bsquare\}}
\end{IEEEeqnarray*} 
\end{minipage}
\begin{minipage}[t]{0.52\linewidth}
\tiny
\begin{IEEEeqnarray*}{*LL*}
S_{2.2.3.2}^\phi=(S_{2.2.3}^\phi-[7])\cup \{\tilde{a}_4\swedge \tilde{a}_2\geql \tilde{a}_5
& [41] \\
\phantom{S_{2.2.3.2}^\phi=(S_{2.2.3}^\phi-[7])\cup \{}
                                            \tilde{a}_5\gle \tilde{a}_4\}
& [42] \\
\IEEEeqnarraymulticol{2}{l}{
\text{\bf Rule (\cref{ceq4hr11})} : [9] : S_{2.2.3.2.1}^\phi : S_{2.2.3.2.2}^\phi :} \\[2mm]
S_{2.2.3.2.1}^\phi=(S_{2.2.3.2}^\phi-[9])\cup \{\tilde{a}_6\gleq \tilde{a}_7\}
& [43] \\
\IEEEeqnarraymulticol{2}{l}{
\text{\bf Rule (\cref{ceq4hr103})} : [43] [10] :} \\
S_{2.2.3.2.1.1}^\phi=(S_{2.2.3.2.1}^\phi-[10])\cup \{\tilde{a}_4\geql \gu\}
& [44] \\
\IEEEeqnarraymulticol{2}{l}{
\text{\bf Rule (\cref{ceq4hr4})} : [44] [41] :} \\
S_{2.2.3.2.1.2}^\phi=(S_{2.2.3.2.1.1}^\phi-[41])\cup \{\tilde{a}_2\geql \tilde{a}_5\}
& [45] \\
\IEEEeqnarraymulticol{2}{l}{
\text{\bf Rule (\cref{ceq4hr11})} : [17] : S_{2.2.3.2.1.2.1}^\phi : S_{2.2.3.2.1.2.2}^\phi :} \\[2mm]
S_{2.2.3.2.1.2.1}^\phi=(S_{2.2.3.2.1.2}^\phi-[17])\cup \{\tilde{a}_8\gleq \tilde{a}_9\}
& [46] \\
\IEEEeqnarraymulticol{2}{l}{
\text{\bf Rule (\cref{ceq4hr103})} : [46] [18] :} \\
S_{2.2.3.2.1.2.1.1}^\phi=(S_{2.2.3.2.1.2.1}^\phi-[18])\cup \{\tilde{a}_5\geql \gu\}
& [47] \\
\IEEEeqnarraymulticol{2}{l}{
\text{\bf Rule (\cref{ceq4hr4})} : [47] [42] :} \\
S_{2.2.3.2.1.2.1.2}^\phi=(S_{2.2.3.2.1.2.1.1}^\phi-[42])\cup \{\gu\gle \tilde{a}_4\} \quad
& [48] \\
\IEEEeqnarraymulticol{2}{l}{
\text{\bf Rule (\cref{ceq4hr2})} : [48] :} \\
\IEEEeqnarraymulticol{2}{l}{
S_{2.2.3.2.1.2.1.3}^\phi=(S_{2.2.3.2.1.2.1.2}^\phi-[48])\cup \{\bsquare\}}
\\[2mm]
S_{2.2.3.2.1.2.2}^\phi=(S_{2.2.3.2.1.2}^\phi-[17])\cup \{\tilde{a}_5\geql \gz
& [49] \\
\phantom{S_{2.2.3.2.1.2.2}^\phi=(S_{2.2.3.2.1.2}^\phi-[17])\cup \{}
                                                         \tilde{a}_9\gle \tilde{a}_8\}
& [50] \\
\IEEEeqnarraymulticol{2}{l}{
\text{\bf Rule (\cref{ceq4hr1})} : b : S_{2.2.3.2.1.2.2.1}^\phi : S_{2.2.3.2.1.2.2.2}^\phi :} \\[2mm]
S_{2.2.3.2.1.2.2.1}^\phi=S_{2.2.3.2.1.2.2}^\phi\cup \{b\geql \gz\}
& [51] \\
\IEEEeqnarraymulticol{2}{l}{
\text{\bf Rule (\cref{ceq4hr3})} : [51] [19] :} \\
\IEEEeqnarraymulticol{2}{l}{
S_{2.2.3.2.1.2.2.1.1}^\phi=(S_{2.2.3.2.1.2.2.1}^\phi-[19])\cup}
\\
\phantom{S_{2.2.3.2.1.2.2.1.1}^\phi=\mbox{}}
                           \{\tilde{a}_8\geql \gz\}
& [52] \\
\IEEEeqnarraymulticol{2}{l}{
\text{\bf Rule (\cref{ceq4hr3})} : [52] [50] :} \\
\IEEEeqnarraymulticol{2}{l}{
S_{2.2.3.2.1.2.2.1.2}^\phi=(S_{2.2.3.2.1.2.2.1.1}^\phi-[50])\cup}
\\
\phantom{S_{2.2.3.2.1.2.2.1.2}^\phi=\mbox{}} 
                           \{\tilde{a}_9\gle \gz\}
& [53] \\
\IEEEeqnarraymulticol{2}{l}{
\text{\bf Rule (\cref{ceq4hr2})} : [53] :} \\
\IEEEeqnarraymulticol{2}{l}{
S_{2.2.3.2.1.2.2.1.3}^\phi=(S_{2.2.3.2.1.2.2.1.2}^\phi-[53])\cup \{\bsquare\}}
\\[2mm]
S_{2.2.3.2.1.2.2.2}^\phi=S_{2.2.3.2.1.2.2}^\phi\cup \{\gz\gle b\}
& [54] \\
\IEEEeqnarraymulticol{2}{l}{
\text{\bf Rule (\cref{ceq4hr100})} : [28] [12] :} \\
S_{2.2.3.2.1.2.2.2.1}^\phi=S_{2.2.3.2.1.2.2.2}^\phi\cup \{\gz\gle \tilde{a}_{10}\}
& [55] \\
\IEEEeqnarraymulticol{2}{l}{
\text{\bf Rule (\cref{ceq4hr100})} : [54] [13] :} \\
S_{2.2.3.2.1.2.2.2.2}^\phi=S_{2.2.3.2.1.2.2.2.1}^\phi\cup \{\gz\gle \tilde{a}_{11}\}
& [56] \\
\IEEEeqnarraymulticol{2}{l}{
\text{\bf Rule (\cref{ceq4hr101})} : [55] [56] [11]:} \\
S_{2.2.3.2.1.2.2.2.3}^\phi=S_{2.2.3.2.1.2.2.2.2}^\phi\cup \{\gz\gle \tilde{a}_6\}
& [57] \\
\IEEEeqnarraymulticol{2}{l}{
\text{\bf Rule (\cref{ceq4hr100})} : [57] [43] :} \\
S_{2.2.3.2.1.2.2.2.4}^\phi=S_{2.2.3.2.1.2.2.2.3}^\phi\cup \{\gz\gle \tilde{a}_7\}
& [58] \\
\IEEEeqnarraymulticol{2}{l}{
\text{\bf Rule (\cref{ceq4hr1})} : c : S_{2.2.3.2.1.2.2.2.4.1}^\phi : S_{2.2.3.2.1.2.2.2.4.2}^\phi :} \\[2mm]
S_{2.2.3.2.1.2.2.2.4.1}^\phi=S_{2.2.3.2.1.2.2.2.4}^\phi\cup \{c\geql \gz\}
& [59] \\
\IEEEeqnarraymulticol{2}{l}{
\text{\bf Rule (\cref{ceq4hr3})} : [59] [16] :} \\
\IEEEeqnarraymulticol{2}{l}{
S_{2.2.3.2.1.2.2.2.4.1.1}^\phi=(S_{2.2.3.2.1.2.2.2.4.1}^\phi-[16])\cup} 
\\
\phantom{S_{2.2.3.2.1.2.2.2.4.1.1}^\phi=\mbox{}}
                               \{\tilde{a}_{13}\geql \gz\}
& [60] \\ 
\IEEEeqnarraymulticol{2}{l}{
\text{\bf Rule (\cref{ceq4hr3})} : [60] [14] :} \\
\IEEEeqnarraymulticol{2}{l}{
S_{2.2.3.2.1.2.2.2.4.1.2}^\phi=(S_{2.2.3.2.1.2.2.2.4.1.1}^\phi-[14])\cup}
\\
\phantom{S_{2.2.3.2.1.2.2.2.4.1.2}^\phi=\mbox{}} 
                               \{\tilde{a}_7\geql \gz\} 
& [61] 
\end{IEEEeqnarray*} 
\end{minipage}}
\begin{minipage}[t]{\linewidth}
\begin{IEEEeqnarray*}{L}
\hline \hline 
\end{IEEEeqnarray*}
\end{minipage}
\end{table}
\begin{table}
\caption{{\it DPLL} derivation from the order clausal theory $S^\phi$}
\label{tab7}  
\begin{minipage}[t]{\linewidth}
\tiny
\begin{IEEEeqnarray*}{LL}
\hline \hline \\[2mm]
\IEEEeqnarraymulticol{2}{l}{
\text{\bf Rule (\cref{ceq4hr3})} : [61] [58] :} \\
S_{2.2.3.2.1.2.2.2.4.1.3}^\phi=(S_{2.2.3.2.1.2.2.2.4.1.2}^\phi-[58])\cup \{\gz\gle \gz\} \quad
& [62] \\
\IEEEeqnarraymulticol{2}{l}{
\text{\bf Rule (\cref{ceq4hr2})} : [62] :} \\
\IEEEeqnarraymulticol{2}{l}{
S_{2.2.3.2.1.2.2.2.4.1.4}^\phi=(S_{2.2.3.2.1.2.2.2.4.1.3}^\phi-[62])\cup \{\bsquare\}}
\\[2mm]
S_{2.2.3.2.1.2.2.2.4.2}^\phi=S_{2.2.3.2.1.2.2.2.4}^\phi\cup \{\gz\gle c\}
& [63] \\ 
\IEEEeqnarraymulticol{2}{l}{
\text{\bf Rule (\cref{ceq4hr100})} : [54] [19] :} \\
S_{2.2.3.2.1.2.2.2.4.2.1}^\phi=S_{2.2.3.2.1.2.2.2.4.2}^\phi\cup \{\gz\gle \tilde{a}_8\}
& [64] \\
\IEEEeqnarraymulticol{2}{l}{
\text{\bf Rule (\cref{ceq4hr100})} : [63] [20] :} \\
S_{2.2.3.2.1.2.2.2.4.2.2}^\phi=S_{2.2.3.2.1.2.2.2.4.2.1}^\phi\cup \{\gz\gle \tilde{a}_9\}
& [65] \\
\IEEEeqnarraymulticol{2}{l}{
\text{\bf Rule (\cref{ceq4hr100})} : [28] [15] :} \\
S_{2.2.3.2.1.2.2.2.4.2.3}^\phi=S_{2.2.3.2.1.2.2.2.4.2.2}^\phi\cup \{\gz\gle \tilde{a}_{12}\}
& [66] \\
\IEEEeqnarraymulticol{2}{l}{
\text{\bf Rule (\cref{ceq4hr100})} : [63] [16] :} \\
S_{2.2.3.2.1.2.2.2.4.2.4}^\phi=S_{2.2.3.2.1.2.2.2.4.2.3}^\phi\cup \{\gz\gle \tilde{a}_{13}\}
& [67] \\
\IEEEeqnarraymulticol{2}{l}{
\text{\bf Rule (\cref{ceq4hr0})} : [28] [54] [63] : [55]-[58] [64]-[67] : [11]-[16] [19] [20] : [43] [50] :} \\
\IEEEeqnarraymulticol{2}{l}{
S_{2.2.3.2.1.2.2.2.4.2.5}^\phi=S_{2.2.3.2.1.2.2.2.4.2.4}^\phi\cup \{\bsquare\}}
\\[2mm]
S_{2.2.3.2.2}^\phi=(S_{2.2.3.2}^\phi-[9])\cup \{\tilde{a}_4\geql \gz
& [68] \\
\phantom{S_{2.2.3.2.2}^\phi=(S_{2.2.3.2}^\phi-[9])\cup \{}
                                                \tilde{a}_7\gle \tilde{a}_6
& [69] \\
\IEEEeqnarraymulticol{2}{l}{
\text{\bf Rule (\cref{ceq4hr3})} : [68] [42] :} \\
S_{2.2.3.2.2.1}^\phi=(S_{2.2.3.2.2}^\phi-[42])\cup \{\tilde{a}_5\gle \gz\}
& [70] \\
\IEEEeqnarraymulticol{2}{l}{
\text{\bf Rule (\cref{ceq4hr2})} : [70] :} \\
\IEEEeqnarraymulticol{2}{l}{
S_{2.2.3.2.2.2}^\phi=(S_{2.2.3.2.2.1}^\phi-[70])\cup \{\bsquare\}} \\[2mm]
\hline \hline 
\end{IEEEeqnarray*} 
\end{minipage}
\end{table}

\begin{figure}
\centering
\vspace*{-15mm}
\hspace*{-36.5mm} 
\includegraphics[width=200mm,height=180mm,angle=0]{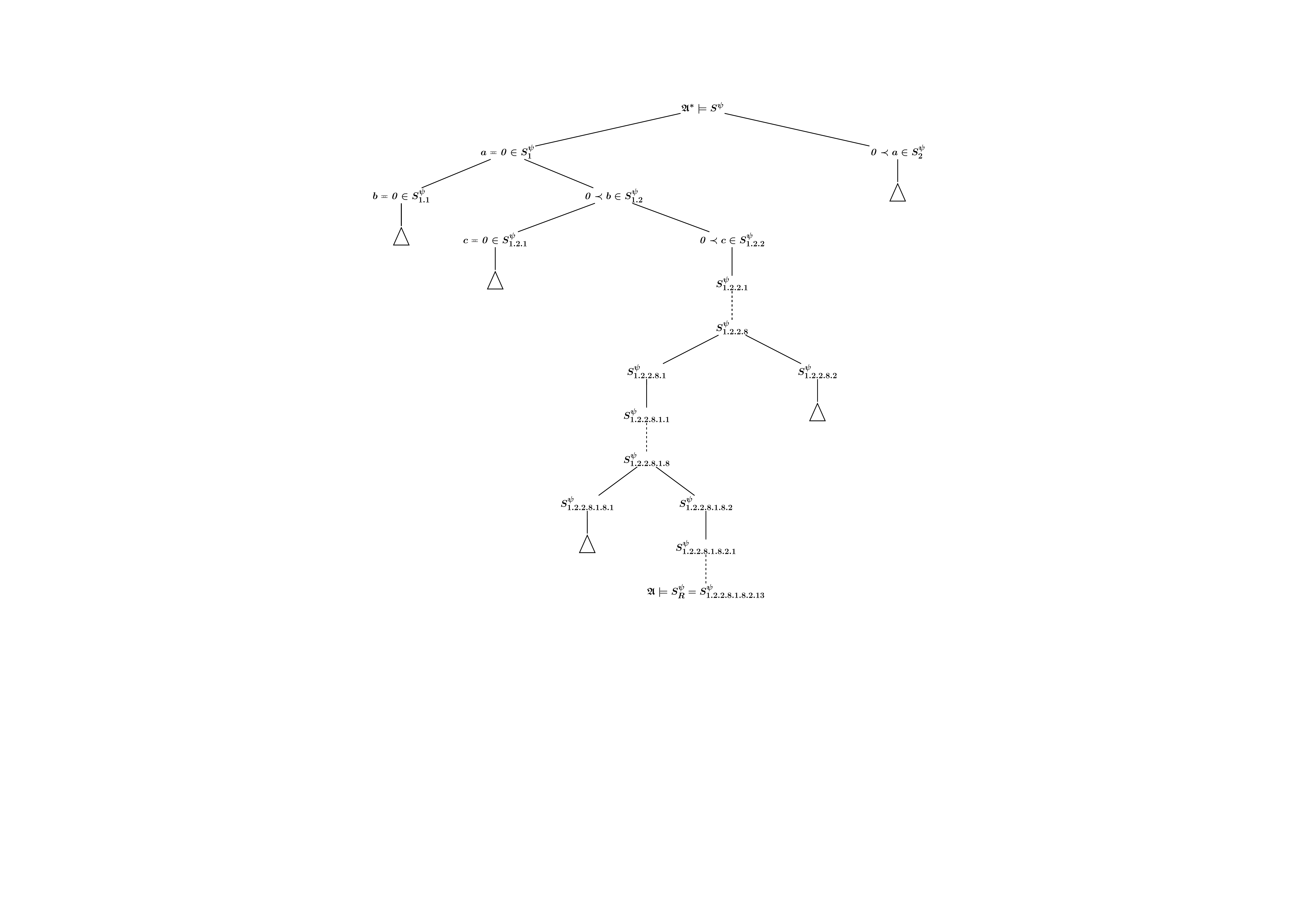}
\vspace{-65mm}
\caption{An open {\it DPLL}-tree $\mi{Tree}$ for $\psi$.}
\label{fig2} 
\end{figure}

We further show that $\psi=a\swedge b\gleq a\swedge c\rightarrow b\gleq c\in \mi{PropForm}_\emptyset$ is not a tautology.
Its translation to clausal form is given in Tables \ref{tab8} and \ref{tab9}.
\begin{table}
\caption{Translation of the formula $\psi$ to clausal form}
\label{tab8}
\begin{minipage}[t]{\linewidth}
\footnotesize
\begin{IEEEeqnarray*}{LL}
\hline \hline \\[2mm]
\Big\{\tilde{a}_0\gle \gu,
      \tilde{a}_0\leftrightarrow (\underbrace{a\swedge b\gleq a\swedge c}_{\tilde{a}_1}\rightarrow \underbrace{b\gleq c}_{\tilde{a}_2})\Big\}
& (\ref{eq0rr3+}) \\
\Big\{\tilde{a}_0\gle \gu,
      \tilde{a}_1\gleq \tilde{a}_2\vee \tilde{a}_1\swedge \tilde{a}_0\geql \tilde{a}_2,
      \tilde{a}_2\gle \tilde{a}_1\vee \tilde{a}_0\geql \gu,
      \tilde{a}_1\leftrightarrow \underbrace{a\swedge b}_{\tilde{a}_3}\gleq \underbrace{a\swedge c}_{\tilde{a}_4},
      \tilde{a}_2\leftrightarrow \underbrace{b}_{\tilde{a}_5}\gleq \underbrace{c}_{\tilde{a}_6}\Big\} \quad
& (\ref{eq0rr77+}) \\
\Big\{\tilde{a}_0\gle \gu,
      \tilde{a}_1\gleq \tilde{a}_2\vee \tilde{a}_1\swedge \tilde{a}_0\geql \tilde{a}_2,
      \tilde{a}_2\gle \tilde{a}_1\vee \tilde{a}_0\geql \gu,
& \\
\phantom{\Big\{}
      \tilde{a}_3\gleq \tilde{a}_4\vee \tilde{a}_1\geql \gz,
      \tilde{a}_4\gle \tilde{a}_3\vee \tilde{a}_1\geql \gu,
      \tilde{a}_3\leftrightarrow \underbrace{a}_{\tilde{a}_7}\swedge \underbrace{b}_{\tilde{a}_8},
      \tilde{a}_4\leftrightarrow \underbrace{a}_{\tilde{a}_9}\swedge \underbrace{c}_{\tilde{a}_{10}},
& \\
\phantom{\Big\{}
      \tilde{a}_5\gleq \tilde{a}_6\vee \tilde{a}_2\geql \gz,
      \tilde{a}_6\gle \tilde{a}_5\vee \tilde{a}_2\geql \gu,
      \tilde{a}_5\geql b,
      \tilde{a}_6\geql c\Big\}
& (\ref{eq0rr11+}) \\[2mm]
\hline \hline 
\end{IEEEeqnarray*}
\end{minipage}
\end{table}
\begin{table}
\caption{Translation of the formula $\psi$ to clausal form}
\label{tab9}
\begin{minipage}[t]{\linewidth}
\begin{IEEEeqnarray*}{L}
\hline \hline 
\end{IEEEeqnarray*}
\end{minipage}
\begin{minipage}[t]{0.49\linewidth}
\footnotesize
\begin{IEEEeqnarray*}{RLL}
S^\psi=\Bigg\{
& \tilde{a}_0\gle \gu
& [1] \\
& \tilde{a}_1\gleq \tilde{a}_2\vee \tilde{a}_1\swedge \tilde{a}_0\geql \tilde{a}_2 \quad
& [2] \\
& \tilde{a}_2\gle \tilde{a}_1\vee \tilde{a}_0\geql \gu
& [3] \\
& \tilde{a}_3\gleq \tilde{a}_4\vee \tilde{a}_1\geql \gz
& [4] \\
& \tilde{a}_4\gle \tilde{a}_3\vee \tilde{a}_1\geql \gu
& [5] \\
& \tilde{a}_3\geql \tilde{a}_7\swedge \tilde{a}_8
& [6] \\
& \tilde{a}_7\geql a
& [7] \\
& \tilde{a}_8\geql b
& [8] 
\end{IEEEeqnarray*} 
\end{minipage}
\begin{minipage}[t]{0.49\linewidth}
\footnotesize
\begin{IEEEeqnarray*}{RLL}
& \tilde{a}_4\geql \tilde{a}_9\swedge \tilde{a}_{10}
& [9] \\
& \tilde{a}_9\geql a
& [10] \\
& \tilde{a}_{10}\geql c
& [11] \\
& \tilde{a}_5\gleq \tilde{a}_6\vee \tilde{a}_2\geql \gz \quad
& [12] \\
& \tilde{a}_6\gle \tilde{a}_5\vee \tilde{a}_2\geql \gu
& [13] \\
& \tilde{a}_5\geql b
& [14] \\
& \tilde{a}_6\geql c\Bigg\}
& [15] 
\end{IEEEeqnarray*} 
\end{minipage}
\begin{minipage}[t]{\linewidth}
\begin{IEEEeqnarray*}{L}
\hline \hline 
\end{IEEEeqnarray*}
\end{minipage}
\end{table}
We have got $S^\psi\subseteq \mi{OrdPropCl}_{\{\tilde{a}_0,\dots,\tilde{a}_{10}\}}$.
During the translation, we have introduced some fresh atoms $\tilde{a}_0,\dots,\tilde{a}_{10}$.
In Figure \ref{fig2}, it is outlined an open {\it DPLL}-tree $\mi{Tree}$ for $\psi$ with the root $S^\psi$.
The corresponding {\it DPLL} derivation at the code level appears in Table \ref{tab10}.
$\mi{Tree}$ contains an open branch ending in a leaf labelled with a unit order clausal theory $S_R^\psi$.
\begin{table}
\caption{{\it DPLL} derivation from the order clausal theory $S^\psi$}
\label{tab10}  
\begin{minipage}[t]{\linewidth}
\begin{IEEEeqnarray*}{L}
\hline \hline 
\end{IEEEeqnarray*}
\end{minipage}
\hspace*{-10mm}
\mbox{
\begin{minipage}[t]{0.5\linewidth}
\tiny
\begin{IEEEeqnarray*}{*LL*}
\IEEEeqnarraymulticol{2}{l}{
\text{\bf Rule (\cref{ceq4hr1})} : a : S_1^\psi : S_2^\psi :} \\[2mm]
S_1^\psi=S^\psi\cup \{a\geql \gz\}
& [16] \\
\IEEEeqnarraymulticol{2}{l}{
\text{\bf Rule (\cref{ceq4hr1})} : b : S_{1.1}^\psi : S_{1.2}^\psi :} \\[2mm]
S_{1.2}^\psi=S_1^\psi\cup \{\gz\gle b\}
& [17] \\
\IEEEeqnarraymulticol{2}{l}{
\text{\bf Rule (\cref{ceq4hr1})} : c : S_{1.2.1}^\psi : S_{1.2.2}^\psi :} \\[2mm]
S_{1.2.2}^\psi=S_{1.2}^\psi\cup \{\gz\gle c\}
& [18] \\
\IEEEeqnarraymulticol{2}{l}{
\text{\bf Rule (\cref{ceq4hr3})} : [16] [7] :} \\
S_{1.2.2.1}^\psi=(S_{1.2.2}^\psi-[7])\cup \{\tilde{a}_7\geql \gz\}
& [19] \\
\IEEEeqnarraymulticol{2}{l}{
\text{\bf Rule (\cref{ceq4hr3})} : [19] [6] :} \\
S_{1.2.2.2}^\psi=(S_{1.2.2.1}^\psi-[6])\cup \{\tilde{a}_3\geql \gz\}
& [20] \\
\IEEEeqnarraymulticol{2}{l}{
\text{\bf Rule (\cref{ceq4hr3})} : [20] [4] :} \\
S_{1.2.2.3}^\psi=(S_{1.2.2.2}^\psi-[4])\cup \{\gz\gleq \tilde{a}_4\vee \tilde{a}_1\geql \gz\}
& [21] \\
\IEEEeqnarraymulticol{2}{l}{
\text{\bf Rule (\cref{ceq4hr22})} : [21] :} \\
\IEEEeqnarraymulticol{2}{l}{
S_{1.2.2.4}^\psi=S_{1.2.2.3}^\psi-[21]}
\\
\IEEEeqnarraymulticol{2}{l}{
\text{\bf Rule (\cref{ceq4hr3})} : [20] [5] :} \\
S_{1.2.2.5}^\psi=(S_{1.2.2.4}^\psi-[5])\cup \{\tilde{a}_4\gle \gz\vee \tilde{a}_1\geql \gu\}
& [22] \\
\IEEEeqnarraymulticol{2}{l}{
\text{\bf Rule (\cref{ceq4hr2})} : [22] :} \\
S_{1.2.2.6}^\psi=(S_{1.2.2.5}^\psi-[22])\cup \{\tilde{a}_1\geql \gu\}
& [23] \\
\IEEEeqnarraymulticol{2}{l}{
\text{\bf Rule (\cref{ceq4hr4})} : [23] : [2] :} \\
S_{1.2.2.7}^\psi=(S_{1.2.2.6}^\psi-[2])\cup \{\gu\gleq \tilde{a}_2\vee \tilde{a}_0\geql \tilde{a}_2\}
& [24] \\
\IEEEeqnarraymulticol{2}{l}{
\text{\bf Rule (\cref{ceq4hr4})} : [23] : [3] :} \\
S_{1.2.2.8}^\psi=(S_{1.2.2.7}^\psi-[3])\cup \{\tilde{a}_2\gle \gu\vee \tilde{a}_0\geql \gu\}
& [25] \\
\IEEEeqnarraymulticol{2}{l}{
\text{\bf Rule (\cref{ceq4hr111})} : [24] : S_{1.2.2.8.1}^\psi : S_{1.2.2.8.2}^\psi :} \\[2mm]
S_{1.2.2.8.1}^\psi=(S_{1.2.2.8}^\psi-[24])\cup \{\tilde{a}_0\geql \tilde{a}_2\}
& [26] \\
\IEEEeqnarraymulticol{2}{l}{
\text{\bf Rule (\cref{ceq4hr3})} : [16] [10] :} \\
S_{1.2.2.8.1.1}^\psi=(S_{1.2.2.8.1}^\psi-[10])\cup \{\tilde{a}_9\geql \gz\}
& [27] \\
\IEEEeqnarraymulticol{2}{l}{
\text{\bf Rule (\cref{ceq4hr3})} : [27] [9] :} \\
S_{1.2.2.8.1.2}^\psi=(S_{1.2.2.8.1.1}^\psi-[9])\cup \{\tilde{a}_4\geql \gz\}
& [28] \\
\IEEEeqnarraymulticol{2}{l}{
\text{\bf Rule (\cref{ceq4hr7})} : [16] :} \\
\IEEEeqnarraymulticol{2}{l}{
S_{1.2.2.8.1.3}^\psi=S_{1.2.2.8.1.2}^\psi-[16]}
\\
\IEEEeqnarraymulticol{2}{l}{
\text{\bf Rule (\cref{ceq4hr7})} : [19] :} \\
\IEEEeqnarraymulticol{2}{l}{
S_{1.2.2.8.1.4}^\psi=S_{1.2.2.8.1.3}^\psi-[19]}
\\
\IEEEeqnarraymulticol{2}{l}{
\text{\bf Rule (\cref{ceq4hr7})} : [20] :} \\
\IEEEeqnarraymulticol{2}{l}{
S_{1.2.2.8.1.5}^\psi=S_{1.2.2.8.1.4}^\psi-[20]}
\\
\IEEEeqnarraymulticol{2}{l}{
\text{\bf Rule (\cref{ceq4hr8})} : [23] :} \\
\IEEEeqnarraymulticol{2}{l}{
S_{1.2.2.8.1.6}^\psi=S_{1.2.2.8.1.5}^\psi-[23]}
\\
\IEEEeqnarraymulticol{2}{l}{
\text{\bf Rule (\cref{ceq4hr7})} : [27] :} \\
\IEEEeqnarraymulticol{2}{l}{
S_{1.2.2.8.1.7}^\psi=S_{1.2.2.8.1.6}^\psi-[27]}
\\
\IEEEeqnarraymulticol{2}{l}{
\text{\bf Rule (\cref{ceq4hr7})} : [28] :} \\
\IEEEeqnarraymulticol{2}{l}{
S_{1.2.2.8.1.8}^\psi=S_{1.2.2.8.1.7}^\psi-[28]}
\\
\IEEEeqnarraymulticol{2}{l}{
\text{\bf Rule (\cref{ceq4hr11})} : [12] : S_{1.2.2.8.1.8.1}^\psi : S_{1.2.2.8.1.8.2}^\psi :} \\[2mm]
S_{1.2.2.8.1.8.2}^\psi=(S_{1.2.2.8.1.8}^\psi-[12])\cup \{\tilde{a}_2\geql \gz
& [29] \\
\phantom{S_{1.2.2.8.1.8.2}^\psi=(S_{1.2.2.8.1.8}^\psi-[12])\cup \{}
                                                         \tilde{a}_6\gle \tilde{a}_5\} \quad
& [30] 
\end{IEEEeqnarray*} 
\end{minipage}
\begin{minipage}[t]{0.5\linewidth}
\tiny
\begin{IEEEeqnarray*}{*LL*}
\IEEEeqnarraymulticol{2}{l}{
\text{\bf Rule (\cref{ceq4hr3})} : [29] [26] :} \\
\IEEEeqnarraymulticol{2}{l}{
S_{1.2.2.8.1.8.2.1}^\psi=(S_{1.2.2.8.1.8.2}^\psi-[26])\cup} 
\\
\phantom{S_{1.2.2.8.1.8.2.1}^\psi=\mbox{}}
                         \{\tilde{a}_0\geql \gz\}
& [31] \\
\IEEEeqnarraymulticol{2}{l}{
\text{\bf Rule (\cref{ceq4hr3})} : [29] [13] :} \\
\IEEEeqnarraymulticol{2}{l}{
S_{1.2.2.8.1.8.2.2}^\psi=(S_{1.2.2.8.1.8.2.1}^\psi-[13])\cup} 
\\
\phantom{S_{1.2.2.8.1.8.2.2}^\psi=\mbox{}}
                         \{\tilde{a}_6\gle \tilde{a}_5\vee \gz\geql \gu\} 
& [32] \\
\IEEEeqnarraymulticol{2}{l}{
\text{\bf Rule (\cref{ceq4hr2})} : [32] :} \\
\IEEEeqnarraymulticol{2}{l}{
S_{1.2.2.8.1.8.2.3}^\psi=S_{1.2.2.8.1.8.2.2}^\psi-[32]} 
\\
\IEEEeqnarraymulticol{2}{l}{
\text{\bf Rule (\cref{ceq4hr3})} : [31] [1] :} \\
\IEEEeqnarraymulticol{2}{l}{
S_{1.2.2.8.1.8.2.4}^\psi=(S_{1.2.2.8.1.8.2.3}^\psi-[1])\cup} 
\\
\phantom{S_{1.2.2.8.1.8.2.4}^\psi=\mbox{}}
                         \{\gz\gle \gu\}
& [33] \\
\IEEEeqnarraymulticol{2}{l}{
\text{\bf Rule (\cref{ceq4hr22})} : [33] :} \\
\IEEEeqnarraymulticol{2}{l}{
S_{1.2.2.8.1.8.2.5}^\psi=S_{1.2.2.8.1.8.2.4}^\psi-[33]} 
\\
\IEEEeqnarraymulticol{2}{l}{
\text{\bf Rule (\cref{ceq4hr3})} : [29] [25] :} \\
\IEEEeqnarraymulticol{2}{l}{
S_{1.2.2.8.1.8.2.6}^\psi=(S_{1.2.2.8.1.8.2.5}^\psi-[25])\cup} 
\\
\phantom{S_{1.2.2.8.1.8.2.6}^\psi=\mbox{}}
                         \{\gz\gle \gu\vee \tilde{a}_0\geql \gu\}
& [34] \\
\IEEEeqnarraymulticol{2}{l}{
\text{\bf Rule (\cref{ceq4hr22})} : [34] :} \\
\IEEEeqnarraymulticol{2}{l}{
S_{1.2.2.8.1.8.2.7}^\psi=S_{1.2.2.8.1.8.2.6}^\psi-[34]} 
\\
\IEEEeqnarraymulticol{2}{l}{
\text{\bf Rule (\cref{ceq4hr7})} : [29] :} \\
\IEEEeqnarraymulticol{2}{l}{
S_{1.2.2.8.1.8.2.8}^\psi=S_{1.2.2.8.1.8.2.7}^\psi-[29]} 
\\
\IEEEeqnarraymulticol{2}{l}{
\text{\bf Rule (\cref{ceq4hr7})} : [31] :} \\
\IEEEeqnarraymulticol{2}{l}{
S_{1.2.2.8.1.8.2.9}^\psi=S_{1.2.2.8.1.8.2.8}^\psi-[31]} 
\\
\IEEEeqnarraymulticol{2}{l}{
\text{\bf Rule (\cref{ceq4hr100})} : [17] [8] :} \\
S_{1.2.2.8.1.8.2.10}^\psi=S_{1.2.2.8.1.8.2.9}^\psi\cup \{\gz\gle \tilde{a}_8\}
& [35] \\
\IEEEeqnarraymulticol{2}{l}{
\text{\bf Rule (\cref{ceq4hr100})} : [17] [14] :} \\
S_{1.2.2.8.1.8.2.11}^\psi=S_{1.2.2.8.1.8.2.10}^\psi\cup \{\gz\gle \tilde{a}_5\}
& [36] \\
\IEEEeqnarraymulticol{2}{l}{
\text{\bf Rule (\cref{ceq4hr100})} : [18] [11] :} \\
S_{1.2.2.8.1.8.2.12}^\psi=S_{1.2.2.8.1.8.2.11}^\psi\cup \{\gz\gle \tilde{a}_{10}\} \quad
& [37] \\ 
\IEEEeqnarraymulticol{2}{l}{
\text{\bf Rule (\cref{ceq4hr100})} : [18] [15] :} \\
\IEEEeqnarraymulticol{2}{l}{
S_R^\psi=S_{1.2.2.8.1.8.2.13}^\psi=S_{1.2.2.8.1.8.2.12}^\psi\cup} 
\\
\phantom{S_R^\psi=S_{1.2.2.8.1.8.2.13}^\psi=\mbox{}}
                                   \{\gz\gle \tilde{a}_6\} 
& [38] 
\end{IEEEeqnarray*} 
\footnotesize
\begin{IEEEeqnarray*}{*RLL*}
\hline \hline \\[2mm]
S_R^\psi=\Bigg\{
& \tilde{a}_8\geql b
& [8] \\
& \tilde{a}_{10}\geql c
& [11] \\
& \tilde{a}_5\geql b
& [14] \\
& \tilde{a}_6\geql c
& [15] \\
& \gz\gle b
& [17] \\
& \gz\gle c
& [18] \\
& \tilde{a}_6\gle \tilde{a}_5
& [30] \\
& \gz\gle \tilde{a}_8
& [35] \\
& \gz\gle \tilde{a}_5
& [36] \\
& \gz\gle \tilde{a}_{10}
& [37] \\
& \gz\gle \tilde{a}_6\Bigg\} \quad
& [38] 
\end{IEEEeqnarray*} 
\end{minipage}}
\begin{minipage}[t]{\linewidth}
\begin{IEEEeqnarray*}{L}
\hline \hline 
\end{IEEEeqnarray*}
\end{minipage}
\end{table}
$S_R^\psi$, $\mi{atoms}(S_R^\psi)=\{b,c,\tilde{a}_5,\tilde{a}_6,\tilde{a}_8,\tilde{a}_{10}\}\subseteq \mi{atoms}(S^\psi)=\{a,b,c,\tilde{a}_0,\dots,\tilde{a}_{10}\}$, can be divided 
into a subset of guards and a subset of remaining unit order clauses:
\begin{alignat*}{1}
\mi{guards}(S_R^\psi)          &= \{\gz\gle b,\gz\gle c,\gz\gle \tilde{a}_5,\gz\gle \tilde{a}_6,\gz\gle \tilde{a}_8,\gz\gle \tilde{a}_{10}\}, \\
S_R^\psi-\mi{guards}(S_R^\psi) &= \{\tilde{a}_5\geql b,\tilde{a}_8\geql b,\tilde{a}_6\geql c,\tilde{a}_{10}\geql c,\tilde{a}_6\gle \tilde{a}_5\}\subseteq \mi{PurOrdPropCl}.
\end{alignat*}
Every atom in $S_R^\psi$ is positively guarded; 
hence, $S_R^\psi$ is a unit positive order clausal theory, and moreover, 
there does not exist an application of Rule (\cref{ceq4hr0}) to $S_R^\psi$.
Using Lemma \ref{le2} for $S_R^\psi$, we can construct a model ${\mf A}$ of $S_R^\psi$:
\begin{equation}
\notag
{\mf A}|_{\mi{atoms}(S_R^\psi)}=\left\{\left(b,\dfrac{1}{2}\right),\left(c,\dfrac{1}{4}\right), 
                                       \left(\tilde{a}_5,\dfrac{1}{2}\right),\left(\tilde{a}_6,\dfrac{1}{4}\right), 
                                       \left(\tilde{a}_8,\dfrac{1}{2}\right),\left(\tilde{a}_{10},\dfrac{1}{4}\right)\right\}.
\end{equation}
Using Lemma \ref{le33}(ii) for $S^\psi$, $S_R^\psi$, and ${\mf A}$, we obtain a model ${\mf A}^*$ of $S^\psi$ related to $\mi{Tree}$:
\begin{alignat*}{1}
{\mf A}^*|_{\mi{atoms}(S^\psi)}
&= {\mf A}|_{\mi{atoms}(S_R^\psi)}\cup \{(a,0),(\tilde{a}_0,0),(\tilde{a}_1,1),(\tilde{a}_2,0), \\
&\phantom{\mbox{}={\mf A}|_{\mi{atoms}(S_R^\psi)}\cup \{}
                                         (\tilde{a}_3,0),(\tilde{a}_4,0),(\tilde{a}_7,0),(\tilde{a}_9,0)\}.
\end{alignat*}
By Theorem \ref{T1}(i) for $\psi$, $S^\psi$, and ${\mf A}^*$, there exists a valuation ${\mf A}^\#$ such that ${\mf A}^\#\not\models \psi$, ${\mf A}^\#|_\mbb{O}={\mf A}^*|_\mbb{O}$;
$\psi$ is not a tautology; 
indeed $\mi{atoms}(\psi)=\{a,b,c\}\subseteq \mbb{O}$, ${\mf A}^*\not\models \psi$,
$\|\psi\|^{{\mf A}^*}=\|a\swedge b\gleq a\swedge c\rightarrow b\gleq c\|^{{\mf A}^*}=
                      {\mf A}^*(a)\fswedge {\mf A}^*(b)\fleq {\mf A}^*(a)\fswedge {\mf A}^*(c)\frightarrow {\mf A}^*(b)\fleq {\mf A}^*(c)=
                      0\fswedge \frac{1}{2}\fleq 0\fswedge \frac{1}{4}\frightarrow \frac{1}{2}\fleq \frac{1}{4}=1\frightarrow 0=0$.

\subsection{Full proof of Lemma \ref{le111}}
\label{S7.1aa}

\begin{proof}
The proof is by induction on the structure of $\theta$ using (\ref{eq0b}), (\ref{eq0kkk}), the obvious simplification identities
over $\bs{\Pi}$ with respect to $0$, $1$, $\fvee$, $\fwedge$, $\fswedge$, $\frightarrow$, $\feql$, $\fleq$, and $\fle$
(e.g. $0\fvee a=a$, $1\fvee a=1$, $a\fvee a=a$, 
      $0\fwedge a=0$, $1\fwedge a=a$, $a\fwedge a=a$,
      $0\fswedge a=0$, $1\fswedge a=a$,
      $0\frightarrow a=1$, $1\frightarrow a=a$, $a\frightarrow 1=1$, $a\frightarrow a=1$,
      $a\feql a=1$, 
      $0\fleq a=1$, $1\fleq a=1\feql a$, $a\fleq 0=a\feql 0$, $a\fleq 1=1$, $a\fleq a=1$,
      $a\fle 0=0$, $1\fle a=0$, $a\fle a=0$, $a\in \bs{\Pi}$, etc.);
the postorder traversal of $\theta$ uses the input $\theta$ and the output $\xi$;
$|\xi|\underset{\text{(b)}}{\leq} 2\cdot |\theta|\in O(|\theta|)$,
$\#{\mc O}(\theta)\in O(|\theta|)$;
by (\ref{eq00t}) for $n_\theta$, $\theta$, $\emptyset$, $\xi$, $q=2$, and $r=1$,
the time complexity of the postorder traversal of $\theta$ is in $O(\#{\mc O}(\theta)\cdot (\log (1+n_\theta)+\log (\#{\mc O}(\theta)+|\theta|)))\subseteq O(|\theta|\cdot (\log (1+n_\theta)+\log |\theta|))$;
by (\ref{eq00s}) for $n_\theta$, $\theta$, $\emptyset$, $\xi$, $q=2$, and $r=1$,
the space complexity of the postorder traversal of $\theta$ is in $O((\#{\mc O}(\theta)+|\theta|)\cdot (\log (1+n_\theta)+\log |\theta|))\subseteq O(|\theta|\cdot (\log (1+n_\theta)+\log |\theta|))$.
%
%
%
\end{proof}

\subsection{Full proof of Lemma \ref{le11}}
\label{S7.1a}

\begin{proof}
We proceed by induction on the structure of $\theta$.

Case 1 (the base case):
$\theta\in \mi{PropAtom}\cup \{\gz,\gu\}$.
We put $n_J=j_\mbbm{i}$ and $J=\{(n_\theta,j) \,|\, j_\mbbm{i}+1\leq j\leq n_J\}\subseteq \{(n_\theta,j) \,|\, j\in \mbb{N}\}\subseteq \mbb{I}$. 
Then $\mbbm{i}\not\in J=\{(n_\theta,j) \,|\, j_\mbbm{i}+1\leq j\leq n_J\}=\{(n_\theta,j) \,|\, j_\mbbm{i}+1\leq j\leq j_\mbbm{i}\}=\emptyset$,
$\theta\not\in \{\gz,\gu\}$, $\theta\in \mi{PropAtom}$, $\mi{atoms}(\theta)\subseteq \mbb{O}$, 
$\tilde{a}_\mbbm{i}\in \tilde{\mbb{A}}$, $\mbb{O}\cap \tilde{\mbb{A}}=\emptyset$, $\tilde{a}_\mbbm{i}\not\in \mbb{O}$,
$\mi{atoms}(\tilde{a}_\mbbm{i}\geql \theta)=\{\tilde{a}_\mbbm{i},\theta\}\subseteq \mbb{O}\cup \{\tilde{a}_\mbbm{i}\}$, 
$\tilde{a}_\mbbm{i}\geql \theta\in \mi{OrdPropLit}_{\{\tilde{a}_\mbbm{i}\}}$.
We put $S=\{\tilde{a}_\mbbm{i}\geql \theta\}\subseteq_{\mc F} \mi{OrdPropCl}_{\{\tilde{a}_\mbbm{i}\}}$.

(a,c,d) can be proved straightforwardly.

Hence, $\mi{atoms}(\theta)\subseteq \mbb{O}$, $\tilde{a}_\mbbm{i}\not\in \mbb{O}$, 
$\mi{atoms}(\tilde{a}_\mbbm{i}\leftrightarrow \theta)=\{\tilde{a}_\mbbm{i},\theta\}\subseteq \mbb{O}\cup \{\tilde{a}_\mbbm{i}\}$,
$\tilde{a}_\mbbm{i}\leftrightarrow \theta\in \mi{PropForm}_{\{\tilde{a}_\mbbm{i}\}}$;
for every valuation ${\mf A}$,
${\mf A}\models \tilde{a}_\mbbm{i}\leftrightarrow \theta$ if and only if
$\|\tilde{a}_\mbbm{i}\leftrightarrow \theta\|^{\mf A}=(\|\tilde{a}_\mbbm{i}\|^{\mf A}\frightarrow \|\theta\|^{\mf A})\fwedge (\|\theta\|^{\mf A}\frightarrow \|\tilde{a}_\mbbm{i}\|^{\mf A})=1$ if and only if
$\|\tilde{a}_\mbbm{i}\|^{\mf A}\frightarrow \|\theta\|^{\mf A}=1$, $\|\theta\|^{\mf A}\frightarrow \|\tilde{a}_\mbbm{i}\|^{\mf A}=1$ if and only if
$\|\tilde{a}_\mbbm{i}\|^{\mf A}\leq \|\theta\|^{\mf A}$, $\|\theta\|^{\mf A}\leq \|\tilde{a}_\mbbm{i}\|^{\mf A}$ if and only if
$\|\tilde{a}_\mbbm{i}\|^{\mf A}=\|\theta\|^{\mf A}$ if and only if
$\|\tilde{a}_\mbbm{i}\geql \theta\|^{\mf A}=\|\tilde{a}_\mbbm{i}\|^{\mf A}\feql \|\theta\|^{\mf A}=1$ if and only if
${\mf A}\models \tilde{a}_\mbbm{i}\geql \theta$ if and only if
${\mf A}\models S=\{\tilde{a}_\mbbm{i}\geql \theta\}$;
there exists a valuation ${\mf A}$ satisfying ${\mf A}\models \tilde{a}_\mbbm{i}\leftrightarrow \theta$ if and only if 
there exists a valuation ${\mf A}'$ satisfying ${\mf A}'\models S$; 
${\mf A}={\mf A}'$, ${\mf A}|_{\mbb{O}\cup \{\tilde{a}_\mbbm{i}\}}={\mf A}'|_{\mbb{O}\cup \{\tilde{a}_\mbbm{i}\}}$; 
(b) holds.

Case 2 (the induction case): 
$\theta\in \mi{PropForm}_\emptyset-(\mi{PropAtom}\cup \{\gz,\gu\})$.
We have that (c,d) of Lemma \ref{le111} hold for $\theta$.
Then $\theta\not\in \mi{PropAtom}$, $\theta\not\in \{\gz,\gu\}$;
$\theta$ does not contain $\neg$ and $\del$; 
either $\theta\in \{\gz,\gu\}$, 
or for every subformula of $\theta$ of the form $\theta_1\diamond \theta_2$, $\theta_i\in \mi{PropForm}_\emptyset$, $\diamond\in \{\wedge,\swedge,\vee,\leftrightarrow,\gleq\}$, 
for both $i$, $\theta_i\neq \gz, \gu$, and
for every subformula of $\theta$ of the form $\theta_1\rightarrow \theta_2$, $\theta_i\in \mi{PropForm}_\emptyset$, $\theta_1\neq \gz, \gu$, $\theta_2\neq \gu$, and
for every subformula of $\theta$ of the form $\theta_1\geql \theta_2$, $\theta_i\in \mi{PropForm}_\emptyset$, $\theta_1\neq \gz, \gu$, and
for every subformula of $\theta$ of the form $\theta_1\gle \theta_2$, $\theta_i\in \mi{PropForm}_\emptyset$, $\theta_1\neq \gu$, $\theta_2\neq \gz$, $\theta_1\gle \theta_2\neq \gz\gle \gu$;
either $\theta=\theta_1\diamond \theta_2$, $\theta_i\in \mi{PropForm}_\emptyset-\{\gz,\gu\}$, $\diamond\in \{\wedge,\swedge,\vee,\rightarrow,\leftrightarrow,\geql,\gleq,\gle\}$, 
or $\theta=\theta_1\rightarrow \gz$ or $\theta=\theta_1\geql \gz$ or $\theta=\theta_1\geql \gu$ or $\theta=\gz\gle \theta_1$ or $\theta=\theta_1\gle \gu$, 
$\theta_1\in \mi{PropForm}_\emptyset-\{\gz,\gu\}$.
We distinguish two cases for $\theta$.

Case 2.1 (the binary interpolation case): 
$\theta=\theta_1\diamond \theta_2$, $\theta_i\in \mi{PropForm}_\emptyset-\{\gz,\gu\}$, $\diamond\in \{\wedge,\swedge,\vee,\rightarrow,\leftrightarrow,\geql,\gleq,\gle\}$.
We have that (c,d) of Lemma \ref{le111} hold for $\theta$.
Then, for both $i$, (c,d) of Lemma \ref{le111} hold for $\theta_i$.
We put $j_{\mbbm{i}_1}=j_\mbbm{i}+1$ and $\mbbm{i}_1=(n_\theta,j_{\mbbm{i}_1})\in \{(n_\theta,j) \,|\, j\in \mbb{N}\}\subseteq \mbb{I}$. 
Hence, $\tilde{a}_{\mbbm{i}_1}\in \tilde{\mbb{A}}$;
by the induction hypothesis for $j_{\mbbm{i}_1}$, $\theta_1$, $\mbbm{i}_1$, and $\tilde{a}_{\mbbm{i}_1}$,
there exist $n_{J_1}\geq j_{\mbbm{i}_1}$, $J_1=\{(n_\theta,j) \,|\, j_{\mbbm{i}_1}+1\leq j\leq n_{J_1}\}\subseteq \{(n_\theta,j) \,|\, j\in \mbb{N}\}\subseteq \mbb{I}$, $\mbbm{i}_1\not\in J_1$,
$S_1\subseteq_{\mc F} \mi{OrdPropCl}_{\{\tilde{a}_{\mbbm{i}_1}\}\cup \{\tilde{a}_\mbbm{j} \,|\, \mbbm{j}\in J_1\}}$ satisfying that
(a--d) hold for $J_1$, $\theta_1$, $\tilde{a}_{\mbbm{i}_1}$, and $S_1$.
We put $j_{\mbbm{i}_2}=n_{J_1}+1$ and $\mbbm{i}_2=(n_\theta,j_{\mbbm{i}_2})\in \{(n_\theta,j) \,|\, j\in \mbb{N}\}\subseteq \mbb{I}$.
Hence, $\tilde{a}_{\mbbm{i}_2}\in \tilde{\mbb{A}}$;
by the induction hypothesis for $j_{\mbbm{i}_2}$, $\theta_2$, $\mbbm{i}_2$, and $\tilde{a}_{\mbbm{i}_2}$,
there exist $n_{J_2}\geq j_{\mbbm{i}_2}$, $J_2=\{(n_\theta,j) \,|\, j_{\mbbm{i}_2}+1\leq j\leq n_{J_2}\}\subseteq \{(n_\theta,j) \,|\, j\in \mbb{N}\}\subseteq \mbb{I}$, $\mbbm{i}_2\not\in J_2$,
$S_2\subseteq_{\mc F} \mi{OrdPropCl}_{\{\tilde{a}_{\mbbm{i}_2}\}\cup \{\tilde{a}_\mbbm{j} \,|\, \mbbm{j}\in J_2\}}$ satisfying that
(a--d) hold for $J_2$, $\theta_2$, $\tilde{a}_{\mbbm{i}_2}$, and $S_2$.
We put $n_J=n_{J_2}$ and $J=\{(n_\theta,j) \,|\, j_\mbbm{i}+1\leq j\leq n_J\}\subseteq \{(n_\theta,j) \,|\, j\in \mbb{N}\}\subseteq \mbb{I}$.
Then $n_J=n_{J_2}\geq j_{\mbbm{i}_2}=n_{J_1}+1>n_{J_1}\geq j_{\mbbm{i}_1}=j_\mbbm{i}+1>j_\mbbm{i}$,
$\mbbm{i}=(n_\theta,j_\mbbm{i})\not\in J=\{(n_\theta,j) \,|\, j_\mbbm{i}+1\leq j\leq n_J\}$,
$j_{\mbbm{i}_1}+1=j_\mbbm{i}+2$, $j_{\mbbm{i}_2}+1=n_{J_1}+2$,
$n_{J_1}+2>n_{J_1}+1>j_\mbbm{i}+1>j_\mbbm{i}$,
\begin{alignat}{1}
\label{eq1a}
& J=\{(n_\theta,j) \,|\, j_\mbbm{i}+1\leq j\leq n_J\}= \\
\notag
& \phantom{J=\mbox{}}
    \{(n_\theta,j_\mbbm{i}+1)\}\cup \{(n_\theta,j) \,|\, j_\mbbm{i}+2\leq j\leq n_{J_1}\}\cup \\
\notag
& \phantom{J=\mbox{}} \quad
    \{(n_\theta,n_{J_1}+1)\}\cup \{(n_\theta,j) \,|\, n_{J_1}+2\leq j\leq n_J\}= \\
\notag
& \phantom{J=\mbox{}}
    \{(n_\theta,j_{\mbbm{i}_1})\}\cup \{(n_\theta,j) \,|\, j_{\mbbm{i}_1}+1\leq j\leq n_{J_1}\}\cup \\
\notag
& \phantom{J=\mbox{}} \quad
    \{(n_\theta,j_{\mbbm{i}_2})\}\cup \{(n_\theta,j) \,|\, j_{\mbbm{i}_2}+1\leq j\leq n_{J_2}\}= \\
\notag
& \phantom{J=\mbox{}}
    \{\mbbm{i}_1\}\cup J_1\cup \{\mbbm{i}_2\}\cup J_2; \\[1mm]
\label{eq1b}
& \{\mbbm{i}\}=\{(n_\theta,j_\mbbm{i})\}, \{\mbbm{i}_1\}=\{(n_\theta,j_{\mbbm{i}_1})\}=\{(n_\theta,j_\mbbm{i}+1)\}, \\
\notag
& J_1=\{(n_\theta,j) \,|\, j_{\mbbm{i}_1}+1\leq j\leq n_{J_1}\}=\{(n_\theta,j) \,|\, j_\mbbm{i}+2\leq j\leq n_{J_1}\}, \\
\notag
& \{\mbbm{i}_2\}=\{(n_\theta,j_{\mbbm{i}_2})\}=\{(n_\theta,n_{J_1}+1)\}, \\
\notag
& J_2=\{(n_\theta,j) \,|\, j_{\mbbm{i}_2}+1\leq j\leq n_{J_2}\}=\{(n_\theta,j) \,|\, n_{J_1}+2\leq j\leq n_{J_2}\} \\ 
\notag
& \quad \text{are pairwise disjoint}.
\end{alignat}   
In Table \ref{tab2}, for every form of $\theta$, a binary interpolation rule of the respective form
\begin{alignat*}{1}
& \dfrac{\tilde{a}_\mbbm{i}\leftrightarrow \theta\in \mi{PropForm}_{\{\tilde{a}_\mbbm{i}\}}}
        {\mi{ClPrefix}\cup \{\tilde{a}_{\mbbm{i}_1}\leftrightarrow \theta_1,\tilde{a}_{\mbbm{i}_2}\leftrightarrow \theta_2\}}, \\
& \mi{ClPrefix}\subseteq_{\mc F} \mi{OrdPropCl}_{\{\tilde{a}_\mbbm{i},\tilde{a}_{\mbbm{i}_1},\tilde{a}_{\mbbm{i}_2}\}}, 
  \tilde{a}_{\mbbm{i}_i}\leftrightarrow \theta_i\in \mi{PropForm}_{\{\tilde{a}_{\mbbm{i}_i}\}}, 
\end{alignat*}
is assigned.
Note that $\mi{atoms}(\theta)\subseteq \mbb{O}$, 
$\tilde{a}_\mbbm{i}, \tilde{a}_{\mbbm{i}_1}, \tilde{a}_{\mbbm{i}_2}\in \tilde{\mbb{A}}$, $\mbb{O}\cap \tilde{\mbb{A}}=\emptyset$, 
$\tilde{a}_\mbbm{i}, \tilde{a}_{\mbbm{i}_1}, \tilde{a}_{\mbbm{i}_2}\not\in \mbb{O}$;
by (\ref{eq1b}), $\tilde{a}_\mbbm{i}$, $\tilde{a}_{\mbbm{i}_1}$, $\tilde{a}_{\mbbm{i}_2}$ are pairwise different;
$\mi{atoms}(\tilde{a}_\mbbm{i}\leftrightarrow \theta)=\mi{atoms}(\tilde{a}_\mbbm{i})\cup \mi{atoms}(\theta)\subseteq \mbb{O}\cup \{\tilde{a}_\mbbm{i}\}$,
$\tilde{a}_\mbbm{i}\leftrightarrow \theta\in \mi{PropForm}_{\{\tilde{a}_\mbbm{i}\}}$;
for both $i$, 
$\mi{atoms}(\theta_i)\subseteq \mbb{O}$, 
$\mi{atoms}(\tilde{a}_{\mbbm{i}_i}\leftrightarrow \theta_i)=\mi{atoms}(\tilde{a}_{\mbbm{i}_i})\cup \mi{atoms}(\theta_i)\subseteq \mbb{O}\cup \{\tilde{a}_{\mbbm{i}_i}\}$,
$\tilde{a}_{\mbbm{i}_i}\leftrightarrow \theta_i\in \mi{PropForm}_{\{\tilde{a}_{\mbbm{i}_i}\}}$;
$\mi{atoms}(\mi{ClPrefix})=\{\tilde{a}_\mbbm{i},\tilde{a}_{\mbbm{i}_1},\tilde{a}_{\mbbm{i}_2}\}$,
$\mi{ClPrefix}\subseteq_{\mc F} \mi{OrdPropCl}_{\{\tilde{a}_\mbbm{i},\tilde{a}_{\mbbm{i}_1},\tilde{a}_{\mbbm{i}_2}\}}$.
We denote $\text{Consequent}=\mi{ClPrefix}\cup \{\tilde{a}_{\mbbm{i}_1}\leftrightarrow \theta_1,\tilde{a}_{\mbbm{i}_2}\leftrightarrow \theta_2\}$.
We put 
\begin{equation}
\notag
S=\mi{ClPrefix}\cup S_1\cup S_2\subseteq_{\mc F} \mi{OrdPropCl}_{\{\tilde{a}_\mbbm{i}\}\cup \{\tilde{a}_\mbbm{j} \,|\, \mbbm{j}\in J\}}.
\end{equation}
Note that for both $i$, $\mi{atoms}(S_i)\subseteq \mbb{O}\cup \{\tilde{a}_{\mbbm{i}_i}\}\cup \{\tilde{a}_\mbbm{j} \,|\, \mbbm{j}\in J_i\}$;
$\mi{atoms}(S)=\mi{atoms}(\mi{ClPrefix}\cup S_1\cup S_2)=\mi{atoms}(\mi{ClPrefix})\cup \mi{atoms}(S_1)\cup \mi{atoms}(S_2)\subseteq
               \{\tilde{a}_\mbbm{i},\tilde{a}_{\mbbm{i}_1},\tilde{a}_{\mbbm{i}_2}\}\cup \mbb{O}\cup \{\tilde{a}_\mbbm{j} \,|\, \mbbm{j}\in J_1\}\cup \{\tilde{a}_\mbbm{j} \,|\, \mbbm{j}\in J_2\}=
               \mbb{O}\cup \{\tilde{a}_\mbbm{i}\}\cup \{\tilde{a}_\mbbm{j} \,|\, \mbbm{j}\in \{\mbbm{i}_1\}\cup J_1\cup \{\mbbm{i}_2\}\cup J_2\}\overset{\text{(\ref{eq1a})}}{=\!\!=}
               \mbb{O}\cup \{\tilde{a}_\mbbm{i}\}\cup \{\tilde{a}_\mbbm{j} \,|\, \mbbm{j}\in J\}$,
$S=\mi{ClPrefix}\cup S_1\cup S_2\subseteq_{\mc F} \mi{OrdPropCl}_{\{\tilde{a}_\mbbm{i}\}\cup \{\tilde{a}_\mbbm{j} \,|\, \mbbm{j}\in J\}}$.
It can be proved that
\begin{equation}
\label{eq1c}
\mi{ClPrefix}, S_1, S_2\ \text{are pairwise disjoint}.
\end{equation}

(a,c,d) can be proved straightforwardly.

Let ${\mf A}$ be a valuation. 
Then $\tilde{a}_\mbbm{i}, \tilde{a}_{\mbbm{i}_1}, \tilde{a}_{\mbbm{i}_2}\not\in \mbb{O}$;
by (\ref{eq1b}), $\tilde{a}_\mbbm{i}$, $\tilde{a}_{\mbbm{i}_1}$, $\tilde{a}_{\mbbm{i}_2}$ are pairwise different.
We define a valuation 
\begin{equation}
\notag
{\mf A}^\#={\mf A}|_{\mbb{O}\cup \{\tilde{a}_\mbbm{i}\}}\cup \{(\tilde{a}_{\mbbm{i}_1},\|\theta_1\|^{\mf A}),(\tilde{a}_{\mbbm{i}_2},\|\theta_2\|^{\mf A})\}\cup 
           {\mf A}|_{\mi{PropAtom}-(\mbb{O}\cup \{\tilde{a}_\mbbm{i},\tilde{a}_{\mbbm{i}_1},\tilde{a}_{\mbbm{i}_2}\})}.
\end{equation}
Hence, for both $i$, 
$\mi{atoms}(\theta_i)\subseteq \mbb{O}$,
${\mf A}^\#|_{\mbb{O}\cup \{\tilde{a}_\mbbm{i}\}}={\mf A}|_{\mbb{O}\cup \{\tilde{a}_\mbbm{i}\}}$,
$\|\tilde{a}_{\mbbm{i}_i}\|^{{\mf A}^\#}=\|\theta_i\|^{\mf A}=\|\theta_i\|^{{\mf A}^\#}$,
\begin{alignat*}{1}
\|\tilde{a}_{\mbbm{i}_i}\leftrightarrow \theta_i\|^{{\mf A}^\#}
&= (\|\tilde{a}_{\mbbm{i}_i}\|^{{\mf A}^\#}\frightarrow \|\theta_i\|^{{\mf A}^\#})\fwedge (\|\theta_i\|^{{\mf A}^\#}\frightarrow \|\tilde{a}_{\mbbm{i}_i}\|^{{\mf A}^\#})= \\
&\phantom{\mbox{}=\mbox{}}
   (\|\theta_i\|^{{\mf A}^\#}\frightarrow \|\theta_i\|^{{\mf A}^\#})\fwedge (\|\theta_i\|^{{\mf A}^\#}\frightarrow \|\theta_i\|^{{\mf A}^\#})=1,
\end{alignat*}
${\mf A}^\#\models \tilde{a}_{\mbbm{i}_i}\leftrightarrow \theta_i$.

Let ${\mf A}\models \tilde{a}_\mbbm{i}\leftrightarrow \theta$.
Then $\mi{atoms}(\tilde{a}_\mbbm{i}\leftrightarrow \theta)\subseteq \mbb{O}\cup \{\tilde{a}_\mbbm{i}\}$,
${\mf A}^\#|_{\mbb{O}\cup \{\tilde{a}_\mbbm{i}\}}={\mf A}|_{\mbb{O}\cup \{\tilde{a}_\mbbm{i}\}}$,
${\mf A}^\#\models \tilde{a}_\mbbm{i}\leftrightarrow \theta$,
\begin{alignat*}{1}
& \|\tilde{a}_\mbbm{i}\leftrightarrow \theta\|^{{\mf A}^\#}=1, \\
& \|\tilde{a}_\mbbm{i}\leftrightarrow (\theta_1\diamond \theta_2)\|^{{\mf A}^\#}=1, \\
& (\|\tilde{a}_\mbbm{i}\|^{{\mf A}^\#}\frightarrow (\|\theta_1\|^{{\mf A}^\#}\fdiamond \|\theta_2\|^{{\mf A}^\#}))\fwedge 
  ((\|\theta_1\|^{{\mf A}^\#}\fdiamond \|\theta_2\|^{{\mf A}^\#})\frightarrow \|\tilde{a}_\mbbm{i}\|^{{\mf A}^\#})=1, \\
& \|\tilde{a}_\mbbm{i}\|^{{\mf A}^\#}\frightarrow (\|\theta_1\|^{{\mf A}^\#}\fdiamond \|\theta_2\|^{{\mf A}^\#})=1,\ 
  (\|\theta_1\|^{{\mf A}^\#}\fdiamond \|\theta_2\|^{{\mf A}^\#})\frightarrow \|\tilde{a}_\mbbm{i}\|^{{\mf A}^\#}=1, \\
& \|\tilde{a}_\mbbm{i}\|^{{\mf A}^\#}\leq \|\theta_1\|^{{\mf A}^\#}\fdiamond \|\theta_2\|^{{\mf A}^\#},\ 
  \|\theta_1\|^{{\mf A}^\#}\fdiamond \|\theta_2\|^{{\mf A}^\#}\leq \|\tilde{a}_\mbbm{i}\|^{{\mf A}^\#}, \\
& \|\tilde{a}_\mbbm{i}\|^{{\mf A}^\#}=\|\theta_1\|^{{\mf A}^\#}\fdiamond \|\theta_2\|^{{\mf A}^\#}, \\
& \|\tilde{a}_\mbbm{i}\|^{{\mf A}^\#}=\|\tilde{a}_{\mbbm{i}_1}\|^{{\mf A}^\#}\fdiamond \|\tilde{a}_{\mbbm{i}_2}\|^{{\mf A}^\#};
\end{alignat*}
concerning Table \ref{tab2}, for every form of $\theta$, ${\mf A}^\#\models \mi{ClPrefix}$;
for both $i$, 
${\mf A}^\#\models \tilde{a}_{\mbbm{i}_i}\leftrightarrow \theta_i$; 
by the induction hypothesis (b) for $\theta_i$, ${\mf A}^\#$, $\tilde{a}_{\mbbm{i}_i}$, and $S_i$, 
there exists a valuation ${\mf A}_i$ satisfying ${\mf A}_i\models S_i$, ${\mf A}_i|_{\mbb{O}\cup \{\tilde{a}_{\mbbm{i}_i}\}}={\mf A}^\#|_{\mbb{O}\cup \{\tilde{a}_{\mbbm{i}_i}\}}$;
$\tilde{a}_\mbbm{i}, \tilde{a}_{\mbbm{i}_1}, \tilde{a}_{\mbbm{i}_2}\not\in \mbb{O}$,
$\{\tilde{a}_\mbbm{j} \,|\, \mbbm{j}\in J_1\}, \{\tilde{a}_\mbbm{j} \,|\, \mbbm{j}\in J_2\}\subseteq \tilde{\mbb{A}}$,
$(\{\tilde{a}_\mbbm{j} \,|\, \mbbm{j}\in J_1\}\cup \{\tilde{a}_\mbbm{j} \,|\, \mbbm{j}\in J_2\})\cap \mbb{O}=\tilde{\mbb{A}}\cap \mbb{O}=\emptyset$;
by (\ref{eq1b}), $\{\tilde{a}_\mbbm{i}\}$, $\{\tilde{a}_{\mbbm{i}_1}\}$, $\{\tilde{a}_\mbbm{j} \,|\, \mbbm{j}\in J_1\}$, $\{\tilde{a}_{\mbbm{i}_2}\}$, $\{\tilde{a}_\mbbm{j} \,|\, \mbbm{j}\in J_2\}$ are pairwise disjoint.
We define a valuation 
\begin{alignat*}{1}
{\mf A}' &= {\mf A}^\#|_{\mbb{O}\cup \{\tilde{a}_\mbbm{i},\tilde{a}_{\mbbm{i}_1},\tilde{a}_{\mbbm{i}_2}\}}\cup 
            {\mf A}_1|_{\{\tilde{a}_\mbbm{j} \,|\, \mbbm{j}\in J_1\}}\cup {\mf A}_2|_{\{\tilde{a}_\mbbm{j} \,|\, \mbbm{j}\in J_2\}}\cup \\
         &\phantom{\mbox{}=\mbox{}}
            {\mf A}^\#|_{\mi{PropAtom}-(\mbb{O}\cup \{\tilde{a}_\mbbm{i}\}\cup 
                                        \{\tilde{a}_\mbbm{j} \,|\, \mbbm{j}\in J\overset{\text{(\ref{eq1a})}}{=\!\!=} \{\mbbm{i}_1\}\cup J_1\cup \{\mbbm{i}_2\}\cup J_2\})}.
\end{alignat*}
We get that
${\mf A}^\#\models \mi{ClPrefix}$,
$\mi{atoms}(\mi{ClPrefix})=\{\tilde{a}_\mbbm{i},\tilde{a}_{\mbbm{i}_1},\tilde{a}_{\mbbm{i}_2}\}$,
${\mf A}'|_{\mbb{O}\cup \{\tilde{a}_\mbbm{i},\tilde{a}_{\mbbm{i}_1},\tilde{a}_{\mbbm{i}_2}\}}={\mf A}^\#|_{\mbb{O}\cup \{\tilde{a}_\mbbm{i},\tilde{a}_{\mbbm{i}_1},\tilde{a}_{\mbbm{i}_2}\}}$, 
${\mf A}'\models \mi{ClPrefix}$;
for both $i$, 
${\mf A}_i\models S_i$,
$\mi{atoms}(S_i)\subseteq \mbb{O}\cup \{\tilde{a}_{\mbbm{i}_i}\}\cup \{\tilde{a}_\mbbm{j} \,|\, \mbbm{j}\in J_i\}$,
\begin{alignat*}{1}
{\mf A}'|_{\mbb{O}\cup \{\tilde{a}_{\mbbm{i}_i}\}\cup \{\tilde{a}_\mbbm{j} \,|\, \mbbm{j}\in J_i\}} 
&= {\mf A}'|_{\mbb{O}\cup \{\tilde{a}_{\mbbm{i}_i}\}}\cup {\mf A}'|_{\{\tilde{a}_\mbbm{j} \,|\, \mbbm{j}\in J_i\}}=
   {\mf A}^\#|_{\mbb{O}\cup \{\tilde{a}_{\mbbm{i}_i}\}}\cup {\mf A}_i|_{\{\tilde{a}_\mbbm{j} \,|\, \mbbm{j}\in J_i\}}= \\
&\phantom{\mbox{}=\mbox{}}
   {\mf A}_i|_{\mbb{O}\cup \{\tilde{a}_{\mbbm{i}_i}\}}\cup {\mf A}_i|_{\{\tilde{a}_\mbbm{j} \,|\, \mbbm{j}\in J_i\}}= 
   {\mf A}_i|_{\mbb{O}\cup \{\tilde{a}_{\mbbm{i}_i}\}\cup \{\tilde{a}_\mbbm{j} \,|\, \mbbm{j}\in J_i\}},
\end{alignat*}
${\mf A}'\models S_i$; 
${\mf A}'\models S=\mi{ClPrefix}\cup S_1\cup S_2$, ${\mf A}'|_{\mbb{O}\cup \{\tilde{a}_\mbbm{i}\}}={\mf A}^\#|_{\mbb{O}\cup \{\tilde{a}_\mbbm{i}\}}={\mf A}|_{\mbb{O}\cup \{\tilde{a}_\mbbm{i}\}}$.

Let ${\mf A}'$ be a valuation such that ${\mf A}'\models S$.
Then $\mi{ClPrefix}, S_1, S_2\subseteq S$, ${\mf A}'\models \mi{ClPrefix}$; 
for both $i$, 
${\mf A}'\models S_i$;
by the induction hypothesis (b) for $\theta_i$, $\tilde{a}_{\mbbm{i}_i}$, and $S_i$, 
there exists a valuation ${\mf A}_i$ satisfying ${\mf A}_i\models \tilde{a}_{\mbbm{i}_i}\leftrightarrow \theta_i$, 
${\mf A}_i|_{\mbb{O}\cup \{\tilde{a}_{\mbbm{i}_i}\}}={\mf A}'|_{\mbb{O}\cup \{\tilde{a}_{\mbbm{i}_i}\}}$;
$\mi{atoms}(\tilde{a}_{\mbbm{i}_i}\leftrightarrow \theta_i)\subseteq \mbb{O}\cup \{\tilde{a}_{\mbbm{i}_i}\}$,
${\mf A}'\models \tilde{a}_{\mbbm{i}_i}\leftrightarrow \theta_i$,
\begin{alignat*}{1}
& \|\tilde{a}_{\mbbm{i}_i}\leftrightarrow \theta_i\|^{{\mf A}'}=1, \\
& (\|\tilde{a}_{\mbbm{i}_i}\|^{{\mf A}'}\frightarrow \|\theta_i\|^{{\mf A}'})\fwedge (\|\theta_i\|^{{\mf A}'}\frightarrow \|\tilde{a}_{\mbbm{i}_i}\|^{{\mf A}'})=1, \\
& \|\tilde{a}_{\mbbm{i}_i}\|^{{\mf A}'}\frightarrow \|\theta_i\|^{{\mf A}'}=1,\ \|\theta_i\|^{{\mf A}'}\frightarrow \|\tilde{a}_{\mbbm{i}_i}\|^{{\mf A}'}=1, \\
& \|\tilde{a}_{\mbbm{i}_i}\|^{{\mf A}'}\leq \|\theta_i\|^{{\mf A}'},\ \|\theta_i\|^{{\mf A}'}\leq \|\tilde{a}_{\mbbm{i}_i}\|^{{\mf A}'}, \\
& \|\tilde{a}_{\mbbm{i}_i}\|^{{\mf A}'}=\|\theta_i\|^{{\mf A}'};  
\end{alignat*}
concerning Table \ref{tab2}, for every form of $\theta$, 
\begin{alignat*}{1}
& \|\tilde{a}_\mbbm{i}\|^{{\mf A}'}=\|\tilde{a}_{\mbbm{i}_1}\|^{{\mf A}'}\fdiamond \|\tilde{a}_{\mbbm{i}_2}\|^{{\mf A}'}, \\
& \|\tilde{a}_\mbbm{i}\|^{{\mf A}'}=\|\theta_1\|^{{\mf A}'}\fdiamond \|\theta_2\|^{{\mf A}'}, \\
& \|\tilde{a}_\mbbm{i}\|^{{\mf A}'}\leq \|\theta_1\|^{{\mf A}'}\fdiamond \|\theta_2\|^{{\mf A}'},\ 
  \|\theta_1\|^{{\mf A}'}\fdiamond \|\theta_2\|^{{\mf A}'}\leq \|\tilde{a}_\mbbm{i}\|^{{\mf A}'}, \\
& \|\tilde{a}_\mbbm{i}\|^{{\mf A}'}\frightarrow (\|\theta_1\|^{{\mf A}'}\fdiamond \|\theta_2\|^{{\mf A}'})=1,\ 
  (\|\theta_1\|^{{\mf A}'}\fdiamond \|\theta_2\|^{{\mf A}'})\frightarrow \|\tilde{a}_\mbbm{i}\|^{{\mf A}'}=1, \\
& (\|\tilde{a}_\mbbm{i}\|^{{\mf A}'}\frightarrow (\|\theta_1\|^{{\mf A}'}\fdiamond \|\theta_2\|^{{\mf A}'}))\fwedge 
  ((\|\theta_1\|^{{\mf A}'}\fdiamond \|\theta_2\|^{{\mf A}'})\frightarrow \|\tilde{a}_\mbbm{i}\|^{{\mf A}'})=1, \\
& \|\tilde{a}_\mbbm{i}\leftrightarrow (\theta_1\diamond \theta_2)\|^{{\mf A}'}=1, \\
& \|\tilde{a}_\mbbm{i}\leftrightarrow \theta\|^{{\mf A}'}=1,
\end{alignat*}
${\mf A}'\models \tilde{a}_\mbbm{i}\leftrightarrow \theta$.
We put ${\mf A}={\mf A}'$.
Hence, ${\mf A}\models \tilde{a}_\mbbm{i}\leftrightarrow \theta$, ${\mf A}|_{\mbb{O}\cup \{\tilde{a}_\mbbm{i}\}}={\mf A}'|_{\mbb{O}\cup \{\tilde{a}_\mbbm{i}\}}$; 
(b) holds.

Case 2.2 (the unary interpolation case):
Either $\theta=\theta_1\rightarrow \gz$ or $\theta=\theta_1\geql \gz$ or $\theta=\theta_1\geql \gu$ or $\theta=\gz\gle \theta_1$ or $\theta=\theta_1\gle \gu$, $\theta_1\in \mi{PropForm}_\emptyset-\{\gz,\gu\}$.
The proof is a straightforward simplification of Case 2.1.

So, in all Cases 1, 2.1, and 2.2, (a--d) hold.
The induction is completed.
%
%
%
\end{proof}

\subsection{Case 2.2 (the unary interpolation case) of Lemma \ref{le11}}
\label{S7.1b}

\begin{proof}
Either $\theta=\theta_1\rightarrow \gz$ or $\theta=\theta_1\geql \gz$ or $\theta=\theta_1\geql \gu$ or $\theta=\gz\gle \theta_1$ or $\theta=\theta_1\gle \gu$, $\theta_1\in \mi{PropForm}_\emptyset-\{\gz,\gu\}$.
We have that (c,d) of Lemma \ref{le111} hold for $\theta$. 
Then (c,d) of Lemma \ref{le111} hold for $\theta_1$.
We put $j_{\mbbm{i}_1}=j_\mbbm{i}+1$ and $\mbbm{i}_1=(n_\theta,j_{\mbbm{i}_1})\in \{(n_\theta,j) \,|\, j\in \mbb{N}\}\subseteq \mbb{I}$. 
Hence, $\tilde{a}_{\mbbm{i}_1}\in \tilde{\mbb{A}}$;
by the induction hypothesis for $j_{\mbbm{i}_1}$, $\theta_1$, $\mbbm{i}_1$, and $\tilde{a}_{\mbbm{i}_1}$,
there exist $n_{J_1}\geq j_{\mbbm{i}_1}$, $J_1=\{(n_\theta,j) \,|\, j_{\mbbm{i}_1}+1\leq j\leq n_{J_1}\}\subseteq \{(n_\theta,j) \,|\, j\in \mbb{N}\}\subseteq \mbb{I}$, $\mbbm{i}_1\not\in J_1$,
$S_1\subseteq_{\mc F} \mi{OrdPropCl}_{\{\tilde{a}_{\mbbm{i}_1}\}\cup \{\tilde{a}_\mbbm{j} \,|\, \mbbm{j}\in J_1\}}$ satisfying that
(a--d) hold for $J_1$, $\theta_1$, $\tilde{a}_{\mbbm{i}_1}$, and $S_1$.
We put $n_J=n_{J_1}$ and $J=\{(n_\theta,j) \,|\, j_\mbbm{i}+1\leq j\leq n_J\}\subseteq \{(n_\theta,j) \,|\, j\in \mbb{N}\}\subseteq \mbb{I}$.
Then $n_J=n_{J_1}\geq j_{\mbbm{i}_1}=j_\mbbm{i}+1>j_\mbbm{i}$,
$\mbbm{i}=(n_\theta,j_\mbbm{i})\not\in J=\{(n_\theta,j) \,|\, j_\mbbm{i}+1\leq j\leq n_J\}$,
$j_{\mbbm{i}_1}+1=j_\mbbm{i}+2$,
\begin{alignat}{1}
\label{eq1f}
& J=\{(n_\theta,j) \,|\, j_\mbbm{i}+1\leq j\leq n_J\}= \\
\notag
& \phantom{J=\mbox{}}
    \{(n_\theta,j_\mbbm{i}+1)\}\cup \{(n_\theta,j) \,|\, j_\mbbm{i}+2\leq j\leq n_J\}= \\
\notag
& \phantom{J=\mbox{}}
    \{(n_\theta,j_{\mbbm{i}_1})\}\cup \{(n_\theta,j) \,|\, j_{\mbbm{i}_1}+1\leq j\leq n_{J_1}\}=\{\mbbm{i}_1\}\cup J_1; \\[1mm]
\label{eq1g}
& \{\mbbm{i}\}=\{(n_\theta,j_\mbbm{i})\}, \{\mbbm{i}_1\}=\{(n_\theta,j_{\mbbm{i}_1})\}=\{(n_\theta,j_\mbbm{i}+1)\}, \\
\notag
& J_1=\{(n_\theta,j) \,|\, j_{\mbbm{i}_1}+1\leq j\leq n_{J_1}\}=\{(n_\theta,j) \,|\, j_\mbbm{i}+2\leq j\leq n_{J_1}\}\ \text{are pairwise disjoint}.
\end{alignat}   
In Table \ref{tab3}, for every form of $\theta$, a unary interpolation rule of the respective form
\begin{alignat*}{1}
& \dfrac{\tilde{a}_\mbbm{i}\leftrightarrow \theta\in \mi{PropForm}_{\{\tilde{a}_\mbbm{i}\}}}
        {\mi{ClPrefix}\cup \{\tilde{a}_{\mbbm{i}_1}\leftrightarrow \theta_1\}}, \\
& \mi{ClPrefix}\subseteq_{\mc F} \mi{OrdPropCl}_{\{\tilde{a}_\mbbm{i},\tilde{a}_{\mbbm{i}_1}\}}, 
  \tilde{a}_{\mbbm{i}_1}\leftrightarrow \theta_1\in \mi{PropForm}_{\{\tilde{a}_{\mbbm{i}_1}\}},
\end{alignat*}
is assigned.
Note that $\mi{atoms}(\theta)\subseteq \mbb{O}$, 
$\tilde{a}_\mbbm{i}, \tilde{a}_{\mbbm{i}_1}\in \tilde{\mbb{A}}$, $\mbb{O}\cap \tilde{\mbb{A}}=\emptyset$, $\tilde{a}_\mbbm{i}, \tilde{a}_{\mbbm{i}_1}\not\in \mbb{O}$;
by (\ref{eq1g}), $\tilde{a}_\mbbm{i}$ and $\tilde{a}_{\mbbm{i}_1}$ are different;
$\mi{atoms}(\tilde{a}_\mbbm{i}\leftrightarrow \theta)=\mi{atoms}(\tilde{a}_\mbbm{i})\cup \mi{atoms}(\theta)\subseteq \mbb{O}\cup \{\tilde{a}_\mbbm{i}\}$,
$\tilde{a}_\mbbm{i}\leftrightarrow \theta\in \mi{PropForm}_{\{\tilde{a}_\mbbm{i}\}}$,
$\mi{atoms}(\theta_1)\subseteq \mbb{O}$, 
$\mi{atoms}(\tilde{a}_{\mbbm{i}_1}\leftrightarrow \theta_1)=\mi{atoms}(\tilde{a}_{\mbbm{i}_1})\cup \mi{atoms}(\theta_1)\subseteq \mbb{O}\cup \{\tilde{a}_{\mbbm{i}_1}\}$,
$\tilde{a}_{\mbbm{i}_1}\leftrightarrow \theta_1\in \mi{PropForm}_{\{\tilde{a}_{\mbbm{i}_1}\}}$,
$\mi{atoms}(\mi{ClPrefix})=\{\tilde{a}_\mbbm{i},\tilde{a}_{\mbbm{i}_1}\}$,
$\mi{ClPrefix}\subseteq_{\mc F} \mi{OrdPropCl}_{\{\tilde{a}_\mbbm{i},\tilde{a}_{\mbbm{i}_1}\}}$.
We denote $\text{Consequent}=\mi{ClPrefix}\cup \{\tilde{a}_{\mbbm{i}_1}\leftrightarrow \theta_1\}$.
We put
\begin{equation}
\notag 
S=\mi{ClPrefix}\cup S_1\subseteq_{\mc F} \mi{OrdPropCl}_{\{\tilde{a}_\mbbm{i}\}\cup \{\tilde{a}_\mbbm{j} \,|\, \mbbm{j}\in J\}}.
\end{equation}
Note that $\mi{atoms}(S_1)\subseteq \mbb{O}\cup \{\tilde{a}_{\mbbm{i}_1}\}\cup \{\tilde{a}_\mbbm{j} \,|\, \mbbm{j}\in J_1\}$,
$\mi{atoms}(S)=\mi{atoms}(\mi{ClPrefix}\cup S_1)=\mi{atoms}(\mi{ClPrefix})\cup \mi{atoms}(S_1)\subseteq
               \{\tilde{a}_\mbbm{i},\tilde{a}_{\mbbm{i}_1}\}\cup \mbb{O}\cup \{\tilde{a}_\mbbm{j} \,|\, \mbbm{j}\in J_1\}=
               \mbb{O}\cup \{\tilde{a}_\mbbm{i}\}\cup \{\tilde{a}_\mbbm{j} \,|\, \mbbm{j}\in \{\mbbm{i}_1\}\cup J_1\}\overset{\text{(\ref{eq1f})}}{=\!\!=}
               \mbb{O}\cup \{\tilde{a}_\mbbm{i}\}\cup \{\tilde{a}_\mbbm{j} \,|\, \mbbm{j}\in J\}$,
$S=\mi{ClPrefix}\cup S_1\subseteq_{\mc F} \mi{OrdPropCl}_{\{\tilde{a}_\mbbm{i}\}\cup \{\tilde{a}_\mbbm{j} \,|\, \mbbm{j}\in J\}}$.
It can be proved that
\begin{equation}
\label{eq1h}
\mi{ClPrefix}\cap S_1=\emptyset.
\end{equation}

(a,c,d) can be proved straightforwardly.

Let ${\mf A}$ be a valuation. 
Then $\tilde{a}_\mbbm{i}, \tilde{a}_{\mbbm{i}_1}\not\in \mbb{O}$;
by (\ref{eq1g}), $\tilde{a}_\mbbm{i}$ and $\tilde{a}_{\mbbm{i}_1}$ are different.
We define a valuation 
\begin{equation}
\notag
{\mf A}^\#={\mf A}|_{\mbb{O}\cup \{\tilde{a}_\mbbm{i}\}}\cup \{(\tilde{a}_{\mbbm{i}_1},\|\theta_1\|^{\mf A})\}\cup
           {\mf A}|_{\mi{PropAtom}-(\mbb{O}\cup \{\tilde{a}_\mbbm{i},\tilde{a}_{\mbbm{i}_1}\})}.
\end{equation}
Hence, $\mi{atoms}(\theta_1)\subseteq \mbb{O}$,
${\mf A}^\#|_{\mbb{O}\cup \{\tilde{a}_\mbbm{i}\}}={\mf A}|_{\mbb{O}\cup \{\tilde{a}_\mbbm{i}\}}$,
$\|\tilde{a}_{\mbbm{i}_1}\|^{{\mf A}^\#}=\|\theta_1\|^{\mf A}=\|\theta_1\|^{{\mf A}^\#}$,
\begin{alignat*}{1}
\|\tilde{a}_{\mbbm{i}_1}\leftrightarrow \theta_1\|^{{\mf A}^\#} 
&= (\|\tilde{a}_{\mbbm{i}_1}\|^{{\mf A}^\#}\frightarrow \|\theta_1\|^{{\mf A}^\#})\fwedge (\|\theta_1\|^{{\mf A}^\#}\frightarrow \|\tilde{a}_{\mbbm{i}_1}\|^{{\mf A}^\#})= \\
&\phantom{\mbox{}=\mbox{}}
   (\|\theta_1\|^{{\mf A}^\#}\frightarrow \|\theta_1\|^{{\mf A}^\#})\fwedge (\|\theta_1\|^{{\mf A}^\#}\frightarrow \|\theta_1\|^{{\mf A}^\#})=1,
\end{alignat*}
${\mf A}^\#\models \tilde{a}_{\mbbm{i}_1}\leftrightarrow \theta_1$.

Let ${\mf A}\models \tilde{a}_\mbbm{i}\leftrightarrow \theta$.
Then $\mi{atoms}(\tilde{a}_\mbbm{i}\leftrightarrow \theta)\subseteq \mbb{O}\cup \{\tilde{a}_\mbbm{i}\}$,
${\mf A}^\#|_{\mbb{O}\cup \{\tilde{a}_\mbbm{i}\}}={\mf A}|_{\mbb{O}\cup \{\tilde{a}_\mbbm{i}\}}$,
${\mf A}^\#\models \tilde{a}_\mbbm{i}\leftrightarrow \theta$,
\begin{alignat*}{1}
& \|\tilde{a}_\mbbm{i}\leftrightarrow \theta\|^{{\mf A}^\#}=1, \\[1mm]
& \text{either}\ \|\tilde{a}_\mbbm{i}\leftrightarrow (\theta_1\rightarrow \gz)\|^{{\mf A}^\#}=1\ \text{or}\ 
                 \|\tilde{a}_\mbbm{i}\leftrightarrow (\theta_1\geql \gz)\|^{{\mf A}^\#}=1\ \text{or} \\
&                \|\tilde{a}_\mbbm{i}\leftrightarrow (\theta_1\geql \gu)\|^{{\mf A}^\#}=1\ \text{or}\
                 \|\tilde{a}_\mbbm{i}\leftrightarrow (\gz\gle \theta_1)\|^{{\mf A}^\#}=1\ \text{or}\
                 \|\tilde{a}_\mbbm{i}\leftrightarrow (\theta_1\gle \gu)\|^{{\mf A}^\#}=1, \\[1mm] 
& \text{either}\ (\|\tilde{a}_\mbbm{i}\|^{{\mf A}^\#}\frightarrow (\|\theta_1\|^{{\mf A}^\#}\frightarrow 0))\fwedge 
                 ((\|\theta_1\|^{{\mf A}^\#}\frightarrow 0)\frightarrow \|\tilde{a}_\mbbm{i}\|^{{\mf A}^\#})=1\ \text{or} \\
&                (\|\tilde{a}_\mbbm{i}\|^{{\mf A}^\#}\frightarrow (\|\theta_1\|^{{\mf A}^\#}\feql 0))\fwedge 
                 ((\|\theta_1\|^{{\mf A}^\#}\feql 0)\frightarrow \|\tilde{a}_\mbbm{i}\|^{{\mf A}^\#})=1\ \text{or} \\
&                (\|\tilde{a}_\mbbm{i}\|^{{\mf A}^\#}\frightarrow (\|\theta_1\|^{{\mf A}^\#}\feql 1))\fwedge 
                 ((\|\theta_1\|^{{\mf A}^\#}\feql 1)\frightarrow \|\tilde{a}_\mbbm{i}\|^{{\mf A}^\#})=1\ \text{or} \\
&                (\|\tilde{a}_\mbbm{i}\|^{{\mf A}^\#}\frightarrow (0\fle \|\theta_1\|^{{\mf A}^\#}))\fwedge 
                 ((0\fle \|\theta_1\|^{{\mf A}^\#})\frightarrow \|\tilde{a}_\mbbm{i}\|^{{\mf A}^\#})=1\ \text{or} \\
&                (\|\tilde{a}_\mbbm{i}\|^{{\mf A}^\#}\frightarrow (\|\theta_1\|^{{\mf A}^\#}\fle 1))\fwedge 
                 ((\|\theta_1\|^{{\mf A}^\#}\fle 1)\frightarrow \|\tilde{a}_\mbbm{i}\|^{{\mf A}^\#})=1, \\[1mm]
& \text{either}\ \|\tilde{a}_\mbbm{i}\|^{{\mf A}^\#}\leq \|\theta_1\|^{{\mf A}^\#}\frightarrow 0,\ \|\theta_1\|^{{\mf A}^\#}\frightarrow 0\leq \|\tilde{a}_\mbbm{i}\|^{{\mf A}^\#}\ \text{or} \\
&                \|\tilde{a}_\mbbm{i}\|^{{\mf A}^\#}\leq \|\theta_1\|^{{\mf A}^\#}\feql 0,\ \|\theta_1\|^{{\mf A}^\#}\feql 0\leq \|\tilde{a}_\mbbm{i}\|^{{\mf A}^\#}\ \text{or} \\
&                \|\tilde{a}_\mbbm{i}\|^{{\mf A}^\#}\leq \|\theta_1\|^{{\mf A}^\#}\feql 1,\ \|\theta_1\|^{{\mf A}^\#}\feql 1\leq \|\tilde{a}_\mbbm{i}\|^{{\mf A}^\#}\ \text{or} \\
&                \|\tilde{a}_\mbbm{i}\|^{{\mf A}^\#}\leq 0\fle \|\theta_1\|^{{\mf A}^\#},\ 0\fle \|\theta_1\|^{{\mf A}^\#}\leq \|\tilde{a}_\mbbm{i}\|^{{\mf A}^\#}\ \text{or} \\
&                \|\tilde{a}_\mbbm{i}\|^{{\mf A}^\#}\leq \|\theta_1\|^{{\mf A}^\#}\fle 1,\ \|\theta_1\|^{{\mf A}^\#}\fle 1\leq \|\tilde{a}_\mbbm{i}\|^{{\mf A}^\#}, \\[1mm]
& \text{either}\ \|\tilde{a}_\mbbm{i}\|^{{\mf A}^\#}=\|\theta_1\|^{{\mf A}^\#}\frightarrow 0\ \text{or}\ 
                 \|\tilde{a}_\mbbm{i}\|^{{\mf A}^\#}=\|\theta_1\|^{{\mf A}^\#}\feql 0\ \text{or}\ \|\tilde{a}_\mbbm{i}\|^{{\mf A}^\#}=\|\theta_1\|^{{\mf A}^\#}\feql 1\ \text{or} \\
&                \|\tilde{a}_\mbbm{i}\|^{{\mf A}^\#}=0\fle \|\theta_1\|^{{\mf A}^\#}\ \text{or}\ \|\tilde{a}_\mbbm{i}\|^{{\mf A}^\#}=\|\theta_1\|^{{\mf A}^\#}\fle 1, \\[1mm]
& \text{either}\ \|\tilde{a}_\mbbm{i}\|^{{\mf A}^\#}=\|\tilde{a}_{\mbbm{i}_1}\|^{{\mf A}^\#}\frightarrow 0\ \text{or}\ 
                 \|\tilde{a}_\mbbm{i}\|^{{\mf A}^\#}=\|\tilde{a}_{\mbbm{i}_1}\|^{{\mf A}^\#}\feql 0\ \text{or}\ \|\tilde{a}_\mbbm{i}\|^{{\mf A}^\#}=\|\tilde{a}_{\mbbm{i}_1}\|^{{\mf A}^\#}\feql 1\ \text{or} \\
&                \|\tilde{a}_\mbbm{i}\|^{{\mf A}^\#}=0\fle \|\tilde{a}_{\mbbm{i}_1}\|^{{\mf A}^\#}\ \text{or}\ \|\tilde{a}_\mbbm{i}\|^{{\mf A}^\#}=\|\tilde{a}_{\mbbm{i}_1}\|^{{\mf A}^\#}\fle 1;
\end{alignat*}
concerning Table \ref{tab3}, for every form of $\theta$, ${\mf A}^\#\models \mi{ClPrefix}$;
${\mf A}^\#\models \tilde{a}_{\mbbm{i}_1}\leftrightarrow \theta_1$; 
by the induction hypothesis (b) for $\theta_1$, ${\mf A}^\#$, $\tilde{a}_{\mbbm{i}_1}$, and $S_1$, 
there exists a valuation ${\mf A}_1$ satisfying ${\mf A}_1\models S_1$, ${\mf A}_1|_{\mbb{O}\cup \{\tilde{a}_{\mbbm{i}_1}\}}={\mf A}^\#|_{\mbb{O}\cup \{\tilde{a}_{\mbbm{i}_1}\}}$;
$\tilde{a}_\mbbm{i}, \tilde{a}_{\mbbm{i}_1}\not\in \mbb{O}$,
$\{\tilde{a}_\mbbm{j} \,|\, \mbbm{j}\in J_1\}\subseteq \tilde{\mbb{A}}$,
$\{\tilde{a}_\mbbm{j} \,|\, \mbbm{j}\in J_1\}\cap \mbb{O}=\tilde{\mbb{A}}\cap \mbb{O}=\emptyset$;
by (\ref{eq1g}), $\{\tilde{a}_\mbbm{i}\}$, $\{\tilde{a}_{\mbbm{i}_1}\}$, $\{\tilde{a}_\mbbm{j} \,|\, \mbbm{j}\in J_1\}$ are pairwise disjoint.
We define a valuation 
\begin{equation}
\notag
{\mf A}'={\mf A}^\#|_{\mbb{O}\cup \{\tilde{a}_\mbbm{i},\tilde{a}_{\mbbm{i}_1}\}}\cup {\mf A}_1|_{\{\tilde{a}_\mbbm{j} \,|\, \mbbm{j}\in J_1\}}\cup 
         {\mf A}^\#|_{\mi{PropAtom}-(\mbb{O}\cup \{\tilde{a}_\mbbm{i}\}\cup \{\tilde{a}_\mbbm{j} \,|\, \mbbm{j}\in J\overset{\text{(\ref{eq1f})}}{=\!\!=} \{\mbbm{i}_1\}\cup J_1\})}.
\end{equation}
We get that
${\mf A}^\#\models \mi{ClPrefix}$,
$\mi{atoms}(\mi{ClPrefix})=\{\tilde{a}_\mbbm{i},\tilde{a}_{\mbbm{i}_1}\}$,
${\mf A}'|_{\mbb{O}\cup \{\tilde{a}_\mbbm{i},\tilde{a}_{\mbbm{i}_1}\}}={\mf A}^\#|_{\mbb{O}\cup \{\tilde{a}_\mbbm{i},\tilde{a}_{\mbbm{i}_1}\}}$, 
${\mf A}'\models \mi{ClPrefix}$;
${\mf A}_1\models S_1$,
$\mi{atoms}(S_1)\subseteq \mbb{O}\cup \{\tilde{a}_{\mbbm{i}_1}\}\cup \{\tilde{a}_\mbbm{j} \,|\, \mbbm{j}\in J_1\}$, 
\begin{alignat*}{1}
{\mf A}'|_{\mbb{O}\cup \{\tilde{a}_{\mbbm{i}_1}\}\cup \{\tilde{a}_\mbbm{j} \,|\, \mbbm{j}\in J_1\}} 
&= {\mf A}'|_{\mbb{O}\cup \{\tilde{a}_{\mbbm{i}_1}\}}\cup {\mf A}'|_{\{\tilde{a}_\mbbm{j} \,|\, \mbbm{j}\in J_1\}}=
   {\mf A}^\#|_{\mbb{O}\cup \{\tilde{a}_{\mbbm{i}_1}\}}\cup {\mf A}_1|_{\{\tilde{a}_\mbbm{j} \,|\, \mbbm{j}\in J_1\}}= \\
&\phantom{\mbox{}=\mbox{}}
   {\mf A}_1|_{\mbb{O}\cup \{\tilde{a}_{\mbbm{i}_1}\}}\cup {\mf A}_1|_{\{\tilde{a}_\mbbm{j} \,|\, \mbbm{j}\in J_1\}}= 
   {\mf A}_1|_{\mbb{O}\cup \{\tilde{a}_{\mbbm{i}_1}\}\cup \{\tilde{a}_\mbbm{j} \,|\, \mbbm{j}\in J_1\}},
\end{alignat*}
${\mf A}'\models S_1$;
${\mf A}'\models S=\mi{ClPrefix}\cup S_1$, ${\mf A}'|_{\mbb{O}\cup \{\tilde{a}_\mbbm{i}\}}={\mf A}^\#|_{\mbb{O}\cup \{\tilde{a}_\mbbm{i}\}}={\mf A}|_{\mbb{O}\cup \{\tilde{a}_\mbbm{i}\}}$.

Let ${\mf A}'$ be a valuation such that ${\mf A}'\models S$.
Then $\mi{ClPrefix}, S_1\subseteq S$, ${\mf A}'\models \mi{ClPrefix}$, ${\mf A}'\models S_1$;
by the induction hypothesis (b) for $\theta_1$, $\tilde{a}_{\mbbm{i}_1}$, and $S_1$,
there exists a valuation ${\mf A}_1$ satisfying ${\mf A}_1\models \tilde{a}_{\mbbm{i}_1}\leftrightarrow \theta_1$, 
${\mf A}_1|_{\mbb{O}\cup \{\tilde{a}_{\mbbm{i}_1}\}}={\mf A}'|_{\mbb{O}\cup \{\tilde{a}_{\mbbm{i}_1}\}}$;
$\mi{atoms}(\tilde{a}_{\mbbm{i}_1}\leftrightarrow \theta_1)\subseteq \mbb{O}\cup \{\tilde{a}_{\mbbm{i}_1}\}$,
${\mf A}'\models \tilde{a}_{\mbbm{i}_1}\leftrightarrow \theta_1$,
\begin{alignat*}{1}
& \|\tilde{a}_{\mbbm{i}_1}\leftrightarrow \theta_1\|^{{\mf A}'}=1, \\
& (\|\tilde{a}_{\mbbm{i}_1}\|^{{\mf A}'}\frightarrow \|\theta_1\|^{{\mf A}'})\fwedge (\|\theta_1\|^{{\mf A}'}\frightarrow \|\tilde{a}_{\mbbm{i}_1}\|^{{\mf A}'})=1, \\
& (\|\tilde{a}_{\mbbm{i}_1}\|^{{\mf A}'}\frightarrow \|\theta_1\|^{{\mf A}'})=1,\ (\|\theta_1\|^{{\mf A}'}\frightarrow \|\tilde{a}_{\mbbm{i}_1}\|^{{\mf A}'})=1, \\
& \|\tilde{a}_{\mbbm{i}_1}\|^{{\mf A}'}\leq \|\theta_1\|^{{\mf A}'},\ \|\theta_1\|^{{\mf A}'}\leq \|\tilde{a}_{\mbbm{i}_1}\|^{{\mf A}'}, \\
& \|\tilde{a}_{\mbbm{i}_1}\|^{{\mf A}'}=\|\theta_1\|^{{\mf A}'};  
\end{alignat*}
concerning Table \ref{tab3}, for every form of $\theta$, 
\begin{alignat*}{1}
& \text{either}\ \|\tilde{a}_\mbbm{i}\|^{{\mf A}'}=\|\tilde{a}_{\mbbm{i}_1}\|^{{\mf A}'}\frightarrow 0\ \text{or}\ 
                 \|\tilde{a}_\mbbm{i}\|^{{\mf A}'}=\|\tilde{a}_{\mbbm{i}_1}\|^{{\mf A}'}\feql 0\ \text{or}\ \|\tilde{a}_\mbbm{i}\|^{{\mf A}'}=\|\tilde{a}_{\mbbm{i}_1}\|^{{\mf A}'}\feql 1\ \text{or} \\
&                \|\tilde{a}_\mbbm{i}\|^{{\mf A}'}=0\fle \|\tilde{a}_{\mbbm{i}_1}\|^{{\mf A}'}\ \text{or}\ \|\tilde{a}_\mbbm{i}\|^{{\mf A}'}=\|\tilde{a}_{\mbbm{i}_1}\|^{{\mf A}'}\fle 1, \\[1mm]
& \text{either}\ \|\tilde{a}_\mbbm{i}\|^{{\mf A}'}=\|\theta_1\|^{{\mf A}'}\frightarrow 0\ \text{or}\ 
                 \|\tilde{a}_\mbbm{i}\|^{{\mf A}'}=\|\theta_1\|^{{\mf A}'}\feql 0\ \text{or}\ \|\tilde{a}_\mbbm{i}\|^{{\mf A}'}=\|\theta_1\|^{{\mf A}'}\feql 1\ \text{or} \\
&                \|\tilde{a}_\mbbm{i}\|^{{\mf A}'}=0\fle \|\theta_1\|^{{\mf A}'}\ \text{or}\ \|\tilde{a}_\mbbm{i}\|^{{\mf A}'}=\|\theta_1\|^{{\mf A}'}\fle 1, \\[1mm]
& \text{either}\ \|\tilde{a}_\mbbm{i}\|^{{\mf A}'}\leq \|\theta_1\|^{{\mf A}'}\frightarrow 0,\ \|\theta_1\|^{{\mf A}'}\frightarrow 0\leq \|\tilde{a}_\mbbm{i}\|^{{\mf A}'}\ \text{or} \\
&                \|\tilde{a}_\mbbm{i}\|^{{\mf A}'}\leq \|\theta_1\|^{{\mf A}'}\feql 0,\ \|\theta_1\|^{{\mf A}'}\feql 0\leq \|\tilde{a}_\mbbm{i}\|^{{\mf A}'}\ \text{or} \\
&                \|\tilde{a}_\mbbm{i}\|^{{\mf A}'}\leq \|\theta_1\|^{{\mf A}'}\feql 1,\ \|\theta_1\|^{{\mf A}'}\feql 1\leq \|\tilde{a}_\mbbm{i}\|^{{\mf A}'}\ \text{or} \\
&                \|\tilde{a}_\mbbm{i}\|^{{\mf A}'}\leq 0\fle \|\theta_1\|^{{\mf A}'},\ 0\fle \|\theta_1\|^{{\mf A}'}\leq \|\tilde{a}_\mbbm{i}\|^{{\mf A}'}\ \text{or} \\
&                \|\tilde{a}_\mbbm{i}\|^{{\mf A}'}\leq \|\theta_1\|^{{\mf A}'}\fle 1,\ \|\theta_1\|^{{\mf A}'}\fle 1\leq \|\tilde{a}_\mbbm{i}\|^{{\mf A}'}, \\[1mm]
& \text{either}\ (\|\tilde{a}_\mbbm{i}\|^{{\mf A}'}\frightarrow (\|\theta_1\|^{{\mf A}'}\frightarrow 0))\fwedge 
                 ((\|\theta_1\|^{{\mf A}'}\frightarrow 0)\frightarrow \|\tilde{a}_\mbbm{i}\|^{{\mf A}'})=1\ \text{or} \\
&                (\|\tilde{a}_\mbbm{i}\|^{{\mf A}'}\frightarrow (\|\theta_1\|^{{\mf A}'}\feql 0))\fwedge 
                 ((\|\theta_1\|^{{\mf A}'}\feql 0)\frightarrow \|\tilde{a}_\mbbm{i}\|^{{\mf A}'})=1\ \text{or} \\
&                (\|\tilde{a}_\mbbm{i}\|^{{\mf A}'}\frightarrow (\|\theta_1\|^{{\mf A}'}\feql 1))\fwedge 
                 ((\|\theta_1\|^{{\mf A}'}\feql 1)\frightarrow \|\tilde{a}_\mbbm{i}\|^{{\mf A}'})=1\ \text{or} \\
&                (\|\tilde{a}_\mbbm{i}\|^{{\mf A}'}\frightarrow (0\fle \|\theta_1\|^{{\mf A}'}))\fwedge 
                 ((0\fle \|\theta_1\|^{{\mf A}'})\frightarrow \|\tilde{a}_\mbbm{i}\|^{{\mf A}'})=1\ \text{or} \\
&                (\|\tilde{a}_\mbbm{i}\|^{{\mf A}'}\frightarrow (\|\theta_1\|^{{\mf A}'}\fle 1))\fwedge 
                 ((\|\theta_1\|^{{\mf A}'}\fle 1)\frightarrow \|\tilde{a}_\mbbm{i}\|^{{\mf A}'})=1, \\[1mm]
& \text{either}\ \|\tilde{a}_\mbbm{i}\leftrightarrow (\theta_1\rightarrow \gz)\|^{{\mf A}'}=1\ \text{or}\ 
                 \|\tilde{a}_\mbbm{i}\leftrightarrow (\theta_1\geql \gz)\|^{{\mf A}'}=1\ \text{or} \\
&                \|\tilde{a}_\mbbm{i}\leftrightarrow (\theta_1\geql \gu)\|^{{\mf A}'}=1\ \text{or}\
                 \|\tilde{a}_\mbbm{i}\leftrightarrow (\gz\gle \theta_1)\|^{{\mf A}'}=1\ \text{or}\
                 \|\tilde{a}_\mbbm{i}\leftrightarrow (\theta_1\gle \gu)\|^{{\mf A}'}=1, \\[1mm] 
& \|\tilde{a}_\mbbm{i}\leftrightarrow \theta\|^{{\mf A}'}=1,
\end{alignat*}
${\mf A}'\models \tilde{a}_\mbbm{i}\leftrightarrow \theta$.
We put ${\mf A}={\mf A}'$.
Hence, ${\mf A}\models \tilde{a}_\mbbm{i}\leftrightarrow \theta$, ${\mf A}|_{\mbb{O}\cup \{\tilde{a}_\mbbm{i}\}}={\mf A}'|_{\mbb{O}\cup \{\tilde{a}_\mbbm{i}\}}$; 
(b) holds.                                                                                                                                                                                 \linebreak[4]
%
%
%
\end{proof}

\subsection{Full proofs of the points (a,c,d) and the statements (\ref{eq1c}) and (\ref{eq1h}) of Lemma \ref{le11}}
\label{S7.1c}

\begin{proof}
Case 1 (the base case):

$\theta\in \mi{PropAtom}$, $|\theta|=1$, $\mi{card}(J)=\mi{card}(\emptyset)=0=|\theta|-1$; 
(a) holds.

$|S|=|\{\tilde{a}_\mbbm{i}\geql \theta\}|=2+|\theta|<25\cdot |\theta|$;
$S=\{\tilde{a}_\mbbm{i}\geql \theta\}$ can be built up from $\theta$ via a trivial preorder traversal of $\theta$ with $\#{\mc O}(\theta)\in O(|S|)=O(|\theta|)$; 
(c) holds.

$\mi{atoms}(\theta)=\{\theta\}\subseteq \mbb{O}$, $\theta\in \mbb{O}$, $\mbb{O}\cap \tilde{\mbb{A}}=\emptyset$, $\theta\not\in \tilde{\mbb{A}}$;
for all $C\in S=\{\tilde{a}_\mbbm{i}\geql \theta\}$,
$C=\tilde{a}_\mbbm{i}\geql \theta$,
$\emptyset\neq \mi{atoms}(C)\cap \tilde{\mbb{A}}=\mi{atoms}(\tilde{a}_\mbbm{i}\geql \theta)\cap \tilde{\mbb{A}}=\{\tilde{a}_\mbbm{i},\theta\}\cap \tilde{\mbb{A}}=\{\tilde{a}_\mbbm{i}\}$;
$\tilde{a}_\mbbm{i}\in \tilde{\mbb{A}}$, $\gu\not\in \mi{PropAtom}$, $\gu\neq \tilde{a}_\mbbm{i}, \theta$, $\tilde{a}_\mbbm{i}\geql \gu, \tilde{a}_\mbbm{i}\gle \gu\neq \tilde{a}_\mbbm{i}\geql \theta$;
$\tilde{a}_\mbbm{i}\geql \gu, \tilde{a}_\mbbm{i}\gle \gu\not\in S=\{\tilde{a}_\mbbm{i}\geql \theta\}$; 
(d) holds.

Case 2.1 (the binary interpolation case):

Concerning Table \ref{tab2}, for every form of $\theta$, for all $C\in \mi{ClPrefix}$, $\tilde{a}_\mbbm{i}\in \mi{atoms}(C)$;
for both $i$, 
by the induction hypothesis (d) for $\theta_i$, $S_i$, $\tilde{a}_{\mbbm{i}_i}$, and $J_i$, 
for all $C\in S_i$, $\emptyset\neq \mi{atoms}(C)\cap \tilde{\mbb{A}}\subseteq \{\tilde{a}_{\mbbm{i}_i}\}\cup \{\tilde{a}_\mbbm{j} \,|\, \mbbm{j}\in J_i\}$;
for all $C_0\in \mi{ClPrefix}$, $C_1\in S_1$, and $C_2\in S_2$,
$\tilde{a}_\mbbm{i}\in \mi{atoms}(C_0)$;
for both $i$, 
$\{\tilde{a}_\mbbm{i}\}\cap (\{\tilde{a}_{\mbbm{i}_i}\}\cup \{\tilde{a}_\mbbm{j} \,|\, \mbbm{j}\in J_i\})\overset{\text{(\ref{eq1b})}}{=\!\!=} \emptyset$,
$\tilde{a}_\mbbm{i}\not\in \mi{atoms}(C_i)\cap \tilde{\mbb{A}}\subseteq \{\tilde{a}_{\mbbm{i}_i}\}\cup \{\tilde{a}_\mbbm{j} \,|\, \mbbm{j}\in J_i\}$,
$\tilde{a}_\mbbm{i}\in \tilde{\mbb{A}}$, $\tilde{a}_\mbbm{i}\not\in \mi{atoms}(C_i)$, $\mi{atoms}(C_0)\neq \mi{atoms}(C_i)$, $C_0\neq C_i$;
$(\mi{atoms}(C_1)\cap \tilde{\mbb{A}})\cap (\mi{atoms}(C_2)\cap \tilde{\mbb{A}})=
 (\{\tilde{a}_{\mbbm{i}_1}\}\cup \{\tilde{a}_\mbbm{j} \,|\, \mbbm{j}\in J_1\})\cap (\{\tilde{a}_{\mbbm{i}_2}\}\cup \{\tilde{a}_\mbbm{j} \,|\, \mbbm{j}\in J_2\})\overset{\text{(\ref{eq1b})}}{=\!\!=} 
 \emptyset$;
for both $i$, $\mi{atoms}(C_i)\cap \tilde{\mbb{A}}\neq \emptyset$;
$\mi{atoms}(C_1)\cap \tilde{\mbb{A}}\neq \mi{atoms}(C_2)\cap \tilde{\mbb{A}}$, $\mi{atoms}(C_1)\neq \mi{atoms}(C_2)$, $C_1\neq C_2$;
\begin{equation}
\notag
\mi{ClPrefix}, S_1, S_2\ \text{are pairwise disjoint};
\end{equation}
(\ref{eq1c}) holds.

$|\theta|=|\theta_1\diamond \theta_2|=1+|\theta_1|+|\theta_2|$;
for both $i$, by the induction hypothesis (a) for $\theta_i$ and $J_i$, $\mi{card}(J_i)\leq |\theta_i|-1$;
$\mi{card}(J)\overset{\text{(\ref{eq1a})}}{=\!\!=} \mi{card}(\{\mbbm{i}_1\}\cup J_1\cup \{\mbbm{i}_2\}\cup J_2)\overset{\text{(\ref{eq1b})}}{=\!\!=} 
                                                   \mi{card}(\{\mbbm{i}_1\})+\mi{card}(J_1)+\mi{card}(\{\mbbm{i}_2\})+\mi{card}(J_2)\leq 2+|\theta_1|-1+|\theta_2|-1=|\theta|-1$; 
(a) holds.

Concerning Table \ref{tab2}, for every form of $\theta$, $1\leq |\mi{ClPrefix}|\leq 25$;
$\mi{ClPrefix}$ can be built up from $\emptyset$ with $\#{\mc O}(\emptyset)\in O(|\mi{ClPrefix}|)=O(1)$;
for both $i$, 
by the induction hypothesis (c) for $\theta_i$ and $S_i$, 
$|S_i|\leq 25\cdot |\theta_i|$;
$S_i$ can be built up from $\theta_i$ via a preorder traversal of $\theta_i$ with $\#{\mc O}(\theta_i)\in O(|\theta_i|)$;
$|S|=|\mi{ClPrefix}\cup S_1\cup S_2|\overset{\text{(\ref{eq1c})}}{=\!\!=} |\mi{ClPrefix}|+|S_1|+|S_2|\leq 25+25\cdot |\theta_1|+25\cdot |\theta_2|=25\cdot (1+|\theta_1|+|\theta_2|)=25\cdot |\theta|$;
$S=\mi{ClPrefix}\cup S_1\cup S_2$ can be built up from $\theta$ via a preorder traversal of $\theta$ with $\#{\mc O}(\theta)\in O(|\theta_1|+|\theta_2|)=O(|\theta|)$; 
(c) holds.

Concerning Table \ref{tab2}, for every form of $\theta$,
for all $C\in \mi{ClPrefix}$, 
$\emptyset\neq \mi{atoms}(C)\cap \tilde{\mbb{A}}=\{\tilde{a}_\mbbm{i},\tilde{a}_{\mbbm{i}_1},\tilde{a}_{\mbbm{i}_2}\}\subseteq
               \{\tilde{a}_\mbbm{i}\}\cup \{\tilde{a}_\mbbm{j} \,|\, \mbbm{j}\in J\overset{\text{(\ref{eq1a})}}{=\!\!=} \{\mbbm{i}_1\}\cup J_1\cup \{\mbbm{i}_2\}\cup J_2\}$;
$\tilde{a}_\mbbm{i}\geql \gu, \tilde{a}_\mbbm{i}\gle \gu\not\in \mi{ClPrefix}$;
for both $i$, 
by the induction hypothesis (d) for $\theta_i$, $S_i$, $\tilde{a}_{\mbbm{i}_i}$, and $J_i$,
for all $C\in S_i$,
$\emptyset\neq \mi{atoms}(C)\cap \tilde{\mbb{A}}\subseteq \{\tilde{a}_{\mbbm{i}_i}\}\cup \{\tilde{a}_\mbbm{j} \,|\, \mbbm{j}\in J_i\}\subseteq
               \{\tilde{a}_\mbbm{i}\}\cup \{\tilde{a}_\mbbm{j} \,|\, \mbbm{j}\in J\overset{\text{(\ref{eq1a})}}{=\!\!=} \{\mbbm{i}_1\}\cup J_1\cup \{\mbbm{i}_2\}\cup J_2\}$,
$\{\tilde{a}_\mbbm{i}\}\cap (\{\tilde{a}_{\mbbm{i}_i}\}\cup \{\tilde{a}_\mbbm{j} \,|\, \mbbm{j}\in J_i\})\overset{\text{(\ref{eq1b})}}{=\!\!=} \emptyset$,
$\tilde{a}_\mbbm{i}\not\in \mi{atoms}(C)\cap \tilde{\mbb{A}}\subseteq \{\tilde{a}_{\mbbm{i}_i}\}\cup \{\tilde{a}_\mbbm{j} \,|\, \mbbm{j}\in J_i\}$,
$\tilde{a}_\mbbm{i}\in \tilde{\mbb{A}}$, $\tilde{a}_\mbbm{i}\not\in \mi{atoms}(C)$;
for both $\diamond^\#\in \{\geql,\gle\}$, 
$\tilde{a}_\mbbm{i}\in \mi{atoms}(\tilde{a}_\mbbm{i}\diamond^\# \gu)$, $\mi{atoms}(\tilde{a}_\mbbm{i}\diamond^\# \gu)\neq \mi{atoms}(C)$, $\tilde{a}_\mbbm{i}\diamond^\# \gu\neq C$;   
$\tilde{a}_\mbbm{i}\geql \gu, \tilde{a}_\mbbm{i}\gle \gu\not\in S_i$;
for all $C\in S=\mi{ClPrefix}\cup S_1\cup S_2$, $\emptyset\neq \mi{atoms}(C)\cap \tilde{\mbb{A}}\subseteq \{\tilde{a}_\mbbm{i}\}\cup \{\tilde{a}_\mbbm{j} \,|\, \mbbm{j}\in J\}$;
$\tilde{a}_\mbbm{i}\geql \gu, \tilde{a}_\mbbm{i}\gle \gu\not\in S=\mi{ClPrefix}\cup S_1\cup S_2$; 
(d) holds.

Case 2.2 (the unary interpolation case):

Concerning Table \ref{tab3}, for every form of $\theta$, for all $C\in \mi{ClPrefix}$, $\tilde{a}_\mbbm{i}\in \mi{atoms}(C)$;
by the induction hypothesis (d) for $\theta_1$, $S_1$, $\tilde{a}_{\mbbm{i}_1}$, and $J_1$, 
for all $C\in S_1$, $\mi{atoms}(C)\cap \tilde{\mbb{A}}\subseteq \{\tilde{a}_{\mbbm{i}_1}\}\cup \{\tilde{a}_\mbbm{j} \,|\, \mbbm{j}\in J_1\}$;
for all $C_0\in \mi{ClPrefix}$ and $C_1\in S_1$,
$\tilde{a}_\mbbm{i}\in \mi{atoms}(C_0)$, 
$\{\tilde{a}_\mbbm{i}\}\cap (\{\tilde{a}_{\mbbm{i}_1}\}\cup \{\tilde{a}_\mbbm{j} \,|\, \mbbm{j}\in J_1\})\overset{\text{(\ref{eq1g})}}{=\!\!=} \emptyset$,
$\tilde{a}_\mbbm{i}\not\in \mi{atoms}(C_1)\cap \tilde{\mbb{A}}\subseteq \{\tilde{a}_{\mbbm{i}_1}\}\cup \{\tilde{a}_\mbbm{j} \,|\, \mbbm{j}\in J_1\}$,
$\tilde{a}_\mbbm{i}\in \tilde{\mbb{A}}$, $\tilde{a}_\mbbm{i}\not\in \mi{atoms}(C_1)$, $\mi{atoms}(C_0)\neq \mi{atoms}(C_1)$, $C_0\neq C_1$;
\begin{equation}
\notag 
\mi{ClPrefix}\cap S_1=\emptyset;
\end{equation}
(\ref{eq1h}) holds.

$|\theta|=|\theta_1\rightarrow \gz|=|\theta_1\geql \gz|=|\theta_1\geql \gu|=|\gz\gle \theta_1|=|\theta_1\gle \gu|=2+|\theta_1|$;
by the induction hypothesis (a) for $\theta_1$ and $J_1$, $\mi{card}(J_1)\leq |\theta_1|-1$;
$\mi{card}(J)\overset{\text{(\ref{eq1f})}}{=\!\!=} \mi{card}(\{\mbbm{i}_1\}\cup J_1)\overset{\text{(\ref{eq1g})}}{=\!\!=} \mi{card}(\{\mbbm{i}_1\})+\mi{card}(J_1)\leq 1+|\theta_1|-1=|\theta|-2<
 |\theta|-1$; 
(a) holds.

Concerning Table \ref{tab3}, for every form of $\theta$, $1\leq |\mi{ClPrefix}|\leq 25$;
$\mi{ClPrefix}$ can be built up from $\emptyset$ with $\#{\mc O}(\emptyset)\in O(|\mi{ClPrefix}|)=O(1)$;
by the induction hypothesis (c) for $\theta_1$ and $S_1$,
$|S_1|\leq 25\cdot |\theta_1|$;
$S_1$ can be built up from $\theta_1$ via a preorder traversal of $\theta_1$ with $\#{\mc O}(\theta_1)\in O(|\theta_1|)$;
$|S|=|\mi{ClPrefix}\cup S_1|\overset{\text{(\ref{eq1h})}}{=\!\!=} |\mi{ClPrefix}|+|S_1|\leq 25+25\cdot |\theta_1|<25\cdot (2+|\theta_1|)=25\cdot |\theta|$;
$S=\mi{ClPrefix}\cup S_1$ can be built up from $\theta$ via a preorder traversal of $\theta$ with $\#{\mc O}(\theta)\in O(|\theta_1|)=O(|\theta|)$; 
(c) holds.

Concerning Table \ref{tab3}, for every form of $\theta$,
for all $C\in \mi{ClPrefix}$, 
$\emptyset\neq \mi{atoms}(C)\cap \tilde{\mbb{A}}=\{\tilde{a}_\mbbm{i},\tilde{a}_{\mbbm{i}_1}\}\subseteq
               \{\tilde{a}_\mbbm{i}\}\cup \{\tilde{a}_\mbbm{j} \,|\, \mbbm{j}\in J\overset{\text{(\ref{eq1f})}}{=\!\!=} \{\mbbm{i}_1\}\cup J_1\}$;
$\tilde{a}_\mbbm{i}\geql \gu, \tilde{a}_\mbbm{i}\gle \gu\not\in \mi{ClPrefix}$;
by the induction hypothesis (d) for $\theta_1$, $S_1$, $\tilde{a}_{\mbbm{i}_1}$, and $J_1$,
for all $C\in S_1$,
$\emptyset\neq \mi{atoms}(C)\cap \tilde{\mbb{A}}\subseteq \{\tilde{a}_{\mbbm{i}_1}\}\cup \{\tilde{a}_\mbbm{j} \,|\, \mbbm{j}\in J_1\}\subseteq
               \{\tilde{a}_\mbbm{i}\}\cup \{\tilde{a}_\mbbm{j} \,|\, \mbbm{j}\in J\overset{\text{(\ref{eq1f})}}{=\!\!=} \{\mbbm{i}_1\}\cup J_1\}$,
$\{\tilde{a}_\mbbm{i}\}\cap (\{\tilde{a}_{\mbbm{i}_1}\}\cup \{\tilde{a}_\mbbm{j} \,|\, \mbbm{j}\in J_1\})\overset{\text{(\ref{eq1g})}}{=\!\!=} \emptyset$,
$\tilde{a}_\mbbm{i}\not\in \mi{atoms}(C)\cap \tilde{\mbb{A}}\subseteq \{\tilde{a}_{\mbbm{i}_1}\}\cup \{\tilde{a}_\mbbm{j} \,|\, \mbbm{j}\in J_1\}$,
$\tilde{a}_\mbbm{i}\in \tilde{\mbb{A}}$, $\tilde{a}_\mbbm{i}\not\in \mi{atoms}(C)$;
for both $\diamond^\#\in \{\geql,\gle\}$, 
$\tilde{a}_\mbbm{i}\in \mi{atoms}(\tilde{a}_\mbbm{i}\diamond^\# \gu)$, $\mi{atoms}(\tilde{a}_\mbbm{i}\diamond^\# \gu)\neq \mi{atoms}(C)$, $\tilde{a}_\mbbm{i}\diamond^\# \gu\neq C$;
$\tilde{a}_\mbbm{i}\geql \gu, \tilde{a}_\mbbm{i}\gle \gu\not\in S_1$;
for all $C\in S=\mi{ClPrefix}\cup S_1$, $\emptyset\neq \mi{atoms}(C)\cap \tilde{\mbb{A}}\subseteq \{\tilde{a}_\mbbm{i}\}\cup \{\tilde{a}_\mbbm{j} \,|\, \mbbm{j}\in J\}$;
$\tilde{a}_\mbbm{i}\geql \gu, \tilde{a}_\mbbm{i}\gle \gu\not\in S=\mi{ClPrefix}\cup S_1$; 
(d) holds.
%
%
%
\end{proof}

\subsection{Full proof of Lemma \ref{le1}}
\label{S7.2a}

\begin{proof}
By Lemma \ref{le111} for $n_\phi$ and $\phi$, there exists $\phi'\in \mi{PropForm}_\emptyset$ such that (a--d) of Lemma \ref{le111} hold for $\phi'$, $\phi$, and $n_\phi$.
We distinguish three cases for $\phi'$.

Case 1:
$\phi'=\gz$.
We put $J_\phi=\emptyset\subseteq_{\mc F} \{(n_\phi,j) \,|\, j\in \mbb{N}\}\subseteq \mbb{I}$ and $S_\phi=\{\square\}\subseteq_{\mc F} \mi{OrdPropCl}_\emptyset$.

(a,b,d,e) can be proved straightforwardly.

Then, for every valuation ${\mf A}^\#$,
${\mf A}^\#\not\models \phi'=\gz$,
${\mf A}^\#\not\models \phi'\overset{\text{Lemma \ref{le111}(a)}}{\eqvl\!\!\eqvl\!\!\eqvl\!\!\eqvl\!\!\eqvl\!\!\eqvl\!\!\eqvl\!\!\eqvl\!\!\eqvl\!\!\eqvl\!\!\eqvl\!\!\eqvl\!\!\eqvl} \phi$,
${\mf A}^\#\not\models S_\phi=\{\square\}$;
trivially, 
there exists a valuation ${\mf A}$ satisfying ${\mf A}\models \phi$ if and only if 
there exists a valuation ${\mf A}'$ satisfying ${\mf A}'\models S_\phi$; 
${\mf A}|_\mbb{O}={\mf A}'|_\mbb{O}$; 
(c) holds.

Case 2:
$\phi'=\gu$.
We put $J_\phi=\emptyset\subseteq_{\mc F} \{(n_\phi,j) \,|\, j\in \mbb{N}\}\subseteq \mbb{I}$ and $S_\phi=\emptyset\subseteq_{\mc F} \mi{OrdPropCl}_\emptyset$.

(a,b,d,e) can be proved straightforwardly.

Then, for every valuation ${\mf A}^\#$,
${\mf A}^\#\models \phi'=\gu$,
${\mf A}^\#\models \phi'\overset{\text{Lemma \ref{le111}(a)}}{\eqvl\!\!\eqvl\!\!\eqvl\!\!\eqvl\!\!\eqvl\!\!\eqvl\!\!\eqvl\!\!\eqvl\!\!\eqvl\!\!\eqvl\!\!\eqvl\!\!\eqvl\!\!\eqvl} \phi$,
${\mf A}^\#\models S_\phi=\emptyset$;
there exists a valuation ${\mf A}$ satisfying ${\mf A}\models \phi$ if and only if 
there exists a valuation ${\mf A}'$ satisfying ${\mf A}'\models S_\phi$; 
${\mf A}={\mf A}'$, ${\mf A}|_\mbb{O}={\mf A}'|_\mbb{O}$; 
(c) holds.

Case 3: 
$\phi'\neq \gz, \gu$. 
We have that (a--d) of Lemma \ref{le111} hold for $\phi'$, $\phi$, and $n_\phi$.
Then $\phi'\in \mi{PropForm}_\emptyset-\{\gz,\gu\}$ and (c,d) of Lemma \ref{le111} hold for $\phi'$.
We put $j_\mbbm{i}=0$ and $\mbbm{i}=(n_\phi,j_\mbbm{i})\in \{(n_\phi,j) \,|\, j\in \mbb{N}\}\subseteq \mbb{I}$.
Hence, $\tilde{a}_\mbbm{i}\in \tilde{\mbb{A}}$;
by Lemma \ref{le11} for $n_\phi$ and $\phi'$, there exist $n_J\geq j_\mbbm{i}$, $J=\{(n_\phi,j) \,|\, j_\mbbm{i}+1\leq j\leq n_J\}\subseteq \{(n_\phi,j) \,|\, j\in \mbb{N}\}\subseteq \mbb{I}$, $\mbbm{i}\not\in J$, 
$S\subseteq_{\mc F} \mi{OrdPropCl}_{\{\tilde{a}_\mbbm{i}\}\cup \{\tilde{a}_\mbbm{j} \,|\, \mbbm{j}\in J\}}$ satisfying that
(a--d) of Lemma \ref{le11} hold for $\phi'$.
We put $n_{J_\phi}=n_J$ and $J_\phi=\{(n_\phi,j) \,|\, j\leq n_{J_\phi}\}\subseteq_{\mc F} \{(n_\phi,j) \,|\, j\in \mbb{N}\}\subseteq \mbb{I}$.
Then 
\begin{alignat}{1}
\label{eq1k}
& J_\phi=\{(n_\phi,j) \,|\, j\leq n_{J_\phi}\}=\{(n_\phi,0)\}\cup \{(n_\phi,j) \,|\, 1\leq j\leq n_{J_\phi}\}= \\
\notag
& \phantom{J_\phi=\mbox{}}
         \{(n_\phi,j_\mbbm{i})\}\cup \{(n_\phi,j) \,|\, j_\mbbm{i}+1\leq j\leq n_J\}=\{\mbbm{i}\}\cup J, \\[1mm]
\label{eq1l}
& \{\mbbm{i}\}\cap J=\{(n_\phi,j_\mbbm{i})\}\cap \{(n_\phi,j) \,|\, j_\mbbm{i}+1\leq j\leq n_J\}=\emptyset.
\end{alignat}
We put $S_\phi=\{\tilde{a}_\mbbm{i}\geql \gu\}\cup S\subseteq_{\mc F} \mi{OrdPropCl}_{\{\tilde{a}_\mbbm{j} \,|\, \mbbm{j}\in J_\phi\}}$.
Note that $\tilde{a}_\mbbm{i}\in \tilde{\mbb{A}}$, $\mbb{O}\cap \tilde{\mbb{A}}=\emptyset$, $\tilde{a}_\mbbm{i}\not\in \mbb{O}$,
$\mi{atoms}(\tilde{a}_\mbbm{i}\geql \gu)=\{\tilde{a}_\mbbm{i}\}\subseteq \mbb{O}\cup \{\tilde{a}_\mbbm{i}\}$, 
$\tilde{a}_\mbbm{i}\geql \gu\in \mi{OrdPropLit}_{\{\tilde{a}_\mbbm{i}\}}$,
$\mi{atoms}(\phi')\subseteq \mbb{O}$,
$\mi{atoms}(\tilde{a}_\mbbm{i}\leftrightarrow \phi')=\mi{atoms}(\tilde{a}_\mbbm{i})\cup \mi{atoms}(\phi')\subseteq \mbb{O}\cup \{\tilde{a}_\mbbm{i}\}$,
$\tilde{a}_\mbbm{i}\leftrightarrow \phi'\in \mi{PropForm}_{\{\tilde{a}_\mbbm{i}\}}$;
$\mi{atoms}(S)\subseteq \mbb{O}\cup \{\tilde{a}_\mbbm{i}\}\cup \{\tilde{a}_\mbbm{j} \,|\, \mbbm{j}\in J\}$,
$\mi{atoms}(S_\phi)=\mi{atoms}(\{\tilde{a}_\mbbm{i}\geql \gu\}\cup S)=\mi{atoms}(\tilde{a}_\mbbm{i}\geql \gu)\cup \mi{atoms}(S)\subseteq
                    \mbb{O}\cup \{\tilde{a}_\mbbm{i}\}\cup \{\tilde{a}_\mbbm{j} \,|\, \mbbm{j}\in J\}=\mbb{O}\cup \{\tilde{a}_\mbbm{j} \,|\, \mbbm{j}\in \{\mbbm{i}\}\cup J\}\overset{\text{(\ref{eq1k})}}{=\!\!=}
                    \mbb{O}\cup \{\tilde{a}_\mbbm{j} \,|\, \mbbm{j}\in J_\phi\}$,
$S_\phi=\{\tilde{a}_\mbbm{i}\geql \gu\}\cup S\subseteq_{\mc F} \mi{OrdPropCl}_{\{\tilde{a}_\mbbm{j} \,|\, \mbbm{j}\in J_\phi\}}$.
\begin{equation}
\label{eq1m}
\{\tilde{a}_\mbbm{i}\geql \gu\}\cap S\overset{\text{Lemma \ref{le11}(d)}}{=\!\!=\!\!=\!\!=\!\!=\!\!=\!\!=\!\!=\!\!=\!\!=\!\!=\!\!=\!\!=} \emptyset.
\end{equation}

(a,b,e) can be proved straightforwardly.

Let ${\mf A}$ be a valuation such that ${\mf A}\models \phi$.
Then $\tilde{a}_\mbbm{i}\not\in \mbb{O}$.
We define a valuation
\begin{equation}
\notag
{\mf A}^\#={\mf A}|_\mbb{O}\cup \{(\tilde{a}_\mbbm{i},\|\phi'\|^{\mf A})\}\cup {\mf A}|_{\mi{PropAtom}-(\mbb{O}\cup \{\tilde{a}_\mbbm{i}\})}.
\end{equation}
Hence, ${\mf A}\models \phi\overset{\text{Lemma \ref{le111}(a)}}{\eqvl\!\!\eqvl\!\!\eqvl\!\!\eqvl\!\!\eqvl\!\!\eqvl\!\!\eqvl\!\!\eqvl\!\!\eqvl\!\!\eqvl\!\!\eqvl\!\!\eqvl\!\!\eqvl} \phi'$,
$\mi{atoms}(\phi')\subseteq \mbb{O}$,
${\mf A}^\#|_\mbb{O}={\mf A}|_\mbb{O}$,
$\|\tilde{a}_\mbbm{i}\|^{{\mf A}^\#}=\|\phi'\|^{\mf A}=\|\phi'\|^{{\mf A}^\#}=1$,
\begin{alignat*}{1}
\|\tilde{a}_\mbbm{i}\geql \gu\|^{{\mf A}^\#}
&= \|\tilde{a}_\mbbm{i}\|^{{\mf A}^\#}\feql \|\gu\|^{{\mf A}^\#}=1\feql 1=1, \\[1mm]
\|\tilde{a}_\mbbm{i}\leftrightarrow \phi'\|^{{\mf A}^\#}
&= (\|\tilde{a}_\mbbm{i}\|^{{\mf A}^\#}\frightarrow \|\phi'\|^{{\mf A}^\#})\fwedge (\|\phi'\|^{{\mf A}^\#}\frightarrow \|\tilde{a}_\mbbm{i}\|^{{\mf A}^\#})= \\
&\phantom{\mbox{}=\mbox{}}
   (\|\phi'\|^{{\mf A}^\#}\frightarrow \|\phi'\|^{{\mf A}^\#})\fwedge (\|\phi'\|^{{\mf A}^\#}\frightarrow \|\phi'\|^{{\mf A}^\#})=1,
\end{alignat*}
${\mf A}^\#\models \tilde{a}_\mbbm{i}\geql \gu$,
${\mf A}^\#\models \tilde{a}_\mbbm{i}\leftrightarrow \phi'$;
by Lemma \ref{le11}(b) for ${\mf A}^\#$, there exists a valuation ${\mf A}'$ satisfying ${\mf A}'\models S$, 
${\mf A}'|_{\mbb{O}\cup \{\tilde{a}_\mbbm{i}\}}={\mf A}^\#|_{\mbb{O}\cup \{\tilde{a}_\mbbm{i}\}}$;
$\mi{atoms}(\tilde{a}_\mbbm{i}\geql \gu)=\{\tilde{a}_\mbbm{i}\}$,
${\mf A}'\models \tilde{a}_\mbbm{i}\geql \gu$; 
${\mf A}'\models S_\phi=\{\tilde{a}_\mbbm{i}\geql \gu\}\cup S$, ${\mf A}'|_\mbb{O}={\mf A}^\#|_\mbb{O}={\mf A}|_\mbb{O}$.

Let ${\mf A}'$ be a valuation such that ${\mf A}'\models S_\phi$.
Then $\{\tilde{a}_\mbbm{i}\geql \gu\}, S\subseteq S_\phi$, ${\mf A}'\models \tilde{a}_\mbbm{i}\geql \gu$, ${\mf A}'\models S$;
by Lemma \ref{le11}(b), there exists a valuation ${\mf A}^\#$ satisfying ${\mf A}^\#\models \tilde{a}_\mbbm{i}\leftrightarrow \phi'$,
${\mf A}^\#|_{\mbb{O}\cup \{\tilde{a}_\mbbm{i}\}}={\mf A}'|_{\mbb{O}\cup \{\tilde{a}_\mbbm{i}\}}$;
$\mi{atoms}(\tilde{a}_\mbbm{i}\leftrightarrow \phi')\subseteq \mbb{O}\cup \{\tilde{a}_\mbbm{i}\}$,
${\mf A}'\models \tilde{a}_\mbbm{i}\leftrightarrow \phi'$,
\begin{equation}
\notag
\|\tilde{a}_\mbbm{i}\geql \gu\|^{{\mf A}'}=\|\tilde{a}_\mbbm{i}\|^{{\mf A}'}\feql \|\gu\|^{{\mf A}'}=\|\tilde{a}_\mbbm{i}\|^{{\mf A}'}\feql 1=1,
\end{equation}
$\|\tilde{a}_\mbbm{i}\|^{{\mf A}'}=1$,
\begin{alignat*}{1}
\|\tilde{a}_\mbbm{i}\leftrightarrow \phi'\|^{{\mf A}'}
&= (\|\tilde{a}_\mbbm{i}\|^{{\mf A}'}\frightarrow \|\phi'\|^{{\mf A}'})\fwedge (\|\phi'\|^{{\mf A}'}\frightarrow \|\tilde{a}_\mbbm{i}\|^{{\mf A}'})= \\
&\phantom{\mbox{}=\mbox{}}
   (1\frightarrow \|\phi'\|^{{\mf A}^\#})\fwedge (\|\phi'\|^{{\mf A}^\#}\frightarrow 1)=1,
\end{alignat*}
$1\frightarrow \|\phi'\|^{{\mf A}'}=1$, $\|\phi'\|^{{\mf A}'}=1$,
${\mf A}'\models \phi'\overset{\text{Lemma \ref{le111}(a)}}{\eqvl\!\!\eqvl\!\!\eqvl\!\!\eqvl\!\!\eqvl\!\!\eqvl\!\!\eqvl\!\!\eqvl\!\!\eqvl\!\!\eqvl\!\!\eqvl\!\!\eqvl\!\!\eqvl} \phi$.
We put ${\mf A}={\mf A}'$.
Hence, ${\mf A}\models \phi$, ${\mf A}|_\mbb{O}={\mf A}'|_\mbb{O}$; 
(c) holds.

$|\{\tilde{a}_\mbbm{i}\geql \gu\}|\in O(1)$, $|\phi'|\underset{\text{Lemma \ref{le111}(b)}}{\in} O(|\phi|)$,
$|S|\leq |S_\phi|=|\{\tilde{a}_\mbbm{i}\geql \gu\}\cup S|\overset{\text{(\ref{eq1m})}}{=\!\!=} |\{\tilde{a}_\mbbm{i}\geql \gu\}|+|S|=3+|S|\underset{\text{Lemma \ref{le11}(c)}}{\leq} 
 3+25\cdot |\phi'|\in O(|\phi'|)\underset{\text{Lemma \ref{le111}(b)}}{\subseteq} O(|\phi|)$;
the translation of $\phi$ to $S_\phi$ uses the input $\phi$, the output $S_\phi$, auxiliary $\phi'$, $\{\tilde{a}_\mbbm{i}\geql \gu\}$, $S$;
by Lemma \ref{le111}(b), $\phi'$ can be built up from $\phi$ via a postorder traversal of $\phi$ with $\#{\mc O}(\phi)\in O(|\phi|)$;
the test $\phi'\neq \gz, \gu$ is with $\#{\mc O}(\phi')\in O(1)$;
$\{\tilde{a}_\mbbm{i}\geql \gu\}$ can be built up from $\emptyset$ with $\#{\mc O}(\emptyset)\in O(|\{\tilde{a}_\mbbm{i}\geql \gu\}|)=O(1)$;
by Lemma \ref{le11}(c), $S$ can be built up from $\phi'$ via a preorder traversal of $\phi'$ with $\#{\mc O}(\phi')\in O(|\phi'|)\underset{\text{Lemma \ref{le111}(b)}}{\subseteq} O(|\phi|)$;
$S_\phi$ can be built up from $\{\tilde{a}_\mbbm{i}\geql \gu\}$ and $S$ by copying and concatenating with $\#{\mc O}(\{\tilde{a}_\mbbm{i}\geql \gu\},S)\in O(|S_\phi|)\subseteq O(|\phi|)$;
the number of all elementary operations of the translation of $\phi$ to $S_\phi$ $\#{\mc O}(\phi)\in O(|\phi|)$;
by (\ref{eq00t}) for $n_\phi$, $\phi$, $\emptyset$, $\phi'$, $\{\tilde{a}_\mbbm{i}\geql \gu\}$, $S$, $S_\phi$, $q=5$, and $r=1$,
the time complexity of the translation of $\phi$ to $S_\phi$ is in $O(\#{\mc O}(\phi)\cdot (\log (1+n_\phi)+\log (\#{\mc O}(\phi)+|\phi|)))\subseteq O(|\phi|\cdot (\log (1+n_\phi)+\log |\phi|))$;
by (\ref{eq00s}) for $n_\phi$, $\phi$, $\emptyset$, $\phi'$, $\{\tilde{a}_\mbbm{i}\geql \gu\}$, $S$, $S_\phi$, $q=5$, and $r=1$,
the space complexity of the translation of $\phi$ to $S_\phi$ is in $O((\#{\mc O}(\phi)+|\phi|)\cdot (\log (1+n_\phi)+\log |\phi|))\subseteq O(|\phi|\cdot (\log (1+n_\phi)+\log |\phi|))$; 
(d) holds.

So, in all Cases 1--3, (a--e) hold.
%
%
%
\end{proof}

\subsection{Full proofs of the points (a,b,d,e) of Lemma \ref{le1}}
\label{S7.2b}

\begin{proof}
Case 1:

$\mi{card}(J_\phi)=\mi{card}(\emptyset)=0\leq 2\cdot |\phi|$; 
(a) holds.

$J_\phi=\emptyset$, $S_\phi=\{\square\}$; 
(b) holds.

$|\phi'|\underset{\text{Lemma \ref{le111}(b)}}{\in} O(|\phi|)$, $|S_\phi|=|\{\square\}|=0\in O(|\phi|)$;
the translation of $\phi$ to $S_\phi$ uses the input $\phi$, the output $S_\phi$, an auxiliary $\phi'$;
by Lemma \ref{le111}(b), $\phi'$ can be built up from $\phi$ via a postorder traversal of $\phi$ with $\#{\mc O}(\phi)\in O(|\phi|)$;
the test $\phi'=\gz$ is with $\#{\mc O}(\phi')\in O(1)$;
$S_\phi$ can be built up from $\emptyset$ with $\#{\mc O}(\emptyset)\in O(1+|S_\phi|)=O(1)$;
the number of all elementary operations of the translation of $\phi$ to $S_\phi$ $\#{\mc O}(\phi)\in O(|\phi|)$;
by (\ref{eq00t}) for $n_\phi$, $\phi$, $\emptyset$, $\phi'$, $S_\phi$, $q=3$, and $r=1$,
the time complexity of the translation of $\phi$ to $S_\phi$ is in $O(\#{\mc O}(\phi)\cdot (\log (1+n_\phi)+\log (\#{\mc O}(\phi)+|\phi|)))\subseteq O(|\phi|\cdot (\log (1+n_\phi)+\log |\phi|))$;
by (\ref{eq00s}) for $n_\phi$, $\phi$, $\emptyset$, $\phi'$, $S_\phi$, $q=3$, and $r=1$,
the space complexity of the translation of $\phi$ to $S_\phi$ is in $O((\#{\mc O}(\phi)+|\phi|)\cdot (\log (1+n_\phi)+\log |\phi|))\subseteq O(|\phi|\cdot (\log (1+n_\phi)+\log |\phi|))$; 
(d) holds.

$S_\phi=\{\square\}$; 
(e) holds trivially.

Case 2:

$\mi{card}(J_\phi)=\mi{card}(\emptyset)=0\leq 2\cdot |\phi|$; 
(a) holds.

$J_\phi=S_\phi=\emptyset$; 
(b) holds.

$|\phi'|\underset{\text{Lemma \ref{le111}(b)}}{\in} O(|\phi|)$, $|S_\phi|=|\emptyset|=0\in O(|\phi|)$;
the translation of $\phi$ to $S_\phi$ uses the input $\phi$, the output $S_\phi$, an auxiliary $\phi'$;
by Lemma \ref{le111}(b), $\phi'$ can be built up from $\phi$ via a postorder traversal of $\phi$ with $\#{\mc O}(\phi)\in O(|\phi|)$;
the test $\phi'=\gu$ is with $\#{\mc O}(\phi')\in O(1)$;
$S_\phi$ can be built up from $\emptyset$ with $\#{\mc O}(\emptyset)\in O(1+|S_\phi|)=O(1)$;
the number of all elementary operations of the translation of $\phi$ to $S_\phi$ $\#{\mc O}(\phi)\in O(|\phi|)$;
by (\ref{eq00t}) for $n_\phi$, $\phi$, $\emptyset$, $\phi'$, $S_\phi$, $q=3$, and $r=1$,
the time complexity of the translation of $\phi$ to $S_\phi$ is in $O(\#{\mc O}(\phi)\cdot (\log (1+n_\phi)+\log (\#{\mc O}(\phi)+|\phi|)))\subseteq O(|\phi|\cdot (\log (1+n_\phi)+\log |\phi|))$;
by (\ref{eq00s}) for $n_\phi$, $\phi$, $\emptyset$, $\phi'$, $S_\phi$, $q=3$, and $r=1$,
the space complexity of the translation of $\phi$ to $S_\phi$ is in $O((\#{\mc O}(\phi)+|\phi|)\cdot (\log (1+n_\phi)+\log |\phi|))\subseteq O(|\phi|\cdot (\log (1+n_\phi)+\log |\phi|))$; 
(d) holds.

$S_\phi=\emptyset$; 
(e) holds trivially.

Case 3:

$\mi{card}(J_\phi)\overset{\text{(\ref{eq1k})}}{=\!\!=} \mi{card}(\{\mbbm{i}\}\cup J)\overset{\text{(\ref{eq1l})}}{=\!\!=} \mi{card}(\{\mbbm{i}\})+\mi{card}(J)=
                                                        1+\mi{card}(J)\underset{\text{Lemma \ref{le11}(a)}}{\leq} 1+|\phi'|-1\underset{\text{Lemma \ref{le111}(b)}}{\leq} 2\cdot |\phi|$; 
(a) holds.

$J_\phi\overset{\text{(\ref{eq1k})}}{=\!\!=} \{\mbbm{i}\}\cup J\neq \emptyset$, $S_\phi=\{\tilde{a}_\mbbm{i}\geql \gu\}\cup S\neq \emptyset$,
$\square\not\in \{\tilde{a}_\mbbm{i}\geql \gu\}$;
for all $C\in S$, by Lemma \ref{le11}(d), 
$\mi{atoms}(C)\cap \tilde{\mbb{A}}\neq \emptyset$, $\mi{atoms}(C)\cap \tilde{\mbb{A}}\subseteq \mi{atoms}(C)$, $\mi{atoms}(C)\neq \mi{atoms}(\square)=\emptyset$, $C\neq \square$;
$\square\not\in S$;
$\square\not\in S_\phi=\{\tilde{a}_\mbbm{i}\geql \gu\}\cup S$; 
(b) holds.

$\emptyset\neq \mi{atoms}(\tilde{a}_\mbbm{i}\geql \gu)\cap \tilde{\mbb{A}}=\{\tilde{a}_\mbbm{i}\}\cap \tilde{\mbb{A}}=\{\tilde{a}_\mbbm{i}\}\subseteq 
                                                                           \{\tilde{a}_\mbbm{j} \,|\, \mbbm{j}\in J_\phi\overset{\text{(\ref{eq1k})}}{=\!\!=} \{\mbbm{i}\}\cup J\}$;
by Lemma \ref{le11}(d), for all $C\in S$, 
$\emptyset\neq \mi{atoms}(C)\cap \tilde{\mbb{A}}\subseteq \{\tilde{a}_\mbbm{i}\}\cup \{\tilde{a}_\mbbm{j} \,|\, \mbbm{j}\in J\}=
                                                          \{\tilde{a}_\mbbm{j} \,|\, \mbbm{j}\in J_\phi\overset{\text{(\ref{eq1k})}}{=\!\!=} \{\mbbm{i}\}\cup J\}$; 
for all $C\in S_\phi=\{\tilde{a}_\mbbm{i}\geql \gu\}\cup S$, $\emptyset\neq \mi{atoms}(C)\cap \tilde{\mbb{A}}\subseteq \{\tilde{a}_\mbbm{j} \,|\, \mbbm{j}\in J_\phi\}$; 
(e) holds.
%
%
%
\end{proof}

\subsection{Full proof of Corollary \ref{cor12}}
\label{S7.3}

\begin{proof}
We put $\Gamma=\{J \,|\, J\subseteq_{\mc F} \mbb{I}\}$ and $\Lambda=\{S \,|\, S\subseteq_{\mc F} \mi{OrdPropCl}\}$.
We have that $\mi{PropAtom}$ and $\mbb{I}$ are countably infinite.
Then $\mi{PropPow}$, $\mi{PropConj}$, $\mi{OrdPropLit}$, $\mi{OrdPropCl}$, $\Lambda$, $\Gamma$, $\Gamma\times \Lambda$ are countably infinite, and 
there exists a well order $\worder\subseteq (\Gamma\times \Lambda)^2$.
Let $\emptyset\neq K\subseteq \Gamma\times \Lambda$.
By $\mi{least}(K)\in K$ we denote the least element of $K$ with respect to $\worder$.
Let $(J,S)\in \Gamma\times \Lambda$, $\theta\in \mi{PropForm}_\emptyset$, $n_\theta\in \mbb{N}$. 
$(J,S)$ is a clausal translation of $\theta$ with respect to $n_\theta$ iff 
either $J=\emptyset$, or there exists $n_J$ satisfying $J=\{(n_\theta,j) \,|\, j\leq n_J\}$; 
$J\subseteq \{(n_\theta,j) \,|\, j\in \mbb{N}\}\subseteq \mbb{I}$, 
$S\subseteq \mi{OrdPropCl}_{\{\tilde{a}_\mbbm{j} \,|\, \mbbm{j}\in J\}}$ satisfying that
(a--e) of Lemma \ref{le1} hold for $J$, $\theta$, $S$, and $n_\theta$.
We put
\begin{equation}
\notag
K_{n_\theta}^\theta=\{(J,S) \,|\, (J,S)\in \Gamma\times \Lambda\ \text{\it is a clausal translation of}\ \theta\ \text{\it with respect to}\ n_\theta\}.
\end{equation}
We have that $\mi{PropForm}$ is countably infinite.
Then $T\subseteq \mi{PropForm}$ is countable, and
there exist $\gamma\leq \omega$ and a sequence $\delta : \gamma\longrightarrow T$ of $T$.
We put $n_\alpha=n_0+\alpha\geq n_0$ and $(J_\alpha,S_\alpha)=\mi{least}(K_{n_\alpha}^{\delta(\alpha)})\in \Gamma\times \Lambda$, $\alpha<\gamma$.
Note that for all $\alpha<\gamma$, 
$\delta(\alpha)\in T\subseteq \mi{PropForm}_\emptyset$;
by Lemma \ref{le1} for $n_\alpha$ and $\delta(\alpha)$, there exist either $J_{\delta(\alpha)}=\emptyset$, or $n_{J_{\delta(\alpha)}}$, $J_{\delta(\alpha)}=\{(n_\alpha,j) \,|\, j\leq n_{J_{\delta(\alpha)}}\}$,
$J_{\delta(\alpha)}\subseteq_{\mc F} \{(n_\alpha,j) \,|\, j\in \mbb{N}\}\subseteq \mbb{I}$, 
$S_{\delta(\alpha)}\subseteq_{\mc F} \mi{OrdPropCl}_{\{\tilde{a}_\mbbm{j} \,|\, \mbbm{j}\in J_{\delta(\alpha)}\}}$ satisfying that
(a--e) of Lemma \ref{le1} hold for $J_{\delta(\alpha)}$, $\delta(\alpha)$, $S_{\delta(\alpha)}$, and $n_\alpha$;
$(J_{\delta(\alpha)},S_{\delta(\alpha)})\in \Gamma\times \Lambda$ is a clausal translation of $\delta(\alpha)$ with respect to $n_\alpha$;
$(J_{\delta(\alpha)},S_{\delta(\alpha)})\in K_{n_\alpha}^{\delta(\alpha)}\neq \emptyset$;
$(J_\alpha,S_\alpha)=\mi{least}(K_{n_\alpha}^{\delta(\alpha)})\in K_{n_\alpha}^{\delta(\alpha)}\subseteq \Gamma\times \Lambda$;
$(J_\alpha,S_\alpha)$ is a clausal translation of $\delta(\alpha)$ with respect to $n_\alpha$;
either $J_\alpha=\emptyset$, or there exists $n_{J_\alpha}$ satisfying $J_\alpha=\{(n_\alpha,j) \,|\, j\leq n_{J_\alpha}\}$;
$J_\alpha\subseteq \{(n_\alpha,j) \,|\, j\in \mbb{N}\}\subseteq \mbb{I}$,
$S_\alpha\subseteq \mi{OrdPropCl}_{\{\tilde{a}_\mbbm{j} \,|\, \mbbm{j}\in J_\alpha\}}$ satisfying that
(a--e) of Lemma \ref{le1} hold for $J_\alpha$, $\delta(\alpha)$, $S_\alpha$, and $n_\alpha$.
Hence, for all $\alpha<\alpha'<\gamma$, $n_\alpha=n_0+\alpha<n_{\alpha'}=n_0+\alpha'$;
\begin{equation}
\label{eq2a}
\text{for all}\ \alpha<\alpha'<\gamma,\ J_\alpha\cap J_{\alpha'}=\{(n_\alpha,j) \,|\, j\in \mbb{N}\}\cap \{(n_{\alpha'},j) \,|\, j\in \mbb{N}\}=\emptyset.
\end{equation}
We put 
\begin{alignat*}{1}
J_T &= \left\{\begin{array}{ll}
              \emptyset                        &\ \text{\it if there exists}\ \alpha^*<\gamma\ \text{\it such that}\ S_{\alpha^*}=\{\square\}, \\[1mm]
              \bigcup_{\alpha<\gamma} J_\alpha &\ \text{\it else};
              \end{array}
       \right. \\[1mm]
S_T &= \left\{\begin{array}{ll}   
              \{\square\}                      &\ \text{\it if there exists}\ \alpha^*<\gamma\ \text{\it such that}\ S_{\alpha^*}=\{\square\}, \\[1mm]
              \bigcup_{\alpha<\gamma} S_\alpha &\ \text{\it else}.
              \end{array}   
       \right.
\end{alignat*}
We distinguish two cases for $\gamma$.

Case 1:
There exists $\alpha^*<\gamma$ such that $S_{\alpha^*}=\{\square\}$.
Then $J_T=\emptyset\subseteq_{\mc F} \{(i,j) \,|\, i\geq n_0, j\in \mbb{N}\}\subseteq \mbb{I}$, $\mi{card}(J_T)=\mi{card}(\emptyset)=0\leq 2\cdot |T|$,
$S_T=\{\square\}\subseteq_{\mc F} \mi{OrdPropCl}_\emptyset$, $|S_T|=|\{\square\}|=0\in O(|T|)$.

$J_T=\emptyset$, $S_T=\{\square\}$; 
(a) holds.

For every valuation ${\mf A}^\#$, ${\mf A}^\#\not\models S_{\alpha^*}=\{\square\}$;
by Lemma \ref{le1}(c) for $\delta(\alpha^*)$ and $S_{\alpha^*}$, 
there exists a valuation ${\mf A}$ satisfying ${\mf A}\models \delta(\alpha^*)$ if and only if 
there exists a valuation ${\mf A}'$ satisfying ${\mf A}'\models S_{\alpha^*}$; 
${\mf A}|_\mbb{O}={\mf A}'|_\mbb{O}$;
for every valuation ${\mf A}^\#$, ${\mf A}^\#\not\models \delta(\alpha^*)$, $\delta(\alpha^*)\in T$, ${\mf A}^\#\not\models T$, ${\mf A}^\#\not\models S_T=\{\square\}$; 
trivially, 
there exists a valuation ${\mf A}$ satisfying ${\mf A}\models T$ if and only if 
there exists a valuation ${\mf A}'$ satisfying ${\mf A}'\models S_T$; 
${\mf A}|_\mbb{O}={\mf A}'|_\mbb{O}$; 
(b) holds.

$S_T=\{\square\}$; 
(d) holds trivially.

Case 2:
For all $\alpha<\gamma$, $S_\alpha\neq \{\square\}$.
Then, for all $\alpha<\gamma$, $J_\alpha\subseteq \{(n_\alpha,j) \,|\, j\in \mbb{N}\}$; 
$J_T=\bigcup_{\alpha<\gamma} J_\alpha\subseteq \bigcup_{\alpha<\gamma} \{(n_\alpha,j) \,|\, j\in \mbb{N}\}\subseteq \{(i,j) \,|\, i\geq n_0, j\in \mbb{N}\}\subseteq \mbb{I}$,
$S_T=\bigcup_{\alpha<\gamma} S_\alpha\subseteq \mi{OrdPropCl}_{\{\tilde{a}_\mbbm{j} \,|\, \mbbm{j}\in J_T\}}$.

For all $\alpha<\gamma$, 
either $\square\not\in S_\alpha=\emptyset$, 
or $S_\alpha\neq \emptyset, \{\square\}$; 
by Lemma \ref{le1}(b), $\square\not\in S_\alpha$; 
$\square\not\in S_\alpha$; 
$\square\not\in S_T=\bigcup_{\alpha<\gamma} S_\alpha$.
We get two cases for $\gamma$.

Case 2.1:
There exists $\alpha^*<\gamma$ such that $S_{\alpha^*}\neq \emptyset$.
Then $S_{\alpha^*}\neq \{\square\}$;
by Lemma \ref{le1}(b) for $J_{\alpha^*}$ and $S_{\alpha^*}$, $J_{\alpha^*}\neq \emptyset$, $J_{\alpha^*}\subseteq J_T$, $J_T\neq \emptyset$, $S_{\alpha^*}\subseteq S_T$, $\square\not\in S_T\neq \emptyset$;
(a) holds.

Case 2.2:
For all $\alpha<\gamma$, $S_\alpha=\emptyset$.
Then, for all $\alpha<\gamma$, $J_\alpha\overset{\text{Lemma \ref{le1}(b)}}{=\!\!=\!\!=\!\!=\!\!=\!\!=\!\!=\!\!=\!\!=\!\!=\!\!=\!\!=\!\!=} \emptyset$;
$J_T=\bigcup_{\alpha<\gamma} J_\alpha=\emptyset$, $S_T=\bigcup_{\alpha<\gamma} S_\alpha=\emptyset$;
(a) holds.

So, in both Cases 2.1 and 2.2, (a) holds.

Let ${\mf A}$ be a valuation such that ${\mf A}\models T$.
Then, for all $\alpha<\gamma$,
$\delta(\alpha)\in T$, ${\mf A}\models \delta(\alpha)$; 
by Lemma \ref{le1}(c), there exists a valuation ${\mf A}_\alpha$ satisfying ${\mf A}_\alpha\models S_\alpha$, ${\mf A}_\alpha|_\mbb{O}={\mf A}|_\mbb{O}$;
for all $\alpha<\gamma$, $\{\tilde{a}_\mbbm{j} \,|\, \mbbm{j}\in J_\alpha\}\subseteq \tilde{\mbb{A}}$;
$(\bigcup_{\alpha<\gamma} \{\tilde{a}_\mbbm{j} \,|\, \mbbm{j}\in J_\alpha\})\cap \mbb{O}=\tilde{\mbb{A}}\cap \mbb{O}=\emptyset$;
for all $\alpha<\alpha'<\gamma$, $\{\tilde{a}_\mbbm{j} \,|\, \mbbm{j}\in J_\alpha\}\cap \{\tilde{a}_\mbbm{j} \,|\, \mbbm{j}\in J_{\alpha'}\}\overset{\text{(\ref{eq2a})}}{=\!\!=} \emptyset$.
We define a valuation
\begin{equation}
\notag
{\mf A}'={\mf A}|_\mbb{O}\cup \left(\bigcup_{\alpha<\gamma} {\mf A}_\alpha|_{\{\tilde{a}_\mbbm{j} \,|\, \mbbm{j}\in J_\alpha\}}\right)\cup         
         {\mf A}|_{\mi{PropAtom}-(\mbb{O}\cup \{\tilde{a}_\mbbm{j} \,|\, \mbbm{j}\in J_T=\bigcup_{\alpha<\gamma} J_\alpha\})}.
\end{equation}
We get that for all $\alpha<\gamma$, 
${\mf A}_\alpha\models S_\alpha$,
$\mi{atoms}(S_\alpha)\subseteq \mbb{O}\cup \{\tilde{a}_\mbbm{j} \,|\, \mbbm{j}\in J_\alpha\}$,
\begin{alignat*}{1}
{\mf A}'|_{\mbb{O}\cup \{\tilde{a}_\mbbm{j} \,|\, \mbbm{j}\in J_\alpha\}}
&= {\mf A}'|_\mbb{O}\cup {\mf A}'|_{\{\tilde{a}_\mbbm{j} \,|\, \mbbm{j}\in J_\alpha\}}=
   {\mf A}|_\mbb{O}\cup {\mf A}_\alpha|_{\{\tilde{a}_\mbbm{j} \,|\, \mbbm{j}\in J_\alpha\}}= \\
&\phantom{\mbox{}=\mbox{}}
   {\mf A}_\alpha|_\mbb{O}\cup {\mf A}_\alpha|_{\{\tilde{a}_\mbbm{j} \,|\, \mbbm{j}\in J_\alpha\}}=
   {\mf A}_\alpha|_{\mbb{O}\cup \{\tilde{a}_\mbbm{j} \,|\, \mbbm{j}\in J_\alpha\}},
\end{alignat*}
${\mf A}'\models S_\alpha$;
${\mf A}'\models S_T=\bigcup_{\alpha<\gamma} S_\alpha$, ${\mf A}'|_\mbb{O}={\mf A}|_\mbb{O}$.

Let ${\mf A}'$ be a valuation such that ${\mf A}'\models S_T$.
Then, for all $\alpha<\gamma$, 
$S_\alpha\subseteq S_T$, ${\mf A}'\models S_\alpha$; 
by Lemma \ref{le1}(c), there exists a valuation ${\mf A}_\alpha$ satisfying ${\mf A}_\alpha\models \delta(\alpha)$, ${\mf A}_\alpha|_\mbb{O}={\mf A}'|_\mbb{O}$;
$\mi{atoms}(\delta(\alpha))\subseteq \mbb{O}$,
${\mf A}'\models \delta(\alpha)$.
We put ${\mf A}={\mf A}'$.
We get that for all $\alpha<\gamma$, ${\mf A}\models \delta(\alpha)$; 
${\mf A}\models T=\{\delta(\alpha) \,|\, \alpha<\gamma\}$, ${\mf A}|_\mbb{O}={\mf A}'|_\mbb{O}$; 
(b) holds.

Let $\alpha<\alpha'<\gamma$.
We get two cases for $S_\alpha$ and $S_{\alpha'}$.

Case 2.3:
$S_\alpha=\emptyset$ or $S_{\alpha'}=\emptyset$.
Then $S_\alpha\cap S_{\alpha'}=\emptyset$.

Case 2.4:
$S_\alpha, S_{\alpha'}\neq \emptyset$.
Then $S_\alpha\neq \{\square\}$;
by Lemma \ref{le1}(e), for all $C\in S_\alpha$, $\emptyset\neq \mi{atoms}(C)\cap \tilde{\mbb{A}}\subseteq \{\tilde{a}_\mbbm{j} \,|\, \mbbm{j}\in J_\alpha\}$;
$S_{\alpha'}\neq \{\square\}$;
by Lemma \ref{le1}(e) for $S_{\alpha'}$ and $J_{\alpha'}$, for all $C\in S_{\alpha'}$, $\emptyset\neq \mi{atoms}(C)\cap \tilde{\mbb{A}}\subseteq \{\tilde{a}_\mbbm{j} \,|\, \mbbm{j}\in J_{\alpha'}\}$;
for all $C\in S_\alpha$ and $C'\in S_{\alpha'}$,
$(\mi{atoms}(C)\cap \tilde{\mbb{A}})\cap (\mi{atoms}(C')\cap \tilde{\mbb{A}})=
 \{\tilde{a}_\mbbm{j} \,|\, \mbbm{j}\in J_\alpha\}\cap \{\tilde{a}_\mbbm{j} \,|\, \mbbm{j}\in J_{\alpha'}\}\overset{\text{(\ref{eq2a})}}{=\!\!=} \emptyset$,
$\mi{atoms}(C)\cap \tilde{\mbb{A}}, \mi{atoms}(C')\cap \tilde{\mbb{A}}\neq \emptyset$, 
$\mi{atoms}(C)\cap \tilde{\mbb{A}}\neq \mi{atoms}(C')\cap \tilde{\mbb{A}}$, $\mi{atoms}(C)\neq \mi{atoms}(C')$, $C\neq C'$;
$S_\alpha\cap S_{\alpha'}=\emptyset$.

So, in both Cases 2.3 and 2.4, $S_\alpha\cap S_{\alpha'}=\emptyset$;
\begin{equation}
\label{eq2b}
\text{for all}\ \alpha<\alpha'<\gamma,\ S_\alpha\cap S_{\alpha'}=\emptyset.
\end{equation}

Let $T\subseteq_{\mc F} \mi{PropForm}_\emptyset$.
Then $\gamma<\omega$,
$J_T=\bigcup_{\alpha<\gamma} J_\alpha\subseteq_{\mc F} \{(i,j) \,|\, i\geq n_0, j\in \mbb{N}\}\subseteq \mbb{I}$,
$\mi{card}(J_T)=\mi{card}(\bigcup_{\alpha<\gamma} J_\alpha)\overset{\text{(\ref{eq2a})}}{=\!\!=} \sum_{\alpha<\gamma} \mi{card}(J_\alpha)\underset{\text{Lemma \ref{le1}(a)}}{\leq} 
                \sum_{\alpha<\gamma} 2\cdot |\delta(\alpha)|=2\cdot \sum_{\alpha<\gamma} |\delta(\alpha)|=2\cdot |T|$,
$S_T=\bigcup_{\alpha<\gamma} S_\alpha\subseteq_{\mc F} \mi{OrdPropCl}_{\{\tilde{a}_\mbbm{j} \,|\, \mbbm{j}\in J_T\}}$,
$|S_T|=|\bigcup_{\alpha<\gamma} S_\alpha|\overset{\text{(\ref{eq2b})}}{=\!\!=} \sum_{\alpha<\gamma} |S_\alpha|\underset{\text{Lemma \ref{le1}(d)}}{\in} O(\sum_{\alpha<\gamma} |\delta(\alpha)|)=O(|T|)$.

Let $S_T=\bigcup_{\alpha<\gamma} S_\alpha\neq \emptyset$.
Then there exists $\alpha^*<\gamma$ satisfying $S_{\alpha^*}\neq \emptyset, \{\square\}$;
by Lemma \ref{le1}(b) for $J_{\alpha^*}$ and $S_{\alpha^*}$, $J_{\alpha^*}\neq \emptyset$, $J_{\alpha^*}\subseteq J_T$, $J_T\neq \emptyset$;
for all $C\in S_T=\bigcup_{\alpha<\gamma} S_\alpha$,
there exists $\alpha^*<\gamma$ satisfying $C\in S_{\alpha^*}\neq \emptyset, \{\square\}$;
by Lemma \ref{le1}(e) for $S_{\alpha^*}$ and $J_{\alpha^*}$, 
$\emptyset\neq \mi{atoms}(C)\cap \tilde{\mbb{A}}\subseteq \{\tilde{a}_\mbbm{j} \,|\, \mbbm{j}\in J_{\alpha^*}\}\subseteq \{\tilde{a}_\mbbm{j} \,|\, \mbbm{j}\in J_T\}$; 
(d) holds.

Let $T\subseteq_{\mc F} \mi{PropForm}_\emptyset$.
Then $\gamma<\omega$;
we get in both Cases 1 and 2 that $J_T\subseteq_{\mc F} \{(i,j) \,|\, i\geq n_0, j\in \mbb{N}\}\subseteq \mbb{I}$, $\mi{card}(J_T)\leq 2\cdot |T|$,
$S_T\subseteq_{\mc F} \mi{OrdPropCl}_{\{\tilde{a}_\mbbm{j} \,|\, \mbbm{j}\in J_T\}}$, $|S_T|\in O(|T|)$;
the translation of $T$ to $S_T$ uses the input $T$ and the output $S_T$;
for all $\alpha<\gamma$, 
by Lemma \ref{le1}(d), the number of all elementary operations of the translation of $\delta(\alpha)$ to $S_\alpha$ is in $O(|\delta(\alpha)|)$;
the time complexity of the translation of $\delta(\alpha)$ to $S_\alpha$ is in $O(|\delta(\alpha)|\cdot (\log (1+n_\alpha)+\log |\delta(\alpha)|))$;
the translation of $T$ to $S_T$ uses a constant number of auxiliary data structures of size in $O(1)$ and data structures of size in $O(\log (1+n_\alpha)+\log |\delta(\alpha)|)$ at the $\alpha$th stage,
including {\it index generator}, values of which are of the form $(n_\alpha,j)\in \mbb{I}$,
$j\leq n_{J_\alpha}=\mi{card}(J_\alpha)-1<\mi{card}(J_\alpha)\underset{\text{Lemma \ref{le1}(a)}}{\leq} 2\cdot |\delta(\alpha)|$;
the translation of $T$ to $S_T$ executes a constant number of all auxiliary elementary operations on auxiliary data structures with the time complexity in $O(\log (1+n_\alpha)+\log |\delta(\alpha)|)$ at the $\alpha$th stage, 
including the test $S_\alpha=\{\square\}$;
the translation of $T$ to $S_T$ also executes the union of the partial output $\bigcup_{\beta<\alpha} S_\beta$, $S_\beta\neq \{\square\}$, and $S_\alpha$ at the $\alpha$th stage, 
including appending $S_\alpha\neq \emptyset, \{\square\}$, $|S_\alpha|\geq 1$, to $\bigcup_{\beta<\alpha} S_\beta$, $S_\beta\neq \{\square\}$, 
which uses the input $S_\alpha$ and the output $S_\alpha$ (a copy) with $\#{\mc O}(S_\alpha)\in O(|S_\alpha|)\underset{\text{Lemma \ref{le1}(d)}}{\subseteq} O(|\delta(\alpha)|)$;
by (\ref{eq00t}) for $n_\alpha$, $S_\alpha$, $\emptyset$, $S_\alpha$, $q=2$, and $r=1$,
the time complexity is in $O(\#{\mc O}(S_\alpha)\cdot (\log (1+n_\alpha)+\log (\#{\mc O}(S_\alpha)+|S_\alpha|)))\subseteq O(|\delta(\alpha)|\cdot (\log (1+n_\alpha)+\log |\delta(\alpha)|))$;
by (\ref{eq00s}) for $n_\alpha$, $S_\alpha$, $\emptyset$, $S_\alpha$, $q=2$, and $r=1$,
the space complexity is in $O((\#{\mc O}(S_\alpha)+|S_\alpha|)\cdot (\log (1+n_\alpha)+\log (1+|S_\alpha|)))\subseteq O(|\delta(\alpha)|\cdot (\log (1+n_\alpha)+\log |\delta(\alpha)|))$;  
the number of all elementary operations of the translation of $T$ to $S_T$ at the $\alpha$th stage is in $O(|\delta(\alpha)|)$;
the time complexity of the translation of $T$ to $S_T$ at the $\alpha$th stage is in $O(|\delta(\alpha)|\cdot (\log (1+n_\alpha)+\log |\delta(\alpha)|))$;
the total number of all elementary operations of the translation of $T$ to $S_T$ is in $O(\sum_{\alpha<\gamma} |\delta(\alpha)|)=O(|T|)$;
$\mi{card}(T)=\gamma=\sum_{\alpha<\gamma} 1\leq |T|=\sum_{\alpha<\gamma} |\delta(\alpha)|$, $n_\alpha=n_0+\alpha<n_0+\gamma\leq n_0+|T|$, $|\delta(\alpha)|\leq |T|=\sum_{\alpha<\gamma} |\delta(\alpha)|$;
the total time complexity of the translation of $T$ to $S_T$ is in
\begin{alignat*}{1} 
& \bigO\left(\sum_{\alpha<\gamma} |\delta(\alpha)|\cdot (\log (1+n_\alpha)+\log |\delta(\alpha)|)\right)\subseteq \\
& \bigO\left(\sum_{\alpha<\gamma} |\delta(\alpha)|\cdot (\log (1+n_0+|T|)+\log (1+|T|))\right)\subseteq \\
& \bigO\left(\left(\sum_{\alpha<\gamma} |\delta(\alpha)|\right)\cdot (\log (1+n_0+|T|)+\log (1+|T|))\right)= \\
& O(|T|\cdot \log (1+n_0+|T|));
\end{alignat*}
(c) holds.

So, in both Cases 1 and 2, (a--d) hold.
%
%
%
\end{proof}

\subsection{Full proof of Theorem \ref{T1}} 
\label{S7.44}

\begin{proof}
By Lemma \ref{le111} for $n_0$ and $\phi$, there exists $\phi'\in \mi{PropForm}_\emptyset$ satisfying that (a--d) of Lemma \ref{le111} hold for $\phi'$, $\phi$, and $n_0$;
by Corollary \ref{cor12} for $n_0+1$, there exist $J_T\subseteq \{(i,j) \,|\, i\geq n_0+1, j\in \mbb{N}\}\subseteq \mbb{I}$ and
$S_T\subseteq \mi{OrdPropCl}_{\{\tilde{a}_\mbbm{j} \,|\, \mbbm{j}\in J_T\}}$ satisfying that (a--d) of Corollary \ref{cor12} hold for $n_0+1$.
We distinguish three cases for $\phi'$.

Case 1: 
$\phi'=\gz$.
We put $J_T^\phi=J_T\subseteq \{(i,j) \,|\, i\geq n_0+1, j\in \mbb{N}\}\subset \{(i,j) \,|\, i\geq n_0, j\in \mbb{N}\}\subseteq \mbb{I}$ and
$S_T^\phi=S_T\subseteq \mi{OrdPropCl}_{\{\tilde{a}_\mbbm{j} \,|\, \mbbm{j}\in J_T^\phi=J_T\}}$.

Then, for every valuation ${\mf A}^\#$,
${\mf A}^\#\not\models \phi'=\gz$,
${\mf A}^\#\not\models \phi'\overset{\text{Lemma \ref{le111}(a)}}{\eqvl\!\!\eqvl\!\!\eqvl\!\!\eqvl\!\!\eqvl\!\!\eqvl\!\!\eqvl\!\!\eqvl\!\!\eqvl\!\!\eqvl\!\!\eqvl\!\!\eqvl\!\!\eqvl} \phi$;
by Corollary \ref{cor12}(b),
there exists a valuation ${\mf A}$ satisfying ${\mf A}\models T$, ${\mf A}\not\models \phi$ if and only if 
there exists a valuation ${\mf A}'$ satisfying ${\mf A}'\models S_T^\phi=S_T$;  
${\mf A}|_\mbb{O}={\mf A}'|_\mbb{O}$; 
(i) holds.

(iii) can be proved straightforwardly.

Case 2:
$\phi'=\gu$.
We put $J_T^\phi=\emptyset\subseteq \{(i,j) \,|\, i\geq n_0, j\in \mbb{N}\}\subseteq \mbb{I}$ and $S_T^\phi=\{\square\}\subseteq \mi{OrdPropCl}_\emptyset$.

Then, for every valuation ${\mf A}^\#$,
${\mf A}^\#\models \phi'=\gu$,
${\mf A}^\#\models \phi'\overset{\text{Lemma \ref{le111}(a)}}{\eqvl\!\!\eqvl\!\!\eqvl\!\!\eqvl\!\!\eqvl\!\!\eqvl\!\!\eqvl\!\!\eqvl\!\!\eqvl\!\!\eqvl\!\!\eqvl\!\!\eqvl\!\!\eqvl} \phi$,
${\mf A}^\#\not\models S_T^\phi=\{\square\}$;
trivially, 
there exists a valuation ${\mf A}$ satisfying ${\mf A}\models T$, ${\mf A}\not\models \phi$ if and only if 
there exists a valuation ${\mf A}'$ satisfying ${\mf A}'\models S_T^\phi$;  
${\mf A}|_\mbb{O}={\mf A}'|_\mbb{O}$; 
(i) holds.

(iii) can be proved straightforwardly.

Case 3: 
$\phi'\neq \gz, \gu$.
We have that (a--d) of Lemma \ref{le111} hold for $\phi'$, $\phi$, and $n_0$.
Then $\phi'\in \mi{PropForm}_\emptyset-\{\gz,\gu\}$ and (c,d) of Lemma \ref{le111} hold for $\phi'$.
We put $j_\mbbm{i}=0$ and $\mbbm{i}=(n_0,j_\mbbm{i})\in \{(n_0,j) \,|\, j\in \mbb{N}\}\subseteq \mbb{I}$.
Hence, $\tilde{a}_\mbbm{i}\in \tilde{\mbb{A}}$;
by Lemma \ref{le11} for $n_0$ and $\phi'$, there exist $n_J\geq j_\mbbm{i}$, $J=\{(n_0,j) \,|\, j_\mbbm{i}+1\leq j\leq n_J\}\subseteq \{(n_0,j) \,|\, j\in \mbb{N}\}\subseteq \mbb{I}$, $\mbbm{i}\not\in J$,
$S\subseteq_{\mc F} \mi{OrdPropCl}_{\{\tilde{a}_\mbbm{i}\}\cup \{\tilde{a}_\mbbm{j} \,|\, \mbbm{j}\in J\}}$ satisfying that
(a--d) of Lemma \ref{le11} hold for $\phi'$.
We put $J_T^\phi=\{\mbbm{i}\}\cup J\cup J_T\subseteq \{(i,j) \,|\, i\geq n_0, j\in \mbb{N}\}\subseteq \mbb{I}$.
Note that $\mbbm{i}\in \{(n_0,j) \,|\, j\in \mbb{N}\}$, $J\subseteq \{(n_0,j) \,|\, j\in \mbb{N}\}$, $J_T\subseteq \{(i,j) \,|\, i\geq n_0+1, j\in \mbb{N}\}\subset \{(i,j) \,|\, i\geq n_0, j\in \mbb{N}\}$,
$J_T^\phi=\{\mbbm{i}\}\cup J\cup J_T\subseteq \{(i,j) \,|\, i\geq n_0, j\in \mbb{N}\}\subseteq \mbb{I}$;
$\mbbm{i}\not\in J$, $(\{\mbbm{i}\}\cup J)\cap J_T=\{(n_0,j) \,|\, j\in \mbb{N}\}\cap \{(i,j) \,|\, i\geq n_0+1, j\in \mbb{N}\}=\emptyset$;
\begin{equation}
\label{eq2c}
\{\mbbm{i}\}, J, J_T\ \text{are pairwise disjoint}.
\end{equation}
We put $S_T^\phi=\{\tilde{a}_\mbbm{i}\gle \gu\}\cup S\cup S_T\subseteq \mi{OrdPropCl}_{\{\tilde{a}_\mbbm{j} \,|\, \mbbm{j}\in J_T^\phi\}}$.
Note that $\tilde{a}_\mbbm{i}\in \tilde{\mbb{A}}$, $\mbb{O}\cap \tilde{\mbb{A}}=\emptyset$, $\tilde{a}_\mbbm{i}\not\in \mbb{O}$,
$\mi{atoms}(\tilde{a}_\mbbm{i}\gle \gu)=\{\tilde{a}_\mbbm{i}\}\subseteq \mbb{O}\cup \{\tilde{a}_\mbbm{i}\}$, 
$\tilde{a}_\mbbm{i}\gle \gu\in \mi{OrdPropLit}_{\{\tilde{a}_\mbbm{i}\}}$,
$\mi{atoms}(\phi')\subseteq \mbb{O}$,
$\mi{atoms}(\tilde{a}_\mbbm{i}\leftrightarrow \phi')=\mi{atoms}(\tilde{a}_\mbbm{i})\cup \mi{atoms}(\phi')\subseteq \mbb{O}\cup \{\tilde{a}_\mbbm{i}\}$,
$\tilde{a}_\mbbm{i}\leftrightarrow \phi'\in \mi{PropForm}_{\{\tilde{a}_\mbbm{i}\}}$;
$\mi{atoms}(S)\subseteq \mbb{O}\cup \{\tilde{a}_\mbbm{i}\}\cup \{\tilde{a}_\mbbm{j} \,|\, \mbbm{j}\in J\}$,
$\mi{atoms}(S_T)\subseteq \mbb{O}\cup \{\tilde{a}_\mbbm{j} \,|\, \mbbm{j}\in J_T\}$,
$\mi{atoms}(S_T^\phi)=\mi{atoms}(\{\tilde{a}_\mbbm{i}\gle \gu\}\cup S\cup S_T)=\mi{atoms}(\tilde{a}_\mbbm{i}\gle \gu)\cup \mi{atoms}(S)\cup \mi{atoms}(S_T)\subseteq
                      \mbb{O}\cup \{\tilde{a}_\mbbm{i}\}\cup \{\tilde{a}_\mbbm{j} \,|\, \mbbm{j}\in J\}\cup \{\tilde{a}_\mbbm{j} \,|\, \mbbm{j}\in J_T\}=
                      \mbb{O}\cup \{\tilde{a}_\mbbm{j} \,|\, \mbbm{j}\in \{\mbbm{i}\}\cup J\cup J_T\}=\mbb{O}\cup \{\tilde{a}_\mbbm{j} \,|\, \mbbm{j}\in J_T^\phi\}$,
$S_T^\phi=\{\tilde{a}_\mbbm{i}\gle \gu\}\cup S\cup S_T\subseteq \mi{OrdPropCl}_{\{\tilde{a}_\mbbm{j} \,|\, \mbbm{j}\in J_T^\phi\}}$.
It can be proved that
\begin{equation}
\label{eq2d}
\{\tilde{a}_\mbbm{i}\gle \gu\}, S, S_T\ \text{are pairwise disjoint}.
\end{equation}

Let ${\mf A}$ be a valuation such that ${\mf A}\models T$ and ${\mf A}\not\models \phi$.
Then, by Corollary \ref{cor12}(b), there exists a valuation ${\mf A}_T$ satisfying ${\mf A}_T\models S_T$, ${\mf A}_T|_\mbb{O}={\mf A}|_\mbb{O}$;
${\mf A}\not\models \phi\overset{\text{Lemma \ref{le111}(a)}}{\eqvl\!\!\eqvl\!\!\eqvl\!\!\eqvl\!\!\eqvl\!\!\eqvl\!\!\eqvl\!\!\eqvl\!\!\eqvl\!\!\eqvl\!\!\eqvl\!\!\eqvl\!\!\eqvl} \phi'$,
$\|\phi'\|^{\mf A}<1$;
$\tilde{a}_\mbbm{i}\not\in \mbb{O}$.
We define a valuation
\begin{equation}
\notag
{\mf A}^\#={\mf A}|_\mbb{O}\cup \{(\tilde{a}_\mbbm{i},\|\phi'\|^{\mf A})\}\cup {\mf A}|_{\mi{PropAtom}-(\mbb{O}\cup \{\tilde{a}_\mbbm{i}\})}.
\end{equation}
Hence, $\mi{atoms}(\phi')\subseteq \mbb{O}$,
${\mf A}^\#|_\mbb{O}={\mf A}|_\mbb{O}$,
$\|\tilde{a}_\mbbm{i}\|^{{\mf A}^\#}=\|\phi'\|^{\mf A}=\|\phi'\|^{{\mf A}^\#}<1$,
\begin{alignat*}{1}
\|\tilde{a}_\mbbm{i}\gle \gu\|^{{\mf A}^\#}
&= \|\tilde{a}_\mbbm{i}\|^{{\mf A}^\#}\fle \|\gu\|^{{\mf A}^\#}=\|\tilde{a}_\mbbm{i}\|^{{\mf A}^\#}\fle 1=1, \\[1mm]
\|\tilde{a}_\mbbm{i}\leftrightarrow \phi'\|^{{\mf A}^\#}
&= (\|\tilde{a}_\mbbm{i}\|^{{\mf A}^\#}\frightarrow \|\phi'\|^{{\mf A}^\#})\fwedge (\|\phi'\|^{{\mf A}^\#}\frightarrow \|\tilde{a}_\mbbm{i}\|^{{\mf A}^\#})= \\
&\phantom{\mbox{}=\mbox{}}
   (\|\phi'\|^{{\mf A}^\#}\frightarrow \|\phi'\|^{{\mf A}^\#})\fwedge (\|\phi'\|^{{\mf A}^\#}\frightarrow \|\phi'\|^{{\mf A}^\#})=1,
\end{alignat*}
${\mf A}^\#\models \tilde{a}_\mbbm{i}\gle \gu$,
${\mf A}^\#\models \tilde{a}_\mbbm{i}\leftrightarrow \phi'$;
by Lemma \ref{le11}(b) for ${\mf A}^\#$, there exists a valuation ${\mf A}_\phi$ satisfying ${\mf A}_\phi\models S$,
${\mf A}_\phi|_{\mbb{O}\cup \{\tilde{a}_\mbbm{i}\}}={\mf A}^\#|_{\mbb{O}\cup \{\tilde{a}_\mbbm{i}\}}$;
$\mi{atoms}(\tilde{a}_\mbbm{i}\gle \gu)=\{\tilde{a}_\mbbm{i}\}$,
${\mf A}_\phi\models \tilde{a}_\mbbm{i}\gle \gu$, 
${\mf A}_\phi|_\mbb{O}={\mf A}^\#|_\mbb{O}={\mf A}|_\mbb{O}$;
$\tilde{a}_\mbbm{i}\not\in \mbb{O}$,
$\{\tilde{a}_\mbbm{j} \,|\, \mbbm{j}\in J\}, \{\tilde{a}_\mbbm{j} \,|\, \mbbm{j}\in J_T\}\subseteq \tilde{\mbb{A}}$,
$(\{\tilde{a}_\mbbm{j} \,|\, \mbbm{j}\in J\}\cup \{\tilde{a}_\mbbm{j} \,|\, \mbbm{j}\in J_T\})\cap \mbb{O}=\tilde{\mbb{A}}\cap \mbb{O}=\emptyset$;
by (\ref{eq2c}), $\{\tilde{a}_\mbbm{i}\}$, $\{\tilde{a}_\mbbm{j} \,|\, \mbbm{j}\in J\}$, $\{\tilde{a}_\mbbm{j} \,|\, \mbbm{j}\in J_T\}$ are pairwise disjoint.
We define a valuation
\begin{equation}
\notag
{\mf A}'={\mf A}|_\mbb{O}\cup {\mf A}_\phi|_{\{\tilde{a}_\mbbm{i}\}\cup \{\tilde{a}_\mbbm{j} \,|\, \mbbm{j}\in J\}}\cup {\mf A}_T|_{\{\tilde{a}_\mbbm{j} \,|\, \mbbm{j}\in J_T\}}\cup 
         {\mf A}|_{\mi{PropAtom}-(\mbb{O}\cup \{\tilde{a}_\mbbm{j} \,|\, \mbbm{j}\in J_T^\phi=\{\mbbm{i}\}\cup J\cup J_T\})}.
\end{equation}
We get that
${\mf A}_\phi\models \tilde{a}_\mbbm{i}\gle \gu$,
$\mi{atoms}(\tilde{a}_\mbbm{i}\gle \gu)=\{\tilde{a}_\mbbm{i}\}$,
${\mf A}_\phi\models S$,
$\mi{atoms}(S)\subseteq \mbb{O}\cup \{\tilde{a}_\mbbm{i}\}\cup \{\tilde{a}_\mbbm{j} \,|\, \mbbm{j}\in J\}$,
\begin{alignat*}{1}
{\mf A}'|_{\mbb{O}\cup \{\tilde{a}_\mbbm{i}\}\cup \{\tilde{a}_\mbbm{j} \,|\, \mbbm{j}\in J\}}
&= {\mf A}|_\mbb{O}\cup {\mf A}_\phi|_{\{\tilde{a}_\mbbm{i}\}\cup \{\tilde{a}_\mbbm{j} \,|\, \mbbm{j}\in J\}}=
   {\mf A}_\phi|_\mbb{O}\cup {\mf A}_\phi|_{\{\tilde{a}_\mbbm{i}\}\cup \{\tilde{a}_\mbbm{j} \,|\, \mbbm{j}\in J\}}= \\
&\phantom{\mbox{}=\mbox{}}
   {\mf A}_\phi|_{\mbb{O}\cup \{\tilde{a}_\mbbm{i}\}\cup \{\tilde{a}_\mbbm{j} \,|\, \mbbm{j}\in J\}},
\end{alignat*}
${\mf A}'\models \tilde{a}_\mbbm{i}\gle \gu$,
${\mf A}'\models S$,
${\mf A}_T\models S_T$,
$\mi{atoms}(S_T)\subseteq \mbb{O}\cup \{\tilde{a}_\mbbm{j} \,|\, \mbbm{j}\in J_T\}$,
\begin{equation}
\notag
{\mf A}'|_{\mbb{O}\cup \{\tilde{a}_\mbbm{j} \,|\, \mbbm{j}\in J_T\}}=                         
{\mf A}|_\mbb{O}\cup {\mf A}_T|_{\{\tilde{a}_\mbbm{j} \,|\, \mbbm{j}\in J_T\}}=
{\mf A}_T|_\mbb{O}\cup {\mf A}_T|_{\{\tilde{a}_\mbbm{j} \,|\, \mbbm{j}\in J_T\}}= 
{\mf A}_T|_{\mbb{O}\cup \{\tilde{a}_\mbbm{j} \,|\, \mbbm{j}\in J_T\}},
\end{equation}
${\mf A}'\models S_T$;
${\mf A}'\models S_T^\phi=\{\tilde{a}_\mbbm{i}\gle \gu\}\cup S\cup S_T$, ${\mf A}'|_\mbb{O}={\mf A}|_\mbb{O}$.

Let ${\mf A}'$ be a valuation such that ${\mf A}'\models S_T^\phi$.
Then $\{\tilde{a}_\mbbm{i}\gle \gu\}, S, S_T\subseteq S_T^\phi$, ${\mf A}'\models \tilde{a}_\mbbm{i}\gle \gu$, ${\mf A}'\models S$, ${\mf A}'\models S_T$;
by Corollary \ref{cor12}(b), there exists a valuation ${\mf A}_T$ satisfying ${\mf A}_T\models T$, ${\mf A}_T|_\mbb{O}={\mf A}'|_\mbb{O}$;
$\mi{atoms}(T)\subseteq \mbb{O}$, 
${\mf A}'\models T$;
by Lemma \ref{le11}(b), there exists a valuation ${\mf A}^\#$ satisfying ${\mf A}^\#\models \tilde{a}_\mbbm{i}\leftrightarrow \phi'$,
${\mf A}^\#|_{\mbb{O}\cup \{\tilde{a}_\mbbm{i}\}}={\mf A}'|_{\mbb{O}\cup \{\tilde{a}_\mbbm{i}\}}$;
$\mi{atoms}(\tilde{a}_\mbbm{i}\leftrightarrow \phi')\subseteq \mbb{O}\cup \{\tilde{a}_\mbbm{i}\}$,
${\mf A}'\models \tilde{a}_\mbbm{i}\leftrightarrow \phi'$,
\begin{equation}
\notag
\|\tilde{a}_\mbbm{i}\gle \gu\|^{{\mf A}'}=\|\tilde{a}_\mbbm{i}\|^{{\mf A}'}\fle \|\gu\|^{{\mf A}'}=\|\tilde{a}_\mbbm{i}\|^{{\mf A}'}\fle 1=1,
\end{equation}
$\|\tilde{a}_\mbbm{i}\|^{{\mf A}'}<1$,
\begin{equation}
\notag
\|\tilde{a}_\mbbm{i}\leftrightarrow \phi'\|^{{\mf A}'}=(\|\tilde{a}_\mbbm{i}\|^{{\mf A}'}\frightarrow \|\phi'\|^{{\mf A}'})\fwedge (\|\phi'\|^{{\mf A}'}\frightarrow \|\tilde{a}_\mbbm{i}\|^{{\mf A}'})=1,
\end{equation}
$\|\phi'\|^{{\mf A}'}\frightarrow \|\tilde{a}_\mbbm{i}\|^{{\mf A}'}=1$,
$\|\phi'\|^{{\mf A}'}\leq \|\tilde{a}_\mbbm{i}\|^{{\mf A}'}<1$;
${\mf A}'\not\models \phi'\overset{\text{Lemma \ref{le111}(a)}}{\eqvl\!\!\eqvl\!\!\eqvl\!\!\eqvl\!\!\eqvl\!\!\eqvl\!\!\eqvl\!\!\eqvl\!\!\eqvl\!\!\eqvl\!\!\eqvl\!\!\eqvl\!\!\eqvl} \phi$.
We put ${\mf A}={\mf A}'$.
Then ${\mf A}\models T$ and ${\mf A}\not\models \phi$, ${\mf A}|_\mbb{O}={\mf A}'|_\mbb{O}$; 
(i) holds.

Let $T\subseteq_{\mc F} \mi{PropForm}_\emptyset$.
Then $\mbbm{i}\in \{(n_0,j) \,|\, j\in \mbb{N}\}$,
$J\subseteq_{\mc F} \{(n_0,j) \,|\, j\in \mbb{N}\}\subseteq \mbb{I}$,
$\mi{card}(J)\underset{\text{Lemma \ref{le11}(a)}}{\leq} |\phi'|-1\in O(|\phi'|)\underset{\text{Lemma \ref{le111}(b)}}{\subseteq} O(|\phi|)$,
$|\{\tilde{a}_\mbbm{i}\gle \gu\}|\in O(1)$, $|S|, |\phi'|\underset{\text{Lemma \ref{le11}(c)}}{\leq} 25\cdot |\phi'|\in O(|\phi'|)\underset{\text{Lemma \ref{le111}(b)}}{\subseteq} O(|\phi|)$;
by Corollary \ref{cor12}(c), $J_T\subseteq_{\mc F} \{(i,j) \,|\, i\geq n_0+1, j\in \mbb{N}\}\subset \{(i,j) \,|\, i\geq n_0, j\in \mbb{N}\}\subseteq \mbb{I}$, 
$\mi{card}(J_T)\leq 2\cdot |T|\in O(|T|)$,
$S_T\subseteq_{\mc F} \mi{OrdPropCl}_{\{\tilde{a}_\mbbm{j} \,|\, \mbbm{j}\in J_T\}}$, $|S_T|\in O(|T|)$;
$J_T^\phi=\{\mbbm{i}\}\cup J\cup J_T\subseteq_{\mc F} \{(i,j) \,|\, i\geq n_0, j\in \mbb{N}\}\subseteq \mbb{I}$,
$\mi{card}(J_T^\phi)=\mi{card}(\{\mbbm{i}\}\cup J\cup J_T)\overset{\text{(\ref{eq2c})}}{=\!\!=} \mi{card}(\{\mbbm{i}\})+\mi{card}(J)+\mi{card}(J_T)=1+\mi{card}(J)+\mi{card}(J_T)\in O(|\phi|+|T|)$,
$S_T^\phi=\{\tilde{a}_\mbbm{i}\gle \gu\}\cup S\cup S_T\subseteq_{\mc F} \mi{OrdPropCl}_{\{\tilde{a}_\mbbm{j} \,|\, \mbbm{j}\in J_T^\phi\}}$,
$|S_T^\phi|=|\{\tilde{a}_\mbbm{i}\gle \gu\}\cup S\cup S_T|\overset{\text{(\ref{eq2d})}}{=\!\!=} |\{\tilde{a}_\mbbm{i}\gle \gu\}|+|S|+|S_T|=3+|S|+|S_T|\in O(|\phi|+|T|)$;
the translation of $\phi$ and $T$ to $S_T^\phi$ uses the input $\phi$, $T$, the output $S_T^\phi$, auxiliary $\phi'$, $\{\tilde{a}_\mbbm{i}\gle \gu\}$, $S$, $S_T$;
by Lemma \ref{le111}(b), $\phi'$ can be built up from $\phi$ via a postorder traversal of $\phi$ with $\#{\mc O}_1(\phi)\in O(|\phi|)$;
the test $\phi'\neq \gz, \gu$ is with $\#{\mc O}_2(\phi')\in O(1)$;
$\{\tilde{a}_\mbbm{i}\gle \gu\}$ can be built up from $\emptyset$ with $\#{\mc O}_3(\emptyset)\in O(|\{\tilde{a}_\mbbm{i}\gle \gu\}|)=O(1)$;
by Lemma \ref{le11}(c), $S$ can be built up from $\phi'$ via a preorder traversal of $\phi'$ with $\#{\mc O}_4(\phi')\in O(|\phi'|)\underset{\text{Lemma \ref{le111}(b)}}{\subseteq} O(|\phi|)$;
by Corollary \ref{cor12}(c), the number of all elementary operations of the translation of $T$ to $S_T$ is in $O(|T|)$;
the time complexity of the translation of $T$ to $S_T$ is in $O(|T|\cdot \log (1+n_0+|T|))$;
$S_T^\phi$ can be built up from $\{\tilde{a}_\mbbm{i}\gle \gu\}$ and $S$ by copying and appending them to $S_T$
with $\#{\mc O}_5(\{\tilde{a}_\mbbm{i}\gle \gu\},S)\in O(|\{\tilde{a}_\mbbm{i}\gle \gu\}|+|S|)=O(1+|S|)\subseteq O(|\phi|)$;
$\sum_{i=1}^5 \#{\mc O}_i\in O(|\phi|)$;
by (\ref{eq00t}) for $n_0$, $\phi$, $\emptyset$, $\phi'$, $\{\tilde{a}_\mbbm{i}\gle \gu\}$, $S$, $\{\tilde{a}_\mbbm{i}\gle \gu\}$ (a copy), $S$ (a copy), $q=6$, and $r=1$,
the time complexity of all elementary operations at the stages $1,\dots,5$ is 
in $O((\sum_{i=1}^5 \#{\mc O}_i)\cdot (\log (1+n_0)+\log ((\sum_{i=1}^5 \#{\mc O}_i)+|\phi|)))\subseteq O(|\phi|\cdot (\log (1+n_0)+\log |\phi|))$;
by (\ref{eq00s}) for $n_0$, $\phi$, $\emptyset$, $\phi'$, $\{\tilde{a}_\mbbm{i}\gle \gu\}$, $S$, $\{\tilde{a}_\mbbm{i}\gle \gu\}$ (a copy), $S$ (a copy), $q=6$, and $r=1$,
the space complexity of all elementary operations at the stages $1,\dots,5$ is
in $O(((\sum_{i=1}^5 \#{\mc O}_i)+|\phi|)\cdot (\log (1+n_0)+\log |\phi|))\subseteq O(|\phi|\cdot (\log (1+n_0)+\log |\phi|))$;
the total number of all elementary operations of the translation of $\phi$ and $T$ to $S_T^\phi$ is in $O(|\phi|+|T|)$;
the total time complexity of the translation of $\phi$ and $T$ to $S_T^\phi$ is in $O(|\phi|\cdot (\log (1+n_0)+\log |\phi|)+|T|\cdot \log (1+n_0+|T|))$; 
(iii) holds.

So, in all Cases 1--3, (i,iii) hold.

Let $T\models \phi$.
Then there does not exist a valuation ${\mf A}$ satisfying ${\mf A}\models T$, ${\mf A}\not\models \phi$;
by (i), there does not exist a valuation ${\mf A}'$ satisfying ${\mf A}'\models S_T^\phi$; 
$S_T^\phi$ is unsatisfiable.
Let $S_T^\phi$ be unsatisfiable.
Then there does not exist a valuation ${\mf A}'$ satisfying ${\mf A}'\models S_T^\phi$;
by (i), there does not exist a valuation ${\mf A}$ satisfying ${\mf A}\models T$, ${\mf A}\not\models \phi$;
$T\models \phi$; 
(ii) holds.
%
%
%
\end{proof}

\subsection{Full proofs of the point (iii) and the statement (\ref{eq2d}) of Theorem \ref{T1}} 
\label{S7.4}

\begin{proof}   
Case 1:
Let $T\subseteq_{\mc F} \mi{PropForm}_\emptyset$. 
Then, by Corollary \ref{cor12}(c), $J_T^\phi=J_T\subseteq_{\mc F} \{(i,j) \,|\, i\geq n_0+1, j\in \mbb{N}\}\subset \{(i,j) \,|\, i\geq n_0, j\in \mbb{N}\}\subseteq \mbb{I}$,
$\mi{card}(J_T^\phi)=\mi{card}(J_T)\leq 2\cdot |T|\in O(|\phi|+|T|)$,
$S_T^\phi=S_T\subseteq_{\mc F} \mi{OrdPropCl}_{\{\tilde{a}_\mbbm{j} \,|\, \mbbm{j}\in J_T^\phi=J_T\}}$, $|S_T^\phi|=|S_T|\in O(|T|)\subseteq O(|\phi|+|T|)$;
the translation of $\phi$ and $T$ to $S_T^\phi$ uses the input $\phi$, $T$, the output $S_T^\phi$, an auxiliary $\phi'$;
by Lemma \ref{le111}(b), $\phi'$ can be built up from $\phi$ via a postorder traversal of $\phi$ with $\#{\mc O}(\phi)\in O(|\phi|)$ and the time complexity in $O(|\phi|\cdot (\log (1+n_0)+\log |\phi|))$;
the test $\phi'=\gz$ is with $\#{\mc O}(\phi')\in O(1)$ and the time complexity in $O(1)$;
by Corollary \ref{cor12}(c), the number of all elementary operations of the translation of $T$ to $S_T^\phi=S_T$ is in $O(|T|)$;
the time complexity of the translation of $T$ to $S_T^\phi=S_T$ is in $O(|T|\cdot \log (1+n_0+|T|))$;
the number of all elementary operations of the translation of $\phi$ and $T$ to $S_T^\phi$ is in $O(|\phi|+|T|)$;
the time complexity of the translation of $\phi$ and $T$ to $S_T^\phi$ is in $O(|\phi|\cdot (\log (1+n_0)+\log |\phi|)+|T|\cdot \log (1+n_0+|T|))$; 
(iii) holds.

Case 2:
Let $T\subseteq_{\mc F} \mi{PropForm}_\emptyset$. 
Then $J_T^\phi=\emptyset\subseteq_{\mc F} \{(i,j) \,|\, i\geq n_0, j\in \mbb{N}\}\subseteq \mbb{I}$, $\mi{card}(J_T^\phi)=\mi{card}(\emptyset)=0\in O(|\phi|+|T|)$,
$S_T^\phi=\{\square\}\subseteq_{\mc F} \mi{OrdPropCl}_\emptyset$, $|S_T^\phi|=|\{\square\}|=0\in O(|\phi|+|T|)$; 
the translation of $\phi$ and $T$ to $S_T^\phi$ uses the input $\phi$, $T$, the output $S_T^\phi$, an auxiliary $\phi'$;
by Lemma \ref{le111}(b), $\phi'$ can be built up from $\phi$ via a postorder traversal of $\phi$ with $\#{\mc O}(\phi)\in O(|\phi|)$ and the time complexity in $O(|\phi|\cdot (\log (1+n_0)+\log |\phi|))$;
the test $\phi'=\gu$ is with $\#{\mc O}(\phi')\in O(1)$ and the time complexity in $O(1)$;
$S_T^\phi$ can be built up from $\emptyset$ with $\#{\mc O}(\emptyset)\in O(1+|S_T^\phi|)=O(1)$ and the time complexity in $O(1)$;
the number of all elementary operations of the translation of $\phi$ and $T$ to $S_T^\phi$ is in $O(|\phi|+|T|)$;
the time complexity of the translation of $\phi$ and $T$ to $S_T^\phi$ is in $O(|\phi|\cdot (\log (1+n_0)+\log |\phi|)+|T|\cdot \log (1+n_0+|T|))$; 
(iii) holds.

Case 3:
For all $C\in S$, 
$\mi{atoms}(C)\cap \tilde{\mbb{A}}\underset{\text{Lemma \ref{le11}(d)}}{\neq} \emptyset$,
$\mi{atoms}(C)\cap \tilde{\mbb{A}}\subseteq \mi{atoms}(C)$, $\mi{atoms}(C)\neq \mi{atoms}(\square)=\emptyset$, $C\neq \square$;
$\square\not\in S$;
$\{\tilde{a}_\mbbm{i}\gle \gu\}\cap S\overset{\text{Lemma \ref{le11}(d)}}{=\!\!=\!\!=\!\!=\!\!=\!\!=\!\!=\!\!=\!\!=\!\!=\!\!=\!\!=\!\!=} \emptyset$.
We distinguish three cases for $S_T$.

Case 3.1:
$S_T=\emptyset$.
Then $(\{\tilde{a}_\mbbm{i}\gle \gu\}\cup S)\cap S_T=\emptyset$.

Case 3.2:
$S_T=\{\square\}$.
Then $\square\not\in \{\tilde{a}_\mbbm{i}\gle \gu\}$, $\square\not\in S$, $(\{\tilde{a}_\mbbm{i}\gle \gu\}\cup S)\cap S_T=\emptyset$.

Case 3.3:
$S_T\neq \emptyset, \{\square\}$.
Then, by Lemma \ref{le11}(d), for all $C\in S$, $\emptyset\neq \mi{atoms}(C)\cap \tilde{\mbb{A}}\subseteq \{\tilde{a}_\mbbm{i}\}\cup \{\tilde{a}_\mbbm{j} \,|\, \mbbm{j}\in J\}$;
by Corollary \ref{cor12}(d), for all $C\in S_T$, $\emptyset\neq \mi{atoms}(C)\cap \tilde{\mbb{A}}\subseteq \{\tilde{a}_\mbbm{j} \,|\, \mbbm{j}\in J_T\}$;
for all $C_1\in \{\tilde{a}_\mbbm{i}\gle \gu\}\cup S$ and $C_2\in S_T$,
$C_1=\tilde{a}_\mbbm{i}\gle \gu$, 
$\emptyset\neq \mi{atoms}(C_1)\cap \tilde{\mbb{A}}=\mi{atoms}(\tilde{a}_\mbbm{i}\gle \gu)\cap \tilde{\mbb{A}}=\{\tilde{a}_\mbbm{i}\}\cap \tilde{\mbb{A}}=\{\tilde{a}_\mbbm{i}\}\subseteq 
                                                   \{\tilde{a}_\mbbm{i}\}\cup \{\tilde{a}_\mbbm{j} \,|\, \mbbm{j}\in J\}$,
or $C_1\in S$, $\emptyset\neq \mi{atoms}(C_1)\cap \tilde{\mbb{A}}\subseteq \{\tilde{a}_\mbbm{i}\}\cup \{\tilde{a}_\mbbm{j} \,|\, \mbbm{j}\in J\}$;
$\emptyset\neq \mi{atoms}(C_1)\cap \tilde{\mbb{A}}\subseteq \{\tilde{a}_\mbbm{i}\}\cup \{\tilde{a}_\mbbm{j} \,|\, \mbbm{j}\in J\}$,
$\emptyset\neq \mi{atoms}(C_2)\cap \tilde{\mbb{A}}\subseteq \{\tilde{a}_\mbbm{j} \,|\, \mbbm{j}\in J_T\}$,
$(\mi{atoms}(C_1)\cap \tilde{\mbb{A}})\cap (\mi{atoms}(C_2)\cap \tilde{\mbb{A}})=
 (\{\tilde{a}_\mbbm{i}\}\cup \{\tilde{a}_\mbbm{j} \,|\, \mbbm{j}\in J\})\cap \{\tilde{a}_\mbbm{j} \,|\, \mbbm{j}\in J_T\}\overset{\text{(\ref{eq2c})}}{=\!\!=} \emptyset$;
for both $i$, $\mi{atoms}(C_i)\cap \tilde{\mbb{A}}\neq \emptyset$; 
$\mi{atoms}(C_1)\cap \tilde{\mbb{A}}\neq \mi{atoms}(C_2)\cap \tilde{\mbb{A}}$, $\mi{atoms}(C_1)\neq \mi{atoms}(C_2)$, $C_1\neq C_2$;
$(\{\tilde{a}_\mbbm{i}\gle \gu\}\cup S)\cap S_T=\emptyset$.

So, $\{\tilde{a}_\mbbm{i}\gle \gu\}\cap S=\emptyset$;
in all Cases 3.1--3.3, $(\{\tilde{a}_\mbbm{i}\gle \gu\}\cup S)\cap S_T=\emptyset$;
\begin{equation}
\notag
\{\tilde{a}_\mbbm{i}\gle \gu\}, S, S_T\ \text{are pairwise disjoint};
\end{equation}
(\ref{eq2d}) holds.
%
%
%
\end{proof}

\subsection{Full proof of Lemma \ref{le2}}
\label{S7.5aa}

\begin{proof}
We have that $S$ is positive and unit.
We denote $A=\mi{atoms}(S)=\{a_1,\dots,a_m\}$.
Then $S\subseteq_{\mc F} \mi{OrdPropCl}_A^{\gz,\gu}$, $S-\mi{guards}(S)\subseteq S$, $S-\mi{guards}(S)\subseteq_{\mc F} \mi{OrdPropCl}_A^\gu$;
$S$ is simplified;
$S-\mi{guards}(S)\subseteq S$ is simplified and unit;
for all $C\in S-\mi{guards}(S)$,
$C$ does not contain contradictions and tautologies;
$C$ is unit;
either $C\in \mi{PurOrdPropCl}_A$, $C=\Cn_1\diamond \Cn_2$, $\Cn_i\in \mi{PropConj}_A$, $\diamond\in \{\geql,\gleq,\gle\}$, $\Cn_1\neq \Cn_2$,
or $C=b_0\swedge\cdots\swedge b_k\gle \gu\vee C^\natural$, $b_j\in A=\mi{atoms}(S)$, $\mi{guards}(S,b_j)=\{\gz\gle b_j\}$, $C^\natural\in \mi{PurOrdPropCl}_A$, 
$C\not\in \mi{guards}(S)$, $k\geq 1$ or $C^\natural\neq \square$, $C^\natural=\square$, $C=b_0\swedge\cdots\swedge b_k\gle \gu$, $k\geq 1$;
$S$ is positively guarded;
for all $C\in \mi{guards}(S)=\bigcup_{a\in A=\mi{atoms}(S)} \mi{guards}(S,a)$,
there exists $a^*\in A=\mi{atoms}(S)$ satisfying $C\in \mi{guards}(S,a^*)$; 
$a^*$ is positively guarded in $S$;
either $C\in \mi{guards}(S,a^*)=\{\gz\gle a^*\}$ or $C\in \mi{guards}(S,a^*)=\{\gz\gle a^*,a^*\gle \gu\}$,
either $C=\gz\gle a^*$ or $C=a^*\gle \gu$;
for all $C\in S=(S-\mi{guards}(S))\cup \mi{guards}(S)$,
either $C=\Cn_1\diamond \Cn_2$, $\Cn_i\in \mi{PropConj}_A$, $\diamond\in \{\geql,\gleq,\gle\}$, $\Cn_1\neq \Cn_2$,
or $C=b_0\swedge\cdots\swedge b_k\gle \gu$, $b_j\in A=\mi{atoms}(S)$, $\mi{guards}(S,b_j)=\{\gz\gle b_j\}$, $k\geq 1$,
or $C=\gz\gle a$ or $C=a\gle \gu$, $a\in A$.

Let $\varepsilon_1, \varepsilon_2\in \{\gu\}\cup \mi{PropConj}_A$ and $\diamond\in \{\geql,\gleq,\gle\}$.
We define six auxiliary operators as follows:
\begin{IEEEeqnarray*}{RL}
\IEEEeqnarraymulticol{2}{l}{
\# : A\times (\{\gu\}\cup \mi{PropLit}_A^\gu)\longrightarrow \mbb{Z};} \\[1mm]
\#(a,\gu)                                           &= 0; \\[1mm]
\#(a,\varepsilon_1\diamond \varepsilon_2)           &= \left\{\begin{array}{ll}
                                                              s-r &\ \text{\it if}\ a^r\in \varepsilon_1\in \mi{PropConj}_A, a^s\in \varepsilon_2\in \mi{PropConj}_A, r, s\geq 1, \\[1mm]
                                                              -r  &\ \text{\it if}\ a^r\in \varepsilon_1\in \mi{PropConj}_A, r\geq 1, a\not\in \mi{atoms}(\varepsilon_2), \\[1mm]
                                                              s   &\ \text{\it if}\ a\not\in \mi{atoms}(\varepsilon_1), a^s\in \varepsilon_2\in \mi{PropConj}_A, s\geq 1, \\[1mm]
                                                              0   &\ \text{\it if}\ a\not\in \mi{atoms}(\varepsilon_1), a\not\in \mi{atoms}(\varepsilon_2);
                                                              \end{array}
                                                       \right. \\[1mm]
\mi{rcd}^-(\gu)                                     &= \emptyset; \\
\mi{rcd}^-(\varepsilon_1\diamond \varepsilon_2)     &= \{a^r \,|\, a\in \mi{atoms}(\varepsilon_1), r=-\#(a,\varepsilon_1\diamond \varepsilon_2)\geq 1\}; \\
                                                    &\subseteq_{\mc F} \mi{PropPow}_A; \\[1mm]
\mi{reduced}^-(\gu)                                 &= \gu; \\[1mm]
\mi{reduced}^-(\varepsilon_1\diamond \varepsilon_2) &= \left\{\begin{array}{ll}
                                                              \gu                                             &\ \text{\it if}\ \mi{rcd}^-(\varepsilon_1\diamond \varepsilon_2)=\emptyset, \\[1mm]
                                                              \mi{rcd}^-(\varepsilon_1\diamond \varepsilon_2) &\ \text{\it if}\ \mi{rcd}^-(\varepsilon_1\diamond \varepsilon_2)\neq \emptyset;
                                                              \end{array}
                                                       \right. \\[1mm]
                                                    &\in \{\gu\}\cup \mi{PropConj}_A; \\[1mm]
\mi{rcd}^+(\gu)                                     &= \emptyset; \\
\mi{rcd}^+(\varepsilon_1\diamond \varepsilon_2)     &= \{a^r \,|\, a\in \mi{atoms}(\varepsilon_2), r=\#(a,\varepsilon_1\diamond \varepsilon_2)\geq 1\}; \\
                                                    &\subseteq_{\mc F} \mi{PropPow}_A; \\[1mm]
\mi{reduced}^+(\gu)                                 &= \gu; \\[1mm]
\mi{reduced}^+(\varepsilon_1\diamond \varepsilon_2) &= \left\{\begin{array}{ll}
                                                              \gu                                             &\ \text{\it if}\ \mi{rcd}^+(\varepsilon_1\diamond \varepsilon_2)=\emptyset, \\[1mm]
                                                              \mi{rcd}^+(\varepsilon_1\diamond \varepsilon_2) &\ \text{\it if}\ \mi{rcd}^+(\varepsilon_1\diamond \varepsilon_2)\neq \emptyset; 
                                                              \end{array}
                                                       \right. \\[1mm]
                                                    &\in \{\gu\}\cup \mi{PropConj}_A; \\[1mm]
\mi{reduced}(\gu)                                   &= \gu; \\
\mi{reduced}(\varepsilon_1\diamond \varepsilon_2)   &= \mi{reduced}^-(\varepsilon_1\diamond \varepsilon_2)\diamond \mi{reduced}^+(\varepsilon_1\diamond \varepsilon_2); \\
                                                    &\in \{\gu\}\cup \mi{PropLit}_A^\gu.
\end{IEEEeqnarray*}
Let $\triangle\in \{-,+\}$.
Note that if $\mi{rcd}^\triangle(\varepsilon_1\diamond \varepsilon_2)=\emptyset$, $\mi{reduced}^\triangle(\varepsilon_1\diamond \varepsilon_2)=\gu$;
if $\mi{rcd}^\triangle(\varepsilon_1\diamond \varepsilon_2)\neq \emptyset$, 
for all $a^r, b^s\in \mi{rcd}^\triangle(\varepsilon_1\diamond \varepsilon_2)\subseteq_{\mc F} \mi{PropPow}_A$, $a, b\in A$, $r, s\geq 1$, satisfying $a^r\neq b^s$, $a\neq b$;
$\mi{reduced}^\triangle(\varepsilon_1\diamond \varepsilon_2)=\mi{rcd}^\triangle(\varepsilon_1\diamond \varepsilon_2)\in \mi{PropConj}_A$;
$\mi{reduced}^\triangle(\varepsilon_1\diamond \varepsilon_2)\in \{\gu\}\cup \mi{PropConj}_A$.
From an input literal not containing $\gz$, we compute a reduced form $\upsilon_1\diamond^\# \upsilon_2$, $\upsilon_i\in \{\gu\}\cup \mi{PropConj}_A$, $\diamond^\#\in \{\geql,\gleq,\gle\}$, of it, 
not containing $\gz$ as well as $\mi{atoms}(\upsilon_1)\cap \mi{atoms}(\upsilon_2)=\emptyset$.
For example, $\mi{reduced}(a^3\swedge b^2\geql a^2\swedge c^3)=a\swedge b^2\geql c^3$, $\mi{reduced}(a^3\gleq a^4\swedge c)=\gu\gleq a\swedge c$, $\mi{reduced}(a^3\gle a^3)=\gu\gle \gu$, $a, b, c\in A$.

Let $\alpha, \alpha'\geq 1$, $\gamma\in \mbb{N}$, $a\in A$, $\varepsilon, \varepsilon_1', \varepsilon_2'\in \{\gu\}\cup \mi{PropConj}_A$, $\diamond^\#\in \{\geql,\gleq,\gle\}$, 
${\mc V}$ be a valuation such that ${\mc V}[A]\subseteq (0,1]$.
The following properties hold:
\begin{alignat}{1}
\label{eq7ax}
& \#(a,\varepsilon_1\diamond \varepsilon_2)=-\#(a,\varepsilon_2\diamond \varepsilon_1); \\
\label{eq7aax}
& \#(a,(\varepsilon_1\diamond \varepsilon_2)\mult (\varepsilon_1'\diamond^\# \varepsilon_2'))=\#(a,\varepsilon_1\diamond \varepsilon_2)+\#(a,\varepsilon_1'\diamond^\# \varepsilon_2'); \\
\label{eq7aaax}
& \#(a,(\varepsilon_1\diamond \varepsilon_2)^\gamma)=\gamma\cdot \#(a,\varepsilon_1\diamond \varepsilon_2); \\
\label{eq7aaaax}
& \#(a,\mi{reduced}(\varepsilon_1\diamond \varepsilon_2))=\#(a,\varepsilon_1\diamond \varepsilon_2); \\
\label{eq7a}
& \mi{atoms}(\mi{reduced}^-(\varepsilon_1\diamond \varepsilon_2))\subseteq \mi{atoms}(\varepsilon_1); \\
\label{eq7b}
& \mi{atoms}(\mi{reduced}^+(\varepsilon_1\diamond \varepsilon_2))\subseteq \mi{atoms}(\varepsilon_2); \\
\label{eq7c}
& \text{if}\ a\not\in \mi{atoms}(\varepsilon_1),\
  \text{then}\ \mi{reduced}^-(\varepsilon_1\diamond (\varepsilon_2\mult a^\alpha))=\mi{reduced}^-(\varepsilon_1\diamond \varepsilon_2); \\
\label{eq7d}
& \text{if}\ a\not\in \mi{atoms}(\varepsilon_2),\
  \text{then}\ \mi{reduced}^+((\varepsilon_1\mult a^\alpha)\diamond \varepsilon_2)=\mi{reduced}^+(\varepsilon_1\diamond \varepsilon_2); \\
\label{eq7e}
& \text{if}\ a\not\in \mi{atoms}(\varepsilon_2),\
  \text{then}\ \mi{reduced}^-((\varepsilon_1\mult a^\alpha)\diamond \varepsilon_2)=\mi{reduced}^-(\varepsilon_1\diamond \varepsilon_2)\mult a^\alpha; \\
\label{eq7f}
& \text{if}\ a\not\in \mi{atoms}(\varepsilon_1),\
  \text{then}\ \mi{reduced}^+(\varepsilon_1\diamond (\varepsilon_2\mult a^\alpha))=\mi{reduced}^+(\varepsilon_1\diamond \varepsilon_2)\mult a^\alpha; \\[1mm]
\label{eq7cc}
& \text{if}\ a\not\in \mi{atoms}(\varepsilon_1), a\not\in \mi{atoms}(\varepsilon_2), \#(a,(\varepsilon_1\mult a^\alpha)\diamond (\varepsilon_2\mult a^{\alpha'}))=\alpha'-\alpha\geq 1, \\
\notag
& \quad \text{then}\ \mi{reduced}^-((\varepsilon_1\mult a^\alpha)\diamond (\varepsilon_2\mult a^{\alpha'}))=\mi{reduced}^-(\varepsilon_1\diamond \varepsilon_2); \\[1mm]
\label{eq7dd}
& \text{if}\ a\not\in \mi{atoms}(\varepsilon_1), a\not\in \mi{atoms}(\varepsilon_2), \#(a,(\varepsilon_1\mult a^\alpha)\diamond (\varepsilon_2\mult a^{\alpha'}))=\alpha'-\alpha\leq -1, \\
\notag
& \quad \text{then}\ \mi{reduced}^+((\varepsilon_1\mult a^\alpha)\diamond (\varepsilon_2\mult a^{\alpha'}))=\mi{reduced}^+(\varepsilon_1\diamond \varepsilon_2); \\[1mm]
\label{eq7ee}
& \text{if}\ a\not\in \mi{atoms}(\varepsilon_1), a\not\in \mi{atoms}(\varepsilon_2), \#(a,(\varepsilon_1\mult a^\alpha)\diamond (\varepsilon_2\mult a^{\alpha'}))=\alpha'-\alpha\leq -1, \\
\notag
& \quad \text{then}\ \mi{reduced}^-((\varepsilon_1\mult a^\alpha)\diamond (\varepsilon_2\mult a^{\alpha'}))=\mi{reduced}^-(\varepsilon_1\diamond \varepsilon_2)\mult a^{\alpha-\alpha'}; \\[1mm]
\label{eq7ff}
& \text{if}\ a\not\in \mi{atoms}(\varepsilon_1), a\not\in \mi{atoms}(\varepsilon_2), \#(a,(\varepsilon_1\mult a^\alpha)\diamond (\varepsilon_2\mult a^{\alpha'}))=\alpha'-\alpha\geq 1, \\
\notag
& \quad \text{then}\ \mi{reduced}^+((\varepsilon_1\mult a^\alpha)\diamond (\varepsilon_2\mult a^{\alpha'}))=\mi{reduced}^+(\varepsilon_1\diamond \varepsilon_2)\mult a^{\alpha'-\alpha}; \\[1mm]
\label{eq7g}
& \mi{atoms}(\mi{reduced}(\varepsilon_1\diamond \varepsilon_2))\subseteq \mi{atoms}(\varepsilon_1\diamond \varepsilon_2); \\
\label{eq7gg}
& \text{if}\ \mi{reduced}(\varepsilon_1\diamond \varepsilon_2)=\varepsilon_1'\diamond \varepsilon_2',\
  \text{then}\ \mi{atoms}(\varepsilon_1')\cap \mi{atoms}(\varepsilon_2')=\emptyset; \\
\label{eq7ggg}
& \text{if}\ \mi{atoms}(\varepsilon_1)\cap \mi{atoms}(\varepsilon_2)=\emptyset,\
  \text{then}\ \mi{reduced}(\varepsilon_1\diamond \varepsilon_2)=\varepsilon_1\diamond \varepsilon_2; \\
\label{eq7h}
& \mi{reduced}((\varepsilon_1\mult \varepsilon)\diamond (\varepsilon_2\mult \varepsilon))=\mi{reduced}(\varepsilon_1\diamond \varepsilon_2); \\[1mm]
\label{eq7i}
& \varepsilon_1\diamond \varepsilon_2=(\mi{reduced}^-(\varepsilon_1\diamond \varepsilon_2)\mult \varepsilon')\diamond (\mi{reduced}^+(\varepsilon_1\diamond \varepsilon_2)\mult \varepsilon'), \\
\notag
& \phantom{\varepsilon_1\diamond \varepsilon_2=\mbox{}}            
                                      \varepsilon'\in \{\gu\}\cup \mi{PropConj}_A; \\[1mm] 
\label{eq7j}
& \mi{reduced}(\mi{reduced}(\varepsilon_1\diamond \varepsilon_2))=\mi{reduced}(\varepsilon_1\diamond \varepsilon_2); \\
\label{eq7k}
& \mi{reduced}(\mi{reduced}(\varepsilon_1\diamond \varepsilon_2)\mult (\varepsilon_1'\diamond^\# \varepsilon_2'))=
  \mi{reduced}((\varepsilon_1\diamond \varepsilon_2)\mult (\varepsilon_1'\diamond^\# \varepsilon_2')); \\
\label{eq7l}
& \mi{reduced}((\varepsilon_1\diamond \varepsilon_2)^\gamma)=\mi{reduced}(\varepsilon_1\diamond \varepsilon_2)^\gamma; \\
\label{eq7m}      
& \|\mi{reduced}(\varepsilon_1\diamond \varepsilon_2)\|^{\mc V}=\|\varepsilon_1\diamond \varepsilon_2\|^{\mc V}; \\
\label{eq7n}
& \dfrac{\|\mi{reduced}^-(\varepsilon_1\diamond \varepsilon_2)\|^{\mc V}}   
        {\|\mi{reduced}^+(\varepsilon_1\diamond \varepsilon_2)\|^{\mc V}}=
  \dfrac{\|\varepsilon_1\|^{\mc V}}
        {\|\varepsilon_2\|^{\mc V}}.
\end{alignat}
The proof is straightforward by definition.

We denote
\begin{alignat*}{1}
\Lambda                &= (S\cap \mi{OrdPropCl}_A^\gu)\cup \{a\gleq \gu \,|\, a\in A, a\gle \gu\not\in \mi{guards}(S)\}\subseteq \mi{OrdPropCl}_A^\gu, \\
\langle \Lambda\rangle &= \{\langle l\rangle \,|\, l\in \Lambda\}=\{\lambda_1,\dots,\lambda_n\}\subseteq \mi{PropLit}_A^\gu, \\[1mm]
\mi{clo}               &= \{l \,|\, \text{\it there exist}\ \{l_0,\dots,l_\kappa\}\subseteq \Lambda, \langle l_0\rangle,\dots,\langle l_\kappa\rangle\in \langle \Lambda\rangle, \\ 
                       &\phantom{\mbox{}=\{l \,|\, \text{\it there exist}\ \mbox{}}
                                                            \alpha_0,\dots,\alpha_\kappa, \alpha_j\geq 1,\ \text{\it such that}\ 
                                                            l=\textstyle{\mi{reduced}(\bigmult_{j=0}^\kappa \langle l_j\rangle^{\alpha_j})}\} \\
                       &\subseteq \mi{PropLit}_A^\gu.
\end{alignat*}
Note that for all $j\leq \kappa$,
$l_j\in \Lambda=(S\cap \mi{OrdPropCl}_A^\gu)\cup \{a\gleq \gu \,|\, a\in A, a\gle \gu\not\in \mi{guards}(S)\}\subseteq \mi{OrdPropCl}_A^\gu$, 
$l_j\in S\cap \mi{OrdPropCl}_A^\gu$ or $l_j\in \{a\gleq \gu \,|\, a\in A, a\gle \gu\not\in \mi{guards}(S)\}$;
either $l_j=\Cn_1\diamond^\natural \Cn_2$, $\Cn_e\in \mi{PropConj}_A$, $\diamond^\natural\in \{\geql,\gleq,\gle\}$, $\Cn_1\neq \Cn_2$,
or $l_j=b_0\swedge\cdots\swedge b_k\gle \gu$, $b_e\in A$, $k\geq 1$,
or $l_j=\gz\gle a$ or $l_j=a\gle \gu$, $a\in A$, $l_j\in \mi{OrdPropCl}_A^\gu$, $\gz\gle a\not\in \mi{OrdPropCl}_A^\gu$, $l_j\neq \gz\gle a$, $l_j=a\gle \gu$,
or $l_j=a\gleq \gu$, $a\in A$;
either $l_j=\Cn_1\diamond^\natural \Cn_2$, $\Cn_e\in \mi{PropConj}_A$, $\diamond^\natural\in \{\geql,\gleq,\gle\}$, $\Cn_1\neq \Cn_2$,
or $l_j=b_0\swedge\cdots\swedge b_k\gle \gu$, $b_e\in A$, 
or $l_j=a\gleq \gu$, $a\in A$;
for all $j<j'\leq \kappa$, $l_j\neq l_{j'}$, $\langle l_j\rangle\neq \langle l_{j'}\rangle$.

We define an auxiliary binary relation $\preceq$ on $\mbb{N}^n$ component-wisely.
Let $(c_1,\dots,c_n), (d_1,\dots,d_n)\in \mbb{N}^n$.
$(c_1,\dots,c_n)\preceq (d_1,\dots,d_n)$ iff, for all $1\leq i\leq n$, $c_i\leq d_i$.
Note that $\preceq$ is reflexive, antisymmetric, transitive, a well-founded order on $\mbb{N}^n$.
$(c_1,\dots,c_n)\prec (d_1,\dots,d_n)$ iff $(c_1,\dots,c_n)\preceq (d_1,\dots,d_n)$ and $(c_1,\dots,c_n)\neq (d_1,\dots,d_n)$.
Note that $\prec$ is irreflexive, transitive, a strict order on $\mbb{N}^n$.
Let $K\subseteq \mbb{N}^n$.
By $\mi{min}(K)\subseteq K$ we denote the set of all minimal elements of $K$ with respect to $\preceq$.
We denote $\bs{0}^n=(\underbrace{0,\dots,0}_n)\in \mbb{N}^n$.
We denote $\lambda^{(c_1,\dots,c_n)}=\bigmult_{i=1}^n \lambda_i^{c_i}\in \{\gu\}\cup \mi{PropLit}_A^\gu$.
Note that $\lambda^{\bs{0}^n}=\gu$.
\begin{alignat}{1}
\label{eq8e}
& \begin{minipage}[t]{\linewidth-15mm}
  Let $K\subseteq \mbb{N}^n$.
  $\mi{min}(K)\subseteq_{\mc F} K$.
  \end{minipage}
\end{alignat}   
An immediate consequence of that $\preceq$ is a well-founded order on $\mbb{N}^n$, using the Dirichlet's Box Principle.

The following properties hold:
\begin{alignat}{1}
\label{eq8a}
& \gu, \gu\gle \gu\not\in \mi{clo}; \\
\label{eq8aa}
& \text{either}\ a\gle \gu\in \mi{clo}\ \text{or}\ a\gleq \gu\in \mi{clo}; \\
\label{eq8aaa}
& \text{if}\ \varepsilon_1\diamond \varepsilon_2\in \mi{clo},\ \text{then}\ \mi{atoms}(\varepsilon_1)\cap \mi{atoms}(\varepsilon_2)=\emptyset; \\
\label{eq8b}
& \text{if}\ \varepsilon_1\geql \varepsilon_2\in \mi{clo},\ \text{then}\ \varepsilon_2\geql \varepsilon_1\in \mi{clo}; \\
\label{eq8c}  
& \text{if}\ \varepsilon_1\diamond \varepsilon_2, \varepsilon_1'\diamond^\# \varepsilon_2'\in \mi{clo},\ 
  \text{then}\ \mi{reduced}((\varepsilon_1\diamond \varepsilon_2)\mult (\varepsilon_1'\diamond^\# \varepsilon_2'))\in \mi{clo}; \\
\label{eq8d}  
& \text{if}\ \varepsilon_1\diamond \varepsilon_2\in \mi{clo},\ \text{then}\ (\varepsilon_1\diamond \varepsilon_2)^\alpha\in \mi{clo}; \\
\label{eq8dd}
& \text{if}\ (c_1,\dots,c_n)\neq \bs{0}^n,\ \text{then}\ \mi{reduced}(\lambda^{(c_1,\dots,c_n)})\in \mi{clo}; \\[1mm]
\label{eq8ddd}
& \text{if}\ l\in \mi{clo},\ \text{then there exists}\ \bs{0}^n\neq (c_1^*,\dots,c_n^*)\in \mbb{N}^n\ \text{such that} \\
\notag 
& \phantom{\text{if}\ l\in \mi{clo},\ \text{then} \mbox{}} \quad
                                        l=\mi{reduced}(\lambda^{(c_1^*,\dots,c_n^*)}); \\[1mm]
\label{eq8dddd}
& \lambda^{(c_1+d_1,\dots,c_n+d_n)}=\lambda^{(c_1,\dots,c_n)}\mult \lambda^{(d_1,\dots,d_n)}; \\
\label{eq8aax}
& \#(a,\lambda^{(c_1,\dots,c_n)}\mult \lambda^{(d_1,\dots,d_n)})=\#(a,\lambda^{(c_1,\dots,c_n)})+\#(a,\lambda^{(d_1,\dots,d_n)}); \\
\label{eq8aaax}
& \#(a,(\lambda^{(c_1,\dots,c_n)})^\gamma)=\gamma\cdot \#(a,\lambda^{(c_1,\dots,c_n)}); \\
\label{eq8aaaax}
& \#(a,\mi{reduced}(\lambda^{(c_1,\dots,c_n)}))=\#(a,\lambda^{(c_1,\dots,c_n)}).
\end{alignat}
The proof is straightforward by definition; (\ref{eq8a}) is an immediate consequence of that there does not exist an application of Rule (\cref{ceq4hr0}) to $S$.

Let $X\subseteq A$.
A partial valuation ${\mc V}$ is a mapping ${\mc V} : X\longrightarrow [0,1]$.
We denote $\mi{dom}({\mc V})=X\subseteq A$.
For all $\iota\leq m$, we define a partial valuation ${\mc V}_\iota$ by recursion on $\iota$ as follows:%
\footnote{For a partial valuation ${\mc V}$, the truth value $\|\varepsilon\|^{\mc V}\in [0,1]$ of $\varepsilon$ in ${\mc V}$ is defined equally as with a valuation.}
\footnote{For $c\in [0,1]$ and $r\geq 1$, $c^{\frac{1}{r}}\in [0,1]$ denotes the unique real $r$th root of $c$ in $[0,1]$.} 
\footnote{Note that $\bigfvee \emptyset=0$ and $\bigfwedge \emptyset=1$.}
\begin{alignat*}{2}
& {\mc V}_0     & &= \emptyset; \\[1mm]
& {\mc V}_\iota & &= {\mc V}_{\iota-1}\cup \{(a_\iota,\delta_\iota)\} \quad (1\leq \iota\leq m), \\[1mm]
& & &                
\begin{alignedat}{2}
& E_{\iota-1}       & &= \{l \,|\, l=(\varepsilon_1\mult a_\iota^\alpha)\geql \varepsilon_2\in \mi{clo}, 
                                   \varepsilon_i\in \{\gu\}\cup \mi{PropConj}_A, \alpha\geq 1, \\
                  & & &\phantom{\mbox{}=\{l \,|\, \mbox{}}
                                   \mi{atoms}(\varepsilon_i)\subseteq \mi{dom}({\mc V}_{\iota-1})\}, \\[1mm]
& \mi{DE}_{\iota-1} & &= \{l \,|\, l=\varepsilon_2\gleq (\varepsilon_1\mult a_\iota^\alpha)\in \mi{clo},
                                   \varepsilon_i\in \{\gu\}\cup \mi{PropConj}_A, \alpha\geq 1, \\
                  & & &\phantom{\mbox{}=\{l \,|\, \mbox{}}
                                   \mi{atoms}(\varepsilon_i)\subseteq \mi{dom}({\mc V}_{\iota-1})\}, \\[1mm]
& D_{\iota-1}       & &= \{l \,|\, l=\varepsilon_2\gle (\varepsilon_1\mult a_\iota^\alpha)\in \mi{clo},
                                   \varepsilon_i\in \{\gu\}\cup \mi{PropConj}_A, \alpha\geq 1, \\
                  & & &\phantom{\mbox{}=\{l \,|\, \mbox{}} 
                                   \mi{atoms}(\varepsilon_i)\subseteq \mi{dom}({\mc V}_{\iota-1})\}, \\[1mm]
& \mi{UE}_{\iota-1} & &= \{l \,|\, l=(\varepsilon_1\mult a_\iota^\alpha)\gleq \varepsilon_2\in \mi{clo},
                                   \varepsilon_i\in \{\gu\}\cup \mi{PropConj}_A, \alpha\geq 1, \\
                  & & &\phantom{\mbox{}=\{l \,|\, \mbox{}}
                                   \mi{atoms}(\varepsilon_i)\subseteq \mi{dom}({\mc V}_{\iota-1})\}, \\[1mm]
& U_{\iota-1}       & &= \{l \,|\, l=(\varepsilon_1\mult a_\iota^\alpha)\gle \varepsilon_2\in \mi{clo},
                                   \varepsilon_i\in \{\gu\}\cup \mi{PropConj}_A, \alpha\geq 1, \\
                  & & &\phantom{\mbox{}=\{l \,|\, \mbox{}}
                                   \mi{atoms}(\varepsilon_i)\subseteq \mi{dom}({\mc V}_{\iota-1})\}, \\[1mm]
& \mi{min}(E_{\iota-1})       & &= \{l \,|\, l=\mi{reduced}(\lambda^{(\gamma_1,\dots,\gamma_n)})\in E_{\iota-1}, 
                                             \bs{0}^n\underset{\text{(\ref{eq8ddd})}}{\neq} (\gamma_1,\dots,\gamma_n)\in \mbb{N}^n, \\
                            & & &\phantom{\mbox{}=\{l \,|\, \mbox{}}
                                             \text{\it for all}\ (\gamma_1',\dots,\gamma_n')\in \mbb{N}^n\ \text{\it such that}\ (\gamma_1',\dots,\gamma_n')\prec (\gamma_1,\dots,\gamma_n), \\
                            & & &\phantom{\mbox{}=\{l \,|\, \mbox{}} \quad
                                             \mi{reduced}(\lambda^{(\gamma_1',\dots,\gamma_n')})\not\in E_{\iota-1}\}\subseteq \mi{clo}, \\[1mm]
& \mi{min}(\mi{DE}_{\iota-1}) & &= \{l \,|\, l=\mi{reduced}(\lambda^{(\gamma_1,\dots,\gamma_n)})\in \mi{DE}_{\iota-1}, 
                                             \bs{0}^n\underset{\text{(\ref{eq8ddd})}}{\neq} (\gamma_1,\dots,\gamma_n)\in \mbb{N}^n, \\
                            & & &\phantom{\mbox{}=\{l \,|\, \mbox{}}
                                             \text{\it for all}\ (\gamma_1',\dots,\gamma_n')\in \mbb{N}^n\ \text{\it such that}\ (\gamma_1',\dots,\gamma_n')\prec (\gamma_1,\dots,\gamma_n), \\
                            & & &\phantom{\mbox{}=\{l \,|\, \mbox{}} \quad
                                             \mi{reduced}(\lambda^{(\gamma_1',\dots,\gamma_n')})\not\in \mi{DE}_{\iota-1}\}\subseteq \mi{clo}, \\[1mm]
& \mi{min}(D_{\iota-1})       & &= \{l \,|\, l=\mi{reduced}(\lambda^{(\gamma_1,\dots,\gamma_n)})\in D_{\iota-1}, 
                                             \bs{0}^n\underset{\text{(\ref{eq8ddd})}}{\neq} (\gamma_1,\dots,\gamma_n)\in \mbb{N}^n, \\
                            & & &\phantom{\mbox{}=\{l \,|\, \mbox{}}
                                             \text{\it for all}\ (\gamma_1',\dots,\gamma_n')\in \mbb{N}^n\ \text{\it such that}\ (\gamma_1',\dots,\gamma_n')\prec (\gamma_1,\dots,\gamma_n), \\
                            & & &\phantom{\mbox{}=\{l \,|\, \mbox{}} \quad
                                             \mi{reduced}(\lambda^{(\gamma_1',\dots,\gamma_n')})\not\in D_{\iota-1}\}\subseteq \mi{clo}, \\[1mm]
& \mi{min}(\mi{UE}_{\iota-1}) & &= \{l \,|\, l=\mi{reduced}(\lambda^{(\gamma_1,\dots,\gamma_n)})\in \mi{UE}_{\iota-1}, 
                                             \bs{0}^n\underset{\text{(\ref{eq8ddd})}}{\neq} (\gamma_1,\dots,\gamma_n)\in \mbb{N}^n, \\
                            & & &\phantom{\mbox{}=\{l \,|\, \mbox{}}
                                             \text{\it for all}\ (\gamma_1',\dots,\gamma_n')\in \mbb{N}^n\ \text{\it such that}\ (\gamma_1',\dots,\gamma_n')\prec (\gamma_1,\dots,\gamma_n), \\
                            & & &\phantom{\mbox{}=\{l \,|\, \mbox{}} \quad
                                             \mi{reduced}(\lambda^{(\gamma_1',\dots,\gamma_n')})\not\in \mi{UE}_{\iota-1}\}\subseteq \mi{clo}, \\[1mm]
& \mi{min}(U_{\iota-1})       & &= \{l \,|\, l=\mi{reduced}(\lambda^{(\gamma_1,\dots,\gamma_n)})\in U_{\iota-1}, 
                                             \bs{0}^n\underset{\text{(\ref{eq8ddd})}}{\neq} (\gamma_1,\dots,\gamma_n)\in \mbb{N}^n, \\
                            & & &\phantom{\mbox{}=\{l \,|\, \mbox{}}
                                             \text{\it for all}\ (\gamma_1',\dots,\gamma_n')\in \mbb{N}^n\ \text{\it such that}\ (\gamma_1',\dots,\gamma_n')\prec (\gamma_1,\dots,\gamma_n), \\
                            & & &\phantom{\mbox{}=\{l \,|\, \mbox{}} \quad
                                             \mi{reduced}(\lambda^{(\gamma_1',\dots,\gamma_n')})\not\in U_{\iota-1}\}\subseteq \mi{clo}, 
\end{alignedat} 
\end{alignat*}
\begin{alignat*}{2}
& \hphantom{{\mc V}_0} & &
\begin{alignedat}{2}
& \hphantom{\mi{min}(\mi{DE}_{\iota-1})} & &\hphantom{\mbox{}=\{l \,|\, \text{\it for all}\ (\gamma_1',\dots,\gamma_n')\in \mbb{N}^n\ \text{\it such that}\ (\gamma_1',\dots,\gamma_n')\prec (\gamma_1,\dots,\gamma_n),} \\[-\baselineskip]
& \mbb{E}_{\iota-1}  & &= \left\{\left(\dfrac{\|\varepsilon_2\|^{{\mc V}_{\iota-1}}}
                                             {\|\varepsilon_1\|^{{\mc V}_{\iota-1}}}\right)^{\dfrac{1}{\alpha}}\ \begin{array}{c} \bigsetver \end{array}\ 
                                 \begin{array}{l}
                                 (\varepsilon_1\mult a_\iota^\alpha)\geql \varepsilon_2\in \mi{min}(E_{\iota-1}), \\
                                 \varepsilon_i\in \{\gu\}\cup \mi{PropConj}_A, \alpha\geq 1, \\
                                 \mi{atoms}(\varepsilon_i)\subseteq \mi{dom}({\mc V}_{\iota-1})
                                 \end{array}\right\}, \\[1mm]
& \mbb{DE}_{\iota-1} & &= \left\{\left(\dfrac{\|\varepsilon_2\|^{{\mc V}_{\iota-1}}}
                                             {\|\varepsilon_1\|^{{\mc V}_{\iota-1}}}\right)^{\dfrac{1}{\alpha}}\ \begin{array}{c} \bigsetver \end{array}\ 
                                 \begin{array}{l}
                                 \varepsilon_2\gleq (\varepsilon_1\mult a_\iota^\alpha)\in \mi{min}(\mi{DE}_{\iota-1}), \\
                                 \varepsilon_i\in \{\gu\}\cup \mi{PropConj}_A, \alpha\geq 1, \\
                                 \mi{atoms}(\varepsilon_i)\subseteq \mi{dom}({\mc V}_{\iota-1})
                                 \end{array}\right\}, \\[1mm]
& \mbb{D}_{\iota-1}  & &= \left\{\left(\dfrac{\|\varepsilon_2\|^{{\mc V}_{\iota-1}}}
                                             {\|\varepsilon_1\|^{{\mc V}_{\iota-1}}}\right)^{\dfrac{1}{\alpha}}\ \begin{array}{c} \bigsetver \end{array}\ 
                                 \begin{array}{l}
                                 \varepsilon_2\gle (\varepsilon_1\mult a_\iota^\alpha)\in \mi{min}(D_{\iota-1}), \\
                                 \varepsilon_i\in \{\gu\}\cup \mi{PropConj}_A, \alpha\geq 1, \\
                                 \mi{atoms}(\varepsilon_i)\subseteq \mi{dom}({\mc V}_{\iota-1})                                        
                                 \end{array}\right\}, \\[1mm]
& \mbb{UE}_{\iota-1} & &= \left\{\mi{min}\left(\left(\dfrac{\|\varepsilon_2\|^{{\mc V}_{\iota-1}}}
                                                           {\|\varepsilon_1\|^{{\mc V}_{\iota-1}}}\right)^{\dfrac{1}{\alpha}},1\right)\ \begin{array}{c} \bigsetver \end{array}\ 
                                 \begin{array}{l}
                                 (\varepsilon_1\mult a_\iota^\alpha)\gleq \varepsilon_2\in \mi{min}(\mi{UE}_{\iota-1}), \\
                                 \varepsilon_i\in \{\gu\}\cup \mi{PropConj}_A, \alpha\geq 1, \\
                                 \mi{atoms}(\varepsilon_i)\subseteq \mi{dom}({\mc V}_{\iota-1})
                                 \end{array}\right\}, \\[1mm]
& \mbb{U}_{\iota-1}  & &= \left\{\mi{min}\left(\left(\dfrac{\|\varepsilon_2\|^{{\mc V}_{\iota-1}}}
                                                           {\|\varepsilon_1\|^{{\mc V}_{\iota-1}}}\right)^{\dfrac{1}{\alpha}},1\right)\ \begin{array}{c} \bigsetver \end{array}\ 
                                 \begin{array}{l}
                                 (\varepsilon_1\mult a_\iota^\alpha)\gle \varepsilon_2\in \mi{min}(U_{\iota-1}), \\
                                 \varepsilon_i\in \{\gu\}\cup \mi{PropConj}_A, \alpha\geq 1, \\
                                 \mi{atoms}(\varepsilon_i)\subseteq \mi{dom}({\mc V}_{\iota-1})
                                 \end{array}\right\}, \\[1mm]
& \delta_\iota       & &= \left\{\begin{array}{ll}
                                 \dfrac{\mi{max}(\bigfvee \mbb{DE}_{\iota-1},\bigfvee \mbb{D}_{\iota-1})+\mi{min}(\bigfwedge \mbb{UE}_{\iota-1},\bigfwedge \mbb{U}_{\iota-1})}{2} 
                                                            &\ \ \text{\it if}\ \mbb{E}_{\iota-1}=\emptyset, \\[1mm]
                                 \bigfvee \mbb{E}_{\iota-1} &\ \ \text{\it if}\ \mbb{E}_{\iota-1}\neq \emptyset.
                                 \end{array}
                          \right. 
\end{alignedat} 
\end{alignat*}
Note that for all $1\leq \iota\leq m$, $E_{\iota-1}$, $\mi{DE}_{\iota-1}$, $D_{\iota-1}$, $\mi{UE}_{\iota-1}$, $U_{\iota-1}$ are pairwise disjoint, and hence,
$\mi{min}(E_{\iota-1})\subseteq E_{\iota-1}$, $\mi{min}(\mi{DE}_{\iota-1})\subseteq \mi{DE}_{\iota-1}$, $\mi{min}(D_{\iota-1})\subseteq D_{\iota-1}$, 
$\mi{min}(\mi{UE}_{\iota-1})\subseteq \mi{UE}_{\iota-1}$, $\mi{min}(U_{\iota-1})\subseteq U_{\iota-1}$ are pairwise disjoint as well.
\begin{alignat}{1}
\label{eq8f}
& \begin{minipage}[t]{\linewidth-15mm}
  For all $\iota\leq m$,
  \begin{enumerate}[\rm (a)]
  \item
  ${\mc V}_\iota$ is a partial valuation ${\mc V}_\iota : \{a_1,\dots,a_\iota\}\longrightarrow (0,1]$, $\mi{dom}({\mc V}_\iota)=\{a_1,\dots,a_\iota\}$;
  \item
  for all $\varsigma\leq \iota$, ${\mc V}_\varsigma\subseteq {\mc V}_\iota$;
  \item[]
  for all $\varepsilon_1, \varepsilon_2\in \{\gu\}\cup \mi{PropConj}_A$ satisfying $\mi{atoms}(\varepsilon_1), \mi{atoms}(\varepsilon_2)\subseteq \mi{dom}({\mc V}_\iota)$,
  \begin{enumerate}[\rm (c)]
  \item[\rm (c)]
  if $\varepsilon_1\geql \varepsilon_2\in \mi{clo}$, then $\|\varepsilon_1\|^{{\mc V}_\iota}=\|\varepsilon_2\|^{{\mc V}_\iota}$;
  \item[\rm (d)]
  if $\varepsilon_1\gleq \varepsilon_2\in \mi{clo}$, then $\|\varepsilon_1\|^{{\mc V}_\iota}\leq \|\varepsilon_2\|^{{\mc V}_\iota}$;
  \item[\rm (e)]
  if $\varepsilon_1\gle \varepsilon_2\in \mi{clo}$, then $\|\varepsilon_1\|^{{\mc V}_\iota}<\|\varepsilon_2\|^{{\mc V}_\iota}$.
  \end{enumerate}
  \end{enumerate}
  \end{minipage}
\end{alignat}
We proceed by induction on $\iota\leq m$.

Case 1 (the base case):
$\iota=0$.
Straightforward.

Case 2 (the induction case):
$1\leq \iota\leq m$.
Then $\iota-1<\iota\leq m$.

At first, we prove the following statements:
\begin{alignat}{1}
\label{eq8ff}
& \begin{minipage}[t]{\linewidth-15mm}
  $\mi{min}(E_{\iota-1}), \mi{min}(\mi{DE}_{\iota-1}), \mi{min}(D_{\iota-1}), \mi{min}(\mi{UE}_{\iota-1}), \mi{min}(U_{\iota-1})\subseteq_{\mc F} \mi{clo}$.
  \end{minipage}
\end{alignat}   
An immediate consequence of (\ref{eq8e}) and (\ref{eq8dd}).

\begin{alignat}{1}
\label{eq8g}
& \begin{minipage}[t]{\linewidth-15mm}
  If $\mbb{E}_{\iota-1}\neq \emptyset$, then there exists $(\varepsilon_1^*\mult a_\iota^{\alpha^*})\geql \varepsilon_2^*\in \mi{min}(E_{\iota-1})$, 
  $\varepsilon_i^*\in \{\gu\}\cup \mi{PropConj}_A$, $\mi{atoms}(\varepsilon_i^*)\subseteq \mi{dom}({\mc V}_{\iota-1})$, $\alpha^*\geq 1$, such that
  $\delta_\iota=\left(\frac{\|\varepsilon_2^*\|^{{\mc V}_{\iota-1}}}
                           {\|\varepsilon_1^*\|^{{\mc V}_{\iota-1}}}\right)^{\frac{1}{\alpha^*}}\in (0,1]$.
  \end{minipage}
\end{alignat}   
The proof is straightforward, using (\ref{eq7ax})--(\ref{eq7n}) and (\ref{eq8a})--(\ref{eq8aaaax}).

\begin{alignat}{1}
\label{eq8h}
& \begin{minipage}[t]{\linewidth-15mm}
  Let $\varepsilon_1, \varepsilon_2\in \{\gu\}\cup \mi{PropConj}_A$ such that 
  $\mi{atoms}(\varepsilon_1), \mi{atoms}(\varepsilon_2)\subseteq \mi{dom}({\mc V}_{\iota-1})\cup \{a_\iota\}$.
  If $a_\iota\in \mi{atoms}(\varepsilon_1)$ or $a_\iota\in \mi{atoms}(\varepsilon_2)$, and $\mbb{E}_{\iota-1}=\emptyset$, then $\varepsilon_1\geql \varepsilon_2\not\in \mi{clo}$.
  \end{minipage}
\end{alignat}   
The proof is straightforward, using (\ref{eq8a})--(\ref{eq8aaaax}).

\begin{alignat}{1}
\label{eq8i}
& \begin{minipage}[t]{\linewidth-15mm}
  Let $\varepsilon_1, \varepsilon_2\in \{\gu\}\cup \mi{PropConj}_A$ such that 
  $\mi{atoms}(\varepsilon_1), \mi{atoms}(\varepsilon_2)\subseteq \mi{dom}({\mc V}_{\iota-1})\cup \{a_\iota\}$.
  If $a_\iota\in \mi{atoms}(\varepsilon_1)$ or $a_\iota\in \mi{atoms}(\varepsilon_2)$, $\mbb{E}_{\iota-1}=\emptyset$, $\varepsilon_1\diamond \varepsilon_2\in \mi{clo}$, $\diamond\in \{\gleq,\gle\}$, 
  then there exist $\kappa_1\leq \kappa_2\leq \kappa_3\leq \kappa_4$,
  \end{minipage} \\
\notag
& \mu_2^j\gleq (\mu_1^j\mult a_\iota^{\beta_j})=\mi{reduced}(\lambda^{(\gamma_1^j,\dots,\gamma_n^j)})\in \mi{min}(\mi{DE}_{\iota-1}), j=1,\dots,\kappa_1, \\ 
\notag
& \mu_2^j\gle (\mu_1^j\mult a_\iota^{\beta_j})=\mi{reduced}(\lambda^{(\gamma_1^j,\dots,\gamma_n^j)})\in \mi{min}(D_{\iota-1}), j=\kappa_1+1,\dots,\kappa_2, \\
\notag
& (\mu_1^j\mult a_\iota^{\beta_j})\gleq \mu_2^j=\mi{reduced}(\lambda^{(\gamma_1^j,\dots,\gamma_n^j)})\in \mi{min}(\mi{UE}_{\iota-1}), j=\kappa_2+1,\dots,\kappa_3, \\
\notag
& (\mu_1^j\mult a_\iota^{\beta_j})\gle \mu_2^j=\mi{reduced}(\lambda^{(\gamma_1^j,\dots,\gamma_n^j)})\in \mi{min}(U_{\iota-1}), j=\kappa_3+1,\dots,\kappa_4, \\ 
\notag
& \mu_e^j\in \{\gu\}\cup \mi{PropConj}_A, \mi{atoms}(\mu_e^j)\subseteq \mi{dom}({\mc V}_{\iota-1}), \beta_j\geq 1, \bs{0}^n\neq (\gamma_1^j,\dots,\gamma_n^j)\in \mbb{N}^n, \\
\notag
& \begin{minipage}[t]{\linewidth-15mm}  
  such that either $\varepsilon_1\diamond \varepsilon_2=\mi{reduced}(\lambda^{(\sum_{j=1}^{\kappa_4} \gamma_1^j,\dots,\sum_{j=1}^{\kappa_4} \gamma_n^j)})$, 
  or there exists $\zeta_1\diamond^\zeta \zeta_2=\mi{reduced}(\lambda^{(\gamma_1^\zeta,\dots,\gamma_n^\zeta)})\in \mi{clo}$,
  $\zeta_e\in \{\gu\}\cup \mi{PropConj}_A$, $\mi{atoms}(\zeta_e)\subseteq \mi{dom}({\mc V}_{\iota-1})$, $\diamond^\zeta\in \{\geql,\gleq,\gle\}$, 
  $\bs{0}^n\neq (\gamma_1^\zeta,\dots,\gamma_n^\zeta)\in \mbb{N}^n$, satisfying
  $\varepsilon_1\diamond \varepsilon_2=\mi{reduced}(\lambda^{((\sum_{j=1}^{\kappa_4} \gamma_1^j)+\gamma_1^\zeta,\dots,(\sum_{j=1}^{\kappa_4} \gamma_n^j)+\gamma_n^\zeta)})$.
  \begin{enumerate}[\rm (a)]
  \item
  If $\varepsilon_1\diamond \varepsilon_2=\varepsilon_1\gleq (\upsilon_2\mult a_\iota^\alpha)$, then $1\leq \kappa_1\leq \kappa_2\leq \kappa_3\leq \kappa_4$;
  \item
  if $\varepsilon_1\diamond \varepsilon_2=\varepsilon_1\gle (\upsilon_2\mult a_\iota^\alpha)$, then $\kappa_1<\kappa_2\leq \kappa_3\leq \kappa_4$; 
  \item
  if $\varepsilon_1\diamond \varepsilon_2=(\upsilon_1\mult a_\iota^\alpha)\gleq \varepsilon_2$, then $\kappa_1\leq \kappa_2<\kappa_3\leq \kappa_4$;
  \item
  if $\varepsilon_1\diamond \varepsilon_2=(\upsilon_1\mult a_\iota^\alpha)\gle \varepsilon_2$, then $\kappa_1\leq \kappa_2\leq \kappa_3<\kappa_4$;
  $\upsilon_e\in \{\gu\}\cup \mi{PropConj}_A$, $\mi{atoms}(\upsilon_e)\subseteq \mi{dom}({\mc V}_{\iota-1})$.
  \end{enumerate}
  \end{minipage}
\end{alignat}
Let $\varepsilon_1\diamond \varepsilon_2\in \mi{clo}$, $\diamond\in \{\gleq,\gle\}$.
Then there exists $\bs{0}^n\neq (\gamma_1^*,\dots,\gamma_n^*)\in \mbb{N}^n$ such that
$\varepsilon_1\diamond \varepsilon_2\overset{\text{(\ref{eq8ddd})}}{=\!\!=} \mi{reduced}(\lambda^{(\gamma_1^*,\dots,\gamma_n^*)})$.
The proof is by complete induction on $\sum_{i=1}^n \gamma_i^*$, using (\ref{eq8h}), (\ref{eq7k}), (\ref{eq8a})--(\ref{eq8aaaax}).

\begin{alignat}{1}
\label{eq8jj}
& \begin{minipage}[t]{\linewidth-15mm}
  \begin{enumerate}[\rm (a)]
  \item
  $\mbb{DE}_{\iota-1}, \mbb{UE}_{\iota-1}, \mbb{U}_{\iota-1}\subseteq_{\mc F} (0,1]$, 
  \item
  $\mbb{D}_{\iota-1}\subseteq_{\mc F} (0,1)$.
  \end{enumerate}
  \end{minipage}
\end{alignat}   
The proof is straightforward, using (\ref{eq8ff}), (\ref{eq7ax})--(\ref{eq7n}), (\ref{eq8a})--(\ref{eq8aaaax}).

\begin{alignat}{1}
\label{eq8j}
& \begin{minipage}[t]{\linewidth-15mm}
  If $\mbb{E}_{\iota-1}=\emptyset$, then
  \begin{enumerate}[\rm (a)]
  \item
  $0\leq \bigfvee \mbb{DE}_{\iota-1}\leq \delta_\iota\leq \bigfwedge \mbb{UE}_{\iota-1}\leq 1$;
  \item
  $0\leq \bigfvee \mbb{D}_{\iota-1}<\delta_\iota\leq \bigfwedge \mbb{U}_{\iota-1}\leq 1$;
  \item
  if $\mbb{U}_{\iota-1}\neq \emptyset$, then $\delta_\iota<\bigfwedge \mbb{U}_{\iota-1}$;
  \item
  $\delta_\iota\in (0,1]$.
  \end{enumerate}
  \end{minipage}
\end{alignat}   
The proof is by case analysis whether $\mbb{DE}_{\iota-1}$, $\mbb{D}_{\iota-1}$, or $\mbb{UE}_{\iota-1}$, $\mbb{U}_{\iota-1}$ are empty, or not,
using (\ref{eq8jj}), (\ref{eq7ax})--(\ref{eq7n}), (\ref{eq8a})--(\ref{eq8aaaax}).

By the induction hypothesis (a) for $\iota-1$, ${\mc V}_{\iota-1}$ is a partial valuation ${\mc V}_{\iota-1} : \{a_1,\dots,a_{\iota-1}\}\longrightarrow (0,1]$, 
$\mi{dom}({\mc V}_{\iota-1})=\{a_1,\dots,a_{\iota-1}\}$;
either $\mbb{E}_{\iota-1}=\emptyset$, $\delta_\iota\underset{\text{(\ref{eq8j}d)}}{\in} (0,1]$,
or $\mbb{E}_{\iota-1}\neq \emptyset$, $\delta_\iota\underset{\text{(\ref{eq8g})}}{\in} (0,1]$;
$\delta_\iota\in (0,1]$,
$a_\iota\not\in \{a_1,\dots,a_{\iota-1}\}$, $\{a_1,\dots,a_{\iota-1}\}\cup \{a_\iota\}=\{a_1,\dots,a_\iota\}$;
${\mc V}_\iota={\mc V}_{\iota-1}\cup \{(a_\iota,\delta_\iota)\}$ is a partial valuation ${\mc V}_\iota : \{a_1,\dots,a_\iota\}\longrightarrow (0,1]$, 
$\mi{dom}({\mc V}_\iota)=\{a_1,\dots,a_\iota\}$;
(a) holds.

For all $\varsigma\leq \iota$, 
either $\varsigma\leq \iota-1$; 
by the induction hypothesis for $\iota-1$, ${\mc V}_\varsigma\underset{\text{(b)}}{\subseteq} {\mc V}_{\iota-1}\subseteq {\mc V}_\iota={\mc V}_{\iota-1}\cup \{(a_\iota,\delta_\iota)\}$;
or $\varsigma=\iota$, ${\mc V}_\varsigma={\mc V}_\iota$;
${\mc V}_\varsigma\subseteq {\mc V}_\iota$;
(b) holds.

Let $\varepsilon_1, \varepsilon_2\in \{\gu\}\cup \mi{PropConj}_A$ such that $\mi{atoms}(\varepsilon_1), \mi{atoms}(\varepsilon_2)\subseteq \mi{dom}({\mc V}_\iota)$.   
Then $a_\iota\not\in \mi{dom}({\mc V}_{\iota-1})=\{a_1,\dots,a_{\iota-1}\}$ and
$\mi{dom}({\mc V}_\iota)=\{a_1,\dots,a_\iota\}=\mi{dom}({\mc V}_{\iota-1})\cup \{a_\iota\}=\{a_1,\dots,a_{\iota-1}\}\cup \{a_\iota\}$.
We distinguish for $\mi{atoms}(\varepsilon_1)$ and $\mi{atoms}(\varepsilon_2)$ whether $a_\iota\in \mi{atoms}(\varepsilon_1)$ or $a_\iota\in \mi{atoms}(\varepsilon_2)$, or not.
(c--e) can be proved straightforwardly for both cases $\mbb{E}_{\iota-1}\neq \emptyset$ and $\mbb{E}_{\iota-1}=\emptyset$, 
using (\ref{eq8g}), (\ref{eq8h}), (\ref{eq8i}), (\ref{eq8j}), (\ref{eq7ax})--(\ref{eq7n}), (\ref{eq8a})--(\ref{eq8aaaax}).

So, in both Cases 1 and 2, (\ref{eq8f}) holds.
The induction is completed.
Thus, (\ref{eq8f}) holds.

We get that
$A=\{a_1,\dots,a_m\}$;
by (\ref{eq8f}a) for $\iota=m$, ${\mc V}_m$ is a partial valuation ${\mc V}_m : A\longrightarrow (0,1]$, $\mi{dom}({\mc V}_m)=A=\{a_1,\dots,a_m\}$.
We define a valuation 
\begin{equation}
\notag
{\mf A}={\mc V}_m\cup \{(a,0) \,|\, a\in \mi{PropAtom}-A\}.
\end{equation}
Let $C\in S$.
Then either $C=\Cn_1\diamond \Cn_2$, $\Cn_i\in \mi{PropConj}_A$, $\diamond\in \{\geql,\gleq,\gle\}$, 
or $C=b_0\swedge\cdots\swedge b_k\gle \gu$, $b_j\in A$, $k\geq 1$,
or $C=\gz\gle a$ or $C=a\gle \gu$, $a\in A$;
either $C=\Cn_1\diamond \Cn_2$, $\Cn_i\in \mi{PropConj}_A$, $\diamond\in \{\geql,\gleq,\gle\}$, 
or $C=b_0\swedge\cdots\swedge b_k\gle \gu$, $b_j\in A$, 
or $C=\gz\gle a$, $a\in A$.
We get three cases for $C$.

Case 1:
$C=\Cn_1\diamond \Cn_2\in S$, $\Cn_i\in \mi{PropConj}_A$, $\diamond\in \{\geql,\gleq,\gle\}$.
Then $\Cn_1\diamond \Cn_2\in \mi{PurOrdPropCl}_A$, $\mi{atoms}(\Cn_1\diamond \Cn_2)\subseteq \mi{dom}({\mc V}_m)=A$,
$\Cn_1\diamond \Cn_2\in \Lambda$, $\langle \Cn_1\diamond \Cn_2\rangle=\Cn_1^\#\diamond \Cn_2^\#\in \langle \Lambda\rangle$, $\Cn_i^\#\in \mi{PropConj}_A$,
$\mi{reduced}(\langle \Cn_1\diamond \Cn_2\rangle)=\mi{reduced}(\Cn_1^\#\diamond \Cn_2^\#)\overset{\text{(\ref{eq7i})}}{=\!\!=} \varepsilon_1\diamond \varepsilon_2\in \mi{clo}$, 
$\varepsilon_i\in \{\gu\}\cup \mi{PropConj}_A$, $\mi{atoms}(\varepsilon_i)\subseteq \mi{dom}({\mc V}_m)=A$,
either $\diamond=\geql$, $\varepsilon_1\diamond \varepsilon_2=\varepsilon_1\geql \varepsilon_2\in \mi{clo}$, 
$\|\varepsilon_1\|^{{\mc V}_m}\overset{\text{(\ref{eq8f}c)}}{=\!\!=} \|\varepsilon_2\|^{{\mc V}_m}$,
$\|\varepsilon_1\diamond \varepsilon_2\|^{{\mc V}_m}=\|\varepsilon_1\geql \varepsilon_2\|^{{\mc V}_m}=\|\varepsilon_1\|^{{\mc V}_m}\feql \|\varepsilon_2\|^{{\mc V}_m}=1$,
or $\diamond=\gleq$, $\varepsilon_1\diamond \varepsilon_2=\varepsilon_1\gleq \varepsilon_2\in \mi{clo}$, 
$\|\varepsilon_1\|^{{\mc V}_m}\underset{\text{(\ref{eq8f}d)}}{\leq} \|\varepsilon_2\|^{{\mc V}_m}$,
$\|\varepsilon_1\diamond \varepsilon_2\|^{{\mc V}_m}=\|\varepsilon_1\gleq \varepsilon_2\|^{{\mc V}_m}=\|\varepsilon_1\|^{{\mc V}_m}\fleq \|\varepsilon_2\|^{{\mc V}_m}=1$,
or $\diamond=\gle$, $\varepsilon_1\diamond \varepsilon_2=\varepsilon_1\gle \varepsilon_2\in \mi{clo}$, 
$\|\varepsilon_1\|^{{\mc V}_m}\underset{\text{(\ref{eq8f}e)}}{<} \|\varepsilon_2\|^{{\mc V}_m}$,
$\|\varepsilon_1\diamond \varepsilon_2\|^{{\mc V}_m}=\|\varepsilon_1\gle \varepsilon_2\|^{{\mc V}_m}=\|\varepsilon_1\|^{{\mc V}_m}\fle \|\varepsilon_2\|^{{\mc V}_m}=1$;
$\|C\|^{\mf A}=\|\Cn_1\diamond \Cn_2\|^{\mf A}=\|\Cn_1\diamond \Cn_2\|^{{\mc V}_m}=\|\langle \Cn_1\diamond \Cn_2\rangle\|^{{\mc V}_m}=\|\Cn_1^\#\diamond \Cn_2^\#\|^{{\mc V}_m}\overset{\text{(\ref{eq7m})}}{=\!\!=} 
               \|\mi{reduced}(\Cn_1^\#\diamond \Cn_2^\#)\|^{{\mc V}_m}=\|\varepsilon_1\diamond \varepsilon_2\|^{{\mc V}_m}=1$,
${\mf A}\models C$.

Case 2:
$C=b_0\swedge\cdots\swedge b_k\gle \gu\in S$, $b_j\in A$.
Then $b_0\swedge\cdots\swedge b_k\gle \gu\in \mi{OrdPropCl}_A^\gu$, 
$\mi{atoms}(b_0\swedge\cdots\swedge b_k\gle \gu)\subseteq \mi{dom}({\mc V}_m)=A$,
$b_0\swedge\cdots\swedge b_k\gle \gu\in \Lambda$, $\langle b_0\swedge\cdots\swedge b_k\gle \gu\rangle\in \langle \Lambda\rangle$, 
$\mi{reduced}(\langle b_0\swedge\cdots\swedge b_k\gle \gu\rangle)=\mi{reduced}(b_0\swedge\cdots\swedge b_k\gle \gu)=b_0\swedge\cdots\swedge b_k\gle \gu\in \mi{clo}$, 
$\|b_0\swedge\cdots\swedge b_k\|^{{\mc V}_m}\underset{\text{(\ref{eq8f}e)}}{<} 1$,
$\|C\|^{\mf A}=\|b_0\swedge\cdots\swedge b_k\gle \gu\|^{\mf A}=\|b_0\swedge\cdots\swedge b_k\gle \gu\|^{{\mc V}_m}=\|b_0\swedge\cdots\swedge b_k\|^{{\mc V}_m}\fle 1=1$,
${\mf A}\models C$.

Case 3:
$C=\gz\gle a\in S$, $a\in A$.
Then $\mi{atoms}(\gz\gle a)\subseteq \mi{dom}({\mc V}_m)=A$,
$0\underset{\text{(\ref{eq8f}a)}}{<} {\mc V}_m(a)$,
$\|C\|^{\mf A}=\|\gz\gle a\|^{\mf A}=\|\gz\gle a\|^{{\mc V}_m}=0\fle {\mc V}_m(a)=1$,
${\mf A}\models C$.

So, in all Cases 1--3, ${\mf A}\models C$; 
${\mf A}\models S$;
$S$ is satisfiable.
%
%
%
\end{proof}

\subsection{Full proof of the statement (\ref{eq8e}) of Lemma \ref{le2}}
\label{S7.5a}

\begin{proof}   
We distinguish two cases for $n$.

Case 1:
$n=0$.
Then $\mi{min}(K)\subseteq_{\mc F} K\subseteq \mbb{N}^n=\mbb{N}^0=\{\emptyset\}$;
(\ref{eq8e}) holds.

Case 2:
$n\geq 1$.
Let $\emptyset\neq R\subseteq \{1,\dots,n\}$, and $a_i\in \mbb{N}$, $i\in R$.
We put 
\begin{equation}
\notag
\mi{min}(K)_{[a_i, i\in R]}=\{(c_1,\dots,c_n) \,|\, (c_1,\dots,c_n)\in \mi{min}(K),\ \text{\it for all}\ i\in R,\ c_i=a_i\}.
\end{equation}
We prove the following statement:
\begin{alignat}{1}
\label{eq8ee}
& \begin{minipage}[t]{\linewidth-15mm}
  Let $\mi{min}(K)$ be infinite.
  For all $1\leq r\leq n$, there exist $\emptyset\neq R^*\subseteq \{1,\dots,n\}$, and $a_i^*\in \mbb{N}$, $i\in R^*$, such that
  $\mi{card}(R^*)=r$ and $\mi{min}(K)_{[a_i^*, i\in R^*]}$ is infinite.  
  \end{minipage}
\end{alignat}   
We proceed by induction on $1\leq r\leq n$.

Case 2.1 (the base case):
$r=1$.
We have that $\mi{min}(K)$ is infinite.
Then there exists $(c_1^*,\dots,c_n^*)\in \mi{min}(K)\subseteq K\subseteq \mbb{N}^n$;
$(c_1^*,\dots,c_n^*)$ is a minimal element of $K$ with respect to $\preceq$;
for all $(c_1,\dots,c_n)\in \mi{min}(K)\subseteq \mbb{N}^n$ satisfying $(c_1,\dots,c_n)\neq (c_1^*,\dots,c_n^*)$,
$(c_1,\dots,c_n)$ is a minimal element of $K$ with respect to $\preceq$;
there exists $1\leq i^*\leq n$ satisfying $c_{i^*}<c_{i^*}^*$;
by the Dirichlet's Box Principle, there exists $1\leq i^*\leq n$ satisfying 
for infinitely many $(c_1,\dots,c_n)\in \mi{min}(K)\subseteq \mbb{N}^n$ that $c_{i^*}<c_{i^*}^*$;
by the Dirichlet's Box Principle, there exists $a_{i^*}^*<c_{i^*}^*$ satisfying 
for infinitely many $(c_1,\dots,c_n)\in \mi{min}(K)\subseteq \mbb{N}^n$ that 
$c_{i^*}=a_{i^*}^*$ and
$(c_1,\dots,c_n)\in \mi{min}(K)_{[a_i^*, i\in \{i^*\}]}=\{(c_1,\dots,c_n) \,|\, (c_1,\dots,c_n)\in \mi{min}(K),\ \text{\it for all}\ i\in \{i^*\},\ c_i=a_i^*\}$;
$\mi{min}(K)_{[a_i^*, i\in \{i^*\}]}$ is infinite.
We put $\emptyset\neq R^*=\{i^*\}\subseteq \{1,\dots,n\}$.
We get that $\mi{card}(R^*)=\mi{card}(\{i^*\})=1$ and $\mi{min}(K)_{[a_i^*, i\in R^*]}=\mi{min}(K)_{[a_i^*, i\in \{i^*\}]}$ is infinite.
Hence, $\mi{card}(R^*)=1$ and $\mi{min}(K)_{[a_i^*, i\in R^*]}$ is infinite.

Case 2.2 (the induction case):
$2\leq r\leq n$.
Then $1\leq r-1<r\leq n$;
by the induction hypothesis for $r-1$, there exist $\emptyset\neq R^\#\subseteq \{1,\dots,n\}$, and $a_i^*\in \mbb{N}$, $i\in R^\#$, satisfying that
$\mi{card}(R^\#)=r-1$ and $\mi{min}(K)_{[a_i^*, i\in R^\#]}$ is infinite;
there exists $(c_1^*,\dots,c_n^*)\in \mi{min}(K)_{[a_i^*, i\in R^\#]}\subseteq \mi{min}(K)\subseteq K\subseteq \mbb{N}^n$;
$(c_1^*,\dots,c_n^*)$ is a minimal element of $K$ with respect to $\preceq$;
for all $i\in R^\#$, $c_i^*=a_i^*$;
for all $(c_1,\dots,c_n)\in \mi{min}(K)_{[a_i^*, i\in R^\#]}\subseteq \mi{min}(K)\subseteq \mbb{N}^n$ satisfying $(c_1,\dots,c_n)\neq (c_1^*,\dots,c_n^*)$,
$(c_1,\dots,c_n)$ is a minimal element of $K$ with respect to $\preceq$;
for all $i\in R^\#$, $c_i=c_i^*=a_i^*$;
there exists $i^*\in \{1,\dots,n\}-R^\#$ satisfying $c_{i^*}<c_{i^*}^*$;
by the Dirichlet's Box Principle, there exists $i^*\in \{1,\dots,n\}-R^\#$ satisfying 
for infinitely many $(c_1,\dots,c_n)\in \mi{min}(K)_{[a_i^*, i\in R^\#]}\subseteq \mbb{N}^n$ that $c_{i^*}<c_{i^*}^*$;
by the Dirichlet's Box Principle, there exists $a_{i^*}^*<c_{i^*}^*$ satisfying 
for infinitely many $(c_1,\dots,c_n)\in \mi{min}(K)_{[a_i^*, i\in R^\#]}\subseteq \mi{min}(K)\subseteq \mbb{N}^n$ that 
$c_{i^*}=a_{i^*}^*$;
for all $i\in R^\#$, $c_i=a_i^*$;
$i^*\not\in R^\#$,
$(c_1,\dots,c_n)\in \mi{min}(K)_{[a_i^*, i\in R^\#\cup \{i^*\}]}=\{(c_1,\dots,c_n) \,|\, (c_1,\dots,c_n)\in \mi{min}(K),\ \text{\it for all}\ i\in R^\#\cup \{i^*\},\ c_i=a_i^*\}$;
$\mi{min}(K)_{[a_i^*, i\in R^\#\cup \{i^*\}]}$ is infinite.
We put $\emptyset\neq R^*=R^\#\cup \{i^*\}\subseteq \{1,\dots,n\}$.
We get that $i^*\not\in R^\#$, $\mi{card}(R^*)=\mi{card}(R^\#\cup \{i^*\})=\mi{card}(R^\#)+\mi{card}(\{i^*\})=r-1+1=r$;
$\mi{min}(K)_{[a_i^*, i\in R^*]}=\mi{min}(K)_{[a_i^*, i\in R^\#\cup \{i^*\}]}$ is infinite.
Hence, $\mi{card}(R^*)=r$ and $\mi{min}(K)_{[a_i^*, i\in R^*]}$ is infinite.

So, in both Cases 2.1 and 2.2, (\ref{eq8ee}) holds.
The induction is completed.
Thus, (\ref{eq8ee}) holds.

Let $\mi{min}(K)$ be infinite.
We get by (\ref{eq8ee}) for $n$ that there exist $\emptyset\neq R^*\subseteq \{1,\dots,n\}$, and $a_i^*\in \mbb{N}$, $i\in R^*$, satisfying that
$\mi{card}(R^*)=n$ and $\mi{min}(K)_{[a_i^*, i\in R^*]}$ is infinite;
$R^*=\{1,\dots,n\}$;
$\mi{min}(K)_{[a_i^*, i\in R^*]}=\mi{min}(K)_{[a_i^*, i\in \{1,\dots,n\}]}=
                                 \{(c_1,\dots,c_n) \,|\, (c_1,\dots,c_n)\in \mi{min}(K),\ \text{\it for all}\ i\in \{1,\dots,n\},\ c_i=a_i^*\}=\{(a_1^*,\dots,a_n^*)\}$ is finite,
which is a contradiction that $\mi{min}(K)_{[a_i^*, i\in R^*]}$ is infinite;
$\mi{min}(K)\subseteq_{\mc F} K$;
(\ref{eq8e}) holds.

So, in both Cases 1 and 2, (\ref{eq8e}) holds.
%
%
%
\end{proof}

\subsection{Full proof of Case 1 (the base case) of the statement (\ref{eq8f}) of Lemma \ref{le2}}
\label{S7.5b}

\begin{proof}   
Case 1 (the base case):
$\iota=0$.

${\mc V}_0=\emptyset$;
${\mc V}_0$ is a partial valuation ${\mc V}_0 : \emptyset\longrightarrow (0,1]$, $\mi{dom}({\mc V}_0)=\{a_1,\dots,a_0\}=\emptyset$;  
(a) holds.

For all $\varsigma\leq \iota=0$, $\varsigma=0$, ${\mc V}_\varsigma={\mc V}_\iota={\mc V}_0$;
(b) holds.

For all $\varepsilon_1, \varepsilon_2\in \{\gu\}\cup \mi{PropConj}_A$ satisfying $\mi{atoms}(\varepsilon_1), \mi{atoms}(\varepsilon_2)\subseteq \mi{dom}({\mc V}_\iota)=\mi{dom}({\mc V}_0)=\emptyset$,
$\varepsilon_1=\varepsilon_2=\gu$;
if $\varepsilon_1\geql \varepsilon_2\in \mi{clo}$, $\|\varepsilon_1\|^{{\mc V}_\iota}=\|\varepsilon_2\|^{{\mc V}_\iota}$;
(c) holds;
if $\varepsilon_1\gleq \varepsilon_2\in \mi{clo}$, $\|\varepsilon_1\|^{{\mc V}_\iota}\leq \|\varepsilon_2\|^{{\mc V}_\iota}$;
(d) holds;
$\varepsilon_1\gle \varepsilon_2=\gu\gle \gu\underset{\text{(\ref{eq8a})}}{\not\in} \mi{clo}$;
(e) holds trivially.
%
%
%
\end{proof}

\subsection{Full proof of the statement (\ref{eq8ff}) of Lemma \ref{le2}}
\label{S7.5c}

\begin{proof}   
Let $X_{\iota-1}\in \{E_{\iota-1},\mi{DE}_{\iota-1},D_{\iota-1},\mi{UE}_{\iota-1},U_{\iota-1}\}$.
We put $\mi{EXP}=\{(\gamma_1,\dots,\gamma_n) \,|\, \bs{0}^n\neq (\gamma_1,\dots,\gamma_n)\in \mbb{N}^n, \mi{reduced}(\lambda^{(\gamma_1,\dots,\gamma_n)})\in X_{\iota-1}\}$.
For all $l\in \mi{min}(X_{\iota-1})$,
$l=\mi{reduced}(\lambda^{(\gamma_1,\dots,\gamma_n)})\in X_{\iota-1}$, $\bs{0}^n\neq (\gamma_1,\dots,\gamma_n)\in \mbb{N}^n$,
$(\gamma_1,\dots,\gamma_n)\in \mi{EXP}$;
for all $(\gamma_1',\dots,\gamma_n')\in \mbb{N}^n$ satisfying $(\gamma_1',\dots,\gamma_n')\prec (\gamma_1,\dots,\gamma_n)$,
$\mi{reduced}(\lambda^{(\gamma_1',\dots,\gamma_n')})\not\in X_{\iota-1}$,
$(\gamma_1',\dots,\gamma_n')\not\in \mi{EXP}$;
$(\gamma_1,\dots,\gamma_n)\in \mi{min}(\mi{EXP})\underset{\text{(\ref{eq8e})}}{\subseteq_{\mc F}} \mi{EXP}$;
for all $\bs{0}^n\neq (\gamma_1,\dots,\gamma_n)\in \mbb{N}^n$, $\mi{reduced}(\lambda^{(\gamma_1,\dots,\gamma_n)})\underset{\text{(\ref{eq8dd})}}{\in} \mi{clo}$;
$\mi{min}(X_{\iota-1})\subseteq \{l \,|\, l=\mi{reduced}(\lambda^{(\gamma_1,\dots,\gamma_n)}), \bs{0}^n\neq (\gamma_1,\dots,\gamma_n)\in \mi{min}(\mi{EXP})\}\subseteq_{\mc F}
                                \{l \,|\, l=\mi{reduced}(\lambda^{(\gamma_1,\dots,\gamma_n)}), \bs{0}^n\neq (\gamma_1,\dots,\gamma_n)\in \mi{EXP}\subseteq \mbb{N}^n\}\subseteq \mi{clo}$;
(\ref{eq8ff}) holds.
%
%
%
\end{proof}

\subsection{Full proof of the statement (\ref{eq8g}) of Lemma \ref{le2}}
\label{S7.5d}

\begin{proof}   
Let $\mbb{E}_{\iota-1}\neq \emptyset$.
Then $\delta_\iota=\bigfvee \mbb{E}_{\iota-1}$;
there exists $(\varepsilon_1^*\mult a_\iota^{\alpha^*})\geql \varepsilon_2^*\in \mi{min}(E_{\iota-1})\subseteq \mi{clo}$, 
$\varepsilon_i^*\in \{\gu\}\cup \mi{PropConj}_A$, $\mi{atoms}(\varepsilon_i^*)\subseteq \mi{dom}({\mc V}_{\iota-1})$, $\alpha^*\geq 1$, satisfying
$\left(\frac{\|\varepsilon_2^*\|^{{\mc V}_{\iota-1}}}
            {\|\varepsilon_1^*\|^{{\mc V}_{\iota-1}}}\right)^{\frac{1}{\alpha^*}}\in \mbb{E}_{\iota-1}$;
$\varepsilon_2^*\geql (\varepsilon_1^*\mult a_\iota^{\alpha^*})\underset{\text{(\ref{eq8b})}}{\in} \mi{clo}$,
$a_\iota\diamond^\# \gu\underset{\text{(\ref{eq8aa})}}{\in} \mi{clo}$, $\diamond^\#\in \{\gleq,\gle\}$,
$(a_\iota\diamond^\# \gu)^{\alpha^*}=a_\iota^{\alpha^*}\diamond^\# \gu\underset{\text{(\ref{eq8d})}}{\in} \mi{clo}$,
$\mi{atoms}(\varepsilon_2^*)\cap \mi{atoms}(\varepsilon_1^*)=\mi{atoms}(\varepsilon_2^*)\cap \mi{atoms}(\varepsilon_1^*\mult a_\iota^{\alpha^*})=
 \mi{atoms}(\varepsilon_2^*)\cap (\mi{atoms}(\varepsilon_1^*)\cup \mi{atoms}(a_\iota^{\alpha^*}))\overset{\text{(\ref{eq8aaa})}}{=\!\!=} \emptyset$,
$\mi{reduced}((\varepsilon_2^*\geql (\varepsilon_1^*\mult a_\iota^{\alpha^*}))\mult (a_\iota^{\alpha^*}\diamond^\# \gu))=
 \mi{reduced}((\varepsilon_2^*\mult a_\iota^{\alpha^*})\diamond^\# (\varepsilon_1^*\mult a_\iota^{\alpha^*}))\overset{\text{(\ref{eq7h})}}{=\!\!=}
 \mi{reduced}(\varepsilon_2^*\diamond^\# \varepsilon_1^*)\overset{\text{(\ref{eq7ggg})}}{=\!\!=} \varepsilon_2^*\diamond^\# \varepsilon_1^*\underset{\text{(\ref{eq8c})}}{\in} \mi{clo}$;
by the induction hypothesis for $\iota-1$, $\|\varepsilon_1^*\|^{{\mc V}_{\iota-1}}, \|\varepsilon_2^*\|^{{\mc V}_{\iota-1}}\underset{\text{(a)}}{\in} (0,1]$, 
either $\diamond^\#=\gleq$, $\|\varepsilon_2^*\|^{{\mc V}_{\iota-1}}\underset{\text{(d)}}{\leq} \|\varepsilon_1^*\|^{{\mc V}_{\iota-1}}$,
or $\diamond^\#=\gle$, $\|\varepsilon_2^*\|^{{\mc V}_{\iota-1}}\underset{\text{(e)}}{<} \|\varepsilon_1^*\|^{{\mc V}_{\iota-1}}$;
$\|\varepsilon_2^*\|^{{\mc V}_{\iota-1}}\leq \|\varepsilon_1^*\|^{{\mc V}_{\iota-1}}$,
$\left(\frac{\|\varepsilon_2^*\|^{{\mc V}_{\iota-1}}}
            {\|\varepsilon_1^*\|^{{\mc V}_{\iota-1}}}\right)^{\frac{1}{\alpha^*}}\in (0,1]$.
Let there exist $(\varepsilon_1\mult a_\iota^\alpha)\geql \varepsilon_2\in \mi{min}(E_{\iota-1})\subseteq \mi{clo}$, 
$\varepsilon_i\in \{\gu\}\cup \mi{PropConj}_A$, $\mi{atoms}(\varepsilon_i)\subseteq \mi{dom}({\mc V}_{\iota-1})$, $\alpha\geq 1$, such that
$\left(\frac{\|\varepsilon_2\|^{{\mc V}_{\iota-1}}}
            {\|\varepsilon_1\|^{{\mc V}_{\iota-1}}}\right)^{\frac{1}{\alpha}}\in \mbb{E}_{\iota-1}$.
Hence, 
$((\varepsilon_1\mult a_\iota^\alpha)\geql \varepsilon_2)^{\alpha^*}=(\varepsilon_1^{\alpha^*}\mult a_\iota^{\alpha\cdot \alpha^*})\geql \varepsilon_2^{\alpha^*}\underset{\text{(\ref{eq8d})}}{\in} \mi{clo}$,
$\varepsilon_2^*\geql (\varepsilon_1^*\mult a_\iota^{\alpha^*})\in \mi{clo}$,
$(\varepsilon_2^*\geql (\varepsilon_1^*\mult a_\iota^{\alpha^*}))^\alpha=(\varepsilon_2^*)^\alpha\geql ((\varepsilon_1^*)^\alpha\mult a_\iota^{\alpha\cdot \alpha^*})\underset{\text{(\ref{eq8d})}}{\in} \mi{clo}$,
\begin{alignat*}{1}
& \mi{reduced}(((\varepsilon_1^{\alpha^*}\mult a_\iota^{\alpha\cdot \alpha^*})\geql \varepsilon_2^{\alpha^*})\mult
               ((\varepsilon_2^*)^\alpha\geql ((\varepsilon_1^*)^\alpha\mult a_\iota^{\alpha\cdot \alpha^*})))= \\
& \mi{reduced}((\varepsilon_1^{\alpha^*}\mult (\varepsilon_2^*)^\alpha\mult a_\iota^{\alpha\cdot \alpha^*})\geql 
               ((\varepsilon_1^*)^\alpha\mult \varepsilon_2^{\alpha^*}\mult a_\iota^{\alpha\cdot \alpha^*}))\overset{\text{(\ref{eq7h})}}{=\!\!=} \\
& \mi{reduced}((\varepsilon_1^{\alpha^*}\mult (\varepsilon_2^*)^\alpha)\geql ((\varepsilon_1^*)^\alpha\mult \varepsilon_2^{\alpha^*}))= \\
& \mi{reduced}^-((\varepsilon_1^{\alpha^*}\mult (\varepsilon_2^*)^\alpha)\geql ((\varepsilon_1^*)^\alpha\mult \varepsilon_2^{\alpha^*}))\geql \\
& \quad \mi{reduced}^+((\varepsilon_1^{\alpha^*}\mult (\varepsilon_2^*)^\alpha)\geql ((\varepsilon_1^*)^\alpha\mult \varepsilon_2^{\alpha^*}))\underset{\text{(\ref{eq8c})}}{\in} \mi{clo},
\end{alignat*}
$\mi{atoms}(\varepsilon_1^*), \mi{atoms}(\varepsilon_2^*)\subseteq \mi{dom}({\mc V}_{\iota-1})$,
$\mi{atoms}(\mi{reduced}^-((\varepsilon_1^{\alpha^*}\mult (\varepsilon_2^*)^\alpha)\geql ((\varepsilon_1^*)^\alpha\mult \varepsilon_2^{\alpha^*})))\underset{\text{(\ref{eq7a})}}{\subseteq}
 \mi{atoms}(\varepsilon_1^{\alpha^*}\mult (\varepsilon_2^*)^\alpha)=\mi{atoms}(\varepsilon_1)\cup \mi{atoms}(\varepsilon_2^*)\subseteq \mi{dom}({\mc V}_{\iota-1})$,
$\mi{atoms}(\mi{reduced}^+((\varepsilon_1^{\alpha^*}\mult (\varepsilon_2^*)^\alpha)\geql ((\varepsilon_1^*)^\alpha\mult \varepsilon_2^{\alpha^*})))\underset{\text{(\ref{eq7b})}}{\subseteq}
 \mi{atoms}((\varepsilon_1^*)^\alpha\mult \varepsilon_2^{\alpha^*})=\mi{atoms}(\varepsilon_1^*)\cup \mi{atoms}(\varepsilon_2)\subseteq \mi{dom}({\mc V}_{\iota-1})$;
by the induction hypothesis for $\iota-1$, 
$\|\varepsilon_1\|^{{\mc V}_{\iota-1}}, \|\varepsilon_1^*\|^{{\mc V}_{\iota-1}}, \|\varepsilon_2\|^{{\mc V}_{\iota-1}}, \|\varepsilon_2^*\|^{{\mc V}_{\iota-1}},
 \|\mi{reduced}^-((\varepsilon_1^{\alpha^*}\mult (\varepsilon_2^*)^\alpha)\geql ((\varepsilon_1^*)^\alpha\mult \varepsilon_2^{\alpha^*}))\|^{{\mc V}_{\iota-1}},
 \|\mi{reduced}^+((\varepsilon_1^{\alpha^*}\mult (\varepsilon_2^*)^\alpha)\geql ((\varepsilon_1^*)^\alpha\mult \varepsilon_2^{\alpha^*}))\|^{{\mc V}_{\iota-1}}\underset{\text{(a)}}{\in} (0,1]$,
$\|\mi{reduced}^-((\varepsilon_1^{\alpha^*}\mult (\varepsilon_2^*)^\alpha)\geql ((\varepsilon_1^*)^\alpha\mult \varepsilon_2^{\alpha^*}))\|^{{\mc V}_{\iota-1}}\overset{\text{(c)}}{=\!\!=}
 \|\mi{reduced}^+((\varepsilon_1^{\alpha^*}\mult (\varepsilon_2^*)^\alpha)\geql ((\varepsilon_1^*)^\alpha\mult \varepsilon_2^{\alpha^*}))\|^{{\mc V}_{\iota-1}}$;
\begin{alignat*}{1}
& \dfrac{\|\mi{reduced}^-((\varepsilon_1^{\alpha^*}\mult (\varepsilon_2^*)^\alpha)\geql ((\varepsilon_1^*)^\alpha\mult \varepsilon_2^{\alpha^*}))\|^{{\mc V}_{\iota-1}}}
        {\|\mi{reduced}^+((\varepsilon_1^{\alpha^*}\mult (\varepsilon_2^*)^\alpha)\geql ((\varepsilon_1^*)^\alpha\mult \varepsilon_2^{\alpha^*}))\|^{{\mc V}_{\iota-1}}}\overset{\text{(\ref{eq7n})}}{=\!\!=} 
  \dfrac{\|\varepsilon_1^{\alpha^*}\mult (\varepsilon_2^*)^\alpha\|^{{\mc V}_{\iota-1}}}
        {\|(\varepsilon_1^*)^\alpha\mult \varepsilon_2^{\alpha^*}\|^{{\mc V}_{\iota-1}}}= \\
& \dfrac{\left(\|\varepsilon_1\|^{{\mc V}_{\iota-1}}\right)^{\alpha^*}\fswedge \left(\|\varepsilon_2^*\|^{{\mc V}_{\iota-1}}\right)^\alpha}
        {\left(\|\varepsilon_1^*\|^{{\mc V}_{\iota-1}}\right)^\alpha\fswedge \left(\|\varepsilon_2\|^{{\mc V}_{\iota-1}}\right)^{\alpha^*}}=1, \\[1mm]
& \left(\dfrac{\|\varepsilon_2\|^{{\mc V}_{\iota-1}}}
              {\|\varepsilon_1\|^{{\mc V}_{\iota-1}}}\right)^{\alpha^*}=
  \left(\dfrac{\|\varepsilon_2^*\|^{{\mc V}_{\iota-1}}}
              {\|\varepsilon_1^*\|^{{\mc V}_{\iota-1}}}\right)^\alpha, \\
& \left(\dfrac{\|\varepsilon_2\|^{{\mc V}_{\iota-1}}}
              {\|\varepsilon_1\|^{{\mc V}_{\iota-1}}}\right)^{\dfrac{1}{\alpha}}=
  \left(\dfrac{\|\varepsilon_2^*\|^{{\mc V}_{\iota-1}}}
              {\|\varepsilon_1^*\|^{{\mc V}_{\iota-1}}}\right)^{\dfrac{1}{\alpha^*}};
\end{alignat*}
$\mbb{E}_{\iota-1}=\left\{\left(\frac{\|\varepsilon_2^*\|^{{\mc V}_{\iota-1}}}
                                     {\|\varepsilon_1^*\|^{{\mc V}_{\iota-1}}}\right)^{\frac{1}{\alpha^*}}\right\}$,
\begin{equation}
\notag
\delta_\iota=\bigfvee \mbb{E}_{\iota-1}=\bigfvee \left\{\left(\dfrac{\|\varepsilon_2^*\|^{{\mc V}_{\iota-1}}}
                                                                    {\|\varepsilon_1^*\|^{{\mc V}_{\iota-1}}}\right)^{\dfrac{1}{\alpha^*}}\right\}=
             \left(\dfrac{\|\varepsilon_2^*\|^{{\mc V}_{\iota-1}}}
                         {\|\varepsilon_1^*\|^{{\mc V}_{\iota-1}}}\right)^{\dfrac{1}{\alpha^*}}\in (0,1];
\end{equation}
(\ref{eq8g}) holds.
%
%
%
\end{proof}

\subsection{Full proof of the statement (\ref{eq8h}) of Lemma \ref{le2}}
\label{S7.5e}

\begin{proof}   
Let $a_\iota\in \mi{atoms}(\varepsilon_1)$ or $a_\iota\in \mi{atoms}(\varepsilon_2)$, $\mbb{E}_{\iota-1}=\emptyset$, $\varepsilon_1\geql \varepsilon_2\in \mi{clo}$.
Then, by the induction hypothesis for $\iota-1$, $a_\iota\not\in \mi{dom}({\mc V}_{\iota-1})\overset{\text{(a)}}{=\!\!=} \{a_1,\dots,a_{\iota-1}\}$;
$\mi{atoms}(\varepsilon_1), \mi{atoms}(\varepsilon_2)\subseteq \mi{dom}({\mc V}_{\iota-1})\cup \{a_\iota\}$,
$\mi{atoms}(\varepsilon_1)\cap \mi{atoms}(\varepsilon_2)\overset{\text{(\ref{eq8aaa})}}{=\!\!=} \emptyset$,
either $a_\iota\in \mi{atoms}(\varepsilon_1)$ or $a_\iota\in \mi{atoms}(\varepsilon_2)$.
We distinguish two cases for $\mi{atoms}(\varepsilon_1)$ and $\mi{atoms}(\varepsilon_2)$.

Case 2.1:
$a_\iota\in \mi{atoms}(\varepsilon_1)$ and $a_\iota\not\in \mi{atoms}(\varepsilon_2)\subseteq \mi{dom}({\mc V}_{\iota-1})$.
Then $\mi{atoms}(\varepsilon_1)\subseteq \mi{dom}({\mc V}_{\iota-1})\cup \{a_\iota\}$,
$\varepsilon_1=\upsilon_1\mult a_\iota^\alpha$, $\upsilon_1\in \{\gu\}\cup \mi{PropConj}_A$, $\alpha\geq 1$, 
$\mi{atoms}(\upsilon_1)\subseteq \mi{atoms}(\varepsilon_1)\subseteq \mi{dom}({\mc V}_{\iota-1})\cup \{a_\iota\}$,
$a_\iota\not\in \mi{atoms}(\upsilon_1)\subseteq \mi{dom}({\mc V}_{\iota-1})$,
$\varepsilon_1\geql \varepsilon_2=(\upsilon_1\mult a_\iota^\alpha)\geql \varepsilon_2\in \mi{clo}$;
there exists $\bs{0}^n\neq (\gamma_1^*,\dots,\gamma_n^*)\in \mbb{N}^n$ satisfying 
$(\upsilon_1\mult a_\iota^\alpha)\geql \varepsilon_2\overset{\text{(\ref{eq8ddd})}}{=\!\!=} \mi{reduced}(\lambda^{(\gamma_1^*,\dots,\gamma_n^*)})$.
We put $\mi{EXP}=\{(\gamma_1,\dots,\gamma_n) \,|\, \bs{0}^n\neq (\gamma_1,\dots,\gamma_n)\in \mbb{N}^n, (\gamma_1,\dots,\gamma_n)\preceq (\gamma_1^*,\dots,\gamma_n^*), 
                                                   \mi{reduced}(\lambda^{(\gamma_1,\dots,\gamma_n)})\in E_{\iota-1}\}$.
Hence, $(\upsilon_1\mult a_\iota^\alpha)\geql \varepsilon_2=\mi{reduced}(\lambda^{(\gamma_1^*,\dots,\gamma_n^*)})\in \mi{clo}$, 
$\mi{atoms}(\upsilon_1), \mi{atoms}(\varepsilon_2)\subseteq \mi{dom}({\mc V}_{\iota-1})$,
$(\upsilon_1\mult a_\iota^\alpha)\geql \varepsilon_2=\mi{reduced}(\lambda^{(\gamma_1^*,\dots,\gamma_n^*)})\in E_{\iota-1}$,
$(\gamma_1^*,\dots,\gamma_n^*)\neq \bs{0}^n$, $(\gamma_1^*,\dots,\gamma_n^*)\preceq (\gamma_1^*,\dots,\gamma_n^*)$, 
$(\gamma_1^*,\dots,\gamma_n^*)\in \mi{EXP}\neq \emptyset$;
we have that $\preceq$ is a well-founded order on $\mbb{N}^n$;
$\mi{min}(\mi{EXP})\neq \emptyset$;
there exists $\bs{0}^n\neq (\gamma_1^\natural,\dots,\gamma_n^\natural)\in \mi{min}(\mi{EXP})\subseteq \mi{EXP}\subseteq \mbb{N}^n$ satisfying 
$(\gamma_1^\natural,\dots,\gamma_n^\natural)\preceq (\gamma_1^*,\dots,\gamma_n^*)$, $\mi{reduced}(\lambda^{(\gamma_1^\natural,\dots,\gamma_n^\natural)})\in E_{\iota-1}$;
for all $(\gamma_1,\dots,\gamma_n)\in \mbb{N}^n$ satisfying $(\gamma_1,\dots,\gamma_n)\prec (\gamma_1^\natural,\dots,\gamma_n^\natural)$, 
$(\gamma_1,\dots,\gamma_n)\not\in \mi{EXP}$,
$(\gamma_1,\dots,\gamma_n)\prec (\gamma_1^\natural,\dots,\gamma_n^\natural)\preceq (\gamma_1^*,\dots,\gamma_n^*)$,
either $(\gamma_1,\dots,\gamma_n)=\bs{0}^n$, $\mi{reduced}(\lambda^{(\gamma_1,\dots,\gamma_n)})=\mi{reduced}(\lambda^{\bs{0}^n})=\gu\underset{\text{(\ref{eq8a})}}{\not\in} \mi{clo}\supseteq E_{\iota-1}$,
or $(\gamma_1,\dots,\gamma_n)\neq \bs{0}^n$, $\mi{reduced}(\lambda^{(\gamma_1,\dots,\gamma_n)})\not\in E_{\iota-1}$;
$\mi{reduced}(\lambda^{(\gamma_1,\dots,\gamma_n)})\not\in E_{\iota-1}$;
$\mi{reduced}(\lambda^{(\gamma_1^\natural,\dots,\gamma_n^\natural)})\in \mi{min}(E_{\iota-1})$, 
$\mi{reduced}(\lambda^{(\gamma_1^\natural,\dots,\gamma_n^\natural)})=(\mu_1^\natural\mult a_\iota^{\beta^\natural})\geql \mu_2^\natural$, 
$\mu_i^\natural\in \{\gu\}\cup \mi{PropConj}_A$, $\mi{atoms}(\mu_i^\natural)\subseteq \mi{dom}({\mc V}_{\iota-1})$, $\beta^\natural\geq 1$,
$\left(\frac{\|\mu_2\|^{{\mc V}_{\iota-1}}}
            {\|\mu_1\|^{{\mc V}_{\iota-1}}}\right)^{\frac{1}{\beta^\natural}}\in \mbb{E}_{\iota-1}\neq \emptyset$,
which is a contradiction with $\mbb{E}_{\iota-1}=\emptyset$;
$\varepsilon_1\geql \varepsilon_2=(\upsilon_1\mult a_\iota^\alpha)\geql \varepsilon_2\not\in \mi{clo}$;
(\ref{eq8h}) holds.

Case 2.2:
$a_\iota\not\in \mi{atoms}(\varepsilon_1)\subseteq \mi{dom}({\mc V}_{\iota-1})$ and $a_\iota\in \mi{atoms}(\varepsilon_2)$.
Then $\mi{atoms}(\varepsilon_2)\subseteq \mi{dom}({\mc V}_{\iota-1})\cup \{a_\iota\}$,
$\varepsilon_2=\upsilon_2\mult a_\iota^\alpha$, $\upsilon_2\in \{\gu\}\cup \mi{PropConj}_A$, $\alpha\geq 1$, 
$\mi{atoms}(\upsilon_2)\subseteq \mi{atoms}(\varepsilon_2)\subseteq \mi{dom}({\mc V}_{\iota-1})\cup \{a_\iota\}$,
$a_\iota\not\in \mi{atoms}(\upsilon_2)\subseteq \mi{dom}({\mc V}_{\iota-1})$,
$\varepsilon_1\geql \varepsilon_2=\varepsilon_1\geql (\upsilon_2\mult a_\iota^\alpha)\in \mi{clo}$,
$(\upsilon_2\mult a_\iota^\alpha)\geql \varepsilon_1\underset{\text{(\ref{eq8b})}}{\in} \mi{clo}$;
there exists $\bs{0}^n\neq (\gamma_1^*,\dots,\gamma_n^*)\in \mbb{N}^n$ satisfying 
$(\upsilon_2\mult a_\iota^\alpha)\geql \varepsilon_1\overset{\text{(\ref{eq8ddd})}}{=\!\!=} \mi{reduced}(\lambda^{(\gamma_1^*,\dots,\gamma_n^*)})$.
We put $\mi{EXP}=\{(\gamma_1,\dots,\gamma_n) \,|\, \bs{0}^n\neq (\gamma_1,\dots,\gamma_n)\in \mbb{N}^n, (\gamma_1,\dots,\gamma_n)\preceq (\gamma_1^*,\dots,\gamma_n^*), 
                                                   \mi{reduced}(\lambda^{(\gamma_1,\dots,\gamma_n)})\in E_{\iota-1}\}$.
Hence, $(\upsilon_2\mult a_\iota^\alpha)\geql \varepsilon_1=\mi{reduced}(\lambda^{(\gamma_1^*,\dots,\gamma_n^*)})\in \mi{clo}$, 
$\mi{atoms}(\varepsilon_1), \mi{atoms}(\upsilon_2)\subseteq \mi{dom}({\mc V}_{\iota-1})$,
$(\upsilon_2\mult a_\iota^\alpha)\geql \varepsilon_1=\mi{reduced}(\lambda^{(\gamma_1^*,\dots,\gamma_n^*)})\in E_{\iota-1}$,
$(\gamma_1^*,\dots,\gamma_n^*)\neq \bs{0}^n$, $(\gamma_1^*,\dots,\gamma_n^*)\preceq (\gamma_1^*,\dots,\gamma_n^*)$, 
$(\gamma_1^*,\dots,\gamma_n^*)\in \mi{EXP}\neq \emptyset$;
we have that $\preceq$ is a well-founded order on $\mbb{N}^n$;
$\mi{min}(\mi{EXP})\neq \emptyset$;
there exists $\bs{0}^n\neq (\gamma_1^\natural,\dots,\gamma_n^\natural)\in \mi{min}(\mi{EXP})\subseteq \mi{EXP}\subseteq \mbb{N}^n$ satisfying 
$(\gamma_1^\natural,\dots,\gamma_n^\natural)\preceq (\gamma_1^*,\dots,\gamma_n^*)$, $\mi{reduced}(\lambda^{(\gamma_1^\natural,\dots,\gamma_n^\natural)})\in E_{\iota-1}$;
for all $(\gamma_1,\dots,\gamma_n)\in \mbb{N}^n$ satisfying $(\gamma_1,\dots,\gamma_n)\prec (\gamma_1^\natural,\dots,\gamma_n^\natural)$, 
$(\gamma_1,\dots,\gamma_n)\not\in \mi{EXP}$,
$(\gamma_1,\dots,\gamma_n)\prec (\gamma_1^\natural,\dots,\gamma_n^\natural)\preceq (\gamma_1^*,\dots,\gamma_n^*)$,
either $(\gamma_1,\dots,\gamma_n)=\bs{0}^n$, $\mi{reduced}(\lambda^{(\gamma_1,\dots,\gamma_n)})=\mi{reduced}(\lambda^{\bs{0}^n})=\gu\underset{\text{(\ref{eq8a})}}{\not\in} \mi{clo}\supseteq E_{\iota-1}$,
or $(\gamma_1,\dots,\gamma_n)\neq \bs{0}^n$, $\mi{reduced}(\lambda^{(\gamma_1,\dots,\gamma_n)})\not\in E_{\iota-1}$;
$\mi{reduced}(\lambda^{(\gamma_1,\dots,\gamma_n)})\not\in E_{\iota-1}$;
$\mi{reduced}(\lambda^{(\gamma_1^\natural,\dots,\gamma_n^\natural)})\in \mi{min}(E_{\iota-1})$, 
$\mi{reduced}(\lambda^{(\gamma_1^\natural,\dots,\gamma_n^\natural)})=(\mu_1^\natural\mult a_\iota^{\beta^\natural})\geql \mu_2^\natural$, 
$\mu_i^\natural\in \{\gu\}\cup \mi{PropConj}_A$, $\mi{atoms}(\mu_i^\natural)\subseteq \mi{dom}({\mc V}_{\iota-1})$, $\beta^\natural\geq 1$,
$\left(\frac{\|\mu_2\|^{{\mc V}_{\iota-1}}}
            {\|\mu_1\|^{{\mc V}_{\iota-1}}}\right)^{\frac{1}{\beta^\natural}}\in \mbb{E}_{\iota-1}\neq \emptyset$,
which is a contradiction with $\mbb{E}_{\iota-1}=\emptyset$;
$\varepsilon_1\geql \varepsilon_2=\varepsilon_1\geql (\upsilon_2\mult a_\iota^\alpha)\not\in \mi{clo}$;
(\ref{eq8h}) holds.

So, in both Cases 2.1 and 2.2, (\ref{eq8h}) holds.
%
%
%
\end{proof}

\subsection{Full proof of the statement (\ref{eq8i}) of Lemma \ref{le2}}
\label{S7.5f}

\begin{proof}   
Let $a_\iota\in \mi{atoms}(\varepsilon_1)$ or $a_\iota\in \mi{atoms}(\varepsilon_2)$, $\mbb{E}_{\iota-1}=\emptyset$, $\varepsilon_1\diamond \varepsilon_2\in \mi{clo}$, $\diamond\in \{\gleq,\gle\}$.
Then there exists $\bs{0}^n\neq (\gamma_1^*,\dots,\gamma_n^*)\in \mbb{N}^n$ satisfying 
$\varepsilon_1\diamond \varepsilon_2\overset{\text{(\ref{eq8ddd})}}{=\!\!=} \mi{reduced}(\lambda^{(\gamma_1^*,\dots,\gamma_n^*)})$;
by the induction hypothesis for $\iota-1$, $\mi{dom}({\mc V}_{\iota-1})\overset{\text{(a)}}{=\!\!=} \{a_1,\dots,a_{\iota-1}\}$;
for all $\iota+1\leq j\leq m$, 
$a_j\not\in \mi{dom}({\mc V}_{\iota-1})\cup \{a_\iota\}=\{a_1,\dots,a_{\iota-1}\}\cup \{a_\iota\}=\{a_1,\dots,a_\iota\}\supseteq \mi{atoms}(\varepsilon_1), \mi{atoms}(\varepsilon_2)$,
$\#(a_j,\varepsilon_1\diamond \varepsilon_2)=0$;
$a_\iota\not\in \mi{dom}({\mc V}_{\iota-1})=\{a_1,\dots,a_{\iota-1}\}$, 
$\mi{atoms}(\varepsilon_1), \mi{atoms}(\varepsilon_2)\subseteq \mi{dom}({\mc V}_{\iota-1})\cup \{a_\iota\}$,
$\mi{atoms}(\varepsilon_1)\cap \mi{atoms}(\varepsilon_2)\overset{\text{(\ref{eq8aaa})}}{=\!\!=} \emptyset$,
either $a_\iota\in \mi{atoms}(\varepsilon_1)$ or $a_\iota\in \mi{atoms}(\varepsilon_2)$.
We proceed by complete induction on $\sum_{i=1}^n \gamma_i^*$.
We distinguish two cases for $\varepsilon_1$ and $\varepsilon_2$.

Case 2.3:
$a_\iota\in \mi{atoms}(\varepsilon_1)$ and $a_\iota\not\in \mi{atoms}(\varepsilon_2)\subseteq \mi{dom}({\mc V}_{\iota-1})$.
Then $\mi{atoms}(\varepsilon_1)\subseteq \mi{dom}({\mc V}_{\iota-1})\cup \{a_\iota\}$,
$\varepsilon_1=\upsilon_1\mult a_\iota^\alpha$, $\upsilon_1\in \{\gu\}\cup \mi{PropConj}_A$, $\alpha\geq 1$, 
$\mi{atoms}(\upsilon_1)\subseteq \mi{atoms}(\varepsilon_1)\subseteq \mi{dom}({\mc V}_{\iota-1})\cup \{a_\iota\}$,
$a_\iota\not\in \mi{atoms}(\upsilon_1)\subseteq \mi{dom}({\mc V}_{\iota-1})$.
We get two cases for $\diamond$.

Case 2.3.1:
$\diamond=\gleq$.
We put $\mi{EXP}=\{(\gamma_1,\dots,\gamma_n) \,|\, \bs{0}^n\neq (\gamma_1,\dots,\gamma_n)\in \mbb{N}^n, (\gamma_1,\dots,\gamma_n)\preceq (\gamma_1^*,\dots,\gamma_n^*), 
                                                   \mi{reduced}(\lambda^{(\gamma_1,\dots,\gamma_n)})\in \mi{UE}_{\iota-1}\}$.
Then $\varepsilon_1\diamond \varepsilon_2=(\upsilon_1\mult a_\iota^\alpha)\gleq \varepsilon_2=\mi{reduced}(\lambda^{(\gamma_1^*,\dots,\gamma_n^*)})\in \mi{clo}$, 
$\mi{atoms}(\upsilon_1), \mi{atoms}(\varepsilon_2)\subseteq \mi{dom}({\mc V}_{\iota-1})$,
$(\upsilon_1\mult a_\iota^\alpha)\gleq \varepsilon_2=\mi{reduced}(\lambda^{(\gamma_1^*,\dots,\gamma_n^*)})\in \mi{UE}_{\iota-1}$,
$(\gamma_1^*,\dots,\gamma_n^*)\neq \bs{0}^n$, $(\gamma_1^*,\dots,\gamma_n^*)\preceq (\gamma_1^*,\dots,\gamma_n^*)$, 
$(\gamma_1^*,\dots,\gamma_n^*)\in \mi{EXP}\neq \emptyset$;
we have that $\preceq$ is a well-founded order on $\mbb{N}^n$;
$\mi{min}(\mi{EXP})\neq \emptyset$;
there exists $\bs{0}^n\neq (\gamma_1^\natural,\dots,\gamma_n^\natural)\in \mi{min}(\mi{EXP})\subseteq \mi{EXP}\subseteq \mbb{N}^n$ satisfying 
$(\gamma_1^\natural,\dots,\gamma_n^\natural)\preceq (\gamma_1^*,\dots,\gamma_n^*)$, $\mi{reduced}(\lambda^{(\gamma_1^\natural,\dots,\gamma_n^\natural)})\in \mi{UE}_{\iota-1}$;
for all $(\gamma_1,\dots,\gamma_n)\in \mbb{N}^n$ satisfying $(\gamma_1,\dots,\gamma_n)\prec (\gamma_1^\natural,\dots,\gamma_n^\natural)$, 
$(\gamma_1,\dots,\gamma_n)\not\in \mi{EXP}$,
$(\gamma_1,\dots,\gamma_n)\prec (\gamma_1^\natural,\dots,\gamma_n^\natural)\preceq (\gamma_1^*,\dots,\gamma_n^*)$,
either $(\gamma_1,\dots,\gamma_n)=\bs{0}^n$, $\mi{reduced}(\lambda^{(\gamma_1,\dots,\gamma_n)})=\mi{reduced}(\lambda^{\bs{0}^n})=\gu\underset{\text{(\ref{eq8a})}}{\not\in} 
                                                                                                \mi{clo}\supseteq \mi{UE}_{\iota-1}$,
or $(\gamma_1,\dots,\gamma_n)\neq \bs{0}^n$, $\mi{reduced}(\lambda^{(\gamma_1,\dots,\gamma_n)})\not\in \mi{UE}_{\iota-1}$;
$\mi{reduced}(\lambda^{(\gamma_1,\dots,\gamma_n)})\not\in \mi{UE}_{\iota-1}$;
$\mi{reduced}(\lambda^{(\gamma_1^\natural,\dots,\gamma_n^\natural)})\in \mi{min}(\mi{UE}_{\iota-1})$,
$\mi{reduced}(\lambda^{(\gamma_1^\natural,\dots,\gamma_n^\natural)})=(\mu_1^\natural\mult a_\iota^{\beta^\natural})\gleq \mu_2^\natural$, 
$\mu_i^\natural\in \{\gu\}\cup \mi{PropConj}_A$, $\mi{atoms}(\mu_i^\natural)\subseteq \mi{dom}({\mc V}_{\iota-1})$, $\beta^\natural\geq 1$,
$\mi{atoms}(\mu_1^\natural\mult a_\iota^{\beta^\natural})=\mi{atoms}(\mu_1^\natural)\cup \mi{atoms}(a_\iota^{\beta^\natural})\subseteq \mi{dom}({\mc V}_{\iota-1})\cup \{a_\iota\}$;
for all $\iota+1\leq j\leq m$, 
$a_j\not\in \mi{dom}({\mc V}_{\iota-1})\cup \{a_\iota\}\supseteq \mi{atoms}(\mu_1^\natural\mult a_\iota^{\beta^\natural}), \mi{atoms}(\mu_2^\natural)$,
$\#(a_j,(\mu_1^\natural\mult a_\iota^{\beta^\natural})\gleq \mu_2^\natural)=0$.
We get two cases for $(\gamma_1^*,\dots,\gamma_n^*)$ and $(\gamma_1^\natural,\dots,\gamma_n^\natural)$.

Case 2.3.1.1:
$(\gamma_1^*,\dots,\gamma_n^*)=(\gamma_1^\natural,\dots,\gamma_n^\natural)$.
We put $\kappa_1=\kappa_2=0$, $\kappa_3=\kappa_4=1$, $\mu_i^1=\mu_i^\natural\in \{\gu\}\cup \mi{PropConj}_A$, $i=1,2$, $\beta_1=\beta^\natural\geq 1$,
$\bs{0}^n\neq (\gamma_1^1,\dots,\gamma_n^1)=(\gamma_1^\natural,\dots,\gamma_n^\natural)\in \mbb{N}^n$.
Then $\kappa_1=\kappa_2=0<\kappa_3=\kappa_4=1$,
\begin{alignat*}{1}
(\upsilon_1\mult a_\iota^\alpha)\gleq \varepsilon_2
&= \mi{reduced}(\lambda^{(\gamma_1^*,\dots,\gamma_n^*)})=\mi{reduced}(\lambda^{(\gamma_1^1,\dots,\gamma_n^1)})= \\
&\phantom{\mbox{}=\mbox{}}                           
   \mi{reduced}(\lambda^{(\gamma_1^\natural,\dots,\gamma_n^\natural)})=\mi{reduced}(\lambda^{(\sum_{j=1}^{\kappa_4} \gamma_1^j,\dots,\sum_{j=1}^{\kappa_4} \gamma_n^j)}),
\end{alignat*}
$(\mu_1^1\mult a_\iota^{\beta_1})\gleq \mu_2^1=(\mu_1^\natural\mult a_\iota^{\beta^\natural})\gleq \mu_2^\natural= 
 \mi{reduced}(\lambda^{(\gamma_1^1,\dots,\gamma_n^1)})=\mi{reduced}(\lambda^{(\gamma_1^\natural,\dots,\gamma_n^\natural)})\in \mi{min}(\mi{UE}_{\iota-1})$;
for both $e$, $\mi{atoms}(\mu_e^1)=\mi{atoms}(\mu_e^\natural)\subseteq \mi{dom}({\mc V}_{\iota-1})$.
Hence, there exist $\kappa_1=\kappa_2<\kappa_3=\kappa_4$,
$(\mu_1^j\mult a_\iota^{\beta_j})\gleq \mu_2^j=\mi{reduced}(\lambda^{(\gamma_1^j,\dots,\gamma_n^j)})\in \mi{min}(\mi{UE}_{\iota-1})$, 
$\mu_e^j\in \{\gu\}\cup \mi{PropConj}_A$, $\mi{atoms}(\mu_e^j)\subseteq \mi{dom}({\mc V}_{\iota-1})$, $\beta_j\geq 1$, $\bs{0}^n\neq (\gamma_1^j,\dots,\gamma_n^j)\in \mbb{N}^n$,
$j=\kappa_2+1,\dots,\kappa_3$, such that 
$\varepsilon_1\diamond \varepsilon_2=(\upsilon_1\mult a_\iota^\alpha)\gleq \varepsilon_2=\mi{reduced}(\lambda^{(\sum_{j=1}^{\kappa_4} \gamma_1^j,\dots,\sum_{j=1}^{\kappa_4} \gamma_n^j)})$;
(\ref{eq8i}c) holds.

Case 2.3.1.2:
$(\gamma_1^*,\dots,\gamma_n^*)\neq (\gamma_1^\natural,\dots,\gamma_n^\natural)$.
Then $(\gamma_1^\natural,\dots,\gamma_n^\natural)\preceq (\gamma_1^*,\dots,\gamma_n^*)$;
for all $i\leq n$, $\gamma_i^\natural\leq \gamma_i^*$;
there exists $i^*\leq n$ satisfying $\gamma_{i^*}^\natural\neq \gamma_{i^*}^*$, $\gamma_{i^*}^\natural<\gamma_{i^*}^*$, $\gamma_{i^*}^*-\gamma_{i^*}^\natural\geq 1$;
$\bs{0}^n\neq (\gamma_1^*-\gamma_1^\natural,\dots,\gamma_n^*-\gamma_n^\natural)\in \mbb{N}^n$,
$\mi{reduced}(\lambda^{(\gamma_1^*-\gamma_1^\natural,\dots,\gamma_n^*-\gamma_n^\natural)})=\mu_1^\#\diamond^\# \mu_2^\#\underset{\text{(\ref{eq8dd})}}{\in} \mi{clo}$, 
$\mu_j^\#\in \{\gu\}\cup \mi{PropConj}_A$, $\diamond^\#\in \{\geql,\gleq,\gle\}$,
$\lambda^{(\gamma_1^*,\dots,\gamma_n^*)}=\lambda^{(\gamma_1^\natural+\gamma_1^*-\gamma_1^\natural,\dots,\gamma_n^\natural+\gamma_n^*-\gamma_n^\natural)}\overset{\text{(\ref{eq8dddd})}}{=\!\!=}
                                         \lambda^{(\gamma_1^\natural,\dots,\gamma_n^\natural)}\mult \lambda^{(\gamma_1^*-\gamma_1^\natural,\dots,\gamma_n^*-\gamma_n^\natural)}$,
$(\upsilon_1\mult a_\iota^\alpha)\gleq \varepsilon_2=\mi{reduced}(\lambda^{(\gamma_1^*,\dots,\gamma_n^*)})=
                                                     \mi{reduced}(\lambda^{(\gamma_1^\natural,\dots,\gamma_n^\natural)}\mult 
                                                                  \lambda^{(\gamma_1^*-\gamma_1^\natural,\dots,\gamma_n^*-\gamma_n^\natural)})$;
for all $\iota+1\leq j\leq m$,
\begin{alignat*}{1}
\#(a_j,(\mu_1^\natural\mult a_\iota^{\beta^\natural})\gleq \mu_2^\natural)
&= \#(a_j,\mi{reduced}(\lambda^{(\gamma_1^\natural,\dots,\gamma_n^\natural)}))\overset{\text{(\ref{eq8aaaax})}}{=\!\!=} \#(a_j,\lambda^{(\gamma_1^\natural,\dots,\gamma_n^\natural)})=0, \\[1mm]
\#(a_j,\varepsilon_1\diamond \varepsilon_2)
&= \#(a_j,(\upsilon_1\mult a_\iota^\alpha)\gleq \varepsilon_2)=\#(a_j,\mi{reduced}(\lambda^{(\gamma_1^*,\dots,\gamma_n^*)}))\overset{\text{(\ref{eq8aaaax})}}{=\!\!=} \\
&\phantom{\mbox{}=\mbox{}}
   \#(a_j,\lambda^{(\gamma_1^*,\dots,\gamma_n^*)})= \\
&\phantom{\mbox{}=\mbox{}}
   \#(a_j,\lambda^{(\gamma_1^\natural,\dots,\gamma_n^\natural)}\mult \lambda^{(\gamma_1^*-\gamma_1^\natural,\dots,\gamma_n^*-\gamma_n^\natural)})\overset{\text{(\ref{eq8aax})}}{=\!\!=} \\
&\phantom{\mbox{}=\mbox{}}
   \#(a_j,\lambda^{(\gamma_1^\natural,\dots,\gamma_n^\natural)})+\#(a_j,\lambda^{(\gamma_1^*-\gamma_1^\natural,\dots,\gamma_n^*-\gamma_n^\natural)})= \\
&\phantom{\mbox{}=\mbox{}}
   \#(a_j,\lambda^{(\gamma_1^*-\gamma_1^\natural,\dots,\gamma_n^*-\gamma_n^\natural)})=0, \\[1mm]
\#(a_j,\mu_1^\#\diamond^\# \mu_2^\#)
&= \#(a_j,\mi{reduced}(\lambda^{(\gamma_1^*-\gamma_1^\natural,\dots,\gamma_n^*-\gamma_n^\natural)}))\overset{\text{(\ref{eq8aaaax})}}{=\!\!=} \\
&\phantom{\mbox{}=\mbox{}}
   \#(a_j,\lambda^{(\gamma_1^*-\gamma_1^\natural,\dots,\gamma_n^*-\gamma_n^\natural)})=0,
\end{alignat*}
$\mi{atoms}(\mu_1^\#)\cap \mi{atoms}(\mu_2^\#)\overset{\text{(\ref{eq8aaa})}}{=\!\!=} \emptyset$;
if $a_j\in \mi{atoms}(\mu_1^\#\diamond^\# \mu_2^\#)=\mi{atoms}(\mu_1^\#)\cup \mi{atoms}(\mu_1^\#)$,
either $a_j\in \mi{atoms}(\mu_1^\#)$ or $a_j\in \mi{atoms}(\mu_2^\#)$,
either $\#(a_j,\mu_1^\#\diamond^\# \mu_2^\#)\leq -1$ or $\#(a_j,\mu_1^\#\diamond^\# \mu_2^\#)\geq 1$;
$a_j\not\in \mi{atoms}(\mu_1^\#\diamond^\# \mu_2^\#)\supseteq \mi{atoms}(\mu_1^\#), \mi{atoms}(\mu_2^\#)$;
$\mi{atoms}(\mu_1^\#), \mi{atoms}(\mu_2^\#)\subseteq A=\{a_1,\dots,a_m\}$,
$\mi{atoms}(\mu_1^\#), \mi{atoms}(\mu_2^\#)\subseteq \{a_1,\dots,a_m\}-\{a_{\iota+1},\dots,a_m\}=\mi{dom}({\mc V}_{\iota-1})\cup \{a_\iota\}=\{a_1,\dots,a_\iota\}$,
$a_\iota\not\in \mi{dom}({\mc V}_{\iota-1})$.
We get two cases for $\mu_1^\#$ and $\mu_2^\#$.

Case 2.3.1.2.1:
$\mi{atoms}(\mu_1^\#), \mi{atoms}(\mu_2^\#)\subseteq \mi{dom}({\mc V}_{\iota-1})$.
We put $\kappa_1=\kappa_2=0$, $\kappa_3=\kappa_4=1$, $\mu_i^1=\mu_i^\natural\in \{\gu\}\cup \mi{PropConj}_A$, $i=1,2$, $\beta_1=\beta^\natural\geq 1$,
$\bs{0}^n\neq (\gamma_1^1,\dots,\gamma_n^1)=(\gamma_1^\natural,\dots,\gamma_n^\natural)\in \mbb{N}^n$,
$\zeta_i=\mu_i^\#\in \{\gu\}\cup \mi{PropConj}_A$, $i=1,2$, $\diamond^\zeta=\diamond^\#\in \{\geql,\gleq,\gle\}$,
$\bs{0}^n\neq (\gamma_1^\zeta,\dots,\gamma_n^\zeta)=(\gamma_1^*-\gamma_1^\natural,\dots,\gamma_n^*-\gamma_n^\natural)\in \mbb{N}^n$.
Then $\kappa_1=\kappa_2=0<\kappa_3=\kappa_4=1$,
\begin{alignat*}{1}
(\upsilon_1\mult a_\iota^\alpha)\gleq \varepsilon_2
&= \mi{reduced}(\lambda^{(\gamma_1^*,\dots,\gamma_n^*)})=\mi{reduced}(\lambda^{(\gamma_1^\natural,\dots,\gamma_n^\natural)}\mult 
                                                                      \lambda^{(\gamma_1^*-\gamma_1^\natural,\dots,\gamma_n^*-\gamma_n^\natural)})= \\
&\phantom{\mbox{}=\mbox{}}
   \mi{reduced}(\lambda^{(\gamma_1^1,\dots,\gamma_n^1)}\mult \lambda^{(\gamma_1^\zeta,\dots,\gamma_n^\zeta)})\overset{\text{(\ref{eq8dddd})}}{=\!\!=} \\
&\phantom{\mbox{}=\mbox{}}
   \mi{reduced}(\lambda^{((\sum_{j=1}^{\kappa_4} \gamma_1^j)+\gamma_1^\zeta,\dots,(\sum_{j=1}^{\kappa_4} \gamma_n^j)+\gamma_n^\zeta)}),
\end{alignat*}
$(\mu_1^1\mult a_\iota^{\beta_1})\gleq \mu_2^1=(\mu_1^\natural\mult a_\iota^{\beta^\natural})\gleq \mu_2^\natural= 
 \mi{reduced}(\lambda^{(\gamma_1^1,\dots,\gamma_n^1)})=\mi{reduced}(\lambda^{(\gamma_1^\natural,\dots,\gamma_n^\natural)})\in \mi{min}(\mi{UE}_{\iota-1})$;
for both $e$, $\mi{atoms}(\mu_e^1)=\mi{atoms}(\mu_e^\natural)\subseteq \mi{dom}({\mc V}_{\iota-1})$;
$\zeta_1\diamond^\zeta \zeta_2=\mu_1^\#\diamond^\# \mu_2^\#=
 \mi{reduced}(\lambda^{(\gamma_1^\zeta,\dots,\gamma_n^\zeta)})=\mi{reduced}(\lambda^{(\gamma_1^*-\gamma_1^\natural,\dots,\gamma_n^*-\gamma_n^\natural)})\in \mi{clo}$;
for both $e$, $\mi{atoms}(\zeta_e)=\mi{atoms}(\mu_e^\#)\subseteq \mi{dom}({\mc V}_{\iota-1})$.
Hence, there exist $\kappa_1=\kappa_2<\kappa_3=\kappa_4$,
$(\mu_1^j\mult a_\iota^{\beta_j})\gleq \mu_2^j=\mi{reduced}(\lambda^{(\gamma_1^j,\dots,\gamma_n^j)})\in \mi{min}(\mi{UE}_{\iota-1})$, 
$\mu_e^j\in \{\gu\}\cup \mi{PropConj}_A$, $\mi{atoms}(\mu_e^j)\subseteq \mi{dom}({\mc V}_{\iota-1})$, $\beta_j\geq 1$, $\bs{0}^n\neq (\gamma_1^j,\dots,\gamma_n^j)\in \mbb{N}^n$,
$j=\kappa_2+1,\dots,\kappa_3$, 
$\zeta_1\diamond^\zeta \zeta_2=\mi{reduced}(\lambda^{(\gamma_1^\zeta,\dots,\gamma_n^\zeta)})\in \mi{clo}$,
$\zeta_e\in \{\gu\}\cup \mi{PropConj}_A$, $\mi{atoms}(\zeta_e)\subseteq \mi{dom}({\mc V}_{\iota-1})$, $\diamond^\zeta\in \{\geql,\gleq,\gle\}$, 
$\bs{0}^n\neq (\gamma_1^\zeta,\dots,\gamma_n^\zeta)\in \mbb{N}^n$, such that 
$\varepsilon_1\diamond \varepsilon_2=(\upsilon_1\mult a_\iota^\alpha)\gleq \varepsilon_2=
                                     \mi{reduced}(\lambda^{((\sum_{j=1}^{\kappa_4} \gamma_1^j)+\gamma_1^\zeta,\dots,(\sum_{j=1}^{\kappa_4} \gamma_n^j)+\gamma_n^\zeta)})$;
(\ref{eq8i}c) holds.

Case 2.3.1.2.2:
$a_\iota\in \mi{atoms}(\mu_1^\#)$ or $a_\iota\in \mi{atoms}(\mu_2^\#)$. 
Then $\mi{atoms}(\mu_1^\#),                                                                                                                                                                \linebreak[4]
                            \mi{atoms}(\mu_2^\#)\subseteq \mi{dom}({\mc V}_{\iota-1})\cup \{a_\iota\}$,
$\mbb{E}_{\iota-1}=\emptyset$, 
$\mu_1^\#\diamond^\# \mu_2^\#\in \mi{clo}$, $\diamond^\#\in \{\geql,\gleq,\gle\}$,
$\mu_1^\#\geql \mu_2^\#\underset{\text{(\ref{eq8h})}}{\not\in} \mi{clo}$,
$\diamond^\#\neq \geql$, $\diamond^\#\in \{\gleq,\gle\}$;
for all $i\leq n$, $\gamma_i^*-\gamma_i^\natural\leq \gamma_i^*$;
$(\gamma_1^\natural,\dots,\gamma_n^\natural)\neq \bs{0}^n$;
there exists $i^{**}\leq n$ satisfying $\gamma_{i^{**}}^\natural\geq 1$, $\gamma_{i^{**}}^*-\gamma_{i^{**}}^\natural<\gamma_{i^{**}}^*$;
$\sum_{i=1}^n \gamma_i^*-\gamma_i^\natural<\sum_{i=1}^n \gamma_i^*$;
by the induction hypothesis for $\mu_1^\#\diamond^\# \mu_2^\#$ and $\sum_{i=1}^n \gamma_i^*-\gamma_i^\natural$, there exist $\kappa_1\leq \kappa_2<\kappa_3\leq \kappa_4$,
\begin{alignat*}{1}
& \mu_2^j\gleq (\mu_1^j\mult a_\iota^{\beta_j})=\mi{reduced}(\lambda^{(\gamma_1^j,\dots,\gamma_n^j)})\in \mi{min}(\mi{DE}_{\iota-1}), j=1,\dots,\kappa_1, \\ 
& \mu_2^j\gle (\mu_1^j\mult a_\iota^{\beta_j})=\mi{reduced}(\lambda^{(\gamma_1^j,\dots,\gamma_n^j)})\in \mi{min}(D_{\iota-1}), j=\kappa_1+1,\dots,\kappa_2, \\
& (\mu_1^j\mult a_\iota^{\beta_j})\gleq \mu_2^j=\mi{reduced}(\lambda^{(\gamma_1^j,\dots,\gamma_n^j)})\in \mi{min}(\mi{UE}_{\iota-1}), j=\kappa_2+2,\dots,\kappa_3, \\
& (\mu_1^j\mult a_\iota^{\beta_j})\gle \mu_2^j=\mi{reduced}(\lambda^{(\gamma_1^j,\dots,\gamma_n^j)})\in \mi{min}(U_{\iota-1}), j=\kappa_3+1,\dots,\kappa_4, \\ 
& \mu_e^j\in \{\gu\}\cup \mi{PropConj}_A, \mi{atoms}(\mu_e^j)\subseteq \mi{dom}({\mc V}_{\iota-1}), \beta_j\geq 1, \bs{0}^n\neq (\gamma_1^j,\dots,\gamma_n^j)\in \mbb{N}^n, 
\end{alignat*}
satisfying either 
\begin{alignat*}{1}
\mu_1^\#\diamond^\# \mu_2^\# &= \mi{reduced}(\lambda^{(\gamma_1^*-\gamma_1^\natural,\dots,\gamma_n^*-\gamma_n^\natural)})= \\
                             &\phantom{\mbox{}=\mbox{}}
                                \mi{reduced}(\lambda^{(\sum_{j=1,\dots,\kappa_2,\kappa_2+2,\dots,\kappa_4} \gamma_1^j,\dots,\sum_{j=1,\dots,\kappa_2,\kappa_2+2,\dots,\kappa_4} \gamma_n^j)}),
\end{alignat*}
or there exists $\zeta_1\diamond^\zeta \zeta_2=\mi{reduced}(\lambda^{(\gamma_1^\zeta,\dots,\gamma_n^\zeta)})\in \mi{clo}$,
$\zeta_e\in \{\gu\}\cup \mi{PropConj}_A$, $\mi{atoms}(\zeta_e)\subseteq \mi{dom}({\mc V}_{\iota-1})$, $\diamond^\zeta\in \{\geql,\gleq,\gle\}$, 
$\bs{0}^n\neq (\gamma_1^\zeta,\dots,\gamma_n^\zeta)\in \mbb{N}^n$, satisfying
\begin{alignat*}{1}
\mu_1^\#\diamond^\# \mu_2^\# &= \mi{reduced}(\lambda^{(\gamma_1^*-\gamma_1^\natural,\dots,\gamma_n^*-\gamma_n^\natural)})= \\
                             &\phantom{\mbox{}=\mbox{}}
                                \mi{reduced}(\lambda^{((\sum_{j=1,\dots,\kappa_2,\kappa_2+2,\dots,\kappa_4} \gamma_1^j)+\gamma_1^\zeta,\dots,(\sum_{j=1,\dots,\kappa_2,\kappa_2+2,\dots,\kappa_4} \gamma_n^j)+\gamma_n^\zeta)}).
\end{alignat*}
We put $\mu_i^{\kappa_2+1}=\mu_i^\natural\in \{\gu\}\cup \mi{PropConj}_A$, $i=1,2$, $\beta_{\kappa_2+1}=\beta^\natural\geq 1$,
$\bs{0}^n\neq (\gamma_1^{\kappa_2+1},\dots,\gamma_n^{\kappa_2+1})=(\gamma_1^\natural,\dots,\gamma_n^\natural)\in \mbb{N}^n$.
Then either
\begin{alignat*}{1}
(\upsilon_1\mult a_\iota^\alpha)\gleq \varepsilon_2 
&= \mi{reduced}(\lambda^{(\gamma_1^\natural,\dots,\gamma_n^\natural)}\mult \lambda^{(\gamma_1^*-\gamma_1^\natural,\dots,\gamma_n^*-\gamma_n^\natural)})\overset{\text{(\ref{eq7k})}}{=\!\!=} \\
&\phantom{\mbox{}=\mbox{}}
   \mi{reduced}(\lambda^{(\gamma_1^\natural,\dots,\gamma_n^\natural)}\mult \mi{reduced}(\lambda^{(\gamma_1^*-\gamma_1^\natural,\dots,\gamma_n^*-\gamma_n^\natural)}))= \\
&\phantom{\mbox{}=\mbox{}}
   \mi{reduced}(\lambda^{(\gamma_1^{\kappa_2+1},\dots,\gamma_n^{\kappa_2+1})}\mult \\
&\phantom{\mbox{}=\mbox{}} \quad
                \mi{reduced}(\lambda^{(\sum_{j=1,\dots,\kappa_2,\kappa_2+2,\dots,\kappa_4} \gamma_1^j,\dots,\sum_{j=1,\dots,\kappa_2,\kappa_2+2,\dots,\kappa_4} \gamma_n^j)}))\overset{\text{(\ref{eq7k})}}{=\!\!=} \\
&\phantom{\mbox{}=\mbox{}}
   \mi{reduced}(\lambda^{(\gamma_1^{\kappa_2+1},\dots,\gamma_n^{\kappa_2+1})}\mult \\
&\phantom{\mbox{}=\mi{reduced}(} 
                \lambda^{(\sum_{j=1,\dots,\kappa_2,\kappa_2+2,\dots,\kappa_4} \gamma_1^j,\dots,\sum_{j=1,\dots,\kappa_2,\kappa_2+2,\dots,\kappa_4} \gamma_n^j)})\overset{\text{(\ref{eq8dddd})}}{=\!\!=} \\
&\phantom{\mbox{}=\mbox{}}
   \mi{reduced}(\lambda^{(\sum_{j=1}^{\kappa_4} \gamma_1^j,\dots,\sum_{j=1}^{\kappa_4} \gamma_n^j)})
\end{alignat*}
or
\begin{alignat*}{1}
& (\upsilon_1\mult a_\iota^\alpha)\gleq \varepsilon_2= 
  \mi{reduced}(\lambda^{(\gamma_1^\natural,\dots,\gamma_n^\natural)}\mult \mi{reduced}(\lambda^{(\gamma_1^*-\gamma_1^\natural,\dots,\gamma_n^*-\gamma_n^\natural)}))= \\
& \quad
  \mi{reduced}(\lambda^{(\gamma_1^{\kappa_2+1},\dots,\gamma_n^{\kappa_2+1})}\mult \\
& \quad \phantom{\mi{reduced}(} 
               \mi{reduced}(\lambda^{((\sum_{j=1,\dots,\kappa_2,\kappa_2+2,\dots,\kappa_4} \gamma_1^j)+\gamma_1^\zeta,\dots,(\sum_{j=1,\dots,\kappa_2,\kappa_2+2,\dots,\kappa_4} \gamma_n^j)+\gamma_n^\zeta)}))\overset{\text{(\ref{eq7k})}}{=\!\!=} \\
& \quad
  \mi{reduced}(\lambda^{(\gamma_1^{\kappa_2+1},\dots,\gamma_n^{\kappa_2+1})}\mult \\
& \quad \phantom{\mi{reduced}(}
               \lambda^{((\sum_{j=1,\dots,\kappa_2,\kappa_2+2,\dots,\kappa_4} \gamma_1^j)+\gamma_1^\zeta,\dots,(\sum_{j=1,\dots,\kappa_2,\kappa_2+2,\dots,\kappa_4} \gamma_n^j)+\gamma_n^\zeta)})\overset{\text{(\ref{eq8dddd})}}{=\!\!=} \\
& \quad
  \mi{reduced}(\lambda^{((\sum_{j=1}^{\kappa_4} \gamma_1^j)+\gamma_1^\zeta,\dots,(\sum_{j=1}^{\kappa_4} \gamma_n^j)+\gamma_n^\zeta)}),
\end{alignat*}
$(\mu_1^{\kappa_2+1}\mult a_\iota^{\beta_{\kappa_2+1}})\gleq \mu_2^{\kappa_2+1}=(\mu_1^\natural\mult a_\iota^{\beta^\natural})\gleq \mu_2^\natural=
 \mi{reduced}(\lambda^{(\gamma_1^{\kappa_2+1},\dots,\gamma_n^{\kappa_2+1})})=\mi{reduced}(\lambda^{(\gamma_1^\natural,\dots,\gamma_n^\natural)})\in \mi{min}(\mi{UE}_{\iota-1})$;
for both $e$, $\mi{atoms}(\mu_e^{\kappa_2+1})=\mi{atoms}(\mu_e^\natural)\subseteq \mi{dom}({\mc V}_{\iota-1})$.
Hence, there exist $\kappa_1\leq \kappa_2<\kappa_3\leq \kappa_4$,
\begin{alignat*}{1}
& \mu_2^j\gleq (\mu_1^j\mult a_\iota^{\beta_j})=\mi{reduced}(\lambda^{(\gamma_1^j,\dots,\gamma_n^j)})\in \mi{min}(\mi{DE}_{\iota-1}), j=1,\dots,\kappa_1, \\ 
& \mu_2^j\gle (\mu_1^j\mult a_\iota^{\beta_j})=\mi{reduced}(\lambda^{(\gamma_1^j,\dots,\gamma_n^j)})\in \mi{min}(D_{\iota-1}), j=\kappa_1+1,\dots,\kappa_2, \\
& (\mu_1^j\mult a_\iota^{\beta_j})\gleq \mu_2^j=\mi{reduced}(\lambda^{(\gamma_1^j,\dots,\gamma_n^j)})\in \mi{min}(\mi{UE}_{\iota-1}), j=\kappa_2+1,\dots,\kappa_3, \\
& (\mu_1^j\mult a_\iota^{\beta_j})\gle \mu_2^j=\mi{reduced}(\lambda^{(\gamma_1^j,\dots,\gamma_n^j)})\in \mi{min}(U_{\iota-1}), j=\kappa_3+1,\dots,\kappa_4, \\ 
& \mu_e^j\in \{\gu\}\cup \mi{PropConj}_A, \mi{atoms}(\mu_e^j)\subseteq \mi{dom}({\mc V}_{\iota-1}), \beta_j\geq 1, \bs{0}^n\neq (\gamma_1^j,\dots,\gamma_n^j)\in \mbb{N}^n, 
\end{alignat*}
such that either $\varepsilon_1\diamond \varepsilon_2=(\upsilon_1\mult a_\iota^\alpha)\gleq \varepsilon_2=
                                                      \mi{reduced}(\lambda^{(\sum_{j=1}^{\kappa_4} \gamma_1^j,\dots,\sum_{j=1}^{\kappa_4} \gamma_n^j)})$, 
or there exists $\zeta_1\diamond^\zeta \zeta_2=\mi{reduced}(\lambda^{(\gamma_1^\zeta,\dots,\gamma_n^\zeta)})\in \mi{clo}$,
$\zeta_e\in \{\gu\}\cup \mi{PropConj}_A$, $\mi{atoms}(\zeta_e)\subseteq \mi{dom}({\mc V}_{\iota-1})$, $\diamond^\zeta\in \{\geql,\gleq,\gle\}$, 
$\bs{0}^n\neq (\gamma_1^\zeta,\dots,\gamma_n^\zeta)\in \mbb{N}^n$, satisfying
$\varepsilon_1\diamond \varepsilon_2=(\upsilon_1\mult a_\iota^\alpha)\gleq \varepsilon_2=
                                     \mi{reduced}(\lambda^{((\sum_{j=1}^{\kappa_4} \gamma_1^j)+\gamma_1^\zeta,\dots,(\sum_{j=1}^{\kappa_4} \gamma_n^j)+\gamma_n^\zeta)})$;
(\ref{eq8i}c) holds.

Case 2.3.2:
$\diamond=\gle$.
We put $\mi{EXP}=\{(\gamma_1,\dots,\gamma_n) \,|\, \bs{0}^n\neq (\gamma_1,\dots,\gamma_n)\in \mbb{N}^n, (\gamma_1,\dots,\gamma_n)\preceq (\gamma_1^*,\dots,\gamma_n^*), 
                                                   \mi{reduced}(\lambda^{(\gamma_1,\dots,\gamma_n)})\in U_{\iota-1}\}$.
Then $\varepsilon_1\diamond \varepsilon_2=(\upsilon_1\mult a_\iota^\alpha)\gle \varepsilon_2=\mi{reduced}(\lambda^{(\gamma_1^*,\dots,\gamma_n^*)})\in \mi{clo}$, 
$\mi{atoms}(\upsilon_1), \mi{atoms}(\varepsilon_2)\subseteq \mi{dom}({\mc V}_{\iota-1})$,
$(\upsilon_1\mult a_\iota^\alpha)\gle \varepsilon_2=\mi{reduced}(\lambda^{(\gamma_1^*,\dots,\gamma_n^*)})\in U_{\iota-1}$,
$(\gamma_1^*,\dots,\gamma_n^*)\neq \bs{0}^n$, $(\gamma_1^*,\dots,\gamma_n^*)\preceq (\gamma_1^*,\dots,\gamma_n^*)$, 
$(\gamma_1^*,\dots,\gamma_n^*)\in \mi{EXP}\neq \emptyset$;
we have that $\preceq$ is a well-founded order on $\mbb{N}^n$;
$\mi{min}(\mi{EXP})\neq \emptyset$;
there exists $\bs{0}^n\neq (\gamma_1^\natural,\dots,\gamma_n^\natural)\in \mi{min}(\mi{EXP})\subseteq \mi{EXP}\subseteq \mbb{N}^n$ satisfying 
$(\gamma_1^\natural,\dots,\gamma_n^\natural)\preceq (\gamma_1^*,\dots,\gamma_n^*)$, $\mi{reduced}(\lambda^{(\gamma_1^\natural,\dots,\gamma_n^\natural)})\in U_{\iota-1}$;
for all $(\gamma_1,\dots,\gamma_n)\in \mbb{N}^n$ satisfying $(\gamma_1,\dots,\gamma_n)\prec (\gamma_1^\natural,\dots,\gamma_n^\natural)$, 
$(\gamma_1,\dots,\gamma_n)\not\in \mi{EXP}$,
$(\gamma_1,\dots,\gamma_n)\prec (\gamma_1^\natural,\dots,\gamma_n^\natural)\preceq (\gamma_1^*,\dots,\gamma_n^*)$,
either $(\gamma_1,\dots,\gamma_n)=\bs{0}^n$, $\mi{reduced}(\lambda^{(\gamma_1,\dots,\gamma_n)})=\mi{reduced}(\lambda^{\bs{0}^n})=\gu\underset{\text{(\ref{eq8a})}}{\not\in}
                                                                                                \mi{clo}\supseteq U_{\iota-1}$,
or $(\gamma_1,\dots,\gamma_n)\neq \bs{0}^n$, $\mi{reduced}(\lambda^{(\gamma_1,\dots,\gamma_n)})\not\in U_{\iota-1}$;
$\mi{reduced}(\lambda^{(\gamma_1,\dots,\gamma_n)})\not\in U_{\iota-1}$;
$\mi{reduced}(\lambda^{(\gamma_1^\natural,\dots,\gamma_n^\natural)})\in \mi{min}(U_{\iota-1})$,
$\mi{reduced}(\lambda^{(\gamma_1^\natural,\dots,\gamma_n^\natural)})=(\mu_1^\natural\mult a_\iota^{\beta^\natural})\gle \mu_2^\natural$, 
$\mu_i^\natural\in \{\gu\}\cup \mi{PropConj}_A$, $\mi{atoms}(\mu_i^\natural)\subseteq \mi{dom}({\mc V}_{\iota-1})$, $\beta^\natural\geq 1$,
$\mi{atoms}(\mu_1^\natural\mult a_\iota^{\beta^\natural})=\mi{atoms}(\mu_1^\natural)\cup \mi{atoms}(a_\iota^{\beta^\natural})\subseteq \mi{dom}({\mc V}_{\iota-1})\cup \{a_\iota\}$;
for all $\iota+1\leq j\leq m$, 
$a_j\not\in \mi{dom}({\mc V}_{\iota-1})\cup \{a_\iota\}\supseteq \mi{atoms}(\mu_1^\natural\mult a_\iota^{\beta^\natural}), \mi{atoms}(\mu_2^\natural)$,
$\#(a_j,(\mu_1^\natural\mult a_\iota^{\beta^\natural})\gle \mu_2^\natural)=0$.
We get two cases for $(\gamma_1^*,\dots,\gamma_n^*)$ and $(\gamma_1^\natural,\dots,\gamma_n^\natural)$.

Case 2.3.2.1:
$(\gamma_1^*,\dots,\gamma_n^*)=(\gamma_1^\natural,\dots,\gamma_n^\natural)$.
We put $\kappa_1=\kappa_2=\kappa_3=0$, $\kappa_4=1$, $\mu_i^1=\mu_i^\natural\in \{\gu\}\cup \mi{PropConj}_A$, $i=1,2$, $\beta_1=\beta^\natural\geq 1$,
$\bs{0}^n\neq (\gamma_1^1,\dots,\gamma_n^1)=(\gamma_1^\natural,\dots,\gamma_n^\natural)\in \mbb{N}^n$.
Then $\kappa_1=\kappa_2=\kappa_3=0<\kappa_4=1$,
\begin{alignat*}{1}
(\upsilon_1\mult a_\iota^\alpha)\gle \varepsilon_2
&= \mi{reduced}(\lambda^{(\gamma_1^*,\dots,\gamma_n^*)})=\mi{reduced}(\lambda^{(\gamma_1^1,\dots,\gamma_n^1)})= \\
&\phantom{\mbox{}=\mbox{}}                           
   \mi{reduced}(\lambda^{(\gamma_1^\natural,\dots,\gamma_n^\natural)})=\mi{reduced}(\lambda^{(\sum_{j=1}^{\kappa_4} \gamma_1^j,\dots,\sum_{j=1}^{\kappa_4} \gamma_n^j)}),
\end{alignat*}
$(\mu_1^1\mult a_\iota^{\beta_1})\gle \mu_2^1=(\mu_1^\natural\mult a_\iota^{\beta^\natural})\gle \mu_2^\natural= 
 \mi{reduced}(\lambda^{(\gamma_1^1,\dots,\gamma_n^1)})=\mi{reduced}(\lambda^{(\gamma_1^\natural,\dots,\gamma_n^\natural)})\in \mi{min}(U_{\iota-1})$;
for both $e$, $\mi{atoms}(\mu_e^1)=\mi{atoms}(\mu_e^\natural)\subseteq \mi{dom}({\mc V}_{\iota-1})$.
Hence, there exist $\kappa_1=\kappa_2=\kappa_3<\kappa_4$,
$(\mu_1^j\mult a_\iota^{\beta_j})\gle \mu_2^j=\mi{reduced}(\lambda^{(\gamma_1^j,\dots,\gamma_n^j)})\in \mi{min}(U_{\iota-1})$, 
$\mu_e^j\in \{\gu\}\cup \mi{PropConj}_A$, $\mi{atoms}(\mu_e^j)\subseteq \mi{dom}({\mc V}_{\iota-1})$, $\beta_j\geq 1$, $\bs{0}^n\neq (\gamma_1^j,\dots,\gamma_n^j)\in \mbb{N}^n$,
$j=\kappa_3+1,\dots,\kappa_4$, such that 
$\varepsilon_1\diamond \varepsilon_2=(\upsilon_1\mult a_\iota^\alpha)\gle \varepsilon_2=\mi{reduced}(\lambda^{(\sum_{j=1}^{\kappa_4} \gamma_1^j,\dots,\sum_{j=1}^{\kappa_4} \gamma_n^j)})$;
(\ref{eq8i}d) holds.

Case 2.3.2.2:
$(\gamma_1^*,\dots,\gamma_n^*)\neq (\gamma_1^\natural,\dots,\gamma_n^\natural)$.
Then $(\gamma_1^\natural,\dots,\gamma_n^\natural)\preceq (\gamma_1^*,\dots,\gamma_n^*)$;
for all $i\leq n$, $\gamma_i^\natural\leq \gamma_i^*$;
there exists $i^*\leq n$ satisfying $\gamma_{i^*}^\natural\neq \gamma_{i^*}^*$, $\gamma_{i^*}^\natural<\gamma_{i^*}^*$, $\gamma_{i^*}^*-\gamma_{i^*}^\natural\geq 1$;
$\bs{0}^n\neq (\gamma_1^*-\gamma_1^\natural,\dots,\gamma_n^*-\gamma_n^\natural)\in \mbb{N}^n$,
$\mi{reduced}(\lambda^{(\gamma_1^*-\gamma_1^\natural,\dots,\gamma_n^*-\gamma_n^\natural)})=\mu_1^\#\diamond^\# \mu_2^\#\underset{\text{(\ref{eq8dd})}}{\in} \mi{clo}$, 
$\mu_j^\#\in \{\gu\}\cup \mi{PropConj}_A$, $\diamond^\#\in \{\geql,\gleq,\gle\}$,
$\lambda^{(\gamma_1^*,\dots,\gamma_n^*)}=\lambda^{(\gamma_1^\natural+\gamma_1^*-\gamma_1^\natural,\dots,\gamma_n^\natural+\gamma_n^*-\gamma_n^\natural)}\overset{\text{(\ref{eq8dddd})}}{=\!\!=}
                                         \lambda^{(\gamma_1^\natural,\dots,\gamma_n^\natural)}\mult \lambda^{(\gamma_1^*-\gamma_1^\natural,\dots,\gamma_n^*-\gamma_n^\natural)}$,
$(\upsilon_1\mult a_\iota^\alpha)\gle \varepsilon_2=\mi{reduced}(\lambda^{(\gamma_1^*,\dots,\gamma_n^*)})=
                                                    \mi{reduced}(\lambda^{(\gamma_1^\natural,\dots,\gamma_n^\natural)}\mult 
                                                                 \lambda^{(\gamma_1^*-\gamma_1^\natural,\dots,\gamma_n^*-\gamma_n^\natural)})$;
for all $\iota+1\leq j\leq m$,
\begin{alignat*}{1}
\#(a_j,(\mu_1^\natural\mult a_\iota^{\beta^\natural})\gle \mu_2^\natural)
&= \#(a_j,\mi{reduced}(\lambda^{(\gamma_1^\natural,\dots,\gamma_n^\natural)}))\overset{\text{(\ref{eq8aaaax})}}{=\!\!=} \#(a_j,\lambda^{(\gamma_1^\natural,\dots,\gamma_n^\natural)})=0, \\[1mm]
\#(a_j,\varepsilon_1\diamond \varepsilon_2)
&= \#(a_j,(\upsilon_1\mult a_\iota^\alpha)\gle \varepsilon_2)=\#(a_j,\mi{reduced}(\lambda^{(\gamma_1^*,\dots,\gamma_n^*)}))\overset{\text{(\ref{eq8aaaax})}}{=\!\!=} \\
&\phantom{\mbox{}=\mbox{}}
   \#(a_j,\lambda^{(\gamma_1^*,\dots,\gamma_n^*)})= \\
&\phantom{\mbox{}=\mbox{}}
   \#(a_j,\lambda^{(\gamma_1^\natural,\dots,\gamma_n^\natural)}\mult \lambda^{(\gamma_1^*-\gamma_1^\natural,\dots,\gamma_n^*-\gamma_n^\natural)})\overset{\text{(\ref{eq8aax})}}{=\!\!=} \\
&\phantom{\mbox{}=\mbox{}}
   \#(a_j,\lambda^{(\gamma_1^\natural,\dots,\gamma_n^\natural)})+\#(a_j,\lambda^{(\gamma_1^*-\gamma_1^\natural,\dots,\gamma_n^*-\gamma_n^\natural)})= \\
&\phantom{\mbox{}=\mbox{}}
   \#(a_j,\lambda^{(\gamma_1^*-\gamma_1^\natural,\dots,\gamma_n^*-\gamma_n^\natural)})=0, \\[1mm]
\#(a_j,\mu_1^\#\diamond^\# \mu_2^\#)
&= \#(a_j,\mi{reduced}(\lambda^{(\gamma_1^*-\gamma_1^\natural,\dots,\gamma_n^*-\gamma_n^\natural)}))\overset{\text{(\ref{eq8aaaax})}}{=\!\!=} \\
&\phantom{\mbox{}=\mbox{}}
   \#(a_j,\lambda^{(\gamma_1^*-\gamma_1^\natural,\dots,\gamma_n^*-\gamma_n^\natural)})=0,
\end{alignat*}
$\mi{atoms}(\mu_1^\#)\cap \mi{atoms}(\mu_2^\#)\overset{\text{(\ref{eq8aaa})}}{=\!\!=} \emptyset$;
if $a_j\in \mi{atoms}(\mu_1^\#\diamond^\# \mu_2^\#)=\mi{atoms}(\mu_1^\#)\cup \mi{atoms}(\mu_1^\#)$,
either $a_j\in \mi{atoms}(\mu_1^\#)$ or $a_j\in \mi{atoms}(\mu_2^\#)$,
either $\#(a_j,\mu_1^\#\diamond^\# \mu_2^\#)\leq -1$ or $\#(a_j,\mu_1^\#\diamond^\# \mu_2^\#)\geq 1$;
$a_j\not\in \mi{atoms}(\mu_1^\#\diamond^\# \mu_2^\#)\supseteq \mi{atoms}(\mu_1^\#), \mi{atoms}(\mu_2^\#)$;
$\mi{atoms}(\mu_1^\#), \mi{atoms}(\mu_2^\#)\subseteq A=\{a_1,\dots,a_m\}$,
$\mi{atoms}(\mu_1^\#), \mi{atoms}(\mu_2^\#)\subseteq \{a_1,\dots,a_m\}-\{a_{\iota+1},\dots,a_m\}=\mi{dom}({\mc V}_{\iota-1})\cup \{a_\iota\}=\{a_1,\dots,a_\iota\}$,
$a_\iota\not\in \mi{dom}({\mc V}_{\iota-1})$.
We get two cases for $\mu_1^\#$ and $\mu_2^\#$.

Case 2.3.2.2.1:
$\mi{atoms}(\mu_1^\#), \mi{atoms}(\mu_2^\#)\subseteq \mi{dom}({\mc V}_{\iota-1})$.
We put $\kappa_1=\kappa_2=\kappa_3=0$, $\kappa_4=1$, $\mu_i^1=\mu_i^\natural\in \{\gu\}\cup \mi{PropConj}_A$, $i=1,2$, $\beta_1=\beta^\natural\geq 1$,
$\bs{0}^n\neq (\gamma_1^1,\dots,\gamma_n^1)=(\gamma_1^\natural,\dots,\gamma_n^\natural)\in \mbb{N}^n$,
$\zeta_i=\mu_i^\#\in \{\gu\}\cup \mi{PropConj}_A$, $i=1,2$, $\diamond^\zeta=\diamond^\#\in \{\geql,\gleq,\gle\}$,
$\bs{0}^n\neq (\gamma_1^\zeta,\dots,\gamma_n^\zeta)=(\gamma_1^*-\gamma_1^\natural,\dots,\gamma_n^*-\gamma_n^\natural)\in \mbb{N}^n$.
Then $\kappa_1=\kappa_2=\kappa_3=0<\kappa_4=1$,
\begin{alignat*}{1}
(\upsilon_1\mult a_\iota^\alpha)\gle \varepsilon_2
&= \mi{reduced}(\lambda^{(\gamma_1^*,\dots,\gamma_n^*)})=\mi{reduced}(\lambda^{(\gamma_1^\natural,\dots,\gamma_n^\natural)}\mult 
                                                                      \lambda^{(\gamma_1^*-\gamma_1^\natural,\dots,\gamma_n^*-\gamma_n^\natural)})= \\
&\phantom{\mbox{}=\mbox{}}
   \mi{reduced}(\lambda^{(\gamma_1^1,\dots,\gamma_n^1)}\mult \lambda^{(\gamma_1^\zeta,\dots,\gamma_n^\zeta)})\overset{\text{(\ref{eq8dddd})}}{=\!\!=} \\
&\phantom{\mbox{}=\mbox{}}
   \mi{reduced}(\lambda^{((\sum_{j=1}^{\kappa_4} \gamma_1^j)+\gamma_1^\zeta,\dots,(\sum_{j=1}^{\kappa_4} \gamma_n^j)+\gamma_n^\zeta)}),
\end{alignat*}
$(\mu_1^1\mult a_\iota^{\beta_1})\gle \mu_2^1=(\mu_1^\natural\mult a_\iota^{\beta^\natural})\gle \mu_2^\natural= 
 \mi{reduced}(\lambda^{(\gamma_1^1,\dots,\gamma_n^1)})=\mi{reduced}(\lambda^{(\gamma_1^\natural,\dots,\gamma_n^\natural)})\in \mi{min}(U_{\iota-1})$;
for both $e$, $\mi{atoms}(\mu_e^1)=\mi{atoms}(\mu_e^\natural)\subseteq \mi{dom}({\mc V}_{\iota-1})$;
$\zeta_1\diamond^\zeta \zeta_2=\mu_1^\#\diamond^\# \mu_2^\#=
 \mi{reduced}(\lambda^{(\gamma_1^\zeta,\dots,\gamma_n^\zeta)})=\mi{reduced}(\lambda^{(\gamma_1^*-\gamma_1^\natural,\dots,\gamma_n^*-\gamma_n^\natural)})\in \mi{clo}$;
for both $e$, $\mi{atoms}(\zeta_e)=\mi{atoms}(\mu_e^\#)\subseteq \mi{dom}({\mc V}_{\iota-1})$.
Hence, there exist $\kappa_1=\kappa_2=\kappa_3<\kappa_4$,
$(\mu_1^j\mult a_\iota^{\beta_j})\gle \mu_2^j=\mi{reduced}(\lambda^{(\gamma_1^j,\dots,\gamma_n^j)})\in \mi{min}(U_{\iota-1})$, 
$\mu_e^j\in \{\gu\}\cup \mi{PropConj}_A$, $\mi{atoms}(\mu_e^j)\subseteq \mi{dom}({\mc V}_{\iota-1})$, $\beta_j\geq 1$, $\bs{0}^n\neq (\gamma_1^j,\dots,\gamma_n^j)\in \mbb{N}^n$,
$j=\kappa_3+1,\dots,\kappa_4$, 
$\zeta_1\diamond^\zeta \zeta_2=\mi{reduced}(\lambda^{(\gamma_1^\zeta,\dots,\gamma_n^\zeta)})\in \mi{clo}$,
$\zeta_e\in \{\gu\}\cup \mi{PropConj}_A$, $\mi{atoms}(\zeta_e)\subseteq \mi{dom}({\mc V}_{\iota-1})$, $\diamond^\zeta\in \{\geql,\gleq,\gle\}$, 
$\bs{0}^n\neq (\gamma_1^\zeta,\dots,\gamma_n^\zeta)\in \mbb{N}^n$, such that 
$\varepsilon_1\diamond \varepsilon_2=(\upsilon_1\mult a_\iota^\alpha)\gle \varepsilon_2=
                                     \mi{reduced}(\lambda^{((\sum_{j=1}^{\kappa_4} \gamma_1^j)+\gamma_1^\zeta,\dots,(\sum_{j=1}^{\kappa_4} \gamma_n^j)+\gamma_n^\zeta)})$;
(\ref{eq8i}d) holds.

Case 2.3.2.2.2:
$a_\iota\in \mi{atoms}(\mu_1^\#)$ or $a_\iota\in \mi{atoms}(\mu_2^\#)$. 
Then $\mi{atoms}(\mu_1^\#),                                                                                                                                                                \linebreak[4]
                            \mi{atoms}(\mu_2^\#)\subseteq \mi{dom}({\mc V}_{\iota-1})\cup \{a_\iota\}$,
$\mbb{E}_{\iota-1}=\emptyset$, 
$\mu_1^\#\diamond^\# \mu_2^\#\in \mi{clo}$, $\diamond^\#\in \{\geql,\gleq,\gle\}$,
$\mu_1^\#\geql \mu_2^\#\underset{\text{(\ref{eq8h})}}{\not\in} \mi{clo}$,
$\diamond^\#\neq \geql$, $\diamond^\#\in \{\gleq,\gle\}$;
for all $i\leq n$, $\gamma_i^*-\gamma_i^\natural\leq \gamma_i^*$;
$(\gamma_1^\natural,\dots,\gamma_n^\natural)\neq \bs{0}^n$;
there exists $i^{**}\leq n$ satisfying $\gamma_{i^{**}}^\natural\geq 1$, $\gamma_{i^{**}}^*-\gamma_{i^{**}}^\natural<\gamma_{i^{**}}^*$;
$\sum_{i=1}^n \gamma_i^*-\gamma_i^\natural<\sum_{i=1}^n \gamma_i^*$;
by the induction hypothesis for $\mu_1^\#\diamond^\# \mu_2^\#$ and $\sum_{i=1}^n \gamma_i^*-\gamma_i^\natural$, there exist $\kappa_1\leq \kappa_2\leq \kappa_3<\kappa_4$,
\begin{alignat*}{1}
& \mu_2^j\gleq (\mu_1^j\mult a_\iota^{\beta_j})=\mi{reduced}(\lambda^{(\gamma_1^j,\dots,\gamma_n^j)})\in \mi{min}(\mi{DE}_{\iota-1}), j=1,\dots,\kappa_1, \\ 
& \mu_2^j\gle (\mu_1^j\mult a_\iota^{\beta_j})=\mi{reduced}(\lambda^{(\gamma_1^j,\dots,\gamma_n^j)})\in \mi{min}(D_{\iota-1}), j=\kappa_1+1,\dots,\kappa_2, \\
& (\mu_1^j\mult a_\iota^{\beta_j})\gleq \mu_2^j=\mi{reduced}(\lambda^{(\gamma_1^j,\dots,\gamma_n^j)})\in \mi{min}(\mi{UE}_{\iota-1}), j=\kappa_2+1,\dots,\kappa_3, \\
& (\mu_1^j\mult a_\iota^{\beta_j})\gle \mu_2^j=\mi{reduced}(\lambda^{(\gamma_1^j,\dots,\gamma_n^j)})\in \mi{min}(U_{\iota-1}), j=\kappa_3+2,\dots,\kappa_4, \\ 
& \mu_e^j\in \{\gu\}\cup \mi{PropConj}_A, \mi{atoms}(\mu_e^j)\subseteq \mi{dom}({\mc V}_{\iota-1}), \beta_j\geq 1, \bs{0}^n\neq (\gamma_1^j,\dots,\gamma_n^j)\in \mbb{N}^n, 
\end{alignat*}
satisfying either 
\begin{alignat*}{1}
\mu_1^\#\diamond^\# \mu_2^\# &= \mi{reduced}(\lambda^{(\gamma_1^*-\gamma_1^\natural,\dots,\gamma_n^*-\gamma_n^\natural)})= \\
                             &\phantom{\mbox{}=\mbox{}}
                                \mi{reduced}(\lambda^{(\sum_{j=1,\dots,\kappa_3,\kappa_3+2,\dots,\kappa_4} \gamma_1^j,\dots,\sum_{j=1,\dots,\kappa_3,\kappa_3+2,\dots,\kappa_4} \gamma_n^j)}),
\end{alignat*}
or there exists $\zeta_1\diamond^\zeta \zeta_2=\mi{reduced}(\lambda^{(\gamma_1^\zeta,\dots,\gamma_n^\zeta)})\in \mi{clo}$,
$\zeta_e\in \{\gu\}\cup \mi{PropConj}_A$, $\mi{atoms}(\zeta_e)\subseteq \mi{dom}({\mc V}_{\iota-1})$, $\diamond^\zeta\in \{\geql,\gleq,\gle\}$, 
$\bs{0}^n\neq (\gamma_1^\zeta,\dots,\gamma_n^\zeta)\in \mbb{N}^n$, satisfying
\begin{alignat*}{1}
\mu_1^\#\diamond^\# \mu_2^\# &= \mi{reduced}(\lambda^{(\gamma_1^*-\gamma_1^\natural,\dots,\gamma_n^*-\gamma_n^\natural)})= \\
                             &\phantom{\mbox{}=\mbox{}}
                                \mi{reduced}(\lambda^{((\sum_{j=1,\dots,\kappa_3,\kappa_3+2,\dots,\kappa_4} \gamma_1^j)+\gamma_1^\zeta,\dots,(\sum_{j=1,\dots,\kappa_3,\kappa_3+2,\dots,\kappa_4} \gamma_n^j)+\gamma_n^\zeta)}).
\end{alignat*}
We put $\mu_i^{\kappa_3+1}=\mu_i^\natural\in \{\gu\}\cup \mi{PropConj}_A$, $i=1,2$, $\beta_{\kappa_3+1}=\beta^\natural\geq 1$,
$\bs{0}^n\neq (\gamma_1^{\kappa_3+1},\dots,\gamma_n^{\kappa_3+1})=(\gamma_1^\natural,\dots,\gamma_n^\natural)\in \mbb{N}^n$.
Then either
\begin{alignat*}{1}
(\upsilon_1\mult a_\iota^\alpha)\gle \varepsilon_2 
&= \mi{reduced}(\lambda^{(\gamma_1^\natural,\dots,\gamma_n^\natural)}\mult \lambda^{(\gamma_1^*-\gamma_1^\natural,\dots,\gamma_n^*-\gamma_n^\natural)})\overset{\text{(\ref{eq7k})}}{=\!\!=} \\
&\phantom{\mbox{}=\mbox{}}
   \mi{reduced}(\lambda^{(\gamma_1^\natural,\dots,\gamma_n^\natural)}\mult \mi{reduced}(\lambda^{(\gamma_1^*-\gamma_1^\natural,\dots,\gamma_n^*-\gamma_n^\natural)}))= \\
&\phantom{\mbox{}=\mbox{}}
   \mi{reduced}(\lambda^{(\gamma_1^{\kappa_3+1},\dots,\gamma_n^{\kappa_3+1})}\mult \\
&\phantom{\mbox{}=\mbox{}} \quad
                \mi{reduced}(\lambda^{(\sum_{j=1,\dots,\kappa_3,\kappa_3+2,\dots,\kappa_4} \gamma_1^j,\dots,\sum_{j=1,\dots,\kappa_3,\kappa_3+2,\dots,\kappa_4} \gamma_n^j)}))\overset{\text{(\ref{eq7k})}}{=\!\!=} \\
&\phantom{\mbox{}=\mbox{}}
   \mi{reduced}(\lambda^{(\gamma_1^{\kappa_3+1},\dots,\gamma_n^{\kappa_3+1})}\mult \\
&\phantom{\mbox{}=\mi{reduced}(}
                \lambda^{(\sum_{j=1,\dots,\kappa_3,\kappa_3+2,\dots,\kappa_4} \gamma_1^j,\dots,\sum_{j=1,\dots,\kappa_3,\kappa_3+2,\dots,\kappa_4} \gamma_n^j)})\overset{\text{(\ref{eq8dddd})}}{=\!\!=} \\
&\phantom{\mbox{}=\mbox{}}
   \mi{reduced}(\lambda^{(\sum_{j=1}^{\kappa_4} \gamma_1^j,\dots,\sum_{j=1}^{\kappa_4} \gamma_n^j)})
\end{alignat*}
or
\begin{alignat*}{1}
& (\upsilon_1\mult a_\iota^\alpha)\gle \varepsilon_2=
  \mi{reduced}(\lambda^{(\gamma_1^\natural,\dots,\gamma_n^\natural)}\mult \mi{reduced}(\lambda^{(\gamma_1^*-\gamma_1^\natural,\dots,\gamma_n^*-\gamma_n^\natural)}))= \\
& \quad
  \mi{reduced}(\lambda^{(\gamma_1^{\kappa_3+1},\dots,\gamma_n^{\kappa_3+1})}\mult \\
& \quad \phantom{\mi{reduced}(}
               \mi{reduced}(\lambda^{((\sum_{j=1,\dots,\kappa_3,\kappa_3+2,\dots,\kappa_4} \gamma_1^j)+\gamma_1^\zeta,\dots,(\sum_{j=1,\dots,\kappa_3,\kappa_3+2,\dots,\kappa_4} \gamma_n^j)+\gamma_n^\zeta)}))\overset{\text{(\ref{eq7k})}}{=\!\!=} \\
& \quad
  \mi{reduced}(\lambda^{(\gamma_1^{\kappa_3+1},\dots,\gamma_n^{\kappa_3+1})}\mult \\
& \quad \phantom{\mi{reduced}(}
               \lambda^{((\sum_{j=1,\dots,\kappa_3,\kappa_3+2,\dots,\kappa_4} \gamma_1^j)+\gamma_1^\zeta,\dots,(\sum_{j=1,\dots,\kappa_3,\kappa_3+2,\dots,\kappa_4} \gamma_n^j)+\gamma_n^\zeta)})\overset{\text{(\ref{eq8dddd})}}{=\!\!=} \\
& \quad
  \mi{reduced}(\lambda^{((\sum_{j=1}^{\kappa_4} \gamma_1^j)+\gamma_1^\zeta,\dots,(\sum_{j=1}^{\kappa_4} \gamma_n^j)+\gamma_n^\zeta)}),
\end{alignat*}
$(\mu_1^{\kappa_3+1}\mult a_\iota^{\beta_{\kappa_3+1}})\gle \mu_2^{\kappa_3+1}=(\mu_1^\natural\mult a_\iota^{\beta^\natural})\gle \mu_2^\natural=
 \mi{reduced}(\lambda^{(\gamma_1^{\kappa_3+1},\dots,\gamma_n^{\kappa_3+1})})=\mi{reduced}(\lambda^{(\gamma_1^\natural,\dots,\gamma_n^\natural)})\in \mi{min}(U_{\iota-1})$;
for both $e$, $\mi{atoms}(\mu_e^{\kappa_3+1})=\mi{atoms}(\mu_e^\natural)\subseteq \mi{dom}({\mc V}_{\iota-1})$.
Hence, there exist $\kappa_1\leq \kappa_2\leq \kappa_3<\kappa_4$,
\begin{alignat*}{1}
& \mu_2^j\gleq (\mu_1^j\mult a_\iota^{\beta_j})=\mi{reduced}(\lambda^{(\gamma_1^j,\dots,\gamma_n^j)})\in \mi{min}(\mi{DE}_{\iota-1}), j=1,\dots,\kappa_1, \\ 
& \mu_2^j\gle (\mu_1^j\mult a_\iota^{\beta_j})=\mi{reduced}(\lambda^{(\gamma_1^j,\dots,\gamma_n^j)})\in \mi{min}(D_{\iota-1}), j=\kappa_1+1,\dots,\kappa_2, \\
& (\mu_1^j\mult a_\iota^{\beta_j})\gleq \mu_2^j=\mi{reduced}(\lambda^{(\gamma_1^j,\dots,\gamma_n^j)})\in \mi{min}(\mi{UE}_{\iota-1}), j=\kappa_2+1,\dots,\kappa_3, \\
& (\mu_1^j\mult a_\iota^{\beta_j})\gle \mu_2^j=\mi{reduced}(\lambda^{(\gamma_1^j,\dots,\gamma_n^j)})\in \mi{min}(U_{\iota-1}), j=\kappa_3+1,\dots,\kappa_4, \\ 
& \mu_e^j\in \{\gu\}\cup \mi{PropConj}_A, \mi{atoms}(\mu_e^j)\subseteq \mi{dom}({\mc V}_{\iota-1}), \beta_j\geq 1, \bs{0}^n\neq (\gamma_1^j,\dots,\gamma_n^j)\in \mbb{N}^n, 
\end{alignat*}
such that either $\varepsilon_1\diamond \varepsilon_2=(\upsilon_1\mult a_\iota^\alpha)\gle \varepsilon_2=
                                                      \mi{reduced}(\lambda^{(\sum_{j=1}^{\kappa_4} \gamma_1^j,\dots,\sum_{j=1}^{\kappa_4} \gamma_n^j)})$, 
or there exists $\zeta_1\diamond^\zeta \zeta_2=\mi{reduced}(\lambda^{(\gamma_1^\zeta,\dots,\gamma_n^\zeta)})\in \mi{clo}$,
$\zeta_e\in \{\gu\}\cup \mi{PropConj}_A$, $\mi{atoms}(\zeta_e)\subseteq \mi{dom}({\mc V}_{\iota-1})$, $\diamond^\zeta\in \{\geql,\gleq,\gle\}$, 
$\bs{0}^n\neq (\gamma_1^\zeta,\dots,\gamma_n^\zeta)\in \mbb{N}^n$, satisfying
$\varepsilon_1\diamond \varepsilon_2=(\upsilon_1\mult a_\iota^\alpha)\gle \varepsilon_2=
                                     \mi{reduced}(\lambda^{((\sum_{j=1}^{\kappa_4} \gamma_1^j)+\gamma_1^\zeta,\dots,(\sum_{j=1}^{\kappa_4} \gamma_n^j)+\gamma_n^\zeta)})$;
(\ref{eq8i}d) holds.

Case 2.4:
$a_\iota\not\in \mi{atoms}(\varepsilon_1)\subseteq \mi{dom}({\mc V}_{\iota-1})$ and $a_\iota\in \mi{atoms}(\varepsilon_2)$.
Then $\mi{atoms}(\varepsilon_2)\subseteq \mi{dom}({\mc V}_{\iota-1})\cup \{a_\iota\}$,
$\varepsilon_2=\upsilon_2\mult a_\iota^\alpha$, $\upsilon_2\in \{\gu\}\cup \mi{PropConj}_A$, $\alpha\geq 1$, 
$\mi{atoms}(\upsilon_2)\subseteq \mi{atoms}(\varepsilon_2)\subseteq \mi{dom}({\mc V}_{\iota-1})\cup \{a_\iota\}$,
$a_\iota\not\in \mi{atoms}(\upsilon_2)\subseteq \mi{dom}({\mc V}_{\iota-1})$.
We get two cases for $\diamond$.

Case 2.4.1:
$\diamond=\gleq$.
We put $\mi{EXP}=\{(\gamma_1,\dots,\gamma_n) \,|\, \bs{0}^n\neq (\gamma_1,\dots,\gamma_n)\in \mbb{N}^n, (\gamma_1,\dots,\gamma_n)\preceq (\gamma_1^*,\dots,\gamma_n^*), 
                                                   \mi{reduced}(\lambda^{(\gamma_1,\dots,\gamma_n)})\in \mi{DE}_{\iota-1}\}$.
Then $\varepsilon_1\diamond \varepsilon_2=\varepsilon_1\gleq (\upsilon_2\mult a_\iota^\alpha)=\mi{reduced}(\lambda^{(\gamma_1^*,\dots,\gamma_n^*)})\in \mi{clo}$, 
$\mi{atoms}(\varepsilon_1), \mi{atoms}(\upsilon_2)\subseteq \mi{dom}({\mc V}_{\iota-1})$,
$\varepsilon_1\gleq (\upsilon_2\mult a_\iota^\alpha)=\mi{reduced}(\lambda^{(\gamma_1^*,\dots,\gamma_n^*)})\in \mi{DE}_{\iota-1}$,
$(\gamma_1^*,\dots,\gamma_n^*)\neq \bs{0}^n$, $(\gamma_1^*,\dots,\gamma_n^*)\preceq (\gamma_1^*,\dots,\gamma_n^*)$, 
$(\gamma_1^*,\dots,\gamma_n^*)\in \mi{EXP}\neq \emptyset$;
we have that $\preceq$ is a well-founded order on $\mbb{N}^n$;
$\mi{min}(\mi{EXP})\neq \emptyset$;
there exists $\bs{0}^n\neq (\gamma_1^\natural,\dots,\gamma_n^\natural)\in \mi{min}(\mi{EXP})\subseteq \mi{EXP}\subseteq \mbb{N}^n$ satisfying 
$(\gamma_1^\natural,\dots,\gamma_n^\natural)\preceq (\gamma_1^*,\dots,\gamma_n^*)$, $\mi{reduced}(\lambda^{(\gamma_1^\natural,\dots,\gamma_n^\natural)})\in \mi{DE}_{\iota-1}$;
for all $(\gamma_1,\dots,\gamma_n)\in \mbb{N}^n$ satisfying $(\gamma_1,\dots,\gamma_n)\prec (\gamma_1^\natural,\dots,\gamma_n^\natural)$, 
$(\gamma_1,\dots,\gamma_n)\not\in \mi{EXP}$,
$(\gamma_1,\dots,\gamma_n)\prec (\gamma_1^\natural,\dots,\gamma_n^\natural)\preceq (\gamma_1^*,\dots,\gamma_n^*)$,
either $(\gamma_1,\dots,\gamma_n)=\bs{0}^n$, $\mi{reduced}(\lambda^{(\gamma_1,\dots,\gamma_n)})=\mi{reduced}(\lambda^{\bs{0}^n})=\gu\underset{\text{(\ref{eq8a})}}{\not\in} 
                                                                                                \mi{clo}\supseteq \mi{DE}_{\iota-1}$,
or $(\gamma_1,\dots,\gamma_n)\neq \bs{0}^n$, $\mi{reduced}(\lambda^{(\gamma_1,\dots,\gamma_n)})\not\in \mi{DE}_{\iota-1}$;
$\mi{reduced}(\lambda^{(\gamma_1,\dots,\gamma_n)})\not\in \mi{DE}_{\iota-1}$;
$\mi{reduced}(\lambda^{(\gamma_1^\natural,\dots,\gamma_n^\natural)})\in \mi{min}(\mi{DE}_{\iota-1})$,
$\mi{reduced}(\lambda^{(\gamma_1^\natural,\dots,\gamma_n^\natural)})=\mu_2^\natural\gleq (\mu_1^\natural\mult a_\iota^{\beta^\natural})$, 
$\mu_i^\natural\in \{\gu\}\cup \mi{PropConj}_A$, $\mi{atoms}(\mu_i^\natural)\subseteq \mi{dom}({\mc V}_{\iota-1})$, $\beta^\natural\geq 1$,
$\mi{atoms}(\mu_1^\natural\mult a_\iota^{\beta^\natural})=\mi{atoms}(\mu_1^\natural)\cup \mi{atoms}(a_\iota^{\beta^\natural})\subseteq \mi{dom}({\mc V}_{\iota-1})\cup \{a_\iota\}$;
for all $\iota+1\leq j\leq m$, 
$a_j\not\in \mi{dom}({\mc V}_{\iota-1})\cup \{a_\iota\}\supseteq \mi{atoms}(\mu_1^\natural\mult a_\iota^{\beta^\natural}), \mi{atoms}(\mu_2^\natural)$, 
$\#(a_j,\mu_2^\natural\gleq (\mu_1^\natural\mult a_\iota^{\beta^\natural}))=0$.
We get two cases for $(\gamma_1^*,\dots,\gamma_n^*)$ and $(\gamma_1^\natural,\dots,\gamma_n^\natural)$.

Case 2.4.1.1:
$(\gamma_1^*,\dots,\gamma_n^*)=(\gamma_1^\natural,\dots,\gamma_n^\natural)$.
We put $\kappa_1=\kappa_2=\kappa_3=\kappa_4=1$, $\mu_i^1=\mu_i^\natural\in \{\gu\}\cup \mi{PropConj}_A$, $i=1,2$, $\beta_1=\beta^\natural\geq 1$,
$\bs{0}^n\neq (\gamma_1^1,\dots,\gamma_n^1)=(\gamma_1^\natural,\dots,\gamma_n^\natural)\in \mbb{N}^n$.
Then $\kappa_1=\kappa_2=\kappa_3=\kappa_4=1$,
\begin{alignat*}{1}
\varepsilon_1\gleq (\upsilon_2\mult a_\iota^\alpha)
&= \mi{reduced}(\lambda^{(\gamma_1^*,\dots,\gamma_n^*)})=\mi{reduced}(\lambda^{(\gamma_1^1,\dots,\gamma_n^1)})= \\
&\phantom{\mbox{}=\mbox{}}
   \mi{reduced}(\lambda^{(\gamma_1^\natural,\dots,\gamma_n^\natural)})=\mi{reduced}(\lambda^{(\sum_{j=1}^{\kappa_4} \gamma_1^j,\dots,\sum_{j=1}^{\kappa_4} \gamma_n^j)}),
\end{alignat*}
$\mu_2^1\gleq (\mu_1^1\mult a_\iota^{\beta_1})=\mu_2^\natural\gleq (\mu_1^\natural\mult a_\iota^{\beta^\natural})=
 \mi{reduced}(\lambda^{(\gamma_1^1,\dots,\gamma_n^1)})=\mi{reduced}(\lambda^{(\gamma_1^\natural,\dots,\gamma_n^\natural)})\in \mi{min}(\mi{DE}_{\iota-1})$;
for both $e$, $\mi{atoms}(\mu_e^1)=\mi{atoms}(\mu_e^\natural)\subseteq \mi{dom}({\mc V}_{\iota-1})$.
Hence, there exist $\kappa_1=\kappa_2=\kappa_3=\kappa_4=1$,
$\mu_2^j\gleq (\mu_1^j\mult a_\iota^{\beta_j})=\mi{reduced}(\lambda^{(\gamma_1^j,\dots,\gamma_n^j)})\in \mi{min}(\mi{DE}_{\iota-1})$, 
$\mu_e^j\in \{\gu\}\cup \mi{PropConj}_A$, $\mi{atoms}(\mu_e^j)\subseteq \mi{dom}({\mc V}_{\iota-1})$, $\beta_j\geq 1$, $\bs{0}^n\neq (\gamma_1^j,\dots,\gamma_n^j)\in \mbb{N}^n$,
$j=1,\dots,\kappa_1$, such that 
$\varepsilon_1\diamond \varepsilon_2=\varepsilon_1\gleq (\upsilon_2\mult a_\iota^\alpha)=\mi{reduced}(\lambda^{(\sum_{j=1}^{\kappa_4} \gamma_1^j,\dots,\sum_{j=1}^{\kappa_4} \gamma_n^j)})$;
(\ref{eq8i}a) holds.

Case 2.4.1.2:
$(\gamma_1^*,\dots,\gamma_n^*)\neq (\gamma_1^\natural,\dots,\gamma_n^\natural)$.
Then $(\gamma_1^\natural,\dots,\gamma_n^\natural)\preceq (\gamma_1^*,\dots,\gamma_n^*)$;
for all $i\leq n$, $\gamma_i^\natural\leq \gamma_i^*$;
there exists $i^*\leq n$ satisfying $\gamma_{i^*}^\natural\neq \gamma_{i^*}^*$, $\gamma_{i^*}^\natural<\gamma_{i^*}^*$, $\gamma_{i^*}^*-\gamma_{i^*}^\natural\geq 1$;
$\bs{0}^n\neq (\gamma_1^*-\gamma_1^\natural,\dots,\gamma_n^*-\gamma_n^\natural)\in \mbb{N}^n$,
$\mi{reduced}(\lambda^{(\gamma_1^*-\gamma_1^\natural,\dots,\gamma_n^*-\gamma_n^\natural)})=\mu_1^\#\diamond^\# \mu_2^\#\underset{\text{(\ref{eq8dd})}}{\in} \mi{clo}$, 
$\mu_j^\#\in \{\gu\}\cup \mi{PropConj}_A$, $\diamond^\#\in \{\geql,\gleq,\gle\}$,
$\lambda^{(\gamma_1^*,\dots,\gamma_n^*)}=\lambda^{(\gamma_1^\natural+\gamma_1^*-\gamma_1^\natural,\dots,\gamma_n^\natural+\gamma_n^*-\gamma_n^\natural)}\overset{\text{(\ref{eq8dddd})}}{=\!\!=}
                                         \lambda^{(\gamma_1^\natural,\dots,\gamma_n^\natural)}\mult \lambda^{(\gamma_1^*-\gamma_1^\natural,\dots,\gamma_n^*-\gamma_n^\natural)}$,
$\varepsilon_1\gleq (\upsilon_2\mult a_\iota^\alpha)=\mi{reduced}(\lambda^{(\gamma_1^*,\dots,\gamma_n^*)})=
                                                     \mi{reduced}(\lambda^{(\gamma_1^\natural,\dots,\gamma_n^\natural)}\mult 
                                                                  \lambda^{(\gamma_1^*-\gamma_1^\natural,\dots,\gamma_n^*-\gamma_n^\natural)})$;
for all $\iota+1\leq j\leq m$,
\begin{alignat*}{1}
\#(a_j,\mu_2^\natural\gleq (\mu_1^\natural\mult a_\iota^{\beta^\natural}))
&= \#(a_j,\mi{reduced}(\lambda^{(\gamma_1^\natural,\dots,\gamma_n^\natural)}))\overset{\text{(\ref{eq8aaaax})}}{=\!\!=} \#(a_j,\lambda^{(\gamma_1^\natural,\dots,\gamma_n^\natural)})=0, \\[1mm]
\#(a_j,\varepsilon_1\diamond \varepsilon_2)
&= \#(a_j,\varepsilon_1\gleq (\upsilon_2\mult a_\iota^\alpha))=\#(a_j,\mi{reduced}(\lambda^{(\gamma_1^*,\dots,\gamma_n^*)}))\overset{\text{(\ref{eq8aaaax})}}{=\!\!=} \\
&\phantom{\mbox{}=\mbox{}}
   \#(a_j,\lambda^{(\gamma_1^*,\dots,\gamma_n^*)})= \\
&\phantom{\mbox{}=\mbox{}}
   \#(a_j,\lambda^{(\gamma_1^\natural,\dots,\gamma_n^\natural)}\mult \lambda^{(\gamma_1^*-\gamma_1^\natural,\dots,\gamma_n^*-\gamma_n^\natural)})\overset{\text{(\ref{eq8aax})}}{=\!\!=} \\
&\phantom{\mbox{}=\mbox{}}
   \#(a_j,\lambda^{(\gamma_1^\natural,\dots,\gamma_n^\natural)})+\#(a_j,\lambda^{(\gamma_1^*-\gamma_1^\natural,\dots,\gamma_n^*-\gamma_n^\natural)})= \\
&\phantom{\mbox{}=\mbox{}}
   \#(a_j,\lambda^{(\gamma_1^*-\gamma_1^\natural,\dots,\gamma_n^*-\gamma_n^\natural)})=0, \\[1mm]
\#(a_j,\mu_1^\#\diamond^\# \mu_2^\#)
&= \#(a_j,\mi{reduced}(\lambda^{(\gamma_1^*-\gamma_1^\natural,\dots,\gamma_n^*-\gamma_n^\natural)}))\overset{\text{(\ref{eq8aaaax})}}{=\!\!=} \\
&\phantom{\mbox{}=\mbox{}}
   \#(a_j,\lambda^{(\gamma_1^*-\gamma_1^\natural,\dots,\gamma_n^*-\gamma_n^\natural)})=0,
\end{alignat*}
$\mi{atoms}(\mu_1^\#)\cap \mi{atoms}(\mu_2^\#)\overset{\text{(\ref{eq8aaa})}}{=\!\!=} \emptyset$;
if $a_j\in \mi{atoms}(\mu_1^\#\diamond^\# \mu_2^\#)=\mi{atoms}(\mu_1^\#)\cup \mi{atoms}(\mu_1^\#)$,
either $a_j\in \mi{atoms}(\mu_1^\#)$ or $a_j\in \mi{atoms}(\mu_2^\#)$,
either $\#(a_j,\mu_1^\#\diamond^\# \mu_2^\#)\leq -1$ or $\#(a_j,\mu_1^\#\diamond^\# \mu_2^\#)\geq 1$;
$a_j\not\in \mi{atoms}(\mu_1^\#\diamond^\# \mu_2^\#)\supseteq \mi{atoms}(\mu_1^\#), \mi{atoms}(\mu_2^\#)$;
$\mi{atoms}(\mu_1^\#), \mi{atoms}(\mu_2^\#)\subseteq A=\{a_1,\dots,a_m\}$,
$\mi{atoms}(\mu_1^\#), \mi{atoms}(\mu_2^\#)\subseteq \{a_1,\dots,a_m\}-\{a_{\iota+1},\dots,a_m\}=\mi{dom}({\mc V}_{\iota-1})\cup \{a_\iota\}=\{a_1,\dots,a_\iota\}$,
$a_\iota\not\in \mi{dom}({\mc V}_{\iota-1})$.
We get two cases for $\mu_1^\#$ and $\mu_2^\#$.

Case 2.4.1.2.1:
$\mi{atoms}(\mu_1^\#), \mi{atoms}(\mu_2^\#)\subseteq \mi{dom}({\mc V}_{\iota-1})$.
We put $\kappa_1=\kappa_2=\kappa_3=\kappa_4=1$, $\mu_i^1=\mu_i^\natural\in \{\gu\}\cup \mi{PropConj}_A$, $i=1,2$, $\beta_1=\beta^\natural\geq 1$,
$\bs{0}^n\neq (\gamma_1^1,\dots,\gamma_n^1)=(\gamma_1^\natural,\dots,\gamma_n^\natural)\in \mbb{N}^n$,
$\zeta_i=\mu_i^\#\in \{\gu\}\cup \mi{PropConj}_A$, $i=1,2$, $\diamond^\zeta=\diamond^\#\in \{\geql,\gleq,\gle\}$,
$\bs{0}^n\neq (\gamma_1^\zeta,\dots,\gamma_n^\zeta)=(\gamma_1^*-\gamma_1^\natural,\dots,\gamma_n^*-\gamma_n^\natural)\in \mbb{N}^n$.
Then $\kappa_1=\kappa_2=\kappa_3=\kappa_4=1$,
\begin{alignat*}{1}
\varepsilon_1\gleq (\upsilon_2\mult a_\iota^\alpha)
&= \mi{reduced}(\lambda^{(\gamma_1^*,\dots,\gamma_n^*)})=\mi{reduced}(\lambda^{(\gamma_1^\natural,\dots,\gamma_n^\natural)}\mult 
                                                                      \lambda^{(\gamma_1^*-\gamma_1^\natural,\dots,\gamma_n^*-\gamma_n^\natural)})= \\
&\phantom{\mbox{}=\mbox{}}
   \mi{reduced}(\lambda^{(\gamma_1^1,\dots,\gamma_n^1)}\mult \lambda^{(\gamma_1^\zeta,\dots,\gamma_n^\zeta)})\overset{\text{(\ref{eq8dddd})}}{=\!\!=} \\
&\phantom{\mbox{}=\mbox{}}
   \mi{reduced}(\lambda^{((\sum_{j=1}^{\kappa_4} \gamma_1^j)+\gamma_1^\zeta,\dots,(\sum_{j=1}^{\kappa_4} \gamma_n^j)+\gamma_n^\zeta)}),
\end{alignat*}
$\mu_2^1\gleq (\mu_1^1\mult a_\iota^{\beta_1})=\mu_2^\natural\gleq (\mu_1^\natural\mult a_\iota^{\beta^\natural})=
 \mi{reduced}(\lambda^{(\gamma_1^1,\dots,\gamma_n^1)})=\mi{reduced}(\lambda^{(\gamma_1^\natural,\dots,\gamma_n^\natural)})\in \mi{min}(\mi{DE}_{\iota-1})$;
for both $e$, $\mi{atoms}(\mu_e^1)=\mi{atoms}(\mu_e^\natural)\subseteq \mi{dom}({\mc V}_{\iota-1})$;
$\zeta_1\diamond^\zeta \zeta_2=\mu_1^\#\diamond^\# \mu_2^\#=
 \mi{reduced}(\lambda^{(\gamma_1^\zeta,\dots,\gamma_n^\zeta)})=\mi{reduced}(\lambda^{(\gamma_1^*-\gamma_1^\natural,\dots,\gamma_n^*-\gamma_n^\natural)})\in \mi{clo}$;
for both $e$, $\mi{atoms}(\zeta_e)=\mi{atoms}(\mu_e^\#)\subseteq \mi{dom}({\mc V}_{\iota-1})$.
Hence, there exist $\kappa_1=\kappa_2=\kappa_3=\kappa_4=1$,
$\mu_2^j\gleq (\mu_1^j\mult a_\iota^{\beta_j})=\mi{reduced}(\lambda^{(\gamma_1^j,\dots,\gamma_n^j)})\in \mi{min}(\mi{DE}_{\iota-1})$, 
$\mu_e^j\in \{\gu\}\cup \mi{PropConj}_A$, $\mi{atoms}(\mu_e^j)\subseteq \mi{dom}({\mc V}_{\iota-1})$, $\beta_j\geq 1$, $\bs{0}^n\neq (\gamma_1^j,\dots,\gamma_n^j)\in \mbb{N}^n$,
$j=1,\dots,\kappa_1$, 
$\zeta_1\diamond^\zeta \zeta_2=\mi{reduced}(\lambda^{(\gamma_1^\zeta,\dots,\gamma_n^\zeta)})\in \mi{clo}$,
$\zeta_e\in \{\gu\}\cup \mi{PropConj}_A$, $\mi{atoms}(\zeta_e)\subseteq \mi{dom}({\mc V}_{\iota-1})$, $\diamond^\zeta\in \{\geql,\gleq,\gle\}$, 
$\bs{0}^n\neq (\gamma_1^\zeta,\dots,\gamma_n^\zeta)\in \mbb{N}^n$, such that 
$\varepsilon_1\diamond \varepsilon_2=\varepsilon_1\gleq (\upsilon_2\mult a_\iota^\alpha)=
                                     \mi{reduced}(\lambda^{((\sum_{j=1}^{\kappa_4} \gamma_1^j)+\gamma_1^\zeta,\dots,(\sum_{j=1}^{\kappa_4} \gamma_n^j)+\gamma_n^\zeta)})$;
(\ref{eq8i}a) holds.

Case 2.4.1.2.2:
$a_\iota\in \mi{atoms}(\mu_1^\#)$ or $a_\iota\in \mi{atoms}(\mu_2^\#)$. 
Then $\mi{atoms}(\mu_1^\#),                                                                                                                                                                \linebreak[4]
                            \mi{atoms}(\mu_2^\#)\subseteq \mi{dom}({\mc V}_{\iota-1})\cup \{a_\iota\}$,
$\mbb{E}_{\iota-1}=\emptyset$, 
$\mu_1^\#\diamond^\# \mu_2^\#\in \mi{clo}$, $\diamond^\#\in \{\geql,\gleq,\gle\}$,
$\mu_1^\#\geql \mu_2^\#\underset{\text{(\ref{eq8h})}}{\not\in} \mi{clo}$,
$\diamond^\#\neq \geql$, $\diamond^\#\in \{\gleq,\gle\}$;
for all $i\leq n$, $\gamma_i^*-\gamma_i^\natural\leq \gamma_i^*$;
$(\gamma_1^\natural,\dots,\gamma_n^\natural)\neq \bs{0}^n$;
there exists $i^{**}\leq n$ satisfying $\gamma_{i^{**}}^\natural\geq 1$, $\gamma_{i^{**}}^*-\gamma_{i^{**}}^\natural<\gamma_{i^{**}}^*$;
$\sum_{i=1}^n \gamma_i^*-\gamma_i^\natural<\sum_{i=1}^n \gamma_i^*$;
by the induction hypothesis for $\mu_1^\#\diamond^\# \mu_2^\#$ and $\sum_{i=1}^n \gamma_i^*-\gamma_i^\natural$, there exist $1\leq \kappa_1\leq \kappa_2\leq \kappa_3\leq \kappa_4$,
\begin{alignat*}{1}
& \mu_2^j\gleq (\mu_1^j\mult a_\iota^{\beta_j})=\mi{reduced}(\lambda^{(\gamma_1^j,\dots,\gamma_n^j)})\in \mi{min}(\mi{DE}_{\iota-1}), j=2,\dots,\kappa_1, \\ 
& \mu_2^j\gle (\mu_1^j\mult a_\iota^{\beta_j})=\mi{reduced}(\lambda^{(\gamma_1^j,\dots,\gamma_n^j)})\in \mi{min}(D_{\iota-1}), j=\kappa_1+1,\dots,\kappa_2, \\
& (\mu_1^j\mult a_\iota^{\beta_j})\gleq \mu_2^j=\mi{reduced}(\lambda^{(\gamma_1^j,\dots,\gamma_n^j)})\in \mi{min}(\mi{UE}_{\iota-1}), j=\kappa_2+1,\dots,\kappa_3, \\
& (\mu_1^j\mult a_\iota^{\beta_j})\gle \mu_2^j=\mi{reduced}(\lambda^{(\gamma_1^j,\dots,\gamma_n^j)})\in \mi{min}(U_{\iota-1}), j=\kappa_3+1,\dots,\kappa_4, \\ 
& \mu_e^j\in \{\gu\}\cup \mi{PropConj}_A, \mi{atoms}(\mu_e^j)\subseteq \mi{dom}({\mc V}_{\iota-1}), \beta_j\geq 1, \bs{0}^n\neq (\gamma_1^j,\dots,\gamma_n^j)\in \mbb{N}^n, 
\end{alignat*}
satisfying either 
\begin{equation}
\notag
\mu_1^\#\diamond^\# \mu_2^\#=\mi{reduced}(\lambda^{(\gamma_1^*-\gamma_1^\natural,\dots,\gamma_n^*-\gamma_n^\natural)})= 
                             \mi{reduced}(\lambda^{(\sum_{j=2}^{\kappa_4} \gamma_1^j,\dots,\sum_{j=2}^{\kappa_4} \gamma_n^j)}),
\end{equation}
or there exists $\zeta_1\diamond^\zeta \zeta_2=\mi{reduced}(\lambda^{(\gamma_1^\zeta,\dots,\gamma_n^\zeta)})\in \mi{clo}$,
$\zeta_e\in \{\gu\}\cup \mi{PropConj}_A$, $\mi{atoms}(\zeta_e)\subseteq \mi{dom}({\mc V}_{\iota-1})$, $\diamond^\zeta\in \{\geql,\gleq,\gle\}$, 
$\bs{0}^n\neq (\gamma_1^\zeta,\dots,\gamma_n^\zeta)\in \mbb{N}^n$, satisfying
\begin{equation}
\notag
\mu_1^\#\diamond^\# \mu_2^\#=\mi{reduced}(\lambda^{(\gamma_1^*-\gamma_1^\natural,\dots,\gamma_n^*-\gamma_n^\natural)})= 
                             \mi{reduced}(\lambda^{((\sum_{j=2}^{\kappa_4} \gamma_1^j)+\gamma_1^\zeta,\dots,(\sum_{j=2}^{\kappa_4} \gamma_n^j)+\gamma_n^\zeta)}).
\end{equation}
We put $\mu_i^1=\mu_i^\natural\in \{\gu\}\cup \mi{PropConj}_A$, $i=1,2$, $\beta_1=\beta^\natural\geq 1$,
$\bs{0}^n\neq (\gamma_1^1,\dots,\gamma_n^1)=(\gamma_1^\natural,\dots,\gamma_n^\natural)\in \mbb{N}^n$.
Then either
\begin{alignat*}{1}
\varepsilon_1\gleq (\upsilon_2\mult a_\iota^\alpha) 
&= \mi{reduced}(\lambda^{(\gamma_1^\natural,\dots,\gamma_n^\natural)}\mult \lambda^{(\gamma_1^*-\gamma_1^\natural,\dots,\gamma_n^*-\gamma_n^\natural)})\overset{\text{(\ref{eq7k})}}{=\!\!=} \\
&\phantom{\mbox{}=\mbox{}}
   \mi{reduced}(\lambda^{(\gamma_1^\natural,\dots,\gamma_n^\natural)}\mult \mi{reduced}(\lambda^{(\gamma_1^*-\gamma_1^\natural,\dots,\gamma_n^*-\gamma_n^\natural)}))= \\
&\phantom{\mbox{}=\mbox{}}
   \mi{reduced}(\lambda^{(\gamma_1^1,\dots,\gamma_n^1)}\mult \mi{reduced}(\lambda^{(\sum_{j=2}^{\kappa_4} \gamma_1^j,\dots,\sum_{j=2}^{\kappa_4} \gamma_n^j)}))\overset{\text{(\ref{eq7k})}}{=\!\!=} \\
&\phantom{\mbox{}=\mbox{}}
   \mi{reduced}(\lambda^{(\gamma_1^1,\dots,\gamma_n^1)}\mult \lambda^{(\sum_{j=2}^{\kappa_4} \gamma_1^j,\dots,\sum_{j=2}^{\kappa_4} \gamma_n^j)})\overset{\text{(\ref{eq8dddd})}}{=\!\!=} \\
&\phantom{\mbox{}=\mbox{}}
   \mi{reduced}(\lambda^{(\sum_{j=1}^{\kappa_4} \gamma_1^j,\dots,\sum_{j=1}^{\kappa_4} \gamma_n^j)})
\end{alignat*}
or
\begin{alignat*}{1}
\varepsilon_1\gleq (\upsilon_2\mult a_\iota^\alpha)
&= \mi{reduced}(\lambda^{(\gamma_1^\natural,\dots,\gamma_n^\natural)}\mult \mi{reduced}(\lambda^{(\gamma_1^*-\gamma_1^\natural,\dots,\gamma_n^*-\gamma_n^\natural)}))= \\
&\phantom{\mbox{}=\mbox{}}
   \mi{reduced}(\lambda^{(\gamma_1^1,\dots,\gamma_n^1)}\mult 
                \mi{reduced}(\lambda^{((\sum_{j=2}^{\kappa_4} \gamma_1^j)+\gamma_1^\zeta,\dots,(\sum_{j=2}^{\kappa_4} \gamma_n^j)+\gamma_n^\zeta)}))\overset{\text{(\ref{eq7k})}}{=\!\!=} \\
&\phantom{\mbox{}=\mbox{}}
   \mi{reduced}(\lambda^{(\gamma_1^1,\dots,\gamma_n^1)}\mult 
                \lambda^{((\sum_{j=2}^{\kappa_4} \gamma_1^j)+\gamma_1^\zeta,\dots,(\sum_{j=2}^{\kappa_4} \gamma_n^j)+\gamma_n^\zeta)})\overset{\text{(\ref{eq8dddd})}}{=\!\!=} \\
&\phantom{\mbox{}=\mbox{}}
   \mi{reduced}(\lambda^{((\sum_{j=1}^{\kappa_4} \gamma_1^j)+\gamma_1^\zeta,\dots,(\sum_{j=1}^{\kappa_4} \gamma_n^j)+\gamma_n^\zeta)}),
\end{alignat*}
$\mu_2^1\gleq (\mu_1^1\mult a_\iota^{\beta_1})=\mu_2^\natural\gleq (\mu_1^\natural\mult a_\iota^{\beta^\natural})=
 \mi{reduced}(\lambda^{(\gamma_1^1,\dots,\gamma_n^1)})=\mi{reduced}(\lambda^{(\gamma_1^\natural,\dots,\gamma_n^\natural)})\in \mi{min}(\mi{DE}_{\iota-1})$;
for both $e$, $\mi{atoms}(\mu_e^1)=\mi{atoms}(\mu_e^\natural)\subseteq \mi{dom}({\mc V}_{\iota-1})$.
Hence, there exist $1\leq \kappa_1\leq \kappa_2\leq \kappa_3\leq \kappa_4$,
\begin{alignat*}{1}
& \mu_2^j\gleq (\mu_1^j\mult a_\iota^{\beta_j})=\mi{reduced}(\lambda^{(\gamma_1^j,\dots,\gamma_n^j)})\in \mi{min}(\mi{DE}_{\iota-1}), j=1,\dots,\kappa_1, \\ 
& \mu_2^j\gle (\mu_1^j\mult a_\iota^{\beta_j})=\mi{reduced}(\lambda^{(\gamma_1^j,\dots,\gamma_n^j)})\in \mi{min}(D_{\iota-1}), j=\kappa_1+1,\dots,\kappa_2, \\
& (\mu_1^j\mult a_\iota^{\beta_j})\gleq \mu_2^j=\mi{reduced}(\lambda^{(\gamma_1^j,\dots,\gamma_n^j)})\in \mi{min}(\mi{UE}_{\iota-1}), j=\kappa_2+1,\dots,\kappa_3, \\
& (\mu_1^j\mult a_\iota^{\beta_j})\gle \mu_2^j=\mi{reduced}(\lambda^{(\gamma_1^j,\dots,\gamma_n^j)})\in \mi{min}(U_{\iota-1}), j=\kappa_3+1,\dots,\kappa_4, \\ 
& \mu_e^j\in \{\gu\}\cup \mi{PropConj}_A, \mi{atoms}(\mu_e^j)\subseteq \mi{dom}({\mc V}_{\iota-1}), \beta_j\geq 1, \bs{0}^n\neq (\gamma_1^j,\dots,\gamma_n^j)\in \mbb{N}^n, 
\end{alignat*}
such that either $\varepsilon_1\diamond \varepsilon_2=\varepsilon_1\gleq (\upsilon_2\mult a_\iota^\alpha)=
                                                      \mi{reduced}(\lambda^{(\sum_{j=1}^{\kappa_4} \gamma_1^j,\dots,\sum_{j=1}^{\kappa_4} \gamma_n^j)})$, 
or there exists $\zeta_1\diamond^\zeta \zeta_2=\mi{reduced}(\lambda^{(\gamma_1^\zeta,\dots,\gamma_n^\zeta)})\in \mi{clo}$,
$\zeta_e\in \{\gu\}\cup \mi{PropConj}_A$, $\mi{atoms}(\zeta_e)\subseteq \mi{dom}({\mc V}_{\iota-1})$, $\diamond^\zeta\in \{\geql,\gleq,\gle\}$, 
$\bs{0}^n\neq (\gamma_1^\zeta,\dots,\gamma_n^\zeta)\in \mbb{N}^n$, satisfying
$\varepsilon_1\diamond \varepsilon_2=\varepsilon_1\gleq (\upsilon_2\mult a_\iota^\alpha)=
                                     \mi{reduced}(\lambda^{((\sum_{j=1}^{\kappa_4} \gamma_1^j)+\gamma_1^\zeta,\dots,(\sum_{j=1}^{\kappa_4} \gamma_n^j)+\gamma_n^\zeta)})$;
(\ref{eq8i}a) holds.

Case 2.4.2:
$\diamond=\gle$.
We put $\mi{EXP}=\{(\gamma_1,\dots,\gamma_n) \,|\, \bs{0}^n\neq (\gamma_1,\dots,\gamma_n)\in \mbb{N}^n, (\gamma_1,\dots,\gamma_n)\preceq (\gamma_1^*,\dots,\gamma_n^*), 
                                                   \mi{reduced}(\lambda^{(\gamma_1,\dots,\gamma_n)})\in D_{\iota-1}\}$.
Then $\varepsilon_1\diamond \varepsilon_2=\varepsilon_1\gle (\upsilon_2\mult a_\iota^\alpha)=\mi{reduced}(\lambda^{(\gamma_1^*,\dots,\gamma_n^*)})\in \mi{clo}$, 
$\mi{atoms}(\varepsilon_1), \mi{atoms}(\upsilon_2)\subseteq \mi{dom}({\mc V}_{\iota-1})$,
$\varepsilon_1\gle (\upsilon_2\mult a_\iota^\alpha)=\mi{reduced}(\lambda^{(\gamma_1^*,\dots,\gamma_n^*)})\in D_{\iota-1}$,
$(\gamma_1^*,\dots,\gamma_n^*)\neq \bs{0}^n$, $(\gamma_1^*,\dots,\gamma_n^*)\preceq (\gamma_1^*,\dots,\gamma_n^*)$, 
$(\gamma_1^*,\dots,\gamma_n^*)\in \mi{EXP}\neq \emptyset$;
we have that $\preceq$ is a well-founded order on $\mbb{N}^n$;
$\mi{min}(\mi{EXP})\neq \emptyset$;
there exists $\bs{0}^n\neq (\gamma_1^\natural,\dots,\gamma_n^\natural)\in \mi{min}(\mi{EXP})\subseteq \mi{EXP}\subseteq \mbb{N}^n$ satisfying 
$(\gamma_1^\natural,\dots,\gamma_n^\natural)\preceq (\gamma_1^*,\dots,\gamma_n^*)$, $\mi{reduced}(\lambda^{(\gamma_1^\natural,\dots,\gamma_n^\natural)})\in D_{\iota-1}$;
for all $(\gamma_1,\dots,\gamma_n)\in \mbb{N}^n$ satisfying $(\gamma_1,\dots,\gamma_n)\prec (\gamma_1^\natural,\dots,\gamma_n^\natural)$, 
$(\gamma_1,\dots,\gamma_n)\not\in \mi{EXP}$,
$(\gamma_1,\dots,\gamma_n)\prec (\gamma_1^\natural,\dots,\gamma_n^\natural)\preceq (\gamma_1^*,\dots,\gamma_n^*)$,
either $(\gamma_1,\dots,\gamma_n)=\bs{0}^n$, $\mi{reduced}(\lambda^{(\gamma_1,\dots,\gamma_n)})=\mi{reduced}(\lambda^{\bs{0}^n})=\gu\underset{\text{(\ref{eq8a})}}{\not\in} 
                                                                                                \mi{clo}\supseteq D_{\iota-1}$,
or $(\gamma_1,\dots,\gamma_n)\neq \bs{0}^n$, $\mi{reduced}(\lambda^{(\gamma_1,\dots,\gamma_n)})\not\in D_{\iota-1}$;
$\mi{reduced}(\lambda^{(\gamma_1,\dots,\gamma_n)})\not\in D_{\iota-1}$;
$\mi{reduced}(\lambda^{(\gamma_1^\natural,\dots,\gamma_n^\natural)})\in \mi{min}(D_{\iota-1})$,
$\mi{reduced}(\lambda^{(\gamma_1^\natural,\dots,\gamma_n^\natural)})=\mu_2^\natural\gle (\mu_1^\natural\mult a_\iota^{\beta^\natural})$, 
$\mu_i^\natural\in \{\gu\}\cup \mi{PropConj}_A$, $\mi{atoms}(\mu_i^\natural)\subseteq \mi{dom}({\mc V}_{\iota-1})$, $\beta^\natural\geq 1$,
$\mi{atoms}(\mu_1^\natural\mult a_\iota^{\beta^\natural})=\mi{atoms}(\mu_1^\natural)\cup \mi{atoms}(a_\iota^{\beta^\natural})\subseteq \mi{dom}({\mc V}_{\iota-1})\cup \{a_\iota\}$;
for all $\iota+1\leq j\leq m$, 
$a_j\not\in \mi{dom}({\mc V}_{\iota-1})\cup \{a_\iota\}\supseteq \mi{atoms}(\mu_1^\natural\mult a_\iota^{\beta^\natural}), \mi{atoms}(\mu_2^\natural)$, 
$\#(a_j,\mu_2^\natural\gle (\mu_1^\natural\mult a_\iota^{\beta^\natural}))=0$.
We get two cases for $(\gamma_1^*,\dots,\gamma_n^*)$ and $(\gamma_1^\natural,\dots,\gamma_n^\natural)$.

Case 2.4.2.1:
$(\gamma_1^*,\dots,\gamma_n^*)=(\gamma_1^\natural,\dots,\gamma_n^\natural)$.
We put $\kappa_1=0$, $\kappa_2=\kappa_3=\kappa_4=1$, $\mu_i^1=\mu_i^\natural\in \{\gu\}\cup \mi{PropConj}_A$, $i=1,2$, $\beta_1=\beta^\natural\geq 1$,
$\bs{0}^n\neq (\gamma_1^1,\dots,\gamma_n^1)=(\gamma_1^\natural,\dots,\gamma_n^\natural)\in \mbb{N}^n$.
Then $\kappa_1=0<\kappa_2=\kappa_3=\kappa_4=1$,
\begin{alignat*}{1}
\varepsilon_1\gle (\upsilon_2\mult a_\iota^\alpha)  
&= \mi{reduced}(\lambda^{(\gamma_1^*,\dots,\gamma_n^*)})=\mi{reduced}(\lambda^{(\gamma_1^1,\dots,\gamma_n^1)})= \\
&\phantom{\mbox{}=\mbox{}}
   \mi{reduced}(\lambda^{(\gamma_1^\natural,\dots,\gamma_n^\natural)})=\mi{reduced}(\lambda^{(\sum_{j=1}^{\kappa_4} \gamma_1^j,\dots,\sum_{j=1}^{\kappa_4} \gamma_n^j)}),
\end{alignat*}
$\mu_2^1\gle (\mu_1^1\mult a_\iota^{\beta_1})=\mu_2^\natural\gle (\mu_1^\natural\mult a_\iota^{\beta^\natural})=
 \mi{reduced}(\lambda^{(\gamma_1^1,\dots,\gamma_n^1)})=\mi{reduced}(\lambda^{(\gamma_1^\natural,\dots,\gamma_n^\natural)})\in \mi{min}(D_{\iota-1})$;
for both $e$, $\mi{atoms}(\mu_e^1)=\mi{atoms}(\mu_e^\natural)\subseteq \mi{dom}({\mc V}_{\iota-1})$.
Hence, there exist $\kappa_1<\kappa_2=\kappa_3=\kappa_4$,
$\mu_2^j\gle (\mu_1^j\mult a_\iota^{\beta_j})=\mi{reduced}(\lambda^{(\gamma_1^j,\dots,\gamma_n^j)})\in \mi{min}(D_{\iota-1})$, 
$\mu_e^j\in \{\gu\}\cup \mi{PropConj}_A$, $\mi{atoms}(\mu_e^j)\subseteq \mi{dom}({\mc V}_{\iota-1})$, $\beta_j\geq 1$, $\bs{0}^n\neq (\gamma_1^j,\dots,\gamma_n^j)\in \mbb{N}^n$,
$j=\kappa_1+1,\dots,\kappa_2$, such that 
$\varepsilon_1\diamond \varepsilon_2=\varepsilon_1\gle (\upsilon_2\mult a_\iota^\alpha)=\mi{reduced}(\lambda^{(\sum_{j=1}^{\kappa_4} \gamma_1^j,\dots,\sum_{j=1}^{\kappa_4} \gamma_n^j)})$;
(\ref{eq8i}b) holds.

Case 2.4.2.2:
$(\gamma_1^*,\dots,\gamma_n^*)\neq (\gamma_1^\natural,\dots,\gamma_n^\natural)$.
Then $(\gamma_1^\natural,\dots,\gamma_n^\natural)\preceq (\gamma_1^*,\dots,\gamma_n^*)$;
for all $i\leq n$, $\gamma_i^\natural\leq \gamma_i^*$;
there exists $i^*\leq n$ satisfying $\gamma_{i^*}^\natural\neq \gamma_{i^*}^*$, $\gamma_{i^*}^\natural<\gamma_{i^*}^*$, $\gamma_{i^*}^*-\gamma_{i^*}^\natural\geq 1$;
$\bs{0}^n\neq (\gamma_1^*-\gamma_1^\natural,\dots,\gamma_n^*-\gamma_n^\natural)\in \mbb{N}^n$,
$\mi{reduced}(\lambda^{(\gamma_1^*-\gamma_1^\natural,\dots,\gamma_n^*-\gamma_n^\natural)})=\mu_1^\#\diamond^\# \mu_2^\#\underset{\text{(\ref{eq8dd})}}{\in} \mi{clo}$, 
$\mu_j^\#\in \{\gu\}\cup \mi{PropConj}_A$, $\diamond^\#\in \{\geql,\gleq,\gle\}$,
$\lambda^{(\gamma_1^*,\dots,\gamma_n^*)}=\lambda^{(\gamma_1^\natural+\gamma_1^*-\gamma_1^\natural,\dots,\gamma_n^\natural+\gamma_n^*-\gamma_n^\natural)}\overset{\text{(\ref{eq8dddd})}}{=\!\!=}
                                         \lambda^{(\gamma_1^\natural,\dots,\gamma_n^\natural)}\mult \lambda^{(\gamma_1^*-\gamma_1^\natural,\dots,\gamma_n^*-\gamma_n^\natural)}$,
$\varepsilon_1\gle (\upsilon_2\mult a_\iota^\alpha)=\mi{reduced}(\lambda^{(\gamma_1^*,\dots,\gamma_n^*)})=
                                                    \mi{reduced}(\lambda^{(\gamma_1^\natural,\dots,\gamma_n^\natural)}\mult 
                                                                 \lambda^{(\gamma_1^*-\gamma_1^\natural,\dots,\gamma_n^*-\gamma_n^\natural)})$;
for all $\iota+1\leq j\leq m$,
\begin{alignat*}{1}
\#(a_j,\mu_2^\natural\gle (\mu_1^\natural\mult a_\iota^{\beta^\natural}))
&= \#(a_j,\mi{reduced}(\lambda^{(\gamma_1^\natural,\dots,\gamma_n^\natural)}))\overset{\text{(\ref{eq8aaaax})}}{=\!\!=} \#(a_j,\lambda^{(\gamma_1^\natural,\dots,\gamma_n^\natural)})=0, \\[1mm]
\#(a_j,\varepsilon_1\diamond \varepsilon_2)
&= \#(a_j,\varepsilon_1\gle (\upsilon_2\mult a_\iota^\alpha))=\#(a_j,\mi{reduced}(\lambda^{(\gamma_1^*,\dots,\gamma_n^*)}))\overset{\text{(\ref{eq8aaaax})}}{=\!\!=} \\
&\phantom{\mbox{}=\mbox{}}
   \#(a_j,\lambda^{(\gamma_1^*,\dots,\gamma_n^*)})= \\
&\phantom{\mbox{}=\mbox{}}
   \#(a_j,\lambda^{(\gamma_1^\natural,\dots,\gamma_n^\natural)}\mult \lambda^{(\gamma_1^*-\gamma_1^\natural,\dots,\gamma_n^*-\gamma_n^\natural)})\overset{\text{(\ref{eq8aax})}}{=\!\!=} \\
&\phantom{\mbox{}=\mbox{}}
   \#(a_j,\lambda^{(\gamma_1^\natural,\dots,\gamma_n^\natural)})+\#(a_j,\lambda^{(\gamma_1^*-\gamma_1^\natural,\dots,\gamma_n^*-\gamma_n^\natural)})= \\
&\phantom{\mbox{}=\mbox{}}
   \#(a_j,\lambda^{(\gamma_1^*-\gamma_1^\natural,\dots,\gamma_n^*-\gamma_n^\natural)})=0, \\[1mm]
\#(a_j,\mu_1^\#\diamond^\# \mu_2^\#)
&= \#(a_j,\mi{reduced}(\lambda^{(\gamma_1^*-\gamma_1^\natural,\dots,\gamma_n^*-\gamma_n^\natural)}))\overset{\text{(\ref{eq8aaaax})}}{=\!\!=} \\
&\phantom{\mbox{}=\mbox{}}
   \#(a_j,\lambda^{(\gamma_1^*-\gamma_1^\natural,\dots,\gamma_n^*-\gamma_n^\natural)})=0,
\end{alignat*}
$\mi{atoms}(\mu_1^\#)\cap \mi{atoms}(\mu_2^\#)\overset{\text{(\ref{eq8aaa})}}{=\!\!=} \emptyset$;
if $a_j\in \mi{atoms}(\mu_1^\#\diamond^\# \mu_2^\#)=\mi{atoms}(\mu_1^\#)\cup \mi{atoms}(\mu_1^\#)$,
either $a_j\in \mi{atoms}(\mu_1^\#)$ or $a_j\in \mi{atoms}(\mu_2^\#)$,
either $\#(a_j,\mu_1^\#\diamond^\# \mu_2^\#)\leq -1$ or $\#(a_j,\mu_1^\#\diamond^\# \mu_2^\#)\geq 1$;
$a_j\not\in \mi{atoms}(\mu_1^\#\diamond^\# \mu_2^\#)\supseteq \mi{atoms}(\mu_1^\#), \mi{atoms}(\mu_2^\#)$;
$\mi{atoms}(\mu_1^\#), \mi{atoms}(\mu_2^\#)\subseteq A=\{a_1,\dots,a_m\}$,
$\mi{atoms}(\mu_1^\#), \mi{atoms}(\mu_2^\#)\subseteq \{a_1,\dots,a_m\}-\{a_{\iota+1},\dots,a_m\}=\mi{dom}({\mc V}_{\iota-1})\cup \{a_\iota\}=\{a_1,\dots,a_\iota\}$,
$a_\iota\not\in \mi{dom}({\mc V}_{\iota-1})$.
We get two cases for $\mu_1^\#$ and $\mu_2^\#$.

Case 2.4.2.2.1:
$\mi{atoms}(\mu_1^\#), \mi{atoms}(\mu_2^\#)\subseteq \mi{dom}({\mc V}_{\iota-1})$.
We put $\kappa_1=0$, $\kappa_2=\kappa_3=\kappa_4=1$, $\mu_i^1=\mu_i^\natural\in \{\gu\}\cup \mi{PropConj}_A$, $i=1,2$, $\beta_1=\beta^\natural\geq 1$,
$\bs{0}^n\neq (\gamma_1^1,\dots,\gamma_n^1)=(\gamma_1^\natural,\dots,\gamma_n^\natural)\in \mbb{N}^n$,
$\zeta_i=\mu_i^\#\in \{\gu\}\cup \mi{PropConj}_A$, $i=1,2$, $\diamond^\zeta=\diamond^\#\in \{\geql,\gleq,\gle\}$,
$\bs{0}^n\neq (\gamma_1^\zeta,\dots,\gamma_n^\zeta)=(\gamma_1^*-\gamma_1^\natural,\dots,\gamma_n^*-\gamma_n^\natural)\in \mbb{N}^n$.
Then $\kappa_1=0<\kappa_2=\kappa_3=\kappa_4=1$,
\begin{alignat*}{1}
\varepsilon_1\gle (\upsilon_2\mult a_\iota^\alpha)
&= \mi{reduced}(\lambda^{(\gamma_1^*,\dots,\gamma_n^*)})=\mi{reduced}(\lambda^{(\gamma_1^\natural,\dots,\gamma_n^\natural)}\mult 
                                                                      \lambda^{(\gamma_1^*-\gamma_1^\natural,\dots,\gamma_n^*-\gamma_n^\natural)})= \\
&\phantom{\mbox{}=\mbox{}}
   \mi{reduced}(\lambda^{(\gamma_1^1,\dots,\gamma_n^1)}\mult \lambda^{(\gamma_1^\zeta,\dots,\gamma_n^\zeta)})\overset{\text{(\ref{eq8dddd})}}{=\!\!=} \\
&\phantom{\mbox{}=\mbox{}}
   \mi{reduced}(\lambda^{((\sum_{j=1}^{\kappa_4} \gamma_1^j)+\gamma_1^\zeta,\dots,(\sum_{j=1}^{\kappa_4} \gamma_n^j)+\gamma_n^\zeta)}),
\end{alignat*}
$\mu_2^1\gle (\mu_1^1\mult a_\iota^{\beta_1})=\mu_2^\natural\gle (\mu_1^\natural\mult a_\iota^{\beta^\natural})=
 \mi{reduced}(\lambda^{(\gamma_1^1,\dots,\gamma_n^1)})=\mi{reduced}(\lambda^{(\gamma_1^\natural,\dots,\gamma_n^\natural)})\in \mi{min}(D_{\iota-1})$;
for both $e$, $\mi{atoms}(\mu_e^1)=\mi{atoms}(\mu_e^\natural)\subseteq \mi{dom}({\mc V}_{\iota-1})$;
$\zeta_1\diamond^\zeta \zeta_2=\mu_1^\#\diamond^\# \mu_2^\#=
 \mi{reduced}(\lambda^{(\gamma_1^\zeta,\dots,\gamma_n^\zeta)})=\mi{reduced}(\lambda^{(\gamma_1^*-\gamma_1^\natural,\dots,\gamma_n^*-\gamma_n^\natural)})\in \mi{clo}$;
for both $e$, $\mi{atoms}(\zeta_e)=\mi{atoms}(\mu_e^\#)\subseteq \mi{dom}({\mc V}_{\iota-1})$.
Hence, there exist $\kappa_1<\kappa_2=\kappa_3=\kappa_4$,
$\mu_2^j\gle (\mu_1^j\mult a_\iota^{\beta_j})=\mi{reduced}(\lambda^{(\gamma_1^j,\dots,\gamma_n^j)})\in \mi{min}(D_{\iota-1})$, 
$\mu_e^j\in \{\gu\}\cup \mi{PropConj}_A$, $\mi{atoms}(\mu_e^j)\subseteq \mi{dom}({\mc V}_{\iota-1})$, $\beta_j\geq 1$, $\bs{0}^n\neq (\gamma_1^j,\dots,\gamma_n^j)\in \mbb{N}^n$,
$j=\kappa_1+1,\dots,\kappa_2$,
$\zeta_1\diamond^\zeta \zeta_2=\mi{reduced}(\lambda^{(\gamma_1^\zeta,\dots,\gamma_n^\zeta)})\in \mi{clo}$,
$\zeta_e\in \{\gu\}\cup \mi{PropConj}_A$, $\mi{atoms}(\zeta_e)\subseteq \mi{dom}({\mc V}_{\iota-1})$, $\diamond^\zeta\in \{\geql,\gleq,\gle\}$, 
$\bs{0}^n\neq (\gamma_1^\zeta,\dots,\gamma_n^\zeta)\in \mbb{N}^n$, such that 
$\varepsilon_1\diamond \varepsilon_2=\varepsilon_1\gle (\upsilon_2\mult a_\iota^\alpha)=
                                     \mi{reduced}(\lambda^{((\sum_{j=1}^{\kappa_4} \gamma_1^j)+\gamma_1^\zeta,\dots,(\sum_{j=1}^{\kappa_4} \gamma_n^j)+\gamma_n^\zeta)})$;
(\ref{eq8i}b) holds.

Case 2.4.2.2.2:
$a_\iota\in \mi{atoms}(\mu_1^\#)$ or $a_\iota\in \mi{atoms}(\mu_2^\#)$. 
Then $\mi{atoms}(\mu_1^\#),                                                                                                                                                                \linebreak[4]
                            \mi{atoms}(\mu_2^\#)\subseteq \mi{dom}({\mc V}_{\iota-1})\cup \{a_\iota\}$,
$\mbb{E}_{\iota-1}=\emptyset$, 
$\mu_1^\#\diamond^\# \mu_2^\#\in \mi{clo}$, $\diamond^\#\in \{\geql,\gleq,\gle\}$,
$\mu_1^\#\geql \mu_2^\#\underset{\text{(\ref{eq8h})}}{\not\in} \mi{clo}$,
$\diamond^\#\neq \geql$, $\diamond^\#\in \{\gleq,\gle\}$;
for all $i\leq n$, $\gamma_i^*-\gamma_i^\natural\leq \gamma_i^*$;
$(\gamma_1^\natural,\dots,\gamma_n^\natural)\neq \bs{0}^n$;
there exists $i^{**}\leq n$ satisfying $\gamma_{i^{**}}^\natural\geq 1$, $\gamma_{i^{**}}^*-\gamma_{i^{**}}^\natural<\gamma_{i^{**}}^*$;
$\sum_{i=1}^n \gamma_i^*-\gamma_i^\natural<\sum_{i=1}^n \gamma_i^*$;
by the induction hypothesis for $\mu_1^\#\diamond^\# \mu_2^\#$ and $\sum_{i=1}^n \gamma_i^*-\gamma_i^\natural$, there exist $\kappa_1<\kappa_2\leq \kappa_3\leq \kappa_4$,
\begin{alignat*}{1}
& \mu_2^j\gleq (\mu_1^j\mult a_\iota^{\beta_j})=\mi{reduced}(\lambda^{(\gamma_1^j,\dots,\gamma_n^j)})\in \mi{min}(\mi{DE}_{\iota-1}), j=1,\dots,\kappa_1, \\ 
& \mu_2^j\gle (\mu_1^j\mult a_\iota^{\beta_j})=\mi{reduced}(\lambda^{(\gamma_1^j,\dots,\gamma_n^j)})\in \mi{min}(D_{\iota-1}), j=\kappa_1+2,\dots,\kappa_2, \\
& (\mu_1^j\mult a_\iota^{\beta_j})\gleq \mu_2^j=\mi{reduced}(\lambda^{(\gamma_1^j,\dots,\gamma_n^j)})\in \mi{min}(\mi{UE}_{\iota-1}), j=\kappa_2+1,\dots,\kappa_3, \\
& (\mu_1^j\mult a_\iota^{\beta_j})\gle \mu_2^j=\mi{reduced}(\lambda^{(\gamma_1^j,\dots,\gamma_n^j)})\in \mi{min}(U_{\iota-1}), j=\kappa_3+1,\dots,\kappa_4, \\ 
& \mu_e^j\in \{\gu\}\cup \mi{PropConj}_A, \mi{atoms}(\mu_e^j)\subseteq \mi{dom}({\mc V}_{\iota-1}), \beta_j\geq 1, \bs{0}^n\neq (\gamma_1^j,\dots,\gamma_n^j)\in \mbb{N}^n, 
\end{alignat*}
satisfying either 
\begin{alignat*}{1}
\mu_1^\#\diamond^\# \mu_2^\# &= \mi{reduced}(\lambda^{(\gamma_1^*-\gamma_1^\natural,\dots,\gamma_n^*-\gamma_n^\natural)})= \\
                             &\phantom{\mbox{}=\mbox{}}
                                \mi{reduced}(\lambda^{(\sum_{j=1,\dots,\kappa_1,\kappa_1+2,\dots,\kappa_4} \gamma_1^j,\dots,\sum_{j=1,\dots,\kappa_1,\kappa_1+2,\dots,\kappa_4} \gamma_n^j)}),
\end{alignat*}
or there exists $\zeta_1\diamond^\zeta \zeta_2=\mi{reduced}(\lambda^{(\gamma_1^\zeta,\dots,\gamma_n^\zeta)})\in \mi{clo}$,
$\zeta_e\in \{\gu\}\cup \mi{PropConj}_A$, $\mi{atoms}(\zeta_e)\subseteq \mi{dom}({\mc V}_{\iota-1})$, $\diamond^\zeta\in \{\geql,\gleq,\gle\}$, 
$\bs{0}^n\neq (\gamma_1^\zeta,\dots,\gamma_n^\zeta)\in \mbb{N}^n$, satisfying
\begin{alignat*}{1}
\mu_1^\#\diamond^\# \mu_2^\# &= \mi{reduced}(\lambda^{(\gamma_1^*-\gamma_1^\natural,\dots,\gamma_n^*-\gamma_n^\natural)})= \\
                             &\phantom{\mbox{}=\mbox{}}
                                \mi{reduced}(\lambda^{((\sum_{j=1,\dots,\kappa_1,\kappa_1+2,\dots,\kappa_4} \gamma_1^j)+\gamma_1^\zeta,\dots,(\sum_{j=1,\dots,\kappa_1,\kappa_1+2,\dots,\kappa_4} \gamma_n^j)+\gamma_n^\zeta)}).
\end{alignat*}
We put $\mu_i^{\kappa_1+1}=\mu_i^\natural\in \{\gu\}\cup \mi{PropConj}_A$, $i=1,2$, $\beta_{\kappa_1+1}=\beta^\natural\geq 1$,
$\bs{0}^n\neq (\gamma_1^{\kappa_1+1},\dots,\gamma_n^{\kappa_1+1})=(\gamma_1^\natural,\dots,\gamma_n^\natural)\in \mbb{N}^n$.
Then either
\begin{alignat*}{1}
\varepsilon_1\gle (\upsilon_2\mult a_\iota^\alpha) 
&= \mi{reduced}(\lambda^{(\gamma_1^\natural,\dots,\gamma_n^\natural)}\mult \lambda^{(\gamma_1^*-\gamma_1^\natural,\dots,\gamma_n^*-\gamma_n^\natural)})\overset{\text{(\ref{eq7k})}}{=\!\!=} \\
&\phantom{\mbox{}=\mbox{}}
   \mi{reduced}(\lambda^{(\gamma_1^\natural,\dots,\gamma_n^\natural)}\mult \mi{reduced}(\lambda^{(\gamma_1^*-\gamma_1^\natural,\dots,\gamma_n^*-\gamma_n^\natural)}))= \\
&\phantom{\mbox{}=\mbox{}}
   \mi{reduced}(\lambda^{(\gamma_1^{\kappa_1+1},\dots,\gamma_n^{\kappa_1+1})}\mult \\
&\phantom{\mbox{}=\mbox{}} \quad
                \mi{reduced}(\lambda^{(\sum_{j=1,\dots,\kappa_1,\kappa_1+2,\dots,\kappa_4} \gamma_1^j,\dots,\sum_{j=1,\dots,\kappa_1,\kappa_1+2,\dots,\kappa_4} \gamma_n^j)}))\overset{\text{(\ref{eq7k})}}{=\!\!=} \\
&\phantom{\mbox{}=\mbox{}}
   \mi{reduced}(\lambda^{(\gamma_1^{\kappa_1+1},\dots,\gamma_n^{\kappa_1+1})}\mult \\
&\phantom{\mbox{}=\mi{reduced}(}
                \lambda^{(\sum_{j=1,\dots,\kappa_1,\kappa_1+2,\dots,\kappa_4} \gamma_1^j,\dots,\sum_{j=1,\dots,\kappa_1,\kappa_1+2,\dots,\kappa_4} \gamma_n^j)})\overset{\text{(\ref{eq8dddd})}}{=\!\!=} \\
&\phantom{\mbox{}=\mbox{}}
   \mi{reduced}(\lambda^{(\sum_{j=1}^{\kappa_4} \gamma_1^j,\dots,\sum_{j=1}^{\kappa_4} \gamma_n^j)})
\end{alignat*}
or
\begin{alignat*}{1}
& \varepsilon_1\gle (\upsilon_2\mult a_\iota^\alpha)=
  \mi{reduced}(\lambda^{(\gamma_1^\natural,\dots,\gamma_n^\natural)}\mult \mi{reduced}(\lambda^{(\gamma_1^*-\gamma_1^\natural,\dots,\gamma_n^*-\gamma_n^\natural)}))= \\
& \quad
  \mi{reduced}(\lambda^{(\gamma_1^{\kappa_1+1},\dots,\gamma_n^{\kappa_1+1})}\mult \\
& \quad \phantom{\mi{reduced}(}
               \mi{reduced}(\lambda^{((\sum_{j=1,\dots,\kappa_1,\kappa_1+2,\dots,\kappa_4} \gamma_1^j)+\gamma_1^\zeta,\dots,(\sum_{j=1,\dots,\kappa_1,\kappa_1+2,\dots,\kappa_4} \gamma_n^j)+\gamma_n^\zeta)}))\overset{\text{(\ref{eq7k})}}{=\!\!=} \\
& \quad
  \mi{reduced}(\lambda^{(\gamma_1^{\kappa_1+1},\dots,\gamma_n^{\kappa_1+1})}\mult \\
& \quad \phantom{\mi{reduced}(}
               \lambda^{((\sum_{j=1,\dots,\kappa_1,\kappa_1+2,\dots,\kappa_4} \gamma_1^j)+\gamma_1^\zeta,\dots,(\sum_{j=1,\dots,\kappa_1,\kappa_1+2,\dots,\kappa_4} \gamma_n^j)+\gamma_n^\zeta)})\overset{\text{(\ref{eq8dddd})}}{=\!\!=} \\
& \quad
  \mi{reduced}(\lambda^{((\sum_{j=1}^{\kappa_4} \gamma_1^j)+\gamma_1^\zeta,\dots,(\sum_{j=1}^{\kappa_4} \gamma_n^j)+\gamma_n^\zeta)}),
\end{alignat*}
$\mu_2^{\kappa_1+1}\gle (\mu_1^{\kappa_1+1}\mult a_\iota^{\beta_{\kappa_1+1}})=\mu_2^\natural\gle (\mu_1^\natural\mult a_\iota^{\beta^\natural})=
 \mi{reduced}(\lambda^{(\gamma_1^{\kappa_1+1},\dots,\gamma_n^{\kappa_1+1})})=\mi{reduced}(\lambda^{(\gamma_1^\natural,\dots,\gamma_n^\natural)})\in \mi{min}(D_{\iota-1})$;
for both $e$, $\mi{atoms}(\mu_e^{\kappa_1+1})=\mi{atoms}(\mu_e^\natural)\subseteq \mi{dom}({\mc V}_{\iota-1})$.
Hence, there exist $\kappa_1<\kappa_2\leq \kappa_3\leq \kappa_4$,
\begin{alignat*}{1}
& \mu_2^j\gleq (\mu_1^j\mult a_\iota^{\beta_j})=\mi{reduced}(\lambda^{(\gamma_1^j,\dots,\gamma_n^j)})\in \mi{min}(\mi{DE}_{\iota-1}), j=1,\dots,\kappa_1, \\ 
& \mu_2^j\gle (\mu_1^j\mult a_\iota^{\beta_j})=\mi{reduced}(\lambda^{(\gamma_1^j,\dots,\gamma_n^j)})\in \mi{min}(D_{\iota-1}), j=\kappa_1+1,\dots,\kappa_2, \\
& (\mu_1^j\mult a_\iota^{\beta_j})\gleq \mu_2^j=\mi{reduced}(\lambda^{(\gamma_1^j,\dots,\gamma_n^j)})\in \mi{min}(\mi{UE}_{\iota-1}), j=\kappa_2+1,\dots,\kappa_3, \\
& (\mu_1^j\mult a_\iota^{\beta_j})\gle \mu_2^j=\mi{reduced}(\lambda^{(\gamma_1^j,\dots,\gamma_n^j)})\in \mi{min}(U_{\iota-1}), j=\kappa_3+1,\dots,\kappa_4, \\ 
& \mu_e^j\in \{\gu\}\cup \mi{PropConj}_A, \mi{atoms}(\mu_e^j)\subseteq \mi{dom}({\mc V}_{\iota-1}), \beta_j\geq 1, \bs{0}^n\neq (\gamma_1^j,\dots,\gamma_n^j)\in \mbb{N}^n, 
\end{alignat*}
such that either $\varepsilon_1\diamond \varepsilon_2=\varepsilon_1\gle (\upsilon_2\mult a_\iota^\alpha)=
                                                      \mi{reduced}(\lambda^{(\sum_{j=1}^{\kappa_4} \gamma_1^j,\dots,\sum_{j=1}^{\kappa_4} \gamma_n^j)})$, 
or there exists $\zeta_1\diamond^\zeta \zeta_2=\mi{reduced}(\lambda^{(\gamma_1^\zeta,\dots,\gamma_n^\zeta)})\in \mi{clo}$,
$\zeta_e\in \{\gu\}\cup \mi{PropConj}_A$, $\mi{atoms}(\zeta_e)\subseteq \mi{dom}({\mc V}_{\iota-1})$, $\diamond^\zeta\in \{\geql,\gleq,\gle\}$, 
$\bs{0}^n\neq (\gamma_1^\zeta,\dots,\gamma_n^\zeta)\in \mbb{N}^n$, satisfying
$\varepsilon_1\diamond \varepsilon_2=\varepsilon_1\gle (\upsilon_2\mult a_\iota^\alpha)=
                                     \mi{reduced}(\lambda^{((\sum_{j=1}^{\kappa_4} \gamma_1^j)+\gamma_1^\zeta,\dots,(\sum_{j=1}^{\kappa_4} \gamma_n^j)+\gamma_n^\zeta)})$;
(\ref{eq8i}b) holds.

So, in both Cases 2.3 and 2.4, (\ref{eq8i}) holds.
The induction is completed.
Thus, (\ref{eq8i}) holds.
%
%
%
\end{proof}

\subsection{Full proof of the statement (\ref{eq8jj}) of Lemma \ref{le2}}
\label{S7.5g}

\begin{proof}   
Let $c\in \mbb{DE}_{\iota-1}$.
Then there exists $\varepsilon_2\gleq (\varepsilon_1\mult a_\iota^\alpha)\in \mi{min}(\mi{DE}_{\iota-1})\subseteq \mi{clo}$, 
$\varepsilon_i\in \{\gu\}\cup \mi{PropConj}_A$, $\mi{atoms}(\varepsilon_i)\subseteq \mi{dom}({\mc V}_{\iota-1})$, $\alpha\geq 1$, satisfying
$c=\left(\frac{\|\varepsilon_2\|^{{\mc V}_{\iota-1}}}
              {\|\varepsilon_1\|^{{\mc V}_{\iota-1}}}\right)^{\frac{1}{\alpha}}$;
$a_\iota\diamond^\# \gu\underset{\text{(\ref{eq8aa})}}{\in} \mi{clo}$, $\diamond^\#\in \{\gleq,\gle\}$,
$(a_\iota\diamond^\# \gu)^\alpha=a_\iota^\alpha\diamond^\# \gu\underset{\text{(\ref{eq8d})}}{\in} \mi{clo}$,
$\mi{atoms}(\varepsilon_2)\cap \mi{atoms}(\varepsilon_1)=\mi{atoms}(\varepsilon_2)\cap (\mi{atoms}(\varepsilon_1)\cup \mi{atoms}(a_\iota^\alpha))=
 \mi{atoms}(\varepsilon_2)\cap \mi{atoms}(\varepsilon_1\mult a_\iota^\alpha)\overset{\text{(\ref{eq8aaa})}}{=\!\!=} \emptyset$,
$\mi{reduced}((\varepsilon_2\gleq (\varepsilon_1\mult a_\iota^\alpha))\mult (a_\iota^\alpha\diamond^\# \gu))=
 \mi{reduced}((\varepsilon_2\mult a_\iota^\alpha)\diamond^\# (\varepsilon_1\mult a_\iota^\alpha))\overset{\text{(\ref{eq7h})}}{=\!\!=}
 \mi{reduced}(\varepsilon_2\diamond^\# \varepsilon_1)\overset{\text{(\ref{eq7ggg})}}{=\!\!=} \varepsilon_2\diamond^\# \varepsilon_1\underset{\text{(\ref{eq8c})}}{\in} \mi{clo}$;
by the induction hypothesis for $\iota-1$, $\|\varepsilon_1\|^{{\mc V}_{\iota-1}}, \|\varepsilon_2\|^{{\mc V}_{\iota-1}}\underset{\text{(a)}}{\in} (0,1]$, 
either $\diamond^\#=\gleq$, $\|\varepsilon_2\|^{{\mc V}_{\iota-1}}\underset{\text{(d)}}{\leq} \|\varepsilon_1\|^{{\mc V}_{\iota-1}}$,
or $\diamond^\#=\gle$, $\|\varepsilon_2\|^{{\mc V}_{\iota-1}}\underset{\text{(e)}}{<} \|\varepsilon_1\|^{{\mc V}_{\iota-1}}$;
$\|\varepsilon_2\|^{{\mc V}_{\iota-1}}\leq \|\varepsilon_1\|^{{\mc V}_{\iota-1}}$,
$c=\left(\frac{\|\varepsilon_2\|^{{\mc V}_{\iota-1}}}
              {\|\varepsilon_1\|^{{\mc V}_{\iota-1}}}\right)^{\frac{1}{\alpha}}\in (0,1]$;
$\mi{min}(\mi{DE}_{\iota-1})\underset{\text{(\ref{eq8ff})}}{\subseteq_{\mc F}} \mi{clo}$,
$\mbb{DE}_{\iota-1}\subseteq_{\mc F} (0,1]$.

Let $X_{\iota-1}\in \{\mi{UE}_{\iota-1},U_{\iota-1}\}$, $\mbb{X}_{\iota-1}\in \{\mbb{UE}_{\iota-1},\mbb{U}_{\iota-1}\}$, $c\in \mbb{X}_{\iota-1}$.
Then there exists $(\varepsilon_1\mult a_\iota^\alpha)\diamond^\# \varepsilon_2\in \mi{min}(X_{\iota-1})\subseteq \mi{clo}$, 
$\varepsilon_i\in \{\gu\}\cup \mi{PropConj}_A$, $\mi{atoms}(\varepsilon_i)\subseteq \mi{dom}({\mc V}_{\iota-1})$, $\alpha\geq 1$, $\diamond^\#\in \{\gleq,\gle\}$, satisfying
$c=\mi{min}\left(\left(\frac{\|\varepsilon_2\|^{{\mc V}_{\iota-1}}}
                            {\|\varepsilon_1\|^{{\mc V}_{\iota-1}}}\right)^{\frac{1}{\alpha}},1\right)$;
by the induction hypothesis for $\iota-1$, $\|\varepsilon_1\|^{{\mc V}_{\iota-1}}, \|\varepsilon_2\|^{{\mc V}_{\iota-1}}\underset{\text{(a)}}{\in} (0,1]$;
$c=\mi{min}\left(\left(\frac{\|\varepsilon_2\|^{{\mc V}_{\iota-1}}}
                            {\|\varepsilon_1\|^{{\mc V}_{\iota-1}}}\right)^{\frac{1}{\alpha}},1\right)\in (0,1]$;
$\mi{min}(X_{\iota-1})\underset{\text{(\ref{eq8ff})}}{\subseteq_{\mc F}} \mi{clo}$,
$\mbb{X}_{\iota-1}\subseteq_{\mc F} (0,1]$;
(a) holds.

Let $c\in \mbb{D}_{\iota-1}$.
Then there exists $\varepsilon_2\gle (\varepsilon_1\mult a_\iota^\alpha)\in \mi{min}(D_{\iota-1})\subseteq \mi{clo}$, 
$\varepsilon_i\in \{\gu\}\cup \mi{PropConj}_A$, $\mi{atoms}(\varepsilon_i)\subseteq \mi{dom}({\mc V}_{\iota-1})$, $\alpha\geq 1$, satisfying
$c=\left(\frac{\|\varepsilon_2\|^{{\mc V}_{\iota-1}}}
              {\|\varepsilon_1\|^{{\mc V}_{\iota-1}}}\right)^{\frac{1}{\alpha}}$;
$a_\iota\diamond^\# \gu\underset{\text{(\ref{eq8aa})}}{\in} \mi{clo}$, $\diamond^\#\in \{\gleq,\gle\}$,
$(a_\iota\diamond^\# \gu)^\alpha=a_\iota^\alpha\diamond^\# \gu\underset{\text{(\ref{eq8d})}}{\in} \mi{clo}$,
$\mi{atoms}(\varepsilon_2)\cap \mi{atoms}(\varepsilon_1)=\mi{atoms}(\varepsilon_2)\cap (\mi{atoms}(\varepsilon_1)\cup \mi{atoms}(a_\iota^\alpha))=
 \mi{atoms}(\varepsilon_2)\cap \mi{atoms}(\varepsilon_1\mult a_\iota^\alpha)\overset{\text{(\ref{eq8aaa})}}{=\!\!=} \emptyset$,
$\mi{reduced}((\varepsilon_2\gle (\varepsilon_1\mult a_\iota^\alpha))\mult (a_\iota^\alpha\diamond^\# \gu))=
 \mi{reduced}((\varepsilon_2\mult a_\iota^\alpha)\gle (\varepsilon_1\mult a_\iota^\alpha))\overset{\text{(\ref{eq7h})}}{=\!\!=}
 \mi{reduced}(\varepsilon_2\gle \varepsilon_1)\overset{\text{(\ref{eq7ggg})}}{=\!\!=} \varepsilon_2\gle \varepsilon_1\underset{\text{(\ref{eq8c})}}{\in} \mi{clo}$;
by the induction hypothesis for $\iota-1$, $\|\varepsilon_1\|^{{\mc V}_{\iota-1}}, \|\varepsilon_2\|^{{\mc V}_{\iota-1}}\underset{\text{(a)}}{\in} (0,1]$, 
$\|\varepsilon_2\|^{{\mc V}_{\iota-1}}\underset{\text{(e)}}{<} \|\varepsilon_1\|^{{\mc V}_{\iota-1}}$;
$c=\left(\frac{\|\varepsilon_2\|^{{\mc V}_{\iota-1}}}
              {\|\varepsilon_1\|^{{\mc V}_{\iota-1}}}\right)^{\frac{1}{\alpha}}\in (0,1)$;
$\mi{min}(D_{\iota-1})\underset{\text{(\ref{eq8ff})}}{\subseteq_{\mc F}} \mi{clo}$,
$\mbb{D}_{\iota-1}\subseteq_{\mc F} (0,1)$;
(b) holds.

So, (\ref{eq8jj}) holds.
%
%
%
\end{proof}

\subsection{Full proof of the statement (\ref{eq8j}) of Lemma \ref{le2}}
\label{S7.5h}

\begin{proof}   
Let $\mbb{E}_{\iota-1}=\emptyset$.
Then $\delta_\iota=\frac{\mi{max}(\bigfvee \mbb{DE}_{\iota-1},\bigfvee \mbb{D}_{\iota-1})+\mi{min}(\bigfwedge \mbb{UE}_{\iota-1},\bigfwedge \mbb{U}_{\iota-1})}{2}$.
We distinguish four cases for $\mbb{DE}_{\iota-1}$, $\mbb{D}_{\iota-1}$, $\mbb{UE}_{\iota-1}$, and $\mbb{U}_{\iota-1}$.

Case 2.5:
$\mbb{DE}_{\iota-1}=\mbb{D}_{\iota-1}=\mbb{UE}_{\iota-1}=\mbb{U}_{\iota-1}=\emptyset$.
Then 
\begin{alignat*}{1}
\delta_\iota &= \dfrac{\mi{max}(\bigfvee \mbb{DE}_{\iota-1},\bigfvee \mbb{D}_{\iota-1})+\mi{min}(\bigfwedge \mbb{UE}_{\iota-1},\bigfwedge \mbb{U}_{\iota-1})}{2}= \\
             &\phantom{\mbox{}=\mbox{}}
                \dfrac{\mi{max}(\bigfvee \emptyset,\bigfvee \emptyset)+\mi{min}(\bigfwedge \emptyset,\bigfwedge \emptyset)}{2}=\dfrac{1}{2}\in (0,1];
\end{alignat*}
(d) holds.

$\bigfvee \mbb{DE}_{\iota-1}=\bigfvee \emptyset=0<\delta_\iota=\frac{1}{2}<\bigfwedge \mbb{UE}_{\iota-1}=\bigfwedge \emptyset=1$;
(a) holds.

$\bigfvee \mbb{D}_{\iota-1}=\bigfvee \emptyset=0<\delta_\iota=\frac{1}{2}<\bigfwedge \mbb{U}_{\iota-1}=\bigfwedge \emptyset=1$;
(b) holds.

(c) holds trivially.

Case 2.6:
$\mbb{DE}_{\iota-1}\neq \emptyset$ or $\mbb{D}_{\iota-1}\neq \emptyset$, and $\mbb{UE}_{\iota-1}=\mbb{U}_{\iota-1}=\emptyset$.
Then $\mbb{DE}_{\iota-1}\underset{\text{(\ref{eq8jj}a)}}{\subseteq_{\mc F}} (0,1]$, $\mbb{D}_{\iota-1}\underset{\text{(\ref{eq8jj}b)}}{\subseteq_{\mc F}} (0,1)$,
$\bigfvee \mbb{DE}_{\iota-1}\in [0,1]$, $\bigfvee \mbb{D}_{\iota-1}\in [0,1)$,
$\mi{max}(\bigfvee \mbb{DE}_{\iota-1},\bigfvee \mbb{D}_{\iota-1})\in (0,1]$,
\begin{alignat*}{1}
\delta_\iota &= \dfrac{\mi{max}(\bigfvee \mbb{DE}_{\iota-1},\bigfvee \mbb{D}_{\iota-1})+\mi{min}(\bigfwedge \mbb{UE}_{\iota-1},\bigfwedge \mbb{U}_{\iota-1})}{2}= \\
             &\phantom{\mbox{}=\mbox{}}
                \dfrac{\mi{max}(\bigfvee \mbb{DE}_{\iota-1},\bigfvee \mbb{D}_{\iota-1})+\mi{min}(\bigfwedge \emptyset,\bigfwedge \emptyset)}{2}= \\
             &\phantom{\mbox{}=\mbox{}}
                \dfrac{\mi{max}(\bigfvee \mbb{DE}_{\iota-1},\bigfvee \mbb{D}_{\iota-1})+1}{2}\in (0,1];
\end{alignat*}
(d) holds.

$0\leq \bigfvee \mbb{DE}_{\iota-1}\leq \mi{max}(\bigfvee \mbb{DE}_{\iota-1},\bigfvee \mbb{D}_{\iota-1})\leq \delta_\iota=\frac{\mi{max}(\bigfvee \mbb{DE}_{\iota-1},\bigfvee \mbb{D}_{\iota-1})+1}{2}\leq 
 \bigfwedge \mbb{UE}_{\iota-1}=\bigfwedge \emptyset=1$;
(a) holds.

Either $\mi{max}(\bigfvee \mbb{DE}_{\iota-1},\bigfvee \mbb{D}_{\iota-1})=1$, 
$\bigfvee \mbb{D}_{\iota-1}<\mi{max}(\bigfvee \mbb{DE}_{\iota-1},\bigfvee \mbb{D}_{\iota-1})=\frac{\mi{max}(\bigfvee \mbb{DE}_{\iota-1},\bigfvee \mbb{D}_{\iota-1})+1}{2}=1$,
or $\mi{max}(\bigfvee \mbb{DE}_{\iota-1},\bigfvee \mbb{D}_{\iota-1})<1$,
$\bigfvee \mbb{D}_{\iota-1}\leq \mi{max}(\bigfvee \mbb{DE}_{\iota-1},\bigfvee \mbb{D}_{\iota-1})<\frac{\mi{max}(\bigfvee \mbb{DE}_{\iota-1},\bigfvee \mbb{D}_{\iota-1})+1}{2}$;
$0\leq \bigfvee \mbb{D}_{\iota-1}<\delta_\iota=\frac{\mi{max}(\bigfvee \mbb{DE}_{\iota-1},\bigfvee \mbb{D}_{\iota-1})+1}{2}\leq \bigfwedge \mbb{U}_{\iota-1}=\bigfwedge \emptyset=1$;
(b) holds.

(c) holds trivially.

Case 2.7:
$\mbb{DE}_{\iota-1}=\mbb{D}_{\iota-1}=\emptyset$, and $\mbb{UE}_{\iota-1}\neq \emptyset$ or $\mbb{U}_{\iota-1}\neq \emptyset$.
Then $\mbb{UE}_{\iota-1}, \mbb{U}_{\iota-1}\underset{\text{(\ref{eq8jj}a)}}{\subseteq_{\mc F}} (0,1]$,
$\bigfwedge \mbb{UE}_{\iota-1}, \bigfwedge \mbb{U}_{\iota-1}, \mi{min}(\bigfwedge \mbb{UE}_{\iota-1},\bigfwedge \mbb{U}_{\iota-1})\in (0,1]$,
\begin{alignat*}{1}
\delta_\iota &= \dfrac{\mi{max}(\bigfvee \mbb{DE}_{\iota-1},\bigfvee \mbb{D}_{\iota-1})+\mi{min}(\bigfwedge \mbb{UE}_{\iota-1},\bigfwedge \mbb{U}_{\iota-1})}{2}= \\
             &\phantom{\mbox{}=\mbox{}}
                \dfrac{\mi{max}(\bigfvee \emptyset,\bigfvee \emptyset)+\mi{min}(\bigfwedge \mbb{UE}_{\iota-1},\bigfwedge \mbb{U}_{\iota-1})}{2}=
                \dfrac{\mi{min}(\bigfwedge \mbb{UE}_{\iota-1},\bigfwedge \mbb{U}_{\iota-1})}{2}\in (0,1];
\end{alignat*}
(d) holds.

$\bigfvee \mbb{DE}_{\iota-1}=\bigfvee \emptyset=0<\delta_\iota=\frac{\mi{min}(\bigfwedge \mbb{UE}_{\iota-1},\bigfwedge \mbb{U}_{\iota-1})}{2}<
 \mi{min}(\bigfwedge \mbb{UE}_{\iota-1},\bigfwedge \mbb{U}_{\iota-1})\leq \bigfwedge \mbb{UE}_{\iota-1}\leq 1$;
(a) holds.

$\bigfvee \mbb{D}_{\iota-1}=\bigfvee \emptyset=0<\delta_\iota=\frac{\mi{min}(\bigfwedge \mbb{UE}_{\iota-1},\bigfwedge \mbb{U}_{\iota-1})}{2}<
 \mi{min}(\bigfwedge \mbb{UE}_{\iota-1},\bigfwedge \mbb{U}_{\iota-1})\leq \bigfwedge \mbb{U}_{\iota-1}\leq 1$;
(b) and (c) hold.

Case 2.8:
$\mbb{DE}_{\iota-1}\neq \emptyset$ or $\mbb{D}_{\iota-1}\neq \emptyset$, and $\mbb{UE}_{\iota-1}\neq \emptyset$ or $\mbb{U}_{\iota-1}\neq \emptyset$.
Then $\mbb{DE}_{\iota-1}, \mbb{UE}_{\iota-1}, \mbb{U}_{\iota-1}\underset{\text{(\ref{eq8jj}a)}}{\subseteq_{\mc F}} (0,1]$, $\mbb{D}_{\iota-1}\underset{\text{(\ref{eq8jj}b)}}{\subseteq_{\mc F}} (0,1)$,
$\bigfvee \mbb{DE}_{\iota-1}\in [0,1]$, $\bigfvee \mbb{D}_{\iota-1}\in [0,1)$,
$\bigfwedge \mbb{UE}_{\iota-1}, \bigfwedge \mbb{U}_{\iota-1}, 
 \mi{max}(\bigfvee \mbb{DE}_{\iota-1},\bigfvee \mbb{D}_{\iota-1}), \mi{min}(\bigfwedge \mbb{UE}_{\iota-1},\bigfwedge \mbb{U}_{\iota-1})\in (0,1]$;
$\mi{min}(\mi{DE}_{\iota-1})\neq \emptyset$ or $\mi{min}(D_{\iota-1})\neq \emptyset$;
there exists $l_d\in \mi{clo}$ satisfying $l_d\in \mi{min}(\mi{DE}_{\iota-1})\subseteq \mi{DE}_{\iota-1}\subseteq \mi{clo}$ or $l_d\in \mi{min}(D_{\iota-1})\subseteq D_{\iota-1}\subseteq \mi{clo}$,
either $l_d=\varepsilon_2^d\gleq (\varepsilon_1^d\mult a_\iota^{\alpha_d})\in \mi{min}(\mi{DE}_{\iota-1})$ or $l_d=\varepsilon_2^d\gle (\varepsilon_1^d\mult a_\iota^{\alpha_d})\in \mi{min}(D_{\iota-1})$,
$\varepsilon_i^d\in \{\gu\}\cup \mi{PropConj}_A$, $\mi{atoms}(\varepsilon_i^d)\subseteq \mi{dom}({\mc V}_{\iota-1})$, $\alpha_d\geq 1$,
either $0\leq \bigfvee \mbb{D}_{\iota-1}<\bigfvee \mbb{DE}_{\iota-1}$, $\bigfvee \emptyset=0$, $\mbb{DE}_{\iota-1}\neq \emptyset$, $l_d\in \mi{min}(\mi{DE}_{\iota-1})\neq \emptyset$,
$\bigfvee \mbb{D}_{\iota-1}<\bigfvee \mbb{DE}_{\iota-1}=\left(\frac{\|\varepsilon_2^d\|^{{\mc V}_{\iota-1}}}
                                                                   {\|\varepsilon_1^d\|^{{\mc V}_{\iota-1}}}\right)^{\frac{1}{\alpha_d}}\in \mbb{DE}_{\iota-1}$,
or $0\leq \bigfvee \mbb{DE}_{\iota-1}\leq \bigfvee \mbb{D}_{\iota-1}$;
if $\mbb{D}_{\iota-1}=\emptyset$, $\bigfvee \mbb{DE}_{\iota-1}=\bigfvee \mbb{D}_{\iota-1}=0$, $\mbb{DE}_{\iota-1}=\emptyset$;
$\mbb{D}_{\iota-1}\neq \emptyset$, $l_d\in \mi{min}(D_{\iota-1})\neq \emptyset$,
$\bigfvee \mbb{DE}_{\iota-1}\leq \bigfvee \mbb{D}_{\iota-1}=\left(\frac{\|\varepsilon_2^d\|^{{\mc V}_{\iota-1}}}
                                                                       {\|\varepsilon_1^d\|^{{\mc V}_{\iota-1}}}\right)^{\frac{1}{\alpha_d}}\in \mbb{D}_{\iota-1}$;
$\mi{max}(\bigfvee \mbb{DE}_{\iota-1},\bigfvee \mbb{D}_{\iota-1})=\left(\frac{\|\varepsilon_2^d\|^{{\mc V}_{\iota-1}}}
                                                                             {\|\varepsilon_1^d\|^{{\mc V}_{\iota-1}}}\right)^{\frac{1}{\alpha_d}}$,
$l_d=\varepsilon_2^d\diamond^d (\varepsilon_1^d\mult a_\iota^{\alpha_d})\in \mi{clo}$, $\diamond^d\in \{\gleq,\gle\}$;
$\mi{min}(\mi{UE}_{\iota-1})\neq \emptyset$ or $\mi{min}(U_{\iota-1})\neq \emptyset$;
there exists $l_u\in \mi{clo}$ satisfying $l_u\in \mi{min}(\mi{UE}_{\iota-1})\subseteq \mi{UE}_{\iota-1}\subseteq \mi{clo}$ or $l_u\in \mi{min}(U_{\iota-1})\subseteq U_{\iota-1}\subseteq \mi{clo}$,
either $l_u=(\varepsilon_1^u\mult a_\iota^{\alpha_u})\gleq \varepsilon_2^u\in \mi{min}(\mi{UE}_{\iota-1})$ or $l_u=(\varepsilon_1^u\mult a_\iota^{\alpha_u})\gle \varepsilon_2^u\in \mi{min}(U_{\iota-1})$,
$\varepsilon_i^u\in \{\gu\}\cup \mi{PropConj}_A$, $\mi{atoms}(\varepsilon_i^u)\subseteq \mi{dom}({\mc V}_{\iota-1})$, $\alpha_u\geq 1$,
either $\mbb{U}_{\iota-1}=\emptyset$, $\bigfwedge \mbb{UE}_{\iota-1}\leq \bigfwedge \mbb{U}_{\iota-1}=\bigfwedge \emptyset=1$, $\mbb{UE}_{\iota-1}\neq \emptyset$,
or $\mbb{U}_{\iota-1}\neq \emptyset$, $\bigfwedge \mbb{UE}_{\iota-1}<\bigfwedge \mbb{U}_{\iota-1}\leq 1$, $\bigfwedge \emptyset=1$, $\mbb{UE}_{\iota-1}\neq \emptyset$;
$\bigfwedge \mbb{UE}_{\iota-1}\leq \bigfwedge \mbb{U}_{\iota-1}$, $\mbb{UE}_{\iota-1}\neq \emptyset$, $l_u\in \mi{min}(\mi{UE}_{\iota-1})\neq \emptyset$,
$\bigfwedge \mbb{U}_{\iota-1}\geq \bigfwedge \mbb{UE}_{\iota-1}=\mi{min}\left(\left(\frac{\|\varepsilon_2^u\|^{{\mc V}_{\iota-1}}}
                                                                                         {\|\varepsilon_1^u\|^{{\mc V}_{\iota-1}}}\right)^{\frac{1}{\alpha_u}},1\right)\in \mbb{UE}_{\iota-1}$,
or $\mbb{U}_{\iota-1}\neq \emptyset$, $\bigfwedge \mbb{U}_{\iota-1}\leq \bigfwedge \mbb{UE}_{\iota-1}$, $l_u\in \mi{min}(U_{\iota-1})\neq \emptyset$,
$\bigfwedge \mbb{UE}_{\iota-1}\geq \bigfwedge \mbb{U}_{\iota-1}=\mi{min}\left(\left(\frac{\|\varepsilon_2^u\|^{{\mc V}_{\iota-1}}}
                                                                                         {\|\varepsilon_1^u\|^{{\mc V}_{\iota-1}}}\right)^{\frac{1}{\alpha_u}},1\right)\in \mbb{U}_{\iota-1}$;
$\mi{min}(\bigfwedge \mbb{UE}_{\iota-1},\bigfwedge \mbb{U}_{\iota-1})=\mi{min}\left(\left(\frac{\|\varepsilon_2^u\|^{{\mc V}_{\iota-1}}}
                                                                                               {\|\varepsilon_1^u\|^{{\mc V}_{\iota-1}}}\right)^{\frac{1}{\alpha_u}},1\right)$,
$l_u=(\varepsilon_1^u\mult a_\iota^{\alpha_u})\diamond^u \varepsilon_2^u\in \mi{clo}$, $\diamond^u\in \{\gleq,\gle\}$;
$(\varepsilon_2^d\diamond^d (\varepsilon_1^d\mult a_\iota^{\alpha_d}))^{\alpha_u}=(\varepsilon_2^d)^{\alpha_u}\diamond^d ((\varepsilon_1^d)^{\alpha_u}\mult a_\iota^{\alpha_d\cdot \alpha_u})\underset{\text{(\ref{eq8d})}}{\in} \mi{clo}$,
$((\varepsilon_1^u\mult a_\iota^{\alpha_u})\diamond^u \varepsilon_2^u)^{\alpha_d}=((\varepsilon_1^u)^{\alpha_d}\mult a_\iota^{\alpha_d\cdot \alpha_u})\diamond^u (\varepsilon_2^u)^{\alpha_d}\underset{\text{(\ref{eq8d})}}{\in} \mi{clo}$, 
\begin{alignat*}{1}
& \mi{reduced}(((\varepsilon_2^d)^{\alpha_u}\diamond^d ((\varepsilon_1^d)^{\alpha_u}\mult a_\iota^{\alpha_d\cdot \alpha_u}))\mult
               (((\varepsilon_1^u)^{\alpha_d}\mult a_\iota^{\alpha_d\cdot \alpha_u})\diamond^u (\varepsilon_2^u)^{\alpha_d}))= \\
& \mi{reduced}(((\varepsilon_1^u)^{\alpha_d}\mult (\varepsilon_2^d)^{\alpha_u}\mult a_\iota^{\alpha_d\cdot \alpha_u})\diamond^\# 
               ((\varepsilon_1^d)^{\alpha_u}\mult (\varepsilon_2^u)^{\alpha_d}\mult a_\iota^{\alpha_d\cdot \alpha_u}))\overset{\text{(\ref{eq7h})}}{=\!\!=} \\
& \mi{reduced}(((\varepsilon_1^u)^{\alpha_d}\mult (\varepsilon_2^d)^{\alpha_u})\diamond^\# ((\varepsilon_1^d)^{\alpha_u}\mult (\varepsilon_2^u)^{\alpha_d}))= \\ 
& \mi{reduced}^-(((\varepsilon_1^u)^{\alpha_d}\mult (\varepsilon_2^d)^{\alpha_u})\diamond^\# ((\varepsilon_1^d)^{\alpha_u}\mult (\varepsilon_2^u)^{\alpha_d}))\diamond^\# \\
& \quad
  \mi{reduced}^+(((\varepsilon_1^u)^{\alpha_d}\mult (\varepsilon_2^d)^{\alpha_u})\diamond^\# ((\varepsilon_1^d)^{\alpha_u}\mult (\varepsilon_2^u)^{\alpha_d}))\underset{\text{(\ref{eq8c})}}{\in} \mi{clo},
\end{alignat*}
$\diamond^\#\in \{\gleq,\gle\}$,
$\mi{atoms}(\mi{reduced}^-(((\varepsilon_1^u)^{\alpha_d}\mult (\varepsilon_2^d)^{\alpha_u})\diamond^\# ((\varepsilon_1^d)^{\alpha_u}\mult (\varepsilon_2^u)^{\alpha_d})))\underset{\text{(\ref{eq7a})}}{\subseteq}
 \mi{atoms}((\varepsilon_1^u)^{\alpha_d}\mult (\varepsilon_2^d)^{\alpha_u})=\mi{atoms}(\varepsilon_1^u)\cup \mi{atoms}(\varepsilon_2^d)\subseteq \mi{dom}({\mc V}_{\iota-1})$,
$\mi{atoms}(\mi{reduced}^+(((\varepsilon_1^u)^{\alpha_d}\mult (\varepsilon_2^d)^{\alpha_u})\diamond^\# ((\varepsilon_1^d)^{\alpha_u}\mult (\varepsilon_2^u)^{\alpha_d}))\underset{\text{(\ref{eq7b})}}{\subseteq}
 \mi{atoms}((\varepsilon_1^d)^{\alpha_u}\mult (\varepsilon_2^u)^{\alpha_d})=\mi{atoms}(\varepsilon_1^d)\cup \mi{atoms}(\varepsilon_2^u)\subseteq \mi{dom}({\mc V}_{\iota-1})$;
by the induction hypothesis for $\iota-1$, 
$\|\varepsilon_1^d\|^{{\mc V}_{\iota-1}}, \|\varepsilon_1^u\|^{{\mc V}_{\iota-1}}, \|\varepsilon_2^d\|^{{\mc V}_{\iota-1}}, \|\varepsilon_2^u\|^{{\mc V}_{\iota-1}},
 \|\mi{reduced}^-(((\varepsilon_1^u)^{\alpha_d}\mult (\varepsilon_2^d)^{\alpha_u})\diamond^\# ((\varepsilon_1^d)^{\alpha_u}\mult (\varepsilon_2^u)^{\alpha_d}))\|^{{\mc V}_{\iota-1}},
 \|\mi{reduced}^+(((\varepsilon_1^u)^{\alpha_d}\mult (\varepsilon_2^d)^{\alpha_u})\diamond^\# ((\varepsilon_1^d)^{\alpha_u}\mult (\varepsilon_2^u)^{\alpha_d}))\|^{{\mc V}_{\iota-1}}\underset{\text{(a)}}{\in} (0,1]$;
either $\diamond^d=\diamond^u=\diamond^\#=\gleq$,
$\mi{reduced}^-(((\varepsilon_1^u)^{\alpha_d}\mult (\varepsilon_2^d)^{\alpha_u})\diamond^\# ((\varepsilon_1^d)^{\alpha_u}\mult (\varepsilon_2^u)^{\alpha_d}))\diamond^\#
 \mi{reduced}^+(((\varepsilon_1^u)^{\alpha_d}\mult (\varepsilon_2^d)^{\alpha_u})\diamond^\# ((\varepsilon_1^d)^{\alpha_u}\mult (\varepsilon_2^u)^{\alpha_d}))=
 \mi{reduced}^-(((\varepsilon_1^u)^{\alpha_d}\mult (\varepsilon_2^d)^{\alpha_u})\diamond^\# ((\varepsilon_1^d)^{\alpha_u}\mult (\varepsilon_2^u)^{\alpha_d}))\gleq
 \mi{reduced}^+(((\varepsilon_1^u)^{\alpha_d}\mult (\varepsilon_2^d)^{\alpha_u})\diamond^\# ((\varepsilon_1^d)^{\alpha_u}\mult (\varepsilon_2^u)^{\alpha_d}))\in \mi{clo}$;
by the induction hypothesis for $\iota-1$,
$\|\mi{reduced}^-(((\varepsilon_1^u)^{\alpha_d}\mult (\varepsilon_2^d)^{\alpha_u})\diamond^\# ((\varepsilon_1^d)^{\alpha_u}\mult (\varepsilon_2^u)^{\alpha_d}))\|^{{\mc V}_{\iota-1}}\underset{\text{(d)}}{\leq}
 \|\mi{reduced}^+(((\varepsilon_1^u)^{\alpha_d}\mult (\varepsilon_2^d)^{\alpha_u})\diamond^\# ((\varepsilon_1^d)^{\alpha_u}\mult (\varepsilon_2^u)^{\alpha_d}))\|^{{\mc V}_{\iota-1}}$;
\begin{alignat*}{1}
& \dfrac{\|\mi{reduced}^-(((\varepsilon_1^u)^{\alpha_d}\mult (\varepsilon_2^d)^{\alpha_u})\diamond^\# ((\varepsilon_1^d)^{\alpha_u}\mult (\varepsilon_2^u)^{\alpha_d}))\|^{{\mc V}_{\iota-1}}}
        {\|\mi{reduced}^+(((\varepsilon_1^u)^{\alpha_d}\mult (\varepsilon_2^d)^{\alpha_u})\diamond^\# ((\varepsilon_1^d)^{\alpha_u}\mult (\varepsilon_2^u)^{\alpha_d}))\|^{{\mc V}_{\iota-1}}}\overset{\text{(\ref{eq7n})}}{=\!\!=} \\
& \dfrac{\|(\varepsilon_1^u)^{\alpha_d}\mult (\varepsilon_2^d)^{\alpha_u}\|^{{\mc V}_{\iota-1}}}
        {\|(\varepsilon_1^d)^{\alpha_u}\mult (\varepsilon_2^u)^{\alpha_d}\|^{{\mc V}_{\iota-1}}}= 
  \dfrac{\left(\|\varepsilon_1^u\|^{{\mc V}_{\iota-1}}\right)^{\alpha_d}\fswedge \left(\|\varepsilon_2^d\|^{{\mc V}_{\iota-1}}\right)^{\alpha_u}}
        {\left(\|\varepsilon_1^d\|^{{\mc V}_{\iota-1}}\right)^{\alpha_u}\fswedge \left(\|\varepsilon_2^u\|^{{\mc V}_{\iota-1}}\right)^{\alpha_d}}\leq 1, \\[1mm]
& \left(\dfrac{\|\varepsilon_2^d\|^{{\mc V}_{\iota-1}}}
              {\|\varepsilon_1^d\|^{{\mc V}_{\iota-1}}}\right)^{\alpha_u}\leq 
  \left(\dfrac{\|\varepsilon_2^u\|^{{\mc V}_{\iota-1}}}
              {\|\varepsilon_1^u\|^{{\mc V}_{\iota-1}}}\right)^{\alpha_d}, \\
& \left(\dfrac{\|\varepsilon_2^d\|^{{\mc V}_{\iota-1}}}
              {\|\varepsilon_1^d\|^{{\mc V}_{\iota-1}}}\right)^{\dfrac{1}{\alpha_d}}\leq
  \left(\dfrac{\|\varepsilon_2^u\|^{{\mc V}_{\iota-1}}}
              {\|\varepsilon_1^u\|^{{\mc V}_{\iota-1}}}\right)^{\dfrac{1}{\alpha_u}}, \\ 
& \left(\dfrac{\|\varepsilon_2^d\|^{{\mc V}_{\iota-1}}}
              {\|\varepsilon_1^d\|^{{\mc V}_{\iota-1}}}\right)^{\dfrac{1}{\alpha_d}}\leq 1, \\[1mm]
& 0\leq \bigfvee \mbb{D}_{\iota-1}<\bigfvee \mbb{DE}_{\iota-1}=\mi{max}\left(\bigfvee \mbb{DE}_{\iota-1},\bigfvee \mbb{D}_{\iota-1}\right)=
  \left(\dfrac{\|\varepsilon_2^d\|^{{\mc V}_{\iota-1}}}
              {\|\varepsilon_1^d\|^{{\mc V}_{\iota-1}}}\right)^{\dfrac{1}{\alpha_d}}\leq \\
& \delta_\iota=\dfrac{\mi{max}(\bigfvee \mbb{DE}_{\iota-1},\bigfvee \mbb{D}_{\iota-1})+\mi{min}(\bigfwedge \mbb{UE}_{\iota-1},\bigfwedge \mbb{U}_{\iota-1})}{2}\leq \\
& \mi{min}\left(\bigfwedge \mbb{UE}_{\iota-1},\bigfwedge \mbb{U}_{\iota-1}\right)=
  \mi{min}\left(\left(\dfrac{\|\varepsilon_2^u\|^{{\mc V}_{\iota-1}}}
                            {\|\varepsilon_1^u\|^{{\mc V}_{\iota-1}}}\right)^{\dfrac{1}{\alpha_u}},1\right)=\bigfwedge \mbb{UE}_{\iota-1}\leq \bigfwedge \mbb{U}_{\iota-1}\leq 1;
\end{alignat*}
if $\mbb{U}_{\iota-1}\neq \emptyset$, $\delta_\iota\leq \bigfwedge \mbb{UE}_{\iota-1}<\bigfwedge \mbb{U}_{\iota-1}$;
$\delta_\iota\in (0,1]$,
or $\diamond^d=\diamond^\#=\gle$,
$\mi{reduced}^-(((\varepsilon_1^u)^{\alpha_d}\mult (\varepsilon_2^d)^{\alpha_u})\diamond^\# ((\varepsilon_1^d)^{\alpha_u}\mult (\varepsilon_2^u)^{\alpha_d}))\diamond^\#
 \mi{reduced}^+(((\varepsilon_1^u)^{\alpha_d}\mult (\varepsilon_2^d)^{\alpha_u})\diamond^\# ((\varepsilon_1^d)^{\alpha_u}\mult (\varepsilon_2^u)^{\alpha_d}))=
 \mi{reduced}^-(((\varepsilon_1^u)^{\alpha_d}\mult (\varepsilon_2^d)^{\alpha_u})\diamond^\# ((\varepsilon_1^d)^{\alpha_u}\mult (\varepsilon_2^u)^{\alpha_d}))\gle
 \mi{reduced}^+(((\varepsilon_1^u)^{\alpha_d}\mult (\varepsilon_2^d)^{\alpha_u})\diamond^\# ((\varepsilon_1^d)^{\alpha_u}\mult (\varepsilon_2^u)^{\alpha_d}))\in \mi{clo}$;
by the induction hypothesis for $\iota-1$,
$\|\mi{reduced}^-(((\varepsilon_1^u)^{\alpha_d}\mult (\varepsilon_2^d)^{\alpha_u})\diamond^\# ((\varepsilon_1^d)^{\alpha_u}\mult (\varepsilon_2^u)^{\alpha_d}))\|^{{\mc V}_{\iota-1}}\underset{\text{(e)}}{<}
 \|\mi{reduced}^+(((\varepsilon_1^u)^{\alpha_d}\mult (\varepsilon_2^d)^{\alpha_u})\diamond^\# ((\varepsilon_1^d)^{\alpha_u}\mult (\varepsilon_2^u)^{\alpha_d}))\|^{{\mc V}_{\iota-1}}$;
\begin{alignat*}{1}
& \dfrac{\|\mi{reduced}^-(((\varepsilon_1^u)^{\alpha_d}\mult (\varepsilon_2^d)^{\alpha_u})\diamond^\# ((\varepsilon_1^d)^{\alpha_u}\mult (\varepsilon_2^u)^{\alpha_d}))\|^{{\mc V}_{\iota-1}}}
        {\|\mi{reduced}^+(((\varepsilon_1^u)^{\alpha_d}\mult (\varepsilon_2^d)^{\alpha_u})\diamond^\# ((\varepsilon_1^d)^{\alpha_u}\mult (\varepsilon_2^u)^{\alpha_d}))\|^{{\mc V}_{\iota-1}}}= \\
& \dfrac{\left(\|\varepsilon_1^u\|^{{\mc V}_{\iota-1}}\right)^{\alpha_d}\fswedge \left(\|\varepsilon_2^d\|^{{\mc V}_{\iota-1}}\right)^{\alpha_u}}
        {\left(\|\varepsilon_1^d\|^{{\mc V}_{\iota-1}}\right)^{\alpha_u}\fswedge \left(\|\varepsilon_2^u\|^{{\mc V}_{\iota-1}}\right)^{\alpha_d}}<1, \\[1mm]
& \left(\dfrac{\|\varepsilon_2^d\|^{{\mc V}_{\iota-1}}}
              {\|\varepsilon_1^d\|^{{\mc V}_{\iota-1}}}\right)^{\alpha_u}< 
  \left(\dfrac{\|\varepsilon_2^u\|^{{\mc V}_{\iota-1}}}
              {\|\varepsilon_1^u\|^{{\mc V}_{\iota-1}}}\right)^{\alpha_d}, \\
& \left(\dfrac{\|\varepsilon_2^d\|^{{\mc V}_{\iota-1}}}
              {\|\varepsilon_1^d\|^{{\mc V}_{\iota-1}}}\right)^{\dfrac{1}{\alpha_d}}<
  \left(\dfrac{\|\varepsilon_2^u\|^{{\mc V}_{\iota-1}}}
              {\|\varepsilon_1^u\|^{{\mc V}_{\iota-1}}}\right)^{\dfrac{1}{\alpha_u}}, \\
& \left(\dfrac{\|\varepsilon_2^d\|^{{\mc V}_{\iota-1}}}
              {\|\varepsilon_1^d\|^{{\mc V}_{\iota-1}}}\right)^{\dfrac{1}{\alpha_d}}<1, \\[1mm] 
& 0\leq \bigfvee \mbb{DE}_{\iota-1}\leq \bigfvee \mbb{D}_{\iota-1}=\mi{max}\left(\bigfvee \mbb{DE}_{\iota-1},\bigfvee \mbb{D}_{\iota-1}\right)=
  \left(\dfrac{\|\varepsilon_2^d\|^{{\mc V}_{\iota-1}}}
              {\|\varepsilon_1^d\|^{{\mc V}_{\iota-1}}}\right)^{\dfrac{1}{\alpha_d}}< \\
& \delta_\iota=\dfrac{\mi{max}(\bigfvee \mbb{DE}_{\iota-1},\bigfvee \mbb{D}_{\iota-1})+\mi{min}(\bigfwedge \mbb{UE}_{\iota-1},\bigfwedge \mbb{U}_{\iota-1})}{2}< \\
& \mi{min}\left(\bigfwedge \mbb{UE}_{\iota-1},\bigfwedge \mbb{U}_{\iota-1}\right)=
  \mi{min}\left(\left(\dfrac{\|\varepsilon_2^u\|^{{\mc V}_{\iota-1}}}
                            {\|\varepsilon_1^u\|^{{\mc V}_{\iota-1}}}\right)^{\dfrac{1}{\alpha_u}},1\right)\leq \bigfwedge \mbb{UE}_{\iota-1}, \bigfwedge \mbb{U}_{\iota-1}\leq 1,
\end{alignat*}
$\delta_\iota\in (0,1]$;
(a--d) hold.

So, in all Cases 2.5--2.8, (\ref{eq8j}) holds.
%
%
%
\end{proof}

\subsection{Full proof of the points (c--e) of Case 2 (the induction case) of the statement (\ref{eq8f}) of Lemma \ref{le2}}
\label{S7.5i}

\begin{proof}   
Let $\varepsilon_1, \varepsilon_2\in \{\gu\}\cup \mi{PropConj}_A$ such that $\mi{atoms}(\varepsilon_1), \mi{atoms}(\varepsilon_2)\subseteq \mi{dom}({\mc V}_\iota)$.   
Then $a_\iota\not\in \mi{dom}({\mc V}_{\iota-1})$ and $\mi{dom}({\mc V}_\iota)=\mi{dom}({\mc V}_{\iota-1})\cup \{a_\iota\}$.
We distinguish two cases for $\mi{atoms}(\varepsilon_1)$ and $\mi{atoms}(\varepsilon_2)$.

Case 2.9:
$\mi{atoms}(\varepsilon_1), \mi{atoms}(\varepsilon_2)\subseteq \mi{dom}({\mc V}_{\iota-1})$.
Then, by the induction hypothesis for $\iota-1<\iota$,
if $\varepsilon_1\geql \varepsilon_2\in \mi{clo}$, 
$\|\varepsilon_1\|^{{\mc V}_\iota}\overset{\text{(b)}}{=\!\!=} \|\varepsilon_1\|^{{\mc V}_{\iota-1}}\overset{\text{(c)}}{=\!\!=} \|\varepsilon_2\|^{{\mc V}_{\iota-1}}\overset{\text{(b)}}{=\!\!=} 
                                                               \|\varepsilon_2\|^{{\mc V}_\iota}$;
(c) holds;
if $\varepsilon_1\gleq \varepsilon_2\in \mi{clo}$, 
$\|\varepsilon_1\|^{{\mc V}_\iota}\overset{\text{(b)}}{=\!\!=} \|\varepsilon_1\|^{{\mc V}_{\iota-1}}\underset{\text{(d)}}{\leq} \|\varepsilon_2\|^{{\mc V}_{\iota-1}}\overset{\text{(b)}}{=\!\!=} 
                                                               \|\varepsilon_2\|^{{\mc V}_\iota}$;
(d) holds;
if $\varepsilon_1\gle \varepsilon_2\in \mi{clo}$,
$\|\varepsilon_1\|^{{\mc V}_\iota}\overset{\text{(b)}}{=\!\!=} \|\varepsilon_1\|^{{\mc V}_{\iota-1}}\underset{\text{(e)}}{<} \|\varepsilon_2\|^{{\mc V}_{\iota-1}}\overset{\text{(b)}}{=\!\!=} 
                                                               \|\varepsilon_2\|^{{\mc V}_\iota}$;
(e) holds.

Case 2.10:
$a_\iota\in \mi{atoms}(\varepsilon_1)$ or $a_\iota\in \mi{atoms}(\varepsilon_2)$.
We get two cases for $\mbb{E}_{\iota-1}$.

Case 2.10.1:
$\mbb{E}_{\iota-1}\neq \emptyset$.
Then, by (\ref{eq8g}), there exists $(\varepsilon_1^*\mult a_\iota^{\alpha^*})\geql \varepsilon_2^*\in \mi{min}(E_{\iota-1})\subseteq \mi{clo}$, 
$\varepsilon_i^*\in \{\gu\}\cup \mi{PropConj}_A$, $\mi{atoms}(\varepsilon_i^*)\subseteq \mi{dom}({\mc V}_{\iota-1}$, $\alpha^*\geq 1$, satisfying
${\mc V}_\iota(a_\iota)=\delta_\iota=\left(\frac{\|\varepsilon_2^*\|^{{\mc V}_{\iota-1}}}
                                                {\|\varepsilon_1^*\|^{{\mc V}_{\iota-1}}}\right)^{\frac{1}{\alpha^*}}\in (0,1]$;
$a_\iota\not\in \mi{dom}({\mc V}_{\iota-1})$, $\mi{atoms}(\varepsilon_1), \mi{atoms}(\varepsilon_2)\subseteq \mi{dom}({\mc V}_\iota)=\mi{dom}({\mc V}_{\iota-1})\cup \{a_\iota\}$,
$a_\iota\in \mi{atoms}(\varepsilon_1)$ or $a_\iota\in \mi{atoms}(\varepsilon_2)$.
We get three cases for $\mi{atoms}(\varepsilon_1)$ and $\mi{atoms}(\varepsilon_2)$.

Case 2.10.1.1:
$a_\iota\in \mi{atoms}(\varepsilon_1)$ and $a_\iota\not\in \mi{atoms}(\varepsilon_2)\subseteq \mi{dom}({\mc V}_{\iota-1})$.
Then $\mi{atoms}(\varepsilon_1)\subseteq \mi{dom}({\mc V}_\iota)=\mi{dom}({\mc V}_{\iota-1})\cup \{a_\iota\}$,
$\varepsilon_1=\upsilon_1\mult a_\iota^\alpha$, $\upsilon_1\in \{\gu\}\cup \mi{PropConj}_A$, $\alpha\geq 1$, $a_\iota\not\in \mi{atoms}(\upsilon_1)\subseteq \mi{dom}({\mc V}_{\iota-1})$.
We get three cases for $\varepsilon_1$ and $\varepsilon_2$.

Case 2.10.1.1.1:
$\varepsilon_1\geql \varepsilon_2\in \mi{clo}$.
Then $\varepsilon_1\geql \varepsilon_2=(\upsilon_1\mult a_\iota^\alpha)\geql \varepsilon_2\in \mi{clo}$,
$((\upsilon_1\mult a_\iota^\alpha)\geql \varepsilon_2)^{\alpha^*}=(\upsilon_1^{\alpha^*}\mult a_\iota^{\alpha\cdot \alpha^*})\geql \varepsilon_2^{\alpha^*}\underset{\text{(\ref{eq8d})}}{\in} \mi{clo}$,
$(\varepsilon_1^*\mult a_\iota^{\alpha^*})\geql \varepsilon_2^*\in \mi{clo}$,
$\varepsilon_2^*\geql (\varepsilon_1^*\mult a_\iota^{\alpha^*})\underset{\text{(\ref{eq8b})}}{\in} \mi{clo}$,
$(\varepsilon_2^*\geql (\varepsilon_1^*\mult a_\iota^{\alpha^*}))^\alpha=(\varepsilon_2^*)^\alpha\geql ((\varepsilon_1^*)^\alpha\mult a_\iota^{\alpha\cdot \alpha^*})\underset{\text{(\ref{eq8d})}}{\in} \mi{clo}$,
\begin{alignat*}{1}
& \mi{reduced}(((\upsilon_1^{\alpha^*}\mult a_\iota^{\alpha\cdot \alpha^*})\geql \varepsilon_2^{\alpha^*})\mult
               ((\varepsilon_2^*)^\alpha\geql ((\varepsilon_1^*)^\alpha\mult a_\iota^{\alpha\cdot \alpha^*})))= \\
& \mi{reduced}((\upsilon_1^{\alpha^*}\mult (\varepsilon_2^*)^\alpha\mult a_\iota^{\alpha\cdot \alpha^*})\geql 
               ((\varepsilon_1^*)^\alpha\mult \varepsilon_2^{\alpha^*}\mult a_\iota^{\alpha\cdot \alpha^*}))\overset{\text{(\ref{eq7h})}}{=\!\!=} \\
& \mi{reduced}((\upsilon_1^{\alpha^*}\mult (\varepsilon_2^*)^\alpha)\geql ((\varepsilon_1^*)^\alpha\mult \varepsilon_2^{\alpha^*}))= \\
& \mi{reduced}^-((\upsilon_1^{\alpha^*}\mult (\varepsilon_2^*)^\alpha)\geql ((\varepsilon_1^*)^\alpha\mult \varepsilon_2^{\alpha^*}))\geql \\
& \quad \mi{reduced}^+((\upsilon_1^{\alpha^*}\mult (\varepsilon_2^*)^\alpha)\geql ((\varepsilon_1^*)^\alpha\mult \varepsilon_2^{\alpha^*}))\underset{\text{(\ref{eq8c})}}{\in} \mi{clo},
\end{alignat*}
$\mi{atoms}(\upsilon_1), \mi{atoms}(\varepsilon_1^*), \mi{atoms}(\varepsilon_2), \mi{atoms}(\varepsilon_2^*)\subseteq \mi{dom}({\mc V}_{\iota-1})$,
$\mi{atoms}(\mi{reduced}^-((\upsilon_1^{\alpha^*}\mult (\varepsilon_2^*)^\alpha)\geql ((\varepsilon_1^*)^\alpha\mult \varepsilon_2^{\alpha^*})))\underset{\text{(\ref{eq7a})}}{\subseteq}
 \mi{atoms}(\upsilon_1^{\alpha^*}\mult (\varepsilon_2^*)^\alpha)=\mi{atoms}(\upsilon_1)\cup \mi{atoms}(\varepsilon_2^*)\subseteq \mi{dom}({\mc V}_{\iota-1})$,
$\mi{atoms}(\mi{reduced}^+((\upsilon_1^{\alpha^*}\mult (\varepsilon_2^*)^\alpha)\geql ((\varepsilon_1^*)^\alpha\mult \varepsilon_2^{\alpha^*})))\underset{\text{(\ref{eq7b})}}{\subseteq}
 \mi{atoms}((\varepsilon_1^*)^\alpha\mult \varepsilon_2^{\alpha^*})=\mi{atoms}(\varepsilon_1^*)\cup \mi{atoms}(\varepsilon_2)\subseteq \mi{dom}({\mc V}_{\iota-1})$;
by the induction hypothesis for $\iota-1<\iota$, 
$\|\upsilon_1\|^{{\mc V}_{\iota-1}}, \|\varepsilon_1^*\|^{{\mc V}_{\iota-1}}, \|\varepsilon_2\|^{{\mc V}_{\iota-1}}, \|\varepsilon_2^*\|^{{\mc V}_{\iota-1}},
 \|\mi{reduced}^-((\upsilon_1^{\alpha^*}\mult (\varepsilon_2^*)^\alpha)\geql ((\varepsilon_1^*)^\alpha\mult \varepsilon_2^{\alpha^*}))\|^{{\mc V}_{\iota-1}},
 \|\mi{reduced}^+((\upsilon_1^{\alpha^*}\mult (\varepsilon_2^*)^\alpha)\geql ((\varepsilon_1^*)^\alpha\mult \varepsilon_2^{\alpha^*}))\|^{{\mc V}_{\iota-1}}\underset{\text{(a)}}{\in} (0,1]$,
$\|\mi{reduced}^-((\upsilon_1^{\alpha^*}\mult (\varepsilon_2^*)^\alpha)\geql ((\varepsilon_1^*)^\alpha\mult \varepsilon_2^{\alpha^*}))\|^{{\mc V}_{\iota-1}}\overset{\text{(c)}}{=\!\!=}
 \|\mi{reduced}^+((\upsilon_1^{\alpha^*}\mult (\varepsilon_2^*)^\alpha)\geql ((\varepsilon_1^*)^\alpha\mult \varepsilon_2^{\alpha^*}))\|^{{\mc V}_{\iota-1}}$;
\begin{alignat*}{1}
& \dfrac{\|\mi{reduced}^-((\upsilon_1^{\alpha^*}\mult (\varepsilon_2^*)^\alpha)\geql ((\varepsilon_1^*)^\alpha\mult \varepsilon_2^{\alpha^*}))\|^{{\mc V}_{\iota-1}}}
        {\|\mi{reduced}^+((\upsilon_1^{\alpha^*}\mult (\varepsilon_2^*)^\alpha)\geql ((\varepsilon_1^*)^\alpha\mult \varepsilon_2^{\alpha^*}))\|^{{\mc V}_{\iota-1}}}\overset{\text{(\ref{eq7n})}}{=\!\!=} 
  \dfrac{\|\upsilon_1^{\alpha^*}\mult (\varepsilon_2^*)^\alpha\|^{{\mc V}_{\iota-1}}}
        {\|(\varepsilon_1^*)^\alpha\mult \varepsilon_2^{\alpha^*}\|^{{\mc V}_{\iota-1}}}= \\
& \dfrac{\left(\|\upsilon_1\|^{{\mc V}_{\iota-1}}\right)^{\alpha^*}\fswedge \left(\|\varepsilon_2^*\|^{{\mc V}_{\iota-1}}\right)^\alpha}
        {\left(\|\varepsilon_1^*\|^{{\mc V}_{\iota-1}}\right)^\alpha\fswedge \left(\|\varepsilon_2\|^{{\mc V}_{\iota-1}}\right)^{\alpha^*}}=1, \\[1mm]
& \left(\|\upsilon_1\|^{{\mc V}_{\iota-1}}\right)^{\alpha^*}\fswedge \left(\dfrac{\|\varepsilon_2^*\|^{{\mc V}_{\iota-1}}}
                                                                                 {\|\varepsilon_1^*\|^{{\mc V}_{\iota-1}}}\right)^\alpha=\left(\|\varepsilon_2\|^{{\mc V}_{\iota-1}}\right)^{\alpha^*}, \\
& \|\upsilon_1\|^{{\mc V}_{\iota-1}}\fswedge \left(\dfrac{\|\varepsilon_2^*\|^{{\mc V}_{\iota-1}}}
                                                         {\|\varepsilon_1^*\|^{{\mc V}_{\iota-1}}}\right)^{\dfrac{\alpha}{\alpha^*}}=\|\varepsilon_2\|^{{\mc V}_{\iota-1}}, \\[1mm]
& \|\varepsilon_1\|^{{\mc V}_\iota}=\|\upsilon_1\mult a_\iota^\alpha\|^{{\mc V}_\iota}=\|\upsilon_1\|^{{\mc V}_\iota}\fswedge {\mc V}_\iota(a_\iota)^\alpha\overset{\text{(b)}}{=\!\!=} 
  \|\upsilon_1\|^{{\mc V}_{\iota-1}}\fswedge {\mc V}_\iota(a_\iota)^\alpha= \\
& \|\upsilon_1\|^{{\mc V}_{\iota-1}}\fswedge \left(\dfrac{\|\varepsilon_2^*\|^{{\mc V}_{\iota-1}}}
                                                         {\|\varepsilon_1^*\|^{{\mc V}_{\iota-1}}}\right)^{\dfrac{\alpha}{\alpha^*}}= 
  \|\varepsilon_2\|^{{\mc V}_{\iota-1}}\overset{\text{(b)}}{=\!\!=} \|\varepsilon_2\|^{{\mc V}_\iota};
\end{alignat*}
(c) holds.

Case 2.10.1.1.2:
$\varepsilon_1\gleq \varepsilon_2\in \mi{clo}$.
Then $\varepsilon_1\gleq \varepsilon_2=(\upsilon_1\mult a_\iota^\alpha)\gleq \varepsilon_2\in \mi{clo}$,
$((\upsilon_1\mult a_\iota^\alpha)\gleq \varepsilon_2)^{\alpha^*}=(\upsilon_1^{\alpha^*}\mult a_\iota^{\alpha\cdot \alpha^*})\gleq \varepsilon_2^{\alpha^*}\underset{\text{(\ref{eq8d})}}{\in} \mi{clo}$,
$(\varepsilon_1^*\mult a_\iota^{\alpha^*})\geql \varepsilon_2^*\in \mi{clo}$,
$\varepsilon_2^*\geql (\varepsilon_1^*\mult a_\iota^{\alpha^*})\underset{\text{(\ref{eq8b})}}{\in} \mi{clo}$,
$(\varepsilon_2^*\geql (\varepsilon_1^*\mult a_\iota^{\alpha^*}))^\alpha=(\varepsilon_2^*)^\alpha\geql ((\varepsilon_1^*)^\alpha\mult a_\iota^{\alpha\cdot \alpha^*})\underset{\text{(\ref{eq8d})}}{\in} \mi{clo}$,
\begin{alignat*}{1}
& \mi{reduced}(((\upsilon_1^{\alpha^*}\mult a_\iota^{\alpha\cdot \alpha^*})\gleq \varepsilon_2^{\alpha^*})\mult
               ((\varepsilon_2^*)^\alpha\geql ((\varepsilon_1^*)^\alpha\mult a_\iota^{\alpha\cdot \alpha^*})))= \\
& \mi{reduced}((\upsilon_1^{\alpha^*}\mult (\varepsilon_2^*)^\alpha\mult a_\iota^{\alpha\cdot \alpha^*})\gleq 
               ((\varepsilon_1^*)^\alpha\mult \varepsilon_2^{\alpha^*}\mult a_\iota^{\alpha\cdot \alpha^*}))\overset{\text{(\ref{eq7h})}}{=\!\!=} \\
& \mi{reduced}((\upsilon_1^{\alpha^*}\mult (\varepsilon_2^*)^\alpha)\gleq ((\varepsilon_1^*)^\alpha\mult \varepsilon_2^{\alpha^*}))= \\
& \mi{reduced}^-((\upsilon_1^{\alpha^*}\mult (\varepsilon_2^*)^\alpha)\gleq ((\varepsilon_1^*)^\alpha\mult \varepsilon_2^{\alpha^*}))\gleq \\
& \quad \mi{reduced}^+((\upsilon_1^{\alpha^*}\mult (\varepsilon_2^*)^\alpha)\gleq ((\varepsilon_1^*)^\alpha\mult \varepsilon_2^{\alpha^*}))\underset{\text{(\ref{eq8c})}}{\in} \mi{clo},
\end{alignat*}
$\mi{atoms}(\upsilon_1), \mi{atoms}(\varepsilon_1^*), \mi{atoms}(\varepsilon_2), \mi{atoms}(\varepsilon_2^*)\subseteq \mi{dom}({\mc V}_{\iota-1})$,
$\mi{atoms}(\mi{reduced}^-((\upsilon_1^{\alpha^*}\mult (\varepsilon_2^*)^\alpha)\gleq ((\varepsilon_1^*)^\alpha\mult \varepsilon_2^{\alpha^*})))\underset{\text{(\ref{eq7a})}}{\subseteq}
 \mi{atoms}(\upsilon_1^{\alpha^*}\mult (\varepsilon_2^*)^\alpha)=\mi{atoms}(\upsilon_1)\cup \mi{atoms}(\varepsilon_2^*)\subseteq \mi{dom}({\mc V}_{\iota-1})$,
$\mi{atoms}(\mi{reduced}^+((\upsilon_1^{\alpha^*}\mult (\varepsilon_2^*)^\alpha)\gleq ((\varepsilon_1^*)^\alpha\mult \varepsilon_2^{\alpha^*})))\underset{\text{(\ref{eq7b})}}{\subseteq}
 \mi{atoms}((\varepsilon_1^*)^\alpha\mult \varepsilon_2^{\alpha^*})=\mi{atoms}(\varepsilon_1^*)\cup \mi{atoms}(\varepsilon_2)\subseteq \mi{dom}({\mc V}_{\iota-1})$;
by the induction hypothesis for $\iota-1<\iota$, 
$\|\upsilon_1\|^{{\mc V}_{\iota-1}}, \|\varepsilon_1^*\|^{{\mc V}_{\iota-1}}, \|\varepsilon_2\|^{{\mc V}_{\iota-1}}, \|\varepsilon_2^*\|^{{\mc V}_{\iota-1}},
 \|\mi{reduced}^-((\upsilon_1^{\alpha^*}\mult (\varepsilon_2^*)^\alpha)\gleq ((\varepsilon_1^*)^\alpha\mult \varepsilon_2^{\alpha^*}))\|^{{\mc V}_{\iota-1}},
 \|\mi{reduced}^+((\upsilon_1^{\alpha^*}\mult (\varepsilon_2^*)^\alpha)\gleq ((\varepsilon_1^*)^\alpha\mult \varepsilon_2^{\alpha^*}))\|^{{\mc V}_{\iota-1}}\underset{\text{(a)}}{\in} (0,1]$,
$\|\mi{reduced}^-((\upsilon_1^{\alpha^*}\mult (\varepsilon_2^*)^\alpha)\gleq ((\varepsilon_1^*)^\alpha\mult \varepsilon_2^{\alpha^*}))\|^{{\mc V}_{\iota-1}}\underset{\text{(d)}}{\leq}
 \|\mi{reduced}^+((\upsilon_1^{\alpha^*}\mult (\varepsilon_2^*)^\alpha)\gleq ((\varepsilon_1^*)^\alpha\mult \varepsilon_2^{\alpha^*}))\|^{{\mc V}_{\iota-1}}$;
\begin{alignat*}{1}
& \dfrac{\|\mi{reduced}^-((\upsilon_1^{\alpha^*}\mult (\varepsilon_2^*)^\alpha)\gleq ((\varepsilon_1^*)^\alpha\mult \varepsilon_2^{\alpha^*}))\|^{{\mc V}_{\iota-1}}}
        {\|\mi{reduced}^+((\upsilon_1^{\alpha^*}\mult (\varepsilon_2^*)^\alpha)\gleq ((\varepsilon_1^*)^\alpha\mult \varepsilon_2^{\alpha^*}))\|^{{\mc V}_{\iota-1}}}\overset{\text{(\ref{eq7n})}}{=\!\!=} 
  \dfrac{\|\upsilon_1^{\alpha^*}\mult (\varepsilon_2^*)^\alpha\|^{{\mc V}_{\iota-1}}}
        {\|(\varepsilon_1^*)^\alpha\mult \varepsilon_2^{\alpha^*}\|^{{\mc V}_{\iota-1}}}= \\
& \dfrac{\left(\|\upsilon_1\|^{{\mc V}_{\iota-1}}\right)^{\alpha^*}\fswedge \left(\|\varepsilon_2^*\|^{{\mc V}_{\iota-1}}\right)^\alpha}
        {\left(\|\varepsilon_1^*\|^{{\mc V}_{\iota-1}}\right)^\alpha\fswedge \left(\|\varepsilon_2\|^{{\mc V}_{\iota-1}}\right)^{\alpha^*}}\leq 1, \\[1mm]
& \left(\|\upsilon_1\|^{{\mc V}_{\iota-1}}\right)^{\alpha^*}\fswedge \left(\dfrac{\|\varepsilon_2^*\|^{{\mc V}_{\iota-1}}}
                                                                                 {\|\varepsilon_1^*\|^{{\mc V}_{\iota-1}}}\right)^\alpha\leq \left(\|\varepsilon_2\|^{{\mc V}_{\iota-1}}\right)^{\alpha^*}, \\
& \|\upsilon_1\|^{{\mc V}_{\iota-1}}\fswedge \left(\dfrac{\|\varepsilon_2^*\|^{{\mc V}_{\iota-1}}}
                                                         {\|\varepsilon_1^*\|^{{\mc V}_{\iota-1}}}\right)^{\dfrac{\alpha}{\alpha^*}}\leq \|\varepsilon_2\|^{{\mc V}_{\iota-1}}, \\[1mm]
& \|\varepsilon_1\|^{{\mc V}_\iota}=\|\upsilon_1\mult a_\iota^\alpha\|^{{\mc V}_\iota}=\|\upsilon_1\|^{{\mc V}_\iota}\fswedge {\mc V}_\iota(a_\iota)^\alpha\overset{\text{(b)}}{=\!\!=} 
  \|\upsilon_1\|^{{\mc V}_{\iota-1}}\fswedge {\mc V}_\iota(a_\iota)^\alpha= \\
& \|\upsilon_1\|^{{\mc V}_{\iota-1}}\fswedge \left(\dfrac{\|\varepsilon_2^*\|^{{\mc V}_{\iota-1}}}
                                                         {\|\varepsilon_1^*\|^{{\mc V}_{\iota-1}}}\right)^{\dfrac{\alpha}{\alpha^*}}\leq 
  \|\varepsilon_2\|^{{\mc V}_{\iota-1}}\overset{\text{(b)}}{=\!\!=} \|\varepsilon_2\|^{{\mc V}_\iota};
\end{alignat*}
(d) holds.

Case 2.10.1.1.3:
$\varepsilon_1\gle \varepsilon_2\in \mi{clo}$.
Then $\varepsilon_1\gle \varepsilon_2=(\upsilon_1\mult a_\iota^\alpha)\gle \varepsilon_2\in \mi{clo}$,
$((\upsilon_1\mult a_\iota^\alpha)\gle \varepsilon_2)^{\alpha^*}=(\upsilon_1^{\alpha^*}\mult a_\iota^{\alpha\cdot \alpha^*})\gle \varepsilon_2^{\alpha^*}\underset{\text{(\ref{eq8d})}}{\in} \mi{clo}$,
$(\varepsilon_1^*\mult a_\iota^{\alpha^*})\geql \varepsilon_2^*\in \mi{clo}$,
$\varepsilon_2^*\geql (\varepsilon_1^*\mult a_\iota^{\alpha^*})\underset{\text{(\ref{eq8b})}}{\in} \mi{clo}$,
$(\varepsilon_2^*\geql (\varepsilon_1^*\mult a_\iota^{\alpha^*}))^\alpha=(\varepsilon_2^*)^\alpha\geql ((\varepsilon_1^*)^\alpha\mult a_\iota^{\alpha\cdot \alpha^*})\underset{\text{(\ref{eq8d})}}{\in} \mi{clo}$,
\begin{alignat*}{1}
& \mi{reduced}(((\upsilon_1^{\alpha^*}\mult a_\iota^{\alpha\cdot \alpha^*})\gle \varepsilon_2^{\alpha^*})\mult
               ((\varepsilon_2^*)^\alpha\geql ((\varepsilon_1^*)^\alpha\mult a_\iota^{\alpha\cdot \alpha^*})))= \\
& \mi{reduced}((\upsilon_1^{\alpha^*}\mult (\varepsilon_2^*)^\alpha\mult a_\iota^{\alpha\cdot \alpha^*})\gle 
               ((\varepsilon_1^*)^\alpha\mult \varepsilon_2^{\alpha^*}\mult a_\iota^{\alpha\cdot \alpha^*}))\overset{\text{(\ref{eq7h})}}{=\!\!=} \\
& \mi{reduced}((\upsilon_1^{\alpha^*}\mult (\varepsilon_2^*)^\alpha)\gle ((\varepsilon_1^*)^\alpha\mult \varepsilon_2^{\alpha^*}))= \\
& \mi{reduced}^-((\upsilon_1^{\alpha^*}\mult (\varepsilon_2^*)^\alpha)\gle ((\varepsilon_1^*)^\alpha\mult \varepsilon_2^{\alpha^*}))\gle \\
& \quad \mi{reduced}^+((\upsilon_1^{\alpha^*}\mult (\varepsilon_2^*)^\alpha)\gle ((\varepsilon_1^*)^\alpha\mult \varepsilon_2^{\alpha^*}))\underset{\text{(\ref{eq8c})}}{\in} \mi{clo},
\end{alignat*}
$\mi{atoms}(\upsilon_1), \mi{atoms}(\varepsilon_1^*), \mi{atoms}(\varepsilon_2), \mi{atoms}(\varepsilon_2^*)\subseteq \mi{dom}({\mc V}_{\iota-1})$,
$\mi{atoms}(\mi{reduced}^-((\upsilon_1^{\alpha^*}\mult (\varepsilon_2^*)^\alpha)\gle ((\varepsilon_1^*)^\alpha\mult \varepsilon_2^{\alpha^*})))\underset{\text{(\ref{eq7a})}}{\subseteq}
 \mi{atoms}(\upsilon_1^{\alpha^*}\mult (\varepsilon_2^*)^\alpha)=\mi{atoms}(\upsilon_1)\cup \mi{atoms}(\varepsilon_2^*)\subseteq \mi{dom}({\mc V}_{\iota-1})$,
$\mi{atoms}(\mi{reduced}^+((\upsilon_1^{\alpha^*}\mult (\varepsilon_2^*)^\alpha)\gle ((\varepsilon_1^*)^\alpha\mult \varepsilon_2^{\alpha^*})))\underset{\text{(\ref{eq7b})}}{\subseteq}
 \mi{atoms}((\varepsilon_1^*)^\alpha\mult \varepsilon_2^{\alpha^*})=\mi{atoms}(\varepsilon_1^*)\cup \mi{atoms}(\varepsilon_2)\subseteq \mi{dom}({\mc V}_{\iota-1})$;
by the induction hypothesis for $\iota-1<\iota$, 
$\|\upsilon_1\|^{{\mc V}_{\iota-1}}, \|\varepsilon_1^*\|^{{\mc V}_{\iota-1}}, \|\varepsilon_2\|^{{\mc V}_{\iota-1}}, \|\varepsilon_2^*\|^{{\mc V}_{\iota-1}},
 \|\mi{reduced}^-((\upsilon_1^{\alpha^*}\mult (\varepsilon_2^*)^\alpha)\gle ((\varepsilon_1^*)^\alpha\mult \varepsilon_2^{\alpha^*}))\|^{{\mc V}_{\iota-1}},
 \|\mi{reduced}^+((\upsilon_1^{\alpha^*}\mult (\varepsilon_2^*)^\alpha)\gle ((\varepsilon_1^*)^\alpha\mult \varepsilon_2^{\alpha^*}))\|^{{\mc V}_{\iota-1}}\underset{\text{(a)}}{\in} (0,1]$,
$\|\mi{reduced}^-((\upsilon_1^{\alpha^*}\mult (\varepsilon_2^*)^\alpha)\gle ((\varepsilon_1^*)^\alpha\mult \varepsilon_2^{\alpha^*}))\|^{{\mc V}_{\iota-1}}\underset{\text{(e)}}{<}
 \|\mi{reduced}^+((\upsilon_1^{\alpha^*}\mult (\varepsilon_2^*)^\alpha)\gle ((\varepsilon_1^*)^\alpha\mult \varepsilon_2^{\alpha^*}))\|^{{\mc V}_{\iota-1}}$;
\begin{alignat*}{1}
& \dfrac{\|\mi{reduced}^-((\upsilon_1^{\alpha^*}\mult (\varepsilon_2^*)^\alpha)\gle ((\varepsilon_1^*)^\alpha\mult \varepsilon_2^{\alpha^*}))\|^{{\mc V}_{\iota-1}}}
        {\|\mi{reduced}^+((\upsilon_1^{\alpha^*}\mult (\varepsilon_2^*)^\alpha)\gle ((\varepsilon_1^*)^\alpha\mult \varepsilon_2^{\alpha^*}))\|^{{\mc V}_{\iota-1}}}\overset{\text{(\ref{eq7n})}}{=\!\!=} 
  \dfrac{\|\upsilon_1^{\alpha^*}\mult (\varepsilon_2^*)^\alpha\|^{{\mc V}_{\iota-1}}}
        {\|(\varepsilon_1^*)^\alpha\mult \varepsilon_2^{\alpha^*}\|^{{\mc V}_{\iota-1}}}= \\
& \dfrac{\left(\|\upsilon_1\|^{{\mc V}_{\iota-1}}\right)^{\alpha^*}\fswedge \left(\|\varepsilon_2^*\|^{{\mc V}_{\iota-1}}\right)^\alpha}
        {\left(\|\varepsilon_1^*\|^{{\mc V}_{\iota-1}}\right)^\alpha\fswedge \left(\|\varepsilon_2\|^{{\mc V}_{\iota-1}}\right)^{\alpha^*}}<1, \\[1mm]
& \left(\|\upsilon_1\|^{{\mc V}_{\iota-1}}\right)^{\alpha^*}\fswedge \left(\dfrac{\|\varepsilon_2^*\|^{{\mc V}_{\iota-1}}}
                                                                                 {\|\varepsilon_1^*\|^{{\mc V}_{\iota-1}}}\right)^\alpha<\left(\|\varepsilon_2\|^{{\mc V}_{\iota-1}}\right)^{\alpha^*}, \\
& \|\upsilon_1\|^{{\mc V}_{\iota-1}}\fswedge \left(\dfrac{\|\varepsilon_2^*\|^{{\mc V}_{\iota-1}}}
                                                         {\|\varepsilon_1^*\|^{{\mc V}_{\iota-1}}}\right)^{\dfrac{\alpha}{\alpha^*}}<\|\varepsilon_2\|^{{\mc V}_{\iota-1}}, \\[1mm]
& \|\varepsilon_1\|^{{\mc V}_\iota}=\|\upsilon_1\mult a_\iota^\alpha\|^{{\mc V}_\iota}=\|\upsilon_1\|^{{\mc V}_\iota}\fswedge {\mc V}_\iota(a_\iota)^\alpha\overset{\text{(b)}}{=\!\!=} 
  \|\upsilon_1\|^{{\mc V}_{\iota-1}}\fswedge {\mc V}_\iota(a_\iota)^\alpha= \\
& \|\upsilon_1\|^{{\mc V}_{\iota-1}}\fswedge \left(\dfrac{\|\varepsilon_2^*\|^{{\mc V}_{\iota-1}}}
                                                         {\|\varepsilon_1^*\|^{{\mc V}_{\iota-1}}}\right)^{\dfrac{\alpha}{\alpha^*}}< 
  \|\varepsilon_2\|^{{\mc V}_{\iota-1}}\overset{\text{(b)}}{=\!\!=} \|\varepsilon_2\|^{{\mc V}_\iota};
\end{alignat*}
(e) holds.

Case 2.10.1.2:
$a_\iota\not\in \mi{atoms}(\varepsilon_1)\subseteq \mi{dom}({\mc V}_{\iota-1})$ and $a_\iota\in \mi{atoms}(\varepsilon_2)$.
Then $\mi{atoms}(\varepsilon_2)\subseteq \mi{dom}({\mc V}_\iota)=\mi{dom}({\mc V}_{\iota-1})\cup \{a_\iota\}$,
$\varepsilon_2=\upsilon_2\mult a_\iota^\alpha$, $\upsilon_2\in \{\gu\}\cup \mi{PropConj}_A$, $\alpha\geq 1$, $a_\iota\not\in \mi{atoms}(\upsilon_2)\subseteq \mi{dom}({\mc V}_{\iota-1})$.
We get three cases for $\varepsilon_1$ and $\varepsilon_2$.

Case 2.10.1.2.1:
$\varepsilon_1\geql \varepsilon_2\in \mi{clo}$.
Then $\varepsilon_2\geql \varepsilon_1\underset{\text{(\ref{eq8b})}}{\in} \mi{clo}$, $a_\iota\in \mi{atoms}(\varepsilon_2)$, $\mi{atoms}(\varepsilon_1)\subseteq \mi{dom}({\mc V}_{\iota-1})$.
We get from Case 2.10.1.1.1 for $\varepsilon_2$ and $\varepsilon_1$ that $\|\varepsilon_1\|^{{\mc V}_\iota}=\|\varepsilon_2\|^{{\mc V}_\iota}$;
(c) holds.

Case 2.10.1.2.2:
$\varepsilon_1\gleq \varepsilon_2\in \mi{clo}$.
Then $\varepsilon_1\gleq \varepsilon_2=\varepsilon_1\gleq (\upsilon_2\mult a_\iota^\alpha)\in \mi{clo}$,
$(\varepsilon_1\gleq (\upsilon_2\mult a_\iota^\alpha))^{\alpha^*}=\varepsilon_1^{\alpha^*}\gleq (\upsilon_2^{\alpha^*}\mult a_\iota^{\alpha\cdot \alpha^*})\underset{\text{(\ref{eq8d})}}{\in} \mi{clo}$,
$(\varepsilon_1^*\mult a_\iota^{\alpha^*})\geql \varepsilon_2^*\in \mi{clo}$,
$((\varepsilon_1^*\mult a_\iota^{\alpha^*})\geql \varepsilon_2^*)^\alpha=((\varepsilon_1^*)^\alpha\mult a_\iota^{\alpha\cdot \alpha^*})\geql (\varepsilon_2^*)^\alpha\underset{\text{(\ref{eq8d})}}{\in} \mi{clo}$,
\begin{alignat*}{1}
& \mi{reduced}((\varepsilon_1^{\alpha^*}\gleq (\upsilon_2^{\alpha^*}\mult a_\iota^{\alpha\cdot \alpha^*}))\mult
               (((\varepsilon_1^*)^\alpha\mult a_\iota^{\alpha\cdot \alpha^*})\geql (\varepsilon_2^*)^\alpha))= \\
& \mi{reduced}((\varepsilon_1^{\alpha^*}\mult (\varepsilon_1^*)^\alpha\mult a_\iota^{\alpha\cdot \alpha^*})\gleq 
               (\upsilon_2^{\alpha^*}\mult (\varepsilon_2^*)^\alpha\mult a_\iota^{\alpha\cdot \alpha^*}))\overset{\text{(\ref{eq7h})}}{=\!\!=} \\
& \mi{reduced}((\varepsilon_1^{\alpha^*}\mult (\varepsilon_1^*)^\alpha)\gleq (\upsilon_2^{\alpha^*}\mult (\varepsilon_2^*)^\alpha))= \\
& \mi{reduced}^-((\varepsilon_1^{\alpha^*}\mult (\varepsilon_1^*)^\alpha)\gleq (\upsilon_2^{\alpha^*}\mult (\varepsilon_2^*)^\alpha))\gleq \\
& \quad \mi{reduced}^+((\varepsilon_1^{\alpha^*}\mult (\varepsilon_1^*)^\alpha)\gleq (\upsilon_2^{\alpha^*}\mult (\varepsilon_2^*)^\alpha))\underset{\text{(\ref{eq8c})}}{\in} \mi{clo},
\end{alignat*}
$\mi{atoms}(\varepsilon_1), \mi{atoms}(\varepsilon_1^*), \mi{atoms}(\upsilon_2), \mi{atoms}(\varepsilon_2^*)\subseteq \mi{dom}({\mc V}_{\iota-1})$,
$\mi{atoms}(\mi{reduced}^-((\varepsilon_1^{\alpha^*}\mult (\varepsilon_1^*)^\alpha)\gleq (\upsilon_2^{\alpha^*}\mult (\varepsilon_2^*)^\alpha)))\underset{\text{(\ref{eq7a})}}{\subseteq}
 \mi{atoms}(\varepsilon_1^{\alpha^*}\mult (\varepsilon_1^*)^\alpha)=\mi{atoms}(\varepsilon_1)\cup \mi{atoms}(\varepsilon_1^*)\subseteq \mi{dom}({\mc V}_{\iota-1})$,
$\mi{atoms}(\mi{reduced}^+((\varepsilon_1^{\alpha^*}\mult (\varepsilon_1^*)^\alpha)\gleq (\upsilon_2^{\alpha^*}\mult (\varepsilon_2^*)^\alpha)))\underset{\text{(\ref{eq7b})}}{\subseteq}
 \mi{atoms}(\upsilon_2^{\alpha^*}\mult (\varepsilon_2^*)^\alpha)=\mi{atoms}(\upsilon_2)\cup \mi{atoms}(\varepsilon_2^*)\subseteq \mi{dom}({\mc V}_{\iota-1})$;
by the induction hypothesis for $\iota-1<\iota$, 
$\|\varepsilon_1\|^{{\mc V}_{\iota-1}}, \|\varepsilon_1^*\|^{{\mc V}_{\iota-1}}, \|\upsilon_2\|^{{\mc V}_{\iota-1}}, \|\varepsilon_2^*\|^{{\mc V}_{\iota-1}},
 \|\mi{reduced}^-((\varepsilon_1^{\alpha^*}\mult (\varepsilon_1^*)^\alpha)\gleq (\upsilon_2^{\alpha^*}\mult (\varepsilon_2^*)^\alpha))\|^{{\mc V}_{\iota-1}},
 \|\mi{reduced}^+((\varepsilon_1^{\alpha^*}\mult (\varepsilon_1^*)^\alpha)\gleq (\upsilon_2^{\alpha^*}\mult (\varepsilon_2^*)^\alpha))\|^{{\mc V}_{\iota-1}}\underset{\text{(a)}}{\in} (0,1]$,
$\|\mi{reduced}^-((\varepsilon_1^{\alpha^*}\mult (\varepsilon_1^*)^\alpha)\gleq (\upsilon_2^{\alpha^*}\mult (\varepsilon_2^*)^\alpha))\|^{{\mc V}_{\iota-1}}\underset{\text{(d)}}{\leq}
 \|\mi{reduced}^+((\varepsilon_1^{\alpha^*}\mult (\varepsilon_1^*)^\alpha)\gleq (\upsilon_2^{\alpha^*}\mult (\varepsilon_2^*)^\alpha))\|^{{\mc V}_{\iota-1}}$;
\begin{alignat*}{1}
& \dfrac{\|\mi{reduced}^-((\varepsilon_1^{\alpha^*}\mult (\varepsilon_1^*)^\alpha)\gleq (\upsilon_2^{\alpha^*}\mult (\varepsilon_2^*)^\alpha))\|^{{\mc V}_{\iota-1}}}
        {\|\mi{reduced}^+((\varepsilon_1^{\alpha^*}\mult (\varepsilon_1^*)^\alpha)\gleq (\upsilon_2^{\alpha^*}\mult (\varepsilon_2^*)^\alpha))\|^{{\mc V}_{\iota-1}}}\overset{\text{(\ref{eq7n})}}{=\!\!=} 
  \dfrac{\|\varepsilon_1^{\alpha^*}\mult (\varepsilon_1^*)^\alpha\|^{{\mc V}_{\iota-1}}}
        {\|\upsilon_2^{\alpha^*}\mult (\varepsilon_2^*)^\alpha\|^{{\mc V}_{\iota-1}}}= \\
& \dfrac{\left(\|\varepsilon_1\|^{{\mc V}_{\iota-1}}\right)^{\alpha^*}\fswedge \left(\|\varepsilon_1^*\|^{{\mc V}_{\iota-1}}\right)^\alpha}
        {\left(\|\upsilon_2\|^{{\mc V}_{\iota-1}}\right)^{\alpha^*}\fswedge \left(\|\varepsilon_2^*\|^{{\mc V}_{\iota-1}}\right)^\alpha}\leq 1, \\[1mm]
& \left(\|\varepsilon_1\|^{{\mc V}_{\iota-1}}\right)^{\alpha^*}\leq \left(\|\upsilon_2\|^{{\mc V}_{\iota-1}}\right)^{\alpha^*}\fswedge \left(\dfrac{\|\varepsilon_2^*\|^{{\mc V}_{\iota-1}}}
                                                                                                                                                   {\|\varepsilon_1^*\|^{{\mc V}_{\iota-1}}}\right)^\alpha, \\
& \|\varepsilon_1\|^{{\mc V}_{\iota-1}}\leq \|\upsilon_2\|^{{\mc V}_{\iota-1}}\fswedge \left(\dfrac{\|\varepsilon_2^*\|^{{\mc V}_{\iota-1}}}
                                                                                                   {\|\varepsilon_1^*\|^{{\mc V}_{\iota-1}}}\right)^{\dfrac{\alpha}{\alpha^*}}, \\[1mm]
& \|\varepsilon_1\|^{{\mc V}_\iota}\overset{\text{(b)}}{=\!\!=} \|\varepsilon_1\|^{{\mc V}_{\iota-1}}\leq 
  \|\upsilon_2\|^{{\mc V}_{\iota-1}}\fswedge \left(\dfrac{\|\varepsilon_2^*\|^{{\mc V}_{\iota-1}}}
                                                         {\|\varepsilon_1^*\|^{{\mc V}_{\iota-1}}}\right)^{\dfrac{\alpha}{\alpha^*}}=
  \|\upsilon_2\|^{{\mc V}_{\iota-1}}\fswedge {\mc V}_\iota(a_\iota)^\alpha\overset{\text{(b)}}{=\!\!=} \\
& \|\upsilon_2\|^{{\mc V}_\iota}\fswedge {\mc V}_\iota(a_\iota)^\alpha=\|\upsilon_2\mult a_\iota^\alpha\|^{{\mc V}_\iota}=\|\varepsilon_2\|^{{\mc V}_\iota};
\end{alignat*}
(d) holds.

Case 2.10.1.2.3:
$\varepsilon_1\gle \varepsilon_2\in \mi{clo}$.
Then $\varepsilon_1\gle \varepsilon_2=\varepsilon_1\gle (\upsilon_2\mult a_\iota^\alpha)\in \mi{clo}$,
$(\varepsilon_1\gle (\upsilon_2\mult a_\iota^\alpha))^{\alpha^*}=\varepsilon_1^{\alpha^*}\gle (\upsilon_2^{\alpha^*}\mult a_\iota^{\alpha\cdot \alpha^*})\underset{\text{(\ref{eq8d})}}{\in} \mi{clo}$,
$(\varepsilon_1^*\mult a_\iota^{\alpha^*})\geql \varepsilon_2^*\in \mi{clo}$,
$((\varepsilon_1^*\mult a_\iota^{\alpha^*})\geql \varepsilon_2^*)^\alpha=((\varepsilon_1^*)^\alpha\mult a_\iota^{\alpha\cdot \alpha^*})\geql (\varepsilon_2^*)^\alpha\underset{\text{(\ref{eq8d})}}{\in} \mi{clo}$,
\begin{alignat*}{1}
& \mi{reduced}((\varepsilon_1^{\alpha^*}\gle (\upsilon_2^{\alpha^*}\mult a_\iota^{\alpha\cdot \alpha^*}))\mult
               (((\varepsilon_1^*)^\alpha\mult a_\iota^{\alpha\cdot \alpha^*})\geql (\varepsilon_2^*)^\alpha))= \\
& \mi{reduced}((\varepsilon_1^{\alpha^*}\mult (\varepsilon_1^*)^\alpha\mult a_\iota^{\alpha\cdot \alpha^*})\gle 
               (\upsilon_2^{\alpha^*}\mult (\varepsilon_2^*)^\alpha\mult a_\iota^{\alpha\cdot \alpha^*}))\overset{\text{(\ref{eq7h})}}{=\!\!=} \\
& \mi{reduced}((\varepsilon_1^{\alpha^*}\mult (\varepsilon_1^*)^\alpha)\gle (\upsilon_2^{\alpha^*}\mult (\varepsilon_2^*)^\alpha))= \\
& \mi{reduced}^-((\varepsilon_1^{\alpha^*}\mult (\varepsilon_1^*)^\alpha)\gle (\upsilon_2^{\alpha^*}\mult (\varepsilon_2^*)^\alpha))\gle \\
& \quad \mi{reduced}^+((\varepsilon_1^{\alpha^*}\mult (\varepsilon_1^*)^\alpha)\gle (\upsilon_2^{\alpha^*}\mult (\varepsilon_2^*)^\alpha))\underset{\text{(\ref{eq8c})}}{\in} \mi{clo},
\end{alignat*}
$\mi{atoms}(\varepsilon_1), \mi{atoms}(\varepsilon_1^*), \mi{atoms}(\upsilon_2), \mi{atoms}(\varepsilon_2^*)\subseteq \mi{dom}({\mc V}_{\iota-1})$,
$\mi{atoms}(\mi{reduced}^-((\varepsilon_1^{\alpha^*}\mult (\varepsilon_1^*)^\alpha)\gle (\upsilon_2^{\alpha^*}\mult (\varepsilon_2^*)^\alpha)))\underset{\text{(\ref{eq7a})}}{\subseteq}
 \mi{atoms}(\varepsilon_1^{\alpha^*}\mult (\varepsilon_1^*)^\alpha)=\mi{atoms}(\varepsilon_1)\cup \mi{atoms}(\varepsilon_1^*)\subseteq \mi{dom}({\mc V}_{\iota-1})$,
$\mi{atoms}(\mi{reduced}^+((\varepsilon_1^{\alpha^*}\mult (\varepsilon_1^*)^\alpha)\gle (\upsilon_2^{\alpha^*}\mult (\varepsilon_2^*)^\alpha)))\underset{\text{(\ref{eq7b})}}{\subseteq}
 \mi{atoms}(\upsilon_2^{\alpha^*}\mult (\varepsilon_2^*)^\alpha)=\mi{atoms}(\upsilon_2)\cup \mi{atoms}(\varepsilon_2^*)\subseteq \mi{dom}({\mc V}_{\iota-1})$;
by the induction hypothesis for $\iota-1<\iota$, 
$\|\varepsilon_1\|^{{\mc V}_{\iota-1}}, \|\varepsilon_1^*\|^{{\mc V}_{\iota-1}}, \|\upsilon_2\|^{{\mc V}_{\iota-1}}, \|\varepsilon_2^*\|^{{\mc V}_{\iota-1}},
 \|\mi{reduced}^-((\varepsilon_1^{\alpha^*}\mult (\varepsilon_1^*)^\alpha)\gle (\upsilon_2^{\alpha^*}\mult (\varepsilon_2^*)^\alpha))\|^{{\mc V}_{\iota-1}},
 \|\mi{reduced}^+((\varepsilon_1^{\alpha^*}\mult (\varepsilon_1^*)^\alpha)\gle (\upsilon_2^{\alpha^*}\mult (\varepsilon_2^*)^\alpha))\|^{{\mc V}_{\iota-1}}\underset{\text{(a)}}{\in} (0,1]$,
$\|\mi{reduced}^-((\varepsilon_1^{\alpha^*}\mult (\varepsilon_1^*)^\alpha)\gle (\upsilon_2^{\alpha^*}\mult (\varepsilon_2^*)^\alpha))\|^{{\mc V}_{\iota-1}}\underset{\text{(e)}}{<}
 \|\mi{reduced}^+((\varepsilon_1^{\alpha^*}\mult (\varepsilon_1^*)^\alpha)\gle (\upsilon_2^{\alpha^*}\mult (\varepsilon_2^*)^\alpha))\|^{{\mc V}_{\iota-1}}$;
\begin{alignat*}{1}
& \dfrac{\|\mi{reduced}^-((\varepsilon_1^{\alpha^*}\mult (\varepsilon_1^*)^\alpha)\gle (\upsilon_2^{\alpha^*}\mult (\varepsilon_2^*)^\alpha))\|^{{\mc V}_{\iota-1}}}
        {\|\mi{reduced}^+((\varepsilon_1^{\alpha^*}\mult (\varepsilon_1^*)^\alpha)\gle (\upsilon_2^{\alpha^*}\mult (\varepsilon_2^*)^\alpha))\|^{{\mc V}_{\iota-1}}}\overset{\text{(\ref{eq7n})}}{=\!\!=} 
  \dfrac{\|\varepsilon_1^{\alpha^*}\mult (\varepsilon_1^*)^\alpha\|^{{\mc V}_{\iota-1}}}
        {\|\upsilon_2^{\alpha^*}\mult (\varepsilon_2^*)^\alpha\|^{{\mc V}_{\iota-1}}}= \\
& \dfrac{\left(\|\varepsilon_1\|^{{\mc V}_{\iota-1}}\right)^{\alpha^*}\fswedge \left(\|\varepsilon_1^*\|^{{\mc V}_{\iota-1}}\right)^\alpha}
        {\left(\|\upsilon_2\|^{{\mc V}_{\iota-1}}\right)^{\alpha^*}\fswedge \left(\|\varepsilon_2^*\|^{{\mc V}_{\iota-1}}\right)^\alpha}<1, \\[1mm]
& \left(\|\varepsilon_1\|^{{\mc V}_{\iota-1}}\right)^{\alpha^*}<\left(\|\upsilon_2\|^{{\mc V}_{\iota-1}}\right)^{\alpha^*}\fswedge \left(\dfrac{\|\varepsilon_2^*\|^{{\mc V}_{\iota-1}}}
                                                                                                                                               {\|\varepsilon_1^*\|^{{\mc V}_{\iota-1}}}\right)^\alpha, \\
& \|\varepsilon_1\|^{{\mc V}_{\iota-1}}<\|\upsilon_2\|^{{\mc V}_{\iota-1}}\fswedge \left(\dfrac{\|\varepsilon_2^*\|^{{\mc V}_{\iota-1}}}
                                                                                               {\|\varepsilon_1^*\|^{{\mc V}_{\iota-1}}}\right)^{\dfrac{\alpha}{\alpha^*}}, \\[1mm]
& \|\varepsilon_1\|^{{\mc V}_\iota}\overset{\text{(b)}}{=\!\!=} \|\varepsilon_1\|^{{\mc V}_{\iota-1}}< 
  \|\upsilon_2\|^{{\mc V}_{\iota-1}}\fswedge \left(\dfrac{\|\varepsilon_2^*\|^{{\mc V}_{\iota-1}}}
                                                         {\|\varepsilon_1^*\|^{{\mc V}_{\iota-1}}}\right)^{\dfrac{\alpha}{\alpha^*}}=
  \|\upsilon_2\|^{{\mc V}_{\iota-1}}\fswedge {\mc V}_\iota(a_\iota)^\alpha\overset{\text{(b)}}{=\!\!=} \\
& \|\upsilon_2\|^{{\mc V}_\iota}\fswedge {\mc V}_\iota(a_\iota)^\alpha=\|\upsilon_2\mult a_\iota^\alpha\|^{{\mc V}_\iota}=\|\varepsilon_2\|^{{\mc V}_\iota};
\end{alignat*}
(e) holds.

Case 2.10.1.3:
$a_\iota\in \mi{atoms}(\varepsilon_1)$ and $a_\iota\in \mi{atoms}(\varepsilon_2)$.
Then $a_\iota\in \mi{atoms}(\varepsilon_1)\cap \mi{atoms}(\varepsilon_2)\neq \emptyset$;
for all $\diamond\in \{\geql,\gleq,\gle\}$, $\varepsilon_1\diamond \varepsilon_2\underset{\text{(\ref{eq8aaa})}}{\not\in} \mi{clo}$; 
(c--e) hold trivially.

Case 2.10.2:
$\mbb{E}_{\iota-1}=\emptyset$.
Then $a_\iota\not\in \mi{dom}({\mc V}_{\iota-1})$, $\mi{atoms}(\varepsilon_1), \mi{atoms}(\varepsilon_2)\subseteq \mi{dom}({\mc V}_\iota)=\mi{dom}({\mc V}_{\iota-1})\cup \{a_\iota\}$,
$a_\iota\in \mi{atoms}(\varepsilon_1)$ or $a_\iota\in \mi{atoms}(\varepsilon_2)$,
$\varepsilon_1\geql \varepsilon_2\underset{\text{(\ref{eq8h})}}{\not\in} \mi{clo}$.
We get three cases for $\mi{atoms}(\varepsilon_1)$ and $\mi{atoms}(\varepsilon_2)$.

Case 2.10.2.1:
$a_\iota\in \mi{atoms}(\varepsilon_1)$ and $a_\iota\not\in \mi{atoms}(\varepsilon_2)\subseteq \mi{dom}({\mc V}_{\iota-1})$.
Then $\mi{atoms}(\varepsilon_1)\subseteq \mi{dom}({\mc V}_\iota)=\mi{dom}({\mc V}_{\iota-1})\cup \{a_\iota\}$,
$\varepsilon_1=\upsilon_1\mult a_\iota^\alpha$, $\upsilon_1\in \{\gu\}\cup \mi{PropConj}_A$, $\alpha\geq 1$, $a_\iota\not\in \mi{atoms}(\upsilon_1)\subseteq \mi{dom}({\mc V}_{\iota-1})$,
$\varepsilon_1\geql \varepsilon_2\not\in \mi{clo}$.
We get two cases for $\varepsilon_1$ and $\varepsilon_2$.

Case 2.10.2.1.1:
$\varepsilon_1\gleq \varepsilon_2\in \mi{clo}$.
Then $\mi{atoms}(\varepsilon_1), \mi{atoms}(\varepsilon_2)\subseteq \mi{dom}({\mc V}_\iota)=\mi{dom}({\mc V}_{\iota-1})\cup \{a_\iota\}$,
$a_\iota\in \mi{atoms}(\varepsilon_1)$ or $a_\iota\in \mi{atoms}(\varepsilon_2)$, $\mbb{E}_{\iota-1}=\emptyset$,
$\varepsilon_1\gleq \varepsilon_2=(\upsilon_1\mult a_\iota^\alpha)\gleq \varepsilon_2\in \mi{clo}$;
by (\ref{eq8i}) for $\gleq$ and (\ref{eq8i}c), there exist $\kappa_1\leq \kappa_2<\kappa_3\leq \kappa_4$,
\begin{alignat*}{1}
& \mu_2^j\gleq (\mu_1^j\mult a_\iota^{\beta_j})=\mi{reduced}(\lambda^{(\gamma_1^j,\dots,\gamma_n^j)})\in \mi{min}(\mi{DE}_{\iota-1}), j=1,\dots,\kappa_1, \\ 
& \mu_2^j\gle (\mu_1^j\mult a_\iota^{\beta_j})=\mi{reduced}(\lambda^{(\gamma_1^j,\dots,\gamma_n^j)})\in \mi{min}(D_{\iota-1}), j=\kappa_1+1,\dots,\kappa_2, \\
& (\mu_1^j\mult a_\iota^{\beta_j})\gleq \mu_2^j=\mi{reduced}(\lambda^{(\gamma_1^j,\dots,\gamma_n^j)})\in \mi{min}(\mi{UE}_{\iota-1}), j=\kappa_2+1,\dots,\kappa_3, \\
& (\mu_1^j\mult a_\iota^{\beta_j})\gle \mu_2^j=\mi{reduced}(\lambda^{(\gamma_1^j,\dots,\gamma_n^j)})\in \mi{min}(U_{\iota-1}), j=\kappa_3+1,\dots,\kappa_4, \\ 
& \mu_e^j\in \{\gu\}\cup \mi{PropConj}_A, \mi{atoms}(\mu_e^j)\subseteq \mi{dom}({\mc V}_{\iota-1}), \beta_j\geq 1, \bs{0}^n\neq (\gamma_1^j,\dots,\gamma_n^j)\in \mbb{N}^n, 
\end{alignat*}
satisfying either $\varepsilon_1\gleq \varepsilon_2=(\upsilon_1\mult a_\iota^\alpha)\gleq \varepsilon_2=
                                                    \mi{reduced}(\lambda^{(\sum_{j=1}^{\kappa_4} \gamma_1^j,\dots,\sum_{j=1}^{\kappa_4} \gamma_n^j)})$, 
or there exists $\zeta_1\diamond^\zeta \zeta_2=\mi{reduced}(\lambda^{(\gamma_1^\zeta,\dots,\gamma_n^\zeta)})\in \mi{clo}$,
$\zeta_e\in \{\gu\}\cup \mi{PropConj}_A$, $\mi{atoms}(\zeta_e)\subseteq \mi{dom}({\mc V}_{\iota-1})$, $\diamond^\zeta\in \{\geql,\gleq,\gle\}$, 
$\bs{0}^n\neq (\gamma_1^\zeta,\dots,\gamma_n^\zeta)\in \mbb{N}^n$, satisfying
$\varepsilon_1\gleq \varepsilon_2=(\upsilon_1\mult a_\iota^\alpha)\gleq \varepsilon_2=
                                  \mi{reduced}(\lambda^{((\sum_{j=1}^{\kappa_4} \gamma_1^j)+\gamma_1^\zeta,\dots,(\sum_{j=1}^{\kappa_4} \gamma_n^j)+\gamma_n^\zeta)})$.
We get two cases for $(\upsilon_1\mult a_\iota^\alpha)\gleq \varepsilon_2$.

Case 2.10.2.1.1.1:
$(\upsilon_1\mult a_\iota^\alpha)\gleq \varepsilon_2=\mi{reduced}(\lambda^{(\sum_{j=1}^{\kappa_4} \gamma_1^j,\dots,\sum_{j=1}^{\kappa_4} \gamma_n^j)})$.
Then 
$\mi{atoms}(\upsilon_1), \mi{atoms}(\varepsilon_2), \mi{atoms}(\mu_1^1),\dots,\mi{atoms}(\mu_1^{\kappa_4}), \mi{atoms}(\mu_2^1),\dots,\mi{atoms}(\mu_2^{\kappa_4})\subseteq \mi{dom}({\mc V}_{\iota-1})$,
\begin{alignat*}{1}
& (\upsilon_1\mult a_\iota^\alpha)\gleq \varepsilon_2=
  \mi{reduced}(\lambda^{(\sum_{j=1}^{\kappa_4} \gamma_1^j,\dots,\sum_{j=1}^{\kappa_4} \gamma_n^j)})\overset{\text{(\ref{eq8dddd})}}{=\!\!=} \\
& \mi{reduced}((\bigmult_{j=1}^{\kappa_1} \lambda^{(\gamma_1^j,\dots,\gamma_n^j)})\mult
               (\bigmult_{j=\kappa_1+1}^{\kappa_2} \lambda^{(\gamma_1^j,\dots,\gamma_n^j)})\mult \\
&\phantom{\mi{reduced}(}
               (\bigmult_{j=\kappa_2+1}^{\kappa_3} \lambda^{(\gamma_1^j,\dots,\gamma_n^j)})\mult
               \bigmult_{j=\kappa_3+1}^{\kappa_4} \lambda^{(\gamma_1^j,\dots,\gamma_n^j)})\overset{\text{(\ref{eq7k})}}{=\!\!=} \\
& \mi{reduced}((\bigmult_{j=1}^{\kappa_1} \mi{reduced}(\lambda^{(\gamma_1^j,\dots,\gamma_n^j)}))\mult
               (\bigmult_{j=\kappa_1+1}^{\kappa_2} \mi{reduced}(\lambda^{(\gamma_1^j,\dots,\gamma_n^j)}))\mult \\
&\phantom{\mi{reduced}(}
               (\bigmult_{j=\kappa_2+1}^{\kappa_3} \mi{reduced}(\lambda^{(\gamma_1^j,\dots,\gamma_n^j)}))\mult
               \bigmult_{j=\kappa_3+1}^{\kappa_4} \mi{reduced}(\lambda^{(\gamma_1^j,\dots,\gamma_n^j)}))= \\
& \mi{reduced}((\bigmult_{j=1}^{\kappa_1} \mu_2^j\gleq (\mu_1^j\mult a_\iota^{\beta_j}))\mult
               (\bigmult_{j=\kappa_1+1}^{\kappa_2} \mu_2^j\gle (\mu_1^j\mult a_\iota^{\beta_j}))\mult \\
&\phantom{\mi{reduced}(}
               (\bigmult_{j=\kappa_2+1}^{\kappa_3} (\mu_1^j\mult a_\iota^{\beta_j})\gleq \mu_2^j)\mult
               \bigmult_{j=\kappa_3+1}^{\kappa_4} (\mu_1^j\mult a_\iota^{\beta_j})\gle \mu_2^j)\overset{\text{(\ref{eq7i})}}{=\!\!=} \\
& \mi{reduced}(((\bigmult_{j=1}^{\kappa_1} \mu_2^j)\mult (\bigmult_{j=\kappa_1+1}^{\kappa_2} \mu_2^j)\mult 
                (\bigmult_{j=\kappa_2+1}^{\kappa_3} \mu_1^j)\mult (\bigmult_{j=\kappa_3+1}^{\kappa_4} \mu_1^j)\mult \\
&\phantom{\mi{reduced}(} \quad
                a_\iota^{(\sum_{j=\kappa_2+1}^{\kappa_3} \beta_j)+\sum_{j=\kappa_3+1}^{\kappa_4} \beta_j})\gleq \\
&\phantom{\mi{reduced}(} \quad
               ((\bigmult_{j=1}^{\kappa_1} \mu_1^j)\mult (\bigmult_{j=\kappa_1+1}^{\kappa_2} \mu_1^j)\mult 
                (\bigmult_{j=\kappa_2+1}^{\kappa_3} \mu_2^j)\mult (\bigmult_{j=\kappa_3+1}^{\kappa_4} \mu_2^j)\mult \\
&\phantom{\mi{reduced}(} \quad \quad
                a_\iota^{(\sum_{j=1}^{\kappa_1} \beta_j)+\sum_{j=\kappa_1+1}^{\kappa_2} \beta_j}))= \\
& \mi{reduced}^-(((\bigmult_{j=1}^{\kappa_1} \mu_2^j)\mult (\bigmult_{j=\kappa_1+1}^{\kappa_2} \mu_2^j)\mult 
                  (\bigmult_{j=\kappa_2+1}^{\kappa_3} \mu_1^j)\mult (\bigmult_{j=\kappa_3+1}^{\kappa_4} \mu_1^j)\mult \\
&\phantom{\mi{reduced}^-(} \quad
                  a_\iota^{(\sum_{j=\kappa_2+1}^{\kappa_3} \beta_j)+\sum_{j=\kappa_3+1}^{\kappa_4} \beta_j})\gleq \\
&\phantom{\mi{reduced}^-(} \quad
                 ((\bigmult_{j=1}^{\kappa_1} \mu_1^j)\mult (\bigmult_{j=\kappa_1+1}^{\kappa_2} \mu_1^j)\mult 
                  (\bigmult_{j=\kappa_2+1}^{\kappa_3} \mu_2^j)\mult (\bigmult_{j=\kappa_3+1}^{\kappa_4} \mu_2^j)\mult \\
&\phantom{\mi{reduced}^-(} \quad \quad
                  a_\iota^{(\sum_{j=1}^{\kappa_1} \beta_j)+\sum_{j=\kappa_1+1}^{\kappa_2} \beta_j}))\gleq \\
& \mi{reduced}^+(((\bigmult_{j=1}^{\kappa_1} \mu_2^j)\mult (\bigmult_{j=\kappa_1+1}^{\kappa_2} \mu_2^j)\mult 
                  (\bigmult_{j=\kappa_2+1}^{\kappa_3} \mu_1^j)\mult (\bigmult_{j=\kappa_3+1}^{\kappa_4} \mu_1^j)\mult \\
&\phantom{\mi{reduced}^+(} \quad
                  a_\iota^{(\sum_{j=\kappa_2+1}^{\kappa_3} \beta_j)+\sum_{j=\kappa_3+1}^{\kappa_4} \beta_j})\gleq \\
&\phantom{\mi{reduced}^+(} \quad
                 ((\bigmult_{j=1}^{\kappa_1} \mu_1^j)\mult (\bigmult_{j=\kappa_1+1}^{\kappa_2} \mu_1^j)\mult 
                  (\bigmult_{j=\kappa_2+1}^{\kappa_3} \mu_2^j)\mult (\bigmult_{j=\kappa_3+1}^{\kappa_4} \mu_2^j)\mult \\
&\phantom{\mi{reduced}^+(} \quad \quad
                  a_\iota^{(\sum_{j=1}^{\kappa_1} \beta_j)+\sum_{j=\kappa_1+1}^{\kappa_2} \beta_j})), \\[1mm]
& \#(a_\iota,(\upsilon_1\mult a_\iota^\alpha)\gleq \varepsilon_2)=-\alpha= \\
& \#(a_\iota,\mi{reduced}(((\bigmult_{j=1}^{\kappa_1} \mu_2^j)\mult (\bigmult_{j=\kappa_1+1}^{\kappa_2} \mu_2^j)\mult 
                           (\bigmult_{j=\kappa_2+1}^{\kappa_3} \mu_1^j)\mult (\bigmult_{j=\kappa_3+1}^{\kappa_4} \mu_1^j)\mult \\
&\phantom{\#(a_\iota,\mi{reduced}(} \quad
                           a_\iota^{(\sum_{j=\kappa_2+1}^{\kappa_3} \beta_j)+\sum_{j=\kappa_3+1}^{\kappa_4} \beta_j})\gleq \\
&\phantom{\#(a_\iota,\mi{reduced}(} \quad
                          ((\bigmult_{j=1}^{\kappa_1} \mu_1^j)\mult (\bigmult_{j=\kappa_1+1}^{\kappa_2} \mu_1^j)\mult 
                           (\bigmult_{j=\kappa_2+1}^{\kappa_3} \mu_2^j)\mult (\bigmult_{j=\kappa_3+1}^{\kappa_4} \mu_2^j)\mult \\
&\phantom{\#(a_\iota,\mi{reduced}(} \quad \quad
                           a_\iota^{(\sum_{j=1}^{\kappa_1} \beta_j)+\sum_{j=\kappa_1+1}^{\kappa_2} \beta_j})))\overset{\text{(\ref{eq7aaaax})}}{=\!\!=} \\
& \#(a_\iota,((\bigmult_{j=1}^{\kappa_1} \mu_2^j)\mult (\bigmult_{j=\kappa_1+1}^{\kappa_2} \mu_2^j)\mult 
              (\bigmult_{j=\kappa_2+1}^{\kappa_3} \mu_1^j)\mult (\bigmult_{j=\kappa_3+1}^{\kappa_4} \mu_1^j)\mult \\
&\phantom{\#(a_\iota,} \quad
              a_\iota^{(\sum_{j=\kappa_2+1}^{\kappa_3} \beta_j)+\sum_{j=\kappa_3+1}^{\kappa_4} \beta_j})\gleq \\
&\phantom{\#(a_\iota,} \quad
             ((\bigmult_{j=1}^{\kappa_1} \mu_1^j)\mult (\bigmult_{j=\kappa_1+1}^{\kappa_2} \mu_1^j)\mult 
              (\bigmult_{j=\kappa_2+1}^{\kappa_3} \mu_2^j)\mult (\bigmult_{j=\kappa_3+1}^{\kappa_4} \mu_2^j)\mult \\
&\phantom{\#(a_\iota,} \quad \quad
              a_\iota^{(\sum_{j=1}^{\kappa_1} \beta_j)+\sum_{j=\kappa_1+1}^{\kappa_2} \beta_j}))= \\
& \left(\sum_{j=1}^{\kappa_1} \beta_j\right)+\left(\sum_{j=\kappa_1+1}^{\kappa_2} \beta_j\right)-\left(\sum_{j=\kappa_2+1}^{\kappa_3} \beta_j\right)-\sum_{j=\kappa_3+1}^{\kappa_4} \beta_j\leq -1, 
\end{alignat*}
$a_\iota\not\in \mi{dom}({\mc V}_{\iota-1})\supseteq \mi{atoms}((\bigmult_{j=1}^{\kappa_1} \mu_2^j)\mult (\bigmult_{j=\kappa_1+1}^{\kappa_2} \mu_2^j)\mult 
                                                                (\bigmult_{j=\kappa_2+1}^{\kappa_3} \mu_1^j)\mult \bigmult_{j=\kappa_3+1}^{\kappa_4} \mu_1^j)=
                                                     (\bigcup_{j=1}^{\kappa_1} \mi{atoms}(\mu_2^j))\cup (\bigcup_{j=\kappa_1+1}^{\kappa_2} \mi{atoms}(\mu_2^j))\cup 
                                                     (\bigcup_{j=\kappa_2+1}^{\kappa_3} \mi{atoms}(\mu_1^j))\cup \bigcup_{j=\kappa_3+1}^{\kappa_4} \mi{atoms}(\mu_1^j)$,
$a_\iota\not\in \mi{dom}({\mc V}_{\iota-1})\supseteq \mi{atoms}((\bigmult_{j=1}^{\kappa_1} \mu_1^j)\mult (\bigmult_{j=\kappa_1+1}^{\kappa_2} \mu_1^j)\mult 
                                                                (\bigmult_{j=\kappa_2+1}^{\kappa_3} \mu_2^j)\mult \bigmult_{j=\kappa_3+1}^{\kappa_4} \mu_2^j)=
                                                     (\bigcup_{j=1}^{\kappa_1} \mi{atoms}(\mu_1^j))\cup (\bigcup_{j=\kappa_1+1}^{\kappa_2} \mi{atoms}(\mu_1^j))\cup 
                                                     (\bigcup_{j=\kappa_2+1}^{\kappa_3} \mi{atoms}(\mu_2^j))\cup \bigcup_{j=\kappa_3+1}^{\kappa_4} \mi{atoms}(\mu_2^j)$,
\begin{alignat*}{1}
& \mi{reduced}^-(((\bigmult_{j=1}^{\kappa_1} \mu_2^j)\mult (\bigmult_{j=\kappa_1+1}^{\kappa_2} \mu_2^j)\mult 
                  (\bigmult_{j=\kappa_2+1}^{\kappa_3} \mu_1^j)\mult (\bigmult_{j=\kappa_3+1}^{\kappa_4} \mu_1^j)\mult \\
&\phantom{\mi{reduced}^-(} \quad
                  a_\iota^{(\sum_{j=\kappa_2+1}^{\kappa_3} \beta_j)+\sum_{j=\kappa_3+1}^{\kappa_4} \beta_j})\gleq \\
&\phantom{\mi{reduced}^-(} \quad
                 ((\bigmult_{j=1}^{\kappa_1} \mu_1^j)\mult (\bigmult_{j=\kappa_1+1}^{\kappa_2} \mu_1^j)\mult 
                  (\bigmult_{j=\kappa_2+1}^{\kappa_3} \mu_2^j)\mult (\bigmult_{j=\kappa_3+1}^{\kappa_4} \mu_2^j)\mult \\
&\phantom{\mi{reduced}^-(} \quad \quad
                  a_\iota^{(\sum_{j=1}^{\kappa_1} \beta_j)+\sum_{j=\kappa_1+1}^{\kappa_2} \beta_j}))\overset{\text{(\ref{eq7ee})}}{=\!\!=} \\
& \mi{reduced}^-(((\bigmult_{j=1}^{\kappa_1} \mu_2^j)\mult (\bigmult_{j=\kappa_1+1}^{\kappa_2} \mu_2^j)\mult 
                  (\bigmult_{j=\kappa_2+1}^{\kappa_3} \mu_1^j)\mult \bigmult_{j=\kappa_3+1}^{\kappa_4} \mu_1^j)\gleq \\
&\phantom{\mi{reduced}^-(} \quad
                 ((\bigmult_{j=1}^{\kappa_1} \mu_1^j)\mult (\bigmult_{j=\kappa_1+1}^{\kappa_2} \mu_1^j)\mult 
                  (\bigmult_{j=\kappa_2+1}^{\kappa_3} \mu_2^j)\mult \bigmult_{j=\kappa_3+1}^{\kappa_4} \mu_2^j))\mult \\
& a_\iota^{(\sum_{j=\kappa_2+1}^{\kappa_3} \beta_j)+(\sum_{j=\kappa_3+1}^{\kappa_4} \beta_j)-(\sum_{j=1}^{\kappa_1} \beta_j)-\sum_{j=\kappa_1+1}^{\kappa_2} \beta_j}= \\
& \mi{reduced}^-(((\bigmult_{j=1}^{\kappa_1} \mu_2^j)\mult (\bigmult_{j=\kappa_1+1}^{\kappa_2} \mu_2^j)\mult 
                  (\bigmult_{j=\kappa_2+1}^{\kappa_3} \mu_1^j)\mult \bigmult_{j=\kappa_3+1}^{\kappa_4} \mu_1^j)\gleq \\
&\phantom{\mi{reduced}^-(} \quad
                 ((\bigmult_{j=1}^{\kappa_1} \mu_1^j)\mult (\bigmult_{j=\kappa_1+1}^{\kappa_2} \mu_1^j)\mult 
                  (\bigmult_{j=\kappa_2+1}^{\kappa_3} \mu_2^j)\mult \bigmult_{j=\kappa_3+1}^{\kappa_4} \mu_2^j))\mult a_\iota^\alpha, \\[1mm]
& \mi{reduced}^+(((\bigmult_{j=1}^{\kappa_1} \mu_2^j)\mult (\bigmult_{j=\kappa_1+1}^{\kappa_2} \mu_2^j)\mult 
                  (\bigmult_{j=\kappa_2+1}^{\kappa_3} \mu_1^j)\mult (\bigmult_{j=\kappa_3+1}^{\kappa_4} \mu_1^j)\mult \\
&\phantom{\mi{reduced}^+(} \quad
                  a_\iota^{(\sum_{j=\kappa_2+1}^{\kappa_3} \beta_j)+\sum_{j=\kappa_3+1}^{\kappa_4} \beta_j})\gleq \\
&\phantom{\mi{reduced}^+(} \quad
                 ((\bigmult_{j=1}^{\kappa_1} \mu_1^j)\mult (\bigmult_{j=\kappa_1+1}^{\kappa_2} \mu_1^j)\mult 
                  (\bigmult_{j=\kappa_2+1}^{\kappa_3} \mu_2^j)\mult (\bigmult_{j=\kappa_3+1}^{\kappa_4} \mu_2^j)\mult \\
&\phantom{\mi{reduced}^+(} \quad \quad
                  a_\iota^{(\sum_{j=1}^{\kappa_1} \beta_j)+\sum_{j=\kappa_1+1}^{\kappa_2} \beta_j}))\overset{\text{(\ref{eq7dd})}}{=\!\!=} \\
& \mi{reduced}^+(((\bigmult_{j=1}^{\kappa_1} \mu_2^j)\mult (\bigmult_{j=\kappa_1+1}^{\kappa_2} \mu_2^j)\mult 
                  (\bigmult_{j=\kappa_2+1}^{\kappa_3} \mu_1^j)\mult \bigmult_{j=\kappa_3+1}^{\kappa_4} \mu_1^j)\gleq \\
&\phantom{\mi{reduced}^+(} \quad
                 ((\bigmult_{j=1}^{\kappa_1} \mu_1^j)\mult (\bigmult_{j=\kappa_1+1}^{\kappa_2} \mu_1^j)\mult 
                  (\bigmult_{j=\kappa_2+1}^{\kappa_3} \mu_2^j)\mult \bigmult_{j=\kappa_3+1}^{\kappa_4} \mu_2^j)), \\[1mm]
& (\upsilon_1\mult a_\iota^\alpha)\gleq \varepsilon_2= \\
& (\mi{reduced}^-(((\bigmult_{j=1}^{\kappa_1} \mu_2^j)\mult (\bigmult_{j=\kappa_1+1}^{\kappa_2} \mu_2^j)\mult 
                   (\bigmult_{j=\kappa_2+1}^{\kappa_3} \mu_1^j)\mult \bigmult_{j=\kappa_3+1}^{\kappa_4} \mu_1^j)\gleq \\
&\phantom{(\mi{reduced}^-(} \quad
                  ((\bigmult_{j=1}^{\kappa_1} \mu_1^j)\mult (\bigmult_{j=\kappa_1+1}^{\kappa_2} \mu_1^j)\mult 
                   (\bigmult_{j=\kappa_2+1}^{\kappa_3} \mu_2^j)\mult \bigmult_{j=\kappa_3+1}^{\kappa_4} \mu_2^j))\mult a_\iota^\alpha)\gleq \\
& \mi{reduced}^+(((\bigmult_{j=1}^{\kappa_1} \mu_2^j)\mult (\bigmult_{j=\kappa_1+1}^{\kappa_2} \mu_2^j)\mult 
                  (\bigmult_{j=\kappa_2+1}^{\kappa_3} \mu_1^j)\mult \bigmult_{j=\kappa_3+1}^{\kappa_4} \mu_1^j)\gleq \\
&\phantom{\mi{reduced}^+(} \quad
                 ((\bigmult_{j=1}^{\kappa_1} \mu_1^j)\mult (\bigmult_{j=\kappa_1+1}^{\kappa_2} \mu_1^j)\mult 
                  (\bigmult_{j=\kappa_2+1}^{\kappa_3} \mu_2^j)\mult \bigmult_{j=\kappa_3+1}^{\kappa_4} \mu_2^j)), \\[1mm]
& \mi{atoms}(\mi{reduced}^-(((\bigmult_{j=1}^{\kappa_1} \mu_2^j)\mult (\bigmult_{j=\kappa_1+1}^{\kappa_2} \mu_2^j)\mult 
                             (\bigmult_{j=\kappa_2+1}^{\kappa_3} \mu_1^j)\mult \bigmult_{j=\kappa_3+1}^{\kappa_4} \mu_1^j)\gleq \\
&\phantom{\mi{atoms}(\mi{reduced}^-(} \quad
                            ((\bigmult_{j=1}^{\kappa_1} \mu_1^j)\mult (\bigmult_{j=\kappa_1+1}^{\kappa_2} \mu_1^j)\mult 
                             (\bigmult_{j=\kappa_2+1}^{\kappa_3} \mu_2^j)\mult \bigmult_{j=\kappa_3+1}^{\kappa_4} \mu_2^j)))\underset{\text{(\ref{eq7a})}}{\subseteq} \\
& \mi{atoms}((\bigmult_{j=1}^{\kappa_1} \mu_2^j)\mult (\bigmult_{j=\kappa_1+1}^{\kappa_2} \mu_2^j)\mult 
             (\bigmult_{j=\kappa_2+1}^{\kappa_3} \mu_1^j)\mult \bigmult_{j=\kappa_3+1}^{\kappa_4} \mu_1^j)\subseteq \mi{dom}({\mc V}_{\iota-1}), \\[1mm]
& \mi{atoms}(\mi{reduced}^+(((\bigmult_{j=1}^{\kappa_1} \mu_2^j)\mult (\bigmult_{j=\kappa_1+1}^{\kappa_2} \mu_2^j)\mult 
                             (\bigmult_{j=\kappa_2+1}^{\kappa_3} \mu_1^j)\mult \bigmult_{j=\kappa_3+1}^{\kappa_4} \mu_1^j)\gleq \\
&\phantom{\mi{atoms}(\mi{reduced}^+(} \quad
                            ((\bigmult_{j=1}^{\kappa_1} \mu_1^j)\mult (\bigmult_{j=\kappa_1+1}^{\kappa_2} \mu_1^j)\mult 
                             (\bigmult_{j=\kappa_2+1}^{\kappa_3} \mu_2^j)\mult \bigmult_{j=\kappa_3+1}^{\kappa_4} \mu_2^j)))\underset{\text{(\ref{eq7b})}}{\subseteq} \\
& \mi{atoms}((\bigmult_{j=1}^{\kappa_1} \mu_1^j)\mult (\bigmult_{j=\kappa_1+1}^{\kappa_2} \mu_1^j)\mult 
             (\bigmult_{j=\kappa_2+1}^{\kappa_3} \mu_2^j)\mult \bigmult_{j=\kappa_3+1}^{\kappa_4} \mu_2^j)\subseteq \mi{dom}({\mc V}_{\iota-1}), 
\end{alignat*}
\begin{alignat*}{1}
& \upsilon_1=\mi{reduced}^-(((\bigmult_{j=1}^{\kappa_1} \mu_2^j)\mult (\bigmult_{j=\kappa_1+1}^{\kappa_2} \mu_2^j)\mult 
                             (\bigmult_{j=\kappa_2+1}^{\kappa_3} \mu_1^j)\mult \bigmult_{j=\kappa_3+1}^{\kappa_4} \mu_1^j)\gleq \\
&\phantom{\upsilon_1=\mi{reduced}^-(} \quad
                            ((\bigmult_{j=1}^{\kappa_1} \mu_1^j)\mult (\bigmult_{j=\kappa_1+1}^{\kappa_2} \mu_1^j)\mult 
                             (\bigmult_{j=\kappa_2+1}^{\kappa_3} \mu_2^j)\mult \bigmult_{j=\kappa_3+1}^{\kappa_4} \mu_2^j)), \\[1mm]
& \varepsilon_2=\mi{reduced}^+(((\bigmult_{j=1}^{\kappa_1} \mu_2^j)\mult (\bigmult_{j=\kappa_1+1}^{\kappa_2} \mu_2^j)\mult 
                                (\bigmult_{j=\kappa_2+1}^{\kappa_3} \mu_1^j)\mult \bigmult_{j=\kappa_3+1}^{\kappa_4} \mu_1^j)\gleq \\
&\phantom{\varepsilon_2=\mi{reduced}^+(} \quad
                               ((\bigmult_{j=1}^{\kappa_1} \mu_1^j)\mult (\bigmult_{j=\kappa_1+1}^{\kappa_2} \mu_1^j)\mult 
                                (\bigmult_{j=\kappa_2+1}^{\kappa_3} \mu_2^j)\mult \bigmult_{j=\kappa_3+1}^{\kappa_4} \mu_2^j));
\end{alignat*}
by the induction hypothesis for $\iota-1<\iota$,
$\|\upsilon_1\|^{{\mc V}_{\iota-1}}, \|\varepsilon_2\|^{{\mc V}_{\iota-1}},                                                                                                                
 \|\mu_1^1\|^{{\mc V}_{\iota-1}},\dots,                                                                                                                                                    \linebreak[4]
                                       \|\mu_1^{\kappa_4}\|^{{\mc V}_{\iota-1}}, \|\mu_2^1\|^{{\mc V}_{\iota-1}},\dots,\|\mu_2^{\kappa_4}\|^{{\mc V}_{\iota-1}},                           
 \|\mi{reduced}^-(((\bigmult_{j=1}^{\kappa_1} \mu_2^j)\mult (\bigmult_{j=\kappa_1+1}^{\kappa_2} \mu_2^j)\mult 
                   (\bigmult_{j=\kappa_2+1}^{\kappa_3} \mu_1^j)\mult \bigmult_{j=\kappa_3+1}^{\kappa_4} \mu_1^j)\gleq
                  ((\bigmult_{j=1}^{\kappa_1} \mu_1^j)\mult (\bigmult_{j=\kappa_1+1}^{\kappa_2} \mu_1^j)\mult 
                   (\bigmult_{j=\kappa_2+1}^{\kappa_3} \mu_2^j)\mult \bigmult_{j=\kappa_3+1}^{\kappa_4} \mu_2^j))\|^{{\mc V}_{\iota-1}}, 
 \|\mi{reduced}^+(((\bigmult_{j=1}^{\kappa_1} \mu_2^j)\mult (\bigmult_{j=\kappa_1+1}^{\kappa_2} \mu_2^j)\mult 
                   (\bigmult_{j=\kappa_2+1}^{\kappa_3} \mu_1^j)\mult \bigmult_{j=\kappa_3+1}^{\kappa_4} \mu_1^j)\gleq
                  ((\bigmult_{j=1}^{\kappa_1} \mu_1^j)\mult (\bigmult_{j=\kappa_1+1}^{\kappa_2} \mu_1^j)\mult 
                   (\bigmult_{j=\kappa_2+1}^{\kappa_3} \mu_2^j)\mult \bigmult_{j=\kappa_3+1}^{\kappa_4} \mu_2^j))\|^{{\mc V}_{\iota-1}}\underset{\text{(a)}}{\in} (0,1]$;
$\mbb{E}_{\iota-1}=\emptyset$;
for all $1\leq j\leq \kappa_1$, 
$\mu_2^j\gleq (\mu_1^j\mult a_\iota^{\beta_j})\in \mi{min}(\mi{DE}_{\iota-1})$, 
$\left(\frac{\|\mu_2^j\|^{{\mc V}_{\iota-1}}}
            {\|\mu_1^j\|^{{\mc V}_{\iota-1}}}\right)^{\frac{1}{\beta_j}}\in \mbb{DE}_{\iota-1}$,
$\left(\frac{\|\mu_2^j\|^{{\mc V}_{\iota-1}}}
            {\|\mu_1^j\|^{{\mc V}_{\iota-1}}}\right)^{\frac{1}{\beta_j}}\leq \bigfvee \mbb{DE}_{\iota-1}\underset{\text{(\ref{eq8j}a)}}{\leq} {\mc V}_\iota(a_\iota)=\delta_\iota$,
$\frac{\|\mu_2^j\|^{{\mc V}_{\iota-1}}}
      {\|\mu_1^j\|^{{\mc V}_{\iota-1}}}\leq {\mc V}_\iota(a_\iota)^{\beta_j}$;
for all $\kappa_1+1\leq j\leq \kappa_2$, 
$\mu_2^j\gle (\mu_1^j\mult a_\iota^{\beta_j})\in \mi{min}(D_{\iota-1})$, 
$\left(\frac{\|\mu_2^j\|^{{\mc V}_{\iota-1}}}
            {\|\mu_1^j\|^{{\mc V}_{\iota-1}}}\right)^{\frac{1}{\beta_j}}\in \mbb{D}_{\iota-1}$,
$\left(\frac{\|\mu_2^j\|^{{\mc V}_{\iota-1}}}
            {\|\mu_1^j\|^{{\mc V}_{\iota-1}}}\right)^{\frac{1}{\beta_j}}\leq \bigfvee \mbb{D}_{\iota-1}\underset{\text{(\ref{eq8j}b)}}{<} {\mc V}_\iota(a_\iota)=\delta_\iota$,
$\frac{\|\mu_2^j\|^{{\mc V}_{\iota-1}}}
      {\|\mu_1^j\|^{{\mc V}_{\iota-1}}}<{\mc V}_\iota(a_\iota)^{\beta_j}$;
for all $\kappa_2+1\leq j\leq \kappa_3$,
$(\mu_1^j\mult a_\iota^{\beta_j})\gleq \mu_2^j\in \mi{min}(\mi{UE}_{\iota-1})$,
$\mi{min}\left(\left(\frac{\|\mu_2^j\|^{{\mc V}_{\iota-1}}}
                          {\|\mu_1^j\|^{{\mc V}_{\iota-1}}}\right)^{\frac{1}{\beta_j}},1\right)\in \mbb{UE}_{\iota-1}$,
${\mc V}_\iota(a_\iota)=\delta_\iota\underset{\text{(\ref{eq8j}a)}}{\leq} \bigfwedge \mbb{UE}_{\iota-1}\leq 
 \mi{min}\left(\left(\frac{\|\mu_2^j\|^{{\mc V}_{\iota-1}}}
                          {\|\mu_1^j\|^{{\mc V}_{\iota-1}}}\right)^{\frac{1}{\beta_j}},1\right)\leq
 \left(\frac{\|\mu_2^j\|^{{\mc V}_{\iota-1}}}
            {\|\mu_1^j\|^{{\mc V}_{\iota-1}}}\right)^{\frac{1}{\beta_j}}$,
${\mc V}_\iota(a_\iota)^{\beta_j}\leq \frac{\|\mu_2^j\|^{{\mc V}_{\iota-1}}}
                                           {\|\mu_1^j\|^{{\mc V}_{\iota-1}}}$,
$\frac{\|\mu_1^j\|^{{\mc V}_{\iota-1}}}
      {\|\mu_2^j\|^{{\mc V}_{\iota-1}}}\leq \frac{1}
                                                 {{\mc V}_\iota(a_\iota)^{\beta_j}}$;
for all $\kappa_3+1\leq j\leq \kappa_4$,
$(\mu_1^j\mult a_\iota^{\beta_j})\gle \mu_2^j\in \mi{min}(U_{\iota-1})$,
$\mi{min}\left(\left(\frac{\|\mu_2^j\|^{{\mc V}_{\iota-1}}}
                          {\|\mu_1^j\|^{{\mc V}_{\iota-1}}}\right)^{\frac{1}{\beta_j}},1\right)\in \mbb{U}_{\iota-1}$,
${\mc V}_\iota(a_\iota)=\delta_\iota\underset{\text{(\ref{eq8j}b)}}{\leq} \bigfwedge \mbb{U}_{\iota-1}\leq 
 \mi{min}\left(\left(\frac{\|\mu_2^j\|^{{\mc V}_{\iota-1}}}
                          {\|\mu_1^j\|^{{\mc V}_{\iota-1}}}\right)^{\frac{1}{\beta_j}},1\right)\leq
 \left(\frac{\|\mu_2^j\|^{{\mc V}_{\iota-1}}}
            {\|\mu_1^j\|^{{\mc V}_{\iota-1}}}\right)^{\frac{1}{\beta_j}}$,
${\mc V}_\iota(a_\iota)^{\beta_j}\leq \frac{\|\mu_2^j\|^{{\mc V}_{\iota-1}}}
                                           {\|\mu_1^j\|^{{\mc V}_{\iota-1}}}$,
$\frac{\|\mu_1^j\|^{{\mc V}_{\iota-1}}}
      {\|\mu_2^j\|^{{\mc V}_{\iota-1}}}\leq \frac{1}
                                                 {{\mc V}_\iota(a_\iota)^{\beta_j}}$;
\begin{alignat*}{1}
& \dfrac{\|\upsilon_1\|^{{\mc V}_{\iota-1}}}
        {\|\varepsilon_2\|^{{\mc V}_{\iota-1}}}= \\
& \dfrac{\left\|\mi{reduced}^-\left(\begin{array}{l}
                                    ((\bigmult_{j=1}^{\kappa_1} \mu_2^j)\mult (\bigmult_{j=\kappa_1+1}^{\kappa_2} \mu_2^j)\mult 
                                     (\bigmult_{j=\kappa_2+1}^{\kappa_3} \mu_1^j)\mult \bigmult_{j=\kappa_3+1}^{\kappa_4} \mu_1^j)\gleq \\
                                    \quad
                                    ((\bigmult_{j=1}^{\kappa_1} \mu_1^j)\mult (\bigmult_{j=\kappa_1+1}^{\kappa_2} \mu_1^j)\mult 
                                     (\bigmult_{j=\kappa_2+1}^{\kappa_3} \mu_2^j)\mult \bigmult_{j=\kappa_3+1}^{\kappa_4} \mu_2^j)
                                    \end{array}\right)\right\|^{{\mc V}_{\iota-1}}}
        {\left\|\mi{reduced}^+\left(\begin{array}{l}
                                    ((\bigmult_{j=1}^{\kappa_1} \mu_2^j)\mult (\bigmult_{j=\kappa_1+1}^{\kappa_2} \mu_2^j)\mult 
                                     (\bigmult_{j=\kappa_2+1}^{\kappa_3} \mu_1^j)\mult \bigmult_{j=\kappa_3+1}^{\kappa_4} \mu_1^j)\gleq \\
                                    \quad
                                    ((\bigmult_{j=1}^{\kappa_1} \mu_1^j)\mult (\bigmult_{j=\kappa_1+1}^{\kappa_2} \mu_1^j)\mult 
                                     (\bigmult_{j=\kappa_2+1}^{\kappa_3} \mu_2^j)\mult \bigmult_{j=\kappa_3+1}^{\kappa_4} \mu_2^j)
                                    \end{array}\right)\right\|^{{\mc V}_{\iota-1}}}\overset{\text{(\ref{eq7n})}}{=\!\!=} \\
& \dfrac{\|(\bigmult_{j=1}^{\kappa_1} \mu_2^j)\mult (\bigmult_{j=\kappa_1+1}^{\kappa_2} \mu_2^j)\mult 
           (\bigmult_{j=\kappa_2+1}^{\kappa_3} \mu_1^j)\mult \bigmult_{j=\kappa_3+1}^{\kappa_4} \mu_1^j\|^{{\mc V}_{\iota-1}}}
        {\|(\bigmult_{j=1}^{\kappa_1} \mu_1^j)\mult (\bigmult_{j=\kappa_1+1}^{\kappa_2} \mu_1^j)\mult 
           (\bigmult_{j=\kappa_2+1}^{\kappa_3} \mu_2^j)\mult \bigmult_{j=\kappa_3+1}^{\kappa_4} \mu_2^j\|^{{\mc V}_{\iota-1}}}= \\
& \dfrac{(\bigfswedge_{j=1}^{\kappa_1} \|\mu_2^j\|^{{\mc V}_{\iota-1}})\fswedge (\bigfswedge_{j=\kappa_1+1}^{\kappa_2} \|\mu_2^j\|^{{\mc V}_{\iota-1}})\fswedge
         (\bigfswedge_{j=\kappa_2+1}^{\kappa_3} \|\mu_1^j\|^{{\mc V}_{\iota-1}})\fswedge \bigfswedge_{j=\kappa_3+1}^{\kappa_4} \|\mu_1^j\|^{{\mc V}_{\iota-1}}}
        {(\bigfswedge_{j=1}^{\kappa_1} \|\mu_1^j\|^{{\mc V}_{\iota-1}})\fswedge (\bigfswedge_{j=\kappa_1+1}^{\kappa_2} \|\mu_1^j\|^{{\mc V}_{\iota-1}})\fswedge
         (\bigfswedge_{j=\kappa_2+1}^{\kappa_3} \|\mu_2^j\|^{{\mc V}_{\iota-1}})\fswedge \bigfswedge_{j=\kappa_3+1}^{\kappa_4} \|\mu_2^j\|^{{\mc V}_{\iota-1}}}= \\
& \left(\bigfswedge_{j=1}^{\kappa_1} \dfrac{\|\mu_2^j\|^{{\mc V}_{\iota-1}}}
                                           {\|\mu_1^j\|^{{\mc V}_{\iota-1}}}\right)\fswedge 
  \left(\bigfswedge_{j=\kappa_1+1}^{\kappa_2} \dfrac{\|\mu_2^j\|^{{\mc V}_{\iota-1}}}
                                                    {\|\mu_1^j\|^{{\mc V}_{\iota-1}}}\right)\fswedge 
  \left(\bigfswedge_{j=\kappa_1+2}^{\kappa_3} \dfrac{\|\mu_1^j\|^{{\mc V}_{\iota-1}}}
                                                    {\|\mu_2^j\|^{{\mc V}_{\iota-1}}}\right)\fswedge
  \bigfswedge_{j=\kappa_1+3}^{\kappa_4} \dfrac{\|\mu_1^j\|^{{\mc V}_{\iota-1}}}
                                              {\|\mu_2^j\|^{{\mc V}_{\iota-1}}}\leq \\
& \left(\bigfswedge_{j=1}^{\kappa_1} {\mc V}_\iota(a_\iota)^{\beta_j}\right)\fswedge \left(\bigfswedge_{j=\kappa_1+1}^{\kappa_2} {\mc V}_\iota(a_\iota)^{\beta_j}\right)\fswedge 
  \left(\bigfswedge_{j=\kappa_1+2}^{\kappa_3} \dfrac{1}
                                                    {{\mc V}_\iota(a_\iota)^{\beta_j}}\right)\fswedge
  \bigfswedge_{j=\kappa_1+3}^{\kappa_4} \dfrac{1}
                                              {{\mc V}_\iota(a_\iota)^{\beta_j}}= \\
& {\mc V}_\iota(a_\iota)^{(\sum_{j=1}^{\kappa_1} \beta_j)+(\sum_{j=\kappa_1+1}^{\kappa_2} \beta_j)-(\sum_{j=\kappa_2+1}^{\kappa_3} \beta_j)-\sum_{j=\kappa_3+1}^{\kappa_4} \beta_j}=
  {\mc V}_\iota(a_\iota)^{-\alpha}=\dfrac{1}
                                         {{\mc V}_\iota(a_\iota)^\alpha}, \\[1mm]
& \|\varepsilon_1\|^{{\mc V}_\iota}=\|\upsilon_1\mult a_\iota^\alpha\|^{{\mc V}_\iota}=\|\upsilon_1\|^{{\mc V}_\iota}\fswedge {\mc V}_\iota(a_\iota)^\alpha\overset{\text{(b)}}{=\!\!=}
  \|\upsilon_1\|^{{\mc V}_{\iota-1}}\fswedge {\mc V}_\iota(a_\iota)^\alpha\leq \|\varepsilon_2\|^{{\mc V}_{\iota-1}}\overset{\text{(b)}}{=\!\!=} \|\varepsilon_2\|^{{\mc V}_\iota};  
\end{alignat*}
(d) holds.

Case 2.10.2.1.1.2:
There exists $\zeta_1\diamond^\zeta \zeta_2=\mi{reduced}(\lambda^{(\gamma_1^\zeta,\dots,\gamma_n^\zeta)})\in \mi{clo}$,
$\zeta_e\in \{\gu\}\cup \mi{PropConj}_A$, $\mi{atoms}(\zeta_e)\subseteq \mi{dom}({\mc V}_{\iota-1})$, $\diamond^\zeta\in \{\geql,\gleq,\gle\}$, 
$\bs{0}^n\neq (\gamma_1^\zeta,\dots,\gamma_n^\zeta)\in \mbb{N}^n$, such that
$\varepsilon_1\gleq \varepsilon_2=(\upsilon_1\mult a_\iota^\alpha)\gleq \varepsilon_2=
                                  \mi{reduced}(\lambda^{((\sum_{j=1}^{\kappa_4} \gamma_1^j)+\gamma_1^\zeta,\dots,(\sum_{j=1}^{\kappa_4} \gamma_n^j)+\gamma_n^\zeta)})$.
Then 
$\mi{atoms}(\upsilon_1), \mi{atoms}(\varepsilon_2), \mi{atoms}(\mu_1^1),\dots,\mi{atoms}(\mu_1^{\kappa_4}), \mi{atoms}(\mu_2^1),\dots,\mi{atoms}(\mu_2^{\kappa_4})\subseteq \mi{dom}({\mc V}_{\iota-1})$,
\begin{alignat*}{1}
& (\upsilon_1\mult a_\iota^\alpha)\gleq \varepsilon_2=
  \mi{reduced}(\lambda^{((\sum_{j=1}^{\kappa_4} \gamma_1^j)+\gamma_1^\zeta,\dots,(\sum_{j=1}^{\kappa_4} \gamma_n^j)+\gamma_n^\zeta)})\overset{\text{(\ref{eq8dddd})}}{=\!\!=} \\
& \mi{reduced}((\bigmult_{j=1}^{\kappa_1} \lambda^{(\gamma_1^j,\dots,\gamma_n^j)})\mult
               (\bigmult_{j=\kappa_1+1}^{\kappa_2} \lambda^{(\gamma_1^j,\dots,\gamma_n^j)})\mult \\
&\phantom{\mi{reduced}(}
               (\bigmult_{j=\kappa_2+1}^{\kappa_3} \lambda^{(\gamma_1^j,\dots,\gamma_n^j)})\mult
               (\bigmult_{j=\kappa_3+1}^{\kappa_4} \lambda^{(\gamma_1^j,\dots,\gamma_n^j)})\mult
               \lambda^{(\gamma_1^\zeta,\dots,\gamma_n^\zeta)})\overset{\text{(\ref{eq7k})}}{=\!\!=} \\
& \mi{reduced}((\bigmult_{j=1}^{\kappa_1} \mi{reduced}(\lambda^{(\gamma_1^j,\dots,\gamma_n^j)}))\mult
               (\bigmult_{j=\kappa_1+1}^{\kappa_2} \mi{reduced}(\lambda^{(\gamma_1^j,\dots,\gamma_n^j)}))\mult \\
&\phantom{\mi{reduced}(}
               (\bigmult_{j=\kappa_2+1}^{\kappa_3} \mi{reduced}(\lambda^{(\gamma_1^j,\dots,\gamma_n^j)}))\mult
               (\bigmult_{j=\kappa_3+1}^{\kappa_4} \mi{reduced}(\lambda^{(\gamma_1^j,\dots,\gamma_n^j)}))\mult \\
&\phantom{\mi{reduced}(}
               \mi{reduced}(\lambda^{(\gamma_1^\zeta,\dots,\gamma_n^\zeta)}))= \\
& \mi{reduced}((\bigmult_{j=1}^{\kappa_1} \mu_2^j\gleq (\mu_1^j\mult a_\iota^{\beta_j}))\mult
               (\bigmult_{j=\kappa_1+1}^{\kappa_2} \mu_2^j\gle (\mu_1^j\mult a_\iota^{\beta_j}))\mult \\
&\phantom{\mi{reduced}(}
               (\bigmult_{j=\kappa_2+1}^{\kappa_3} (\mu_1^j\mult a_\iota^{\beta_j})\gleq \mu_2^j)\mult
               (\bigmult_{j=\kappa_3+1}^{\kappa_4} (\mu_1^j\mult a_\iota^{\beta_j})\gle \mu_2^j)\mult
               (\zeta_1\diamond^\zeta \zeta_2))\overset{\text{(\ref{eq7i})}}{=\!\!=} \\
& \mi{reduced}(((\bigmult_{j=1}^{\kappa_1} \mu_2^j)\mult (\bigmult_{j=\kappa_1+1}^{\kappa_2} \mu_2^j)\mult 
                (\bigmult_{j=\kappa_2+1}^{\kappa_3} \mu_1^j)\mult (\bigmult_{j=\kappa_3+1}^{\kappa_4} \mu_1^j)\mult \zeta_1\mult \\
&\phantom{\mi{reduced}(} \quad
                a_\iota^{(\sum_{j=\kappa_2+1}^{\kappa_3} \beta_j)+\sum_{j=\kappa_3+1}^{\kappa_4} \beta_j})\gleq \\
&\phantom{\mi{reduced}(} \quad
               ((\bigmult_{j=1}^{\kappa_1} \mu_1^j)\mult (\bigmult_{j=\kappa_1+1}^{\kappa_2} \mu_1^j)\mult 
                (\bigmult_{j=\kappa_2+1}^{\kappa_3} \mu_2^j)\mult (\bigmult_{j=\kappa_3+1}^{\kappa_4} \mu_2^j)\mult \zeta_2\mult \\
&\phantom{\mi{reduced}(} \quad \quad
                a_\iota^{(\sum_{j=1}^{\kappa_1} \beta_j)+\sum_{j=\kappa_1+1}^{\kappa_2} \beta_j}))= \\
& \mi{reduced}^-(((\bigmult_{j=1}^{\kappa_1} \mu_2^j)\mult (\bigmult_{j=\kappa_1+1}^{\kappa_2} \mu_2^j)\mult 
                  (\bigmult_{j=\kappa_2+1}^{\kappa_3} \mu_1^j)\mult (\bigmult_{j=\kappa_3+1}^{\kappa_4} \mu_1^j)\mult \zeta_1\mult \\
&\phantom{\mi{reduced}^-(} \quad
                  a_\iota^{(\sum_{j=\kappa_2+1}^{\kappa_3} \beta_j)+\sum_{j=\kappa_3+1}^{\kappa_4} \beta_j})\gleq \\
&\phantom{\mi{reduced}^-(} \quad
                 ((\bigmult_{j=1}^{\kappa_1} \mu_1^j)\mult (\bigmult_{j=\kappa_1+1}^{\kappa_2} \mu_1^j)\mult 
                  (\bigmult_{j=\kappa_2+1}^{\kappa_3} \mu_2^j)\mult (\bigmult_{j=\kappa_3+1}^{\kappa_4} \mu_2^j)\mult \zeta_2\mult \\
&\phantom{\mi{reduced}^-(} \quad \quad
                  a_\iota^{(\sum_{j=1}^{\kappa_1} \beta_j)+\sum_{j=\kappa_1+1}^{\kappa_2} \beta_j}))\gleq \\
& \mi{reduced}^+(((\bigmult_{j=1}^{\kappa_1} \mu_2^j)\mult (\bigmult_{j=\kappa_1+1}^{\kappa_2} \mu_2^j)\mult 
                  (\bigmult_{j=\kappa_2+1}^{\kappa_3} \mu_1^j)\mult (\bigmult_{j=\kappa_3+1}^{\kappa_4} \mu_1^j)\mult \zeta_1\mult \\
&\phantom{\mi{reduced}^+(} \quad
                  a_\iota^{(\sum_{j=\kappa_2+1}^{\kappa_3} \beta_j)+\sum_{j=\kappa_3+1}^{\kappa_4} \beta_j})\gleq \\
&\phantom{\mi{reduced}^+(} \quad
                 ((\bigmult_{j=1}^{\kappa_1} \mu_1^j)\mult (\bigmult_{j=\kappa_1+1}^{\kappa_2} \mu_1^j)\mult 
                  (\bigmult_{j=\kappa_2+1}^{\kappa_3} \mu_2^j)\mult (\bigmult_{j=\kappa_3+1}^{\kappa_4} \mu_2^j)\mult \zeta_2\mult \\
&\phantom{\mi{reduced}^+(} \quad \quad 
                  a_\iota^{(\sum_{j=1}^{\kappa_1} \beta_j)+\sum_{j=\kappa_1+1}^{\kappa_2} \beta_j})), \\[1mm]
& \#(a_\iota,(\upsilon_1\mult a_\iota^\alpha)\gleq \varepsilon_2)=-\alpha= \\
& \#(a_\iota,\mi{reduced}(((\bigmult_{j=1}^{\kappa_1} \mu_2^j)\mult (\bigmult_{j=\kappa_1+1}^{\kappa_2} \mu_2^j)\mult 
                           (\bigmult_{j=\kappa_2+1}^{\kappa_3} \mu_1^j)\mult (\bigmult_{j=\kappa_3+1}^{\kappa_4} \mu_1^j)\mult \zeta_1\mult \\
&\phantom{\#(a_\iota,\mi{reduced}(} \quad
                           a_\iota^{(\sum_{j=\kappa_2+1}^{\kappa_3} \beta_j)+\sum_{j=\kappa_3+1}^{\kappa_4} \beta_j})\gleq \\
&\phantom{\#(a_\iota,\mi{reduced}(} \quad
                          ((\bigmult_{j=1}^{\kappa_1} \mu_1^j)\mult (\bigmult_{j=\kappa_1+1}^{\kappa_2} \mu_1^j)\mult 
                           (\bigmult_{j=\kappa_2+1}^{\kappa_3} \mu_2^j)\mult (\bigmult_{j=\kappa_3+1}^{\kappa_4} \mu_2^j)\mult \zeta_2\mult \\
&\phantom{\#(a_\iota,\mi{reduced}(} \quad \quad
                           a_\iota^{(\sum_{j=1}^{\kappa_1} \beta_j)+\sum_{j=\kappa_1+1}^{\kappa_2} \beta_j})))\overset{\text{(\ref{eq7aaaax})}}{=\!\!=} \\
& \#(a_\iota,((\bigmult_{j=1}^{\kappa_1} \mu_2^j)\mult (\bigmult_{j=\kappa_1+1}^{\kappa_2} \mu_2^j)\mult 
              (\bigmult_{j=\kappa_2+1}^{\kappa_3} \mu_1^j)\mult (\bigmult_{j=\kappa_3+1}^{\kappa_4} \mu_1^j)\mult \zeta_1\mult \\
&\phantom{\#(a_\iota,} \quad
              a_\iota^{(\sum_{j=\kappa_2+1}^{\kappa_3} \beta_j)+\sum_{j=\kappa_3+1}^{\kappa_4} \beta_j})\gleq \\
&\phantom{\#(a_\iota,} \quad
             ((\bigmult_{j=1}^{\kappa_1} \mu_1^j)\mult (\bigmult_{j=\kappa_1+1}^{\kappa_2} \mu_1^j)\mult 
              (\bigmult_{j=\kappa_2+1}^{\kappa_3} \mu_2^j)\mult (\bigmult_{j=\kappa_3+1}^{\kappa_4} \mu_2^j)\mult \zeta_2\mult \\
&\phantom{\#(a_\iota,} \quad \quad
              a_\iota^{(\sum_{j=1}^{\kappa_1} \beta_j)+\sum_{j=\kappa_1+1}^{\kappa_2} \beta_j}))= \\
& \left(\sum_{j=1}^{\kappa_1} \beta_j\right)+\left(\sum_{j=\kappa_1+1}^{\kappa_2} \beta_j\right)-\left(\sum_{j=\kappa_2+1}^{\kappa_3} \beta_j\right)-\sum_{j=\kappa_3+1}^{\kappa_4} \beta_j\leq -1, 
\end{alignat*}
$a_\iota\not\in \mi{dom}({\mc V}_{\iota-1})\supseteq \mi{atoms}((\bigmult_{j=1}^{\kappa_1} \mu_2^j)\mult (\bigmult_{j=\kappa_1+1}^{\kappa_2} \mu_2^j)\mult 
                                                                (\bigmult_{j=\kappa_2+1}^{\kappa_3} \mu_1^j)\mult                                                                          \linebreak[4]
                                                                                                                  (\bigmult_{j=\kappa_3+1}^{\kappa_4} \mu_1^j)\mult \zeta_1)=
                                                     (\bigcup_{j=1}^{\kappa_1} \mi{atoms}(\mu_2^j))\cup (\bigcup_{j=\kappa_1+1}^{\kappa_2} \mi{atoms}(\mu_2^j))\cup                        \linebreak[4]
                                                     (\bigcup_{j=\kappa_2+1}^{\kappa_3} \mi{atoms}(\mu_1^j))\cup (\bigcup_{j=\kappa_3+1}^{\kappa_4} \mi{atoms}(\mu_1^j))\cup \mi{atoms}(\zeta_1)$,
$a_\iota\not\in \mi{dom}({\mc V}_{\iota-1})\supseteq \mi{atoms}((\bigmult_{j=1}^{\kappa_1} \mu_1^j)\mult (\bigmult_{j=\kappa_1+1}^{\kappa_2} \mu_1^j)\mult 
                                                                (\bigmult_{j=\kappa_2+1}^{\kappa_3} \mu_2^j)\mult (\bigmult_{j=\kappa_3+1}^{\kappa_4} \mu_2^j)\mult \zeta_2)=              \linebreak[4]
                                                     (\bigcup_{j=1}^{\kappa_1} \mi{atoms}(\mu_1^j))\cup (\bigcup_{j=\kappa_1+1}^{\kappa_2} \mi{atoms}(\mu_1^j))\cup 
                                                     (\bigcup_{j=\kappa_2+1}^{\kappa_3} \mi{atoms}(\mu_2^j))\cup (\bigcup_{j=\kappa_3+1}^{\kappa_4} \mi{atoms}(\mu_2^j))\cup \mi{atoms}(\zeta_2)$,
\begin{alignat*}{1}
& \mi{reduced}^-(((\bigmult_{j=1}^{\kappa_1} \mu_2^j)\mult (\bigmult_{j=\kappa_1+1}^{\kappa_2} \mu_2^j)\mult 
                  (\bigmult_{j=\kappa_2+1}^{\kappa_3} \mu_1^j)\mult (\bigmult_{j=\kappa_3+1}^{\kappa_4} \mu_1^j)\mult \zeta_1\mult \\
&\phantom{\mi{reduced}^-(} \quad
                  a_\iota^{(\sum_{j=\kappa_2+1}^{\kappa_3} \beta_j)+\sum_{j=\kappa_3+1}^{\kappa_4} \beta_j})\gleq \\
&\phantom{\mi{reduced}^-(} \quad
                 ((\bigmult_{j=1}^{\kappa_1} \mu_1^j)\mult (\bigmult_{j=\kappa_1+1}^{\kappa_2} \mu_1^j)\mult 
                  (\bigmult_{j=\kappa_2+1}^{\kappa_3} \mu_2^j)\mult (\bigmult_{j=\kappa_3+1}^{\kappa_4} \mu_2^j)\mult \zeta_2\mult \\
&\phantom{\mi{reduced}^-(} \quad \quad
                  a_\iota^{(\sum_{j=1}^{\kappa_1} \beta_j)+\sum_{j=\kappa_1+1}^{\kappa_2} \beta_j}))\overset{\text{(\ref{eq7ee})}}{=\!\!=} \\
& \mi{reduced}^-(((\bigmult_{j=1}^{\kappa_1} \mu_2^j)\mult (\bigmult_{j=\kappa_1+1}^{\kappa_2} \mu_2^j)\mult 
                  (\bigmult_{j=\kappa_2+1}^{\kappa_3} \mu_1^j)\mult (\bigmult_{j=\kappa_3+1}^{\kappa_4} \mu_1^j)\mult \zeta_1)\gleq \\
&\phantom{\mi{reduced}^-(} \quad
                 ((\bigmult_{j=1}^{\kappa_1} \mu_1^j)\mult (\bigmult_{j=\kappa_1+1}^{\kappa_2} \mu_1^j)\mult 
                  (\bigmult_{j=\kappa_2+1}^{\kappa_3} \mu_2^j)\mult (\bigmult_{j=\kappa_3+1}^{\kappa_4} \mu_2^j)\mult \zeta_2))\mult \\
& a_\iota^{(\sum_{j=\kappa_2+1}^{\kappa_3} \beta_j)+(\sum_{j=\kappa_3+1}^{\kappa_4} \beta_j)-(\sum_{j=1}^{\kappa_1} \beta_j)-\sum_{j=\kappa_1+1}^{\kappa_2} \beta_j}= \\
& \mi{reduced}^-(((\bigmult_{j=1}^{\kappa_1} \mu_2^j)\mult (\bigmult_{j=\kappa_1+1}^{\kappa_2} \mu_2^j)\mult 
                  (\bigmult_{j=\kappa_2+1}^{\kappa_3} \mu_1^j)\mult (\bigmult_{j=\kappa_3+1}^{\kappa_4} \mu_1^j)\mult \zeta_1)\gleq \\
&\phantom{\mi{reduced}^-(} \quad
                 ((\bigmult_{j=1}^{\kappa_1} \mu_1^j)\mult (\bigmult_{j=\kappa_1+1}^{\kappa_2} \mu_1^j)\mult 
                  (\bigmult_{j=\kappa_2+1}^{\kappa_3} \mu_2^j)\mult (\bigmult_{j=\kappa_3+1}^{\kappa_4} \mu_2^j)\mult \zeta_2))\mult a_\iota^\alpha, \\[1mm]
& \mi{reduced}^+(((\bigmult_{j=1}^{\kappa_1} \mu_2^j)\mult (\bigmult_{j=\kappa_1+1}^{\kappa_2} \mu_2^j)\mult 
                  (\bigmult_{j=\kappa_2+1}^{\kappa_3} \mu_1^j)\mult (\bigmult_{j=\kappa_3+1}^{\kappa_4} \mu_1^j)\mult \zeta_1\mult \\
&\phantom{\mi{reduced}^+(} \quad
                  a_\iota^{(\sum_{j=\kappa_2+1}^{\kappa_3} \beta_j)+\sum_{j=\kappa_3+1}^{\kappa_4} \beta_j})\gleq \\
&\phantom{\mi{reduced}^+(} \quad
                 ((\bigmult_{j=1}^{\kappa_1} \mu_1^j)\mult (\bigmult_{j=\kappa_1+1}^{\kappa_2} \mu_1^j)\mult 
                  (\bigmult_{j=\kappa_2+1}^{\kappa_3} \mu_2^j)\mult (\bigmult_{j=\kappa_3+1}^{\kappa_4} \mu_2^j)\mult \zeta_2\mult \\
&\phantom{\mi{reduced}^+(} \quad \quad
                  a_\iota^{(\sum_{j=1}^{\kappa_1} \beta_j)+\sum_{j=\kappa_1+1}^{\kappa_2} \beta_j}))\overset{\text{(\ref{eq7dd})}}{=\!\!=} \\
& \mi{reduced}^+(((\bigmult_{j=1}^{\kappa_1} \mu_2^j)\mult (\bigmult_{j=\kappa_1+1}^{\kappa_2} \mu_2^j)\mult 
                  (\bigmult_{j=\kappa_2+1}^{\kappa_3} \mu_1^j)\mult (\bigmult_{j=\kappa_3+1}^{\kappa_4} \mu_1^j)\mult \zeta_1)\gleq \\
&\phantom{\mi{reduced}^+(} \quad
                 ((\bigmult_{j=1}^{\kappa_1} \mu_1^j)\mult (\bigmult_{j=\kappa_1+1}^{\kappa_2} \mu_1^j)\mult 
                  (\bigmult_{j=\kappa_2+1}^{\kappa_3} \mu_2^j)\mult (\bigmult_{j=\kappa_3+1}^{\kappa_4} \mu_2^j)\mult \zeta_2)), \\[1mm]
& (\upsilon_1\mult a_\iota^\alpha)\gleq \varepsilon_2= \\
& (\mi{reduced}^-(((\bigmult_{j=1}^{\kappa_1} \mu_2^j)\mult (\bigmult_{j=\kappa_1+1}^{\kappa_2} \mu_2^j)\mult 
                   (\bigmult_{j=\kappa_2+1}^{\kappa_3} \mu_1^j)\mult (\bigmult_{j=\kappa_3+1}^{\kappa_4} \mu_1^j)\mult \zeta_1)\gleq \\
&\phantom{(\mi{reduced}^-(} \quad
                  ((\bigmult_{j=1}^{\kappa_1} \mu_1^j)\mult (\bigmult_{j=\kappa_1+1}^{\kappa_2} \mu_1^j)\mult 
                   (\bigmult_{j=\kappa_2+1}^{\kappa_3} \mu_2^j)\mult (\bigmult_{j=\kappa_3+1}^{\kappa_4} \mu_2^j)\mult \zeta_2))\mult a_\iota^\alpha)\gleq \\
& \mi{reduced}^+(((\bigmult_{j=1}^{\kappa_1} \mu_2^j)\mult (\bigmult_{j=\kappa_1+1}^{\kappa_2} \mu_2^j)\mult 
                  (\bigmult_{j=\kappa_2+1}^{\kappa_3} \mu_1^j)\mult (\bigmult_{j=\kappa_3+1}^{\kappa_4} \mu_1^j)\mult \zeta_1)\gleq \\
&\phantom{\mi{reduced}^+(} \quad
                 ((\bigmult_{j=1}^{\kappa_1} \mu_1^j)\mult (\bigmult_{j=\kappa_1+1}^{\kappa_2} \mu_1^j)\mult 
                  (\bigmult_{j=\kappa_2+1}^{\kappa_3} \mu_2^j)\mult (\bigmult_{j=\kappa_3+1}^{\kappa_4} \mu_2^j)\mult \zeta_2)), \\[1mm]
& \mi{atoms}(\mi{reduced}^-(((\bigmult_{j=1}^{\kappa_1} \mu_2^j)\mult (\bigmult_{j=\kappa_1+1}^{\kappa_2} \mu_2^j)\mult 
                             (\bigmult_{j=\kappa_2+1}^{\kappa_3} \mu_1^j)\mult (\bigmult_{j=\kappa_3+1}^{\kappa_4} \mu_1^j)\mult \zeta_1)\gleq \\
&\phantom{\mi{atoms}(\mi{reduced}^-(} \quad
                            ((\bigmult_{j=1}^{\kappa_1} \mu_1^j)\mult (\bigmult_{j=\kappa_1+1}^{\kappa_2} \mu_1^j)\mult 
                             (\bigmult_{j=\kappa_2+1}^{\kappa_3} \mu_2^j)\mult (\bigmult_{j=\kappa_3+1}^{\kappa_4} \mu_2^j)\mult \zeta_2)))\underset{\text{(\ref{eq7a})}}{\subseteq} \\
& \mi{atoms}((\bigmult_{j=1}^{\kappa_1} \mu_2^j)\mult (\bigmult_{j=\kappa_1+1}^{\kappa_2} \mu_2^j)\mult 
             (\bigmult_{j=\kappa_2+1}^{\kappa_3} \mu_1^j)\mult (\bigmult_{j=\kappa_3+1}^{\kappa_4} \mu_1^j)\mult \zeta_1)\subseteq \mi{dom}({\mc V}_{\iota-1}), \\[1mm]
& \mi{atoms}(\mi{reduced}^+(((\bigmult_{j=1}^{\kappa_1} \mu_2^j)\mult (\bigmult_{j=\kappa_1+1}^{\kappa_2} \mu_2^j)\mult 
                             (\bigmult_{j=\kappa_2+1}^{\kappa_3} \mu_1^j)\mult (\bigmult_{j=\kappa_3+1}^{\kappa_4} \mu_1^j)\mult \zeta_1)\gleq \\
&\phantom{\mi{atoms}(\mi{reduced}^+(} \quad
                            ((\bigmult_{j=1}^{\kappa_1} \mu_1^j)\mult (\bigmult_{j=\kappa_1+1}^{\kappa_2} \mu_1^j)\mult 
                             (\bigmult_{j=\kappa_2+1}^{\kappa_3} \mu_2^j)\mult (\bigmult_{j=\kappa_3+1}^{\kappa_4} \mu_2^j)\mult \zeta_2)))\underset{\text{(\ref{eq7b})}}{\subseteq} \\
& \mi{atoms}((\bigmult_{j=1}^{\kappa_1} \mu_1^j)\mult (\bigmult_{j=\kappa_1+1}^{\kappa_2} \mu_1^j)\mult 
             (\bigmult_{j=\kappa_2+1}^{\kappa_3} \mu_2^j)\mult (\bigmult_{j=\kappa_3+1}^{\kappa_4} \mu_2^j)\mult \zeta_2)\subseteq \mi{dom}({\mc V}_{\iota-1}), 
\end{alignat*}
\begin{alignat*}{1}
& \upsilon_1=\mi{reduced}^-(((\bigmult_{j=1}^{\kappa_1} \mu_2^j)\mult (\bigmult_{j=\kappa_1+1}^{\kappa_2} \mu_2^j)\mult 
                             (\bigmult_{j=\kappa_2+1}^{\kappa_3} \mu_1^j)\mult (\bigmult_{j=\kappa_3+1}^{\kappa_4} \mu_1^j)\mult \zeta_1)\gleq \\
&\phantom{\upsilon_1=\mi{reduced}^-(} \quad
                            ((\bigmult_{j=1}^{\kappa_1} \mu_1^j)\mult (\bigmult_{j=\kappa_1+1}^{\kappa_2} \mu_1^j)\mult 
                             (\bigmult_{j=\kappa_2+1}^{\kappa_3} \mu_2^j)\mult (\bigmult_{j=\kappa_3+1}^{\kappa_4} \mu_2^j)\mult \zeta_2)), \\[1mm]
& \varepsilon_2=\mi{reduced}^+(((\bigmult_{j=1}^{\kappa_1} \mu_2^j)\mult (\bigmult_{j=\kappa_1+1}^{\kappa_2} \mu_2^j)\mult 
                                (\bigmult_{j=\kappa_2+1}^{\kappa_3} \mu_1^j)\mult (\bigmult_{j=\kappa_3+1}^{\kappa_4} \mu_1^j)\mult \zeta_1)\gleq \\
&\phantom{\varepsilon_2=\mi{reduced}^+(} \quad
                               ((\bigmult_{j=1}^{\kappa_1} \mu_1^j)\mult (\bigmult_{j=\kappa_1+1}^{\kappa_2} \mu_1^j)\mult 
                                (\bigmult_{j=\kappa_2+1}^{\kappa_3} \mu_2^j)\mult (\bigmult_{j=\kappa_3+1}^{\kappa_4} \mu_2^j)\mult \zeta_2));
\end{alignat*}
by the induction hypothesis for $\iota-1<\iota$,
$\|\upsilon_1\|^{{\mc V}_{\iota-1}}, \|\varepsilon_2\|^{{\mc V}_{\iota-1}},                                                                                                                
 \|\mu_1^1\|^{{\mc V}_{\iota-1}},\dots,                                                                                                                                                    \linebreak[4]
                                       \|\mu_1^{\kappa_4}\|^{{\mc V}_{\iota-1}}, \|\mu_2^1\|^{{\mc V}_{\iota-1}},\dots,\|\mu_2^{\kappa_4}\|^{{\mc V}_{\iota-1}},
 \|\zeta_1\|^{{\mc V}_{\iota-1}}, \|\zeta_2\|^{{\mc V}_{\iota-1}},
 \|\mi{reduced}^-(((\bigmult_{j=1}^{\kappa_1} \mu_2^j)\mult (\bigmult_{j=\kappa_1+1}^{\kappa_2} \mu_2^j)\mult 
                   (\bigmult_{j=\kappa_2+1}^{\kappa_3} \mu_1^j)\mult (\bigmult_{j=\kappa_3+1}^{\kappa_4} \mu_1^j)\mult \zeta_1)\gleq
                  ((\bigmult_{j=1}^{\kappa_1} \mu_1^j)\mult (\bigmult_{j=\kappa_1+1}^{\kappa_2} \mu_1^j)\mult 
                   (\bigmult_{j=\kappa_2+1}^{\kappa_3} \mu_2^j)\mult (\bigmult_{j=\kappa_3+1}^{\kappa_4} \mu_2^j)\mult \zeta_2))\|^{{\mc V}_{\iota-1}}, 
 \|\mi{reduced}^+(((\bigmult_{j=1}^{\kappa_1} \mu_2^j)\mult (\bigmult_{j=\kappa_1+1}^{\kappa_2} \mu_2^j)\mult 
                   (\bigmult_{j=\kappa_2+1}^{\kappa_3} \mu_1^j)\mult (\bigmult_{j=\kappa_3+1}^{\kappa_4} \mu_1^j)\mult \zeta_1)\gleq
                  ((\bigmult_{j=1}^{\kappa_1} \mu_1^j)\mult (\bigmult_{j=\kappa_1+1}^{\kappa_2} \mu_1^j)\mult 
                   (\bigmult_{j=\kappa_2+1}^{\kappa_3} \mu_2^j)\mult (\bigmult_{j=\kappa_3+1}^{\kappa_4} \mu_2^j)\mult \zeta_2))\|^{{\mc V}_{\iota-1}}\underset{\text{(a)}}{\in} (0,1]$,
either $\diamond^\zeta=\geql$, $\zeta_1\diamond^\zeta \zeta_2=\zeta_1\geql \zeta_2\in \mi{clo}$, $\|\zeta_1\|^{{\mc V}_{\iota-1}}\overset{\text{(c)}}{=\!\!=} \|\zeta_2\|^{{\mc V}_{\iota-1}}$, 
or $\diamond^\zeta=\gleq$, $\zeta_1\diamond^\zeta \zeta_2=\zeta_1\gleq \zeta_2\in \mi{clo}$, $\|\zeta_1\|^{{\mc V}_{\iota-1}}\underset{\text{(d)}}{\leq} \|\zeta_2\|^{{\mc V}_{\iota-1}}$,
or $\diamond^\zeta=\gle$, $\zeta_1\diamond^\zeta \zeta_2=\zeta_1\gle \zeta_2\in \mi{clo}$, $\|\zeta_1\|^{{\mc V}_{\iota-1}}\underset{\text{(e)}}{<} \|\zeta_2\|^{{\mc V}_{\iota-1}}$;
$\|\zeta_1\|^{{\mc V}_{\iota-1}}\leq \|\zeta_2\|^{{\mc V}_{\iota-1}}$,
$\frac{\|\zeta_1\|^{{\mc V}_{\iota-1}}}
      {\|\zeta_2\|^{{\mc V}_{\iota-1}}}\leq 1$,
$\mbb{E}_{\iota-1}=\emptyset$;
for all $1\leq j\leq \kappa_1$, 
$\mu_2^j\gleq (\mu_1^j\mult a_\iota^{\beta_j})\in \mi{min}(\mi{DE}_{\iota-1})$, 
$\left(\frac{\|\mu_2^j\|^{{\mc V}_{\iota-1}}}
            {\|\mu_1^j\|^{{\mc V}_{\iota-1}}}\right)^{\frac{1}{\beta_j}}\in \mbb{DE}_{\iota-1}$,
$\left(\frac{\|\mu_2^j\|^{{\mc V}_{\iota-1}}}
            {\|\mu_1^j\|^{{\mc V}_{\iota-1}}}\right)^{\frac{1}{\beta_j}}\leq \bigfvee \mbb{DE}_{\iota-1}\underset{\text{(\ref{eq8j}a)}}{\leq} {\mc V}_\iota(a_\iota)=\delta_\iota$,
$\frac{\|\mu_2^j\|^{{\mc V}_{\iota-1}}}
      {\|\mu_1^j\|^{{\mc V}_{\iota-1}}}\leq {\mc V}_\iota(a_\iota)^{\beta_j}$;
for all $\kappa_1+1\leq j\leq \kappa_2$, 
$\mu_2^j\gle (\mu_1^j\mult a_\iota^{\beta_j})\in \mi{min}(D_{\iota-1})$, 
$\left(\frac{\|\mu_2^j\|^{{\mc V}_{\iota-1}}}
            {\|\mu_1^j\|^{{\mc V}_{\iota-1}}}\right)^{\frac{1}{\beta_j}}\in \mbb{D}_{\iota-1}$,
$\left(\frac{\|\mu_2^j\|^{{\mc V}_{\iota-1}}}
            {\|\mu_1^j\|^{{\mc V}_{\iota-1}}}\right)^{\frac{1}{\beta_j}}\leq \bigfvee \mbb{D}_{\iota-1}\underset{\text{(\ref{eq8j}b)}}{<} {\mc V}_\iota(a_\iota)=\delta_\iota$,
$\frac{\|\mu_2^j\|^{{\mc V}_{\iota-1}}}
      {\|\mu_1^j\|^{{\mc V}_{\iota-1}}}<{\mc V}_\iota(a_\iota)^{\beta_j}$;
for all $\kappa_2+1\leq j\leq \kappa_3$,
$(\mu_1^j\mult a_\iota^{\beta_j})\gleq \mu_2^j\in \mi{min}(\mi{UE}_{\iota-1})$,
$\mi{min}\left(\left(\frac{\|\mu_2^j\|^{{\mc V}_{\iota-1}}}
                          {\|\mu_1^j\|^{{\mc V}_{\iota-1}}}\right)^{\frac{1}{\beta_j}},1\right)\in \mbb{UE}_{\iota-1}$,
${\mc V}_\iota(a_\iota)=\delta_\iota\underset{\text{(\ref{eq8j}a)}}{\leq} \bigfwedge \mbb{UE}_{\iota-1}\leq 
 \mi{min}\left(\left(\frac{\|\mu_2^j\|^{{\mc V}_{\iota-1}}}
                          {\|\mu_1^j\|^{{\mc V}_{\iota-1}}}\right)^{\frac{1}{\beta_j}},1\right)\leq
 \left(\frac{\|\mu_2^j\|^{{\mc V}_{\iota-1}}}
            {\|\mu_1^j\|^{{\mc V}_{\iota-1}}}\right)^{\frac{1}{\beta_j}}$,
${\mc V}_\iota(a_\iota)^{\beta_j}\leq \frac{\|\mu_2^j\|^{{\mc V}_{\iota-1}}}
                                           {\|\mu_1^j\|^{{\mc V}_{\iota-1}}}$,
$\frac{\|\mu_1^j\|^{{\mc V}_{\iota-1}}}
      {\|\mu_2^j\|^{{\mc V}_{\iota-1}}}\leq \frac{1}
                                                 {{\mc V}_\iota(a_\iota)^{\beta_j}}$;
for all $\kappa_3+1\leq j\leq \kappa_4$,
$(\mu_1^j\mult a_\iota^{\beta_j})\gle \mu_2^j\in \mi{min}(U_{\iota-1})$,
$\mi{min}\left(\left(\frac{\|\mu_2^j\|^{{\mc V}_{\iota-1}}}
                          {\|\mu_1^j\|^{{\mc V}_{\iota-1}}}\right)^{\frac{1}{\beta_j}},1\right)\in \mbb{U}_{\iota-1}$,
${\mc V}_\iota(a_\iota)=\delta_\iota\underset{\text{(\ref{eq8j}b)}}{\leq} \bigfwedge \mbb{U}_{\iota-1}\leq 
 \mi{min}\left(\left(\frac{\|\mu_2^j\|^{{\mc V}_{\iota-1}}}
                          {\|\mu_1^j\|^{{\mc V}_{\iota-1}}}\right)^{\frac{1}{\beta_j}},1\right)\leq
 \left(\frac{\|\mu_2^j\|^{{\mc V}_{\iota-1}}}
            {\|\mu_1^j\|^{{\mc V}_{\iota-1}}}\right)^{\frac{1}{\beta_j}}$,
${\mc V}_\iota(a_\iota)^{\beta_j}\leq \frac{\|\mu_2^j\|^{{\mc V}_{\iota-1}}}
                                           {\|\mu_1^j\|^{{\mc V}_{\iota-1}}}$,
$\frac{\|\mu_1^j\|^{{\mc V}_{\iota-1}}}
      {\|\mu_2^j\|^{{\mc V}_{\iota-1}}}\leq \frac{1}
                                                 {{\mc V}_\iota(a_\iota)^{\beta_j}}$;
\begin{alignat*}{1}
& \dfrac{\|\upsilon_1\|^{{\mc V}_{\iota-1}}}
        {\|\varepsilon_2\|^{{\mc V}_{\iota-1}}}= \\
& \dfrac{\left\|\mi{reduced}^-\left(\begin{array}{l}
                                    ((\bigmult_{j=1}^{\kappa_1} \mu_2^j)\mult (\bigmult_{j=\kappa_1+1}^{\kappa_2} \mu_2^j)\mult 
                                     (\bigmult_{j=\kappa_2+1}^{\kappa_3} \mu_1^j)\mult \\
                                    \quad
                                     (\bigmult_{j=\kappa_3+1}^{\kappa_4} \mu_1^j)\mult \zeta_1)\gleq \\
                                    \quad
                                    ((\bigmult_{j=1}^{\kappa_1} \mu_1^j)\mult (\bigmult_{j=\kappa_1+1}^{\kappa_2} \mu_1^j)\mult 
                                     (\bigmult_{j=\kappa_2+1}^{\kappa_3} \mu_2^j)\mult \\
                                    \quad \quad
                                     (\bigmult_{j=\kappa_3+1}^{\kappa_4} \mu_2^j)\mult \zeta_2)
                                    \end{array}\right)\right\|^{{\mc V}_{\iota-1}}}
        {\left\|\mi{reduced}^+\left(\begin{array}{l}
                                    ((\bigmult_{j=1}^{\kappa_1} \mu_2^j)\mult (\bigmult_{j=\kappa_1+1}^{\kappa_2} \mu_2^j)\mult 
                                     (\bigmult_{j=\kappa_2+1}^{\kappa_3} \mu_1^j)\mult \\
                                    \quad
                                     (\bigmult_{j=\kappa_3+1}^{\kappa_4} \mu_1^j)\mult \zeta_1)\gleq \\
                                    \quad
                                    ((\bigmult_{j=1}^{\kappa_1} \mu_1^j)\mult (\bigmult_{j=\kappa_1+1}^{\kappa_2} \mu_1^j)\mult 
                                     (\bigmult_{j=\kappa_2+1}^{\kappa_3} \mu_2^j)\mult \\
                                    \quad \quad
                                     (\bigmult_{j=\kappa_3+1}^{\kappa_4} \mu_2^j)\mult \zeta_2)
                                    \end{array}\right)\right\|^{{\mc V}_{\iota-1}}}\overset{\text{(\ref{eq7n})}}{=\!\!=} \\
& \dfrac{\|(\bigmult_{j=1}^{\kappa_1} \mu_2^j)\mult (\bigmult_{j=\kappa_1+1}^{\kappa_2} \mu_2^j)\mult 
           (\bigmult_{j=\kappa_2+1}^{\kappa_3} \mu_1^j)\mult (\bigmult_{j=\kappa_3+1}^{\kappa_4} \mu_1^j)\mult \zeta_1\|^{{\mc V}_{\iota-1}}}
        {\|(\bigmult_{j=1}^{\kappa_1} \mu_1^j)\mult (\bigmult_{j=\kappa_1+1}^{\kappa_2} \mu_1^j)\mult 
           (\bigmult_{j=\kappa_2+1}^{\kappa_3} \mu_2^j)\mult (\bigmult_{j=\kappa_3+1}^{\kappa_4} \mu_2^j)\mult \zeta_2\|^{{\mc V}_{\iota-1}}}= \\
& \dfrac{\begin{array}{l}
         (\bigfswedge_{j=1}^{\kappa_1} \|\mu_2^j\|^{{\mc V}_{\iota-1}})\fswedge (\bigfswedge_{j=\kappa_1+1}^{\kappa_2} \|\mu_2^j\|^{{\mc V}_{\iota-1}})\fswedge
         (\bigfswedge_{j=\kappa_2+1}^{\kappa_3} \|\mu_1^j\|^{{\mc V}_{\iota-1}})\fswedge (\bigfswedge_{j=\kappa_3+1}^{\kappa_4} \|\mu_1^j\|^{{\mc V}_{\iota-1}})\fswedge \\
         \quad
         \|\zeta_1\|^{{\mc V}_{\iota-1}}
         \end{array}}
        {\begin{array}{l}
         (\bigfswedge_{j=1}^{\kappa_1} \|\mu_1^j\|^{{\mc V}_{\iota-1}})\fswedge (\bigfswedge_{j=\kappa_1+1}^{\kappa_2} \|\mu_1^j\|^{{\mc V}_{\iota-1}})\fswedge
         (\bigfswedge_{j=\kappa_2+1}^{\kappa_3} \|\mu_2^j\|^{{\mc V}_{\iota-1}})\fswedge (\bigfswedge_{j=\kappa_3+1}^{\kappa_4} \|\mu_2^j\|^{{\mc V}_{\iota-1}})\fswedge \\
         \quad
         \|\zeta_2\|^{{\mc V}_{\iota-1}}
         \end{array}}= \\
& \left(\bigfswedge_{j=1}^{\kappa_1} \dfrac{\|\mu_2^j\|^{{\mc V}_{\iota-1}}}
                                           {\|\mu_1^j\|^{{\mc V}_{\iota-1}}}\right)\fswedge 
  \left(\bigfswedge_{j=\kappa_1+1}^{\kappa_2} \dfrac{\|\mu_2^j\|^{{\mc V}_{\iota-1}}}
                                                    {\|\mu_1^j\|^{{\mc V}_{\iota-1}}}\right)\fswedge 
  \left(\bigfswedge_{j=\kappa_1+2}^{\kappa_3} \dfrac{\|\mu_1^j\|^{{\mc V}_{\iota-1}}}
                                                    {\|\mu_2^j\|^{{\mc V}_{\iota-1}}}\right)\fswedge \\
& \quad
  \left(\bigfswedge_{j=\kappa_1+3}^{\kappa_4} \dfrac{\|\mu_1^j\|^{{\mc V}_{\iota-1}}}
                                                    {\|\mu_2^j\|^{{\mc V}_{\iota-1}}}\right)\fswedge
  \dfrac{\|\zeta_1\|^{{\mc V}_{\iota-1}}}
        {\|\zeta_2\|^{{\mc V}_{\iota-1}}}\leq \\
& \left(\bigfswedge_{j=1}^{\kappa_1} {\mc V}_\iota(a_\iota)^{\beta_j}\right)\fswedge \left(\bigfswedge_{j=\kappa_1+1}^{\kappa_2} {\mc V}_\iota(a_\iota)^{\beta_j}\right)\fswedge 
  \left(\bigfswedge_{j=\kappa_1+2}^{\kappa_3} \dfrac{1}
                                                    {{\mc V}_\iota(a_\iota)^{\beta_j}}\right)\fswedge
  \bigfswedge_{j=\kappa_1+3}^{\kappa_4} \dfrac{1}
                                              {{\mc V}_\iota(a_\iota)^{\beta_j}}= \\
& {\mc V}_\iota(a_\iota)^{(\sum_{j=1}^{\kappa_1} \beta_j)+(\sum_{j=\kappa_1+1}^{\kappa_2} \beta_j)-(\sum_{j=\kappa_2+1}^{\kappa_3} \beta_j)-\sum_{j=\kappa_3+1}^{\kappa_4} \beta_j}=
  {\mc V}_\iota(a_\iota)^{-\alpha}=\dfrac{1}
                                         {{\mc V}_\iota(a_\iota)^\alpha}, \\[1mm]
& \|\varepsilon_1\|^{{\mc V}_\iota}=\|\upsilon_1\mult a_\iota^\alpha\|^{{\mc V}_\iota}=\|\upsilon_1\|^{{\mc V}_\iota}\fswedge {\mc V}_\iota(a_\iota)^\alpha\overset{\text{(b)}}{=\!\!=}
  \|\upsilon_1\|^{{\mc V}_{\iota-1}}\fswedge {\mc V}_\iota(a_\iota)^\alpha\leq \|\varepsilon_2\|^{{\mc V}_{\iota-1}}\overset{\text{(b)}}{=\!\!=} \|\varepsilon_2\|^{{\mc V}_\iota};  
\end{alignat*}
(d) holds.

Case 2.10.2.1.2:
$\varepsilon_1\gle \varepsilon_2\in \mi{clo}$.
Then $\mi{atoms}(\varepsilon_1), \mi{atoms}(\varepsilon_2)\subseteq \mi{dom}({\mc V}_\iota)=\mi{dom}({\mc V}_{\iota-1})\cup \{a_\iota\}$,
$a_\iota\in \mi{atoms}(\varepsilon_1)$ or $a_\iota\in \mi{atoms}(\varepsilon_2)$, $\mbb{E}_{\iota-1}=\emptyset$,
$\varepsilon_1\gle \varepsilon_2=(\upsilon_1\mult a_\iota^\alpha)\gle \varepsilon_2\in \mi{clo}$;
by (\ref{eq8i}) for $\gle$ and (\ref{eq8i}d), there exist $\kappa_1\leq \kappa_2\leq \kappa_3<\kappa_4$,
\begin{alignat*}{1}
& \mu_2^j\gleq (\mu_1^j\mult a_\iota^{\beta_j})=\mi{reduced}(\lambda^{(\gamma_1^j,\dots,\gamma_n^j)})\in \mi{min}(\mi{DE}_{\iota-1}), j=1,\dots,\kappa_1, \\ 
& \mu_2^j\gle (\mu_1^j\mult a_\iota^{\beta_j})=\mi{reduced}(\lambda^{(\gamma_1^j,\dots,\gamma_n^j)})\in \mi{min}(D_{\iota-1}), j=\kappa_1+1,\dots,\kappa_2, \\
& (\mu_1^j\mult a_\iota^{\beta_j})\gleq \mu_2^j=\mi{reduced}(\lambda^{(\gamma_1^j,\dots,\gamma_n^j)})\in \mi{min}(\mi{UE}_{\iota-1}), j=\kappa_2+1,\dots,\kappa_3, \\
& (\mu_1^j\mult a_\iota^{\beta_j})\gle \mu_2^j=\mi{reduced}(\lambda^{(\gamma_1^j,\dots,\gamma_n^j)})\in \mi{min}(U_{\iota-1}), j=\kappa_3+1,\dots,\kappa_4, \\ 
& \mu_e^j\in \{\gu\}\cup \mi{PropConj}_A, \mi{atoms}(\mu_e^j)\subseteq \mi{dom}({\mc V}_{\iota-1}), \beta_j\geq 1, \bs{0}^n\neq (\gamma_1^j,\dots,\gamma_n^j)\in \mbb{N}^n, 
\end{alignat*}
satisfying either $\varepsilon_1\gle \varepsilon_2=(\upsilon_1\mult a_\iota^\alpha)\gle \varepsilon_2=
                                                   \mi{reduced}(\lambda^{(\sum_{j=1}^{\kappa_4} \gamma_1^j,\dots,\sum_{j=1}^{\kappa_4} \gamma_n^j)})$, 
or there exists $\zeta_1\diamond^\zeta \zeta_2=\mi{reduced}(\lambda^{(\gamma_1^\zeta,\dots,\gamma_n^\zeta)})\in \mi{clo}$,
$\zeta_e\in \{\gu\}\cup \mi{PropConj}_A$, $\mi{atoms}(\zeta_e)\subseteq \mi{dom}({\mc V}_{\iota-1})$, $\diamond^\zeta\in \{\geql,\gleq,\gle\}$, 
$\bs{0}^n\neq (\gamma_1^\zeta,\dots,\gamma_n^\zeta)\in \mbb{N}^n$, satisfying
$\varepsilon_1\gle \varepsilon_2=(\upsilon_1\mult a_\iota^\alpha)\gle \varepsilon_2=
                                 \mi{reduced}(\lambda^{((\sum_{j=1}^{\kappa_4} \gamma_1^j)+\gamma_1^\zeta,\dots,(\sum_{j=1}^{\kappa_4} \gamma_n^j)+\gamma_n^\zeta)})$.
We get two cases for $(\upsilon_1\mult a_\iota^\alpha)\gle \varepsilon_2$.

Case 2.10.2.1.2.1:
$(\upsilon_1\mult a_\iota^\alpha)\gle \varepsilon_2=\mi{reduced}(\lambda^{(\sum_{j=1}^{\kappa_4} \gamma_1^j,\dots,\sum_{j=1}^{\kappa_4} \gamma_n^j)})$.
Then 
$\mi{atoms}(\upsilon_1), \mi{atoms}(\varepsilon_2), \mi{atoms}(\mu_1^1),\dots,\mi{atoms}(\mu_1^{\kappa_4}), \mi{atoms}(\mu_2^1),\dots,\mi{atoms}(\mu_2^{\kappa_4})\subseteq \mi{dom}({\mc V}_{\iota-1})$,
\begin{alignat*}{1}
& (\upsilon_1\mult a_\iota^\alpha)\gle \varepsilon_2=
  \mi{reduced}(\lambda^{(\sum_{j=1}^{\kappa_4} \gamma_1^j,\dots,\sum_{j=1}^{\kappa_4} \gamma_n^j)})\overset{\text{(\ref{eq8dddd})}}{=\!\!=} \\
& \mi{reduced}((\bigmult_{j=1}^{\kappa_1} \lambda^{(\gamma_1^j,\dots,\gamma_n^j)})\mult
               (\bigmult_{j=\kappa_1+1}^{\kappa_2} \lambda^{(\gamma_1^j,\dots,\gamma_n^j)})\mult \\
&\phantom{\mi{reduced}(}
               (\bigmult_{j=\kappa_2+1}^{\kappa_3} \lambda^{(\gamma_1^j,\dots,\gamma_n^j)})\mult
               \bigmult_{j=\kappa_3+1}^{\kappa_4} \lambda^{(\gamma_1^j,\dots,\gamma_n^j)})\overset{\text{(\ref{eq7k})}}{=\!\!=} \\
& \mi{reduced}((\bigmult_{j=1}^{\kappa_1} \mi{reduced}(\lambda^{(\gamma_1^j,\dots,\gamma_n^j)}))\mult
               (\bigmult_{j=\kappa_1+1}^{\kappa_2} \mi{reduced}(\lambda^{(\gamma_1^j,\dots,\gamma_n^j)}))\mult \\
&\phantom{\mi{reduced}(}
               (\bigmult_{j=\kappa_2+1}^{\kappa_3} \mi{reduced}(\lambda^{(\gamma_1^j,\dots,\gamma_n^j)}))\mult
               \bigmult_{j=\kappa_3+1}^{\kappa_4} \mi{reduced}(\lambda^{(\gamma_1^j,\dots,\gamma_n^j)}))= \\
& \mi{reduced}((\bigmult_{j=1}^{\kappa_1} \mu_2^j\gleq (\mu_1^j\mult a_\iota^{\beta_j}))\mult
               (\bigmult_{j=\kappa_1+1}^{\kappa_2} \mu_2^j\gle (\mu_1^j\mult a_\iota^{\beta_j}))\mult \\
&\phantom{\mi{reduced}(}
               (\bigmult_{j=\kappa_2+1}^{\kappa_3} (\mu_1^j\mult a_\iota^{\beta_j})\gleq \mu_2^j)\mult
               \bigmult_{j=\kappa_3+1}^{\kappa_4} (\mu_1^j\mult a_\iota^{\beta_j})\gle \mu_2^j)\overset{\text{(\ref{eq7i})}}{=\!\!=} \\
& \mi{reduced}(((\bigmult_{j=1}^{\kappa_1} \mu_2^j)\mult (\bigmult_{j=\kappa_1+1}^{\kappa_2} \mu_2^j)\mult 
                (\bigmult_{j=\kappa_2+1}^{\kappa_3} \mu_1^j)\mult (\bigmult_{j=\kappa_3+1}^{\kappa_4} \mu_1^j)\mult \\
&\phantom{\mi{reduced}(} \quad
                a_\iota^{(\sum_{j=\kappa_2+1}^{\kappa_3} \beta_j)+\sum_{j=\kappa_3+1}^{\kappa_4} \beta_j})\gle \\
&\phantom{\mi{reduced}(} \quad
               ((\bigmult_{j=1}^{\kappa_1} \mu_1^j)\mult (\bigmult_{j=\kappa_1+1}^{\kappa_2} \mu_1^j)\mult 
                (\bigmult_{j=\kappa_2+1}^{\kappa_3} \mu_2^j)\mult (\bigmult_{j=\kappa_3+1}^{\kappa_4} \mu_2^j)\mult \\
&\phantom{\mi{reduced}(} \quad \quad
                a_\iota^{(\sum_{j=1}^{\kappa_1} \beta_j)+\sum_{j=\kappa_1+1}^{\kappa_2} \beta_j}))= \\
& \mi{reduced}^-(((\bigmult_{j=1}^{\kappa_1} \mu_2^j)\mult (\bigmult_{j=\kappa_1+1}^{\kappa_2} \mu_2^j)\mult 
                  (\bigmult_{j=\kappa_2+1}^{\kappa_3} \mu_1^j)\mult (\bigmult_{j=\kappa_3+1}^{\kappa_4} \mu_1^j)\mult \\
&\phantom{\mi{reduced}^-(} \quad
                  a_\iota^{(\sum_{j=\kappa_2+1}^{\kappa_3} \beta_j)+\sum_{j=\kappa_3+1}^{\kappa_4} \beta_j})\gle \\
&\phantom{\mi{reduced}^-(} \quad
                 ((\bigmult_{j=1}^{\kappa_1} \mu_1^j)\mult (\bigmult_{j=\kappa_1+1}^{\kappa_2} \mu_1^j)\mult 
                  (\bigmult_{j=\kappa_2+1}^{\kappa_3} \mu_2^j)\mult (\bigmult_{j=\kappa_3+1}^{\kappa_4} \mu_2^j)\mult \\
&\phantom{\mi{reduced}^-(} \quad \quad
                  a_\iota^{(\sum_{j=1}^{\kappa_1} \beta_j)+\sum_{j=\kappa_1+1}^{\kappa_2} \beta_j}))\gle \\
& \mi{reduced}^+(((\bigmult_{j=1}^{\kappa_1} \mu_2^j)\mult (\bigmult_{j=\kappa_1+1}^{\kappa_2} \mu_2^j)\mult 
                  (\bigmult_{j=\kappa_2+1}^{\kappa_3} \mu_1^j)\mult (\bigmult_{j=\kappa_3+1}^{\kappa_4} \mu_1^j)\mult \\
&\phantom{\mi{reduced}^+(} \quad
                  a_\iota^{(\sum_{j=\kappa_2+1}^{\kappa_3} \beta_j)+\sum_{j=\kappa_3+1}^{\kappa_4} \beta_j})\gle \\
&\phantom{\mi{reduced}^+(} \quad
                 ((\bigmult_{j=1}^{\kappa_1} \mu_1^j)\mult (\bigmult_{j=\kappa_1+1}^{\kappa_2} \mu_1^j)\mult 
                  (\bigmult_{j=\kappa_2+1}^{\kappa_3} \mu_2^j)\mult (\bigmult_{j=\kappa_3+1}^{\kappa_4} \mu_2^j)\mult \\
&\phantom{\mi{reduced}^+(} \quad \quad
                  a_\iota^{(\sum_{j=1}^{\kappa_1} \beta_j)+\sum_{j=\kappa_1+1}^{\kappa_2} \beta_j})), \\[1mm]
& \#(a_\iota,(\upsilon_1\mult a_\iota^\alpha)\gle \varepsilon_2)=-\alpha= \\
& \#(a_\iota,\mi{reduced}(((\bigmult_{j=1}^{\kappa_1} \mu_2^j)\mult (\bigmult_{j=\kappa_1+1}^{\kappa_2} \mu_2^j)\mult 
                           (\bigmult_{j=\kappa_2+1}^{\kappa_3} \mu_1^j)\mult (\bigmult_{j=\kappa_3+1}^{\kappa_4} \mu_1^j)\mult \\
&\phantom{\#(a_\iota,\mi{reduced}(} \quad
                           a_\iota^{(\sum_{j=\kappa_2+1}^{\kappa_3} \beta_j)+\sum_{j=\kappa_3+1}^{\kappa_4} \beta_j})\gle \\
&\phantom{\#(a_\iota,\mi{reduced}(} \quad
                          ((\bigmult_{j=1}^{\kappa_1} \mu_1^j)\mult (\bigmult_{j=\kappa_1+1}^{\kappa_2} \mu_1^j)\mult 
                           (\bigmult_{j=\kappa_2+1}^{\kappa_3} \mu_2^j)\mult (\bigmult_{j=\kappa_3+1}^{\kappa_4} \mu_2^j)\mult \\
&\phantom{\#(a_\iota,\mi{reduced}(} \quad \quad
                           a_\iota^{(\sum_{j=1}^{\kappa_1} \beta_j)+\sum_{j=\kappa_1+1}^{\kappa_2} \beta_j})))\overset{\text{(\ref{eq7aaaax})}}{=\!\!=} \\
& \#(a_\iota,((\bigmult_{j=1}^{\kappa_1} \mu_2^j)\mult (\bigmult_{j=\kappa_1+1}^{\kappa_2} \mu_2^j)\mult 
              (\bigmult_{j=\kappa_2+1}^{\kappa_3} \mu_1^j)\mult (\bigmult_{j=\kappa_3+1}^{\kappa_4} \mu_1^j)\mult \\
&\phantom{\#(a_\iota,} \quad
              a_\iota^{(\sum_{j=\kappa_2+1}^{\kappa_3} \beta_j)+\sum_{j=\kappa_3+1}^{\kappa_4} \beta_j})\gle \\
&\phantom{\#(a_\iota,} \quad
             ((\bigmult_{j=1}^{\kappa_1} \mu_1^j)\mult (\bigmult_{j=\kappa_1+1}^{\kappa_2} \mu_1^j)\mult 
              (\bigmult_{j=\kappa_2+1}^{\kappa_3} \mu_2^j)\mult (\bigmult_{j=\kappa_3+1}^{\kappa_4} \mu_2^j)\mult \\
&\phantom{\#(a_\iota,} \quad \quad
              a_\iota^{(\sum_{j=1}^{\kappa_1} \beta_j)+\sum_{j=\kappa_1+1}^{\kappa_2} \beta_j}))= \\
& \left(\sum_{j=1}^{\kappa_1} \beta_j\right)+\left(\sum_{j=\kappa_1+1}^{\kappa_2} \beta_j\right)-\left(\sum_{j=\kappa_2+1}^{\kappa_3} \beta_j\right)-\sum_{j=\kappa_3+1}^{\kappa_4} \beta_j\leq -1, 
\end{alignat*}
$a_\iota\not\in \mi{dom}({\mc V}_{\iota-1})\supseteq \mi{atoms}((\bigmult_{j=1}^{\kappa_1} \mu_2^j)\mult (\bigmult_{j=\kappa_1+1}^{\kappa_2} \mu_2^j)\mult 
                                                                (\bigmult_{j=\kappa_2+1}^{\kappa_3} \mu_1^j)\mult \bigmult_{j=\kappa_3+1}^{\kappa_4} \mu_1^j)=
                                                     (\bigcup_{j=1}^{\kappa_1} \mi{atoms}(\mu_2^j))\cup (\bigcup_{j=\kappa_1+1}^{\kappa_2} \mi{atoms}(\mu_2^j))\cup 
                                                     (\bigcup_{j=\kappa_2+1}^{\kappa_3} \mi{atoms}(\mu_1^j))\cup \bigcup_{j=\kappa_3+1}^{\kappa_4} \mi{atoms}(\mu_1^j)$,
$a_\iota\not\in \mi{dom}({\mc V}_{\iota-1})\supseteq \mi{atoms}((\bigmult_{j=1}^{\kappa_1} \mu_1^j)\mult (\bigmult_{j=\kappa_1+1}^{\kappa_2} \mu_1^j)\mult 
                                                                (\bigmult_{j=\kappa_2+1}^{\kappa_3} \mu_2^j)\mult \bigmult_{j=\kappa_3+1}^{\kappa_4} \mu_2^j)=
                                                     (\bigcup_{j=1}^{\kappa_1} \mi{atoms}(\mu_1^j))\cup (\bigcup_{j=\kappa_1+1}^{\kappa_2} \mi{atoms}(\mu_1^j))\cup 
                                                     (\bigcup_{j=\kappa_2+1}^{\kappa_3} \mi{atoms}(\mu_2^j))\cup \bigcup_{j=\kappa_3+1}^{\kappa_4} \mi{atoms}(\mu_2^j)$,
\begin{alignat*}{1}
& \mi{reduced}^-(((\bigmult_{j=1}^{\kappa_1} \mu_2^j)\mult (\bigmult_{j=\kappa_1+1}^{\kappa_2} \mu_2^j)\mult 
                  (\bigmult_{j=\kappa_2+1}^{\kappa_3} \mu_1^j)\mult (\bigmult_{j=\kappa_3+1}^{\kappa_4} \mu_1^j)\mult \\
&\phantom{\mi{reduced}^-(} \quad
                  a_\iota^{(\sum_{j=\kappa_2+1}^{\kappa_3} \beta_j)+\sum_{j=\kappa_3+1}^{\kappa_4} \beta_j})\gle \\
&\phantom{\mi{reduced}^-(} \quad
                 ((\bigmult_{j=1}^{\kappa_1} \mu_1^j)\mult (\bigmult_{j=\kappa_1+1}^{\kappa_2} \mu_1^j)\mult 
                  (\bigmult_{j=\kappa_2+1}^{\kappa_3} \mu_2^j)\mult (\bigmult_{j=\kappa_3+1}^{\kappa_4} \mu_2^j)\mult \\
&\phantom{\mi{reduced}^-(} \quad \quad
                  a_\iota^{(\sum_{j=1}^{\kappa_1} \beta_j)+\sum_{j=\kappa_1+1}^{\kappa_2} \beta_j}))\overset{\text{(\ref{eq7ee})}}{=\!\!=} \\
& \mi{reduced}^-(((\bigmult_{j=1}^{\kappa_1} \mu_2^j)\mult (\bigmult_{j=\kappa_1+1}^{\kappa_2} \mu_2^j)\mult 
                  (\bigmult_{j=\kappa_2+1}^{\kappa_3} \mu_1^j)\mult \bigmult_{j=\kappa_3+1}^{\kappa_4} \mu_1^j)\gle \\
&\phantom{\mi{reduced}^-(} \quad
                 ((\bigmult_{j=1}^{\kappa_1} \mu_1^j)\mult (\bigmult_{j=\kappa_1+1}^{\kappa_2} \mu_1^j)\mult 
                  (\bigmult_{j=\kappa_2+1}^{\kappa_3} \mu_2^j)\mult \bigmult_{j=\kappa_3+1}^{\kappa_4} \mu_2^j))\mult \\
& a_\iota^{(\sum_{j=\kappa_2+1}^{\kappa_3} \beta_j)+(\sum_{j=\kappa_3+1}^{\kappa_4} \beta_j)-(\sum_{j=1}^{\kappa_1} \beta_j)-\sum_{j=\kappa_1+1}^{\kappa_2} \beta_j}= \\
& \mi{reduced}^-(((\bigmult_{j=1}^{\kappa_1} \mu_2^j)\mult (\bigmult_{j=\kappa_1+1}^{\kappa_2} \mu_2^j)\mult 
                  (\bigmult_{j=\kappa_2+1}^{\kappa_3} \mu_1^j)\mult \bigmult_{j=\kappa_3+1}^{\kappa_4} \mu_1^j)\gle \\
&\phantom{\mi{reduced}^-(} \quad
                 ((\bigmult_{j=1}^{\kappa_1} \mu_1^j)\mult (\bigmult_{j=\kappa_1+1}^{\kappa_2} \mu_1^j)\mult 
                  (\bigmult_{j=\kappa_2+1}^{\kappa_3} \mu_2^j)\mult \bigmult_{j=\kappa_3+1}^{\kappa_4} \mu_2^j))\mult a_\iota^\alpha, \\[1mm]
& \mi{reduced}^+(((\bigmult_{j=1}^{\kappa_1} \mu_2^j)\mult (\bigmult_{j=\kappa_1+1}^{\kappa_2} \mu_2^j)\mult 
                  (\bigmult_{j=\kappa_2+1}^{\kappa_3} \mu_1^j)\mult (\bigmult_{j=\kappa_3+1}^{\kappa_4} \mu_1^j)\mult \\
&\phantom{\mi{reduced}^+(} \quad
                  a_\iota^{(\sum_{j=\kappa_2+1}^{\kappa_3} \beta_j)+\sum_{j=\kappa_3+1}^{\kappa_4} \beta_j})\gle \\
&\phantom{\mi{reduced}^+(} \quad
                 ((\bigmult_{j=1}^{\kappa_1} \mu_1^j)\mult (\bigmult_{j=\kappa_1+1}^{\kappa_2} \mu_1^j)\mult 
                  (\bigmult_{j=\kappa_2+1}^{\kappa_3} \mu_2^j)\mult (\bigmult_{j=\kappa_3+1}^{\kappa_4} \mu_2^j)\mult \\
&\phantom{\mi{reduced}^+(} \quad \quad
                  a_\iota^{(\sum_{j=1}^{\kappa_1} \beta_j)+\sum_{j=\kappa_1+1}^{\kappa_2} \beta_j}))\overset{\text{(\ref{eq7dd})}}{=\!\!=} \\
& \mi{reduced}^+(((\bigmult_{j=1}^{\kappa_1} \mu_2^j)\mult (\bigmult_{j=\kappa_1+1}^{\kappa_2} \mu_2^j)\mult 
                  (\bigmult_{j=\kappa_2+1}^{\kappa_3} \mu_1^j)\mult \bigmult_{j=\kappa_3+1}^{\kappa_4} \mu_1^j)\gle \\
&\phantom{\mi{reduced}^+(} \quad
                 ((\bigmult_{j=1}^{\kappa_1} \mu_1^j)\mult (\bigmult_{j=\kappa_1+1}^{\kappa_2} \mu_1^j)\mult 
                  (\bigmult_{j=\kappa_2+1}^{\kappa_3} \mu_2^j)\mult \bigmult_{j=\kappa_3+1}^{\kappa_4} \mu_2^j)), \\[1mm]
& (\upsilon_1\mult a_\iota^\alpha)\gle \varepsilon_2= \\
& (\mi{reduced}^-(((\bigmult_{j=1}^{\kappa_1} \mu_2^j)\mult (\bigmult_{j=\kappa_1+1}^{\kappa_2} \mu_2^j)\mult 
                   (\bigmult_{j=\kappa_2+1}^{\kappa_3} \mu_1^j)\mult \bigmult_{j=\kappa_3+1}^{\kappa_4} \mu_1^j)\gle \\
&\phantom{(\mi{reduced}^-(} \quad
                  ((\bigmult_{j=1}^{\kappa_1} \mu_1^j)\mult (\bigmult_{j=\kappa_1+1}^{\kappa_2} \mu_1^j)\mult 
                   (\bigmult_{j=\kappa_2+1}^{\kappa_3} \mu_2^j)\mult \bigmult_{j=\kappa_3+1}^{\kappa_4} \mu_2^j))\mult a_\iota^\alpha)\gle \\
& \mi{reduced}^+(((\bigmult_{j=1}^{\kappa_1} \mu_2^j)\mult (\bigmult_{j=\kappa_1+1}^{\kappa_2} \mu_2^j)\mult 
                  (\bigmult_{j=\kappa_2+1}^{\kappa_3} \mu_1^j)\mult \bigmult_{j=\kappa_3+1}^{\kappa_4} \mu_1^j)\gle \\
&\phantom{\mi{reduced}^+(} \quad
                 ((\bigmult_{j=1}^{\kappa_1} \mu_1^j)\mult (\bigmult_{j=\kappa_1+1}^{\kappa_2} \mu_1^j)\mult 
                  (\bigmult_{j=\kappa_2+1}^{\kappa_3} \mu_2^j)\mult \bigmult_{j=\kappa_3+1}^{\kappa_4} \mu_2^j)), \\[1mm]
& \mi{atoms}(\mi{reduced}^-(((\bigmult_{j=1}^{\kappa_1} \mu_2^j)\mult (\bigmult_{j=\kappa_1+1}^{\kappa_2} \mu_2^j)\mult 
                             (\bigmult_{j=\kappa_2+1}^{\kappa_3} \mu_1^j)\mult \bigmult_{j=\kappa_3+1}^{\kappa_4} \mu_1^j)\gle \\
&\phantom{\mi{atoms}(\mi{reduced}^-(} \quad
                            ((\bigmult_{j=1}^{\kappa_1} \mu_1^j)\mult (\bigmult_{j=\kappa_1+1}^{\kappa_2} \mu_1^j)\mult 
                             (\bigmult_{j=\kappa_2+1}^{\kappa_3} \mu_2^j)\mult \bigmult_{j=\kappa_3+1}^{\kappa_4} \mu_2^j)))\underset{\text{(\ref{eq7a})}}{\subseteq} \\
& \mi{atoms}((\bigmult_{j=1}^{\kappa_1} \mu_2^j)\mult (\bigmult_{j=\kappa_1+1}^{\kappa_2} \mu_2^j)\mult 
             (\bigmult_{j=\kappa_2+1}^{\kappa_3} \mu_1^j)\mult \bigmult_{j=\kappa_3+1}^{\kappa_4} \mu_1^j)\subseteq \mi{dom}({\mc V}_{\iota-1}), \\[1mm]
& \mi{atoms}(\mi{reduced}^+(((\bigmult_{j=1}^{\kappa_1} \mu_2^j)\mult (\bigmult_{j=\kappa_1+1}^{\kappa_2} \mu_2^j)\mult 
                             (\bigmult_{j=\kappa_2+1}^{\kappa_3} \mu_1^j)\mult \bigmult_{j=\kappa_3+1}^{\kappa_4} \mu_1^j)\gle \\
&\phantom{\mi{atoms}(\mi{reduced}^+(} \quad
                            ((\bigmult_{j=1}^{\kappa_1} \mu_1^j)\mult (\bigmult_{j=\kappa_1+1}^{\kappa_2} \mu_1^j)\mult 
                             (\bigmult_{j=\kappa_2+1}^{\kappa_3} \mu_2^j)\mult \bigmult_{j=\kappa_3+1}^{\kappa_4} \mu_2^j)))\underset{\text{(\ref{eq7b})}}{\subseteq} \\
& \mi{atoms}((\bigmult_{j=1}^{\kappa_1} \mu_1^j)\mult (\bigmult_{j=\kappa_1+1}^{\kappa_2} \mu_1^j)\mult 
             (\bigmult_{j=\kappa_2+1}^{\kappa_3} \mu_2^j)\mult \bigmult_{j=\kappa_3+1}^{\kappa_4} \mu_2^j)\subseteq \mi{dom}({\mc V}_{\iota-1}), 
\end{alignat*}
\begin{alignat*}{1}
& \upsilon_1=\mi{reduced}^-(((\bigmult_{j=1}^{\kappa_1} \mu_2^j)\mult (\bigmult_{j=\kappa_1+1}^{\kappa_2} \mu_2^j)\mult 
                             (\bigmult_{j=\kappa_2+1}^{\kappa_3} \mu_1^j)\mult \bigmult_{j=\kappa_3+1}^{\kappa_4} \mu_1^j)\gle \\
&\phantom{\upsilon_1=\mi{reduced}^-(} \quad
                            ((\bigmult_{j=1}^{\kappa_1} \mu_1^j)\mult (\bigmult_{j=\kappa_1+1}^{\kappa_2} \mu_1^j)\mult 
                             (\bigmult_{j=\kappa_2+1}^{\kappa_3} \mu_2^j)\mult \bigmult_{j=\kappa_3+1}^{\kappa_4} \mu_2^j)), \\[1mm]
& \varepsilon_2=\mi{reduced}^+(((\bigmult_{j=1}^{\kappa_1} \mu_2^j)\mult (\bigmult_{j=\kappa_1+1}^{\kappa_2} \mu_2^j)\mult 
                                (\bigmult_{j=\kappa_2+1}^{\kappa_3} \mu_1^j)\mult \bigmult_{j=\kappa_3+1}^{\kappa_4} \mu_1^j)\gle \\
&\phantom{\varepsilon_2=\mi{reduced}^+(} \quad
                               ((\bigmult_{j=1}^{\kappa_1} \mu_1^j)\mult (\bigmult_{j=\kappa_1+1}^{\kappa_2} \mu_1^j)\mult 
                                (\bigmult_{j=\kappa_2+1}^{\kappa_3} \mu_2^j)\mult \bigmult_{j=\kappa_3+1}^{\kappa_4} \mu_2^j));
\end{alignat*}
by the induction hypothesis for $\iota-1<\iota$,
$\|\upsilon_1\|^{{\mc V}_{\iota-1}}, \|\varepsilon_2\|^{{\mc V}_{\iota-1}},
 \|\mu_1^1\|^{{\mc V}_{\iota-1}},\dots,                                                                                                                                                    \linebreak[4]
                                       \|\mu_1^{\kappa_4}\|^{{\mc V}_{\iota-1}}, \|\mu_2^1\|^{{\mc V}_{\iota-1}},\dots,\|\mu_2^{\kappa_4}\|^{{\mc V}_{\iota-1}},
 \|\mi{reduced}^-(((\bigmult_{j=1}^{\kappa_1} \mu_2^j)\mult (\bigmult_{j=\kappa_1+1}^{\kappa_2} \mu_2^j)\mult 
                   (\bigmult_{j=\kappa_2+1}^{\kappa_3} \mu_1^j)\mult \bigmult_{j=\kappa_3+1}^{\kappa_4} \mu_1^j)\gle
                  ((\bigmult_{j=1}^{\kappa_1} \mu_1^j)\mult (\bigmult_{j=\kappa_1+1}^{\kappa_2} \mu_1^j)\mult 
                   (\bigmult_{j=\kappa_2+1}^{\kappa_3} \mu_2^j)\mult \bigmult_{j=\kappa_3+1}^{\kappa_4} \mu_2^j))\|^{{\mc V}_{\iota-1}}, 
 \|\mi{reduced}^+(((\bigmult_{j=1}^{\kappa_1} \mu_2^j)\mult (\bigmult_{j=\kappa_1+1}^{\kappa_2} \mu_2^j)\mult 
                   (\bigmult_{j=\kappa_2+1}^{\kappa_3} \mu_1^j)\mult \bigmult_{j=\kappa_3+1}^{\kappa_4} \mu_1^j)\gle
                  ((\bigmult_{j=1}^{\kappa_1} \mu_1^j)\mult (\bigmult_{j=\kappa_1+1}^{\kappa_2} \mu_1^j)\mult 
                   (\bigmult_{j=\kappa_2+1}^{\kappa_3} \mu_2^j)\mult \bigmult_{j=\kappa_3+1}^{\kappa_4} \mu_2^j))\|^{{\mc V}_{\iota-1}}\underset{\text{(a)}}{\in} (0,1]$;
$\mbb{E}_{\iota-1}=\emptyset$;
for all $1\leq j\leq \kappa_1$, 
$\mu_2^j\gleq (\mu_1^j\mult a_\iota^{\beta_j})\in \mi{min}(\mi{DE}_{\iota-1})$, 
$\left(\frac{\|\mu_2^j\|^{{\mc V}_{\iota-1}}}
            {\|\mu_1^j\|^{{\mc V}_{\iota-1}}}\right)^{\frac{1}{\beta_j}}\in \mbb{DE}_{\iota-1}$,
$\left(\frac{\|\mu_2^j\|^{{\mc V}_{\iota-1}}}
            {\|\mu_1^j\|^{{\mc V}_{\iota-1}}}\right)^{\frac{1}{\beta_j}}\leq \bigfvee \mbb{DE}_{\iota-1}\underset{\text{(\ref{eq8j}a)}}{\leq} {\mc V}_\iota(a_\iota)=\delta_\iota$,
$\frac{\|\mu_2^j\|^{{\mc V}_{\iota-1}}}
      {\|\mu_1^j\|^{{\mc V}_{\iota-1}}}\leq {\mc V}_\iota(a_\iota)^{\beta_j}$;
for all $\kappa_1+1\leq j\leq \kappa_2$, 
$\mu_2^j\gle (\mu_1^j\mult a_\iota^{\beta_j})\in \mi{min}(D_{\iota-1})$, 
$\left(\frac{\|\mu_2^j\|^{{\mc V}_{\iota-1}}}
            {\|\mu_1^j\|^{{\mc V}_{\iota-1}}}\right)^{\frac{1}{\beta_j}}\in \mbb{D}_{\iota-1}$,
$\left(\frac{\|\mu_2^j\|^{{\mc V}_{\iota-1}}}
            {\|\mu_1^j\|^{{\mc V}_{\iota-1}}}\right)^{\frac{1}{\beta_j}}\leq \bigfvee \mbb{D}_{\iota-1}\underset{\text{(\ref{eq8j}b)}}{<} {\mc V}_\iota(a_\iota)=\delta_\iota$,
$\frac{\|\mu_2^j\|^{{\mc V}_{\iota-1}}}
      {\|\mu_1^j\|^{{\mc V}_{\iota-1}}}<{\mc V}_\iota(a_\iota)^{\beta_j}$;
for all $\kappa_2+1\leq j\leq \kappa_3$,
$(\mu_1^j\mult a_\iota^{\beta_j})\gleq \mu_2^j\in \mi{min}(\mi{UE}_{\iota-1})$,
$\mi{min}\left(\left(\frac{\|\mu_2^j\|^{{\mc V}_{\iota-1}}}
                          {\|\mu_1^j\|^{{\mc V}_{\iota-1}}}\right)^{\frac{1}{\beta_j}},1\right)\in \mbb{UE}_{\iota-1}$,
${\mc V}_\iota(a_\iota)=\delta_\iota\underset{\text{(\ref{eq8j}a)}}{\leq} \bigfwedge \mbb{UE}_{\iota-1}\leq 
 \mi{min}\left(\left(\frac{\|\mu_2^j\|^{{\mc V}_{\iota-1}}}
                          {\|\mu_1^j\|^{{\mc V}_{\iota-1}}}\right)^{\frac{1}{\beta_j}},1\right)\leq
 \left(\frac{\|\mu_2^j\|^{{\mc V}_{\iota-1}}}
            {\|\mu_1^j\|^{{\mc V}_{\iota-1}}}\right)^{\frac{1}{\beta_j}}$,
${\mc V}_\iota(a_\iota)^{\beta_j}\leq \frac{\|\mu_2^j\|^{{\mc V}_{\iota-1}}}
                                           {\|\mu_1^j\|^{{\mc V}_{\iota-1}}}$,
$\frac{\|\mu_1^j\|^{{\mc V}_{\iota-1}}}
      {\|\mu_2^j\|^{{\mc V}_{\iota-1}}}\leq \frac{1}
                                                 {{\mc V}_\iota(a_\iota)^{\beta_j}}$;
$\kappa_3<\kappa_4$, $\kappa_3+1\leq \kappa_4$,
$\{\kappa_3+1,\dots,\kappa_4\}\neq \emptyset$,
$(\mu_1^{\kappa_3+1}\mult a_\iota^{\beta_{\kappa_3+1}})\gle \mu_2^{\kappa_3+1}\in \mi{min}(U_{\iota-1})$,
$\mi{min}\left(\left(\frac{\|\mu_2^{\kappa_3+1}\|^{{\mc V}_{\iota-1}}}
                          {\|\mu_1^{\kappa_3+1}\|^{{\mc V}_{\iota-1}}}\right)^{\frac{1}{\beta_{\kappa_3+1}}},1\right)\in \mbb{U}_{\iota-1}\neq \emptyset$;
for all $\kappa_3+1\leq j\leq \kappa_4$,
$(\mu_1^j\mult a_\iota^{\beta_j})\gle \mu_2^j\in \mi{min}(U_{\iota-1})$,
$\mi{min}\left(\left(\frac{\|\mu_2^j\|^{{\mc V}_{\iota-1}}}
                          {\|\mu_1^j\|^{{\mc V}_{\iota-1}}}\right)^{\frac{1}{\beta_j}},1\right)\in \mbb{U}_{\iota-1}$,
${\mc V}_\iota(a_\iota)=\delta_\iota\underset{\text{(\ref{eq8j}c)}}{<} \bigfwedge \mbb{U}_{\iota-1}\leq 
 \mi{min}\left(\left(\frac{\|\mu_2^j\|^{{\mc V}_{\iota-1}}}
                          {\|\mu_1^j\|^{{\mc V}_{\iota-1}}}\right)^{\frac{1}{\beta_j}},1\right)\leq
 \left(\frac{\|\mu_2^j\|^{{\mc V}_{\iota-1}}}
            {\|\mu_1^j\|^{{\mc V}_{\iota-1}}}\right)^{\frac{1}{\beta_j}}$,
${\mc V}_\iota(a_\iota)^{\beta_j}<\frac{\|\mu_2^j\|^{{\mc V}_{\iota-1}}}
                                       {\|\mu_1^j\|^{{\mc V}_{\iota-1}}}$,
$\frac{\|\mu_1^j\|^{{\mc V}_{\iota-1}}}
      {\|\mu_2^j\|^{{\mc V}_{\iota-1}}}<\frac{1}
                                             {{\mc V}_\iota(a_\iota)^{\beta_j}}$;
\begin{alignat*}{1}
& \dfrac{\|\upsilon_1\|^{{\mc V}_{\iota-1}}}
        {\|\varepsilon_2\|^{{\mc V}_{\iota-1}}}= \\
& \dfrac{\left\|\mi{reduced}^-\left(\begin{array}{l}
                                    ((\bigmult_{j=1}^{\kappa_1} \mu_2^j)\mult (\bigmult_{j=\kappa_1+1}^{\kappa_2} \mu_2^j)\mult 
                                     (\bigmult_{j=\kappa_2+1}^{\kappa_3} \mu_1^j)\mult \bigmult_{j=\kappa_3+1}^{\kappa_4} \mu_1^j)\gle \\
                                    \quad
                                    ((\bigmult_{j=1}^{\kappa_1} \mu_1^j)\mult (\bigmult_{j=\kappa_1+1}^{\kappa_2} \mu_1^j)\mult 
                                     (\bigmult_{j=\kappa_2+1}^{\kappa_3} \mu_2^j)\mult \bigmult_{j=\kappa_3+1}^{\kappa_4} \mu_2^j)
                                    \end{array}\right)\right\|^{{\mc V}_{\iota-1}}}
        {\left\|\mi{reduced}^+\left(\begin{array}{l}
                                    ((\bigmult_{j=1}^{\kappa_1} \mu_2^j)\mult (\bigmult_{j=\kappa_1+1}^{\kappa_2} \mu_2^j)\mult 
                                     (\bigmult_{j=\kappa_2+1}^{\kappa_3} \mu_1^j)\mult \bigmult_{j=\kappa_3+1}^{\kappa_4} \mu_1^j)\gle \\
                                    \quad
                                    ((\bigmult_{j=1}^{\kappa_1} \mu_1^j)\mult (\bigmult_{j=\kappa_1+1}^{\kappa_2} \mu_1^j)\mult 
                                     (\bigmult_{j=\kappa_2+1}^{\kappa_3} \mu_2^j)\mult \bigmult_{j=\kappa_3+1}^{\kappa_4} \mu_2^j)
                                    \end{array}\right)\right\|^{{\mc V}_{\iota-1}}}\overset{\text{(\ref{eq7n})}}{=\!\!=} \\
& \dfrac{\|(\bigmult_{j=1}^{\kappa_1} \mu_2^j)\mult (\bigmult_{j=\kappa_1+1}^{\kappa_2} \mu_2^j)\mult 
           (\bigmult_{j=\kappa_2+1}^{\kappa_3} \mu_1^j)\mult \bigmult_{j=\kappa_3+1}^{\kappa_4} \mu_1^j\|^{{\mc V}_{\iota-1}}}
        {\|(\bigmult_{j=1}^{\kappa_1} \mu_1^j)\mult (\bigmult_{j=\kappa_1+1}^{\kappa_2} \mu_1^j)\mult 
           (\bigmult_{j=\kappa_2+1}^{\kappa_3} \mu_2^j)\mult \bigmult_{j=\kappa_3+1}^{\kappa_4} \mu_2^j\|^{{\mc V}_{\iota-1}}}= \\
& \dfrac{(\bigfswedge_{j=1}^{\kappa_1} \|\mu_2^j\|^{{\mc V}_{\iota-1}})\fswedge (\bigfswedge_{j=\kappa_1+1}^{\kappa_2} \|\mu_2^j\|^{{\mc V}_{\iota-1}})\fswedge
         (\bigfswedge_{j=\kappa_2+1}^{\kappa_3} \|\mu_1^j\|^{{\mc V}_{\iota-1}})\fswedge \bigfswedge_{j=\kappa_3+1}^{\kappa_4} \|\mu_1^j\|^{{\mc V}_{\iota-1}}}
        {(\bigfswedge_{j=1}^{\kappa_1} \|\mu_1^j\|^{{\mc V}_{\iota-1}})\fswedge (\bigfswedge_{j=\kappa_1+1}^{\kappa_2} \|\mu_1^j\|^{{\mc V}_{\iota-1}})\fswedge
         (\bigfswedge_{j=\kappa_2+1}^{\kappa_3} \|\mu_2^j\|^{{\mc V}_{\iota-1}})\fswedge \bigfswedge_{j=\kappa_3+1}^{\kappa_4} \|\mu_2^j\|^{{\mc V}_{\iota-1}}}= \\
& \left(\bigfswedge_{j=1}^{\kappa_1} \dfrac{\|\mu_2^j\|^{{\mc V}_{\iota-1}}}
                                           {\|\mu_1^j\|^{{\mc V}_{\iota-1}}}\right)\fswedge 
  \left(\bigfswedge_{j=\kappa_1+1}^{\kappa_2} \dfrac{\|\mu_2^j\|^{{\mc V}_{\iota-1}}}
                                                    {\|\mu_1^j\|^{{\mc V}_{\iota-1}}}\right)\fswedge 
  \left(\bigfswedge_{j=\kappa_1+2}^{\kappa_3} \dfrac{\|\mu_1^j\|^{{\mc V}_{\iota-1}}}
                                                    {\|\mu_2^j\|^{{\mc V}_{\iota-1}}}\right)\fswedge
  \bigfswedge_{j=\kappa_1+3}^{\kappa_4} \dfrac{\|\mu_1^j\|^{{\mc V}_{\iota-1}}}
                                              {\|\mu_2^j\|^{{\mc V}_{\iota-1}}}< \\
& \left(\bigfswedge_{j=1}^{\kappa_1} {\mc V}_\iota(a_\iota)^{\beta_j}\right)\fswedge \left(\bigfswedge_{j=\kappa_1+1}^{\kappa_2} {\mc V}_\iota(a_\iota)^{\beta_j}\right)\fswedge 
  \left(\bigfswedge_{j=\kappa_1+2}^{\kappa_3} \dfrac{1}
                                                    {{\mc V}_\iota(a_\iota)^{\beta_j}}\right)\fswedge
  \bigfswedge_{j=\kappa_1+3}^{\kappa_4} \dfrac{1}
                                              {{\mc V}_\iota(a_\iota)^{\beta_j}}= \\
& {\mc V}_\iota(a_\iota)^{(\sum_{j=1}^{\kappa_1} \beta_j)+(\sum_{j=\kappa_1+1}^{\kappa_2} \beta_j)-(\sum_{j=\kappa_2+1}^{\kappa_3} \beta_j)-\sum_{j=\kappa_3+1}^{\kappa_4} \beta_j}=
  {\mc V}_\iota(a_\iota)^{-\alpha}=\dfrac{1}
                                         {{\mc V}_\iota(a_\iota)^\alpha}, \\[1mm]
& \|\varepsilon_1\|^{{\mc V}_\iota}=\|\upsilon_1\mult a_\iota^\alpha\|^{{\mc V}_\iota}=\|\upsilon_1\|^{{\mc V}_\iota}\fswedge {\mc V}_\iota(a_\iota)^\alpha\overset{\text{(b)}}{=\!\!=}
  \|\upsilon_1\|^{{\mc V}_{\iota-1}}\fswedge {\mc V}_\iota(a_\iota)^\alpha<\|\varepsilon_2\|^{{\mc V}_{\iota-1}}\overset{\text{(b)}}{=\!\!=} \|\varepsilon_2\|^{{\mc V}_\iota};  
\end{alignat*}
(e) holds.

Case 2.10.2.1.2.2:
There exists $\zeta_1\diamond^\zeta \zeta_2=\mi{reduced}(\lambda^{(\gamma_1^\zeta,\dots,\gamma_n^\zeta)})\in \mi{clo}$,
$\zeta_e\in \{\gu\}\cup \mi{PropConj}_A$, $\mi{atoms}(\zeta_e)\subseteq \mi{dom}({\mc V}_{\iota-1})$, $\diamond^\zeta\in \{\geql,\gleq,\gle\}$, 
$\bs{0}^n\neq (\gamma_1^\zeta,\dots,\gamma_n^\zeta)\in \mbb{N}^n$, such that
$\varepsilon_1\gle \varepsilon_2=(\upsilon_1\mult a_\iota^\alpha)\gle \varepsilon_2=
                                 \mi{reduced}(\lambda^{((\sum_{j=1}^{\kappa_4} \gamma_1^j)+\gamma_1^\zeta,\dots,(\sum_{j=1}^{\kappa_4} \gamma_n^j)+\gamma_n^\zeta)})$.
Then 
$\mi{atoms}(\upsilon_1), \mi{atoms}(\varepsilon_2), \mi{atoms}(\mu_1^1),\dots,\mi{atoms}(\mu_1^{\kappa_4}), \mi{atoms}(\mu_2^1),\dots,\mi{atoms}(\mu_2^{\kappa_4})\subseteq \mi{dom}({\mc V}_{\iota-1})$,
\begin{alignat*}{1}
& (\upsilon_1\mult a_\iota^\alpha)\gle \varepsilon_2=
  \mi{reduced}(\lambda^{((\sum_{j=1}^{\kappa_4} \gamma_1^j)+\gamma_1^\zeta,\dots,(\sum_{j=1}^{\kappa_4} \gamma_n^j)+\gamma_n^\zeta)})\overset{\text{(\ref{eq8dddd})}}{=\!\!=} \\
& \mi{reduced}((\bigmult_{j=1}^{\kappa_1} \lambda^{(\gamma_1^j,\dots,\gamma_n^j)})\mult
               (\bigmult_{j=\kappa_1+1}^{\kappa_2} \lambda^{(\gamma_1^j,\dots,\gamma_n^j)})\mult \\
&\phantom{\mi{reduced}(}
               (\bigmult_{j=\kappa_2+1}^{\kappa_3} \lambda^{(\gamma_1^j,\dots,\gamma_n^j)})\mult
               (\bigmult_{j=\kappa_3+1}^{\kappa_4} \lambda^{(\gamma_1^j,\dots,\gamma_n^j)})\mult
               \lambda^{(\gamma_1^\zeta,\dots,\gamma_n^\zeta)})\overset{\text{(\ref{eq7k})}}{=\!\!=} \\
& \mi{reduced}((\bigmult_{j=1}^{\kappa_1} \mi{reduced}(\lambda^{(\gamma_1^j,\dots,\gamma_n^j)}))\mult
               (\bigmult_{j=\kappa_1+1}^{\kappa_2} \mi{reduced}(\lambda^{(\gamma_1^j,\dots,\gamma_n^j)}))\mult \\
&\phantom{\mi{reduced}(}
               (\bigmult_{j=\kappa_2+1}^{\kappa_3} \mi{reduced}(\lambda^{(\gamma_1^j,\dots,\gamma_n^j)}))\mult
               (\bigmult_{j=\kappa_3+1}^{\kappa_4} \mi{reduced}(\lambda^{(\gamma_1^j,\dots,\gamma_n^j)}))\mult \\
&\phantom{\mi{reduced}(}
               \mi{reduced}(\lambda^{(\gamma_1^\zeta,\dots,\gamma_n^\zeta)}))= \\
& \mi{reduced}((\bigmult_{j=1}^{\kappa_1} \mu_2^j\gleq (\mu_1^j\mult a_\iota^{\beta_j}))\mult
               (\bigmult_{j=\kappa_1+1}^{\kappa_2} \mu_2^j\gle (\mu_1^j\mult a_\iota^{\beta_j}))\mult \\
&\phantom{\mi{reduced}(}
               (\bigmult_{j=\kappa_2+1}^{\kappa_3} (\mu_1^j\mult a_\iota^{\beta_j})\gleq \mu_2^j)\mult
               (\bigmult_{j=\kappa_3+1}^{\kappa_4} (\mu_1^j\mult a_\iota^{\beta_j})\gle \mu_2^j)\mult
               (\zeta_1\diamond^\zeta \zeta_2))\overset{\text{(\ref{eq7i})}}{=\!\!=} \\
& \mi{reduced}(((\bigmult_{j=1}^{\kappa_1} \mu_2^j)\mult (\bigmult_{j=\kappa_1+1}^{\kappa_2} \mu_2^j)\mult 
                (\bigmult_{j=\kappa_2+1}^{\kappa_3} \mu_1^j)\mult (\bigmult_{j=\kappa_3+1}^{\kappa_4} \mu_1^j)\mult \zeta_1\mult \\
&\phantom{\mi{reduced}(} \quad
                a_\iota^{(\sum_{j=\kappa_2+1}^{\kappa_3} \beta_j)+\sum_{j=\kappa_3+1}^{\kappa_4} \beta_j})\gle \\
&\phantom{\mi{reduced}(} \quad
               ((\bigmult_{j=1}^{\kappa_1} \mu_1^j)\mult (\bigmult_{j=\kappa_1+1}^{\kappa_2} \mu_1^j)\mult 
                (\bigmult_{j=\kappa_2+1}^{\kappa_3} \mu_2^j)\mult (\bigmult_{j=\kappa_3+1}^{\kappa_4} \mu_2^j)\mult \zeta_2\mult \\
&\phantom{\mi{reduced}(} \quad \quad
                a_\iota^{(\sum_{j=1}^{\kappa_1} \beta_j)+\sum_{j=\kappa_1+1}^{\kappa_2} \beta_j}))= \\
& \mi{reduced}^-(((\bigmult_{j=1}^{\kappa_1} \mu_2^j)\mult (\bigmult_{j=\kappa_1+1}^{\kappa_2} \mu_2^j)\mult 
                  (\bigmult_{j=\kappa_2+1}^{\kappa_3} \mu_1^j)\mult (\bigmult_{j=\kappa_3+1}^{\kappa_4} \mu_1^j)\mult \zeta_1\mult \\
&\phantom{\mi{reduced}^-(} \quad
                  a_\iota^{(\sum_{j=\kappa_2+1}^{\kappa_3} \beta_j)+\sum_{j=\kappa_3+1}^{\kappa_4} \beta_j})\gle \\
&\phantom{\mi{reduced}^-(} \quad
                 ((\bigmult_{j=1}^{\kappa_1} \mu_1^j)\mult (\bigmult_{j=\kappa_1+1}^{\kappa_2} \mu_1^j)\mult 
                  (\bigmult_{j=\kappa_2+1}^{\kappa_3} \mu_2^j)\mult (\bigmult_{j=\kappa_3+1}^{\kappa_4} \mu_2^j)\mult \zeta_2\mult \\
&\phantom{\mi{reduced}^-(} \quad \quad
                  a_\iota^{(\sum_{j=1}^{\kappa_1} \beta_j)+\sum_{j=\kappa_1+1}^{\kappa_2} \beta_j}))\gle \\
& \mi{reduced}^+(((\bigmult_{j=1}^{\kappa_1} \mu_2^j)\mult (\bigmult_{j=\kappa_1+1}^{\kappa_2} \mu_2^j)\mult 
                  (\bigmult_{j=\kappa_2+1}^{\kappa_3} \mu_1^j)\mult (\bigmult_{j=\kappa_3+1}^{\kappa_4} \mu_1^j)\mult \zeta_1\mult \\
&\phantom{\mi{reduced}^+(} \quad
                  a_\iota^{(\sum_{j=\kappa_2+1}^{\kappa_3} \beta_j)+\sum_{j=\kappa_3+1}^{\kappa_4} \beta_j})\gle \\
&\phantom{\mi{reduced}^+(} \quad
                 ((\bigmult_{j=1}^{\kappa_1} \mu_1^j)\mult (\bigmult_{j=\kappa_1+1}^{\kappa_2} \mu_1^j)\mult 
                  (\bigmult_{j=\kappa_2+1}^{\kappa_3} \mu_2^j)\mult (\bigmult_{j=\kappa_3+1}^{\kappa_4} \mu_2^j)\mult \zeta_2\mult \\
&\phantom{\mi{reduced}^+(} \quad \quad
                  a_\iota^{(\sum_{j=1}^{\kappa_1} \beta_j)+\sum_{j=\kappa_1+1}^{\kappa_2} \beta_j})), \\[1mm]
& \#(a_\iota,(\upsilon_1\mult a_\iota^\alpha)\gle \varepsilon_2)=-\alpha= \\
& \#(a_\iota,\mi{reduced}(((\bigmult_{j=1}^{\kappa_1} \mu_2^j)\mult (\bigmult_{j=\kappa_1+1}^{\kappa_2} \mu_2^j)\mult 
                           (\bigmult_{j=\kappa_2+1}^{\kappa_3} \mu_1^j)\mult (\bigmult_{j=\kappa_3+1}^{\kappa_4} \mu_1^j)\mult \zeta_1\mult \\
&\phantom{\#(a_\iota,\mi{reduced}(} \quad
                           a_\iota^{(\sum_{j=\kappa_2+1}^{\kappa_3} \beta_j)+\sum_{j=\kappa_3+1}^{\kappa_4} \beta_j})\gle \\
&\phantom{\#(a_\iota,\mi{reduced}(} \quad
                          ((\bigmult_{j=1}^{\kappa_1} \mu_1^j)\mult (\bigmult_{j=\kappa_1+1}^{\kappa_2} \mu_1^j)\mult 
                           (\bigmult_{j=\kappa_2+1}^{\kappa_3} \mu_2^j)\mult (\bigmult_{j=\kappa_3+1}^{\kappa_4} \mu_2^j)\mult \zeta_2\mult \\
&\phantom{\#(a_\iota,\mi{reduced}(} \quad \quad
                           a_\iota^{(\sum_{j=1}^{\kappa_1} \beta_j)+\sum_{j=\kappa_1+1}^{\kappa_2} \beta_j})))\overset{\text{(\ref{eq7aaaax})}}{=\!\!=} \\
& \#(a_\iota,((\bigmult_{j=1}^{\kappa_1} \mu_2^j)\mult (\bigmult_{j=\kappa_1+1}^{\kappa_2} \mu_2^j)\mult 
              (\bigmult_{j=\kappa_2+1}^{\kappa_3} \mu_1^j)\mult (\bigmult_{j=\kappa_3+1}^{\kappa_4} \mu_1^j)\mult \zeta_1\mult \\
&\phantom{\#(a_\iota,} \quad
              a_\iota^{(\sum_{j=\kappa_2+1}^{\kappa_3} \beta_j)+\sum_{j=\kappa_3+1}^{\kappa_4} \beta_j})\gle \\
&\phantom{\#(a_\iota,} \quad
             ((\bigmult_{j=1}^{\kappa_1} \mu_1^j)\mult (\bigmult_{j=\kappa_1+1}^{\kappa_2} \mu_1^j)\mult 
              (\bigmult_{j=\kappa_2+1}^{\kappa_3} \mu_2^j)\mult (\bigmult_{j=\kappa_3+1}^{\kappa_4} \mu_2^j)\mult \zeta_2\mult \\
&\phantom{\#(a_\iota,} \quad \quad
              a_\iota^{(\sum_{j=1}^{\kappa_1} \beta_j)+\sum_{j=\kappa_1+1}^{\kappa_2} \beta_j}))= \\
& \left(\sum_{j=1}^{\kappa_1} \beta_j\right)+\left(\sum_{j=\kappa_1+1}^{\kappa_2} \beta_j\right)-\left(\sum_{j=\kappa_2+1}^{\kappa_3} \beta_j\right)-\sum_{j=\kappa_3+1}^{\kappa_4} \beta_j\leq -1, 
\end{alignat*}
$a_\iota\not\in \mi{dom}({\mc V}_{\iota-1})\supseteq \mi{atoms}((\bigmult_{j=1}^{\kappa_1} \mu_2^j)\mult (\bigmult_{j=\kappa_1+1}^{\kappa_2} \mu_2^j)\mult 
                                                                (\bigmult_{j=\kappa_2+1}^{\kappa_3} \mu_1^j)\mult                                                                          \linebreak[4]
                                                                                                                  (\bigmult_{j=\kappa_3+1}^{\kappa_4} \mu_1^j)\mult \zeta_1)=
                                                     (\bigcup_{j=1}^{\kappa_1} \mi{atoms}(\mu_2^j))\cup (\bigcup_{j=\kappa_1+1}^{\kappa_2} \mi{atoms}(\mu_2^j))\cup                        \linebreak[4]
                                                     (\bigcup_{j=\kappa_2+1}^{\kappa_3} \mi{atoms}(\mu_1^j))\cup (\bigcup_{j=\kappa_3+1}^{\kappa_4} \mi{atoms}(\mu_1^j))\cup \mi{atoms}(\zeta_1)$,
$a_\iota\not\in \mi{dom}({\mc V}_{\iota-1})\supseteq \mi{atoms}((\bigmult_{j=1}^{\kappa_1} \mu_1^j)\mult (\bigmult_{j=\kappa_1+1}^{\kappa_2} \mu_1^j)\mult 
                                                                (\bigmult_{j=\kappa_2+1}^{\kappa_3} \mu_2^j)\mult (\bigmult_{j=\kappa_3+1}^{\kappa_4} \mu_2^j)\mult \zeta_2)=              \linebreak[4]
                                                     (\bigcup_{j=1}^{\kappa_1} \mi{atoms}(\mu_1^j))\cup (\bigcup_{j=\kappa_1+1}^{\kappa_2} \mi{atoms}(\mu_1^j))\cup 
                                                     (\bigcup_{j=\kappa_2+1}^{\kappa_3} \mi{atoms}(\mu_2^j))\cup (\bigcup_{j=\kappa_3+1}^{\kappa_4} \mi{atoms}(\mu_2^j))\cup \mi{atoms}(\zeta_2)$,
\begin{alignat*}{1}
& \mi{reduced}^-(((\bigmult_{j=1}^{\kappa_1} \mu_2^j)\mult (\bigmult_{j=\kappa_1+1}^{\kappa_2} \mu_2^j)\mult 
                  (\bigmult_{j=\kappa_2+1}^{\kappa_3} \mu_1^j)\mult (\bigmult_{j=\kappa_3+1}^{\kappa_4} \mu_1^j)\mult \zeta_1\mult \\
&\phantom{\mi{reduced}^-(} \quad
                  a_\iota^{(\sum_{j=\kappa_2+1}^{\kappa_3} \beta_j)+\sum_{j=\kappa_3+1}^{\kappa_4} \beta_j})\gle \\
&\phantom{\mi{reduced}^-(} \quad
                 ((\bigmult_{j=1}^{\kappa_1} \mu_1^j)\mult (\bigmult_{j=\kappa_1+1}^{\kappa_2} \mu_1^j)\mult 
                  (\bigmult_{j=\kappa_2+1}^{\kappa_3} \mu_2^j)\mult (\bigmult_{j=\kappa_3+1}^{\kappa_4} \mu_2^j)\mult \zeta_2\mult \\
&\phantom{\mi{reduced}^-(} \quad \quad
                  a_\iota^{(\sum_{j=1}^{\kappa_1} \beta_j)+\sum_{j=\kappa_1+1}^{\kappa_2} \beta_j}))\overset{\text{(\ref{eq7ee})}}{=\!\!=} \\
& \mi{reduced}^-(((\bigmult_{j=1}^{\kappa_1} \mu_2^j)\mult (\bigmult_{j=\kappa_1+1}^{\kappa_2} \mu_2^j)\mult 
                  (\bigmult_{j=\kappa_2+1}^{\kappa_3} \mu_1^j)\mult (\bigmult_{j=\kappa_3+1}^{\kappa_4} \mu_1^j)\mult \zeta_1)\gle \\
&\phantom{\mi{reduced}^-(} \quad
                 ((\bigmult_{j=1}^{\kappa_1} \mu_1^j)\mult (\bigmult_{j=\kappa_1+1}^{\kappa_2} \mu_1^j)\mult 
                  (\bigmult_{j=\kappa_2+1}^{\kappa_3} \mu_2^j)\mult (\bigmult_{j=\kappa_3+1}^{\kappa_4} \mu_2^j)\mult \zeta_2))\mult \\
& a_\iota^{(\sum_{j=\kappa_2+1}^{\kappa_3} \beta_j)+(\sum_{j=\kappa_3+1}^{\kappa_4} \beta_j)-(\sum_{j=1}^{\kappa_1} \beta_j)-\sum_{j=\kappa_1+1}^{\kappa_2} \beta_j}= \\
& \mi{reduced}^-(((\bigmult_{j=1}^{\kappa_1} \mu_2^j)\mult (\bigmult_{j=\kappa_1+1}^{\kappa_2} \mu_2^j)\mult 
                  (\bigmult_{j=\kappa_2+1}^{\kappa_3} \mu_1^j)\mult (\bigmult_{j=\kappa_3+1}^{\kappa_4} \mu_1^j)\mult \zeta_1)\gle \\
&\phantom{\mi{reduced}^-(} \quad
                 ((\bigmult_{j=1}^{\kappa_1} \mu_1^j)\mult (\bigmult_{j=\kappa_1+1}^{\kappa_2} \mu_1^j)\mult 
                  (\bigmult_{j=\kappa_2+1}^{\kappa_3} \mu_2^j)\mult (\bigmult_{j=\kappa_3+1}^{\kappa_4} \mu_2^j)\mult \zeta_2))\mult a_\iota^\alpha, \\[1mm]
& \mi{reduced}^+(((\bigmult_{j=1}^{\kappa_1} \mu_2^j)\mult (\bigmult_{j=\kappa_1+1}^{\kappa_2} \mu_2^j)\mult 
                  (\bigmult_{j=\kappa_2+1}^{\kappa_3} \mu_1^j)\mult (\bigmult_{j=\kappa_3+1}^{\kappa_4} \mu_1^j)\mult \zeta_1\mult \\
&\phantom{\mi{reduced}^+(} \quad
                  a_\iota^{(\sum_{j=\kappa_2+1}^{\kappa_3} \beta_j)+\sum_{j=\kappa_3+1}^{\kappa_4} \beta_j})\gle \\
&\phantom{\mi{reduced}^+(} \quad
                 ((\bigmult_{j=1}^{\kappa_1} \mu_1^j)\mult (\bigmult_{j=\kappa_1+1}^{\kappa_2} \mu_1^j)\mult 
                  (\bigmult_{j=\kappa_2+1}^{\kappa_3} \mu_2^j)\mult (\bigmult_{j=\kappa_3+1}^{\kappa_4} \mu_2^j)\mult \zeta_2\mult \\
&\phantom{\mi{reduced}^+(} \quad \quad
                  a_\iota^{(\sum_{j=1}^{\kappa_1} \beta_j)+\sum_{j=\kappa_1+1}^{\kappa_2} \beta_j}))\overset{\text{(\ref{eq7dd})}}{=\!\!=} \\
& \mi{reduced}^+(((\bigmult_{j=1}^{\kappa_1} \mu_2^j)\mult (\bigmult_{j=\kappa_1+1}^{\kappa_2} \mu_2^j)\mult 
                  (\bigmult_{j=\kappa_2+1}^{\kappa_3} \mu_1^j)\mult (\bigmult_{j=\kappa_3+1}^{\kappa_4} \mu_1^j)\mult \zeta_1)\gle \\
&\phantom{\mi{reduced}^+(} \quad
                 ((\bigmult_{j=1}^{\kappa_1} \mu_1^j)\mult (\bigmult_{j=\kappa_1+1}^{\kappa_2} \mu_1^j)\mult 
                  (\bigmult_{j=\kappa_2+1}^{\kappa_3} \mu_2^j)\mult (\bigmult_{j=\kappa_3+1}^{\kappa_4} \mu_2^j)\mult \zeta_2)), \\[1mm]
& (\upsilon_1\mult a_\iota^\alpha)\gle \varepsilon_2= \\
& (\mi{reduced}^-(((\bigmult_{j=1}^{\kappa_1} \mu_2^j)\mult (\bigmult_{j=\kappa_1+1}^{\kappa_2} \mu_2^j)\mult 
                   (\bigmult_{j=\kappa_2+1}^{\kappa_3} \mu_1^j)\mult (\bigmult_{j=\kappa_3+1}^{\kappa_4} \mu_1^j)\mult \zeta_1)\gle \\
&\phantom{(\mi{reduced}^-(} \quad
                  ((\bigmult_{j=1}^{\kappa_1} \mu_1^j)\mult (\bigmult_{j=\kappa_1+1}^{\kappa_2} \mu_1^j)\mult 
                   (\bigmult_{j=\kappa_2+1}^{\kappa_3} \mu_2^j)\mult (\bigmult_{j=\kappa_3+1}^{\kappa_4} \mu_2^j)\mult \zeta_2))\mult a_\iota^\alpha)\gle \\
& \mi{reduced}^+(((\bigmult_{j=1}^{\kappa_1} \mu_2^j)\mult (\bigmult_{j=\kappa_1+1}^{\kappa_2} \mu_2^j)\mult 
                  (\bigmult_{j=\kappa_2+1}^{\kappa_3} \mu_1^j)\mult (\bigmult_{j=\kappa_3+1}^{\kappa_4} \mu_1^j)\mult \zeta_1)\gle \\
&\phantom{\mi{reduced}^+(} \quad
                 ((\bigmult_{j=1}^{\kappa_1} \mu_1^j)\mult (\bigmult_{j=\kappa_1+1}^{\kappa_2} \mu_1^j)\mult 
                  (\bigmult_{j=\kappa_2+1}^{\kappa_3} \mu_2^j)\mult (\bigmult_{j=\kappa_3+1}^{\kappa_4} \mu_2^j)\mult \zeta_2)), \\[1mm]
& \mi{atoms}(\mi{reduced}^-(((\bigmult_{j=1}^{\kappa_1} \mu_2^j)\mult (\bigmult_{j=\kappa_1+1}^{\kappa_2} \mu_2^j)\mult 
                             (\bigmult_{j=\kappa_2+1}^{\kappa_3} \mu_1^j)\mult (\bigmult_{j=\kappa_3+1}^{\kappa_4} \mu_1^j)\mult \zeta_1)\gle \\
&\phantom{\mi{atoms}(\mi{reduced}^-(} \quad
                            ((\bigmult_{j=1}^{\kappa_1} \mu_1^j)\mult (\bigmult_{j=\kappa_1+1}^{\kappa_2} \mu_1^j)\mult 
                             (\bigmult_{j=\kappa_2+1}^{\kappa_3} \mu_2^j)\mult (\bigmult_{j=\kappa_3+1}^{\kappa_4} \mu_2^j)\mult \zeta_2)))\underset{\text{(\ref{eq7a})}}{\subseteq} \\
& \mi{atoms}((\bigmult_{j=1}^{\kappa_1} \mu_2^j)\mult (\bigmult_{j=\kappa_1+1}^{\kappa_2} \mu_2^j)\mult 
             (\bigmult_{j=\kappa_2+1}^{\kappa_3} \mu_1^j)\mult (\bigmult_{j=\kappa_3+1}^{\kappa_4} \mu_1^j)\mult \zeta_1)\subseteq \mi{dom}({\mc V}_{\iota-1}), \\[1mm]
& \mi{atoms}(\mi{reduced}^+(((\bigmult_{j=1}^{\kappa_1} \mu_2^j)\mult (\bigmult_{j=\kappa_1+1}^{\kappa_2} \mu_2^j)\mult 
                             (\bigmult_{j=\kappa_2+1}^{\kappa_3} \mu_1^j)\mult (\bigmult_{j=\kappa_3+1}^{\kappa_4} \mu_1^j)\mult \zeta_1)\gle \\
&\phantom{\mi{atoms}(\mi{reduced}^+(} \quad
                            ((\bigmult_{j=1}^{\kappa_1} \mu_1^j)\mult (\bigmult_{j=\kappa_1+1}^{\kappa_2} \mu_1^j)\mult 
                             (\bigmult_{j=\kappa_2+1}^{\kappa_3} \mu_2^j)\mult (\bigmult_{j=\kappa_3+1}^{\kappa_4} \mu_2^j)\mult \zeta_2)))\underset{\text{(\ref{eq7b})}}{\subseteq} \\
& \mi{atoms}((\bigmult_{j=1}^{\kappa_1} \mu_1^j)\mult (\bigmult_{j=\kappa_1+1}^{\kappa_2} \mu_1^j)\mult 
             (\bigmult_{j=\kappa_2+1}^{\kappa_3} \mu_2^j)\mult (\bigmult_{j=\kappa_3+1}^{\kappa_4} \mu_2^j)\mult \zeta_2)\subseteq \mi{dom}({\mc V}_{\iota-1}), 
\end{alignat*}
\begin{alignat*}{1}
& \upsilon_1=\mi{reduced}^-(((\bigmult_{j=1}^{\kappa_1} \mu_2^j)\mult (\bigmult_{j=\kappa_1+1}^{\kappa_2} \mu_2^j)\mult 
                             (\bigmult_{j=\kappa_2+1}^{\kappa_3} \mu_1^j)\mult (\bigmult_{j=\kappa_3+1}^{\kappa_4} \mu_1^j)\mult \zeta_1)\gle \\
&\phantom{\upsilon_1=\mi{reduced}^-(} \quad
                            ((\bigmult_{j=1}^{\kappa_1} \mu_1^j)\mult (\bigmult_{j=\kappa_1+1}^{\kappa_2} \mu_1^j)\mult 
                             (\bigmult_{j=\kappa_2+1}^{\kappa_3} \mu_2^j)\mult (\bigmult_{j=\kappa_3+1}^{\kappa_4} \mu_2^j)\mult \zeta_2)), \\[1mm]
& \varepsilon_2=\mi{reduced}^+(((\bigmult_{j=1}^{\kappa_1} \mu_2^j)\mult (\bigmult_{j=\kappa_1+1}^{\kappa_2} \mu_2^j)\mult 
                                (\bigmult_{j=\kappa_2+1}^{\kappa_3} \mu_1^j)\mult (\bigmult_{j=\kappa_3+1}^{\kappa_4} \mu_1^j)\mult \zeta_1)\gle \\
&\phantom{\varepsilon_2=\mi{reduced}^+(} \quad
                               ((\bigmult_{j=1}^{\kappa_1} \mu_1^j)\mult (\bigmult_{j=\kappa_1+1}^{\kappa_2} \mu_1^j)\mult 
                                (\bigmult_{j=\kappa_2+1}^{\kappa_3} \mu_2^j)\mult (\bigmult_{j=\kappa_3+1}^{\kappa_4} \mu_2^j)\mult \zeta_2));
\end{alignat*}
by the induction hypothesis for $\iota-1<\iota$,
$\|\upsilon_1\|^{{\mc V}_{\iota-1}}, \|\varepsilon_2\|^{{\mc V}_{\iota-1}},
 \|\mu_1^1\|^{{\mc V}_{\iota-1}},\dots,                                                                                                                                                    \linebreak[4]
                                       \|\mu_1^{\kappa_4}\|^{{\mc V}_{\iota-1}}, \|\mu_2^1\|^{{\mc V}_{\iota-1}},\dots,\|\mu_2^{\kappa_4}\|^{{\mc V}_{\iota-1}},
 \|\zeta_1\|^{{\mc V}_{\iota-1}}, \|\zeta_2\|^{{\mc V}_{\iota-1}},
 \|\mi{reduced}^-(((\bigmult_{j=1}^{\kappa_1} \mu_2^j)\mult (\bigmult_{j=\kappa_1+1}^{\kappa_2} \mu_2^j)\mult 
                   (\bigmult_{j=\kappa_2+1}^{\kappa_3} \mu_1^j)\mult (\bigmult_{j=\kappa_3+1}^{\kappa_4} \mu_1^j)\mult \zeta_1)\gle
                  ((\bigmult_{j=1}^{\kappa_1} \mu_1^j)\mult (\bigmult_{j=\kappa_1+1}^{\kappa_2} \mu_1^j)\mult 
                   (\bigmult_{j=\kappa_2+1}^{\kappa_3} \mu_2^j)\mult (\bigmult_{j=\kappa_3+1}^{\kappa_4} \mu_2^j)\mult \zeta_2))\|^{{\mc V}_{\iota-1}}, 
 \|\mi{reduced}^+(((\bigmult_{j=1}^{\kappa_1} \mu_2^j)\mult (\bigmult_{j=\kappa_1+1}^{\kappa_2} \mu_2^j)\mult 
                   (\bigmult_{j=\kappa_2+1}^{\kappa_3} \mu_1^j)\mult (\bigmult_{j=\kappa_3+1}^{\kappa_4} \mu_1^j)\mult \zeta_1)\gle
                  ((\bigmult_{j=1}^{\kappa_1} \mu_1^j)\mult (\bigmult_{j=\kappa_1+1}^{\kappa_2} \mu_1^j)\mult 
                   (\bigmult_{j=\kappa_2+1}^{\kappa_3} \mu_2^j)\mult (\bigmult_{j=\kappa_3+1}^{\kappa_4} \mu_2^j)\mult \zeta_2))\|^{{\mc V}_{\iota-1}}\underset{\text{(a)}}{\in} (0,1]$,
either $\diamond^\zeta=\geql$, $\zeta_1\diamond^\zeta \zeta_2=\zeta_1\geql \zeta_2\in \mi{clo}$, $\|\zeta_1\|^{{\mc V}_{\iota-1}}\overset{\text{(c)}}{=\!\!=} \|\zeta_2\|^{{\mc V}_{\iota-1}}$, 
or $\diamond^\zeta=\gleq$, $\zeta_1\diamond^\zeta \zeta_2=\zeta_1\gleq \zeta_2\in \mi{clo}$, $\|\zeta_1\|^{{\mc V}_{\iota-1}}\underset{\text{(d)}}{\leq} \|\zeta_2\|^{{\mc V}_{\iota-1}}$,
or $\diamond^\zeta=\gle$, $\zeta_1\diamond^\zeta \zeta_2=\zeta_1\gle \zeta_2\in \mi{clo}$, $\|\zeta_1\|^{{\mc V}_{\iota-1}}\underset{\text{(e)}}{<} \|\zeta_2\|^{{\mc V}_{\iota-1}}$;
$\|\zeta_1\|^{{\mc V}_{\iota-1}}\leq \|\zeta_2\|^{{\mc V}_{\iota-1}}$,
$\frac{\|\zeta_1\|^{{\mc V}_{\iota-1}}}
      {\|\zeta_2\|^{{\mc V}_{\iota-1}}}\leq 1$,
$\mbb{E}_{\iota-1}=\emptyset$;
for all $1\leq j\leq \kappa_1$, 
$\mu_2^j\gleq (\mu_1^j\mult a_\iota^{\beta_j})\in \mi{min}(\mi{DE}_{\iota-1})$, 
$\left(\frac{\|\mu_2^j\|^{{\mc V}_{\iota-1}}}
            {\|\mu_1^j\|^{{\mc V}_{\iota-1}}}\right)^{\frac{1}{\beta_j}}\in \mbb{DE}_{\iota-1}$,
$\left(\frac{\|\mu_2^j\|^{{\mc V}_{\iota-1}}}
            {\|\mu_1^j\|^{{\mc V}_{\iota-1}}}\right)^{\frac{1}{\beta_j}}\leq \bigfvee \mbb{DE}_{\iota-1}\underset{\text{(\ref{eq8j}a)}}{\leq} {\mc V}_\iota(a_\iota)=\delta_\iota$,
$\frac{\|\mu_2^j\|^{{\mc V}_{\iota-1}}}
      {\|\mu_1^j\|^{{\mc V}_{\iota-1}}}\leq {\mc V}_\iota(a_\iota)^{\beta_j}$;
for all $\kappa_1+1\leq j\leq \kappa_2$, 
$\mu_2^j\gle (\mu_1^j\mult a_\iota^{\beta_j})\in \mi{min}(D_{\iota-1})$, 
$\left(\frac{\|\mu_2^j\|^{{\mc V}_{\iota-1}}}
            {\|\mu_1^j\|^{{\mc V}_{\iota-1}}}\right)^{\frac{1}{\beta_j}}\in \mbb{D}_{\iota-1}$,
$\left(\frac{\|\mu_2^j\|^{{\mc V}_{\iota-1}}}
            {\|\mu_1^j\|^{{\mc V}_{\iota-1}}}\right)^{\frac{1}{\beta_j}}\leq \bigfvee \mbb{D}_{\iota-1}\underset{\text{(\ref{eq8j}b)}}{<} {\mc V}_\iota(a_\iota)=\delta_\iota$,
$\frac{\|\mu_2^j\|^{{\mc V}_{\iota-1}}}
      {\|\mu_1^j\|^{{\mc V}_{\iota-1}}}<{\mc V}_\iota(a_\iota)^{\beta_j}$;
for all $\kappa_2+1\leq j\leq \kappa_3$,
$(\mu_1^j\mult a_\iota^{\beta_j})\gleq \mu_2^j\in \mi{min}(\mi{UE}_{\iota-1})$,
$\mi{min}\left(\left(\frac{\|\mu_2^j\|^{{\mc V}_{\iota-1}}}
                          {\|\mu_1^j\|^{{\mc V}_{\iota-1}}}\right)^{\frac{1}{\beta_j}},1\right)\in \mbb{UE}_{\iota-1}$,
${\mc V}_\iota(a_\iota)=\delta_\iota\underset{\text{(\ref{eq8j}a)}}{\leq} \bigfwedge \mbb{UE}_{\iota-1}\leq 
 \mi{min}\left(\left(\frac{\|\mu_2^j\|^{{\mc V}_{\iota-1}}}
                          {\|\mu_1^j\|^{{\mc V}_{\iota-1}}}\right)^{\frac{1}{\beta_j}},1\right)\leq
 \left(\frac{\|\mu_2^j\|^{{\mc V}_{\iota-1}}}
            {\|\mu_1^j\|^{{\mc V}_{\iota-1}}}\right)^{\frac{1}{\beta_j}}$,
${\mc V}_\iota(a_\iota)^{\beta_j}\leq \frac{\|\mu_2^j\|^{{\mc V}_{\iota-1}}}
                                           {\|\mu_1^j\|^{{\mc V}_{\iota-1}}}$,
$\frac{\|\mu_1^j\|^{{\mc V}_{\iota-1}}}
      {\|\mu_2^j\|^{{\mc V}_{\iota-1}}}\leq \frac{1}
                                                 {{\mc V}_\iota(a_\iota)^{\beta_j}}$;
$\kappa_3<\kappa_4$, $\kappa_3+1\leq \kappa_4$,
$\{\kappa_3+1,\dots,\kappa_4\}\neq \emptyset$,
$(\mu_1^{\kappa_3+1}\mult a_\iota^{\beta_{\kappa_3+1}})\gle \mu_2^{\kappa_3+1}\in \mi{min}(U_{\iota-1})$,
$\mi{min}\left(\left(\frac{\|\mu_2^{\kappa_3+1}\|^{{\mc V}_{\iota-1}}}
                          {\|\mu_1^{\kappa_3+1}\|^{{\mc V}_{\iota-1}}}\right)^{\frac{1}{\beta_{\kappa_3+1}}},1\right)\in \mbb{U}_{\iota-1}\neq \emptyset$;
for all $\kappa_3+1\leq j\leq \kappa_4$,
$(\mu_1^j\mult a_\iota^{\beta_j})\gle \mu_2^j\in \mi{min}(U_{\iota-1})$,
$\mi{min}\left(\left(\frac{\|\mu_2^j\|^{{\mc V}_{\iota-1}}}
                          {\|\mu_1^j\|^{{\mc V}_{\iota-1}}}\right)^{\frac{1}{\beta_j}},1\right)\in \mbb{U}_{\iota-1}$,
${\mc V}_\iota(a_\iota)=\delta_\iota\underset{\text{(\ref{eq8j}c)}}{<} \bigfwedge \mbb{U}_{\iota-1}\leq 
 \mi{min}\left(\left(\frac{\|\mu_2^j\|^{{\mc V}_{\iota-1}}}
                          {\|\mu_1^j\|^{{\mc V}_{\iota-1}}}\right)^{\frac{1}{\beta_j}},1\right)\leq
 \left(\frac{\|\mu_2^j\|^{{\mc V}_{\iota-1}}}
            {\|\mu_1^j\|^{{\mc V}_{\iota-1}}}\right)^{\frac{1}{\beta_j}}$,
${\mc V}_\iota(a_\iota)^{\beta_j}<\frac{\|\mu_2^j\|^{{\mc V}_{\iota-1}}}
                                       {\|\mu_1^j\|^{{\mc V}_{\iota-1}}}$,
$\frac{\|\mu_1^j\|^{{\mc V}_{\iota-1}}}
      {\|\mu_2^j\|^{{\mc V}_{\iota-1}}}<\frac{1}
                                             {{\mc V}_\iota(a_\iota)^{\beta_j}}$;
\begin{alignat*}{1}
& \dfrac{\|\upsilon_1\|^{{\mc V}_{\iota-1}}}
        {\|\varepsilon_2\|^{{\mc V}_{\iota-1}}}= \\
& \dfrac{\left\|\mi{reduced}^-\left(\begin{array}{l}
                                    ((\bigmult_{j=1}^{\kappa_1} \mu_2^j)\mult (\bigmult_{j=\kappa_1+1}^{\kappa_2} \mu_2^j)\mult 
                                     (\bigmult_{j=\kappa_2+1}^{\kappa_3} \mu_1^j)\mult \\
                                    \quad
                                     (\bigmult_{j=\kappa_3+1}^{\kappa_4} \mu_1^j)\mult \zeta_1)\gle \\
                                    \quad
                                    ((\bigmult_{j=1}^{\kappa_1} \mu_1^j)\mult (\bigmult_{j=\kappa_1+1}^{\kappa_2} \mu_1^j)\mult 
                                     (\bigmult_{j=\kappa_2+1}^{\kappa_3} \mu_2^j)\mult \\
                                    \quad \quad
                                     (\bigmult_{j=\kappa_3+1}^{\kappa_4} \mu_2^j)\mult \zeta_2)
                                    \end{array}\right)\right\|^{{\mc V}_{\iota-1}}}
        {\left\|\mi{reduced}^+\left(\begin{array}{l}
                                    ((\bigmult_{j=1}^{\kappa_1} \mu_2^j)\mult (\bigmult_{j=\kappa_1+1}^{\kappa_2} \mu_2^j)\mult 
                                     (\bigmult_{j=\kappa_2+1}^{\kappa_3} \mu_1^j)\mult \\
                                    \quad
                                     (\bigmult_{j=\kappa_3+1}^{\kappa_4} \mu_1^j)\mult \zeta_1)\gle \\
                                    \quad
                                    ((\bigmult_{j=1}^{\kappa_1} \mu_1^j)\mult (\bigmult_{j=\kappa_1+1}^{\kappa_2} \mu_1^j)\mult 
                                     (\bigmult_{j=\kappa_2+1}^{\kappa_3} \mu_2^j)\mult \\
                                    \quad \quad
                                     (\bigmult_{j=\kappa_3+1}^{\kappa_4} \mu_2^j)\mult \zeta_2)
                                    \end{array}\right)\right\|^{{\mc V}_{\iota-1}}}\overset{\text{(\ref{eq7n})}}{=\!\!=} \\
& \dfrac{\|(\bigmult_{j=1}^{\kappa_1} \mu_2^j)\mult (\bigmult_{j=\kappa_1+1}^{\kappa_2} \mu_2^j)\mult 
           (\bigmult_{j=\kappa_2+1}^{\kappa_3} \mu_1^j)\mult (\bigmult_{j=\kappa_3+1}^{\kappa_4} \mu_1^j)\mult \zeta_1\|^{{\mc V}_{\iota-1}}}
        {\|(\bigmult_{j=1}^{\kappa_1} \mu_1^j)\mult (\bigmult_{j=\kappa_1+1}^{\kappa_2} \mu_1^j)\mult 
           (\bigmult_{j=\kappa_2+1}^{\kappa_3} \mu_2^j)\mult (\bigmult_{j=\kappa_3+1}^{\kappa_4} \mu_2^j)\mult \zeta_2\|^{{\mc V}_{\iota-1}}}= \\
& \dfrac{\begin{array}{l}
         (\bigfswedge_{j=1}^{\kappa_1} \|\mu_2^j\|^{{\mc V}_{\iota-1}})\fswedge (\bigfswedge_{j=\kappa_1+1}^{\kappa_2} \|\mu_2^j\|^{{\mc V}_{\iota-1}})\fswedge
         (\bigfswedge_{j=\kappa_2+1}^{\kappa_3} \|\mu_1^j\|^{{\mc V}_{\iota-1}})\fswedge (\bigfswedge_{j=\kappa_3+1}^{\kappa_4} \|\mu_1^j\|^{{\mc V}_{\iota-1}})\fswedge \\
         \quad
         \|\zeta_1\|^{{\mc V}_{\iota-1}}
         \end{array}}
        {\begin{array}{l}
         (\bigfswedge_{j=1}^{\kappa_1} \|\mu_1^j\|^{{\mc V}_{\iota-1}})\fswedge (\bigfswedge_{j=\kappa_1+1}^{\kappa_2} \|\mu_1^j\|^{{\mc V}_{\iota-1}})\fswedge
         (\bigfswedge_{j=\kappa_2+1}^{\kappa_3} \|\mu_2^j\|^{{\mc V}_{\iota-1}})\fswedge (\bigfswedge_{j=\kappa_3+1}^{\kappa_4} \|\mu_2^j\|^{{\mc V}_{\iota-1}})\fswedge \\
         \quad
         \|\zeta_2\|^{{\mc V}_{\iota-1}}
         \end{array}}= \\
& \left(\bigfswedge_{j=1}^{\kappa_1} \dfrac{\|\mu_2^j\|^{{\mc V}_{\iota-1}}}
                                           {\|\mu_1^j\|^{{\mc V}_{\iota-1}}}\right)\fswedge 
  \left(\bigfswedge_{j=\kappa_1+1}^{\kappa_2} \dfrac{\|\mu_2^j\|^{{\mc V}_{\iota-1}}}
                                                    {\|\mu_1^j\|^{{\mc V}_{\iota-1}}}\right)\fswedge 
  \left(\bigfswedge_{j=\kappa_1+2}^{\kappa_3} \dfrac{\|\mu_1^j\|^{{\mc V}_{\iota-1}}}
                                                    {\|\mu_2^j\|^{{\mc V}_{\iota-1}}}\right)\fswedge \\
& \quad
  \left(\bigfswedge_{j=\kappa_1+3}^{\kappa_4} \dfrac{\|\mu_1^j\|^{{\mc V}_{\iota-1}}}
                                                    {\|\mu_2^j\|^{{\mc V}_{\iota-1}}}\right)\fswedge
  \dfrac{\|\zeta_1\|^{{\mc V}_{\iota-1}}}
        {\|\zeta_2\|^{{\mc V}_{\iota-1}}}< \\
& \left(\bigfswedge_{j=1}^{\kappa_1} {\mc V}_\iota(a_\iota)^{\beta_j}\right)\fswedge \left(\bigfswedge_{j=\kappa_1+1}^{\kappa_2} {\mc V}_\iota(a_\iota)^{\beta_j}\right)\fswedge 
  \left(\bigfswedge_{j=\kappa_1+2}^{\kappa_3} \dfrac{1}
                                                    {{\mc V}_\iota(a_\iota)^{\beta_j}}\right)\fswedge
  \bigfswedge_{j=\kappa_1+3}^{\kappa_4} \dfrac{1}
                                              {{\mc V}_\iota(a_\iota)^{\beta_j}}= \\
& {\mc V}_\iota(a_\iota)^{(\sum_{j=1}^{\kappa_1} \beta_j)+(\sum_{j=\kappa_1+1}^{\kappa_2} \beta_j)-(\sum_{j=\kappa_2+1}^{\kappa_3} \beta_j)-\sum_{j=\kappa_3+1}^{\kappa_4} \beta_j}=
  {\mc V}_\iota(a_\iota)^{-\alpha}=\dfrac{1}
                                         {{\mc V}_\iota(a_\iota)^\alpha}, \\[1mm]
& \|\varepsilon_1\|^{{\mc V}_\iota}=\|\upsilon_1\mult a_\iota^\alpha\|^{{\mc V}_\iota}=\|\upsilon_1\|^{{\mc V}_\iota}\fswedge {\mc V}_\iota(a_\iota)^\alpha\overset{\text{(b)}}{=\!\!=}
  \|\upsilon_1\|^{{\mc V}_{\iota-1}}\fswedge {\mc V}_\iota(a_\iota)^\alpha<\|\varepsilon_2\|^{{\mc V}_{\iota-1}}\overset{\text{(b)}}{=\!\!=} \|\varepsilon_2\|^{{\mc V}_\iota};  
\end{alignat*}
(e) holds.

Case 2.10.2.2:
$a_\iota\not\in \mi{atoms}(\varepsilon_1)\subseteq \mi{dom}({\mc V}_{\iota-1})$ and $a_\iota\in \mi{atoms}(\varepsilon_2)$.
Then $\mi{atoms}(\varepsilon_2)\subseteq \mi{dom}({\mc V}_\iota)=\mi{dom}({\mc V}_{\iota-1})\cup \{a_\iota\}$,
$\varepsilon_2=\upsilon_2\mult a_\iota^\alpha$, $\upsilon_2\in \{\gu\}\cup \mi{PropConj}_A$, $\alpha\geq 1$, $a_\iota\not\in \mi{atoms}(\upsilon_2)\subseteq \mi{dom}({\mc V}_{\iota-1})$,
$\varepsilon_1\geql \varepsilon_2\not\in \mi{clo}$.
We get two cases for $\varepsilon_1$ and $\varepsilon_2$.

Case 2.10.2.2.1:
$\varepsilon_1\gleq \varepsilon_2\in \mi{clo}$.
Then $\mi{atoms}(\varepsilon_1), \mi{atoms}(\varepsilon_2)\subseteq \mi{dom}({\mc V}_\iota)=\mi{dom}({\mc V}_{\iota-1})\cup \{a_\iota\}$,
$a_\iota\in \mi{atoms}(\varepsilon_1)$ or $a_\iota\in \mi{atoms}(\varepsilon_2)$, $\mbb{E}_{\iota-1}=\emptyset$,
$\varepsilon_1\gleq \varepsilon_2=\varepsilon_1\gleq (\upsilon_2\mult a_\iota^\alpha)\in \mi{clo}$;
by (\ref{eq8i}) for $\gleq$ and (\ref{eq8i}a), there exist $1\leq \kappa_1\leq \kappa_2\leq \kappa_3\leq \kappa_4$,
\begin{alignat*}{1}
& \mu_2^j\gleq (\mu_1^j\mult a_\iota^{\beta_j})=\mi{reduced}(\lambda^{(\gamma_1^j,\dots,\gamma_n^j)})\in \mi{min}(\mi{DE}_{\iota-1}), j=1,\dots,\kappa_1, \\ 
& \mu_2^j\gle (\mu_1^j\mult a_\iota^{\beta_j})=\mi{reduced}(\lambda^{(\gamma_1^j,\dots,\gamma_n^j)})\in \mi{min}(D_{\iota-1}), j=\kappa_1+1,\dots,\kappa_2, \\
& (\mu_1^j\mult a_\iota^{\beta_j})\gleq \mu_2^j=\mi{reduced}(\lambda^{(\gamma_1^j,\dots,\gamma_n^j)})\in \mi{min}(\mi{UE}_{\iota-1}), j=\kappa_2+1,\dots,\kappa_3, \\
& (\mu_1^j\mult a_\iota^{\beta_j})\gle \mu_2^j=\mi{reduced}(\lambda^{(\gamma_1^j,\dots,\gamma_n^j)})\in \mi{min}(U_{\iota-1}), j=\kappa_3+1,\dots,\kappa_4, \\ 
& \mu_e^j\in \{\gu\}\cup \mi{PropConj}_A, \mi{atoms}(\mu_e^j)\subseteq \mi{dom}({\mc V}_{\iota-1}), \beta_j\geq 1, \bs{0}^n\neq (\gamma_1^j,\dots,\gamma_n^j)\in \mbb{N}^n, 
\end{alignat*}
satisfying either $\varepsilon_1\gleq \varepsilon_2=\varepsilon_1\gleq (\upsilon_2\mult a_\iota^\alpha)=
                                                    \mi{reduced}(\lambda^{(\sum_{j=1}^{\kappa_4} \gamma_1^j,\dots,\sum_{j=1}^{\kappa_4} \gamma_n^j)})$, 
or there exists $\zeta_1\diamond^\zeta \zeta_2=\mi{reduced}(\lambda^{(\gamma_1^\zeta,\dots,\gamma_n^\zeta)})\in \mi{clo}$,
$\zeta_e\in \{\gu\}\cup \mi{PropConj}_A$, $\mi{atoms}(\zeta_e)\subseteq \mi{dom}({\mc V}_{\iota-1})$, $\diamond^\zeta\in \{\geql,\gleq,\gle\}$, 
$\bs{0}^n\neq (\gamma_1^\zeta,\dots,\gamma_n^\zeta)\in \mbb{N}^n$, satisfying
$\varepsilon_1\gleq \varepsilon_2=\varepsilon_1\gleq (\upsilon_2\mult a_\iota^\alpha)=
                                  \mi{reduced}(\lambda^{((\sum_{j=1}^{\kappa_4} \gamma_1^j)+\gamma_1^\zeta,\dots,(\sum_{j=1}^{\kappa_4} \gamma_n^j)+\gamma_n^\zeta)})$.
We get two cases for $\varepsilon_1\gleq (\upsilon_2\mult a_\iota^\alpha)$.

Case 2.10.2.2.1.1:
$\varepsilon_1\gleq (\upsilon_2\mult a_\iota^\alpha)=\mi{reduced}(\lambda^{(\sum_{j=1}^{\kappa_4} \gamma_1^j,\dots,\sum_{j=1}^{\kappa_4} \gamma_n^j)})$.
Then 
$\mi{atoms}(\varepsilon_1), \mi{atoms}(\upsilon_2), \mi{atoms}(\mu_1^1),\dots,\mi{atoms}(\mu_1^{\kappa_4}), \mi{atoms}(\mu_2^1),\dots,\mi{atoms}(\mu_2^{\kappa_4})\subseteq \mi{dom}({\mc V}_{\iota-1})$,
\begin{alignat*}{1}
& \varepsilon_1\gleq (\upsilon_2\mult a_\iota^\alpha)=
  \mi{reduced}(\lambda^{(\sum_{j=1}^{\kappa_4} \gamma_1^j,\dots,\sum_{j=1}^{\kappa_4} \gamma_n^j)})\overset{\text{(\ref{eq8dddd})}}{=\!\!=} \\
& \mi{reduced}((\bigmult_{j=1}^{\kappa_1} \lambda^{(\gamma_1^j,\dots,\gamma_n^j)})\mult
               (\bigmult_{j=\kappa_1+1}^{\kappa_2} \lambda^{(\gamma_1^j,\dots,\gamma_n^j)})\mult \\
&\phantom{\mi{reduced}(}
               (\bigmult_{j=\kappa_2+1}^{\kappa_3} \lambda^{(\gamma_1^j,\dots,\gamma_n^j)})\mult
               \bigmult_{j=\kappa_3+1}^{\kappa_4} \lambda^{(\gamma_1^j,\dots,\gamma_n^j)})\overset{\text{(\ref{eq7k})}}{=\!\!=} \\
& \mi{reduced}((\bigmult_{j=1}^{\kappa_1} \mi{reduced}(\lambda^{(\gamma_1^j,\dots,\gamma_n^j)}))\mult
               (\bigmult_{j=\kappa_1+1}^{\kappa_2} \mi{reduced}(\lambda^{(\gamma_1^j,\dots,\gamma_n^j)}))\mult \\
&\phantom{\mi{reduced}(}
               (\bigmult_{j=\kappa_2+1}^{\kappa_3} \mi{reduced}(\lambda^{(\gamma_1^j,\dots,\gamma_n^j)}))\mult
               \bigmult_{j=\kappa_3+1}^{\kappa_4} \mi{reduced}(\lambda^{(\gamma_1^j,\dots,\gamma_n^j)}))= \\
& \mi{reduced}((\bigmult_{j=1}^{\kappa_1} \mu_2^j\gleq (\mu_1^j\mult a_\iota^{\beta_j}))\mult
               (\bigmult_{j=\kappa_1+1}^{\kappa_2} \mu_2^j\gle (\mu_1^j\mult a_\iota^{\beta_j}))\mult \\
&\phantom{\mi{reduced}(}
               (\bigmult_{j=\kappa_2+1}^{\kappa_3} (\mu_1^j\mult a_\iota^{\beta_j})\gleq \mu_2^j)\mult
               \bigmult_{j=\kappa_3+1}^{\kappa_4} (\mu_1^j\mult a_\iota^{\beta_j})\gle \mu_2^j)\overset{\text{(\ref{eq7i})}}{=\!\!=} \\
& \mi{reduced}(((\bigmult_{j=1}^{\kappa_1} \mu_2^j)\mult (\bigmult_{j=\kappa_1+1}^{\kappa_2} \mu_2^j)\mult 
                (\bigmult_{j=\kappa_2+1}^{\kappa_3} \mu_1^j)\mult (\bigmult_{j=\kappa_3+1}^{\kappa_4} \mu_1^j)\mult \\
&\phantom{\mi{reduced}(} \quad
                a_\iota^{(\sum_{j=\kappa_2+1}^{\kappa_3} \beta_j)+\sum_{j=\kappa_3+1}^{\kappa_4} \beta_j})\gleq \\
&\phantom{\mi{reduced}(} \quad
               ((\bigmult_{j=1}^{\kappa_1} \mu_1^j)\mult (\bigmult_{j=\kappa_1+1}^{\kappa_2} \mu_1^j)\mult 
                (\bigmult_{j=\kappa_2+1}^{\kappa_3} \mu_2^j)\mult (\bigmult_{j=\kappa_3+1}^{\kappa_4} \mu_2^j)\mult \\
&\phantom{\mi{reduced}(} \quad \quad
                a_\iota^{(\sum_{j=1}^{\kappa_1} \beta_j)+\sum_{j=\kappa_1+1}^{\kappa_2} \beta_j}))= \\
& \mi{reduced}^-(((\bigmult_{j=1}^{\kappa_1} \mu_2^j)\mult (\bigmult_{j=\kappa_1+1}^{\kappa_2} \mu_2^j)\mult 
                  (\bigmult_{j=\kappa_2+1}^{\kappa_3} \mu_1^j)\mult (\bigmult_{j=\kappa_3+1}^{\kappa_4} \mu_1^j)\mult \\
&\phantom{\mi{reduced}^-(} \quad
                  a_\iota^{(\sum_{j=\kappa_2+1}^{\kappa_3} \beta_j)+\sum_{j=\kappa_3+1}^{\kappa_4} \beta_j})\gleq \\
&\phantom{\mi{reduced}^-(} \quad
                 ((\bigmult_{j=1}^{\kappa_1} \mu_1^j)\mult (\bigmult_{j=\kappa_1+1}^{\kappa_2} \mu_1^j)\mult 
                  (\bigmult_{j=\kappa_2+1}^{\kappa_3} \mu_2^j)\mult (\bigmult_{j=\kappa_3+1}^{\kappa_4} \mu_2^j)\mult \\
&\phantom{\mi{reduced}^-(} \quad \quad
                  a_\iota^{(\sum_{j=1}^{\kappa_1} \beta_j)+\sum_{j=\kappa_1+1}^{\kappa_2} \beta_j}))\gleq \\
& \mi{reduced}^+(((\bigmult_{j=1}^{\kappa_1} \mu_2^j)\mult (\bigmult_{j=\kappa_1+1}^{\kappa_2} \mu_2^j)\mult 
                  (\bigmult_{j=\kappa_2+1}^{\kappa_3} \mu_1^j)\mult (\bigmult_{j=\kappa_3+1}^{\kappa_4} \mu_1^j)\mult \\
&\phantom{\mi{reduced}^+(} \quad
                  a_\iota^{(\sum_{j=\kappa_2+1}^{\kappa_3} \beta_j)+\sum_{j=\kappa_3+1}^{\kappa_4} \beta_j})\gleq \\
&\phantom{\mi{reduced}^+(} \quad
                 ((\bigmult_{j=1}^{\kappa_1} \mu_1^j)\mult (\bigmult_{j=\kappa_1+1}^{\kappa_2} \mu_1^j)\mult 
                  (\bigmult_{j=\kappa_2+1}^{\kappa_3} \mu_2^j)\mult (\bigmult_{j=\kappa_3+1}^{\kappa_4} \mu_2^j)\mult \\
&\phantom{\mi{reduced}^+(} \quad \quad
                  a_\iota^{(\sum_{j=1}^{\kappa_1} \beta_j)+\sum_{j=\kappa_1+1}^{\kappa_2} \beta_j})), \\[1mm]
& \#(a_\iota,\varepsilon_1\gleq (\upsilon_2\mult a_\iota^\alpha))=\alpha= \\
& \#(a_\iota,\mi{reduced}(((\bigmult_{j=1}^{\kappa_1} \mu_2^j)\mult (\bigmult_{j=\kappa_1+1}^{\kappa_2} \mu_2^j)\mult 
                           (\bigmult_{j=\kappa_2+1}^{\kappa_3} \mu_1^j)\mult (\bigmult_{j=\kappa_3+1}^{\kappa_4} \mu_1^j)\mult \\
&\phantom{\#(a_\iota,\mi{reduced}(} \quad
                           a_\iota^{(\sum_{j=\kappa_2+1}^{\kappa_3} \beta_j)+\sum_{j=\kappa_3+1}^{\kappa_4} \beta_j})\gleq \\
&\phantom{\#(a_\iota,\mi{reduced}(} \quad
                          ((\bigmult_{j=1}^{\kappa_1} \mu_1^j)\mult (\bigmult_{j=\kappa_1+1}^{\kappa_2} \mu_1^j)\mult 
                           (\bigmult_{j=\kappa_2+1}^{\kappa_3} \mu_2^j)\mult (\bigmult_{j=\kappa_3+1}^{\kappa_4} \mu_2^j)\mult \\
&\phantom{\#(a_\iota,\mi{reduced}(} \quad \quad
                           a_\iota^{(\sum_{j=1}^{\kappa_1} \beta_j)+\sum_{j=\kappa_1+1}^{\kappa_2} \beta_j})))\overset{\text{(\ref{eq7aaaax})}}{=\!\!=} \\
& \#(a_\iota,((\bigmult_{j=1}^{\kappa_1} \mu_2^j)\mult (\bigmult_{j=\kappa_1+1}^{\kappa_2} \mu_2^j)\mult 
              (\bigmult_{j=\kappa_2+1}^{\kappa_3} \mu_1^j)\mult (\bigmult_{j=\kappa_3+1}^{\kappa_4} \mu_1^j)\mult \\
&\phantom{\#(a_\iota,} \quad
              a_\iota^{(\sum_{j=\kappa_2+1}^{\kappa_3} \beta_j)+\sum_{j=\kappa_3+1}^{\kappa_4} \beta_j})\gleq \\
&\phantom{\#(a_\iota,} \quad
             ((\bigmult_{j=1}^{\kappa_1} \mu_1^j)\mult (\bigmult_{j=\kappa_1+1}^{\kappa_2} \mu_1^j)\mult 
              (\bigmult_{j=\kappa_2+1}^{\kappa_3} \mu_2^j)\mult (\bigmult_{j=\kappa_3+1}^{\kappa_4} \mu_2^j)\mult \\
&\phantom{\#(a_\iota,} \quad \quad
              a_\iota^{(\sum_{j=1}^{\kappa_1} \beta_j)+\sum_{j=\kappa_1+1}^{\kappa_2} \beta_j}))= \\
& \left(\sum_{j=1}^{\kappa_1} \beta_j\right)+\left(\sum_{j=\kappa_1+1}^{\kappa_2} \beta_j\right)-\left(\sum_{j=\kappa_2+1}^{\kappa_3} \beta_j\right)-\sum_{j=\kappa_3+1}^{\kappa_4} \beta_j\geq 1, 
\end{alignat*}
$a_\iota\not\in \mi{dom}({\mc V}_{\iota-1})\supseteq \mi{atoms}((\bigmult_{j=1}^{\kappa_1} \mu_2^j)\mult (\bigmult_{j=\kappa_1+1}^{\kappa_2} \mu_2^j)\mult 
                                                                (\bigmult_{j=\kappa_2+1}^{\kappa_3} \mu_1^j)\mult \bigmult_{j=\kappa_3+1}^{\kappa_4} \mu_1^j)=
                                                     (\bigcup_{j=1}^{\kappa_1} \mi{atoms}(\mu_2^j))\cup (\bigcup_{j=\kappa_1+1}^{\kappa_2} \mi{atoms}(\mu_2^j))\cup 
                                                     (\bigcup_{j=\kappa_2+1}^{\kappa_3} \mi{atoms}(\mu_1^j))\cup \bigcup_{j=\kappa_3+1}^{\kappa_4} \mi{atoms}(\mu_1^j)$,
$a_\iota\not\in \mi{dom}({\mc V}_{\iota-1})\supseteq \mi{atoms}((\bigmult_{j=1}^{\kappa_1} \mu_1^j)\mult (\bigmult_{j=\kappa_1+1}^{\kappa_2} \mu_1^j)\mult 
                                                                (\bigmult_{j=\kappa_2+1}^{\kappa_3} \mu_2^j)\mult \bigmult_{j=\kappa_3+1}^{\kappa_4} \mu_2^j)=
                                                     (\bigcup_{j=1}^{\kappa_1} \mi{atoms}(\mu_1^j))\cup (\bigcup_{j=\kappa_1+1}^{\kappa_2} \mi{atoms}(\mu_1^j))\cup 
                                                     (\bigcup_{j=\kappa_2+1}^{\kappa_3} \mi{atoms}(\mu_2^j))\cup \bigcup_{j=\kappa_3+1}^{\kappa_4} \mi{atoms}(\mu_2^j)$,
\begin{alignat*}{1}
& \mi{reduced}^-(((\bigmult_{j=1}^{\kappa_1} \mu_2^j)\mult (\bigmult_{j=\kappa_1+1}^{\kappa_2} \mu_2^j)\mult 
                  (\bigmult_{j=\kappa_2+1}^{\kappa_3} \mu_1^j)\mult (\bigmult_{j=\kappa_3+1}^{\kappa_4} \mu_1^j)\mult \\
&\phantom{\mi{reduced}^-(} \quad
                  a_\iota^{(\sum_{j=\kappa_2+1}^{\kappa_3} \beta_j)+\sum_{j=\kappa_3+1}^{\kappa_4} \beta_j})\gleq \\
&\phantom{\mi{reduced}^-(} \quad
                 ((\bigmult_{j=1}^{\kappa_1} \mu_1^j)\mult (\bigmult_{j=\kappa_1+1}^{\kappa_2} \mu_1^j)\mult 
                  (\bigmult_{j=\kappa_2+1}^{\kappa_3} \mu_2^j)\mult (\bigmult_{j=\kappa_3+1}^{\kappa_4} \mu_2^j)\mult \\
&\phantom{\mi{reduced}^-(} \quad \quad
                  a_\iota^{(\sum_{j=1}^{\kappa_1} \beta_j)+\sum_{j=\kappa_1+1}^{\kappa_2} \beta_j}))\overset{\text{(\ref{eq7cc})}}{=\!\!=} \\
& \mi{reduced}^-(((\bigmult_{j=1}^{\kappa_1} \mu_2^j)\mult (\bigmult_{j=\kappa_1+1}^{\kappa_2} \mu_2^j)\mult 
                  (\bigmult_{j=\kappa_2+1}^{\kappa_3} \mu_1^j)\mult \bigmult_{j=\kappa_3+1}^{\kappa_4} \mu_1^j)\gleq \\
&\phantom{\mi{reduced}^-(} \quad
                 ((\bigmult_{j=1}^{\kappa_1} \mu_1^j)\mult (\bigmult_{j=\kappa_1+1}^{\kappa_2} \mu_1^j)\mult 
                  (\bigmult_{j=\kappa_2+1}^{\kappa_3} \mu_2^j)\mult \bigmult_{j=\kappa_3+1}^{\kappa_4} \mu_2^j)), \\[1mm]
& \mi{reduced}^+(((\bigmult_{j=1}^{\kappa_1} \mu_2^j)\mult (\bigmult_{j=\kappa_1+1}^{\kappa_2} \mu_2^j)\mult 
                  (\bigmult_{j=\kappa_2+1}^{\kappa_3} \mu_1^j)\mult (\bigmult_{j=\kappa_3+1}^{\kappa_4} \mu_1^j)\mult \\
&\phantom{\mi{reduced}^+(} \quad
                  a_\iota^{(\sum_{j=\kappa_2+1}^{\kappa_3} \beta_j)+\sum_{j=\kappa_3+1}^{\kappa_4} \beta_j})\gleq \\
&\phantom{\mi{reduced}^+(} \quad
                 ((\bigmult_{j=1}^{\kappa_1} \mu_1^j)\mult (\bigmult_{j=\kappa_1+1}^{\kappa_2} \mu_1^j)\mult 
                  (\bigmult_{j=\kappa_2+1}^{\kappa_3} \mu_2^j)\mult (\bigmult_{j=\kappa_3+1}^{\kappa_4} \mu_2^j)\mult \\
&\phantom{\mi{reduced}^+(} \quad \quad
                  a_\iota^{(\sum_{j=1}^{\kappa_1} \beta_j)+\sum_{j=\kappa_1+1}^{\kappa_2} \beta_j}))\overset{\text{(\ref{eq7ff})}}{=\!\!=} \\
& \mi{reduced}^+(((\bigmult_{j=1}^{\kappa_1} \mu_2^j)\mult (\bigmult_{j=\kappa_1+1}^{\kappa_2} \mu_2^j)\mult 
                  (\bigmult_{j=\kappa_2+1}^{\kappa_3} \mu_1^j)\mult \bigmult_{j=\kappa_3+1}^{\kappa_4} \mu_1^j)\gleq \\
&\phantom{\mi{reduced}^+(} \quad
                 ((\bigmult_{j=1}^{\kappa_1} \mu_1^j)\mult (\bigmult_{j=\kappa_1+1}^{\kappa_2} \mu_1^j)\mult 
                  (\bigmult_{j=\kappa_2+1}^{\kappa_3} \mu_2^j)\mult \bigmult_{j=\kappa_3+1}^{\kappa_4} \mu_2^j))\mult \\
& a_\iota^{(\sum_{j=1}^{\kappa_1} \beta_j)+(\sum_{j=\kappa_1+1}^{\kappa_2} \beta_j)-(\sum_{j=\kappa_2+1}^{\kappa_3} \beta_j)-\sum_{j=\kappa_3+1}^{\kappa_4} \beta_j}= \\
& \mi{reduced}^+(((\bigmult_{j=1}^{\kappa_1} \mu_2^j)\mult (\bigmult_{j=\kappa_1+1}^{\kappa_2} \mu_2^j)\mult 
                  (\bigmult_{j=\kappa_2+1}^{\kappa_3} \mu_1^j)\mult \bigmult_{j=\kappa_3+1}^{\kappa_4} \mu_1^j)\gleq \\
&\phantom{\mi{reduced}^+(} \quad
                 ((\bigmult_{j=1}^{\kappa_1} \mu_1^j)\mult (\bigmult_{j=\kappa_1+1}^{\kappa_2} \mu_1^j)\mult 
                  (\bigmult_{j=\kappa_2+1}^{\kappa_3} \mu_2^j)\mult \bigmult_{j=\kappa_3+1}^{\kappa_4} \mu_2^j))\mult a_\iota^\alpha, \\[1mm]
& \varepsilon_1\gleq (\upsilon_2\mult a_\iota^\alpha)= \\
& \mi{reduced}^-(((\bigmult_{j=1}^{\kappa_1} \mu_2^j)\mult (\bigmult_{j=\kappa_1+1}^{\kappa_2} \mu_2^j)\mult 
                  (\bigmult_{j=\kappa_2+1}^{\kappa_3} \mu_1^j)\mult \bigmult_{j=\kappa_3+1}^{\kappa_4} \mu_1^j)\gleq \\
&\phantom{\mi{reduced}^-(} \quad
                 ((\bigmult_{j=1}^{\kappa_1} \mu_1^j)\mult (\bigmult_{j=\kappa_1+1}^{\kappa_2} \mu_1^j)\mult 
                  (\bigmult_{j=\kappa_2+1}^{\kappa_3} \mu_2^j)\mult \bigmult_{j=\kappa_3+1}^{\kappa_4} \mu_2^j))\gleq \\
& (\mi{reduced}^+(((\bigmult_{j=1}^{\kappa_1} \mu_2^j)\mult (\bigmult_{j=\kappa_1+1}^{\kappa_2} \mu_2^j)\mult 
                   (\bigmult_{j=\kappa_2+1}^{\kappa_3} \mu_1^j)\mult \bigmult_{j=\kappa_3+1}^{\kappa_4} \mu_1^j)\gleq \\
&\phantom{(\mi{reduced}^+(} \quad
                  ((\bigmult_{j=1}^{\kappa_1} \mu_1^j)\mult (\bigmult_{j=\kappa_1+1}^{\kappa_2} \mu_1^j)\mult 
                   (\bigmult_{j=\kappa_2+1}^{\kappa_3} \mu_2^j)\mult \bigmult_{j=\kappa_3+1}^{\kappa_4} \mu_2^j))\mult a_\iota^\alpha), \\[1mm]
& \mi{atoms}(\mi{reduced}^-(((\bigmult_{j=1}^{\kappa_1} \mu_2^j)\mult (\bigmult_{j=\kappa_1+1}^{\kappa_2} \mu_2^j)\mult 
                             (\bigmult_{j=\kappa_2+1}^{\kappa_3} \mu_1^j)\mult \bigmult_{j=\kappa_3+1}^{\kappa_4} \mu_1^j)\gleq \\
&\phantom{\mi{atoms}(\mi{reduced}^-(} \quad
                            ((\bigmult_{j=1}^{\kappa_1} \mu_1^j)\mult (\bigmult_{j=\kappa_1+1}^{\kappa_2} \mu_1^j)\mult 
                             (\bigmult_{j=\kappa_2+1}^{\kappa_3} \mu_2^j)\mult \bigmult_{j=\kappa_3+1}^{\kappa_4} \mu_2^j)))\underset{\text{(\ref{eq7a})}}{\subseteq} \\
& \mi{atoms}((\bigmult_{j=1}^{\kappa_1} \mu_2^j)\mult (\bigmult_{j=\kappa_1+1}^{\kappa_2} \mu_2^j)\mult 
             (\bigmult_{j=\kappa_2+1}^{\kappa_3} \mu_1^j)\mult \bigmult_{j=\kappa_3+1}^{\kappa_4} \mu_1^j)\subseteq \mi{dom}({\mc V}_{\iota-1}), \\[1mm]
& \mi{atoms}(\mi{reduced}^+(((\bigmult_{j=1}^{\kappa_1} \mu_2^j)\mult (\bigmult_{j=\kappa_1+1}^{\kappa_2} \mu_2^j)\mult 
                             (\bigmult_{j=\kappa_2+1}^{\kappa_3} \mu_1^j)\mult \bigmult_{j=\kappa_3+1}^{\kappa_4} \mu_1^j)\gleq \\
&\phantom{\mi{atoms}(\mi{reduced}^+(} \quad
                            ((\bigmult_{j=1}^{\kappa_1} \mu_1^j)\mult (\bigmult_{j=\kappa_1+1}^{\kappa_2} \mu_1^j)\mult 
                             (\bigmult_{j=\kappa_2+1}^{\kappa_3} \mu_2^j)\mult \bigmult_{j=\kappa_3+1}^{\kappa_4} \mu_2^j)))\underset{\text{(\ref{eq7b})}}{\subseteq} \\
& \mi{atoms}((\bigmult_{j=1}^{\kappa_1} \mu_1^j)\mult (\bigmult_{j=\kappa_1+1}^{\kappa_2} \mu_1^j)\mult 
             (\bigmult_{j=\kappa_2+1}^{\kappa_3} \mu_2^j)\mult \bigmult_{j=\kappa_3+1}^{\kappa_4} \mu_2^j)\subseteq \mi{dom}({\mc V}_{\iota-1}), 
\end{alignat*}
\begin{alignat*}{1}
& \varepsilon_1=\mi{reduced}^-(((\bigmult_{j=1}^{\kappa_1} \mu_2^j)\mult (\bigmult_{j=\kappa_1+1}^{\kappa_2} \mu_2^j)\mult 
                                (\bigmult_{j=\kappa_2+1}^{\kappa_3} \mu_1^j)\mult \bigmult_{j=\kappa_3+1}^{\kappa_4} \mu_1^j)\gleq \\
&\phantom{\varepsilon_1=\mi{reduced}^-(} \quad
                               ((\bigmult_{j=1}^{\kappa_1} \mu_1^j)\mult (\bigmult_{j=\kappa_1+1}^{\kappa_2} \mu_1^j)\mult 
                                (\bigmult_{j=\kappa_2+1}^{\kappa_3} \mu_2^j)\mult \bigmult_{j=\kappa_3+1}^{\kappa_4} \mu_2^j)), \\[1mm]
& \upsilon_2=\mi{reduced}^+(((\bigmult_{j=1}^{\kappa_1} \mu_2^j)\mult (\bigmult_{j=\kappa_1+1}^{\kappa_2} \mu_2^j)\mult 
                             (\bigmult_{j=\kappa_2+1}^{\kappa_3} \mu_1^j)\mult \bigmult_{j=\kappa_3+1}^{\kappa_4} \mu_1^j)\gleq \\
&\phantom{\upsilon_2=\mi{reduced}^+(} \quad
                            ((\bigmult_{j=1}^{\kappa_1} \mu_1^j)\mult (\bigmult_{j=\kappa_1+1}^{\kappa_2} \mu_1^j)\mult 
                             (\bigmult_{j=\kappa_2+1}^{\kappa_3} \mu_2^j)\mult \bigmult_{j=\kappa_3+1}^{\kappa_4} \mu_2^j));
\end{alignat*}
by the induction hypothesis for $\iota-1<\iota$,
$\|\varepsilon_1\|^{{\mc V}_{\iota-1}}, \|\upsilon_2\|^{{\mc V}_{\iota-1}},
 \|\mu_1^1\|^{{\mc V}_{\iota-1}},\dots,                                                                                                                                                    \linebreak[4]
                                       \|\mu_1^{\kappa_4}\|^{{\mc V}_{\iota-1}}, \|\mu_2^1\|^{{\mc V}_{\iota-1}},\dots,\|\mu_2^{\kappa_4}\|^{{\mc V}_{\iota-1}},
 \|\mi{reduced}^-(((\bigmult_{j=1}^{\kappa_1} \mu_2^j)\mult (\bigmult_{j=\kappa_1+1}^{\kappa_2} \mu_2^j)\mult 
                   (\bigmult_{j=\kappa_2+1}^{\kappa_3} \mu_1^j)\mult \bigmult_{j=\kappa_3+1}^{\kappa_4} \mu_1^j)\gleq
                  ((\bigmult_{j=1}^{\kappa_1} \mu_1^j)\mult (\bigmult_{j=\kappa_1+1}^{\kappa_2} \mu_1^j)\mult 
                   (\bigmult_{j=\kappa_2+1}^{\kappa_3} \mu_2^j)\mult \bigmult_{j=\kappa_3+1}^{\kappa_4} \mu_2^j))\|^{{\mc V}_{\iota-1}}, 
 \|\mi{reduced}^+(((\bigmult_{j=1}^{\kappa_1} \mu_2^j)\mult (\bigmult_{j=\kappa_1+1}^{\kappa_2} \mu_2^j)\mult 
                   (\bigmult_{j=\kappa_2+1}^{\kappa_3} \mu_1^j)\mult \bigmult_{j=\kappa_3+1}^{\kappa_4} \mu_1^j)\gleq
                  ((\bigmult_{j=1}^{\kappa_1} \mu_1^j)\mult (\bigmult_{j=\kappa_1+1}^{\kappa_2} \mu_1^j)\mult 
                   (\bigmult_{j=\kappa_2+1}^{\kappa_3} \mu_2^j)\mult \bigmult_{j=\kappa_3+1}^{\kappa_4} \mu_2^j))\|^{{\mc V}_{\iota-1}}\underset{\text{(a)}}{\in} (0,1]$;
$\mbb{E}_{\iota-1}=\emptyset$;
for all $1\leq j\leq \kappa_1$, 
$\mu_2^j\gleq (\mu_1^j\mult a_\iota^{\beta_j})\in \mi{min}(\mi{DE}_{\iota-1})$, 
$\left(\frac{\|\mu_2^j\|^{{\mc V}_{\iota-1}}}
            {\|\mu_1^j\|^{{\mc V}_{\iota-1}}}\right)^{\frac{1}{\beta_j}}\in \mbb{DE}_{\iota-1}$,
$\left(\frac{\|\mu_2^j\|^{{\mc V}_{\iota-1}}}
            {\|\mu_1^j\|^{{\mc V}_{\iota-1}}}\right)^{\frac{1}{\beta_j}}\leq \bigfvee \mbb{DE}_{\iota-1}\underset{\text{(\ref{eq8j}a)}}{\leq} {\mc V}_\iota(a_\iota)=\delta_\iota$,
$\frac{\|\mu_2^j\|^{{\mc V}_{\iota-1}}}
      {\|\mu_1^j\|^{{\mc V}_{\iota-1}}}\leq {\mc V}_\iota(a_\iota)^{\beta_j}$;
for all $\kappa_1+1\leq j\leq \kappa_2$, 
$\mu_2^j\gle (\mu_1^j\mult a_\iota^{\beta_j})\in \mi{min}(D_{\iota-1})$, 
$\left(\frac{\|\mu_2^j\|^{{\mc V}_{\iota-1}}}
            {\|\mu_1^j\|^{{\mc V}_{\iota-1}}}\right)^{\frac{1}{\beta_j}}\in \mbb{D}_{\iota-1}$,
$\left(\frac{\|\mu_2^j\|^{{\mc V}_{\iota-1}}}
            {\|\mu_1^j\|^{{\mc V}_{\iota-1}}}\right)^{\frac{1}{\beta_j}}\leq \bigfvee \mbb{D}_{\iota-1}\underset{\text{(\ref{eq8j}b)}}{<} {\mc V}_\iota(a_\iota)=\delta_\iota$,
$\frac{\|\mu_2^j\|^{{\mc V}_{\iota-1}}}
      {\|\mu_1^j\|^{{\mc V}_{\iota-1}}}<{\mc V}_\iota(a_\iota)^{\beta_j}$;
for all $\kappa_2+1\leq j\leq \kappa_3$,
$(\mu_1^j\mult a_\iota^{\beta_j})\gleq \mu_2^j\in \mi{min}(\mi{UE}_{\iota-1})$,
$\mi{min}\left(\left(\frac{\|\mu_2^j\|^{{\mc V}_{\iota-1}}}
                          {\|\mu_1^j\|^{{\mc V}_{\iota-1}}}\right)^{\frac{1}{\beta_j}},1\right)\in \mbb{UE}_{\iota-1}$,
${\mc V}_\iota(a_\iota)=\delta_\iota\underset{\text{(\ref{eq8j}a)}}{\leq} \bigfwedge \mbb{UE}_{\iota-1}\leq 
 \mi{min}\left(\left(\frac{\|\mu_2^j\|^{{\mc V}_{\iota-1}}}
                          {\|\mu_1^j\|^{{\mc V}_{\iota-1}}}\right)^{\frac{1}{\beta_j}},1\right)\leq
 \left(\frac{\|\mu_2^j\|^{{\mc V}_{\iota-1}}}
            {\|\mu_1^j\|^{{\mc V}_{\iota-1}}}\right)^{\frac{1}{\beta_j}}$,
${\mc V}_\iota(a_\iota)^{\beta_j}\leq \frac{\|\mu_2^j\|^{{\mc V}_{\iota-1}}}
                                           {\|\mu_1^j\|^{{\mc V}_{\iota-1}}}$,
$\frac{\|\mu_1^j\|^{{\mc V}_{\iota-1}}}
      {\|\mu_2^j\|^{{\mc V}_{\iota-1}}}\leq \frac{1}
                                                 {{\mc V}_\iota(a_\iota)^{\beta_j}}$;
for all $\kappa_3+1\leq j\leq \kappa_4$,
$(\mu_1^j\mult a_\iota^{\beta_j})\gle \mu_2^j\in \mi{min}(U_{\iota-1})$,
$\mi{min}\left(\left(\frac{\|\mu_2^j\|^{{\mc V}_{\iota-1}}}
                          {\|\mu_1^j\|^{{\mc V}_{\iota-1}}}\right)^{\frac{1}{\beta_j}},1\right)\in \mbb{U}_{\iota-1}$,
${\mc V}_\iota(a_\iota)=\delta_\iota\underset{\text{(\ref{eq8j}b)}}{\leq} \bigfwedge \mbb{U}_{\iota-1}\leq 
 \mi{min}\left(\left(\frac{\|\mu_2^j\|^{{\mc V}_{\iota-1}}}
                          {\|\mu_1^j\|^{{\mc V}_{\iota-1}}}\right)^{\frac{1}{\beta_j}},1\right)\leq
 \left(\frac{\|\mu_2^j\|^{{\mc V}_{\iota-1}}}
            {\|\mu_1^j\|^{{\mc V}_{\iota-1}}}\right)^{\frac{1}{\beta_j}}$,
${\mc V}_\iota(a_\iota)^{\beta_j}\leq \frac{\|\mu_2^j\|^{{\mc V}_{\iota-1}}}
                                           {\|\mu_1^j\|^{{\mc V}_{\iota-1}}}$,
$\frac{\|\mu_1^j\|^{{\mc V}_{\iota-1}}}
      {\|\mu_2^j\|^{{\mc V}_{\iota-1}}}\leq \frac{1}
                                                 {{\mc V}_\iota(a_\iota)^{\beta_j}}$;
\begin{alignat*}{1}
& \dfrac{\|\varepsilon_1\|^{{\mc V}_{\iota-1}}}
        {\|\upsilon_2\|^{{\mc V}_{\iota-1}}}= \\
& \dfrac{\left\|\mi{reduced}^-\left(\begin{array}{l}
                                    ((\bigmult_{j=1}^{\kappa_1} \mu_2^j)\mult (\bigmult_{j=\kappa_1+1}^{\kappa_2} \mu_2^j)\mult 
                                     (\bigmult_{j=\kappa_2+1}^{\kappa_3} \mu_1^j)\mult \bigmult_{j=\kappa_3+1}^{\kappa_4} \mu_1^j)\gleq \\
                                    \quad
                                    ((\bigmult_{j=1}^{\kappa_1} \mu_1^j)\mult (\bigmult_{j=\kappa_1+1}^{\kappa_2} \mu_1^j)\mult 
                                     (\bigmult_{j=\kappa_2+1}^{\kappa_3} \mu_2^j)\mult \bigmult_{j=\kappa_3+1}^{\kappa_4} \mu_2^j)
                                    \end{array}\right)\right\|^{{\mc V}_{\iota-1}}}
        {\left\|\mi{reduced}^+\left(\begin{array}{l}
                                    ((\bigmult_{j=1}^{\kappa_1} \mu_2^j)\mult (\bigmult_{j=\kappa_1+1}^{\kappa_2} \mu_2^j)\mult 
                                     (\bigmult_{j=\kappa_2+1}^{\kappa_3} \mu_1^j)\mult \bigmult_{j=\kappa_3+1}^{\kappa_4} \mu_1^j)\gleq \\
                                    \quad
                                    ((\bigmult_{j=1}^{\kappa_1} \mu_1^j)\mult (\bigmult_{j=\kappa_1+1}^{\kappa_2} \mu_1^j)\mult 
                                     (\bigmult_{j=\kappa_2+1}^{\kappa_3} \mu_2^j)\mult \bigmult_{j=\kappa_3+1}^{\kappa_4} \mu_2^j)
                                    \end{array}\right)\right\|^{{\mc V}_{\iota-1}}}\overset{\text{(\ref{eq7n})}}{=\!\!=} \\
& \dfrac{\|(\bigmult_{j=1}^{\kappa_1} \mu_2^j)\mult (\bigmult_{j=\kappa_1+1}^{\kappa_2} \mu_2^j)\mult 
           (\bigmult_{j=\kappa_2+1}^{\kappa_3} \mu_1^j)\mult \bigmult_{j=\kappa_3+1}^{\kappa_4} \mu_1^j\|^{{\mc V}_{\iota-1}}}
        {\|(\bigmult_{j=1}^{\kappa_1} \mu_1^j)\mult (\bigmult_{j=\kappa_1+1}^{\kappa_2} \mu_1^j)\mult 
           (\bigmult_{j=\kappa_2+1}^{\kappa_3} \mu_2^j)\mult \bigmult_{j=\kappa_3+1}^{\kappa_4} \mu_2^j\|^{{\mc V}_{\iota-1}}}= \\
& \dfrac{(\bigfswedge_{j=1}^{\kappa_1} \|\mu_2^j\|^{{\mc V}_{\iota-1}})\fswedge (\bigfswedge_{j=\kappa_1+1}^{\kappa_2} \|\mu_2^j\|^{{\mc V}_{\iota-1}})\fswedge
         (\bigfswedge_{j=\kappa_2+1}^{\kappa_3} \|\mu_1^j\|^{{\mc V}_{\iota-1}})\fswedge \bigfswedge_{j=\kappa_3+1}^{\kappa_4} \|\mu_1^j\|^{{\mc V}_{\iota-1}}}
        {(\bigfswedge_{j=1}^{\kappa_1} \|\mu_1^j\|^{{\mc V}_{\iota-1}})\fswedge (\bigfswedge_{j=\kappa_1+1}^{\kappa_2} \|\mu_1^j\|^{{\mc V}_{\iota-1}})\fswedge
         (\bigfswedge_{j=\kappa_2+1}^{\kappa_3} \|\mu_2^j\|^{{\mc V}_{\iota-1}})\fswedge \bigfswedge_{j=\kappa_3+1}^{\kappa_4} \|\mu_2^j\|^{{\mc V}_{\iota-1}}}= \\
& \left(\bigfswedge_{j=1}^{\kappa_1} \dfrac{\|\mu_2^j\|^{{\mc V}_{\iota-1}}}
                                           {\|\mu_1^j\|^{{\mc V}_{\iota-1}}}\right)\fswedge 
  \left(\bigfswedge_{j=\kappa_1+1}^{\kappa_2} \dfrac{\|\mu_2^j\|^{{\mc V}_{\iota-1}}}
                                                    {\|\mu_1^j\|^{{\mc V}_{\iota-1}}}\right)\fswedge 
  \left(\bigfswedge_{j=\kappa_1+2}^{\kappa_3} \dfrac{\|\mu_1^j\|^{{\mc V}_{\iota-1}}}
                                                    {\|\mu_2^j\|^{{\mc V}_{\iota-1}}}\right)\fswedge
  \bigfswedge_{j=\kappa_1+3}^{\kappa_4} \dfrac{\|\mu_1^j\|^{{\mc V}_{\iota-1}}}
                                              {\|\mu_2^j\|^{{\mc V}_{\iota-1}}}\leq \\
& \left(\bigfswedge_{j=1}^{\kappa_1} {\mc V}_\iota(a_\iota)^{\beta_j}\right)\fswedge \left(\bigfswedge_{j=\kappa_1+1}^{\kappa_2} {\mc V}_\iota(a_\iota)^{\beta_j}\right)\fswedge 
  \left(\bigfswedge_{j=\kappa_1+2}^{\kappa_3} \dfrac{1}
                                                    {{\mc V}_\iota(a_\iota)^{\beta_j}}\right)\fswedge
  \bigfswedge_{j=\kappa_1+3}^{\kappa_4} \dfrac{1}
                                              {{\mc V}_\iota(a_\iota)^{\beta_j}}= \\
& {\mc V}_\iota(a_\iota)^{(\sum_{j=1}^{\kappa_1} \beta_j)+(\sum_{j=\kappa_1+1}^{\kappa_2} \beta_j)-(\sum_{j=\kappa_2+1}^{\kappa_3} \beta_j)-\sum_{j=\kappa_3+1}^{\kappa_4} \beta_j}=
  {\mc V}_\iota(a_\iota)^\alpha, \\[1mm]
& \|\varepsilon_1\|^{{\mc V}_\iota}\overset{\text{(b)}}{=\!\!=} \|\varepsilon_1\|^{{\mc V}_{\iota-1}}\leq 
  \|\upsilon_2\|^{{\mc V}_{\iota-1}}\fswedge {\mc V}_\iota(a_\iota)^\alpha\overset{\text{(b)}}{=\!\!=}
  \|\upsilon_2\|^{{\mc V}_\iota}\fswedge {\mc V}_\iota(a_\iota)^\alpha=\|\upsilon_2\mult a_\iota^\alpha\|^{{\mc V}_\iota}=\|\varepsilon_2\|^{{\mc V}_\iota};
\end{alignat*}
(d) holds.

Case 2.10.2.2.1.2:
There exists $\zeta_1\diamond^\zeta \zeta_2=\mi{reduced}(\lambda^{(\gamma_1^\zeta,\dots,\gamma_n^\zeta)})\in \mi{clo}$,
$\zeta_e\in \{\gu\}\cup \mi{PropConj}_A$, $\mi{atoms}(\zeta_e)\subseteq \mi{dom}({\mc V}_{\iota-1})$, $\diamond^\zeta\in \{\geql,\gleq,\gle\}$, 
$\bs{0}^n\neq (\gamma_1^\zeta,\dots,\gamma_n^\zeta)\in \mbb{N}^n$, such that
$\varepsilon_1\gleq \varepsilon_2=\varepsilon_1\gleq (\upsilon_2\mult a_\iota^\alpha)=
                                  \mi{reduced}(\lambda^{((\sum_{j=1}^{\kappa_4} \gamma_1^j)+\gamma_1^\zeta,\dots,(\sum_{j=1}^{\kappa_4} \gamma_n^j)+\gamma_n^\zeta)})$.
Then 
$\mi{atoms}(\varepsilon_1), \mi{atoms}(\upsilon_2), \mi{atoms}(\mu_1^1),\dots,\mi{atoms}(\mu_1^{\kappa_4}), \mi{atoms}(\mu_2^1),\dots,\mi{atoms}(\mu_2^{\kappa_4})\subseteq \mi{dom}({\mc V}_{\iota-1})$,
\begin{alignat*}{1}
& \varepsilon_1\gleq (\upsilon_2\mult a_\iota^\alpha)=
  \mi{reduced}(\lambda^{((\sum_{j=1}^{\kappa_4} \gamma_1^j)+\gamma_1^\zeta,\dots,(\sum_{j=1}^{\kappa_4} \gamma_n^j)+\gamma_n^\zeta)})\overset{\text{(\ref{eq8dddd})}}{=\!\!=} \\
& \mi{reduced}((\bigmult_{j=1}^{\kappa_1} \lambda^{(\gamma_1^j,\dots,\gamma_n^j)})\mult
               (\bigmult_{j=\kappa_1+1}^{\kappa_2} \lambda^{(\gamma_1^j,\dots,\gamma_n^j)})\mult \\
&\phantom{\mi{reduced}(}
               (\bigmult_{j=\kappa_2+1}^{\kappa_3} \lambda^{(\gamma_1^j,\dots,\gamma_n^j)})\mult
               (\bigmult_{j=\kappa_3+1}^{\kappa_4} \lambda^{(\gamma_1^j,\dots,\gamma_n^j)})\mult
               \lambda^{(\gamma_1^\zeta,\dots,\gamma_n^\zeta)})\overset{\text{(\ref{eq7k})}}{=\!\!=} \\
& \mi{reduced}((\bigmult_{j=1}^{\kappa_1} \mi{reduced}(\lambda^{(\gamma_1^j,\dots,\gamma_n^j)}))\mult
               (\bigmult_{j=\kappa_1+1}^{\kappa_2} \mi{reduced}(\lambda^{(\gamma_1^j,\dots,\gamma_n^j)}))\mult \\
&\phantom{\mi{reduced}(}
               (\bigmult_{j=\kappa_2+1}^{\kappa_3} \mi{reduced}(\lambda^{(\gamma_1^j,\dots,\gamma_n^j)}))\mult
               (\bigmult_{j=\kappa_3+1}^{\kappa_4} \mi{reduced}(\lambda^{(\gamma_1^j,\dots,\gamma_n^j)}))\mult \\
&\phantom{\mi{reduced}(}
               \mi{reduced}(\lambda^{(\gamma_1^\zeta,\dots,\gamma_n^\zeta)}))= \\
& \mi{reduced}((\bigmult_{j=1}^{\kappa_1} \mu_2^j\gleq (\mu_1^j\mult a_\iota^{\beta_j}))\mult
               (\bigmult_{j=\kappa_1+1}^{\kappa_2} \mu_2^j\gle (\mu_1^j\mult a_\iota^{\beta_j}))\mult \\
&\phantom{\mi{reduced}(}
               (\bigmult_{j=\kappa_2+1}^{\kappa_3} (\mu_1^j\mult a_\iota^{\beta_j})\gleq \mu_2^j)\mult
               (\bigmult_{j=\kappa_3+1}^{\kappa_4} (\mu_1^j\mult a_\iota^{\beta_j})\gle \mu_2^j)\mult
               (\zeta_1\diamond^\zeta \zeta_2))\overset{\text{(\ref{eq7i})}}{=\!\!=} \\
& \mi{reduced}(((\bigmult_{j=1}^{\kappa_1} \mu_2^j)\mult (\bigmult_{j=\kappa_1+1}^{\kappa_2} \mu_2^j)\mult 
                (\bigmult_{j=\kappa_2+1}^{\kappa_3} \mu_1^j)\mult (\bigmult_{j=\kappa_3+1}^{\kappa_4} \mu_1^j)\mult \zeta_1\mult \\
&\phantom{\mi{reduced}(} \quad
                a_\iota^{(\sum_{j=\kappa_2+1}^{\kappa_3} \beta_j)+\sum_{j=\kappa_3+1}^{\kappa_4} \beta_j})\gleq \\
&\phantom{\mi{reduced}(} \quad
               ((\bigmult_{j=1}^{\kappa_1} \mu_1^j)\mult (\bigmult_{j=\kappa_1+1}^{\kappa_2} \mu_1^j)\mult 
                (\bigmult_{j=\kappa_2+1}^{\kappa_3} \mu_2^j)\mult (\bigmult_{j=\kappa_3+1}^{\kappa_4} \mu_2^j)\mult \zeta_2\mult \\
&\phantom{\mi{reduced}(} \quad \quad
                a_\iota^{(\sum_{j=1}^{\kappa_1} \beta_j)+\sum_{j=\kappa_1+1}^{\kappa_2} \beta_j}))= \\
& \mi{reduced}^-(((\bigmult_{j=1}^{\kappa_1} \mu_2^j)\mult (\bigmult_{j=\kappa_1+1}^{\kappa_2} \mu_2^j)\mult 
                  (\bigmult_{j=\kappa_2+1}^{\kappa_3} \mu_1^j)\mult (\bigmult_{j=\kappa_3+1}^{\kappa_4} \mu_1^j)\mult \zeta_1\mult \\
&\phantom{\mi{reduced}^-(} \quad
                  a_\iota^{(\sum_{j=\kappa_2+1}^{\kappa_3} \beta_j)+\sum_{j=\kappa_3+1}^{\kappa_4} \beta_j})\gleq \\
&\phantom{\mi{reduced}^-(} \quad
                 ((\bigmult_{j=1}^{\kappa_1} \mu_1^j)\mult (\bigmult_{j=\kappa_1+1}^{\kappa_2} \mu_1^j)\mult 
                  (\bigmult_{j=\kappa_2+1}^{\kappa_3} \mu_2^j)\mult (\bigmult_{j=\kappa_3+1}^{\kappa_4} \mu_2^j)\mult \zeta_2\mult \\
&\phantom{\mi{reduced}^-(} \quad \quad
                  a_\iota^{(\sum_{j=1}^{\kappa_1} \beta_j)+\sum_{j=\kappa_1+1}^{\kappa_2} \beta_j}))\gleq \\
& \mi{reduced}^+(((\bigmult_{j=1}^{\kappa_1} \mu_2^j)\mult (\bigmult_{j=\kappa_1+1}^{\kappa_2} \mu_2^j)\mult 
                  (\bigmult_{j=\kappa_2+1}^{\kappa_3} \mu_1^j)\mult (\bigmult_{j=\kappa_3+1}^{\kappa_4} \mu_1^j)\mult \zeta_1\mult \\
&\phantom{\mi{reduced}^+(} \quad
                  a_\iota^{(\sum_{j=\kappa_2+1}^{\kappa_3} \beta_j)+\sum_{j=\kappa_3+1}^{\kappa_4} \beta_j})\gleq \\
&\phantom{\mi{reduced}^+(} \quad
                 ((\bigmult_{j=1}^{\kappa_1} \mu_1^j)\mult (\bigmult_{j=\kappa_1+1}^{\kappa_2} \mu_1^j)\mult 
                  (\bigmult_{j=\kappa_2+1}^{\kappa_3} \mu_2^j)\mult (\bigmult_{j=\kappa_3+1}^{\kappa_4} \mu_2^j)\mult \zeta_2\mult \\
&\phantom{\mi{reduced}^+(} \quad \quad
                  a_\iota^{(\sum_{j=1}^{\kappa_1} \beta_j)+\sum_{j=\kappa_1+1}^{\kappa_2} \beta_j})), \\[1mm]
& \#(a_\iota,\varepsilon_1\gleq (\upsilon_2\mult a_\iota^\alpha))=\alpha= \\
& \#(a_\iota,\mi{reduced}(((\bigmult_{j=1}^{\kappa_1} \mu_2^j)\mult (\bigmult_{j=\kappa_1+1}^{\kappa_2} \mu_2^j)\mult 
                           (\bigmult_{j=\kappa_2+1}^{\kappa_3} \mu_1^j)\mult (\bigmult_{j=\kappa_3+1}^{\kappa_4} \mu_1^j)\mult \zeta_1\mult \\
&\phantom{\#(a_\iota,\mi{reduced}(} \quad
                           a_\iota^{(\sum_{j=\kappa_2+1}^{\kappa_3} \beta_j)+\sum_{j=\kappa_3+1}^{\kappa_4} \beta_j})\gleq \\
&\phantom{\#(a_\iota,\mi{reduced}(} \quad
                          ((\bigmult_{j=1}^{\kappa_1} \mu_1^j)\mult (\bigmult_{j=\kappa_1+1}^{\kappa_2} \mu_1^j)\mult 
                           (\bigmult_{j=\kappa_2+1}^{\kappa_3} \mu_2^j)\mult (\bigmult_{j=\kappa_3+1}^{\kappa_4} \mu_2^j)\mult \zeta_2\mult \\
&\phantom{\#(a_\iota,\mi{reduced}(} \quad \quad
                           a_\iota^{(\sum_{j=1}^{\kappa_1} \beta_j)+\sum_{j=\kappa_1+1}^{\kappa_2} \beta_j})))\overset{\text{(\ref{eq7aaaax})}}{=\!\!=} \\
& \#(a_\iota,((\bigmult_{j=1}^{\kappa_1} \mu_2^j)\mult (\bigmult_{j=\kappa_1+1}^{\kappa_2} \mu_2^j)\mult 
              (\bigmult_{j=\kappa_2+1}^{\kappa_3} \mu_1^j)\mult (\bigmult_{j=\kappa_3+1}^{\kappa_4} \mu_1^j)\mult \zeta_1\mult \\
&\phantom{\#(a_\iota,} \quad
              a_\iota^{(\sum_{j=\kappa_2+1}^{\kappa_3} \beta_j)+\sum_{j=\kappa_3+1}^{\kappa_4} \beta_j})\gleq \\
&\phantom{\#(a_\iota,} \quad
             ((\bigmult_{j=1}^{\kappa_1} \mu_1^j)\mult (\bigmult_{j=\kappa_1+1}^{\kappa_2} \mu_1^j)\mult 
              (\bigmult_{j=\kappa_2+1}^{\kappa_3} \mu_2^j)\mult (\bigmult_{j=\kappa_3+1}^{\kappa_4} \mu_2^j)\mult \zeta_2\mult \\
&\phantom{\#(a_\iota,} \quad \quad
              a_\iota^{(\sum_{j=1}^{\kappa_1} \beta_j)+\sum_{j=\kappa_1+1}^{\kappa_2} \beta_j}))= \\
& \left(\sum_{j=1}^{\kappa_1} \beta_j\right)+\left(\sum_{j=\kappa_1+1}^{\kappa_2} \beta_j\right)-\left(\sum_{j=\kappa_2+1}^{\kappa_3} \beta_j\right)-\sum_{j=\kappa_3+1}^{\kappa_4} \beta_j\geq 1, 
\end{alignat*}
$a_\iota\not\in \mi{dom}({\mc V}_{\iota-1})\supseteq \mi{atoms}((\bigmult_{j=1}^{\kappa_1} \mu_2^j)\mult (\bigmult_{j=\kappa_1+1}^{\kappa_2} \mu_2^j)\mult 
                                                                (\bigmult_{j=\kappa_2+1}^{\kappa_3} \mu_1^j)\mult                                                                          \linebreak[4]
                                                                                                                  (\bigmult_{j=\kappa_3+1}^{\kappa_4} \mu_1^j)\mult \zeta_1)=
                                                     (\bigcup_{j=1}^{\kappa_1} \mi{atoms}(\mu_2^j))\cup (\bigcup_{j=\kappa_1+1}^{\kappa_2} \mi{atoms}(\mu_2^j))\cup                        \linebreak[4]
                                                     (\bigcup_{j=\kappa_2+1}^{\kappa_3} \mi{atoms}(\mu_1^j))\cup (\bigcup_{j=\kappa_3+1}^{\kappa_4} \mi{atoms}(\mu_1^j))\cup \mi{atoms}(\zeta_1)$,
$a_\iota\not\in \mi{dom}({\mc V}_{\iota-1})\supseteq \mi{atoms}((\bigmult_{j=1}^{\kappa_1} \mu_1^j)\mult (\bigmult_{j=\kappa_1+1}^{\kappa_2} \mu_1^j)\mult 
                                                                (\bigmult_{j=\kappa_2+1}^{\kappa_3} \mu_2^j)\mult (\bigmult_{j=\kappa_3+1}^{\kappa_4} \mu_2^j)\mult \zeta_2)=              \linebreak[4]
                                                     (\bigcup_{j=1}^{\kappa_1} \mi{atoms}(\mu_1^j))\cup (\bigcup_{j=\kappa_1+1}^{\kappa_2} \mi{atoms}(\mu_1^j))\cup 
                                                     (\bigcup_{j=\kappa_2+1}^{\kappa_3} \mi{atoms}(\mu_2^j))\cup (\bigcup_{j=\kappa_3+1}^{\kappa_4} \mi{atoms}(\mu_2^j))\cup \mi{atoms}(\zeta_2)$,
\begin{alignat*}{1}
& \mi{reduced}^-(((\bigmult_{j=1}^{\kappa_1} \mu_2^j)\mult (\bigmult_{j=\kappa_1+1}^{\kappa_2} \mu_2^j)\mult 
                  (\bigmult_{j=\kappa_2+1}^{\kappa_3} \mu_1^j)\mult (\bigmult_{j=\kappa_3+1}^{\kappa_4} \mu_1^j)\mult \zeta_1\mult \\
&\phantom{\mi{reduced}^-(} \quad
                  a_\iota^{(\sum_{j=\kappa_2+1}^{\kappa_3} \beta_j)+\sum_{j=\kappa_3+1}^{\kappa_4} \beta_j})\gleq \\
&\phantom{\mi{reduced}^-(} \quad
                 ((\bigmult_{j=1}^{\kappa_1} \mu_1^j)\mult (\bigmult_{j=\kappa_1+1}^{\kappa_2} \mu_1^j)\mult 
                  (\bigmult_{j=\kappa_2+1}^{\kappa_3} \mu_2^j)\mult (\bigmult_{j=\kappa_3+1}^{\kappa_4} \mu_2^j)\mult \zeta_2\mult \\
&\phantom{\mi{reduced}^-(} \quad \quad
                  a_\iota^{(\sum_{j=1}^{\kappa_1} \beta_j)+\sum_{j=\kappa_1+1}^{\kappa_2} \beta_j}))\overset{\text{(\ref{eq7cc})}}{=\!\!=} \\
& \mi{reduced}^-(((\bigmult_{j=1}^{\kappa_1} \mu_2^j)\mult (\bigmult_{j=\kappa_1+1}^{\kappa_2} \mu_2^j)\mult 
                  (\bigmult_{j=\kappa_2+1}^{\kappa_3} \mu_1^j)\mult (\bigmult_{j=\kappa_3+1}^{\kappa_4} \mu_1^j)\mult \zeta_1)\gleq \\
&\phantom{\mi{reduced}^-(} \quad
                 ((\bigmult_{j=1}^{\kappa_1} \mu_1^j)\mult (\bigmult_{j=\kappa_1+1}^{\kappa_2} \mu_1^j)\mult 
                  (\bigmult_{j=\kappa_2+1}^{\kappa_3} \mu_2^j)\mult (\bigmult_{j=\kappa_3+1}^{\kappa_4} \mu_2^j)\mult \zeta_2)), \\[1mm]
& \mi{reduced}^+(((\bigmult_{j=1}^{\kappa_1} \mu_2^j)\mult (\bigmult_{j=\kappa_1+1}^{\kappa_2} \mu_2^j)\mult 
                  (\bigmult_{j=\kappa_2+1}^{\kappa_3} \mu_1^j)\mult (\bigmult_{j=\kappa_3+1}^{\kappa_4} \mu_1^j)\mult \zeta_1\mult \\
&\phantom{\mi{reduced}^+(} \quad
                  a_\iota^{(\sum_{j=\kappa_2+1}^{\kappa_3} \beta_j)+\sum_{j=\kappa_3+1}^{\kappa_4} \beta_j})\gleq \\
&\phantom{\mi{reduced}^+(} \quad
                 ((\bigmult_{j=1}^{\kappa_1} \mu_1^j)\mult (\bigmult_{j=\kappa_1+1}^{\kappa_2} \mu_1^j)\mult 
                  (\bigmult_{j=\kappa_2+1}^{\kappa_3} \mu_2^j)\mult (\bigmult_{j=\kappa_3+1}^{\kappa_4} \mu_2^j)\mult \zeta_2\mult \\
&\phantom{\mi{reduced}^+(} \quad \quad
                  a_\iota^{(\sum_{j=1}^{\kappa_1} \beta_j)+\sum_{j=\kappa_1+1}^{\kappa_2} \beta_j}))\overset{\text{(\ref{eq7ff})}}{=\!\!=} \\
& \mi{reduced}^+(((\bigmult_{j=1}^{\kappa_1} \mu_2^j)\mult (\bigmult_{j=\kappa_1+1}^{\kappa_2} \mu_2^j)\mult 
                  (\bigmult_{j=\kappa_2+1}^{\kappa_3} \mu_1^j)\mult (\bigmult_{j=\kappa_3+1}^{\kappa_4} \mu_1^j)\mult \zeta_1)\gleq \\
&\phantom{\mi{reduced}^+(} \quad
                 ((\bigmult_{j=1}^{\kappa_1} \mu_1^j)\mult (\bigmult_{j=\kappa_1+1}^{\kappa_2} \mu_1^j)\mult 
                  (\bigmult_{j=\kappa_2+1}^{\kappa_3} \mu_2^j)\mult (\bigmult_{j=\kappa_3+1}^{\kappa_4} \mu_2^j)\mult \zeta_2))\mult \\
& a_\iota^{(\sum_{j=1}^{\kappa_1} \beta_j)+(\sum_{j=\kappa_1+1}^{\kappa_2} \beta_j)-(\sum_{j=\kappa_2+1}^{\kappa_3} \beta_j)-\sum_{j=\kappa_3+1}^{\kappa_4} \beta_j}= \\
& \mi{reduced}^+(((\bigmult_{j=1}^{\kappa_1} \mu_2^j)\mult (\bigmult_{j=\kappa_1+1}^{\kappa_2} \mu_2^j)\mult 
                  (\bigmult_{j=\kappa_2+1}^{\kappa_3} \mu_1^j)\mult (\bigmult_{j=\kappa_3+1}^{\kappa_4} \mu_1^j)\mult \zeta_1)\gleq \\
&\phantom{\mi{reduced}^+(} \quad
                 ((\bigmult_{j=1}^{\kappa_1} \mu_1^j)\mult (\bigmult_{j=\kappa_1+1}^{\kappa_2} \mu_1^j)\mult 
                  (\bigmult_{j=\kappa_2+1}^{\kappa_3} \mu_2^j)\mult (\bigmult_{j=\kappa_3+1}^{\kappa_4} \mu_2^j)\mult \zeta_2))\mult a_\iota^\alpha, \\[1mm]
& \varepsilon_1\gleq (\upsilon_2\mult a_\iota^\alpha)= \\
& \mi{reduced}^-(((\bigmult_{j=1}^{\kappa_1} \mu_2^j)\mult (\bigmult_{j=\kappa_1+1}^{\kappa_2} \mu_2^j)\mult 
                  (\bigmult_{j=\kappa_2+1}^{\kappa_3} \mu_1^j)\mult (\bigmult_{j=\kappa_3+1}^{\kappa_4} \mu_1^j)\mult \zeta_1)\gleq \\
&\phantom{\mi{reduced}^-(} \quad
                 ((\bigmult_{j=1}^{\kappa_1} \mu_1^j)\mult (\bigmult_{j=\kappa_1+1}^{\kappa_2} \mu_1^j)\mult 
                  (\bigmult_{j=\kappa_2+1}^{\kappa_3} \mu_2^j)\mult (\bigmult_{j=\kappa_3+1}^{\kappa_4} \mu_2^j)\mult \zeta_2))\gleq \\
& (\mi{reduced}^+(((\bigmult_{j=1}^{\kappa_1} \mu_2^j)\mult (\bigmult_{j=\kappa_1+1}^{\kappa_2} \mu_2^j)\mult 
                   (\bigmult_{j=\kappa_2+1}^{\kappa_3} \mu_1^j)\mult (\bigmult_{j=\kappa_3+1}^{\kappa_4} \mu_1^j)\mult \zeta_1)\gleq \\
&\phantom{(\mi{reduced}^+(} \quad
                  ((\bigmult_{j=1}^{\kappa_1} \mu_1^j)\mult (\bigmult_{j=\kappa_1+1}^{\kappa_2} \mu_1^j)\mult 
                   (\bigmult_{j=\kappa_2+1}^{\kappa_3} \mu_2^j)\mult (\bigmult_{j=\kappa_3+1}^{\kappa_4} \mu_2^j)\mult \zeta_2))\mult a_\iota^\alpha), \\[1mm]
& \mi{atoms}(\mi{reduced}^-(((\bigmult_{j=1}^{\kappa_1} \mu_2^j)\mult (\bigmult_{j=\kappa_1+1}^{\kappa_2} \mu_2^j)\mult 
                             (\bigmult_{j=\kappa_2+1}^{\kappa_3} \mu_1^j)\mult (\bigmult_{j=\kappa_3+1}^{\kappa_4} \mu_1^j)\mult \zeta_1)\gleq \\
&\phantom{\mi{atoms}(\mi{reduced}^-(} \quad
                            ((\bigmult_{j=1}^{\kappa_1} \mu_1^j)\mult (\bigmult_{j=\kappa_1+1}^{\kappa_2} \mu_1^j)\mult 
                             (\bigmult_{j=\kappa_2+1}^{\kappa_3} \mu_2^j)\mult (\bigmult_{j=\kappa_3+1}^{\kappa_4} \mu_2^j)\mult \zeta_2)))\underset{\text{(\ref{eq7a})}}{\subseteq} \\
& \mi{atoms}((\bigmult_{j=1}^{\kappa_1} \mu_2^j)\mult (\bigmult_{j=\kappa_1+1}^{\kappa_2} \mu_2^j)\mult 
             (\bigmult_{j=\kappa_2+1}^{\kappa_3} \mu_1^j)\mult (\bigmult_{j=\kappa_3+1}^{\kappa_4} \mu_1^j)\mult \zeta_1)\subseteq \mi{dom}({\mc V}_{\iota-1}), \\[1mm]
& \mi{atoms}(\mi{reduced}^+(((\bigmult_{j=1}^{\kappa_1} \mu_2^j)\mult (\bigmult_{j=\kappa_1+1}^{\kappa_2} \mu_2^j)\mult 
                             (\bigmult_{j=\kappa_2+1}^{\kappa_3} \mu_1^j)\mult (\bigmult_{j=\kappa_3+1}^{\kappa_4} \mu_1^j)\mult \zeta_1)\gleq \\
&\phantom{\mi{atoms}(\mi{reduced}^+(} \quad
                            ((\bigmult_{j=1}^{\kappa_1} \mu_1^j)\mult (\bigmult_{j=\kappa_1+1}^{\kappa_2} \mu_1^j)\mult 
                             (\bigmult_{j=\kappa_2+1}^{\kappa_3} \mu_2^j)\mult (\bigmult_{j=\kappa_3+1}^{\kappa_4} \mu_2^j)\mult \zeta_2)))\underset{\text{(\ref{eq7b})}}{\subseteq} \\
& \mi{atoms}((\bigmult_{j=1}^{\kappa_1} \mu_1^j)\mult (\bigmult_{j=\kappa_1+1}^{\kappa_2} \mu_1^j)\mult 
             (\bigmult_{j=\kappa_2+1}^{\kappa_3} \mu_2^j)\mult (\bigmult_{j=\kappa_3+1}^{\kappa_4} \mu_2^j)\mult \zeta_2)\subseteq \mi{dom}({\mc V}_{\iota-1}), 
\end{alignat*}
\begin{alignat*}{1}
& \varepsilon_1=\mi{reduced}^-(((\bigmult_{j=1}^{\kappa_1} \mu_2^j)\mult (\bigmult_{j=\kappa_1+1}^{\kappa_2} \mu_2^j)\mult 
                                (\bigmult_{j=\kappa_2+1}^{\kappa_3} \mu_1^j)\mult (\bigmult_{j=\kappa_3+1}^{\kappa_4} \mu_1^j)\mult \zeta_1)\gleq \\
&\phantom{\varepsilon_1=\mi{reduced}^-(} \quad
                               ((\bigmult_{j=1}^{\kappa_1} \mu_1^j)\mult (\bigmult_{j=\kappa_1+1}^{\kappa_2} \mu_1^j)\mult 
                                (\bigmult_{j=\kappa_2+1}^{\kappa_3} \mu_2^j)\mult (\bigmult_{j=\kappa_3+1}^{\kappa_4} \mu_2^j)\mult \zeta_2)), \\[1mm]
& \upsilon_2=\mi{reduced}^+(((\bigmult_{j=1}^{\kappa_1} \mu_2^j)\mult (\bigmult_{j=\kappa_1+1}^{\kappa_2} \mu_2^j)\mult 
                             (\bigmult_{j=\kappa_2+1}^{\kappa_3} \mu_1^j)\mult (\bigmult_{j=\kappa_3+1}^{\kappa_4} \mu_1^j)\mult \zeta_1)\gleq \\
&\phantom{\upsilon_2=\mi{reduced}^+(} \quad
                            ((\bigmult_{j=1}^{\kappa_1} \mu_1^j)\mult (\bigmult_{j=\kappa_1+1}^{\kappa_2} \mu_1^j)\mult 
                             (\bigmult_{j=\kappa_2+1}^{\kappa_3} \mu_2^j)\mult (\bigmult_{j=\kappa_3+1}^{\kappa_4} \mu_2^j)\mult \zeta_2));
\end{alignat*}
by the induction hypothesis for $\iota-1<\iota$,
$\|\varepsilon_1\|^{{\mc V}_{\iota-1}}, \|\upsilon_2\|^{{\mc V}_{\iota-1}},
 \|\mu_1^1\|^{{\mc V}_{\iota-1}},\dots,                                                                                                                                                    \linebreak[4]
                                       \|\mu_1^{\kappa_4}\|^{{\mc V}_{\iota-1}}, \|\mu_2^1\|^{{\mc V}_{\iota-1}},\dots,\|\mu_2^{\kappa_4}\|^{{\mc V}_{\iota-1}},
 \|\zeta_1\|^{{\mc V}_{\iota-1}}, \|\zeta_2\|^{{\mc V}_{\iota-1}},
 \|\mi{reduced}^-(((\bigmult_{j=1}^{\kappa_1} \mu_2^j)\mult (\bigmult_{j=\kappa_1+1}^{\kappa_2} \mu_2^j)\mult 
                   (\bigmult_{j=\kappa_2+1}^{\kappa_3} \mu_1^j)\mult (\bigmult_{j=\kappa_3+1}^{\kappa_4} \mu_1^j)\mult \zeta_1)\gleq
                  ((\bigmult_{j=1}^{\kappa_1} \mu_1^j)\mult (\bigmult_{j=\kappa_1+1}^{\kappa_2} \mu_1^j)\mult 
                   (\bigmult_{j=\kappa_2+1}^{\kappa_3} \mu_2^j)\mult (\bigmult_{j=\kappa_3+1}^{\kappa_4} \mu_2^j)\mult \zeta_2))\|^{{\mc V}_{\iota-1}}, 
 \|\mi{reduced}^+(((\bigmult_{j=1}^{\kappa_1} \mu_2^j)\mult (\bigmult_{j=\kappa_1+1}^{\kappa_2} \mu_2^j)\mult 
                   (\bigmult_{j=\kappa_2+1}^{\kappa_3} \mu_1^j)\mult (\bigmult_{j=\kappa_3+1}^{\kappa_4} \mu_1^j)\mult \zeta_1)\gleq
                  ((\bigmult_{j=1}^{\kappa_1} \mu_1^j)\mult (\bigmult_{j=\kappa_1+1}^{\kappa_2} \mu_1^j)\mult 
                   (\bigmult_{j=\kappa_2+1}^{\kappa_3} \mu_2^j)\mult (\bigmult_{j=\kappa_3+1}^{\kappa_4} \mu_2^j)\mult \zeta_2))\|^{{\mc V}_{\iota-1}}\underset{\text{(a)}}{\in} (0,1]$,
either $\diamond^\zeta=\geql$, $\zeta_1\diamond^\zeta \zeta_2=\zeta_1\geql \zeta_2\in \mi{clo}$, $\|\zeta_1\|^{{\mc V}_{\iota-1}}\overset{\text{(c)}}{=\!\!=} \|\zeta_2\|^{{\mc V}_{\iota-1}}$, 
or $\diamond^\zeta=\gleq$, $\zeta_1\diamond^\zeta \zeta_2=\zeta_1\gleq \zeta_2\in \mi{clo}$, $\|\zeta_1\|^{{\mc V}_{\iota-1}}\underset{\text{(d)}}{\leq} \|\zeta_2\|^{{\mc V}_{\iota-1}}$,
or $\diamond^\zeta=\gle$, $\zeta_1\diamond^\zeta \zeta_2=\zeta_1\gle \zeta_2\in \mi{clo}$, $\|\zeta_1\|^{{\mc V}_{\iota-1}}\underset{\text{(e)}}{<} \|\zeta_2\|^{{\mc V}_{\iota-1}}$;
$\|\zeta_1\|^{{\mc V}_{\iota-1}}\leq \|\zeta_2\|^{{\mc V}_{\iota-1}}$,
$\frac{\|\zeta_1\|^{{\mc V}_{\iota-1}}}
      {\|\zeta_2\|^{{\mc V}_{\iota-1}}}\leq 1$,
$\mbb{E}_{\iota-1}=\emptyset$;
for all $1\leq j\leq \kappa_1$, 
$\mu_2^j\gleq (\mu_1^j\mult a_\iota^{\beta_j})\in \mi{min}(\mi{DE}_{\iota-1})$, 
$\left(\frac{\|\mu_2^j\|^{{\mc V}_{\iota-1}}}
            {\|\mu_1^j\|^{{\mc V}_{\iota-1}}}\right)^{\frac{1}{\beta_j}}\in \mbb{DE}_{\iota-1}$,
$\left(\frac{\|\mu_2^j\|^{{\mc V}_{\iota-1}}}
            {\|\mu_1^j\|^{{\mc V}_{\iota-1}}}\right)^{\frac{1}{\beta_j}}\leq \bigfvee \mbb{DE}_{\iota-1}\underset{\text{(\ref{eq8j}a)}}{\leq} {\mc V}_\iota(a_\iota)=\delta_\iota$,
$\frac{\|\mu_2^j\|^{{\mc V}_{\iota-1}}}
      {\|\mu_1^j\|^{{\mc V}_{\iota-1}}}\leq {\mc V}_\iota(a_\iota)^{\beta_j}$;
for all $\kappa_1+1\leq j\leq \kappa_2$, 
$\mu_2^j\gle (\mu_1^j\mult a_\iota^{\beta_j})\in \mi{min}(D_{\iota-1})$, 
$\left(\frac{\|\mu_2^j\|^{{\mc V}_{\iota-1}}}
            {\|\mu_1^j\|^{{\mc V}_{\iota-1}}}\right)^{\frac{1}{\beta_j}}\in \mbb{D}_{\iota-1}$,
$\left(\frac{\|\mu_2^j\|^{{\mc V}_{\iota-1}}}
            {\|\mu_1^j\|^{{\mc V}_{\iota-1}}}\right)^{\frac{1}{\beta_j}}\leq \bigfvee \mbb{D}_{\iota-1}\underset{\text{(\ref{eq8j}b)}}{<} {\mc V}_\iota(a_\iota)=\delta_\iota$,
$\frac{\|\mu_2^j\|^{{\mc V}_{\iota-1}}}
      {\|\mu_1^j\|^{{\mc V}_{\iota-1}}}<{\mc V}_\iota(a_\iota)^{\beta_j}$;
for all $\kappa_2+1\leq j\leq \kappa_3$,
$(\mu_1^j\mult a_\iota^{\beta_j})\gleq \mu_2^j\in \mi{min}(\mi{UE}_{\iota-1})$,
$\mi{min}\left(\left(\frac{\|\mu_2^j\|^{{\mc V}_{\iota-1}}}
                          {\|\mu_1^j\|^{{\mc V}_{\iota-1}}}\right)^{\frac{1}{\beta_j}},1\right)\in \mbb{UE}_{\iota-1}$,
${\mc V}_\iota(a_\iota)=\delta_\iota\underset{\text{(\ref{eq8j}a)}}{\leq} \bigfwedge \mbb{UE}_{\iota-1}\leq 
 \mi{min}\left(\left(\frac{\|\mu_2^j\|^{{\mc V}_{\iota-1}}}
                          {\|\mu_1^j\|^{{\mc V}_{\iota-1}}}\right)^{\frac{1}{\beta_j}},1\right)\leq
 \left(\frac{\|\mu_2^j\|^{{\mc V}_{\iota-1}}}
            {\|\mu_1^j\|^{{\mc V}_{\iota-1}}}\right)^{\frac{1}{\beta_j}}$,
${\mc V}_\iota(a_\iota)^{\beta_j}\leq \frac{\|\mu_2^j\|^{{\mc V}_{\iota-1}}}
                                           {\|\mu_1^j\|^{{\mc V}_{\iota-1}}}$,
$\frac{\|\mu_1^j\|^{{\mc V}_{\iota-1}}}
      {\|\mu_2^j\|^{{\mc V}_{\iota-1}}}\leq \frac{1}
                                                 {{\mc V}_\iota(a_\iota)^{\beta_j}}$;
for all $\kappa_3+1\leq j\leq \kappa_4$,
$(\mu_1^j\mult a_\iota^{\beta_j})\gle \mu_2^j\in \mi{min}(U_{\iota-1})$,
$\mi{min}\left(\left(\frac{\|\mu_2^j\|^{{\mc V}_{\iota-1}}}
                          {\|\mu_1^j\|^{{\mc V}_{\iota-1}}}\right)^{\frac{1}{\beta_j}},1\right)\in \mbb{U}_{\iota-1}$,
${\mc V}_\iota(a_\iota)=\delta_\iota\underset{\text{(\ref{eq8j}b)}}{\leq} \bigfwedge \mbb{U}_{\iota-1}\leq 
 \mi{min}\left(\left(\frac{\|\mu_2^j\|^{{\mc V}_{\iota-1}}}
                          {\|\mu_1^j\|^{{\mc V}_{\iota-1}}}\right)^{\frac{1}{\beta_j}},1\right)\leq
 \left(\frac{\|\mu_2^j\|^{{\mc V}_{\iota-1}}}
            {\|\mu_1^j\|^{{\mc V}_{\iota-1}}}\right)^{\frac{1}{\beta_j}}$,
${\mc V}_\iota(a_\iota)^{\beta_j}\leq \frac{\|\mu_2^j\|^{{\mc V}_{\iota-1}}}
                                           {\|\mu_1^j\|^{{\mc V}_{\iota-1}}}$,
$\frac{\|\mu_1^j\|^{{\mc V}_{\iota-1}}}
      {\|\mu_2^j\|^{{\mc V}_{\iota-1}}}\leq \frac{1}
                                                 {{\mc V}_\iota(a_\iota)^{\beta_j}}$;
\begin{alignat*}{1}
& \dfrac{\|\varepsilon_1\|^{{\mc V}_{\iota-1}}}
        {\|\upsilon_2\|^{{\mc V}_{\iota-1}}}= \\
& \dfrac{\left\|\mi{reduced}^-\left(\begin{array}{l}
                                    ((\bigmult_{j=1}^{\kappa_1} \mu_2^j)\mult (\bigmult_{j=\kappa_1+1}^{\kappa_2} \mu_2^j)\mult 
                                     (\bigmult_{j=\kappa_2+1}^{\kappa_3} \mu_1^j)\mult \\
                                    \quad
                                     (\bigmult_{j=\kappa_3+1}^{\kappa_4} \mu_1^j)\mult \zeta_1)\gleq \\
                                    \quad
                                    ((\bigmult_{j=1}^{\kappa_1} \mu_1^j)\mult (\bigmult_{j=\kappa_1+1}^{\kappa_2} \mu_1^j)\mult 
                                     (\bigmult_{j=\kappa_2+1}^{\kappa_3} \mu_2^j)\mult \\
                                    \quad \quad
                                     (\bigmult_{j=\kappa_3+1}^{\kappa_4} \mu_2^j)\mult \zeta_2)
                                    \end{array}\right)\right\|^{{\mc V}_{\iota-1}}}
        {\left\|\mi{reduced}^+\left(\begin{array}{l}
                                    ((\bigmult_{j=1}^{\kappa_1} \mu_2^j)\mult (\bigmult_{j=\kappa_1+1}^{\kappa_2} \mu_2^j)\mult 
                                     (\bigmult_{j=\kappa_2+1}^{\kappa_3} \mu_1^j)\mult \\
                                    \quad
                                     (\bigmult_{j=\kappa_3+1}^{\kappa_4} \mu_1^j)\mult \zeta_1)\gleq \\
                                    \quad
                                    ((\bigmult_{j=1}^{\kappa_1} \mu_1^j)\mult (\bigmult_{j=\kappa_1+1}^{\kappa_2} \mu_1^j)\mult 
                                     (\bigmult_{j=\kappa_2+1}^{\kappa_3} \mu_2^j)\mult \\
                                    \quad \quad
                                     (\bigmult_{j=\kappa_3+1}^{\kappa_4} \mu_2^j)\mult \zeta_2)
                                    \end{array}\right)\right\|^{{\mc V}_{\iota-1}}}\overset{\text{(\ref{eq7n})}}{=\!\!=} \\
& \dfrac{\|(\bigmult_{j=1}^{\kappa_1} \mu_2^j)\mult (\bigmult_{j=\kappa_1+1}^{\kappa_2} \mu_2^j)\mult 
           (\bigmult_{j=\kappa_2+1}^{\kappa_3} \mu_1^j)\mult (\bigmult_{j=\kappa_3+1}^{\kappa_4} \mu_1^j)\mult \zeta_1\|^{{\mc V}_{\iota-1}}}
        {\|(\bigmult_{j=1}^{\kappa_1} \mu_1^j)\mult (\bigmult_{j=\kappa_1+1}^{\kappa_2} \mu_1^j)\mult 
           (\bigmult_{j=\kappa_2+1}^{\kappa_3} \mu_2^j)\mult (\bigmult_{j=\kappa_3+1}^{\kappa_4} \mu_2^j)\mult \zeta_2\|^{{\mc V}_{\iota-1}}}= \\
& \dfrac{\begin{array}{l}
         (\bigfswedge_{j=1}^{\kappa_1} \|\mu_2^j\|^{{\mc V}_{\iota-1}})\fswedge (\bigfswedge_{j=\kappa_1+1}^{\kappa_2} \|\mu_2^j\|^{{\mc V}_{\iota-1}})\fswedge
         (\bigfswedge_{j=\kappa_2+1}^{\kappa_3} \|\mu_1^j\|^{{\mc V}_{\iota-1}})\fswedge (\bigfswedge_{j=\kappa_3+1}^{\kappa_4} \|\mu_1^j\|^{{\mc V}_{\iota-1}})\fswedge \\
         \quad
         \|\zeta_1\|^{{\mc V}_{\iota-1}}
         \end{array}}
        {\begin{array}{l}
         (\bigfswedge_{j=1}^{\kappa_1} \|\mu_1^j\|^{{\mc V}_{\iota-1}})\fswedge (\bigfswedge_{j=\kappa_1+1}^{\kappa_2} \|\mu_1^j\|^{{\mc V}_{\iota-1}})\fswedge
         (\bigfswedge_{j=\kappa_2+1}^{\kappa_3} \|\mu_2^j\|^{{\mc V}_{\iota-1}})\fswedge (\bigfswedge_{j=\kappa_3+1}^{\kappa_4} \|\mu_2^j\|^{{\mc V}_{\iota-1}})\fswedge \\
         \quad
         \|\zeta_2\|^{{\mc V}_{\iota-1}}
         \end{array}}= \\
& \left(\bigfswedge_{j=1}^{\kappa_1} \dfrac{\|\mu_2^j\|^{{\mc V}_{\iota-1}}}
                                           {\|\mu_1^j\|^{{\mc V}_{\iota-1}}}\right)\fswedge 
  \left(\bigfswedge_{j=\kappa_1+1}^{\kappa_2} \dfrac{\|\mu_2^j\|^{{\mc V}_{\iota-1}}}
                                                    {\|\mu_1^j\|^{{\mc V}_{\iota-1}}}\right)\fswedge 
  \left(\bigfswedge_{j=\kappa_1+2}^{\kappa_3} \dfrac{\|\mu_1^j\|^{{\mc V}_{\iota-1}}}
                                                    {\|\mu_2^j\|^{{\mc V}_{\iota-1}}}\right)\fswedge \\
& \quad
  \left(\bigfswedge_{j=\kappa_1+3}^{\kappa_4} \dfrac{\|\mu_1^j\|^{{\mc V}_{\iota-1}}}
                                                    {\|\mu_2^j\|^{{\mc V}_{\iota-1}}}\right)\fswedge
  \dfrac{\|\zeta_1\|^{{\mc V}_{\iota-1}}}
        {\|\zeta_2\|^{{\mc V}_{\iota-1}}}\leq \\
& \left(\bigfswedge_{j=1}^{\kappa_1} {\mc V}_\iota(a_\iota)^{\beta_j}\right)\fswedge \left(\bigfswedge_{j=\kappa_1+1}^{\kappa_2} {\mc V}_\iota(a_\iota)^{\beta_j}\right)\fswedge 
  \left(\bigfswedge_{j=\kappa_1+2}^{\kappa_3} \dfrac{1}
                                                    {{\mc V}_\iota(a_\iota)^{\beta_j}}\right)\fswedge
  \bigfswedge_{j=\kappa_1+3}^{\kappa_4} \dfrac{1}
                                              {{\mc V}_\iota(a_\iota)^{\beta_j}}= \\
& {\mc V}_\iota(a_\iota)^{(\sum_{j=1}^{\kappa_1} \beta_j)+(\sum_{j=\kappa_1+1}^{\kappa_2} \beta_j)-(\sum_{j=\kappa_2+1}^{\kappa_3} \beta_j)-\sum_{j=\kappa_3+1}^{\kappa_4} \beta_j}=
  {\mc V}_\iota(a_\iota)^\alpha, \\[1mm]
& \|\varepsilon_1\|^{{\mc V}_\iota}\overset{\text{(b)}}{=\!\!=} \|\varepsilon_1\|^{{\mc V}_{\iota-1}}\leq 
  \|\upsilon_2\|^{{\mc V}_{\iota-1}}\fswedge {\mc V}_\iota(a_\iota)^\alpha\overset{\text{(b)}}{=\!\!=}
  \|\upsilon_2\|^{{\mc V}_\iota}\fswedge {\mc V}_\iota(a_\iota)^\alpha=\|\upsilon_2\mult a_\iota^\alpha\|^{{\mc V}_\iota}=\|\varepsilon_2\|^{{\mc V}_\iota};
\end{alignat*}
(d) holds.

Case 2.10.2.2.2:
$\varepsilon_1\gle \varepsilon_2\in \mi{clo}$.
Then $\mi{atoms}(\varepsilon_1), \mi{atoms}(\varepsilon_2)\subseteq \mi{dom}({\mc V}_\iota)=\mi{dom}({\mc V}_{\iota-1})\cup \{a_\iota\}$,
$a_\iota\in \mi{atoms}(\varepsilon_1)$ or $a_\iota\in \mi{atoms}(\varepsilon_2)$, $\mbb{E}_{\iota-1}=\emptyset$,
$\varepsilon_1\gle \varepsilon_2=\varepsilon_1\gle (\upsilon_2\mult a_\iota^\alpha)\in \mi{clo}$;
by (\ref{eq8i}) for $\gle$ and (\ref{eq8i}b), there exist $\kappa_1<\kappa_2\leq \kappa_3\leq \kappa_4$,
\begin{alignat*}{1}
& \mu_2^j\gleq (\mu_1^j\mult a_\iota^{\beta_j})=\mi{reduced}(\lambda^{(\gamma_1^j,\dots,\gamma_n^j)})\in \mi{min}(\mi{DE}_{\iota-1}), j=1,\dots,\kappa_1, \\ 
& \mu_2^j\gle (\mu_1^j\mult a_\iota^{\beta_j})=\mi{reduced}(\lambda^{(\gamma_1^j,\dots,\gamma_n^j)})\in \mi{min}(D_{\iota-1}), j=\kappa_1+1,\dots,\kappa_2, \\
& (\mu_1^j\mult a_\iota^{\beta_j})\gleq \mu_2^j=\mi{reduced}(\lambda^{(\gamma_1^j,\dots,\gamma_n^j)})\in \mi{min}(\mi{UE}_{\iota-1}), j=\kappa_2+1,\dots,\kappa_3, \\
& (\mu_1^j\mult a_\iota^{\beta_j})\gle \mu_2^j=\mi{reduced}(\lambda^{(\gamma_1^j,\dots,\gamma_n^j)})\in \mi{min}(U_{\iota-1}), j=\kappa_3+1,\dots,\kappa_4, \\ 
& \mu_e^j\in \{\gu\}\cup \mi{PropConj}_A, \mi{atoms}(\mu_e^j)\subseteq \mi{dom}({\mc V}_{\iota-1}), \beta_j\geq 1, \bs{0}^n\neq (\gamma_1^j,\dots,\gamma_n^j)\in \mbb{N}^n, 
\end{alignat*}
satisfying either $\varepsilon_1\gle \varepsilon_2=\varepsilon_1\gle (\upsilon_2\mult a_\iota^\alpha)=
                                                   \mi{reduced}(\lambda^{(\sum_{j=1}^{\kappa_4} \gamma_1^j,\dots,\sum_{j=1}^{\kappa_4} \gamma_n^j)})$, 
or there exists $\zeta_1\diamond^\zeta \zeta_2=\mi{reduced}(\lambda^{(\gamma_1^\zeta,\dots,\gamma_n^\zeta)})\in \mi{clo}$,
$\zeta_e\in \{\gu\}\cup \mi{PropConj}_A$, $\mi{atoms}(\zeta_e)\subseteq \mi{dom}({\mc V}_{\iota-1})$, $\diamond^\zeta\in \{\geql,\gleq,\gle\}$, 
$\bs{0}^n\neq (\gamma_1^\zeta,\dots,\gamma_n^\zeta)\in \mbb{N}^n$, satisfying
$\varepsilon_1\gle \varepsilon_2=\varepsilon_1\gle (\upsilon_2\mult a_\iota^\alpha)=
                                 \mi{reduced}(\lambda^{((\sum_{j=1}^{\kappa_4} \gamma_1^j)+\gamma_1^\zeta,\dots,(\sum_{j=1}^{\kappa_4} \gamma_n^j)+\gamma_n^\zeta)})$.
We get two cases for $\varepsilon_1\gle (\upsilon_2\mult a_\iota^\alpha)$.

Case 2.10.2.2.2.1:
$\varepsilon_1\gle (\upsilon_2\mult a_\iota^\alpha)=\mi{reduced}(\lambda^{(\sum_{j=1}^{\kappa_4} \gamma_1^j,\dots,\sum_{j=1}^{\kappa_4} \gamma_n^j)})$.
Then 
$\mi{atoms}(\varepsilon_1), \mi{atoms}(\upsilon_2), \mi{atoms}(\mu_1^1),\dots,\mi{atoms}(\mu_1^{\kappa_4}), \mi{atoms}(\mu_2^1),\dots,\mi{atoms}(\mu_2^{\kappa_4})\subseteq \mi{dom}({\mc V}_{\iota-1})$,
\begin{alignat*}{1}
& \varepsilon_1\gle (\upsilon_2\mult a_\iota^\alpha)=
  \mi{reduced}(\lambda^{(\sum_{j=1}^{\kappa_4} \gamma_1^j,\dots,\sum_{j=1}^{\kappa_4} \gamma_n^j)})\overset{\text{(\ref{eq8dddd})}}{=\!\!=} \\
& \mi{reduced}((\bigmult_{j=1}^{\kappa_1} \lambda^{(\gamma_1^j,\dots,\gamma_n^j)})\mult
               (\bigmult_{j=\kappa_1+1}^{\kappa_2} \lambda^{(\gamma_1^j,\dots,\gamma_n^j)})\mult \\
&\phantom{\mi{reduced}(}
               (\bigmult_{j=\kappa_2+1}^{\kappa_3} \lambda^{(\gamma_1^j,\dots,\gamma_n^j)})\mult
               \bigmult_{j=\kappa_3+1}^{\kappa_4} \lambda^{(\gamma_1^j,\dots,\gamma_n^j)})\overset{\text{(\ref{eq7k})}}{=\!\!=} \\
& \mi{reduced}((\bigmult_{j=1}^{\kappa_1} \mi{reduced}(\lambda^{(\gamma_1^j,\dots,\gamma_n^j)}))\mult
               (\bigmult_{j=\kappa_1+1}^{\kappa_2} \mi{reduced}(\lambda^{(\gamma_1^j,\dots,\gamma_n^j)}))\mult \\
&\phantom{\mi{reduced}(}
               (\bigmult_{j=\kappa_2+1}^{\kappa_3} \mi{reduced}(\lambda^{(\gamma_1^j,\dots,\gamma_n^j)}))\mult
               \bigmult_{j=\kappa_3+1}^{\kappa_4} \mi{reduced}(\lambda^{(\gamma_1^j,\dots,\gamma_n^j)}))= \\
& \mi{reduced}((\bigmult_{j=1}^{\kappa_1} \mu_2^j\gleq (\mu_1^j\mult a_\iota^{\beta_j}))\mult
               (\bigmult_{j=\kappa_1+1}^{\kappa_2} \mu_2^j\gle (\mu_1^j\mult a_\iota^{\beta_j}))\mult \\
&\phantom{\mi{reduced}(}
               (\bigmult_{j=\kappa_2+1}^{\kappa_3} (\mu_1^j\mult a_\iota^{\beta_j})\gleq \mu_2^j)\mult
               \bigmult_{j=\kappa_3+1}^{\kappa_4} (\mu_1^j\mult a_\iota^{\beta_j})\gle \mu_2^j)\overset{\text{(\ref{eq7i})}}{=\!\!=} \\
& \mi{reduced}(((\bigmult_{j=1}^{\kappa_1} \mu_2^j)\mult (\bigmult_{j=\kappa_1+1}^{\kappa_2} \mu_2^j)\mult 
                (\bigmult_{j=\kappa_2+1}^{\kappa_3} \mu_1^j)\mult (\bigmult_{j=\kappa_3+1}^{\kappa_4} \mu_1^j)\mult \\
&\phantom{\mi{reduced}(} \quad
                a_\iota^{(\sum_{j=\kappa_2+1}^{\kappa_3} \beta_j)+\sum_{j=\kappa_3+1}^{\kappa_4} \beta_j})\gle \\
&\phantom{\mi{reduced}(} \quad
               ((\bigmult_{j=1}^{\kappa_1} \mu_1^j)\mult (\bigmult_{j=\kappa_1+1}^{\kappa_2} \mu_1^j)\mult 
                (\bigmult_{j=\kappa_2+1}^{\kappa_3} \mu_2^j)\mult (\bigmult_{j=\kappa_3+1}^{\kappa_4} \mu_2^j)\mult \\
&\phantom{\mi{reduced}(} \quad \quad
                a_\iota^{(\sum_{j=1}^{\kappa_1} \beta_j)+\sum_{j=\kappa_1+1}^{\kappa_2} \beta_j}))= \\
& \mi{reduced}^-(((\bigmult_{j=1}^{\kappa_1} \mu_2^j)\mult (\bigmult_{j=\kappa_1+1}^{\kappa_2} \mu_2^j)\mult 
                  (\bigmult_{j=\kappa_2+1}^{\kappa_3} \mu_1^j)\mult (\bigmult_{j=\kappa_3+1}^{\kappa_4} \mu_1^j)\mult \\
&\phantom{\mi{reduced}^-(} \quad
                  a_\iota^{(\sum_{j=\kappa_2+1}^{\kappa_3} \beta_j)+\sum_{j=\kappa_3+1}^{\kappa_4} \beta_j})\gle \\
&\phantom{\mi{reduced}^-(} \quad
                 ((\bigmult_{j=1}^{\kappa_1} \mu_1^j)\mult (\bigmult_{j=\kappa_1+1}^{\kappa_2} \mu_1^j)\mult 
                  (\bigmult_{j=\kappa_2+1}^{\kappa_3} \mu_2^j)\mult (\bigmult_{j=\kappa_3+1}^{\kappa_4} \mu_2^j)\mult \\
&\phantom{\mi{reduced}^-(} \quad \quad
                  a_\iota^{(\sum_{j=1}^{\kappa_1} \beta_j)+\sum_{j=\kappa_1+1}^{\kappa_2} \beta_j}))\gle \\
& \mi{reduced}^+(((\bigmult_{j=1}^{\kappa_1} \mu_2^j)\mult (\bigmult_{j=\kappa_1+1}^{\kappa_2} \mu_2^j)\mult 
                  (\bigmult_{j=\kappa_2+1}^{\kappa_3} \mu_1^j)\mult (\bigmult_{j=\kappa_3+1}^{\kappa_4} \mu_1^j)\mult \\
&\phantom{\mi{reduced}^+(} \quad
                  a_\iota^{(\sum_{j=\kappa_2+1}^{\kappa_3} \beta_j)+\sum_{j=\kappa_3+1}^{\kappa_4} \beta_j})\gle \\
&\phantom{\mi{reduced}^+(} \quad
                 ((\bigmult_{j=1}^{\kappa_1} \mu_1^j)\mult (\bigmult_{j=\kappa_1+1}^{\kappa_2} \mu_1^j)\mult 
                  (\bigmult_{j=\kappa_2+1}^{\kappa_3} \mu_2^j)\mult (\bigmult_{j=\kappa_3+1}^{\kappa_4} \mu_2^j)\mult \\
&\phantom{\mi{reduced}^+(} \quad \quad
                  a_\iota^{(\sum_{j=1}^{\kappa_1} \beta_j)+\sum_{j=\kappa_1+1}^{\kappa_2} \beta_j})), \\[1mm]
& \#(a_\iota,\varepsilon_1\gle (\upsilon_2\mult a_\iota^\alpha))=\alpha= \\
& \#(a_\iota,\mi{reduced}(((\bigmult_{j=1}^{\kappa_1} \mu_2^j)\mult (\bigmult_{j=\kappa_1+1}^{\kappa_2} \mu_2^j)\mult 
                           (\bigmult_{j=\kappa_2+1}^{\kappa_3} \mu_1^j)\mult (\bigmult_{j=\kappa_3+1}^{\kappa_4} \mu_1^j)\mult \\
&\phantom{\#(a_\iota,\mi{reduced}(} \quad
                           a_\iota^{(\sum_{j=\kappa_2+1}^{\kappa_3} \beta_j)+\sum_{j=\kappa_3+1}^{\kappa_4} \beta_j})\gle \\
&\phantom{\#(a_\iota,\mi{reduced}(} \quad
                          ((\bigmult_{j=1}^{\kappa_1} \mu_1^j)\mult (\bigmult_{j=\kappa_1+1}^{\kappa_2} \mu_1^j)\mult 
                           (\bigmult_{j=\kappa_2+1}^{\kappa_3} \mu_2^j)\mult (\bigmult_{j=\kappa_3+1}^{\kappa_4} \mu_2^j)\mult \\
&\phantom{\#(a_\iota,\mi{reduced}(} \quad \quad
                           a_\iota^{(\sum_{j=1}^{\kappa_1} \beta_j)+\sum_{j=\kappa_1+1}^{\kappa_2} \beta_j})))\overset{\text{(\ref{eq7aaaax})}}{=\!\!=} \\
& \#(a_\iota,((\bigmult_{j=1}^{\kappa_1} \mu_2^j)\mult (\bigmult_{j=\kappa_1+1}^{\kappa_2} \mu_2^j)\mult 
              (\bigmult_{j=\kappa_2+1}^{\kappa_3} \mu_1^j)\mult (\bigmult_{j=\kappa_3+1}^{\kappa_4} \mu_1^j)\mult \\
&\phantom{\#(a_\iota,} \quad
              a_\iota^{(\sum_{j=\kappa_2+1}^{\kappa_3} \beta_j)+\sum_{j=\kappa_3+1}^{\kappa_4} \beta_j})\gle \\
&\phantom{\#(a_\iota,} \quad
             ((\bigmult_{j=1}^{\kappa_1} \mu_1^j)\mult (\bigmult_{j=\kappa_1+1}^{\kappa_2} \mu_1^j)\mult 
              (\bigmult_{j=\kappa_2+1}^{\kappa_3} \mu_2^j)\mult (\bigmult_{j=\kappa_3+1}^{\kappa_4} \mu_2^j)\mult \\
&\phantom{\#(a_\iota,} \quad \quad
              a_\iota^{(\sum_{j=1}^{\kappa_1} \beta_j)+\sum_{j=\kappa_1+1}^{\kappa_2} \beta_j}))= \\
& \left(\sum_{j=1}^{\kappa_1} \beta_j\right)+\left(\sum_{j=\kappa_1+1}^{\kappa_2} \beta_j\right)-\left(\sum_{j=\kappa_2+1}^{\kappa_3} \beta_j\right)-\sum_{j=\kappa_3+1}^{\kappa_4} \beta_j\geq 1, 
\end{alignat*}
$a_\iota\not\in \mi{dom}({\mc V}_{\iota-1})\supseteq \mi{atoms}((\bigmult_{j=1}^{\kappa_1} \mu_2^j)\mult (\bigmult_{j=\kappa_1+1}^{\kappa_2} \mu_2^j)\mult 
                                                                (\bigmult_{j=\kappa_2+1}^{\kappa_3} \mu_1^j)\mult \bigmult_{j=\kappa_3+1}^{\kappa_4} \mu_1^j)=
                                                     (\bigcup_{j=1}^{\kappa_1} \mi{atoms}(\mu_2^j))\cup (\bigcup_{j=\kappa_1+1}^{\kappa_2} \mi{atoms}(\mu_2^j))\cup 
                                                     (\bigcup_{j=\kappa_2+1}^{\kappa_3} \mi{atoms}(\mu_1^j))\cup \bigcup_{j=\kappa_3+1}^{\kappa_4} \mi{atoms}(\mu_1^j)$,
$a_\iota\not\in \mi{dom}({\mc V}_{\iota-1})\supseteq \mi{atoms}((\bigmult_{j=1}^{\kappa_1} \mu_1^j)\mult (\bigmult_{j=\kappa_1+1}^{\kappa_2} \mu_1^j)\mult 
                                                                (\bigmult_{j=\kappa_2+1}^{\kappa_3} \mu_2^j)\mult \bigmult_{j=\kappa_3+1}^{\kappa_4} \mu_2^j)=
                                                     (\bigcup_{j=1}^{\kappa_1} \mi{atoms}(\mu_1^j))\cup (\bigcup_{j=\kappa_1+1}^{\kappa_2} \mi{atoms}(\mu_1^j))\cup 
                                                     (\bigcup_{j=\kappa_2+1}^{\kappa_3} \mi{atoms}(\mu_2^j))\cup \bigcup_{j=\kappa_3+1}^{\kappa_4} \mi{atoms}(\mu_2^j)$,
\begin{alignat*}{1}
& \mi{reduced}^-(((\bigmult_{j=1}^{\kappa_1} \mu_2^j)\mult (\bigmult_{j=\kappa_1+1}^{\kappa_2} \mu_2^j)\mult 
                  (\bigmult_{j=\kappa_2+1}^{\kappa_3} \mu_1^j)\mult (\bigmult_{j=\kappa_3+1}^{\kappa_4} \mu_1^j)\mult \\
&\phantom{\mi{reduced}^-(} \quad
                  a_\iota^{(\sum_{j=\kappa_2+1}^{\kappa_3} \beta_j)+\sum_{j=\kappa_3+1}^{\kappa_4} \beta_j})\gle \\
&\phantom{\mi{reduced}^-(} \quad
                 ((\bigmult_{j=1}^{\kappa_1} \mu_1^j)\mult (\bigmult_{j=\kappa_1+1}^{\kappa_2} \mu_1^j)\mult 
                  (\bigmult_{j=\kappa_2+1}^{\kappa_3} \mu_2^j)\mult (\bigmult_{j=\kappa_3+1}^{\kappa_4} \mu_2^j)\mult \\
&\phantom{\mi{reduced}^-(} \quad \quad
                  a_\iota^{(\sum_{j=1}^{\kappa_1} \beta_j)+\sum_{j=\kappa_1+1}^{\kappa_2} \beta_j}))\overset{\text{(\ref{eq7cc})}}{=\!\!=} \\
& \mi{reduced}^-(((\bigmult_{j=1}^{\kappa_1} \mu_2^j)\mult (\bigmult_{j=\kappa_1+1}^{\kappa_2} \mu_2^j)\mult 
                  (\bigmult_{j=\kappa_2+1}^{\kappa_3} \mu_1^j)\mult \bigmult_{j=\kappa_3+1}^{\kappa_4} \mu_1^j)\gle \\
&\phantom{\mi{reduced}^-(} \quad
                 ((\bigmult_{j=1}^{\kappa_1} \mu_1^j)\mult (\bigmult_{j=\kappa_1+1}^{\kappa_2} \mu_1^j)\mult 
                  (\bigmult_{j=\kappa_2+1}^{\kappa_3} \mu_2^j)\mult \bigmult_{j=\kappa_3+1}^{\kappa_4} \mu_2^j)), \\[1mm]
& \mi{reduced}^+(((\bigmult_{j=1}^{\kappa_1} \mu_2^j)\mult (\bigmult_{j=\kappa_1+1}^{\kappa_2} \mu_2^j)\mult 
                  (\bigmult_{j=\kappa_2+1}^{\kappa_3} \mu_1^j)\mult (\bigmult_{j=\kappa_3+1}^{\kappa_4} \mu_1^j)\mult \\
&\phantom{\mi{reduced}^+(} \quad
                  a_\iota^{(\sum_{j=\kappa_2+1}^{\kappa_3} \beta_j)+\sum_{j=\kappa_3+1}^{\kappa_4} \beta_j})\gle \\
&\phantom{\mi{reduced}^+(} \quad
                 ((\bigmult_{j=1}^{\kappa_1} \mu_1^j)\mult (\bigmult_{j=\kappa_1+1}^{\kappa_2} \mu_1^j)\mult 
                  (\bigmult_{j=\kappa_2+1}^{\kappa_3} \mu_2^j)\mult (\bigmult_{j=\kappa_3+1}^{\kappa_4} \mu_2^j)\mult \\
&\phantom{\mi{reduced}^+(} \quad \quad
                  a_\iota^{(\sum_{j=1}^{\kappa_1} \beta_j)+\sum_{j=\kappa_1+1}^{\kappa_2} \beta_j}))\overset{\text{(\ref{eq7ff})}}{=\!\!=} \\
& \mi{reduced}^+(((\bigmult_{j=1}^{\kappa_1} \mu_2^j)\mult (\bigmult_{j=\kappa_1+1}^{\kappa_2} \mu_2^j)\mult 
                  (\bigmult_{j=\kappa_2+1}^{\kappa_3} \mu_1^j)\mult \bigmult_{j=\kappa_3+1}^{\kappa_4} \mu_1^j)\gle \\
&\phantom{\mi{reduced}^+(} \quad
                 ((\bigmult_{j=1}^{\kappa_1} \mu_1^j)\mult (\bigmult_{j=\kappa_1+1}^{\kappa_2} \mu_1^j)\mult 
                  (\bigmult_{j=\kappa_2+1}^{\kappa_3} \mu_2^j)\mult \bigmult_{j=\kappa_3+1}^{\kappa_4} \mu_2^j))\mult \\
& a_\iota^{(\sum_{j=1}^{\kappa_1} \beta_j)+(\sum_{j=\kappa_1+1}^{\kappa_2} \beta_j)-(\sum_{j=\kappa_2+1}^{\kappa_3} \beta_j)-\sum_{j=\kappa_3+1}^{\kappa_4} \beta_j}= \\
& \mi{reduced}^+(((\bigmult_{j=1}^{\kappa_1} \mu_2^j)\mult (\bigmult_{j=\kappa_1+1}^{\kappa_2} \mu_2^j)\mult 
                  (\bigmult_{j=\kappa_2+1}^{\kappa_3} \mu_1^j)\mult \bigmult_{j=\kappa_3+1}^{\kappa_4} \mu_1^j)\gle \\
&\phantom{\mi{reduced}^+(} \quad
                 ((\bigmult_{j=1}^{\kappa_1} \mu_1^j)\mult (\bigmult_{j=\kappa_1+1}^{\kappa_2} \mu_1^j)\mult 
                  (\bigmult_{j=\kappa_2+1}^{\kappa_3} \mu_2^j)\mult \bigmult_{j=\kappa_3+1}^{\kappa_4} \mu_2^j))\mult a_\iota^\alpha, \\[1mm]
& \varepsilon_1\gle (\upsilon_2\mult a_\iota^\alpha)= \\
& \mi{reduced}^-(((\bigmult_{j=1}^{\kappa_1} \mu_2^j)\mult (\bigmult_{j=\kappa_1+1}^{\kappa_2} \mu_2^j)\mult 
                  (\bigmult_{j=\kappa_2+1}^{\kappa_3} \mu_1^j)\mult \bigmult_{j=\kappa_3+1}^{\kappa_4} \mu_1^j)\gle \\
&\phantom{\mi{reduced}^-(} \quad
                 ((\bigmult_{j=1}^{\kappa_1} \mu_1^j)\mult (\bigmult_{j=\kappa_1+1}^{\kappa_2} \mu_1^j)\mult 
                  (\bigmult_{j=\kappa_2+1}^{\kappa_3} \mu_2^j)\mult \bigmult_{j=\kappa_3+1}^{\kappa_4} \mu_2^j))\gle \\
& (\mi{reduced}^+(((\bigmult_{j=1}^{\kappa_1} \mu_2^j)\mult (\bigmult_{j=\kappa_1+1}^{\kappa_2} \mu_2^j)\mult 
                   (\bigmult_{j=\kappa_2+1}^{\kappa_3} \mu_1^j)\mult \bigmult_{j=\kappa_3+1}^{\kappa_4} \mu_1^j)\gle \\
&\phantom{(\mi{reduced}^+(} \quad
                  ((\bigmult_{j=1}^{\kappa_1} \mu_1^j)\mult (\bigmult_{j=\kappa_1+1}^{\kappa_2} \mu_1^j)\mult 
                   (\bigmult_{j=\kappa_2+1}^{\kappa_3} \mu_2^j)\mult \bigmult_{j=\kappa_3+1}^{\kappa_4} \mu_2^j))\mult a_\iota^\alpha), \\[1mm]
& \mi{atoms}(\mi{reduced}^-(((\bigmult_{j=1}^{\kappa_1} \mu_2^j)\mult (\bigmult_{j=\kappa_1+1}^{\kappa_2} \mu_2^j)\mult 
                             (\bigmult_{j=\kappa_2+1}^{\kappa_3} \mu_1^j)\mult \bigmult_{j=\kappa_3+1}^{\kappa_4} \mu_1^j)\gle \\
&\phantom{\mi{atoms}(\mi{reduced}^-(} \quad
                            ((\bigmult_{j=1}^{\kappa_1} \mu_1^j)\mult (\bigmult_{j=\kappa_1+1}^{\kappa_2} \mu_1^j)\mult 
                             (\bigmult_{j=\kappa_2+1}^{\kappa_3} \mu_2^j)\mult \bigmult_{j=\kappa_3+1}^{\kappa_4} \mu_2^j)))\underset{\text{(\ref{eq7a})}}{\subseteq} \\
& \mi{atoms}((\bigmult_{j=1}^{\kappa_1} \mu_2^j)\mult (\bigmult_{j=\kappa_1+1}^{\kappa_2} \mu_2^j)\mult 
             (\bigmult_{j=\kappa_2+1}^{\kappa_3} \mu_1^j)\mult \bigmult_{j=\kappa_3+1}^{\kappa_4} \mu_1^j)\subseteq \mi{dom}({\mc V}_{\iota-1}), \\[1mm]
& \mi{atoms}(\mi{reduced}^+(((\bigmult_{j=1}^{\kappa_1} \mu_2^j)\mult (\bigmult_{j=\kappa_1+1}^{\kappa_2} \mu_2^j)\mult 
                             (\bigmult_{j=\kappa_2+1}^{\kappa_3} \mu_1^j)\mult \bigmult_{j=\kappa_3+1}^{\kappa_4} \mu_1^j)\gle \\
&\phantom{\mi{atoms}(\mi{reduced}^+(} \quad
                            ((\bigmult_{j=1}^{\kappa_1} \mu_1^j)\mult (\bigmult_{j=\kappa_1+1}^{\kappa_2} \mu_1^j)\mult 
                             (\bigmult_{j=\kappa_2+1}^{\kappa_3} \mu_2^j)\mult \bigmult_{j=\kappa_3+1}^{\kappa_4} \mu_2^j)))\underset{\text{(\ref{eq7b})}}{\subseteq} \\
& \mi{atoms}((\bigmult_{j=1}^{\kappa_1} \mu_1^j)\mult (\bigmult_{j=\kappa_1+1}^{\kappa_2} \mu_1^j)\mult 
             (\bigmult_{j=\kappa_2+1}^{\kappa_3} \mu_2^j)\mult \bigmult_{j=\kappa_3+1}^{\kappa_4} \mu_2^j)\subseteq \mi{dom}({\mc V}_{\iota-1}), 
\end{alignat*}
\begin{alignat*}{1}
& \varepsilon_1=\mi{reduced}^-(((\bigmult_{j=1}^{\kappa_1} \mu_2^j)\mult (\bigmult_{j=\kappa_1+1}^{\kappa_2} \mu_2^j)\mult 
                                (\bigmult_{j=\kappa_2+1}^{\kappa_3} \mu_1^j)\mult \bigmult_{j=\kappa_3+1}^{\kappa_4} \mu_1^j)\gle \\
&\phantom{\varepsilon_1=\mi{reduced}^-(} \quad
                               ((\bigmult_{j=1}^{\kappa_1} \mu_1^j)\mult (\bigmult_{j=\kappa_1+1}^{\kappa_2} \mu_1^j)\mult 
                                (\bigmult_{j=\kappa_2+1}^{\kappa_3} \mu_2^j)\mult \bigmult_{j=\kappa_3+1}^{\kappa_4} \mu_2^j)), \\[1mm]
& \upsilon_2=\mi{reduced}^+(((\bigmult_{j=1}^{\kappa_1} \mu_2^j)\mult (\bigmult_{j=\kappa_1+1}^{\kappa_2} \mu_2^j)\mult 
                             (\bigmult_{j=\kappa_2+1}^{\kappa_3} \mu_1^j)\mult \bigmult_{j=\kappa_3+1}^{\kappa_4} \mu_1^j)\gle \\
&\phantom{\upsilon_2=\mi{reduced}^+(} \quad
                            ((\bigmult_{j=1}^{\kappa_1} \mu_1^j)\mult (\bigmult_{j=\kappa_1+1}^{\kappa_2} \mu_1^j)\mult 
                             (\bigmult_{j=\kappa_2+1}^{\kappa_3} \mu_2^j)\mult \bigmult_{j=\kappa_3+1}^{\kappa_4} \mu_2^j));
\end{alignat*}
by the induction hypothesis for $\iota-1<\iota$,
$\|\varepsilon_1\|^{{\mc V}_{\iota-1}}, \|\upsilon_2\|^{{\mc V}_{\iota-1}},
 \|\mu_1^1\|^{{\mc V}_{\iota-1}},\dots,                                                                                                                                                    \linebreak[4]
                                       \|\mu_1^{\kappa_4}\|^{{\mc V}_{\iota-1}}, \|\mu_2^1\|^{{\mc V}_{\iota-1}},\dots,\|\mu_2^{\kappa_4}\|^{{\mc V}_{\iota-1}},
 \|\mi{reduced}^-(((\bigmult_{j=1}^{\kappa_1} \mu_2^j)\mult (\bigmult_{j=\kappa_1+1}^{\kappa_2} \mu_2^j)\mult 
                   (\bigmult_{j=\kappa_2+1}^{\kappa_3} \mu_1^j)\mult \bigmult_{j=\kappa_3+1}^{\kappa_4} \mu_1^j)\gle
                  ((\bigmult_{j=1}^{\kappa_1} \mu_1^j)\mult (\bigmult_{j=\kappa_1+1}^{\kappa_2} \mu_1^j)\mult 
                   (\bigmult_{j=\kappa_2+1}^{\kappa_3} \mu_2^j)\mult \bigmult_{j=\kappa_3+1}^{\kappa_4} \mu_2^j))\|^{{\mc V}_{\iota-1}}, 
 \|\mi{reduced}^+(((\bigmult_{j=1}^{\kappa_1} \mu_2^j)\mult (\bigmult_{j=\kappa_1+1}^{\kappa_2} \mu_2^j)\mult 
                   (\bigmult_{j=\kappa_2+1}^{\kappa_3} \mu_1^j)\mult \bigmult_{j=\kappa_3+1}^{\kappa_4} \mu_1^j)\gle
                  ((\bigmult_{j=1}^{\kappa_1} \mu_1^j)\mult (\bigmult_{j=\kappa_1+1}^{\kappa_2} \mu_1^j)\mult 
                   (\bigmult_{j=\kappa_2+1}^{\kappa_3} \mu_2^j)\mult \bigmult_{j=\kappa_3+1}^{\kappa_4} \mu_2^j))\|^{{\mc V}_{\iota-1}}\underset{\text{(a)}}{\in} (0,1]$;
$\mbb{E}_{\iota-1}=\emptyset$;
for all $1\leq j\leq \kappa_1$, 
$\mu_2^j\gleq (\mu_1^j\mult a_\iota^{\beta_j})\in \mi{min}(\mi{DE}_{\iota-1})$, 
$\left(\frac{\|\mu_2^j\|^{{\mc V}_{\iota-1}}}
            {\|\mu_1^j\|^{{\mc V}_{\iota-1}}}\right)^{\frac{1}{\beta_j}}\in \mbb{DE}_{\iota-1}$,
$\left(\frac{\|\mu_2^j\|^{{\mc V}_{\iota-1}}}
            {\|\mu_1^j\|^{{\mc V}_{\iota-1}}}\right)^{\frac{1}{\beta_j}}\leq \bigfvee \mbb{DE}_{\iota-1}\underset{\text{(\ref{eq8j}a)}}{\leq} {\mc V}_\iota(a_\iota)=\delta_\iota$,
$\frac{\|\mu_2^j\|^{{\mc V}_{\iota-1}}}
      {\|\mu_1^j\|^{{\mc V}_{\iota-1}}}\leq {\mc V}_\iota(a_\iota)^{\beta_j}$;
$\kappa_1<\kappa_2$, $\kappa_1+1\leq \kappa_2$,
$\{\kappa_1+1,\dots,\kappa_2\}\neq \emptyset$;
for all $\kappa_1+1\leq j\leq \kappa_2$, 
$\mu_2^j\gle (\mu_1^j\mult a_\iota^{\beta_j})\in \mi{min}(D_{\iota-1})$, 
$\left(\frac{\|\mu_2^j\|^{{\mc V}_{\iota-1}}}
            {\|\mu_1^j\|^{{\mc V}_{\iota-1}}}\right)^{\frac{1}{\beta_j}}\in \mbb{D}_{\iota-1}$,
$\left(\frac{\|\mu_2^j\|^{{\mc V}_{\iota-1}}}
            {\|\mu_1^j\|^{{\mc V}_{\iota-1}}}\right)^{\frac{1}{\beta_j}}\leq \bigfvee \mbb{D}_{\iota-1}\underset{\text{(\ref{eq8j}b)}}{<} {\mc V}_\iota(a_\iota)=\delta_\iota$,
$\frac{\|\mu_2^j\|^{{\mc V}_{\iota-1}}}
      {\|\mu_1^j\|^{{\mc V}_{\iota-1}}}<{\mc V}_\iota(a_\iota)^{\beta_j}$;
for all $\kappa_2+1\leq j\leq \kappa_3$,
$(\mu_1^j\mult a_\iota^{\beta_j})\gleq \mu_2^j\in \mi{min}(\mi{UE}_{\iota-1})$,
$\mi{min}\left(\left(\frac{\|\mu_2^j\|^{{\mc V}_{\iota-1}}}
                          {\|\mu_1^j\|^{{\mc V}_{\iota-1}}}\right)^{\frac{1}{\beta_j}},1\right)\in \mbb{UE}_{\iota-1}$,
${\mc V}_\iota(a_\iota)=\delta_\iota\underset{\text{(\ref{eq8j}a)}}{\leq} \bigfwedge \mbb{UE}_{\iota-1}\leq 
 \mi{min}\left(\left(\frac{\|\mu_2^j\|^{{\mc V}_{\iota-1}}}
                          {\|\mu_1^j\|^{{\mc V}_{\iota-1}}}\right)^{\frac{1}{\beta_j}},1\right)\leq
 \left(\frac{\|\mu_2^j\|^{{\mc V}_{\iota-1}}}
            {\|\mu_1^j\|^{{\mc V}_{\iota-1}}}\right)^{\frac{1}{\beta_j}}$,
${\mc V}_\iota(a_\iota)^{\beta_j}\leq \frac{\|\mu_2^j\|^{{\mc V}_{\iota-1}}}
                                           {\|\mu_1^j\|^{{\mc V}_{\iota-1}}}$,
$\frac{\|\mu_1^j\|^{{\mc V}_{\iota-1}}}
      {\|\mu_2^j\|^{{\mc V}_{\iota-1}}}\leq \frac{1}
                                                 {{\mc V}_\iota(a_\iota)^{\beta_j}}$;
for all $\kappa_3+1\leq j\leq \kappa_4$,
$(\mu_1^j\mult a_\iota^{\beta_j})\gle \mu_2^j\in \mi{min}(U_{\iota-1})$,
$\mi{min}\left(\left(\frac{\|\mu_2^j\|^{{\mc V}_{\iota-1}}}
                          {\|\mu_1^j\|^{{\mc V}_{\iota-1}}}\right)^{\frac{1}{\beta_j}},1\right)\in \mbb{U}_{\iota-1}$,
${\mc V}_\iota(a_\iota)=\delta_\iota\underset{\text{(\ref{eq8j}b)}}{\leq} \bigfwedge \mbb{U}_{\iota-1}\leq 
 \mi{min}\left(\left(\frac{\|\mu_2^j\|^{{\mc V}_{\iota-1}}}
                          {\|\mu_1^j\|^{{\mc V}_{\iota-1}}}\right)^{\frac{1}{\beta_j}},1\right)\leq
 \left(\frac{\|\mu_2^j\|^{{\mc V}_{\iota-1}}}
            {\|\mu_1^j\|^{{\mc V}_{\iota-1}}}\right)^{\frac{1}{\beta_j}}$,
${\mc V}_\iota(a_\iota)^{\beta_j}\leq \frac{\|\mu_2^j\|^{{\mc V}_{\iota-1}}}
                                           {\|\mu_1^j\|^{{\mc V}_{\iota-1}}}$,
$\frac{\|\mu_1^j\|^{{\mc V}_{\iota-1}}}
      {\|\mu_2^j\|^{{\mc V}_{\iota-1}}}\leq \frac{1}
                                                 {{\mc V}_\iota(a_\iota)^{\beta_j}}$;
\begin{alignat*}{1}
& \dfrac{\|\varepsilon_1\|^{{\mc V}_{\iota-1}}}
        {\|\upsilon_2\|^{{\mc V}_{\iota-1}}}= \\
& \dfrac{\left\|\mi{reduced}^-\left(\begin{array}{l}
                                    ((\bigmult_{j=1}^{\kappa_1} \mu_2^j)\mult (\bigmult_{j=\kappa_1+1}^{\kappa_2} \mu_2^j)\mult 
                                     (\bigmult_{j=\kappa_2+1}^{\kappa_3} \mu_1^j)\mult \bigmult_{j=\kappa_3+1}^{\kappa_4} \mu_1^j)\gle \\
                                    \quad
                                    ((\bigmult_{j=1}^{\kappa_1} \mu_1^j)\mult (\bigmult_{j=\kappa_1+1}^{\kappa_2} \mu_1^j)\mult 
                                     (\bigmult_{j=\kappa_2+1}^{\kappa_3} \mu_2^j)\mult \bigmult_{j=\kappa_3+1}^{\kappa_4} \mu_2^j)
                                    \end{array}\right)\right\|^{{\mc V}_{\iota-1}}}
        {\left\|\mi{reduced}^+\left(\begin{array}{l}
                                    ((\bigmult_{j=1}^{\kappa_1} \mu_2^j)\mult (\bigmult_{j=\kappa_1+1}^{\kappa_2} \mu_2^j)\mult 
                                     (\bigmult_{j=\kappa_2+1}^{\kappa_3} \mu_1^j)\mult \bigmult_{j=\kappa_3+1}^{\kappa_4} \mu_1^j)\gle \\
                                    \quad
                                    ((\bigmult_{j=1}^{\kappa_1} \mu_1^j)\mult (\bigmult_{j=\kappa_1+1}^{\kappa_2} \mu_1^j)\mult 
                                     (\bigmult_{j=\kappa_2+1}^{\kappa_3} \mu_2^j)\mult \bigmult_{j=\kappa_3+1}^{\kappa_4} \mu_2^j)
                                    \end{array}\right)\right\|^{{\mc V}_{\iota-1}}}\overset{\text{(\ref{eq7n})}}{=\!\!=} \\
& \dfrac{\|(\bigmult_{j=1}^{\kappa_1} \mu_2^j)\mult (\bigmult_{j=\kappa_1+1}^{\kappa_2} \mu_2^j)\mult 
           (\bigmult_{j=\kappa_2+1}^{\kappa_3} \mu_1^j)\mult \bigmult_{j=\kappa_3+1}^{\kappa_4} \mu_1^j\|^{{\mc V}_{\iota-1}}}
        {\|(\bigmult_{j=1}^{\kappa_1} \mu_1^j)\mult (\bigmult_{j=\kappa_1+1}^{\kappa_2} \mu_1^j)\mult 
           (\bigmult_{j=\kappa_2+1}^{\kappa_3} \mu_2^j)\mult \bigmult_{j=\kappa_3+1}^{\kappa_4} \mu_2^j\|^{{\mc V}_{\iota-1}}}= \\
& \dfrac{(\bigfswedge_{j=1}^{\kappa_1} \|\mu_2^j\|^{{\mc V}_{\iota-1}})\fswedge (\bigfswedge_{j=\kappa_1+1}^{\kappa_2} \|\mu_2^j\|^{{\mc V}_{\iota-1}})\fswedge
         (\bigfswedge_{j=\kappa_2+1}^{\kappa_3} \|\mu_1^j\|^{{\mc V}_{\iota-1}})\fswedge \bigfswedge_{j=\kappa_3+1}^{\kappa_4} \|\mu_1^j\|^{{\mc V}_{\iota-1}}}
        {(\bigfswedge_{j=1}^{\kappa_1} \|\mu_1^j\|^{{\mc V}_{\iota-1}})\fswedge (\bigfswedge_{j=\kappa_1+1}^{\kappa_2} \|\mu_1^j\|^{{\mc V}_{\iota-1}})\fswedge
         (\bigfswedge_{j=\kappa_2+1}^{\kappa_3} \|\mu_2^j\|^{{\mc V}_{\iota-1}})\fswedge \bigfswedge_{j=\kappa_3+1}^{\kappa_4} \|\mu_2^j\|^{{\mc V}_{\iota-1}}}= \\
& \left(\bigfswedge_{j=1}^{\kappa_1} \dfrac{\|\mu_2^j\|^{{\mc V}_{\iota-1}}}
                                           {\|\mu_1^j\|^{{\mc V}_{\iota-1}}}\right)\fswedge 
  \left(\bigfswedge_{j=\kappa_1+1}^{\kappa_2} \dfrac{\|\mu_2^j\|^{{\mc V}_{\iota-1}}}
                                                    {\|\mu_1^j\|^{{\mc V}_{\iota-1}}}\right)\fswedge 
  \left(\bigfswedge_{j=\kappa_1+2}^{\kappa_3} \dfrac{\|\mu_1^j\|^{{\mc V}_{\iota-1}}}
                                                    {\|\mu_2^j\|^{{\mc V}_{\iota-1}}}\right)\fswedge
  \bigfswedge_{j=\kappa_1+3}^{\kappa_4} \dfrac{\|\mu_1^j\|^{{\mc V}_{\iota-1}}}
                                              {\|\mu_2^j\|^{{\mc V}_{\iota-1}}}< \\
& \left(\bigfswedge_{j=1}^{\kappa_1} {\mc V}_\iota(a_\iota)^{\beta_j}\right)\fswedge \left(\bigfswedge_{j=\kappa_1+1}^{\kappa_2} {\mc V}_\iota(a_\iota)^{\beta_j}\right)\fswedge 
  \left(\bigfswedge_{j=\kappa_1+2}^{\kappa_3} \dfrac{1}
                                                    {{\mc V}_\iota(a_\iota)^{\beta_j}}\right)\fswedge
  \bigfswedge_{j=\kappa_1+3}^{\kappa_4} \dfrac{1}
                                              {{\mc V}_\iota(a_\iota)^{\beta_j}}= \\
& {\mc V}_\iota(a_\iota)^{(\sum_{j=1}^{\kappa_1} \beta_j)+(\sum_{j=\kappa_1+1}^{\kappa_2} \beta_j)-(\sum_{j=\kappa_2+1}^{\kappa_3} \beta_j)-\sum_{j=\kappa_3+1}^{\kappa_4} \beta_j}=
  {\mc V}_\iota(a_\iota)^\alpha, \\[1mm]
& \|\varepsilon_1\|^{{\mc V}_\iota}\overset{\text{(b)}}{=\!\!=} \|\varepsilon_1\|^{{\mc V}_{\iota-1}}<
  \|\upsilon_2\|^{{\mc V}_{\iota-1}}\fswedge {\mc V}_\iota(a_\iota)^\alpha\overset{\text{(b)}}{=\!\!=}
  \|\upsilon_2\|^{{\mc V}_\iota}\fswedge {\mc V}_\iota(a_\iota)^\alpha=\|\upsilon_2\mult a_\iota^\alpha\|^{{\mc V}_\iota}=\|\varepsilon_2\|^{{\mc V}_\iota};
\end{alignat*}
(e) holds.

Case 2.10.2.2.2.2:
There exists $\zeta_1\diamond^\zeta \zeta_2=\mi{reduced}(\lambda^{(\gamma_1^\zeta,\dots,\gamma_n^\zeta)})\in \mi{clo}$,
$\zeta_e\in \{\gu\}\cup \mi{PropConj}_A$, $\mi{atoms}(\zeta_e)\subseteq \mi{dom}({\mc V}_{\iota-1})$, $\diamond^\zeta\in \{\geql,\gleq,\gle\}$, 
$\bs{0}^n\neq (\gamma_1^\zeta,\dots,\gamma_n^\zeta)\in \mbb{N}^n$, such that
$\varepsilon_1\gle \varepsilon_2=\varepsilon_1\gle (\upsilon_2\mult a_\iota^\alpha)=
                                 \mi{reduced}(\lambda^{((\sum_{j=1}^{\kappa_4} \gamma_1^j)+\gamma_1^\zeta,\dots,(\sum_{j=1}^{\kappa_4} \gamma_n^j)+\gamma_n^\zeta)})$.
Then 
$\mi{atoms}(\varepsilon_1), \mi{atoms}(\upsilon_2), \mi{atoms}(\mu_1^1),\dots,\mi{atoms}(\mu_1^{\kappa_4}), \mi{atoms}(\mu_2^1),\dots,\mi{atoms}(\mu_2^{\kappa_4})\subseteq \mi{dom}({\mc V}_{\iota-1})$,
\begin{alignat*}{1}
& \varepsilon_1\gle (\upsilon_2\mult a_\iota^\alpha)=
  \mi{reduced}(\lambda^{((\sum_{j=1}^{\kappa_4} \gamma_1^j)+\gamma_1^\zeta,\dots,(\sum_{j=1}^{\kappa_4} \gamma_n^j)+\gamma_n^\zeta)})\overset{\text{(\ref{eq8dddd})}}{=\!\!=} \\
& \mi{reduced}((\bigmult_{j=1}^{\kappa_1} \lambda^{(\gamma_1^j,\dots,\gamma_n^j)})\mult
               (\bigmult_{j=\kappa_1+1}^{\kappa_2} \lambda^{(\gamma_1^j,\dots,\gamma_n^j)})\mult \\
&\phantom{\mi{reduced}(}
               (\bigmult_{j=\kappa_2+1}^{\kappa_3} \lambda^{(\gamma_1^j,\dots,\gamma_n^j)})\mult
               (\bigmult_{j=\kappa_3+1}^{\kappa_4} \lambda^{(\gamma_1^j,\dots,\gamma_n^j)})\mult
               \lambda^{(\gamma_1^\zeta,\dots,\gamma_n^\zeta)})\overset{\text{(\ref{eq7k})}}{=\!\!=} \\
& \mi{reduced}((\bigmult_{j=1}^{\kappa_1} \mi{reduced}(\lambda^{(\gamma_1^j,\dots,\gamma_n^j)}))\mult
               (\bigmult_{j=\kappa_1+1}^{\kappa_2} \mi{reduced}(\lambda^{(\gamma_1^j,\dots,\gamma_n^j)}))\mult \\
&\phantom{\mi{reduced}(}
               (\bigmult_{j=\kappa_2+1}^{\kappa_3} \mi{reduced}(\lambda^{(\gamma_1^j,\dots,\gamma_n^j)}))\mult
               (\bigmult_{j=\kappa_3+1}^{\kappa_4} \mi{reduced}(\lambda^{(\gamma_1^j,\dots,\gamma_n^j)}))\mult \\
&\phantom{\mi{reduced}(}
               \mi{reduced}(\lambda^{(\gamma_1^\zeta,\dots,\gamma_n^\zeta)}))= \\
& \mi{reduced}((\bigmult_{j=1}^{\kappa_1} \mu_2^j\gleq (\mu_1^j\mult a_\iota^{\beta_j}))\mult
               (\bigmult_{j=\kappa_1+1}^{\kappa_2} \mu_2^j\gle (\mu_1^j\mult a_\iota^{\beta_j}))\mult \\
&\phantom{\mi{reduced}(}
               (\bigmult_{j=\kappa_2+1}^{\kappa_3} (\mu_1^j\mult a_\iota^{\beta_j})\gleq \mu_2^j)\mult
               (\bigmult_{j=\kappa_3+1}^{\kappa_4} (\mu_1^j\mult a_\iota^{\beta_j})\gle \mu_2^j)\mult
               (\zeta_1\diamond^\zeta \zeta_2))\overset{\text{(\ref{eq7i})}}{=\!\!=} \\
& \mi{reduced}(((\bigmult_{j=1}^{\kappa_1} \mu_2^j)\mult (\bigmult_{j=\kappa_1+1}^{\kappa_2} \mu_2^j)\mult 
                (\bigmult_{j=\kappa_2+1}^{\kappa_3} \mu_1^j)\mult (\bigmult_{j=\kappa_3+1}^{\kappa_4} \mu_1^j)\mult \zeta_1\mult \\
&\phantom{\mi{reduced}(} \quad
                a_\iota^{(\sum_{j=\kappa_2+1}^{\kappa_3} \beta_j)+\sum_{j=\kappa_3+1}^{\kappa_4} \beta_j})\gle \\
&\phantom{\mi{reduced}(} \quad
               ((\bigmult_{j=1}^{\kappa_1} \mu_1^j)\mult (\bigmult_{j=\kappa_1+1}^{\kappa_2} \mu_1^j)\mult 
                (\bigmult_{j=\kappa_2+1}^{\kappa_3} \mu_2^j)\mult (\bigmult_{j=\kappa_3+1}^{\kappa_4} \mu_2^j)\mult \zeta_2\mult \\
&\phantom{\mi{reduced}(} \quad \quad
                a_\iota^{(\sum_{j=1}^{\kappa_1} \beta_j)+\sum_{j=\kappa_1+1}^{\kappa_2} \beta_j}))= \\
& \mi{reduced}^-(((\bigmult_{j=1}^{\kappa_1} \mu_2^j)\mult (\bigmult_{j=\kappa_1+1}^{\kappa_2} \mu_2^j)\mult 
                  (\bigmult_{j=\kappa_2+1}^{\kappa_3} \mu_1^j)\mult (\bigmult_{j=\kappa_3+1}^{\kappa_4} \mu_1^j)\mult \zeta_1\mult \\
&\phantom{\mi{reduced}^-(} \quad
                  a_\iota^{(\sum_{j=\kappa_2+1}^{\kappa_3} \beta_j)+\sum_{j=\kappa_3+1}^{\kappa_4} \beta_j})\gle \\
&\phantom{\mi{reduced}^-(} \quad
                 ((\bigmult_{j=1}^{\kappa_1} \mu_1^j)\mult (\bigmult_{j=\kappa_1+1}^{\kappa_2} \mu_1^j)\mult 
                  (\bigmult_{j=\kappa_2+1}^{\kappa_3} \mu_2^j)\mult (\bigmult_{j=\kappa_3+1}^{\kappa_4} \mu_2^j)\mult \zeta_2\mult \\
&\phantom{\mi{reduced}^-(} \quad \quad
                  a_\iota^{(\sum_{j=1}^{\kappa_1} \beta_j)+\sum_{j=\kappa_1+1}^{\kappa_2} \beta_j}))\gle \\
& \mi{reduced}^+(((\bigmult_{j=1}^{\kappa_1} \mu_2^j)\mult (\bigmult_{j=\kappa_1+1}^{\kappa_2} \mu_2^j)\mult 
                  (\bigmult_{j=\kappa_2+1}^{\kappa_3} \mu_1^j)\mult (\bigmult_{j=\kappa_3+1}^{\kappa_4} \mu_1^j)\mult \zeta_1\mult \\
&\phantom{\mi{reduced}^+(} \quad
                  a_\iota^{(\sum_{j=\kappa_2+1}^{\kappa_3} \beta_j)+\sum_{j=\kappa_3+1}^{\kappa_4} \beta_j})\gle \\
&\phantom{\mi{reduced}^+(} \quad
                 ((\bigmult_{j=1}^{\kappa_1} \mu_1^j)\mult (\bigmult_{j=\kappa_1+1}^{\kappa_2} \mu_1^j)\mult 
                  (\bigmult_{j=\kappa_2+1}^{\kappa_3} \mu_2^j)\mult (\bigmult_{j=\kappa_3+1}^{\kappa_4} \mu_2^j)\mult \zeta_2\mult \\
&\phantom{\mi{reduced}^+(} \quad \quad
                  a_\iota^{(\sum_{j=1}^{\kappa_1} \beta_j)+\sum_{j=\kappa_1+1}^{\kappa_2} \beta_j})), \\[1mm]
& \#(a_\iota,\varepsilon_1\gle (\upsilon_2\mult a_\iota^\alpha))=\alpha= \\
& \#(a_\iota,\mi{reduced}(((\bigmult_{j=1}^{\kappa_1} \mu_2^j)\mult (\bigmult_{j=\kappa_1+1}^{\kappa_2} \mu_2^j)\mult 
                           (\bigmult_{j=\kappa_2+1}^{\kappa_3} \mu_1^j)\mult (\bigmult_{j=\kappa_3+1}^{\kappa_4} \mu_1^j)\mult \zeta_1\mult \\
&\phantom{\#(a_\iota,\mi{reduced}(} \quad
                           a_\iota^{(\sum_{j=\kappa_2+1}^{\kappa_3} \beta_j)+\sum_{j=\kappa_3+1}^{\kappa_4} \beta_j})\gle \\
&\phantom{\#(a_\iota,\mi{reduced}(} \quad
                          ((\bigmult_{j=1}^{\kappa_1} \mu_1^j)\mult (\bigmult_{j=\kappa_1+1}^{\kappa_2} \mu_1^j)\mult 
                           (\bigmult_{j=\kappa_2+1}^{\kappa_3} \mu_2^j)\mult (\bigmult_{j=\kappa_3+1}^{\kappa_4} \mu_2^j)\mult \zeta_2\mult \\
&\phantom{\#(a_\iota,\mi{reduced}(} \quad \quad
                           a_\iota^{(\sum_{j=1}^{\kappa_1} \beta_j)+\sum_{j=\kappa_1+1}^{\kappa_2} \beta_j})))\overset{\text{(\ref{eq7aaaax})}}{=\!\!=} \\
& \#(a_\iota,((\bigmult_{j=1}^{\kappa_1} \mu_2^j)\mult (\bigmult_{j=\kappa_1+1}^{\kappa_2} \mu_2^j)\mult 
              (\bigmult_{j=\kappa_2+1}^{\kappa_3} \mu_1^j)\mult (\bigmult_{j=\kappa_3+1}^{\kappa_4} \mu_1^j)\mult \zeta_1\mult \\
&\phantom{\#(a_\iota,} \quad
              a_\iota^{(\sum_{j=\kappa_2+1}^{\kappa_3} \beta_j)+\sum_{j=\kappa_3+1}^{\kappa_4} \beta_j})\gle \\
&\phantom{\#(a_\iota,} \quad
             ((\bigmult_{j=1}^{\kappa_1} \mu_1^j)\mult (\bigmult_{j=\kappa_1+1}^{\kappa_2} \mu_1^j)\mult 
              (\bigmult_{j=\kappa_2+1}^{\kappa_3} \mu_2^j)\mult (\bigmult_{j=\kappa_3+1}^{\kappa_4} \mu_2^j)\mult \zeta_2\mult \\
&\phantom{\#(a_\iota,} \quad \quad
              a_\iota^{(\sum_{j=1}^{\kappa_1} \beta_j)+\sum_{j=\kappa_1+1}^{\kappa_2} \beta_j}))= \\
& \left(\sum_{j=1}^{\kappa_1} \beta_j\right)+\left(\sum_{j=\kappa_1+1}^{\kappa_2} \beta_j\right)-\left(\sum_{j=\kappa_2+1}^{\kappa_3} \beta_j\right)-\sum_{j=\kappa_3+1}^{\kappa_4} \beta_j\geq 1, 
\end{alignat*}
$a_\iota\not\in \mi{dom}({\mc V}_{\iota-1})\supseteq \mi{atoms}((\bigmult_{j=1}^{\kappa_1} \mu_2^j)\mult (\bigmult_{j=\kappa_1+1}^{\kappa_2} \mu_2^j)\mult 
                                                                (\bigmult_{j=\kappa_2+1}^{\kappa_3} \mu_1^j)\mult                                                                          \linebreak[4]
                                                                                                                  (\bigmult_{j=\kappa_3+1}^{\kappa_4} \mu_1^j)\mult \zeta_1)=
                                                     (\bigcup_{j=1}^{\kappa_1} \mi{atoms}(\mu_2^j))\cup (\bigcup_{j=\kappa_1+1}^{\kappa_2} \mi{atoms}(\mu_2^j))\cup                        \linebreak[4]
                                                     (\bigcup_{j=\kappa_2+1}^{\kappa_3} \mi{atoms}(\mu_1^j))\cup (\bigcup_{j=\kappa_3+1}^{\kappa_4} \mi{atoms}(\mu_1^j))\cup \mi{atoms}(\zeta_1)$,
$a_\iota\not\in \mi{dom}({\mc V}_{\iota-1})\supseteq \mi{atoms}((\bigmult_{j=1}^{\kappa_1} \mu_1^j)\mult (\bigmult_{j=\kappa_1+1}^{\kappa_2} \mu_1^j)\mult 
                                                                (\bigmult_{j=\kappa_2+1}^{\kappa_3} \mu_2^j)\mult (\bigmult_{j=\kappa_3+1}^{\kappa_4} \mu_2^j)\mult \zeta_2)=              \linebreak[4]
                                                     (\bigcup_{j=1}^{\kappa_1} \mi{atoms}(\mu_1^j))\cup (\bigcup_{j=\kappa_1+1}^{\kappa_2} \mi{atoms}(\mu_1^j))\cup 
                                                     (\bigcup_{j=\kappa_2+1}^{\kappa_3} \mi{atoms}(\mu_2^j))\cup (\bigcup_{j=\kappa_3+1}^{\kappa_4} \mi{atoms}(\mu_2^j))\cup \mi{atoms}(\zeta_2)$,
\begin{alignat*}{1}
& \mi{reduced}^-(((\bigmult_{j=1}^{\kappa_1} \mu_2^j)\mult (\bigmult_{j=\kappa_1+1}^{\kappa_2} \mu_2^j)\mult 
                  (\bigmult_{j=\kappa_2+1}^{\kappa_3} \mu_1^j)\mult (\bigmult_{j=\kappa_3+1}^{\kappa_4} \mu_1^j)\mult \zeta_1\mult \\
&\phantom{\mi{reduced}^-(} \quad
                  a_\iota^{(\sum_{j=\kappa_2+1}^{\kappa_3} \beta_j)+\sum_{j=\kappa_3+1}^{\kappa_4} \beta_j})\gle \\
&\phantom{\mi{reduced}^-(} \quad
                 ((\bigmult_{j=1}^{\kappa_1} \mu_1^j)\mult (\bigmult_{j=\kappa_1+1}^{\kappa_2} \mu_1^j)\mult 
                  (\bigmult_{j=\kappa_2+1}^{\kappa_3} \mu_2^j)\mult (\bigmult_{j=\kappa_3+1}^{\kappa_4} \mu_2^j)\mult \zeta_2\mult \\
&\phantom{\mi{reduced}^-(} \quad \quad
                  a_\iota^{(\sum_{j=1}^{\kappa_1} \beta_j)+\sum_{j=\kappa_1+1}^{\kappa_2} \beta_j}))\overset{\text{(\ref{eq7cc})}}{=\!\!=} \\
& \mi{reduced}^-(((\bigmult_{j=1}^{\kappa_1} \mu_2^j)\mult (\bigmult_{j=\kappa_1+1}^{\kappa_2} \mu_2^j)\mult 
                  (\bigmult_{j=\kappa_2+1}^{\kappa_3} \mu_1^j)\mult (\bigmult_{j=\kappa_3+1}^{\kappa_4} \mu_1^j)\mult \zeta_1)\gle \\
&\phantom{\mi{reduced}^-(} \quad
                 ((\bigmult_{j=1}^{\kappa_1} \mu_1^j)\mult (\bigmult_{j=\kappa_1+1}^{\kappa_2} \mu_1^j)\mult 
                  (\bigmult_{j=\kappa_2+1}^{\kappa_3} \mu_2^j)\mult (\bigmult_{j=\kappa_3+1}^{\kappa_4} \mu_2^j)\mult \zeta_2)), \\[1mm]
& \mi{reduced}^+(((\bigmult_{j=1}^{\kappa_1} \mu_2^j)\mult (\bigmult_{j=\kappa_1+1}^{\kappa_2} \mu_2^j)\mult 
                  (\bigmult_{j=\kappa_2+1}^{\kappa_3} \mu_1^j)\mult (\bigmult_{j=\kappa_3+1}^{\kappa_4} \mu_1^j)\mult \zeta_1\mult \\
&\phantom{\mi{reduced}^+(} \quad
                  a_\iota^{(\sum_{j=\kappa_2+1}^{\kappa_3} \beta_j)+\sum_{j=\kappa_3+1}^{\kappa_4} \beta_j})\gle \\
&\phantom{\mi{reduced}^+(} \quad
                 ((\bigmult_{j=1}^{\kappa_1} \mu_1^j)\mult (\bigmult_{j=\kappa_1+1}^{\kappa_2} \mu_1^j)\mult 
                  (\bigmult_{j=\kappa_2+1}^{\kappa_3} \mu_2^j)\mult (\bigmult_{j=\kappa_3+1}^{\kappa_4} \mu_2^j)\mult \zeta_2\mult \\
&\phantom{\mi{reduced}^+(} \quad \quad
                  a_\iota^{(\sum_{j=1}^{\kappa_1} \beta_j)+\sum_{j=\kappa_1+1}^{\kappa_2} \beta_j}))\overset{\text{(\ref{eq7ff})}}{=\!\!=} \\
& \mi{reduced}^+(((\bigmult_{j=1}^{\kappa_1} \mu_2^j)\mult (\bigmult_{j=\kappa_1+1}^{\kappa_2} \mu_2^j)\mult 
                  (\bigmult_{j=\kappa_2+1}^{\kappa_3} \mu_1^j)\mult (\bigmult_{j=\kappa_3+1}^{\kappa_4} \mu_1^j)\mult \zeta_1)\gle \\
&\phantom{\mi{reduced}^+(} \quad
                 ((\bigmult_{j=1}^{\kappa_1} \mu_1^j)\mult (\bigmult_{j=\kappa_1+1}^{\kappa_2} \mu_1^j)\mult 
                  (\bigmult_{j=\kappa_2+1}^{\kappa_3} \mu_2^j)\mult (\bigmult_{j=\kappa_3+1}^{\kappa_4} \mu_2^j)\mult \zeta_2))\mult \\
& a_\iota^{(\sum_{j=1}^{\kappa_1} \beta_j)+(\sum_{j=\kappa_1+1}^{\kappa_2} \beta_j)-(\sum_{j=\kappa_2+1}^{\kappa_3} \beta_j)-\sum_{j=\kappa_3+1}^{\kappa_4} \beta_j}= \\
& \mi{reduced}^+(((\bigmult_{j=1}^{\kappa_1} \mu_2^j)\mult (\bigmult_{j=\kappa_1+1}^{\kappa_2} \mu_2^j)\mult 
                  (\bigmult_{j=\kappa_2+1}^{\kappa_3} \mu_1^j)\mult (\bigmult_{j=\kappa_3+1}^{\kappa_4} \mu_1^j)\mult \zeta_1)\gle \\
&\phantom{\mi{reduced}^+(} \quad
                 ((\bigmult_{j=1}^{\kappa_1} \mu_1^j)\mult (\bigmult_{j=\kappa_1+1}^{\kappa_2} \mu_1^j)\mult 
                  (\bigmult_{j=\kappa_2+1}^{\kappa_3} \mu_2^j)\mult (\bigmult_{j=\kappa_3+1}^{\kappa_4} \mu_2^j)\mult \zeta_2))\mult a_\iota^\alpha, \\[1mm]
& \varepsilon_1\gle (\upsilon_2\mult a_\iota^\alpha)= \\
& \mi{reduced}^-(((\bigmult_{j=1}^{\kappa_1} \mu_2^j)\mult (\bigmult_{j=\kappa_1+1}^{\kappa_2} \mu_2^j)\mult 
                  (\bigmult_{j=\kappa_2+1}^{\kappa_3} \mu_1^j)\mult (\bigmult_{j=\kappa_3+1}^{\kappa_4} \mu_1^j)\mult \zeta_1)\gle \\
&\phantom{\mi{reduced}^-(} \quad
                 ((\bigmult_{j=1}^{\kappa_1} \mu_1^j)\mult (\bigmult_{j=\kappa_1+1}^{\kappa_2} \mu_1^j)\mult 
                  (\bigmult_{j=\kappa_2+1}^{\kappa_3} \mu_2^j)\mult (\bigmult_{j=\kappa_3+1}^{\kappa_4} \mu_2^j)\mult \zeta_2))\gle \\
& (\mi{reduced}^+(((\bigmult_{j=1}^{\kappa_1} \mu_2^j)\mult (\bigmult_{j=\kappa_1+1}^{\kappa_2} \mu_2^j)\mult 
                   (\bigmult_{j=\kappa_2+1}^{\kappa_3} \mu_1^j)\mult (\bigmult_{j=\kappa_3+1}^{\kappa_4} \mu_1^j)\mult \zeta_1)\gle \\
&\phantom{(\mi{reduced}^+(} \quad
                  ((\bigmult_{j=1}^{\kappa_1} \mu_1^j)\mult (\bigmult_{j=\kappa_1+1}^{\kappa_2} \mu_1^j)\mult 
                   (\bigmult_{j=\kappa_2+1}^{\kappa_3} \mu_2^j)\mult (\bigmult_{j=\kappa_3+1}^{\kappa_4} \mu_2^j)\mult \zeta_2))\mult a_\iota^\alpha), \\[1mm]
& \mi{atoms}(\mi{reduced}^-(((\bigmult_{j=1}^{\kappa_1} \mu_2^j)\mult (\bigmult_{j=\kappa_1+1}^{\kappa_2} \mu_2^j)\mult 
                             (\bigmult_{j=\kappa_2+1}^{\kappa_3} \mu_1^j)\mult (\bigmult_{j=\kappa_3+1}^{\kappa_4} \mu_1^j)\mult \zeta_1)\gle \\
&\phantom{\mi{atoms}(\mi{reduced}^-(} \quad
                            ((\bigmult_{j=1}^{\kappa_1} \mu_1^j)\mult (\bigmult_{j=\kappa_1+1}^{\kappa_2} \mu_1^j)\mult 
                             (\bigmult_{j=\kappa_2+1}^{\kappa_3} \mu_2^j)\mult (\bigmult_{j=\kappa_3+1}^{\kappa_4} \mu_2^j)\mult \zeta_2)))\underset{\text{(\ref{eq7a})}}{\subseteq} \\
& \mi{atoms}((\bigmult_{j=1}^{\kappa_1} \mu_2^j)\mult (\bigmult_{j=\kappa_1+1}^{\kappa_2} \mu_2^j)\mult 
             (\bigmult_{j=\kappa_2+1}^{\kappa_3} \mu_1^j)\mult (\bigmult_{j=\kappa_3+1}^{\kappa_4} \mu_1^j)\mult \zeta_1)\subseteq \mi{dom}({\mc V}_{\iota-1}), \\[1mm]
& \mi{atoms}(\mi{reduced}^+(((\bigmult_{j=1}^{\kappa_1} \mu_2^j)\mult (\bigmult_{j=\kappa_1+1}^{\kappa_2} \mu_2^j)\mult 
                             (\bigmult_{j=\kappa_2+1}^{\kappa_3} \mu_1^j)\mult (\bigmult_{j=\kappa_3+1}^{\kappa_4} \mu_1^j)\mult \zeta_1)\gle \\
&\phantom{\mi{atoms}(\mi{reduced}^+(} \quad
                            ((\bigmult_{j=1}^{\kappa_1} \mu_1^j)\mult (\bigmult_{j=\kappa_1+1}^{\kappa_2} \mu_1^j)\mult 
                             (\bigmult_{j=\kappa_2+1}^{\kappa_3} \mu_2^j)\mult (\bigmult_{j=\kappa_3+1}^{\kappa_4} \mu_2^j)\mult \zeta_2)))\underset{\text{(\ref{eq7b})}}{\subseteq} \\
& \mi{atoms}((\bigmult_{j=1}^{\kappa_1} \mu_1^j)\mult (\bigmult_{j=\kappa_1+1}^{\kappa_2} \mu_1^j)\mult 
             (\bigmult_{j=\kappa_2+1}^{\kappa_3} \mu_2^j)\mult (\bigmult_{j=\kappa_3+1}^{\kappa_4} \mu_2^j)\mult \zeta_2)\subseteq \mi{dom}({\mc V}_{\iota-1}), 
\end{alignat*}
\begin{alignat*}{1}
& \varepsilon_1=\mi{reduced}^-(((\bigmult_{j=1}^{\kappa_1} \mu_2^j)\mult (\bigmult_{j=\kappa_1+1}^{\kappa_2} \mu_2^j)\mult 
                                (\bigmult_{j=\kappa_2+1}^{\kappa_3} \mu_1^j)\mult (\bigmult_{j=\kappa_3+1}^{\kappa_4} \mu_1^j)\mult \zeta_1)\gle \\
&\phantom{\varepsilon_1=\mi{reduced}^-(} \quad
                               ((\bigmult_{j=1}^{\kappa_1} \mu_1^j)\mult (\bigmult_{j=\kappa_1+1}^{\kappa_2} \mu_1^j)\mult 
                                (\bigmult_{j=\kappa_2+1}^{\kappa_3} \mu_2^j)\mult (\bigmult_{j=\kappa_3+1}^{\kappa_4} \mu_2^j)\mult \zeta_2)), \\[1mm]
& \upsilon_2=\mi{reduced}^+(((\bigmult_{j=1}^{\kappa_1} \mu_2^j)\mult (\bigmult_{j=\kappa_1+1}^{\kappa_2} \mu_2^j)\mult 
                             (\bigmult_{j=\kappa_2+1}^{\kappa_3} \mu_1^j)\mult (\bigmult_{j=\kappa_3+1}^{\kappa_4} \mu_1^j)\mult \zeta_1)\gle \\
&\phantom{\upsilon_2=\mi{reduced}^+(} \quad
                            ((\bigmult_{j=1}^{\kappa_1} \mu_1^j)\mult (\bigmult_{j=\kappa_1+1}^{\kappa_2} \mu_1^j)\mult 
                             (\bigmult_{j=\kappa_2+1}^{\kappa_3} \mu_2^j)\mult (\bigmult_{j=\kappa_3+1}^{\kappa_4} \mu_2^j)\mult \zeta_2));
\end{alignat*}
by the induction hypothesis for $\iota-1<\iota$,
$\|\varepsilon_1\|^{{\mc V}_{\iota-1}}, \|\upsilon_2\|^{{\mc V}_{\iota-1}},
 \|\mu_1^1\|^{{\mc V}_{\iota-1}},\dots,                                                                                                                                                    \linebreak[4]
                                       \|\mu_1^{\kappa_4}\|^{{\mc V}_{\iota-1}}, \|\mu_2^1\|^{{\mc V}_{\iota-1}},\dots,\|\mu_2^{\kappa_4}\|^{{\mc V}_{\iota-1}},
 \|\zeta_1\|^{{\mc V}_{\iota-1}}, \|\zeta_2\|^{{\mc V}_{\iota-1}},
 \|\mi{reduced}^-(((\bigmult_{j=1}^{\kappa_1} \mu_2^j)\mult (\bigmult_{j=\kappa_1+1}^{\kappa_2} \mu_2^j)\mult 
                   (\bigmult_{j=\kappa_2+1}^{\kappa_3} \mu_1^j)\mult (\bigmult_{j=\kappa_3+1}^{\kappa_4} \mu_1^j)\mult \zeta_1)\gle
                  ((\bigmult_{j=1}^{\kappa_1} \mu_1^j)\mult (\bigmult_{j=\kappa_1+1}^{\kappa_2} \mu_1^j)\mult 
                   (\bigmult_{j=\kappa_2+1}^{\kappa_3} \mu_2^j)\mult (\bigmult_{j=\kappa_3+1}^{\kappa_4} \mu_2^j)\mult \zeta_2))\|^{{\mc V}_{\iota-1}}, 
 \|\mi{reduced}^+(((\bigmult_{j=1}^{\kappa_1} \mu_2^j)\mult (\bigmult_{j=\kappa_1+1}^{\kappa_2} \mu_2^j)\mult 
                   (\bigmult_{j=\kappa_2+1}^{\kappa_3} \mu_1^j)\mult (\bigmult_{j=\kappa_3+1}^{\kappa_4} \mu_1^j)\mult \zeta_1)\gle
                  ((\bigmult_{j=1}^{\kappa_1} \mu_1^j)\mult (\bigmult_{j=\kappa_1+1}^{\kappa_2} \mu_1^j)\mult 
                   (\bigmult_{j=\kappa_2+1}^{\kappa_3} \mu_2^j)\mult (\bigmult_{j=\kappa_3+1}^{\kappa_4} \mu_2^j)\mult \zeta_2))\|^{{\mc V}_{\iota-1}}\underset{\text{(a)}}{\in} (0,1]$,
either $\diamond^\zeta=\geql$, $\zeta_1\diamond^\zeta \zeta_2=\zeta_1\geql \zeta_2\in \mi{clo}$, $\|\zeta_1\|^{{\mc V}_{\iota-1}}\overset{\text{(c)}}{=\!\!=} \|\zeta_2\|^{{\mc V}_{\iota-1}}$, 
or $\diamond^\zeta=\gleq$, $\zeta_1\diamond^\zeta \zeta_2=\zeta_1\gleq \zeta_2\in \mi{clo}$, $\|\zeta_1\|^{{\mc V}_{\iota-1}}\underset{\text{(d)}}{\leq} \|\zeta_2\|^{{\mc V}_{\iota-1}}$,
or $\diamond^\zeta=\gle$, $\zeta_1\diamond^\zeta \zeta_2=\zeta_1\gle \zeta_2\in \mi{clo}$, $\|\zeta_1\|^{{\mc V}_{\iota-1}}\underset{\text{(e)}}{<} \|\zeta_2\|^{{\mc V}_{\iota-1}}$;
$\|\zeta_1\|^{{\mc V}_{\iota-1}}\leq \|\zeta_2\|^{{\mc V}_{\iota-1}}$,
$\frac{\|\zeta_1\|^{{\mc V}_{\iota-1}}}
      {\|\zeta_2\|^{{\mc V}_{\iota-1}}}\leq 1$,
$\mbb{E}_{\iota-1}=\emptyset$;
for all $1\leq j\leq \kappa_1$, 
$\mu_2^j\gleq (\mu_1^j\mult a_\iota^{\beta_j})\in \mi{min}(\mi{DE}_{\iota-1})$, 
$\left(\frac{\|\mu_2^j\|^{{\mc V}_{\iota-1}}}
            {\|\mu_1^j\|^{{\mc V}_{\iota-1}}}\right)^{\frac{1}{\beta_j}}\in \mbb{DE}_{\iota-1}$,
$\left(\frac{\|\mu_2^j\|^{{\mc V}_{\iota-1}}}
            {\|\mu_1^j\|^{{\mc V}_{\iota-1}}}\right)^{\frac{1}{\beta_j}}\leq \bigfvee \mbb{DE}_{\iota-1}\underset{\text{(\ref{eq8j}a)}}{\leq} {\mc V}_\iota(a_\iota)=\delta_\iota$,
$\frac{\|\mu_2^j\|^{{\mc V}_{\iota-1}}}
      {\|\mu_1^j\|^{{\mc V}_{\iota-1}}}\leq {\mc V}_\iota(a_\iota)^{\beta_j}$;
$\kappa_1<\kappa_2$, $\kappa_1+1\leq \kappa_2$,
$\{\kappa_1+1,\dots,\kappa_2\}\neq \emptyset$;
for all $\kappa_1+1\leq j\leq \kappa_2$, 
$\mu_2^j\gle (\mu_1^j\mult a_\iota^{\beta_j})\in \mi{min}(D_{\iota-1})$, 
$\left(\frac{\|\mu_2^j\|^{{\mc V}_{\iota-1}}}
            {\|\mu_1^j\|^{{\mc V}_{\iota-1}}}\right)^{\frac{1}{\beta_j}}\in \mbb{D}_{\iota-1}$,
$\left(\frac{\|\mu_2^j\|^{{\mc V}_{\iota-1}}}
            {\|\mu_1^j\|^{{\mc V}_{\iota-1}}}\right)^{\frac{1}{\beta_j}}\leq \bigfvee \mbb{D}_{\iota-1}\underset{\text{(\ref{eq8j}b)}}{<} {\mc V}_\iota(a_\iota)=\delta_\iota$,
$\frac{\|\mu_2^j\|^{{\mc V}_{\iota-1}}}
      {\|\mu_1^j\|^{{\mc V}_{\iota-1}}}<{\mc V}_\iota(a_\iota)^{\beta_j}$;
for all $\kappa_2+1\leq j\leq \kappa_3$,
$(\mu_1^j\mult a_\iota^{\beta_j})\gleq \mu_2^j\in \mi{min}(\mi{UE}_{\iota-1})$,
$\mi{min}\left(\left(\frac{\|\mu_2^j\|^{{\mc V}_{\iota-1}}}
                          {\|\mu_1^j\|^{{\mc V}_{\iota-1}}}\right)^{\frac{1}{\beta_j}},1\right)\in \mbb{UE}_{\iota-1}$,
${\mc V}_\iota(a_\iota)=\delta_\iota\underset{\text{(\ref{eq8j}a)}}{\leq} \bigfwedge \mbb{UE}_{\iota-1}\leq 
 \mi{min}\left(\left(\frac{\|\mu_2^j\|^{{\mc V}_{\iota-1}}}
                          {\|\mu_1^j\|^{{\mc V}_{\iota-1}}}\right)^{\frac{1}{\beta_j}},1\right)\leq
 \left(\frac{\|\mu_2^j\|^{{\mc V}_{\iota-1}}}
            {\|\mu_1^j\|^{{\mc V}_{\iota-1}}}\right)^{\frac{1}{\beta_j}}$,
${\mc V}_\iota(a_\iota)^{\beta_j}\leq \frac{\|\mu_2^j\|^{{\mc V}_{\iota-1}}}
                                           {\|\mu_1^j\|^{{\mc V}_{\iota-1}}}$,
$\frac{\|\mu_1^j\|^{{\mc V}_{\iota-1}}}
      {\|\mu_2^j\|^{{\mc V}_{\iota-1}}}\leq \frac{1}
                                                 {{\mc V}_\iota(a_\iota)^{\beta_j}}$;
for all $\kappa_3+1\leq j\leq \kappa_4$,
$(\mu_1^j\mult a_\iota^{\beta_j})\gle \mu_2^j\in \mi{min}(U_{\iota-1})$,
$\mi{min}\left(\left(\frac{\|\mu_2^j\|^{{\mc V}_{\iota-1}}}
                          {\|\mu_1^j\|^{{\mc V}_{\iota-1}}}\right)^{\frac{1}{\beta_j}},1\right)\in \mbb{U}_{\iota-1}$,
${\mc V}_\iota(a_\iota)=\delta_\iota\underset{\text{(\ref{eq8j}b)}}{\leq} \bigfwedge \mbb{U}_{\iota-1}\leq 
 \mi{min}\left(\left(\frac{\|\mu_2^j\|^{{\mc V}_{\iota-1}}}
                          {\|\mu_1^j\|^{{\mc V}_{\iota-1}}}\right)^{\frac{1}{\beta_j}},1\right)\leq
 \left(\frac{\|\mu_2^j\|^{{\mc V}_{\iota-1}}}
            {\|\mu_1^j\|^{{\mc V}_{\iota-1}}}\right)^{\frac{1}{\beta_j}}$,
${\mc V}_\iota(a_\iota)^{\beta_j}\leq \frac{\|\mu_2^j\|^{{\mc V}_{\iota-1}}}
                                           {\|\mu_1^j\|^{{\mc V}_{\iota-1}}}$,
$\frac{\|\mu_1^j\|^{{\mc V}_{\iota-1}}}
      {\|\mu_2^j\|^{{\mc V}_{\iota-1}}}\leq \frac{1}
                                                 {{\mc V}_\iota(a_\iota)^{\beta_j}}$;
\begin{alignat*}{1}
& \dfrac{\|\varepsilon_1\|^{{\mc V}_{\iota-1}}}
        {\|\upsilon_2\|^{{\mc V}_{\iota-1}}}= \\
& \dfrac{\left\|\mi{reduced}^-\left(\begin{array}{l}
                                    ((\bigmult_{j=1}^{\kappa_1} \mu_2^j)\mult (\bigmult_{j=\kappa_1+1}^{\kappa_2} \mu_2^j)\mult 
                                     (\bigmult_{j=\kappa_2+1}^{\kappa_3} \mu_1^j)\mult \\
                                    \quad
                                     (\bigmult_{j=\kappa_3+1}^{\kappa_4} \mu_1^j)\mult \zeta_1)\gle \\
                                    \quad
                                    ((\bigmult_{j=1}^{\kappa_1} \mu_1^j)\mult (\bigmult_{j=\kappa_1+1}^{\kappa_2} \mu_1^j)\mult 
                                     (\bigmult_{j=\kappa_2+1}^{\kappa_3} \mu_2^j)\mult \\
                                    \quad \quad
                                     (\bigmult_{j=\kappa_3+1}^{\kappa_4} \mu_2^j)\mult \zeta_2)
                                    \end{array}\right)\right\|^{{\mc V}_{\iota-1}}}
        {\left\|\mi{reduced}^+\left(\begin{array}{l}
                                    ((\bigmult_{j=1}^{\kappa_1} \mu_2^j)\mult (\bigmult_{j=\kappa_1+1}^{\kappa_2} \mu_2^j)\mult 
                                     (\bigmult_{j=\kappa_2+1}^{\kappa_3} \mu_1^j)\mult \\
                                    \quad
                                     (\bigmult_{j=\kappa_3+1}^{\kappa_4} \mu_1^j)\mult \zeta_1)\gle \\
                                    \quad
                                    ((\bigmult_{j=1}^{\kappa_1} \mu_1^j)\mult (\bigmult_{j=\kappa_1+1}^{\kappa_2} \mu_1^j)\mult 
                                     (\bigmult_{j=\kappa_2+1}^{\kappa_3} \mu_2^j)\mult \\
                                    \quad \quad
                                     (\bigmult_{j=\kappa_3+1}^{\kappa_4} \mu_2^j)\mult \zeta_2)
                                    \end{array}\right)\right\|^{{\mc V}_{\iota-1}}}\overset{\text{(\ref{eq7n})}}{=\!\!=} \\
& \dfrac{\|(\bigmult_{j=1}^{\kappa_1} \mu_2^j)\mult (\bigmult_{j=\kappa_1+1}^{\kappa_2} \mu_2^j)\mult 
           (\bigmult_{j=\kappa_2+1}^{\kappa_3} \mu_1^j)\mult (\bigmult_{j=\kappa_3+1}^{\kappa_4} \mu_1^j)\mult \zeta_1\|^{{\mc V}_{\iota-1}}}
        {\|(\bigmult_{j=1}^{\kappa_1} \mu_1^j)\mult (\bigmult_{j=\kappa_1+1}^{\kappa_2} \mu_1^j)\mult 
           (\bigmult_{j=\kappa_2+1}^{\kappa_3} \mu_2^j)\mult (\bigmult_{j=\kappa_3+1}^{\kappa_4} \mu_2^j)\mult \zeta_2\|^{{\mc V}_{\iota-1}}}= \\
& \dfrac{\begin{array}{l}
         (\bigfswedge_{j=1}^{\kappa_1} \|\mu_2^j\|^{{\mc V}_{\iota-1}})\fswedge (\bigfswedge_{j=\kappa_1+1}^{\kappa_2} \|\mu_2^j\|^{{\mc V}_{\iota-1}})\fswedge
         (\bigfswedge_{j=\kappa_2+1}^{\kappa_3} \|\mu_1^j\|^{{\mc V}_{\iota-1}})\fswedge (\bigfswedge_{j=\kappa_3+1}^{\kappa_4} \|\mu_1^j\|^{{\mc V}_{\iota-1}})\fswedge \\
         \quad
         \|\zeta_1\|^{{\mc V}_{\iota-1}}
         \end{array}}
        {\begin{array}{l}
         (\bigfswedge_{j=1}^{\kappa_1} \|\mu_1^j\|^{{\mc V}_{\iota-1}})\fswedge (\bigfswedge_{j=\kappa_1+1}^{\kappa_2} \|\mu_1^j\|^{{\mc V}_{\iota-1}})\fswedge
         (\bigfswedge_{j=\kappa_2+1}^{\kappa_3} \|\mu_2^j\|^{{\mc V}_{\iota-1}})\fswedge (\bigfswedge_{j=\kappa_3+1}^{\kappa_4} \|\mu_2^j\|^{{\mc V}_{\iota-1}})\fswedge \\
         \quad
         \|\zeta_2\|^{{\mc V}_{\iota-1}}
         \end{array}}= \\
& \left(\bigfswedge_{j=1}^{\kappa_1} \dfrac{\|\mu_2^j\|^{{\mc V}_{\iota-1}}}
                                           {\|\mu_1^j\|^{{\mc V}_{\iota-1}}}\right)\fswedge 
  \left(\bigfswedge_{j=\kappa_1+1}^{\kappa_2} \dfrac{\|\mu_2^j\|^{{\mc V}_{\iota-1}}}
                                                    {\|\mu_1^j\|^{{\mc V}_{\iota-1}}}\right)\fswedge 
  \left(\bigfswedge_{j=\kappa_1+2}^{\kappa_3} \dfrac{\|\mu_1^j\|^{{\mc V}_{\iota-1}}}
                                                    {\|\mu_2^j\|^{{\mc V}_{\iota-1}}}\right)\fswedge \\
& \quad
  \left(\bigfswedge_{j=\kappa_1+3}^{\kappa_4} \dfrac{\|\mu_1^j\|^{{\mc V}_{\iota-1}}}
                                                    {\|\mu_2^j\|^{{\mc V}_{\iota-1}}}\right)\fswedge
  \dfrac{\|\zeta_1\|^{{\mc V}_{\iota-1}}}
        {\|\zeta_2\|^{{\mc V}_{\iota-1}}}< \\
& \left(\bigfswedge_{j=1}^{\kappa_1} {\mc V}_\iota(a_\iota)^{\beta_j}\right)\fswedge \left(\bigfswedge_{j=\kappa_1+1}^{\kappa_2} {\mc V}_\iota(a_\iota)^{\beta_j}\right)\fswedge 
  \left(\bigfswedge_{j=\kappa_1+2}^{\kappa_3} \dfrac{1}
                                                    {{\mc V}_\iota(a_\iota)^{\beta_j}}\right)\fswedge
  \bigfswedge_{j=\kappa_1+3}^{\kappa_4} \dfrac{1}
                                              {{\mc V}_\iota(a_\iota)^{\beta_j}}= \\
& {\mc V}_\iota(a_\iota)^{(\sum_{j=1}^{\kappa_1} \beta_j)+(\sum_{j=\kappa_1+1}^{\kappa_2} \beta_j)-(\sum_{j=\kappa_2+1}^{\kappa_3} \beta_j)-\sum_{j=\kappa_3+1}^{\kappa_4} \beta_j}=
  {\mc V}_\iota(a_\iota)^\alpha, \\[1mm]
& \|\varepsilon_1\|^{{\mc V}_\iota}\overset{\text{(b)}}{=\!\!=} \|\varepsilon_1\|^{{\mc V}_{\iota-1}}<
  \|\upsilon_2\|^{{\mc V}_{\iota-1}}\fswedge {\mc V}_\iota(a_\iota)^\alpha\overset{\text{(b)}}{=\!\!=}
  \|\upsilon_2\|^{{\mc V}_\iota}\fswedge {\mc V}_\iota(a_\iota)^\alpha=\|\upsilon_2\mult a_\iota^\alpha\|^{{\mc V}_\iota}=\|\varepsilon_2\|^{{\mc V}_\iota};
\end{alignat*}
(e) holds.

Case 2.10.2.3:
$a_\iota\in \mi{atoms}(\varepsilon_1)$ and $a_\iota\in \mi{atoms}(\varepsilon_2)$.
Then $a_\iota\in \mi{atoms}(\varepsilon_1)\cap \mi{atoms}(\varepsilon_2)\neq \emptyset$;
for all $\diamond\in \{\geql,\gleq,\gle\}$, $\varepsilon_1\diamond \varepsilon_2\underset{\text{(\ref{eq8aaa})}}{\not\in} \mi{clo}$; 
(c--e) hold trivially.

So, in Case 2.10, (c--e) hold.
%
%
%
\end{proof}

\subsection{Full proof of Lemma \ref{le44}}
\label{S7.6a}

\begin{proof}
We have that $S$ is positive.
Then $a^*\gle \gu\in \mi{guards}(a^*)$, $a^*\gle \gu\not\in \mi{guards}(S,a^*)=\{\gz\gle a^*\}=S\cap \mi{guards}(a^*)$, $a^*\gle \gu\not\in S\supseteq \mi{guards}(S)$,
$S\cup \{a^*\gle \gu\}\subseteq_{\mc F} \mi{OrdPropCl}$,
$a^*\in \mi{atoms}(S)$,
$a^*\in \mi{atoms}(S\cup \{a^*\gle \gu\})=\mi{atoms}(S)\cup \mi{atoms}(a^*\gle \gu)=\mi{atoms}(S)\cup \{a^*\}=\mi{atoms}(S)$;
$S$ is simplified; 
$a^*\gle \gu\neq \square$ does not contain contradictions and tautologies;
$S\cup \{a^*\gle \gu\}$ is simplified;
$\mi{guards}(S\cup \{a^*\gle \gu\},a^*)=(S\cup \{a^*\gle \gu\})\cap \mi{guards}(a^*)=(S\cap \mi{guards}(a^*))\cup (\{a^*\gle \gu\}\cap \mi{guards}(a^*))=\mi{guards}(S,a^*)\cup \{a^*\gle \gu\}=
                                        \{\gz\gle a^*,a^*\gle \gu\}$;
$a^*$ is positively guarded in $S\cup \{a^*\gle \gu\}$;
for all $a\in \mi{atoms}(S\cup \{a^*\gle \gu\})-\{a^*\}=\mi{atoms}(S)-\{a^*\}$,
$a$ is positively guarded in $S$;
$a\neq a^*$,
$\mi{guards}(a)\cap \{a^*\gle \gu\}=\mi{guards}(a)\cap \mi{guards}(a^*)=\emptyset$,
$\mi{guards}(S\cup \{a^*\gle \gu\},a)=(S\cup \{a^*\gle \gu\})\cap \mi{guards}(a)=(S\cap \mi{guards}(a))\cup (\{a^*\gle \gu\}\cap \mi{guards}(a))=\mi{guards}(S,a)$;
$a$ is positively guarded in $S\cup \{a^*\gle \gu\}$;
$S\cup \{a^*\gle \gu\}$ is positively guarded; 
$\mi{guards}(S\cup \{a^*\gle \gu\})=\mi{guards}(S)\cup \{a^*\gle \gu\}$.
We define a binary measure operator $\mi{undeleted} : \mi{PropAtom}\times {\mc P}(\mi{OrdPropCl}^\gu)\longrightarrow {\mc P}(\mi{OrdPropCl}^\gu)$,
$\mi{undeleted}(a^F,S^F)=\{C \,|\, C=a_0\swedge\cdots\swedge a_n\gle \gu\vee C^\natural\in S^F, a_i\in \mi{PropAtom}, C^\natural\in \mi{OrdPropCl}^\gu, a^F\in \{a_0,\dots,a_n\}\}$.
We prove the following statement:
\begin{alignat}{1}
\label{eq9a}
& \begin{minipage}[t]{\linewidth-15mm}
  Let $S_r\subseteq S-\mi{guards}(S)$.
  $S_r\cup \mi{guards}(S)\cup \{a^*\gle \gu\}\subseteq_{\mc F} \mi{OrdPropCl}$ is simplified, and
  there exists a finite linear {\it DPLL}-tree with the root $S_r\cup \mi{guards}(S)\cup \{a^*\gle \gu\}$ constructed using Rule (\cref{ceq4hr66}) such that
  for its only leaf $S'$, $S'\subseteq_{\mc F} \mi{OrdPropCl}$ is simplified, 
  $S'-\mi{guards}(S')\subseteq S_r$, $\mi{atoms}(S')=\mi{atoms}(S)$, $\mi{guards}(S')=\mi{guards}(S)\cup \{a^*\gle \gu\}$, $a^*\gle \gu\not\in S$, $\mi{undeleted}(a^*,S'-\mi{guards}(S'))=\emptyset$.
  \end{minipage}
\end{alignat}   
We have that $S\cup \{a^*\gle \gu\}$ is simplified.
Then $S_r\subseteq S-\mi{guards}(S)$;
$S_r\cup \mi{guards}(S)\cup \{a^*\gle \gu\}\subseteq (S-\mi{guards}(S))\cup \mi{guards}(S)\cup \{a^*\gle \gu\}=S\cup \{a^*\gle \gu\}\subseteq_{\mc F} \mi{OrdPropCl}$ is simplified;
we have that $S$ is positive;
$S_r\subseteq S-\mi{guards}(S)\subseteq_{\mc F} \mi{OrdPropCl}^\gu$.
We proceed by induction on $\mi{undeleted}(a^*,S_r)\subseteq S_r\subseteq_{\mc F} \mi{OrdPropCl}^\gu$.

Case 1 (the base case):
$\mi{undeleted}(a^*,S_r)=\emptyset$.
Then $S_r\subseteq S-\mi{guards}(S)$, $\mi{guards}(S_r)=\mi{guards}(S-\mi{guards}(S))=\emptyset$, $a^*\gle \gu\not\in \mi{guards}(S)$,
$\mi{guards}(S_r\cup \mi{guards}(S)\cup \{a^*\gle \gu\})=\mi{guards}(S_r)\cup \mi{guards}(\mi{guards}(S))\cup \mi{guards}(\{a^*\gle \gu\})=\mi{guards}(S)\cup \{a^*\gle \gu\}$,
$S_r\cap \mi{guards}(S)=(S-\mi{guards}(S))\cap \mi{guards}(S)=\emptyset$,
$a^*\gle \gu\not\in S\supseteq S-\mi{guards}(S)\supseteq S_r$,
$(S_r\cup \mi{guards}(S)\cup \{a^*\gle \gu\})-\mi{guards}(S_r\cup \mi{guards}(S)\cup \{a^*\gle \gu\})=
 (S_r\cup \mi{guards}(S)\cup \{a^*\gle \gu\})-(\mi{guards}(S)\cup \{a^*\gle \gu\})=
 (S_r-(\mi{guards}(S)\cup \{a^*\gle \gu\}))\cup ((\mi{guards}(S)\cup \{a^*\gle \gu\})-(\mi{guards}(S)\cup \{a^*\gle \gu\}))=
 (S_r-\mi{guards}(S))-\{a^*\gle \gu\}=S_r$;
$S$ is positively guarded;
$\mi{atoms}(S_r)\subseteq \mi{atoms}(S-\mi{guards}(S))\subseteq \mi{atoms}(S)$, $a^*\in \mi{atoms}(S)$,
$\mi{atoms}(S_r\cup \mi{guards}(S)\cup \{a^*\gle \gu\})=\mi{atoms}(S_r)\cup \mi{atoms}(\mi{guards}(S))\cup \mi{atoms}(a^*\gle \gu)=\mi{atoms}(S_r)\cup \mi{atoms}(S)\cup \{a^*\}=\mi{atoms}(S)$,
$\mi{undeleted}(a^*,(S_r\cup \mi{guards}(S)\cup \{a^*\gle \gu\})-\mi{guards}(S_r\cup \mi{guards}(S)\cup \{a^*\gle \gu\}))=\mi{undeleted}(a^*,S_r)=\emptyset$.
We put $\mi{Tree}=S_r\cup \mi{guards}(S)\cup \{a^*\gle \gu\}$.
Hence, $\mi{Tree}$ is a finite linear {\it DPLL}-tree with the root $S_r\cup \mi{guards}(S)\cup \{a^*\gle \gu\}$ constructed using Rule (\cref{ceq4hr66}) such that
for its only leaf $S_r\cup \mi{guards}(S)\cup \{a^*\gle \gu\}$, $S_r\cup \mi{guards}(S)\cup \{a^*\gle \gu\}\subseteq_{\mc F} \mi{OrdPropCl}$ is simplified,
$(S_r\cup \mi{guards}(S)\cup \{a^*\gle \gu\})-\mi{guards}(S_r\cup \mi{guards}(S)\cup \{a^*\gle \gu\})=S_r$,
$\mi{atoms}(S_r\cup \mi{guards}(S)\cup \{a^*\gle \gu\})=\mi{atoms}(S)$,
$\mi{guards}(S_r\cup \mi{guards}(S)\cup \{a^*\gle \gu\})=\mi{guards}(S)\cup \{a^*\gle \gu\}$, $a^*\gle \gu\not\in S$, 
$\mi{undeleted}(a^*,(S_r\cup \mi{guards}(S)\cup \{a^*\gle \gu\})-\mi{guards}(S_r\cup \mi{guards}(S)\cup \{a^*\gle \gu\}))=\emptyset$.

Case 2 (the induction case):
$\emptyset\neq \mi{undeleted}(a^*,S_r)\subseteq_{\mc F} \mi{OrdPropCl}^\gu$.
Then there exist $C^*=a_0\swedge\cdots\swedge a_n\gle \gu\vee C^\natural\in \mi{undeleted}(a^*,S_r)\subseteq S_r$, $a_i\in \mi{PropAtom}$, $C^\natural\in \mi{OrdPropCl}^\gu$, and $i^*\leq n$ satisfying 
$a^*\in \{a_0,\dots,a_n\}$, $a_{i^*}=a^*$;
$a_{i^*}\gle \gu=a^*\gle \gu\in \mi{guards}(S_r\cup \mi{guards}(S)\cup \{a^*\gle \gu\})=\mi{guards}(S)\cup \{a^*\gle \gu\}$,
$C^*\in (S_r\cup \mi{guards}(S)\cup \{a^*\gle \gu\})-\mi{guards}(S_r\cup \mi{guards}(S)\cup \{a^*\gle \gu\})=S_r$,
applying Rule (\cref{ceq4hr66}) to $S_r\cup \mi{guards}(S)\cup \{a^*\gle \gu\}$ and $C^*$, we derive
\begin{equation}
\notag
\dfrac{S_r\cup \mi{guards}(S)\cup \{a^*\gle \gu\}}
      {(S_r\cup \mi{guards}(S)\cup \{a^*\gle \gu\})-\{C^*\}}.
\end{equation}
We have that $S_r\cup \mi{guards}(S)\cup \{a^*\gle \gu\}$ is simplified.
We get that
$C^*\not\in \mi{guards}(S_r\cup \mi{guards}(S)\cup \{a^*\gle \gu\})=\mi{guards}(S)\cup \{a^*\gle \gu\}$;
$(S_r\cup \mi{guards}(S)\cup \{a^*\gle \gu\})-\{C^*\}=(S_r-\{C^*\})\cup ((\mi{guards}(S)\cup \{a^*\gle \gu\})-\{C^*\})=(S_r-\{C^*\})\cup \mi{guards}(S)\cup \{a^*\gle \gu\}\subseteq
 S_r\cup \mi{guards}(S)\cup \{a^*\gle \gu\}\subseteq_{\mc F} \mi{OrdPropCl}$ is simplified;
$S_r-\{C^*\}\subseteq S_r\subseteq S-\mi{guards}(S)$,
$C^*\in \mi{undeleted}(a^*,S_r)$,
$\mi{undeleted}(a^*,S_r-\{C^*\})=\{C \,|\, C=b_0\swedge\cdots\swedge b_k\gle \gu\vee C^\flat\in S_r-\{C^*\}, b_j\in \mi{PropAtom}, C^\flat\in \mi{OrdPropCl}^\gu, a^*\in \{b_0,\dots,b_k\}\}=
                                 \mi{undeleted}(a^*,S_r)-\{C^*\}=
                                 \{C \,|\, C=b_0\swedge\cdots\swedge b_k\gle \gu\vee C^\flat\in S_r, b_j\in \mi{PropAtom}, C^\flat\in \mi{OrdPropCl}^\gu, a^*\in \{b_0,\dots,b_k\}\}-\{C^*\}\subset
                                 \mi{undeleted}(a^*,S_r)$;
by the induction hypothesis for $S_r-\{C^*\}$, there exists a finite linear {\it DPLL}-tree $\mi{Tree}'$ with the root $(S_r-\{C^*\})\cup \mi{guards}(S)\cup \{a^*\gle \gu\}$ constructed using Rule (\cref{ceq4hr66}) satisfying
for its only leaf $S'$ that $S'\subseteq_{\mc F} \mi{OrdPropCl}$ is simplified, 
$S'-\mi{guards}(S')\subseteq S_r-\{C^*\}\subseteq S_r$, $\mi{atoms}(S')=\mi{atoms}(S)$, $\mi{guards}(S')=\mi{guards}(S)\cup \{a^*\gle \gu\}$, $a^*\gle \gu\not\in S$,
$\mi{undeleted}(a^*,S'-\mi{guards}(S'))=\emptyset$.
We put
\begin{equation}
\notag
\mi{Tree}=\dfrac{S_r\cup \mi{guards}(S)\cup \{a^*\gle \gu\}}
                {\mi{Tree}'}.
\end{equation}
Hence, $\mi{Tree}$ is a finite linear {\it DPLL}-tree with the root $S_r\cup \mi{guards}(S)\cup \{a^*\gle \gu\}$ constructed using Rule (\cref{ceq4hr66}) such that
for its only leaf $S'$, $S'\subseteq_{\mc F} \mi{OrdPropCl}$ is simplified, 
$S'-\mi{guards}(S')\subseteq S_r$, $\mi{atoms}(S')=\mi{atoms}(S)$, $\mi{guards}(S')=\mi{guards}(S)\cup \{a^*\gle \gu\}$, $a^*\gle \gu\not\in S$, $\mi{undeleted}(a^*,S'-\mi{guards}(S'))=\emptyset$.

So, in both Cases 1 and 2, (\ref{eq9a}) holds.
The induction is completed.
Thus, (\ref{eq9a}) holds.

We get by (\ref{eq9a}) for $S-\mi{guards}(S)$ that 
there exists a finite linear {\it DPLL}-tree $\mi{Tree}$ with the root $S\cup \{a^*\gle \gu\}=(S-\mi{guards}(S))\cup \mi{guards}(S)\cup \{a^*\gle \gu\}$ constructed using Rule (\cref{ceq4hr66}) satisfying
for its only leaf $S'$ that $S'\subseteq_{\mc F} \mi{OrdPropCl}$ is simplified, 
$S'-\mi{guards}(S')\subseteq S-\mi{guards}(S)$, $\mi{atoms}(S')=\mi{atoms}(S\cup \{a^*\gle \gu\})=\mi{atoms}(S)$, 
$\mi{guards}(S')=\mi{guards}(S\cup \{a^*\gle \gu\})=\mi{guards}(S)\cup \{a^*\gle \gu\}$, $a^*\gle \gu\not\in S$, $\mi{undeleted}(a^*,S'-\mi{guards}(S'))=\emptyset$;
we have that $S\cup \{a^*\gle \gu\}$ is positively guarded;
$S'$ is positively guarded;
we have that $S$ is positive;
for all $C\in S'-\mi{guards}(S')\subseteq S-\mi{guards}(S)$, 
either $C\in \mi{PurOrdPropCl}$, 
or $C=a_0\swedge\cdots\swedge a_n\gle \gu\vee C^\natural$, $a_i\in \mi{PropAtom}$, $\mi{guards}(S,a_i)=\{\gz\gle a_i\}$, $C^\natural\in \mi{PurOrdPropCl}$,
$C\not\in \mi{undeleted}(a^*,S'-\mi{guards}(S'))=\emptyset$, $a^*\not\in \{a_0,\dots,a_n\}$;
for all $i\leq n$, 
$a_i\in \mi{atoms}(C)\subseteq \mi{atoms}(S')$, $a_i\neq a^*$, 
$a_i\in \mi{atoms}(S')-\{a^*\}=\mi{atoms}(S\cup \{a^*\gle \gu\})-\{a^*\}=\mi{atoms}(S)-\{a^*\}$,
$\mi{guards}(S',a_i)=\mi{guards}(S\cup \{a^*\gle \gu\},a_i)=\mi{guards}(S,a_i)=\{\gz\gle a_i\}$;
$S'$ is positive.
%
%
%
\end{proof}

\subsection{Full proof of Lemma \ref{le4}}
\label{S7.6b}

\begin{proof}
We exploit the excess literal technique \cite{ANBL70} to construct a finite {\it DPLL}-tree $\mi{Tree}$ with the root $S$ using Rules (\cref{ceq4hr0}), (\cref{ceq4hr11}), (\cref{ceq4hr111}), (\cref{ceq4hr66}).
Let $l^F\in \mi{OrdPropLit}$, $C^F\in \mi{OrdPropCl}$, $\square\not\in S^F\subseteq_{\mc F} \mi{OrdPropCl}$.
We define three measure operators as follows:
\begin{alignat*}{1}
\mi{count}(l^F)     &= \left\{\begin{array}{ll}
                              2 &\ \text{\it if}\ l^F=\varepsilon_1\geql \varepsilon_2, \varepsilon_i\in \{\gz,\gu\}\cup \mi{PropConj}, \\[1mm]
                              1 &\ \text{\it if}\ l^F=\varepsilon_1\diamond \varepsilon_2, \varepsilon_i\in \{\gz,\gu\}\cup \mi{PropConj}, \diamond\in \{\gleq,\gle\};
                              \end{array}
                       \right. \\[1mm]
\mi{count}(C^F)     &= \sum_{l\in C^F} \mi{count}(l); \\
\mi{elmeasure}(S^F) &= \sum_{C\in S^F} \mi{count}(C)-1.
\end{alignat*}
Note that if $C^F\neq \square$, then $\mi{count}(C^F)\geq 1$.
We have that $S$ is positive.
Then $S$ is simplified and $\square\not\in S$.
We proceed by induction on $\mi{elmeasure}(S)$.

Case 1 (the base case):
$\mi{elmeasure}(S)=0$.
Then $\square\not\in S$,
$\mi{elmeasure}(S)=\sum_{C\in S} \mi{count}(C)-1=0$;
for all $C\in S$, $C\neq \square$, $\mi{count}(C)=1$, $\mi{card}(C)=1$; 
$S$ is unit.
We get two cases for $S$.

Case 1.1: 
There exists an application of Rule (\cref{ceq4hr0}) to $S$.
Then, applying Rule (\cref{ceq4hr0}), we derive 
$\dfrac{S}
       {S\cup \{\square\}}$.
We put $\mi{Tree}=\dfrac{S}
                        {S\cup \{\square\}}$.
We get that
$\mi{Tree}$ is a closed {\it DPLL}-tree with the root $S$ and its only leaf $\square\in S\cup \{\square\}$ constructed using Rule (\cref{ceq4hr0}); 
by Lemma \ref{le3}, $S$ is unsatisfiable.
Hence, $\mi{Tree}$ is a closed {\it DPLL}-tree with the root $S$ constructed using Rules (\cref{ceq4hr0}), (\cref{ceq4hr11}), (\cref{ceq4hr111}), (\cref{ceq4hr66});
(\ref{eq5a}) holds and (\ref{eq5b}) holds trivially.

Case 1.2:
There does not exist an application of Rule (\cref{ceq4hr0}) to $S$.
Then, by Lemma \ref{le2}, $S$ is satisfiable, and we can construct a model ${\mf A}$ of $S$.
We put $\mi{Tree}=S$. 
We get that
$\mi{Tree}$ is a finite open {\it DPLL}-tree with the root $S$ and its only leaf $\square\not\in S$;
${\mf A}$ is a model of $S$ related to $\mi{Tree}$.
Hence, $\mi{Tree}$ is a finite open {\it DPLL}-tree with the root $S$ constructed using Rules (\cref{ceq4hr0}), (\cref{ceq4hr11}), (\cref{ceq4hr111}), (\cref{ceq4hr66});
(\ref{eq5a}) holds trivially and (\ref{eq5b}) holds.

Case 2 (the induction case):
$\mi{elmeasure}(S)\geq 1$.
We have that $S$ is positive.
We get two cases for $S$.

Case 2.1:
$S$ is unit.
This case is the same as Case 1.

Case 2.2:
$S$ is not unit.
Then there exists $C^*\in S$ satisfying that $C^*$ is not unit;
$C^*$ is not a guard;
$C^*\not\in \mi{guards}(S)$,
$C^*\in S-\mi{guards}(S)$;
$S$ is simplified;
$\square\not\in S$, $C^*\neq \square$;
we have that $S$ is positive;
either $C^*\in \mi{PurOrdPropCl}$, 
or $C^*=a_0\swedge\cdots\swedge a_n\gle \gu\vee C^\natural$, $a_i\in \mi{PropAtom}$, $\mi{guards}(S,a_i)=\{\gz\gle a_i\}$, $C^\natural\in \mi{PurOrdPropCl}$.
We get two cases for $C^*$.

Case 2.2.1:
$C^*\in \mi{PurOrdPropCl}$.
We have that $C^*$ is not unit.
Then $C^*\neq \square$,
$C^*=\Cn_1\diamond^* \Cn_2\vee C^\natural$, $\Cn_i\in \mi{PropConj}$, $\diamond^*\in \{\geql,\gleq,\gle\}$, $\square\neq C^\natural\in \mi{PurOrdPropCl}$;
$S$ is simplified;
$\Cn_1\diamond^* \Cn_2\in C^*\in S$ is not a contradiction or tautology;
$\Cn_1\neq \Cn_2$.
We get two cases for $\diamond^*$.

Case 2.2.1.1:
$\diamond^*=\geql$.
Then $C^*=\Cn_1\diamond^* \Cn_2\vee C^\natural=\Cn_1\geql \Cn_2\vee C^\natural\in S-\mi{guards}(S)$, $C^\natural\neq \square$, $\Cn_1\neq \Cn_2$,
applying Rule (\cref{ceq4hr111}) to $C^*$, $\Cn_1\geql \Cn_2$, and $C^\natural$, we derive
\begin{equation}
\notag
\dfrac{S}
      {(S-\{C^*\})\cup \{\Cn_1\geql \Cn_2\}\ \big|\ (S-\{C^*\})\cup \{C^\natural\}\cup \{\Cn_1\gle \Cn_2\vee \Cn_2\gle \Cn_1\}}.
\end{equation}
Hence, $C^*\in S$, $\Cn_1\neq \Cn_2$, $\Cn_1\gle \Cn_2\neq \Cn_2\gle \Cn_1$.
We put 
$S_1=(S-\{C^*\})\cup \{\Cn_1\geql \Cn_2\}\subseteq_{\mc F} \mi{OrdPropCl}$ and
$S_2=(S-\{C^*\})\cup \{C^\natural\}\cup \{\Cn_1\gle \Cn_2\vee \Cn_2\gle \Cn_1\}\subseteq_{\mc F} \mi{OrdPropCl}$.
We get that
$S$ is simplified; 
$S-\{C^*\}\subseteq S$ is simplified;
$\Cn_1\neq \Cn_2$;
$\Cn_1\geql \Cn_2$, $\Cn_1\gle \Cn_2$, $\Cn_2\gle \Cn_1$ are not contradictions or tautologies;
$\Cn_1\geql \Cn_2, \Cn_1\gle \Cn_2\vee \Cn_2\gle \Cn_1\neq \square$ do not contain contradictions and tautologies;
$C^*\in S$ does not contain contradictions and tautologies;
$\square\neq C^\natural\subseteq C^*$ does not contain contradictions and tautologies;
$S_1=(S-\{C^*\})\cup \{\Cn_1\geql \Cn_2\}$ and
$S_2=(S-\{C^*\})\cup \{C^\natural\}\cup \{\Cn_1\gle \Cn_2\vee \Cn_2\gle \Cn_1\}$ are simplified;
$\Cn_1\geql \Cn_2$, $C^\natural$, $\Cn_1\gle \Cn_2\vee \Cn_2\gle \Cn_1$ are not guards;
$C^*\not\in \mi{guards}(S)$,
$\mi{guards}(S_1)=\{C \,|\, C\in (S-\{C^*\})\cup \{\Cn_1\geql \Cn_2\}\ \text{\it is a guard}\}=\{C \,|\, C\in S-\{C^*\}\ \text{\it is a guard}\}=\{C \,|\, C\in S\ \text{\it is a guard}\}-\{C^*\}=
                  \mi{guards}(S)-\{C^*\}=\mi{guards}(S)$,
$\mi{guards}(S_2)=\{C \,|\, C\in (S-\{C^*\})\cup \{C^\natural\}\cup \{\Cn_1\gle \Cn_2\vee \Cn_2\gle \Cn_1\}\ \text{\it is a guard}\}=\{C \,|\, C\in S-\{C^*\}\ \text{\it is a guard}\}=\mi{guards}(S)$;
$\mi{atoms}(\Cn_1\geql \Cn_2), \mi{atoms}(C^\natural)\subseteq \mi{atoms}(C^*)\subseteq \mi{atoms}(S)$, 
$\mi{atoms}(\Cn_1\gle \Cn_2\vee \Cn_2\gle \Cn_1)=\mi{atoms}(\Cn_1\geql \Cn_2)\subseteq \mi{atoms}(S)$,
$\mi{atoms}(S_1)=\mi{atoms}((S-\{C^*\})\cup \{\Cn_1\geql \Cn_2\})=\mi{atoms}(S-\{C^*\})\cup \mi{atoms}(\Cn_1\geql \Cn_2)\subseteq \mi{atoms}(S)$,
$\mi{atoms}(S_2)=\mi{atoms}((S-\{C^*\})\cup \{C^\natural\}\cup \{\Cn_1\gle \Cn_2\vee \Cn_2\gle \Cn_1\})=
                 \mi{atoms}(S-\{C^*\})\cup \mi{atoms}(C^\natural)\cup \mi{atoms}(\Cn_1\gle \Cn_2\vee \Cn_2\gle \Cn_1)\subseteq \mi{atoms}(S)$;
$S$ is positively guarded;
for both $i$,
$\mi{atoms}(S_i-\mi{guards}(S_i))\subseteq \mi{atoms}(S_i)\subseteq \mi{atoms}(S)$,
$\mi{atoms}(S_i)=\mi{atoms}((S_i-\mi{guards}(S_i))\cup \mi{guards}(S_i))=\mi{atoms}(S_i-\mi{guards}(S_i))\cup \mi{atoms}(\mi{guards}(S_i))=
                 \mi{atoms}(S_i-\mi{guards}(S_i))\cup \mi{atoms}(\mi{guards}(S))=\mi{atoms}(S_i-\mi{guards}(S_i))\cup \mi{atoms}(S)=\mi{atoms}(S)$;
$S_1$ and $S_2$ are positively guarded;
we have that $S$ is positive;
$\Cn_1\geql \Cn_2\not\in \mi{guards}(S)$;
for all $C\in S_1-\mi{guards}(S_1)=((S-\{C^*\})\cup \{\Cn_1\geql \Cn_2\})-\mi{guards}(S)=((S-\{C^*\})-\mi{guards}(S))\cup (\{\Cn_1\geql \Cn_2\}-\mi{guards}(S))=
              ((S-\{C^*\})-\mi{guards}(S))\cup \{\Cn_1\geql \Cn_2\}\subseteq (S-\mi{guards}(S))\cup \{\Cn_1\geql \Cn_2\}$, 
either $C\in \mi{PurOrdPropCl}$ or $C=\Cn_1\geql \Cn_2\in \mi{PurOrdPropCl}$, 
or $C=b_0\swedge\cdots\swedge b_k\gle \gu\vee C^\flat$, $b_j\in \mi{PropAtom}$, $\mi{guards}(S,b_j)=\{\gz\gle b_j\}$, $C^\flat\in \mi{PurOrdPropCl}$;
for all $j\leq k$, 
$b_j\in \mi{atoms}(C)\subseteq \mi{atoms}(S_1)=\mi{atoms}(S)$,
$\mi{guards}(S_1,b_j)=\mi{guards}(S,b_j)=\{\gz\gle b_j\}$;
$C^\natural, \Cn_1\gle \Cn_2\vee \Cn_2\gle \Cn_1\not\in \mi{guards}(S)$;
for all $C\in S_2-\mi{guards}(S_2)=((S-\{C^*\})\cup \{C^\natural\}\cup \{\Cn_1\gle \Cn_2\vee \Cn_2\gle \Cn_1\})-\mi{guards}(S)=
                                   ((S-\{C^*\})-\mi{guards}(S))\cup (\{C^\natural\}-\mi{guards}(S))\cup (\{\Cn_1\gle \Cn_2\vee \Cn_2\gle \Cn_1\}-\mi{guards}(S))=
                                   ((S-\{C^*\})-\mi{guards}(S))\cup \{C^\natural\}\cup \{\Cn_1\gle \Cn_2\vee \Cn_2\gle \Cn_1\}\subseteq 
                                   (S-\mi{guards}(S))\cup \{C^\natural\}\cup \{\Cn_1\gle \Cn_2\vee \Cn_2\gle \Cn_1\}$, 
either $C\in \mi{PurOrdPropCl}$ or $C=C^\natural\in \mi{PurOrdPropCl}$ or $C=\Cn_1\gle \Cn_2\vee \Cn_2\gle \Cn_1\in \mi{PurOrdPropCl}$, 
or $C=b_0\swedge\cdots\swedge b_k\gle \gu\vee C^\flat$, $b_j\in \mi{PropAtom}$, $\mi{guards}(S,b_j)=\{\gz\gle b_j\}$, $C^\flat\in \mi{PurOrdPropCl}$;
for all $j\leq k$, 
$b_j\in \mi{atoms}(C)\subseteq \mi{atoms}(S_2)=\mi{atoms}(S)$,
$\mi{guards}(S_2,b_j)=\mi{guards}(S,b_j)=\{\gz\gle b_j\}$;
$S_1$ and $S_2$ are positive;
$\square\not\in S_1$,
$\mi{count}(C^\natural)\geq 1$,
$\mi{count}(C^*)=\mi{count}(\Cn_1\geql \Cn_2\vee C^\natural)=\mi{count}(\Cn_1\geql \Cn_2)+\mi{count}(C^\natural)=\mi{count}(C^\natural)+2$, 
$\mi{elmeasure}(S_1)=\sum_{C\in (S-\{C^*\})\cup \{\Cn_1\geql \Cn_2\}} \mi{count}(C)-1\leq (\sum_{C\in S-\{C^*\}} \mi{count}(C)-1)+\mi{count}(\Cn_1\geql \Cn_2)-1=
                     (\sum_{C\in S-\{C^*\}} \mi{count}(C)-1)+1<(\sum_{C\in S-\{C^*\}} \mi{count}(C)-1)+\mi{count}(C^\natural)+1=
                     (\sum_{C\in S-\{C^*\}} \mi{count}(C)-1)+\mi{count}(C^*)-1=\sum_{C\in (S-\{C^*\})\cup \{C^*\}} \mi{count}(C)-1=\mi{elmeasure}(S)=\sum_{C\in S} \mi{count}(C)-1$,
$\square\not\in S_2$,
$\mi{elmeasure}(S_2)=\sum_{C\in (S-\{C^*\})\cup \{C^\natural\}\cup \{\Cn_1\gle \Cn_2\vee \Cn_2\gle \Cn_1\}} \mi{count}(C)-1\leq 
                     (\sum_{C\in S-\{C^*\}} \mi{count}(C)-1)+\mi{count}(C^\natural)-1+\mi{count}(\Cn_1\gle \Cn_2\vee \Cn_2\gle \Cn_1)-1=
                     (\sum_{C\in S-\{C^*\}} \mi{count}(C)-1)+\mi{count}(C^\natural)<(\sum_{C\in S-\{C^*\}} \mi{count}(C)-1)+\mi{count}(C^\natural)+1=\mi{elmeasure}(S)$.

Case 2.2.1.2:
$\diamond^*\in \{\gleq,\gle\}$.
We put
\begin{equation}
\notag
\diamond^{**}=\left\{\begin{array}{ll}
                     \gle  &\ \text{\it if}\ \diamond^*=\gleq, \\[1mm]
                     \gleq &\ \text{\it if}\ \diamond^*=\gle.
                     \end{array}
              \right. 
\end{equation}
Then $C^*=\Cn_1\diamond^* \Cn_2\vee C^\natural\in S-\mi{guards}(S)$, 
$\Cn_1\diamond^* \Cn_2, \Cn_2\diamond^{**} \Cn_1\in \mi{PurOrdPropLit}$, $C^\natural\neq \square$, $\Cn_1\neq \Cn_2$;
$\Cn_1\diamond^* \Cn_2\vee \Cn_2\diamond^{**} \Cn_1$ is a pure dichotomy;
applying Rule (\cref{ceq4hr11}) to $C^*$, $\Cn_1\diamond^* \Cn_2$, and $C^\natural$, we derive
\begin{equation}
\notag
\dfrac{S}
      {(S-\{C^*\})\cup \{\Cn_1\diamond^* \Cn_2\}\ \big|\ (S-\{C^*\})\cup \{C^\natural\}\cup \{\Cn_2\diamond^{**} \Cn_1\}}.
\end{equation}
Hence, $C^*\in S$.
We put 
$S_1=(S-\{C^*\})\cup \{\Cn_1\diamond^* \Cn_2\}\subseteq_{\mc F} \mi{OrdPropCl}$ and
$S_2=(S-\{C^*\})\cup \{C^\natural\}\cup \{\Cn_2\diamond^{**} \Cn_1\}\subseteq_{\mc F} \mi{OrdPropCl}$.
We get that
$S$ is simplified; 
$S-\{C^*\}\subseteq S$ is simplified;
$\Cn_1\neq \Cn_2$;
$\Cn_1\diamond^* \Cn_2$ and $\Cn_2\diamond^{**} \Cn_1$ are not contradictions or tautologies;
$\Cn_1\diamond^* \Cn_2, \Cn_2\diamond^{**} \Cn_1\neq \square$ do not contain contradictions and tautologies;
$C^*\in S$ does not contain contradictions and tautologies;
$\square\neq C^\natural\subseteq C^*$ does not contain contradictions and tautologies;
$S_1=(S-\{C^*\})\cup \{\Cn_1\diamond^* \Cn_2\}$ and
$S_2=(S-\{C^*\})\cup \{C^\natural\}\cup \{\Cn_2\diamond^{**} \Cn_1\}$ are simplified;
$\Cn_1\diamond^* \Cn_2$, $C^\natural$, $\Cn_2\diamond^{**} \Cn_1$ are not guards;
$C^*\not\in \mi{guards}(S)$,
$\mi{guards}(S_1)=\{C \,|\, C\in (S-\{C^*\})\cup \{\Cn_1\diamond^* \Cn_2\}\ \text{\it is a guard}\}=\{C \,|\, C\in S-\{C^*\}\ \text{\it is a guard}\}=\{C \,|\, C\in S\ \text{\it is a guard}\}-\{C^*\}=
                  \mi{guards}(S)-\{C^*\}=\mi{guards}(S)$,
$\mi{guards}(S_2)=\{C \,|\, C\in (S-\{C^*\})\cup \{C^\natural\}\cup \{\Cn_2\diamond^{**} \Cn_1\}\ \text{\it is a guard}\}=\{C \,|\, C\in S-\{C^*\}\ \text{\it is a guard}\}=\mi{guards}(S)$;
$\mi{atoms}(\Cn_1\diamond^* \Cn_2), \mi{atoms}(C^\natural)\subseteq \mi{atoms}(C^*)\subseteq \mi{atoms}(S)$, 
$\mi{atoms}(\Cn_2\diamond^{**} \Cn_1)=\mi{atoms}(\Cn_1\diamond^* \Cn_2)\subseteq \mi{atoms}(S)$,
$\mi{atoms}(S_1)=\mi{atoms}((S-\{C^*\})\cup \{\Cn_1\diamond^* \Cn_2\})=\mi{atoms}(S-\{C^*\})\cup \mi{atoms}(\Cn_1\diamond^* \Cn_2)\subseteq \mi{atoms}(S)$,
$\mi{atoms}(S_2)=\mi{atoms}((S-\{C^*\})\cup \{C^\natural\}\cup \{\Cn_2\diamond^{**} \Cn_1\})=\mi{atoms}(S-\{C^*\})\cup \mi{atoms}(C^\natural)\cup \mi{atoms}(\Cn_2\diamond^{**} \Cn_1)\subseteq 
                 \mi{atoms}(S)$;
$S$ is positively guarded;
for both $i$,
$\mi{atoms}(S_i-\mi{guards}(S_i))\subseteq \mi{atoms}(S_i)\subseteq \mi{atoms}(S)$,
$\mi{atoms}(S_i)=\mi{atoms}((S_i-\mi{guards}(S_i))\cup \mi{guards}(S_i))=\mi{atoms}(S_i-\mi{guards}(S_i))\cup \mi{atoms}(\mi{guards}(S_i))=
                 \mi{atoms}(S_i-\mi{guards}(S_i))\cup \mi{atoms}(\mi{guards}(S))=\mi{atoms}(S_i-\mi{guards}(S_i))\cup \mi{atoms}(S)=\mi{atoms}(S)$;
$S_1$ and $S_2$ are positively guarded;
we have that $S$ is positive;
$\Cn_1\diamond^* \Cn_2\not\in \mi{guards}(S)$;
for all $C\in S_1-\mi{guards}(S_1)=((S-\{C^*\})\cup \{\Cn_1\diamond^* \Cn_2\})-\mi{guards}(S)=((S-\{C^*\})-\mi{guards}(S))\cup (\{\Cn_1\diamond^* \Cn_2\}-\mi{guards}(S))=
              ((S-\{C^*\})-\mi{guards}(S))\cup \{\Cn_1\diamond^* \Cn_2\}\subseteq (S-\mi{guards}(S))\cup \{\Cn_1\diamond^* \Cn_2\}$, 
either $C\in \mi{PurOrdPropCl}$ or $C=\Cn_1\diamond^* \Cn_2\in \mi{PurOrdPropCl}$, 
or $C=b_0\swedge\cdots\swedge b_k\gle \gu\vee C^\flat$, $b_j\in \mi{PropAtom}$, $\mi{guards}(S,b_j)=\{\gz\gle b_j\}$, $C^\flat\in \mi{PurOrdPropCl}$;
for all $j\leq k$, 
$b_j\in \mi{atoms}(C)\subseteq \mi{atoms}(S_1)=\mi{atoms}(S)$,
$\mi{guards}(S_1,b_j)=\mi{guards}(S,b_j)=\{\gz\gle b_j\}$;
$C^\natural, \Cn_2\diamond^{**} \Cn_1\not\in \mi{guards}(S)$;
for all $C\in S_2-\mi{guards}(S_2)=((S-\{C^*\})\cup \{C^\natural\}\cup \{\Cn_2\diamond^{**} \Cn_1\})-\mi{guards}(S)=
                                   ((S-\{C^*\})-\mi{guards}(S))\cup (\{C^\natural\}-\mi{guards}(S))\cup (\{\Cn_2\diamond^{**} \Cn_1\}-\mi{guards}(S))=
                                   ((S-\{C^*\})-\mi{guards}(S))\cup \{C^\natural\}\cup \{\Cn_2\diamond^{**} \Cn_1\}\subseteq 
                                   (S-\mi{guards}(S))\cup \{C^\natural\}\cup \{\Cn_2\diamond^{**} \Cn_1\}$, 
either $C\in \mi{PurOrdPropCl}$ or $C=C^\natural\in \mi{PurOrdPropCl}$ or $C=\Cn_2\diamond^{**} \Cn_1\in \mi{PurOrdPropCl}$, 
or $C=b_0\swedge\cdots\swedge b_k\gle \gu\vee C^\flat$, $b_j\in \mi{PropAtom}$, $\mi{guards}(S,b_j)=\{\gz\gle b_j\}$, $C^\flat\in \mi{PurOrdPropCl}$;
for all $j\leq k$, 
$b_j\in \mi{atoms}(C)\subseteq \mi{atoms}(S_2)=\mi{atoms}(S)$,
$\mi{guards}(S_2,b_j)=\mi{guards}(S,b_j)=\{\gz\gle b_j\}$;
$S_1$ and $S_2$ are positive;
$\square\not\in S_1$,
$\diamond^*, \diamond^{**}\in \{\gleq,\gle\}$,
$\mi{count}(\Cn_1\diamond^* \Cn_2)=1$, $\mi{count}(C^\natural)\geq 1$, 
$\mi{count}(C^*)=\mi{count}(\Cn_1\diamond^* \Cn_2\vee C^\natural)=\mi{count}(\Cn_1\diamond^* \Cn_2)+\mi{count}(C^\natural)=\mi{count}(C^\natural)+1$, 
$\mi{elmeasure}(S_1)=\sum_{C\in (S-\{C^*\})\cup \{\Cn_1\diamond^* \Cn_2\}} \mi{count}(C)-1=(\sum_{C\in S-\{C^*\}} \mi{count}(C)-1)+\mi{count}(\Cn_1\diamond^* \Cn_2)-1=
                     \sum_{C\in S-\{C^*\}} \mi{count}(C)-1<(\sum_{C\in S-\{C^*\}} \mi{count}(C)-1)+\mi{count}(C^\natural)=(\sum_{C\in S-\{C^*\}} \mi{count}(C)-1)+\mi{count}(C^*)-1=
                     \sum_{C\in (S-\{C^*\})\cup \{C^*\}} \mi{count}(C)-1=\mi{elmeasure}(S)=\sum_{C\in S} \mi{count}(C)-1$,
$\square\not\in S_2$,
$\mi{count}(\Cn_2\diamond^{**} \Cn_1)=1$, 
$\mi{elmeasure}(S_2)=\sum_{C\in (S-\{C^*\})\cup \{C^\natural\}\cup \{\Cn_2\diamond^{**} \Cn_1\}} \mi{count}(C)-1\leq 
                     (\sum_{C\in S-\{C^*\}} \mi{count}(C)-1)+\mi{count}(C^\natural)-1+\mi{count}(\Cn_2\diamond^{**} \Cn_1)-1=
                     (\sum_{C\in S-\{C^*\}} \mi{count}(C)-1)+\mi{count}(C^\natural)-1<(\sum_{C\in S-\{C^*\}} \mi{count}(C)-1)+\mi{count}(C^\natural)=\mi{elmeasure}(S)$.

Case 2.2.2:
$C^*=a_0\swedge\cdots\swedge a_n\gle \gu\vee C^\natural$, $a_i\in \mi{PropAtom}$, $\mi{guards}(S,a_i)=\{\gz\gle a_i\}$, $C^\natural\in \mi{PurOrdPropCl}$.
We have that $C^*$ is not unit.
Then $C^\natural\neq \square$;
there exists $\Cn_1\diamond^* \Cn_2\in C^\natural$, $\Cn_i\in \mi{PropConj}$, $\diamond^*\in \{\geql,\gleq,\gle\}$;
$a_0\swedge\cdots\swedge a_n\gle \gu\not\in C^\natural\supseteq C^\natural-\{\Cn_1\diamond^* \Cn_2\}$,
$C^*=a_0\swedge\cdots\swedge a_n\gle \gu\vee C^\natural=a_0\swedge\cdots\swedge a_n\gle \gu\vee ((C^\natural-\{\Cn_1\diamond^* \Cn_2\})\cup \{\Cn_1\diamond^* \Cn_2\})=
     \Cn_1\diamond^* \Cn_2\vee a_0\swedge\cdots\swedge a_n\gle \gu\vee (C^\natural-\{\Cn_1\diamond^* \Cn_2\})$;
$S$ is simplified;
$\Cn_1\diamond^* \Cn_2\in C^*\in S$ is not a contradiction or tautology;
$\Cn_1\neq \Cn_2$.
We get two cases for $n$ and $C^\natural-\{\Cn_1\diamond^* \Cn_2\}$.

Case 2.2.2.1:
$n=0$ and $C^\natural-\{\Cn_1\diamond^* \Cn_2\}=\square$.
Then $C^*=\Cn_1\diamond^* \Cn_2\vee a_0\swedge\cdots\swedge a_n\gle \gu\vee (C^\natural-\{\Cn_1\diamond^* \Cn_2\})=\Cn_1\diamond^* \Cn_2\vee a_0\gle \gu$.
We get two cases for $\diamond^*$.

Case 2.2.2.1.1:
$\diamond^*=\geql$.
Then $C^*=\Cn_1\diamond^* \Cn_2\vee a_0\gle \gu=\Cn_1\geql \Cn_2\vee a_0\gle \gu\in S-\mi{guards}(S)$, $\square\neq a_0\gle \gu\in \mi{OrdPropCl}^\gu$, $\Cn_1\neq \Cn_2$,
applying Rule (\cref{ceq4hr111}) to $C^*$, $\Cn_1\geql \Cn_2$, and $a_0\gle \gu$, we derive
\begin{equation}
\notag
\dfrac{S}
      {(S-\{C^*\})\cup \{\Cn_1\geql \Cn_2\}\ \big|\ (S-\{C^*\})\cup \{a_0\gle \gu\}\cup \{\Cn_1\gle \Cn_2\vee \Cn_2\gle \Cn_1\}}.
\end{equation}
Hence, $C^*\in S$, $\Cn_1\neq \Cn_2$, $\Cn_1\gle \Cn_2\neq \Cn_2\gle \Cn_1$.
We put 
$S_1=(S-\{C^*\})\cup \{\Cn_1\geql \Cn_2\}\subseteq_{\mc F} \mi{OrdPropCl}$ and
$S_2'=(S-\{C^*\})\cup \{\Cn_1\gle \Cn_2\vee \Cn_2\gle \Cn_1\}\subseteq_{\mc F} \mi{OrdPropCl}$.
We get that
$a_0\gle \gu\in \mi{guards}(a_0)$, $a_0\gle \gu\not\in \mi{guards}(S,a_0)=\{\gz\gle a_0\}=S\cap \mi{guards}(a_0)$, $a_0\gle \gu\not\in S$,
$a_0\gle \gu\neq \Cn_1\gle \Cn_2\vee \Cn_2\gle \Cn_1$,
$a_0\gle \gu\not\in S_2'=(S-\{C^*\})\cup \{\Cn_1\gle \Cn_2\vee \Cn_2\gle \Cn_1\}$,
$S_2'\cup \{a_0\gle \gu\}=(S-\{C^*\})\cup \{a_0\gle \gu\}\cup \{\Cn_1\gle \Cn_2\vee \Cn_2\gle \Cn_1\}$;
$S$ is simplified; 
$S-\{C^*\}\subseteq S$ is simplified;
$\Cn_1\neq \Cn_2$;
$\Cn_1\geql \Cn_2$, $\Cn_1\gle \Cn_2$, $\Cn_2\gle \Cn_1$ are not contradictions or tautologies;
$\Cn_1\geql \Cn_2, Cn_1\gle \Cn_2\vee \Cn_2\gle \Cn_1\neq \square$ do not contain contradictions and tautologies;
$S_1=(S-\{C^*\})\cup \{\Cn_1\geql \Cn_2\}$ and $S_2'=(S-\{C^*\})\cup \{\Cn_1\gle \Cn_2\vee \Cn_2\gle \Cn_1\}$ are simplified;
$\Cn_1\geql \Cn_2$ and $\Cn_1\gle \Cn_2\vee \Cn_2\gle \Cn_1$ are not guards;
$C^*\not\in \mi{guards}(S)$,
$\mi{guards}(S_1)=\{C \,|\, C\in (S-\{C^*\})\cup \{\Cn_1\geql \Cn_2\}\ \text{\it is a guard}\}=\{C \,|\, C\in S-\{C^*\}\ \text{\it is a guard}\}=\{C \,|\, C\in S\ \text{\it is a guard}\}-\{C^*\}=
                  \mi{guards}(S)-\{C^*\}=\mi{guards}(S)$,
$\mi{guards}(S_2')=\{C \,|\, C\in (S-\{C^*\})\cup \{\Cn_1\gle \Cn_2\vee \Cn_2\gle \Cn_1\}\ \text{\it is a guard}\}=\{C \,|\, C\in S-\{C^*\}\ \text{\it is a guard}\}=\mi{guards}(S)$;
$C^*\in S$, $\mi{atoms}(\Cn_1\gle \Cn_2\vee \Cn_2\gle \Cn_1)=\mi{atoms}(\Cn_1\geql \Cn_2)\subseteq \mi{atoms}(C^*)\subseteq \mi{atoms}(S)$,
$\mi{atoms}(S_1)=\mi{atoms}((S-\{C^*\})\cup \{\Cn_1\geql \Cn_2\})=\mi{atoms}(S-\{C^*\})\cup \mi{atoms}(\Cn_1\geql \Cn_2)\subseteq \mi{atoms}(S)$,
$\mi{atoms}(S_2')=\mi{atoms}((S-\{C^*\})\cup \{\Cn_1\gle \Cn_2\vee \Cn_2\gle \Cn_1\})=\mi{atoms}(S-\{C^*\})\cup \mi{atoms}(\Cn_1\gle \Cn_2\vee \Cn_2\gle \Cn_1)\subseteq \mi{atoms}(S)$;
$S$ is positively guarded;
$\mi{atoms}(S_1-\mi{guards}(S_1))\subseteq \mi{atoms}(S_1)\subseteq \mi{atoms}(S)$,
$\mi{atoms}(S_1)=\mi{atoms}((S_1-\mi{guards}(S_1))\cup \mi{guards}(S_1))=\mi{atoms}(S_1-\mi{guards}(S_1))\cup \mi{atoms}(\mi{guards}(S_1))=
                 \mi{atoms}(S_1-\mi{guards}(S_1))\cup \mi{atoms}(\mi{guards}(S))=\mi{atoms}(S_1-\mi{guards}(S_1))\cup \mi{atoms}(S)=\mi{atoms}(S)$;
$\mi{atoms}(S_2'-\mi{guards}(S_2'))\subseteq \mi{atoms}(S_2')\subseteq \mi{atoms}(S)$,
$\mi{atoms}(S_2')=\mi{atoms}((S_2'-\mi{guards}(S_2'))\cup \mi{guards}(S_2'))=\mi{atoms}(S_2'-\mi{guards}(S_2'))\cup \mi{atoms}(\mi{guards}(S_2'))=
                  \mi{atoms}(S_2'-\mi{guards}(S_2'))\cup \mi{atoms}(\mi{guards}(S))=\mi{atoms}(S_2'-\mi{guards}(S_2'))\cup \mi{atoms}(S)=\mi{atoms}(S)$;
$S_1$ and $S_2'$ are positively guarded;
we have that $S$ is positive;
$\Cn_1\geql \Cn_2\not\in \mi{guards}(S)$;
for all $C\in S_1-\mi{guards}(S_1)=((S-\{C^*\})\cup \{\Cn_1\geql \Cn_2\})-\mi{guards}(S)=((S-\{C^*\})-\mi{guards}(S))\cup (\{\Cn_1\geql \Cn_2\}-\mi{guards}(S))=
              ((S-\{C^*\})-\mi{guards}(S))\cup \{\Cn_1\geql \Cn_2\}\subseteq (S-\mi{guards}(S))\cup \{\Cn_1\geql \Cn_2\}$, 
either $C\in \mi{PurOrdPropCl}$ or $C=\Cn_1\geql \Cn_2\in \mi{PurOrdPropCl}$, 
or $C=b_0\swedge\cdots\swedge b_k\gle \gu\vee C^\flat$, $b_j\in \mi{PropAtom}$, $\mi{guards}(S,b_j)=\{\gz\gle b_j\}$, $C^\flat\in \mi{PurOrdPropCl}$;
for all $j\leq k$, 
$b_j\in \mi{atoms}(C)\subseteq \mi{atoms}(S_1)=\mi{atoms}(S)$,
$\mi{guards}(S_1,b_j)=\mi{guards}(S,b_j)=\{\gz\gle b_j\}$;
$\Cn_1\gle \Cn_2\vee \Cn_2\gle \Cn_1\not\in \mi{guards}(S)$;
for all $C\in S_2'-\mi{guards}(S_2')=((S-\{C^*\})\cup \{\Cn_1\gle \Cn_2\vee \Cn_2\gle \Cn_1\})-\mi{guards}(S)=((S-\{C^*\})-\mi{guards}(S))\cup (\{\Cn_1\gle \Cn_2\vee \Cn_2\gle \Cn_1\}-\mi{guards}(S))=
              ((S-\{C^*\})-\mi{guards}(S))\cup \{\Cn_1\gle \Cn_2\vee \Cn_2\gle \Cn_1\}\subseteq (S-\mi{guards}(S))\cup \{\Cn_1\gle \Cn_2\vee \Cn_2\gle \Cn_1\}$, 
either $C\in \mi{PurOrdPropCl}$ or $C=\Cn_1\gle \Cn_2\vee \Cn_2\gle \Cn_1\in \mi{PurOrdPropCl}$, 
or $C=b_0\swedge\cdots\swedge b_k\gle \gu\vee C^\flat$, $b_j\in \mi{PropAtom}$, $\mi{guards}(S,b_j)=\{\gz\gle b_j\}$, $C^\flat\in \mi{PurOrdPropCl}$;
for all $j\leq k$, 
$b_j\in \mi{atoms}(C)\subseteq \mi{atoms}(S_2')=\mi{atoms}(S)$,
$\mi{guards}(S_2',b_j)=\mi{guards}(S,b_j)=\{\gz\gle b_j\}$;
$S_1$ and $S_2'$ are positive;
$a_0\in \mi{atoms}(C^*)\subseteq \mi{atoms}(S)=\mi{atoms}(S_2')$, $\mi{guards}(S_2',a_0)=\mi{guards}(S,a_0)=\{\gz\gle a_0\}$;
by Lemma \ref{le44} for $S_2'$ and $a_0$, there exists a finite linear {\it DPLL}-tree $\mi{Tree}_2'$ with the root $S_2'\cup \{a_0\gle \gu\}$ constructed using Rule (\cref{ceq4hr66}) satisfying
for its only leaf $S_2$ that $S_2\subseteq_{\mc F} \mi{OrdPropCl}$ is positive, 
$S_2-\mi{guards}(S_2)\subseteq S_2'-\mi{guards}(S_2')$, $\mi{guards}(S_2)=\mi{guards}(S_2')\cup \{a_0\gle \gu\}$, $a_0\gle \gu\not\in S_2'\supseteq \mi{guards}(S_2')$;
$\square\not\in S_1$,
$\mi{count}(C^*)=\mi{count}(\Cn_1\geql \Cn_2\vee a_0\gle \gu)=3$, 
$\mi{elmeasure}(S_1)=\sum_{C\in (S-\{C^*\})\cup \{\Cn_1\geql \Cn_2\}} \mi{count}(C)-1\leq (\sum_{C\in S-\{C^*\}} \mi{count}(C)-1)+\mi{count}(\Cn_1\geql \Cn_2)-1=
                     (\sum_{C\in S-\{C^*\}} \mi{count}(C)-1)+1<(\sum_{C\in S-\{C^*\}} \mi{count}(C)-1)+2=(\sum_{C\in S-\{C^*\}} \mi{count}(C)-1)+\mi{count}(C^*)-1=
                     \sum_{C\in (S-\{C^*\})\cup \{C^*\}} \mi{count}(C)-1=\mi{elmeasure}(S)=\sum_{C\in S} \mi{count}(C)-1$;
$S_2$ is simplified;
$\square\not\in S_2$,
$\square\not\in S_2'\supseteq S_2'-\mi{guards}(S_2')\supseteq S_2-\mi{guards}(S_2)$, 
$\mi{elmeasure}(S_2)=\sum_{C\in S_2} \mi{count}(C)-1=\sum_{C\in (S_2-\mi{guards}(S_2))\cup \mi{guards}(S_2)} \mi{count}(C)-1=
                     (\sum_{C\in S_2-\mi{guards}(S_2)} \mi{count}(C)-1)+\sum_{C\in \mi{guards}(S_2)} \mi{count}(C)-1\leq 
                     (\sum_{C\in S_2'-\mi{guards}(S_2')} \mi{count}(C)-1)+                                                                                                                 \linebreak[4]
                                                                          \sum_{C\in \mi{guards}(S_2')\cup \{a_0\gle \gu\}} \mi{count}(C)-1=
                     (\sum_{C\in S_2'-\mi{guards}(S_2')} \mi{count}(C)-1)+                                                                                                                 \linebreak[4]
                                                                          (\sum_{C\in \mi{guards}(S_2')} \mi{count}(C)-1)+\mi{count}(a_0\gle \gu)-1=                                       \linebreak[4]
                     \sum_{C\in (S_2'-\mi{guards}(S_2'))\cup \mi{guards}(S_2')} \mi{count}(C)-1=\sum_{C\in S_2'} \mi{count}(C)-1=
                     \sum_{C\in (S-\{C^*\})\cup \{\Cn_1\gle \Cn_2\vee \Cn_2\gle \Cn_1\}} \mi{count}(C)-1\leq (\sum_{C\in S-\{C^*\}} \mi{count}(C)-1)+\mi{count}(\Cn_1\gle \Cn_2\vee \Cn_2\gle \Cn_1)-1=
                     (\sum_{C\in S-\{C^*\}} \mi{count}(C)-1)+1<\mi{elmeasure}(S)$.

Case 2.2.2.1.2:
$\diamond^*\in \{\gleq,\gle\}$.
We put
\begin{equation}
\notag
\diamond^{**}=\left\{\begin{array}{ll}
                     \gle  &\ \text{\it if}\ \diamond^*=\gleq, \\[1mm]
                     \gleq &\ \text{\it if}\ \diamond^*=\gle.
                     \end{array}
              \right. 
\end{equation}
Then $C^*=\Cn_1\diamond^* \Cn_2\vee a_0\gle \gu\in S-\mi{guards}(S)$, $\Cn_1\diamond^* \Cn_2, \Cn_2\diamond^{**} \Cn_1\in \mi{PurOrdPropLit}$, $\square\neq a_0\gle \gu\in \mi{OrdPropCl}^\gu$,
$\Cn_1\neq \Cn_2$;
$\Cn_1\diamond^* \Cn_2\vee \Cn_2\diamond^{**} \Cn_1$ is a pure dichotomy;
applying Rule (\cref{ceq4hr11}) to $C^*$, $\Cn_1\diamond^* \Cn_2$, and $a_0\gle \gu$, we derive
\begin{equation}
\notag
\dfrac{S}
      {(S-\{C^*\})\cup \{\Cn_1\diamond^* \Cn_2\}\ \big|\ (S-\{C^*\})\cup \{a_0\gle \gu\}\cup \{\Cn_2\diamond^{**} \Cn_1\}}.
\end{equation}
Hence, $C^*\in S$.
We put 
$S_1=(S-\{C^*\})\cup \{\Cn_1\diamond^* \Cn_2\}\subseteq_{\mc F} \mi{OrdPropCl}$ and
$S_2'=(S-\{C^*\})\cup \{\Cn_2\diamond^{**} \Cn_1\}\subseteq_{\mc F} \mi{OrdPropCl}$.
We get that
$a_0\gle \gu\in \mi{guards}(a_0)$, $a_0\gle \gu\not\in \mi{guards}(S,a_0)=\{\gz\gle a_0\}=S\cap \mi{guards}(a_0)$, $a_0\gle \gu\not\in S$,
$a_0\gle \gu\neq \Cn_2\diamond^{**} \Cn_1$,
$a_0\gle \gu\not\in S_2'=(S-\{C^*\})\cup \{\Cn_2\diamond^{**} \Cn_1\}$,
$S_2'\cup \{a_0\gle \gu\}=(S-\{C^*\})\cup \{a_0\gle \gu\}\cup \{\Cn_2\diamond^{**} \Cn_1\}$;
$S$ is simplified; 
$S-\{C^*\}\subseteq S$ is simplified;
$\Cn_1\neq \Cn_2$;
$\Cn_1\diamond^* \Cn_2$ and $\Cn_2\diamond^{**} \Cn_1$ are not contradictions or tautologies;
$\Cn_1\diamond^* \Cn_2, \Cn_2\diamond^{**} \Cn_1\neq \square$ do not contain contradictions and tautologies;
$S_1=(S-\{C^*\})\cup \{\Cn_1\diamond^* \Cn_2\}$ and $S_2'=(S-\{C^*\})\cup \{\Cn_2\diamond^{**} \Cn_1\}$ are simplified;
$\Cn_1\diamond^* \Cn_2$ and $\Cn_2\diamond^{**} \Cn_1$ are not guards;
$C^*\not\in \mi{guards}(S)$,
$\mi{guards}(S_1)=\{C \,|\, C\in (S-\{C^*\})\cup \{\Cn_1\diamond^* \Cn_2\}\ \text{\it is a guard}\}=\{C \,|\, C\in S-\{C^*\}\ \text{\it is a guard}\}=\{C \,|\, C\in S\ \text{\it is a guard}\}-\{C^*\}=
                  \mi{guards}(S)-\{C^*\}=\mi{guards}(S)$,
$\mi{guards}(S_2')=\{C \,|\, C\in (S-\{C^*\})\cup \{\Cn_2\diamond^{**} \Cn_1\}\ \text{\it is a guard}\}=\{C \,|\, C\in S-\{C^*\}\ \text{\it is a guard}\}=\mi{guards}(S)$;
$C^*\in S$, $\mi{atoms}(\Cn_2\diamond^{**} \Cn_1)=\mi{atoms}(\Cn_1\diamond^* \Cn_2)\subseteq \mi{atoms}(C^*)\subseteq \mi{atoms}(S)$,
$\mi{atoms}(S_1)=\mi{atoms}((S-\{C^*\})\cup \{\Cn_1\diamond^* \Cn_2\})=\mi{atoms}(S-\{C^*\})\cup \mi{atoms}(\Cn_1\diamond^* \Cn_2)\subseteq \mi{atoms}(S)$,
$\mi{atoms}(S_2')=\mi{atoms}((S-\{C^*\})\cup \{\Cn_2\diamond^{**} \Cn_1\})=\mi{atoms}(S-\{C^*\})\cup \mi{atoms}(\Cn_2\diamond^{**} \Cn_1)\subseteq \mi{atoms}(S)$;
$S$ is positively guarded;
$\mi{atoms}(S_1-\mi{guards}(S_1))\subseteq \mi{atoms}(S_1)\subseteq \mi{atoms}(S)$,
$\mi{atoms}(S_1)=\mi{atoms}((S_1-\mi{guards}(S_1))\cup \mi{guards}(S_1))=\mi{atoms}(S_1-\mi{guards}(S_1))\cup \mi{atoms}(\mi{guards}(S_1))=
                 \mi{atoms}(S_1-\mi{guards}(S_1))\cup \mi{atoms}(\mi{guards}(S))=\mi{atoms}(S_1-\mi{guards}(S_1))\cup \mi{atoms}(S)=\mi{atoms}(S)$;
$\mi{atoms}(S_2'-\mi{guards}(S_2'))\subseteq \mi{atoms}(S_2')\subseteq \mi{atoms}(S)$,
$\mi{atoms}(S_2')=\mi{atoms}((S_2'-\mi{guards}(S_2'))\cup \mi{guards}(S_2'))=\mi{atoms}(S_2'-\mi{guards}(S_2'))\cup \mi{atoms}(\mi{guards}(S_2'))=
                  \mi{atoms}(S_2'-\mi{guards}(S_2'))\cup \mi{atoms}(\mi{guards}(S))=\mi{atoms}(S_2'-\mi{guards}(S_2'))\cup \mi{atoms}(S)=\mi{atoms}(S)$;
$S_1$ and $S_2'$ are positively guarded;
we have that $S$ is positive;
$\Cn_1\diamond^* \Cn_2\not\in \mi{guards}(S)$;
for all $C\in S_1-\mi{guards}(S_1)=((S-\{C^*\})\cup \{\Cn_1\diamond^* \Cn_2\})-\mi{guards}(S)=((S-\{C^*\})-\mi{guards}(S))\cup (\{\Cn_1\diamond^* \Cn_2\}-\mi{guards}(S))=
              ((S-\{C^*\})-\mi{guards}(S))\cup \{\Cn_1\diamond^* \Cn_2\}\subseteq (S-\mi{guards}(S))\cup \{\Cn_1\diamond^* \Cn_2\}$, 
either $C\in \mi{PurOrdPropCl}$ or $C=\Cn_1\diamond^* \Cn_2\in \mi{PurOrdPropCl}$, 
or $C=b_0\swedge\cdots\swedge b_k\gle \gu\vee C^\flat$, $b_j\in \mi{PropAtom}$, $\mi{guards}(S,b_j)=\{\gz\gle b_j\}$, $C^\flat\in \mi{PurOrdPropCl}$;
for all $j\leq k$, 
$b_j\in \mi{atoms}(C)\subseteq \mi{atoms}(S_1)=\mi{atoms}(S)$,
$\mi{guards}(S_1,b_j)=\mi{guards}(S,b_j)=\{\gz\gle b_j\}$;
$\Cn_2\diamond^{**} \Cn_1\not\in \mi{guards}(S)$;
for all $C\in S_2'-\mi{guards}(S_2')=((S-\{C^*\})\cup \{\Cn_2\diamond^{**} \Cn_1\})-\mi{guards}(S)=((S-\{C^*\})-\mi{guards}(S))\cup (\{\Cn_2\diamond^{**} \Cn_1\}-\mi{guards}(S))=
              ((S-\{C^*\})-\mi{guards}(S))\cup \{\Cn_2\diamond^{**} \Cn_1\}\subseteq (S-\mi{guards}(S))\cup \{\Cn_2\diamond^{**} \Cn_1\}$, 
either $C\in \mi{PurOrdPropCl}$ or $C=\Cn_2\diamond^{**} \Cn_1\in \mi{PurOrdPropCl}$, 
or $C=b_0\swedge\cdots\swedge b_k\gle \gu\vee C^\flat$, $b_j\in \mi{PropAtom}$, $\mi{guards}(S,b_j)=\{\gz\gle b_j\}$, $C^\flat\in \mi{PurOrdPropCl}$;
for all $j\leq k$, 
$b_j\in \mi{atoms}(C)\subseteq \mi{atoms}(S_2')=\mi{atoms}(S)$,
$\mi{guards}(S_2',b_j)=\mi{guards}(S,b_j)=\{\gz\gle b_j\}$;
$S_1$ and $S_2'$ are positive;
$a_0\in \mi{atoms}(C^*)\subseteq \mi{atoms}(S)=\mi{atoms}(S_2')$, $\mi{guards}(S_2',a_0)=\mi{guards}(S,a_0)=\{\gz\gle a_0\}$;
by Lemma \ref{le44} for $S_2'$ and $a_0$, there exists a finite linear {\it DPLL}-tree $\mi{Tree}_2'$ with the root $S_2'\cup \{a_0\gle \gu\}$ constructed using Rule (\cref{ceq4hr66}) satisfying
for its only leaf $S_2$ that $S_2\subseteq_{\mc F} \mi{OrdPropCl}$ is positive, 
$S_2-\mi{guards}(S_2)\subseteq S_2'-\mi{guards}(S_2')$, $\mi{guards}(S_2)=\mi{guards}(S_2')\cup \{a_0\gle \gu\}$, $a_0\gle \gu\not\in S_2'\supseteq \mi{guards}(S_2')$;
$\square\not\in S_1$,
$\diamond^*, \diamond^{**}\in \{\gleq,\gle\}$,
$\mi{count}(\Cn_1\diamond^* \Cn_2)=1$, $\mi{count}(C^*)=\mi{count}(\Cn_1\diamond^* \Cn_2\vee a_0\gle \gu)=2$, 
$\mi{elmeasure}(S_1)=\sum_{C\in (S-\{C^*\})\cup \{\Cn_1\diamond^* \Cn_2\}} \mi{count}(C)-1=(\sum_{C\in S-\{C^*\}} \mi{count}(C)-1)+\mi{count}(\Cn_1\diamond^* \Cn_2)-1=
                     \sum_{C\in S-\{C^*\}} \mi{count}(C)-1<(\sum_{C\in S-\{C^*\}} \mi{count}(C)-1)+1=(\sum_{C\in S-\{C^*\}} \mi{count}(C)-1)+\mi{count}(C^*)-1=
                     \sum_{C\in (S-\{C^*\})\cup \{C^*\}} \mi{count}(C)-1=\mi{elmeasure}(S)=\sum_{C\in S} \mi{count}(C)-1$;
$S_2$ is simplified;
$\square\not\in S_2$,
$\square\not\in S_2'\supseteq S_2'-\mi{guards}(S_2')\supseteq S_2-\mi{guards}(S_2)$, $\mi{count}(\Cn_2\diamond^{**} \Cn_1)=1$, 
$\mi{elmeasure}(S_2)=\sum_{C\in S_2} \mi{count}(C)-1=\sum_{C\in (S_2-\mi{guards}(S_2))\cup \mi{guards}(S_2)} \mi{count}(C)-1=
                     (\sum_{C\in S_2-\mi{guards}(S_2)} \mi{count}(C)-1)+\sum_{C\in \mi{guards}(S_2)} \mi{count}(C)-1\leq 
                     (\sum_{C\in S_2'-\mi{guards}(S_2')} \mi{count}(C)-1)+\sum_{C\in \mi{guards}(S_2')\cup \{a_0\gle \gu\}} \mi{count}(C)-1=
                     (\sum_{C\in S_2'-\mi{guards}(S_2')} \mi{count}(C)-1)+(\sum_{C\in \mi{guards}(S_2')} \mi{count}(C)-1)+\mi{count}(a_0\gle \gu)-1=
                     \sum_{C\in (S_2'-\mi{guards}(S_2'))\cup \mi{guards}(S_2')} \mi{count}(C)-1=\sum_{C\in S_2'} \mi{count}(C)-1=
                     \sum_{C\in (S-\{C^*\})\cup \{\Cn_2\diamond^{**} \Cn_1\}} \mi{count}(C)-1=(\sum_{C\in S-\{C^*\}} \mi{count}(C)-1)+\mi{count}(\Cn_2\diamond^{**} \Cn_1)-1=
                     \sum_{C\in S-\{C^*\}} \mi{count}(C)-1<\mi{elmeasure}(S)$.

Case 2.2.2.2:
$n\geq 1$ or $C^\natural-\{\Cn_1\diamond^* \Cn_2\}\neq \square$.
Then $a_0\swedge\cdots\swedge a_n\gle \gu\vee (C^\natural-\{\Cn_1\diamond^* \Cn_2\})$ is not a guard.
We get two cases for $\diamond^*$.

Case 2.2.2.2.1:
$\diamond^*=\geql$.
Then $C^*=\Cn_1\diamond^* \Cn_2\vee a_0\swedge\cdots\swedge a_n\gle \gu\vee (C^\natural-\{\Cn_1\diamond^* \Cn_2\})=
          \Cn_1\geql \Cn_2\vee a_0\swedge\cdots\swedge a_n\gle \gu\vee (C^\natural-\{\Cn_1\geql \Cn_2\})\in S-\mi{guards}(S)$, 
$\square\neq a_0\swedge\cdots\swedge a_n\gle \gu\vee (C^\natural-\{\Cn_1\geql \Cn_2\})\in \mi{OrdPropCl}^\gu$, $\Cn_1\neq \Cn_2$,
applying Rule (\cref{ceq4hr111}) to $C^*$, $\Cn_1\geql \Cn_2$, and $a_0\swedge\cdots\swedge a_n\gle \gu\vee (C^\natural-\{\Cn_1\geql \Cn_2\})$, we derive
\begin{equation}
\notag
\dfrac{S}
      {\begin{array}{l}
       (S-\{C^*\})\cup \{\Cn_1\geql \Cn_2\}\ \big| \\
       (S-\{C^*\})\cup \{a_0\swedge\cdots\swedge a_n\gle \gu\vee (C^\natural-\{\Cn_1\geql \Cn_2\})\}\cup \\
       \quad \{\Cn_1\gle \Cn_2\vee \Cn_2\gle \Cn_1\}
       \end{array}}.
\end{equation}
Hence, $C^*\in S$, $\Cn_1\neq \Cn_2$, $\Cn_1\gle \Cn_2\neq \Cn_2\gle \Cn_1$.
We put 
$S_1=(S-\{C^*\})\cup \{\Cn_1\geql \Cn_2\}\subseteq_{\mc F} \mi{OrdPropCl}$ and
$S_2=(S-\{C^*\})\cup \{a_0\swedge\cdots\swedge a_n\gle \gu\vee (C^\natural-\{\Cn_1\geql \Cn_2\})\}\cup \{\Cn_1\gle \Cn_2\vee \Cn_2\gle \Cn_1\}\subseteq_{\mc F} \mi{OrdPropCl}$.
We get that
$S$ is simplified; 
$S-\{C^*\}\subseteq S$ is simplified;
$\Cn_1\neq \Cn_2$;
$\Cn_1\geql \Cn_2$, $\Cn_1\gle \Cn_2$, $\Cn_2\gle \Cn_1$ are not contradictions or tautologies;
$\Cn_1\geql \Cn_2, \Cn_1\gle \Cn_2\vee \Cn_2\gle \Cn_1\neq \square$ do not contain contradictions and tautologies;
$a_0\swedge\cdots\swedge a_n\gle \gu$ is not a contradiction or tautology;
$C^*\in S$ does not contain contradictions and tautologies;
$C^\natural-\{\Cn_1\geql \Cn_2\}\subseteq C^*$ does not contain contradictions and tautologies;
$a_0\swedge\cdots\swedge a_n\gle \gu\vee (C^\natural-\{\Cn_1\geql \Cn_2\})\neq \square$ does not contain contradictions and tautologies;
$S_1=(S-\{C^*\})\cup \{\Cn_1\geql \Cn_2\}$ and
$S_2=(S-\{C^*\})\cup \{a_0\swedge\cdots\swedge a_n\gle \gu\vee (C^\natural-\{\Cn_1\geql \Cn_2\})\}\cup \{\Cn_1\gle \Cn_2\vee \Cn_2\gle \Cn_1\}$ are simplified;
$\Cn_1\geql \Cn_2$ and $\Cn_1\gle \Cn_2\vee \Cn_2\gle \Cn_1$ are not guards;
$a_0\swedge\cdots\swedge a_n\gle \gu\vee (C^\natural-\{\Cn_1\diamond^* \Cn_2\})=a_0\swedge\cdots\swedge a_n\gle \gu\vee (C^\natural-\{\Cn_1\geql \Cn_2\})$ is not a guard;
$C^*\not\in \mi{guards}(S)$,
$\mi{guards}(S_1)=\{C \,|\, C\in (S-\{C^*\})\cup \{\Cn_1\geql \Cn_2\}\ \text{\it is a guard}\}=\{C \,|\, C\in S-\{C^*\}\ \text{\it is a guard}\}=\{C \,|\, C\in S\ \text{\it is a guard}\}-\{C^*\}=
                  \mi{guards}(S)-\{C^*\}=\mi{guards}(S)$,
$\mi{guards}(S_2)=\{C \,|\, C\in (S-\{C^*\})\cup \{a_0\swedge\cdots\swedge a_n\gle \gu\vee (C^\natural-\{\Cn_1\geql \Cn_2\})\}\cup \{\Cn_1\gle \Cn_2\vee \Cn_2\gle \Cn_1\}\ \text{\it is a guard}\}=
                  \{C \,|\, C\in S-\{C^*\}\ \text{\it is a guard}\}=\mi{guards}(S)$;
$\mi{atoms}(\Cn_1\geql \Cn_2), \mi{atoms}(a_0\swedge\cdots\swedge a_n\gle \gu\vee (C^\natural-\{\Cn_1\geql \Cn_2\}))\subseteq \mi{atoms}(C^*)\subseteq \mi{atoms}(S)$, 
$\mi{atoms}(\Cn_1\gle \Cn_2\vee \Cn_2\gle \Cn_1)=\mi{atoms}(\Cn_1\geql \Cn_2)\subseteq \mi{atoms}(S)$,
$\mi{atoms}(S_1)=\mi{atoms}((S-\{C^*\})\cup \{\Cn_1\geql \Cn_2\})=\mi{atoms}(S-\{C^*\})\cup \mi{atoms}(\Cn_1\geql \Cn_2)\subseteq \mi{atoms}(S)$,
$\mi{atoms}(S_2)=\mi{atoms}((S-\{C^*\})\cup \{a_0\swedge\cdots\swedge a_n\gle \gu\vee (C^\natural-\{\Cn_1\geql \Cn_2\})\}\cup \{\Cn_1\gle \Cn_2\vee \Cn_2\gle \Cn_1\})=
                 \mi{atoms}(S-\{C^*\})\cup \mi{atoms}(a_0\swedge\cdots\swedge a_n\gle \gu\vee (C^\natural-\{\Cn_1\geql \Cn_2\}))\cup \mi{atoms}(\Cn_1\gle \Cn_2\vee \Cn_2\gle \Cn_1)\subseteq 
                 \mi{atoms}(S)$;
$S$ is positively guarded;
for both $i$,
$\mi{atoms}(S_i-\mi{guards}(S_i))\subseteq \mi{atoms}(S_i)\subseteq \mi{atoms}(S)$,
$\mi{atoms}(S_i)=\mi{atoms}((S_i-\mi{guards}(S_i))\cup \mi{guards}(S_i))=\mi{atoms}(S_i-\mi{guards}(S_i))\cup \mi{atoms}(\mi{guards}(S_i))=
                 \mi{atoms}(S_i-\mi{guards}(S_i))\cup \mi{atoms}(\mi{guards}(S))=\mi{atoms}(S_i-\mi{guards}(S_i))\cup \mi{atoms}(S)=\mi{atoms}(S)$;
$S_1$ and $S_2$ are positively guarded;
we have that $S$ is positive;
$\Cn_1\geql \Cn_2\not\in \mi{guards}(S)$;
for all $C\in S_1-\mi{guards}(S_1)=((S-\{C^*\})\cup \{\Cn_1\geql \Cn_2\})-\mi{guards}(S)=((S-\{C^*\})-\mi{guards}(S))\cup (\{\Cn_1\geql \Cn_2\}-\mi{guards}(S))=
              ((S-\{C^*\})-\mi{guards}(S))\cup \{\Cn_1\geql \Cn_2\}\subseteq (S-\mi{guards}(S))\cup \{\Cn_1\geql \Cn_2\}$, 
either $C\in \mi{PurOrdPropCl}$ or $C=\Cn_1\geql \Cn_2\in \mi{PurOrdPropCl}$, 
or $C=b_0\swedge\cdots\swedge b_k\gle \gu\vee C^\flat$, $b_j\in \mi{PropAtom}$, $\mi{guards}(S,b_j)=\{\gz\gle b_j\}$, $C^\flat\in \mi{PurOrdPropCl}$;
for all $j\leq k$, 
$b_j\in \mi{atoms}(C)\subseteq \mi{atoms}(S_1)=\mi{atoms}(S)$,
$\mi{guards}(S_1,b_j)=\mi{guards}(S,b_j)=\{\gz\gle b_j\}$;
$a_0\swedge\cdots\swedge a_n\gle \gu\vee (C^\natural-\{\Cn_1\geql \Cn_2\}), \Cn_1\gle \Cn_2\vee \Cn_2\gle \Cn_1\not\in \mi{guards}(S)$;
for all $C\in S_2-\mi{guards}(S_2)=((S-\{C^*\})\cup \{a_0\swedge\cdots\swedge a_n\gle \gu\vee (C^\natural-\{\Cn_1\geql \Cn_2\})\}\cup \{\Cn_1\gle \Cn_2\vee \Cn_2\gle \Cn_1\})-\mi{guards}(S)=
                                   ((S-\{C^*\})-\mi{guards}(S))\cup (\{a_0\swedge\cdots\swedge a_n\gle \gu\vee (C^\natural-\{\Cn_1\geql \Cn_2\})\}-\mi{guards}(S))\cup 
                                   (\{\Cn_1\gle \Cn_2\vee \Cn_2\gle \Cn_1\}-\mi{guards}(S))=
                                   ((S-\{C^*\})-\mi{guards}(S))\cup \{a_0\swedge\cdots\swedge a_n\gle \gu\vee (C^\natural-\{\Cn_1\geql \Cn_2\})\}\cup \{\Cn_1\gle \Cn_2\vee \Cn_2\gle \Cn_1\}\subseteq 
                                   (S-\mi{guards}(S))\cup \{a_0\swedge\cdots\swedge a_n\gle \gu\vee (C^\natural-\{\Cn_1\geql \Cn_2\})\}\cup \{\Cn_1\gle \Cn_2\vee \Cn_2\gle \Cn_1\}$, 
either $C\in \mi{PurOrdPropCl}$ or $C=\Cn_1\gle \Cn_2\vee \Cn_2\gle \Cn_1\in \mi{PurOrdPropCl}$, 
or $C=b_0\swedge\cdots\swedge b_k\gle \gu\vee C^\flat$, $b_j\in \mi{PropAtom}$, $\mi{guards}(S,b_j)=\{\gz\gle b_j\}$, $C^\flat\in \mi{PurOrdPropCl}$;
for all $j\leq k$, 
$b_j\in \mi{atoms}(C)\subseteq \mi{atoms}(S_2)=\mi{atoms}(S)$,
$\mi{guards}(S_2,b_j)=\mi{guards}(S,b_j)=\{\gz\gle b_j\}$;
or $C=a_0\swedge\cdots\swedge a_n\gle \gu\vee (C^\natural-\{\Cn_1\geql \Cn_2\})$, $\mi{guards}(S,a_i)=\{\gz\gle a_i\}$, $C^\natural-\{\Cn_1\geql \Cn_2\}\in \mi{PurOrdPropCl}$;
for all $i\leq n$, 
$a_i\in \mi{atoms}(C)\subseteq \mi{atoms}(S_2)=\mi{atoms}(S)$,
$\mi{guards}(S_2,a_i)=\mi{guards}(S,a_i)=\{\gz\gle a_i\}$;
$S_1$ and $S_2$ are positive;
$\square\not\in S_1$,
$\mi{count}(C^*)=\mi{count}(\Cn_1\geql \Cn_2\vee a_0\swedge\cdots\swedge a_n\gle \gu\vee (C^\natural-\{\Cn_1\geql \Cn_2\}))=
                 \mi{count}(\Cn_1\geql \Cn_2)+\mi{count}(a_0\swedge\cdots\swedge a_n\gle \gu)+\mi{count}(C^\natural-\{\Cn_1\geql \Cn_2\})=
                 \mi{count}(C^\natural-\{\Cn_1\geql \Cn_2\})+3$, 
$\mi{elmeasure}(S_1)=\sum_{C\in (S-\{C^*\})\cup \{\Cn_1\geql \Cn_2\}} \mi{count}(C)-1\leq (\sum_{C\in S-\{C^*\}} \mi{count}(C)-1)+\mi{count}(\Cn_1\geql \Cn_2)-1=
                     (\sum_{C\in S-\{C^*\}} \mi{count}(C)-1)+1<(\sum_{C\in S-\{C^*\}} \mi{count}(C)-1)+\mi{count}(C^\natural-\{\Cn_1\geql \Cn_2\})+2=
                     (\sum_{C\in S-\{C^*\}} \mi{count}(C)-1)+\mi{count}(C^*)-1=\sum_{C\in (S-\{C^*\})\cup \{C^*\}} \mi{count}(C)-1=\mi{elmeasure}(S)=\sum_{C\in S} \mi{count}(C)-1$,
$\square\not\in S_2$,
$\mi{count}(a_0\swedge\cdots\swedge a_n\gle \gu\vee (C^\natural-\{\Cn_1\geql \Cn_2\}))=\mi{count}(a_0\swedge\cdots\swedge a_n\gle \gu)+\mi{count}(C^\natural-\{\Cn_1\geql \Cn_2\})=
                                                                                       \mi{count}(C^\natural-\{\Cn_1\geql \Cn_2\})+1$, 
$\mi{elmeasure}(S_2)=\sum_{C\in (S-\{C^*\})\cup \{a_0\swedge\cdots\swedge a_n\gle \gu\vee (C^\natural-\{\Cn_1\geql \Cn_2\})\}\cup \{\Cn_1\gle \Cn_2\vee \Cn_2\gle \Cn_1\}} \mi{count}(C)-1\leq 
                     (\sum_{C\in S-\{C^*\}} \mi{count}(C)-1)+\mi{count}(a_0\swedge\cdots\swedge a_n\gle \gu\vee (C^\natural-\{\Cn_1\geql \Cn_2\}))-1+\mi{count}(\Cn_1\gle \Cn_2\vee \Cn_2\gle \Cn_1)-1=
                     (\sum_{C\in S-\{C^*\}} \mi{count}(C)-1)+\mi{count}(C^\natural-\{\Cn_1\geql \Cn_2\})+1<(\sum_{C\in S-\{C^*\}} \mi{count}(C)-1)+\mi{count}(C^\natural-\{\Cn_1\geql \Cn_2\})+2=
                     \mi{elmeasure}(S)$.

Case 2.2.2.2.2:
$\diamond^*\in \{\gleq,\gle\}$.
We put
\begin{equation}
\notag
\diamond^{**}=\left\{\begin{array}{ll}
                     \gle  &\ \text{\it if}\ \diamond^*=\gleq, \\[1mm]
                     \gleq &\ \text{\it if}\ \diamond^*=\gle.
                     \end{array}
              \right. 
\end{equation}
Then $C^*=\Cn_1\diamond^* \Cn_2\vee a_0\swedge\cdots\swedge a_n\gle \gu\vee (C^\natural-\{\Cn_1\diamond^* \Cn_2\})\in S-\mi{guards}(S)$, 
$\Cn_1\diamond^* \Cn_2, \Cn_2\diamond^{**} \Cn_1\in \mi{PurOrdPropLit}$, $\square\neq a_0\swedge\cdots\swedge a_n\gle \gu\vee (C^\natural-\{\Cn_1\diamond^* \Cn_2\})\in \mi{OrdPropCl}^\gu$,
$\Cn_1\neq \Cn_2$;
$\Cn_1\diamond^* \Cn_2\vee \Cn_2\diamond^{**} \Cn_1$ is a pure dichotomy;
applying Rule (\cref{ceq4hr11}) to $C^*$, $\Cn_1\diamond^* \Cn_2$, and $a_0\swedge\cdots\swedge a_n\gle \gu\vee (C^\natural-\{\Cn_1\diamond^* \Cn_2\})$, we derive
\begin{equation}
\notag
\dfrac{S}
      {\begin{array}{l}
       (S-\{C^*\})\cup \{\Cn_1\diamond^* \Cn_2\}\ \big| \\
       (S-\{C^*\})\cup \{a_0\swedge\cdots\swedge a_n\gle \gu\vee (C^\natural-\{\Cn_1\diamond^* \Cn_2\})\}\cup \{\Cn_2\diamond^{**} \Cn_1\}
       \end{array}}.
\end{equation}
Hence, $C^*\in S$.
We put 
$S_1=(S-\{C^*\})\cup \{\Cn_1\diamond^* \Cn_2\}\subseteq_{\mc F} \mi{OrdPropCl}$ and
$S_2=(S-\{C^*\})\cup \{a_0\swedge\cdots\swedge a_n\gle \gu\vee (C^\natural-\{\Cn_1\diamond^* \Cn_2\})\}\cup \{\Cn_2\diamond^{**} \Cn_1\}\subseteq_{\mc F} \mi{OrdPropCl}$.
We get that
$S$ is simplified; 
$S-\{C^*\}\subseteq S$ is simplified;
$\Cn_1\neq \Cn_2$;
$\Cn_1\diamond^* \Cn_2$ and $\Cn_2\diamond^{**} \Cn_1$ are not contradictions or tautologies;
$\Cn_1\diamond^* \Cn_2, \Cn_2\diamond^{**} \Cn_1\neq \square$ do not contain contradictions and tautologies;
$a_0\swedge\cdots\swedge a_n\gle \gu$ is not a contradiction or tautology;
$C^*\in S$ does not contain contradictions and tautologies;
$C^\natural-\{\Cn_1\diamond^* \Cn_2\}\subseteq C^*$ does not contain contradictions and tautologies;
$a_0\swedge\cdots\swedge a_n\gle \gu\vee (C^\natural-\{\Cn_1\diamond^* \Cn_2\})\neq \square$ does not contain contradictions and tautologies;
$S_1=(S-\{C^*\})\cup \{\Cn_1\diamond^* \Cn_2\}$ and
$S_2=(S-\{C^*\})\cup \{a_0\swedge\cdots\swedge a_n\gle \gu\vee (C^\natural-\{\Cn_1\diamond^* \Cn_2\})\}\cup \{\Cn_2\diamond^{**} \Cn_1\}$ are simplified;
$\Cn_1\diamond^* \Cn_2$ and $\Cn_2\diamond^{**} \Cn_1$ are not guards;
we have that $a_0\swedge\cdots\swedge a_n\gle \gu\vee (C^\natural-\{\Cn_1\diamond^* \Cn_2\})$ is not a guard;
$C^*\not\in \mi{guards}(S)$,
$\mi{guards}(S_1)=\{C \,|\, C\in (S-\{C^*\})\cup \{\Cn_1\diamond^* \Cn_2\}\ \text{\it is a guard}\}=\{C \,|\, C\in S-\{C^*\}\ \text{\it is a guard}\}=\{C \,|\, C\in S\ \text{\it is a guard}\}-\{C^*\}=
                  \mi{guards}(S)-\{C^*\}=\mi{guards}(S)$,
$\mi{guards}(S_2)=\{C \,|\, C\in (S-\{C^*\})\cup \{a_0\swedge\cdots\swedge a_n\gle \gu\vee (C^\natural-\{\Cn_1\diamond^* \Cn_2\})\}\cup \{\Cn_2\diamond^{**} \Cn_1\}\ \text{\it is a guard}\}=
                  \{C \,|\, C\in S-\{C^*\}\ \text{\it is a guard}\}=\mi{guards}(S)$;
$\mi{atoms}(\Cn_1\diamond^* \Cn_2), \mi{atoms}(a_0\swedge\cdots\swedge a_n\gle \gu\vee (C^\natural-\{\Cn_1\diamond^* \Cn_2\}))\subseteq \mi{atoms}(C^*)\subseteq \mi{atoms}(S)$, 
$\mi{atoms}(\Cn_2\diamond^{**} \Cn_1)=\mi{atoms}(\Cn_1\diamond^* \Cn_2)\subseteq \mi{atoms}(S)$,
$\mi{atoms}(S_1)=\mi{atoms}((S-\{C^*\})\cup \{\Cn_1\diamond^* \Cn_2\})=\mi{atoms}(S-\{C^*\})\cup \mi{atoms}(\Cn_1\diamond^* \Cn_2)\subseteq \mi{atoms}(S)$,
$\mi{atoms}(S_2)=\mi{atoms}((S-\{C^*\})\cup \{a_0\swedge\cdots\swedge a_n\gle \gu\vee (C^\natural-\{\Cn_1\diamond^* \Cn_2\})\}\cup \{\Cn_2\diamond^{**} \Cn_1\})=
                 \mi{atoms}(S-\{C^*\})\cup \mi{atoms}(a_0\swedge\cdots\swedge a_n\gle \gu\vee (C^\natural-\{\Cn_1\diamond^* \Cn_2\}))\cup \mi{atoms}(\Cn_2\diamond^{**} \Cn_1)\subseteq 
                 \mi{atoms}(S)$;
$S$ is positively guarded;
for both $i$,
$\mi{atoms}(S_i-\mi{guards}(S_i))\subseteq \mi{atoms}(S_i)\subseteq \mi{atoms}(S)$,
$\mi{atoms}(S_i)=\mi{atoms}((S_i-\mi{guards}(S_i))\cup \mi{guards}(S_i))=\mi{atoms}(S_i-\mi{guards}(S_i))\cup \mi{atoms}(\mi{guards}(S_i))=
                 \mi{atoms}(S_i-\mi{guards}(S_i))\cup \mi{atoms}(\mi{guards}(S))=\mi{atoms}(S_i-\mi{guards}(S_i))\cup \mi{atoms}(S)=\mi{atoms}(S)$;
$S_1$ and $S_2$ are positively guarded;
we have that $S$ is positive;
$\Cn_1\diamond^* \Cn_2\not\in \mi{guards}(S)$;
for all $C\in S_1-\mi{guards}(S_1)=((S-\{C^*\})\cup \{\Cn_1\diamond^* \Cn_2\})-\mi{guards}(S)=((S-\{C^*\})-\mi{guards}(S))\cup (\{\Cn_1\diamond^* \Cn_2\}-\mi{guards}(S))=
              ((S-\{C^*\})-\mi{guards}(S))\cup \{\Cn_1\diamond^* \Cn_2\}\subseteq (S-\mi{guards}(S))\cup \{\Cn_1\diamond^* \Cn_2\}$, 
either $C\in \mi{PurOrdPropCl}$ or $C=\Cn_1\diamond^* \Cn_2\in \mi{PurOrdPropCl}$, 
or $C=b_0\swedge\cdots\swedge b_k\gle \gu\vee C^\flat$, $b_j\in \mi{PropAtom}$, $\mi{guards}(S,b_j)=\{\gz\gle b_j\}$, $C^\flat\in \mi{PurOrdPropCl}$;
for all $j\leq k$, 
$b_j\in \mi{atoms}(C)\subseteq \mi{atoms}(S_1)=\mi{atoms}(S)$,
$\mi{guards}(S_1,b_j)=\mi{guards}(S,b_j)=\{\gz\gle b_j\}$;
$a_0\swedge\cdots\swedge a_n\gle \gu\vee (C^\natural-\{\Cn_1\diamond^* \Cn_2\}), \Cn_2\diamond^{**} \Cn_1\not\in \mi{guards}(S)$;
for all $C\in S_2-\mi{guards}(S_2)=((S-\{C^*\})\cup \{a_0\swedge\cdots\swedge a_n\gle \gu\vee (C^\natural-\{\Cn_1\diamond^* \Cn_2\})\}\cup \{\Cn_2\diamond^{**} \Cn_1\})-\mi{guards}(S)=
                                   ((S-\{C^*\})-\mi{guards}(S))\cup (\{a_0\swedge\cdots\swedge a_n\gle \gu\vee (C^\natural-\{\Cn_1\diamond^* \Cn_2\})\}-\mi{guards}(S))\cup 
                                   (\{\Cn_2\diamond^{**} \Cn_1\}-\mi{guards}(S))=
                                   ((S-\{C^*\})-\mi{guards}(S))\cup \{a_0\swedge\cdots\swedge a_n\gle \gu\vee (C^\natural-\{\Cn_1\diamond^* \Cn_2\})\}\cup \{\Cn_2\diamond^{**} \Cn_1\}\subseteq 
                                   (S-\mi{guards}(S))\cup \{a_0\swedge\cdots\swedge a_n\gle \gu\vee (C^\natural-\{\Cn_1\diamond^* \Cn_2\})\}\cup \{\Cn_2\diamond^{**} \Cn_1\}$, 
either $C\in \mi{PurOrdPropCl}$ or $C=\Cn_2\diamond^{**} \Cn_1\in \mi{PurOrdPropCl}$, 
or $C=b_0\swedge\cdots\swedge b_k\gle \gu\vee C^\flat$, $b_j\in \mi{PropAtom}$, $\mi{guards}(S,b_j)=\{\gz\gle b_j\}$, $C^\flat\in \mi{PurOrdPropCl}$;
for all $j\leq k$, 
$b_j\in \mi{atoms}(C)\subseteq \mi{atoms}(S_2)=\mi{atoms}(S)$,
$\mi{guards}(S_2,b_j)=\mi{guards}(S,b_j)=\{\gz\gle b_j\}$;
or $C=a_0\swedge\cdots\swedge a_n\gle \gu\vee (C^\natural-\{\Cn_1\diamond^* \Cn_2\})$, $\mi{guards}(S,a_i)=\{\gz\gle a_i\}$, $C^\natural-\{\Cn_1\diamond^* \Cn_2\}\in \mi{PurOrdPropCl}$;
for all $i\leq n$, 
$a_i\in \mi{atoms}(C)\subseteq \mi{atoms}(S_2)=\mi{atoms}(S)$,
$\mi{guards}(S_2,a_i)=\mi{guards}(S,a_i)=\{\gz\gle a_i\}$;
$S_1$ and $S_2$ are positive;
$\square\not\in S_1$,
$\diamond^*, \diamond^{**}\in \{\gleq,\gle\}$,
$\mi{count}(\Cn_1\diamond^* \Cn_2)=1$, 
$\mi{count}(C^*)=\mi{count}(\Cn_1\diamond^* \Cn_2\vee a_0\swedge\cdots\swedge a_n\gle \gu\vee (C^\natural-\{\Cn_1\diamond^* \Cn_2\}))=
                 \mi{count}(\Cn_1\diamond^* \Cn_2)+\mi{count}(a_0\swedge\cdots\swedge a_n\gle \gu)+\mi{count}(C^\natural-\{\Cn_1\diamond^* \Cn_2\})=
                 \mi{count}(C^\natural-\{\Cn_1\diamond^* \Cn_2\})+2$, 
$\mi{elmeasure}(S_1)=\sum_{C\in (S-\{C^*\})\cup \{\Cn_1\diamond^* \Cn_2\}} \mi{count}(C)-1=(\sum_{C\in S-\{C^*\}} \mi{count}(C)-1)+\mi{count}(\Cn_1\diamond^* \Cn_2)-1=
                     \sum_{C\in S-\{C^*\}} \mi{count}(C)-1<(\sum_{C\in S-\{C^*\}} \mi{count}(C)-1)+\mi{count}(C^\natural-\{\Cn_1\diamond^* \Cn_2\})+1=
                     (\sum_{C\in S-\{C^*\}} \mi{count}(C)-1)+\mi{count}(C^*)-1=\sum_{C\in (S-\{C^*\})\cup \{C^*\}} \mi{count}(C)-1=\mi{elmeasure}(S)=\sum_{C\in S} \mi{count}(C)-1$,
$\square\not\in S_2$,
$\mi{count}(a_0\swedge\cdots\swedge a_n\gle \gu\vee (C^\natural-\{\Cn_1\diamond^* \Cn_2\}))=\mi{count}(a_0\swedge\cdots\swedge a_n\gle \gu)+\mi{count}(C^\natural-\{\Cn_1\diamond^* \Cn_2\})=
                                                                                            \mi{count}(C^\natural-\{\Cn_1\diamond^* \Cn_2\})+1$, 
$\mi{count}(\Cn_2\diamond^{**} \Cn_1)=1$, 
$\mi{elmeasure}(S_2)=\sum_{C\in (S-\{C^*\})\cup \{a_0\swedge\cdots\swedge a_n\gle \gu\vee (C^\natural-\{\Cn_1\diamond^* \Cn_2\})\}\cup \{\Cn_2\diamond^{**} \Cn_1\}} \mi{count}(C)-1\leq 
                     (\sum_{C\in S-\{C^*\}} \mi{count}(C)-1)+\mi{count}(a_0\swedge\cdots\swedge a_n\gle \gu\vee (C^\natural-\{\Cn_1\diamond^* \Cn_2\}))-1+\mi{count}(\Cn_2\diamond^{**} \Cn_1)-1=
                     (\sum_{C\in S-\{C^*\}} \mi{count}(C)-1)+\mi{count}(C^\natural-\{\Cn_1\diamond^* \Cn_2\})<(\sum_{C\in S-\{C^*\}} \mi{count}(C)-1)+\mi{count}(C^\natural-\{\Cn_1\diamond^* \Cn_2\})+1=
                     \mi{elmeasure}(S)$.

We get in all Cases 2.2.1.1, 2.2.1.2, 2.2.2.1.1, 2.2.2.1.2, 2.2.2.2.1, and 2.2.2.2.2 that
for both $i$, 
$S_i$ is positive; 
$\mi{elmeasure}(S_i)<\mi{elmeasure}(S)$;
by the induction hypothesis for $S_i$, there exists a finite {\it DPLL}-tree $\mi{Tree}_i$ with the root $S_i$ constructed using Rules (\cref{ceq4hr0}), (\cref{ceq4hr11}), (\cref{ceq4hr111}), (\cref{ceq4hr66}) satisfying that
(\ref{eq5a}) and (\ref{eq5b}) hold for $S_i$ and $\mi{Tree}_i$.
We put
\begin{alignat*}{1}
\mi{Tree}_2^* &= \left\{\begin{array}{ll}
                        \dfrac{\mi{Tree}_2'}
                              {\mi{Tree}_2} &\ \ \text{\it in {\rm Cases 2.2.2.1.1} and {\rm 2.2.2.1.2}}, \\[1mm]
                        \mi{Tree}_2         &\ \ \text{\it in {\rm Cases 2.2.1.1, 2.2.1.2, 2.2.2.2.1,} and {\rm 2.2.2.2.2}}; 
                        \end{array}
                 \right. \\[1mm]
\mi{Tree}     &= \dfrac{S}
                       {\mi{Tree}_1\ \big|\ \mi{Tree}_2^*}.
\end{alignat*}
Note that in Cases 2.2.2.1.1 and 2.2.2.1.2, $\mi{Tree}_2^*$ is a finite {\it DPLL}-tree with the root $S_2'\cup \{a_0\gle \gu\}$ constructed using Rules (\cref{ceq4hr0}), (\cref{ceq4hr11}), (\cref{ceq4hr111}), (\cref{ceq4hr66});
in Cases 2.2.1.1, 2.2.1.2, 2.2.2.2.1, and 2.2.2.2.2, $\mi{Tree}_2^*$ is a finite {\it DPLL}-tree with the root $S_2$ constructed using Rules (\cref{ceq4hr0}), (\cref{ceq4hr11}), (\cref{ceq4hr111}), (\cref{ceq4hr66});
$\mi{Tree}$ is a finite {\it DPLL}-tree with the root $S$ constructed using Rules (\cref{ceq4hr0}), (\cref{ceq4hr11}), (\cref{ceq4hr111}), (\cref{ceq4hr66}).
We get two cases for $S$.

Case 2.2.3:
$S$ is unsatisfiable.
Then, in Cases 2.2.2.1.1 and 2.2.2.1.2,
by Lemma \ref{le33333} for $n=2$, $S_2'\cup \{a_0\gle \gu\}$, and $\dfrac{S}
                                                                         {S_1\ \big|\ S_2'\cup \{a_0\gle \gu\}}$,
$S_1$ and $S_2'\cup \{a_0\gle \gu\}$ are unsatisfiable;
by Lemma \ref{le33}(ii) for $S_2'\cup \{a_0\gle \gu\}$, $\mi{Tree}_2'$, and $S_2$, $S_2$ is unsatisfiable;
in Cases 2.2.1.1, 2.2.1.2, 2.2.2.2.1, and 2.2.2.2.2,
by Lemma \ref{le33333} for $n=2$ and $\dfrac{S}
                                            {S_1\ \big|\ S_2}$,
$S_1$ and $S_2$ are unsatisfiable;
$S_1$ and $S_2$ are unsatisfiable;
for both $i$, by (\ref{eq5a}) for $S_i$ and $\mi{Tree}_i$, $\mi{Tree}_i$ is closed;
$\mi{Tree}_2$ is closed;
in Cases 2.2.2.1.1 and 2.2.2.1.2, 
$\mi{Tree}_2^*=\dfrac{\mi{Tree}_2'}
                     {\mi{Tree}_2}$ is closed;
in Cases 2.2.1.1, 2.2.1.2, 2.2.2.2.1, and 2.2.2.2.2, $\mi{Tree}_2^*=\mi{Tree}_2$ is closed;
$\mi{Tree}_2^*$ is closed;
$\mi{Tree}_1$ is closed;
$\mi{Tree}=\dfrac{S}
                 {\mi{Tree}_1\ \big|\ \mi{Tree}_2^*}$ is closed;
(\ref{eq5a}) holds and (\ref{eq5b}) holds trivially.

Case 2.2.4: 
$S$ is satisfiable.
Then, in Cases 2.2.2.1.1 and 2.2.2.1.2,
by Lemma \ref{le33333} for $n=2$, $S_2'\cup \{a_0\gle \gu\}$, and $\dfrac{S}
                                                                         {S_1\ \big|\ S_2'\cup \{a_0\gle \gu\}}$,
$S_1$ or $S_2'\cup \{a_0\gle \gu\}$ is satisfiable;
if $S_2'\cup \{a_0\gle \gu\}$ is satisfiable,
by Lemma \ref{le33}(i) for $S_2'\cup \{a_0\gle \gu\}$, $\mi{Tree}_2'$ with its only leaf $S_2$, and $S_2$, $S_2$ is satisfiable;
$S_1$ or $S_2$ is satisfiable;
in Cases 2.2.1.1, 2.2.1.2, 2.2.2.2.1, and 2.2.2.2.2,
by Lemma \ref{le33333} for $n=2$ and $\dfrac{S}
                                            {S_1\ \big|\ S_2}$,
$S_1$ or $S_2$ is satisfiable;
$S_1$ or $S_2$ is satisfiable;
there exists $1\leq i^*\leq 2$ satisfying that $S_{i^*}$ is satisfiable;
by (\ref{eq5b}) for $S_{i^*}$ and $\mi{Tree}_{i^*}$, 
$\mi{Tree}_{i^*}$ is open; 
there exists a model ${\mf A}_{i^*}$ of $S_{i^*}$ related to $\mi{Tree}_{i^*}$.
We get two cases for $i^*$.

Case 2.2.4.1: 
$i^*=1$.
Then $\mi{Tree}_{i^*}=\mi{Tree}_1$ is open, and there exists a model ${\mf A}_{i^*}={\mf A}_1$ of $S_{i^*}=S_1$ related to $\mi{Tree}_{i^*}=\mi{Tree}_1$.
We put ${\mf A}={\mf A}_1$.
Hence, $\mi{Tree}=\dfrac{S}
                        {\mi{Tree}_1\ \big|\ \mi{Tree}_2^*}$ is open;
${\mf A}\models S_1$;
in Cases 2.2.2.1.1 and 2.2.2.1.2,
by Lemma \ref{le333}(ii) for $n=2$, $S_2'\cup \{a_0\gle \gu\}$, and $\dfrac{S}
                                                                           {S_1\ \big|\ S_2'\cup \{a_0\gle \gu\}}$,
${\mf A}\models S$; 
${\mf A}$ is a model of $S_1$ related to $\mi{Tree}_1$;
${\mf A}$ is a model of $S$ related to $\mi{Tree}=\dfrac{S}
                                                        {\mi{Tree}_1\ \big|\ \mi{Tree}_2^*}$;
in Cases 2.2.1.1, 2.2.1.2, 2.2.2.2.1, and 2.2.2.2.2,
by Lemma \ref{le333}(ii) for $n=2$ and $\dfrac{S}
                                              {S_1\ \big|\ S_2}$,
${\mf A}\models S$; 
${\mf A}$ is a model of $S_1$ related to $\mi{Tree}_1$;
${\mf A}$ is a model of $S$ related to $\mi{Tree}=\dfrac{S}
                                                        {\mi{Tree}_1\ \big|\ \mi{Tree}_2^*}$;
${\mf A}$ is a model of $S$ related to $\mi{Tree}$;
(\ref{eq5a}) holds trivially and (\ref{eq5b}) holds.

Case 2.2.4.2: 
$i^*=2$.
Then $\mi{Tree}_{i^*}=\mi{Tree}_2$ is open; 
there exists a model ${\mf A}_{i^*}={\mf A}_2$ of $S_{i^*}=S_2$ related to $\mi{Tree}_{i^*}=\mi{Tree}_2$;
in Cases 2.2.2.1.1 and 2.2.2.1.2, 
$\mi{Tree}_2^*=\dfrac{\mi{Tree}_2'}
                     {\mi{Tree}_2}$ is open;
in Cases 2.2.1.1, 2.2.1.2, 2.2.2.2.1, and 2.2.2.2.2, $\mi{Tree}_2^*=\mi{Tree}_2$ is open;
$\mi{Tree}_2^*$ is open;
$\mi{Tree}=\dfrac{S}
                 {\mi{Tree}_1\ \big|\ \mi{Tree}_2^*}$ is open;
in Cases 2.2.2.1.1 and 2.2.2.1.2,
by Lemma \ref{le33}(ii) for $S_2'\cup \{a_0\gle \gu\}$, $\mi{Tree}_2'$, $S_2$, and ${\mf A}_2$, there exists a model ${\mf A}$ of $S_2'\cup \{a_0\gle \gu\}$ 
related to $\mi{Tree}_2'$ and $\mi{Tree}_2^*=\dfrac{\mi{Tree}_2'}
                                                   {\mi{Tree}_2}$;
by Lemma \ref{le333}(ii) for $n=2$, $S_2'\cup \{a_0\gle \gu\}$, and $\dfrac{S}
                                                                           {S_1\ \big|\ S_2'\cup \{a_0\gle \gu\}}$,
${\mf A}\models S$;
${\mf A}$ is a model of $S$ related to $\mi{Tree}=\dfrac{S}
                                                        {\mi{Tree}_1\ \big|\ \mi{Tree}_2^*}$;
in Cases 2.2.1.1, 2.2.1.2, 2.2.2.2.1, and 2.2.2.2.2, 
we put ${\mf A}={\mf A}_2$;
${\mf A}\models S_2$;
by Lemma \ref{le333}(ii) for $n=2$ and $\dfrac{S}
                                              {S_1\ \big|\ S_2}$,
${\mf A}\models S$; 
${\mf A}$ is a model of $S_2$ related to $\mi{Tree}_2^*=\mi{Tree}_2$;
${\mf A}$ is a model of $S$ related to $\mi{Tree}=\dfrac{S}
                                                        {\mi{Tree}_1\ \big|\ \mi{Tree}_2^*}$;
${\mf A}$ is a model of $S$ related to $\mi{Tree}$;
(\ref{eq5a}) holds trivially and (\ref{eq5b}) holds.

So, in both Cases 1 and 2, (\ref{eq5a}) and (\ref{eq5b}) hold.
The induction is completed.
%
%
%
\end{proof}

\subsection{Full proof of Lemma \ref{le555}}
\label{S7.7a}

\begin{proof}
We distinguish two cases for $S$.

Case 1:
$\square\in S$.
We put $\mi{Tree}=S$.
Hence, $\mi{Tree}$ is a finite linear {\it DPLL}-tree with the root $S$ constructed using Rules (\cref{ceq4hr2}) and (\cref{ceq4hr22}) such that
for its only leaf $S$, $\square\in S$.

Case 2:
$\square\not\in S$.
Let $C^F\in \mi{OrdPropCl}$.
We define a measure operator $\mi{unsimplified}(C^F)=\{l \,|\, l\in C^F\ \text{\it is either a contradiction or tautology}\}$.
Then $C^*\in S\subseteq_{\mc F} \mi{OrdPropCl}$.
We proceed by induction on $\mi{unsimplified}(C^*)\subseteq C^*\subseteq_{\mc F} \mi{OrdPropLit}$.

Case 2.1 (the base case):
$\mi{unsimplified}(C^*)=\emptyset$.
Then $C^*\in S$, $\square\not\in S$;
for all $l\in C^*$, $l$ is not a contradiction or tautology;
$C^*\neq \square$ does not contain contradictions and tautologies.
We put $\mi{Tree}=S$.
Hence, $\mi{Tree}$ is a finite linear {\it DPLL}-tree with the root $S$ constructed using Rules (\cref{ceq4hr2}) and (\cref{ceq4hr22}) such that
for its only leaf $S$, $\square\not\in S\subseteq_{\mc F} \mi{OrdPropCl}$ and $C^*\neq \square$ does not contain contradictions and tautologies.

Case 2.2 (the induction case):
$\emptyset\neq \mi{unsimplified}(C^*)\subseteq_{\mc F} \mi{OrdPropLit}$.
Then there exists $l^*\in \mi{unsimplified}(C^*)\subseteq C^*\subseteq_{\mc F} \mi{OrdPropLit}$ satisfying that $l^*$ is either a contradiction or tautology.
We get two cases for $l^*$.

Case 2.2.1:
$l^*$ is a contradiction.
Then $C^*\in S$;
$l^*\in C^*$ is not a guard;
$C^*\not\in \mi{guards}(S)$, 
$l^*\in C^*\in S-\mi{guards}(S)$,
applying Rule (\cref{ceq4hr2}) to $C^*$ and $l^*$, we derive
\begin{equation}
\notag
\dfrac{S}
      {(S-\{C^*\})\cup \{C^*-\{l^*\}\}}.
\end{equation}
We put $S'=(S-\{C^*\})\cup \{C^*-\{l^*\}\}\subseteq_{\mc F} \mi{OrdPropCl}$.
We get two cases for $C^*-\{l^*\}$.

Case 2.2.1.1:
$C^*-\{l^*\}=\square$.
Then $C^*-\{l^*\}=\square\in S'$.
We put 
\begin{equation}
\notag
\mi{Tree}=\dfrac{S}
                {S'}.
\end{equation}
Hence, $\mi{Tree}$ is a finite linear {\it DPLL}-tree with the root $S$ constructed using Rules (\cref{ceq4hr2}) and (\cref{ceq4hr22}) such that
for its only leaf $S'$, $\square\in S'$.

Case 2.2.1.2:
$C^*-\{l^*\}\neq \square$.
We get two cases for $C^*-\{l^*\}$.

Case 2.2.1.2.1:
$C^*-\{l^*\}\in S$.
Then $l^*\in C^*$, $C^*-\{l^*\}\neq C^*$, $C^*-\{l^*\}\in S-\{C^*\}$,
$S'=(S-\{C^*\})\cup \{C^*-\{l^*\}\}=S-\{C^*\}$,
$\square\not\in S\supseteq S'=S-\{C^*\}$.
We put
\begin{equation}
\notag
\mi{Tree}=\dfrac{S}
                {S'}.
\end{equation}
Hence, $\mi{Tree}$ is a finite linear {\it DPLL}-tree with the root $S$ constructed using Rules (\cref{ceq4hr2}) and (\cref{ceq4hr22}) such that
for its only leaf $S'$, $\square\not\in S'\subseteq_{\mc F} \mi{OrdPropCl}$ and $S'=S-\{C^*\}$.

Case 2.2.1.2.2:
$C^*-\{l^*\}\not\in S$.
Then $C^*-\{l^*\}\in S'$, $l^*\in \mi{unsimplified}(C^*)$,
$\mi{unsimplified}(C^*-\{l^*\})=\{l \,|\, l\in C^*-\{l^*\}\ \text{\it is either a contradiction or tautology}\}=
                                \mi{unsimplified}(C^*)-\{l^*\}=\{l \,|\, l\in C^*\ \text{\it is either a contradiction or tautology}\}-\{l^*\}\subset \mi{unsimplified}(C^*)$;
by the induction hypothesis for $S'$ and $C^*-\{l^*\}$, there exists a finite linear {\it DPLL}-tree $\mi{Tree}'$ with the root $S'$ constructed using Rules (\cref{ceq4hr2}) and (\cref{ceq4hr22}) satisfying
for its only leaf $S''$ that either $\square\in S''$, or $\square\not\in S''\subseteq_{\mc F} \mi{OrdPropCl}$ and exactly one of the following points holds.
\begin{enumerate}[\rm (a)]
\item
$S''=S'$, $C^*-\{l^*\}\neq \square$ does not contain contradictions and tautologies;
\item
$S''=S'-\{C^*-\{l^*\}\}$;
\item
there exists $C^{**}\in \mi{OrdPropCl}$ satisfying that 
$S''=(S'-\{C^*-\{l^*\}\})\cup \{C^{**}\}$, $C^{**}\not\in S'$, $\square\neq C^{**}\subset C^*-\{l^*\}$ does not contain contradictions and tautologies.
\end{enumerate}
We get that 
$C^*-\{l^*\}\not\in S$,
$S'-\{C^*-\{l^*\}\}=((S-\{C^*\})\cup \{C^*-\{l^*\}\})-\{C^*-\{l^*\}\}=((S-\{C^*\})-\{C^*-\{l^*\}\})\cup (\{C^*-\{l^*\}\}-\{C^*-\{l^*\}\})=S-\{C^*\}$;
\begin{enumerate}[\rm (a)]
\item
$l^*\in C^*$;
we put $C^{**}=C^*-\{l^*\}\in \mi{OrdPropCl}$;
$C^{**}=C^*-\{l^*\}\not\in S$,
$S''=S'=(S-\{C^*\})\cup \{C^*-\{l^*\}\}=(S-\{C^*\})\cup \{C^{**}\}$;
$\square\neq C^{**}=C^*-\{l^*\}\subset C^*$ does not contain contradictions and tautologies;
$C^{**}\in S''$, $S''\neq S$;
\item
$C^*\in S$, $S''=S'-\{C^*-\{l^*\}\}=S-\{C^*\}\neq S$;
\item
there exists $C^{**}\in \mi{OrdPropCl}$ satisfying that 
$S''=(S'-\{C^*-\{l^*\}\})\cup \{C^{**}\}=(S-\{C^*\})\cup \{C^{**}\}$ and
$\square\neq C^{**}\subset C^*-\{l^*\}\subset C^*$ does not contain contradictions and tautologies;
we have that $l^*$ is a contradiction;
$l^*\not\in C^{**}$, $C^{**}\neq C^*$,
$C^{**}\not\in S'\supseteq S-\{C^*\}$, $C^{**}\not\in S$, $C^{**}\in S''$, $S''\neq S$.
\end{enumerate}
We put
\begin{equation}
\notag
\mi{Tree}=\dfrac{S}
                {\mi{Tree}'}.
\end{equation}
Hence, $\mi{Tree}$ is a finite linear {\it DPLL}-tree with the root $S$ constructed using Rules (\cref{ceq4hr2}) and (\cref{ceq4hr22}) such that
for its only leaf $S''$, either $\square\in S''$, or $\square\not\in S''\subseteq_{\mc F} \mi{OrdPropCl}$, $S''\neq S$, and exactly one of the following points holds.
\begin{enumerate}[\rm (a)]
\item
$S''=S-\{C^*\}$;
\item
there exists $C^{**}\in \mi{OrdPropCl}$ satisfying that $S''=(S-\{C^*\})\cup \{C^{**}\}$, $C^{**}\not\in S$, $\square\neq C^{**}\subset C^*$ does not contain contradictions and tautologies.
\end{enumerate}

Case 2.2.2:
$l^*$ is a tautology.
Then $C^*\in S$;
$l^*\in C^*$ is not a guard;
$C^*\not\in \mi{guards}(S)$, 
$l^*\in C^*\in S-\mi{guards}(S)$,
applying Rule (\cref{ceq4hr22}) to $C^*$ and $l^*$, we derive
\begin{equation}
\notag
\dfrac{S}
      {S-\{C^*\}}.
\end{equation}
We put $S'=S-\{C^*\}\subseteq_{\mc F} \mi{OrdPropCl}$.
Hence, this case is the same as Case 2.2.1.2.1.

So, in both Cases 2.1 and 2.2, the statement holds.
The induction is completed.
Thus, in both Cases 1 and 2, the statement holds.
%
%
%
\end{proof}

\subsection{Full proof of Lemma \ref{le55}}
\label{S7.7b}

\begin{proof}
We distinguish two cases for $S$.

Case 1:
$\square\in S$.
We put $\mi{Tree}=S$.
Hence, $\mi{Tree}$ is a finite linear {\it DPLL}-tree with the root $S$ constructed using Rules (\cref{ceq4hr2}) and (\cref{ceq4hr22}) such that
for its only leaf $S$, $\square\in S$.

Case 2:
$\square\not\in S$.
Let $S^F\subseteq \mi{OrdPropCl}$.
We define a measure operator $\mi{unsimplified}(S^F)=\{C \,|\, C\in S^F\ \text{\it contains a contradiction or tautology}\}$.
We proceed by induction on $\mi{unsimplified}(S)\subseteq S\subseteq_{\mc F} \mi{OrdPropCl}$.

Case 2.1 (the base case):
$\mi{unsimplified}(S)=\emptyset$.
Then $\square\not\in S$;
for all $C\in S$, $C$ does not contain contradictions and tautologies;
$S$ is simplified.
We put $\mi{Tree}=S$.
Hence, $\mi{Tree}$ is a finite linear {\it DPLL}-tree with the root $S$ constructed using Rules (\cref{ceq4hr2}) and (\cref{ceq4hr22}) such that
for its only leaf $S$, $S\subseteq_{\mc F} \mi{OrdPropCl}$ is simplified.

Case 2.2 (the induction case):
$\emptyset\neq \mi{unsimplified}(S)\subseteq_{\mc F} \mi{OrdPropCl}$.
Then there exists $C^*\in \mi{unsimplified}(S)\subseteq S$ satisfying that $C^*$ contains a contradiction or tautology;
by Lemma \ref{le555}, there exists a finite linear {\it DPLL}-tree $\mi{Tree}'$ with the root $S$ constructed using Rules (\cref{ceq4hr2}) and (\cref{ceq4hr22}) satisfying
for its only leaf $S'$ that either $\square\in S'$, or $\square\not\in S'\subseteq_{\mc F} \mi{OrdPropCl}$ and exactly one of the following points holds.
\begin{enumerate}[\rm (a)]
\item
$S'=S$, $C^*\neq \square$ does not contain contradictions and tautologies;
\item
$S'=S-\{C^*\}$;
\item
there exists $C^{**}\in \mi{OrdPropCl}$ satisfying that $S'=(S-\{C^*\})\cup \{C^{**}\}$, $C^{**}\not\in S$, $\square\neq C^{**}\subset C^*$ does not contain contradictions and tautologies.
\end{enumerate}
We get four cases for $S'$.

Case 2.2.1:
$\square\in S'$.
We put $\mi{Tree}=\mi{Tree}'$.
Hence, $\mi{Tree}$ is a finite linear {\it DPLL}-tree with the root $S$ constructed using Rules (\cref{ceq4hr2}) and (\cref{ceq4hr22}) such that
for its only leaf $S'$, $\square\in S'$.

Case 2.2.2:
$S'=S$ and $C^*\neq \square$ does not contain contradictions and tautologies.
We have that $C^*$ contains a contradiction or tautology,
which is a contradiction.

Case 2.2.3:
$S'=S-\{C^*\}$.
Then $\square\not\in S'$, $C^*\in \mi{unsimplified}(S)$,
$\mi{unsimplified}(S')=\{C \,|\, C\in S-\{C^*\}\ \text{\it contains a contradiction or tautology}\}=
                       \mi{unsimplified}(S)-\{C^*\}=\{C \,|\, C\in S\ \text{\it contains a contradiction or tautology}\}-\{C^*\}\subset \mi{unsimplified}(S)$;
by the induction hypothesis for $S'$, there exists a finite linear {\it DPLL}-tree $\mi{Tree}''$ with the root $S'$ constructed using Rules (\cref{ceq4hr2}) and (\cref{ceq4hr22}) satisfying 
for its only leaf $S''$ that either $\square\in S''$ or $S''\subseteq_{\mc F} \mi{OrdPropCl}$ is simplified. 
We put
\begin{equation}
\notag
\mi{Tree}=\dfrac{\mi{Tree}'} 
                {\mi{Tree}''}.
\end{equation}
Hence, $\mi{Tree}$ is a finite linear {\it DPLL}-tree with the root $S$ constructed using Rules (\cref{ceq4hr2}) and (\cref{ceq4hr22}) such that
for its only leaf $S''$, either $\square\in S''$ or $S''\subseteq_{\mc F} \mi{OrdPropCl}$ is simplified.

Case 2.2.4:
There exists $C^{**}\in \mi{OrdPropCl}$ such that $S'=(S-\{C^*\})\cup \{C^{**}\}$, $C^{**}\not\in S$, $\square\neq C^{**}\subset C^*$ does not contain contradictions and tautologies.
Then $\square\not\in S'$, $C^*\in \mi{unsimplified}(S)$,
$\mi{unsimplified}(S')=\{C \,|\, C\in (S-\{C^*\})\cup \{C^{**}\}\ \text{\it contains a contradiction or tautology}\}=\{C \,|\, C\in S-\{C^*\}\ \text{\it contains a contradiction or tautology}\}=
                       \mi{unsimplified}(S)-\{C^*\}=\{C \,|\, C\in S\ \text{\it contains a contradiction or tautology}\}-\{C^*\}\subset \mi{unsimplified}(S)$;
by the induction hypothesis for $S'$, there exists a finite linear {\it DPLL}-tree $\mi{Tree}''$ with the root $S'$ constructed using Rules (\cref{ceq4hr2}) and (\cref{ceq4hr22}) satisfying 
for its only leaf $S''$ that either $\square\in S''$ or $S''\subseteq_{\mc F} \mi{OrdPropCl}$ is simplified. 
We put
\begin{equation}
\notag
\mi{Tree}=\dfrac{\mi{Tree}'} 
                {\mi{Tree}''}.
\end{equation}
Hence, $\mi{Tree}$ is a finite linear {\it DPLL}-tree with the root $S$ constructed using Rules (\cref{ceq4hr2}) and (\cref{ceq4hr22}) such that
for its only leaf $S''$, either $\square\in S''$ or $S''\subseteq_{\mc F} \mi{OrdPropCl}$ is simplified.

So, in both Cases 2.1 and 2.2, the statement holds.
The induction is completed.
Thus, in both Cases 1 and 2, the statement holds.
%
%
%
\end{proof}

\subsection{Full proof of Lemma \ref{le5}}
\label{S7.7c}

\begin{proof}
Let $S^F\subseteq \mi{OrdPropCl}$.
We define a measure operator $\mi{unguarded}(S^F)=\{a \,|\, a\in \mi{atoms}(S^F)\ \text{\it is not $\gz$-guarded in}\ S^F\}$.
Then $S\subseteq_{\mc F} \mi{OrdPropCl}$.
We proceed by induction on $\mi{unguarded}(S)\subseteq \mi{atoms}(S)\subseteq_{\mc F} \mi{PropAtom}$.

Case 1 (the base case):
$\mi{unguarded}(S)=\emptyset$.
We have that $S$ is simplified.
Then, for all $a\in \mi{atoms}(S)$, $a$ is $\gz$-guarded in $S$;
$S$ is $\gz$-guarded.
We put $\mi{Tree}=S$.
Hence, $\mi{Tree}$ is a finite {\it DPLL}-tree with the root $S$ constructed using Rules (\cref{ceq4hr1}), (\cref{ceq4hr1111111})--(\cref{ceq4hr4}) such that
for its only leaf $S$, $S\subseteq_{\mc F} \mi{OrdPropCl}$ is $\gz$-guarded.

Case 2 (the induction case):
$\emptyset\neq \mi{unguarded}(S)\subseteq_{\mc F} \mi{PropAtom}$.
Then there exists $a^*\in \mi{unguarded}(S)\subseteq \mi{atoms}(S)$ not satisfying that $a^*$ is $\gz$-guarded in $S$.
We distinguish two cases for $S$.

Case 2.1:
$S$ is a contradictory set of guards for $a^*$. 
Then $\mi{guards}(S,a^*)\supseteq \{a^*\geql \gz,\gz\gle a^*\}$ or $\mi{guards}(S,a^*)\supseteq \{a^*\gleq \gz,\gz\gle a^*\}$ or
$\mi{guards}(S,a^*)\supseteq \{a^*\geql \gz,a^*\geql \gu\}$ or $\mi{guards}(S,a^*)\supseteq \{a^*\gleq \gz,a^*\geql \gu\}$ or $\mi{guards}(S,a^*)\supseteq \{a^*\geql \gz,\gu\gleq a^*\}$ or 
$\mi{guards}(S,a^*)\supseteq \{a^*\gleq \gz,\gu\gleq a^*\}$ or
$\mi{guards}(S,a^*)\supseteq \{a^*\gle \gu,a^*\geql \gu\}$ or $\mi{guards}(S,a^*)\supseteq \{a^*\gle \gu,\gu\gleq a^*\}$.
We get eight cases for $\mi{guards}(S,a^*)$.

Case 2.1.1:
$\mi{guards}(S,a^*)\supseteq \{a^*\geql \gz,\gz\gle a^*\}$.
Then $S\supseteq \mi{guards}(S)\supseteq \mi{guards}(S,a^*)\supseteq \{a^*\geql \gz,\gz\gle a^*\}$,
$a^*\geql \gz\in \mi{guards}(S)$, $\gz\gle a^*\in S$, 
$a^*\in \mi{atoms}(\gz\gle a^*)$, $a^*\geql \gz\neq \gz\gle a^*$, $\mi{simplify}(\gz\gle a^*,a^*,\gz)=\gz\gle \gz$,
applying Rule (\cref{ceq4hr3}) to $a^*\geql \gz$ and $\gz\gle a^*$, we derive
\begin{equation} \notag
\dfrac{S}
      {(S-\{\gz\gle a^*\})\cup \{\gz\gle \gz\}};
\end{equation}
$\gz\gle \gz\in (S-\{\gz\gle a^*\})\cup \{\gz\gle \gz\}$;
$\gz\gle \gz\in \mi{OrdPropLit}$ is a contradiction;
$\gz\gle \gz$ is not a guard;
$\gz\gle \gz\not\in \mi{guards}((S-\{\gz\gle a^*\})\cup \{\gz\gle \gz\})$,
$\gz\gle \gz\in ((S-\{\gz\gle a^*\})\cup \{\gz\gle \gz\})-\mi{guards}((S-\{\gz\gle a^*\})\cup \{\gz\gle \gz\})$,
applying Rule (\cref{ceq4hr2}) to $(S-\{\gz\gle a^*\})\cup \{\gz\gle \gz\}$ and $\gz\gle \gz$, we derive
\begin{equation} \notag
\dfrac{(S-\{\gz\gle a^*\})\cup \{\gz\gle \gz\}}
      {(((S-\{\gz\gle a^*\})\cup \{\gz\gle \gz\})-\{\gz\gle \gz\})\cup \{\square\}}.
\end{equation}
We put 
\begin{equation} \notag
\mi{Tree}=\begin{array}[c]{c}
          S \\[0.4mm]
          \hline \\[-3.8mm]
          (S-\{\gz\gle a^*\})\cup \{\gz\gle \gz\} \\[0.4mm]
          \hline \\[-3.8mm]
          S'=(((S-\{\gz\gle a^*\})\cup \{\gz\gle \gz\})-\{\gz\gle \gz\})\cup \{\square\}.
          \end{array} 
\end{equation}
Hence, $\mi{Tree}$ is a finite {\it DPLL}-tree with the root $S$ constructed using Rules (\cref{ceq4hr1}), (\cref{ceq4hr1111111})--(\cref{ceq4hr4}) such that
for its only leaf $S'$, $\square\in S'$.

Case 2.1.2:
$\mi{guards}(S,a^*)\supseteq \{a^*\gleq \gz,\gz\gle a^*\}$.
Then $\mi{guards}(S)\supseteq \mi{guards}(S,a^*)\supseteq \{a^*\gleq \gz,\gz\gle a^*\}$,
$a^*\gleq \gz\in \mi{guards}(S)$, applying Rule (\cref{ceq4hr1111111}) to $a^*\gleq \gz$, we derive
\begin{equation} \notag
\dfrac{S}
      {(S-\{a^*\gleq \gz\})\cup \{a^*\geql \gz\}};
\end{equation}
$a^*\in \mi{atoms}((S-\{a^*\gleq \gz\})\cup \{a^*\geql \gz\})$,
$(S-\{a^*\gleq \gz\})\cup \{a^*\geql \gz\}\supseteq \mi{guards}((S-\{a^*\gleq \gz\})\cup \{a^*\geql \gz\})\supseteq 
 \mi{guards}((S-\{a^*\gleq \gz\})\cup \{a^*\geql \gz\},a^*)=((S-\{a^*\gleq \gz\})\cup \{a^*\geql \gz\})\cap \mi{guards}(a^*)=
 ((S-\{a^*\gleq \gz\})\cap \mi{guards}(a^*))\cup (\{a^*\geql \gz\}\cap \mi{guards}(a^*))=((S\cap \mi{guards}(a^*))-\{a^*\gleq \gz\})\cup \{a^*\geql \gz\}=
 (\mi{guards}(S,a^*)-\{a^*\gleq \gz\})\cup \{a^*\geql \gz\}\supseteq (\{a^*\gleq \gz,\gz\gle a^*\}-\{a^*\gleq \gz\})\cup \{a^*\geql \gz\}=\{a^*\geql \gz,\gz\gle a^*\}$,
$a^*\geql \gz\in \mi{guards}((S-\{a^*\gleq \gz\})\cup \{a^*\geql \gz\})$, $\gz\gle a^*\in (S-\{a^*\gleq \gz\})\cup \{a^*\geql \gz\}$, 
$a^*\in \mi{atoms}(\gz\gle a^*)$, $a^*\geql \gz\neq \gz\gle a^*$, $\mi{simplify}(\gz\gle a^*,a^*,\gz)=\gz\gle \gz$,
applying Rule (\cref{ceq4hr3}) to $(S-\{a^*\gleq \gz\})\cup \{a^*\geql \gz\}$, $a^*\geql \gz$, and $\gz\gle a^*$, we derive
\begin{equation} \notag
\dfrac{(S-\{a^*\gleq \gz\})\cup \{a^*\geql \gz\}}
      {(((S-\{a^*\gleq \gz\})\cup \{a^*\geql \gz\})-\{\gz\gle a^*\})\cup \{\gz\gle \gz\}};
\end{equation}
$\gz\gle \gz\in (((S-\{a^*\gleq \gz\})\cup \{a^*\geql \gz\})-\{\gz\gle a^*\})\cup \{\gz\gle \gz\}$;
$\gz\gle \gz\in \mi{OrdPropLit}$ is a contradiction;
$\gz\gle \gz$ is not a guard;
$\gz\gle \gz\not\in \mi{guards}((((S-\{a^*\gleq \gz\})\cup \{a^*\geql \gz\})-\{\gz\gle a^*\})\cup \{\gz\gle \gz\})$,
$\gz\gle \gz\in ((((S-\{a^*\gleq \gz\})\cup \{a^*\geql \gz\})-\{\gz\gle a^*\})\cup \{\gz\gle \gz\})-\mi{guards}((((S-\{a^*\gleq \gz\})\cup \{a^*\geql \gz\})-\{\gz\gle a^*\})\cup \{\gz\gle \gz\})$,
applying Rule (\cref{ceq4hr2}) to $(((S-\{a^*\gleq \gz\})\cup \{a^*\geql \gz\})-\{\gz\gle a^*\})\cup \{\gz\gle \gz\}$ and $\gz\gle \gz$, we derive
\begin{equation} \notag
\dfrac{(((S-\{a^*\gleq \gz\})\cup \{a^*\geql \gz\})-\{\gz\gle a^*\})\cup \{\gz\gle \gz\}}
      {(((((S-\{a^*\gleq \gz\})\cup \{a^*\geql \gz\})-\{\gz\gle a^*\})\cup \{\gz\gle \gz\})-\{\gz\gle \gz\})\cup \{\square\}}.
\end{equation}
We put 
\begin{equation} \notag
\mi{Tree}=\begin{array}[c]{c}
          S \\[0.4mm]
          \hline \\[-3.8mm]
          (S-\{a^*\gleq \gz\})\cup \{a^*\geql \gz\} \\[0.4mm]
          \hline \\[-3.8mm]
          (((S-\{a^*\gleq \gz\})\cup \{a^*\geql \gz\})-\{\gz\gle a^*\})\cup \{\gz\gle \gz\} \\[0.4mm]
          \hline \\[-3.8mm]
          S'=(((((S-\{a^*\gleq \gz\})\cup \{a^*\geql \gz\})-\{\gz\gle a^*\})\cup \{\gz\gle \gz\})- \\
          \hfill \{\gz\gle \gz\})\cup \{\square\}.
          \end{array} 
\end{equation}
Hence, $\mi{Tree}$ is a finite {\it DPLL}-tree with the root $S$ constructed using Rules (\cref{ceq4hr1}), (\cref{ceq4hr1111111})--(\cref{ceq4hr4}) such that
for its only leaf $S'$, $\square\in S'$.

Case 2.1.3:
$\mi{guards}(S,a^*)\supseteq \{a^*\geql \gz,a^*\geql \gu\}$.
Then $S\supseteq \mi{guards}(S)\supseteq \mi{guards}(S,a^*)\supseteq \{a^*\geql \gz,a^*\geql \gu\}$,
$a^*\geql \gz\in \mi{guards}(S)$, $a^*\geql \gu\in S$,
$a^*\in \mi{atoms}(a^*\geql \gu)$, $a^*\geql \gz\neq a^*\geql \gu$, $\mi{simplify}(a^*\geql \gu,a^*,\gz)=\gz\geql \gu$,
applying Rule (\cref{ceq4hr3}) to $a^*\geql \gz$ and $a^*\geql \gu$, we derive
\begin{equation} \notag
\dfrac{S}
      {(S-\{a^*\geql \gu\})\cup \{\gz\geql \gu\}};
\end{equation}
$\gz\geql \gu\in (S-\{a^*\geql \gu\})\cup \{\gz\geql \gu\}$;
$\gz\geql \gu\in \mi{OrdPropLit}$ is a contradiction;
$\gz\geql \gu$ is not a guard;
$\gz\geql \gu\not\in \mi{guards}((S-\{a^*\geql \gu\})\cup \{\gz\geql \gu\})$,
$\gz\geql \gu\in ((S-\{a^*\geql \gu\})\cup \{\gz\geql \gu\})-\mi{guards}((S-\{a^*\geql \gu\})\cup \{\gz\geql \gu\})$,
applying Rule (\cref{ceq4hr2}) to $(S-\{a^*\geql \gu\})\cup \{\gz\geql \gu\}$ and $\gz\geql \gu$, we derive
\begin{equation} \notag
\dfrac{(S-\{a^*\geql \gu\})\cup \{\gz\geql \gu\}}
      {(((S-\{a^*\geql \gu\})\cup \{\gz\geql \gu\})-\{\gz\geql \gu\})\cup \{\square\}}.
\end{equation}
We put 
\begin{equation} \notag
\mi{Tree}=\begin{array}[c]{c}
          S \\[0.4mm]
          \hline \\[-3.8mm]
          (S-\{a^*\geql \gu\})\cup \{\gz\geql \gu\} \\[0.4mm]
          \hline \\[-3.8mm]
          S'=(((S-\{a^*\geql \gu\})\cup \{\gz\geql \gu\})-\{\gz\geql \gu\})\cup \{\square\}.
          \end{array} 
\end{equation}
Hence, $\mi{Tree}$ is a finite {\it DPLL}-tree with the root $S$ constructed using Rules (\cref{ceq4hr1}), (\cref{ceq4hr1111111})--(\cref{ceq4hr4}) such that
for its only leaf $S'$, $\square\in S'$.

Case 2.1.4:
$\mi{guards}(S,a^*)\supseteq \{a^*\gleq \gz,a^*\geql \gu\}$.
Then $S\supseteq \mi{guards}(S)\supseteq \mi{guards}(S,a^*)\supseteq \{a^*\gleq \gz,a^*\geql \gu\}$,
$a^*\geql \gu\in \mi{guards}(S)$, $a^*\gleq \gz\in S$,
$a^*\in \mi{atoms}(a^*\gleq \gz)$, $a^*\geql \gu\neq a^*\gleq \gz$, $\mi{simplify}(a^*\gleq \gz,a^*,\gu)=\gu\gleq \gz$,
applying Rule (\cref{ceq4hr4}) to $a^*\geql \gu$ and $a^*\gleq \gz$, we derive
\begin{equation} \notag
\dfrac{S}
      {(S-\{a^*\gleq \gz\})\cup \{\gu\gleq \gz\}};
\end{equation}
$\gu\gleq \gz\in (S-\{a^*\gleq \gz\})\cup \{\gu\gleq \gz\}$;
$\gu\gleq \gz\in \mi{OrdPropLit}$ is a contradiction;
$\gu\gleq \gz$ is not a guard;
$\gu\gleq \gz\not\in \mi{guards}((S-\{a^*\gleq \gz\})\cup \{\gu\gleq \gz\})$,
$\gu\gleq \gz\in ((S-\{a^*\gleq \gz\})\cup \{\gu\gleq \gz\})-\mi{guards}((S-\{a^*\gleq \gz\})\cup \{\gu\gleq \gz\})$,
applying Rule (\cref{ceq4hr2}) to $(S-\{a^*\gleq \gz\})\cup \{\gu\gleq \gz\}$ and $\gu\gleq \gz$, we derive
\begin{equation} \notag
\dfrac{(S-\{a^*\gleq \gz\})\cup \{\gu\gleq \gz\}}
      {(((S-\{a^*\gleq \gz\})\cup \{\gu\gleq \gz\})-\{\gu\gleq \gz\})\cup \{\square\}}.
\end{equation}
We put 
\begin{equation} \notag
\mi{Tree}=\begin{array}[c]{c}
          S \\[0.4mm]
          \hline \\[-3.8mm]
          (S-\{a^*\gleq \gz\})\cup \{\gu\gleq \gz\} \\[0.4mm]
          \hline \\[-3.8mm]
          S'=(((S-\{a^*\gleq \gz\})\cup \{\gu\gleq \gz\})-\{\gu\gleq \gz\})\cup \{\square\}.
          \end{array} 
\end{equation}
Hence, $\mi{Tree}$ is a finite {\it DPLL}-tree with the root $S$ constructed using Rules (\cref{ceq4hr1}), (\cref{ceq4hr1111111})--(\cref{ceq4hr4}) such that
for its only leaf $S'$, $\square\in S'$.

Case 2.1.5:
$\mi{guards}(S,a^*)\supseteq \{a^*\geql \gz,\gu\gleq a^*\}$.
Then $S\supseteq \mi{guards}(S)\supseteq \mi{guards}(S,a^*)\supseteq \{a^*\geql \gz,\gu\gleq a^*\}$,
$a^*\geql \gz\in \mi{guards}(S)$, $\gu\gleq a^*\in S$,
$a^*\in \mi{atoms}(\gu\gleq a^*)$, $a^*\geql \gz\neq \gu\gleq a^*$, $\mi{simplify}(\gu\gleq a^*,a^*,\gz)=\gu\gleq \gz$,
applying Rule (\cref{ceq4hr3}) to $a^*\geql \gz$ and $\gu\gleq a^*$, we derive
\begin{equation} \notag
\dfrac{S}
      {(S-\{\gu\gleq a^*\})\cup \{\gu\gleq \gz\}};
\end{equation}
$\gu\gleq \gz\in (S-\{\gu\gleq a^*\})\cup \{\gu\gleq \gz\}$;
$\gu\gleq \gz\in \mi{OrdPropLit}$ is a contradiction;
$\gu\gleq \gz$ is not a guard;
$\gu\gleq \gz\not\in \mi{guards}((S-\{\gu\gleq a^*\})\cup \{\gu\gleq \gz\})$,
$\gu\gleq \gz\in ((S-\{\gu\gleq a^*\})\cup \{\gu\gleq \gz\})-\mi{guards}((S-\{\gu\gleq a^*\})\cup \{\gu\gleq \gz\})$,
applying Rule (\cref{ceq4hr2}) to $(S-\{\gu\gleq a^*\})\cup \{\gu\gleq \gz\}$ and $\gu\gleq \gz$, we derive
\begin{equation} \notag
\dfrac{(S-\{\gu\gleq a^*\})\cup \{\gu\gleq \gz\}}
      {(((S-\{\gu\gleq a^*\})\cup \{\gu\gleq \gz\})-\{\gu\gleq \gz\})\cup \{\square\}}.
\end{equation}
We put 
\begin{equation} \notag
\mi{Tree}=\begin{array}[c]{c}
          S \\[0.4mm]
          \hline \\[-3.8mm]
          (S-\{\gu\gleq a^*\})\cup \{\gu\gleq \gz\} \\[0.4mm]
          \hline \\[-3.8mm]
          S'=(((S-\{\gu\gleq a^*\})\cup \{\gu\gleq \gz\})-\{\gu\gleq \gz\})\cup \{\square\}.
          \end{array} 
\end{equation}
Hence, $\mi{Tree}$ is a finite {\it DPLL}-tree with the root $S$ constructed using Rules (\cref{ceq4hr1}), (\cref{ceq4hr1111111})--(\cref{ceq4hr4}) such that
for its only leaf $S'$, $\square\in S'$.

Case 2.1.6:
$\mi{guards}(S,a^*)\supseteq \{a^*\gleq \gz,\gu\gleq a^*\}$.
Then $\mi{guards}(S)\supseteq \mi{guards}(S,a^*)\supseteq \{a^*\gleq \gz,\gu\gleq a^*\}$,
$a^*\gleq \gz\in \mi{guards}(S)$, applying Rule (\cref{ceq4hr1111111}) to $a^*\gleq \gz$, we derive
\begin{equation} \notag
\dfrac{S}
      {(S-\{a^*\gleq \gz\})\cup \{a^*\geql \gz\}};
\end{equation}
$a^*\in \mi{atoms}((S-\{a^*\gleq \gz\})\cup \{a^*\geql \gz\})$,
$(S-\{a^*\gleq \gz\})\cup \{a^*\geql \gz\}\supseteq \mi{guards}((S-\{a^*\gleq \gz\})\cup \{a^*\geql \gz\})\supseteq 
 \mi{guards}((S-\{a^*\gleq \gz\})\cup \{a^*\geql \gz\},a^*)=((S-\{a^*\gleq \gz\})\cup \{a^*\geql \gz\})\cap \mi{guards}(a^*)=
 ((S-\{a^*\gleq \gz\})\cap \mi{guards}(a^*))\cup (\{a^*\geql \gz\}\cap \mi{guards}(a^*))=((S\cap \mi{guards}(a^*))-\{a^*\gleq \gz\})\cup \{a^*\geql \gz\}=
 (\mi{guards}(S,a^*)-\{a^*\gleq \gz\})\cup \{a^*\geql \gz\}\supseteq (\{a^*\gleq \gz,\gu\gleq a^*\}-\{a^*\gleq \gz\})\cup \{a^*\geql \gz\}=\{a^*\geql \gz,\gu\gleq a^*\}$,
$a^*\geql \gz\in \mi{guards}((S-\{a^*\gleq \gz\})\cup \{a^*\geql \gz\})$, $\gu\gleq a^*\in (S-\{a^*\gleq \gz\})\cup \{a^*\geql \gz\}$,
$a^*\in \mi{atoms}(\gu\gleq a^*)$, $a^*\geql \gz\neq \gu\gleq a^*$, $\mi{simplify}(\gu\gleq a^*,a^*,\gz)=\gu\gleq \gz$,
applying Rule (\cref{ceq4hr3}) to $(S-\{a^*\gleq \gz\})\cup \{a^*\geql \gz\}$, $a^*\geql \gz$, and $\gu\gleq a^*$, we derive
\begin{equation} \notag
\dfrac{(S-\{a^*\gleq \gz\})\cup \{a^*\geql \gz\}}
      {(((S-\{a^*\gleq \gz\})\cup \{a^*\geql \gz\})-\{\gu\gleq a^*\})\cup \{\gu\gleq \gz\}};
\end{equation}
$\gu\gleq \gz\in (((S-\{a^*\gleq \gz\})\cup \{a^*\geql \gz\})-\{\gu\gleq a^*\})\cup \{\gu\gleq \gz\}$;
$\gu\gleq \gz\in \mi{OrdPropLit}$ is a contradiction;
$\gu\gleq \gz$ is not a guard;
$\gu\gleq \gz\not\in \mi{guards}((((S-\{a^*\gleq \gz\})\cup \{a^*\geql \gz\})-\{\gu\gleq a^*\})\cup \{\gu\gleq \gz\})$,
$\gu\gleq \gz\in ((((S-\{a^*\gleq \gz\})\cup \{a^*\geql \gz\})-\{\gu\gleq a^*\})\cup \{\gu\gleq \gz\})-\mi{guards}((((S-\{a^*\gleq \gz\})\cup \{a^*\geql \gz\})-\{\gu\gleq a^*\})\cup \{\gu\gleq \gz\})$,
applying Rule (\cref{ceq4hr2}) to $(((S-\{a^*\gleq \gz\})\cup \{a^*\geql \gz\})-\{\gu\gleq a^*\})\cup \{\gu\gleq \gz\}$ and $\gu\gleq \gz$, we derive
\begin{equation} \notag
\dfrac{(((S-\{a^*\gleq \gz\})\cup \{a^*\geql \gz\})-\{\gu\gleq a^*\})\cup \{\gu\gleq \gz\}}
      {(((((S-\{a^*\gleq \gz\})\cup \{a^*\geql \gz\})-\{\gu\gleq a^*\})\cup \{\gu\gleq \gz\})-\{\gu\gleq \gz\})\cup \{\square\}}.
\end{equation}
We put 
\begin{equation} \notag
\mi{Tree}=\begin{array}[c]{c}
          S \\[0.4mm]
          \hline \\[-3.8mm]
          (S-\{a^*\gleq \gz\})\cup \{a^*\geql \gz\} \\[0.4mm]
          \hline \\[-3.8mm]
          (((S-\{a^*\gleq \gz\})\cup \{a^*\geql \gz\})-\{\gu\gleq a^*\})\cup \{\gu\gleq \gz\} \\[0.4mm]
          \hline \\[-3.8mm]
          S'=(((((S-\{a^*\gleq \gz\})\cup \{a^*\geql \gz\})-\{\gu\gleq a^*\})\cup \{\gu\gleq \gz\})- \\
          \hfill \{\gu\gleq \gz\})\cup \{\square\}.
          \end{array} 
\end{equation}
Hence, $\mi{Tree}$ is a finite {\it DPLL}-tree with the root $S$ constructed using Rules (\cref{ceq4hr1}), (\cref{ceq4hr1111111})--(\cref{ceq4hr4}) such that
for its only leaf $S'$, $\square\in S'$.

Case 2.1.7:
$\mi{guards}(S,a^*)\supseteq \{a^*\gle \gu,a^*\geql \gu\}$.
Then $S\supseteq \mi{guards}(S)\supseteq \mi{guards}(S,a^*)\supseteq \{a^*\gle \gu,a^*\geql \gu\}$,
$a^*\geql \gu\in \mi{guards}(S)$, $a^*\gle \gu\in S$,
$a^*\in \mi{atoms}(a^*\gle \gu)$, $a^*\geql \gu\neq a^*\gle \gu$, $\mi{simplify}(a^*\gle \gu,a^*,\gu)=\gu\gle \gu$,
applying Rule (\cref{ceq4hr4}) to $a^*\geql \gu$ and $a^*\gle \gu$, we derive
\begin{equation} \notag
\dfrac{S}
      {(S-\{a^*\gle \gu\})\cup \{\gu\gle \gu\}};
\end{equation}
$\gu\gle \gu\in (S-\{a^*\gle \gu\})\cup \{\gu\gle \gu\}$;
$\gu\gle \gu\in \mi{OrdPropLit}$ is a contradiction;
$\gu\gle \gu$ is not a guard;
$\gu\gle \gu\not\in \mi{guards}((S-\{a^*\gle \gu\})\cup \{\gu\gle \gu\})$,
$\gu\gle \gu\in ((S-\{a^*\gle \gu\})\cup \{\gu\gle \gu\})-\mi{guards}((S-\{a^*\gle \gu\})\cup \{\gu\gle \gu\})$,
applying Rule (\cref{ceq4hr2}) to $(S-\{a^*\gle \gu\})\cup \{\gu\gle \gu\}$ and $\gu\gle \gu$, we derive
\begin{equation} \notag
\dfrac{(S-\{a^*\gle \gu\})\cup \{\gu\gle \gu\}}
      {(((S-\{a^*\gle \gu\})\cup \{\gu\gle \gu\})-\{\gu\gle \gu\})\cup \{\square\}}.
\end{equation}
We put 
\begin{equation} \notag
\mi{Tree}=\begin{array}[c]{c}
          S \\[0.4mm]
          \hline \\[-3.8mm]
          (S-\{a^*\gle \gu\})\cup \{\gu\gle \gu\} \\[0.4mm]
          \hline \\[-3.8mm]
          S'=(((S-\{a^*\gle \gu\})\cup \{\gu\gle \gu\})-\{\gu\gle \gu\})\cup \{\square\}.
          \end{array} 
\end{equation}
Hence, $\mi{Tree}$ is a finite {\it DPLL}-tree with the root $S$ constructed using Rules (\cref{ceq4hr1}), (\cref{ceq4hr1111111})--(\cref{ceq4hr4}) such that
for its only leaf $S'$, $\square\in S'$.

Case 2.1.8:
$\mi{guards}(S,a^*)\supseteq \{a^*\gle \gu,\gu\gleq a^*\}$.
Then $\mi{guards}(S)\supseteq \mi{guards}(S,a^*)\supseteq \{a^*\gle \gu,\gu\gleq a^*\}$,
$\gu\gleq a^*\in \mi{guards}(S)$, applying Rule (\cref{ceq4hr11111111}) to $\gu\gleq a^*$, we derive
\begin{equation} \notag
\dfrac{S}
      {(S-\{\gu\gleq a^*\})\cup \{a^*\geql \gu\}};
\end{equation}
$a^*\in \mi{atoms}((S-\{\gu\gleq a^*\})\cup \{a^*\geql \gu\})$,
$(S-\{\gu\gleq a^*\})\cup \{a^*\geql \gu\}\supseteq \mi{guards}((S-\{\gu\gleq a^*\})\cup \{a^*\geql \gu\})\supseteq 
 \mi{guards}((S-\{\gu\gleq a^*\})\cup \{a^*\geql \gu\},a^*)=((S-\{\gu\gleq a^*\})\cup \{a^*\geql \gu\})\cap \mi{guards}(a^*)=
 ((S-\{\gu\gleq a^*\})\cap \mi{guards}(a^*))\cup (\{a^*\geql \gu\}\cap \mi{guards}(a^*))=((S\cap \mi{guards}(a^*))-\{\gu\gleq a^*\})\cup \{a^*\geql \gu\}=
 (\mi{guards}(S,a^*)-\{\gu\gleq a^*\})\cup \{a^*\geql \gu\}\supseteq (\{a^*\gle \gu,\gu\gleq a^*\}-\{\gu\gleq a^*\})\cup \{a^*\geql \gu\}=\{a^*\gle \gu,a^*\geql \gu\}$,
$a^*\geql \gu\in \mi{guards}((S-\{\gu\gleq a^*\})\cup \{a^*\geql \gu\})$, $a^*\gle \gu\in (S-\{\gu\gleq a^*\})\cup \{a^*\geql \gu\}$,
$a^*\in \mi{atoms}(a^*\gle \gu)$, $a^*\geql \gu\neq a^*\gle \gu$, $\mi{simplify}(a^*\gle \gu,a^*,\gu)=\gu\gle \gu$,
applying Rule (\cref{ceq4hr4}) to $(S-\{\gu\gleq a^*\})\cup \{a^*\geql \gu\}$, $a^*\geql \gu$, and $a^*\gle \gu$, we derive
\begin{equation} \notag
\dfrac{(S-\{\gu\gleq a^*\})\cup \{a^*\geql \gu\}}
      {(((S-\{\gu\gleq a^*\})\cup \{a^*\geql \gu\})-\{a^*\gle \gu\})\cup \{\gu\gle \gu\}};
\end{equation}
$\gu\gle \gu\in (((S-\{\gu\gleq a^*\})\cup \{a^*\geql \gu\})-\{a^*\gle \gu\})\cup \{\gu\gle \gu\}$;
$\gu\gle \gu\in \mi{OrdPropLit}$ is a contradiction;
$\gu\gle \gu$ is not a guard;
$\gu\gle \gu\not\in \mi{guards}((((S-\{\gu\gleq a^*\})\cup \{a^*\geql \gu\})-\{a^*\gle \gu\})\cup \{\gu\gle \gu\})$,
$\gu\gle \gu\in ((((S-\{\gu\gleq a^*\})\cup \{a^*\geql \gu\})-\{a^*\gle \gu\})\cup \{\gu\gle \gu\})-\mi{guards}((((S-\{\gu\gleq a^*\})\cup \{a^*\geql \gu\})-\{a^*\gle \gu\})\cup \{\gu\gle \gu\})$,
applying Rule (\cref{ceq4hr2}) to $(((S-\{\gu\gleq a^*\})\cup \{a^*\geql \gu\})-\{a^*\gle \gu\})\cup \{\gu\gle \gu\}$ and $\gu\gle \gu$, we derive
\begin{equation} \notag
\dfrac{(((S-\{\gu\gleq a^*\})\cup \{a^*\geql \gu\})-\{a^*\gle \gu\})\cup \{\gu\gle \gu\}}
      {(((((S-\{\gu\gleq a^*\})\cup \{a^*\geql \gu\})-\{a^*\gle \gu\})\cup \{\gu\gle \gu\})-\{\gu\gle \gu\})\cup \{\square\}}.
\end{equation}
We put 
\begin{equation} \notag
\mi{Tree}=\begin{array}[c]{c}
          S \\[0.4mm]
          \hline \\[-3.8mm]
          (S-\{\gu\gleq a^*\})\cup \{a^*\geql \gu\} \\[0.4mm]
          \hline \\[-3.8mm]
          (((S-\{\gu\gleq a^*\})\cup \{a^*\geql \gu\})-\{a^*\gle \gu\})\cup \{\gu\gle \gu\} \\[0.4mm]
          \hline \\[-3.8mm]
          S'=(((((S-\{\gu\gleq a^*\})\cup \{a^*\geql \gu\})-\{a^*\gle \gu\})\cup \{\gu\gle \gu\})- \\
          \hfill \{\gu\gle \gu\})\cup \{\square\}.
          \end{array} 
\end{equation}
Hence, $\mi{Tree}$ is a finite {\it DPLL}-tree with the root $S$ constructed using Rules (\cref{ceq4hr1}), (\cref{ceq4hr1111111})--(\cref{ceq4hr4}) such that
for its only leaf $S'$, $\square\in S'$.

Case 2.2:
$S$ is not a contradictory set of guards for $a^*$.
We have that $S$ is simplified.
Then 
$\mi{guards}(S,a^*)\not\supseteq \{a^*\geql \gz,\gz\gle a^*\}$, 
$\mi{guards}(S,a^*)\not\supseteq \{a^*\gleq \gz,\gz\gle a^*\}$,
$\mi{guards}(S,a^*)\not\supseteq \{a^*\geql \gz,a^*\geql \gu\}$,
$\mi{guards}(S,a^*)\not\supseteq \{a^*\gleq \gz,a^*\geql \gu\}$,
$\mi{guards}(S,a^*)\not\supseteq \{a^*\geql \gz,\gu\gleq a^*\}$,
$\mi{guards}(S,a^*)\not\supseteq \{a^*\gleq \gz,\gu\gleq a^*\}$,
$\mi{guards}(S,a^*)\not\supseteq \{a^*\gle \gu,a^*\geql \gu\}$,                                                                                                                            
$\mi{guards}(S,a^*)\not\supseteq \{a^*\gle \gu,\gu\gleq a^*\}$;
we have that $a^*$ is not $\gz$-guarded in $S$;
$\mi{guards}(S,a^*)\neq \{a^*\geql \gz\}$, 
$\mi{guards}(S,a^*)\neq \{\gz\gle a^*\}$,
$\mi{guards}(S,a^*)\neq \{\gz\gle a^*,a^*\gle \gu\}$, 
$\mi{guards}(S,a^*)\neq \{a^*\geql \gu\}$,
$\mi{guards}(S,a^*)\subseteq \mi{guards}(a^*)=\{a^*\geql \gz,a^*\gleq \gz,\gz\gle a^*,a^*\gle \gu,a^*\geql \gu,\gu\gleq a^*\}$;
either $\mi{guards}(S,a^*)=\emptyset$ or $\mi{guards}(S,a^*)=\{a^*\gleq \gz\}$ or $\mi{guards}(S,a^*)=\{a^*\gle \gu\}$ or $\mi{guards}(S,a^*)=\{\gu\gleq a^*\}$ 
or $\mi{guards}(S,a^*)=\{a^*\geql \gz,a^*\gleq \gz\}$ 
or $\mi{guards}(S,a^*)=\{a^*\geql \gz,a^*\gle \gu\}$ or $\mi{guards}(S,a^*)=\{a^*\gleq \gz,a^*\gle \gu\}$ 
or $\mi{guards}(S,a^*)=\{\gz\gle a^*,a^*\geql \gu\}$ or $\mi{guards}(S,a^*)=\{\gz\gle a^*,\gu\gleq a^*\}$ 
or $\mi{guards}(S,a^*)=\{a^*\geql \gu,\gu\gleq a^*\}$ 
or $\mi{guards}(S,a^*)=\{a^*\geql \gz,a^*\gleq \gz,a^*\gle \gu\}$ or $\mi{guards}(S,a^*)=\{\gz\gle a^*,a^*\geql \gu,\gu\gleq a^*\}$.
We get twelve cases for $\mi{guards}(S,a^*)$.

Case 2.2.1:
$\mi{guards}(S,a^*)=\emptyset$.
Then $a^*\in \mi{atoms}(S)$, applying Rule (\cref{ceq4hr1}) to $a^*$, we derive
\begin{equation} \notag
\dfrac{S}
      {S\cup \{a^*\geql \gz\}\ \big|\ S\cup \{\gz\gle a^*\}}.
\end{equation}
Hence, $a^*\geql \gz, \gz\gle a^*\in \mi{guards}(a^*)$, $a^*\geql \gz, \gz\gle a^*\not\in \mi{guards}(S,a^*)=S\cap \mi{guards}(a^*)=\emptyset$, $a^*\geql \gz, \gz\gle a^*\not\in S$.
We put 
$S_1=S\cup \{a^*\geql \gz\}\subseteq_{\mc F} \mi{OrdPropCl}$ and
$S_2=S\cup \{\gz\gle a^*\}\subseteq_{\mc F} \mi{OrdPropCl}$.
We get that
$a^*\in \mi{atoms}(S)$, 
$a^*\in \mi{atoms}(S_1)=\mi{atoms}(S\cup \{a^*\geql \gz\})=\mi{atoms}(S)\cup \mi{atoms}(a^*\geql \gz)=\mi{atoms}(S)\cup \{a^*\}=\mi{atoms}(S)$,
$a^*\in \mi{atoms}(S_2)=\mi{atoms}(S\cup \{\gz\gle a^*\})=\mi{atoms}(S)\cup \mi{atoms}(\gz\gle a^*)=\mi{atoms}(S)\cup \{a^*\}=\mi{atoms}(S)$;
we have that $S$ is simplified;
$a^*\geql \gz, \gz\gle a^*\neq \square$ do not contain contradictions and tautologies;
$S_1=S\cup \{a^*\geql \gz\}$ and $S_2=S\cup \{\gz\gle a^*\}$ are simplified;
$\mi{guards}(S_1,a^*)=(S\cup \{a^*\geql \gz\})\cap \mi{guards}(a^*)=(S\cap \mi{guards}(a^*))\cup (\{a^*\geql \gz\}\cap \mi{guards}(a^*))=\mi{guards}(S,a^*)\cup \{a^*\geql \gz\}=\{a^*\geql \gz\}$,
$\mi{guards}(S_2,a^*)=(S\cup \{\gz\gle a^*\})\cap \mi{guards}(a^*)=(S\cap \mi{guards}(a^*))\cup (\{\gz\gle a^*\}\cap \mi{guards}(a^*))=\mi{guards}(S,a^*)\cup \{\gz\gle a^*\}=\{\gz\gle a^*\}$;
for both $i$, 
$a^*\in \mi{atoms}(S_i)=\mi{atoms}(S)$; 
$a^*$ is $\gz$-guarded in $S_i$;
for all $a\in \mi{atoms}(S_i)-\{a^*\}=\mi{atoms}(S)-\{a^*\}$,
$a\neq a^*$,
$\mi{guards}(a)\cap \{a^*\geql \gz,\gz\gle a^*\}=\mi{guards}(a)\cap \mi{guards}(a^*)=\emptyset$,
$\mi{guards}(S_1,a)=(S\cup \{a^*\geql \gz\})\cap \mi{guards}(a)=(S\cap \mi{guards}(a))\cup (\{a^*\geql \gz\}\cap \mi{guards}(a))=\mi{guards}(S,a)$,
$\mi{guards}(S_2,a)=(S\cup \{\gz\gle a^*\})\cap \mi{guards}(a)=(S\cap \mi{guards}(a))\cup (\{\gz\gle a^*\}\cap \mi{guards}(a))=\mi{guards}(S,a)$;
$a$ is $\gz$-guarded in $S_i$ if and only if $a$ is $\gz$-guarded in $S$;
$a^*\in \mi{unguarded}(S)$,
$\mi{unguarded}(S_i)=\{a \,|\, a\in \mi{atoms}(S_i)\ \text{\it is not $\gz$-guarded in}\ S_i\}=\{a \,|\, a\in (\mi{atoms}(S_i)-\{a^*\})\cup \{a^*\}\ \text{\it is not $\gz$-guarded in}\ S_i\}=
                     \{a \,|\, a\in \mi{atoms}(S_i)-\{a^*\}\ \text{\it is not $\gz$-guarded in}\ S_i\}=\{a \,|\, a\in \mi{atoms}(S)-\{a^*\}\ \text{\it is not $\gz$-guarded in}\ S\}=  
                     \mi{unguarded}(S)-\{a^*\}=\{a \,|\, a\in \mi{atoms}(S)\ \text{\it is not $\gz$-guarded in}\ S\}-\{a^*\}\subset \mi{unguarded}(S)$;
by the induction hypothesis for $S_i$, there exists a finite {\it DPLL}-tree $\mi{Tree}_i$ with the root $S_i$ constructed using Rules (\cref{ceq4hr1}), (\cref{ceq4hr1111111})--(\cref{ceq4hr4}) satisfying
for every leaf $S'$ that either $\square\in S'$ or $S'\subseteq_{\mc F} \mi{OrdPropCl}$ is $\gz$-guarded.
We put
\begin{equation} \notag
\mi{Tree}=\dfrac{S}
                {\mi{Tree}_1\ \big|\ \mi{Tree}_2}.
\end{equation}
Hence, $\mi{Tree}$ is a finite {\it DPLL}-tree with the root $S$ constructed using Rules (\cref{ceq4hr1}), (\cref{ceq4hr1111111})--(\cref{ceq4hr4}) such that
for every leaf $S'$, either $\square\in S'$ or $S'\subseteq_{\mc F} \mi{OrdPropCl}$ is $\gz$-guarded.

Case 2.2.2:
$\mi{guards}(S,a^*)=\{a^*\gleq \gz\}$.
Then $\mi{guards}(S)\supseteq \mi{guards}(S,a^*)=\{a^*\gleq \gz\}$,
$a^*\gleq \gz\in \mi{guards}(S)$, applying Rule (\cref{ceq4hr1111111}) to $a^*\gleq \gz$, we derive
\begin{equation} \notag
\dfrac{S}
      {(S-\{a^*\gleq \gz\})\cup \{a^*\geql \gz\}}.
\end{equation}
Hence, $a^*\geql \gz\in \mi{guards}(a^*)$, $a^*\geql \gz\not\in \mi{guards}(S,a^*)=S\cap \mi{guards}(a^*)=\{a^*\gleq \gz\}$, $a^*\gleq \gz\in S$, $a^*\geql \gz\not\in S$.
We put $S'=(S-\{a^*\gleq \gz\})\cup \{a^*\geql \gz\}\subseteq_{\mc F} \mi{OrdPropCl}$.
We get that
$a^*\in \mi{atoms}(S)$, $\mi{atoms}(a^*\geql \gz,a^*\gleq \gz)=\{a^*\}$,
$a^*\in \mi{atoms}(S')=\mi{atoms}((S-\{a^*\gleq \gz\})\cup \{a^*\geql \gz\})=\mi{atoms}(S)$;
we have that $S$ is simplified;
$S-\{a^*\gleq \gz\}\subseteq S$ is simplified;
$a^*\geql \gz\neq \square$ does not contain contradictions and tautologies;
$S'=(S-\{a^*\gleq \gz\})\cup \{a^*\geql \gz\}$ is simplified;
$\mi{guards}(S',a^*)=((S-\{a^*\gleq \gz\})\cup \{a^*\geql \gz\})\cap \mi{guards}(a^*)=((S-\{a^*\gleq \gz\})\cap \mi{guards}(a^*))\cup (\{a^*\geql \gz\}\cap \mi{guards}(a^*))=
                     ((S\cap \mi{guards}(a^*))-\{a^*\gleq \gz\})\cup \{a^*\geql \gz\}=(\mi{guards}(S,a^*)-\{a^*\gleq \gz\})\cup \{a^*\geql \gz\}=
                     (\{a^*\gleq \gz\}-\{a^*\gleq \gz\})\cup \{a^*\geql \gz\}=\{a^*\geql \gz\}$;
$a^*$ is $\gz$-guarded in $S'$;
for all $a\in \mi{atoms}(S')-\{a^*\}=\mi{atoms}(S)-\{a^*\}$,
$a\neq a^*$,
$\mi{guards}(a)\cap \{a^*\geql \gz,a^*\gleq \gz\}=\mi{guards}(a)\cap \mi{guards}(a^*)=\emptyset$,
$\mi{guards}(S',a)=((S-\{a^*\gleq \gz\})\cup \{a^*\geql \gz\})\cap \mi{guards}(a)=((S-\{a^*\gleq \gz\})\cap \mi{guards}(a))\cup (\{a^*\geql \gz\}\cap \mi{guards}(a))=
                   (S\cap \mi{guards}(a))-\{a^*\gleq \gz\}=S\cap \mi{guards}(a)=\mi{guards}(S,a)$;
$a$ is $\gz$-guarded in $S'$ if and only if $a$ is $\gz$-guarded in $S$;
$a^*\in \mi{unguarded}(S)$,
$\mi{unguarded}(S')=\{a \,|\, a\in \mi{atoms}(S')\ \text{\it is not $\gz$-guarded in}\ S'\}=\{a \,|\, a\in (\mi{atoms}(S')-\{a^*\})\cup \{a^*\}\ \text{\it is not $\gz$-guarded in}\ S'\}=
                    \{a \,|\, a\in \mi{atoms}(S')-\{a^*\}\ \text{\it is not $\gz$-guarded in}\ S'\}=\{a \,|\, a\in \mi{atoms}(S)-\{a^*\}\ \text{\it is not $\gz$-guarded in}\ S\}=  
                    \mi{unguarded}(S)-\{a^*\}=\{a \,|\, a\in \mi{atoms}(S)\ \text{\it is not $\gz$-guarded in}\ S\}-\{a^*\}\subset \mi{unguarded}(S)$;
by the induction hypothesis for $S'$, there exists a finite {\it DPLL}-tree $\mi{Tree}'$ with the root $S'$ constructed using Rules (\cref{ceq4hr1}), (\cref{ceq4hr1111111})--(\cref{ceq4hr4}) satisfying
for every leaf $S''$ that either $\square\in S''$ or $S''\subseteq_{\mc F} \mi{OrdPropCl}$ is $\gz$-guarded.
We put
\begin{equation} \notag
\mi{Tree}=\dfrac{S}
                {\mi{Tree}'}.
\end{equation}
Hence, $\mi{Tree}$ is a finite {\it DPLL}-tree with the root $S$ constructed using Rules (\cref{ceq4hr1}), (\cref{ceq4hr1111111})--(\cref{ceq4hr4}) such that
for every leaf $S''$, either $\square\in S''$ or $S''\subseteq_{\mc F} \mi{OrdPropCl}$ is $\gz$-guarded.

Case 2.2.3:
$\mi{guards}(S,a^*)=\{a^*\gle \gu\}$.
Then $a^*\in \mi{atoms}(S)$, applying Rule (\cref{ceq4hr1}) to $a^*$, we derive
\begin{equation} \notag
\dfrac{S}
      {S\cup \{a^*\geql \gz\}\ \big|\ S\cup \{\gz\gle a^*\}}.
\end{equation}
Hence, $a^*\geql \gz, \gz\gle a^*\in \mi{guards}(a^*)$, $a^*\geql \gz, \gz\gle a^*\not\in \mi{guards}(S,a^*)=S\cap \mi{guards}(a^*)=\{a^*\gle \gu\}$, 
$a^*\gle \gu\in S$, $a^*\geql \gz, \gz\gle a^*\not\in S$.
We put 
$S_1=S\cup \{a^*\geql \gz\}\subseteq_{\mc F} \mi{OrdPropCl}$ and
$S_2=S\cup \{\gz\gle a^*\}\subseteq_{\mc F} \mi{OrdPropCl}$.
We get from Case 2.2.1 that $a^*\in \mi{atoms}(S_1)=\mi{atoms}(S_2)=\mi{atoms}(S)$, and $S_1$ and $S_2$ are simplified.
Then $S_1\supseteq \mi{guards}(S_1)\supseteq \mi{guards}(S_1,a^*)=(S\cup \{a^*\geql \gz\})\cap \mi{guards}(a^*)=(S\cap \mi{guards}(a^*))\cup (\{a^*\geql \gz\}\cap \mi{guards}(a^*))=
                                                                  \mi{guards}(S,a^*)\cup \{a^*\geql \gz\}=\{a^*\geql \gz,a^*\gle \gu\}$,
$a^*\geql \gz\in \mi{guards}(S_1)$, $a^*\gle \gu\in S_1$,
$a^*\in \mi{atoms}(a^*\gle \gu)$, $a^*\geql \gz\neq a^*\gle \gu$, $\mi{simplify}(a^*\gle \gu,a^*,\gz)=\gz\gle \gu$,
applying Rule (\cref{ceq4hr3}) to $S_1$, $a^*\geql \gz$, and $a^*\gle \gu$, we derive
\begin{equation} \notag
\dfrac{S_1}
      {(S_1-\{a^*\gle \gu\})\cup \{\gz\gle \gu\}};
\end{equation}
$\gz\gle \gu\in (S_1-\{a^*\gle \gu\})\cup \{\gz\gle \gu\}$;
$\gz\gle \gu\in \mi{OrdPropLit}$ is a tautology;
$\gz\gle \gu$ is not a guard;
$\gz\gle \gu\not\in \mi{guards}((S_1-\{a^*\gle \gu\})\cup \{\gz\gle \gu\})$,
$\gz\gle \gu\in ((S_1-\{a^*\gle \gu\})\cup \{\gz\gle \gu\})-\mi{guards}((S_1-\{a^*\gle \gu\})\cup \{\gz\gle \gu\})$,
applying Rule (\cref{ceq4hr22}) to $(S_1-\{a^*\gle \gu\})\cup \{\gz\gle \gu\}$ and $\gz\gle \gu$, we derive
\begin{equation} \notag
\dfrac{(S_1-\{a^*\gle \gu\})\cup \{\gz\gle \gu\}}
      {((S_1-\{a^*\gle \gu\})\cup \{\gz\gle \gu\})-\{\gz\gle \gu\}}.
\end{equation}
We have that $S_1$ is simplified.
Hence, $\gz\gle \gu\not\in S_1$, $a^*\gle \gu\in S$, $a^*\geql \gz\not\in S$,
$((S_1-\{a^*\gle \gu\})\cup \{\gz\gle \gu\})-\{\gz\gle \gu\}=((S_1-\{a^*\gle \gu\})-\{\gz\gle \gu\})\cup (\{\gz\gle \gu\}-\{\gz\gle \gu\})=S_1-\{a^*\gle \gu\}=
 (S\cup \{a^*\geql \gz\})-\{a^*\gle \gu\}=(S-\{a^*\gle \gu\})\cup (\{a^*\geql \gz\}-\{a^*\gle \gu\})=(S-\{a^*\gle \gu\})\cup \{a^*\geql \gz\}$.
We put $S_1'=(S-\{a^*\gle \gu\})\cup \{a^*\geql \gz\}\subseteq_{\mc F} \mi{OrdPropCl}$.
We get that
$S_1'=(S-\{a^*\gle \gu\})\cup \{a^*\geql \gz\}=S_1-\{a^*\gle \gu\}$,
$\mi{atoms}(a^*\gle \gu)=\{a^*\}$,
$a^*\in \mi{atoms}(S_1')=\mi{atoms}(S_1-\{a^*\gle \gu\})=\mi{atoms}(S_1)=\mi{atoms}(S)$;
we have that $S_1$ is simplified;
$S_1'=S_1-\{a^*\gle \gu\}\subseteq S_1$ is simplified;
$\mi{guards}(S_1',a^*)=((S-\{a^*\gle \gu\})\cup \{a^*\geql \gz\})\cap \mi{guards}(a^*)=((S-\{a^*\gle \gu\})\cap \mi{guards}(a^*))\cup (\{a^*\geql \gz\}\cap \mi{guards}(a^*))=
                       ((S\cap \mi{guards}(a^*))-\{a^*\gle \gu\})\cup \{a^*\geql \gz\}=(\mi{guards}(S,a^*)-\{a^*\gle \gu\})\cup \{a^*\geql \gz\}=(\{a^*\gle \gu\}-\{a^*\gle \gu\})\cup \{a^*\geql \gz\}=
                       \{a^*\geql \gz\}$,
$\mi{guards}(S_2,a^*)=(S\cup \{\gz\gle a^*\})\cap \mi{guards}(a^*)=(S\cap \mi{guards}(a^*))\cup (\{\gz\gle a^*\}\cap \mi{guards}(a^*))=\mi{guards}(S,a^*)\cup \{\gz\gle a^*\}=
                      \{\gz\gle a^*,a^*\gle \gu\}$;
$a^*$ is $\gz$-guarded in $S_1'$ and $S_2$;
for all $a\in \mi{atoms}(S_1')-\{a^*\}=\mi{atoms}(S)-\{a^*\}$,
$a\neq a^*$,
$\mi{guards}(a)\cap \{a^*\geql \gz,a^*\gle \gu\}=\mi{guards}(a)\cap \mi{guards}(a^*)=\emptyset$,
$\mi{guards}(S_1',a)=((S-\{a^*\gle \gu\})\cup \{a^*\geql \gz\})\cap \mi{guards}(a)=((S-\{a^*\gle \gu\})\cap \mi{guards}(a))\cup (\{a^*\geql \gz\}\cap \mi{guards}(a))=
                     (S\cap \mi{guards}(a))-\{a^*\gle \gu\}=S\cap \mi{guards}(a)=\mi{guards}(S,a)$;
$a$ is $\gz$-guarded in $S_1'$ if and only if $a$ is $\gz$-guarded in $S$;
$a^*\in \mi{unguarded}(S)$,
$\mi{unguarded}(S_1')=\{a \,|\, a\in \mi{atoms}(S_1')\ \text{\it is not $\gz$-guarded in}\ S_1'\}=\{a \,|\, a\in (\mi{atoms}(S_1')-\{a^*\})\cup \{a^*\}\ \text{\it is not $\gz$-guarded in}\ S_1'\}=
                      \{a \,|\, a\in \mi{atoms}(S_1')-\{a^*\}\ \text{\it is not $\gz$-guarded in}\ S_1'\}=\{a \,|\, a\in \mi{atoms}(S)-\{a^*\}\ \text{\it is not $\gz$-guarded in}\ S\}=  
                      \mi{unguarded}(S)-\{a^*\}=\{a \,|\, a\in \mi{atoms}(S)\ \text{\it is not $\gz$-guarded in}\ S\}-\{a^*\}\subset \mi{unguarded}(S)$;
by the induction hypothesis for $S_1'$, there exists a finite {\it DPLL}-tree $\mi{Tree}_1'$ with the root $S_1'$ constructed using Rules (\cref{ceq4hr1}), (\cref{ceq4hr1111111})--(\cref{ceq4hr4}) satisfying
for every leaf $S'$ that either $\square\in S'$ or $S'\subseteq_{\mc F} \mi{OrdPropCl}$ is $\gz$-guarded.
We get from Case 2.2.1 that there exists a finite {\it DPLL}-tree $\mi{Tree}_2$ with the root $S_2$ constructed using Rules (\cref{ceq4hr1}), (\cref{ceq4hr1111111})--(\cref{ceq4hr4}) satisfying
for every leaf $S'$ that either $\square\in S'$ or $S'\subseteq_{\mc F} \mi{OrdPropCl}$ is $\gz$-guarded.
We put
\begin{equation} \notag
\mi{Tree}=\dfrac{S}
                {\left.
                 \begin{array}[t]{c}
                 \\[-9.5mm]
                 S_1 \\[0.4mm]
                 \hline \\[-3.8mm]
                 (S_1-\{a^*\gle \gu\})\cup \{\gz\gle \gu\} \\[0.4mm]
                 \hline \\[-3.8mm]
                 \mi{Tree}_1'
                 \end{array}\ \right|\ 
                 \begin{array}[t]{c}
                 \\[-9.5mm]
                 \mi{Tree}_2
                 \end{array}}.
\end{equation}
Hence, $\mi{Tree}$ is a finite {\it DPLL}-tree with the root $S$ constructed using Rules (\cref{ceq4hr1}), (\cref{ceq4hr1111111})--(\cref{ceq4hr4}) such that
for every leaf $S'$, either $\square\in S'$ or $S'\subseteq_{\mc F} \mi{OrdPropCl}$ is $\gz$-guarded.

Case 2.2.4:
$\mi{guards}(S,a^*)=\{\gu\gleq a^*\}$.
Then $\mi{guards}(S)\supseteq \mi{guards}(S,a^*)=\{\gu\gleq a^*\}$,
$\gu\gleq a^*\in \mi{guards}(S)$, applying Rule (\cref{ceq4hr11111111}) to $\gu\gleq a^*$, we derive
\begin{equation} \notag
\dfrac{S}
      {(S-\{\gu\gleq a^*\})\cup \{a^*\geql \gu\}}.
\end{equation}
Hence, $a^*\geql \gu\in \mi{guards}(a^*)$, $a^*\geql \gu\not\in \mi{guards}(S,a^*)=S\cap \mi{guards}(a^*)=\{\gu\gleq a^*\}$, $\gu\gleq a^*\in S$, $a^*\geql \gu\not\in S$.
We put $S'=(S-\{\gu\gleq a^*\})\cup \{a^*\geql \gu\}\subseteq_{\mc F} \mi{OrdPropCl}$.
We get that
$a^*\in \mi{atoms}(S)$, $\mi{atoms}(a^*\geql \gu,\gu\gleq a^*)=\{a^*\}$,
$a^*\in \mi{atoms}(S')=\mi{atoms}((S-\{\gu\gleq a^*\})\cup \{a^*\geql \gu\})=\mi{atoms}(S)$;
we have that $S$ is simplified;
$S-\{\gu\gleq a^*\}\subseteq S$ is simplified;
$a^*\geql \gu\neq \square$ does not contain contradictions and tautologies;
$S'=(S-\{\gu\gleq a^*\})\cup \{a^*\geql \gu\}$ is simplified;
$\mi{guards}(S',a^*)=((S-\{\gu\gleq a^*\})\cup \{a^*\geql \gu\})\cap \mi{guards}(a^*)=((S-\{\gu\gleq a^*\})\cap \mi{guards}(a^*))\cup (\{a^*\geql \gu\}\cap \mi{guards}(a^*))=
                     ((S\cap \mi{guards}(a^*))-\{\gu\gleq a^*\})\cup \{a^*\geql \gu\}=(\mi{guards}(S,a^*)-\{\gu\gleq a^*\})\cup \{a^*\geql \gu\}=
                     (\{\gu\gleq a^*\}-\{\gu\gleq a^*\})\cup \{a^*\geql \gu\}=\{a^*\geql \gu\}$;
$a^*$ is $\gz$-guarded in $S'$;
for all $a\in \mi{atoms}(S')-\{a^*\}=\mi{atoms}(S)-\{a^*\}$,
$a\neq a^*$,
$\mi{guards}(a)\cap \{a^*\geql \gu,\gu\gleq a^*\}=\mi{guards}(a)\cap \mi{guards}(a^*)=\emptyset$,
$\mi{guards}(S',a)=((S-\{\gu\gleq a^*\})\cup \{a^*\geql \gu\})\cap \mi{guards}(a)=((S-\{\gu\gleq a^*\})\cap \mi{guards}(a))\cup (\{a^*\geql \gu\}\cap \mi{guards}(a))=
                   (S\cap \mi{guards}(a))-\{\gu\gleq a^*\}=S\cap \mi{guards}(a)=\mi{guards}(S,a)$;
$a$ is $\gz$-guarded in $S'$ if and only if $a$ is $\gz$-guarded in $S$;
$a^*\in \mi{unguarded}(S)$,
$\mi{unguarded}(S')=\{a \,|\, a\in \mi{atoms}(S')\ \text{\it is not $\gz$-guarded in}\ S'\}=\{a \,|\, a\in (\mi{atoms}(S')-\{a^*\})\cup \{a^*\}\ \text{\it is not $\gz$-guarded in}\ S'\}=
                    \{a \,|\, a\in \mi{atoms}(S')-\{a^*\}\ \text{\it is not $\gz$-guarded in}\ S'\}=\{a \,|\, a\in \mi{atoms}(S)-\{a^*\}\ \text{\it is not $\gz$-guarded in}\ S\}=  
                    \mi{unguarded}(S)-\{a^*\}=\{a \,|\, a\in \mi{atoms}(S)\ \text{\it is not $\gz$-guarded in}\ S\}-\{a^*\}\subset \mi{unguarded}(S)$;
by the induction hypothesis for $S'$, there exists a finite {\it DPLL}-tree $\mi{Tree}'$ with the root $S'$ constructed using Rules (\cref{ceq4hr1}), (\cref{ceq4hr1111111})--(\cref{ceq4hr4}) satisfying
for every leaf $S''$ that either $\square\in S''$ or $S''\subseteq_{\mc F} \mi{OrdPropCl}$ is $\gz$-guarded.
We put
\begin{equation} \notag
\mi{Tree}=\dfrac{S}
                {\mi{Tree}'}.
\end{equation}
Hence, $\mi{Tree}$ is a finite {\it DPLL}-tree with the root $S$ constructed using Rules (\cref{ceq4hr1}), (\cref{ceq4hr1111111})--(\cref{ceq4hr4}) such that
for every leaf $S''$, either $\square\in S''$ or $S''\subseteq_{\mc F} \mi{OrdPropCl}$ is $\gz$-guarded.

Case 2.2.5:
$\mi{guards}(S,a^*)=\{a^*\geql \gz,a^*\gleq \gz\}$.
Then $S\supseteq \mi{guards}(S)\supseteq \mi{guards}(S,a^*)=\{a^*\geql \gz,a^*\gleq \gz\}$, $a^*\geql \gz, a^*\gleq \gz\in S$,
$a^*\gleq \gz\in \mi{guards}(S)$, applying Rule (\cref{ceq4hr1111111}) to $a^*\gleq \gz$, we derive
\begin{equation} \notag
\dfrac{S}
      {(S-\{a^*\gleq \gz\})\cup \{a^*\geql \gz\}}.
\end{equation}
Hence, $a^*\geql \gz, a^*\gleq \gz\in S$, $a^*\geql \gz\neq a^*\gleq \gz$, $a^*\geql \gz\in S-\{a^*\gleq \gz\}$,
$(S-\{a^*\gleq \gz\})\cup \{a^*\geql \gz\}=S-\{a^*\gleq \gz\}$.
We put $S'=S-\{a^*\gleq \gz\}\subseteq_{\mc F} \mi{OrdPropCl}$.
We get that
$a^*\geql \gz\in S'=S-\{a^*\gleq \gz\}$, $\mi{atoms}(a^*\gleq \gz)=\{a^*\}$,
$a^*\in \mi{atoms}(a^*\geql \gz)\subseteq \mi{atoms}(S')=\mi{atoms}(S-\{a^*\gleq \gz\})=\mi{atoms}(S)$;
we have that $S$ is simplified;
$S'=S-\{a^*\gleq \gz\}\subseteq S$ is simplified;
$\mi{guards}(S',a^*)=(S-\{a^*\gleq \gz\})\cap \mi{guards}(a^*)=(S\cap \mi{guards}(a^*))-\{a^*\gleq \gz\}=\mi{guards}(S,a^*)-\{a^*\gleq \gz\}=\{a^*\geql \gz,a^*\gleq \gz\}-\{a^*\gleq \gz\}=
                     \{a^*\geql \gz\}$;
$a^*$ is $\gz$-guarded in $S'$;
for all $a\in \mi{atoms}(S')-\{a^*\}=\mi{atoms}(S)-\{a^*\}$,
$a\neq a^*$,
$\mi{guards}(a)\cap \{a^*\gleq \gz\}=\mi{guards}(a)\cap \mi{guards}(a^*)=\emptyset$,
$\mi{guards}(S',a)=(S-\{a^*\gleq \gz\})\cap \mi{guards}(a)=(S\cap \mi{guards}(a))-\{a^*\gleq \gz\}=S\cap \mi{guards}(a)=\mi{guards}(S,a)$;
$a$ is $\gz$-guarded in $S'$ if and only if $a$ is $\gz$-guarded in $S$;
$a^*\in \mi{unguarded}(S)$,
$\mi{unguarded}(S')=\{a \,|\, a\in \mi{atoms}(S')\ \text{\it is not $\gz$-guarded in}\ S'\}=\{a \,|\, a\in (\mi{atoms}(S')-\{a^*\})\cup \{a^*\}\ \text{\it is not $\gz$-guarded in}\ S'\}=
                    \{a \,|\, a\in \mi{atoms}(S')-\{a^*\}\ \text{\it is not $\gz$-guarded in}\ S'\}=\{a \,|\, a\in \mi{atoms}(S)-\{a^*\}\ \text{\it is not $\gz$-guarded in}\ S\}=  
                    \mi{unguarded}(S)-\{a^*\}=\{a \,|\, a\in \mi{atoms}(S)\ \text{\it is not $\gz$-guarded in}\ S\}-\{a^*\}\subset \mi{unguarded}(S)$;
by the induction hypothesis for $S'$, there exists a finite {\it DPLL}-tree $\mi{Tree}'$ with the root $S'$ constructed using Rules (\cref{ceq4hr1}), (\cref{ceq4hr1111111})--(\cref{ceq4hr4}) satisfying
for every leaf $S''$ that either $\square\in S''$ or $S''\subseteq_{\mc F} \mi{OrdPropCl}$ is $\gz$-guarded.
We put
\begin{equation} \notag
\mi{Tree}=\dfrac{S}
                {\mi{Tree}'}.
\end{equation}
Hence, $\mi{Tree}$ is a finite {\it DPLL}-tree with the root $S$ constructed using Rules (\cref{ceq4hr1}), (\cref{ceq4hr1111111})--(\cref{ceq4hr4}) such that
for every leaf $S''$, either $\square\in S''$ or $S''\subseteq_{\mc F} \mi{OrdPropCl}$ is $\gz$-guarded.

Case 2.2.6:
$\mi{guards}(S,a^*)=\{a^*\geql \gz,a^*\gle \gu\}$.
Then $S\supseteq \mi{guards}(S)\supseteq \mi{guards}(S,a^*)=\{a^*\geql \gz,a^*\gle \gu\}$,
$a^*\geql \gz\in \mi{guards}(S)$, $a^*\geql \gz, a^*\gle \gu\in S$,
$a^*\in \mi{atoms}(a^*\gle \gu)$, $a^*\geql \gz\neq a^*\gle \gu$, $\mi{simplify}(a^*\gle \gu,a^*,\gz)=\gz\gle \gu$,
applying Rule (\cref{ceq4hr3}) to $a^*\geql \gz$ and $a^*\gle \gu$, we derive
\begin{equation} \notag
\dfrac{S}
      {(S-\{a^*\gle \gu\})\cup \{\gz\gle \gu\}};
\end{equation}
$\gz\gle \gu\in (S-\{a^*\gle \gu\})\cup \{\gz\gle \gu\}$;
$\gz\gle \gu\in \mi{OrdPropLit}$ is a tautology;
$\gz\gle \gu$ is not a guard;
$\gz\gle \gu\not\in \mi{guards}((S-\{a^*\gle \gu\})\cup \{\gz\gle \gu\})$,
$\gz\gle \gu\in ((S-\{a^*\gle \gu\})\cup \{\gz\gle \gu\})-\mi{guards}((S-\{a^*\gle \gu\})\cup \{\gz\gle \gu\})$,
applying Rule (\cref{ceq4hr22}) to $(S-\{a^*\gle \gu\})\cup \{\gz\gle \gu\}$ and $\gz\gle \gu$, we derive
\begin{equation} \notag
\dfrac{(S-\{a^*\gle \gu\})\cup \{\gz\gle \gu\}}
      {((S-\{a^*\gle \gu\})\cup \{\gz\gle \gu\})-\{\gz\gle \gu\}}.
\end{equation}
We have that $S$ is simplified.
Hence, $\gz\gle \gu\not\in S$, $a^*\gle \gu\in S$,
$((S-\{a^*\gle \gu\})\cup \{\gz\gle \gu\})-\{\gz\gle \gu\}=((S-\{a^*\gle \gu\})-\{\gz\gle \gu\})\cup (\{\gz\gle \gu\}-\{\gz\gle \gu\})=S-\{a^*\gle \gu\}$.
We put $S'=S-\{a^*\gle \gu\}\subseteq_{\mc F} \mi{OrdPropCl}$.
We get that
$a^*\geql \gz\in S$, $a^*\geql \gz\neq a^*\gle \gu$, $a^*\geql \gz\in S'=S-\{a^*\gle \gu\}$, $\mi{atoms}(a^*\gle \gu)=\{a^*\}$, 
$a^*\in \mi{atoms}(a^*\geql \gz)\subseteq \mi{atoms}(S')=\mi{atoms}(S-\{a^*\gle \gu\})=\mi{atoms}(S)$;
we have that $S$ is simplified;
$S'=S-\{a^*\gle \gu\}\subseteq S$ is simplified;
$\mi{guards}(S',a^*)=(S-\{a^*\gle \gu\})\cap \mi{guards}(a^*)=(S\cap \mi{guards}(a^*))-\{a^*\gle \gu\}=\mi{guards}(S,a^*)-\{a^*\gle \gu\}=\{a^*\geql \gz,a^*\gle \gu\}-\{a^*\gle \gu\}=
                     \{a^*\geql \gz\}$;
$a^*$ is $\gz$-guarded in $S'$;
for all $a\in \mi{atoms}(S')-\{a^*\}=\mi{atoms}(S)-\{a^*\}$,
$a\neq a^*$,
$\mi{guards}(a)\cap \{a^*\gle \gu\}=\mi{guards}(a)\cap \mi{guards}(a^*)=\emptyset$,
$\mi{guards}(S',a)=(S-\{a^*\gle \gu\})\cap \mi{guards}(a)=(S\cap \mi{guards}(a))-\{a^*\gle \gu\}=S\cap \mi{guards}(a)=\mi{guards}(S,a)$;
$a$ is $\gz$-guarded in $S'$ if and only if $a$ is $\gz$-guarded in $S$;
$a^*\in \mi{unguarded}(S)$,
$\mi{unguarded}(S')=\{a \,|\, a\in \mi{atoms}(S')\ \text{\it is not $\gz$-guarded in}\ S'\}=\{a \,|\, a\in (\mi{atoms}(S')-\{a^*\})\cup \{a^*\}\ \text{\it is not $\gz$-guarded in}\ S'\}=
                    \{a \,|\, a\in \mi{atoms}(S')-\{a^*\}\ \text{\it is not $\gz$-guarded in}\ S'\}=\{a \,|\, a\in \mi{atoms}(S)-\{a^*\}\ \text{\it is not $\gz$-guarded in}\ S\}=  
                    \mi{unguarded}(S)-\{a^*\}=\{a \,|\, a\in \mi{atoms}(S)\ \text{\it is not $\gz$-guarded in}\ S\}-\{a^*\}\subset \mi{unguarded}(S)$;
by the induction hypothesis for $S'$, there exists a finite {\it DPLL}-tree $\mi{Tree}'$ with the root $S'$ constructed using Rules (\cref{ceq4hr1}), (\cref{ceq4hr1111111})--(\cref{ceq4hr4}) satisfying
for every leaf $S''$ that either $\square\in S''$ or $S''\subseteq_{\mc F} \mi{OrdPropCl}$ is $\gz$-guarded.
We put
\begin{equation} \notag
\mi{Tree}=\begin{array}[c]{c}
          S \\[0.4mm]
          \hline \\[-3.8mm]
          (S-\{a^*\gle \gu\})\cup \{\gz\gle \gu\} \\[0.4mm]
          \hline \\[-3.8mm]
          \mi{Tree}'.
          \end{array}
\end{equation}
Hence, $\mi{Tree}$ is a finite {\it DPLL}-tree with the root $S$ constructed using Rules (\cref{ceq4hr1}), (\cref{ceq4hr1111111})--(\cref{ceq4hr4}) such that
for every leaf $S''$, either $\square\in S''$ or $S''\subseteq_{\mc F} \mi{OrdPropCl}$ is $\gz$-guarded.

Case 2.2.7:
$\mi{guards}(S,a^*)=\{a^*\gleq \gz,a^*\gle \gu\}$.
Then $\mi{guards}(S)\supseteq \mi{guards}(S,a^*)=\{a^*\gleq \gz,a^*\gle \gu\}$,
$a^*\gleq \gz\in \mi{guards}(S)$, applying Rule (\cref{ceq4hr1111111}) to $a^*\gleq \gz$, we derive
\begin{equation} \notag
\dfrac{S}
      {(S-\{a^*\gleq \gz\})\cup \{a^*\geql \gz\}};
\end{equation}
$a^*\in \mi{atoms}((S-\{a^*\gleq \gz\})\cup \{a^*\geql \gz\})$,
$(S-\{a^*\gleq \gz\})\cup \{a^*\geql \gz\}\supseteq \mi{guards}((S-\{a^*\gleq \gz\})\cup \{a^*\geql \gz\})\supseteq 
 \mi{guards}((S-\{a^*\gleq \gz\})\cup \{a^*\geql \gz\},a^*)=((S-\{a^*\gleq \gz\})\cup \{a^*\geql \gz\})\cap \mi{guards}(a^*)=
 ((S-\{a^*\gleq \gz\})\cap \mi{guards}(a^*))\cup (\{a^*\geql \gz\}\cap \mi{guards}(a^*))=((S\cap \mi{guards}(a^*))-\{a^*\gleq \gz\})\cup \{a^*\geql \gz\}=
 (\mi{guards}(S,a^*)-\{a^*\gleq \gz\})\cup \{a^*\geql \gz\}=(\{a^*\gleq \gz,a^*\gle \gu\}-\{a^*\gleq \gz\})\cup \{a^*\geql \gz\}=\{a^*\geql \gz,a^*\gle \gu\}$,
$a^*\geql \gz\in \mi{guards}((S-\{a^*\gleq \gz\})\cup \{a^*\geql \gz\})$, $a^*\gle \gu\in (S-\{a^*\gleq \gz\})\cup \{a^*\geql \gz\}$,
$a^*\in \mi{atoms}(a^*\gle \gu)$, $a^*\geql \gz\neq a^*\gle \gu$, $\mi{simplify}(a^*\gle \gu,a^*,\gz)=\gz\gle \gu$,
applying Rule (\cref{ceq4hr3}) to $(S-\{a^*\gleq \gz\})\cup \{a^*\geql \gz\}$, $a^*\geql \gz$, and $a^*\gle \gu$, we derive
\begin{equation} \notag
\dfrac{(S-\{a^*\gleq \gz\})\cup \{a^*\geql \gz\}}
      {(((S-\{a^*\gleq \gz\})\cup \{a^*\geql \gz\})-\{a^*\gle \gu\})\cup \{\gz\gle \gu\}};
\end{equation}
$\gz\gle \gu\in (((S-\{a^*\gleq \gz\})\cup \{a^*\geql \gz\})-\{a^*\gle \gu\})\cup \{\gz\gle \gu\}$;
$\gz\gle \gu\in \mi{OrdPropLit}$ is a tautology;
$\gz\gle \gu$ is not a guard;
$\gz\gle \gu\not\in \mi{guards}((((S-\{a^*\gleq \gz\})\cup \{a^*\geql \gz\})-\{a^*\gle \gu\})\cup \{\gz\gle \gu\})$,
$\gz\gle \gu\in ((((S-\{a^*\gleq \gz\})\cup \{a^*\geql \gz\})-\{a^*\gle \gu\})\cup \{\gz\gle \gu\})-\mi{guards}((((S-\{a^*\gleq \gz\})\cup \{a^*\geql \gz\})-\{a^*\gle \gu\})\cup \{\gz\gle \gu\})$,
applying Rule (\cref{ceq4hr22}) to $(((S-\{a^*\gleq \gz\})\cup \{a^*\geql \gz\})-\{a^*\gle \gu\})\cup \{\gz\gle \gu\}$ and $\gz\gle \gu$, we derive
\begin{equation} \notag
\dfrac{(((S-\{a^*\gleq \gz\})\cup \{a^*\geql \gz\})-\{a^*\gle \gu\})\cup \{\gz\gle \gu\}}
      {((((S-\{a^*\gleq \gz\})\cup \{a^*\geql \gz\})-\{a^*\gle \gu\})\cup \{\gz\gle \gu\})-\{\gz\gle \gu\}}.
\end{equation}
We have that $S$ is simplified.
Hence, $\gz\gle \gu\not\in S$, 
$a^*\geql \gz\in \mi{guards}(a^*)$, $a^*\geql \gz\not\in \mi{guards}(S,a^*)=S\cap \mi{guards}(a^*)=\{a^*\gleq \gz,a^*\gle \gu\}$, $a^*\gleq \gz, a^*\gle \gu\in S$, $a^*\geql \gz\not\in S$,
$((((S-\{a^*\gleq \gz\})\cup \{a^*\geql \gz\})-\{a^*\gle \gu\})\cup \{\gz\gle \gu\})-\{\gz\gle \gu\}=
 ((((S-\{a^*\gleq \gz\})\cup \{a^*\geql \gz\})-\{a^*\gle \gu\})-\{\gz\gle \gu\})\cup (\{\gz\gle \gu\}-\{\gz\gle \gu\})=
 (((S-\{a^*\gleq \gz\})\cup \{a^*\geql \gz\})-\{\gz\gle \gu\})-\{a^*\gle \gu\}=
 (((S-\{a^*\gleq \gz\})-\{\gz\gle \gu\})\cup (\{a^*\geql \gz\}-\{\gz\gle \gu\}))-\{a^*\gle \gu\}=
 ((S-\{a^*\gleq \gz\})\cup \{a^*\geql \gz\})-\{a^*\gle \gu\}=
 ((S-\{a^*\gleq \gz\})-\{a^*\gle \gu\})\cup (\{a^*\geql \gz\}-\{a^*\gle \gu\})=
 (S-\{a^*\gleq \gz,a^*\gle \gu\})\cup \{a^*\geql \gz\}$.
We put $S'=(S-\{a^*\gleq \gz,a^*\gle \gu\})\cup \{a^*\geql \gz\}\subseteq_{\mc F} \mi{OrdPropCl}$.
We get that
$a^*\in \mi{atoms}(S)$, $\mi{atoms}(a^*\geql \gz,a^*\gleq \gz,a^*\gle \gu)=\{a^*\}$,
$a^*\in \mi{atoms}(S')=\mi{atoms}((S-\{a^*\gleq \gz,a^*\gle \gu\})\cup \{a^*\geql \gz\})=\mi{atoms}(S)$;
we have that $S$ is simplified;
$S-\{a^*\gleq \gz,a^*\gle \gu\}\subseteq S$ is simplified;
$a^*\geql \gz\neq \square$ does not contain contradictions and tautologies;
$S'=(S-\{a^*\gleq \gz,a^*\gle \gu\})\cup \{a^*\geql \gz\}$ is simplified;
$\mi{guards}(S',a^*)=((S-\{a^*\gleq \gz,a^*\gle \gu\})\cup \{a^*\geql \gz\})\cap \mi{guards}(a^*)=((S-\{a^*\gleq \gz,a^*\gle \gu\})\cap \mi{guards}(a^*))\cup (\{a^*\geql \gz\}\cap \mi{guards}(a^*))=
                     ((S\cap \mi{guards}(a^*))-\{a^*\gleq \gz,a^*\gle \gu\})\cup \{a^*\geql \gz\}=(\mi{guards}(S,a^*)-\{a^*\gleq \gz,a^*\gle \gu\})\cup \{a^*\geql \gz\}=
                     (\{a^*\gleq \gz,a^*\gle \gu\}-\{a^*\gleq \gz,a^*\gle \gu\})\cup \{a^*\geql \gz\}=\{a^*\geql \gz\}$;
$a^*$ is $\gz$-guarded in $S'$;
for all $a\in \mi{atoms}(S')-\{a^*\}=\mi{atoms}(S)-\{a^*\}$,
$a\neq a^*$,
$\mi{guards}(a)\cap \{a^*\geql \gz,a^*\gleq \gz,a^*\gle \gu\}=\mi{guards}(a)\cap \mi{guards}(a^*)=\emptyset$,
$\mi{guards}(S',a)=((S-\{a^*\gleq \gz,a^*\gle \gu\})\cup \{a^*\geql \gz\})\cap \mi{guards}(a)=((S-\{a^*\gleq \gz,a^*\gle \gu\})\cap \mi{guards}(a))\cup (\{a^*\geql \gz\}\cap \mi{guards}(a))=
                   (S\cap \mi{guards}(a))-\{a^*\gleq \gz,a^*\gle \gu\}=S\cap \mi{guards}(a)=\mi{guards}(S,a)$;
$a$ is $\gz$-guarded in $S'$ if and only if $a$ is $\gz$-guarded in $S$;
$a^*\in \mi{unguarded}(S)$,
$\mi{unguarded}(S')=\{a \,|\, a\in \mi{atoms}(S')\ \text{\it is not $\gz$-guarded in}\ S'\}=\{a \,|\, a\in (\mi{atoms}(S')-\{a^*\})\cup \{a^*\}\ \text{\it is not $\gz$-guarded in}\ S'\}=
                    \{a \,|\, a\in \mi{atoms}(S')-\{a^*\}\ \text{\it is not $\gz$-guarded in}\ S'\}=\{a \,|\, a\in \mi{atoms}(S)-\{a^*\}\ \text{\it is not $\gz$-guarded in}\ S\}=  
                    \mi{unguarded}(S)-\{a^*\}=\{a \,|\, a\in \mi{atoms}(S)\ \text{\it is not $\gz$-guarded in}\ S\}-\{a^*\}\subset \mi{unguarded}(S)$;
by the induction hypothesis for $S'$, there exists a finite {\it DPLL}-tree $\mi{Tree}'$ with the root $S'$ constructed using Rules (\cref{ceq4hr1}), (\cref{ceq4hr1111111})--(\cref{ceq4hr4}) satisfying
for every leaf $S''$ that either $\square\in S''$ or $S''\subseteq_{\mc F} \mi{OrdPropCl}$ is $\gz$-guarded.
We put
\begin{equation} \notag
\mi{Tree}=\begin{array}[c]{c}
          S \\[0.4mm]
          \hline \\[-3.8mm]
          (S-\{a^*\gleq \gz\})\cup \{a^*\geql \gz\} \\[0.4mm]
          \hline \\[-3.8mm]
          (((S-\{a^*\gleq \gz\})\cup \{a^*\geql \gz\})-\{a^*\gle \gu\})\cup \{\gz\gle \gu\} \\[0.4mm]
          \hline \\[-3.8mm]
          \mi{Tree}'.
          \end{array}
\end{equation}
Hence, $\mi{Tree}$ is a finite {\it DPLL}-tree with the root $S$ constructed using Rules (\cref{ceq4hr1}), (\cref{ceq4hr1111111})--(\cref{ceq4hr4}) such that
for every leaf $S''$, either $\square\in S''$ or $S''\subseteq_{\mc F} \mi{OrdPropCl}$ is $\gz$-guarded.

Case 2.2.8:
$\mi{guards}(S,a^*)=\{\gz\gle a^*,a^*\geql \gu\}$.
Then $S\supseteq \mi{guards}(S)\supseteq \mi{guards}(S,a^*)=\{\gz\gle a^*,a^*\geql \gu\}$,
$a^*\geql \gu\in \mi{guards}(S)$, $\gz\gle a^*, a^*\geql \gu\in S$,
$a^*\in \mi{atoms}(\gz\gle a^*)$, $a^*\geql \gu\neq \gz\gle a^*$, $\mi{simplify}(\gz\gle a^*,a^*,\gu)=\gz\gle \gu$,
applying Rule (\cref{ceq4hr4}) to $a^*\geql \gu$ and $\gz\gle a^*$, we derive
\begin{equation} \notag
\dfrac{S}
      {(S-\{\gz\gle a^*\})\cup \{\gz\gle \gu\}};
\end{equation}
$\gz\gle \gu\in (S-\{\gz\gle a^*\})\cup \{\gz\gle \gu\}$;
$\gz\gle \gu\in \mi{OrdPropLit}$ is a tautology;
$\gz\gle \gu$ is not a guard;
$\gz\gle \gu\not\in \mi{guards}((S-\{\gz\gle a^*\})\cup \{\gz\gle \gu\})$,
$\gz\gle \gu\in ((S-\{\gz\gle a^*\})\cup \{\gz\gle \gu\})-\mi{guards}((S-\{\gz\gle a^*\})\cup \{\gz\gle \gu\})$,
applying Rule (\cref{ceq4hr22}) to $(S-\{\gz\gle a^*\})\cup \{\gz\gle \gu\}$ and $\gz\gle \gu$, we derive
\begin{equation} \notag
\dfrac{(S-\{\gz\gle a^*\})\cup \{\gz\gle \gu\}}
      {((S-\{\gz\gle a^*\})\cup \{\gz\gle \gu\})-\{\gz\gle \gu\}}.
\end{equation}
We have that $S$ is simplified.
Hence, $\gz\gle \gu\not\in S$, $\gz\gle a^*\in S$,
$((S-\{\gz\gle a^*\})\cup \{\gz\gle \gu\})-\{\gz\gle \gu\}=((S-\{\gz\gle a^*\})-\{\gz\gle \gu\})\cup (\{\gz\gle \gu\}-\{\gz\gle \gu\})=S-\{\gz\gle a^*\}$.
We put $S'=S-\{\gz\gle a^*\}\subseteq_{\mc F} \mi{OrdPropCl}$.
We get that
$a^*\geql \gu\in S$, $a^*\geql \gu\neq \gz\gle a^*$, $a^*\geql \gu\in S'=S-\{\gz\gle a^*\}$, $\mi{atoms}(\gz\gle a^*)=\{a^*\}$,
$a^*\in \mi{atoms}(a^*\geql \gu)\subseteq \mi{atoms}(S')=\mi{atoms}(S-\{\gz\gle a^*\})=\mi{atoms}(S)$;   
we have that $S$ is simplified;
$S'=S-\{\gz\gle a^*\}\subseteq S$ is simplified;
$\mi{guards}(S',a^*)=(S-\{\gz\gle a^*\})\cap \mi{guards}(a^*)=(S\cap \mi{guards}(a^*))-\{\gz\gle a^*\}=\mi{guards}(S,a^*)-\{\gz\gle a^*\}=\{\gz\gle a^*,a^*\geql \gu\}-\{\gz\gle a^*\}=
                     \{a^*\geql \gu\}$;
$a^*$ is $\gz$-guarded in $S'$;
for all $a\in \mi{atoms}(S')-\{a^*\}=\mi{atoms}(S)-\{a^*\}$,
$a\neq a^*$,
$\mi{guards}(a)\cap \{\gz\gle a^*\}=\mi{guards}(a)\cap \mi{guards}(a^*)=\emptyset$,
$\mi{guards}(S',a)=(S-\{\gz\gle a^*\})\cap \mi{guards}(a)=(S\cap \mi{guards}(a))-\{\gz\gle a^*\}=S\cap \mi{guards}(a)=\mi{guards}(S,a)$;
$a$ is $\gz$-guarded in $S'$ if and only if $a$ is $\gz$-guarded in $S$;
$a^*\in \mi{unguarded}(S)$,
$\mi{unguarded}(S')=\{a \,|\, a\in \mi{atoms}(S')\ \text{\it is not $\gz$-guarded in}\ S'\}=\{a \,|\, a\in (\mi{atoms}(S')-\{a^*\})\cup \{a^*\}\ \text{\it is not $\gz$-guarded in}\ S'\}=
                    \{a \,|\, a\in \mi{atoms}(S')-\{a^*\}\ \text{\it is not $\gz$-guarded in}\ S'\}=\{a \,|\, a\in \mi{atoms}(S)-\{a^*\}\ \text{\it is not $\gz$-guarded in}\ S\}=  
                    \mi{unguarded}(S)-\{a^*\}=\{a \,|\, a\in \mi{atoms}(S)\ \text{\it is not $\gz$-guarded in}\ S\}-\{a^*\}\subset \mi{unguarded}(S)$;
by the induction hypothesis for $S'$, there exists a finite {\it DPLL}-tree $\mi{Tree}'$ with the root $S'$ constructed using Rules (\cref{ceq4hr1}), (\cref{ceq4hr1111111})--(\cref{ceq4hr4}) satisfying
for every leaf $S''$ that either $\square\in S''$ or $S''\subseteq_{\mc F} \mi{OrdPropCl}$ is $\gz$-guarded.
We put
\begin{equation} \notag
\mi{Tree}=\begin{array}[c]{c}
          S \\[0.4mm]
          \hline \\[-3.8mm]
          (S-\{\gz\gle a^*\})\cup \{\gz\gle \gu\} \\[0.4mm]
          \hline \\[-3.8mm]
          \mi{Tree}'.
          \end{array}
\end{equation}
Hence, $\mi{Tree}$ is a finite {\it DPLL}-tree with the root $S$ constructed using Rules (\cref{ceq4hr1}), (\cref{ceq4hr1111111})--(\cref{ceq4hr4}) such that
for every leaf $S''$, either $\square\in S''$ or $S''\subseteq_{\mc F} \mi{OrdPropCl}$ is $\gz$-guarded.

Case 2.2.9:
$\mi{guards}(S,a^*)=\{\gz\gle a^*,\gu\gleq a^*\}$.
Then $\mi{guards}(S)\supseteq \mi{guards}(S,a^*)=\{\gz\gle a^*,\gu\gleq a^*\}$,
$\gu\gleq a^*\in \mi{guards}(S)$, applying Rule (\cref{ceq4hr11111111}) to $\gu\gleq a^*$, we derive
\begin{equation} \notag
\dfrac{S}
      {(S-\{\gu\gleq a^*\})\cup \{a^*\geql \gu\}};
\end{equation}
$a^*\in \mi{atoms}((S-\{\gu\gleq a^*\})\cup \{a^*\geql \gu\})$,
$(S-\{\gu\gleq a^*\})\cup \{a^*\geql \gu\}\supseteq \mi{guards}((S-\{\gu\gleq a^*\})\cup \{a^*\geql \gu\})\supseteq 
 \mi{guards}((S-\{\gu\gleq a^*\})\cup \{a^*\geql \gu\},a^*)=((S-\{\gu\gleq a^*\})\cup \{a^*\geql \gu\})\cap \mi{guards}(a^*)=
 ((S-\{\gu\gleq a^*\})\cap \mi{guards}(a^*))\cup (\{a^*\geql \gu\}\cap \mi{guards}(a^*))=((S\cap \mi{guards}(a^*))-\{\gu\gleq a^*\})\cup \{a^*\geql \gu\}=
 (\mi{guards}(S,a^*)-\{\gu\gleq a^*\})\cup \{a^*\geql \gu\}=(\{\gz\gle a^*,\gu\gleq a^*\}-\{\gu\gleq a^*\})\cup \{a^*\geql \gu\}=\{\gz\gle a^*,a^*\geql \gu\}$,
$a^*\geql \gu\in \mi{guards}((S-\{\gu\gleq a^*\})\cup \{a^*\geql \gu\})$, $\gz\gle a^*\in (S-\{\gu\gleq a^*\})\cup \{a^*\geql \gu\}$,
$a^*\in \mi{atoms}(\gz\gle a^*)$, $a^*\geql \gu\neq \gz\gle a^*$, $\mi{simplify}(\gz\gle a^*,a^*,\gu)=\gz\gle \gu$,
applying Rule (\cref{ceq4hr4}) to $(S-\{\gu\gleq a^*\})\cup \{a^*\geql \gu\}$, $a^*\geql \gu$, and $\gz\gle a^*$, we derive
\begin{equation} \notag
\dfrac{(S-\{\gu\gleq a^*\})\cup \{a^*\geql \gu\}}
      {(((S-\{\gu\gleq a^*\})\cup \{a^*\geql \gu\})-\{\gz\gle a^*\})\cup \{\gz\gle \gu\}};
\end{equation}
$\gz\gle \gu\in (((S-\{\gu\gleq a^*\})\cup \{a^*\geql \gu\})-\{\gz\gle a^*\})\cup \{\gz\gle \gu\}$;
$\gz\gle \gu\in \mi{OrdPropLit}$ is a tautology;
$\gz\gle \gu$ is not a guard;
$\gz\gle \gu\not\in \mi{guards}((((S-\{\gu\gleq a^*\})\cup \{a^*\geql \gu\})-\{\gz\gle a^*\})\cup \{\gz\gle \gu\})$,
$\gz\gle \gu\in ((((S-\{\gu\gleq a^*\})\cup \{a^*\geql \gu\})-\{\gz\gle a^*\})\cup \{\gz\gle \gu\})-\mi{guards}((((S-\{\gu\gleq a^*\})\cup \{a^*\geql \gu\})-\{\gz\gle a^*\})\cup \{\gz\gle \gu\})$,
applying Rule (\cref{ceq4hr22}) to $(((S-\{\gu\gleq a^*\})\cup \{a^*\geql \gu\})-\{\gz\gle a^*\})\cup \{\gz\gle \gu\}$ and $\gz\gle \gu$, we derive
\begin{equation} \notag
\dfrac{(((S-\{\gu\gleq a^*\})\cup \{a^*\geql \gu\})-\{\gz\gle a^*\})\cup \{\gz\gle \gu\}}
      {((((S-\{\gu\gleq a^*\})\cup \{a^*\geql \gu\})-\{\gz\gle a^*\})\cup \{\gz\gle \gu\})-\{\gz\gle \gu\}}.
\end{equation}
We have that $S$ is simplified.
Hence, $\gz\gle \gu\not\in S$,
$a^*\geql \gu\in \mi{guards}(a^*)$, $a^*\geql \gu\not\in \mi{guards}(S,a^*)=S\cap \mi{guards}(a^*)=\{\gz\gle a^*,\gu\gleq a^*\}$, $\gz\gle a^*, \gu\gleq a^*\in S$, $a^*\geql \gu\not\in S$,
$((((S-\{\gu\gleq a^*\})\cup \{a^*\geql \gu\})-\{\gz\gle a^*\})\cup \{\gz\gle \gu\})-\{\gz\gle \gu\}=
 ((((S-\{\gu\gleq a^*\})\cup \{a^*\geql \gu\})-\{\gz\gle a^*\})-\{\gz\gle \gu\})\cup (\{\gz\gle \gu\}-\{\gz\gle \gu\})=
 (((S-\{\gu\gleq a^*\})\cup \{a^*\geql \gu\})-\{\gz\gle \gu\})-\{\gz\gle a^*\}=
 (((S-\{\gu\gleq a^*\})-\{\gz\gle \gu\})\cup (\{a^*\geql \gu\}-\{\gz\gle \gu\}))-\{\gz\gle a^*\}=
 ((S-\{\gu\gleq a^*\})\cup \{a^*\geql \gu\})-\{\gz\gle a^*\}=
 ((S-\{\gu\gleq a^*\})-\{\gz\gle a^*\})\cup (\{a^*\geql \gu\}-\{\gz\gle a^*\})=
 (S-\{\gz\gle a^*,\gu\gleq a^*\})\cup \{a^*\geql \gu\}$.
We put $S'=(S-\{\gz\gle a^*,\gu\gleq a^*\})\cup \{a^*\geql \gu\}\subseteq_{\mc F} \mi{OrdPropCl}$.
We get that
$a^*\in \mi{atoms}(S)$, $\mi{atoms}(\gz\gle a^*,a^*\geql \gu,\gu\gleq a^*)=\{a^*\}$,
$a^*\in \mi{atoms}(S')=\mi{atoms}((S-\{\gz\gle a^*,\gu\gleq a^*\})\cup \{a^*\geql \gu\})=\mi{atoms}(S)$;
we have that $S$ is simplified;
$S-\{\gz\gle a^*,\gu\gleq a^*\}\subseteq S$ is simplified;
$a^*\geql \gu\neq \square$ does not contain contradictions and tautologies;
$S'=(S-\{\gz\gle a^*,\gu\gleq a^*\})\cup \{a^*\geql \gu\}$ is simplified;
$\mi{guards}(S',a^*)=((S-\{\gz\gle a^*,\gu\gleq a^*\})\cup \{a^*\geql \gu\})\cap \mi{guards}(a^*)=((S-\{\gz\gle a^*,\gu\gleq a^*\})\cap \mi{guards}(a^*))\cup (\{a^*\geql \gu\}\cap \mi{guards}(a^*))=
                     ((S\cap \mi{guards}(a^*))-\{\gz\gle a^*,\gu\gleq a^*\})\cup \{a^*\geql \gu\}=(\mi{guards}(S,a^*)-\{\gz\gle a^*,\gu\gleq a^*\})\cup \{a^*\geql \gu\}=
                     (\{\gz\gle a^*,\gu\gleq a^*\}-\{\gz\gle a^*,\gu\gleq a^*\})\cup \{a^*\geql \gu\}=\{a^*\geql \gu\}$;
$a^*$ is $\gz$-guarded in $S'$;
for all $a\in \mi{atoms}(S')-\{a^*\}=\mi{atoms}(S)-\{a^*\}$,
$a\neq a^*$,
$\mi{guards}(a)\cap \{\gz\gle a^*,a^*\geql \gu,\gu\gleq a^*\}=\mi{guards}(a)\cap \mi{guards}(a^*)=\emptyset$,
$\mi{guards}(S',a)=((S-\{\gz\gle a^*,\gu\gleq a^*\})\cup \{a^*\geql \gu\})\cap \mi{guards}(a)=((S-\{\gz\gle a^*,\gu\gleq a^*\})\cap \mi{guards}(a))\cup (\{a^*\geql \gu\}\cap \mi{guards}(a))=
                   (S\cap \mi{guards}(a))-\{\gz\gle a^*,\gu\gleq a^*\}=S\cap \mi{guards}(a)=\mi{guards}(S,a)$;
$a$ is $\gz$-guarded in $S'$ if and only if $a$ is $\gz$-guarded in $S$;
$a^*\in \mi{unguarded}(S)$,
$\mi{unguarded}(S')=\{a \,|\, a\in \mi{atoms}(S')\ \text{\it is not $\gz$-guarded in}\ S'\}=\{a \,|\, a\in (\mi{atoms}(S')-\{a^*\})\cup \{a^*\}\ \text{\it is not $\gz$-guarded in}\ S'\}=
                    \{a \,|\, a\in \mi{atoms}(S')-\{a^*\}\ \text{\it is not $\gz$-guarded in}\ S'\}=\{a \,|\, a\in \mi{atoms}(S)-\{a^*\}\ \text{\it is not $\gz$-guarded in}\ S\}=  
                    \mi{unguarded}(S)-\{a^*\}=\{a \,|\, a\in \mi{atoms}(S)\ \text{\it is not $\gz$-guarded in}\ S\}-\{a^*\}\subset \mi{unguarded}(S)$;
by the induction hypothesis for $S'$, there exists a finite {\it DPLL}-tree $\mi{Tree}'$ with the root $S'$ constructed using Rules (\cref{ceq4hr1}), (\cref{ceq4hr1111111})--(\cref{ceq4hr4}) satisfying
for every leaf $S''$ that either $\square\in S''$ or $S''\subseteq_{\mc F} \mi{OrdPropCl}$ is $\gz$-guarded.
We put
\begin{equation} \notag
\mi{Tree}=\begin{array}[c]{c}
          S \\[0.4mm]
          \hline \\[-3.8mm]
          (S-\{\gu\gleq a^*\})\cup \{a^*\geql \gu\} \\[0.4mm]
          \hline \\[-3.8mm]
          (((S-\{\gu\gleq a^*\})\cup \{a^*\geql \gu\})-\{\gz\gle a^*\})\cup \{\gz\gle \gu\} \\[0.4mm]
          \hline \\[-3.8mm]
          \mi{Tree}'.
          \end{array}
\end{equation}
Hence, $\mi{Tree}$ is a finite {\it DPLL}-tree with the root $S$ constructed using Rules (\cref{ceq4hr1}), (\cref{ceq4hr1111111})--(\cref{ceq4hr4}) such that
for every leaf $S''$, either $\square\in S''$ or $S''\subseteq_{\mc F} \mi{OrdPropCl}$ is $\gz$-guarded.

Case 2.2.10:
$\mi{guards}(S,a^*)=\{a^*\geql \gu,\gu\gleq a^*\}$.
Then $S\supseteq \mi{guards}(S)\supseteq \mi{guards}(S,a^*)=\{a^*\geql \gu,\gu\gleq a^*\}$, $a^*\geql \gu, \gu\gleq a^*\in S$,
$\gu\gleq a^*\in \mi{guards}(S)$, applying Rule (\cref{ceq4hr11111111}) to $\gu\gleq a^*$, we derive
\begin{equation} \notag
\dfrac{S}
      {(S-\{\gu\gleq a^*\})\cup \{a^*\geql \gu\}}.
\end{equation}
Hence, $a^*\geql \gu, \gu\gleq a^*\in S$, $a^*\geql \gu\neq \gu\gleq a^*$, $a^*\geql \gu\in S-\{\gu\gleq a^*\}$,
$(S-\{\gu\gleq a^*\})\cup \{a^*\geql \gu\}=S-\{\gu\gleq a^*\}$.
We put $S'=S-\{\gu\gleq a^*\}\subseteq_{\mc F} \mi{OrdPropCl}$.
We get that
$a^*\geql \gu\in S'=S-\{\gu\gleq a^*\}$, $\mi{atoms}(\gu\gleq a^*)=\{a^*\}$,
$a^*\in \mi{atoms}(a^*\geql \gu)\subseteq \mi{atoms}(S')=\mi{atoms}(S-\{\gu\gleq a^*\})=\mi{atoms}(S)$;
we have that $S$ is simplified;
$S'=S-\{\gu\gleq a^*\}\subseteq S$ is simplified;
$\mi{guards}(S',a^*)=(S-\{\gu\gleq a^*\})\cap \mi{guards}(a^*)=(S\cap \mi{guards}(a^*))-\{\gu\gleq a^*\}=\mi{guards}(S,a^*)-\{\gu\gleq a^*\}=\{a^*\geql \gu,\gu\gleq a^*\}-\{\gu\gleq a^*\}=
                     \{a^*\geql \gu\}$;
$a^*$ is $\gz$-guarded in $S'$;
for all $a\in \mi{atoms}(S')-\{a^*\}=\mi{atoms}(S)-\{a^*\}$,
$a\neq a^*$,
$\mi{guards}(a)\cap \{\gu\gleq a^*\}=\mi{guards}(a)\cap \mi{guards}(a^*)=\emptyset$,
$\mi{guards}(S',a)=(S-\{\gu\gleq a^*\})\cap \mi{guards}(a)=(S\cap \mi{guards}(a))-\{\gu\gleq a^*\}=S\cap \mi{guards}(a)=\mi{guards}(S,a)$;
$a$ is $\gz$-guarded in $S'$ if and only if $a$ is $\gz$-guarded in $S$;
$a^*\in \mi{unguarded}(S)$,
$\mi{unguarded}(S')=\{a \,|\, a\in \mi{atoms}(S')\ \text{\it is not $\gz$-guarded in}\ S'\}=\{a \,|\, a\in (\mi{atoms}(S')-\{a^*\})\cup \{a^*\}\ \text{\it is not $\gz$-guarded in}\ S'\}=
                    \{a \,|\, a\in \mi{atoms}(S')-\{a^*\}\ \text{\it is not $\gz$-guarded in}\ S'\}=\{a \,|\, a\in \mi{atoms}(S)-\{a^*\}\ \text{\it is not $\gz$-guarded in}\ S\}=  
                    \mi{unguarded}(S)-\{a^*\}=\{a \,|\, a\in \mi{atoms}(S)\ \text{\it is not $\gz$-guarded in}\ S\}-\{a^*\}\subset \mi{unguarded}(S)$;
by the induction hypothesis for $S'$, there exists a finite {\it DPLL}-tree $\mi{Tree}'$ with the root $S'$ constructed using Rules (\cref{ceq4hr1}), (\cref{ceq4hr1111111})--(\cref{ceq4hr4}) satisfying
for every leaf $S''$ that either $\square\in S''$ or $S''\subseteq_{\mc F} \mi{OrdPropCl}$ is $\gz$-guarded.
We put
\begin{equation} \notag
\mi{Tree}=\dfrac{S}
                {\mi{Tree}'}.
\end{equation}
Hence, $\mi{Tree}$ is a finite {\it DPLL}-tree with the root $S$ constructed using Rules (\cref{ceq4hr1}), (\cref{ceq4hr1111111})--(\cref{ceq4hr4}) such that
for every leaf $S''$, either $\square\in S''$ or $S''\subseteq_{\mc F} \mi{OrdPropCl}$ is $\gz$-guarded.

Case 2.2.11:
$\mi{guards}(S,a^*)=\{a^*\geql \gz,a^*\gleq \gz,a^*\gle \gu\}$.
Then $S\supseteq \mi{guards}(S)\supseteq \mi{guards}(S,a^*)=\{a^*\geql \gz,a^*\gleq \gz,a^*\gle \gu\}$,
$a^*\gleq \gz\in \mi{guards}(S)$, applying Rule (\cref{ceq4hr1111111}) to $a^*\gleq \gz$, we derive
\begin{equation} \notag
\dfrac{S}
      {(S-\{a^*\gleq \gz\})\cup \{a^*\geql \gz\}};
\end{equation}
$a^*\geql \gz, a^*\gleq \gz\in S$, $a^*\geql \gz\neq a^*\gleq \gz$, $a^*\geql \gz\in S-\{a^*\gleq \gz\}$,
$(S-\{a^*\gleq \gz\})\cup \{a^*\geql \gz\}=S-\{a^*\gleq \gz\}$;
$a^*\in \mi{atoms}(a^*\geql \gz)\subseteq \mi{atoms}(S-\{a^*\gleq \gz\})$,
$S-\{a^*\gleq \gz\}\supseteq \mi{guards}(S-\{a^*\gleq \gz\})\supseteq 
 \mi{guards}(S-\{a^*\gleq \gz\},a^*)=(S-\{a^*\gleq \gz\})\cap \mi{guards}(a^*)=(S\cap \mi{guards}(a^*))-\{a^*\gleq \gz\}=\mi{guards}(S,a^*)-\{a^*\gleq \gz\}=
                                     \{a^*\geql \gz,a^*\gleq \gz,a^*\gle \gu\}-\{a^*\gleq \gz\}=\{a^*\geql \gz,a^*\gle \gu\}$,
$a^*\geql \gz\in \mi{guards}(S-\{a^*\gleq \gz\})$, $a^*\gle \gu\in S-\{a^*\gleq \gz\}$,
$a^*\in \mi{atoms}(a^*\gle \gu)$, $a^*\geql \gz\neq a^*\gle \gu$, $\mi{simplify}(a^*\gle \gu,a^*,\gz)=\gz\gle \gu$,
applying Rule (\cref{ceq4hr3}) to $S-\{a^*\gleq \gz\}$, $a^*\geql \gz$, and $a^*\gle \gu$, we derive
\begin{equation} \notag
\dfrac{S-\{a^*\gleq \gz\}}
      {((S-\{a^*\gleq \gz\})-\{a^*\gle \gu\})\cup \{\gz\gle \gu\}};
\end{equation}
$\gz\gle \gu\in ((S-\{a^*\gleq \gz\})-\{a^*\gle \gu\})\cup \{\gz\gle \gu\}$;
$\gz\gle \gu\in \mi{OrdPropLit}$ is a tautology;
$\gz\gle \gu$ is not a guard;
$\gz\gle \gu\not\in \mi{guards}(((S-\{a^*\gleq \gz\})-\{a^*\gle \gu\})\cup \{\gz\gle \gu\})$,
$\gz\gle \gu\in (((S-\{a^*\gleq \gz\})-\{a^*\gle \gu\})\cup \{\gz\gle \gu\})-\mi{guards}(((S-\{a^*\gleq \gz\})-\{a^*\gle \gu\})\cup \{\gz\gle \gu\})$,
applying Rule (\cref{ceq4hr22}) to $((S-\{a^*\gleq \gz\})-\{a^*\gle \gu\})\cup \{\gz\gle \gu\}$ and $\gz\gle \gu$, we derive
\begin{equation} \notag
\dfrac{((S-\{a^*\gleq \gz\})-\{a^*\gle \gu\})\cup \{\gz\gle \gu\}}
      {(((S-\{a^*\gleq \gz\})-\{a^*\gle \gu\})\cup \{\gz\gle \gu\})-\{\gz\gle \gu\}}.
\end{equation}
We have that $S$ is simplified.
Hence, $\gz\gle \gu\not\in S$, $a^*\gleq \gz, a^*\gle \gu\in S$,
$(((S-\{a^*\gleq \gz\})-\{a^*\gle \gu\})\cup \{\gz\gle \gu\})-\{\gz\gle \gu\}=(((S-\{a^*\gleq \gz\})-\{a^*\gle \gu\})-\{\gz\gle \gu\})\cup (\{\gz\gle \gu\}-\{\gz\gle \gu\})=
 (S-\{a^*\gleq \gz,a^*\gle \gu\})-\{\gz\gle \gu\}=S-\{a^*\gleq \gz,a^*\gle \gu\}$.
We put $S'=S-\{a^*\gleq \gz,a^*\gle \gu\}\subseteq_{\mc F} \mi{OrdPropCl}$.
We get that
$a^*\geql \gz\in S$, $a^*\geql \gz\neq a^*\gleq \gz, a^*\gle \gu$, $a^*\geql \gz\in S'=S-\{a^*\gleq \gz,a^*\gle \gu\}$, $\mi{atoms}(a^*\gleq \gz,a^*\gle \gu)=\{a^*\}$, 
$a^*\in \mi{atoms}(a^*\geql \gz)\subseteq \mi{atoms}(S')=\mi{atoms}(S-\{a^*\gleq \gz,a^*\gle \gu\})=\mi{atoms}(S)$;
we have that $S$ is simplified;
$S'=S-\{a^*\gleq \gz,a^*\gle \gu\}\subseteq S$ is simplified;
$\mi{guards}(S',a^*)=(S-\{a^*\gleq \gz,a^*\gle \gu\})\cap \mi{guards}(a^*)=(S\cap \mi{guards}(a^*))-\{a^*\gleq \gz,a^*\gle \gu\}=\mi{guards}(S,a^*)-\{a^*\gleq \gz,a^*\gle \gu\}=
                     \{a^*\geql \gz,a^*\gleq \gz,a^*\gle \gu\}-\{a^*\gleq \gz,a^*\gle \gu\}=\{a^*\geql \gz\}$;
$a^*$ is $\gz$-guarded in $S'$;
for all $a\in \mi{atoms}(S')-\{a^*\}=\mi{atoms}(S)-\{a^*\}$,
$a\neq a^*$,
$\mi{guards}(a)\cap \{a^*\gleq \gz,a^*\gle \gu\}=\mi{guards}(a)\cap \mi{guards}(a^*)=\emptyset$,
$\mi{guards}(S',a)=(S-\{a^*\gleq \gz,a^*\gle \gu\})\cap \mi{guards}(a)=(S\cap \mi{guards}(a))-\{a^*\gleq \gz,a^*\gle \gu\}=S\cap \mi{guards}(a)=\mi{guards}(S,a)$;
$a$ is $\gz$-guarded in $S'$ if and only if $a$ is $\gz$-guarded in $S$;
$a^*\in \mi{unguarded}(S)$,
$\mi{unguarded}(S')=\{a \,|\, a\in \mi{atoms}(S')\ \text{\it is not $\gz$-guarded in}\ S'\}=\{a \,|\, a\in (\mi{atoms}(S')-\{a^*\})\cup \{a^*\}\ \text{\it is not $\gz$-guarded in}\ S'\}=
                    \{a \,|\, a\in \mi{atoms}(S')-\{a^*\}\ \text{\it is not $\gz$-guarded in}\ S'\}=\{a \,|\, a\in \mi{atoms}(S)-\{a^*\}\ \text{\it is not $\gz$-guarded in}\ S\}=  
                    \mi{unguarded}(S)-\{a^*\}=\{a \,|\, a\in \mi{atoms}(S)\ \text{\it is not $\gz$-guarded in}\ S\}-\{a^*\}\subset \mi{unguarded}(S)$;
by the induction hypothesis for $S'$, there exists a finite {\it DPLL}-tree $\mi{Tree}'$ with the root $S'$ constructed using Rules (\cref{ceq4hr1}), (\cref{ceq4hr1111111})--(\cref{ceq4hr4}) satisfying
for every leaf $S''$ that either $\square\in S''$ or $S''\subseteq_{\mc F} \mi{OrdPropCl}$ is $\gz$-guarded.
We put
\begin{equation} \notag
\mi{Tree}=\begin{array}[c]{c}
          S \\[0.4mm]
          \hline \\[-3.8mm]
          (S-\{a^*\gleq \gz\})\cup \{a^*\geql \gz\} \\[0.4mm]
          \hline \\[-3.8mm]
          ((S-\{a^*\gleq \gz\})-\{a^*\gle \gu\})\cup \{\gz\gle \gu\} \\[0.4mm]
          \hline \\[-3.8mm]
          \mi{Tree}'.
          \end{array}
\end{equation}
Hence, $\mi{Tree}$ is a finite {\it DPLL}-tree with the root $S$ constructed using Rules (\cref{ceq4hr1}), (\cref{ceq4hr1111111})--(\cref{ceq4hr4}) such that
for every leaf $S''$, either $\square\in S''$ or $S''\subseteq_{\mc F} \mi{OrdPropCl}$ is $\gz$-guarded.

Case 2.2.12:
$\mi{guards}(S,a^*)=\{\gz\gle a^*,a^*\geql \gu,\gu\gleq a^*\}$.
Then $S\supseteq \mi{guards}(S)\supseteq \mi{guards}(S,a^*)=\{\gz\gle a^*,a^*\geql \gu,\gu\gleq a^*\}$,
$\gu\gleq a^*\in \mi{guards}(S)$, applying Rule (\cref{ceq4hr11111111}) to $\gu\gleq a^*$, we derive
\begin{equation} \notag
\dfrac{S}
      {(S-\{\gu\gleq a^*\})\cup \{a^*\geql \gu\}};
\end{equation}
$a^*\geql \gu, \gu\gleq a^*\in S$, $a^*\geql \gu\neq \gu\gleq a^*$, $a^*\geql \gu\in S-\{\gu\gleq a^*\}$,
$(S-\{\gu\gleq a^*\})\cup \{a^*\geql \gu\}=S-\{\gu\gleq a^*\}$;
$a^*\in \mi{atoms}(a^*\geql \gu)\subseteq \mi{atoms}(S-\{\gu\gleq a^*\})$,
$S-\{\gu\gleq a^*\}\supseteq \mi{guards}(S-\{\gu\gleq a^*\})\supseteq 
 \mi{guards}(S-\{\gu\gleq a^*\},a^*)=(S-\{\gu\gleq a^*\})\cap \mi{guards}(a^*)=(S\cap \mi{guards}(a^*))-\{\gu\gleq a^*\}=\mi{guards}(S,a^*)-\{\gu\gleq a^*\}=
                                     \{\gz\gle a^*,a^*\geql \gu,\gu\gleq a^*\}-\{\gu\gleq a^*\}=\{\gz\gle a^*,a^*\geql \gu\}$,
$a^*\geql \gu\in \mi{guards}(S-\{\gu\gleq a^*\})$, $\gz\gle a^*\in S-\{\gu\gleq a^*\}$,
$a^*\in \mi{atoms}(\gz\gle a^*)$, $a^*\geql \gu\neq \gz\gle a^*$, $\mi{simplify}(\gz\gle a^*,a^*,\gu)=\gz\gle \gu$,
applying Rule (\cref{ceq4hr4}) to $S-\{\gu\gleq a^*\}$, $a^*\geql \gu$, and $\gz\gle a^*$, we derive
\begin{equation} \notag
\dfrac{S-\{\gu\gleq a^*\}}
      {((S-\{\gu\gleq a^*\})-\{\gz\gle a^*\})\cup \{\gz\gle \gu\}};
\end{equation}
$\gz\gle \gu\in ((S-\{\gu\gleq a^*\})-\{\gz\gle a^*\})\cup \{\gz\gle \gu\}$;
$\gz\gle \gu\in \mi{OrdPropLit}$ is a tautology;
$\gz\gle \gu$ is not a guard;
$\gz\gle \gu\not\in \mi{guards}(((S-\{\gu\gleq a^*\})-\{\gz\gle a^*\})\cup \{\gz\gle \gu\})$,
$\gz\gle \gu\in (((S-\{\gu\gleq a^*\})-\{\gz\gle a^*\})\cup \{\gz\gle \gu\})-\mi{guards}(((S-\{\gu\gleq a^*\})-\{\gz\gle a^*\})\cup \{\gz\gle \gu\})$,
applying Rule (\cref{ceq4hr22}) to $((S-\{\gu\gleq a^*\})-\{\gz\gle a^*\})\cup \{\gz\gle \gu\}$ and $\gz\gle \gu$, we derive
\begin{equation} \notag
\dfrac{((S-\{\gu\gleq a^*\})-\{\gz\gle a^*\})\cup \{\gz\gle \gu\}}
      {(((S-\{\gu\gleq a^*\})-\{\gz\gle a^*\})\cup \{\gz\gle \gu\})-\{\gz\gle \gu\}}.
\end{equation}
We have that $S$ is simplified.
Hence, $\gz\gle \gu\not\in S$, $\gz\gle a^*, \gu\gleq a^*\in S$,
$(((S-\{\gu\gleq a^*\})-\{\gz\gle a^*\})\cup \{\gz\gle \gu\})-\{\gz\gle \gu\}=(((S-\{\gu\gleq a^*\})-\{\gz\gle a^*\})-\{\gz\gle \gu\})\cup (\{\gz\gle \gu\}-\{\gz\gle \gu\})=
 (S-\{\gz\gle a^*,\gu\gleq a^*\})-\{\gz\gle \gu\}=S-\{\gz\gle a^*,\gu\gleq a^*\}$.
We put $S'=S-\{\gz\gle a^*,\gu\gleq a^*\}\subseteq_{\mc F} \mi{OrdPropCl}$.
We get that
$a^*\geql \gu\in S$, $a^*\geql \gu\neq \gz\gle a^*, \gu\gleq a^*$, $a^*\geql \gu\in S'=S-\{\gz\gle a^*,\gu\gleq a^*\}$, $\mi{atoms}(\gz\gle a^*,\gu\gleq a^*)=\{a^*\}$, 
$a^*\in \mi{atoms}(a^*\geql \gu)\subseteq \mi{atoms}(S')=\mi{atoms}(S-\{\gz\gle a^*,\gu\gleq a^*\})=\mi{atoms}(S)$;
we have that $S$ is simplified;
$S'=S-\{\gz\gle a^*,\gu\gleq a^*\}\subseteq S$ is simplified;
$\mi{guards}(S',a^*)=(S-\{\gz\gle a^*,\gu\gleq a^*\})\cap \mi{guards}(a^*)=(S\cap \mi{guards}(a^*))-\{\gz\gle a^*,\gu\gleq a^*\}=\mi{guards}(S,a^*)-\{\gz\gle a^*,\gu\gleq a^*\}=
                     \{\gz\gle a^*,a^*\geql \gu,\gu\gleq a^*\}-\{\gz\gle a^*,\gu\gleq a^*\}=\{a^*\geql \gu\}$;
$a^*$ is $\gz$-guarded in $S'$;
for all $a\in \mi{atoms}(S')-\{a^*\}=\mi{atoms}(S)-\{a^*\}$,
$a\neq a^*$,
$\mi{guards}(a)\cap \{\gz\gle a^*,\gu\gleq a^*\}=\mi{guards}(a)\cap \mi{guards}(a^*)=\emptyset$,
$\mi{guards}(S',a)=(S-\{\gz\gle a^*,\gu\gleq a^*\})\cap \mi{guards}(a)=(S\cap \mi{guards}(a))-\{\gz\gle a^*,\gu\gleq a^*\}=S\cap \mi{guards}(a)=\mi{guards}(S,a)$;
$a$ is $\gz$-guarded in $S'$ if and only if $a$ is $\gz$-guarded in $S$;
$a^*\in \mi{unguarded}(S)$,
$\mi{unguarded}(S')=\{a \,|\, a\in \mi{atoms}(S')\ \text{\it is not $\gz$-guarded in}\ S'\}=\{a \,|\, a\in (\mi{atoms}(S')-\{a^*\})\cup \{a^*\}\ \text{\it is not $\gz$-guarded in}\ S'\}=
                    \{a \,|\, a\in \mi{atoms}(S')-\{a^*\}\ \text{\it is not $\gz$-guarded in}\ S'\}=\{a \,|\, a\in \mi{atoms}(S)-\{a^*\}\ \text{\it is not $\gz$-guarded in}\ S\}=  
                    \mi{unguarded}(S)-\{a^*\}=\{a \,|\, a\in \mi{atoms}(S)\ \text{\it is not $\gz$-guarded in}\ S\}-\{a^*\}\subset \mi{unguarded}(S)$;
by the induction hypothesis for $S'$, there exists a finite {\it DPLL}-tree $\mi{Tree}'$ with the root $S'$ constructed using Rules (\cref{ceq4hr1}), (\cref{ceq4hr1111111})--(\cref{ceq4hr4}) satisfying
for every leaf $S''$ that either $\square\in S''$ or $S''\subseteq_{\mc F} \mi{OrdPropCl}$ is $\gz$-guarded.
We put
\begin{equation} \notag
\mi{Tree}=\begin{array}[c]{c}
          S \\[0.4mm]
          \hline \\[-3.8mm]
          (S-\{\gu\gleq a^*\})\cup \{a^*\geql \gu\} \\[0.4mm]
          \hline \\[-3.8mm]
          ((S-\{\gu\gleq a^*\})-\{\gz\gle a^*\})\cup \{\gz\gle \gu\} \\[0.4mm]
          \hline \\[-3.8mm]
          \mi{Tree}'.
          \end{array}
\end{equation}
Hence, $\mi{Tree}$ is a finite {\it DPLL}-tree with the root $S$ constructed using Rules (\cref{ceq4hr1}), (\cref{ceq4hr1111111})--(\cref{ceq4hr4}) such that
for every leaf $S''$, either $\square\in S''$ or $S''\subseteq_{\mc F} \mi{OrdPropCl}$ is $\gz$-guarded.

So, in both Cases 1 and 2, the statement holds.
The induction is completed.
%
%
%
\end{proof}

\subsection{Full proof of Lemma \ref{le6}}
\label{S7.8a}

\begin{proof}
We distinguish two cases for $\mi{guards}(S,a^*)$.

Case 1:
$\mi{guards}(S,a^*)=\{a^*\geql \gz\}$.
Let $S^F\subseteq \mi{OrdPropCl}$.
We define a measure operator $\mi{restricted}(S^F)=\{C \,|\, C\in S^F, a^*\in \mi{atoms}(C), C\neq a^*\geql \gz\}$.
We proceed by induction on $\mi{restricted}(S)\subseteq S\subseteq_{\mc F} \mi{OrdPropCl}$.

Case 1.1 (the base case):
$\mi{restricted}(S)=\emptyset$.
We have that $S$ is $\gz$-guarded.
Then $S\supseteq \mi{guards}(S)\supseteq \mi{guards}(S,a^*)=\{a^*\geql \gz\}$, $a^*\geql \gz\in \mi{guards}(S)$, $a^*\geql \gz\in S$;
for all $C\in S$ satisfying $C\neq a^*\geql \gz$, $a^*\not\in \mi{atoms}(C)$;
$a^*\not\in \mi{atoms}(S-\{a^*\geql \gz\})$, applying Rule (\cref{ceq4hr7}) to $a^*\geql \gz$, we derive
\begin{equation} \notag
\dfrac{S}
      {S-\{a^*\geql \gz\}}.
\end{equation}
We put $S'=S-\{a^*\geql \gz\}\subseteq_{\mc F} \mi{OrdPropCl}$.
We get that
$a^*\in \mi{atoms}(S)$, $\mi{atoms}(a^*\geql \gz)=\{a^*\}$,
$a^*\not\in \mi{atoms}(S')=\mi{atoms}(S-\{a^*\geql \gz\})=\mi{atoms}(S)-\{a^*\}$;
$S$ is simplified;
$S'=S-\{a^*\geql \gz\}\subseteq S$ is simplified;
for all $a\in \mi{atoms}(S')=\mi{atoms}(S)-\{a^*\}$,
$a$ is $\gz$-guarded in $S$;
$a\neq a^*$, 
$\mi{guards}(a)\cap \{a^*\geql \gz\}=\mi{guards}(a)\cap \mi{guards}(a^*)=\emptyset$,
$\mi{guards}(S',a)=(S-\{a^*\geql \gz\})\cap \mi{guards}(a)=(S\cap \mi{guards}(a))-\{a^*\geql \gz\}=S\cap \mi{guards}(a)=\mi{guards}(S,a)$;
$a$ is $\gz$-guarded in $S'$;
$S'$ is $\gz$-guarded.
We put
\begin{equation} \notag
\mi{Tree}=\dfrac{S} 
                {S'}.
\end{equation}
Hence, $\mi{Tree}$ is a finite linear {\it DPLL}-tree with the root $S$ constructed using Rules (\cref{ceq4hr1111111})--(\cref{ceq4hr4}), (\cref{ceq4hr7}) such that
for its only leaf $S'$, $S'\subseteq_{\mc F} \mi{OrdPropCl}$ is $\gz$-guarded, $\mi{atoms}(S')=\mi{atoms}(S)-\{a^*\}$;
(i) holds.

Case 1.2 (the induction case):
$\emptyset\neq \mi{restricted}(S)\subseteq_{\mc F} \mi{OrdPropCl}$.
We have that $S$ is $\gz$-guarded.
Then $S\supseteq \mi{guards}(S)\supseteq \mi{guards}(S,a^*)=\{a^*\geql \gz\}$, $a^*\geql \gz\in \mi{guards}(S)$, $a^*\geql \gz\in S$;
there exists $C^*\in \mi{restricted}(S)\subseteq S$ satisfying $a^*\in \mi{atoms}(C^*)$, $a^*\geql \gz\neq C^*$;
applying Rule (\cref{ceq4hr3}) to $a^*\geql \gz$ and $C^*$, we derive
\begin{equation} \notag
\dfrac{S}
      {(S-\{C^*\})\cup \{\mi{simplify}(C^*,a^*,\gz)\}}.
\end{equation}
We put $S'=(S-\{C^*\})\cup \{\mi{simplify}(C^*,a^*,\gz)\}\subseteq_{\mc F} \mi{OrdPropCl}$.
We get that
$a^*\geql \gz\in S$, $a^*\geql \gz\neq C^*$, $a^*\geql \gz\in S-\{C^*\}\subseteq S'$, $\mi{atoms}(\mi{simplify}(C^*,a^*,\gz))\subseteq \mi{atoms}(C^*)$, $C^*\in S$,
$a^*\in \mi{atoms}(a^*\geql \gz)\subseteq \mi{atoms}(S')=\mi{atoms}((S-\{C^*\})\cup \{\mi{simplify}(C^*,a^*,\gz)\})=\mi{atoms}(S-\{C^*\})\cup \mi{atoms}(\mi{simplify}(C^*,a^*,\gz))\subseteq 
                                                         \mi{atoms}(S-\{C^*\})\cup \mi{atoms}(C^*)=\mi{atoms}((S-\{C^*\})\cup \{C^*\})=\mi{atoms}(S)$;
$C^*\not\in \mi{guards}(S,a^*)=S\cap \mi{guards}(a^*)=\{a^*\geql \gz\}$, $C^*\not\in \mi{guards}(a^*)$;
for all $a\in \mi{atoms}(S)-\{a^*\}$,
$a\neq a^*$;
for all $C\in \mi{guards}(a)$, $a^*\in \mi{atoms}(C^*)$, $a^*\not\in \mi{atoms}(C)=\{a\}$, $\mi{atoms}(C^*)\neq \mi{atoms}(C)$, $C^*\neq C$;
$C^*\not\in \mi{guards}(a)$;
for all $a\in \mi{atoms}(C^*)\subseteq \mi{atoms}(S)$, $C^*\not\in \mi{guards}(a)$;
$C^*$ is not a guard; 
$C^*\not\in \mi{guards}(S)$.
We get two cases for $\mi{simplify}(C^*,a^*,\gz)$.

Case 1.2.1:
$\mi{simplify}(C^*,a^*,\gz)\in S$.
Then $a^*\not\in \mi{atoms}(\mi{simplify}(C^*,a^*,\gz))$, $a^*\in \mi{atoms}(C^*)$, $\mi{atoms}(\mi{simplify}(C^*,a^*,\gz))\neq \mi{atoms}(C^*)$, $\mi{simplify}(C^*,a^*,\gz)\neq C^*$,
$\mi{simplify}(C^*,a^*,\gz)\in S-\{C^*\}$,
$S'=(S-\{C^*\})\cup \{\mi{simplify}(C^*,a^*,\gz)\}=S-\{C^*\}$;
$S$ is simplified;
$S'=S-\{C^*\}\subseteq S$ is simplified;
$C^*\not\in \mi{guards}(S)$,
$\mi{guards}(S')=\{C \,|\, C\in S-\{C^*\}\ \text{\it is a guard}\}=\mi{guards}(S)-\{C^*\}=\{C \,|\, C\in S\ \text{\it is a guard}\}-\{C^*\}=\mi{guards}(S)$;
we have that $S$ is $\gz$-guarded;
$\mi{atoms}(S'-\mi{guards}(S'))\subseteq \mi{atoms}(S')\subseteq \mi{atoms}(S)$,
$\mi{atoms}(S')=\mi{atoms}((S'-\mi{guards}(S'))\cup \mi{guards}(S'))=\mi{atoms}(S'-\mi{guards}(S'))\cup \mi{atoms}(\mi{guards}(S'))=
                \mi{atoms}(S'-\mi{guards}(S'))\cup \mi{atoms}(\mi{guards}(S))=\mi{atoms}(S'-\mi{guards}(S'))\cup \mi{atoms}(S)=\mi{atoms}(S)$;
$S'$ is $\gz$-guarded;
$a^*\in \mi{atoms}(S')$, $\mi{guards}(S',a^*)=\mi{guards}(S,a^*)=\{a^*\geql \gz\}$,
$C^*\in \mi{restricted}(S)$,
$\mi{restricted}(S')=\{C \,|\, C\in S-\{C^*\}, a^*\in \mi{atoms}(C), C\neq a^*\geql \gz\}=
                     \mi{restricted}(S)-\{C^*\}=\{C \,|\, C\in S, a^*\in \mi{atoms}(C), C\neq a^*\geql \gz\}-\{C^*\}\subset \mi{restricted}(S)$;
by the induction hypothesis for $S'$, there exists a finite linear {\it DPLL}-tree $\mi{Tree}'$ with the root $S'$ constructed using Rules (\cref{ceq4hr1111111})--(\cref{ceq4hr4}), (\cref{ceq4hr7}) satisfying
for its only leaf $S''$ that either $\square\in S''$, or $S''\subseteq_{\mc F} \mi{OrdPropCl}$ is $\gz$-guarded, $\mi{atoms}(S'')\subseteq \mi{atoms}(S')-\{a^*\}=\mi{atoms}(S)-\{a^*\}$.
We put
\begin{equation} \notag
\mi{Tree}=\dfrac{S}
                {\mi{Tree}'}.
\end{equation}
Hence, $\mi{Tree}$ is a finite linear {\it DPLL}-tree with the root $S$ constructed using Rules (\cref{ceq4hr1111111})--(\cref{ceq4hr4}), (\cref{ceq4hr7}) such that
for its only leaf $S''$, either $\square\in S''$, or $S''\subseteq_{\mc F} \mi{OrdPropCl}$ is $\gz$-guarded, $\mi{atoms}(S'')\subseteq \mi{atoms}(S)-\{a^*\}$;
(i) holds.

Case 1.2.2:
$\mi{simplify}(C^*,a^*,\gz)\not\in S$.
Then $\mi{simplify}(C^*,a^*,\gz)\in S'$; 
by Lemma \ref{le555} for $S'$ and $\mi{simplify}(C^*,a^*,\gz)$, there exists a finite linear {\it DPLL}-tree $\mi{Tree}'$ with the root $S'$ constructed using Rules (\cref{ceq4hr2}) and (\cref{ceq4hr22}) satisfying 
for its only leaf $S''$ that either $\square\in S''$, or $\square\not\in S''\subseteq_{\mc F} \mi{OrdPropCl}$ and exactly one of the following points holds.
\begin{enumerate}[\rm (a)]
\item
$S''=S'$, $\mi{simplify}(C^*,a^*,\gz)\neq \square$ does not contain contradictions and tautologies;
\item
$S''=S'-\{\mi{simplify}(C^*,a^*,\gz)\}$;
\item
there exists $C^{**}\in \mi{OrdPropCl}$ satisfying that $S''=(S'-\{\mi{simplify}(C^*,a^*,\gz)\})\cup \{C^{**}\}$, $C^{**}\not\in S'$, 
$\square\neq C^{**}\subset \mi{simplify}(C^*,a^*,\gz)$ does not contain contradictions and tautologies.
\end{enumerate}
We get four cases for $S''$.

Case 1.2.2.1:
$\square\in S''$.
We put
\begin{equation} \notag
\mi{Tree}=\dfrac{S}
                {\mi{Tree}'}.
\end{equation}
Hence, $\mi{Tree}$ is a finite linear {\it DPLL}-tree with the root $S$ constructed using Rules (\cref{ceq4hr1111111})--(\cref{ceq4hr4}), (\cref{ceq4hr7}) such that
for its only leaf $S''$, $\square\in S''$;
(i) holds.

Case 1.2.2.2:
$S''=S'$ and $\mi{simplify}(C^*,a^*,\gz)\neq \square$ does not contain contradictions and tautologies.

Case 1.2.2.3:
$S''=S'-\{\mi{simplify}(C^*,a^*,\gz)\}$.
Then $\mi{simplify}(C^*,a^*,\gz)\not\in S$,
$S''=S'-\{\mi{simplify}(C^*,a^*,\gz)\}=((S-\{C^*\})\cup \{\mi{simplify}(C^*,a^*,\gz)\})-                                                                                                   \linebreak[4]
                                                                                        \{\mi{simplify}(C^*,a^*,\gz)\}=
     ((S-\{C^*\})-\{\mi{simplify}(C^*,a^*,\gz)\})\cup (\{\mi{simplify}(C^*,a^*,\gz)\}-\{\mi{simplify}(C^*,a^*,\gz)\})=S-\{C^*\}$,
$a^*\geql \gz\in S''=S-\{C^*\}$, $a^*\in \mi{atoms}(a^*\geql \gz)\subseteq \mi{atoms}(S'')=\mi{atoms}(S-\{C^*\})\subseteq \mi{atoms}(S)$.
We get from Case 1.2.1 for $S''$ that there exists a finite linear {\it DPLL}-tree $\mi{Tree}''$ with the root $S''$ constructed using Rules (\cref{ceq4hr1111111})--(\cref{ceq4hr4}), (\cref{ceq4hr7}) satisfying
for its only leaf $S'''$ that either $\square\in S'''$, or $S'''\subseteq_{\mc F} \mi{OrdPropCl}$ is $\gz$-guarded, $\mi{atoms}(S''')\subseteq \mi{atoms}(S'')-\{a^*\}=\mi{atoms}(S)-\{a^*\}$.
We put
\begin{equation} \notag
\mi{Tree}=\begin{array}[c]{c}
          S \\[0.4mm]
          \hline \\[-3.8mm]
          \mi{Tree}' \\[0.4mm]
          \hline \\[-3.8mm]
          \mi{Tree}''.    
          \end{array}    
\end{equation}
Hence, $\mi{Tree}$ is a finite linear {\it DPLL}-tree with the root $S$ constructed using Rules (\cref{ceq4hr1111111})--(\cref{ceq4hr4}), (\cref{ceq4hr7}) such that
for its only leaf $S'''$, either $\square\in S'''$, or $S'''\subseteq_{\mc F} \mi{OrdPropCl}$ is $\gz$-guarded, $\mi{atoms}(S''')\subseteq \mi{atoms}(S)-\{a^*\}$;
(i) holds.

Case 1.2.2.4:
There exists $C^{**}\in \mi{OrdPropCl}$ such that $S''=(S'-\{\mi{simplify}(C^*,a^*,\gz)\})\cup \{C^{**}\}$, $C^{**}\not\in S'$, 
$\square\neq C^{**}\subset \mi{simplify}(C^*,a^*,\gz)$ does not contain contradictions and tautologies.

In Cases 1.2.2.2 and 1.2.2.4, we put 
\begin{alignat*}{1}
C^\natural &= \left\{\begin{array}{ll}
                     \mi{simplify}(C^*,a^*,\gz) &\ \text{\it in {\rm Case 1.2.2.2}}, \\[1mm]
                     C^{**}                     &\ \text{\it in {\rm Case 1.2.2.4}};
                     \end{array}
              \right. \\[1mm]
           &\in \mi{OrdPropCl}.
\end{alignat*}
Then, in Case 1.2.2.2, 
$C^\natural=\mi{simplify}(C^*,a^*,\gz)\not\in S$, $S''=S'=(S-\{C^*\})\cup \{\mi{simplify}(C^*,a^*,\gz)\}=(S-\{C^*\})\cup \{C^\natural\}$;
$C^\natural=\mi{simplify}(C^*,a^*,\gz)\neq \square$ does not contain contradictions and tautologies;
$a^*\not\in \mi{atoms}(C^\natural)=\mi{atoms}(\mi{simplify}(C^*,a^*,\gz))\subseteq \mi{atoms}(C^*)$;
in Case 1.2.2.4, 
$C^{**}\subset \mi{simplify}(C^*,a^*,\gz)$, 
$\mi{atoms}(C^\natural)=\mi{atoms}(C^{**})\subseteq \mi{atoms}(\mi{simplify}(C^*,a^*,\gz))\subseteq \mi{atoms}(C^*)$,
$a^*\not\in \mi{atoms}(\mi{simplify}(C^*,a^*,\gz))\supseteq \mi{atoms}(C^\natural)=\mi{atoms}(C^{**})$, $a^*\in \mi{atoms}(C^*)$, $\mi{atoms}(C^{**})\neq \mi{atoms}(C^*)$, $C^{**}\neq C^*$,
$C^{**}\not\in S'\supseteq S-\{C^*\}$, $C^\natural=C^{**}\not\in S$,
$S''=(S'-\{\mi{simplify}(C^*,a^*,\gz)\})\cup \{C^{**}\}=(((S-\{C^*\})\cup \{\mi{simplify}(C^*,a^*,\gz)\})-\{\mi{simplify}(C^*,a^*,\gz)\})\cup \{C^{**}\}=
     ((S-\{C^*\})-\{\mi{simplify}(C^*,a^*,\gz)\})\cup (\{\mi{simplify}(C^*,a^*,\gz)\}-\{\mi{simplify}(C^*,a^*,\gz)\})\cup \{C^{**}\}=(S-\{C^*\})\cup \{C^{**}\}=(S-\{C^*\})\cup \{C^\natural\}$;
$C^\natural=C^{**}\neq \square$ does not contain contradictions and tautologies;
in both Cases 1.2.2.2 and 1.2.2.4,
$C^\natural\not\in S$, $S''=(S-\{C^*\})\cup \{C^\natural\}$;
$C^\natural\neq \square$ does not contain contradictions and tautologies;
$a^*\not\in \mi{atoms}(C^\natural)\subseteq \mi{atoms}(C^*)$;
$a^*\geql \gz\in S-\{C^*\}\subseteq S''$, $C^*\in S$,
$a^*\in \mi{atoms}(a^*\geql \gz)\subseteq \mi{atoms}(S'')=\mi{atoms}((S-\{C^*\})\cup \{C^\natural\})=\mi{atoms}(S-\{C^*\})\cup \mi{atoms}(C^\natural)\subseteq 
                                                          \mi{atoms}(S-\{C^*\})\cup \mi{atoms}(C^*)=\mi{atoms}((S-\{C^*\})\cup \{C^*\})=\mi{atoms}(S)$,
$\mi{atoms}(C^\natural)\subseteq \mi{atoms}(S'')$, $\mi{atoms}(C^\natural)\subseteq \mi{atoms}(S'')-\{a^*\}$;
$S$ is simplified;
$S-\{C^*\}\subseteq S$ is simplified;
$S''=(S-\{C^*\})\cup \{C^\natural\}$ is simplified;
for all $C\in \mi{guards}(a^*)$, $a^*\in \mi{atoms}(C)=\{a^*\}$, $\mi{atoms}(C^\natural)\neq \mi{atoms}(C)$, $C^\natural\neq C$;
$C^\natural, C^*\not\in \mi{guards}(a^*)$,
$\mi{guards}(S'',a^*)=((S-\{C^*\})\cup \{C^\natural\})\cap \mi{guards}(a^*)=((S-\{C^*\})\cap \mi{guards}(a^*))\cup (\{C^\natural\}\cap \mi{guards}(a^*))=
                      (S\cap \mi{guards}(a^*))-\{C^*\}=S\cap \mi{guards}(a^*)=\mi{guards}(S,a^*)=\{a^*\geql \gz\}$;
$a^*$ is $\gz$-guarded in $S''$.
We get two cases for $C^\natural$.

Cases 1.2.2.2.1 and 1.2.2.4.1:
There exists $a^{**}\in \mi{atoms}(S'')-\{a^*\}\subseteq \mi{atoms}(S)-\{a^*\}$ such that $C^\natural\in \mi{guards}(a^{**})$.
Then $a^{**}\neq a^*$, $C^*\not\in \mi{guards}(a^{**})$,
$\mi{guards}(S'',a^{**})=((S-\{C^*\})\cup \{C^\natural\})\cap \mi{guards}(a^{**})=((S-\{C^*\})\cap \mi{guards}(a^{**}))\cup (\{C^\natural\}\cap \mi{guards}(a^{**}))=
                         ((S\cap \mi{guards}(a^{**}))-\{C^*\})\cup \{C^\natural\}=(S\cap \mi{guards}(a^{**}))\cup \{C^\natural\}=\mi{guards}(S,a^{**})\cup \{C^\natural\}$;
$a^{**}$ is $\gz$-guarded in $S$.
We get six cases for $C^\natural$.

Cases 1.2.2.2.1.1 and 1.2.2.4.1.1:
$C^\natural=a^{**}\geql \gz$.
We have that $a^{**}$ is $\gz$-guarded in $S$.
We get three cases for $\mi{guards}(S,a^{**})$.

Cases 1.2.2.2.1.1.1 and 1.2.2.4.1.1.1:
$\mi{guards}(S,a^{**})=\{a^{**}\geql \gz\}$.
Then $a^{**}\geql \gz\in \mi{guards}(S,a^{**})\subseteq S$,
which is a contradiction with $C^\natural=a^{**}\geql \gz\not\in S$.

Cases 1.2.2.2.1.1.2 and 1.2.2.4.1.1.2:
Either $\mi{guards}(S,a^{**})=\{\gz\gle a^{**}\}$ or $\mi{guards}(S,a^{**})=\{\gz\gle a^{**},a^{**}\gle \gu\}$.
Then $S''\supseteq \mi{guards}(S'')\supseteq \mi{guards}(S'',a^{**})=\mi{guards}(S,a^{**})\cup \{C^\natural\}=\mi{guards}(S,a^{**})\cup \{a^{**}\geql \gz\}\supseteq \{a^{**}\geql \gz,\gz\gle a^{**}\}$,
$a^{**}\geql \gz\in \mi{guards}(S'')$, $\gz\gle a^{**}\in S''$,
$a^{**}\in \mi{atoms}(\gz\gle a^{**})$, $a^{**}\geql \gz\neq \gz\gle a^{**}$, $\mi{simplify}(\gz\gle a^{**},a^{**},\gz)=\gz\gle \gz$,
applying Rule (\cref{ceq4hr3}) to $S''$, $a^{**}\geql \gz$, and $\gz\gle a^{**}$, we derive
\begin{equation} \notag
\dfrac{S''}
      {(S''-\{\gz\gle a^{**}\})\cup \{\gz\gle \gz\}};
\end{equation}
$\gz\gle \gz\in (S''-\{\gz\gle a^{**}\})\cup \{\gz\gle \gz\}$;
$\gz\gle \gz\in \mi{OrdPropLit}$ is a contradiction;
$\gz\gle \gz$ is not a guard;
$\gz\gle \gz\not\in \mi{guards}((S''-\{\gz\gle a^{**}\})\cup \{\gz\gle \gz\})$,
$\gz\gle \gz\in ((S''-\{\gz\gle a^{**}\})\cup \{\gz\gle \gz\})-\mi{guards}((S''-\{\gz\gle a^{**}\})\cup \{\gz\gle \gz\})$,
applying Rule (\cref{ceq4hr2}) to $(S''-\{\gz\gle a^{**}\})\cup \{\gz\gle \gz\}$ and $\gz\gle \gz$, we derive
\begin{equation} \notag
\dfrac{(S''-\{\gz\gle a^{**}\})\cup \{\gz\gle \gz\}}
      {(((S''-\{\gz\gle a^{**}\})\cup \{\gz\gle \gz\})-\{\gz\gle \gz\})\cup \{\square\}}.
\end{equation}
We put
\begin{equation} \notag
\mi{Tree}=\begin{array}[c]{c}
          S \\[0.4mm]
          \hline \\[-3.8mm]
          \mi{Tree}' \\[0.4mm]
          \hline \\[-3.8mm]
          (S''-\{\gz\gle a^{**}\})\cup \{\gz\gle \gz\} \\[0.4mm]
          \hline \\[-3.8mm]
          S'''=(((S''-\{\gz\gle a^{**}\})\cup \{\gz\gle \gz\})-\{\gz\gle \gz\})\cup \{\square\}.
          \end{array} 
\end{equation}
Hence, $\mi{Tree}$ is a finite linear {\it DPLL}-tree with the root $S$ constructed using Rules (\cref{ceq4hr1111111})--(\cref{ceq4hr4}), (\cref{ceq4hr7}) such that
for its only leaf $S'''$, $\square\in S'''$;
(i) holds.

Cases 1.2.2.2.1.1.3 and 1.2.2.4.1.1.3:
$\mi{guards}(S,a^{**})=\{a^{**}\geql \gu\}$.
Then $S''\supseteq \mi{guards}(S'')\supseteq \mi{guards}(S'',a^{**})=\mi{guards}(S,a^{**})\cup \{C^\natural\}=\{a^{**}\geql \gz,a^{**}\geql \gu\}$,
$a^{**}\geql \gz\in \mi{guards}(S'')$, $a^{**}\geql \gu\in S''$,
$a^{**}\in \mi{atoms}(a^{**}\geql \gu)$, $a^{**}\geql \gz\neq a^{**}\geql \gu$, $\mi{simplify}(a^{**}\geql \gu,a^{**},\gz)=\gz\geql \gu$,
applying Rule (\cref{ceq4hr3}) to $S''$, $a^{**}\geql \gz$, and $a^{**}\geql \gu$, we derive
\begin{equation} \notag
\dfrac{S''}
      {(S''-\{a^{**}\geql \gu\})\cup \{\gz\geql \gu\}};
\end{equation}
$\gz\geql \gu\in (S''-\{a^{**}\geql \gu\})\cup \{\gz\geql \gu\}$;
$\gz\geql \gu\in \mi{OrdPropLit}$ is a contradiction;
$\gz\geql \gu$ is not a guard;
$\gz\geql \gu\not\in \mi{guards}((S''-\{a^{**}\geql \gu\})\cup \{\gz\geql \gu\})$,
$\gz\geql \gu\in ((S''-\{a^{**}\geql \gu\})\cup \{\gz\geql \gu\})-\mi{guards}((S''-\{a^{**}\geql \gu\})\cup \{\gz\geql \gu\})$,
applying Rule (\cref{ceq4hr2}) to $(S''-\{a^{**}\geql \gu\})\cup \{\gz\geql \gu\}$ and $\gz\geql \gu$, we derive
\begin{equation} \notag
\dfrac{(S''-\{a^{**}\geql \gu\})\cup \{\gz\geql \gu\}}
      {(((S''-\{a^{**}\geql \gu\})\cup \{\gz\geql \gu\})-\{\gz\geql \gu\})\cup \{\square\}}.
\end{equation}
We put 
\begin{equation} \notag
\mi{Tree}=\begin{array}[c]{c}
          S \\[0.4mm]
          \hline \\[-3.8mm]
          \mi{Tree}' \\[0.4mm]
          \hline \\[-3.8mm]
          (S''-\{a^{**}\geql \gu\})\cup \{\gz\geql \gu\} \\[0.4mm]
          \hline \\[-3.8mm]
          S'''=(((S''-\{a^{**}\geql \gu\})\cup \{\gz\geql \gu\})-\{\gz\geql \gu\})\cup \{\square\}.
          \end{array} 
\end{equation}
Hence, $\mi{Tree}$ is a finite linear {\it DPLL}-tree with the root $S$ constructed using Rules (\cref{ceq4hr1111111})--(\cref{ceq4hr4}), (\cref{ceq4hr7}) such that
for its only leaf $S'''$, $\square\in S'''$;
(i) holds.

Cases 1.2.2.2.1.2 and 1.2.2.4.1.2:
$C^\natural=a^{**}\gleq \gz$.
We have that $a^{**}$ is $\gz$-guarded in $S$.
We get three cases for $\mi{guards}(S,a^{**})$.

Cases 1.2.2.2.1.2.1 and 1.2.2.4.1.2.1:
$\mi{guards}(S,a^{**})=\{a^{**}\geql \gz\}$.
Then $C^\natural=a^{**}\gleq \gz\not\in S$,
$S''=(S-\{C^*\})\cup \{C^\natural\}=(S-\{C^*\})\cup \{a^{**}\gleq \gz\}$;
$S''\supseteq \mi{guards}(S'')\supseteq \mi{guards}(S'',a^{**})=\mi{guards}(S,a^{**})\cup \{C^\natural\}=\{a^{**}\geql \gz,a^{**}\gleq \gz\}$,
$a^{**}\gleq \gz\in \mi{guards}(S'')$, applying Rule (\cref{ceq4hr1111111}) to $S''$ and $a^{**}\gleq \gz$, we derive
\begin{equation} \notag
\dfrac{S''}
      {(S''-\{a^{**}\gleq \gz\})\cup \{a^{**}\geql \gz\}}.
\end{equation}
Hence, $a^{**}\geql \gz\in S''$, $a^{**}\geql \gz\neq a^{**}\gleq \gz$, $a^{**}\geql \gz\in S''-\{a^{**}\gleq \gz\}$, $a^{**}\gleq \gz\not\in S$,
$(S''-\{a^{**}\gleq \gz\})\cup \{a^{**}\geql \gz\}=S''-\{a^{**}\gleq \gz\}=((S-\{C^*\})\cup \{a^{**}\gleq \gz\})-\{a^{**}\gleq \gz\}=
 ((S-\{C^*\})-\{a^{**}\gleq \gz\})\cup (\{a^{**}\gleq \gz\}-\{a^{**}\gleq \gz\})=S-\{C^*\}$,
$a^*\geql \gz\in S-\{C^*\}$, $a^*\in \mi{atoms}(a^*\geql \gz)\subseteq \mi{atoms}(S-\{C^*\})\subseteq \mi{atoms}(S)$.
We get from Case 1.2.1 for $S-\{C^*\}$ that there exists a finite linear {\it DPLL}-tree $\mi{Tree}''$ with the root $S-\{C^*\}$ constructed using Rules (\cref{ceq4hr1111111})--(\cref{ceq4hr4}), (\cref{ceq4hr7}) satisfying
for its only leaf $S'''$ that either $\square\in S'''$, or $S'''\subseteq_{\mc F} \mi{OrdPropCl}$ is $\gz$-guarded, $\mi{atoms}(S''')\subseteq \mi{atoms}(S-\{C^*\})-\{a^*\}=\mi{atoms}(S)-\{a^*\}$.
We put
\begin{equation} \notag
\mi{Tree}=\begin{array}[c]{c}
          S \\[0.4mm]
          \hline \\[-3.8mm]
          \mi{Tree}' \\[0.4mm]
          \hline \\[-3.8mm]
          \mi{Tree}''.
          \end{array}
\end{equation}
Hence, $\mi{Tree}$ is a finite linear {\it DPLL}-tree with the root $S$ constructed using Rules (\cref{ceq4hr1111111})--(\cref{ceq4hr4}), (\cref{ceq4hr7}) such that
for its only leaf $S'''$, either $\square\in S'''$, or $S'''\subseteq_{\mc F} \mi{OrdPropCl}$ is $\gz$-guarded, $\mi{atoms}(S''')\subseteq \mi{atoms}(S)-\{a^*\}$;
(i) holds.

Cases 1.2.2.2.1.2.2 and 1.2.2.4.1.2.2:
Either $\mi{guards}(S,a^{**})=\{\gz\gle a^{**}\}$ or $\mi{guards}(S,a^{**})=\{\gz\gle a^{**},a^{**}\gle \gu\}$.
Then $\mi{guards}(S'')\supseteq \mi{guards}(S'',a^{**})=\mi{guards}(S,a^{**})\cup \{C^\natural\}=\mi{guards}(S,a^{**})\cup \{a^{**}\gleq \gz\}$,
$a^{**}\gleq \gz\in \mi{guards}(S'')$, applying Rule (\cref{ceq4hr1111111}) to $S''$ and $a^{**}\gleq \gz$, we derive
\begin{equation} \notag
\dfrac{S''}
      {(S''-\{a^{**}\gleq \gz\})\cup \{a^{**}\geql \gz\}};
\end{equation}
$a^{**}\in \mi{atoms}((S''-\{a^{**}\gleq \gz\})\cup \{a^{**}\geql \gz\})$,
$(S''-\{a^{**}\gleq \gz\})\cup \{a^{**}\geql \gz\}\supseteq \mi{guards}((S''-\{a^{**}\gleq \gz\})\cup \{a^{**}\geql \gz\})\supseteq 
 \mi{guards}((S''-\{a^{**}\gleq \gz\})\cup \{a^{**}\geql \gz\},a^{**})=((S''-\{a^{**}\gleq \gz\})\cup \{a^{**}\geql \gz\})\cap \mi{guards}(a^{**})=
 ((S''-\{a^{**}\gleq \gz\})\cap \mi{guards}(a^{**}))\cup (\{a^{**}\geql \gz\}\cap \mi{guards}(a^{**}))=((S''\cap \mi{guards}(a^{**}))-\{a^{**}\gleq \gz\})\cup \{a^{**}\geql \gz\}=
 (\mi{guards}(S'',a^{**})-\{a^{**}\gleq \gz\})\cup \{a^{**}\geql \gz\}=((\mi{guards}(S,a^{**})\cup \{a^{**}\gleq \gz\})-\{a^{**}\gleq \gz\})\cup \{a^{**}\geql \gz\}=
 (\mi{guards}(S,a^{**})-\{a^{**}\gleq \gz\})\cup (\{a^{**}\gleq \gz\}-\{a^{**}\gleq \gz\})\cup \{a^{**}\geql \gz\}=(\mi{guards}(S,a^{**})-\{a^{**}\gleq \gz\})\cup \{a^{**}\geql \gz\}\supseteq
 (\{\gz\gle a^{**}\}-\{a^{**}\gleq \gz\})\cup \{a^{**}\geql \gz\}=\{a^{**}\geql \gz,\gz\gle a^{**}\}$,
$a^{**}\geql \gz\in \mi{guards}((S''-\{a^{**}\gleq \gz\})\cup \{a^{**}\geql \gz\})$, $\gz\gle a^{**}\in (S''-\{a^{**}\gleq \gz\})\cup \{a^{**}\geql \gz\}$, 
$a^{**}\in \mi{atoms}(\gz\gle a^{**})$, $a^{**}\geql \gz\neq \gz\gle a^{**}$, $\mi{simplify}(\gz\gle a^{**},a^{**},\gz)=\gz\gle \gz$,
applying Rule (\cref{ceq4hr3}) to $(S''-\{a^{**}\gleq \gz\})\cup \{a^{**}\geql \gz\}$, $a^{**}\geql \gz$, and $\gz\gle a^{**}$, we derive
\begin{equation} \notag
\dfrac{(S''-\{a^{**}\gleq \gz\})\cup \{a^{**}\geql \gz\}}
      {(((S''-\{a^{**}\gleq \gz\})\cup \{a^{**}\geql \gz\})-\{\gz\gle a^{**}\})\cup \{\gz\gle \gz\}};
\end{equation}
$\gz\gle \gz\in (((S''-\{a^{**}\gleq \gz\})\cup \{a^{**}\geql \gz\})-\{\gz\gle a^{**}\})\cup \{\gz\gle \gz\}$;
$\gz\gle \gz\in \mi{OrdPropLit}$ is a contradiction;
$\gz\gle \gz$ is not a guard;
$\gz\gle \gz\not\in \mi{guards}((((S''-\{a^{**}\gleq \gz\})\cup \{a^{**}\geql \gz\})-\{\gz\gle a^{**}\})\cup \{\gz\gle \gz\})$,
$\gz\gle \gz\in ((((S''-\{a^{**}\gleq \gz\})\cup \{a^{**}\geql \gz\})-\{\gz\gle a^{**}\})\cup \{\gz\gle \gz\})-\mi{guards}((((S''-\{a^{**}\gleq \gz\})\cup \{a^{**}\geql \gz\})-\{\gz\gle a^{**}\})\cup \{\gz\gle \gz\})$,
applying Rule (\cref{ceq4hr2}) to $(((S''-\{a^{**}\gleq \gz\})\cup \{a^{**}\geql \gz\})-\{\gz\gle a^{**}\})\cup \{\gz\gle \gz\}$ and $\gz\gle \gz$, we derive
\begin{equation} \notag
\dfrac{(((S''-\{a^{**}\gleq \gz\})\cup \{a^{**}\geql \gz\})-\{\gz\gle a^{**}\})\cup \{\gz\gle \gz\}}
      {(((((S''-\{a^{**}\gleq \gz\})\cup \{a^{**}\geql \gz\})-\{\gz\gle a^{**}\})\cup \{\gz\gle \gz\})-\{\gz\gle \gz\})\cup \{\square\}}.
\end{equation}
We put
\begin{equation} \notag
\mi{Tree}=\begin{array}[c]{c}
          S \\[0.4mm]
          \hline \\[-3.8mm]
          \mi{Tree}' \\[0.4mm]
          \hline \\[-3.8mm]
          (S''-\{a^{**}\gleq \gz\})\cup \{a^{**}\geql \gz\} \\[0.4mm]
          \hline \\[-3.8mm]
          (((S''-\{a^{**}\gleq \gz\})\cup \{a^{**}\geql \gz\})-\{\gz\gle a^{**}\})\cup \{\gz\gle \gz\} \\[0.4mm]
          \hline \\[-3.8mm]
          S'''=(((((S''-\{a^{**}\gleq \gz\})\cup \{a^{**}\geql \gz\})-\{\gz\gle a^{**}\})\cup \{\gz\gle \gz\})- \\
          \hfill \{\gz\gle \gz\})\cup \{\square\}.
          \end{array} 
\end{equation}
Hence, $\mi{Tree}$ is a finite linear {\it DPLL}-tree with the root $S$ constructed using Rules (\cref{ceq4hr1111111})--(\cref{ceq4hr4}), (\cref{ceq4hr7}) such that
for its only leaf $S'''$, $\square\in S'''$;
(i) holds.

Cases 1.2.2.2.1.2.3 and 1.2.2.4.1.2.3:
$\mi{guards}(S,a^{**})=\{a^{**}\geql \gu\}$.
Then $S''\supseteq \mi{guards}(S'')\supseteq \mi{guards}(S'',a^{**})=\mi{guards}(S,a^{**})\cup \{C^\natural\}=\{a^{**}\gleq \gz,a^{**}\geql \gu\}$,
$a^{**}\geql \gu\in \mi{guards}(S'')$, $a^{**}\gleq \gz\in S''$,
$a^{**}\in \mi{atoms}(a^{**}\gleq \gz)$, $a^{**}\geql \gu\neq a^{**}\gleq \gz$, $\mi{simplify}(a^{**}\gleq \gz,a^{**},\gu)=\gu\gleq \gz$,
applying Rule (\cref{ceq4hr4}) to $S''$, $a^{**}\geql \gu$, and $a^{**}\gleq \gz$, we derive
\begin{equation} \notag
\dfrac{S''}
      {(S''-\{a^{**}\gleq \gz\})\cup \{\gu\gleq \gz\}};
\end{equation}
$\gu\gleq \gz\in (S''-\{a^{**}\gleq \gz\})\cup \{\gu\gleq \gz\}$;
$\gu\gleq \gz\in \mi{OrdPropLit}$ is a contradiction;
$\gu\gleq \gz$ is not a guard;
$\gu\gleq \gz\not\in \mi{guards}((S''-\{a^{**}\gleq \gz\})\cup \{\gu\gleq \gz\})$,
$\gu\gleq \gz\in ((S''-\{a^{**}\gleq \gz\})\cup \{\gu\gleq \gz\})-\mi{guards}((S''-\{a^{**}\gleq \gz\})\cup \{\gu\gleq \gz\})$,
applying Rule (\cref{ceq4hr2}) to $(S''-\{a^{**}\gleq \gz\})\cup \{\gu\gleq \gz\}$ and $\gu\gleq \gz$, we derive
\begin{equation} \notag
\dfrac{(S''-\{a^{**}\gleq \gz\})\cup \{\gu\gleq \gz\}}
      {(((S''-\{a^{**}\gleq \gz\})\cup \{\gu\gleq \gz\})-\{\gu\gleq \gz\})\cup \{\square\}}.
\end{equation}
We put 
\begin{equation} \notag
\mi{Tree}=\begin{array}[c]{c}
          S \\[0.4mm]
          \hline \\[-3.8mm]
          \mi{Tree}' \\[0.4mm]
          \hline \\[-3.8mm]
          (S''-\{a^{**}\gleq \gz\})\cup \{\gu\gleq \gz\} \\[0.4mm]
          \hline \\[-3.8mm]
          S'''=(((S''-\{a^{**}\gleq \gz\})\cup \{\gu\gleq \gz\})-\{\gu\gleq \gz\})\cup \{\square\}.
          \end{array} 
\end{equation}
Hence, $\mi{Tree}$ is a finite linear {\it DPLL}-tree with the root $S$ constructed using Rules (\cref{ceq4hr1111111})--(\cref{ceq4hr4}), (\cref{ceq4hr7}) such that
for its only leaf $S'''$, $\square\in S'''$;
(i) holds.

Cases 1.2.2.2.1.3 and 1.2.2.4.1.3:
$C^\natural=\gz\gle a^{**}$.
We have that $a^{**}$ is $\gz$-guarded in $S$.
We get three cases for $\mi{guards}(S,a^{**})$.

Cases 1.2.2.2.1.3.1 and 1.2.2.4.1.3.1:
$\mi{guards}(S,a^{**})=\{a^{**}\geql \gz\}$.
Then $S''\supseteq \mi{guards}(S'')\supseteq \mi{guards}(S'',a^{**})=\mi{guards}(S,a^{**})\cup \{C^\natural\}=\{a^{**}\geql \gz,\gz\gle a^{**}\}$;
these cases are the same as Cases 1.2.2.2.1.1.2 and 1.2.2.4.1.1.2.

Cases 1.2.2.2.1.3.2 and 1.2.2.4.1.3.2:
Either $\mi{guards}(S,a^{**})=\{\gz\gle a^{**}\}$ or $\mi{guards}(S,a^{**})=\{\gz\gle a^{**},a^{**}\gle \gu\}$.
Then $\gz\gle a^{**}\in \mi{guards}(S,a^{**})\subseteq S$,
which is a contradiction with $C^\natural=\gz\gle a^{**}\not\in S$.

Cases 1.2.2.2.1.3.3 and 1.2.2.4.1.3.3:
$\mi{guards}(S,a^{**})=\{a^{**}\geql \gu\}$.
Then $C^\natural=\gz\gle a^{**}\not\in S$,
$S''=(S-\{C^*\})\cup \{C^\natural\}=(S-\{C^*\})\cup \{\gz\gle a^{**}\}$;
$S''\supseteq \mi{guards}(S'')\supseteq \mi{guards}(S'',a^{**})=\mi{guards}(S,a^{**})\cup \{C^\natural\}=\{\gz\gle a^{**},a^{**}\geql \gu\}$,
$a^{**}\geql \gu\in \mi{guards}(S'')$, $\gz\gle a^{**}\in S''$,
$a^{**}\in \mi{atoms}(\gz\gle a^{**})$, $a^{**}\geql \gu\neq \gz\gle a^{**}$, $\mi{simplify}(\gz\gle a^{**},a^{**},\gu)=\gz\gle \gu$,
applying Rule (\cref{ceq4hr4}) to $S''$, $a^{**}\geql \gu$, and $\gz\gle a^{**}$, we derive
\begin{equation} \notag
\dfrac{S''}
      {(S''-\{\gz\gle a^{**}\})\cup \{\gz\gle \gu\}};
\end{equation}
$\gz\gle \gu\in (S''-\{\gz\gle a^{**}\})\cup \{\gz\gle \gu\}$;
$\gz\gle \gu\in \mi{OrdPropLit}$ is a tautology;
$\gz\gle \gu$ is not a guard;
$\gz\gle \gu\not\in \mi{guards}((S''-\{\gz\gle a^{**}\})\cup \{\gz\gle \gu\})$,
$\gz\gle \gu\in ((S''-\{\gz\gle a^{**}\})\cup \{\gz\gle \gu\})-\mi{guards}((S''-\{\gz\gle a^{**}\})\cup \{\gz\gle \gu\})$,
applying Rule (\cref{ceq4hr22}) to $(S''-\{\gz\gle a^{**}\})\cup \{\gz\gle \gu\}$ and $\gz\gle \gu$, we derive
\begin{equation} \notag
\dfrac{(S''-\{\gz\gle a^{**}\})\cup \{\gz\gle \gu\}}
      {((S''-\{\gz\gle a^{**}\})\cup \{\gz\gle \gu\})-\{\gz\gle \gu\}}.
\end{equation}
We have that $S''$ is simplified.
Hence, $\gz\gle \gu\not\in S''$, $\gz\gle a^{**}\not\in S$,
$((S''-\{\gz\gle a^{**}\})\cup \{\gz\gle \gu\})-\{\gz\gle \gu\}=((S''-\{\gz\gle a^{**}\})-\{\gz\gle \gu\})\cup (\{\gz\gle \gu\}-\{\gz\gle \gu\})=S''-\{\gz\gle a^{**}\}=
 ((S-\{C^*\})\cup \{\gz\gle a^{**}\})-\{\gz\gle a^{**}\}=((S-\{C^*\})-\{\gz\gle a^{**}\})\cup (\{\gz\gle a^{**}\}-\{\gz\gle a^{**}\})=S-\{C^*\}$,
$a^*\geql \gz\in S-\{C^*\}$, $a^*\in \mi{atoms}(a^*\geql \gz)\subseteq \mi{atoms}(S-\{C^*\})\subseteq \mi{atoms}(S)$.
We get from Case 1.2.1 for $S-\{C^*\}$ that there exists a finite linear {\it DPLL}-tree $\mi{Tree}''$ with the root $S-\{C^*\}$ constructed using Rules (\cref{ceq4hr1111111})--(\cref{ceq4hr4}), (\cref{ceq4hr7}) satisfying
for its only leaf $S'''$ that either $\square\in S'''$, or $S'''\subseteq_{\mc F} \mi{OrdPropCl}$ is $\gz$-guarded, $\mi{atoms}(S''')\subseteq \mi{atoms}(S-\{C^*\})-\{a^*\}=\mi{atoms}(S)-\{a^*\}$.
We put
\begin{equation} \notag
\mi{Tree}=\begin{array}[c]{c}
          S \\[0.4mm]
          \hline \\[-3.8mm]
          \mi{Tree}' \\[0.4mm]
          \hline \\[-3.8mm]
          (S''-\{\gz\gle a^{**}\})\cup \{\gz\gle \gu\} \\[0.4mm] 
          \hline \\[-3.8mm]
          \mi{Tree}''.
          \end{array}
\end{equation}
Hence, $\mi{Tree}$ is a finite linear {\it DPLL}-tree with the root $S$ constructed using Rules (\cref{ceq4hr1111111})--(\cref{ceq4hr4}), (\cref{ceq4hr7}) such that
for its only leaf $S'''$, either $\square\in S'''$, or $S'''\subseteq_{\mc F} \mi{OrdPropCl}$ is $\gz$-guarded, $\mi{atoms}(S''')\subseteq \mi{atoms}(S)-\{a^*\}$;
(i) holds.

Cases 1.2.2.2.1.4 and 1.2.2.4.1.4:
$C^\natural=a^{**}\gle \gu$.
We have that $a^{**}$ is $\gz$-guarded in $S$.
We get four cases for $\mi{guards}(S,a^{**})$.

Cases 1.2.2.2.1.4.1 and 1.2.2.4.1.4.1:
$\mi{guards}(S,a^{**})=\{a^{**}\geql \gz\}$.
Then $C^\natural=a^{**}\gle \gu\not\in S$,
$S''=(S-\{C^*\})\cup \{C^\natural\}=(S-\{C^*\})\cup \{a^{**}\gle \gu\}$;
$S''\supseteq \mi{guards}(S'')\supseteq \mi{guards}(S'',a^{**})=\mi{guards}(S,a^{**})\cup \{C^\natural\}=\{a^{**}\geql \gz,a^{**}\gle \gu\}$,
$a^{**}\geql \gz\in \mi{guards}(S'')$, $a^{**}\gle \gu\in S''$,
$a^{**}\in \mi{atoms}(a^{**}\gle \gu)$, $a^{**}\geql \gz\neq a^{**}\gle \gu$, $\mi{simplify}(a^{**}\gle \gu,a^{**},\gz)=\gz\gle \gu$,
applying Rule (\cref{ceq4hr3}) to $S''$, $a^{**}\geql \gz$, and $a^{**}\gle \gu$, we derive
\begin{equation} \notag
\dfrac{S''}
      {(S''-\{a^{**}\gle \gu\})\cup \{\gz\gle \gu\}};
\end{equation}
$\gz\gle \gu\in (S''-\{a^{**}\gle \gu\})\cup \{\gz\gle \gu\}$;
$\gz\gle \gu\in \mi{OrdPropLit}$ is a tautology;
$\gz\gle \gu$ is not a guard;
$\gz\gle \gu\not\in \mi{guards}((S''-\{a^{**}\gle \gu\})\cup \{\gz\gle \gu\})$,
$\gz\gle \gu\in ((S''-\{a^{**}\gle \gu\})\cup \{\gz\gle \gu\})-\mi{guards}((S''-\{a^{**}\gle \gu\})\cup \{\gz\gle \gu\})$,
applying Rule (\cref{ceq4hr22}) to $(S''-\{a^{**}\gle \gu\})\cup \{\gz\gle \gu\}$ and $\gz\gle \gu$, we derive
\begin{equation} \notag
\dfrac{(S''-\{a^{**}\gle \gu\})\cup \{\gz\gle \gu\}}
      {((S''-\{a^{**}\gle \gu\})\cup \{\gz\gle \gu\})-\{\gz\gle \gu\}}.
\end{equation}
We have that $S''$ is simplified.
Hence, $\gz\gle \gu\not\in S''$, $a^{**}\gle \gu\not\in S$,
$((S''-\{a^{**}\gle \gu\})\cup \{\gz\gle \gu\})-\{\gz\gle \gu\}=((S''-\{a^{**}\gle \gu\})-\{\gz\gle \gu\})\cup (\{\gz\gle \gu\}-\{\gz\gle \gu\})=S''-\{a^{**}\gle \gu\}=
 ((S-\{C^*\})\cup \{a^{**}\gle \gu\})-\{a^{**}\gle \gu\}=((S-\{C^*\})-\{a^{**}\gle \gu\})\cup (\{a^{**}\gle \gu\}-\{a^{**}\gle \gu\})=S-\{C^*\}$,
$a^*\geql \gz\in S-\{C^*\}$, $a^*\in \mi{atoms}(a^*\geql \gz)\subseteq \mi{atoms}(S-\{C^*\})\subseteq \mi{atoms}(S)$.
We get from Case 1.2.1 for $S-\{C^*\}$ that there exists a finite linear {\it DPLL}-tree $\mi{Tree}''$ with the root $S-\{C^*\}$ constructed using Rules (\cref{ceq4hr1111111})--(\cref{ceq4hr4}), (\cref{ceq4hr7}) satisfying
for its only leaf $S'''$ that either $\square\in S'''$, or $S'''\subseteq_{\mc F} \mi{OrdPropCl}$ is $\gz$-guarded, $\mi{atoms}(S''')\subseteq \mi{atoms}(S-\{C^*\})-\{a^*\}=\mi{atoms}(S)-\{a^*\}$.
We put
\begin{equation} \notag
\mi{Tree}=\begin{array}[c]{c}
          S \\[0.4mm]
          \hline \\[-3.8mm]
          \mi{Tree}' \\[0.4mm]
          \hline \\[-3.8mm]
          (S''-\{a^{**}\gle \gu\})\cup \{\gz\gle \gu\} \\[0.4mm] 
          \hline \\[-3.8mm]
          \mi{Tree}''.
          \end{array}
\end{equation}
Hence, $\mi{Tree}$ is a finite linear {\it DPLL}-tree with the root $S$ constructed using Rules (\cref{ceq4hr1111111})--(\cref{ceq4hr4}), (\cref{ceq4hr7}) such that
for its only leaf $S'''$, either $\square\in S'''$, or $S'''\subseteq_{\mc F} \mi{OrdPropCl}$ is $\gz$-guarded, $\mi{atoms}(S''')\subseteq \mi{atoms}(S)-\{a^*\}$;
(i) holds.

Cases 1.2.2.2.1.4.2 and 1.2.2.4.1.4.2:
$\mi{guards}(S,a^{**})=\{\gz\gle a^{**}\}$.
Then $S''=(S-\{C^*\})\cup \{C^\natural\}=(S-\{C^*\})\cup \{a^{**}\gle \gu\}$,
$\mi{guards}(S'',a^{**})=\mi{guards}(S,a^{**})\cup \{C^\natural\}=\{\gz\gle a^{**},a^{**}\gle \gu\}$;
$a^{**}$ is $\gz$-guarded in $S''$;  
$\mi{atoms}(S'')\subseteq \mi{atoms}(S)$;
for all $a\in \mi{atoms}(S'')-\{a^*,a^{**}\}\subseteq \mi{atoms}(S)-\{a^*,a^{**}\}$,
$a$ is $\gz$-guarded in $S$;
$a\neq a^{**}$,
$\mi{guards}(a)\cap \{a^{**}\gle \gu\}=\mi{guards}(a)\cap \mi{guards}(a^{**})=\emptyset$,
$C^*\not\in \mi{guards}(a)$,
$\mi{guards}(S'',a)=((S-\{C^*\})\cup \{a^{**}\gle \gu\})\cap \mi{guards}(a)=((S-\{C^*\})\cap \mi{guards}(a))\cup (\{a^{**}\gle \gu\}\cap \mi{guards}(a))=(S\cap \mi{guards}(a))-\{C^*\}=
                    S\cap \mi{guards}(a)=\mi{guards}(S,a)$;
$a$ is $\gz$-guarded in $S''$;
$S''$ is $\gz$-guarded;
$a^*\in \mi{atoms}(S'')$, $\mi{guards}(S'',a^*)=\{a^*\geql \gz\}$,
$a^*\neq a^{**}$, $a^*\not\in \mi{atoms}(a^{**}\gle \gu)$, $C^*\in \mi{restricted}(S)$,
$\mi{restricted}(S'')=\{C \,|\, C\in (S-\{C^*\})\cup \{a^{**}\gle \gu\}, a^*\in \mi{atoms}(C), C\neq a^*\geql \gz\}=\{C \,|\, C\in S-\{C^*\}, a^*\in \mi{atoms}(C), C\neq a^*\geql \gz\}=
                      \mi{restricted}(S)-\{C^*\}=\{C \,|\, C\in S, a^*\in \mi{atoms}(C), C\neq a^*\geql \gz\}-\{C^*\}\subset \mi{restricted}(S)$;
by the induction hypothesis for $S''$, there exists a finite linear {\it DPLL}-tree $\mi{Tree}''$ with the root $S''$ constructed using Rules (\cref{ceq4hr1111111})--(\cref{ceq4hr4}), (\cref{ceq4hr7}) satisfying
for its only leaf $S'''$ that either $\square\in S'''$, or $S'''\subseteq_{\mc F} \mi{OrdPropCl}$ is $\gz$-guarded, $\mi{atoms}(S''')\subseteq \mi{atoms}(S'')-\{a^*\}\subseteq \mi{atoms}(S)-\{a^*\}$.
We put
\begin{equation} \notag
\mi{Tree}=\begin{array}[c]{c}
          S \\[0.4mm]
          \hline \\[-3.8mm]
          \mi{Tree}' \\[0.4mm]
          \hline \\[-3.8mm]
          \mi{Tree}''.
          \end{array}
\end{equation}
Hence, $\mi{Tree}$ is a finite linear {\it DPLL}-tree with the root $S$ constructed using Rules (\cref{ceq4hr1111111})--(\cref{ceq4hr4}), (\cref{ceq4hr7}) such that
for its only leaf $S'''$, either $\square\in S'''$, or $S'''\subseteq_{\mc F} \mi{OrdPropCl}$ is $\gz$-guarded, $\mi{atoms}(S''')\subseteq \mi{atoms}(S)-\{a^*\}$;
(i) holds.

Cases 1.2.2.2.1.4.3 and 1.2.2.4.1.4.3:
$\mi{guards}(S,a^{**})=\{\gz\gle a^{**},a^{**}\gle \gu\}$.
Then $a^{**}\gle \gu\in \mi{guards}(S,a^{**})\subseteq S$,
which is a contradiction with $C^\natural=a^{**}\gle \gu\not\in S$.

Cases 1.2.2.2.1.4.4 and 1.2.2.4.1.4.4:
$\mi{guards}(S,a^{**})=\{a^{**}\geql \gu\}$.
Then $S''\supseteq \mi{guards}(S'')\supseteq \mi{guards}(S'',a^{**})=\mi{guards}(S,a^{**})\cup \{C^\natural\}=\{a^{**}\gle \gu,a^{**}\geql \gu\}$,
$a^{**}\geql \gu\in \mi{guards}(S'')$, $a^{**}\gle \gu\in S''$,
$a^{**}\in \mi{atoms}(a^{**}\gle \gu)$, $a^{**}\geql \gu\neq a^{**}\gle \gu$, $\mi{simplify}(a^{**}\gle \gu,a^{**},\gu)=\gu\gle \gu$,
applying Rule (\cref{ceq4hr4}) to $S''$, $a^{**}\geql \gu$, and $a^{**}\gle \gu$, we derive
\begin{equation} \notag
\dfrac{S''}
      {(S''-\{a^{**}\gle \gu\})\cup \{\gu\gle \gu\}};
\end{equation}
$\gu\gle \gu\in (S''-\{a^{**}\gle \gu\})\cup \{\gu\gle \gu\}$;
$\gu\gle \gu\in \mi{OrdPropLit}$ is a contradiction;
$\gu\gle \gu$ is not a guard;
$\gu\gle \gu\not\in \mi{guards}((S''-\{a^{**}\gle \gu\})\cup \{\gu\gle \gu\})$,
$\gu\gle \gu\in ((S''-\{a^{**}\gle \gu\})\cup \{\gu\gle \gu\})-\mi{guards}((S''-\{a^{**}\gle \gu\})\cup \{\gu\gle \gu\})$,
applying Rule (\cref{ceq4hr2}) to $(S''-\{a^{**}\gle \gu\})\cup \{\gu\gle \gu\}$ and $\gu\gle \gu$, we derive
\begin{equation} \notag
\dfrac{(S''-\{a^{**}\gle \gu\})\cup \{\gu\gle \gu\}}
      {(((S''-\{a^{**}\gle \gu\})\cup \{\gu\gle \gu\})-\{\gu\gle \gu\})\cup \{\square\}}.
\end{equation}
We put
\begin{equation} \notag
\mi{Tree}=\begin{array}[c]{c}
          S \\[0.4mm]
          \hline \\[-3.8mm]
          \mi{Tree}' \\[0.4mm]
          \hline \\[-3.8mm]
          (S''-\{a^{**}\gle \gu\})\cup \{\gu\gle \gu\} \\[0.4mm]
          \hline \\[-3.8mm]
          S'''=(((S''-\{a^{**}\gle \gu\})\cup \{\gu\gle \gu\})-\{\gu\gle \gu\})\cup \{\square\}.
          \end{array} 
\end{equation}
Hence, $\mi{Tree}$ is a finite linear {\it DPLL}-tree with the root $S$ constructed using Rules (\cref{ceq4hr1111111})--(\cref{ceq4hr4}), (\cref{ceq4hr7}) such that
for its only leaf $S'''$, $\square\in S'''$;
(i) holds.

Cases 1.2.2.2.1.5 and 1.2.2.4.1.5:
$C^\natural=a^{**}\geql \gu$.
We have that $a^{**}$ is $\gz$-guarded in $S$.
We get four cases for $\mi{guards}(S,a^{**})$.

Cases 1.2.2.2.1.5.1 and 1.2.2.4.1.5.1:
$\mi{guards}(S,a^{**})=\{a^{**}\geql \gz\}$.
Then $S''\supseteq \mi{guards}(S'')\supseteq \mi{guards}(S'',a^{**})=\mi{guards}(S,a^{**})\cup \{C^\natural\}=\{a^{**}\geql \gz,a^{**}\geql \gu\}$;
these cases are the same as Cases 1.2.2.2.1.1.3 and 1.2.2.4.1.1.3.

Cases 1.2.2.2.1.5.2 and 1.2.2.4.1.5.2:
$\mi{guards}(S,a^{**})=\{\gz\gle a^{**}\}$.
Then $C^\natural=a^{**}\geql \gu\not\in S$,
$S''=(S-\{C^*\})\cup \{C^\natural\}=(S-\{C^*\})\cup \{a^{**}\geql \gu\}$,
$S\supseteq \mi{guards}(S,a^{**})=\{\gz\gle a^{**}\}$;
$S''\supseteq \mi{guards}(S'')\supseteq \mi{guards}(S'',a^{**})=\mi{guards}(S,a^{**})\cup \{C^\natural\}=\{\gz\gle a^{**},a^{**}\geql \gu\}$,
$a^{**}\geql \gu\in \mi{guards}(S'')$, $\gz\gle a^{**}\in S''$,
$a^{**}\in \mi{atoms}(\gz\gle a^{**})$, $a^{**}\geql \gu\neq \gz\gle a^{**}$, $\mi{simplify}(\gz\gle a^{**},a^{**},\gu)=\gz\gle \gu$,
applying Rule (\cref{ceq4hr4}) to $S''$, $a^{**}\geql \gu$, and $\gz\gle a^{**}$, we derive
\begin{equation} \notag
\dfrac{S''}
      {(S''-\{\gz\gle a^{**}\})\cup \{\gz\gle \gu\}};
\end{equation}
$\gz\gle \gu\in (S''-\{\gz\gle a^{**}\})\cup \{\gz\gle \gu\}$;
$\gz\gle \gu\in \mi{OrdPropLit}$ is a tautology;
$\gz\gle \gu$ is not a guard;
$\gz\gle \gu\not\in \mi{guards}((S''-\{\gz\gle a^{**}\})\cup \{\gz\gle \gu\})$,
$\gz\gle \gu\in ((S''-\{\gz\gle a^{**}\})\cup \{\gz\gle \gu\})-\mi{guards}((S''-\{\gz\gle a^{**}\})\cup \{\gz\gle \gu\})$,
applying Rule (\cref{ceq4hr22}) to $(S''-\{\gz\gle a^{**}\})\cup \{\gz\gle \gu\}$ and $\gz\gle \gu$, we derive
\begin{equation} \notag
\dfrac{(S''-\{\gz\gle a^{**}\})\cup \{\gz\gle \gu\}}
      {((S''-\{\gz\gle a^{**}\})\cup \{\gz\gle \gu\})-\{\gz\gle \gu\}}.
\end{equation}
We have that $S''$ is simplified.
Hence, $\gz\gle \gu\not\in S''$, $C^*, \gz\gle a^{**}\in S$, $C^*\not\in \mi{guards}(a^{**})$, $\gz\gle a^{**}\in \mi{guards}(a^{**})$, $C^*\neq \gz\gle a^{**}$, $a^{**}\geql \gu\not\in S$,
$((S''-\{\gz\gle a^{**}\})\cup \{\gz\gle \gu\})-\{\gz\gle \gu\}=((S''-\{\gz\gle a^{**}\})-\{\gz\gle \gu\})\cup (\{\gz\gle \gu\}-\{\gz\gle \gu\})=S''-\{\gz\gle a^{**}\}=
 ((S-\{C^*\})\cup \{a^{**}\geql \gu\})-\{\gz\gle a^{**}\}=((S-\{C^*\})-\{\gz\gle a^{**}\})\cup (\{a^{**}\geql \gu\}-\{\gz\gle a^{**}\})=(S-\{C^*,\gz\gle a^{**}\})\cup \{a^{**}\geql \gu\}$.
We put $S'''=(S-\{C^*,\gz\gle a^{**}\})\cup \{a^{**}\geql \gu\}\subseteq_{\mc F} \mi{OrdPropCl}$.
We get that
$S'''=(S-\{C^*,\gz\gle a^{**}\})\cup \{a^{**}\geql \gu\}=((S-\{C^*\})-\{\gz\gle a^{**}\})\cup \{a^{**}\geql \gu\}$,
$a^*\geql \gz\in S$, $a^*\geql \gz\neq C^*, \gz\gle a^{**}$, 
$a^*\geql \gz\in S'''=(S-\{C^*,\gz\gle a^{**}\})\cup \{a^{**}\geql \gu\}$,
$a^*\in \mi{atoms}(a^*\geql \gz)\subseteq \mi{atoms}(S''')$,
$a^{**}\in \mi{atoms}(S)$,
$a^{**}\in \mi{atoms}(S''')=\mi{atoms}((S-\{C^*,\gz\gle a^{**}\})\cup \{a^{**}\geql \gu\})=\mi{atoms}(S-\{C^*,\gz\gle a^{**}\})\cup \mi{atoms}(a^{**}\geql \gu)=
                            \mi{atoms}(S-\{C^*,\gz\gle a^{**}\})\cup \{a^{**}\}\subseteq \mi{atoms}(S)$;
$S$ is simplified;
$S-\{C^*,\gz\gle a^{**}\}\subseteq S$ is simplified;
$a^{**}\geql \gu\neq \square$ does not contain contradictions and tautologies;
$S'''=(S-\{C^*,\gz\gle a^{**}\})\cup \{a^{**}\geql \gu\}$ is simplified;
$a^*\neq a^{**}$,
$\mi{guards}(a^*)\cap \{\gz\gle a^{**},a^{**}\geql \gu\}=\mi{guards}(a^*)\cap \mi{guards}(a^{**})=\emptyset$,
$C^*\not\in \mi{guards}(a^*)$,
$\mi{guards}(S''',a^*)=(((S-\{C^*\})-\{\gz\gle a^{**}\})\cup \{a^{**}\geql \gu\})\cap \mi{guards}(a^*)=
                       (((S-\{C^*\})-\{\gz\gle a^{**}\})\cap \mi{guards}(a^*))\cup (\{a^{**}\geql \gu\}\cap \mi{guards}(a^*))=
                       ((S\cap \mi{guards}(a^*))-\{C^*\})-\{\gz\gle a^{**}\}=S\cap \mi{guards}(a^*)=\mi{guards}(S,a^*)=\{a^*\geql \gz\}$;
$a^*$ is $\gz$-guarded in $S'''$;
$C^*\not\in \mi{guards}(a^{**})$,
$\mi{guards}(S''',a^{**})=(((S-\{C^*\})-\{\gz\gle a^{**}\})\cup \{a^{**}\geql \gu\})\cap \mi{guards}(a^{**})=
                          (((S-\{C^*\})-\{\gz\gle a^{**}\})\cap \mi{guards}(a^{**}))\cup (\{a^{**}\geql \gu\}\cap \mi{guards}(a^{**}))=
                          (((S\cap \mi{guards}(a^{**}))-\{\gz\gle a^{**}\})-\{C^*\})\cup \{a^{**}\geql \gu\}=((S\cap \mi{guards}(a^{**}))-\{\gz\gle a^{**}\})\cup \{a^{**}\geql \gu\}=
                          (\mi{guards}(S,a^{**})-\{\gz\gle a^{**}\})\cup \{a^{**}\geql \gu\}=(\{\gz\gle a^{**}\}-\{\gz\gle a^{**}\})\cup \{a^{**}\geql \gu\}=\{a^{**}\geql \gu\}$;
$a^{**}$ is $\gz$-guarded in $S'''$;
for all $a\in \mi{atoms}(S''')-\{a^*,a^{**}\}\subseteq \mi{atoms}(S)-\{a^*,a^{**}\}$,
$a$ is $\gz$-guarded in $S$;
$a\neq a^{**}$,
$\mi{guards}(a)\cap \{\gz\gle a^{**},a^{**}\geql \gu\}=\mi{guards}(a)\cap \mi{guards}(a^{**})=\emptyset$,
$C^*\not\in \mi{guards}(a)$,
$\mi{guards}(S''',a)=(((S-\{C^*\})-\{\gz\gle a^{**}\})\cup \{a^{**}\geql \gu\})\cap \mi{guards}(a)=
                     (((S-\{C^*\})-\{\gz\gle a^{**}\})\cap \mi{guards}(a))\cup (\{a^{**}\geql \gu\}\cap \mi{guards}(a))=
                     ((S\cap \mi{guards}(a))-\{C^*\})-\{\gz\gle a^{**}\}=S\cap \mi{guards}(a)=\mi{guards}(S,a)$;
$a$ is $\gz$-guarded in $S'''$;
$S'''$ is $\gz$-guarded;
$a^*\not\in \mi{atoms}(\gz\gle a^{**})$, $a^*\not\in \mi{atoms}(a^{**}\geql \gu)$, $C^*\in \mi{restricted}(S)$,
$\mi{restricted}(S''')=\{C \,|\, C\in ((S-\{C^*\})-\{\gz\gle a^{**}\})\cup \{a^{**}\geql \gu\}, a^*\in \mi{atoms}(C), C\neq a^*\geql \gz\}= 
                       \{C \,|\, C\in S-\{C^*\}, a^*\in \mi{atoms}(C), C\neq a^*\geql \gz\}=
                       \mi{restricted}(S)-\{C^*\}=\{C \,|\, C\in S, a^*\in \mi{atoms}(C), C\neq a^*\geql \gz\}-\{C^*\}\subset \mi{restricted}(S)$;
by the induction hypothesis for $S'''$, there exists a finite linear {\it DPLL}-tree $\mi{Tree}'''$ with the root $S'''$ constructed using Rules (\cref{ceq4hr1111111})--(\cref{ceq4hr4}), (\cref{ceq4hr7}) satisfying
for its only leaf $S''''$ that either $\square\in S''''$, or $S''''\subseteq_{\mc F} \mi{OrdPropCl}$ is $\gz$-guarded, $\mi{atoms}(S'''')\subseteq \mi{atoms}(S''')-\{a^*\}\subseteq \mi{atoms}(S)-\{a^*\}$.
We put
\begin{equation} \notag
\mi{Tree}=\begin{array}[c]{c}
          S \\[0.4mm]
          \hline \\[-3.8mm]
          \mi{Tree}' \\[0.4mm]
          \hline \\[-3.8mm]
          (S''-\{\gz\gle a^{**}\})\cup \{\gz\gle \gu\} \\[0.4mm]
          \hline \\[-3.8mm]
          \mi{Tree}'''.    
          \end{array}    
\end{equation}
Hence, $\mi{Tree}$ is a finite linear {\it DPLL}-tree with the root $S$ constructed using Rules (\cref{ceq4hr1111111})--(\cref{ceq4hr4}), (\cref{ceq4hr7}) such that
for its only leaf $S''''$, either $\square\in S''''$, or $S''''\subseteq_{\mc F} \mi{OrdPropCl}$ is $\gz$-guarded, $\mi{atoms}(S'''')\subseteq \mi{atoms}(S)-\{a^*\}$;
(i) holds.

Cases 1.2.2.2.1.5.3 and 1.2.2.4.1.5.3:
$\mi{guards}(S,a^{**})=\{\gz\gle a^{**},a^{**}\gle \gu\}$.
Then $S''\supseteq \mi{guards}(S'')\supseteq \mi{guards}(S'',a^{**})=\mi{guards}(S,a^{**})\cup \{C^\natural\}=\{\gz\gle a^{**},a^{**}\gle \gu,a^{**}\geql \gu\}\supset \{a^{**}\gle \gu,a^{**}\geql \gu\}$;
these cases are the same as Cases 1.2.2.2.1.4.4 and 1.2.2.4.1.4.4.

Cases 1.2.2.2.1.5.4 and 1.2.2.4.1.5.4:
$\mi{guards}(S,a^{**})=\{a^{**}\geql \gu\}$.
Then $a^{**}\geql \gu\in \mi{guards}(S,a^{**})\subseteq S$,
which is a contradiction with $C^\natural=a^{**}\geql \gu\not\in S$.

Cases 1.2.2.2.1.6 and 1.2.2.4.1.6:
$C^\natural=\gu\gleq a^{**}$.
We have that $a^{**}$ is $\gz$-guarded in $S$.
We get four cases for $\mi{guards}(S,a^{**})$.

Cases 1.2.2.2.1.6.1 and 1.2.2.4.1.6.1:
$\mi{guards}(S,a^{**})=\{a^{**}\geql \gz\}$.
Then $S''\supseteq \mi{guards}(S'')\supseteq \mi{guards}(S'',a^{**})=\mi{guards}(S,a^{**})\cup \{C^\natural\}=\{a^{**}\geql \gz,\gu\gleq a^{**}\}$,
$a^{**}\geql \gz\in \mi{guards}(S'')$, $\gu\gleq a^{**}\in S''$,
$a^{**}\in \mi{atoms}(\gu\gleq a^*)$, $a^{**}\geql \gz\neq \gu\gleq a^{**}$, $\mi{simplify}(\gu\gleq a^{**},a^{**},\gz)=\gu\gleq \gz$,
applying Rule (\cref{ceq4hr3}) to $S''$, $a^{**}\geql \gz$, and $\gu\gleq a^{**}$, we derive
\begin{equation} \notag
\dfrac{S''}
      {(S''-\{\gu\gleq a^{**}\})\cup \{\gu\gleq \gz\}};
\end{equation}
$\gu\gleq \gz\in (S''-\{\gu\gleq a^{**}\})\cup \{\gu\gleq \gz\}$;
$\gu\gleq \gz\in \mi{OrdPropLit}$ is a contradiction;
$\gu\gleq \gz$ is not a guard;
$\gu\gleq \gz\not\in \mi{guards}((S''-\{\gu\gleq a^{**}\})\cup \{\gu\gleq \gz\})$,
$\gu\gleq \gz\in ((S''-\{\gu\gleq a^{**}\})\cup \{\gu\gleq \gz\})-\mi{guards}((S''-\{\gu\gleq a^{**}\})\cup \{\gu\gleq \gz\})$,
applying Rule (\cref{ceq4hr2}) to $(S''-\{\gu\gleq a^{**}\})\cup \{\gu\gleq \gz\}$ and $\gu\gleq \gz$, we derive
\begin{equation} \notag
\dfrac{(S''-\{\gu\gleq a^{**}\})\cup \{\gu\gleq \gz\}}
      {(((S''-\{\gu\gleq a^{**}\})\cup \{\gu\gleq \gz\})-\{\gu\gleq \gz\})\cup \{\square\}}.
\end{equation}
We put
\begin{equation} \notag
\mi{Tree}=\begin{array}[c]{c}
          S \\[0.4mm]
          \hline \\[-3.8mm]
          \mi{Tree}' \\[0.4mm]
          \hline \\[-3.8mm]
          (S''-\{\gu\gleq a^{**}\})\cup \{\gu\gleq \gz\} \\[0.4mm]
          \hline \\[-3.8mm]
          S'''=(((S''-\{\gu\gleq a^{**}\})\cup \{\gu\gleq \gz\})-\{\gu\gleq \gz\})\cup \{\square\}.
          \end{array} 
\end{equation}
Hence, $\mi{Tree}$ is a finite linear {\it DPLL}-tree with the root $S$ constructed using Rules (\cref{ceq4hr1111111})--(\cref{ceq4hr4}), (\cref{ceq4hr7}) such that
for its only leaf $S'''$, $\square\in S'''$;
(i) holds.

Cases 1.2.2.2.1.6.2 and 1.2.2.4.1.6.2:
$\mi{guards}(S,a^{**})=\{\gz\gle a^{**}\}$.
Then $C^\natural=\gu\gleq a^{**}\not\in S$,
$S''=(S-\{C^*\})\cup \{C^\natural\}=(S-\{C^*\})\cup \{\gu\gleq a^{**}\}$,
$S\supseteq \mi{guards}(S,a^{**})=\{\gz\gle a^{**}\}$;
$\mi{guards}(S'')\supseteq \mi{guards}(S'',a^{**})=\mi{guards}(S,a^{**})\cup \{C^\natural\}=\{\gz\gle a^{**},\gu\gleq a^{**}\}$,
$\gu\gleq a^{**}\in \mi{guards}(S'')$, applying Rule (\cref{ceq4hr11111111}) to $S''$ and $\gu\gleq a^{**}$, we derive
\begin{equation} \notag
\dfrac{S''}
      {(S''-\{\gu\gleq a^{**}\})\cup \{a^{**}\geql \gu\}};
\end{equation}
$a^{**}\in \mi{atoms}((S''-\{\gu\gleq a^{**}\})\cup \{a^{**}\geql \gu\})$,
$(S''-\{\gu\gleq a^{**}\})\cup \{a^{**}\geql \gu\}\supseteq \mi{guards}((S''-\{\gu\gleq a^{**}\})\cup \{a^{**}\geql \gu\})\supseteq 
 \mi{guards}((S''-\{\gu\gleq a^{**}\})\cup \{a^{**}\geql \gu\},a^{**})=((S''-\{\gu\gleq a^{**}\})\cup \{a^{**}\geql \gu\})\cap \mi{guards}(a^{**})=
 ((S''-\{\gu\gleq a^{**}\})\cap \mi{guards}(a^{**}))\cup (\{a^{**}\geql \gu\}\cap \mi{guards}(a^{**}))=((S''\cap \mi{guards}(a^{**}))-\{\gu\gleq a^{**}\})\cup \{a^{**}\geql \gu\}=
 (\mi{guards}(S'',a^{**})-\{\gu\gleq a^{**}\})\cup \{a^{**}\geql \gu\}=(\{\gz\gle a^{**},\gu\gleq a^{**}\}-\{\gu\gleq a^{**}\})\cup \{a^{**}\geql \gu\}=\{\gz\gle a^{**},a^{**}\geql \gu\}$,
$a^{**}\geql \gu\in \mi{guards}((S''-\{\gu\gleq a^{**}\})\cup \{a^{**}\geql \gu\})$, $\gz\gle a^{**}\in (S''-\{\gu\gleq a^{**}\})\cup \{a^{**}\geql \gu\}$,
$a^{**}\in \mi{atoms}(\gz\gle a^{**})$, $a^{**}\geql \gu\neq \gz\gle a^{**}$, $\mi{simplify}(\gz\gle a^{**},a^{**},\gu)=\gz\gle \gu$,
applying Rule (\cref{ceq4hr4}) to $(S''-\{\gu\gleq a^{**}\})\cup \{a^{**}\geql \gu\}$, $a^{**}\geql \gu$, and $\gz\gle a^{**}$, we derive
\begin{equation} \notag
\dfrac{(S''-\{\gu\gleq a^{**}\})\cup \{a^{**}\geql \gu\}}
      {(((S''-\{\gu\gleq a^{**}\})\cup \{a^{**}\geql \gu\})-\{\gz\gle a^{**}\})\cup \{\gz\gle \gu\}};
\end{equation}
$\gz\gle \gu\in (((S''-\{\gu\gleq a^{**}\})\cup \{a^{**}\geql \gu\})-\{\gz\gle a^{**}\})\cup \{\gz\gle \gu\}$;
$\gz\gle \gu\in \mi{OrdPropLit}$ is a tautology;
$\gz\gle \gu$ is not a guard;
$\gz\gle \gu\not\in \mi{guards}((((S''-\{\gu\gleq a^{**}\})\cup \{a^{**}\geql \gu\})-\{\gz\gle a^{**}\})\cup \{\gz\gle \gu\})$,
$\gz\gle \gu\in ((((S''-\{\gu\gleq a^{**}\})\cup \{a^{**}\geql \gu\})-\{\gz\gle a^{**}\})\cup \{\gz\gle \gu\})-\mi{guards}((((S''-\{\gu\gleq a^{**}\})\cup \{a^{**}\geql \gu\})-\{\gz\gle a^{**}\})\cup \{\gz\gle \gu\})$,
applying Rule (\cref{ceq4hr22}) to $(((S''-\{\gu\gleq a^{**}\})\cup \{a^{**}\geql \gu\})-\{\gz\gle a^{**}\})\cup \{\gz\gle \gu\}$ and $\gz\gle \gu$, we derive
\begin{equation} \notag
\dfrac{(((S''-\{\gu\gleq a^{**}\})\cup \{a^{**}\geql \gu\})-\{\gz\gle a^{**}\})\cup \{\gz\gle \gu\}}
      {((((S''-\{\gu\gleq a^{**}\})\cup \{a^{**}\geql \gu\})-\{\gz\gle a^{**}\})\cup \{\gz\gle \gu\})-\{\gz\gle \gu\}}.
\end{equation}
We have that $S''$ is simplified.
Hence, $\gz\gle \gu\not\in S''$, $\gu\gleq a^{**}\not\in S$, 
$C^*, \gz\gle a^{**}\in S$, $C^*\not\in \mi{guards}(a^{**})$, $\gz\gle a^{**}, a^{**}\geql \gu\in \mi{guards}(a^{**})$, $C^*\neq \gz\gle a^{**}$, 
$a^{**}\geql \gu\not\in \mi{guards}(S,a^{**})=\{\gz\gle a^{**}\}=S\cap \mi{guards}(a^{**})$, $a^{**}\geql \gu\not\in S$,
$((((S''-\{\gu\gleq a^{**}\})\cup \{a^{**}\geql \gu\})-\{\gz\gle a^{**}\})\cup \{\gz\gle \gu\})-\{\gz\gle \gu\}=
 ((((S''-\{\gu\gleq a^{**}\})\cup \{a^{**}\geql \gu\})-\{\gz\gle a^{**}\})-\{\gz\gle \gu\})\cup (\{\gz\gle \gu\}-\{\gz\gle \gu\})=
 (((S''-\{\gu\gleq a^{**}\})\cup \{a^{**}\geql \gu\})-\{\gz\gle \gu\})-\{\gz\gle a^{**}\}=
 (((S''-\{\gu\gleq a^{**}\})-\{\gz\gle \gu\})\cup (\{a^{**}\geql \gu\}-\{\gz\gle \gu\}))-\{\gz\gle a^{**}\}=
 ((S''-\{\gu\gleq a^{**}\})\cup \{a^{**}\geql \gu\})-\{\gz\gle a^{**}\}=
 ((S''-\{\gu\gleq a^{**}\})-\{\gz\gle a^{**}\})\cup (\{a^{**}\geql \gu\}-\{\gz\gle a^{**}\})=
 ((S''-\{\gu\gleq a^{**}\})-\{\gz\gle a^{**}\})\cup \{a^{**}\geql \gu\}=
 ((((S-\{C^*\})\cup \{\gu\gleq a^{**}\})-\{\gu\gleq a^{**}\})-\{\gz\gle a^{**}\})\cup \{a^{**}\geql \gu\}=
 ((((S-\{C^*\})-\{\gu\gleq a^{**}\})\cup (\{\gu\gleq a^{**}\}-\{\gu\gleq a^{**}\}))-\{\gz\gle a^{**}\})\cup \{a^{**}\geql \gu\}=
 ((S-\{C^*\})-\{\gz\gle a^{**}\})\cup \{a^{**}\geql \gu\}=(S-\{C^*,\gz\gle a^{**}\})\cup \{a^{**}\geql \gu\}$.
We put $S'''=(S-\{C^*,\gz\gle a^{**}\})\cup \{a^{**}\geql \gu\}\subseteq_{\mc F} \mi{OrdPropCl}$.
We get from Cases 1.2.2.2.1.5.2 and 1.2.2.4.1.5.2 that 
there exists a finite linear {\it DPLL}-tree $\mi{Tree}'''$ with the root $S'''$ constructed using Rules (\cref{ceq4hr1111111})--(\cref{ceq4hr4}), (\cref{ceq4hr7}) satisfying
for its only leaf $S''''$ that either $\square\in S''''$, or $S''''\subseteq_{\mc F} \mi{OrdPropCl}$ is $\gz$-guarded, $\mi{atoms}(S'''')\subseteq \mi{atoms}(S''')-\{a^*\}\subseteq \mi{atoms}(S)-\{a^*\}$.
We put
\begin{equation} \notag
\mi{Tree}=\begin{array}[c]{c}
          S \\[0.4mm]
          \hline \\[-3.8mm]
          \mi{Tree}' \\[0.4mm]
          \hline \\[-3.8mm]
          (S''-\{\gu\gleq a^{**}\})\cup \{a^{**}\geql \gu\} \\[0.4mm]
          \hline \\[-3.8mm]
          (((S''-\{\gu\gleq a^{**}\})\cup \{a^{**}\geql \gu\})-\{\gz\gle a^{**}\})\cup \{\gz\gle \gu\} \\[0.4mm]
          \hline \\[-3.8mm]
          \mi{Tree}'''.    
          \end{array}    
\end{equation}
Hence, $\mi{Tree}$ is a finite linear {\it DPLL}-tree with the root $S$ constructed using Rules (\cref{ceq4hr1111111})--(\cref{ceq4hr4}), (\cref{ceq4hr7}) such that
for its only leaf $S''''$, either $\square\in S''''$, or $S''''\subseteq_{\mc F} \mi{OrdPropCl}$ is $\gz$-guarded, $\mi{atoms}(S'''')\subseteq \mi{atoms}(S)-\{a^*\}$;
(i) holds.

Cases 1.2.2.2.1.6.3 and 1.2.2.4.1.6.3:
$\mi{guards}(S,a^{**})=\{\gz\gle a^{**},a^{**}\gle \gu\}$.
Then $\mi{guards}(S'')\supseteq \mi{guards}(S'',a^{**})=\mi{guards}(S,a^{**})\cup \{C^\natural\}=\{\gz\gle a^{**},a^{**}\gle \gu,\gu\gleq a^{**}\}$,
$\gu\gleq a^{**}\in \mi{guards}(S'')$, applying Rule (\cref{ceq4hr11111111}) to $S''$ and $\gu\gleq a^{**}$, we derive
\begin{equation} \notag
\dfrac{S''}
      {(S''-\{\gu\gleq a^{**}\})\cup \{a^{**}\geql \gu\}};
\end{equation}
$a^{**}\in \mi{atoms}((S''-\{\gu\gleq a^{**}\})\cup \{a^{**}\geql \gu\})$,
$(S''-\{\gu\gleq a^{**}\})\cup \{a^{**}\geql \gu\}\supseteq \mi{guards}((S''-\{\gu\gleq a^{**}\})\cup \{a^{**}\geql \gu\})\supseteq 
 \mi{guards}((S''-\{\gu\gleq a^{**}\})\cup \{a^{**}\geql \gu\},a^{**})=((S''-\{\gu\gleq a^{**}\})\cup \{a^{**}\geql \gu\})\cap \mi{guards}(a^{**})=
 ((S''-\{\gu\gleq a^{**}\})\cap \mi{guards}(a^{**}))\cup (\{a^{**}\geql \gu\}\cap \mi{guards}(a^{**}))=((S''\cap \mi{guards}(a^{**}))-\{\gu\gleq a^{**}\})\cup \{a^{**}\geql \gu\}=
 (\mi{guards}(S'',a^{**})-\{\gu\gleq a^{**}\})\cup \{a^{**}\geql \gu\}=(\{\gz\gle a^{**},a^{**}\gle \gu,\gu\gleq a^{**}\}-\{\gu\gleq a^{**}\})\cup \{a^{**}\geql \gu\}\supseteq 
 \{a^{**}\gle \gu,a^{**}\geql \gu\}$,
$a^{**}\geql \gu\in \mi{guards}((S''-\{\gu\gleq a^{**}\})\cup \{a^{**}\geql \gu\})$, $a^{**}\gle \gu\in (S''-\{\gu\gleq a^{**}\})\cup \{a^{**}\geql \gu\}$,
$a^{**}\in \mi{atoms}(a^{**}\gle \gu)$, $a^{**}\geql \gu\neq a^{**}\gle \gu$, $\mi{simplify}(a^{**}\gle \gu,a^{**},\gu)=\gu\gle \gu$,
applying Rule (\cref{ceq4hr4}) to $(S''-\{\gu\gleq a^{**}\})\cup \{a^{**}\geql \gu\}$, $a^{**}\geql \gu$, and $a^{**}\gle \gu$, we derive
\begin{equation} \notag
\dfrac{(S''-\{\gu\gleq a^{**}\})\cup \{a^{**}\geql \gu\}}
      {(((S''-\{\gu\gleq a^{**}\})\cup \{a^{**}\geql \gu\})-\{a^{**}\gle \gu\})\cup \{\gu\gle \gu\}};
\end{equation}
$\gu\gle \gu\in (((S''-\{\gu\gleq a^{**}\})\cup \{a^{**}\geql \gu\})-\{a^{**}\gle \gu\})\cup \{\gu\gle \gu\}$;
$\gu\gle \gu\in \mi{OrdPropLit}$ is a contradiction;
$\gu\gle \gu$ is not a guard;
$\gu\gle \gu\not\in \mi{guards}((((S''-\{\gu\gleq a^{**}\})\cup \{a^{**}\geql \gu\})-\{a^{**}\gle \gu\})\cup \{\gu\gle \gu\})$,
$\gu\gle \gu\in ((((S''-\{\gu\gleq a^{**}\})\cup \{a^{**}\geql \gu\})-\{a^{**}\gle \gu\})\cup \{\gu\gle \gu\})-\mi{guards}((((S''-\{\gu\gleq a^{**}\})\cup \{a^{**}\geql \gu\})-\{a^{**}\gle \gu\})\cup \{\gu\gle \gu\})$,
applying Rule (\cref{ceq4hr2}) to $(((S''-\{\gu\gleq a^{**}\})\cup \{a^{**}\geql \gu\})-\{a^{**}\gle \gu\})\cup \{\gu\gle \gu\}$ and $\gu\gle \gu$, we derive
\begin{equation} \notag
\dfrac{(((S''-\{\gu\gleq a^{**}\})\cup \{a^{**}\geql \gu\})-\{a^{**}\gle \gu\})\cup \{\gu\gle \gu\}}
      {(((((S''-\{\gu\gleq a^{**}\})\cup \{a^{**}\geql \gu\})-\{a^{**}\gle \gu\})\cup \{\gu\gle \gu\})-\{\gu\gle \gu\})\cup \{\square\}}.
\end{equation}
We put
\begin{equation} \notag
\mi{Tree}=\begin{array}[c]{c}
          S \\[0.4mm]
          \hline \\[-3.8mm]
          \mi{Tree}' \\[0.4mm]
          \hline \\[-3.8mm]
          (S''-\{\gu\gleq a^{**}\})\cup \{a^{**}\geql \gu\} \\[0.4mm]
          \hline \\[-3.8mm]
          (((S''-\{\gu\gleq a^{**}\})\cup \{a^{**}\geql \gu\})-\{a^{**}\gle \gu\})\cup \{\gu\gle \gu\} \\[0.4mm]
          \hline \\[-3.8mm]
          S'''=(((((S''-\{\gu\gleq a^{**}\})\cup \{a^{**}\geql \gu\})-\{a^{**}\gle \gu\})\cup \{\gu\gle \gu\})- \\
          \hfill \{\gu\gle \gu\})\cup \{\square\}.
          \end{array} 
\end{equation}
Hence, $\mi{Tree}$ is a finite linear {\it DPLL}-tree with the root $S$ constructed using Rules (\cref{ceq4hr1111111})--(\cref{ceq4hr4}), (\cref{ceq4hr7}) such that
for its only leaf $S'''$, $\square\in S'''$;
(i) holds.

Cases 1.2.2.2.1.6.4 and 1.2.2.4.1.6.4:
$\mi{guards}(S,a^{**})=\{a^{**}\geql \gu\}$.
Then $C^\natural=\gu\gleq a^{**}\not\in S$,
$S''=(S-\{C^*\})\cup \{C^\natural\}=(S-\{C^*\})\cup \{\gu\gleq a^{**}\}$;
$S''\supseteq \mi{guards}(S'')\supseteq \mi{guards}(S'',a^{**})=\mi{guards}(S,a^{**})\cup \{C^\natural\}=\{a^{**}\geql \gu,\gu\gleq a^{**}\}$,
$\gu\gleq a^{**}\in \mi{guards}(S'')$, applying Rule (\cref{ceq4hr11111111}) to $S''$ and $\gu\gleq a^{**}$, we derive
\begin{equation} \notag
\dfrac{S''}
      {(S''-\{\gu\gleq a^{**}\})\cup \{a^{**}\geql \gu\}}.
\end{equation}
Hence, $a^{**}\geql \gu\in S''$, $a^{**}\geql \gu\neq \gu\gleq a^{**}$, $a^{**}\geql \gu\in S''-\{\gu\gleq a^{**}\}$, $\gu\gleq a^{**}\not\in S$,
$(S''-\{\gu\gleq a^{**}\})\cup \{a^{**}\geql \gu\}=S''-\{\gu\gleq a^{**}\}=((S-\{C^*\})\cup \{\gu\gleq a^{**}\})-\{\gu\gleq a^{**}\}=
 ((S-\{C^*\})-\{\gu\gleq a^{**}\})\cup (\{\gu\gleq a^{**}\}-\{\gu\gleq a^{**}\})=S-\{C^*\}$,
$a^*\geql \gz\in S-\{C^*\}$, $a^*\in \mi{atoms}(a^*\geql \gz)\subseteq \mi{atoms}(S-\{C^*\})\subseteq \mi{atoms}(S)$.
We get from Case 1.2.1 for $S-\{C^*\}$ that there exists a finite linear {\it DPLL}-tree $\mi{Tree}''$ with the root $S-\{C^*\}$ constructed using Rules (\cref{ceq4hr1111111})--(\cref{ceq4hr4}), (\cref{ceq4hr7}) satisfying
for its only leaf $S'''$ that either $\square\in S'''$, or $S'''\subseteq_{\mc F} \mi{OrdPropCl}$ is $\gz$-guarded, $\mi{atoms}(S''')\subseteq \mi{atoms}(S-\{C^*\})-\{a^*\}=\mi{atoms}(S)-\{a^*\}$.
We put
\begin{equation} \notag
\mi{Tree}=\begin{array}[c]{c}
          S \\[0.4mm]
          \hline \\[-3.8mm]
          \mi{Tree}' \\[0.4mm]
          \hline \\[-3.8mm]
          \mi{Tree}''.
          \end{array}
\end{equation}
Hence, $\mi{Tree}$ is a finite linear {\it DPLL}-tree with the root $S$ constructed using Rules (\cref{ceq4hr1111111})--(\cref{ceq4hr4}), (\cref{ceq4hr7}) such that
for its only leaf $S'''$, either $\square\in S'''$, or $S'''\subseteq_{\mc F} \mi{OrdPropCl}$ is $\gz$-guarded, $\mi{atoms}(S''')\subseteq \mi{atoms}(S)-\{a^*\}$;
(i) holds.

Cases 1.2.2.2.2 and 1.2.2.4.2:
For all $a\in \mi{atoms}(S'')-\{a^*\}$, $C^\natural\not\in \mi{guards}(a)$.
Then, for all $a\in \mi{atoms}(C^\natural)\subseteq \mi{atoms}(S'')-\{a^*\}$, $C^\natural\not\in \mi{guards}(a)$;
$C^\natural$ is not a guard;
$C^*\not\in \mi{guards}(S)$,
$\mi{guards}(S'')=\{C \,|\, C\in (S-\{C^*\})\cup \{C^\natural\}\ \text{\it is a guard}\}=\{C \,|\, C\in S-\{C^*\}\ \text{\it is a guard}\}=
                  \mi{guards}(S)-\{C^*\}=\{C \,|\, C\in S\ \text{\it is a guard}\}-\{C^*\}=\mi{guards}(S)$;
we have that $S$ is $\gz$-guarded;
$\mi{atoms}(S''-\mi{guards}(S''))\subseteq \mi{atoms}(S'')\subseteq \mi{atoms}(S)$,
$\mi{atoms}(S'')=\mi{atoms}((S''-\mi{guards}(S''))\cup \mi{guards}(S''))=\mi{atoms}(S''-\mi{guards}(S''))\cup \mi{atoms}(\mi{guards}(S''))=
                 \mi{atoms}(S''-\mi{guards}(S''))\cup \mi{atoms}(\mi{guards}(S))=\mi{atoms}(S''-\mi{guards}(S''))\cup \mi{atoms}(S)=\mi{atoms}(S)$;
$S''$ is $\gz$-guarded; 
$a^*\in \mi{atoms}(S'')$, $\mi{guards}(S'',a^*)=\{a^*\geql \gz\}$,
$a^*\not\in \mi{atoms}(C^\natural)$, $C^*\in \mi{restricted}(S)$,
$\mi{restricted}(S'')=\{C \,|\, C\in (S-\{C^*\})\cup \{C^\natural\}, a^*\in \mi{atoms}(C), C\neq a^*\geql \gz\}=\{C \,|\, C\in S-\{C^*\}, a^*\in \mi{atoms}(C), C\neq a^*\geql \gz\}=
                      \mi{restricted}(S)-\{C^*\}=\{C \,|\, C\in S, a^*\in \mi{atoms}(C), C\neq a^*\geql \gz\}-\{C^*\}\subset \mi{restricted}(S)$;
by the induction hypothesis for $S''$, there exists a finite linear {\it DPLL}-tree $\mi{Tree}''$ with the root $S''$ constructed using Rules (\cref{ceq4hr1111111})--(\cref{ceq4hr4}), (\cref{ceq4hr7}) satisfying
for its only leaf $S'''$ that either $\square\in S'''$, or $S'''\subseteq_{\mc F} \mi{OrdPropCl}$ is $\gz$-guarded, $\mi{atoms}(S''')\subseteq \mi{atoms}(S'')-\{a^*\}=\mi{atoms}(S)-\{a^*\}$.
We put
\begin{equation} \notag
\mi{Tree}=\begin{array}[c]{c}
          S \\[0.4mm]
          \hline \\[-3.8mm]
          \mi{Tree}' \\[0.4mm]
          \hline \\[-3.8mm]
          \mi{Tree}''.
          \end{array}
\end{equation}
Hence, $\mi{Tree}$ is a finite linear {\it DPLL}-tree with the root $S$ constructed using Rules (\cref{ceq4hr1111111})--(\cref{ceq4hr4}), (\cref{ceq4hr7}) such that
for its only leaf $S'''$, either $\square\in S'''$, or $S'''\subseteq_{\mc F} \mi{OrdPropCl}$ is $\gz$-guarded, $\mi{atoms}(S''')\subseteq \mi{atoms}(S)-\{a^*\}$;
(i) holds.

So, in both Cases 1.1 and 1.2, (i) holds.
The induction is completed.
Thus, (i) holds.

Case 2:
$\mi{guards}(S,a^*)=\{a^*\geql \gu\}$.
Let $S^F\subseteq \mi{OrdPropCl}$.
We define a measure operator $\mi{restricted}(S^F)=\{C \,|\, C\in S^F, a^*\in \mi{atoms}(C), C\neq a^*\geql \gu\}$.
We proceed by induction on $\mi{restricted}(S)\subseteq S\subseteq_{\mc F} \mi{OrdPropCl}$.

Case 2.1 (the base case):
$\mi{restricted}(S)=\emptyset$.
We have that $S$ is $\gz$-guarded.
Then $S\supseteq \mi{guards}(S)\supseteq \mi{guards}(S,a^*)=\{a^*\geql \gu\}$, $a^*\geql \gu\in \mi{guards}(S)$, $a^*\geql \gu\in S$;
for all $C\in S$ satisfying $C\neq a^*\geql \gu$, $a^*\not\in \mi{atoms}(C)$;
$a^*\not\in \mi{atoms}(S-\{a^*\geql \gu\})$, applying Rule (\cref{ceq4hr8}) to $a^*\geql \gu$, we derive
\begin{equation} \notag
\dfrac{S}
      {S-\{a^*\geql \gu\}}.
\end{equation}
We put $S'=S-\{a^*\geql \gu\}\subseteq_{\mc F} \mi{OrdPropCl}$.
We get that
$a^*\in \mi{atoms}(S)$, $\mi{atoms}(a^*\geql \gu)=\{a^*\}$,
$a^*\not\in \mi{atoms}(S')=\mi{atoms}(S-\{a^*\geql \gu\})=\mi{atoms}(S)-\{a^*\}$;
$S$ is simplified;
$S'=S-\{a^*\geql \gu\}\subseteq S$ is simplified;
for all $a\in \mi{atoms}(S')=\mi{atoms}(S)-\{a^*\}$,
$a$ is $\gz$-guarded in $S$;
$a\neq a^*$, 
$\mi{guards}(a)\cap \{a^*\geql \gu\}=\mi{guards}(a)\cap \mi{guards}(a^*)=\emptyset$,
$\mi{guards}(S',a)=(S-\{a^*\geql \gu\})\cap \mi{guards}(a)=(S\cap \mi{guards}(a))-\{a^*\geql \gu\}=S\cap \mi{guards}(a)=\mi{guards}(S,a)$;
$a$ is $\gz$-guarded in $S'$;
$S'$ is $\gz$-guarded.
We put
\begin{equation} \notag
\mi{Tree}=\dfrac{S} 
                {S'}.
\end{equation}
Hence, $\mi{Tree}$ is a finite linear {\it DPLL}-tree with the root $S$ constructed using Rules (\cref{ceq4hr1111111})--(\cref{ceq4hr4}), (\cref{ceq4hr8}) such that
for its only leaf $S'$, $S'\subseteq_{\mc F} \mi{OrdPropCl}$ is $\gz$-guarded, $\mi{atoms}(S')=\mi{atoms}(S)-\{a^*\}$;
(ii) holds.

Case 2.2 (the induction case):
$\emptyset\neq \mi{restricted}(S)\subseteq_{\mc F} \mi{OrdPropCl}$.
We have that $S$ is $\gz$-guarded.
Then $S\supseteq \mi{guards}(S)\supseteq \mi{guards}(S,a^*)=\{a^*\geql \gu\}$, $a^*\geql \gu\in \mi{guards}(S)$, $a^*\geql \gu\in S$;
there exists $C^*\in \mi{restricted}(S)\subseteq S$ satisfying $a^*\in \mi{atoms}(C^*)$, $a^*\geql \gu\neq C^*$;
applying Rule (\cref{ceq4hr4}) to $a^*\geql \gu$ and $C^*$, we derive
\begin{equation} \notag
\dfrac{S}
      {(S-\{C^*\})\cup \{\mi{simplify}(C^*,a^*,\gu)\}}.
\end{equation}
We put $S'=(S-\{C^*\})\cup \{\mi{simplify}(C^*,a^*,\gu)\}\subseteq_{\mc F} \mi{OrdPropCl}$.
We get that
$a^*\geql \gu\in S$, $a^*\geql \gu\neq C^*$, $a^*\geql \gu\in S-\{C^*\}\subseteq S'$, $\mi{atoms}(\mi{simplify}(C^*,a^*,\gu))\subseteq \mi{atoms}(C^*)$, $C^*\in S$,
$a^*\in \mi{atoms}(a^*\geql \gu)\subseteq \mi{atoms}(S')=\mi{atoms}((S-\{C^*\})\cup \{\mi{simplify}(C^*,a^*,\gu)\})=\mi{atoms}(S-\{C^*\})\cup \mi{atoms}(\mi{simplify}(C^*,a^*,\gu))\subseteq 
                                                         \mi{atoms}(S-\{C^*\})\cup \mi{atoms}(C^*)=\mi{atoms}((S-\{C^*\})\cup \{C^*\})=\mi{atoms}(S)$;
$C^*\not\in \mi{guards}(S,a^*)=S\cap \mi{guards}(a^*)=\{a^*\geql \gu\}$, $C^*\not\in \mi{guards}(a^*)$;
for all $a\in \mi{atoms}(S)-\{a^*\}$,
$a\neq a^*$;
for all $C\in \mi{guards}(a)$, $a^*\in \mi{atoms}(C^*)$, $a^*\not\in \mi{atoms}(C)=\{a\}$, $\mi{atoms}(C^*)\neq \mi{atoms}(C)$, $C^*\neq C$;
$C^*\not\in \mi{guards}(a)$;
for all $a\in \mi{atoms}(C^*)\subseteq \mi{atoms}(S)$, $C^*\not\in \mi{guards}(a)$;
$C^*$ is not a guard; 
$C^*\not\in \mi{guards}(S)$.
We get two cases for $\mi{simplify}(C^*,a^*,\gu)$.

Case 2.2.1:
$\mi{simplify}(C^*,a^*,\gu)\in S$.
Then $a^*\not\in \mi{atoms}(\mi{simplify}(C^*,a^*,\gu))$, $a^*\in \mi{atoms}(C^*)$, $\mi{atoms}(\mi{simplify}(C^*,a^*,\gu))\neq \mi{atoms}(C^*)$, $\mi{simplify}(C^*,a^*,\gu)\neq C^*$,
$\mi{simplify}(C^*,a^*,\gu)\in S-\{C^*\}$,
$S'=(S-\{C^*\})\cup \{\mi{simplify}(C^*,a^*,\gu)\}=S-\{C^*\}$;
$S$ is simplified;
$S'=S-\{C^*\}\subseteq S$ is simplified;
$C^*\not\in \mi{guards}(S)$,
$\mi{guards}(S')=\{C \,|\, C\in S-\{C^*\}\ \text{\it is a guard}\}=\mi{guards}(S)-\{C^*\}=\{C \,|\, C\in S\ \text{\it is a guard}\}-\{C^*\}=\mi{guards}(S)$;
we have that $S$ is $\gz$-guarded;
$\mi{atoms}(S'-\mi{guards}(S'))\subseteq \mi{atoms}(S')\subseteq \mi{atoms}(S)$,
$\mi{atoms}(S')=\mi{atoms}((S'-\mi{guards}(S'))\cup \mi{guards}(S'))=\mi{atoms}(S'-\mi{guards}(S'))\cup \mi{atoms}(\mi{guards}(S'))=
                \mi{atoms}(S'-\mi{guards}(S'))\cup \mi{atoms}(\mi{guards}(S))=\mi{atoms}(S'-\mi{guards}(S'))\cup \mi{atoms}(S)=\mi{atoms}(S)$;
$S'$ is $\gz$-guarded;
$a^*\in \mi{atoms}(S')$, $\mi{guards}(S',a^*)=\mi{guards}(S,a^*)=\{a^*\geql \gu\}$,
$C^*\in \mi{restricted}(S)$,
$\mi{restricted}(S')=\{C \,|\, C\in S-\{C^*\}, a^*\in \mi{atoms}(C), C\neq a^*\geql \gu\}=
                     \mi{restricted}(S)-\{C^*\}=\{C \,|\, C\in S, a^*\in \mi{atoms}(C), C\neq a^*\geql \gu\}-\{C^*\}\subset \mi{restricted}(S)$;
by the induction hypothesis for $S'$, there exists a finite linear {\it DPLL}-tree $\mi{Tree}'$ with the root $S'$ constructed using Rules (\cref{ceq4hr1111111})--(\cref{ceq4hr4}), (\cref{ceq4hr8}) satisfying
for its only leaf $S''$ that either $\square\in S''$, or $S''\subseteq_{\mc F} \mi{OrdPropCl}$ is $\gz$-guarded, $\mi{atoms}(S'')\subseteq \mi{atoms}(S')-\{a^*\}=\mi{atoms}(S)-\{a^*\}$.
We put
\begin{equation} \notag
\mi{Tree}=\dfrac{S}
                {\mi{Tree}'}.
\end{equation}
Hence, $\mi{Tree}$ is a finite linear {\it DPLL}-tree with the root $S$ constructed using Rules (\cref{ceq4hr1111111})--(\cref{ceq4hr4}), (\cref{ceq4hr8}) such that
for its only leaf $S''$, either $\square\in S''$, or $S''\subseteq_{\mc F} \mi{OrdPropCl}$ is $\gz$-guarded, $\mi{atoms}(S'')\subseteq \mi{atoms}(S)-\{a^*\}$;
(ii) holds.

Case 2.2.2:
$\mi{simplify}(C^*,a^*,\gu)\not\in S$.
Then $\mi{simplify}(C^*,a^*,\gu)\in S'$; 
by Lemma \ref{le555} for $S'$ and $\mi{simplify}(C^*,a^*,\gu)$, there exists a finite linear {\it DPLL}-tree $\mi{Tree}'$ with the root $S'$ constructed using Rules (\cref{ceq4hr2}) and (\cref{ceq4hr22}) satisfying 
for its only leaf $S''$ that either $\square\in S''$, or $\square\not\in S''\subseteq_{\mc F} \mi{OrdPropCl}$ and exactly one of the following points holds.
\begin{enumerate}[\rm (a)]
\item
$S''=S'$, $\mi{simplify}(C^*,a^*,\gu)\neq \square$ does not contain contradictions and tautologies;
\item
$S''=S'-\{\mi{simplify}(C^*,a^*,\gu)\}$;
\item
there exists $C^{**}\in \mi{OrdPropCl}$ satisfying that $S''=(S'-\{\mi{simplify}(C^*,a^*,\gu)\})\cup \{C^{**}\}$, $C^{**}\not\in S'$, 
$\square\neq C^{**}\subset \mi{simplify}(C^*,a^*,\gu)$ does not contain contradictions and tautologies.
\end{enumerate}
We get four cases for $S''$.

Case 2.2.2.1:
$\square\in S''$.
We put
\begin{equation} \notag
\mi{Tree}=\dfrac{S}
                {\mi{Tree}'}.
\end{equation}
Hence, $\mi{Tree}$ is a finite linear {\it DPLL}-tree with the root $S$ constructed using Rules (\cref{ceq4hr1111111})--(\cref{ceq4hr4}), (\cref{ceq4hr8}) such that
for its only leaf $S''$, $\square\in S''$;
(ii) holds.

Case 2.2.2.2:
$S''=S'$ and $\mi{simplify}(C^*,a^*,\gu)\neq \square$ does not contain contradictions and tautologies.

Case 2.2.2.3:
$S''=S'-\{\mi{simplify}(C^*,a^*,\gu)\}$.
Then $\mi{simplify}(C^*,a^*,\gu)\not\in S$,
$S''=S'-\{\mi{simplify}(C^*,a^*,\gu)\}=((S-\{C^*\})\cup \{\mi{simplify}(C^*,a^*,\gu)\})-                                                                                                   \linebreak[4]
                                                                                        \{\mi{simplify}(C^*,a^*,\gu)\}=
     ((S-\{C^*\})-\{\mi{simplify}(C^*,a^*,\gu)\})\cup (\{\mi{simplify}(C^*,a^*,\gu)\}-\{\mi{simplify}(C^*,a^*,\gu)\})=S-\{C^*\}$,
$a^*\geql \gu\in S''=S-\{C^*\}$, $a^*\in \mi{atoms}(a^*\geql \gu)\subseteq \mi{atoms}(S'')=\mi{atoms}(S-\{C^*\})\subseteq \mi{atoms}(S)$.
We get from Case 2.2.1 for $S''$ that there exists a finite linear {\it DPLL}-tree $\mi{Tree}''$ with the root $S''$ constructed using Rules (\cref{ceq4hr1111111})--(\cref{ceq4hr4}), (\cref{ceq4hr8}) satisfying
for its only leaf $S'''$ that either $\square\in S'''$, or $S'''\subseteq_{\mc F} \mi{OrdPropCl}$ is $\gz$-guarded, $\mi{atoms}(S''')\subseteq \mi{atoms}(S'')-\{a^*\}=\mi{atoms}(S)-\{a^*\}$.
We put
\begin{equation} \notag
\mi{Tree}=\begin{array}[c]{c}
          S \\[0.4mm]
          \hline \\[-3.8mm]
          \mi{Tree}' \\[0.4mm]
          \hline \\[-3.8mm]
          \mi{Tree}''.    
          \end{array}    
\end{equation}
Hence, $\mi{Tree}$ is a finite linear {\it DPLL}-tree with the root $S$ constructed using Rules (\cref{ceq4hr1111111})--(\cref{ceq4hr4}), (\cref{ceq4hr8}) such that
for its only leaf $S'''$, either $\square\in S'''$, or $S'''\subseteq_{\mc F} \mi{OrdPropCl}$ is $\gz$-guarded, $\mi{atoms}(S''')\subseteq \mi{atoms}(S)-\{a^*\}$;
(ii) holds.

Case 2.2.2.4:
There exists $C^{**}\in \mi{OrdPropCl}$ such that $S''=(S'-\{\mi{simplify}(C^*,a^*,\gu)\})\cup \{C^{**}\}$, $C^{**}\not\in S'$, 
$\square\neq C^{**}\subset \mi{simplify}(C^*,a^*,\gu)$ does not contain contradictions and tautologies.

In Cases 2.2.2.2 and 2.2.2.4, we put 
\begin{alignat*}{1}
C^\natural &= \left\{\begin{array}{ll}
                     \mi{simplify}(C^*,a^*,\gu) &\ \text{\it in {\rm Case 2.2.2.2}}, \\[1mm]
                     C^{**}                     &\ \text{\it in {\rm Case 2.2.2.4}};
                     \end{array}
              \right. \\[1mm]
           &\in \mi{OrdPropCl}.
\end{alignat*}
Then, in Case 2.2.2.2, 
$C^\natural=\mi{simplify}(C^*,a^*,\gu)\not\in S$, $S''=S'=(S-\{C^*\})\cup \{\mi{simplify}(C^*,a^*,\gu)\}=(S-\{C^*\})\cup \{C^\natural\}$;
$C^\natural=\mi{simplify}(C^*,a^*,\gu)\neq \square$ does not contain contradictions and tautologies;
$a^*\not\in \mi{atoms}(C^\natural)=\mi{atoms}(\mi{simplify}(C^*,a^*,\gu))\subseteq \mi{atoms}(C^*)$;
in Case 2.2.2.4, 
$C^{**}\subset \mi{simplify}(C^*,a^*,\gu)$, 
$\mi{atoms}(C^\natural)=\mi{atoms}(C^{**})\subseteq \mi{atoms}(\mi{simplify}(C^*,a^*,\gu))\subseteq \mi{atoms}(C^*)$,
$a^*\not\in \mi{atoms}(\mi{simplify}(C^*,a^*,\gu))\supseteq \mi{atoms}(C^\natural)=\mi{atoms}(C^{**})$, $a^*\in \mi{atoms}(C^*)$, $\mi{atoms}(C^{**})\neq \mi{atoms}(C^*)$, $C^{**}\neq C^*$,
$C^{**}\not\in S'\supseteq S-\{C^*\}$, $C^\natural=C^{**}\not\in S$,
$S''=(S'-\{\mi{simplify}(C^*,a^*,\gu)\})\cup \{C^{**}\}=(((S-\{C^*\})\cup \{\mi{simplify}(C^*,a^*,\gu)\})-\{\mi{simplify}(C^*,a^*,\gu)\})\cup \{C^{**}\}=
     ((S-\{C^*\})-\{\mi{simplify}(C^*,a^*,\gu)\})\cup (\{\mi{simplify}(C^*,a^*,\gu)\}-\{\mi{simplify}(C^*,a^*,\gu)\})\cup \{C^{**}\}=(S-\{C^*\})\cup \{C^{**}\}=(S-\{C^*\})\cup \{C^\natural\}$;
$C^\natural=C^{**}\neq \square$ does not contain contradictions and tautologies;
in both Cases 2.2.2.2 and 2.2.2.4,
$C^\natural\not\in S$, $S''=(S-\{C^*\})\cup \{C^\natural\}$;
$C^\natural\neq \square$ does not contain contradictions and tautologies;
$a^*\not\in \mi{atoms}(C^\natural)\subseteq \mi{atoms}(C^*)$;
$a^*\geql \gu\in S-\{C^*\}\subseteq S''$, $C^*\in S$,
$a^*\in \mi{atoms}(a^*\geql \gu)\subseteq \mi{atoms}(S'')=\mi{atoms}((S-\{C^*\})\cup \{C^\natural\})=\mi{atoms}(S-\{C^*\})\cup \mi{atoms}(C^\natural)\subseteq 
                                                          \mi{atoms}(S-\{C^*\})\cup \mi{atoms}(C^*)=\mi{atoms}((S-\{C^*\})\cup \{C^*\})=\mi{atoms}(S)$,
$\mi{atoms}(C^\natural)\subseteq \mi{atoms}(S'')$, $\mi{atoms}(C^\natural)\subseteq \mi{atoms}(S'')-\{a^*\}$;
$S$ is simplified;
$S-\{C^*\}\subseteq S$ is simplified;
$S''=(S-\{C^*\})\cup \{C^\natural\}$ is simplified;
for all $C\in \mi{guards}(a^*)$, $a^*\in \mi{atoms}(C)=\{a^*\}$, $\mi{atoms}(C^\natural)\neq \mi{atoms}(C)$, $C^\natural\neq C$;
$C^\natural, C^*\not\in \mi{guards}(a^*)$,
$\mi{guards}(S'',a^*)=((S-\{C^*\})\cup \{C^\natural\})\cap \mi{guards}(a^*)=((S-\{C^*\})\cap \mi{guards}(a^*))\cup (\{C^\natural\}\cap \mi{guards}(a^*))=
                      (S\cap \mi{guards}(a^*))-\{C^*\}=S\cap \mi{guards}(a^*)=\mi{guards}(S,a^*)=\{a^*\geql \gu\}$;
$a^*$ is $\gz$-guarded in $S''$.
We get two cases for $C^\natural$.

Cases 2.2.2.2.1 and 2.2.2.4.1:
There exists $a^{**}\in \mi{atoms}(S'')-\{a^*\}\subseteq \mi{atoms}(S)-\{a^*\}$ such that $C^\natural\in \mi{guards}(a^{**})$.
Then $a^{**}\neq a^*$, $C^*\not\in \mi{guards}(a^{**})$,
$\mi{guards}(S'',a^{**})=((S-\{C^*\})\cup \{C^\natural\})\cap \mi{guards}(a^{**})=((S-\{C^*\})\cap \mi{guards}(a^{**}))\cup (\{C^\natural\}\cap \mi{guards}(a^{**}))=
                         ((S\cap \mi{guards}(a^{**}))-\{C^*\})\cup \{C^\natural\}=(S\cap \mi{guards}(a^{**}))\cup \{C^\natural\}=\mi{guards}(S,a^{**})\cup \{C^\natural\}$;
$a^{**}$ is $\gz$-guarded in $S$.
We get six cases for $C^\natural$.

Cases 2.2.2.2.1.1 and 2.2.2.4.1.1:
$C^\natural=a^{**}\geql \gz$.
We have that $a^{**}$ is $\gz$-guarded in $S$.
We get three cases for $\mi{guards}(S,a^{**})$.

Cases 2.2.2.2.1.1.1 and 2.2.2.4.1.1.1:
$\mi{guards}(S,a^{**})=\{a^{**}\geql \gz\}$.
Then $a^{**}\geql \gz\in \mi{guards}(S,a^{**})\subseteq S$,
which is a contradiction with $C^\natural=a^{**}\geql \gz\not\in S$.

Cases 2.2.2.2.1.1.2 and 2.2.2.4.1.1.2:
Either $\mi{guards}(S,a^{**})=\{\gz\gle a^{**}\}$ or $\mi{guards}(S,a^{**})=\{\gz\gle a^{**},a^{**}\gle \gu\}$.
Then $S''\supseteq \mi{guards}(S'')\supseteq \mi{guards}(S'',a^{**})=\mi{guards}(S,a^{**})\cup \{C^\natural\}=\mi{guards}(S,a^{**})\cup \{a^{**}\geql \gz\}\supseteq \{a^{**}\geql \gz,\gz\gle a^{**}\}$,
$a^{**}\geql \gz\in \mi{guards}(S'')$, $\gz\gle a^{**}\in S''$,
$a^{**}\in \mi{atoms}(\gz\gle a^{**})$, $a^{**}\geql \gz\neq \gz\gle a^{**}$, $\mi{simplify}(\gz\gle a^{**},a^{**},\gz)=\gz\gle \gz$,
applying Rule (\cref{ceq4hr3}) to $S''$, $a^{**}\geql \gz$, and $\gz\gle a^{**}$, we derive
\begin{equation} \notag
\dfrac{S''}
      {(S''-\{\gz\gle a^{**}\})\cup \{\gz\gle \gz\}};
\end{equation}
$\gz\gle \gz\in (S''-\{\gz\gle a^{**}\})\cup \{\gz\gle \gz\}$;
$\gz\gle \gz\in \mi{OrdPropLit}$ is a contradiction;
$\gz\gle \gz$ is not a guard;
$\gz\gle \gz\not\in \mi{guards}((S''-\{\gz\gle a^{**}\})\cup \{\gz\gle \gz\})$,
$\gz\gle \gz\in ((S''-\{\gz\gle a^{**}\})\cup \{\gz\gle \gz\})-\mi{guards}((S''-\{\gz\gle a^{**}\})\cup \{\gz\gle \gz\})$,
applying Rule (\cref{ceq4hr2}) to $(S''-\{\gz\gle a^{**}\})\cup \{\gz\gle \gz\}$ and $\gz\gle \gz$, we derive
\begin{equation} \notag
\dfrac{(S''-\{\gz\gle a^{**}\})\cup \{\gz\gle \gz\}}
      {(((S''-\{\gz\gle a^{**}\})\cup \{\gz\gle \gz\})-\{\gz\gle \gz\})\cup \{\square\}}.
\end{equation}
We put
\begin{equation} \notag
\mi{Tree}=\begin{array}[c]{c}
          S \\[0.4mm]
          \hline \\[-3.8mm]
          \mi{Tree}' \\[0.4mm]
          \hline \\[-3.8mm]
          (S''-\{\gz\gle a^{**}\})\cup \{\gz\gle \gz\} \\[0.4mm]
          \hline \\[-3.8mm]
          S'''=(((S''-\{\gz\gle a^{**}\})\cup \{\gz\gle \gz\})-\{\gz\gle \gz\})\cup \{\square\}.
          \end{array} 
\end{equation}
Hence, $\mi{Tree}$ is a finite linear {\it DPLL}-tree with the root $S$ constructed using Rules (\cref{ceq4hr1111111})--(\cref{ceq4hr4}), (\cref{ceq4hr8}) such that
for its only leaf $S'''$, $\square\in S'''$;
(ii) holds.

Cases 2.2.2.2.1.1.3 and 2.2.2.4.1.1.3:
$\mi{guards}(S,a^{**})=\{a^{**}\geql \gu\}$.
Then $S''\supseteq \mi{guards}(S'')\supseteq \mi{guards}(S'',a^{**})=\mi{guards}(S,a^{**})\cup \{C^\natural\}=\{a^{**}\geql \gz,a^{**}\geql \gu\}$,
$a^{**}\geql \gz\in \mi{guards}(S'')$, $a^{**}\geql \gu\in S''$,
$a^{**}\in \mi{atoms}(a^{**}\geql \gu)$, $a^{**}\geql \gz\neq a^{**}\geql \gu$, $\mi{simplify}(a^{**}\geql \gu,a^{**},\gz)=\gz\geql \gu$,
applying Rule (\cref{ceq4hr3}) to $S''$, $a^{**}\geql \gz$, and $a^{**}\geql \gu$, we derive
\begin{equation} \notag
\dfrac{S''}
      {(S''-\{a^{**}\geql \gu\})\cup \{\gz\geql \gu\}};
\end{equation}
$\gz\geql \gu\in (S''-\{a^{**}\geql \gu\})\cup \{\gz\geql \gu\}$;
$\gz\geql \gu\in \mi{OrdPropLit}$ is a contradiction;
$\gz\geql \gu$ is not a guard;
$\gz\geql \gu\not\in \mi{guards}((S''-\{a^{**}\geql \gu\})\cup \{\gz\geql \gu\})$,
$\gz\geql \gu\in ((S''-\{a^{**}\geql \gu\})\cup \{\gz\geql \gu\})-\mi{guards}((S''-\{a^{**}\geql \gu\})\cup \{\gz\geql \gu\})$,
applying Rule (\cref{ceq4hr2}) to $(S''-\{a^{**}\geql \gu\})\cup \{\gz\geql \gu\}$ and $\gz\geql \gu$, we derive
\begin{equation} \notag
\dfrac{(S''-\{a^{**}\geql \gu\})\cup \{\gz\geql \gu\}}
      {(((S''-\{a^{**}\geql \gu\})\cup \{\gz\geql \gu\})-\{\gz\geql \gu\})\cup \{\square\}}.
\end{equation}
We put 
\begin{equation} \notag
\mi{Tree}=\begin{array}[c]{c}
          S \\[0.4mm]
          \hline \\[-3.8mm]
          \mi{Tree}' \\[0.4mm]
          \hline \\[-3.8mm]
          (S''-\{a^{**}\geql \gu\})\cup \{\gz\geql \gu\} \\[0.4mm]
          \hline \\[-3.8mm]
          S'''=(((S''-\{a^{**}\geql \gu\})\cup \{\gz\geql \gu\})-\{\gz\geql \gu\})\cup \{\square\}.
          \end{array} 
\end{equation}
Hence, $\mi{Tree}$ is a finite linear {\it DPLL}-tree with the root $S$ constructed using Rules (\cref{ceq4hr1111111})--(\cref{ceq4hr4}), (\cref{ceq4hr8}) such that
for its only leaf $S'''$, $\square\in S'''$;
(ii) holds.

Cases 2.2.2.2.1.2 and 2.2.2.4.1.2:
$C^\natural=a^{**}\gleq \gz$.
We have that $a^{**}$ is $\gz$-guarded in $S$.
We get three cases for $\mi{guards}(S,a^{**})$.

Cases 2.2.2.2.1.2.1 and 2.2.2.4.1.2.1:
$\mi{guards}(S,a^{**})=\{a^{**}\geql \gz\}$.
Then $C^\natural=a^{**}\gleq \gz\not\in S$,
$S''=(S-\{C^*\})\cup \{C^\natural\}=(S-\{C^*\})\cup \{a^{**}\gleq \gz\}$;
$S''\supseteq \mi{guards}(S'')\supseteq \mi{guards}(S'',a^{**})=\mi{guards}(S,a^{**})\cup \{C^\natural\}=\{a^{**}\geql \gz,a^{**}\gleq \gz\}$,
$a^{**}\gleq \gz\in \mi{guards}(S'')$, applying Rule (\cref{ceq4hr1111111}) to $S''$ and $a^{**}\gleq \gz$, we derive
\begin{equation} \notag
\dfrac{S''}
      {(S''-\{a^{**}\gleq \gz\})\cup \{a^{**}\geql \gz\}}.
\end{equation}
Hence, $a^{**}\geql \gz\in S''$, $a^{**}\geql \gz\neq a^{**}\gleq \gz$, $a^{**}\geql \gz\in S''-\{a^{**}\gleq \gz\}$, $a^{**}\gleq \gz\not\in S$,
$(S''-\{a^{**}\gleq \gz\})\cup \{a^{**}\geql \gz\}=S''-\{a^{**}\gleq \gz\}=((S-\{C^*\})\cup \{a^{**}\gleq \gz\})-\{a^{**}\gleq \gz\}=
 ((S-\{C^*\})-\{a^{**}\gleq \gz\})\cup (\{a^{**}\gleq \gz\}-\{a^{**}\gleq \gz\})=S-\{C^*\}$,
$a^*\geql \gu\in S-\{C^*\}$, $a^*\in \mi{atoms}(a^*\geql \gu)\subseteq \mi{atoms}(S-\{C^*\})\subseteq \mi{atoms}(S)$.
We get from Case 2.2.1 for $S-\{C^*\}$ that there exists a finite linear {\it DPLL}-tree $\mi{Tree}''$ with the root $S-\{C^*\}$ constructed using Rules (\cref{ceq4hr1111111})--(\cref{ceq4hr4}), (\cref{ceq4hr8}) satisfying
for its only leaf $S'''$ that either $\square\in S'''$, or $S'''\subseteq_{\mc F} \mi{OrdPropCl}$ is $\gz$-guarded, $\mi{atoms}(S''')\subseteq \mi{atoms}(S-\{C^*\})-\{a^*\}=\mi{atoms}(S)-\{a^*\}$.
We put
\begin{equation} \notag
\mi{Tree}=\begin{array}[c]{c}
          S \\[0.4mm]
          \hline \\[-3.8mm]
          \mi{Tree}' \\[0.4mm]
          \hline \\[-3.8mm]
          \mi{Tree}''.
          \end{array}
\end{equation}
Hence, $\mi{Tree}$ is a finite linear {\it DPLL}-tree with the root $S$ constructed using Rules (\cref{ceq4hr1111111})--(\cref{ceq4hr4}), (\cref{ceq4hr8}) such that
for its only leaf $S'''$, either $\square\in S'''$, or $S'''\subseteq_{\mc F} \mi{OrdPropCl}$ is $\gz$-guarded, $\mi{atoms}(S''')\subseteq \mi{atoms}(S)-\{a^*\}$;
(ii) holds.

Cases 2.2.2.2.1.2.2 and 2.2.2.4.1.2.2:
Either $\mi{guards}(S,a^{**})=\{\gz\gle a^{**}\}$ or $\mi{guards}(S,a^{**})=\{\gz\gle a^{**},a^{**}\gle \gu\}$.
Then $\mi{guards}(S'')\supseteq \mi{guards}(S'',a^{**})=\mi{guards}(S,a^{**})\cup \{C^\natural\}=\mi{guards}(S,a^{**})\cup \{a^{**}\gleq \gz\}$,
$a^{**}\gleq \gz\in \mi{guards}(S'')$, applying Rule (\cref{ceq4hr1111111}) to $S''$ and $a^{**}\gleq \gz$, we derive
\begin{equation} \notag
\dfrac{S''}
      {(S''-\{a^{**}\gleq \gz\})\cup \{a^{**}\geql \gz\}};
\end{equation}
$a^{**}\in \mi{atoms}((S''-\{a^{**}\gleq \gz\})\cup \{a^{**}\geql \gz\})$,
$(S''-\{a^{**}\gleq \gz\})\cup \{a^{**}\geql \gz\}\supseteq \mi{guards}((S''-\{a^{**}\gleq \gz\})\cup \{a^{**}\geql \gz\})\supseteq 
 \mi{guards}((S''-\{a^{**}\gleq \gz\})\cup \{a^{**}\geql \gz\},a^{**})=((S''-\{a^{**}\gleq \gz\})\cup \{a^{**}\geql \gz\})\cap \mi{guards}(a^{**})=
 ((S''-\{a^{**}\gleq \gz\})\cap \mi{guards}(a^{**}))\cup (\{a^{**}\geql \gz\}\cap \mi{guards}(a^{**}))=((S''\cap \mi{guards}(a^{**}))-\{a^{**}\gleq \gz\})\cup \{a^{**}\geql \gz\}=
 (\mi{guards}(S'',a^{**})-\{a^{**}\gleq \gz\})\cup \{a^{**}\geql \gz\}=((\mi{guards}(S,a^{**})\cup \{a^{**}\gleq \gz\})-\{a^{**}\gleq \gz\})\cup \{a^{**}\geql \gz\}=
 (\mi{guards}(S,a^{**})-\{a^{**}\gleq \gz\})\cup (\{a^{**}\gleq \gz\}-\{a^{**}\gleq \gz\})\cup \{a^{**}\geql \gz\}=(\mi{guards}(S,a^{**})-\{a^{**}\gleq \gz\})\cup \{a^{**}\geql \gz\}\supseteq
 (\{\gz\gle a^{**}\}-\{a^{**}\gleq \gz\})\cup \{a^{**}\geql \gz\}=\{a^{**}\geql \gz,\gz\gle a^{**}\}$,
$a^{**}\geql \gz\in \mi{guards}((S''-\{a^{**}\gleq \gz\})\cup \{a^{**}\geql \gz\})$, $\gz\gle a^{**}\in (S''-\{a^{**}\gleq \gz\})\cup \{a^{**}\geql \gz\}$, 
$a^{**}\in \mi{atoms}(\gz\gle a^{**})$, $a^{**}\geql \gz\neq \gz\gle a^{**}$, $\mi{simplify}(\gz\gle a^{**},a^{**},\gz)=\gz\gle \gz$,
applying Rule (\cref{ceq4hr3}) to $(S''-\{a^{**}\gleq \gz\})\cup \{a^{**}\geql \gz\}$, $a^{**}\geql \gz$, and $\gz\gle a^{**}$, we derive
\begin{equation} \notag
\dfrac{(S''-\{a^{**}\gleq \gz\})\cup \{a^{**}\geql \gz\}}
      {(((S''-\{a^{**}\gleq \gz\})\cup \{a^{**}\geql \gz\})-\{\gz\gle a^{**}\})\cup \{\gz\gle \gz\}};
\end{equation}
$\gz\gle \gz\in (((S''-\{a^{**}\gleq \gz\})\cup \{a^{**}\geql \gz\})-\{\gz\gle a^{**}\})\cup \{\gz\gle \gz\}$;
$\gz\gle \gz\in \mi{OrdPropLit}$ is a contradiction;
$\gz\gle \gz$ is not a guard;
$\gz\gle \gz\not\in \mi{guards}((((S''-\{a^{**}\gleq \gz\})\cup \{a^{**}\geql \gz\})-\{\gz\gle a^{**}\})\cup \{\gz\gle \gz\})$,
$\gz\gle \gz\in ((((S''-\{a^{**}\gleq \gz\})\cup \{a^{**}\geql \gz\})-\{\gz\gle a^{**}\})\cup \{\gz\gle \gz\})-\mi{guards}((((S''-\{a^{**}\gleq \gz\})\cup \{a^{**}\geql \gz\})-\{\gz\gle a^{**}\})\cup \{\gz\gle \gz\})$,
applying Rule (\cref{ceq4hr2}) to $(((S''-\{a^{**}\gleq \gz\})\cup \{a^{**}\geql \gz\})-\{\gz\gle a^{**}\})\cup \{\gz\gle \gz\}$ and $\gz\gle \gz$, we derive
\begin{equation} \notag
\dfrac{(((S''-\{a^{**}\gleq \gz\})\cup \{a^{**}\geql \gz\})-\{\gz\gle a^{**}\})\cup \{\gz\gle \gz\}}
      {(((((S''-\{a^{**}\gleq \gz\})\cup \{a^{**}\geql \gz\})-\{\gz\gle a^{**}\})\cup \{\gz\gle \gz\})-\{\gz\gle \gz\})\cup \{\square\}}.
\end{equation}
We put
\begin{equation} \notag
\mi{Tree}=\begin{array}[c]{c}
          S \\[0.4mm]
          \hline \\[-3.8mm]
          \mi{Tree}' \\[0.4mm]
          \hline \\[-3.8mm]
          (S''-\{a^{**}\gleq \gz\})\cup \{a^{**}\geql \gz\} \\[0.4mm]
          \hline \\[-3.8mm]
          (((S''-\{a^{**}\gleq \gz\})\cup \{a^{**}\geql \gz\})-\{\gz\gle a^{**}\})\cup \{\gz\gle \gz\} \\[0.4mm]
          \hline \\[-3.8mm]
          S'''=(((((S''-\{a^{**}\gleq \gz\})\cup \{a^{**}\geql \gz\})-\{\gz\gle a^{**}\})\cup \{\gz\gle \gz\})- \\
          \hfill \{\gz\gle \gz\})\cup \{\square\}.
          \end{array} 
\end{equation}
Hence, $\mi{Tree}$ is a finite linear {\it DPLL}-tree with the root $S$ constructed using Rules (\cref{ceq4hr1111111})--(\cref{ceq4hr4}), (\cref{ceq4hr8}) such that
for its only leaf $S'''$, $\square\in S'''$;
(ii) holds.

Cases 2.2.2.2.1.2.3 and 2.2.2.4.1.2.3:
$\mi{guards}(S,a^{**})=\{a^{**}\geql \gu\}$.
Then $S''\supseteq \mi{guards}(S'')\supseteq \mi{guards}(S'',a^{**})=\mi{guards}(S,a^{**})\cup \{C^\natural\}=\{a^{**}\gleq \gz,a^{**}\geql \gu\}$,
$a^{**}\geql \gu\in \mi{guards}(S'')$, $a^{**}\gleq \gz\in S''$,
$a^{**}\in \mi{atoms}(a^{**}\gleq \gz)$, $a^{**}\geql \gu\neq a^{**}\gleq \gz$, $\mi{simplify}(a^{**}\gleq \gz,a^{**},\gu)=\gu\gleq \gz$,
applying Rule (\cref{ceq4hr4}) to $S''$, $a^{**}\geql \gu$, and $a^{**}\gleq \gz$, we derive
\begin{equation} \notag
\dfrac{S''}
      {(S''-\{a^{**}\gleq \gz\})\cup \{\gu\gleq \gz\}};
\end{equation}
$\gu\gleq \gz\in (S''-\{a^{**}\gleq \gz\})\cup \{\gu\gleq \gz\}$;
$\gu\gleq \gz\in \mi{OrdPropLit}$ is a contradiction;
$\gu\gleq \gz$ is not a guard;
$\gu\gleq \gz\not\in \mi{guards}((S''-\{a^{**}\gleq \gz\})\cup \{\gu\gleq \gz\})$,
$\gu\gleq \gz\in ((S''-\{a^{**}\gleq \gz\})\cup \{\gu\gleq \gz\})-\mi{guards}((S''-\{a^{**}\gleq \gz\})\cup \{\gu\gleq \gz\})$,
applying Rule (\cref{ceq4hr2}) to $(S''-\{a^{**}\gleq \gz\})\cup \{\gu\gleq \gz\}$ and $\gu\gleq \gz$, we derive
\begin{equation} \notag
\dfrac{(S''-\{a^{**}\gleq \gz\})\cup \{\gu\gleq \gz\}}
      {(((S''-\{a^{**}\gleq \gz\})\cup \{\gu\gleq \gz\})-\{\gu\gleq \gz\})\cup \{\square\}}.
\end{equation}
We put 
\begin{equation} \notag
\mi{Tree}=\begin{array}[c]{c}
          S \\[0.4mm]
          \hline \\[-3.8mm]
          \mi{Tree}' \\[0.4mm]
          \hline \\[-3.8mm]
          (S''-\{a^{**}\gleq \gz\})\cup \{\gu\gleq \gz\} \\[0.4mm]
          \hline \\[-3.8mm]
          S'''=(((S''-\{a^{**}\gleq \gz\})\cup \{\gu\gleq \gz\})-\{\gu\gleq \gz\})\cup \{\square\}.
          \end{array} 
\end{equation}
Hence, $\mi{Tree}$ is a finite linear {\it DPLL}-tree with the root $S$ constructed using Rules (\cref{ceq4hr1111111})--(\cref{ceq4hr4}), (\cref{ceq4hr8}) such that
for its only leaf $S'''$, $\square\in S'''$;
(ii) holds.

Cases 2.2.2.2.1.3 and 2.2.2.4.1.3:
$C^\natural=\gz\gle a^{**}$.
We have that $a^{**}$ is $\gz$-guarded in $S$.
We get three cases for $\mi{guards}(S,a^{**})$.

Cases 2.2.2.2.1.3.1 and 2.2.2.4.1.3.1:
$\mi{guards}(S,a^{**})=\{a^{**}\geql \gz\}$.
Then $S''\supseteq \mi{guards}(S'')\supseteq \mi{guards}(S'',a^{**})=\mi{guards}(S,a^{**})\cup \{C^\natural\}=\{a^{**}\geql \gz,\gz\gle a^{**}\}$;
these cases are the same as Cases 2.2.2.2.1.1.2 and 2.2.2.4.1.1.2.

Cases 2.2.2.2.1.3.2 and 2.2.2.4.1.3.2:
Either $\mi{guards}(S,a^{**})=\{\gz\gle a^{**}\}$ or $\mi{guards}(S,a^{**})=\{\gz\gle a^{**},a^{**}\gle \gu\}$.
Then $\gz\gle a^{**}\in \mi{guards}(S,a^{**})\subseteq S$,
which is a contradiction with $C^\natural=\gz\gle a^{**}\not\in S$.

Cases 2.2.2.2.1.3.3 and 2.2.2.4.1.3.3:
$\mi{guards}(S,a^{**})=\{a^{**}\geql \gu\}$.
Then $C^\natural=\gz\gle a^{**}\not\in S$,
$S''=(S-\{C^*\})\cup \{C^\natural\}=(S-\{C^*\})\cup \{\gz\gle a^{**}\}$;
$S''\supseteq \mi{guards}(S'')\supseteq \mi{guards}(S'',a^{**})=\mi{guards}(S,a^{**})\cup \{C^\natural\}=\{\gz\gle a^{**},a^{**}\geql \gu\}$,
$a^{**}\geql \gu\in \mi{guards}(S'')$, $\gz\gle a^{**}\in S''$,
$a^{**}\in \mi{atoms}(\gz\gle a^{**})$, $a^{**}\geql \gu\neq \gz\gle a^{**}$, $\mi{simplify}(\gz\gle a^{**},a^{**},\gu)=\gz\gle \gu$,
applying Rule (\cref{ceq4hr4}) to $S''$, $a^{**}\geql \gu$, and $\gz\gle a^{**}$, we derive
\begin{equation} \notag
\dfrac{S''}
      {(S''-\{\gz\gle a^{**}\})\cup \{\gz\gle \gu\}};
\end{equation}
$\gz\gle \gu\in (S''-\{\gz\gle a^{**}\})\cup \{\gz\gle \gu\}$;
$\gz\gle \gu\in \mi{OrdPropLit}$ is a tautology;
$\gz\gle \gu$ is not a guard;
$\gz\gle \gu\not\in \mi{guards}((S''-\{\gz\gle a^{**}\})\cup \{\gz\gle \gu\})$,
$\gz\gle \gu\in ((S''-\{\gz\gle a^{**}\})\cup \{\gz\gle \gu\})-\mi{guards}((S''-\{\gz\gle a^{**}\})\cup \{\gz\gle \gu\})$,
applying Rule (\cref{ceq4hr22}) to $(S''-\{\gz\gle a^{**}\})\cup \{\gz\gle \gu\}$ and $\gz\gle \gu$, we derive
\begin{equation} \notag
\dfrac{(S''-\{\gz\gle a^{**}\})\cup \{\gz\gle \gu\}}
      {((S''-\{\gz\gle a^{**}\})\cup \{\gz\gle \gu\})-\{\gz\gle \gu\}}.
\end{equation}
We have that $S''$ is simplified.
Hence, $\gz\gle \gu\not\in S''$, $\gz\gle a^{**}\not\in S$,
$((S''-\{\gz\gle a^{**}\})\cup \{\gz\gle \gu\})-\{\gz\gle \gu\}=((S''-\{\gz\gle a^{**}\})-\{\gz\gle \gu\})\cup (\{\gz\gle \gu\}-\{\gz\gle \gu\})=S''-\{\gz\gle a^{**}\}=
 ((S-\{C^*\})\cup \{\gz\gle a^{**}\})-\{\gz\gle a^{**}\}=((S-\{C^*\})-\{\gz\gle a^{**}\})\cup (\{\gz\gle a^{**}\}-\{\gz\gle a^{**}\})=S-\{C^*\}$,
$a^*\geql \gu\in S-\{C^*\}$, $a^*\in \mi{atoms}(a^*\geql \gu)\subseteq \mi{atoms}(S-\{C^*\})\subseteq \mi{atoms}(S)$.
We get from Case 2.2.1 for $S-\{C^*\}$ that there exists a finite linear {\it DPLL}-tree $\mi{Tree}''$ with the root $S-\{C^*\}$ constructed using Rules (\cref{ceq4hr1111111})--(\cref{ceq4hr4}), (\cref{ceq4hr8}) satisfying
for its only leaf $S'''$ that either $\square\in S'''$, or $S'''\subseteq_{\mc F} \mi{OrdPropCl}$ is $\gz$-guarded, $\mi{atoms}(S''')\subseteq \mi{atoms}(S-\{C^*\})-\{a^*\}=\mi{atoms}(S)-\{a^*\}$.
We put
\begin{equation} \notag
\mi{Tree}=\begin{array}[c]{c}
          S \\[0.4mm]
          \hline \\[-3.8mm]
          \mi{Tree}' \\[0.4mm]
          \hline \\[-3.8mm]
          (S''-\{\gz\gle a^{**}\})\cup \{\gz\gle \gu\} \\[0.4mm] 
          \hline \\[-3.8mm]
          \mi{Tree}''.
          \end{array}
\end{equation}
Hence, $\mi{Tree}$ is a finite linear {\it DPLL}-tree with the root $S$ constructed using Rules (\cref{ceq4hr1111111})--(\cref{ceq4hr4}), (\cref{ceq4hr8}) such that
for its only leaf $S'''$, either $\square\in S'''$, or $S'''\subseteq_{\mc F} \mi{OrdPropCl}$ is $\gz$-guarded, $\mi{atoms}(S''')\subseteq \mi{atoms}(S)-\{a^*\}$;
(ii) holds.

Cases 2.2.2.2.1.4 and 2.2.2.4.1.4:
$C^\natural=a^{**}\gle \gu$.
We have that $a^{**}$ is $\gz$-guarded in $S$.
We get four cases for $\mi{guards}(S,a^{**})$.

Cases 2.2.2.2.1.4.1 and 2.2.2.4.1.4.1:
$\mi{guards}(S,a^{**})=\{a^{**}\geql \gz\}$.
Then $C^\natural=a^{**}\gle \gu\not\in S$,
$S''=(S-\{C^*\})\cup \{C^\natural\}=(S-\{C^*\})\cup \{a^{**}\gle \gu\}$;
$S''\supseteq \mi{guards}(S'')\supseteq \mi{guards}(S'',a^{**})=\mi{guards}(S,a^{**})\cup \{C^\natural\}=\{a^{**}\geql \gz,a^{**}\gle \gu\}$,
$a^{**}\geql \gz\in \mi{guards}(S'')$, $a^{**}\gle \gu\in S''$,
$a^{**}\in \mi{atoms}(a^{**}\gle \gu)$, $a^{**}\geql \gz\neq a^{**}\gle \gu$, $\mi{simplify}(a^{**}\gle \gu,a^{**},\gz)=\gz\gle \gu$,
applying Rule (\cref{ceq4hr3}) to $S''$, $a^{**}\geql \gz$, and $a^{**}\gle \gu$, we derive
\begin{equation} \notag
\dfrac{S''}
      {(S''-\{a^{**}\gle \gu\})\cup \{\gz\gle \gu\}};
\end{equation}
$\gz\gle \gu\in (S''-\{a^{**}\gle \gu\})\cup \{\gz\gle \gu\}$;
$\gz\gle \gu\in \mi{OrdPropLit}$ is a tautology;
$\gz\gle \gu$ is not a guard;
$\gz\gle \gu\not\in \mi{guards}((S''-\{a^{**}\gle \gu\})\cup \{\gz\gle \gu\})$,
$\gz\gle \gu\in ((S''-\{a^{**}\gle \gu\})\cup \{\gz\gle \gu\})-\mi{guards}((S''-\{a^{**}\gle \gu\})\cup \{\gz\gle \gu\})$,
applying Rule (\cref{ceq4hr22}) to $(S''-\{a^{**}\gle \gu\})\cup \{\gz\gle \gu\}$ and $\gz\gle \gu$, we derive
\begin{equation} \notag
\dfrac{(S''-\{a^{**}\gle \gu\})\cup \{\gz\gle \gu\}}
      {((S''-\{a^{**}\gle \gu\})\cup \{\gz\gle \gu\})-\{\gz\gle \gu\}}.
\end{equation}
We have that $S''$ is simplified.
Hence, $\gz\gle \gu\not\in S''$, $a^{**}\gle \gu\not\in S$,
$((S''-\{a^{**}\gle \gu\})\cup \{\gz\gle \gu\})-\{\gz\gle \gu\}=((S''-\{a^{**}\gle \gu\})-\{\gz\gle \gu\})\cup (\{\gz\gle \gu\}-\{\gz\gle \gu\})=S''-\{a^{**}\gle \gu\}=
 ((S-\{C^*\})\cup \{a^{**}\gle \gu\})-\{a^{**}\gle \gu\}=((S-\{C^*\})-\{a^{**}\gle \gu\})\cup (\{a^{**}\gle \gu\}-\{a^{**}\gle \gu\})=S-\{C^*\}$,
$a^*\geql \gu\in S-\{C^*\}$, $a^*\in \mi{atoms}(a^*\geql \gu)\subseteq \mi{atoms}(S-\{C^*\})\subseteq \mi{atoms}(S)$.
We get from Case 2.2.1 for $S-\{C^*\}$ that there exists a finite linear {\it DPLL}-tree $\mi{Tree}''$ with the root $S-\{C^*\}$ constructed using Rules (\cref{ceq4hr1111111})--(\cref{ceq4hr4}), (\cref{ceq4hr8}) satisfying
for its only leaf $S'''$ that either $\square\in S'''$, or $S'''\subseteq_{\mc F} \mi{OrdPropCl}$ is $\gz$-guarded, $\mi{atoms}(S''')\subseteq \mi{atoms}(S-\{C^*\})-\{a^*\}=\mi{atoms}(S)-\{a^*\}$.
We put
\begin{equation} \notag
\mi{Tree}=\begin{array}[c]{c}
          S \\[0.4mm]
          \hline \\[-3.8mm]
          \mi{Tree}' \\[0.4mm]
          \hline \\[-3.8mm]
          (S''-\{a^{**}\gle \gu\})\cup \{\gz\gle \gu\} \\[0.4mm] 
          \hline \\[-3.8mm]
          \mi{Tree}''.
          \end{array}
\end{equation}
Hence, $\mi{Tree}$ is a finite linear {\it DPLL}-tree with the root $S$ constructed using Rules (\cref{ceq4hr1111111})--(\cref{ceq4hr4}), (\cref{ceq4hr8}) such that
for its only leaf $S'''$, either $\square\in S'''$, or $S'''\subseteq_{\mc F} \mi{OrdPropCl}$ is $\gz$-guarded, $\mi{atoms}(S''')\subseteq \mi{atoms}(S)-\{a^*\}$;
(ii) holds.

Cases 2.2.2.2.1.4.2 and 2.2.2.4.1.4.2:
$\mi{guards}(S,a^{**})=\{\gz\gle a^{**}\}$.
Then $S''=(S-\{C^*\})\cup \{C^\natural\}=(S-\{C^*\})\cup \{a^{**}\gle \gu\}$,
$\mi{guards}(S'',a^{**})=\mi{guards}(S,a^{**})\cup \{C^\natural\}=\{\gz\gle a^{**},a^{**}\gle \gu\}$;
$a^{**}$ is $\gz$-guarded in $S''$;  
$\mi{atoms}(S'')\subseteq \mi{atoms}(S)$;
for all $a\in \mi{atoms}(S'')-\{a^*,a^{**}\}\subseteq \mi{atoms}(S)-\{a^*,a^{**}\}$,
$a$ is $\gz$-guarded in $S$;
$a\neq a^{**}$,
$\mi{guards}(a)\cap \{a^{**}\gle \gu\}=\mi{guards}(a)\cap \mi{guards}(a^{**})=\emptyset$,
$C^*\not\in \mi{guards}(a)$,
$\mi{guards}(S'',a)=((S-\{C^*\})\cup \{a^{**}\gle \gu\})\cap \mi{guards}(a)=((S-\{C^*\})\cap \mi{guards}(a))\cup (\{a^{**}\gle \gu\}\cap \mi{guards}(a))=(S\cap \mi{guards}(a))-\{C^*\}=
                    S\cap \mi{guards}(a)=\mi{guards}(S,a)$;
$a$ is $\gz$-guarded in $S''$;
$S''$ is $\gz$-guarded;
$a^*\in \mi{atoms}(S'')$, $\mi{guards}(S'',a^*)=\{a^*\geql \gu\}$,
$a^*\neq a^{**}$, $a^*\not\in \mi{atoms}(a^{**}\gle \gu)$, $C^*\in \mi{restricted}(S)$,
$\mi{restricted}(S'')=\{C \,|\, C\in (S-\{C^*\})\cup \{a^{**}\gle \gu\}, a^*\in \mi{atoms}(C), C\neq a^*\geql \gu\}=\{C \,|\, C\in S-\{C^*\}, a^*\in \mi{atoms}(C), C\neq a^*\geql \gu\}=
                      \mi{restricted}(S)-\{C^*\}=\{C \,|\, C\in S, a^*\in \mi{atoms}(C), C\neq a^*\geql \gu\}-\{C^*\}\subset \mi{restricted}(S)$;
by the induction hypothesis for $S''$, there exists a finite linear {\it DPLL}-tree $\mi{Tree}''$ with the root $S''$ constructed using Rules (\cref{ceq4hr1111111})--(\cref{ceq4hr4}), (\cref{ceq4hr8}) satisfying
for its only leaf $S'''$ that either $\square\in S'''$, or $S'''\subseteq_{\mc F} \mi{OrdPropCl}$ is $\gz$-guarded, $\mi{atoms}(S''')\subseteq \mi{atoms}(S'')-\{a^*\}\subseteq \mi{atoms}(S)-\{a^*\}$.
We put
\begin{equation} \notag
\mi{Tree}=\begin{array}[c]{c}
          S \\[0.4mm]
          \hline \\[-3.8mm]
          \mi{Tree}' \\[0.4mm]
          \hline \\[-3.8mm]
          \mi{Tree}''.
          \end{array}
\end{equation}
Hence, $\mi{Tree}$ is a finite linear {\it DPLL}-tree with the root $S$ constructed using Rules (\cref{ceq4hr1111111})--(\cref{ceq4hr4}), (\cref{ceq4hr8}) such that
for its only leaf $S'''$, either $\square\in S'''$, or $S'''\subseteq_{\mc F} \mi{OrdPropCl}$ is $\gz$-guarded, $\mi{atoms}(S''')\subseteq \mi{atoms}(S)-\{a^*\}$;
(ii) holds.

Cases 2.2.2.2.1.4.3 and 2.2.2.4.1.4.3:
$\mi{guards}(S,a^{**})=\{\gz\gle a^{**},a^{**}\gle \gu\}$.
Then $a^{**}\gle \gu\in \mi{guards}(S,a^{**})\subseteq S$,
which is a contradiction with $C^\natural=a^{**}\gle \gu\not\in S$.

Cases 2.2.2.2.1.4.4 and 2.2.2.4.1.4.4:
$\mi{guards}(S,a^{**})=\{a^{**}\geql \gu\}$.
Then $S''\supseteq \mi{guards}(S'')\supseteq \mi{guards}(S'',a^{**})=\mi{guards}(S,a^{**})\cup \{C^\natural\}=\{a^{**}\gle \gu,a^{**}\geql \gu\}$,
$a^{**}\geql \gu\in \mi{guards}(S'')$, $a^{**}\gle \gu\in S''$,
$a^{**}\in \mi{atoms}(a^{**}\gle \gu)$, $a^{**}\geql \gu\neq a^{**}\gle \gu$, $\mi{simplify}(a^{**}\gle \gu,a^{**},\gu)=\gu\gle \gu$,
applying Rule (\cref{ceq4hr4}) to $S''$, $a^{**}\geql \gu$, and $a^{**}\gle \gu$, we derive
\begin{equation} \notag
\dfrac{S''}
      {(S''-\{a^{**}\gle \gu\})\cup \{\gu\gle \gu\}};
\end{equation}
$\gu\gle \gu\in (S''-\{a^{**}\gle \gu\})\cup \{\gu\gle \gu\}$;
$\gu\gle \gu\in \mi{OrdPropLit}$ is a contradiction;
$\gu\gle \gu$ is not a guard;
$\gu\gle \gu\not\in \mi{guards}((S''-\{a^{**}\gle \gu\})\cup \{\gu\gle \gu\})$,
$\gu\gle \gu\in ((S''-\{a^{**}\gle \gu\})\cup \{\gu\gle \gu\})-\mi{guards}((S''-\{a^{**}\gle \gu\})\cup \{\gu\gle \gu\})$,
applying Rule (\cref{ceq4hr2}) to $(S''-\{a^{**}\gle \gu\})\cup \{\gu\gle \gu\}$ and $\gu\gle \gu$, we derive
\begin{equation} \notag
\dfrac{(S''-\{a^{**}\gle \gu\})\cup \{\gu\gle \gu\}}
      {(((S''-\{a^{**}\gle \gu\})\cup \{\gu\gle \gu\})-\{\gu\gle \gu\})\cup \{\square\}}.
\end{equation}
We put
\begin{equation} \notag
\mi{Tree}=\begin{array}[c]{c}
          S \\[0.4mm]
          \hline \\[-3.8mm]
          \mi{Tree}' \\[0.4mm]
          \hline \\[-3.8mm]
          (S''-\{a^{**}\gle \gu\})\cup \{\gu\gle \gu\} \\[0.4mm]
          \hline \\[-3.8mm]
          S'''=(((S''-\{a^{**}\gle \gu\})\cup \{\gu\gle \gu\})-\{\gu\gle \gu\})\cup \{\square\}.
          \end{array} 
\end{equation}
Hence, $\mi{Tree}$ is a finite linear {\it DPLL}-tree with the root $S$ constructed using Rules (\cref{ceq4hr1111111})--(\cref{ceq4hr4}), (\cref{ceq4hr8}) such that
for its only leaf $S'''$, $\square\in S'''$;
(ii) holds.

Cases 2.2.2.2.1.5 and 2.2.2.4.1.5:
$C^\natural=a^{**}\geql \gu$.
We have that $a^{**}$ is $\gz$-guarded in $S$.
We get four cases for $\mi{guards}(S,a^{**})$.

Cases 2.2.2.2.1.5.1 and 2.2.2.4.1.5.1:
$\mi{guards}(S,a^{**})=\{a^{**}\geql \gz\}$.
Then $S''\supseteq \mi{guards}(S'')\supseteq \mi{guards}(S'',a^{**})=\mi{guards}(S,a^{**})\cup \{C^\natural\}=\{a^{**}\geql \gz,a^{**}\geql \gu\}$;
these cases are the same as Cases 2.2.2.2.1.1.3 and 2.2.2.4.1.1.3.

Cases 2.2.2.2.1.5.2 and 2.2.2.4.1.5.2:
$\mi{guards}(S,a^{**})=\{\gz\gle a^{**}\}$.
Then $C^\natural=a^{**}\geql \gu\not\in S$,
$S''=(S-\{C^*\})\cup \{C^\natural\}=(S-\{C^*\})\cup \{a^{**}\geql \gu\}$,
$S\supseteq \mi{guards}(S,a^{**})=\{\gz\gle a^{**}\}$;
$S''\supseteq \mi{guards}(S'')\supseteq \mi{guards}(S'',a^{**})=\mi{guards}(S,a^{**})\cup \{C^\natural\}=\{\gz\gle a^{**},a^{**}\geql \gu\}$,
$a^{**}\geql \gu\in \mi{guards}(S'')$, $\gz\gle a^{**}\in S''$,
$a^{**}\in \mi{atoms}(\gz\gle a^{**})$, $a^{**}\geql \gu\neq \gz\gle a^{**}$, $\mi{simplify}(\gz\gle a^{**},a^{**},\gu)=\gz\gle \gu$,
applying Rule (\cref{ceq4hr4}) to $S''$, $a^{**}\geql \gu$, and $\gz\gle a^{**}$, we derive
\begin{equation} \notag
\dfrac{S''}
      {(S''-\{\gz\gle a^{**}\})\cup \{\gz\gle \gu\}};
\end{equation}
$\gz\gle \gu\in (S''-\{\gz\gle a^{**}\})\cup \{\gz\gle \gu\}$;
$\gz\gle \gu\in \mi{OrdPropLit}$ is a tautology;
$\gz\gle \gu$ is not a guard;
$\gz\gle \gu\not\in \mi{guards}((S''-\{\gz\gle a^{**}\})\cup \{\gz\gle \gu\})$,
$\gz\gle \gu\in ((S''-\{\gz\gle a^{**}\})\cup \{\gz\gle \gu\})-\mi{guards}((S''-\{\gz\gle a^{**}\})\cup \{\gz\gle \gu\})$,
applying Rule (\cref{ceq4hr22}) to $(S''-\{\gz\gle a^{**}\})\cup \{\gz\gle \gu\}$ and $\gz\gle \gu$, we derive
\begin{equation} \notag
\dfrac{(S''-\{\gz\gle a^{**}\})\cup \{\gz\gle \gu\}}
      {((S''-\{\gz\gle a^{**}\})\cup \{\gz\gle \gu\})-\{\gz\gle \gu\}}.
\end{equation}
We have that $S''$ is simplified.
Hence, $\gz\gle \gu\not\in S''$, $C^*, \gz\gle a^{**}\in S$, $C^*\not\in \mi{guards}(a^{**})$, $\gz\gle a^{**}\in \mi{guards}(a^{**})$, $C^*\neq \gz\gle a^{**}$, $a^{**}\geql \gu\not\in S$,
$((S''-\{\gz\gle a^{**}\})\cup \{\gz\gle \gu\})-\{\gz\gle \gu\}=((S''-\{\gz\gle a^{**}\})-\{\gz\gle \gu\})\cup (\{\gz\gle \gu\}-\{\gz\gle \gu\})=S''-\{\gz\gle a^{**}\}=
 ((S-\{C^*\})\cup \{a^{**}\geql \gu\})-\{\gz\gle a^{**}\}=((S-\{C^*\})-\{\gz\gle a^{**}\})\cup (\{a^{**}\geql \gu\}-\{\gz\gle a^{**}\})=(S-\{C^*,\gz\gle a^{**}\})\cup \{a^{**}\geql \gu\}$.
We put $S'''=(S-\{C^*,\gz\gle a^{**}\})\cup \{a^{**}\geql \gu\}\subseteq_{\mc F} \mi{OrdPropCl}$.
We get that
$S'''=(S-\{C^*,\gz\gle a^{**}\})\cup \{a^{**}\geql \gu\}=((S-\{C^*\})-\{\gz\gle a^{**}\})\cup \{a^{**}\geql \gu\}$,
$a^*\geql \gu\in S$, $a^*\geql \gu\neq C^*, \gz\gle a^{**}$, 
$a^*\geql \gu\in S'''=(S-\{C^*,\gz\gle a^{**}\})\cup \{a^{**}\geql \gu\}$,
$a^*\in \mi{atoms}(a^*\geql \gu)\subseteq \mi{atoms}(S''')$,
$a^{**}\in \mi{atoms}(S)$,
$a^{**}\in \mi{atoms}(S''')=\mi{atoms}((S-\{C^*,\gz\gle a^{**}\})\cup \{a^{**}\geql \gu\})=\mi{atoms}(S-\{C^*,\gz\gle a^{**}\})\cup \mi{atoms}(a^{**}\geql \gu)=
                            \mi{atoms}(S-\{C^*,\gz\gle a^{**}\})\cup \{a^{**}\}\subseteq \mi{atoms}(S)$;
$S$ is simplified;
$S-\{C^*,\gz\gle a^{**}\}\subseteq S$ is simplified;
$a^{**}\geql \gu\neq \square$ does not contain contradictions and tautologies;
$S'''=(S-\{C^*,\gz\gle a^{**}\})\cup \{a^{**}\geql \gu\}$ is simplified;
$a^*\neq a^{**}$,
$\mi{guards}(a^*)\cap \{\gz\gle a^{**},a^{**}\geql \gu\}=\mi{guards}(a^*)\cap \mi{guards}(a^{**})=\emptyset$,
$C^*\not\in \mi{guards}(a^*)$,
$\mi{guards}(S''',a^*)=(((S-\{C^*\})-\{\gz\gle a^{**}\})\cup \{a^{**}\geql \gu\})\cap \mi{guards}(a^*)=
                       (((S-\{C^*\})-\{\gz\gle a^{**}\})\cap \mi{guards}(a^*))\cup (\{a^{**}\geql \gu\}\cap \mi{guards}(a^*))=
                       ((S\cap \mi{guards}(a^*))-\{C^*\})-\{\gz\gle a^{**}\}=S\cap \mi{guards}(a^*)=\mi{guards}(S,a^*)=\{a^*\geql \gu\}$;
$a^*$ is $\gz$-guarded in $S'''$;
$C^*\not\in \mi{guards}(a^{**})$,
$\mi{guards}(S''',a^{**})=(((S-\{C^*\})-\{\gz\gle a^{**}\})\cup \{a^{**}\geql \gu\})\cap \mi{guards}(a^{**})=
                          (((S-\{C^*\})-\{\gz\gle a^{**}\})\cap \mi{guards}(a^{**}))\cup (\{a^{**}\geql \gu\}\cap \mi{guards}(a^{**}))=
                          (((S\cap \mi{guards}(a^{**}))-\{\gz\gle a^{**}\})-\{C^*\})\cup \{a^{**}\geql \gu\}=((S\cap \mi{guards}(a^{**}))-\{\gz\gle a^{**}\})\cup \{a^{**}\geql \gu\}=
                          (\mi{guards}(S,a^{**})-\{\gz\gle a^{**}\})\cup \{a^{**}\geql \gu\}=(\{\gz\gle a^{**}\}-\{\gz\gle a^{**}\})\cup \{a^{**}\geql \gu\}=\{a^{**}\geql \gu\}$;
$a^{**}$ is $\gz$-guarded in $S'''$;
for all $a\in \mi{atoms}(S''')-\{a^*,a^{**}\}\subseteq \mi{atoms}(S)-\{a^*,a^{**}\}$,
$a$ is $\gz$-guarded in $S$;
$a\neq a^{**}$,
$\mi{guards}(a)\cap \{\gz\gle a^{**},a^{**}\geql \gu\}=\mi{guards}(a)\cap \mi{guards}(a^{**})=\emptyset$,
$C^*\not\in \mi{guards}(a)$,
$\mi{guards}(S''',a)=(((S-\{C^*\})-\{\gz\gle a^{**}\})\cup \{a^{**}\geql \gu\})\cap \mi{guards}(a)=
                     (((S-\{C^*\})-\{\gz\gle a^{**}\})\cap \mi{guards}(a))\cup (\{a^{**}\geql \gu\}\cap \mi{guards}(a))=
                     ((S\cap \mi{guards}(a))-\{C^*\})-\{\gz\gle a^{**}\}=S\cap \mi{guards}(a)=\mi{guards}(S,a)$;
$a$ is $\gz$-guarded in $S'''$;
$S'''$ is $\gz$-guarded;
$a^*\not\in \mi{atoms}(\gz\gle a^{**})$, $a^*\not\in \mi{atoms}(a^{**}\geql \gu)$, $C^*\in \mi{restricted}(S)$,
$\mi{restricted}(S''')=\{C \,|\, C\in ((S-\{C^*\})-\{\gz\gle a^{**}\})\cup \{a^{**}\geql \gu\}, a^*\in \mi{atoms}(C), C\neq a^*\geql \gu\}= 
                       \{C \,|\, C\in S-\{C^*\}, a^*\in \mi{atoms}(C), C\neq a^*\geql \gu\}=
                       \mi{restricted}(S)-\{C^*\}=\{C \,|\, C\in S, a^*\in \mi{atoms}(C), C\neq a^*\geql \gu\}-\{C^*\}\subset \mi{restricted}(S)$;
by the induction hypothesis for $S'''$, there exists a finite linear {\it DPLL}-tree $\mi{Tree}'''$ with the root $S'''$ constructed using Rules (\cref{ceq4hr1111111})--(\cref{ceq4hr4}), (\cref{ceq4hr8}) satisfying
for its only leaf $S''''$ that either $\square\in S''''$, or $S''''\subseteq_{\mc F} \mi{OrdPropCl}$ is $\gz$-guarded, $\mi{atoms}(S'''')\subseteq \mi{atoms}(S''')-\{a^*\}\subseteq \mi{atoms}(S)-\{a^*\}$.
We put
\begin{equation} \notag
\mi{Tree}=\begin{array}[c]{c}
          S \\[0.4mm]
          \hline \\[-3.8mm]
          \mi{Tree}' \\[0.4mm]
          \hline \\[-3.8mm]
          (S''-\{\gz\gle a^{**}\})\cup \{\gz\gle \gu\} \\[0.4mm]
          \hline \\[-3.8mm]
          \mi{Tree}'''.    
          \end{array}    
\end{equation}
Hence, $\mi{Tree}$ is a finite linear {\it DPLL}-tree with the root $S$ constructed using Rules (\cref{ceq4hr1111111})--(\cref{ceq4hr4}), (\cref{ceq4hr8}) such that
for its only leaf $S''''$, either $\square\in S''''$, or $S''''\subseteq_{\mc F} \mi{OrdPropCl}$ is $\gz$-guarded, $\mi{atoms}(S'''')\subseteq \mi{atoms}(S)-\{a^*\}$;
(ii) holds.

Cases 2.2.2.2.1.5.3 and 2.2.2.4.1.5.3:
$\mi{guards}(S,a^{**})=\{\gz\gle a^{**},a^{**}\gle \gu\}$.
Then $S''\supseteq \mi{guards}(S'')\supseteq \mi{guards}(S'',a^{**})=\mi{guards}(S,a^{**})\cup \{C^\natural\}=\{\gz\gle a^{**},a^{**}\gle \gu,a^{**}\geql \gu\}\supset \{a^{**}\gle \gu,a^{**}\geql \gu\}$;
these cases are the same as Cases 2.2.2.2.1.4.4 and 2.2.2.4.1.4.4.

Cases 2.2.2.2.1.5.4 and 2.2.2.4.1.5.4:
$\mi{guards}(S,a^{**})=\{a^{**}\geql \gu\}$.
Then $a^{**}\geql \gu\in \mi{guards}(S,a^{**})\subseteq S$,
which is a contradiction with $C^\natural=a^{**}\geql \gu\not\in S$.

Cases 2.2.2.2.1.6 and 2.2.2.4.1.6:
$C^\natural=\gu\gleq a^{**}$.
We have that $a^{**}$ is $\gz$-guarded in $S$.
We get four cases for $\mi{guards}(S,a^{**})$.

Cases 2.2.2.2.1.6.1 and 2.2.2.4.1.6.1:
$\mi{guards}(S,a^{**})=\{a^{**}\geql \gz\}$.
Then $S''\supseteq \mi{guards}(S'')\supseteq \mi{guards}(S'',a^{**})=\mi{guards}(S,a^{**})\cup \{C^\natural\}=\{a^{**}\geql \gz,\gu\gleq a^{**}\}$,
$a^{**}\geql \gz\in \mi{guards}(S'')$, $\gu\gleq a^{**}\in S''$,
$a^{**}\in \mi{atoms}(\gu\gleq a^*)$, $a^{**}\geql \gz\neq \gu\gleq a^{**}$, $\mi{simplify}(\gu\gleq a^{**},a^{**},\gz)=\gu\gleq \gz$,
applying Rule (\cref{ceq4hr3}) to $S''$, $a^{**}\geql \gz$, and $\gu\gleq a^{**}$, we derive
\begin{equation} \notag
\dfrac{S''}
      {(S''-\{\gu\gleq a^{**}\})\cup \{\gu\gleq \gz\}};
\end{equation}
$\gu\gleq \gz\in (S''-\{\gu\gleq a^{**}\})\cup \{\gu\gleq \gz\}$;
$\gu\gleq \gz\in \mi{OrdPropLit}$ is a contradiction;
$\gu\gleq \gz$ is not a guard;
$\gu\gleq \gz\not\in \mi{guards}((S''-\{\gu\gleq a^{**}\})\cup \{\gu\gleq \gz\})$,
$\gu\gleq \gz\in ((S''-\{\gu\gleq a^{**}\})\cup \{\gu\gleq \gz\})-\mi{guards}((S''-\{\gu\gleq a^{**}\})\cup \{\gu\gleq \gz\})$,
applying Rule (\cref{ceq4hr2}) to $(S''-\{\gu\gleq a^{**}\})\cup \{\gu\gleq \gz\}$ and $\gu\gleq \gz$, we derive
\begin{equation} \notag
\dfrac{(S''-\{\gu\gleq a^{**}\})\cup \{\gu\gleq \gz\}}
      {(((S''-\{\gu\gleq a^{**}\})\cup \{\gu\gleq \gz\})-\{\gu\gleq \gz\})\cup \{\square\}}.
\end{equation}
We put
\begin{equation} \notag
\mi{Tree}=\begin{array}[c]{c}
          S \\[0.4mm]
          \hline \\[-3.8mm]
          \mi{Tree}' \\[0.4mm]
          \hline \\[-3.8mm]
          (S''-\{\gu\gleq a^{**}\})\cup \{\gu\gleq \gz\} \\[0.4mm]
          \hline \\[-3.8mm]
          S'''=(((S''-\{\gu\gleq a^{**}\})\cup \{\gu\gleq \gz\})-\{\gu\gleq \gz\})\cup \{\square\}.
          \end{array} 
\end{equation}
Hence, $\mi{Tree}$ is a finite linear {\it DPLL}-tree with the root $S$ constructed using Rules (\cref{ceq4hr1111111})--(\cref{ceq4hr4}), (\cref{ceq4hr8}) such that
for its only leaf $S'''$, $\square\in S'''$;
(ii) holds.

Cases 2.2.2.2.1.6.2 and 2.2.2.4.1.6.2:
$\mi{guards}(S,a^{**})=\{\gz\gle a^{**}\}$.
Then $C^\natural=\gu\gleq a^{**}\not\in S$,
$S''=(S-\{C^*\})\cup \{C^\natural\}=(S-\{C^*\})\cup \{\gu\gleq a^{**}\}$,
$S\supseteq \mi{guards}(S,a^{**})=\{\gz\gle a^{**}\}$;
$\mi{guards}(S'')\supseteq \mi{guards}(S'',a^{**})=\mi{guards}(S,a^{**})\cup \{C^\natural\}=\{\gz\gle a^{**},\gu\gleq a^{**}\}$,
$\gu\gleq a^{**}\in \mi{guards}(S'')$, applying Rule (\cref{ceq4hr11111111}) to $S''$ and $\gu\gleq a^{**}$, we derive
\begin{equation} \notag
\dfrac{S''}
      {(S''-\{\gu\gleq a^{**}\})\cup \{a^{**}\geql \gu\}};
\end{equation}
$a^{**}\in \mi{atoms}((S''-\{\gu\gleq a^{**}\})\cup \{a^{**}\geql \gu\})$,
$(S''-\{\gu\gleq a^{**}\})\cup \{a^{**}\geql \gu\}\supseteq \mi{guards}((S''-\{\gu\gleq a^{**}\})\cup \{a^{**}\geql \gu\})\supseteq 
 \mi{guards}((S''-\{\gu\gleq a^{**}\})\cup \{a^{**}\geql \gu\},a^{**})=((S''-\{\gu\gleq a^{**}\})\cup \{a^{**}\geql \gu\})\cap \mi{guards}(a^{**})=
 ((S''-\{\gu\gleq a^{**}\})\cap \mi{guards}(a^{**}))\cup (\{a^{**}\geql \gu\}\cap \mi{guards}(a^{**}))=((S''\cap \mi{guards}(a^{**}))-\{\gu\gleq a^{**}\})\cup \{a^{**}\geql \gu\}=
 (\mi{guards}(S'',a^{**})-\{\gu\gleq a^{**}\})\cup \{a^{**}\geql \gu\}=(\{\gz\gle a^{**},\gu\gleq a^{**}\}-\{\gu\gleq a^{**}\})\cup \{a^{**}\geql \gu\}=\{\gz\gle a^{**},a^{**}\geql \gu\}$,
$a^{**}\geql \gu\in \mi{guards}((S''-\{\gu\gleq a^{**}\})\cup \{a^{**}\geql \gu\})$, $\gz\gle a^{**}\in (S''-\{\gu\gleq a^{**}\})\cup \{a^{**}\geql \gu\}$,
$a^{**}\in \mi{atoms}(\gz\gle a^{**})$, $a^{**}\geql \gu\neq \gz\gle a^{**}$, $\mi{simplify}(\gz\gle a^{**},a^{**},\gu)=\gz\gle \gu$,
applying Rule (\cref{ceq4hr4}) to $(S''-\{\gu\gleq a^{**}\})\cup \{a^{**}\geql \gu\}$, $a^{**}\geql \gu$, and $\gz\gle a^{**}$, we derive
\begin{equation} \notag
\dfrac{(S''-\{\gu\gleq a^{**}\})\cup \{a^{**}\geql \gu\}}
      {(((S''-\{\gu\gleq a^{**}\})\cup \{a^{**}\geql \gu\})-\{\gz\gle a^{**}\})\cup \{\gz\gle \gu\}};
\end{equation}
$\gz\gle \gu\in (((S''-\{\gu\gleq a^{**}\})\cup \{a^{**}\geql \gu\})-\{\gz\gle a^{**}\})\cup \{\gz\gle \gu\}$;
$\gz\gle \gu\in \mi{OrdPropLit}$ is a tautology;
$\gz\gle \gu$ is not a guard;
$\gz\gle \gu\not\in \mi{guards}((((S''-\{\gu\gleq a^{**}\})\cup \{a^{**}\geql \gu\})-\{\gz\gle a^{**}\})\cup \{\gz\gle \gu\})$,
$\gz\gle \gu\in ((((S''-\{\gu\gleq a^{**}\})\cup \{a^{**}\geql \gu\})-\{\gz\gle a^{**}\})\cup \{\gz\gle \gu\})-\mi{guards}((((S''-\{\gu\gleq a^{**}\})\cup \{a^{**}\geql \gu\})-\{\gz\gle a^{**}\})\cup \{\gz\gle \gu\})$,
applying Rule (\cref{ceq4hr22}) to $(((S''-\{\gu\gleq a^{**}\})\cup \{a^{**}\geql \gu\})-\{\gz\gle a^{**}\})\cup \{\gz\gle \gu\}$ and $\gz\gle \gu$, we derive
\begin{equation} \notag
\dfrac{(((S''-\{\gu\gleq a^{**}\})\cup \{a^{**}\geql \gu\})-\{\gz\gle a^{**}\})\cup \{\gz\gle \gu\}}
      {((((S''-\{\gu\gleq a^{**}\})\cup \{a^{**}\geql \gu\})-\{\gz\gle a^{**}\})\cup \{\gz\gle \gu\})-\{\gz\gle \gu\}}.
\end{equation}
We have that $S''$ is simplified.
Hence, $\gz\gle \gu\not\in S''$, $\gu\gleq a^{**}\not\in S$, 
$C^*, \gz\gle a^{**}\in S$, $C^*\not\in \mi{guards}(a^{**})$, $\gz\gle a^{**}, a^{**}\geql \gu\in \mi{guards}(a^{**})$, $C^*\neq \gz\gle a^{**}$, 
$a^{**}\geql \gu\not\in \mi{guards}(S,a^{**})=\{\gz\gle a^{**}\}=S\cap \mi{guards}(a^{**})$, $a^{**}\geql \gu\not\in S$,
$((((S''-\{\gu\gleq a^{**}\})\cup \{a^{**}\geql \gu\})-\{\gz\gle a^{**}\})\cup \{\gz\gle \gu\})-\{\gz\gle \gu\}=
 ((((S''-\{\gu\gleq a^{**}\})\cup \{a^{**}\geql \gu\})-\{\gz\gle a^{**}\})-\{\gz\gle \gu\})\cup (\{\gz\gle \gu\}-\{\gz\gle \gu\})=
 (((S''-\{\gu\gleq a^{**}\})\cup \{a^{**}\geql \gu\})-\{\gz\gle \gu\})-\{\gz\gle a^{**}\}=
 (((S''-\{\gu\gleq a^{**}\})-\{\gz\gle \gu\})\cup (\{a^{**}\geql \gu\}-\{\gz\gle \gu\}))-\{\gz\gle a^{**}\}=
 ((S''-\{\gu\gleq a^{**}\})\cup \{a^{**}\geql \gu\})-\{\gz\gle a^{**}\}=
 ((S''-\{\gu\gleq a^{**}\})-\{\gz\gle a^{**}\})\cup (\{a^{**}\geql \gu\}-\{\gz\gle a^{**}\})=
 ((S''-\{\gu\gleq a^{**}\})-\{\gz\gle a^{**}\})\cup \{a^{**}\geql \gu\}=
 ((((S-\{C^*\})\cup \{\gu\gleq a^{**}\})-\{\gu\gleq a^{**}\})-\{\gz\gle a^{**}\})\cup \{a^{**}\geql \gu\}=
 ((((S-\{C^*\})-\{\gu\gleq a^{**}\})\cup (\{\gu\gleq a^{**}\}-\{\gu\gleq a^{**}\}))-\{\gz\gle a^{**}\})\cup \{a^{**}\geql \gu\}=
 ((S-\{C^*\})-\{\gz\gle a^{**}\})\cup \{a^{**}\geql \gu\}=(S-\{C^*,\gz\gle a^{**}\})\cup \{a^{**}\geql \gu\}$.
We put $S'''=(S-\{C^*,\gz\gle a^{**}\})\cup \{a^{**}\geql \gu\}\subseteq_{\mc F} \mi{OrdPropCl}$.
We get from Cases 2.2.2.2.1.5.2 and 2.2.2.4.1.5.2 that 
there exists a finite linear {\it DPLL}-tree $\mi{Tree}'''$ with the root $S'''$ constructed using Rules (\cref{ceq4hr1111111})--(\cref{ceq4hr4}), (\cref{ceq4hr8}) satisfying
for its only leaf $S''''$ that either $\square\in S''''$, or $S''''\subseteq_{\mc F} \mi{OrdPropCl}$ is $\gz$-guarded, $\mi{atoms}(S'''')\subseteq \mi{atoms}(S''')-\{a^*\}\subseteq \mi{atoms}(S)-\{a^*\}$.
We put
\begin{equation} \notag
\mi{Tree}=\begin{array}[c]{c}
          S \\[0.4mm]
          \hline \\[-3.8mm]
          \mi{Tree}' \\[0.4mm]
          \hline \\[-3.8mm]
          (S''-\{\gu\gleq a^{**}\})\cup \{a^{**}\geql \gu\} \\[0.4mm]
          \hline \\[-3.8mm]
          (((S''-\{\gu\gleq a^{**}\})\cup \{a^{**}\geql \gu\})-\{\gz\gle a^{**}\})\cup \{\gz\gle \gu\} \\[0.4mm]
          \hline \\[-3.8mm]
          \mi{Tree}'''.    
          \end{array}    
\end{equation}
Hence, $\mi{Tree}$ is a finite linear {\it DPLL}-tree with the root $S$ constructed using Rules (\cref{ceq4hr1111111})--(\cref{ceq4hr4}), (\cref{ceq4hr8}) such that
for its only leaf $S''''$, either $\square\in S''''$, or $S''''\subseteq_{\mc F} \mi{OrdPropCl}$ is $\gz$-guarded, $\mi{atoms}(S'''')\subseteq \mi{atoms}(S)-\{a^*\}$;
(ii) holds.

Cases 2.2.2.2.1.6.3 and 2.2.2.4.1.6.3:
$\mi{guards}(S,a^{**})=\{\gz\gle a^{**},a^{**}\gle \gu\}$.
Then $\mi{guards}(S'')\supseteq \mi{guards}(S'',a^{**})=\mi{guards}(S,a^{**})\cup \{C^\natural\}=\{\gz\gle a^{**},a^{**}\gle \gu,\gu\gleq a^{**}\}$,
$\gu\gleq a^{**}\in \mi{guards}(S'')$, applying Rule (\cref{ceq4hr11111111}) to $S''$ and $\gu\gleq a^{**}$, we derive
\begin{equation} \notag
\dfrac{S''}
      {(S''-\{\gu\gleq a^{**}\})\cup \{a^{**}\geql \gu\}};
\end{equation}
$a^{**}\in \mi{atoms}((S''-\{\gu\gleq a^{**}\})\cup \{a^{**}\geql \gu\})$,
$(S''-\{\gu\gleq a^{**}\})\cup \{a^{**}\geql \gu\}\supseteq \mi{guards}((S''-\{\gu\gleq a^{**}\})\cup \{a^{**}\geql \gu\})\supseteq 
 \mi{guards}((S''-\{\gu\gleq a^{**}\})\cup \{a^{**}\geql \gu\},a^{**})=((S''-\{\gu\gleq a^{**}\})\cup \{a^{**}\geql \gu\})\cap \mi{guards}(a^{**})=
 ((S''-\{\gu\gleq a^{**}\})\cap \mi{guards}(a^{**}))\cup (\{a^{**}\geql \gu\}\cap \mi{guards}(a^{**}))=((S''\cap \mi{guards}(a^{**}))-\{\gu\gleq a^{**}\})\cup \{a^{**}\geql \gu\}=
 (\mi{guards}(S'',a^{**})-\{\gu\gleq a^{**}\})\cup \{a^{**}\geql \gu\}=(\{\gz\gle a^{**},a^{**}\gle \gu,\gu\gleq a^{**}\}-\{\gu\gleq a^{**}\})\cup \{a^{**}\geql \gu\}\supseteq 
 \{a^{**}\gle \gu,a^{**}\geql \gu\}$,
$a^{**}\geql \gu\in \mi{guards}((S''-\{\gu\gleq a^{**}\})\cup \{a^{**}\geql \gu\})$, $a^{**}\gle \gu\in (S''-\{\gu\gleq a^{**}\})\cup \{a^{**}\geql \gu\}$,
$a^{**}\in \mi{atoms}(a^{**}\gle \gu)$, $a^{**}\geql \gu\neq a^{**}\gle \gu$, $\mi{simplify}(a^{**}\gle \gu,a^{**},\gu)=\gu\gle \gu$,
applying Rule (\cref{ceq4hr4}) to $(S''-\{\gu\gleq a^{**}\})\cup \{a^{**}\geql \gu\}$, $a^{**}\geql \gu$, and $a^{**}\gle \gu$, we derive
\begin{equation} \notag
\dfrac{(S''-\{\gu\gleq a^{**}\})\cup \{a^{**}\geql \gu\}}
      {(((S''-\{\gu\gleq a^{**}\})\cup \{a^{**}\geql \gu\})-\{a^{**}\gle \gu\})\cup \{\gu\gle \gu\}};
\end{equation}
$\gu\gle \gu\in (((S''-\{\gu\gleq a^{**}\})\cup \{a^{**}\geql \gu\})-\{a^{**}\gle \gu\})\cup \{\gu\gle \gu\}$;
$\gu\gle \gu\in \mi{OrdPropLit}$ is a contradiction;
$\gu\gle \gu$ is not a guard;
$\gu\gle \gu\not\in \mi{guards}((((S''-\{\gu\gleq a^{**}\})\cup \{a^{**}\geql \gu\})-\{a^{**}\gle \gu\})\cup \{\gu\gle \gu\})$,
$\gu\gle \gu\in ((((S''-\{\gu\gleq a^{**}\})\cup \{a^{**}\geql \gu\})-\{a^{**}\gle \gu\})\cup \{\gu\gle \gu\})-\mi{guards}((((S''-\{\gu\gleq a^{**}\})\cup \{a^{**}\geql \gu\})-\{a^{**}\gle \gu\})\cup \{\gu\gle \gu\})$,
applying Rule (\cref{ceq4hr2}) to $(((S''-\{\gu\gleq a^{**}\})\cup \{a^{**}\geql \gu\})-\{a^{**}\gle \gu\})\cup \{\gu\gle \gu\}$ and $\gu\gle \gu$, we derive
\begin{equation} \notag
\dfrac{(((S''-\{\gu\gleq a^{**}\})\cup \{a^{**}\geql \gu\})-\{a^{**}\gle \gu\})\cup \{\gu\gle \gu\}}
      {(((((S''-\{\gu\gleq a^{**}\})\cup \{a^{**}\geql \gu\})-\{a^{**}\gle \gu\})\cup \{\gu\gle \gu\})-\{\gu\gle \gu\})\cup \{\square\}}.
\end{equation}
We put
\begin{equation} \notag
\mi{Tree}=\begin{array}[c]{c}
          S \\[0.4mm]
          \hline \\[-3.8mm]
          \mi{Tree}' \\[0.4mm]
          \hline \\[-3.8mm]
          (S''-\{\gu\gleq a^{**}\})\cup \{a^{**}\geql \gu\} \\[0.4mm]
          \hline \\[-3.8mm]
          (((S''-\{\gu\gleq a^{**}\})\cup \{a^{**}\geql \gu\})-\{a^{**}\gle \gu\})\cup \{\gu\gle \gu\} \\[0.4mm]
          \hline \\[-3.8mm]
          S'''=(((((S''-\{\gu\gleq a^{**}\})\cup \{a^{**}\geql \gu\})-\{a^{**}\gle \gu\})\cup \{\gu\gle \gu\})- \\
          \hfill \{\gu\gle \gu\})\cup \{\square\}.
          \end{array} 
\end{equation}
Hence, $\mi{Tree}$ is a finite linear {\it DPLL}-tree with the root $S$ constructed using Rules (\cref{ceq4hr1111111})--(\cref{ceq4hr4}), (\cref{ceq4hr8}) such that
for its only leaf $S'''$, $\square\in S'''$;
(ii) holds.

Cases 2.2.2.2.1.6.4 and 2.2.2.4.1.6.4:
$\mi{guards}(S,a^{**})=\{a^{**}\geql \gu\}$.
Then $C^\natural=\gu\gleq a^{**}\not\in S$,
$S''=(S-\{C^*\})\cup \{C^\natural\}=(S-\{C^*\})\cup \{\gu\gleq a^{**}\}$;
$S''\supseteq \mi{guards}(S'')\supseteq \mi{guards}(S'',a^{**})=\mi{guards}(S,a^{**})\cup \{C^\natural\}=\{a^{**}\geql \gu,\gu\gleq a^{**}\}$,
$\gu\gleq a^{**}\in \mi{guards}(S'')$, applying Rule (\cref{ceq4hr11111111}) to $S''$ and $\gu\gleq a^{**}$, we derive
\begin{equation} \notag
\dfrac{S''}
      {(S''-\{\gu\gleq a^{**}\})\cup \{a^{**}\geql \gu\}}.
\end{equation}
Hence, $a^{**}\geql \gu\in S''$, $a^{**}\geql \gu\neq \gu\gleq a^{**}$, $a^{**}\geql \gu\in S''-\{\gu\gleq a^{**}\}$, $\gu\gleq a^{**}\not\in S$,
$(S''-\{\gu\gleq a^{**}\})\cup \{a^{**}\geql \gu\}=S''-\{\gu\gleq a^{**}\}=((S-\{C^*\})\cup \{\gu\gleq a^{**}\})-\{\gu\gleq a^{**}\}=
 ((S-\{C^*\})-\{\gu\gleq a^{**}\})\cup (\{\gu\gleq a^{**}\}-\{\gu\gleq a^{**}\})=S-\{C^*\}$,
$a^*\geql \gu\in S-\{C^*\}$, $a^*\in \mi{atoms}(a^*\geql \gu)\subseteq \mi{atoms}(S-\{C^*\})\subseteq \mi{atoms}(S)$.
We get from Case 2.2.1 for $S-\{C^*\}$ that there exists a finite linear {\it DPLL}-tree $\mi{Tree}''$ with the root $S-\{C^*\}$ constructed using Rules (\cref{ceq4hr1111111})--(\cref{ceq4hr4}), (\cref{ceq4hr8}) satisfying
for its only leaf $S'''$ that either $\square\in S'''$, or $S'''\subseteq_{\mc F} \mi{OrdPropCl}$ is $\gz$-guarded, $\mi{atoms}(S''')\subseteq \mi{atoms}(S-\{C^*\})-\{a^*\}=\mi{atoms}(S)-\{a^*\}$.
We put
\begin{equation} \notag
\mi{Tree}=\begin{array}[c]{c}
          S \\[0.4mm]
          \hline \\[-3.8mm]
          \mi{Tree}' \\[0.4mm]
          \hline \\[-3.8mm]
          \mi{Tree}''.
          \end{array}
\end{equation}
Hence, $\mi{Tree}$ is a finite linear {\it DPLL}-tree with the root $S$ constructed using Rules (\cref{ceq4hr1111111})--(\cref{ceq4hr4}), (\cref{ceq4hr8}) such that
for its only leaf $S'''$, either $\square\in S'''$, or $S'''\subseteq_{\mc F} \mi{OrdPropCl}$ is $\gz$-guarded, $\mi{atoms}(S''')\subseteq \mi{atoms}(S)-\{a^*\}$;
(ii) holds.

Cases 2.2.2.2.2 and 2.2.2.4.2:
For all $a\in \mi{atoms}(S'')-\{a^*\}$, $C^\natural\not\in \mi{guards}(a)$.
Then, for all $a\in \mi{atoms}(C^\natural)\subseteq \mi{atoms}(S'')-\{a^*\}$, $C^\natural\not\in \mi{guards}(a)$;
$C^\natural$ is not a guard;
$C^*\not\in \mi{guards}(S)$,
$\mi{guards}(S'')=\{C \,|\, C\in (S-\{C^*\})\cup \{C^\natural\}\ \text{\it is a guard}\}=\{C \,|\, C\in S-\{C^*\}\ \text{\it is a guard}\}=
                  \mi{guards}(S)-\{C^*\}=\{C \,|\, C\in S\ \text{\it is a guard}\}-\{C^*\}=\mi{guards}(S)$;
we have that $S$ is $\gz$-guarded;
$\mi{atoms}(S''-\mi{guards}(S''))\subseteq \mi{atoms}(S'')\subseteq \mi{atoms}(S)$,
$\mi{atoms}(S'')=\mi{atoms}((S''-\mi{guards}(S''))\cup \mi{guards}(S''))=\mi{atoms}(S''-\mi{guards}(S''))\cup \mi{atoms}(\mi{guards}(S''))=
                 \mi{atoms}(S''-\mi{guards}(S''))\cup \mi{atoms}(\mi{guards}(S))=\mi{atoms}(S''-\mi{guards}(S''))\cup \mi{atoms}(S)=\mi{atoms}(S)$;
$S''$ is $\gz$-guarded; 
$a^*\in \mi{atoms}(S'')$, $\mi{guards}(S'',a^*)=\{a^*\geql \gu\}$,
$a^*\not\in \mi{atoms}(C^\natural)$, $C^*\in \mi{restricted}(S)$,
$\mi{restricted}(S'')=\{C \,|\, C\in (S-\{C^*\})\cup \{C^\natural\}, a^*\in \mi{atoms}(C), C\neq a^*\geql \gu\}=\{C \,|\, C\in S-\{C^*\}, a^*\in \mi{atoms}(C), C\neq a^*\geql \gu\}=
                      \mi{restricted}(S)-\{C^*\}=\{C \,|\, C\in S, a^*\in \mi{atoms}(C), C\neq a^*\geql \gu\}-\{C^*\}\subset \mi{restricted}(S)$;
by the induction hypothesis for $S''$, there exists a finite linear {\it DPLL}-tree $\mi{Tree}''$ with the root $S''$ constructed using Rules (\cref{ceq4hr1111111})--(\cref{ceq4hr4}), (\cref{ceq4hr8}) satisfying
for its only leaf $S'''$ that either $\square\in S'''$, or $S'''\subseteq_{\mc F} \mi{OrdPropCl}$ is $\gz$-guarded, $\mi{atoms}(S''')\subseteq \mi{atoms}(S'')-\{a^*\}=\mi{atoms}(S)-\{a^*\}$.
We put
\begin{equation} \notag
\mi{Tree}=\begin{array}[c]{c}
          S \\[0.4mm]
          \hline \\[-3.8mm]
          \mi{Tree}' \\[0.4mm]
          \hline \\[-3.8mm]
          \mi{Tree}''.
          \end{array}
\end{equation}
Hence, $\mi{Tree}$ is a finite linear {\it DPLL}-tree with the root $S$ constructed using Rules (\cref{ceq4hr1111111})--(\cref{ceq4hr4}), (\cref{ceq4hr8}) such that
for its only leaf $S'''$, either $\square\in S'''$, or $S'''\subseteq_{\mc F} \mi{OrdPropCl}$ is $\gz$-guarded, $\mi{atoms}(S''')\subseteq \mi{atoms}(S)-\{a^*\}$;
(ii) holds.

So, in both Cases 2.1 and 2.2, (ii) holds.
The induction is completed.
Thus, (ii) holds. 
%
%
%
\end{proof}

\subsection{Full proof of Lemma \ref{le66}}
\label{S7.8b}

\begin{proof}
Let $S^F\subseteq \mi{OrdPropCl}$.
We define a measure operator $\mi{restricted}(S^F)=\{C \,|\, C\in S^F, a^*\in \mi{atoms}(C), C\neq a^*\geql \gu\}$.
We proceed by induction on $\mi{restricted}(S)\subseteq S\subseteq_{\mc F} \mi{OrdPropCl}$.

Case 1 (the base case):
$\mi{restricted}(S)=\emptyset$.
We have that $S$ is semi-positively guarded.
Then $S\supseteq \mi{guards}(S)\supseteq \mi{guards}(S,a^*)=\{a^*\geql \gu\}$, $a^*\geql \gu\in \mi{guards}(S)$, $a^*\geql \gu\in S$;
for all $C\in S$ satisfying $C\neq a^*\geql \gu$, $a^*\not\in \mi{atoms}(C)$;
$a^*\not\in \mi{atoms}(S-\{a^*\geql \gu\})$, applying Rule (\cref{ceq4hr8}) to $a^*\geql \gu$, we derive
\begin{equation} \notag
\dfrac{S}
      {S-\{a^*\geql \gu\}}.
\end{equation}
We put $S'=S-\{a^*\geql \gu\}\subseteq_{\mc F} \mi{OrdPropCl}$.
We get that
$a^*\in \mi{atoms}(S)$, $\mi{atoms}(a^*\geql \gu)=\{a^*\}$,
$a^*\not\in \mi{atoms}(S')=\mi{atoms}(S-\{a^*\geql \gu\})=\mi{atoms}(S)-\{a^*\}$;
$S$ is simplified;
$S'=S-\{a^*\geql \gu\}\subseteq S$ is simplified;
for all $a\in \mi{atoms}(S')=\mi{atoms}(S)-\{a^*\}$,
$a$ is semi-positively guarded in $S$;
$a\neq a^*$, 
$\mi{guards}(a)\cap \{a^*\geql \gu\}=\mi{guards}(a)\cap \mi{guards}(a^*)=\emptyset$,
$\mi{guards}(S',a)=(S-\{a^*\geql \gu\})\cap \mi{guards}(a)=(S\cap \mi{guards}(a))-\{a^*\geql \gu\}=S\cap \mi{guards}(a)=\mi{guards}(S,a)$;
$a$ is semi-positively guarded in $S'$;
$S'$ is semi-positively guarded.
We put
\begin{equation} \notag
\mi{Tree}=\dfrac{S} 
                {S'}.
\end{equation}
Hence, $\mi{Tree}$ is a finite linear {\it DPLL}-tree with the root $S$ constructed using Rules (\cref{ceq4hr1111111})--(\cref{ceq4hr4}), (\cref{ceq4hr8}) such that
for its only leaf $S'$, $S'\subseteq_{\mc F} \mi{OrdPropCl}$ is semi-positively guarded, $\mi{atoms}(S')=\mi{atoms}(S)-\{a^*\}$.

Case 2 (the induction case):
$\emptyset\neq \mi{restricted}(S)\subseteq_{\mc F} \mi{OrdPropCl}$.
We have that $S$ is semi-positively guarded.
Then $S\supseteq \mi{guards}(S)\supseteq \mi{guards}(S,a^*)=\{a^*\geql \gu\}$, $a^*\geql \gu\in \mi{guards}(S)$, $a^*\geql \gu\in S$;
there exists $C^*\in \mi{restricted}(S)\subseteq S$ satisfying $a^*\in \mi{atoms}(C^*)$, $a^*\geql \gu\neq C^*$;
applying Rule (\cref{ceq4hr4}) to $a^*\geql \gu$ and $C^*$, we derive
\begin{equation} \notag
\dfrac{S}
      {(S-\{C^*\})\cup \{\mi{simplify}(C^*,a^*,\gu)\}}.
\end{equation}
We put $S'=(S-\{C^*\})\cup \{\mi{simplify}(C^*,a^*,\gu)\}\subseteq_{\mc F} \mi{OrdPropCl}$.
We get that
$a^*\geql \gu\in S$, $a^*\geql \gu\neq C^*$, $a^*\geql \gu\in S-\{C^*\}\subseteq S'$, $\mi{atoms}(\mi{simplify}(C^*,a^*,\gu))\subseteq \mi{atoms}(C^*)$, $C^*\in S$,
$a^*\in \mi{atoms}(a^*\geql \gu)\subseteq \mi{atoms}(S')=\mi{atoms}((S-\{C^*\})\cup \{\mi{simplify}(C^*,a^*,\gu)\})=\mi{atoms}(S-\{C^*\})\cup \mi{atoms}(\mi{simplify}(C^*,a^*,\gu))\subseteq 
                                                         \mi{atoms}(S-\{C^*\})\cup \mi{atoms}(C^*)=\mi{atoms}((S-\{C^*\})\cup \{C^*\})=\mi{atoms}(S)$;
$C^*\not\in \mi{guards}(S,a^*)=S\cap \mi{guards}(a^*)=\{a^*\geql \gu\}$, $C^*\not\in \mi{guards}(a^*)$;
for all $a\in \mi{atoms}(S)-\{a^*\}$,
$a\neq a^*$;
for all $C\in \mi{guards}(a)$, $a^*\in \mi{atoms}(C^*)$, $a^*\not\in \mi{atoms}(C)=\{a\}$, $\mi{atoms}(C^*)\neq \mi{atoms}(C)$, $C^*\neq C$;
$C^*\not\in \mi{guards}(a)$;
for all $a\in \mi{atoms}(C^*)\subseteq \mi{atoms}(S)$, $C^*\not\in \mi{guards}(a)$;
$C^*$ is not a guard; 
$C^*\not\in \mi{guards}(S)$.
We get two cases for $\mi{simplify}(C^*,a^*,\gu)$.

Case 2.1:
$\mi{simplify}(C^*,a^*,\gu)\in S$.
Then $a^*\not\in \mi{atoms}(\mi{simplify}(C^*,a^*,\gu))$, $a^*\in \mi{atoms}(C^*)$, $\mi{atoms}(\mi{simplify}(C^*,a^*,\gu))\neq \mi{atoms}(C^*)$, $\mi{simplify}(C^*,a^*,\gu)\neq C^*$,
$\mi{simplify}(C^*,a^*,\gu)\in S-\{C^*\}$,
$S'=(S-\{C^*\})\cup \{\mi{simplify}(C^*,a^*,\gu)\}=S-\{C^*\}$;
$S$ is simplified;
$S'=S-\{C^*\}\subseteq S$ is simplified;
$C^*\not\in \mi{guards}(S)$,
$\mi{guards}(S')=\{C \,|\, C\in S-\{C^*\}\ \text{\it is a guard}\}=\mi{guards}(S)-\{C^*\}=\{C \,|\, C\in S\ \text{\it is a guard}\}-\{C^*\}=\mi{guards}(S)$;
we have that $S$ is semi-positively guarded;
$\mi{atoms}(S'-\mi{guards}(S'))\subseteq \mi{atoms}(S')\subseteq \mi{atoms}(S)$,
$\mi{atoms}(S')=\mi{atoms}((S'-\mi{guards}(S'))\cup \mi{guards}(S'))=\mi{atoms}(S'-\mi{guards}(S'))\cup \mi{atoms}(\mi{guards}(S'))=
                \mi{atoms}(S'-\mi{guards}(S'))\cup \mi{atoms}(\mi{guards}(S))=\mi{atoms}(S'-\mi{guards}(S'))\cup \mi{atoms}(S)=\mi{atoms}(S)$;
$S'$ is semi-positively guarded;
$a^*\in \mi{atoms}(S')$, $\mi{guards}(S',a^*)=\mi{guards}(S,a^*)=\{a^*\geql \gu\}$,
$C^*\in \mi{restricted}(S)$,
$\mi{restricted}(S')=\{C \,|\, C\in S-\{C^*\}, a^*\in \mi{atoms}(C), C\neq a^*\geql \gu\}=
                     \mi{restricted}(S)-\{C^*\}=\{C \,|\, C\in S, a^*\in \mi{atoms}(C), C\neq a^*\geql \gu\}-\{C^*\}\subset \mi{restricted}(S)$;
by the induction hypothesis for $S'$, there exists a finite linear {\it DPLL}-tree $\mi{Tree}'$ with the root $S'$ constructed using Rules (\cref{ceq4hr1111111})--(\cref{ceq4hr4}), (\cref{ceq4hr8}) satisfying
for its only leaf $S''$ that either $\square\in S''$, or $S''\subseteq_{\mc F} \mi{OrdPropCl}$ is semi-positively guarded, $\mi{atoms}(S'')\subseteq \mi{atoms}(S')-\{a^*\}=\mi{atoms}(S)-\{a^*\}$.
We put
\begin{equation} \notag
\mi{Tree}=\dfrac{S}
                {\mi{Tree}'}.
\end{equation}
Hence, $\mi{Tree}$ is a finite linear {\it DPLL}-tree with the root $S$ constructed using Rules (\cref{ceq4hr1111111})--(\cref{ceq4hr4}), (\cref{ceq4hr8}) such that
for its only leaf $S''$, either $\square\in S''$, or $S''\subseteq_{\mc F} \mi{OrdPropCl}$ is semi-positively guarded, $\mi{atoms}(S'')\subseteq \mi{atoms}(S)-\{a^*\}$.

Case 2.2:
$\mi{simplify}(C^*,a^*,\gu)\not\in S$.
Then $\mi{simplify}(C^*,a^*,\gu)\in S'$; 
by Lemma \ref{le555} for $S'$ and $\mi{simplify}(C^*,a^*,\gu)$, there exists a finite linear {\it DPLL}-tree $\mi{Tree}'$ with the root $S'$ constructed using Rules (\cref{ceq4hr2}) and (\cref{ceq4hr22}) satisfying 
for its only leaf $S''$ that either $\square\in S''$, or $\square\not\in S''\subseteq_{\mc F} \mi{OrdPropCl}$ and exactly one of the following points holds.
\begin{enumerate}[\rm (a)]
\item
$S''=S'$, $\mi{simplify}(C^*,a^*,\gu)\neq \square$ does not contain contradictions and tautologies;
\item
$S''=S'-\{\mi{simplify}(C^*,a^*,\gu)\}$;
\item
there exists $C^{**}\in \mi{OrdPropCl}$ satisfying that $S''=(S'-\{\mi{simplify}(C^*,a^*,\gu)\})\cup \{C^{**}\}$, $C^{**}\not\in S'$, 
$\square\neq C^{**}\subset \mi{simplify}(C^*,a^*,\gu)$ does not contain contradictions and tautologies.
\end{enumerate}
We get four cases for $S''$.

Case 2.2.1:
$\square\in S''$.
We put
\begin{equation} \notag
\mi{Tree}=\dfrac{S}
                {\mi{Tree}'}.
\end{equation}
Hence, $\mi{Tree}$ is a finite linear {\it DPLL}-tree with the root $S$ constructed using Rules (\cref{ceq4hr1111111})--(\cref{ceq4hr4}), (\cref{ceq4hr8}) such that
for its only leaf $S''$, $\square\in S''$.

Case 2.2.2:
$S''=S'$ and $\mi{simplify}(C^*,a^*,\gu)\neq \square$ does not contain contradictions and tautologies.

Case 2.2.3:
$S''=S'-\{\mi{simplify}(C^*,a^*,\gu)\}$.
Then $\mi{simplify}(C^*,a^*,\gu)\not\in S$,
$S''=S'-\{\mi{simplify}(C^*,a^*,\gu)\}=((S-\{C^*\})\cup \{\mi{simplify}(C^*,a^*,\gu)\})-                                                                                                   \linebreak[4]
                                                                                        \{\mi{simplify}(C^*,a^*,\gu)\}=
     ((S-\{C^*\})-\{\mi{simplify}(C^*,a^*,\gu)\})\cup                                                                                                                                      \linebreak[4]
                                                      (\{\mi{simplify}(C^*,a^*,\gu)\}-\{\mi{simplify}(C^*,a^*,\gu)\})=S-\{C^*\}$,
$a^*\geql \gu\in S''=S-\{C^*\}$, $a^*\in \mi{atoms}(a^*\geql \gu)\subseteq \mi{atoms}(S'')=\mi{atoms}(S-\{C^*\})\subseteq \mi{atoms}(S)$.
We get from Case 2.1 for $S''$ that there exists a finite linear {\it DPLL}-tree $\mi{Tree}''$ with the root $S''$ constructed using Rules (\cref{ceq4hr1111111})--(\cref{ceq4hr4}), (\cref{ceq4hr8}) satisfying
for its only leaf $S'''$ that either $\square\in S'''$, or $S'''\subseteq_{\mc F} \mi{OrdPropCl}$ is semi-positively guarded, $\mi{atoms}(S''')\subseteq \mi{atoms}(S'')-\{a^*\}=\mi{atoms}(S)-\{a^*\}$.
We put
\begin{equation} \notag
\mi{Tree}=\begin{array}[c]{c}
          S \\[0.4mm]
          \hline \\[-3.8mm]
          \mi{Tree}' \\[0.4mm]
          \hline \\[-3.8mm]
          \mi{Tree}''.    
          \end{array}    
\end{equation}
Hence, $\mi{Tree}$ is a finite linear {\it DPLL}-tree with the root $S$ constructed using Rules (\cref{ceq4hr1111111})--(\cref{ceq4hr4}), (\cref{ceq4hr8}) such that
for its only leaf $S'''$, either $\square\in S'''$, or $S'''\subseteq_{\mc F} \mi{OrdPropCl}$ is semi-positively guarded, $\mi{atoms}(S''')\subseteq \mi{atoms}(S)-\{a^*\}$.

Case 2.2.4:
There exists $C^{**}\in \mi{OrdPropCl}$ such that $S''=(S'-\{\mi{simplify}(C^*,a^*,\gu)\})\cup \{C^{**}\}$, $C^{**}\not\in S'$, 
$\square\neq C^{**}\subset \mi{simplify}(C^*,a^*,\gu)$ does not contain contradictions and tautologies.

In Cases 2.2.2 and 2.2.4, we put 
\begin{alignat*}{1}
C^\natural &= \left\{\begin{array}{ll}
                     \mi{simplify}(C^*,a^*,\gu) &\ \text{\it in {\rm Case 2.2.2}}, \\[1mm]
                     C^{**}                     &\ \text{\it in {\rm Case 2.2.4}};
                     \end{array}
              \right. \\[1mm]
           &\in \mi{OrdPropCl}.
\end{alignat*}
Then, in Case 2.2.2, 
$C^\natural=\mi{simplify}(C^*,a^*,\gu)\not\in S$, $S''=S'=(S-\{C^*\})\cup \{\mi{simplify}(C^*,a^*,\gu)\}=(S-\{C^*\})\cup \{C^\natural\}$;
$C^\natural=\mi{simplify}(C^*,a^*,\gu)\neq \square$ does not contain contradictions and tautologies;
$a^*\not\in \mi{atoms}(C^\natural)=\mi{atoms}(\mi{simplify}(C^*,a^*,\gu))\subseteq \mi{atoms}(C^*)$;
in Case 2.2.4, 
$C^{**}\subset \mi{simplify}(C^*,a^*,\gu)$, 
$\mi{atoms}(C^\natural)=\mi{atoms}(C^{**})\subseteq \mi{atoms}(\mi{simplify}(C^*,a^*,\gu))\subseteq \mi{atoms}(C^*)$,
$a^*\not\in \mi{atoms}(\mi{simplify}(C^*,a^*,\gu))\supseteq \mi{atoms}(C^\natural)=\mi{atoms}(C^{**})$, $a^*\in \mi{atoms}(C^*)$, $\mi{atoms}(C^{**})\neq \mi{atoms}(C^*)$, $C^{**}\neq C^*$,
$C^{**}\not\in S'\supseteq S-\{C^*\}$, $C^\natural=C^{**}\not\in S$,
$S''=(S'-\{\mi{simplify}(C^*,a^*,\gu)\})\cup \{C^{**}\}=(((S-\{C^*\})\cup \{\mi{simplify}(C^*,a^*,\gu)\})-\{\mi{simplify}(C^*,a^*,\gu)\})\cup \{C^{**}\}=
     ((S-\{C^*\})-\{\mi{simplify}(C^*,a^*,\gu)\})\cup (\{\mi{simplify}(C^*,a^*,\gu)\}-\{\mi{simplify}(C^*,a^*,\gu)\})\cup \{C^{**}\}=(S-\{C^*\})\cup \{C^{**}\}=(S-\{C^*\})\cup \{C^\natural\}$;
$C^\natural=C^{**}\neq \square$ does not contain contradictions and tautologies;
in both Cases 2.2.2 and 2.2.4,
$C^\natural\not\in S$, $S''=(S-\{C^*\})\cup \{C^\natural\}$;
$C^\natural\neq \square$ does not contain contradictions and tautologies;
$a^*\not\in \mi{atoms}(C^\natural)\subseteq \mi{atoms}(C^*)$;
$a^*\geql \gu\in S-\{C^*\}\subseteq S''$, $C^*\in S$,
$a^*\in \mi{atoms}(a^*\geql \gu)\subseteq \mi{atoms}(S'')=\mi{atoms}((S-\{C^*\})\cup \{C^\natural\})=\mi{atoms}(S-\{C^*\})\cup \mi{atoms}(C^\natural)\subseteq 
                                                          \mi{atoms}(S-\{C^*\})\cup \mi{atoms}(C^*)=\mi{atoms}((S-\{C^*\})\cup \{C^*\})=\mi{atoms}(S)$,
$\mi{atoms}(C^\natural)\subseteq \mi{atoms}(S'')$, $\mi{atoms}(C^\natural)\subseteq \mi{atoms}(S'')-\{a^*\}$;
$S$ is simplified;
$S-\{C^*\}\subseteq S$ is simplified;
$S''=(S-\{C^*\})\cup \{C^\natural\}$ is simplified;
for all $C\in \mi{guards}(a^*)$, $a^*\in \mi{atoms}(C)=\{a^*\}$, $\mi{atoms}(C^\natural)\neq \mi{atoms}(C)$, $C^\natural\neq C$;
$C^\natural, C^*\not\in \mi{guards}(a^*)$,
$\mi{guards}(S'',a^*)=((S-\{C^*\})\cup \{C^\natural\})\cap \mi{guards}(a^*)=((S-\{C^*\})\cap \mi{guards}(a^*))\cup (\{C^\natural\}\cap \mi{guards}(a^*))=
                      (S\cap \mi{guards}(a^*))-\{C^*\}=S\cap \mi{guards}(a^*)=\mi{guards}(S,a^*)=\{a^*\geql \gu\}$;
$a^*$ is semi-positively guarded in $S''$.
We get two cases for $C^\natural$.

Cases 2.2.2.1 and 2.2.4.1:
There exists $a^{**}\in \mi{atoms}(S'')-\{a^*\}\subseteq \mi{atoms}(S)-\{a^*\}$ such that $C^\natural\in \mi{guards}(a^{**})$.
Then $a^{**}\neq a^*$, $C^*\not\in \mi{guards}(a^{**})$,
$\mi{guards}(S'',a^{**})=((S-\{C^*\})\cup \{C^\natural\})\cap \mi{guards}(a^{**})=((S-\{C^*\})\cap \mi{guards}(a^{**}))\cup (\{C^\natural\}\cap \mi{guards}(a^{**}))=
                         ((S\cap \mi{guards}(a^{**}))-\{C^*\})\cup \{C^\natural\}=(S\cap \mi{guards}(a^{**}))\cup \{C^\natural\}=\mi{guards}(S,a^{**})\cup \{C^\natural\}$;
$a^{**}$ is semi-positively guarded in $S$.
We get six cases for $C^\natural$.

Cases 2.2.2.1.1 and 2.2.4.1.1:
$C^\natural=a^{**}\geql \gz$.
We have that $a^{**}$ is semi-positively guarded in $S$.
We get two cases for $\mi{guards}(S,a^{**})$.

Cases 2.2.2.1.1.1 and 2.2.4.1.1.1:
Either $\mi{guards}(S,a^{**})=\{\gz\gle a^{**}\}$ or $\mi{guards}(S,a^{**})=\{\gz\gle a^{**},a^{**}\gle \gu\}$.
Then $S''\supseteq \mi{guards}(S'')\supseteq \mi{guards}(S'',a^{**})=\mi{guards}(S,a^{**})\cup \{C^\natural\}=\mi{guards}(S,a^{**})\cup \{a^{**}\geql \gz\}\supseteq \{a^{**}\geql \gz,\gz\gle a^{**}\}$,
$a^{**}\geql \gz\in \mi{guards}(S'')$, $\gz\gle a^{**}\in S''$,
$a^{**}\in \mi{atoms}(\gz\gle a^{**})$, $a^{**}\geql \gz\neq \gz\gle a^{**}$, $\mi{simplify}(\gz\gle a^{**},a^{**},\gz)=\gz\gle \gz$,
applying Rule (\cref{ceq4hr3}) to $S''$, $a^{**}\geql \gz$, and $\gz\gle a^{**}$, we derive
\begin{equation} \notag
\dfrac{S''}
      {(S''-\{\gz\gle a^{**}\})\cup \{\gz\gle \gz\}};
\end{equation}
$\gz\gle \gz\in (S''-\{\gz\gle a^{**}\})\cup \{\gz\gle \gz\}$;
$\gz\gle \gz\in \mi{OrdPropLit}$ is a contradiction;
$\gz\gle \gz$ is not a guard;
$\gz\gle \gz\not\in \mi{guards}((S''-\{\gz\gle a^{**}\})\cup \{\gz\gle \gz\})$,
$\gz\gle \gz\in ((S''-\{\gz\gle a^{**}\})\cup \{\gz\gle \gz\})-\mi{guards}((S''-\{\gz\gle a^{**}\})\cup \{\gz\gle \gz\})$,
applying Rule (\cref{ceq4hr2}) to $(S''-\{\gz\gle a^{**}\})\cup \{\gz\gle \gz\}$ and $\gz\gle \gz$, we derive
\begin{equation} \notag
\dfrac{(S''-\{\gz\gle a^{**}\})\cup \{\gz\gle \gz\}}
      {(((S''-\{\gz\gle a^{**}\})\cup \{\gz\gle \gz\})-\{\gz\gle \gz\})\cup \{\square\}}.
\end{equation}
We put
\begin{equation} \notag
\mi{Tree}=\begin{array}[c]{c}
          S \\[0.4mm]
          \hline \\[-3.8mm]
          \mi{Tree}' \\[0.4mm]
          \hline \\[-3.8mm]
          (S''-\{\gz\gle a^{**}\})\cup \{\gz\gle \gz\} \\[0.4mm]
          \hline \\[-3.8mm]
          S'''=(((S''-\{\gz\gle a^{**}\})\cup \{\gz\gle \gz\})-\{\gz\gle \gz\})\cup \{\square\}.
          \end{array} 
\end{equation}
Hence, $\mi{Tree}$ is a finite linear {\it DPLL}-tree with the root $S$ constructed using Rules (\cref{ceq4hr1111111})--(\cref{ceq4hr4}), (\cref{ceq4hr8}) such that
for its only leaf $S'''$, $\square\in S'''$.

Cases 2.2.2.1.1.2 and 2.2.4.1.1.2:
$\mi{guards}(S,a^{**})=\{a^{**}\geql \gu\}$.
Then $S''\supseteq \mi{guards}(S'')\supseteq \mi{guards}(S'',a^{**})=\mi{guards}(S,a^{**})\cup \{C^\natural\}=\{a^{**}\geql \gz,a^{**}\geql \gu\}$,
$a^{**}\geql \gz\in \mi{guards}(S'')$, $a^{**}\geql \gu\in S''$,
$a^{**}\in \mi{atoms}(a^{**}\geql \gu)$, $a^{**}\geql \gz\neq a^{**}\geql \gu$, $\mi{simplify}(a^{**}\geql \gu,a^{**},\gz)=\gz\geql \gu$,
applying Rule (\cref{ceq4hr3}) to $S''$, $a^{**}\geql \gz$, and $a^{**}\geql \gu$, we derive
\begin{equation} \notag
\dfrac{S''}
      {(S''-\{a^{**}\geql \gu\})\cup \{\gz\geql \gu\}};
\end{equation}
$\gz\geql \gu\in (S''-\{a^{**}\geql \gu\})\cup \{\gz\geql \gu\}$;
$\gz\geql \gu\in \mi{OrdPropLit}$ is a contradiction;
$\gz\geql \gu$ is not a guard;
$\gz\geql \gu\not\in \mi{guards}((S''-\{a^{**}\geql \gu\})\cup \{\gz\geql \gu\})$,
$\gz\geql \gu\in ((S''-\{a^{**}\geql \gu\})\cup \{\gz\geql \gu\})-\mi{guards}((S''-\{a^{**}\geql \gu\})\cup \{\gz\geql \gu\})$,
applying Rule (\cref{ceq4hr2}) to $(S''-\{a^{**}\geql \gu\})\cup \{\gz\geql \gu\}$ and $\gz\geql \gu$, we derive
\begin{equation} \notag
\dfrac{(S''-\{a^{**}\geql \gu\})\cup \{\gz\geql \gu\}}
      {(((S''-\{a^{**}\geql \gu\})\cup \{\gz\geql \gu\})-\{\gz\geql \gu\})\cup \{\square\}}.
\end{equation}
We put 
\begin{equation} \notag
\mi{Tree}=\begin{array}[c]{c}
          S \\[0.4mm]
          \hline \\[-3.8mm]
          \mi{Tree}' \\[0.4mm]
          \hline \\[-3.8mm]
          (S''-\{a^{**}\geql \gu\})\cup \{\gz\geql \gu\} \\[0.4mm]
          \hline \\[-3.8mm]
          S'''=(((S''-\{a^{**}\geql \gu\})\cup \{\gz\geql \gu\})-\{\gz\geql \gu\})\cup \{\square\}.
          \end{array} 
\end{equation}
Hence, $\mi{Tree}$ is a finite linear {\it DPLL}-tree with the root $S$ constructed using Rules (\cref{ceq4hr1111111})--(\cref{ceq4hr4}), (\cref{ceq4hr8}) such that
for its only leaf $S'''$, $\square\in S'''$.

Cases 2.2.2.1.2 and 2.2.4.1.2:
$C^\natural=a^{**}\gleq \gz$.
We have that $a^{**}$ is semi-positively guarded in $S$.
We get two cases for $\mi{guards}(S,a^{**})$.

Cases 2.2.2.1.2.1 and 2.2.4.1.2.1:
Either $\mi{guards}(S,a^{**})=\{\gz\gle a^{**}\}$ or $\mi{guards}(S,a^{**})=\{\gz\gle a^{**},a^{**}\gle \gu\}$.
Then $\mi{guards}(S'')\supseteq \mi{guards}(S'',a^{**})=\mi{guards}(S,a^{**})\cup \{C^\natural\}=\mi{guards}(S,a^{**})\cup \{a^{**}\gleq \gz\}$,
$a^{**}\gleq \gz\in \mi{guards}(S'')$, applying Rule (\cref{ceq4hr1111111}) to $S''$ and $a^{**}\gleq \gz$, we derive
\begin{equation} \notag
\dfrac{S''}
      {(S''-\{a^{**}\gleq \gz\})\cup \{a^{**}\geql \gz\}};
\end{equation}
$a^{**}\in \mi{atoms}((S''-\{a^{**}\gleq \gz\})\cup \{a^{**}\geql \gz\})$,
$(S''-\{a^{**}\gleq \gz\})\cup \{a^{**}\geql \gz\}\supseteq \mi{guards}((S''-\{a^{**}\gleq \gz\})\cup \{a^{**}\geql \gz\})\supseteq 
 \mi{guards}((S''-\{a^{**}\gleq \gz\})\cup \{a^{**}\geql \gz\},a^{**})=((S''-\{a^{**}\gleq \gz\})\cup \{a^{**}\geql \gz\})\cap \mi{guards}(a^{**})=
 ((S''-\{a^{**}\gleq \gz\})\cap \mi{guards}(a^{**}))\cup (\{a^{**}\geql \gz\}\cap \mi{guards}(a^{**}))=((S''\cap \mi{guards}(a^{**}))-\{a^{**}\gleq \gz\})\cup \{a^{**}\geql \gz\}=
 (\mi{guards}(S'',a^{**})-\{a^{**}\gleq \gz\})\cup \{a^{**}\geql \gz\}=((\mi{guards}(S,a^{**})\cup \{a^{**}\gleq \gz\})-\{a^{**}\gleq \gz\})\cup \{a^{**}\geql \gz\}=
 (\mi{guards}(S,a^{**})-\{a^{**}\gleq \gz\})\cup (\{a^{**}\gleq \gz\}-\{a^{**}\gleq \gz\})\cup \{a^{**}\geql \gz\}=(\mi{guards}(S,a^{**})-\{a^{**}\gleq \gz\})\cup \{a^{**}\geql \gz\}\supseteq
 (\{\gz\gle a^{**}\}-\{a^{**}\gleq \gz\})\cup \{a^{**}\geql \gz\}=\{a^{**}\geql \gz,\gz\gle a^{**}\}$,
$a^{**}\geql \gz\in \mi{guards}((S''-\{a^{**}\gleq \gz\})\cup \{a^{**}\geql \gz\})$, $\gz\gle a^{**}\in (S''-\{a^{**}\gleq \gz\})\cup \{a^{**}\geql \gz\}$, 
$a^{**}\in \mi{atoms}(\gz\gle a^{**})$, $a^{**}\geql \gz\neq \gz\gle a^{**}$, $\mi{simplify}(\gz\gle a^{**},a^{**},\gz)=\gz\gle \gz$,
applying Rule (\cref{ceq4hr3}) to $(S''-\{a^{**}\gleq \gz\})\cup \{a^{**}\geql \gz\}$, $a^{**}\geql \gz$, and $\gz\gle a^{**}$, we derive
\begin{equation} \notag
\dfrac{(S''-\{a^{**}\gleq \gz\})\cup \{a^{**}\geql \gz\}}
      {(((S''-\{a^{**}\gleq \gz\})\cup \{a^{**}\geql \gz\})-\{\gz\gle a^{**}\})\cup \{\gz\gle \gz\}};
\end{equation}
$\gz\gle \gz\in (((S''-\{a^{**}\gleq \gz\})\cup \{a^{**}\geql \gz\})-\{\gz\gle a^{**}\})\cup \{\gz\gle \gz\}$;
$\gz\gle \gz\in \mi{OrdPropLit}$ is a contradiction;
$\gz\gle \gz$ is not a guard;
$\gz\gle \gz\not\in \mi{guards}((((S''-\{a^{**}\gleq \gz\})\cup \{a^{**}\geql \gz\})-\{\gz\gle a^{**}\})\cup \{\gz\gle \gz\})$,
$\gz\gle \gz\in ((((S''-\{a^{**}\gleq \gz\})\cup \{a^{**}\geql \gz\})-\{\gz\gle a^{**}\})\cup \{\gz\gle \gz\})-\mi{guards}((((S''-\{a^{**}\gleq \gz\})\cup \{a^{**}\geql \gz\})-\{\gz\gle a^{**}\})\cup \{\gz\gle \gz\})$,
applying Rule (\cref{ceq4hr2}) to $(((S''-\{a^{**}\gleq \gz\})\cup \{a^{**}\geql \gz\})-\{\gz\gle a^{**}\})\cup \{\gz\gle \gz\}$ and $\gz\gle \gz$, we derive
\begin{equation} \notag
\dfrac{(((S''-\{a^{**}\gleq \gz\})\cup \{a^{**}\geql \gz\})-\{\gz\gle a^{**}\})\cup \{\gz\gle \gz\}}
      {(((((S''-\{a^{**}\gleq \gz\})\cup \{a^{**}\geql \gz\})-\{\gz\gle a^{**}\})\cup \{\gz\gle \gz\})-\{\gz\gle \gz\})\cup \{\square\}}.
\end{equation}
We put
\begin{equation} \notag
\mi{Tree}=\begin{array}[c]{c}
          S \\[0.4mm]
          \hline \\[-3.8mm]
          \mi{Tree}' \\[0.4mm]
          \hline \\[-3.8mm]
          (S''-\{a^{**}\gleq \gz\})\cup \{a^{**}\geql \gz\} \\[0.4mm]
          \hline \\[-3.8mm]
          (((S''-\{a^{**}\gleq \gz\})\cup \{a^{**}\geql \gz\})-\{\gz\gle a^{**}\})\cup \{\gz\gle \gz\} \\[0.4mm]
          \hline \\[-3.8mm]
          S'''=(((((S''-\{a^{**}\gleq \gz\})\cup \{a^{**}\geql \gz\})-\{\gz\gle a^{**}\})\cup \{\gz\gle \gz\})- \\
          \hfill \{\gz\gle \gz\})\cup \{\square\}.
          \end{array} 
\end{equation}
Hence, $\mi{Tree}$ is a finite linear {\it DPLL}-tree with the root $S$ constructed using Rules (\cref{ceq4hr1111111})--(\cref{ceq4hr4}), (\cref{ceq4hr8}) such that
for its only leaf $S'''$, $\square\in S'''$.

Cases 2.2.2.1.2.2 and 2.2.4.1.2.2:
$\mi{guards}(S,a^{**})=\{a^{**}\geql \gu\}$.
Then $S''\supseteq \mi{guards}(S'')\supseteq \mi{guards}(S'',a^{**})=\mi{guards}(S,a^{**})\cup \{C^\natural\}=\{a^{**}\gleq \gz,a^{**}\geql \gu\}$,
$a^{**}\geql \gu\in \mi{guards}(S'')$, $a^{**}\gleq \gz\in S''$,
$a^{**}\in \mi{atoms}(a^{**}\gleq \gz)$, $a^{**}\geql \gu\neq a^{**}\gleq \gz$, $\mi{simplify}(a^{**}\gleq \gz,a^{**},\gu)=\gu\gleq \gz$,
applying Rule (\cref{ceq4hr4}) to $S''$, $a^{**}\geql \gu$, and $a^{**}\gleq \gz$, we derive
\begin{equation} \notag
\dfrac{S''}
      {(S''-\{a^{**}\gleq \gz\})\cup \{\gu\gleq \gz\}};
\end{equation}
$\gu\gleq \gz\in (S''-\{a^{**}\gleq \gz\})\cup \{\gu\gleq \gz\}$;
$\gu\gleq \gz\in \mi{OrdPropLit}$ is a contradiction;
$\gu\gleq \gz$ is not a guard;
$\gu\gleq \gz\not\in \mi{guards}((S''-\{a^{**}\gleq \gz\})\cup \{\gu\gleq \gz\})$,
$\gu\gleq \gz\in ((S''-\{a^{**}\gleq \gz\})\cup \{\gu\gleq \gz\})-\mi{guards}((S''-\{a^{**}\gleq \gz\})\cup \{\gu\gleq \gz\})$,
applying Rule (\cref{ceq4hr2}) to $(S''-\{a^{**}\gleq \gz\})\cup \{\gu\gleq \gz\}$ and $\gu\gleq \gz$, we derive
\begin{equation} \notag
\dfrac{(S''-\{a^{**}\gleq \gz\})\cup \{\gu\gleq \gz\}}
      {(((S''-\{a^{**}\gleq \gz\})\cup \{\gu\gleq \gz\})-\{\gu\gleq \gz\})\cup \{\square\}}.
\end{equation}
We put 
\begin{equation} \notag
\mi{Tree}=\begin{array}[c]{c}
          S \\[0.4mm]
          \hline \\[-3.8mm]
          \mi{Tree}' \\[0.4mm]
          \hline \\[-3.8mm]
          (S''-\{a^{**}\gleq \gz\})\cup \{\gu\gleq \gz\} \\[0.4mm]
          \hline \\[-3.8mm]
          S'''=(((S''-\{a^{**}\gleq \gz\})\cup \{\gu\gleq \gz\})-\{\gu\gleq \gz\})\cup \{\square\}.
          \end{array} 
\end{equation}
Hence, $\mi{Tree}$ is a finite linear {\it DPLL}-tree with the root $S$ constructed using Rules (\cref{ceq4hr1111111})--(\cref{ceq4hr4}), (\cref{ceq4hr8}) such that
for its only leaf $S'''$, $\square\in S'''$.

Cases 2.2.2.1.3 and 2.2.4.1.3:
$C^\natural=\gz\gle a^{**}$.
We have that $a^{**}$ is semi-positively guarded in $S$.
We get two cases for $\mi{guards}(S,a^{**})$.

Cases 2.2.2.1.3.1 and 2.2.4.1.3.1:
Either $\mi{guards}(S,a^{**})=\{\gz\gle a^{**}\}$ or $\mi{guards}(S,a^{**})=\{\gz\gle a^{**},a^{**}\gle \gu\}$.
Then $\gz\gle a^{**}\in \mi{guards}(S,a^{**})\subseteq S$,
which is a contradiction with $C^\natural=\gz\gle a^{**}\not\in S$.

Cases 2.2.2.1.3.2 and 2.2.4.1.3.2:
$\mi{guards}(S,a^{**})=\{a^{**}\geql \gu\}$.
Then $C^\natural=\gz\gle a^{**}\not\in S$,
$S''=(S-\{C^*\})\cup \{C^\natural\}=(S-\{C^*\})\cup \{\gz\gle a^{**}\}$;
$S''\supseteq \mi{guards}(S'')\supseteq \mi{guards}(S'',a^{**})=\mi{guards}(S,a^{**})\cup \{C^\natural\}=\{\gz\gle a^{**},a^{**}\geql \gu\}$,
$a^{**}\geql \gu\in \mi{guards}(S'')$, $\gz\gle a^{**}\in S''$,
$a^{**}\in \mi{atoms}(\gz\gle a^{**})$, $a^{**}\geql \gu\neq \gz\gle a^{**}$, $\mi{simplify}(\gz\gle a^{**},a^{**},\gu)=\gz\gle \gu$,
applying Rule (\cref{ceq4hr4}) to $S''$, $a^{**}\geql \gu$, and $\gz\gle a^{**}$, we derive
\begin{equation} \notag
\dfrac{S''}
      {(S''-\{\gz\gle a^{**}\})\cup \{\gz\gle \gu\}};
\end{equation}
$\gz\gle \gu\in (S''-\{\gz\gle a^{**}\})\cup \{\gz\gle \gu\}$;
$\gz\gle \gu\in \mi{OrdPropLit}$ is a tautology;
$\gz\gle \gu$ is not a guard;
$\gz\gle \gu\not\in \mi{guards}((S''-\{\gz\gle a^{**}\})\cup \{\gz\gle \gu\})$,
$\gz\gle \gu\in ((S''-\{\gz\gle a^{**}\})\cup \{\gz\gle \gu\})-\mi{guards}((S''-\{\gz\gle a^{**}\})\cup \{\gz\gle \gu\})$,
applying Rule (\cref{ceq4hr22}) to $(S''-\{\gz\gle a^{**}\})\cup \{\gz\gle \gu\}$ and $\gz\gle \gu$, we derive
\begin{equation} \notag
\dfrac{(S''-\{\gz\gle a^{**}\})\cup \{\gz\gle \gu\}}
      {((S''-\{\gz\gle a^{**}\})\cup \{\gz\gle \gu\})-\{\gz\gle \gu\}}.
\end{equation}
We have that $S''$ is simplified.
Hence, $\gz\gle \gu\not\in S''$, $\gz\gle a^{**}\not\in S$,
$((S''-\{\gz\gle a^{**}\})\cup \{\gz\gle \gu\})-\{\gz\gle \gu\}=((S''-\{\gz\gle a^{**}\})-\{\gz\gle \gu\})\cup (\{\gz\gle \gu\}-\{\gz\gle \gu\})=S''-\{\gz\gle a^{**}\}=
 ((S-\{C^*\})\cup \{\gz\gle a^{**}\})-\{\gz\gle a^{**}\}=((S-\{C^*\})-\{\gz\gle a^{**}\})\cup (\{\gz\gle a^{**}\}-\{\gz\gle a^{**}\})=S-\{C^*\}$,
$a^*\geql \gu\in S-\{C^*\}$, $a^*\in \mi{atoms}(a^*\geql \gu)\subseteq \mi{atoms}(S-\{C^*\})\subseteq \mi{atoms}(S)$.
We get from Case 2.1 for $S-\{C^*\}$ that there exists a finite linear {\it DPLL}-tree $\mi{Tree}''$ with the root $S-\{C^*\}$ constructed using Rules (\cref{ceq4hr1111111})--(\cref{ceq4hr4}), (\cref{ceq4hr8}) satisfying
for its only leaf $S'''$ that either $\square\in S'''$, or $S'''\subseteq_{\mc F} \mi{OrdPropCl}$ is semi-positively guarded, $\mi{atoms}(S''')\subseteq \mi{atoms}(S-\{C^*\})-\{a^*\}=\mi{atoms}(S)-\{a^*\}$.
We put
\begin{equation} \notag
\mi{Tree}=\begin{array}[c]{c}
          S \\[0.4mm]
          \hline \\[-3.8mm]
          \mi{Tree}' \\[0.4mm]
          \hline \\[-3.8mm]
          (S''-\{\gz\gle a^{**}\})\cup \{\gz\gle \gu\} \\[0.4mm] 
          \hline \\[-3.8mm]
          \mi{Tree}''.
          \end{array}
\end{equation}
Hence, $\mi{Tree}$ is a finite linear {\it DPLL}-tree with the root $S$ constructed using Rules (\cref{ceq4hr1111111})--(\cref{ceq4hr4}), (\cref{ceq4hr8}) such that
for its only leaf $S'''$, either $\square\in S'''$, or $S'''\subseteq_{\mc F} \mi{OrdPropCl}$ is semi-positively guarded, $\mi{atoms}(S''')\subseteq \mi{atoms}(S)-\{a^*\}$.

Cases 2.2.2.1.4 and 2.2.4.1.4:
$C^\natural=a^{**}\gle \gu$.
We have that $a^{**}$ is semi-positively guarded in $S$.
We get three cases for $\mi{guards}(S,a^{**})$.

Cases 2.2.2.1.4.1 and 2.2.4.1.4.1:
$\mi{guards}(S,a^{**})=\{\gz\gle a^{**}\}$.
Then $S''=(S-\{C^*\})\cup \{C^\natural\}=(S-\{C^*\})\cup \{a^{**}\gle \gu\}$,
$\mi{guards}(S'',a^{**})=\mi{guards}(S,a^{**})\cup \{C^\natural\}=\{\gz\gle a^{**},a^{**}\gle \gu\}$;
$a^{**}$ is semi-positively guarded in $S''$;  
$\mi{atoms}(S'')\subseteq \mi{atoms}(S)$;
for all $a\in \mi{atoms}(S'')-\{a^*,a^{**}\}\subseteq \mi{atoms}(S)-\{a^*,a^{**}\}$,
$a$ is semi-positively guarded in $S$;
$a\neq a^{**}$,
$\mi{guards}(a)\cap \{a^{**}\gle \gu\}=\mi{guards}(a)\cap \mi{guards}(a^{**})=\emptyset$,
$C^*\not\in \mi{guards}(a)$,
$\mi{guards}(S'',a)=((S-\{C^*\})\cup \{a^{**}\gle \gu\})\cap \mi{guards}(a)=((S-\{C^*\})\cap \mi{guards}(a))\cup (\{a^{**}\gle \gu\}\cap \mi{guards}(a))=(S\cap \mi{guards}(a))-\{C^*\}=
                    S\cap \mi{guards}(a)=\mi{guards}(S,a)$;
$a$ is semi-positively guarded in $S''$;
$S''$ is semi-positively guarded;
$a^*\in \mi{atoms}(S'')$, $\mi{guards}(S'',a^*)=\{a^*\geql \gu\}$,
$a^*\neq a^{**}$, $a^*\not\in \mi{atoms}(a^{**}\gle \gu)$, $C^*\in \mi{restricted}(S)$,
$\mi{restricted}(S'')=\{C \,|\, C\in (S-\{C^*\})\cup \{a^{**}\gle \gu\}, a^*\in \mi{atoms}(C), C\neq a^*\geql \gu\}=\{C \,|\, C\in S-\{C^*\}, a^*\in \mi{atoms}(C), C\neq a^*\geql \gu\}=
                      \mi{restricted}(S)-\{C^*\}=\{C \,|\, C\in S, a^*\in \mi{atoms}(C), C\neq a^*\geql \gu\}-\{C^*\}\subset \mi{restricted}(S)$;
by the induction hypothesis for $S''$, there exists a finite linear {\it DPLL}-tree $\mi{Tree}''$ with the root $S''$ constructed using Rules (\cref{ceq4hr1111111})--(\cref{ceq4hr4}), (\cref{ceq4hr8}) satisfying
for its only leaf $S'''$ that either $\square\in S'''$, or $S'''\subseteq_{\mc F} \mi{OrdPropCl}$ is semi-positively guarded, $\mi{atoms}(S''')\subseteq \mi{atoms}(S'')-\{a^*\}\subseteq \mi{atoms}(S)-\{a^*\}$.
We put
\begin{equation} \notag
\mi{Tree}=\begin{array}[c]{c}
          S \\[0.4mm]
          \hline \\[-3.8mm]
          \mi{Tree}' \\[0.4mm]
          \hline \\[-3.8mm]
          \mi{Tree}''.
          \end{array}
\end{equation}
Hence, $\mi{Tree}$ is a finite linear {\it DPLL}-tree with the root $S$ constructed using Rules (\cref{ceq4hr1111111})--(\cref{ceq4hr4}), (\cref{ceq4hr8}) such that
for its only leaf $S'''$, either $\square\in S'''$, or $S'''\subseteq_{\mc F} \mi{OrdPropCl}$ is semi-positively guarded, $\mi{atoms}(S''')\subseteq \mi{atoms}(S)-\{a^*\}$.

Cases 2.2.2.1.4.2 and 2.2.4.1.4.2:
$\mi{guards}(S,a^{**})=\{\gz\gle a^{**},a^{**}\gle \gu\}$.
Then $a^{**}\gle \gu\in \mi{guards}(S,a^{**})\subseteq S$,
which is a contradiction with $C^\natural=a^{**}\gle \gu\not\in S$.

Cases 2.2.2.1.4.3 and 2.2.4.1.4.3:
$\mi{guards}(S,a^{**})=\{a^{**}\geql \gu\}$.
Then $S''\supseteq \mi{guards}(S'')\supseteq \mi{guards}(S'',a^{**})=\mi{guards}(S,a^{**})\cup \{C^\natural\}=\{a^{**}\gle \gu,a^{**}\geql \gu\}$,
$a^{**}\geql \gu\in \mi{guards}(S'')$, $a^{**}\gle \gu\in S''$,
$a^{**}\in \mi{atoms}(a^{**}\gle \gu)$, $a^{**}\geql \gu\neq a^{**}\gle \gu$, $\mi{simplify}(a^{**}\gle \gu,a^{**},\gu)=\gu\gle \gu$,
applying Rule (\cref{ceq4hr4}) to $S''$, $a^{**}\geql \gu$, and $a^{**}\gle \gu$, we derive
\begin{equation} \notag
\dfrac{S''}
      {(S''-\{a^{**}\gle \gu\})\cup \{\gu\gle \gu\}};
\end{equation}
$\gu\gle \gu\in (S''-\{a^{**}\gle \gu\})\cup \{\gu\gle \gu\}$;
$\gu\gle \gu\in \mi{OrdPropLit}$ is a contradiction;
$\gu\gle \gu$ is not a guard;
$\gu\gle \gu\not\in \mi{guards}((S''-\{a^{**}\gle \gu\})\cup \{\gu\gle \gu\})$,
$\gu\gle \gu\in ((S''-\{a^{**}\gle \gu\})\cup \{\gu\gle \gu\})-\mi{guards}((S''-\{a^{**}\gle \gu\})\cup \{\gu\gle \gu\})$,
applying Rule (\cref{ceq4hr2}) to $(S''-\{a^{**}\gle \gu\})\cup \{\gu\gle \gu\}$ and $\gu\gle \gu$, we derive
\begin{equation} \notag
\dfrac{(S''-\{a^{**}\gle \gu\})\cup \{\gu\gle \gu\}}
      {(((S''-\{a^{**}\gle \gu\})\cup \{\gu\gle \gu\})-\{\gu\gle \gu\})\cup \{\square\}}.
\end{equation}
We put
\begin{equation} \notag
\mi{Tree}=\begin{array}[c]{c}
          S \\[0.4mm]
          \hline \\[-3.8mm]
          \mi{Tree}' \\[0.4mm]
          \hline \\[-3.8mm]
          (S''-\{a^{**}\gle \gu\})\cup \{\gu\gle \gu\} \\[0.4mm]
          \hline \\[-3.8mm]
          S'''=(((S''-\{a^{**}\gle \gu\})\cup \{\gu\gle \gu\})-\{\gu\gle \gu\})\cup \{\square\}.
          \end{array} 
\end{equation}
Hence, $\mi{Tree}$ is a finite linear {\it DPLL}-tree with the root $S$ constructed using Rules (\cref{ceq4hr1111111})--(\cref{ceq4hr4}), (\cref{ceq4hr8}) such that
for its only leaf $S'''$, $\square\in S'''$.

Cases 2.2.2.1.5 and 2.2.4.1.5:
$C^\natural=a^{**}\geql \gu$.
We have that $a^{**}$ is semi-positively guarded in $S$.
We get three cases for $\mi{guards}(S,a^{**})$.

Cases 2.2.2.1.5.1 and 2.2.4.1.5.1:
$\mi{guards}(S,a^{**})=\{\gz\gle a^{**}\}$.
Then $C^\natural=a^{**}\geql \gu\not\in S$,
$S''=(S-\{C^*\})\cup \{C^\natural\}=(S-\{C^*\})\cup \{a^{**}\geql \gu\}$,
$S\supseteq \mi{guards}(S,a^{**})=\{\gz\gle a^{**}\}$;
$S''\supseteq \mi{guards}(S'')\supseteq \mi{guards}(S'',a^{**})=\mi{guards}(S,a^{**})\cup \{C^\natural\}=\{\gz\gle a^{**},a^{**}\geql \gu\}$,
$a^{**}\geql \gu\in \mi{guards}(S'')$, $\gz\gle a^{**}\in S''$,
$a^{**}\in \mi{atoms}(\gz\gle a^{**})$, $a^{**}\geql \gu\neq \gz\gle a^{**}$, $\mi{simplify}(\gz\gle a^{**},a^{**},\gu)=\gz\gle \gu$,
applying Rule (\cref{ceq4hr4}) to $S''$, $a^{**}\geql \gu$, and $\gz\gle a^{**}$, we derive
\begin{equation} \notag
\dfrac{S''}
      {(S''-\{\gz\gle a^{**}\})\cup \{\gz\gle \gu\}};
\end{equation}
$\gz\gle \gu\in (S''-\{\gz\gle a^{**}\})\cup \{\gz\gle \gu\}$;
$\gz\gle \gu\in \mi{OrdPropLit}$ is a tautology;
$\gz\gle \gu$ is not a guard;
$\gz\gle \gu\not\in \mi{guards}((S''-\{\gz\gle a^{**}\})\cup \{\gz\gle \gu\})$,
$\gz\gle \gu\in ((S''-\{\gz\gle a^{**}\})\cup \{\gz\gle \gu\})-\mi{guards}((S''-\{\gz\gle a^{**}\})\cup \{\gz\gle \gu\})$,
applying Rule (\cref{ceq4hr22}) to $(S''-\{\gz\gle a^{**}\})\cup \{\gz\gle \gu\}$ and $\gz\gle \gu$, we derive
\begin{equation} \notag
\dfrac{(S''-\{\gz\gle a^{**}\})\cup \{\gz\gle \gu\}}
      {((S''-\{\gz\gle a^{**}\})\cup \{\gz\gle \gu\})-\{\gz\gle \gu\}}.
\end{equation}
We have that $S''$ is simplified.
Hence, $\gz\gle \gu\not\in S''$, $C^*, \gz\gle a^{**}\in S$, $C^*\not\in \mi{guards}(a^{**})$, $\gz\gle a^{**}\in \mi{guards}(a^{**})$, $C^*\neq \gz\gle a^{**}$, $a^{**}\geql \gu\not\in S$,
$((S''-\{\gz\gle a^{**}\})\cup \{\gz\gle \gu\})-\{\gz\gle \gu\}=((S''-\{\gz\gle a^{**}\})-\{\gz\gle \gu\})\cup (\{\gz\gle \gu\}-\{\gz\gle \gu\})=S''-\{\gz\gle a^{**}\}=
 ((S-\{C^*\})\cup \{a^{**}\geql \gu\})-\{\gz\gle a^{**}\}=((S-\{C^*\})-\{\gz\gle a^{**}\})\cup (\{a^{**}\geql \gu\}-\{\gz\gle a^{**}\})=(S-\{C^*,\gz\gle a^{**}\})\cup \{a^{**}\geql \gu\}$.
We put $S'''=(S-\{C^*,\gz\gle a^{**}\})\cup \{a^{**}\geql \gu\}\subseteq_{\mc F} \mi{OrdPropCl}$.
We get that
$S'''=(S-\{C^*,\gz\gle a^{**}\})\cup \{a^{**}\geql \gu\}=((S-\{C^*\})-\{\gz\gle a^{**}\})\cup \{a^{**}\geql \gu\}$,
$a^*\geql \gu\in S$, $a^*\geql \gu\neq C^*, \gz\gle a^{**}$, 
$a^*\geql \gu\in S'''=(S-\{C^*,\gz\gle a^{**}\})\cup \{a^{**}\geql \gu\}$,
$a^*\in \mi{atoms}(a^*\geql \gu)\subseteq \mi{atoms}(S''')$,
$a^{**}\in \mi{atoms}(S)$,
$a^{**}\in \mi{atoms}(S''')=\mi{atoms}((S-\{C^*,\gz\gle a^{**}\})\cup \{a^{**}\geql \gu\})=\mi{atoms}(S-\{C^*,\gz\gle a^{**}\})\cup \mi{atoms}(a^{**}\geql \gu)=
                            \mi{atoms}(S-\{C^*,\gz\gle a^{**}\})\cup \{a^{**}\}\subseteq \mi{atoms}(S)$;
$S$ is simplified;
$S-\{C^*,\gz\gle a^{**}\}\subseteq S$ is simplified;
$a^{**}\geql \gu\neq \square$ does not contain contradictions and tautologies;
$S'''=(S-\{C^*,\gz\gle a^{**}\})\cup \{a^{**}\geql \gu\}$ is simplified;
$a^*\neq a^{**}$,
$\mi{guards}(a^*)\cap \{\gz\gle a^{**},a^{**}\geql \gu\}=\mi{guards}(a^*)\cap \mi{guards}(a^{**})=\emptyset$,
$C^*\not\in \mi{guards}(a^*)$,
$\mi{guards}(S''',a^*)=(((S-\{C^*\})-\{\gz\gle a^{**}\})\cup \{a^{**}\geql \gu\})\cap \mi{guards}(a^*)=
                       (((S-\{C^*\})-\{\gz\gle a^{**}\})\cap \mi{guards}(a^*))\cup (\{a^{**}\geql \gu\}\cap \mi{guards}(a^*))=
                       ((S\cap \mi{guards}(a^*))-\{C^*\})-\{\gz\gle a^{**}\}=S\cap \mi{guards}(a^*)=\mi{guards}(S,a^*)=\{a^*\geql \gu\}$;
$a^*$ is semi-positively guarded in $S'''$;
$C^*\not\in \mi{guards}(a^{**})$,
$\mi{guards}(S''',a^{**})=(((S-\{C^*\})-\{\gz\gle a^{**}\})\cup \{a^{**}\geql \gu\})\cap \mi{guards}(a^{**})=
                          (((S-\{C^*\})-\{\gz\gle a^{**}\})\cap \mi{guards}(a^{**}))\cup (\{a^{**}\geql \gu\}\cap \mi{guards}(a^{**}))=
                          (((S\cap \mi{guards}(a^{**}))-\{\gz\gle a^{**}\})-\{C^*\})\cup \{a^{**}\geql \gu\}=((S\cap \mi{guards}(a^{**}))-\{\gz\gle a^{**}\})\cup \{a^{**}\geql \gu\}=
                          (\mi{guards}(S,a^{**})-\{\gz\gle a^{**}\})\cup \{a^{**}\geql \gu\}=(\{\gz\gle a^{**}\}-\{\gz\gle a^{**}\})\cup \{a^{**}\geql \gu\}=\{a^{**}\geql \gu\}$;
$a^{**}$ is semi-positively guarded in $S'''$;
for all $a\in \mi{atoms}(S''')-\{a^*,a^{**}\}\subseteq \mi{atoms}(S)-\{a^*,a^{**}\}$,
$a$ is semi-positively guarded in $S$;
$a\neq a^{**}$,
$\mi{guards}(a)\cap \{\gz\gle a^{**},a^{**}\geql \gu\}=\mi{guards}(a)\cap \mi{guards}(a^{**})=\emptyset$,
$C^*\not\in \mi{guards}(a)$,
$\mi{guards}(S''',a)=(((S-\{C^*\})-\{\gz\gle a^{**}\})\cup \{a^{**}\geql \gu\})\cap \mi{guards}(a)=
                     (((S-\{C^*\})-\{\gz\gle a^{**}\})\cap \mi{guards}(a))\cup (\{a^{**}\geql \gu\}\cap \mi{guards}(a))=
                     ((S\cap \mi{guards}(a))-\{C^*\})-\{\gz\gle a^{**}\}=S\cap \mi{guards}(a)=\mi{guards}(S,a)$;
$a$ is semi-positively guarded in $S'''$;
$S'''$ is semi-positively guarded;
$a^*\not\in \mi{atoms}(\gz\gle a^{**})$, $a^*\not\in \mi{atoms}(a^{**}\geql \gu)$, $C^*\in \mi{restricted}(S)$,
$\mi{restricted}(S''')=\{C \,|\, C\in ((S-\{C^*\})-\{\gz\gle a^{**}\})\cup \{a^{**}\geql \gu\}, a^*\in \mi{atoms}(C), C\neq a^*\geql \gu\}= 
                       \{C \,|\, C\in S-\{C^*\}, a^*\in \mi{atoms}(C), C\neq a^*\geql \gu\}=
                       \mi{restricted}(S)-\{C^*\}=\{C \,|\, C\in S, a^*\in \mi{atoms}(C), C\neq a^*\geql \gu\}-\{C^*\}\subset \mi{restricted}(S)$;
by the induction hypothesis for $S'''$, there exists a finite linear {\it DPLL}-tree $\mi{Tree}'''$ with the root $S'''$ constructed using Rules (\cref{ceq4hr1111111})--(\cref{ceq4hr4}), (\cref{ceq4hr8}) satisfying
for its only leaf $S''''$ that either $\square\in S''''$, or $S''''\subseteq_{\mc F} \mi{OrdPropCl}$ is semi-positively guarded, $\mi{atoms}(S'''')\subseteq \mi{atoms}(S''')-\{a^*\}\subseteq \mi{atoms}(S)-\{a^*\}$.
We put
\begin{equation} \notag
\mi{Tree}=\begin{array}[c]{c}
          S \\[0.4mm]
          \hline \\[-3.8mm]
          \mi{Tree}' \\[0.4mm]
          \hline \\[-3.8mm]
          (S''-\{\gz\gle a^{**}\})\cup \{\gz\gle \gu\} \\[0.4mm]
          \hline \\[-3.8mm]
          \mi{Tree}'''.    
          \end{array}    
\end{equation}
Hence, $\mi{Tree}$ is a finite linear {\it DPLL}-tree with the root $S$ constructed using Rules (\cref{ceq4hr1111111})--(\cref{ceq4hr4}), (\cref{ceq4hr8}) such that
for its only leaf $S''''$, either $\square\in S''''$, or $S''''\subseteq_{\mc F} \mi{OrdPropCl}$ is semi-positively guarded, $\mi{atoms}(S'''')\subseteq \mi{atoms}(S)-\{a^*\}$.

Cases 2.2.2.1.5.2 and 2.2.4.1.5.2:
$\mi{guards}(S,a^{**})=\{\gz\gle a^{**},a^{**}\gle \gu\}$.
Then $S''\supseteq \mi{guards}(S'')\supseteq \mi{guards}(S'',a^{**})=\mi{guards}(S,a^{**})\cup \{C^\natural\}=\{\gz\gle a^{**},a^{**}\gle \gu,a^{**}\geql \gu\}\supset \{a^{**}\gle \gu,a^{**}\geql \gu\}$;
these cases are the same as Cases 2.2.2.1.4.3 and 2.2.4.1.4.3.

Cases 2.2.2.1.5.3 and 2.2.4.1.5.3:
$\mi{guards}(S,a^{**})=\{a^{**}\geql \gu\}$.
Then $a^{**}\geql \gu\in \mi{guards}(S,a^{**})\subseteq S$,
which is a contradiction with $C^\natural=a^{**}\geql \gu\not\in S$.

Cases 2.2.2.1.6 and 2.2.4.1.6:
$C^\natural=\gu\gleq a^{**}$.
We have that $a^{**}$ is semi-positively guarded in $S$.
We get three cases for $\mi{guards}(S,a^{**})$.

Cases 2.2.2.1.6.1 and 2.2.4.1.6.1:
$\mi{guards}(S,a^{**})=\{\gz\gle a^{**}\}$.
Then $C^\natural=\gu\gleq a^{**}\not\in S$,
$S''=(S-\{C^*\})\cup \{C^\natural\}=(S-\{C^*\})\cup \{\gu\gleq a^{**}\}$,
$S\supseteq \mi{guards}(S,a^{**})=\{\gz\gle a^{**}\}$;
$\mi{guards}(S'')\supseteq \mi{guards}(S'',a^{**})=\mi{guards}(S,a^{**})\cup \{C^\natural\}=\{\gz\gle a^{**},\gu\gleq a^{**}\}$,
$\gu\gleq a^{**}\in \mi{guards}(S'')$, applying Rule (\cref{ceq4hr11111111}) to $S''$ and $\gu\gleq a^{**}$, we derive
\begin{equation} \notag
\dfrac{S''}
      {(S''-\{\gu\gleq a^{**}\})\cup \{a^{**}\geql \gu\}};
\end{equation}
$a^{**}\in \mi{atoms}((S''-\{\gu\gleq a^{**}\})\cup \{a^{**}\geql \gu\})$,
$(S''-\{\gu\gleq a^{**}\})\cup \{a^{**}\geql \gu\}\supseteq \mi{guards}((S''-\{\gu\gleq a^{**}\})\cup \{a^{**}\geql \gu\})\supseteq 
 \mi{guards}((S''-\{\gu\gleq a^{**}\})\cup \{a^{**}\geql \gu\},a^{**})=((S''-\{\gu\gleq a^{**}\})\cup \{a^{**}\geql \gu\})\cap \mi{guards}(a^{**})=
 ((S''-\{\gu\gleq a^{**}\})\cap \mi{guards}(a^{**}))\cup (\{a^{**}\geql \gu\}\cap \mi{guards}(a^{**}))=((S''\cap \mi{guards}(a^{**}))-\{\gu\gleq a^{**}\})\cup \{a^{**}\geql \gu\}=
 (\mi{guards}(S'',a^{**})-\{\gu\gleq a^{**}\})\cup \{a^{**}\geql \gu\}=(\{\gz\gle a^{**},\gu\gleq a^{**}\}-\{\gu\gleq a^{**}\})\cup \{a^{**}\geql \gu\}=\{\gz\gle a^{**},a^{**}\geql \gu\}$,
$a^{**}\geql \gu\in \mi{guards}((S''-\{\gu\gleq a^{**}\})\cup \{a^{**}\geql \gu\})$, $\gz\gle a^{**}\in (S''-\{\gu\gleq a^{**}\})\cup \{a^{**}\geql \gu\}$,
$a^{**}\in \mi{atoms}(\gz\gle a^{**})$, $a^{**}\geql \gu\neq \gz\gle a^{**}$, $\mi{simplify}(\gz\gle a^{**},a^{**},\gu)=\gz\gle \gu$,
applying Rule (\cref{ceq4hr4}) to $(S''-\{\gu\gleq a^{**}\})\cup \{a^{**}\geql \gu\}$, $a^{**}\geql \gu$, and $\gz\gle a^{**}$, we derive
\begin{equation} \notag
\dfrac{(S''-\{\gu\gleq a^{**}\})\cup \{a^{**}\geql \gu\}}
      {(((S''-\{\gu\gleq a^{**}\})\cup \{a^{**}\geql \gu\})-\{\gz\gle a^{**}\})\cup \{\gz\gle \gu\}};
\end{equation}
$\gz\gle \gu\in (((S''-\{\gu\gleq a^{**}\})\cup \{a^{**}\geql \gu\})-\{\gz\gle a^{**}\})\cup \{\gz\gle \gu\}$;
$\gz\gle \gu\in \mi{OrdPropLit}$ is a tautology;
$\gz\gle \gu$ is not a guard;
$\gz\gle \gu\not\in \mi{guards}((((S''-\{\gu\gleq a^{**}\})\cup \{a^{**}\geql \gu\})-\{\gz\gle a^{**}\})\cup \{\gz\gle \gu\})$,
$\gz\gle \gu\in ((((S''-\{\gu\gleq a^{**}\})\cup \{a^{**}\geql \gu\})-\{\gz\gle a^{**}\})\cup \{\gz\gle \gu\})-\mi{guards}((((S''-\{\gu\gleq a^{**}\})\cup \{a^{**}\geql \gu\})-\{\gz\gle a^{**}\})\cup \{\gz\gle \gu\})$,
applying Rule (\cref{ceq4hr22}) to $(((S''-\{\gu\gleq a^{**}\})\cup \{a^{**}\geql \gu\})-\{\gz\gle a^{**}\})\cup \{\gz\gle \gu\}$ and $\gz\gle \gu$, we derive
\begin{equation} \notag
\dfrac{(((S''-\{\gu\gleq a^{**}\})\cup \{a^{**}\geql \gu\})-\{\gz\gle a^{**}\})\cup \{\gz\gle \gu\}}
      {((((S''-\{\gu\gleq a^{**}\})\cup \{a^{**}\geql \gu\})-\{\gz\gle a^{**}\})\cup \{\gz\gle \gu\})-\{\gz\gle \gu\}}.
\end{equation}
We have that $S''$ is simplified.
Hence, $\gz\gle \gu\not\in S''$, $\gu\gleq a^{**}\not\in S$, 
$C^*, \gz\gle a^{**}\in S$, $C^*\not\in \mi{guards}(a^{**})$, $\gz\gle a^{**}, a^{**}\geql \gu\in \mi{guards}(a^{**})$, $C^*\neq \gz\gle a^{**}$, 
$a^{**}\geql \gu\not\in \mi{guards}(S,a^{**})=\{\gz\gle a^{**}\}=S\cap \mi{guards}(a^{**})$, $a^{**}\geql \gu\not\in S$,
$((((S''-\{\gu\gleq a^{**}\})\cup \{a^{**}\geql \gu\})-\{\gz\gle a^{**}\})\cup \{\gz\gle \gu\})-\{\gz\gle \gu\}=
 ((((S''-\{\gu\gleq a^{**}\})\cup \{a^{**}\geql \gu\})-\{\gz\gle a^{**}\})-\{\gz\gle \gu\})\cup (\{\gz\gle \gu\}-\{\gz\gle \gu\})=
 (((S''-\{\gu\gleq a^{**}\})\cup \{a^{**}\geql \gu\})-\{\gz\gle \gu\})-\{\gz\gle a^{**}\}=
 (((S''-\{\gu\gleq a^{**}\})-\{\gz\gle \gu\})\cup (\{a^{**}\geql \gu\}-\{\gz\gle \gu\}))-\{\gz\gle a^{**}\}=
 ((S''-\{\gu\gleq a^{**}\})\cup \{a^{**}\geql \gu\})-\{\gz\gle a^{**}\}=
 ((S''-\{\gu\gleq a^{**}\})-\{\gz\gle a^{**}\})\cup (\{a^{**}\geql \gu\}-\{\gz\gle a^{**}\})=
 ((S''-\{\gu\gleq a^{**}\})-\{\gz\gle a^{**}\})\cup \{a^{**}\geql \gu\}=
 ((((S-\{C^*\})\cup \{\gu\gleq a^{**}\})-\{\gu\gleq a^{**}\})-\{\gz\gle a^{**}\})\cup \{a^{**}\geql \gu\}=
 ((((S-\{C^*\})-\{\gu\gleq a^{**}\})\cup (\{\gu\gleq a^{**}\}-\{\gu\gleq a^{**}\}))-\{\gz\gle a^{**}\})\cup \{a^{**}\geql \gu\}=
 ((S-\{C^*\})-\{\gz\gle a^{**}\})\cup \{a^{**}\geql \gu\}=(S-\{C^*,\gz\gle a^{**}\})\cup \{a^{**}\geql \gu\}$.
We put $S'''=(S-\{C^*,\gz\gle a^{**}\})\cup \{a^{**}\geql \gu\}\subseteq_{\mc F} \mi{OrdPropCl}$.
We get from Cases 2.2.2.1.5.1 and 2.2.4.1.5.1 that 
there exists a finite linear {\it DPLL}-tree $\mi{Tree}'''$ with the root $S'''$ constructed using Rules (\cref{ceq4hr1111111})--(\cref{ceq4hr4}), (\cref{ceq4hr8}) satisfying
for its only leaf $S''''$ that either $\square\in S''''$, or $S''''\subseteq_{\mc F} \mi{OrdPropCl}$ is semi-positively guarded, $\mi{atoms}(S'''')\subseteq \mi{atoms}(S''')-\{a^*\}\subseteq \mi{atoms}(S)-\{a^*\}$.
We put
\begin{equation} \notag
\mi{Tree}=\begin{array}[c]{c}
          S \\[0.4mm]
          \hline \\[-3.8mm]
          \mi{Tree}' \\[0.4mm]
          \hline \\[-3.8mm]
          (S''-\{\gu\gleq a^{**}\})\cup \{a^{**}\geql \gu\} \\[0.4mm]
          \hline \\[-3.8mm]
          (((S''-\{\gu\gleq a^{**}\})\cup \{a^{**}\geql \gu\})-\{\gz\gle a^{**}\})\cup \{\gz\gle \gu\} \\[0.4mm]
          \hline \\[-3.8mm]
          \mi{Tree}'''.    
          \end{array}    
\end{equation}
Hence, $\mi{Tree}$ is a finite linear {\it DPLL}-tree with the root $S$ constructed using Rules (\cref{ceq4hr1111111})--(\cref{ceq4hr4}), (\cref{ceq4hr8}) such that
for its only leaf $S''''$, either $\square\in S''''$, or $S''''\subseteq_{\mc F} \mi{OrdPropCl}$ is semi-positively guarded, $\mi{atoms}(S'''')\subseteq \mi{atoms}(S)-\{a^*\}$.

Cases 2.2.2.1.6.2 and 2.2.4.1.6.2:
$\mi{guards}(S,a^{**})=\{\gz\gle a^{**},a^{**}\gle \gu\}$.
Then $\mi{guards}(S'')\supseteq \mi{guards}(S'',a^{**})=\mi{guards}(S,a^{**})\cup \{C^\natural\}=\{\gz\gle a^{**},a^{**}\gle \gu,\gu\gleq a^{**}\}$,
$\gu\gleq a^{**}\in \mi{guards}(S'')$, applying Rule (\cref{ceq4hr11111111}) to $S''$ and $\gu\gleq a^{**}$, we derive
\begin{equation} \notag
\dfrac{S''}
      {(S''-\{\gu\gleq a^{**}\})\cup \{a^{**}\geql \gu\}};
\end{equation}
$a^{**}\in \mi{atoms}((S''-\{\gu\gleq a^{**}\})\cup \{a^{**}\geql \gu\})$,
$(S''-\{\gu\gleq a^{**}\})\cup \{a^{**}\geql \gu\}\supseteq \mi{guards}((S''-\{\gu\gleq a^{**}\})\cup \{a^{**}\geql \gu\})\supseteq 
 \mi{guards}((S''-\{\gu\gleq a^{**}\})\cup \{a^{**}\geql \gu\},a^{**})=((S''-\{\gu\gleq a^{**}\})\cup \{a^{**}\geql \gu\})\cap \mi{guards}(a^{**})=
 ((S''-\{\gu\gleq a^{**}\})\cap \mi{guards}(a^{**}))\cup (\{a^{**}\geql \gu\}\cap \mi{guards}(a^{**}))=((S''\cap \mi{guards}(a^{**}))-\{\gu\gleq a^{**}\})\cup \{a^{**}\geql \gu\}=
 (\mi{guards}(S'',a^{**})-\{\gu\gleq a^{**}\})\cup \{a^{**}\geql \gu\}=(\{\gz\gle a^{**},a^{**}\gle \gu,\gu\gleq a^{**}\}-\{\gu\gleq a^{**}\})\cup \{a^{**}\geql \gu\}\supseteq 
 \{a^{**}\gle \gu,a^{**}\geql \gu\}$,
$a^{**}\geql \gu\in \mi{guards}((S''-\{\gu\gleq a^{**}\})\cup \{a^{**}\geql \gu\})$, $a^{**}\gle \gu\in (S''-\{\gu\gleq a^{**}\})\cup \{a^{**}\geql \gu\}$,
$a^{**}\in \mi{atoms}(a^{**}\gle \gu)$, $a^{**}\geql \gu\neq a^{**}\gle \gu$, $\mi{simplify}(a^{**}\gle \gu,a^{**},\gu)=\gu\gle \gu$,
applying Rule (\cref{ceq4hr4}) to $(S''-\{\gu\gleq a^{**}\})\cup \{a^{**}\geql \gu\}$, $a^{**}\geql \gu$, and $a^{**}\gle \gu$, we derive
\begin{equation} \notag
\dfrac{(S''-\{\gu\gleq a^{**}\})\cup \{a^{**}\geql \gu\}}
      {(((S''-\{\gu\gleq a^{**}\})\cup \{a^{**}\geql \gu\})-\{a^{**}\gle \gu\})\cup \{\gu\gle \gu\}};
\end{equation}
$\gu\gle \gu\in (((S''-\{\gu\gleq a^{**}\})\cup \{a^{**}\geql \gu\})-\{a^{**}\gle \gu\})\cup \{\gu\gle \gu\}$;
$\gu\gle \gu\in \mi{OrdPropLit}$ is a contradiction;
$\gu\gle \gu$ is not a guard;
$\gu\gle \gu\not\in \mi{guards}((((S''-\{\gu\gleq a^{**}\})\cup \{a^{**}\geql \gu\})-\{a^{**}\gle \gu\})\cup \{\gu\gle \gu\})$,
$\gu\gle \gu\in ((((S''-\{\gu\gleq a^{**}\})\cup \{a^{**}\geql \gu\})-\{a^{**}\gle \gu\})\cup \{\gu\gle \gu\})-\mi{guards}((((S''-\{\gu\gleq a^{**}\})\cup \{a^{**}\geql \gu\})-\{a^{**}\gle \gu\})\cup \{\gu\gle \gu\})$,
applying Rule (\cref{ceq4hr2}) to $(((S''-\{\gu\gleq a^{**}\})\cup \{a^{**}\geql \gu\})-\{a^{**}\gle \gu\})\cup \{\gu\gle \gu\}$ and $\gu\gle \gu$, we derive
\begin{equation} \notag
\dfrac{(((S''-\{\gu\gleq a^{**}\})\cup \{a^{**}\geql \gu\})-\{a^{**}\gle \gu\})\cup \{\gu\gle \gu\}}
      {(((((S''-\{\gu\gleq a^{**}\})\cup \{a^{**}\geql \gu\})-\{a^{**}\gle \gu\})\cup \{\gu\gle \gu\})-\{\gu\gle \gu\})\cup \{\square\}}.
\end{equation}
We put
\begin{equation} \notag
\mi{Tree}=\begin{array}[c]{c}
          S \\[0.4mm]
          \hline \\[-3.8mm]
          \mi{Tree}' \\[0.4mm]
          \hline \\[-3.8mm]
          (S''-\{\gu\gleq a^{**}\})\cup \{a^{**}\geql \gu\} \\[0.4mm]
          \hline \\[-3.8mm]
          (((S''-\{\gu\gleq a^{**}\})\cup \{a^{**}\geql \gu\})-\{a^{**}\gle \gu\})\cup \{\gu\gle \gu\} \\[0.4mm]
          \hline \\[-3.8mm]
          S'''=(((((S''-\{\gu\gleq a^{**}\})\cup \{a^{**}\geql \gu\})-\{a^{**}\gle \gu\})\cup \{\gu\gle \gu\})- \\
          \hfill \{\gu\gle \gu\})\cup \{\square\}.
          \end{array} 
\end{equation}
Hence, $\mi{Tree}$ is a finite linear {\it DPLL}-tree with the root $S$ constructed using Rules (\cref{ceq4hr1111111})--(\cref{ceq4hr4}), (\cref{ceq4hr8}) such that
for its only leaf $S'''$, $\square\in S'''$.

Cases 2.2.2.1.6.3 and 2.2.4.1.6.3:
$\mi{guards}(S,a^{**})=\{a^{**}\geql \gu\}$.
Then $C^\natural=\gu\gleq a^{**}\not\in S$,
$S''=(S-\{C^*\})\cup \{C^\natural\}=(S-\{C^*\})\cup \{\gu\gleq a^{**}\}$;
$S''\supseteq \mi{guards}(S'')\supseteq \mi{guards}(S'',a^{**})=\mi{guards}(S,a^{**})\cup \{C^\natural\}=\{a^{**}\geql \gu,\gu\gleq a^{**}\}$,
$\gu\gleq a^{**}\in \mi{guards}(S'')$, applying Rule (\cref{ceq4hr11111111}) to $S''$ and $\gu\gleq a^{**}$, we derive
\begin{equation} \notag
\dfrac{S''}
      {(S''-\{\gu\gleq a^{**}\})\cup \{a^{**}\geql \gu\}}.
\end{equation}
Hence, $a^{**}\geql \gu\in S''$, $a^{**}\geql \gu\neq \gu\gleq a^{**}$, $a^{**}\geql \gu\in S''-\{\gu\gleq a^{**}\}$, $\gu\gleq a^{**}\not\in S$,
$(S''-\{\gu\gleq a^{**}\})\cup \{a^{**}\geql \gu\}=S''-\{\gu\gleq a^{**}\}=((S-\{C^*\})\cup \{\gu\gleq a^{**}\})-\{\gu\gleq a^{**}\}=
 ((S-\{C^*\})-\{\gu\gleq a^{**}\})\cup (\{\gu\gleq a^{**}\}-\{\gu\gleq a^{**}\})=S-\{C^*\}$,
$a^*\geql \gu\in S-\{C^*\}$, $a^*\in \mi{atoms}(a^*\geql \gu)\subseteq \mi{atoms}(S-\{C^*\})\subseteq \mi{atoms}(S)$.
We get from Case 2.1 for $S-\{C^*\}$ that there exists a finite linear {\it DPLL}-tree $\mi{Tree}''$ with the root $S-\{C^*\}$ constructed using Rules (\cref{ceq4hr1111111})--(\cref{ceq4hr4}), (\cref{ceq4hr8}) satisfying
for its only leaf $S'''$ that either $\square\in S'''$, or $S'''\subseteq_{\mc F} \mi{OrdPropCl}$ is semi-positively guarded, $\mi{atoms}(S''')\subseteq \mi{atoms}(S-\{C^*\})-\{a^*\}=\mi{atoms}(S)-\{a^*\}$.
We put
\begin{equation} \notag
\mi{Tree}=\begin{array}[c]{c}
          S \\[0.4mm]
          \hline \\[-3.8mm]
          \mi{Tree}' \\[0.4mm]
          \hline \\[-3.8mm]
          \mi{Tree}''.
          \end{array}
\end{equation}
Hence, $\mi{Tree}$ is a finite linear {\it DPLL}-tree with the root $S$ constructed using Rules (\cref{ceq4hr1111111})--(\cref{ceq4hr4}), (\cref{ceq4hr8}) such that
for its only leaf $S'''$, either $\square\in S'''$, or $S'''\subseteq_{\mc F} \mi{OrdPropCl}$ is semi-positively guarded, $\mi{atoms}(S''')\subseteq \mi{atoms}(S)-\{a^*\}$.

Cases 2.2.2.2 and 2.2.4.2:
For all $a\in \mi{atoms}(S'')-\{a^*\}$, $C^\natural\not\in \mi{guards}(a)$.
Then, for all $a\in \mi{atoms}(C^\natural)\subseteq \mi{atoms}(S'')-\{a^*\}$, $C^\natural\not\in \mi{guards}(a)$;
$C^\natural$ is not a guard;
$C^*\not\in \mi{guards}(S)$,
$\mi{guards}(S'')=\{C \,|\, C\in (S-\{C^*\})\cup \{C^\natural\}\ \text{\it is a guard}\}=\{C \,|\, C\in S-\{C^*\}\ \text{\it is a guard}\}=
                  \mi{guards}(S)-\{C^*\}=\{C \,|\, C\in S\ \text{\it is a guard}\}-\{C^*\}=\mi{guards}(S)$;
we have that $S$ is semi-positively guarded;
$\mi{atoms}(S''-\mi{guards}(S''))\subseteq \mi{atoms}(S'')\subseteq \mi{atoms}(S)$,
$\mi{atoms}(S'')=\mi{atoms}((S''-\mi{guards}(S''))\cup \mi{guards}(S''))=\mi{atoms}(S''-\mi{guards}(S''))\cup \mi{atoms}(\mi{guards}(S''))=
                 \mi{atoms}(S''-\mi{guards}(S''))\cup \mi{atoms}(\mi{guards}(S))=\mi{atoms}(S''-\mi{guards}(S''))\cup \mi{atoms}(S)=\mi{atoms}(S)$;
$S''$ is semi-positively guarded; 
$a^*\in \mi{atoms}(S'')$, $\mi{guards}(S'',a^*)=\{a^*\geql \gu\}$,
$a^*\not\in \mi{atoms}(C^\natural)$, $C^*\in \mi{restricted}(S)$,
$\mi{restricted}(S'')=\{C \,|\, C\in (S-\{C^*\})\cup \{C^\natural\}, a^*\in \mi{atoms}(C), C\neq a^*\geql \gu\}=\{C \,|\, C\in S-\{C^*\}, a^*\in \mi{atoms}(C), C\neq a^*\geql \gu\}=
                      \mi{restricted}(S)-\{C^*\}=\{C \,|\, C\in S, a^*\in \mi{atoms}(C), C\neq a^*\geql \gu\}-\{C^*\}\subset \mi{restricted}(S)$;
by the induction hypothesis for $S''$, there exists a finite linear {\it DPLL}-tree $\mi{Tree}''$ with the root $S''$ constructed using Rules (\cref{ceq4hr1111111})--(\cref{ceq4hr4}), (\cref{ceq4hr8}) satisfying
for its only leaf $S'''$ that either $\square\in S'''$, or $S'''\subseteq_{\mc F} \mi{OrdPropCl}$ is semi-positively guarded, $\mi{atoms}(S''')\subseteq \mi{atoms}(S'')-\{a^*\}=\mi{atoms}(S)-\{a^*\}$.
We put
\begin{equation} \notag
\mi{Tree}=\begin{array}[c]{c}
          S \\[0.4mm]
          \hline \\[-3.8mm]
          \mi{Tree}' \\[0.4mm]
          \hline \\[-3.8mm]
          \mi{Tree}''.
          \end{array}
\end{equation}
Hence, $\mi{Tree}$ is a finite linear {\it DPLL}-tree with the root $S$ constructed using Rules (\cref{ceq4hr1111111})--(\cref{ceq4hr4}), (\cref{ceq4hr8}) such that
for its only leaf $S'''$, either $\square\in S'''$, or $S'''\subseteq_{\mc F} \mi{OrdPropCl}$ is semi-positively guarded, $\mi{atoms}(S''')\subseteq \mi{atoms}(S)-\{a^*\}$.

So, in both Cases 1 and 2, the statement holds.
The induction is completed.
%
%
%
\end{proof}

\subsection{Full proof of Lemma \ref{le7}}
\label{S7.9}

\begin{proof}
We have $S\subseteq_{\mc F} \mi{OrdPropCl}$.
We proceed by induction on $\mi{atoms}(S)\subseteq_{\mc F} \mi{PropAtom}$.

Case 1 (the base case):
$\mi{atoms}(S)=\emptyset$.
We have that $S$ is $\gz$-guarded.
Then $S$ is simplified;
trivially, for all $a\in \mi{atoms}(S)=\emptyset$, $a$ is positively guarded in $S$;
$S$ is positively guarded.
We put $\mi{Tree}=S$.
Hence, $\mi{Tree}$ is a finite linear {\it DPLL}-tree with the root $S$ constructed using Rules (\cref{ceq4hr1111111})--(\cref{ceq4hr4}), (\cref{ceq4hr7}), (\cref{ceq4hr8}) such that
for its only leaf $S$, $S\subseteq_{\mc F} \mi{OrdPropCl}$ is positively guarded.

Case 2 (the induction case):
$\emptyset\neq \mi{atoms}(S)\subseteq_{\mc F} \mi{PropAtom}$.
We have that $S$ is $\gz$-guarded.
We distinguish two cases for $S$.

Case 2.1:
For all $a\in \mi{atoms}(S)$, either $\mi{guards}(S,a)=\{\gz\gle a\}$ or $\mi{guards}(S,a)=\{\gz\gle a,a\gle \gu\}$.
Then $S$ is simplified;
for all $a\in \mi{atoms}(S)$, $a$ is positively guarded in $S$;
$S$ is positively guarded.
We put $\mi{Tree}=S$.
Hence, $\mi{Tree}$ is a finite linear {\it DPLL}-tree with the root $S$ constructed using Rules (\cref{ceq4hr1111111})--(\cref{ceq4hr4}), (\cref{ceq4hr7}), (\cref{ceq4hr8}) such that
for its only leaf $S$, $S\subseteq_{\mc F} \mi{OrdPropCl}$ is positively guarded.

Case 2.2:
There exists $a^*\in \mi{atoms}(S)$ such that either $\mi{guards}(S,a^*)=\{a^*\geql \gz\}$ or $\mi{guards}(S,a^*)=\{a^*\geql \gu\}$.
Then, by Lemma \ref{le6}(i,ii), there exists a finite linear {\it DPLL}-tree $\mi{Tree}'$ with the root $S$ constructed using Rules (\cref{ceq4hr1111111})--(\cref{ceq4hr4}), (\cref{ceq4hr7}), (\cref{ceq4hr8}) such that
for its only leaf $S'$, either $\square\in S'$, or $S'\subseteq_{\mc F} \mi{OrdPropCl}$ is $\gz$-guarded, $\mi{atoms}(S')\subseteq \mi{atoms}(S)-\{a^*\}$.
We get two cases for $S'$.

Case 2.2.1:
$\square\in S'$.
We put $\mi{Tree}=\mi{Tree}'$.
Hence, $\mi{Tree}$ is a finite linear {\it DPLL}-tree with the root $S$ constructed using Rules (\cref{ceq4hr1111111})--(\cref{ceq4hr4}), (\cref{ceq4hr7}), (\cref{ceq4hr8}) such that
for its only leaf $S'$, $\square\in S'$.

Case 2.2.2:
$S'\subseteq_{\mc F} \mi{OrdPropCl}$ is $\gz$-guarded, $\mi{atoms}(S')\subseteq \mi{atoms}(S)-\{a^*\}$.
Then $a^*\in \mi{atoms}(S)$, $\mi{atoms}(S')\subseteq \mi{atoms}(S)-\{a^*\}\subset \mi{atoms}(S)$;
by the induction hypothesis for $S'$, there exists a finite linear {\it DPLL}-tree $\mi{Tree}''$ with the root $S'$ constructed using Rules (\cref{ceq4hr1111111})--(\cref{ceq4hr4}), (\cref{ceq4hr7}), (\cref{ceq4hr8}) satisfying
for its only leaf $S''$ that either $\square\in S''$ or $S''\subseteq_{\mc F} \mi{OrdPropCl}$ is positively guarded.
We put
\begin{equation} \notag
\mi{Tree}=\dfrac{\mi{Tree}'} 
                {\mi{Tree}''}.    
\end{equation}
Hence, $\mi{Tree}$ is a finite linear {\it DPLL}-tree with the root $S$ constructed using Rules (\cref{ceq4hr1111111})--(\cref{ceq4hr4}), (\cref{ceq4hr7}), (\cref{ceq4hr8}) such that
for its only leaf $S''$, either $\square\in S''$ or $S''\subseteq_{\mc F} \mi{OrdPropCl}$ is positively guarded.

So, in both Cases 1 and 2, the statement holds.
The induction is completed.
%
%
%
\end{proof}

\subsection{Full proof of Lemma \ref{le88}}
\label{S7.10a}

\begin{proof}
We have that $S$ is simplified.
We proceed by induction on $n$.

Case 1 (the base case):
$n=0$.
We put $\mi{Tree}=S$.
Hence, $\mi{Tree}$ is a finite linear {\it DPLL}-tree with the root $S$ constructed using Rules (\cref{ceq4hr22}) and (\cref{ceq4hr4}) such that
for its only leaf $S$, $S=S-\{\gz\gle a_1,\dots,\gz\gle a_0\}\subseteq_{\mc F} \mi{OrdPropCl}$ is simplified.

Case 2 (the induction case):
$n\geq 1$.
Then $S\supseteq \mi{guards}(S)\supseteq \mi{guards}(S,a_n)=\{\gz\gle a_n,a_n\geql \gu\}$,
$a_n\geql \gu\in \mi{guards}(S)$, $\gz\gle a_n\in S$,
$a_n\in \mi{atoms}(\gz\gle a_n)$, $a_n\geql \gu\neq \gz\gle a_n$, $\mi{simplify}(\gz\gle a_n,a_n,\gu)=\gz\gle \gu$,
applying Rule (\cref{ceq4hr4}) to $a_n\geql \gu$ and $\gz\gle a_n$, we derive
\begin{equation} \notag
\dfrac{S}
      {(S-\{\gz\gle a_n\})\cup \{\gz\gle \gu\}};
\end{equation}
$\gz\gle \gu\in (S-\{\gz\gle a_n\})\cup \{\gz\gle \gu\}$;
$\gz\gle \gu\in \mi{OrdPropLit}$ is a tautology;
$\gz\gle \gu$ is not a guard;
$\gz\gle \gu\not\in \mi{guards}((S-\{\gz\gle a_n\})\cup \{\gz\gle \gu\})$,
$\gz\gle \gu\in ((S-\{\gz\gle a_n\})\cup \{\gz\gle \gu\})-\mi{guards}((S-\{\gz\gle a_n\})\cup \{\gz\gle \gu\})$,
applying Rule (\cref{ceq4hr22}) to $(S-\{\gz\gle a_n\})\cup \{\gz\gle \gu\}$ and $\gz\gle \gu$, we derive
\begin{equation} \notag
\dfrac{(S-\{\gz\gle a_n\})\cup \{\gz\gle \gu\}}
      {((S-\{\gz\gle a_n\})\cup \{\gz\gle \gu\})-\{\gz\gle \gu\}}.
\end{equation}
We have that $S$ is simplified.
Hence, $\gz\gle \gu\not\in S$, $\gz\gle a_n\in S$,
$((S-\{\gz\gle a_n\})\cup \{\gz\gle \gu\})-\{\gz\gle \gu\}=((S-\{\gz\gle a_n\})-\{\gz\gle \gu\})\cup (\{\gz\gle \gu\}-\{\gz\gle \gu\})=S-\{\gz\gle a_n\}$.
We put $S'=S-\{\gz\gle a_n\}\subseteq_{\mc F} \mi{OrdPropCl}$.
We get that
$\mi{atoms}(S')=\mi{atoms}(S-\{\gz\gle a_n\})\subseteq \mi{atoms}(S)$;
we have that $S$ is simplified;
$S'=S-\{\gz\gle a_n\}\subseteq S$ is simplified;
for all $i\leq n-1$, 
$a_i\geql \gu\in \mi{guards}(S,a_i)=\{\gz\gle a_i,a_i\geql \gu\}\subseteq S$, $a_i\geql \gu\neq \gz\gle a_n$, $a_i\geql \gu\in S-\{\gz\gle a_n\}$,
$a_i\in \mi{atoms}(a_i\geql \gu)\subseteq \mi{atoms}(S-\{\gz\gle a_n\})=\mi{atoms}(S')$;
$a_i\neq a_n$,
$\mi{guards}(a_i)\cap \{\gz\gle a_n\}=\mi{guards}(a_i)\cap \mi{guards}(a_n)=\emptyset$,
$\mi{guards}(S',a_i)=(S-\{\gz\gle a_n\})\cap \mi{guards}(a_i)=(S\cap \mi{guards}(a_i))-\{\gz\gle a_n\}=S\cap \mi{guards}(a_i)=\mi{guards}(S,a_i)=\{\gz\gle a_i,a_i\geql \gu\}$;
$\{a_1,\dots,a_{n-1}\}\subseteq \mi{atoms}(S')$;
by the induction hypothesis for $n-1$ and $S'$, there exists a finite linear {\it DPLL}-tree $\mi{Tree}'$ with the root $S'$ constructed using Rules (\cref{ceq4hr22}) and (\cref{ceq4hr4}) satisfying
for its only leaf $S''$ that 
$S''=S'-\{\gz\gle a_1,\dots,\gz\gle a_{n-1}\}=(S-\{\gz\gle a_n\})-\{\gz\gle a_1,\dots,\gz\gle a_{n-1}\}=S-\{\gz\gle a_1,\dots,\gz\gle a_n\}\subseteq_{\mc F} \mi{OrdPropCl}$ is simplified.
We put
\begin{equation} \notag
\mi{Tree}=\begin{array}[c]{c}
          S \\[0.4mm]
          \hline \\[-3.8mm]
          (S-\{\gz\gle a_n\})\cup \{\gz\gle \gu\} \\[0.4mm]
          \hline \\[-3.8mm]
          \mi{Tree}'.
          \end{array}
\end{equation}
Hence, $\mi{Tree}$ is a finite linear {\it DPLL}-tree with the root $S$ constructed using Rules (\cref{ceq4hr22}) and (\cref{ceq4hr4}) such that
for its only leaf $S''$, $S''=S-\{\gz\gle a_1,\dots,\gz\gle a_n\}\subseteq_{\mc F} \mi{OrdPropCl}$ is simplified.

So, in both Cases 1 and 2, the statement holds.
The induction is completed.
%
%
%
\end{proof}

\subsection{Full proof of Lemma \ref{le8}}
\label{S7.10b}

\begin{proof}
Let $S^F\subseteq \mi{OrdPropCl}$ and $C^F\in S^F-\mi{guards}(S^F)$.
We define a binary measure partial operator $\mi{invalid} : \mi{OrdPropCl}\times {\mc P}(\mi{OrdPropCl})\longrightarrow {\mc P}(\mi{OrdPropLit})$,
$\mi{invalid}(C^F,S^F)=\{l \,|\, l\in C^F,\ \text{\it not}\ \mi{valid}(l,S^F)\}$.
Then $C^*\in S\subseteq_{\mc F} \mi{OrdPropCl}$.
We proceed by induction on $\mi{invalid}(C^*,S)\subseteq C^*\subseteq_{\mc F} \mi{OrdPropLit}$.

Case 1 (the base case):
$\mi{invalid}(C^*,S)=\emptyset$.
We have that $S$ is positively guarded.
Then, for all $l\in C^*$, $\mi{valid}(l,S)$, $l\in \mi{OrdPropLit}^\gu$;
$C^*\in \mi{OrdPropCl}^\gu$.
We get two cases for $C^*$.

Case 1.1:
$C^*=(\bigvee_{i=0}^n \Cn_i\gle \gu)\vee C^\natural$, $\Cn_i\in \mi{PropConj}$, $n\geq 1$, 
$C^\natural\in \mi{OrdPropCl}^\gu$ does not contain an order literal of the form $\Cn\gle \gu$, $\Cn\in \mi{PropConj}$.
Then $C^*\in S-\mi{guards}(S)$, applying Rule (\cref{ceq4hr1111}) to $C^*$ and $C^\natural$, we derive
\begin{equation} \notag
\dfrac{S}
      {(S-\{C^*\})\cup \{(\bigmult_{i=0}^n \Cn_i)\gle \gu\vee C^\natural\}}.
\end{equation}
Hence, $C^*\in S$;
we have that $C^\natural$ does not contain an order literal of the form $\Cn\gle \gu$, $\Cn\in \mi{PropConj}$;
$(\bigmult_{i=0}^n \Cn_i)\gle \gu\not\in C^\natural$.
We put $S'=(S-\{C^*\})\cup \{(\bigmult_{i=0}^n \Cn_i)\gle \gu\vee C^\natural\}\subseteq_{\mc F} \mi{OrdPropCl}$.
We get two cases for $(\bigmult_{i=0}^n \Cn_i)\gle \gu\vee C^\natural$.

Case 1.1.1:
$(\bigmult_{i=0}^n \Cn_i)\gle \gu\vee C^\natural\in S$.
Then $n\geq 1$, $(\bigmult_{i=0}^n \Cn_i)\gle \gu\neq \bigvee_{i=0}^n \Cn_i\gle \gu$,
$(\bigmult_{i=0}^n \Cn_i)\gle \gu\vee C^\natural\neq C^*=(\bigvee_{i=0}^n \Cn_i\gle \gu)\vee C^\natural$,
$(\bigmult_{i=0}^n \Cn_i)\gle \gu\vee C^\natural\in S-\{C^*\}$,
$S'=(S-\{C^*\})\cup \{(\bigmult_{i=0}^n \Cn_i)\gle \gu\vee C^\natural\}=S-\{C^*\}$;
$S$ is simplified;
$S'=S-\{C^*\}\subseteq S$ is simplified;
$C^*\not\in \mi{guards}(S)$,
$\mi{guards}(S')=\{C \,|\, C\in S-\{C^*\}\ \text{\it is a guard}\}=\mi{guards}(S)-\{C^*\}=\{C \,|\, C\in S\ \text{\it is a guard}\}-\{C^*\}=\mi{guards}(S)$;
we have that $S$ is positively guarded;
$\mi{atoms}(S'-\mi{guards}(S'))\subseteq \mi{atoms}(S')=\mi{atoms}(S-\{C^*\})\subseteq \mi{atoms}(S)$,
$\mi{atoms}(S')=\mi{atoms}((S'-\mi{guards}(S'))\cup \mi{guards}(S'))=\mi{atoms}(S'-\mi{guards}(S'))\cup \mi{atoms}(\mi{guards}(S'))=\mi{atoms}(S'-\mi{guards}(S'))\cup \mi{atoms}(\mi{guards}(S))=
                \mi{atoms}(S'-\mi{guards}(S'))\cup \mi{atoms}(S)=\mi{atoms}(S)$;
$S'$ is positively guarded.
We put
\begin{equation} \notag
\mi{Tree}=\dfrac{S}
                {S'}.
\end{equation}
Hence, $\mi{Tree}$ is a finite linear {\it DPLL}-tree with the root $S$ constructed using Rules (\cref{ceq4hr1111})--(\cref{ceq4hr66}) such that
for its only leaf $S'$, $S'\subseteq_{\mc F} \mi{OrdPropCl}$ is positively guarded, $S'=S-\{C^*\}$, $\mi{guards}(S')=\mi{guards}(S)$.

Case 1.1.2:
$(\bigmult_{i=0}^n \Cn_i)\gle \gu\vee C^\natural\not\in S$.
We get two cases for $\bigmult_{i=0}^n \Cn_i$.

Case 1.1.2.1:
$\bigmult_{i=0}^n \Cn_i=a_0^{\alpha_0}\swedge\cdots\swedge a_m^{\alpha_m}$, $a_j\in \mi{PropAtom}$, $\alpha_j\geq 1$, and there exists $j^*\leq m$ such that $\alpha_{j^*}\geq 2$.
Then $(\bigmult_{i=0}^n \Cn_i)\gle \gu=a_0^{\alpha_0}\swedge\cdots\swedge a_m^{\alpha_m}\gle \gu\not\in C^\natural$,
$(\bigmult_{i=0}^n \Cn_i)\gle \gu\vee C^\natural=a_0^{\alpha_0}\swedge\cdots\swedge a_m^{\alpha_m}\gle \gu\vee C^\natural\not\in S$,
$S'=(S-\{C^*\})\cup \{(\bigmult_{i=0}^n \Cn_i)\gle \gu\vee C^\natural\}=(S-\{C^*\})\cup \{a_0^{\alpha_0}\swedge\cdots\swedge a_m^{\alpha_m}\gle \gu\vee C^\natural\}$,
$a_0^{\alpha_0}\swedge\cdots\swedge a_m^{\alpha_m}\gle \gu\vee C^\natural\in S'$;
$a_0^{\alpha_0}\swedge\cdots\swedge a_m^{\alpha_m}\gle \gu\vee C^\natural$ is not a guard;
$a_0^{\alpha_0}\swedge\cdots\swedge a_m^{\alpha_m}\gle \gu\vee C^\natural\not\in \mi{guards}(S')$,
$a_0^{\alpha_0}\swedge\cdots\swedge a_m^{\alpha_m}\gle \gu\vee C^\natural\in S'-\mi{guards}(S')$,
applying Rule (\cref{ceq4hr11111}) to $S'$, $a_0^{\alpha_0}\swedge\cdots\swedge a_m^{\alpha_m}\gle \gu\vee C^\natural$, $a_0^{\alpha_0}\swedge\cdots\swedge a_m^{\alpha_m}\gle \gu$,
                                      $C^\natural$, and $j^*$, we derive
\begin{equation} \notag
\dfrac{S'}
      {(S'-\{a_0^{\alpha_0}\swedge\cdots\swedge a_m^{\alpha_m}\gle \gu\vee C^\natural\})\cup \{C^\natural\cup \{a_0\swedge\cdots\swedge a_m\gle \gu\}\}}.
\end{equation}
Hence, $a_0^{\alpha_0}\swedge\cdots\swedge a_m^{\alpha_m}\gle \gu\vee C^\natural\not\in S$, $C^*\in S$;
we have that $C^\natural$ does not contain an order literal of the form $\Cn\gle \gu$, $\Cn\in \mi{PropConj}$;
$a_0\swedge\cdots\swedge a_m\gle \gu\not\in C^\natural$, $C^\natural\cup \{a_0\swedge\cdots\swedge a_m\gle \gu\}=a_0\swedge\cdots\swedge a_m\gle \gu\vee C^\natural$,
$(S'-\{a_0^{\alpha_0}\swedge\cdots\swedge a_m^{\alpha_m}\gle \gu\vee C^\natural\})\cup \{C^\natural\cup \{a_0\swedge\cdots\swedge a_m\gle \gu\}\}=
 (S'-\{a_0^{\alpha_0}\swedge\cdots\swedge a_m^{\alpha_m}\gle \gu\vee C^\natural\})\cup \{a_0\swedge\cdots\swedge a_m\gle \gu\vee C^\natural\}=
 (((S-\{C^*\})\cup \{a_0^{\alpha_0}\swedge\cdots\swedge a_m^{\alpha_m}\gle \gu\vee C^\natural\})-\{a_0^{\alpha_0}\swedge\cdots\swedge a_m^{\alpha_m}\gle \gu\vee C^\natural\})\cup 
 \{a_0\swedge\cdots\swedge a_m\gle \gu\vee C^\natural\}=
 ((S-\{C^*\})-\{a_0^{\alpha_0}\swedge\cdots\swedge a_m^{\alpha_m}\gle \gu\vee C^\natural\})\cup 
 (\{a_0^{\alpha_0}\swedge\cdots\swedge a_m^{\alpha_m}\gle \gu\vee C^\natural\}-\{a_0^{\alpha_0}\swedge\cdots\swedge a_m^{\alpha_m}\gle \gu\vee C^\natural\})\cup 
 \{a_0\swedge\cdots\swedge a_m\gle \gu\vee C^\natural\}=
 (S-\{C^*\})\cup \{a_0\swedge\cdots\swedge a_m\gle \gu\vee C^\natural\}$.
We put $S''=(S-\{C^*\})\cup \{a_0\swedge\cdots\swedge a_m\gle \gu\vee C^\natural\}\subseteq_{\mc F} \mi{OrdPropCl}$.
We get two cases for $a_0\swedge\cdots\swedge a_m\gle \gu\vee C^\natural$.

Case 1.1.2.1.1:
$a_0\swedge\cdots\swedge a_m\gle \gu\vee C^\natural\in S$.
Then $n\geq 1$, $a_0\swedge\cdots\swedge a_m\gle \gu\neq \bigvee_{i=0}^n \Cn_i\gle \gu$,
$a_0\swedge\cdots\swedge a_m\gle \gu\vee C^\natural\neq C^*=(\bigvee_{i=0}^n \Cn_i\gle \gu)\vee C^\natural$,
$a_0\swedge\cdots\swedge a_m\gle \gu\vee C^\natural\in S-\{C^*\}$,
$S''=(S-\{C^*\})\cup \{a_0\swedge\cdots\swedge a_m\gle \gu\vee C^\natural\}=S-\{C^*\}$.
We get from Case 1.1.1 for $S''$ that $\mi{guards}(S'')=\mi{guards}(S)$ and $S''$ is positively guarded.
We put
\begin{equation} \notag
\mi{Tree}=\begin{array}[c]{c}
          S \\[0.4mm]
          \hline \\[-3.8mm]
          S' \\[0.4mm]
          \hline \\[-3.8mm]
          S''.
          \end{array}
\end{equation}
Hence, $\mi{Tree}$ is a finite linear {\it DPLL}-tree with the root $S$ constructed using Rules (\cref{ceq4hr1111})--(\cref{ceq4hr66}) such that
for its only leaf $S''$, $S''\subseteq_{\mc F} \mi{OrdPropCl}$ is positively guarded, $S''=S-\{C^*\}$, $\mi{guards}(S'')=\mi{guards}(S)$.

Case 1.1.2.1.2:
$a_0\swedge\cdots\swedge a_m\gle \gu\vee C^\natural\not\in S$.
Then 
$\mi{atoms}(a_0\swedge\cdots\swedge a_m\gle \gu\vee C^\natural)=\mi{atoms}(a_0^{\alpha_0}\swedge\cdots\swedge a_m^{\alpha_m}\gle \gu\vee C^\natural)=
                                                                \mi{atoms}((\bigmult_{i=0}^n \Cn_i)\gle \gu\vee C^\natural)=\mi{atoms}((\bigvee_{i=0}^n \Cn_i\gle \gu)\vee C^\natural)=\mi{atoms}(C^*)$,
$C^*\in S$,
$\mi{atoms}(S'')=\mi{atoms}((S-\{C^*\})\cup \{a_0\swedge\cdots\swedge a_m\gle \gu\vee C^\natural\})=\mi{atoms}(S-\{C^*\})\cup \mi{atoms}(a_0\swedge\cdots\swedge a_m\gle \gu\vee C^\natural)=
                 \mi{atoms}(S-\{C^*\})\cup \mi{atoms}(C^*)=\mi{atoms}((S-\{C^*\})\cup \{C^*\})=\mi{atoms}(S)$;
$S$ is simplified; 
$S-\{C^*\}\subseteq S$ is simplified;
$C^*\in S$ does not contain contradictions and tautologies;
$C^\natural\subseteq C^*$ does not contain contradictions and tautologies;
$a_0\swedge\cdots\swedge a_m\gle \gu$ is not a contradiction or tautology;
$a_0\swedge\cdots\swedge a_m\gle \gu\vee C^\natural\neq \square$ does not contain contradictions and tautologies;
$S''=(S-\{C^*\})\cup \{a_0\swedge\cdots\swedge a_m\gle \gu\vee C^\natural\}$ is simplified.
We get two cases for $m$ and $C^\natural$.

Case 1.1.2.1.2.1:
$m=0$ and $C^\natural=\square$.
Then $a_0\swedge\cdots\swedge a_m\gle \gu\vee C^\natural=a_0\gle \gu\not\in S\supseteq \mi{guards}(S)$, 
$S''=(S-\{C^*\})\cup \{a_0\swedge\cdots\swedge a_m\gle \gu\vee C^\natural\}=(S-\{C^*\})\cup \{a_0\gle \gu\}$, $a_0\in \mi{atoms}(S'')=\mi{atoms}(S)$,
$a_0\in \mi{atoms}(a_0\swedge\cdots\swedge a_m\gle \gu\vee C^\natural)=\mi{atoms}(C^*)$; 
$C^*\in S$, $C^*\not\in \mi{guards}(S)\supseteq \mi{guards}(S,a_0)=S\cap \mi{guards}(a_0)$, $C^*\not\in \mi{guards}(a_0)$; 
$a_0$ is positively guarded in $S$;
$\mi{guards}(S'',a_0)=((S-\{C^*\})\cup \{a_0\gle \gu\})\cap \mi{guards}(a_0)=((S-\{C^*\})\cap \mi{guards}(a_0))\cup (\{a_0\gle \gu\}\cap \mi{guards}(a_0))=
                      ((S\cap \mi{guards}(a_0))-\{C^*\})\cup \{a_0\gle \gu\}=(S\cap \mi{guards}(a_0))\cup \{a_0\gle \gu\}=\mi{guards}(S,a_0)\cup \{a_0\gle \gu\}=\{\gz\gle a_0,a_0\gle \gu\}$;
$a_0$ is positively guarded in $S''$;
for all $a\in \mi{atoms}(S'')-\{a_0\}=\mi{atoms}(S)-\{a_0\}$,
$a$ is positively guarded in $S$;
$a\neq a_0$,
$\mi{guards}(a)\cap \{a_0\gle \gu\}=\mi{guards}(a)\cap \mi{guards}(a_0)=\emptyset$,
$C^*\not\in \mi{guards}(S)\supseteq \mi{guards}(S,a)=S\cap \mi{guards}(a)$, $C^*\not\in \mi{guards}(a)$,
$\mi{guards}(S'',a)=((S-\{C^*\})\cup \{a_0\gle \gu\})\cap \mi{guards}(a)=((S-\{C^*\})\cap \mi{guards}(a))\cup (\{a_0\gle \gu\}\cap \mi{guards}(a))=
                    (S\cap \mi{guards}(a))-\{C^*\}=S\cap \mi{guards}(a)=\mi{guards}(S,a)$;
$a$ is positively guarded in $S''$;
$S''$ is positively guarded; 
$\mi{guards}(S'')=\mi{guards}(S)\cup \{a_0\gle \gu\}$.
We put
\begin{equation} \notag
\mi{Tree}=\begin{array}[c]{c}
          S \\[0.4mm]
          \hline \\[-3.8mm]
          S' \\[0.4mm]
          \hline \\[-3.8mm]
          S''.
          \end{array}
\end{equation}
Hence, $\mi{Tree}$ is a finite linear {\it DPLL}-tree with the root $S$ constructed using Rules (\cref{ceq4hr1111})--(\cref{ceq4hr66}) such that
for its only leaf $S''$, $S''\subseteq_{\mc F} \mi{OrdPropCl}$ is positively guarded, and there exists $a_0\gle \gu\in \mi{OrdPropCl}^\gu$, $a_0\in \mi{atoms}(C^*)$, satisfying 
$S''=(S-\{C^*\})\cup \{a_0\gle \gu\}$, $\mi{guards}(S'')=\mi{guards}(S)\cup \{a_0\gle \gu\}$, $a_0\gle \gu\not\in S$.

Case 1.1.2.1.2.2:
$m\geq 1$ or $C^\natural\neq \square$.
Then $a_0\swedge\cdots\swedge a_m\gle \gu\vee C^\natural$ is not a guard;
$C^*\not\in \mi{guards}(S)$,
$\mi{guards}(S'')=\{C \,|\, C\in (S-\{C^*\})\cup \{a_0\swedge\cdots\swedge a_m\gle \gu\vee C^\natural\}\ \text{\it is a guard}\}=
                  \{C \,|\, C\in S-\{C^*\}\ \text{\it is a guard}\}=\mi{guards}(S)-\{C^*\}=\{C \,|\, C\in S\ \text{\it is a guard}\}-\{C^*\}=\mi{guards}(S)$;
$S''$ is positively guarded;
$a_0\swedge\cdots\swedge a_m\gle \gu\in \mi{OrdPropLit}^\gu$; 
for all $j\leq m$,
$a_j\in \mi{atoms}(S'')=\mi{atoms}(S)$,
$a_j\in \mi{atoms}(a_0^{\alpha_0}\swedge\cdots\swedge a_m^{\alpha_m})=\mi{atoms}(\bigmult_{i=0}^n \Cn_i)$; 
there exists $i^*\leq n$ satisfying $a_j\in \mi{atoms}(\Cn_{i^*})$;
$\Cn_{i^*}\gle \gu\in C^*$, $\mi{valid}(\Cn_{i^*}\gle \gu,S)$, $C^*\in S$;
for all $a\in \mi{atoms}(\Cn_{i^*})\subseteq \mi{atoms}(C^*)\subseteq \mi{atoms}(S)$, $\mi{guards}(S,a)=\{\gz\gle a\}$;
$\mi{guards}(S'',a_j)=\mi{guards}(S,a_j)=\{\gz\gle a_j\}$; 
$\mi{valid}(a_0\swedge\cdots\swedge a_m\gle \gu,S'')$;
for all $l\in C^\natural\subseteq C^*$, 
$\mi{valid}(l,S)$, $l\in \mi{OrdPropLit}^\gu$;
if either $l=\Cn\diamond \gu$ or $l=\gu\gleq \Cn$, $\Cn\in \mi{PropConj}$, and $\diamond\in \{\geql,\gle\}$, 
$\Cn=b_0\swedge\cdots\swedge b_k$, $b_j\in \mi{PropAtom}$; 
for all $j\leq k$, 
$b_j\in \mi{atoms}(\Cn)=\mi{atoms}(l)\subseteq \mi{atoms}(C^\natural)\subseteq \mi{atoms}(S'')=\mi{atoms}(S)$,
$\mi{guards}(S'',b_j)=\mi{guards}(S,b_j)=\{\gz\gle b_j\}$;
$\mi{valid}(l,S'')$.
We put $C^{**}=a_0\swedge\cdots\swedge a_m\gle \gu\vee C^\natural\in \mi{OrdPropCl}^\gu$.
We get that 
$C^{**}=a_0\swedge\cdots\swedge a_m\gle \gu\vee C^\natural\not\in S$,
$S''=(S-\{C^*\})\cup \{a_0\swedge\cdots\swedge a_m\gle \gu\vee C^\natural\}=(S-\{C^*\})\cup \{C^{**}\}$,
$\mi{atoms}(C^{**})=\mi{atoms}(a_0\swedge\cdots\swedge a_m\gle \gu\vee C^\natural)=\mi{atoms}(C^*)$;
$a_0\swedge\cdots\swedge a_m\gle \gu\vee C^\natural\neq \square$;
for all $l\in a_0\swedge\cdots\swedge a_m\gle \gu\vee C^\natural$, $\mi{valid}(l,S'')$;
we have that $C^\natural$ does not contain an order literal of the form $\Cn\gle \gu$, $\Cn\in \mi{PropConj}$;
$\mi{valid}(a_0\swedge\cdots\swedge a_m\gle \gu\vee C^\natural,S'')$,
$\mi{valid}(C^{**},S'')$.
We put
\begin{equation} \notag
\mi{Tree}=\begin{array}[c]{c}
          S \\[0.4mm]
          \hline \\[-3.8mm]
          S' \\[0.4mm]
          \hline \\[-3.8mm]
          S''.
          \end{array}
\end{equation}
Hence, $\mi{Tree}$ is a finite linear {\it DPLL}-tree with the root $S$ constructed using Rules (\cref{ceq4hr1111})--(\cref{ceq4hr66}) such that
for its only leaf $S''$, $S''\subseteq_{\mc F} \mi{OrdPropCl}$ is positively guarded, and there exists $C^{**}\in \mi{OrdPropCl}^\gu$, $\mi{atoms}(C^{**})=\mi{atoms}(C^*)$, satisfying 
$S''=(S-\{C^*\})\cup \{C^{**}\}$, $\mi{guards}(S'')=\mi{guards}(S)$, $C^{**}\not\in S$, $\mi{valid}(C^{**},S'')$.

Case 1.1.2.2:
$\bigmult_{i=0}^n \Cn_i=a_0\swedge\cdots\swedge a_m$, $a_j\in \mi{PropAtom}$.
Then $(\bigmult_{i=0}^n \Cn_i)\gle \gu=a_0\swedge\cdots\swedge a_m\gle \gu\not\in C^\natural$,
$(\bigmult_{i=0}^n \Cn_i)\gle \gu\vee C^\natural=a_0\swedge\cdots\swedge a_m\gle \gu\vee C^\natural\not\in S$,
$S'=(S-\{C^*\})\cup \{(\bigmult_{i=0}^n \Cn_i)\gle \gu\vee C^\natural\}=(S-\{C^*\})\cup \{a_0\swedge\cdots\swedge a_m\gle \gu\vee C^\natural\}$,
$\mi{atoms}(a_0\swedge\cdots\swedge a_m\gle \gu\vee C^\natural)=\mi{atoms}((\bigmult_{i=0}^n \Cn_i)\gle \gu\vee C^\natural)=\mi{atoms}((\bigvee_{i=0}^n \Cn_i\gle \gu)\vee C^\natural)=\mi{atoms}(C^*)$.
We get from Case 1.1.2.1.2 for $S'$ that $\mi{atoms}(S')=\mi{atoms}(S)$ and $S'$ is simplified.
We get two cases for $m$ and $C^\natural$.

Case 1.1.2.2.1:
$m=0$ and $C^\natural=\square$.
Then $a_0\swedge\cdots\swedge a_m\gle \gu\vee C^\natural=a_0\gle \gu\not\in S$, 
$S'=(S-\{C^*\})\cup \{a_0\swedge\cdots\swedge a_m\gle \gu\vee C^\natural\}=(S-\{C^*\})\cup \{a_0\gle \gu\}$,
$a_0\in \mi{atoms}(a_0\swedge\cdots\swedge a_m\gle \gu\vee C^\natural)=\mi{atoms}(C^*)$.
We get from Case 1.1.2.1.2.1 for $S'$ that $S'$ is positively guarded and $\mi{guards}(S')=\mi{guards}(S)\cup \{a_0\gle \gu\}$. 
We put
\begin{equation} \notag
\mi{Tree}=\dfrac{S}
                {S'}.
\end{equation}
Hence, $\mi{Tree}$ is a finite linear {\it DPLL}-tree with the root $S$ constructed using Rules (\cref{ceq4hr1111})--(\cref{ceq4hr66}) such that
for its only leaf $S'$, $S'\subseteq_{\mc F} \mi{OrdPropCl}$ is positively guarded, and there exists $a_0\gle \gu\in \mi{OrdPropCl}^\gu$, $a_0\in \mi{atoms}(C^*)$, satisfying 
$S'=(S-\{C^*\})\cup \{a_0\gle \gu\}$, $\mi{guards}(S')=\mi{guards}(S)\cup \{a_0\gle \gu\}$, $a_0\gle \gu\not\in S$.

Case 1.1.2.2.2:
$m\geq 1$ or $C^\natural\neq \square$.
We put $C^{**}=a_0\swedge\cdots\swedge a_m\gle \gu\vee C^\natural\in \mi{OrdPropCl}^\gu$.
Then $C^{**}=a_0\swedge\cdots\swedge a_m\gle \gu\vee C^\natural\not\in S$,
$S'=(S-\{C^*\})\cup \{a_0\swedge\cdots\swedge a_m\gle \gu\vee C^\natural\}=(S-\{C^*\})\cup \{C^{**}\}$,
$\mi{atoms}(C^{**})=\mi{atoms}(a_0\swedge\cdots\swedge a_m\gle \gu\vee C^\natural)=\mi{atoms}(C^*)$.
We get from Case 1.1.2.1.2.2 for $S'$ that $\mi{guards}(S')=\mi{guards}(S)$ and $S'$ is positively guarded.
Then $a_0\swedge\cdots\swedge a_m\gle \gu\in \mi{OrdPropLit}^\gu$; 
for all $j\leq m$,
$a_j\in \mi{atoms}(S')=\mi{atoms}(S)$,
$a_j\in \mi{atoms}(a_0\swedge\cdots\swedge a_m)=\mi{atoms}(\bigmult_{i=0}^n \Cn_i)$; 
there exists $i^*\leq n$ satisfying $a_j\in \mi{atoms}(\Cn_{i^*})$;
$\Cn_{i^*}\gle \gu\in C^*$, $\mi{valid}(\Cn_{i^*}\gle \gu,S)$, $C^*\in S$;
for all $a\in \mi{atoms}(\Cn_{i^*})\subseteq \mi{atoms}(C^*)\subseteq \mi{atoms}(S)$, $\mi{guards}(S,a)=\{\gz\gle a\}$;
$\mi{guards}(S',a_j)=\mi{guards}(S,a_j)=\{\gz\gle a_j\}$; 
$\mi{valid}(a_0\swedge\cdots\swedge a_m\gle \gu,S')$.
We get from Case 1.1.2.1.2.2 for $S'$ that $\mi{valid}(C^{**},S')$. 
We put
\begin{equation} \notag
\mi{Tree}=\dfrac{S}
                {S'}.
\end{equation}
Hence, $\mi{Tree}$ is a finite linear {\it DPLL}-tree with the root $S$ constructed using Rules (\cref{ceq4hr1111})--(\cref{ceq4hr66}) such that
for its only leaf $S'$, $S'\subseteq_{\mc F} \mi{OrdPropCl}$ is positively guarded, and there exists $C^{**}\in \mi{OrdPropCl}^\gu$, $\mi{atoms}(C^{**})=\mi{atoms}(C^*)$, satisfying 
$S'=(S-\{C^*\})\cup \{C^{**}\}$, $\mi{guards}(S')=\mi{guards}(S)$, $C^{**}\not\in S$, $\mi{valid}(C^{**},S')$.

Case 1.2:
Either $C^*=\Cn_0\gle \gu\vee C^\natural$ or $C^*=C^\natural$, $\Cn_0\in \mi{PropConj}$, 
$C^\natural\in \mi{OrdPropCl}^\gu$ does not contain an order literal of the form $\Cn\gle \gu$, $\Cn\in \mi{PropConj}$.
Then $C^*\in S$;
$S$ is simplified; 
$\square\not\in S$, $C^*\neq \square$;
for all $l\in C^*$, $\mi{valid}(l,S)$;
$\mi{valid}(C^*,S)$.
We put $\mi{Tree}=S$.
Hence, $\mi{Tree}$ is a finite linear {\it DPLL}-tree with the root $S$ constructed using Rules (\cref{ceq4hr1111})--(\cref{ceq4hr66}) such that
for its only leaf $S$, $S\subseteq_{\mc F} \mi{OrdPropCl}$ is positively guarded and $\mi{valid}(C^*,S)$.

Case 2 (the induction case):
$\emptyset\neq \mi{invalid}(C^*,S)\subseteq_{\mc F} \mi{OrdPropLit}$.
We have that $S$ is positively guarded.
Then there exists $l^*\in \mi{invalid}(C^*,S)\subseteq C^*$ not satisfying $\mi{valid}(l^*,S)$;
$S$ is simplified;
$l^*\in C^*\in S$ is not a contradiction or tautology;
either $l^*\not\in \mi{OrdPropLit}^\gu$, 
or $l^*=\Cn^*\diamond^* \gu$ or $l^*=\gu\gleq \Cn^*$, $\Cn^*\in \mi{PropConj}$, $\diamond^*\in \{\geql,\gle\}$, 
$\Cn^*=a_0^{\alpha_0}\swedge\cdots\swedge a_n^{\alpha_n}$, $a_i\in \mi{PropAtom}$, $\alpha_i\geq 1$, there exists $i^*\leq n$ satisfying that
$\alpha_{i^*}\geq 2$, 
or $a_{i^*}\in \mi{atoms}(\Cn^*)=\mi{atoms}(l^*)\subseteq \mi{atoms}(C^*)\subseteq \mi{atoms}(S)$, $a_{i^*}$ is positively guarded in $S$, $\mi{guards}(S,a_{i^*})=\{\gz\gle a_{i^*},a_{i^*}\gle \gu\}$.
We distinguish six cases for $l^*$.

Case 2.1:
Either $l^*=a_0^{\alpha_0}\swedge\cdots\swedge a_n^{\alpha_n}\geql \gz$ or $l^*=a_0^{\alpha_0}\swedge\cdots\swedge a_n^{\alpha_n}\gleq \gz$, $a_i\in \mi{PropAtom}$, $\alpha_i\geq 1$.
Then, for all $i\leq n$,
$a_i\in \mi{atoms}(l^*)\subseteq \mi{atoms}(S)$;
$a_i$ is positively guarded in $S$;
$\gz\gle a_i\in \mi{guards}(S,a_i)\subseteq \mi{guards}(S)$;
$l^*\in C^*\in S-\mi{guards}(S)$, applying Rule (\cref{ceq4hr5}) to $C^*$ and $l^*$, we derive
\begin{equation} \notag
\dfrac{S}
      {(S-\{C^*\})\cup \{C^*-\{l^*\}\}}.
\end{equation}

Case 2.2:
Either $l^*=a_0^{\alpha_0}\swedge\cdots\swedge a_n^{\alpha_n}\geql \gu$ or $l^*=\gu\gleq a_0^{\alpha_0}\swedge\cdots\swedge a_n^{\alpha_n}$, $a_i\in \mi{PropAtom}$, $\alpha_i\geq 1$, and 
there exists $i^*\leq n$ such that $a_{i^*}\in \mi{atoms}(l^*)\subseteq \mi{atoms}(S)$, $a_{i^*}$ is positively guarded in $S$, $\mi{guards}(S,a_{i^*})=\{\gz\gle a_{i^*},a_{i^*}\gle \gu\}$.
Then $a_{i^*}\gle \gu\in \mi{guards}(S,a_{i^*})\subseteq \mi{guards}(S)$, $l^*\in C^*\in S-\mi{guards}(S)$,
applying Rule (\cref{ceq4hr6}) to $C^*$ and $l^*$, we derive
\begin{equation} \notag
\dfrac{S}
      {(S-\{C^*\})\cup \{C^*-\{l^*\}\}}.
\end{equation}

In Cases 2.1 and 2.2, we put $S'=(S-\{C^*\})\cup \{C^*-\{l^*\}\}\subseteq_{\mc F} \mi{OrdPropCl}$.
We distinguish two cases for $C^*-\{l^*\}$.

Cases 2.1.1 and 2.2.1:
$C^*-\{l^*\}\in S$.
Then $l^*\in C^*$, $C^*-\{l^*\}\neq C^*$,
$C^*-\{l^*\}\in S-\{C^*\}$,
$S'=(S-\{C^*\})\cup \{C^*-\{l^*\}\}=S-\{C^*\}$;
these cases are the same as Case 1.1.1.

Cases 2.1.2 and 2.2.2:
$C^*-\{l^*\}\not\in S$.
We get two cases for $C^*-\{l^*\}$.

Cases 2.1.2.1 and 2.2.2.1:
$C^*-\{l^*\}=\square$.
Then $S'=(S-\{C^*\})\cup \{C^*-\{l^*\}\}=(S-\{C^*\})\cup \{\square\}$.
We put
\begin{equation} \notag
\mi{Tree}=\dfrac{S}
                {S'}.
\end{equation}
Hence, $\mi{Tree}$ is a finite linear {\it DPLL}-tree with the root $S$ constructed using Rules (\cref{ceq4hr1111})--(\cref{ceq4hr66}) such that
for its only leaf $S'$, $\square\in S'$.

Cases 2.1.2.2 and 2.2.2.2:
$C^*-\{l^*\}\neq \square$.
Then $C^*\in S$, $\mi{atoms}(C^*-\{l^*\})\subseteq \mi{atoms}(C^*)\subseteq \mi{atoms}(S)$,
$\mi{atoms}(S')=\mi{atoms}((S-\{C^*\})\cup \{C^*-\{l^*\}\})=\mi{atoms}(S-\{C^*\})\cup \mi{atoms}(C^*-\{l^*\})\subseteq \mi{atoms}(S)$;
$S$ is simplified; 
$S-\{C^*\}\subseteq S$ is simplified;
$C^*\in S$ does not contain contradictions and tautologies;
$\square\neq C^*-\{l^*\}\subseteq C^*$ does not contain contradictions and tautologies;
$S'=(S-\{C^*\})\cup \{C^*-\{l^*\}\}$ is simplified.
We get two cases for $C^*-\{l^*\}$.

Cases 2.1.2.2.1 and 2.2.2.2.1:
There exists $a^*\in \mi{atoms}(S')\subseteq \mi{atoms}(S)$ such that $C^*-\{l^*\}\in \mi{guards}(a^*)$.  
Then $a^*\in \mi{atoms}(C^*-\{l^*\})=\{a^*\}\subseteq \mi{atoms}(C^*)$,
$C^*\in S$, $C^*\not\in \mi{guards}(S)\supseteq \mi{guards}(S,a^*)=S\cap \mi{guards}(a^*)$, $C^*\not\in \mi{guards}(a^*)$,
$\mi{guards}(S',a^*)=((S-\{C^*\})\cup \{C^*-\{l^*\}\})\cap \mi{guards}(a^*)=((S-\{C^*\})\cap \mi{guards}(a^*))\cup (\{C^*-\{l^*\}\}\cap \mi{guards}(a^*))=
                     ((S\cap \mi{guards}(a^*))-\{C^*\})\cup \{C^*-\{l^*\}\}=(S\cap \mi{guards}(a^*))\cup \{C^*-\{l^*\}\}=\mi{guards}(S,a^*)\cup \{C^*-\{l^*\}\}$;
$a^*$ is positively guarded in $S$.
We get six cases for $C^*-\{l^*\}$.

Cases 2.1.2.2.1.1 and 2.2.2.2.1.1:
$C^*-\{l^*\}=a^*\geql \gz$.
We have that $a^*$ is positively guarded in $S$.
Then $S'\supseteq \mi{guards}(S')\supseteq \mi{guards}(S',a^*)=\mi{guards}(S,a^*)\cup \{C^*-\{l^*\}\}=\mi{guards}(S,a^*)\cup \{a^*\geql \gz\}\supseteq \{a^*\geql \gz,\gz\gle a^*\}$,
$a^*\geql \gz\in \mi{guards}(S')$, $\gz\gle a^*\in S'$,
$a^*\in \mi{atoms}(\gz\gle a^*)$, $a^*\geql \gz\neq \gz\gle a^*$, $\mi{simplify}(\gz\gle a^*,a^*,\gz)=\gz\gle \gz$,
applying Rule (\cref{ceq4hr3}) to $S'$, $a^*\geql \gz$, and $\gz\gle a^*$, we derive
\begin{equation} \notag
\dfrac{S'}
      {(S'-\{\gz\gle a^*\})\cup \{\gz\gle \gz\}};
\end{equation}
$\gz\gle \gz\in (S'-\{\gz\gle a^*\})\cup \{\gz\gle \gz\}$;
$\gz\gle \gz\in \mi{OrdPropLit}$ is a contradiction;
$\gz\gle \gz$ is not a guard;
$\gz\gle \gz\not\in \mi{guards}((S'-\{\gz\gle a^*\})\cup \{\gz\gle \gz\})$,
$\gz\gle \gz\in ((S'-\{\gz\gle a^*\})\cup \{\gz\gle \gz\})-\mi{guards}((S'-\{\gz\gle a^*\})\cup \{\gz\gle \gz\})$,
applying Rule (\cref{ceq4hr2}) to $(S'-\{\gz\gle a^*\})\cup \{\gz\gle \gz\}$ and $\gz\gle \gz$, we derive
\begin{equation} \notag
\dfrac{(S'-\{\gz\gle a^*\})\cup \{\gz\gle \gz\}}
      {(((S'-\{\gz\gle a^*\})\cup \{\gz\gle \gz\})-\{\gz\gle \gz\})\cup \{\square\}}.
\end{equation}
We put
\begin{equation} \notag
\mi{Tree}=\begin{array}[c]{c}
          S \\[0.4mm]
          \hline \\[-3.8mm]
          S' \\[0.4mm]
          \hline \\[-3.8mm]
          (S'-\{\gz\gle a^*\})\cup \{\gz\gle \gz\} \\[0.4mm]
          \hline \\[-3.8mm]
          S''=(((S'-\{\gz\gle a^*\})\cup \{\gz\gle \gz\})-\{\gz\gle \gz\})\cup \{\square\}.
          \end{array} 
\end{equation}
Hence, $\mi{Tree}$ is a finite linear {\it DPLL}-tree with the root $S$ constructed using Rules (\cref{ceq4hr1111})--(\cref{ceq4hr66}) such that
for its only leaf $S''$, $\square\in S''$.

Cases 2.1.2.2.1.2 and 2.2.2.2.1.2:
$C^*-\{l^*\}=a^*\gleq \gz$.
Then $\mi{guards}(S')\supseteq \mi{guards}(S',a^*)=\mi{guards}(S,a^*)\cup \{C^*-\{l^*\}\}=\mi{guards}(S,a^*)\cup \{a^*\gleq \gz\}$,
$a^*\gleq \gz\in \mi{guards}(S')$, applying Rule (\cref{ceq4hr1111111}) to $S'$ and $a^*\gleq \gz$, we derive
\begin{equation} \notag
\dfrac{S'}
      {(S'-\{a^*\gleq \gz\})\cup \{a^*\geql \gz\}};
\end{equation}
$a^*\in \mi{atoms}((S'-\{a^*\gleq \gz\})\cup \{a^*\geql \gz\})$;
we have that $a^*$ is positively guarded in $S$;
$(S'-\{a^*\gleq \gz\})\cup \{a^*\geql \gz\}\supseteq \mi{guards}((S'-\{a^*\gleq \gz\})\cup \{a^*\geql \gz\})\supseteq 
 \mi{guards}((S'-\{a^*\gleq \gz\})\cup \{a^*\geql \gz\},a^*)=((S'-\{a^*\gleq \gz\})\cup \{a^*\geql \gz\})\cap \mi{guards}(a^*)=
 ((S'-\{a^*\gleq \gz\})\cap \mi{guards}(a^*))\cup (\{a^*\geql \gz\}\cap \mi{guards}(a^*))=((S'\cap \mi{guards}(a^*))-\{a^*\gleq \gz\})\cup \{a^*\geql \gz\}=
 (\mi{guards}(S',a^*)-\{a^*\gleq \gz\})\cup \{a^*\geql \gz\}=((\mi{guards}(S,a^*)\cup \{a^*\gleq \gz\})-\{a^*\gleq \gz\})\cup \{a^*\geql \gz\}=
 (\mi{guards}(S,a^*)-\{a^*\gleq \gz\})\cup (\{a^*\gleq \gz\}-\{a^*\gleq \gz\})\cup \{a^*\geql \gz\}=(\mi{guards}(S,a^*)-\{a^*\gleq \gz\})\cup \{a^*\geql \gz\}\supseteq
 (\{\gz\gle a^*\}-\{a^*\gleq \gz\})\cup \{a^*\geql \gz\}=\{a^*\geql \gz,\gz\gle a^*\}$,
$a^*\geql \gz\in \mi{guards}((S'-\{a^*\gleq \gz\})\cup \{a^*\geql \gz\})$, $\gz\gle a^*\in (S'-\{a^*\gleq \gz\})\cup \{a^*\geql \gz\}$, 
$a^*\in \mi{atoms}(\gz\gle a^*)$, $a^*\geql \gz\neq \gz\gle a^*$, $\mi{simplify}(\gz\gle a^*,a^*,\gz)=\gz\gle \gz$,
applying Rule (\cref{ceq4hr3}) to $(S'-\{a^*\gleq \gz\})\cup \{a^*\geql \gz\}$, $a^*\geql \gz$, and $\gz\gle a^*$, we derive
\begin{equation} \notag
\dfrac{(S'-\{a^*\gleq \gz\})\cup \{a^*\geql \gz\}}
      {(((S'-\{a^*\gleq \gz\})\cup \{a^*\geql \gz\})-\{\gz\gle a^*\})\cup \{\gz\gle \gz\}};
\end{equation}
$\gz\gle \gz\in (((S'-\{a^*\gleq \gz\})\cup \{a^*\geql \gz\})-\{\gz\gle a^*\})\cup \{\gz\gle \gz\}$;
$\gz\gle \gz\in \mi{OrdPropLit}$ is a contradiction;
$\gz\gle \gz$ is not a guard;
$\gz\gle \gz\not\in \mi{guards}((((S'-\{a^*\gleq \gz\})\cup \{a^*\geql \gz\})-\{\gz\gle a^*\})\cup \{\gz\gle \gz\})$,
$\gz\gle \gz\in ((((S'-\{a^*\gleq \gz\})\cup \{a^*\geql \gz\})-\{\gz\gle a^*\})\cup \{\gz\gle \gz\})-\mi{guards}((((S'-\{a^*\gleq \gz\})\cup \{a^*\geql \gz\})-\{\gz\gle a^*\})\cup \{\gz\gle \gz\})$,
applying Rule (\cref{ceq4hr2}) to $(((S'-\{a^*\gleq \gz\})\cup \{a^*\geql \gz\})-\{\gz\gle a^*\})\cup \{\gz\gle \gz\}$ and $\gz\gle \gz$, we derive
\begin{equation} \notag
\dfrac{(((S'-\{a^*\gleq \gz\})\cup \{a^*\geql \gz\})-\{\gz\gle a^*\})\cup \{\gz\gle \gz\}}
      {(((((S'-\{a^*\gleq \gz\})\cup \{a^*\geql \gz\})-\{\gz\gle a^*\})\cup \{\gz\gle \gz\})-\{\gz\gle \gz\})\cup \{\square\}}.
\end{equation}
We put
\begin{equation} \notag
\mi{Tree}=\begin{array}[c]{c}
          S \\[0.4mm]
          \hline \\[-3.8mm]
          S' \\[0.4mm]
          \hline \\[-3.8mm]
          (S'-\{a^*\gleq \gz\})\cup \{a^*\geql \gz\} \\[0.4mm]
          \hline \\[-3.8mm]
          (((S'-\{a^*\gleq \gz\})\cup \{a^*\geql \gz\})-\{\gz\gle a^*\})\cup \{\gz\gle \gz\} \\[0.4mm]
          \hline \\[-3.8mm]
          S''=(((((S'-\{a^*\gleq \gz\})\cup \{a^*\geql \gz\})-\{\gz\gle a^*\})\cup \{\gz\gle \gz\})- \\
          \hfill \{\gz\gle \gz\})\cup \{\square\}.
          \end{array} 
\end{equation}
Hence, $\mi{Tree}$ is a finite linear {\it DPLL}-tree with the root $S$ constructed using Rules (\cref{ceq4hr1111})--(\cref{ceq4hr66}) such that
for its only leaf $S''$, $\square\in S''$.

Cases 2.1.2.2.1.3 and 2.2.2.2.1.3:
$C^*-\{l^*\}=\gz\gle a^*$.
We have that $a^*$ is positively guarded in $S$.
Then $\gz\gle a^*\in \mi{guards}(S,a^*)\subseteq S$,
which is a contradiction with $C^*-\{l^*\}=\gz\gle a^*\not\in S$.

Cases 2.1.2.2.1.4 and 2.2.2.2.1.4:
$C^*-\{l^*\}=a^*\gle \gu$.
We have that $a^*$ is positively guarded in $S$.
We get two cases for $\mi{guards}(S,a^*)$.

Cases 2.1.2.2.1.4.1 and 2.2.2.2.1.4.1:
$\mi{guards}(S,a^*)=\{\gz\gle a^*\}$.
Then $C^*-\{l^*\}=a^*\gle \gu\not\in S\supseteq \mi{guards}(S)$, $S'=(S-\{C^*\})\cup \{C^*-\{l^*\}\}=(S-\{C^*\})\cup \{a^*\gle \gu\}$, 
$\mi{guards}(S',a^*)=\mi{guards}(S,a^*)\cup \{C^*-\{l^*\}\}=\{\gz\gle a^*,a^*\gle \gu\}$;
$a^*$ is positively guarded in $S'$;
$\mi{atoms}(S')\subseteq \mi{atoms}(S)$;
for all $a\in \mi{atoms}(S')-\{a^*\}\subseteq \mi{atoms}(S)-\{a^*\}$,
$a$ is positively guarded in $S$;
$a\neq a^*$,
$\mi{guards}(a)\cap \{a^*\gle \gu\}=\mi{guards}(a)\cap \mi{guards}(a^*)=\emptyset$,
$C^*\in S$, $C^*\not\in \mi{guards}(S)\supseteq \mi{guards}(S,a)=S\cap \mi{guards}(a)$, $C^*\not\in \mi{guards}(a)$,
$\mi{guards}(S',a)=((S-\{C^*\})\cup \{a^*\gle \gu\})\cap \mi{guards}(a)=((S-\{C^*\})\cap \mi{guards}(a))\cup (\{a^*\gle \gu\}\cap \mi{guards}(a))=
                   (S\cap \mi{guards}(a))-\{C^*\}=S\cap \mi{guards}(a)=\mi{guards}(S,a)$;
$a$ is positively guarded in $S'$;
$S'$ is positively guarded; 
$\mi{guards}(S')=\mi{guards}(S)\cup \{a^*\gle \gu\}$.
We put
\begin{equation} \notag
\mi{Tree}=\dfrac{S}
                {S'}.
\end{equation}
Hence, $\mi{Tree}$ is a finite linear {\it DPLL}-tree with the root $S$ constructed using Rules (\cref{ceq4hr1111})--(\cref{ceq4hr66}) such that
for its only leaf $S'$, $S'\subseteq_{\mc F} \mi{OrdPropCl}$ is positively guarded, and there exists $a^*\gle \gu\in \mi{OrdPropCl}^\gu$, $a^*\in \mi{atoms}(C^*)$, satisfying 
$S'=(S-\{C^*\})\cup \{a^*\gle \gu\}$, $\mi{guards}(S')=\mi{guards}(S)\cup \{a^*\gle \gu\}$, $a^*\gle \gu\not\in S$.

Cases 2.1.2.2.1.4.2 and 2.2.2.2.1.4.2:
$\mi{guards}(S,a^*)=\{\gz\gle a^*,a^*\gle \gu\}$.
Then $a^*\gle \gu\in \mi{guards}(S,a^*)\subseteq S$,
which is a contradiction with $C^*-\{l^*\}=a^*\gle \gu\not\in S$.

Cases 2.1.2.2.1.5 and 2.2.2.2.1.5:
$C^*-\{l^*\}=a^*\geql \gu$.
We have that $a^*$ is positively guarded in $S$.
We get two cases for $\mi{guards}(S,a^*)$.

Cases 2.1.2.2.1.5.1 and 2.2.2.2.1.5.1:
$\mi{guards}(S,a^*)=\{\gz\gle a^*\}$.
Then $C^*-\{l^*\}=a^*\geql \gu\not\in S\supseteq \mi{guards}(S)$,
$S'=(S-\{C^*\})\cup \{C^*-\{l^*\}\}=(S-\{C^*\})\cup \{a^*\geql \gu\}$,
$S\supseteq \mi{guards}(S)\supseteq \mi{guards}(S,a^*)=\{\gz\gle a^*\}$;
$S'\supseteq \mi{guards}(S')\supseteq \mi{guards}(S',a^*)=\mi{guards}(S,a^*)\cup \{C^*-\{l^*\}\}=\{\gz\gle a^*,a^*\geql \gu\}$,
$a^*\geql \gu\in \mi{guards}(S')$, $\gz\gle a^*\in S'$,
$a^*\in \mi{atoms}(\gz\gle a^*)$, $a^*\geql \gu\neq \gz\gle a^*$, $\mi{simplify}(\gz\gle a^*,a^*,\gu)=\gz\gle \gu$,
applying Rule (\cref{ceq4hr4}) to $S'$, $a^*\geql \gu$, and $\gz\gle a^*$, we derive
\begin{equation} \notag
\dfrac{S'}
      {(S'-\{\gz\gle a^*\})\cup \{\gz\gle \gu\}};
\end{equation}
$\gz\gle \gu\in (S'-\{\gz\gle a^*\})\cup \{\gz\gle \gu\}$;
$\gz\gle \gu\in \mi{OrdPropLit}$ is a tautology;
$\gz\gle \gu$ is not a guard;
$\gz\gle \gu\not\in \mi{guards}((S'-\{\gz\gle a^*\})\cup \{\gz\gle \gu\})$,
$\gz\gle \gu\in ((S'-\{\gz\gle a^*\})\cup \{\gz\gle \gu\})-\mi{guards}((S'-\{\gz\gle a^*\})\cup \{\gz\gle \gu\})$,
applying Rule (\cref{ceq4hr22}) to $(S'-\{\gz\gle a^*\})\cup \{\gz\gle \gu\}$ and $\gz\gle \gu$, we derive
\begin{equation} \notag
\dfrac{(S'-\{\gz\gle a^*\})\cup \{\gz\gle \gu\}}
      {((S'-\{\gz\gle a^*\})\cup \{\gz\gle \gu\})-\{\gz\gle \gu\}}.
\end{equation}
We put $S^{**}=\{a^*\geql \gu\}\subseteq \mi{OrdPropCl}^\gu$.
We have that $S'$ is simplified.
Hence, $\gz\gle \gu\not\in S'$, $C^*, \gz\gle a^*\in S$, $C^*\not\in \mi{guards}(a^*)$, $\gz\gle a^*\in \mi{guards}(a^*)$, $C^*\neq \gz\gle a^*$,
$a^*\geql \gu\not\in S\supseteq \mi{guards}(S)$, $S^{**}\cap S=S^{**}\cap \mi{guards}(S)=\{a^*\geql \gu\}\cap S=\{a^*\geql \gu\}\cap \mi{guards}(S)=\emptyset$,
$((S'-\{\gz\gle a^*\})\cup \{\gz\gle \gu\})-\{\gz\gle \gu\}=((S'-\{\gz\gle a^*\})-\{\gz\gle \gu\})\cup (\{\gz\gle \gu\}-\{\gz\gle \gu\})=S'-\{\gz\gle a^*\}=
 ((S-\{C^*\})\cup \{a^*\geql \gu\})-\{\gz\gle a^*\}=((S-\{C^*\})-\{\gz\gle a^*\})\cup (\{a^*\geql \gu\}-\{\gz\gle a^*\})=((S-\{C^*\})-\{\gz\gle a^*\})\cup \{a^*\geql \gu\}=
 (S-(\{C^*\}\cup \{\gz\gle a^*\}))\cup S^{**}$.
We put $S''=(S-(\{C^*\}\cup \{\gz\gle a^*\}))\cup S^{**}\subseteq_{\mc F} \mi{OrdPropCl}$.
We get that
$S''=(S-(\{C^*\}\cup \{\gz\gle a^*\}))\cup S^{**}=((S-\{C^*\})-\{\gz\gle a^*\})\cup \{a^*\geql \gu\}$,
$a^*\in \mi{atoms}(S)$,
$a^*\in \mi{atoms}(S'')=\mi{atoms}(((S-\{C^*\})-\{\gz\gle a^*\})\cup \{a^*\geql \gu\})=\mi{atoms}((S-\{C^*\})-\{\gz\gle a^*\})\cup \mi{atoms}(a^*\geql \gu)=
                        \mi{atoms}((S-\{C^*\})-\{\gz\gle a^*\})\cup \{a^*\}\subseteq \mi{atoms}(S)$;
$S$ is simplified;
$(S-\{C^*\})-\{\gz\gle a^*\}\subseteq S$ is simplified;
$a^*\geql \gu\neq \square$ does not contain contradictions and tautologies;
$S''=((S-\{C^*\})-\{\gz\gle a^*\})\cup \{a^*\geql \gu\}$ is simplified;
$C^*\in S$, $C^*\not\in \mi{guards}(S)\supseteq \mi{guards}(S,a^*)=S\cap \mi{guards}(a^*)$, $C^*\not\in \mi{guards}(a^*)$,
$\mi{guards}(S'',a^*)=(((S-\{C^*\})-\{\gz\gle a^*\})\cup \{a^*\geql \gu\})\cap \mi{guards}(a^*)=(((S-\{C^*\})-\{\gz\gle a^*\})\cap \mi{guards}(a^*))\cup (\{a^*\geql \gu\}\cap \mi{guards}(a^*))=
                      (((S\cap \mi{guards}(a^*))-\{\gz\gle a^*\})-\{C^*\})\cup \{a^*\geql \gu\}=((S\cap \mi{guards}(a^*))-\{\gz\gle a^*\})\cup \{a^*\geql \gu\}=
                      (\mi{guards}(S,a^*)-\{\gz\gle a^*\})\cup \{a^*\geql \gu\}=(\{\gz\gle a^*\}-\{\gz\gle a^*\})\cup \{a^*\geql \gu\}=\{a^*\geql \gu\}$;
$a^*$ is semi-positively guarded in $S''$; 
for all $a\in \mi{atoms}(S'')-\{a^*\}\subseteq \mi{atoms}(S)-\{a^*\}$,
$a$ is positively guarded in $S$;
$a\neq a^*$,
$\mi{guards}(a)\cap \{\gz\gle a^*,a^*\geql \gu\}=\mi{guards}(a)\cap \mi{guards}(a^*)=\emptyset$,
$C^*\not\in \mi{guards}(S)\supseteq \mi{guards}(S,a)=S\cap \mi{guards}(a)$, $C^*\not\in \mi{guards}(a)$,
$\mi{guards}(S'',a)=(((S-\{C^*\})-\{\gz\gle a^*\})\cup \{a^*\geql \gu\})\cap \mi{guards}(a)=(((S-\{C^*\})-\{\gz\gle a^*\})\cap \mi{guards}(a))\cup (\{a^*\geql \gu\}\cap \mi{guards}(a))=
                    ((S\cap \mi{guards}(a))-\{C^*\})-\{\gz\gle a^*\}=S\cap \mi{guards}(a)=\mi{guards}(S,a)$;
$a$ is positively guarded in $S''$;
$S''$ is semi-positively guarded; 
$\gz\gle a^*\in \mi{guards}(S)$, $S^{**}\cap \mi{guards}(S)=\emptyset$, 
$\mi{guards}(S'')=(\mi{guards}(S)-\{\gz\gle a^*\})\cup \{a^*\geql \gu\}=(\mi{guards}(S)-\{\gz\gle a^*\})\cup S^{**}$.
We put
\begin{equation} \notag
\mi{Tree}=\begin{array}[c]{c}
          S \\[0.4mm]
          \hline \\[-3.8mm]
          S' \\[0.4mm]
          \hline \\[-3.8mm]
          (S'-\{\gz\gle a^*\})\cup \{\gz\gle \gu\} \\[0.4mm]
          \hline \\[-3.8mm]
          S''.
          \end{array}
\end{equation}
Hence, $\mi{Tree}$ is a finite linear {\it DPLL}-tree with the root $S$ constructed using Rules (\cref{ceq4hr1111})--(\cref{ceq4hr66}) such that
for its only leaf $S''$, $S''\subseteq_{\mc F} \mi{OrdPropCl}$ is semi-positively guarded, and
there exists $S^{**}=\{a^*\geql \gu\}\subseteq \mi{OrdPropCl}^\gu$, $a^*\in \mi{atoms}(C^*)$, $\{\gz\gle a^*\}\subseteq \mi{guards}(S)$, satisfying 
$S''=(S-(\{C^*\}\cup \{\gz\gle a^*\}))\cup S^{**}$, $\mi{guards}(S'')=(\mi{guards}(S)-\{\gz\gle a^*\})\cup S^{**}$, $S^{**}\cap S=\emptyset$.

Cases 2.1.2.2.1.5.2 and 2.2.2.2.1.5.2:
$\mi{guards}(S,a^*)=\{\gz\gle a^*,a^*\gle \gu\}$.
Then $S'\supseteq \mi{guards}(S')\supseteq \mi{guards}(S',a^*)=\mi{guards}(S,a^*)\cup \{C^*-\{l^*\}\}=\{\gz\gle a^*,a^*\gle \gu,a^*\geql \gu\}$,
$a^*\geql \gu\in \mi{guards}(S')$, $a^*\gle \gu\in S'$,
$a^*\in \mi{atoms}(a^*\gle \gu)$, $a^*\geql \gu\neq a^*\gle \gu$, $\mi{simplify}(a^*\gle \gu,a^*,\gu)=\gu\gle \gu$,
applying Rule (\cref{ceq4hr4}) to $S'$, $a^*\geql \gu$, and $a^*\gle \gu$, we derive
\begin{equation} \notag
\dfrac{S'}
      {(S'-\{a^*\gle \gu\})\cup \{\gu\gle \gu\}};
\end{equation}
$\gu\gle \gu\in (S'-\{a^*\gle \gu\})\cup \{\gu\gle \gu\}$;
$\gu\gle \gu\in \mi{OrdPropLit}$ is a contradiction;
$\gu\gle \gu$ is not a guard;
$\gu\gle \gu\not\in \mi{guards}((S'-\{a^*\gle \gu\})\cup \{\gu\gle \gu\})$,
$\gu\gle \gu\in ((S'-\{a^*\gle \gu\})\cup \{\gu\gle \gu\})-\mi{guards}((S'-\{a^*\gle \gu\})\cup \{\gu\gle \gu\})$,
applying Rule (\cref{ceq4hr2}) to $(S'-\{a^*\gle \gu\})\cup \{\gu\gle \gu\}$ and $\gu\gle \gu$, we derive
\begin{equation} \notag
\dfrac{(S'-\{a^*\gle \gu\})\cup \{\gu\gle \gu\}}
      {(((S'-\{a^*\gle \gu\})\cup \{\gu\gle \gu\})-\{\gu\gle \gu\})\cup \{\square\}}.
\end{equation}
We put
\begin{equation} \notag
\mi{Tree}=\begin{array}[c]{c}
          S \\[0.4mm]
          \hline \\[-3.8mm]
          S' \\[0.4mm]
          \hline \\[-3.8mm]
          (S'-\{a^*\gle \gu\})\cup \{\gu\gle \gu\} \\[0.4mm]
          \hline \\[-3.8mm]
          S''=(((S'-\{a^*\gle \gu\})\cup \{\gu\gle \gu\})-\{\gu\gle \gu\})\cup \{\square\}.
          \end{array} 
\end{equation}
Hence, $\mi{Tree}$ is a finite linear {\it DPLL}-tree with the root $S$ constructed using Rules (\cref{ceq4hr1111})--(\cref{ceq4hr66}) such that
for its only leaf $S''$, $\square\in S''$.

Cases 2.1.2.2.1.6 and 2.2.2.2.1.6:
$C^*-\{l^*\}=\gu\gleq a^*$.
We have that $a^*$ is positively guarded in $S$.
We get two cases for $\mi{guards}(S,a^*)$.

Cases 2.1.2.2.1.6.1 and 2.2.2.2.1.6.1:
$\mi{guards}(S,a^*)=\{\gz\gle a^*\}$.
Then $C^*-\{l^*\}=\gu\gleq a^*\not\in S$,
$S'=(S-\{C^*\})\cup \{C^*-\{l^*\}\}=(S-\{C^*\})\cup \{\gu\gleq a^*\}$,
$S\supseteq \mi{guards}(S)\supseteq \mi{guards}(S,a^*)=\{\gz\gle a^*\}$;
$\mi{guards}(S')\supseteq \mi{guards}(S',a^*)=\mi{guards}(S,a^*)\cup \{C^*-\{l^*\}\}=\{\gz\gle a^*,\gu\gleq a^*\}$,
$\gu\gleq a^*\in \mi{guards}(S')$, applying Rule (\cref{ceq4hr11111111}) to $S'$ and $\gu\gleq a^*$, we derive
\begin{equation} \notag
\dfrac{S'}
      {(S'-\{\gu\gleq a^*\})\cup \{a^*\geql \gu\}};
\end{equation}
$a^*\in \mi{atoms}((S'-\{\gu\gleq a^*\})\cup \{a^*\geql \gu\})$,
$(S'-\{\gu\gleq a^*\})\cup \{a^*\geql \gu\}\supseteq \mi{guards}((S'-\{\gu\gleq a^*\})\cup \{a^*\geql \gu\})\supseteq 
 \mi{guards}((S'-\{\gu\gleq a^*\})\cup \{a^*\geql \gu\},a^*)=((S'-\{\gu\gleq a^*\})\cup \{a^*\geql \gu\})\cap \mi{guards}(a^*)=
 ((S'-\{\gu\gleq a^*\})\cap \mi{guards}(a^*))\cup (\{a^*\geql \gu\}\cap \mi{guards}(a^*))=((S'\cap \mi{guards}(a^*))-\{\gu\gleq a^*\})\cup \{a^*\geql \gu\}=
 (\mi{guards}(S',a^*)-\{\gu\gleq a^*\})\cup \{a^*\geql \gu\}=(\{\gz\gle a^*,\gu\gleq a^*\}-\{\gu\gleq a^*\})\cup \{a^*\geql \gu\}=\{\gz\gle a^*,a^*\geql \gu\}$,
$a^*\geql \gu\in \mi{guards}((S'-\{\gu\gleq a^*\})\cup \{a^*\geql \gu\})$, $\gz\gle a^*\in (S'-\{\gu\gleq a^*\})\cup \{a^*\geql \gu\}$,
$a^*\in \mi{atoms}(\gz\gle a^*)$, $a^*\geql \gu\neq \gz\gle a^*$, $\mi{simplify}(\gz\gle a^*,a^*,\gu)=\gz\gle \gu$,
applying Rule (\cref{ceq4hr4}) to $(S'-\{\gu\gleq a^*\})\cup \{a^*\geql \gu\}$, $a^*\geql \gu$, and $\gz\gle a^*$, we derive
\begin{equation} \notag
\dfrac{(S'-\{\gu\gleq a^*\})\cup \{a^*\geql \gu\}}
      {(((S'-\{\gu\gleq a^*\})\cup \{a^*\geql \gu\})-\{\gz\gle a^*\})\cup \{\gz\gle \gu\}};
\end{equation}
$\gz\gle \gu\in (((S'-\{\gu\gleq a^*\})\cup \{a^*\geql \gu\})-\{\gz\gle a^*\})\cup \{\gz\gle \gu\}$;
$\gz\gle \gu\in \mi{OrdPropLit}$ is a tautology;
$\gz\gle \gu$ is not a guard;
$\gz\gle \gu\not\in \mi{guards}((((S'-\{\gu\gleq a^*\})\cup \{a^*\geql \gu\})-\{\gz\gle a^*\})\cup \{\gz\gle \gu\})$,
$\gz\gle \gu\in ((((S'-\{\gu\gleq a^*\})\cup \{a^*\geql \gu\})-\{\gz\gle a^*\})\cup \{\gz\gle \gu\})-\mi{guards}((((S'-\{\gu\gleq a^*\})\cup \{a^*\geql \gu\})-\{\gz\gle a^*\})\cup \{\gz\gle \gu\})$,
applying Rule (\cref{ceq4hr22}) to $(((S'-\{\gu\gleq a^*\})\cup \{a^*\geql \gu\})-\{\gz\gle a^*\})\cup \{\gz\gle \gu\}$ and $\gz\gle \gu$, we derive
\begin{equation} \notag
\dfrac{(((S'-\{\gu\gleq a^*\})\cup \{a^*\geql \gu\})-\{\gz\gle a^*\})\cup \{\gz\gle \gu\}}
      {((((S'-\{\gu\gleq a^*\})\cup \{a^*\geql \gu\})-\{\gz\gle a^*\})\cup \{\gz\gle \gu\})-\{\gz\gle \gu\}}.
\end{equation}
We put $S^{**}=\{a^*\geql \gu\}\subseteq \mi{OrdPropCl}^\gu$.
We have that $S'$ is simplified.
Hence, $\gz\gle \gu\not\in S'$, $\gu\gleq a^*\not\in S$, $C^*, \gz\gle a^*\in S$, $C^*\not\in \mi{guards}(a^*)$, $\gz\gle a^*, a^*\geql \gu\in \mi{guards}(a^*)$, $C^*\neq \gz\gle a^*$,
$a^*\geql \gu\not\in \mi{guards}(S,a^*)=\{\gz\gle a^*\}=S\cap \mi{guards}(a^*)$, $a^*\geql \gu\not\in S\supseteq \mi{guards}(S)$,
$S^{**}\cap S=S^{**}\cap \mi{guards}(S)=\{a^*\geql \gu\}\cap S=\{a^*\geql \gu\}\cap \mi{guards}(S)=\emptyset$,
$((((S'-\{\gu\gleq a^*\})\cup \{a^*\geql \gu\})-\{\gz\gle a^*\})\cup \{\gz\gle \gu\})-\{\gz\gle \gu\}=
 ((((S'-\{\gu\gleq a^*\})\cup \{a^*\geql \gu\})-\{\gz\gle a^*\})-\{\gz\gle \gu\})\cup (\{\gz\gle \gu\}-\{\gz\gle \gu\})=
 (((S'-\{\gu\gleq a^*\})\cup \{a^*\geql \gu\})-\{\gz\gle \gu\})-\{\gz\gle a^*\}=
 (((S'-\{\gu\gleq a^*\})-\{\gz\gle \gu\})\cup (\{a^*\geql \gu\}-\{\gz\gle \gu\}))-\{\gz\gle a^*\}=
 ((S'-\{\gu\gleq a^*\})\cup \{a^*\geql \gu\})-\{\gz\gle a^*\}=
 ((S'-\{\gu\gleq a^*\})-\{\gz\gle a^*\})\cup (\{a^*\geql \gu\}-\{\gz\gle a^*\})=
 ((S'-\{\gu\gleq a^*\})-\{\gz\gle a^*\})\cup \{a^*\geql \gu\}=
 ((((S-\{C^*\})\cup \{\gu\gleq a^*\})-\{\gu\gleq a^*\})-\{\gz\gle a^*\})\cup \{a^*\geql \gu\}=
 ((((S-\{C^*\})-\{\gu\gleq a^*\})\cup (\{\gu\gleq a^*\}-\{\gu\gleq a^*\}))-\{\gz\gle a^*\})\cup \{a^*\geql \gu\}=
 ((S-\{C^*\})-\{\gz\gle a^*\})\cup \{a^*\geql \gu\}=(S-(\{C^*\}\cup \{\gz\gle a^*\}))\cup S^{**}$.
We put $S''=(S-(\{C^*\}\cup \{\gz\gle a^*\}))\cup S^{**}\subseteq_{\mc F} \mi{OrdPropCl}$.
Then $S''=(S-(\{C^*\}\cup \{\gz\gle a^*\}))\cup S^{**}=((S-\{C^*\})-\{\gz\gle a^*\})\cup \{a^*\geql \gu\}$, $\gz\gle a^*\in \mi{guards}(S)$, $S^{**}\cap \mi{guards}(S)=\emptyset$.
We get from Cases 2.1.2.2.1.5.1 and 2.2.2.2.1.5.1 that $S''$ is semi-positively guarded and $\mi{guards}(S'')=(\mi{guards}(S)-\{\gz\gle a^*\})\cup S^{**}$.
We put
\begin{equation} \notag
\mi{Tree}=\begin{array}[c]{c}
          S \\[0.4mm]
          \hline \\[-3.8mm]
          S' \\[0.4mm]
          \hline \\[-3.8mm]
          (S'-\{\gu\gleq a^*\})\cup \{a^*\geql \gu\} \\[0.4mm]
          \hline \\[-3.8mm]
          (((S'-\{\gu\gleq a^*\})\cup \{a^*\geql \gu\})-\{\gz\gle a^*\})\cup \{\gz\gle \gu\} \\[0.4mm]
          \hline \\[-3.8mm]
          S''.
          \end{array}
\end{equation}
Hence, $\mi{Tree}$ is a finite linear {\it DPLL}-tree with the root $S$ constructed using Rules (\cref{ceq4hr1111})--(\cref{ceq4hr66}) such that
for its only leaf $S''$, $S''\subseteq_{\mc F} \mi{OrdPropCl}$ is semi-positively guarded, and
there exists $S^{**}=\{a^*\geql \gu\}\subseteq \mi{OrdPropCl}^\gu$, $a^*\in \mi{atoms}(C^*)$, $\{\gz\gle a^*\}\subseteq \mi{guards}(S)$, satisfying 
$S''=(S-(\{C^*\}\cup \{\gz\gle a^*\}))\cup S^{**}$, $\mi{guards}(S'')=(\mi{guards}(S)-\{\gz\gle a^*\})\cup S^{**}$, $S^{**}\cap S=\emptyset$.

Cases 2.1.2.2.1.6.2 and 2.2.2.2.1.6.2:
$\mi{guards}(S,a^*)=\{\gz\gle a^*,a^*\gle \gu\}$.
Then $\mi{guards}(S')\supseteq \mi{guards}(S',a^*)=\mi{guards}(S,a^*)\cup \{C^*-\{l^*\}\}=\{\gz\gle a^*,a^*\gle \gu,\gu\gleq a^*\}$,
$\gu\gleq a^*\in \mi{guards}(S')$, applying Rule (\cref{ceq4hr11111111}) to $S'$ and $\gu\gleq a^*$, we derive
\begin{equation} \notag
\dfrac{S'}
      {(S'-\{\gu\gleq a^*\})\cup \{a^*\geql \gu\}};
\end{equation}
$a^*\in \mi{atoms}((S'-\{\gu\gleq a^*\})\cup \{a^*\geql \gu\})$,
$(S'-\{\gu\gleq a^*\})\cup \{a^*\geql \gu\}\supseteq \mi{guards}((S'-\{\gu\gleq a^*\})\cup \{a^*\geql \gu\})\supseteq 
 \mi{guards}((S'-\{\gu\gleq a^*\})\cup \{a^*\geql \gu\},a^*)=((S'-\{\gu\gleq a^*\})\cup \{a^*\geql \gu\})\cap \mi{guards}(a^*)=
 ((S'-\{\gu\gleq a^*\})\cap \mi{guards}(a^*))\cup (\{a^*\geql \gu\}\cap \mi{guards}(a^*))=((S'\cap \mi{guards}(a^*))-\{\gu\gleq a^*\})\cup \{a^*\geql \gu\}=
 (\mi{guards}(S',a^*)-\{\gu\gleq a^*\})\cup \{a^*\geql \gu\}=(\{\gz\gle a^*,a^*\gle \gu,\gu\gleq a^*\}-\{\gu\gleq a^*\})\cup \{a^*\geql \gu\}=\{\gz\gle a^*,a^*\gle \gu,a^*\geql \gu\}$,
$a^*\geql \gu\in \mi{guards}((S'-\{\gu\gleq a^*\})\cup \{a^*\geql \gu\})$, $a^*\gle \gu\in (S'-\{\gu\gleq a^*\})\cup \{a^*\geql \gu\}$,
$a^*\in \mi{atoms}(a^*\gle \gu)$, $a^*\geql \gu\neq a^*\gle \gu$, $\mi{simplify}(a^*\gle \gu,a^*,\gu)=\gu\gle \gu$,
applying Rule (\cref{ceq4hr4}) to $(S'-\{\gu\gleq a^*\})\cup \{a^*\geql \gu\}$, $a^*\geql \gu$, and $a^*\gle \gu$, we derive
\begin{equation} \notag
\dfrac{(S'-\{\gu\gleq a^*\})\cup \{a^*\geql \gu\}}
      {(((S'-\{\gu\gleq a^*\})\cup \{a^*\geql \gu\})-\{a^*\gle \gu\})\cup \{\gu\gle \gu\}};
\end{equation}
$\gu\gle \gu\in (((S'-\{\gu\gleq a^*\})\cup \{a^*\geql \gu\})-\{a^*\gle \gu\})\cup \{\gu\gle \gu\}$;
$\gu\gle \gu\in \mi{OrdPropLit}$ is a contradiction;
$\gu\gle \gu$ is not a guard;
$\gu\gle \gu\not\in \mi{guards}((((S'-\{\gu\gleq a^*\})\cup \{a^*\geql \gu\})-\{a^*\gle \gu\})\cup \{\gu\gle \gu\})$,
$\gu\gle \gu\in ((((S'-\{\gu\gleq a^*\})\cup \{a^*\geql \gu\})-\{a^*\gle \gu\})\cup \{\gu\gle \gu\})-\mi{guards}((((S'-\{\gu\gleq a^*\})\cup \{a^*\geql \gu\})-\{a^*\gle \gu\})\cup \{\gu\gle \gu\})$,
applying Rule (\cref{ceq4hr2}) to $(((S'-\{\gu\gleq a^*\})\cup \{a^*\geql \gu\})-\{a^*\gle \gu\})\cup \{\gu\gle \gu\}$ and $\gu\gle \gu$, we derive
\begin{equation} \notag
\dfrac{(((S'-\{\gu\gleq a^*\})\cup \{a^*\geql \gu\})-\{a^*\gle \gu\})\cup \{\gu\gle \gu\}}
      {(((((S'-\{\gu\gleq a^*\})\cup \{a^*\geql \gu\})-\{a^*\gle \gu\})\cup \{\gu\gle \gu\})-\{\gu\gle \gu\})\cup \{\square\}}.
\end{equation}
We put
\begin{equation} \notag
\mi{Tree}=\begin{array}[c]{c}
          S \\[0.4mm]
          \hline \\[-3.8mm]
          S' \\[0.4mm]
          \hline \\[-3.8mm]
          (S'-\{\gu\gleq a^*\})\cup \{a^*\geql \gu\} \\[0.4mm]
          \hline \\[-3.8mm]
          (((S'-\{\gu\gleq a^*\})\cup \{a^*\geql \gu\})-\{a^*\gle \gu\})\cup \{\gu\gle \gu\} \\[0.4mm]
          \hline \\[-3.8mm]
          S''=(((((S'-\{\gu\gleq a^*\})\cup \{a^*\geql \gu\})-\{a^*\gle \gu\})\cup \{\gu\gle \gu\})- \\
          \hfill \{\gu\gle \gu\})\cup \{\square\}.
          \end{array} 
\end{equation}
Hence, $\mi{Tree}$ is a finite linear {\it DPLL}-tree with the root $S$ constructed using Rules (\cref{ceq4hr1111})--(\cref{ceq4hr66}) such that
for its only leaf $S''$, $\square\in S''$.

Cases 2.1.2.2.2 and 2.2.2.2.2:
For all $a\in \mi{atoms}(S')\subseteq \mi{atoms}(S)$, $C^*-\{l^*\}\not\in \mi{guards}(a)$.
Then $C^*-\{l^*\}\in S'$;
for all $a\in \mi{atoms}(C^*-\{l^*\})\subseteq \mi{atoms}(S')$, $C^*-\{l^*\}\not\in \mi{guards}(a)$;
$C^*-\{l^*\}$ is not a guard;
$C^*-\{l^*\}\not\in \mi{guards}(S')$,
$C^*-\{l^*\}\in S'-\mi{guards}(S')$;
$C^*\not\in \mi{guards}(S)$,
$\mi{guards}(S')=\{C \,|\, C\in (S-\{C^*\})\cup \{C^*-\{l^*\}\}\ \text{\it is a guard}\}=\{C \,|\, C\in S-\{C^*\}\ \text{\it is a guard}\}=
                 \mi{guards}(S)-\{C^*\}=\{C \,|\, C\in S\ \text{\it is a guard}\}-\{C^*\}=\mi{guards}(S)$;
we have that $S$ is positively guarded;
$\mi{atoms}(S'-\mi{guards}(S'))\subseteq \mi{atoms}(S')\subseteq \mi{atoms}(S)$,
$\mi{atoms}(S')=\mi{atoms}((S'-\mi{guards}(S'))\cup \mi{guards}(S'))=\mi{atoms}(S'-\mi{guards}(S'))\cup \mi{atoms}(\mi{guards}(S'))=\mi{atoms}(S'-\mi{guards}(S'))\cup \mi{atoms}(\mi{guards}(S))=
                \mi{atoms}(S'-\mi{guards}(S'))\cup \mi{atoms}(S)=\mi{atoms}(S)$;
$S'$ is positively guarded;
for all $l\in C^*$, 
$\mi{valid}(l,S)$ if and only if
$l\in \mi{OrdPropLit}^\gu$;
if either $l=\Cn\diamond \gu$ or $l=\gu\gleq \Cn$, $\Cn\in \mi{PropConj}$, and $\diamond\in \{\geql,\gle\}$, 
$\Cn=b_0\swedge\cdots\swedge b_k$, $b_j\in \mi{PropAtom}$; 
for all $j\leq k$, 
$C^*\in S$, $b_j\in \mi{atoms}(\Cn)=\mi{atoms}(l)\subseteq \mi{atoms}(C^*)\subseteq \mi{atoms}(S)=\mi{atoms}(S')$,
$\mi{guards}(S,b_j)=\mi{guards}(S',b_j)=\{\gz\gle b_j\}$ if and only if
$\mi{valid}(l,S')$;
$l^*\in \mi{invalid}(C^*,S)$,
$\mi{invalid}(C^*-\{l^*\},S')=\{l \,|\, l\in C^*-\{l^*\},\ \text{\it not}\ \mi{valid}(l,S')\}=\{l \,|\, l\in C^*-\{l^*\},\ \text{\it not}\ \mi{valid}(l,S)\}=
                              \mi{invalid}(C^*,S)-\{l^*\}=\{l \,|\, l\in C^*,\ \text{\it not}\ \mi{valid}(l,S)\}-\{l^*\}\subset \mi{invalid}(C^*,S)$;
by the induction hypothesis for $S'$ and $C^*-\{l^*\}$, there exists a finite linear {\it DPLL}-tree $\mi{Tree}'$ with the root $S'$ constructed using Rules (\cref{ceq4hr1111})--(\cref{ceq4hr66}) satisfying
for its only leaf $S''$ that either $\square\in S''$, or $S''\subseteq_{\mc F} \mi{OrdPropCl}$ is semi-positively guarded, and exactly one of the following points holds.
\begin{enumerate}[\rm (a)]
\item
$S''=S'$, $\mi{valid}(C^*-\{l^*\},S')$;
\item
$S''=S'-\{C^*-\{l^*\}\}$, $\mi{guards}(S'')=\mi{guards}(S')$;
\item
there exists $C^{**}\in \mi{OrdPropCl}^\gu$, $\mi{atoms}(C^{**})\subseteq \mi{atoms}(C^*-\{l^*\})$, satisfying 
$S''=(S'-\{C^*-\{l^*\}\})\cup \{C^{**}\}$, $\mi{guards}(S'')=\mi{guards}(S')$, $C^{**}\not\in S'$, $\mi{valid}(C^{**},S'')$;
\item
there exists $b^*\gle \gu\in \mi{OrdPropCl}^\gu$, $b^*\in \mi{atoms}(C^*-\{l^*\})$, satisfying 
$S''=(S'-\{C^*-\{l^*\}\})\cup \{b^*\gle \gu\}$, $\mi{guards}(S'')=\mi{guards}(S')\cup \{b^*\gle \gu\}$, $b^*\gle \gu\not\in S'$;
\item
there exists $S^{**}=\{b_0\geql \gu,\dots,b_k\geql \gu\}\subseteq \mi{OrdPropCl}^\gu$, $\{b_0,\dots,b_k\}\subseteq \mi{atoms}(C^*-\{l^*\})$, 
$\{\gz\gle b_0,\dots,\gz\gle b_k\}\subseteq \mi{guards}(S')$, satisfying
$S''=(S'-(\{C^*-\{l^*\}\}\cup \{\gz\gle b_0,\dots,\gz\gle b_k\}))\cup S^{**}$, $\mi{guards}(S'')=(\mi{guards}(S')-\{\gz\gle b_0,\dots,\gz\gle b_k\})\cup S^{**}$, $S^{**}\cap S'=\emptyset$. 
\end{enumerate}
We get that 
$C^*-\{l^*\}\not\in S$,
$S'-\{C^*-\{l^*\}\}=((S-\{C^*\})\cup \{C^*-\{l^*\}\})-\{C^*-\{l^*\}\}=((S-\{C^*\})-\{C^*-\{l^*\}\})\cup (\{C^*-\{l^*\}\}-\{C^*-\{l^*\}\})=S-\{C^*\}$;
\begin{enumerate}[\rm (a)]
\item
$\mi{valid}(C^*-\{l^*\},S')$; 
for all $l\in C^*-\{l^*\}$, $\mi{valid}(l,S')$, $l\in \mi{OrdPropLit}^\gu$;
$l^*\in C^*$;
we put $C^{**}=C^*-\{l^*\}\in \mi{OrdPropCl}^\gu$;
$\mi{atoms}(C^{**})=\mi{atoms}(C^*-\{l^*\})\subseteq \mi{atoms}(C^*)$,
$C^{**}=C^*-\{l^*\}\not\in S$,
$S''=S'=(S-\{C^*\})\cup \{C^*-\{l^*\}\}=(S-\{C^*\})\cup \{C^{**}\}$,
$\mi{guards}(S'')=\mi{guards}(S')=\mi{guards}(S)$,
$\mi{valid}(C^{**},S'')$;
$C^{**}\in S''$, $S''\neq S$;
\item
$C^*\in S$, $S''=S'-\{C^*-\{l^*\}\}=S-\{C^*\}\neq S$, $\mi{guards}(S'')=\mi{guards}(S')=\mi{guards}(S)$;
\item
there exists $C^{**}\in \mi{OrdPropCl}^\gu$, $\mi{atoms}(C^{**})\subseteq \mi{atoms}(C^*-\{l^*\})\subseteq \mi{atoms}(C^*)$, satisfying 
$S''=(S'-\{C^*-\{l^*\}\})\cup \{C^{**}\}=(S-\{C^*\})\cup \{C^{**}\}$, $\mi{guards}(S'')=\mi{guards}(S')=\mi{guards}(S)$;
we have that $S''$ is semi-positively guarded, and $S$ is positively guarded;
$\mi{atoms}(S'')=\mi{atoms}(\mi{guards}(S''))=\mi{atoms}(\mi{guards}(S))=\mi{atoms}(S)$;
for all $l\in C^*$, 
$\mi{valid}(l,S)$ if and only if
$l\in \mi{OrdPropLit}^\gu$;
if either $l=\Cn\diamond \gu$ or $l=\gu\gleq \Cn$, $\Cn\in \mi{PropConj}$, and $\diamond\in \{\geql,\gle\}$, 
$\Cn=b_0\swedge\cdots\swedge b_k$, $b_j\in \mi{PropAtom}$; 
for all $j\leq k$, 
$b_j\in \mi{atoms}(\Cn)=\mi{atoms}(l)\subseteq \mi{atoms}(C^*)\subseteq \mi{atoms}(S)=\mi{atoms}(S'')$,
$\mi{guards}(S,b_j)=\mi{guards}(S'',b_j)=\{\gz\gle b_j\}$ if and only if
$\mi{valid}(l,S'')$;
$\mi{valid}(C^{**},S'')$;
for all $l\in C^{**}$, $\mi{valid}(l,S'')$;
not $\mi{valid}(l^*,S)$, not $\mi{valid}(l^*,S'')$, $l^*\not\in C^{**}$, $C^{**}\neq C^*$,
$C^{**}\not\in S'\supseteq S-\{C^*\}$, $C^{**}\not\in S$, $C^{**}\in S''$, $S''\neq S$;  
\item
there exists $b^*\gle \gu\in \mi{OrdPropCl}^\gu$, $b^*\in \mi{atoms}(C^*-\{l^*\})\subseteq \mi{atoms}(C^*)\subseteq \mi{atoms}(S)$, satisfying 
$S''=(S'-\{C^*-\{l^*\}\})\cup \{b^*\gle \gu\}=(S-\{C^*\})\cup \{b^*\gle \gu\}$, $\mi{guards}(S'')=\mi{guards}(S')\cup \{b^*\gle \gu\}=\mi{guards}(S)\cup \{b^*\gle \gu\}$; 
$b^*\gle \gu\in \mi{guards}(b^*)$, $C^*\not\in \mi{guards}(S)\supseteq \mi{guards}(S,b^*)=S\cap \mi{guards}(b^*)$, $C^*\not\in \mi{guards}(b^*)$, $b^*\gle \gu\neq C^*$, 
$b^*\gle \gu\not\in S'\supseteq S-\{C^*\}$, $b^*\gle \gu\not\in S$, $b^*\gle \gu\in S''$, $S''\neq S$;
\item
there exists $S^{**}=\{b_0\geql \gu,\dots,b_k\geql \gu\}\subseteq \mi{OrdPropCl}^\gu$, $\{b_0,\dots,b_k\}\subseteq \mi{atoms}(C^*-\{l^*\})\subseteq \mi{atoms}(C^*)\subseteq \mi{atoms}(S)$, 
$\{\gz\gle b_0,\dots,\gz\gle b_k\}\subseteq \mi{guards}(S')=\mi{guards}(S)$, satisfying
$S''=(S'-(\{C^*-\{l^*\}\}\cup \{\gz\gle b_0,\dots,\gz\gle b_k\}))\cup S^{**}=((S'-\{C^*-\{l^*\}\})-\{\gz\gle b_0,\dots,\gz\gle b_k\})\cup S^{**}=
     ((S-\{C^*\})-\{\gz\gle b_0,\dots,\gz\gle b_k\})\cup S^{**}=(S-(\{C^*\}\cup \{\gz\gle b_0,\dots,\gz\gle b_k\}))\cup S^{**}$, 
$\mi{guards}(S'')=(\mi{guards}(S')-\{\gz\gle b_0,\dots,\gz\gle b_k\})\cup S^{**}=(\mi{guards}(S)-\{\gz\gle b_0,\dots,\gz\gle b_k\})\cup S^{**}$;
for all $j\leq k$, $b_j\geql \gu\in \mi{guards}(b_j)$, $C^*\not\in \mi{guards}(S)\supseteq \mi{guards}(S,b_j)=S\cap \mi{guards}(b_j)$, $C^*\not\in \mi{guards}(b_j)$, $b_j\geql \gu\neq C^*$;
$S^{**}\cap S=\{b_0\geql \gu,\dots,b_k\geql \gu\}\cap S=\{b_0\geql \gu,\dots,b_k\geql \gu\}\cap (S-\{C^*\})=S^{**}\cap (S-\{C^*\})=S^{**}\cap S'=\emptyset$, 
$S^{**}=\{b_0\geql \gu,\dots,b_k\geql \gu\}\neq \emptyset$, $S^{**}\not\subseteq S$, $S^{**}\subseteq S''$, $S''\neq S$. 
\end{enumerate}
We put
\begin{equation} \notag
\mi{Tree}=\dfrac{S}
                {\mi{Tree}'}.
\end{equation}
Hence, $\mi{Tree}$ is a finite linear {\it DPLL}-tree with the root $S$ constructed using Rules (\cref{ceq4hr1111})--(\cref{ceq4hr66}) such that
for its only leaf $S''$, either $\square\in S''$, or $S''\subseteq_{\mc F} \mi{OrdPropCl}$ is semi-positively guarded, $S''\neq S$, and exactly one of the following points holds.
\begin{enumerate}[\rm (a)]
\item
$S''=S-\{C^*\}$, $\mi{guards}(S'')=\mi{guards}(S)$;
\item
there exists $C^{**}\in \mi{OrdPropCl}^\gu$, $\mi{atoms}(C^{**})\subseteq \mi{atoms}(C^*)$, satisfying 
$S''=(S-\{C^*\})\cup \{C^{**}\}$, $\mi{guards}(S'')=\mi{guards}(S)$, $C^{**}\not\in S$, $\mi{valid}(C^{**},S'')$;
\item
there exists $b^*\gle \gu\in \mi{OrdPropCl}^\gu$, $b^*\in \mi{atoms}(C^*)$, satisfying 
$S''=(S-\{C^*\})\cup \{b^*\gle \gu\}$, $\mi{guards}(S'')=\mi{guards}(S)\cup \{b^*\gle \gu\}$, $b^*\gle \gu\not\in S$;
\item
there exists $S^{**}=\{b_0\geql \gu,\dots,b_k\geql \gu\}\subseteq \mi{OrdPropCl}^\gu$, $\{b_0,\dots,b_k\}\subseteq \mi{atoms}(C^*)$, $\{\gz\gle b_0,\dots,\gz\gle b_k\}\subseteq \mi{guards}(S)$, satisfying
$S''=(S-(\{C^*\}\cup \{\gz\gle b_0,\dots,\gz\gle b_k\}))\cup S^{**}$, $\mi{guards}(S'')=(\mi{guards}(S)-\{\gz\gle b_0,\dots,\gz\gle b_k\})\cup S^{**}$, $S^{**}\cap S=\emptyset$. 
\end{enumerate}

Case 2.3:
$l^*=\gz\gle a_0^{\alpha_0}\swedge\cdots\swedge a_n^{\alpha_n}$, $a_i\in \mi{PropAtom}$, $\alpha_i\geq 1$.
Then, for all $i\leq n$,
$a_i\in \mi{atoms}(l^*)\subseteq \mi{atoms}(S)$;
$a_i$ is positively guarded in $S$;
$\gz\gle a_i\in \mi{guards}(S,a_i)\subseteq \mi{guards}(S)$; 
$l^*\in C^*\in S-\mi{guards}(S)$, applying Rule (\cref{ceq4hr55}) to $C^*$ and $l^*$, we derive
\begin{equation} \notag
\dfrac{S}
      {S-\{C^*\}}.
\end{equation}

Case 2.4:
$l^*=a_0^{\alpha_0}\swedge\cdots\swedge a_n^{\alpha_n}\gle \gu$, $a_i\in \mi{PropAtom}$, $\alpha_i\geq 1$, and there exists $i^*\leq n$ such that 
$a_{i^*}\in \mi{atoms}(l^*)\subseteq \mi{atoms}(S)$, $a_{i^*}$ is positively guarded in $S$, $\mi{guards}(S,a_{i^*})=\{\gz\gle a_{i^*},a_{i^*}\gle \gu\}$.
Then $a_{i^*}\gle \gu\in \mi{guards}(S,a_{i^*})\subseteq \mi{guards}(S)$, $l^*\in C^*\in S-\mi{guards}(S)$,
applying Rule (\cref{ceq4hr66}) to $C^*$ and $l^*$, we derive
\begin{equation} \notag
\dfrac{S}
      {S-\{C^*\}}.
\end{equation}

In Cases 2.3 and 2.4, we put $S'=S-\{C^*\}\subseteq_{\mc F} \mi{OrdPropCl}$.
Hence, these cases are the same as Case 1.1.1.

Case 2.5:
$l^*\not\in \mi{OrdPropLit}^\gu$.
We have that $l^*$ is not a contradiction or tautology.
Then $l^*$ is not a contradiction of the form either $l^*=\gz\geql \gu$ or $l^*=\gu\gleq \gz$ or $l^*=\varepsilon\gle \gz$, and
$l^*$ is not a tautology of the form either $l^*=\gz\geql \gz$ or $l^*=\gz\gleq \varepsilon$ or $l^*=\gz\gle \gu$, $\varepsilon\in \{\gz,\gu\}\cup \mi{PropConj}$.
We get two cases for $l^*$.

Case 2.5.1:
Either $l^*=a_0^{\alpha_0}\swedge\cdots\swedge a_n^{\alpha_n}\geql \gz$ or $l^*=a_0^{\alpha_0}\swedge\cdots\swedge a_n^{\alpha_n}\gleq \gz$, $a_i\in \mi{PropAtom}$, $\alpha_i\geq 1$.
This case is the same as Case 2.1.

Case 2.5.2:
$l^*=\gz\gle a_0^{\alpha_0}\swedge\cdots\swedge a_n^{\alpha_n}$, $a_i\in \mi{PropAtom}$, $\alpha_i\geq 1$.
This case is the same as Case 2.3.

Case 2.6:
Either $l^*=\Cn^*\diamond^* \gu$ or $l^*=\gu\gleq \Cn^*$, $\Cn^*\in \mi{PropConj}$, $\diamond^*\in \{\geql,\gle\}$, 
$\Cn^*=a_0^{\alpha_0}\swedge\cdots\swedge a_n^{\alpha_n}$, $a_i\in \mi{PropAtom}$, $\alpha_i\geq 1$, there exists $i^*\leq n$ such that
$\alpha_{i^*}\geq 2$, 
or $a_{i^*}\in \mi{atoms}(\Cn^*)=\mi{atoms}(l^*)\subseteq \mi{atoms}(C^*)\subseteq \mi{atoms}(S)$, $a_{i^*}$ is positively guarded in $S$, $\mi{guards}(S,a_{i^*})=\{\gz\gle a_{i^*},a_{i^*}\gle \gu\}$.
We get two cases for $i^*$.

Case 2.6.1:
$\alpha_{i^*}\geq 2$, and for all $i\leq n$, $a_i\in \mi{atoms}(\Cn^*)\subseteq \mi{atoms}(S)$, $a_i$ is positively guarded in $S$, $\mi{guards}(S,a_i)=\{\gz\gle a_i\}$.
We put 
\begin{alignat*}{1}
l^{**}        &= a_0\swedge\cdots\swedge a_n\diamond^{**} \gu\in \mi{OrdPropLit}^\gu, \\[1mm]
\diamond^{**} &= \left\{\begin{array}{ll}
                        \geql &\ \text{\it if either}\ l^*=\Cn^*\geql \gu\ \text{\it or}\ l^*=\gu\gleq \Cn^*, \\[1mm]
                        \gle  &\ \text{\it if}\ l^*=\Cn^*\gle \gu.
                        \end{array}
                 \right. 
\end{alignat*}
Then $l^*\in C^*\in S-\mi{guards}(S)$, applying Rule (\cref{ceq4hr11111}) to $C^*$, $l^*$, and $l^{**}$, we derive
\begin{equation} \notag
\dfrac{S}
      {(S-\{C^*\})\cup \{(C^*-\{l^*\})\cup \{l^{**}\}\}}.
\end{equation}
We put $S'=(S-\{C^*\})\cup \{(C^*-\{l^*\})\cup \{l^{**}\}\}\subseteq_{\mc F} \mi{OrdPropCl}$.
We get two cases for $l^{**}$.

Case 2.6.1.1:
$l^{**}\in C^*$.
Then $i^*\leq n$, $\alpha_{i^*}\geq 2$, $a_0\swedge\cdots\swedge a_n\neq a_0^{\alpha_0}\swedge\cdots\swedge a_n^{\alpha_n}$,
either $\diamond^{**}=\geql$, either $l^*=\Cn^*\geql \gu$ or $l^*=\gu\gleq \Cn^*$,
either $l^{**}=a_0\swedge\cdots\swedge a_n\diamond^{**} \gu=a_0\swedge\cdots\swedge a_n\geql \gu\neq l^*=\Cn^*\geql \gu=a_0^{\alpha_0}\swedge\cdots\swedge a_n^{\alpha_n}\geql \gu$
or $l^{**}=a_0\swedge\cdots\swedge a_n\geql \gu\neq l^*=\gu\gleq \Cn^*$,
or $\diamond^{**}=\gle$, $l^*=\Cn^*\gle \gu$,
$l^{**}=a_0\swedge\cdots\swedge a_n\diamond^{**} \gu=a_0\swedge\cdots\swedge a_n\gle \gu\neq l^*=\Cn^*\gle \gu=a_0^{\alpha_0}\swedge\cdots\swedge a_n^{\alpha_n}\gle \gu$;
$l^{**}\neq l^*$, $l^{**}\in C^*-\{l^*\}$,
$S'=(S-\{C^*\})\cup \{(C^*-\{l^*\})\cup \{l^{**}\}\}=(S-\{C^*\})\cup \{C^*-\{l^*\}\}$;
this case is the same as Cases 2.1 and 2.2.

Case 2.6.1.2:
$l^{**}\not\in C^*$.
We get two cases for $(C^*-\{l^*\})\cup \{l^{**}\}$.

Case 2.6.1.2.1:
$(C^*-\{l^*\})\cup \{l^{**}\}\in S$.
Then $l^{**}\in (C^*-\{l^*\})\cup \{l^{**}\}$, $l^{**}\not\in C^*$, $(C^*-\{l^*\})\cup \{l^{**}\}\neq C^*$,
$(C^*-\{l^*\})\cup \{l^{**}\}\in S-\{C^*\}$,
$S'=(S-\{C^*\})\cup \{(C^*-\{l^*\})\cup \{l^{**}\}\}=S-\{C^*\}$;
this case is the same as Case 1.1.1.

Case 2.6.1.2.2:
$(C^*-\{l^*\})\cup \{l^{**}\}\not\in S$.
Then $\mi{atoms}(l^{**})=\mi{atoms}(l^*)=\mi{atoms}(\Cn^*)=\{a_0,\dots,a_n\}$, $l^*\in C^*$,
$\mi{atoms}((C^*-\{l^*\})\cup \{l^{**}\})=\mi{atoms}(C^*-\{l^*\})\cup \mi{atoms}(l^{**})=\mi{atoms}(C^*-\{l^*\})\cup \mi{atoms}(l^*)=\mi{atoms}((C^*-\{l^*\})\cup \{l^*\})=\mi{atoms}(C^*)$,
$C^*\in S$,
$\mi{atoms}(S')=\mi{atoms}((S-\{C^*\})\cup \{(C^*-\{l^*\})\cup \{l^{**}\}\})=\mi{atoms}(S-\{C^*\})\cup \mi{atoms}((C^*-\{l^*\})\cup \{l^{**}\})=\mi{atoms}(S-\{C^*\})\cup \mi{atoms}(C^*)=
                \mi{atoms}((S-\{C^*\})\cup \{C^*\})=\mi{atoms}(S)$;
$S$ is simplified; 
$S-\{C^*\}\subseteq S$ is simplified;
$C^*\in S$ does not contain contradictions and tautologies;
$C^*-\{l^*\}\subseteq C^*$ does not contain contradictions and tautologies;
$l^{**}=a_0\swedge\cdots\swedge a_n\diamond^{**} \gu$, $\diamond^{**}\in \{\geql,\gle\}$, is not a contradiction or tautology;
$(C^*-\{l^*\})\cup \{l^{**}\}\neq \square$ does not contain contradictions and tautologies;
$S'=(S-\{C^*\})\cup \{(C^*-\{l^*\})\cup \{l^{**}\}\}$ is simplified.
We get two cases for $C^*-\{l^*\}$.

Case 2.6.1.2.2.1:
$C^*-\{l^*\}=\square$.
Then $l^*\in C^*$, $C^*=(C^*-\{l^*\})\cup \{l^*\}=l^*$, $(C^*-\{l^*\})\cup \{l^{**}\}=l^{**}\not\in S$, $S'=(S-\{C^*\})\cup \{(C^*-\{l^*\})\cup \{l^{**}\}\}=(S-\{C^*\})\cup \{l^{**}\}$.
We get two cases for $l^{**}$.

Case 2.6.1.2.2.1.1:
$\diamond^{**}=\geql$.
Then $l^{**}=a_0\swedge\cdots\swedge a_n\diamond^{**} \gu=a_0\swedge\cdots\swedge a_n\geql \gu$, either $l^*=\Cn^*\geql \gu$ or $l^*=\gu\gleq \Cn^*$.
We get two cases for $n$.

Case 2.6.1.2.2.1.1.1:
$n=0$.
Then $l^{**}=a_0\swedge\cdots\swedge a_n\geql \gu=a_0\geql \gu\not\in S$, $S'=(S-\{C^*\})\cup \{l^{**}\}=(S-\{C^*\})\cup \{a_0\geql \gu\}$, $a_0\in \mi{atoms}(S')$, 
$S\supseteq \mi{guards}(S)\supseteq \mi{guards}(S,a_0)=\{\gz\gle a_0\}$;
$C^*\in S$, $C^*\not\in \mi{guards}(S)\supseteq \mi{guards}(S,a_0)=S\cap \mi{guards}(a_0)$, $C^*\not\in \mi{guards}(a_0)$, 
$S'\supseteq \mi{guards}(S')\supseteq \mi{guards}(S',a_0)=((S-\{C^*\})\cup \{a_0\geql \gu\})\cap \mi{guards}(a_0)=((S-\{C^*\})\cap \mi{guards}(a_0))\cup (\{a_0\geql \gu\}\cap \mi{guards}(a_0))=
 ((S\cap \mi{guards}(a_0))-\{C^*\})\cup \{a_0\geql \gu\}=(S\cap \mi{guards}(a_0))\cup \{a_0\geql \gu\}=\mi{guards}(S,a_0)\cup \{a_0\geql \gu\}=\{\gz\gle a_0,a_0\geql \gu\}$,
$a_0\geql \gu\in \mi{guards}(S')$, $\gz\gle a_0\in S'$,
$a_0\in \mi{atoms}(\gz\gle a_0)$, $a_0\geql \gu\neq \gz\gle a_0$, $\mi{simplify}(\gz\gle a_0,a_0,\gu)=\gz\gle \gu$,
applying Rule (\cref{ceq4hr4}) to $S'$, $a_0\geql \gu$, and $\gz\gle a_0$, we derive
\begin{equation} \notag
\dfrac{S'}
      {(S'-\{\gz\gle a_0\})\cup \{\gz\gle \gu\}};
\end{equation}
$\gz\gle \gu\in (S'-\{\gz\gle a_0\})\cup \{\gz\gle \gu\}$;
$\gz\gle \gu\in \mi{OrdPropLit}$ is a tautology;
$\gz\gle \gu$ is not a guard;
$\gz\gle \gu\not\in \mi{guards}((S'-\{\gz\gle a_0\})\cup \{\gz\gle \gu\})$,
$\gz\gle \gu\in ((S'-\{\gz\gle a_0\})\cup \{\gz\gle \gu\})-\mi{guards}((S'-\{\gz\gle a_0\})\cup \{\gz\gle \gu\})$,
applying Rule (\cref{ceq4hr22}) to $(S'-\{\gz\gle a_0\})\cup \{\gz\gle \gu\}$ and $\gz\gle \gu$, we derive
\begin{equation} \notag
\dfrac{(S'-\{\gz\gle a_0\})\cup \{\gz\gle \gu\}}
      {((S'-\{\gz\gle a_0\})\cup \{\gz\gle \gu\})-\{\gz\gle \gu\}}.
\end{equation}
We put $S^{**}=\{a_0\geql \gu\}\subseteq \mi{OrdPropCl}^\gu$.
We have that $S'$ is simplified.
Hence, $\gz\gle \gu\not\in S'$, $C^*, \gz\gle a_0\in S$, either $C^*=l^*=\Cn^*\geql \gu$ or $C^*=l^*=\gu\gleq \Cn^*$, $C^*\neq \gz\gle a_0$, 
$a_0\geql \gu\not\in S\supseteq \mi{guards}(S)$, $S^{**}\cap S=S^{**}\cap \mi{guards}(S)=\{a_0\geql \gu\}\cap S=\{a_0\geql \gu\}\cap \mi{guards}(S)=\emptyset$,
$((S'-\{\gz\gle a_0\})\cup \{\gz\gle \gu\})-\{\gz\gle \gu\}=((S'-\{\gz\gle a_0\})-\{\gz\gle \gu\})\cup (\{\gz\gle \gu\}-\{\gz\gle \gu\})=S'-\{\gz\gle a_0\}=
 ((S-\{C^*\})\cup \{a_0\geql \gu\})-\{\gz\gle a_0\}=((S-\{C^*\})-\{\gz\gle a_0\})\cup (\{a_0\geql \gu\}-\{\gz\gle a_0\})=((S-\{C^*\})-\{\gz\gle a_0\})\cup \{a_0\geql \gu\}=
 (S-(\{C^*\}\cup \{\gz\gle a_0\}))\cup S^{**}$.
We put $S''=(S-(\{C^*\}\cup \{\gz\gle a_0\}))\cup S^{**}\subseteq_{\mc F} \mi{OrdPropCl}$.
We get that
$S''=(S-(\{C^*\}\cup \{\gz\gle a_0\}))\cup S^{**}=((S-\{C^*\})-\{\gz\gle a_0\})\cup \{a_0\geql \gu\}=S'-\{\gz\gle a_0\}$,
$\mi{atoms}(\gz\gle a_0)=\{a_0\}$, 
$a_0\in \mi{atoms}(S'')=\mi{atoms}(S'-\{\gz\gle a_0\})=\mi{atoms}(S')=\mi{atoms}(S)$;
we have that $S'$ is simplified;
$S''=S'-\{\gz\gle a_0\}\subseteq S'$ is simplified;
$C^*\not\in \mi{guards}(a_0)$,
$\mi{guards}(S'',a_0)=(((S-\{C^*\})-\{\gz\gle a_0\})\cup \{a_0\geql \gu\})\cap \mi{guards}(a_0)=(((S-\{C^*\})-\{\gz\gle a_0\})\cap \mi{guards}(a_0))\cup (\{a_0\geql \gu\}\cap \mi{guards}(a_0))=
                      (((S\cap \mi{guards}(a_0))-\{\gz\gle a_0\})-\{C^*\})\cup \{a_0\geql \gu\}=((S\cap \mi{guards}(a_0))-\{\gz\gle a_0\})\cup \{a_0\geql \gu\}=
                      (\mi{guards}(S,a_0)-\{\gz\gle a_0\})\cup \{a_0\geql \gu\}=(\{\gz\gle a_0\}-\{\gz\gle a_0\})\cup \{a_0\geql \gu\}=\{a_0\geql \gu\}$;
$a_0$ is semi-positively guarded in $S''$;
for all $a\in \mi{atoms}(S'')-\{a_0\}=\mi{atoms}(S)-\{a_0\}$,
$a$ is positively guarded in $S$;
$a\neq a_0$,
$\mi{guards}(a)\cap \{\gz\gle a_0,a_0\geql \gu\}=\mi{guards}(a)\cap \mi{guards}(a_0)=\emptyset$,
$C^*\in S$, $C^*\not\in \mi{guards}(S)\supseteq \mi{guards}(S,a)=S\cap \mi{guards}(a)$, $C^*\not\in \mi{guards}(a)$,
$\mi{guards}(S'',a)=(((S-\{C^*\})-\{\gz\gle a_0\})\cup \{a_0\geql \gu\})\cap \mi{guards}(a)=(((S-\{C^*\})-\{\gz\gle a_0\})\cap \mi{guards}(a))\cup (\{a_0\geql \gu\}\cap \mi{guards}(a))=
                    ((S\cap \mi{guards}(a))-\{C^*\})-\{\gz\gle a_0\}=S\cap \mi{guards}(a)=\mi{guards}(S,a)$;
$a$ is positively guarded in $S''$;
$S''$ is semi-positively guarded; 
$\gz\gle a_0\in \mi{guards}(S)$, $S^{**}\cap \mi{guards}(S)=\emptyset$, 
$\mi{guards}(S'')=(\mi{guards}(S)-\{\gz\gle a_0\})\cup \{a_0\geql \gu\}=(\mi{guards}(S)-\{\gz\gle a_0\})\cup S^{**}$.
We put
\begin{equation} \notag
\mi{Tree}=\begin{array}[c]{c}
          S \\[0.4mm]
          \hline \\[-3.8mm]
          S' \\[0.4mm]
          \hline \\[-3.8mm]
          (S'-\{\gz\gle a_0\})\cup \{\gz\gle \gu\} \\[0.4mm]
          \hline \\[-3.8mm]
          S''.
          \end{array}
\end{equation}
Hence, $\mi{Tree}$ is a finite linear {\it DPLL}-tree with the root $S$ constructed using Rules (\cref{ceq4hr1111})--(\cref{ceq4hr66}) such that
for its only leaf $S''$, $S''\subseteq_{\mc F} \mi{OrdPropCl}$ is semi-positively guarded, and
there exists $S^{**}=\{a_0\geql \gu\}\subseteq \mi{OrdPropCl}^\gu$, $a_0\in \mi{atoms}(C^*)$, $\{\gz\gle a_0\}\subseteq \mi{guards}(S)$, satisfying 
$S''=(S-(\{C^*\}\cup \{\gz\gle a_0\}))\cup S^{**}$, $\mi{guards}(S'')=(\mi{guards}(S)-\{\gz\gle a_0\})\cup S^{**}$, $S^{**}\cap S=\emptyset$.

Case 2.6.1.2.2.1.1.2:
$n\geq 1$.
Then $l^{**}\in S'$;
$l^{**}=a_0\swedge\cdots\swedge a_n\geql \gu$ is not a guard;
$l^{**}\not\in \mi{guards}(S')$,
$l^{**}\in S'-\mi{guards}(S')$,
applying Rule (\cref{ceq4hr111111}) to $S'$ and $l^{**}$, we derive
\begin{equation} \notag
\dfrac{S'}
      {(S'-\{l^{**}\})\cup \{a_0\geql \gu,\dots,a_n\geql \gu\}}.
\end{equation}
We put $S^{**}=\{a_0\geql \gu,\dots,a_n\geql \gu\}\subseteq \mi{OrdPropCl}^\gu$.
Hence, for all $i\leq n$, 
$a_i\geql \gu\in \mi{guards}(a_i)$, $a_i\geql \gu\not\in \mi{guards}(S,a_i)=\{\gz\gle a_i\}=S\cap \mi{guards}(a_i)\subseteq \mi{guards}(S)$, $a_i\geql \gu\not\in S\supseteq \mi{guards}(S)$;
$\{\gz\gle a_0,\dots,\gz\gle a_n\}\subseteq \mi{guards}(S)\subseteq S$,
$l^{**}\not\in S$, $C^*\in S$,
$S^{**}\cap S=S^{**}\cap \mi{guards}(S)=\{a_0\geql \gu,\dots,a_n\geql \gu\}\cap S=\{a_0\geql \gu,\dots,a_n\geql \gu\}\cap \mi{guards}(S)=\emptyset$,
$(S'-\{l^{**}\})\cup \{a_0\geql \gu,\dots,a_n\geql \gu\}=(((S-\{C^*\})\cup \{l^{**}\})-\{l^{**}\})\cup \{a_0\geql \gu,\dots,a_n\geql \gu\}=
 ((S-\{C^*\})-\{l^{**}\})\cup (\{l^{**}\}-\{l^{**}\})\cup \{a_0\geql \gu,\dots,a_n\geql \gu\}=(S-\{C^*\})\cup \{a_0\geql \gu,\dots,a_n\geql \gu\}=(S-\{C^*\})\cup S^{**}$.
We put $S''=(S-\{C^*\})\cup S^{**}\subseteq_{\mc F} \mi{OrdPropCl}$.
We get that
$S''=(S-\{C^*\})\cup S^{**}=(S-\{C^*\})\cup \{a_0\geql \gu,\dots,a_n\geql \gu\}$,
$\mi{atoms}(S^{**})=\mi{atoms}(l^{**})=\{a_0,\dots,a_n\}\subseteq \mi{atoms}(S'')$,
$\mi{atoms}(S'')=\mi{atoms}((S-\{C^*\})\cup S^{**})=\mi{atoms}(S-\{C^*\})\cup \mi{atoms}(S^{**})=\mi{atoms}(S-\{C^*\})\cup \mi{atoms}(l^{**})=\mi{atoms}((S-\{C^*\})\cup \{l^{**}\})=\mi{atoms}(S')$;
$S$ is simplified; 
$S-\{C^*\}\subseteq S$ is simplified;
for all $i\leq n$, $a_i\geql \gu\neq \square$ does not contain contradictions and tautologies;
$S^{**}=\{a_0\geql \gu,\dots,a_n\geql \gu\}$ is simplified;
$S''=(S-\{C^*\})\cup S^{**}$ is simplified;
for all $i\leq n$, 
for all $j\leq n$ and $j\neq i$, $a_i\neq a_j$, $\mi{guards}(a_i)\cap \{\gz\gle a_j,a_j\geql \gu\}=\mi{guards}(a_i)\cap \mi{guards}(a_j)=\emptyset$;
$C^*\in S$, $C^*\not\in \mi{guards}(S)\supseteq \mi{guards}(S,a_i)=S\cap \mi{guards}(a_i)$, $C^*\not\in \mi{guards}(a_i)$,
$\mi{guards}(S'',a_i)=((S-\{C^*\})\cup \{a_0\geql \gu,\dots,a_n\geql \gu\})\cap \mi{guards}(a_i)=
                      ((S-\{C^*\})\cap \mi{guards}(a_i))\cup (\{a_0\geql \gu,\dots,a_n\geql \gu\}\cap \mi{guards}(a_i))=
                      ((S\cap \mi{guards}(a_i))-\{C^*\})\cup \{a_i\geql \gu\}=(S\cap \mi{guards}(a_i))\cup \{a_i\geql \gu\}=\mi{guards}(S,a_i)\cup \{a_i\geql \gu\}=\{\gz\gle a_i,a_i\geql \gu\}$;
by Lemma \ref{le88} for $S''$ and $\{a_0,\dots,a_n\}$, there exists a finite linear {\it DPLL}-tree $\mi{Tree}''$ with the root $S''$ constructed using Rules (\cref{ceq4hr22}) and (\cref{ceq4hr4}) satisfying
for its only leaf $S'''$ that $S'''=S''-\{\gz\gle a_0,\dots,\gz\gle a_n\}\subseteq_{\mc F} \mi{OrdPropCl}$ is simplified;  
$S^{**}\cap \{\gz\gle a_0,\dots,\gz\gle a_n\}=\{a_0\geql \gu,\dots,a_n\geql \gu\}\cap \{\gz\gle a_0,\dots,\gz\gle a_n\}=\emptyset$,
$\{\gz\gle a_0,\dots,\gz\gle a_n\}\subseteq S$;
for all $i\leq n$, either $C^*=l^*=\Cn^*\geql \gu$ or $C^*=l^*=\gu\gleq \Cn^*$, $C^*\neq \gz\gle a_i$;
$\{C^*\}\cap \{\gz\gle a_0,\dots,\gz\gle a_n\}=\emptyset$, 
$S^{**}\cap S=\emptyset$,
$S'''=S''-\{\gz\gle a_0,\dots,\gz\gle a_n\}=((S-\{C^*\})\cup S^{**})-\{\gz\gle a_0,\dots,\gz\gle a_n\}=((S-\{C^*\})-\{\gz\gle a_0,\dots,\gz\gle a_n\})\cup (S^{**}-\{\gz\gle a_0,\dots,\gz\gle a_n\})=
      ((S-\{C^*\})-\{\gz\gle a_0,\dots,\gz\gle a_n\})\cup S^{**}=(S-(\{C^*\}\cup \{\gz\gle a_0,\dots,\gz\gle a_n\}))\cup S^{**}=
      ((S-\{C^*\})-\{\gz\gle a_0,\dots,\gz\gle a_n\})\cup \{a_0\geql \gu,\dots,a_n\geql \gu\}$,
$\mi{atoms}(\gz\gle a_0,\dots,\gz\gle a_n)=\{a_0,\dots,a_n\}$,
$\{a_0,\dots,a_n\}=\mi{atoms}(S^{**})\subseteq \mi{atoms}(S''')=\mi{atoms}(S''-\{\gz\gle a_0,\dots,\gz\gle a_n\})=\mi{atoms}(S'')=\mi{atoms}(S')=\mi{atoms}(S)$;
for all $i\leq n$,
$\mi{guards}(S''',a_i)=(((S-\{C^*\})-\{\gz\gle a_0,\dots,\gz\gle a_n\})\cup \{a_0\geql \gu,\dots,a_n\geql \gu\})\cap \mi{guards}(a_i)=
                       (((S-\{C^*\})-\{\gz\gle a_0,\dots,\gz\gle a_n\})\cap \mi{guards}(a_i))\cup (\{a_0\geql \gu,\dots,a_n\geql \gu\}\cap \mi{guards}(a_i))=
                       (((S\cap \mi{guards}(a_i))-\{\gz\gle a_0,\dots,\gz\gle a_n\})-\{C^*\})\cup \{a_i\geql \gu\}=((S\cap \mi{guards}(a_i))-\{\gz\gle a_i\})\cup \{a_i\geql \gu\}=
                       (\mi{guards}(S,a_i)-\{\gz\gle a_i\})\cup \{a_i\geql \gu\}=(\{\gz\gle a_i\}-\{\gz\gle a_i\})\cup \{a_i\geql \gu\}=\{a_i\geql \gu\}$;
$a_i$ is semi-positively guarded in $S'''$;
for all $a\in \mi{atoms}(S''')-\{a_0,\dots,a_n\}=\mi{atoms}(S)-\{a_0,\dots,a_n\}$,
$a$ is positively guarded in $S$;
for all $i\leq n$, $a\neq a_i$, $\mi{guards}(a)\cap \{\gz\gle a_i,a_i\geql \gu\}=\mi{guards}(a)\cap \mi{guards}(a_i)=\emptyset$;
$C^*\not\in \mi{guards}(S)\supseteq \mi{guards}(S,a)=S\cap \mi{guards}(a)$, $C^*\not\in \mi{guards}(a)$,
$\mi{guards}(S''',a)=(((S-\{C^*\})-\{\gz\gle a_0,\dots,\gz\gle a_n\})\cup \{a_0\geql \gu,\dots,a_n\geql \gu\})\cap \mi{guards}(a)=
                     (((S-\{C^*\})-\{\gz\gle a_0,\dots,\gz\gle a_n\})\cap \mi{guards}(a))\cup (\{a_0\geql \gu,\dots,a_n\geql \gu\}\cap \mi{guards}(a))=
                     ((S\cap \mi{guards}(a))-\{C^*\})-\{\gz\gle a_0,\dots,\gz\gle a_n\}=S\cap \mi{guards}(a)=\mi{guards}(S,a)$;
$a$ is positively guarded in $S'''$;
$S'''$ is semi-positively guarded; 
$\{\gz\gle a_0,\dots,\gz\gle a_n\}\subseteq \mi{guards}(S)$, $S^{**}\cap \mi{guards}(S)=\emptyset$,
$\mi{guards}(S''')=(\mi{guards}(S)-\{\gz\gle a_0,\dots,\gz\gle a_n\})\cup \{a_0\geql \gu,\dots,a_n\geql \gu\}=(\mi{guards}(S)-\{\gz\gle a_0,\dots,\gz\gle a_n\})\cup S^{**}$.
We put
\begin{equation} \notag
\mi{Tree}=\begin{array}[c]{c}
          S \\[0.4mm]
          \hline \\[-3.8mm]
          S' \\[0.4mm]
          \hline \\[-3.8mm]
          \mi{Tree}''.
          \end{array}
\end{equation}
Hence, $\mi{Tree}$ is a finite linear {\it DPLL}-tree with the root $S$ constructed using Rules (\cref{ceq4hr1111})--(\cref{ceq4hr66}) such that
for its only leaf $S'''$, $S'''\subseteq_{\mc F} \mi{OrdPropCl}$ is semi-positively guarded, and
there exists $S^{**}=\{a_0\geql \gu,\dots,a_n\geql \gu\}\subseteq \mi{OrdPropCl}^\gu$, $\{a_0,\dots,a_n\}\subseteq \mi{atoms}(C^*)$, $\{\gz\gle a_0,\dots,\gz\gle a_n\}\subseteq \mi{guards}(S)$, satisfying 
$S'''=(S-(\{C^*\}\cup \{\gz\gle a_0,\dots,\gz\gle a_n\}))\cup S^{**}$, $\mi{guards}(S''')=(\mi{guards}(S)-\{\gz\gle a_0,\dots,\gz\gle a_n\})\cup S^{**}$, $S^{**}\cap S=\emptyset$.

Case 2.6.1.2.2.1.2:
$\diamond^{**}=\gle$.
Then $l^{**}=a_0\swedge\cdots\swedge a_n\diamond^{**} \gu=a_0\swedge\cdots\swedge a_n\gle \gu$.
We get two cases for $n$.

Case 2.6.1.2.2.1.2.1:
$n=0$.
Then $l^{**}=a_0\swedge\cdots\swedge a_n\gle \gu=a_0\gle \gu\not\in S\supseteq \mi{guards}(S)$, $S'=(S-\{C^*\})\cup \{l^{**}\}=(S-\{C^*\})\cup \{a_0\gle \gu\}$, $a_0\in \mi{atoms}(S')$; 
$C^*\in S$, $C^*\not\in \mi{guards}(S)\supseteq \mi{guards}(S,a_0)=S\cap \mi{guards}(a_0)$, $C^*\not\in \mi{guards}(a_0)$,
$\mi{guards}(S',a_0)=((S-\{C^*\})\cup \{a_0\gle \gu\})\cap \mi{guards}(a_0)=((S-\{C^*\})\cap \mi{guards}(a_0))\cup (\{a_0\gle \gu\}\cap \mi{guards}(a_0))=
                     ((S\cap \mi{guards}(a_0))-\{C^*\})\cup \{a_0\gle \gu\}=(S\cap \mi{guards}(a_0))\cup \{a_0\gle \gu\}=\mi{guards}(S,a_0)\cup \{a_0\gle \gu\}=\{\gz\gle a_0,a_0\gle \gu\}$;
$a_0$ is positively guarded in $S'$;
for all $a\in \mi{atoms}(S')-\{a_0\}=\mi{atoms}(S)-\{a_0\}$,
$a$ is positively guarded in $S$;
$a\neq a_0$,
$\mi{guards}(a)\cap \{a_0\gle \gu\}=\mi{guards}(a)\cap \mi{guards}(a_0)=\emptyset$,
$C^*\not\in \mi{guards}(S)\supseteq \mi{guards}(S,a)=S\cap \mi{guards}(a)$, $C^*\not\in \mi{guards}(a)$,
$\mi{guards}(S',a)=((S-\{C^*\})\cup \{a_0\gle \gu\})\cap \mi{guards}(a)=((S-\{C^*\})\cap \mi{guards}(a))\cup (\{a_0\gle \gu\}\cap \mi{guards}(a))=
                   (S\cap \mi{guards}(a))-\{C^*\}=S\cap \mi{guards}(a)=\mi{guards}(S,a)$;
$a$ is positively guarded in $S'$;
$S'$ is positively guarded; 
$\mi{guards}(S')=\mi{guards}(S)\cup \{a_0\gle \gu\}$.
We put
\begin{equation} \notag
\mi{Tree}=\dfrac{S}
                {S'}.
\end{equation}
Hence, $\mi{Tree}$ is a finite linear {\it DPLL}-tree with the root $S$ constructed using Rules (\cref{ceq4hr1111})--(\cref{ceq4hr66}) such that
for its only leaf $S'$, $S'\subseteq_{\mc F} \mi{OrdPropCl}$ is positively guarded, and there exists $a_0\gle \gu\in \mi{OrdPropCl}^\gu$, $a_0\in \mi{atoms}(C^*)$, satisfying 
$S'=(S-\{C^*\})\cup \{a_0\gle \gu\}$, $\mi{guards}(S')=\mi{guards}(S)\cup \{a_0\gle \gu\}$, $a_0\gle \gu\not\in S$.

Case 2.6.1.2.2.1.2.2:
$n\geq 1$.
Then $l^{**}=a_0\swedge\cdots\swedge a_n\gle \gu\not\in S$, $S'=(S-\{C^*\})\cup \{l^{**}\}=(S-\{C^*\})\cup \{a_0\swedge\cdots\swedge a_n\gle \gu\}$, $\{a_0,\dots,a_n\}\subseteq \mi{atoms}(S')$;
$a_0\swedge\cdots\swedge a_n\gle \gu$ is not a guard;
$C^*\not\in \mi{guards}(S)$,
$\mi{guards}(S')=\{C \,|\, C\in (S-\{C^*\})\cup \{a_0\swedge\cdots\swedge a_n\gle \gu\}\ \text{\it is a guard}\}=
                 \{C \,|\, C\in S-\{C^*\}\ \text{\it is a guard}\}=\mi{guards}(S)-\{C^*\}=\{C \,|\, C\in S\ \text{\it is a guard}\}-\{C^*\}=\mi{guards}(S)$;
$S'$ is positively guarded.
We put $C^{**}=a_0\swedge\cdots\swedge a_n\gle \gu\in \mi{OrdPropCl}^\gu$.
We get that
$C^{**}=a_0\swedge\cdots\swedge a_n\gle \gu\not\in S$,
$S'=(S-\{C^*\})\cup \{a_0\swedge\cdots\swedge a_n\gle \gu\}=(S-\{C^*\})\cup \{C^{**}\}$,
$\mi{atoms}(C^{**})=\{a_0,\dots,a_n\}=\mi{atoms}(l^*)=\mi{atoms}(C^*)$;
$C^{**}=a_0\swedge\cdots\swedge a_n\gle \gu\neq \square$;
$a_0\swedge\cdots\swedge a_n\gle \gu\in \mi{OrdPropLit}^\gu$;
for all $i\leq n$, $\mi{guards}(S',a_i)=\mi{guards}(S,a_i)=\{\gz\gle a_i\}$;
$\mi{valid}(a_0\swedge\cdots\swedge a_n\gle \gu,S')$;
$\mi{valid}(C^{**},S')$.
We put
\begin{equation} \notag
\mi{Tree}=\dfrac{S}
                {S'}.
\end{equation}
Hence, $\mi{Tree}$ is a finite linear {\it DPLL}-tree with the root $S$ constructed using Rules (\cref{ceq4hr1111})--(\cref{ceq4hr66}) such that
for its only leaf $S'$, $S'\subseteq_{\mc F} \mi{OrdPropCl}$ is positively guarded, and there exists $C^{**}\in \mi{OrdPropCl}^\gu$, $\mi{atoms}(C^{**})=\mi{atoms}(C^*)$, satisfying 
$S'=(S-\{C^*\})\cup \{C^{**}\}$, $\mi{guards}(S')=\mi{guards}(S)$, $C^{**}\not\in S$, $\mi{valid}(C^{**},S')$.

Case 2.6.1.2.2.2:
$C^*-\{l^*\}\neq \square$.
Then $(C^*-\{l^*\})\cup \{l^{**}\}\in S'$,
$l^{**}\not\in C^*\supseteq C^*-\{l^*\}$;
$(C^*-\{l^*\})\cup \{l^{**}\}$ is not unit;
$(C^*-\{l^*\})\cup \{l^{**}\}$ is not a guard;
$(C^*-\{l^*\})\cup \{l^{**}\}\not\in \mi{guards}(S')$,
$(C^*-\{l^*\})\cup \{l^{**}\}\in S'-\mi{guards}(S')$;
$C^*\not\in \mi{guards}(S)$,
$\mi{guards}(S')=\{C \,|\, C\in (S-\{C^*\})\cup \{(C^*-\{l^*\})\cup \{l^{**}\}\}\ \text{\it is a guard}\}=\{C \,|\, C\in S-\{C^*\}\ \text{\it is a guard}\}=
                 \mi{guards}(S)-\{C^*\}=\{C \,|\, C\in S\ \text{\it is a guard}\}-\{C^*\}=\mi{guards}(S)$;
$S'$ is positively guarded;
$l^{**}=a_0\swedge\cdots\swedge a_n\diamond^{**} \gu\in \mi{OrdPropLit}^\gu$, $\diamond^{**}\in \{\geql,\gle\}$;
for all $i\leq n$, $a_i\in \mi{atoms}(l^{**})\subseteq \mi{atoms}(S')$, $\mi{guards}(S',a_i)=\mi{guards}(S,a_i)=\{\gz\gle a_i\}$;
$\mi{valid}(l^{**},S')$;
for all $l\in C^*$, 
$\mi{valid}(l,S)$ if and only if
$l\in \mi{OrdPropLit}^\gu$;
if either $l=\Cn\diamond \gu$ or $l=\gu\gleq \Cn$, $\Cn\in \mi{PropConj}$, and $\diamond\in \{\geql,\gle\}$, 
$\Cn=b_0\swedge\cdots\swedge b_k$, $b_j\in \mi{PropAtom}$;
for all $j\leq k$, 
$C^*\in S$, $b_j\in \mi{atoms}(\Cn)=\mi{atoms}(l)\subseteq \mi{atoms}(C^*)\subseteq \mi{atoms}(S)=\mi{atoms}(S')$,
$\mi{guards}(S,b_j)=\mi{guards}(S',b_j)=\{\gz\gle b_j\}$ if and only if
$\mi{valid}(l,S')$;
$l^*\in \mi{invalid}(C^*,S)$,
$\mi{invalid}((C^*-\{l^*\})\cup \{l^{**}\},S')=\{l \,|\, l\in (C^*-\{l^*\})\cup \{l^{**}\},\ \text{\it not}\ \mi{valid}(l,S')\}=\{l \,|\, l\in C^*-\{l^*\},\ \text{\it not}\ \mi{valid}(l,S')\}=
                                               \{l \,|\, l\in C^*-\{l^*\},\ \text{\it not}\ \mi{valid}(l,S)\}=
                                               \mi{invalid}(C^*,S)-\{l^*\}=\{l \,|\, l\in C^*,\ \text{\it not}\ \mi{valid}(l,S)\}-\{l^*\}\subset \mi{invalid}(C^*,S)$;
by the induction hypothesis for $S'$ and $(C^*-\{l^*\})\cup \{l^{**}\}$, there exists a finite linear {\it DPLL}-tree $\mi{Tree}'$ with the root $S'$ constructed using Rules (\cref{ceq4hr1111})--(\cref{ceq4hr66}) satisfying
for its only leaf $S''$ that either $\square\in S''$, or $S''\subseteq_{\mc F} \mi{OrdPropCl}$ is semi-positively guarded, and exactly one of the following points holds.
\begin{enumerate}[\rm (a)]
\item
$S''=S'$, $\mi{valid}((C^*-\{l^*\})\cup \{l^{**}\},S')$;
\item
$S''=S'-\{(C^*-\{l^*\})\cup \{l^{**}\}\}$, $\mi{guards}(S'')=\mi{guards}(S')$;
\item
there exists $C^{**}\in \mi{OrdPropCl}^\gu$, $\mi{atoms}(C^{**})\subseteq \mi{atoms}((C^*-\{l^*\})\cup \{l^{**}\})$, satisfying 
$S''=(S'-\{(C^*-\{l^*\})\cup \{l^{**}\}\})\cup \{C^{**}\}$, $\mi{guards}(S'')=\mi{guards}(S')$, $C^{**}\not\in S'$, $\mi{valid}(C^{**},S'')$;
\item
there exists $b^*\gle \gu\in \mi{OrdPropCl}^\gu$, $b^*\in \mi{atoms}((C^*-\{l^*\})\cup \{l^{**}\})$, satisfying 
$S''=(S'-\{(C^*-\{l^*\})\cup \{l^{**}\}\})\cup \{b^*\gle \gu\}$, $\mi{guards}(S'')=\mi{guards}(S')\cup \{b^*\gle \gu\}$, $b^*\gle \gu\not\in S'$;
\item
there exists $S^{**}=\{b_0\geql \gu,\dots,b_k\geql \gu\}\subseteq \mi{OrdPropCl}^\gu$, $\{b_0,\dots,b_k\}\subseteq \mi{atoms}((C^*-\{l^*\})\cup \{l^{**}\})$, 
$\{\gz\gle b_0,\dots,\gz\gle b_k\}\subseteq \mi{guards}(S')$, satisfying
$S''=(S'-(\{(C^*-\{l^*\})\cup \{l^{**}\}\}\cup \{\gz\gle b_0,\dots,\gz\gle b_k\}))\cup S^{**}$, $\mi{guards}(S'')=(\mi{guards}(S')-\{\gz\gle b_0,\dots,\gz\gle b_k\})\cup S^{**}$,
$S^{**}\cap S'=\emptyset$. 
\end{enumerate}
We get that 
$(C^*-\{l^*\})\cup \{l^{**}\}\not\in S$,
$S'-\{(C^*-\{l^*\})\cup \{l^{**}\}\}=((S-\{C^*\})\cup \{(C^*-\{l^*\})\cup \{l^{**}\}\})-\{(C^*-\{l^*\})\cup \{l^{**}\}\}=
                                     ((S-\{C^*\})-\{(C^*-\{l^*\})\cup \{l^{**}\}\})\cup (\{(C^*-\{l^*\})\cup \{l^{**}\}\}-\{(C^*-\{l^*\})\cup \{l^{**}\}\})=S-\{C^*\}$;
\begin{enumerate}[\rm (a)]
\item
$\mi{valid}((C^*-\{l^*\})\cup \{l^{**}\},S')$; 
for all $l\in (C^*-\{l^*\})\cup \{l^{**}\}$, $\mi{valid}(l,S')$, $l\in \mi{OrdPropLit}^\gu$;
$l^*\in C^*$, $l^{**}\not\in C^*$;
we put $C^{**}=(C^*-\{l^*\})\cup \{l^{**}\}\in \mi{OrdPropCl}^\gu$;
$\mi{atoms}(C^{**})=\mi{atoms}((C^*-\{l^*\})\cup \{l^{**}\})=\mi{atoms}(C^*)$,
$C^{**}=(C^*-\{l^*\})\cup \{l^{**}\}\not\in S$,
$S''=S'=(S-\{C^*\})\cup \{(C^*-\{l^*\})\cup \{l^{**}\}\}=(S-\{C^*\})\cup \{C^{**}\}$,
$\mi{guards}(S'')=\mi{guards}(S')=\mi{guards}(S)$,
$\mi{valid}(C^{**},S'')$;
$C^{**}\in S''$, $S''\neq S$;
\item
$C^*\in S$, $S''=S'-\{(C^*-\{l^*\})\cup \{l^{**}\}\}=S-\{C^*\}\neq S$, $\mi{guards}(S'')=\mi{guards}(S')=\mi{guards}(S)$;
\item
there exists $C^{**}\in \mi{OrdPropCl}^\gu$, $\mi{atoms}(C^{**})\subseteq \mi{atoms}((C^*-\{l^*\})\cup \{l^{**}\})=\mi{atoms}(C^*)$, satisfying 
$S''=(S'-\{(C^*-\{l^*\})\cup \{l^{**}\}\})\cup \{C^{**}\}=(S-\{C^*\})\cup \{C^{**}\}$, $\mi{guards}(S'')=\mi{guards}(S')=\mi{guards}(S)$;
we have that $S''$ is semi-positively guarded, and $S$ is positively guarded;
$\mi{atoms}(S'')=\mi{atoms}(\mi{guards}(S''))=\mi{atoms}(\mi{guards}(S))=\mi{atoms}(S)$;
for all $l\in C^*$, 
$\mi{valid}(l,S)$ if and only if
$l\in \mi{OrdPropLit}^\gu$;
if either $l=\Cn\diamond \gu$ or $l=\gu\gleq \Cn$, $\Cn\in \mi{PropConj}$, and $\diamond\in \{\geql,\gle\}$, 
$\Cn=b_0\swedge\cdots\swedge b_k$, $b_j\in \mi{PropAtom}$;
for all $j\leq k$, 
$b_j\in \mi{atoms}(\Cn)=\mi{atoms}(l)\subseteq \mi{atoms}(C^*)\subseteq \mi{atoms}(S)=\mi{atoms}(S'')$,
$\mi{guards}(S,b_j)=\mi{guards}(S'',b_j)=\{\gz\gle b_j\}$ if and only if
$\mi{valid}(l,S'')$;
$\mi{valid}(C^{**},S'')$;
for all $l\in C^{**}$, $\mi{valid}(l,S'')$;
not $\mi{valid}(l^*,S)$, not $\mi{valid}(l^*,S'')$, $l^*\not\in C^{**}$, $C^{**}\neq C^*$,
$C^{**}\not\in S'\supseteq S-\{C^*\}$, $C^{**}\not\in S$, $C^{**}\in S''$, $S''\neq S$;  
\item
there exists $b^*\gle \gu\in \mi{OrdPropCl}^\gu$, $b^*\in \mi{atoms}((C^*-\{l^*\})\cup \{l^{**}\})=\mi{atoms}(C^*)\subseteq \mi{atoms}(S)$, satisfying 
$S''=(S'-\{(C^*-\{l^*\})\cup \{l^{**}\}\})\cup \{b^*\gle \gu\}=(S-\{C^*\})\cup \{b^*\gle \gu\}$, $\mi{guards}(S'')=\mi{guards}(S')\cup \{b^*\gle \gu\}=\mi{guards}(S)\cup \{b^*\gle \gu\}$; 
$b^*\gle \gu\in \mi{guards}(b^*)$, $C^*\not\in \mi{guards}(S)\supseteq \mi{guards}(S,b^*)=S\cap \mi{guards}(b^*)$, $C^*\not\in \mi{guards}(b^*)$, $b^*\gle \gu\neq C^*$,
$b^*\gle \gu\not\in S'\supseteq S-\{C^*\}$, $b^*\gle \gu\not\in S$, $b^*\gle \gu\in S''$, $S''\neq S$;
\item
there exists $S^{**}=\{b_0\geql \gu,\dots,b_k\geql \gu\}\subseteq \mi{OrdPropCl}^\gu$, $\{b_0,\dots,b_k\}\subseteq \mi{atoms}((C^*-\{l^*\})\cup \{l^{**}\})=\mi{atoms}(C^*)\subseteq \mi{atoms}(S)$, 
$\{\gz\gle b_0,\dots,\gz\gle b_k\}\subseteq \mi{guards}(S')=\mi{guards}(S)$, satisfying
$S''=(S'-(\{(C^*-\{l^*\})\cup \{l^{**}\}\}\cup \{\gz\gle b_0,\dots,\gz\gle b_k\}))\cup S^{**}=((S'-\{(C^*-\{l^*\})\cup \{l^{**}\}\})-\{\gz\gle b_0,\dots,\gz\gle b_k\})\cup S^{**}=
     ((S-\{C^*\})-\{\gz\gle b_0,\dots,\gz\gle b_k\})\cup S^{**}=(S-(\{C^*\}\cup \{\gz\gle b_0,\dots,\gz\gle b_k\}))\cup S^{**}$, 
$\mi{guards}(S'')=(\mi{guards}(S')-\{\gz\gle b_0,\dots,\gz\gle b_k\})\cup S^{**}=(\mi{guards}(S)-\{\gz\gle b_0,\dots,\gz\gle b_k\})\cup S^{**}$;
for all $j\leq k$, $b_j\geql \gu\in \mi{guards}(b_j)$, $C^*\not\in \mi{guards}(S)\supseteq \mi{guards}(S,b_j)=S\cap \mi{guards}(b_j)$, $C^*\not\in \mi{guards}(b_j)$, $b_j\geql \gu\neq C^*$;
$S^{**}\cap S=\{b_0\geql \gu,\dots,b_k\geql \gu\}\cap S=\{b_0\geql \gu,\dots,b_k\geql \gu\}\cap (S-\{C^*\})=S^{**}\cap (S-\{C^*\})=S^{**}\cap S'=\emptyset$, 
$S^{**}=\{b_0\geql \gu,\dots,b_k\geql \gu\}\neq \emptyset$, $S^{**}\not\subseteq S$, $S^{**}\subseteq S''$, $S''\neq S$. 
\end{enumerate}
We put
\begin{equation} \notag
\mi{Tree}=\dfrac{S}
                {\mi{Tree}'}.
\end{equation}
Hence, $\mi{Tree}$ is a finite linear {\it DPLL}-tree with the root $S$ constructed using Rules (\cref{ceq4hr1111})--(\cref{ceq4hr66}) such that
for its only leaf $S''$, either $\square\in S''$, or $S''\subseteq_{\mc F} \mi{OrdPropCl}$ is semi-positively guarded, $S''\neq S$, and exactly one of the following points holds.
\begin{enumerate}[\rm (a)]
\item
$S''=S-\{C^*\}$, $\mi{guards}(S'')=\mi{guards}(S)$;
\item
there exists $C^{**}\in \mi{OrdPropCl}^\gu$, $\mi{atoms}(C^{**})\subseteq \mi{atoms}(C^*)$, satisfying 
$S''=(S-\{C^*\})\cup \{C^{**}\}$, $\mi{guards}(S'')=\mi{guards}(S)$, $C^{**}\not\in S$, $\mi{valid}(C^{**},S'')$;
\item
there exists $b^*\gle \gu\in \mi{OrdPropCl}^\gu$, $b^*\in \mi{atoms}(C^*)$, satisfying 
$S''=(S-\{C^*\})\cup \{b^*\gle \gu\}$, $\mi{guards}(S'')=\mi{guards}(S)\cup \{b^*\gle \gu\}$, $b^*\gle \gu\not\in S$;
\item
there exists $S^{**}=\{b_0\geql \gu,\dots,b_k\geql \gu\}\subseteq \mi{OrdPropCl}^\gu$, $\{b_0,\dots,b_k\}\subseteq \mi{atoms}(C^*)$, $\{\gz\gle b_0,\dots,\gz\gle b_k\}\subseteq \mi{guards}(S)$, satisfying
$S''=(S-(\{C^*\}\cup \{\gz\gle b_0,\dots,\gz\gle b_k\}))\cup S^{**}$, $\mi{guards}(S'')=(\mi{guards}(S)-\{\gz\gle b_0,\dots,\gz\gle b_k\})\cup S^{**}$, $S^{**}\cap S=\emptyset$. 
\end{enumerate}

Case 2.6.2:
$\mi{guards}(S,a_{i^*})=\{\gz\gle a_{i^*},a_{i^*}\gle \gu\}$.
We get two cases for $l^*$.

Case 2.6.2.1:
Either $l^*=\Cn^*\geql \gu$ or $l^*=\gu\gleq \Cn^*$.
Then either $l^*=\Cn^*\geql \gu=a_0^{\alpha_0}\swedge\cdots\swedge a_n^{\alpha_n}\geql \gu$ or $l^*=\gu\gleq \Cn^*=\gu\gleq a_0^{\alpha_0}\swedge\cdots\swedge a_n^{\alpha_n}$,
$\mi{guards}(S,a_{i^*})=\{\gz\gle a_{i^*},a_{i^*}\gle \gu\}$;
this case is the same as Case 2.2.

Case 2.6.2.2:
$l^*=\Cn^*\gle \gu$.
Then $l^*=\Cn^*\gle \gu=a_0^{\alpha_0}\swedge\cdots\swedge a_n^{\alpha_n}\gle \gu$, $\mi{guards}(S,a_{i^*})=\{\gz\gle a_{i^*},a_{i^*}\gle \gu\}$;
this case is the same as Case 2.4.

So, in both Cases 1 and 2, the statement holds.
The induction is completed.
%
%
%
\end{proof}

\subsection{Full proof of Lemma \ref{le9}}
\label{S7.11}

\begin{proof}
Let $l^F\in \mi{OrdPropLit}$, $C^F\in \mi{OrdPropCl}$, $S^F\subseteq_{\mc F} \mi{OrdPropCl}$.
We define six measure operators and two auxiliary binary relations as follows:
\begin{IEEEeqnarray*}{RL}
\mi{count}(l^F)       &= \left\{\begin{array}{ll}
                                1 &\ \text{\it if either}\ l^F=\Cn\geql \gu\ \text{\it or}\ l^F=\gu\gleq \Cn, \\
                                  &\ \phantom{\text{\it if}\ \mbox{}}
                                                  \Cn\in \mi{PropConj}, \\[1mm]
                                0 &\ \text{\it else};
                                \end{array}
                         \right. \\[1mm]
\mi{count}(C^F)       &= \sum_{l\in C^F} \mi{count}(l); \\
\mi{count}(S^F)       &= \sum_{C\in S^F-\mi{guards}(S^F)} \mi{count}(C); \\
\mi{invalid}(S^F)     &= \{C \,|\, C\in S^F-\mi{guards}(S^F),\ \text{\it not}\ \mi{valid}(C,S^F)\}; \\
\mi{unsaturated}(S^F) &= \{a \,|\, a\in \mi{atoms}(S^F), \mi{guards}(S^F,a)=\{\gz\gle a\}\}; \\[1mm]
\IEEEeqnarraymulticol{2}{l}{
\mi{measure} : {\mc P}_{\mc F}(\mi{OrdPropCl})\longrightarrow {\mc P}_{\mc F}(\mi{PropAtom})\times {\mc P}_{\mc F}(\mi{PropAtom})\times} \\
\IEEEeqnarraymulticol{2}{l}{
\phantom{\mi{measure} : {\mc P}_{\mc F}(\mi{OrdPropCl})\longrightarrow \mbox{}} \quad 
                                                              {\mc P}_{\mc F}(\mi{OrdPropCl})\times \mbb{N};} \\
\IEEEeqnarraymulticol{2}{l}{
\mi{measure}(S^F)=(\mi{atoms}(S^F),\mi{unsaturated}(S^F),\mi{invalid}(S^F),\mi{count}(S^F));} \\[1mm]
\IEEEeqnarraymulticol{2}{l}{
\preceq, \prec\ \subseteq ({\mc P}_{\mc F}(\mi{PropAtom})\times {\mc P}_{\mc F}(\mi{PropAtom})\times {\mc P}_{\mc F}(\mi{OrdPropCl})\times \mbb{N})^2;} \\[1mm]
(x_1,x_2,x_3,x_4)\preceq (y_1,y_2,y_3,y_4) &\longleftrightarrow \text{\it either}\ x_1\subset y_1,\ \text{\it or}\ x_1=y_1, x_2\subset y_2,\ \text{\it or} \\ 
                                           &\phantom{\mbox{}\longleftrightarrow \mbox{}} \quad
                                                                x_1=y_1, x_2=y_2, x_3\subset y_3,\ \text{\it or} \\ 
                                           &\phantom{\mbox{}\longleftrightarrow \mbox{}} \quad 
                                                                x_1=y_1, x_2=y_2, x_3=y_3, x_4\leq y_4; \\[1mm]
(x_1,x_2,x_3,x_4)\prec (y_1,y_2,y_3,y_4)   &\longleftrightarrow (x_1,x_2,x_3,x_4)\preceq (y_1,y_2,y_3,y_4), \\
                                           &\phantom{\mbox{}\longleftrightarrow \mbox{}} \quad
                                                                (x_1,x_2,x_3,x_4)\neq (y_1,y_2,y_3,y_4).
\end{IEEEeqnarray*}
Note that $\preceq$ is reflexive, antisymmetric, transitive, a well-founded order, which arranges tuples in the lexicographic manner, and 
$\prec$ is irreflexive, transitive, a strict order.
We proceed by induction on $\mi{measure}(S)\in {\mc P}_{\mc F}(\mi{PropAtom})\times {\mc P}_{\mc F}(\mi{PropAtom})\times {\mc P}_{\mc F}(\mi{OrdPropCl})\times \mbb{N}$.

Case 1 (the base case):
$\mi{measure}(S)=(\emptyset,\emptyset,\emptyset,0)$.
We have that $S$ is semi-positively guarded.
Then $S$ is simplified;
$\mi{atoms}(S)=\emptyset$;
trivially, for all $a\in \mi{atoms}(S)=\emptyset$, $a$ is positively guarded in $S$;
$S$ is positively guarded;
$\square\not\in S$; 
for all $C\in S$, 
$\mi{atoms}(C)=\mi{atoms}(S)=\emptyset$;
$C$ does not contain contradictions and tautologies;
$S=\emptyset$;
trivially, for all $C\in S-\mi{guards}(S)=S=\emptyset$, 
either $C\in \mi{PurOrdPropCl}$, 
or $C=a_0\swedge\cdots\swedge a_n\gle \gu\vee C^\natural$, $a_i\in \mi{PropAtom}$, $\mi{guards}(S,a_i)=\{\gz\gle a_i\}$, $C^\natural\in \mi{PurOrdPropCl}$;
$S$ is positive.
We put $\mi{Tree}=S$.
Hence, $\mi{Tree}$ is a finite {\it DPLL}-tree with the root $S$ constructed using Rules (\cref{ceq4hr1x}), (\cref{ceq4hr1111})--(\cref{ceq4hr66}), (\cref{ceq4hr8}) such that
for its only leaf $S$, $S\subseteq_{\mc F} \mi{OrdPropCl}$ is positive.

Case 2 (the induction case):
$(\emptyset,\emptyset,\emptyset,0)\neq \mi{measure}(S)\in {\mc P}_{\mc F}(\mi{PropAtom})\times {\mc P}_{\mc F}(\mi{PropAtom})\times {\mc P}_{\mc F}(\mi{OrdPropCl})\times \mbb{N}$.
We have that $S$ is semi-positively guarded.
Then $S$ is simplified;
$\mi{atoms}(S)\neq \emptyset$ or $\mi{unsaturated}(S)\neq \emptyset$ or $\mi{invalid}(S)\neq \emptyset$ or $\mi{count}(S)\geq 1$;
if $\mi{atoms}(S)=\emptyset$, 
$\square\not\in S$; 
for all $C\in S$, 
$\mi{atoms}(C)=\mi{atoms}(S)=\emptyset$;
$C$ does not contain contradictions and tautologies;
$S=S-\mi{guards}(S)=\mi{unsaturated}(S)=\{a \,|\, a\in \mi{atoms}(S), \mi{guards}(S,a)=\{\gz\gle a\}\}=\mi{invalid}(S)=\{C \,|\, C\in S-\mi{guards}(S),\ \text{\it not}\ \mi{valid}(C,S)\}=\emptyset$, 
$\mi{count}(S)=\sum_{C\in S-\mi{guards}(S)} \mi{count}(C)=0$;
$\mi{atoms}(S)\neq \emptyset$.
We distinguish two cases for $\mi{unsaturated}(S)$.

Case 2.1:
$\mi{unsaturated}(S)=\emptyset$.
We have that $S$ is semi-positively guarded.
Then, for all $a\in \mi{atoms}(S)$, 
either $\mi{guards}(S,a)=\{\gz\gle a\}$ or $\mi{guards}(S,a)=\{\gz\gle a,a\gle \gu\}$ or $\mi{guards}(S,a)=\{a\geql \gu\}$,
$a\not\in \mi{unsaturated}(S)=\emptyset$, $\mi{guards}(S,a)\neq \{\gz\gle a\}$,
either $\mi{guards}(S,a)=\{\gz\gle a,a\gle \gu\}$ or $\mi{guards}(S,a)=\{a\geql \gu\}$.
We get two cases for $S$.

Case 2.1.1:
There exists $a^*\in \mi{atoms}(S)$ such that $\mi{guards}(S,a^*)=\{a^*\geql \gu\}$.
Then, by Lemma \ref{le66}, there exists a finite linear {\it DPLL}-tree $\mi{Tree}'$ with the root $S$ constructed using Rules (\cref{ceq4hr1111111})--(\cref{ceq4hr4}), (\cref{ceq4hr8}) such that
for its only leaf $S'$, either $\square\in S'$, or $S'\subseteq_{\mc F} \mi{OrdPropCl}$ is semi-positively guarded, $\mi{atoms}(S')\subseteq \mi{atoms}(S)-\{a^*\}$.
We get two cases for $S'$.

Case 2.1.1.1:
$\square\in S'$.
We put $\mi{Tree}=\mi{Tree}'$.
Hence, $\mi{Tree}$ is a finite {\it DPLL}-tree with the root $S$ constructed using Rules (\cref{ceq4hr1x}), (\cref{ceq4hr1111})--(\cref{ceq4hr66}), (\cref{ceq4hr8}) such that
for its only leaf $S'$, $\square\in S'$.

Case 2.1.1.2:
$S'\subseteq_{\mc F} \mi{OrdPropCl}$ is semi-positively guarded, $\mi{atoms}(S')\subseteq \mi{atoms}(S)-\{a^*\}$.
Then $a^*\in \mi{atoms}(S)$, $\mi{atoms}(S')\subseteq \mi{atoms}(S)-\{a^*\}\subset \mi{atoms}(S)$, 
$\mi{measure}(S')=(\mi{atoms}(S'),\mi{unsaturated}(S'),\mi{invalid}(S'),\mi{count}(S'))\prec \mi{measure}(S)=(\mi{atoms}(S),\mi{unsaturated}(S),\mi{invalid}(S),\mi{count}(S))$;
by the induction hypothesis for $S'$, there exists a finite {\it DPLL}-tree $\mi{Tree}''$ with the root $S'$ constructed using Rules (\cref{ceq4hr1x}), (\cref{ceq4hr1111})--(\cref{ceq4hr66}), (\cref{ceq4hr8}) satisfying
for every leaf $S''$ that either $\square\in S''$ or $S''\subseteq_{\mc F} \mi{OrdPropCl}$ is positive.
We put
\begin{equation} \notag
\mi{Tree}=\dfrac{\mi{Tree}'} 
                {\mi{Tree}''}.
\end{equation}
Hence, $\mi{Tree}$ is a finite {\it DPLL}-tree with the root $S$ constructed using Rules (\cref{ceq4hr1x}), (\cref{ceq4hr1111})--(\cref{ceq4hr66}), (\cref{ceq4hr8}) such that
for every leaf $S''$, either $\square\in S''$ or $S''\subseteq_{\mc F} \mi{OrdPropCl}$ is positive.

Case 2.1.2:
For all $a\in \mi{atoms}(S)$, $\mi{guards}(S,a)=\{\gz\gle a,a\gle \gu\}$.
Then $S$ is simplified;
for all $a\in \mi{atoms}(S)$, $a$ is positively guarded in $S$;
$S$ is positively guarded.
We get two cases for $\mi{invalid}(S)$.

Case 2.1.2.1:
$\mi{invalid}(S)=\emptyset$.
Then $S$ is simplified;
for all $C\in S-\mi{guards}(S)$,
$C\not\in \mi{invalid}(S)=\emptyset$,
$\mi{valid}(C,S)$;
$C$ does not contain contradictions and tautologies;
for all $l\in C$, 
$\mi{valid}(l,S)$;
$l$ is not a contradiction or tautology;
$l\in \mi{OrdPropLit}^\gu$;
if either $l=\Cn\diamond \gu$ or $l=\gu\gleq \Cn$, $\Cn\in \mi{PropConj}$, and $\diamond\in \{\geql,\gle\}$,
$\Cn=a_0\swedge\cdots\swedge a_n$, $a_i\in \mi{PropAtom}$, $\mi{guards}(S,a_i)=\{\gz\gle a_i\}$;
either $l\in \mi{PurOrdPropLit}$,
or $l=\Cn^*\diamond^* \gu$ or $l=\gu\gleq \Cn^*$, $\Cn^*\in \mi{PropConj}$, $\diamond^*\in \{\geql,\gle\}$,
$\Cn^*=a_0\swedge\cdots\swedge a_n$, $a_i\in \mi{PropAtom}$, $\mi{guards}(S,a_i)=\{\gz\gle a_i\}$;
for all $i\leq n$,
$a_i\in \mi{atoms}(\Cn^*)=\mi{atoms}(l)\subseteq \mi{atoms}(C)\subseteq \mi{atoms}(S)$,
$a_i\not\in \mi{unsaturated}(S)=\emptyset$, $\mi{guards}(S,a_i)\neq \{\gz\gle a_i\}$;
$l\in \mi{PurOrdPropLit}$;
$C\in \mi{PurOrdPropCl}$;
$S$ is positive.
We put $\mi{Tree}=S$.
Hence, $\mi{Tree}$ is a finite {\it DPLL}-tree with the root $S$ constructed using Rules (\cref{ceq4hr1x}), (\cref{ceq4hr1111})--(\cref{ceq4hr66}), (\cref{ceq4hr8}) such that
for its only leaf $S$, $S\subseteq_{\mc F} \mi{OrdPropCl}$ is positive.

Case 2.1.2.2:
$\emptyset\neq \mi{invalid}(S)\subseteq_{\mc F} \mi{OrdPropCl}$.
Then there exists $C^*\in \mi{invalid}(S)\subseteq S-\mi{guards}(S)$ not satisfying $\mi{valid}(C^*,S)$;
by Lemma \ref{le8}, there exists a finite linear {\it DPLL}-tree $\mi{Tree}'$ with the root $S$ constructed using Rules (\cref{ceq4hr1111})--(\cref{ceq4hr66}) satisfying
for its only leaf $S'$ that either $\square\in S'$, or $S'\subseteq_{\mc F} \mi{OrdPropCl}$ is semi-positively guarded, and exactly one of the following points holds.
\begin{enumerate}[\rm (a)]
\item
$S'=S$, $\mi{valid}(C^*,S)$;
\item
$S'=S-\{C^*\}$, $\mi{guards}(S')=\mi{guards}(S)$;
\item
there exists $C^{**}\in \mi{OrdPropCl}^\gu$, $\mi{atoms}(C^{**})\subseteq \mi{atoms}(C^*)$, satisfying 
$S'=(S-\{C^*\})\cup \{C^{**}\}$, $\mi{guards}(S')=\mi{guards}(S)$, $C^{**}\not\in S$, $\mi{valid}(C^{**},S')$;
\item
there exists $a^*\gle \gu\in \mi{OrdPropCl}^\gu$, $a^*\in \mi{atoms}(C^*)$, satisfying 
$S'=(S-\{C^*\})\cup \{a^*\gle \gu\}$, $\mi{guards}(S')=\mi{guards}(S)\cup \{a^*\gle \gu\}$, $a^*\gle \gu\not\in S$;
\item
there exists $S^{**}=\{a_0\geql \gu,\dots,a_n\geql \gu\}\subseteq \mi{OrdPropCl}^\gu$, $\{a_0,\dots,a_n\}\subseteq \mi{atoms}(C^*)$, $\{\gz\gle a_0,\dots,\gz\gle a_n\}\subseteq \mi{guards}(S)$, satisfying 
$S'=(S-(\{C^*\}\cup \{\gz\gle a_0,\dots,\gz\gle a_n\}))\cup S^{**}$, $\mi{guards}(S')=(\mi{guards}(S)-\{\gz\gle a_0,\dots,\gz\gle a_n\})\cup S^{**}$, $S^{**}\cap S=\emptyset$. 
\end{enumerate}
We get five cases for $S'$.

Case 2.1.2.2.1:
(a) $S'=S$ and $\mi{valid}(C^*,S)$.
This case is a contradiction with not $\mi{valid}(C^*,S)$.

Case 2.1.2.2.2:
(b) $S'=S-\{C^*\}$ and $\mi{guards}(S')=\mi{guards}(S)$.
We have that $S'$ is semi-positively guarded, and $S$ is positively guarded.
Then $\mi{atoms}(S')=\mi{atoms}(\mi{guards}(S'))=\mi{atoms}(\mi{guards}(S))=\mi{atoms}(S)$,
$S'-\mi{guards}(S')=(S-\{C^*\})-\mi{guards}(S)\subseteq S-\mi{guards}(S)$;
for all $C\in S'-\mi{guards}(S')\subseteq S-\mi{guards}(S)$,
for all $l\in C$, 
$\mi{valid}(l,S)$ if and only if
$l\in \mi{OrdPropLit}^\gu$;
if either $l=\Cn\diamond \gu$ or $l=\gu\gleq \Cn$, $\Cn\in \mi{PropConj}$, and $\diamond\in \{\geql,\gle\}$, 
$\Cn=a_0\swedge\cdots\swedge a_n$, $a_i\in \mi{PropAtom}$;
for all $i\leq n$, 
$a_i\in \mi{atoms}(\Cn)=\mi{atoms}(l)\subseteq \mi{atoms}(C)\subseteq \mi{atoms}(S)=\mi{atoms}(S')$,
$\mi{guards}(S,a_i)=\mi{guards}(S',a_i)=\{\gz\gle a_i\}$ if and only if
$\mi{valid}(l,S')$;
$\mi{valid}(C,S)$ if and only if
$C\neq \square$; 
for all $l\in C$, $\mi{valid}(l,S)$;
if $C=\Cn\gle \gu\vee C^\natural$, $\Cn\in \mi{PropConj}$, and $\square\neq C^\natural\in \mi{OrdPropCl}$,
$C^\natural$ does not contain an order literal of the form $\Cn'\gle \gu$, $\Cn'\in \mi{PropConj}$, if and only if
$C\neq \square$; 
for all $l\in C$, $\mi{valid}(l,S')$;
if $C=\Cn\gle \gu\vee C^\natural$, $\Cn\in \mi{PropConj}$, and $\square\neq C^\natural\in \mi{OrdPropCl}$,
$C^\natural$ does not contain an order literal of the form $\Cn'\gle \gu$, $\Cn'\in \mi{PropConj}$, if and only if
$\mi{valid}(C,S')$;
for all $a\in \mi{atoms}(S')=\mi{atoms}(S)$, $\mi{guards}(S',a)=\mi{guards}(S,a)=\{\gz\gle a,a\gle \gu\}\neq \{\gz\gle a\}$;
$\mi{unsaturated}(S')=\{a \,|\, a\in \mi{atoms}(S'), \mi{guards}(S',a)=\{\gz\gle a\}\}=\{a \,|\, a\in \mi{atoms}(S'), \mi{guards}(S',a)=\{\gz\gle a,a\gle \gu\}=\{\gz\gle a\}\}=\emptyset$;
$C^*\not\in S'=S-\{C^*\}\supseteq S'-\mi{guards}(S')$, $C^*\in \mi{invalid}(S)$,
$C^*\not\in \mi{invalid}(S')=\{C \,|\, C\in S'-\mi{guards}(S'),\ \text{\it not}\ \mi{valid}(C,S')\}=\{C \,|\, C\in S'-\mi{guards}(S'),\ \text{\it not}\ \mi{valid}(C,S')\}-\{C^*\}=
                             \{C \,|\, C\in S'-\mi{guards}(S'),\ \text{\it not}\ \mi{valid}(C,S)\}-\{C^*\}\subseteq 
                             \mi{invalid}(S)-\{C^*\}=\{C \,|\, C\in S-\mi{guards}(S),\ \text{\it not}\ \mi{valid}(C,S)\}-\{C^*\}\subset \mi{invalid}(S)$;
$\mi{measure}(S')=(\mi{atoms}(S'),\mi{unsaturated}(S'),\mi{invalid}(S'),\mi{count}(S'))=(\mi{atoms}(S),\mi{unsaturated}(S),\mi{invalid}(S'),\mi{count}(S'))=
                  (\mi{atoms}(S),\emptyset,\mi{invalid}(S'),                                                                                                                               \linebreak[4]
                                                            \mi{count}(S'))\prec \mi{measure}(S)=(\mi{atoms}(S),\mi{unsaturated}(S),\mi{invalid}(S),\mi{count}(S))$;
by the induction hypothesis for $S'$, there exists a finite {\it DPLL}-tree $\mi{Tree}''$ with the root $S'$ constructed using Rules (\cref{ceq4hr1x}), (\cref{ceq4hr1111})--(\cref{ceq4hr66}), (\cref{ceq4hr8}) satisfying
for every leaf $S''$ that either $\square\in S''$ or $S''\subseteq_{\mc F} \mi{OrdPropCl}$ is positive.
We put
\begin{equation} \notag
\mi{Tree}=\dfrac{\mi{Tree}'} 
                {\mi{Tree}''}.
\end{equation}
Hence, $\mi{Tree}$ is a finite {\it DPLL}-tree with the root $S$ constructed using Rules (\cref{ceq4hr1x}), (\cref{ceq4hr1111})--(\cref{ceq4hr66}), (\cref{ceq4hr8}) such that
for every leaf $S''$, either $\square\in S''$ or $S''\subseteq_{\mc F} \mi{OrdPropCl}$ is positive.

Case 2.1.2.2.3:
(c) There exists $C^{**}\in \mi{OrdPropCl}^\gu$, $\mi{atoms}(C^{**})\subseteq \mi{atoms}(C^*)$, such that
$S'=(S-\{C^*\})\cup \{C^{**}\}$, $\mi{guards}(S')=\mi{guards}(S)$, $C^{**}\not\in S$, $\mi{valid}(C^{**},S')$.
We have that $S'$ is semi-positively guarded, and $S$ is positively guarded.
Then $\mi{atoms}(S')=\mi{atoms}(\mi{guards}(S'))=\mi{atoms}(\mi{guards}(S))=\mi{atoms}(S)$,
$(S-\{C^*\})-\mi{guards}(S')=(S-\{C^*\})-\mi{guards}(S)\subseteq S-\mi{guards}(S)$;
for all $C\in (S-\{C^*\})-\mi{guards}(S')\subseteq S-\mi{guards}(S)$,
for all $l\in C$, 
$\mi{valid}(l,S)$ if and only if
$l\in \mi{OrdPropLit}^\gu$;
if either $l=\Cn\diamond \gu$ or $l=\gu\gleq \Cn$, $\Cn\in \mi{PropConj}$, and $\diamond\in \{\geql,\gle\}$, 
$\Cn=a_0\swedge\cdots\swedge a_n$, $a_i\in \mi{PropAtom}$;
for all $i\leq n$, 
$a_i\in \mi{atoms}(\Cn)=\mi{atoms}(l)\subseteq \mi{atoms}(C)\subseteq \mi{atoms}(S)=\mi{atoms}(S')$,
$\mi{guards}(S,a_i)=\mi{guards}(S',a_i)=\{\gz\gle a_i\}$ if and only if
$\mi{valid}(l,S')$;
$\mi{valid}(C,S)$ if and only if
$C\neq \square$; 
for all $l\in C$, $\mi{valid}(l,S)$;
if $C=\Cn\gle \gu\vee C^\natural$, $\Cn\in \mi{PropConj}$, and $\square\neq C^\natural\in \mi{OrdPropCl}$,
$C^\natural$ does not contain an order literal of the form $\Cn'\gle \gu$, $\Cn'\in \mi{PropConj}$, if and only if
$C\neq \square$; 
for all $l\in C$, $\mi{valid}(l,S')$;
if $C=\Cn\gle \gu\vee C^\natural$, $\Cn\in \mi{PropConj}$, and $\square\neq C^\natural\in \mi{OrdPropCl}$,
$C^\natural$ does not contain an order literal of the form $\Cn'\gle \gu$, $\Cn'\in \mi{PropConj}$, if and only if
$\mi{valid}(C,S')$;
for all $a\in \mi{atoms}(S')=\mi{atoms}(S)$, $\mi{guards}(S',a)=\mi{guards}(S,a)=\{\gz\gle a,a\gle \gu\}\neq \{\gz\gle a\}$;
$\mi{unsaturated}(S')=\{a \,|\, a\in \mi{atoms}(S'), \mi{guards}(S',a)=\{\gz\gle a\}\}=\{a \,|\, a\in \mi{atoms}(S'), \mi{guards}(S',a)=\{\gz\gle a,a\gle \gu\}=\{\gz\gle a\}\}=\emptyset$;
$C^*\not\in S-\{C^*\}$,
$C^*\in S$, $C^*\neq C^{**}$,
$C^*\not\in S'=(S-\{C^*\})\cup \{C^{**}\}\supseteq S'-\mi{guards}(S')$,
$C^{**}\not\in S\supseteq \mi{guards}(S)=\mi{guards}(S')$, 
$((S-\{C^*\})\cup \{C^{**}\})-\mi{guards}(S')=((S-\{C^*\})-\mi{guards}(S'))\cup (\{C^{**}\}-\mi{guards}(S'))=((S-\{C^*\})-\mi{guards}(S'))\cup \{C^{**}\}$, $C^*\in \mi{invalid}(S)$,
$C^*\not\in \mi{invalid}(S')=\{C \,|\, C\in S'-\mi{guards}(S'),\ \text{\it not}\ \mi{valid}(C,S')\}=\{C \,|\, C\in S'-\mi{guards}(S'),\ \text{\it not}\ \mi{valid}(C,S')\}-\{C^*\}=
                             \{C \,|\, C\in ((S-\{C^*\})\cup \{C^{**}\})-\mi{guards}(S'),\ \text{\it not}\ \mi{valid}(C,S')\}-\{C^*\}=
                             \{C \,|\, C\in ((S-\{C^*\})-\mi{guards}(S'))\cup \{C^{**}\},\ \text{\it not}\ \mi{valid}(C,S')\}-\{C^*\}=
                             \{C \,|\, C\in (S-\{C^*\})-\mi{guards}(S'),\ \text{\it not}\ \mi{valid}(C,S')\}-\{C^*\}=
                             \{C \,|\, C\in (S-\{C^*\})-\mi{guards}(S'),\ \text{\it not}\ \mi{valid}(C,S)\}-\{C^*\}\subseteq 
                             \mi{invalid}(S)-\{C^*\}=\{C \,|\, C\in S-\mi{guards}(S),                                                                                                      \linebreak[4] 
                                                                                      \text{\it not}\ \mi{valid}(C,S)\}-\{C^*\}\subset \mi{invalid}(S)$;
$\mi{measure}(S')=(\mi{atoms}(S'),\mi{unsaturated}(S'),                                                                                                                                    \linebreak[4]
                                                       \mi{invalid}(S'),\mi{count}(S'))=(\mi{atoms}(S),\mi{unsaturated}(S),\mi{invalid}(S'),\mi{count}(S'))=
                  (\mi{atoms}(S),\emptyset,\mi{invalid}(S'),\mi{count}(S'))\prec \mi{measure}(S)=(\mi{atoms}(S),\mi{unsaturated}(S),                                                       \linebreak[4]
                                                                                                                                    \mi{invalid}(S),\mi{count}(S))$;
by the induction hypothesis for $S'$, there exists a finite {\it DPLL}-tree $\mi{Tree}''$ with the root $S'$ constructed using Rules (\cref{ceq4hr1x}), (\cref{ceq4hr1111})--(\cref{ceq4hr66}), (\cref{ceq4hr8}) satisfying
for every leaf $S''$ that either $\square\in S''$ or $S''\subseteq_{\mc F} \mi{OrdPropCl}$ is positive.
We put
\begin{equation} \notag
\mi{Tree}=\dfrac{\mi{Tree}'} 
                {\mi{Tree}''}.
\end{equation}
Hence, $\mi{Tree}$ is a finite {\it DPLL}-tree with the root $S$ constructed using Rules (\cref{ceq4hr1x}), (\cref{ceq4hr1111})--(\cref{ceq4hr66}), (\cref{ceq4hr8}) such that
for every leaf $S''$, either $\square\in S''$ or $S''\subseteq_{\mc F} \mi{OrdPropCl}$ is positive.

Case 2.1.2.2.4:
(d) There exists $a^*\gle \gu\in \mi{OrdPropCl}^\gu$, $a^*\in \mi{atoms}(C^*)$, such that
$S'=(S-\{C^*\})\cup \{a^*\gle \gu\}$, $\mi{guards}(S')=\mi{guards}(S)\cup \{a^*\gle \gu\}$, $a^*\gle \gu\not\in S$.
Then $C^*\in S$, $a^*\in \mi{atoms}(C^*)\subseteq \mi{atoms}(S)$;
$a^*$ is positively guarded in $S$;
either $\mi{guards}(S,a^*)=\{\gz\gle a^*\}$ or $\mi{guards}(S,a^*)=\{\gz\gle a^*,a^*\gle \gu\}$, 
$a^*\not\in \mi{unsaturated}(S)=\emptyset$, $\mi{guards}(S,a^*)\neq \{\gz\gle a^*\}$,
$a^*\gle \gu\not\in S\supseteq \mi{guards}(S,a^*)$, $a^*\gle \gu\in \{\gz\gle a^*,a^*\gle \gu\}$, $\mi{guards}(S,a^*)\neq \{\gz\gle a^*,a^*\gle \gu\}$,
which is a contradiction.

Case 2.1.2.2.5:
(e) There exists $S^{**}=\{a_0\geql \gu,\dots,a_n\geql \gu\}\subseteq \mi{OrdPropCl}^\gu$, $\{a_0,\dots,a_n\}\subseteq \mi{atoms}(C^*)$, $\{\gz\gle a_0,\dots,\gz\gle a_n\}\subseteq \mi{guards}(S)$, such that
$S'=(S-(\{C^*\}\cup \{\gz\gle a_0,\dots,\gz\gle a_n\}))\cup S^{**}$, $\mi{guards}(S')=(\mi{guards}(S)-\{\gz\gle a_0,\dots,\gz\gle a_n\})\cup S^{**}$, $S^{**}\cap S=\emptyset$.
Then $C^*\in S$,
$\mi{atoms}(S')=\mi{atoms}((S-(\{C^*\}\cup \{\gz\gle a_0,\dots,\gz\gle a_n\}))\cup S^{**})=\mi{atoms}(S-(\{C^*\}\cup \{\gz\gle a_0,\dots,\gz\gle a_n\}))\cup \mi{atoms}(S^{**})\subseteq
                \mi{atoms}(S-\{C^*\})\cup \{a_0,\dots,a_n\}\subseteq \mi{atoms}(S-\{C^*\})\cup \mi{atoms}(C^*)=\mi{atoms}((S-\{C^*\})\cup \{C^*\})=\mi{atoms}(S)$, 
$a_0\geql \gu\in S^{**}\subseteq S'$, $a_0\geql \gu\in \mi{guards}(a_0)$, $a_0\geql \gu\in \mi{guards}(S',a_0)=S'\cap \mi{guards}(a_0)$,
$a_0\in \mi{atoms}(a_0\geql \gu)\subseteq \mi{atoms}(S')$;
$a_0$ is semi-positively guarded in $S'$;
either $\mi{guards}(S',a_0)=\{\gz\gle a_0\}$ or $\mi{guards}(S',a_0)=\{\gz\gle a_0,a_0\gle \gu\}$ or $\mi{guards}(S',a_0)=\{a_0\geql \gu\}$,
$a_0\geql \gu\not\in \{\gz\gle a_0\}$, $\mi{guards}(S',a_0)\neq \{\gz\gle a_0\}$,
$a_0\geql \gu\not\in \{\gz\gle a_0,a_0\gle \gu\}$, $\mi{guards}(S',a_0)\neq \{\gz\gle a_0,a_0\gle \gu\}$,
$\mi{guards}(S',a_0)=\{a_0\geql \gu\}$;
by Lemma \ref{le66} for $S'$ and $a_0$, there exists a finite linear {\it DPLL}-tree $\mi{Tree}''$ with the root $S'$ constructed using Rules (\cref{ceq4hr1111111})--(\cref{ceq4hr4}), (\cref{ceq4hr8}) satisfying
for its only leaf $S''$ that either $\square\in S''$, or $S''\subseteq_{\mc F} \mi{OrdPropCl}$ is semi-positively guarded, $\mi{atoms}(S'')\subseteq \mi{atoms}(S')-\{a_0\}$.
We get two cases for $S''$.

Case 2.1.2.2.5.1:
$\square\in S''$.
We put 
\begin{equation} \notag
\mi{Tree}=\dfrac{\mi{Tree}'} 
                {\mi{Tree}''}.
\end{equation}
Hence, $\mi{Tree}$ is a finite {\it DPLL}-tree with the root $S$ constructed using Rules (\cref{ceq4hr1x}), (\cref{ceq4hr1111})--(\cref{ceq4hr66}), (\cref{ceq4hr8}) such that
for its only leaf $S''$, $\square\in S''$.

Case 2.1.2.2.5.2:
$S''\subseteq_{\mc F} \mi{OrdPropCl}$ is semi-positively guarded, $\mi{atoms}(S'')\subseteq \mi{atoms}(S')-\{a_0\}$.
Then $a_0\in \mi{atoms}(S')$, $\mi{atoms}(S'')\subseteq \mi{atoms}(S')-\{a_0\}\subset \mi{atoms}(S')\subseteq \mi{atoms}(S)$, 
$\mi{measure}(S'')=(\mi{atoms}(S''),\mi{unsaturated}(S''),\mi{invalid}(S''),                                                                                                               \linebreak[4]
                                                                            \mi{count}(S''))\prec \mi{measure}(S)=(\mi{atoms}(S),\mi{unsaturated}(S),\mi{invalid}(S),\mi{count}(S))$;
by the induction hypothesis for $S''$, there exists a finite {\it DPLL}-tree $\mi{Tree}'''$ with the root $S''$ constructed using Rules (\cref{ceq4hr1x}), (\cref{ceq4hr1111})--(\cref{ceq4hr66}), (\cref{ceq4hr8}) satisfying
for every leaf $S'''$ that either $\square\in S'''$ or $S'''\subseteq_{\mc F} \mi{OrdPropCl}$ is positive.
We put
\begin{equation} \notag
\mi{Tree}=\begin{array}[c]{c}
          \mi{Tree}' \\[0.4mm]
          \hline \\[-3.8mm]
          \mi{Tree}'' \\[0.4mm]
          \hline \\[-3.8mm]
          \mi{Tree}'''.
          \end{array}
\end{equation}
Hence, $\mi{Tree}$ is a finite {\it DPLL}-tree with the root $S$ constructed using Rules (\cref{ceq4hr1x}), (\cref{ceq4hr1111})--(\cref{ceq4hr66}), (\cref{ceq4hr8}) such that
for every leaf $S'''$, either $\square\in S'''$ or $S'''\subseteq_{\mc F} \mi{OrdPropCl}$ is positive.

Case 2.2:
$\emptyset\neq \mi{unsaturated}(S)\subseteq_{\mc F} \mi{PropAtom}$.
We have that $S$ is semi-positively guarded.
Then, for all $a\in \mi{atoms}(S)$, either $\mi{guards}(S,a)=\{\gz\gle a\}$ or $\mi{guards}(S,a)=\{\gz\gle a,a\gle \gu\}$ or $\mi{guards}(S,a)=\{a\geql \gu\}$.
We get two cases for $S$.

Case 2.2.1:
There exists $a^*\in \mi{atoms}(S)$ such that $\mi{guards}(S,a^*)=\{a^*\geql \gu\}$.
This case is the same as Case 2.1.1.

Case 2.2.2:
For all $a\in \mi{atoms}(S)$, either $\mi{guards}(S,a)=\{\gz\gle a\}$ or $\mi{guards}(S,a)=\{\gz\gle a,a\gle \gu\}$.
Then $S$ is simplified;
for all $a\in \mi{atoms}(S)$, $a$ is positively guarded in $S$;
$S$ is positively guarded.
We get two cases for $\mi{invalid}(S)$.

Case 2.2.2.1:
$\mi{invalid}(S)=\emptyset$.
We get two cases for $\mi{count}(S)$.

Case 2.2.2.1.1:
$\mi{count}(S)=0$.
Then $S$ is simplified;
for all $C\in S-\mi{guards}(S)$,
$C\not\in \mi{invalid}(S)=\emptyset$,
$\mi{valid}(C,S)$,
$\mi{count}(S)=0=\sum_{C\in S-\mi{guards}(S)} \mi{count}(C)$,
$\mi{count}(C)=0$;
$C$ does not contain contradictions and tautologies;
for all $l\in C$, 
$\mi{valid}(l,S)$,
$\mi{count}(C)=0=\sum_{l\in C} \mi{count}(l)$, 
$\mi{count}(l)=0$;
$l$ is not a contradiction or tautology;
$l\in \mi{OrdPropLit}^\gu$;
if either $l=\Cn\diamond \gu$ or $l=\gu\gleq \Cn$, $\Cn\in \mi{PropConj}$, and $\diamond\in \{\geql,\gle\}$,
$\Cn=a_0\swedge\cdots\swedge a_n$, $a_i\in \mi{PropAtom}$, $\mi{guards}(S,a_i)=\{\gz\gle a_i\}$;
either $l\in \mi{PurOrdPropLit}$,
or $l=\Cn^*\diamond^* \gu$ or $l=\gu\gleq \Cn^*$, $\Cn^*\in \mi{PropConj}$, $\diamond^*\in \{\geql,\gle\}$,
$\Cn^*=a_0\swedge\cdots\swedge a_n$, $a_i\in \mi{PropAtom}$, $\mi{guards}(S,a_i)=\{\gz\gle a_i\}$,
$\mi{count}(\Cn^*\geql \gu)=\mi{count}(\gu\gleq \Cn^*)=1$, $l\neq \Cn^*\geql \gu, \gu\gleq \Cn^*$, 
$l=\Cn^*\gle \gu$,
either $l\in \mi{PurOrdPropLit}$ or $l=\Cn^*\gle \gu$;
$C\in \mi{OrdPropCl}^\gu$;
if $C=\Cn\gle \gu\vee C^\natural$, $\Cn\in \mi{PropConj}$, and $C^\natural\in \mi{OrdPropCl}^\gu$,
$C^\natural$ does not contain an order literal of the form $\Cn'\gle \gu$, $\Cn'\in \mi{PropConj}$;
either $C\in \mi{PurOrdPropCl}$,
or $C=\Cn^{**}\gle \gu\vee C^\flat$, $\Cn^{**}\in \mi{PropConj}$, $C^\flat\in \mi{OrdPropCl}^\gu$,
$\Cn^{**}=a_0\swedge\cdots\swedge a_n$, $a_i\in \mi{PropAtom}$, $\mi{guards}(S,a_i)=\{\gz\gle a_i\}$;
$C^\flat$ does not contain an order literal of the form $\Cn'\gle \gu$, $\Cn'\in \mi{PropConj}$;
for all $l\in C^\flat\subseteq C$, 
either $l\in \mi{PurOrdPropLit}$ or $l=\Cn^*\gle \gu$, 
$l\neq \Cn^*\gle \gu$,
$l\in \mi{PurOrdPropLit}$;
$C^\flat\in \mi{PurOrdPropCl}$;
$S$ is positive.
We put $\mi{Tree}=S$.
Hence, $\mi{Tree}$ is a finite {\it DPLL}-tree with the root $S$ constructed using Rules (\cref{ceq4hr1x}), (\cref{ceq4hr1111})--(\cref{ceq4hr66}), (\cref{ceq4hr8}) such that
for its only leaf $S$, $S\subseteq_{\mc F} \mi{OrdPropCl}$ is positive.

Case 2.2.2.1.2:
$\mi{count}(S)\geq 1$.
Then $\mi{count}(S)=\sum_{C\in S-\mi{guards}(S)} \mi{count}(C)\geq 1$;
there exists $C^*\in S-\mi{guards}(S)$ satisfying $\mi{count}(C^*)=\sum_{l\in C^*} \mi{count}(l)\geq 1$;
$C^*\not\in \mi{invalid}(S)=\emptyset$,
$\mi{valid}(C^*,S)$;
for all $l\in C^*$, $\mi{valid}(l,S)$, $l\in \mi{OrdPropLit}^\gu$;
$C^*\in \mi{OrdPropCl}^\gu$;
there exists $l^*\in C^*$ satisfying $\mi{count}(l^*)=1$, either $l^*=\Cn^*\geql \gu$ or $l^*=\gu\gleq \Cn^*$, $\Cn^*\in \mi{PropConj}$;
$\mi{valid}(l^*,S)$,
$\Cn^*=a_0\swedge\cdots\swedge a_n$, $a_i\in \mi{PropAtom}$;
for all $i\leq n$, $a_i\in \mi{atoms}(\Cn^*)=\mi{atoms}(l^*)\subseteq \mi{atoms}(C^*)\subseteq \mi{atoms}(S)$, $\mi{guards}(S,a_i)=\{\gz\gle a_i\}$;
$C^*-\{l^*\}\subseteq C^*$, $C^*-\{l^*\}\in \mi{OrdPropCl}^\gu$.
We get two cases for $C^*-\{l^*\}$.

Case 2.2.2.1.2.1:
$C^*-\{l^*\}=\square$.
Then $l^*\in C^*$, $C^*=(C^*-\{l^*\})\cup \{l^*\}=l^*\in S-\mi{guards}(S)$; 
$l^*$ is not a guard; 
either $l^*=\Cn^*\geql \gu$ or $l^*=\gu\gleq \Cn^*$, $\Cn^*=a_0\swedge\cdots\swedge a_n$, $n\geq 1$,
applying Rule (\cref{ceq4hr111111}) to $l^*$, we derive
\begin{equation} \notag
\dfrac{S}
      {(S-\{l^*\})\cup \{a_0\geql \gu,\dots,a_n\geql \gu\}}.
\end{equation}
Hence, $l^*\in S$;
for all $i\leq n$, $a_i\geql \gu\in \mi{guards}(a_i)$, $a_i\geql \gu\not\in \mi{guards}(S,a_i)=\{\gz\gle a_i\}=S\cap \mi{guards}(a_i)$, $\gz\gle a_i\in S$, $a_i\geql \gu\not\in S$;
$\{\gz\gle a_0,\dots,\gz\gle a_n\}\subseteq S$,
$\{a_0\geql \gu,\dots,a_n\geql \gu\}\cap S=\emptyset$.
We put $S'=(S-\{l^*\})\cup \{a_0\geql \gu,\dots,a_n\geql \gu\}\subseteq_{\mc F} \mi{OrdPropCl}$.
We get that
$\mi{atoms}(a_0\geql \gu,\dots,a_n\geql \gu)=\mi{atoms}(l^*)=\mi{atoms}(\Cn^*)=\{a_0,\dots,a_n\}\subseteq \mi{atoms}(S')$, 
$l^*\in S$,
$\mi{atoms}(S')=\mi{atoms}((S-\{l^*\})\cup \{a_0\geql \gu,\dots,a_n\geql \gu\})=\mi{atoms}(S-\{l^*\})\cup \mi{atoms}(a_0\geql \gu,\dots,a_n\geql \gu)=\mi{atoms}(S-\{l^*\})\cup \mi{atoms}(l^*)=
                \mi{atoms}((S-\{l^*\})\cup \{l^*\})=\mi{atoms}(S)$;
$S$ is simplified; 
$S-\{l^*\}\subseteq S$ is simplified;
for all $i\leq n$, $a_i\geql \gu\neq \square$ does not contain contradictions and tautologies;
$\{a_0\geql \gu,\dots,a_n\geql \gu\}$ is simplified;
$S'=(S-\{l^*\})\cup \{a_0\geql \gu,\dots,a_n\geql \gu\}$ is simplified;
for all $i\leq n$, 
for all $j\leq n$ and $j\neq i$, $a_i\neq a_j$, $\mi{guards}(a_i)\cap \{\gz\gle a_j,a_j\geql \gu\}=\mi{guards}(a_i)\cap \mi{guards}(a_j)=\emptyset$;
we have that $l^*$ is not a guard;
$l^*\not\in \mi{guards}(a_i)$,
$\mi{guards}(S',a_i)=((S-\{l^*\})\cup \{a_0\geql \gu,\dots,a_n\geql \gu\})\cap \mi{guards}(a_i)=((S-\{l^*\})\cap \mi{guards}(a_i))\cup (\{a_0\geql \gu,\dots,a_n\geql \gu\}\cap \mi{guards}(a_i))=
                     ((S\cap \mi{guards}(a_i))-\{l^*\})\cup \{a_i\geql \gu\}=(S\cap \mi{guards}(a_i))\cup \{a_i\geql \gu\}=\mi{guards}(S,a_i)\cup \{a_i\geql \gu\}=\{\gz\gle a_i,a_i\geql \gu\}$;
by Lemma \ref{le88} for $S'$ and $\{a_0,\dots,a_n\}$, there exists a finite linear {\it DPLL}-tree $\mi{Tree}'$ with the root $S'$ constructed using Rules (\cref{ceq4hr22}) and (\cref{ceq4hr4}) satisfying
for its only leaf $S''$ that $S''=S'-\{\gz\gle a_0,\dots,\gz\gle a_n\}\subseteq_{\mc F} \mi{OrdPropCl}$ is simplified;  
$\{\gz\gle a_0,\dots,\gz\gle a_n\}\subseteq S$;
for all $i\leq n$, either $l^*=\Cn^*\geql \gu$ or $l^*=\gu\gleq \Cn^*$, $l^*\neq \gz\gle a_i$;
$\{l^*\}\cap \{\gz\gle a_0,\dots,\gz\gle a_n\}=\emptyset$, 
$\{a_0\geql \gu,\dots,a_n\geql \gu\}\cap S=\emptyset$,
$S''=S'-\{\gz\gle a_0,\dots,\gz\gle a_n\}=((S-\{l^*\})\cup \{a_0\geql \gu,\dots,a_n\geql \gu\})-\{\gz\gle a_0,\dots,\gz\gle a_n\}=
     ((S-\{l^*\})-\{\gz\gle a_0,\dots,\gz\gle a_n\})\cup (\{a_0\geql \gu,\dots,a_n\geql \gu\}-\{\gz\gle a_0,\dots,\gz\gle a_n\})=
     ((S-\{l^*\})-\{\gz\gle a_0,\dots,\gz\gle a_n\})\cup \{a_0\geql \gu,\dots,a_n\geql \gu\}$,
$\mi{atoms}(\gz\gle a_0,\dots,\gz\gle a_n)=\{a_0,\dots,a_n\}$,
$\{a_0,\dots,a_n\}\subseteq \mi{atoms}(S'')=\mi{atoms}(S'-\{\gz\gle a_0,\dots,\gz\gle a_n\})=\mi{atoms}(S')=\mi{atoms}(S)$;
for all $i\leq n$,
$\mi{guards}(S'',a_i)=(((S-\{l^*\})-\{\gz\gle a_0,\dots,\gz\gle a_n\})\cup \{a_0\geql \gu,\dots,a_n\geql \gu\})\cap \mi{guards}(a_i)=
                      (((S-\{l^*\})-\{\gz\gle a_0,\dots,\gz\gle a_n\})\cap \mi{guards}(a_i))\cup (\{a_0\geql \gu,\dots,a_n\geql \gu\}\cap \mi{guards}(a_i))=
                      (((S\cap \mi{guards}(a_i))-\{\gz\gle a_0,\dots,\gz\gle a_n\})-\{l^*\})\cup \{a_i\geql \gu\}=((S\cap \mi{guards}(a_i))-\{\gz\gle a_i\})\cup \{a_i\geql \gu\}=
                      (\mi{guards}(S,a_i)-\{\gz\gle a_i\})\cup \{a_i\geql \gu\}=(\{\gz\gle a_i\}-\{\gz\gle a_i\})\cup \{a_i\geql \gu\}=\{a_i\geql \gu\}$;
$a_i$ is semi-positively guarded in $S''$;
for all $a\in \mi{atoms}(S'')-\{a_0,\dots,a_n\}=\mi{atoms}(S)-\{a_0,\dots,a_n\}$,
$a$ is positively guarded in $S$;
for all $i\leq n$, $a\neq a_i$, $\mi{guards}(a)\cap \{\gz\gle a_i,a_i\geql \gu\}=\mi{guards}(a)\cap \mi{guards}(a_i)=\emptyset$;
$l^*\not\in \mi{guards}(a)$,
$\mi{guards}(S'',a)=(((S-\{l^*\})-\{\gz\gle a_0,\dots,\gz\gle a_n\})\cup \{a_0\geql \gu,\dots,a_n\geql \gu\})\cap \mi{guards}(a)=
                    (((S-\{l^*\})-\{\gz\gle a_0,\dots,\gz\gle a_n\})\cap \mi{guards}(a))\cup (\{a_0\geql \gu,\dots,a_n\geql \gu\}\cap \mi{guards}(a))=
                    ((S\cap \mi{guards}(a))-\{l^*\})-\{\gz\gle a_0,\dots,\gz\gle a_n\}=S\cap \mi{guards}(a)=\mi{guards}(S,a)$;
$a$ is positively guarded in $S''$;
$S''$ is semi-positively guarded; 
$a_0\in \mi{atoms}(S'')$, $\mi{guards}(S'',a_0)=\{a_0\geql \gu\}$;
by Lemma \ref{le66} for $S''$ and $a_0$, there exists a finite linear {\it DPLL}-tree $\mi{Tree}''$ with the root $S''$ constructed using Rules (\cref{ceq4hr1111111})--(\cref{ceq4hr4}), (\cref{ceq4hr8}) satisfying 
for its only leaf $S'''$ that either $\square\in S'''$, or $S'''\subseteq_{\mc F} \mi{OrdPropCl}$ is semi-positively guarded, $\mi{atoms}(S''')\subseteq \mi{atoms}(S'')-\{a_0\}$.
We get two cases for $S'''$.

Case 2.2.2.1.2.1.1:
$\square\in S'''$.
We put 
\begin{equation} \notag
\mi{Tree}=\begin{array}[c]{c}
          S \\[0.4mm]
          \hline \\[-3.8mm]
          \mi{Tree}' \\[0.4mm]
          \hline \\[-3.8mm]
          \mi{Tree}''.
          \end{array}
\end{equation}
Hence, $\mi{Tree}$ is a finite {\it DPLL}-tree with the root $S$ constructed using Rules (\cref{ceq4hr1x}), (\cref{ceq4hr1111})--(\cref{ceq4hr66}), (\cref{ceq4hr8}) such that
for its only leaf $S'''$, $\square\in S'''$.

Case 2.2.2.1.2.1.2:
$S'''\subseteq_{\mc F} \mi{OrdPropCl}$ is semi-positively guarded, $\mi{atoms}(S''')\subseteq                                                                                              \linebreak[4]
                                                                                              \mi{atoms}(S'')-\{a_0\}$.
Then $a_0\in \mi{atoms}(S'')$, $\mi{atoms}(S''')\subseteq \mi{atoms}(S'')-\{a_0\}\subset \mi{atoms}(S'')\subseteq \mi{atoms}(S)$, 
$\mi{measure}(S''')=(\mi{atoms}(S'''),\mi{unsaturated}(S'''),                                                                                                                              \linebreak[4]
                                                             \mi{invalid}(S'''),\mi{count}(S'''))\prec \mi{measure}(S)=(\mi{atoms}(S),\mi{unsaturated}(S),\mi{invalid}(S),                 \linebreak[4]
                                                                                                                                                                          \mi{count}(S))$;
by the induction hypothesis for $S'''$, there exists a finite {\it DPLL}-tree $\mi{Tree}'''$ with the root $S'''$ constructed using Rules (\cref{ceq4hr1x}), (\cref{ceq4hr1111})--(\cref{ceq4hr66}), (\cref{ceq4hr8}) satisfying
for every leaf $S''''$ that either $\square\in S''''$ or $S''''\subseteq_{\mc F} \mi{OrdPropCl}$ is positive.
We put
\begin{equation} \notag
\mi{Tree}=\begin{array}[c]{c}
          S \\[0.4mm]
          \hline \\[-3.8mm]
          \mi{Tree}' \\[0.4mm]
          \hline \\[-3.8mm]
          \mi{Tree}'' \\[0.4mm]
          \hline \\[-3.8mm]
          \mi{Tree}'''.
          \end{array}
\end{equation}
Hence, $\mi{Tree}$ is a finite {\it DPLL}-tree with the root $S$ constructed using Rules (\cref{ceq4hr1x}), (\cref{ceq4hr1111})--(\cref{ceq4hr66}), (\cref{ceq4hr8}) such that
for every leaf $S''''$, either $\square\in S''''$ or $S''''\subseteq_{\mc F} \mi{OrdPropCl}$ is positive.

Case 2.2.2.1.2.2:
$C^*-\{l^*\}\neq \square$.
Then $l^*\in C^*\in S-\mi{guards}(S)$, either $l^*=\Cn^*\geql \gu$ or $l^*=\gu\gleq \Cn^*$, $\Cn^*=a_0\swedge\cdots\swedge a_n$,
applying Rule (\cref{ceq4hr1x}) to $C^*$, $l^*$, $C^*-\{l^*\}$, and $\Cn^*$, we derive
\begin{equation} \notag
\dfrac{S}
      {(S-\{C^*\})\cup \{\Cn^*\geql \gu\}\ \big|\ (S-\{C^*\})\cup \{C^*-\{l^*\}\}\cup \{\Cn^*\gle \gu\}}.
\end{equation}
We put 
$S_1=(S-\{C^*\})\cup \{\Cn^*\geql \gu\}\subseteq_{\mc F} \mi{OrdPropCl}$ and
$S_2=(S-\{C^*\})\cup \{C^*-\{l^*\}\}\cup \{\Cn^*\gle \gu\}\subseteq_{\mc F} \mi{OrdPropCl}$.
We get that
$\mi{atoms}(\Cn^*\geql \gu)=\mi{atoms}(\Cn^*\gle \gu)=\mi{atoms}(l^*)=\mi{atoms}(\Cn^*)=\{a_0,\dots,a_n\}$,
$l^*\in C^*$, $\mi{atoms}(l^*)\subseteq \mi{atoms}(C^*)$, $C^*\in S$,
$\mi{atoms}(S_1)=\mi{atoms}((S-\{C^*\})\cup \{\Cn^*\geql \gu\})=\mi{atoms}(S-\{C^*\})\cup \mi{atoms}(\Cn^*\geql \gu)=\mi{atoms}(S-\{C^*\})\cup \mi{atoms}(l^*)\subseteq 
                 \mi{atoms}(S-\{C^*\})\cup \mi{atoms}(C^*)=\mi{atoms}((S-\{C^*\})\cup \{C^*\})=\mi{atoms}(S)$,
$\mi{atoms}(S_2)=\mi{atoms}((S-\{C^*\})\cup \{C^*-\{l^*\}\}\cup \{\Cn^*\gle \gu\})=\mi{atoms}(S-\{C^*\})\cup \mi{atoms}(C^*-\{l^*\})\cup \mi{atoms}(\Cn^*\gle \gu)=
                 \mi{atoms}(S-\{C^*\})\cup \mi{atoms}(C^*-\{l^*\})\cup \mi{atoms}(l^*)=\mi{atoms}(S-\{C^*\})\cup \mi{atoms}((C^*-\{l^*\})\cup \{l^*\})=                 
                 \mi{atoms}(S-\{C^*\})\cup \mi{atoms}(C^*)=\mi{atoms}((S-\{C^*\})\cup \{C^*\})=\mi{atoms}(S)$;
$S$ is simplified; 
$S-\{C^*\}\subseteq S$ is simplified;
$C^*\in S$ does not contain contradictions and tautologies;
$\square\neq C^*-\{l^*\}\subseteq C^*$ does not contain contradictions and tautologies;
$\Cn^*\geql \gu, \Cn^*\gle \gu\neq \square$ do not contain contradictions and tautologies;
$S_1=(S-\{C^*\})\cup \{\Cn^*\geql \gu\}$ and $S_2=(S-\{C^*\})\cup \{C^*-\{l^*\}\}\cup \{\Cn^*\gle \gu\}$ are simplified.
We get two cases for $n$.

Case 2.2.2.1.2.2.1:
$n=0$.
Then $\Cn^*=a_0\swedge\cdots\swedge a_n=a_0$, 
$a_0\gle \gu, a_0\geql \gu\in \mi{guards}(a_0)$, $a_0\gle \gu, a_0\geql \gu\not\in \mi{guards}(S,a_0)=\{\gz\gle a_0\}=S\cap \mi{guards}(a_0)$, $\gz\gle a_0\in S$, $a_0\gle \gu, a_0\geql \gu\not\in S$,
$C^*\in S$, $C^*\not\in \mi{guards}(S)\supseteq \mi{guards}(S,a_0)=S\cap \mi{guards}(a_0)$, $C^*\not\in \mi{guards}(a_0)$.
We get two cases for $S_1$ and $S_2$.

Case 2.2.2.1.2.2.1.1:
$S_1$.
Then $S_1=(S-\{C^*\})\cup \{\Cn^*\geql \gu\}=(S-\{C^*\})\cup \{a_0\geql \gu\}$, $a_0\in \mi{atoms}(S_1)$, $C^*\not\in \mi{guards}(a_0)$, 
$S_1\supseteq \mi{guards}(S_1)\supseteq \mi{guards}(S_1,a_0)=((S-\{C^*\})\cup \{a_0\geql \gu\})\cap \mi{guards}(a_0)=((S-\{C^*\})\cap \mi{guards}(a_0))\cup (\{a_0\geql \gu\}\cap \mi{guards}(a_0))=
 ((S\cap \mi{guards}(a_0))-\{C^*\})\cup \{a_0\geql \gu\}=(S\cap \mi{guards}(a_0))\cup \{a_0\geql \gu\}=\mi{guards}(S,a_0)\cup \{a_0\geql \gu\}=\{\gz\gle a_0,a_0\geql \gu\}$,
$a_0\geql \gu\in \mi{guards}(S_1)$, $\gz\gle a_0\in S_1$,
$a_0\in \mi{atoms}(\gz\gle a_0)$, $a_0\geql \gu\neq \gz\gle a_0$, $\mi{simplify}(\gz\gle a_0,a_0,\gu)=\gz\gle \gu$,
applying Rule (\cref{ceq4hr4}) to $S_1$, $a_0\geql \gu$, and $\gz\gle a_0$, we derive
\begin{equation} \notag
\dfrac{S_1}
      {(S_1-\{\gz\gle a_0\})\cup \{\gz\gle \gu\}};
\end{equation}
$\gz\gle \gu\in (S_1-\{\gz\gle a_0\})\cup \{\gz\gle \gu\}$;
$\gz\gle \gu\in \mi{OrdPropLit}$ is a tautology;
$\gz\gle \gu$ is not a guard;
$\gz\gle \gu\not\in \mi{guards}((S_1-\{\gz\gle a_0\})\cup \{\gz\gle \gu\})$,
$\gz\gle \gu\in ((S_1-\{\gz\gle a_0\})\cup \{\gz\gle \gu\})-\mi{guards}((S_1-\{\gz\gle a_0\})\cup \{\gz\gle \gu\})$,
applying Rule (\cref{ceq4hr22}) to $(S_1-\{\gz\gle a_0\})\cup \{\gz\gle \gu\}$ and $\gz\gle \gu$, we derive
\begin{equation} \notag
\dfrac{(S_1-\{\gz\gle a_0\})\cup \{\gz\gle \gu\}}
      {((S_1-\{\gz\gle a_0\})\cup \{\gz\gle \gu\})-\{\gz\gle \gu\}}.
\end{equation}
We have that $S_1$ is simplified.
Hence, $\gz\gle \gu\not\in S_1$, $C^*, \gz\gle a_0\in S$, $C^*\not\in \mi{guards}(a_0)$, $\gz\gle a_0\in \mi{guards}(a_0)$, $C^*\neq \gz\gle a_0$, $a_0\geql \gu\not\in S$, 
$((S_1-\{\gz\gle a_0\})\cup \{\gz\gle \gu\})-\{\gz\gle \gu\}=((S_1-\{\gz\gle a_0\})-\{\gz\gle \gu\})\cup (\{\gz\gle \gu\}-\{\gz\gle \gu\})=S_1-\{\gz\gle a_0\}=
 ((S-\{C^*\})\cup \{a_0\geql \gu\})-\{\gz\gle a_0\}=((S-\{C^*\})-\{\gz\gle a_0\})\cup (\{a_0\geql \gu\}-\{\gz\gle a_0\})=((S-\{C^*\})-\{\gz\gle a_0\})\cup \{a_0\geql \gu\}$.
We put $S_1'=((S-\{C^*\})-\{\gz\gle a_0\})\cup \{a_0\geql \gu\}\subseteq_{\mc F} \mi{OrdPropCl}$.
We get that
$S_1'=((S-\{C^*\})-\{\gz\gle a_0\})\cup \{a_0\geql \gu\}=S_1-\{\gz\gle a_0\}$,
$\mi{atoms}(\gz\gle a_0)=\{a_0\}$, 
$a_0\in \mi{atoms}(S_1')=\mi{atoms}(S_1-\{\gz\gle a_0\})=\mi{atoms}(S_1)\subseteq \mi{atoms}(S)$;
we have that $S_1$ is simplified;
$S_1'=S_1-\{\gz\gle a_0\}\subseteq S_1$ is simplified;
$C^*\not\in \mi{guards}(a_0)$,
$\mi{guards}(S_1',a_0)=(((S-\{C^*\})-\{\gz\gle a_0\})\cup \{a_0\geql \gu\})\cap \mi{guards}(a_0)=(((S-\{C^*\})-\{\gz\gle a_0\})\cap \mi{guards}(a_0))\cup (\{a_0\geql \gu\}\cap \mi{guards}(a_0))=
                       (((S\cap \mi{guards}(a_0))-\{\gz\gle a_0\})-\{C^*\})\cup \{a_0\geql \gu\}=((S\cap \mi{guards}(a_0))-\{\gz\gle a_0\})\cup \{a_0\geql \gu\}=
                       (\mi{guards}(S,a_0)-\{\gz\gle a_0\})\cup \{a_0\geql \gu\}=(\{\gz\gle a_0\}-\{\gz\gle a_0\})\cup \{a_0\geql \gu\}=\{a_0\geql \gu\}$;
$a_0$ is semi-positively guarded in $S_1'$;
for all $a\in \mi{atoms}(S_1')-\{a_0\}\subseteq \mi{atoms}(S)-\{a_0\}$,
$a$ is positively guarded in $S$;
$a\neq a_0$,
$\mi{guards}(a)\cap \{\gz\gle a_0,a_0\geql \gu\}=\mi{guards}(a)\cap \mi{guards}(a_0)=\emptyset$,
$C^*\in S$, $C^*\not\in \mi{guards}(S)\supseteq \mi{guards}(S,a)=S\cap \mi{guards}(a)$, $C^*\not\in \mi{guards}(a)$,
$\mi{guards}(S_1',a)=(((S-\{C^*\})-\{\gz\gle a_0\})\cup \{a_0\geql \gu\})\cap \mi{guards}(a)=(((S-\{C^*\})-\{\gz\gle a_0\})\cap \mi{guards}(a))\cup (\{a_0\geql \gu\}\cap \mi{guards}(a))=
                     ((S\cap \mi{guards}(a))-\{C^*\})-\{\gz\gle a_0\}=S\cap \mi{guards}(a)=\mi{guards}(S,a)$;
$a$ is positively guarded in $S_1'$;
$S_1'$ is semi-positively guarded;
by Lemma \ref{le66} for $S_1'$ and $a_0$, there exists a finite linear {\it DPLL}-tree $\mi{Tree}_1'$ with the root $S_1'$ constructed using Rules (\cref{ceq4hr1111111})--(\cref{ceq4hr4}), (\cref{ceq4hr8}) satisfying 
for its only leaf $S_1''$ that either $\square\in S_1''$, or $S_1''\subseteq_{\mc F} \mi{OrdPropCl}$ is semi-positively guarded, $\mi{atoms}(S_1'')\subseteq \mi{atoms}(S_1')-\{a_0\}$.
We get two cases for $S_1''$.

Case 2.2.2.1.2.2.1.1.1:
$\square\in S_1''$.
We put 
\begin{equation} \notag
\mi{Tree}_1=\begin{array}[c]{c}
            S_1 \\[0.4mm]
            \hline \\[-3.8mm]
            (S_1-\{\gz\gle a_0\})\cup \{\gz\gle \gu\} \\[0.4mm]
            \hline \\[-3.8mm]
            \mi{Tree}_1'.
            \end{array}
\end{equation}
Hence, $\mi{Tree}_1$ is a finite {\it DPLL}-tree with the root $S_1$ constructed using Rules (\cref{ceq4hr1x}), (\cref{ceq4hr1111})--(\cref{ceq4hr66}), (\cref{ceq4hr8}) such that
for its only leaf $S_1''$, $\square\in S_1''$.

Case 2.2.2.1.2.2.1.1.2:
$S_1''\subseteq_{\mc F} \mi{OrdPropCl}$ is semi-positively guarded,                                                                                                                        \linebreak[4]
                                                                    $\mi{atoms}(S_1'')\subseteq \mi{atoms}(S_1')-\{a_0\}$.
Then $a_0\in \mi{atoms}(S_1')$, $\mi{atoms}(S_1'')\subseteq \mi{atoms}(S_1')-\{a_0\}\subset \mi{atoms}(S_1')\subseteq \mi{atoms}(S)$, 
$\mi{measure}(S_1'')=(\mi{atoms}(S_1''),\mi{unsaturated}(S_1''),                                                                                                                           \linebreak[4]
                                                                \mi{invalid}(S_1''),\mi{count}(S_1''))\prec \mi{measure}(S)=(\mi{atoms}(S),\mi{unsaturated}(S),\mi{invalid}(S),            \linebreak[4]
                                                                                                                                                                               \mi{count}(S))$;
by the induction hypothesis for $S_1''$, there exists a finite {\it DPLL}-tree $\mi{Tree}_1''$ with the root $S_1''$ constructed using Rules (\cref{ceq4hr1x}), (\cref{ceq4hr1111})--(\cref{ceq4hr66}), (\cref{ceq4hr8}) satisfying
for every leaf $S_1'''$ that either $\square\in S_1'''$ or $S_1'''\subseteq_{\mc F} \mi{OrdPropCl}$ is positive.
We put
\begin{equation} \notag
\mi{Tree}_1=\begin{array}[c]{c}
            S_1 \\[0.4mm]
            \hline \\[-3.8mm]
            (S_1-\{\gz\gle a_0\})\cup \{\gz\gle \gu\} \\[0.4mm]
            \hline \\[-3.8mm]
            \mi{Tree}_1' \\[0.4mm]
            \hline \\[-3.8mm]
            \mi{Tree}_1''.
            \end{array}
\end{equation}
Hence, $\mi{Tree}_1$ is a finite {\it DPLL}-tree with the root $S_1$ constructed using Rules (\cref{ceq4hr1x}), (\cref{ceq4hr1111})--(\cref{ceq4hr66}), (\cref{ceq4hr8}) such that
for every leaf $S_1'''$, either $\square\in S_1'''$ or $S_1'''\subseteq_{\mc F} \mi{OrdPropCl}$ is positive.

Case 2.2.2.1.2.2.1.2:
$S_2$.
Then $S_2=(S-\{C^*\})\cup \{C^*-\{l^*\}\}\cup \{\Cn^*\gle \gu\}=(S-\{C^*\})\cup \{C^*-\{l^*\}\}\cup \{a_0\gle \gu\}$ and $a_0\in \mi{atoms}(S_2)$.
We get two cases for $C^*-\{l^*\}$.

Case 2.2.2.1.2.2.1.2.1:
There exists $a^*\in \mi{atoms}(S_2)=\mi{atoms}(S)$ such that $C^*-\{l^*\}\in \mi{guards}(a^*)$.  
Then $C^*\in S$, $C^*\not\in \mi{guards}(S)\supseteq \mi{guards}(S,a^*)=S\cap \mi{guards}(a^*)$, $C^*\not\in \mi{guards}(a^*)$,
$\mi{guards}(S_2,a^*)=((S-\{C^*\})\cup \{C^*-\{l^*\}\}\cup \{a_0\gle \gu\})\cap \mi{guards}(a^*)=
                      ((S-\{C^*\})\cap \mi{guards}(a^*))\cup (\{C^*-\{l^*\}\}\cap \mi{guards}(a^*))\cup (\{a_0\gle \gu\}\cap \mi{guards}(a^*))=
                      ((S\cap \mi{guards}(a^*))-\{C^*\})\cup \{C^*-\{l^*\}\}\cup (\{a_0\gle \gu\}\cap \mi{guards}(a^*))=
                      (S\cap \mi{guards}(a^*))\cup \{C^*-\{l^*\}\}\cup (\{a_0\gle \gu\}\cap \mi{guards}(a^*))=\mi{guards}(S,a^*)\cup \{C^*-\{l^*\}\}\cup (\{a_0\gle \gu\}\cap \mi{guards}(a^*))$.
We get two cases for $a^*$.

Case 2.2.2.1.2.2.1.2.1.1:
$a^*=a_0$.
Then $C^*-\{l^*\}\in \mi{guards}(a^*)=\mi{guards}(a_0)$,
$\mi{guards}(S_2,a^*)=\mi{guards}(S_2,a_0)=\mi{guards}(S,a^*)\cup \{C^*-\{l^*\}\}\cup (\{a_0\gle \gu\}\cap \mi{guards}(a^*))=
                                           \mi{guards}(S,a_0)\cup \{C^*-\{l^*\}\}\cup (\{a_0\gle \gu\}\cap \mi{guards}(a_0))=\{C^*-\{l^*\}\}\cup \{\gz\gle a_0,a_0\gle \gu\}$,
$C^*-\{l^*\}\in \mi{OrdPropCl}^\gu$.
We get three cases for $C^*-\{l^*\}$.

Case 2.2.2.1.2.2.1.2.1.1.1:
$C^*-\{l^*\}=a_0\gle \gu$.
Then $S_2=(S-\{C^*\})\cup \{C^*-\{l^*\}\}\cup \{a_0\gle \gu\}=(S-\{C^*\})\cup \{a_0\gle \gu\}$,
$\mi{guards}(S_2,a_0)=\{C^*-\{l^*\}\}\cup \{\gz\gle a_0,a_0\gle \gu\}=\{\gz\gle a_0,a_0\gle \gu\}$;
$a_0$ is positively guarded in $S_2$;
for all $a\in \mi{atoms}(S_2)-\{a_0\}=\mi{atoms}(S)-\{a_0\}$,
$a$ is positively guarded in $S$;
$a\neq a_0$,
$\mi{guards}(a)\cap \{a_0\gle \gu\}=\mi{guards}(a)\cap \mi{guards}(a_0)=\emptyset$,
$C^*\in S$, $C^*\not\in \mi{guards}(S)\supseteq \mi{guards}(S,a)=S\cap \mi{guards}(a)$, $C^*\not\in \mi{guards}(a)$,
$\mi{guards}(S_2,a)=((S-\{C^*\})\cup \{a_0\gle \gu\})\cap \mi{guards}(a)=((S-\{C^*\})\cap \mi{guards}(a))\cup (\{a_0\gle \gu\}\cap \mi{guards}(a))=
                    (S\cap \mi{guards}(a))-\{C^*\}=S\cap \mi{guards}(a)=\mi{guards}(S,a)$;
$a$ is positively guarded in $S_2$;
$S_2$ is positively guarded; 
$\mi{guards}(S_2,a_0)=\{\gz\gle a_0,a_0\gle \gu\}\neq \{\gz\gle a_0\}$, 
$a_0\in \mi{atoms}(S)$, $\mi{guards}(S,a_0)=\{\gz\gle a_0\}$, $a_0\in \mi{unsaturated}(S)=\{a \,|\, a\in \mi{atoms}(S), \mi{guards}(S,a)=\{\gz\gle a\}\}$,
$a_0\not\in \mi{unsaturated}(S_2)=\{a \,|\, a\in \mi{atoms}(S_2), \mi{guards}(S_2,a)=\{\gz\gle a\}\}=\{a \,|\, a\in \mi{atoms}(S_2), \mi{guards}(S_2,a)=\{\gz\gle a\}\}-\{a_0\}=
                                  \{a \,|\, a\in \mi{atoms}(S_2)-\{a_0\}, \mi{guards}(S_2,a)=\{\gz\gle a\}\}=\{a \,|\, a\in \mi{atoms}(S)-\{a_0\}, \mi{guards}(S,a)=\{\gz\gle a\}\}=
                                  \mi{unsaturated}(S)-\{a_0\}=\{a \,|\, a\in \mi{atoms}(S), \mi{guards}(S,a)=\{\gz\gle a\}\}-\{a_0\}\subset \mi{unsaturated}(S)$, 
$\mi{measure}(S_2)=(\mi{atoms}(S_2),\mi{unsaturated}(S_2),\mi{invalid}(S_2),\mi{count}(S_2))=(\mi{atoms}(S),                                                                               \linebreak[4]
                                                                                                            \mi{unsaturated}(S_2),\mi{invalid}(S_2),\mi{count}(S_2))\prec 
                   \mi{measure}(S)=(\mi{atoms}(S),                                                                                                                                         \linebreak[4]
                                                  \mi{unsaturated}(S),\mi{invalid}(S),\mi{count}(S))$;
by the induction hypothesis for $S_2$, there exists a finite {\it DPLL}-tree $\mi{Tree}_2$ with the root $S_2$ constructed using Rules (\cref{ceq4hr1x}), (\cref{ceq4hr1111})--(\cref{ceq4hr66}), (\cref{ceq4hr8}) satisfying
for every leaf $S_2'$ that either $\square\in S_2'$ or $S_2'\subseteq_{\mc F} \mi{OrdPropCl}$ is positive.

Case 2.2.2.1.2.2.1.2.1.1.2:
$C^*-\{l^*\}=a_0\geql \gu$.
Then $S_2\supseteq \mi{guards}(S_2)\supseteq \mi{guards}(S_2,a_0)=\{C^*-\{l^*\}\}\cup \{\gz\gle a_0,a_0\gle \gu\}=\{\gz\gle a_0,a_0\gle \gu,a_0\geql \gu\}$,
$a_0\geql \gu\in \mi{guards}(S_2)$, $a_0\gle \gu\in S_2$,
$a_0\in \mi{atoms}(a_0\gle \gu)$, $a_0\geql \gu\neq a_0\gle \gu$, $\mi{simplify}(a_0\gle \gu,a_0,\gu)=\gu\gle \gu$,
applying Rule (\cref{ceq4hr4}) to $S_2$, $a_0\geql \gu$, and $a_0\gle \gu$, we derive
\begin{equation} \notag
\dfrac{S_2}
      {(S_2-\{a_0\gle \gu\})\cup \{\gu\gle \gu\}};
\end{equation}
$\gu\gle \gu\in (S_2-\{a_0\gle \gu\})\cup \{\gu\gle \gu\}$;
$\gu\gle \gu\in \mi{OrdPropLit}$ is a contradiction;
$\gu\gle \gu$ is not a guard;
$\gu\gle \gu\not\in \mi{guards}((S_2-\{a_0\gle \gu\})\cup \{\gu\gle \gu\})$,
$\gu\gle \gu\in ((S_2-\{a_0\gle \gu\})\cup \{\gu\gle \gu\})-\mi{guards}((S_2-\{a_0\gle \gu\})\cup \{\gu\gle \gu\})$,
applying Rule (\cref{ceq4hr2}) to $(S_2-\{a_0\gle \gu\})\cup \{\gu\gle \gu\}$ and $\gu\gle \gu$, we derive
\begin{equation} \notag
\dfrac{(S_2-\{a_0\gle \gu\})\cup \{\gu\gle \gu\}}
      {(((S_2-\{a_0\gle \gu\})\cup \{\gu\gle \gu\})-\{\gu\gle \gu\})\cup \{\square\}}.
\end{equation}
We put
\begin{equation} \notag
\mi{Tree}_2=\begin{array}[c]{c}
            S_2 \\[0.4mm]
            \hline \\[-3.8mm]
            (S_2-\{a_0\gle \gu\})\cup \{\gu\gle \gu\} \\[0.4mm]
            \hline \\[-3.8mm]
            S_2'=(((S_2-\{a_0\gle \gu\})\cup \{\gu\gle \gu\})-\{\gu\gle \gu\})\cup \{\square\}.
            \end{array} 
\end{equation}
Hence, $\mi{Tree}_2$ is a finite {\it DPLL}-tree with the root $S_2$ constructed using Rules (\cref{ceq4hr1x}), (\cref{ceq4hr1111})--(\cref{ceq4hr66}), (\cref{ceq4hr8}) such that
for its only leaf $S_2'$, $\square\in S_2'$.

Case 2.2.2.1.2.2.1.2.1.1.3:
$C^*-\{l^*\}=\gu\gleq a_0$.
Then $\mi{guards}(S_2)\supseteq \mi{guards}(S_2,a_0)=\{C^*-\{l^*\}\}\cup \{\gz\gle a_0,a_0\gle \gu\}=\{\gz\gle a_0,a_0\gle \gu,\gu\gleq a_0\}$,
$\gu\gleq a_0\in \mi{guards}(S_2)$, applying Rule (\cref{ceq4hr11111111}) to $S_2$ and $\gu\gleq a_0$, we derive
\begin{equation} \notag
\dfrac{S_2}
      {(S_2-\{\gu\gleq a_0\})\cup \{a_0\geql \gu\}};
\end{equation}
$a_0\in \mi{atoms}((S_2-\{\gu\gleq a_0\})\cup \{a_0\geql \gu\})$,
$(S_2-\{\gu\gleq a_0\})\cup \{a_0\geql \gu\}\supseteq \mi{guards}((S_2-\{\gu\gleq a_0\})\cup \{a_0\geql \gu\})\supseteq 
 \mi{guards}((S_2-\{\gu\gleq a_0\})\cup \{a_0\geql \gu\},a_0)=((S_2-\{\gu\gleq a_0\})\cup \{a_0\geql \gu\})\cap \mi{guards}(a_0)=
 ((S_2-\{\gu\gleq a_0\})\cap \mi{guards}(a_0))\cup (\{a_0\geql \gu\}\cap \mi{guards}(a_0))=((S_2\cap \mi{guards}(a_0))-\{\gu\gleq a_0\})\cup \{a_0\geql \gu\}=
 (\mi{guards}(S_2,a_0)-\{\gu\gleq a_0\})\cup \{a_0\geql \gu\}=(\{\gz\gle a_0,a_0\gle \gu,\gu\gleq a_0\}-\{\gu\gleq a_0\})\cup \{a_0\geql \gu\}=\{\gz\gle a_0,a_0\gle \gu,a_0\geql \gu\}$,
$a_0\geql \gu\in \mi{guards}((S_2-\{\gu\gleq a_0\})\cup \{a_0\geql \gu\})$, $a_0\gle \gu\in (S_2-\{\gu\gleq a_0\})\cup \{a_0\geql \gu\}$,
$a_0\in \mi{atoms}(a_0\gle \gu)$, $a_0\geql \gu\neq a_0\gle \gu$, $\mi{simplify}(a_0\gle \gu,a_0,\gu)=\gu\gle \gu$,
applying Rule (\cref{ceq4hr4}) to $(S_2-\{\gu\gleq a_0\})\cup \{a_0\geql \gu\}$, $a_0\geql \gu$, and $a_0\gle \gu$, we derive
\begin{equation} \notag
\dfrac{(S_2-\{\gu\gleq a_0\})\cup \{a_0\geql \gu\}}
      {(((S_2-\{\gu\gleq a_0\})\cup \{a_0\geql \gu\})-\{a_0\gle \gu\})\cup \{\gu\gle \gu\}};
\end{equation}
$\gu\gle \gu\in (((S_2-\{\gu\gleq a_0\})\cup \{a_0\geql \gu\})-\{a_0\gle \gu\})\cup \{\gu\gle \gu\}$;
$\gu\gle \gu\in \mi{OrdPropLit}$ is a contradiction;
$\gu\gle \gu$ is not a guard;
$\gu\gle \gu\not\in \mi{guards}((((S_2-\{\gu\gleq a_0\})\cup \{a_0\geql \gu\})-\{a_0\gle \gu\})\cup \{\gu\gle \gu\})$,
$\gu\gle \gu\in ((((S_2-\{\gu\gleq a_0\})\cup \{a_0\geql \gu\})-\{a_0\gle \gu\})\cup \{\gu\gle \gu\})-\mi{guards}((((S_2-\{\gu\gleq a_0\})\cup \{a_0\geql \gu\})-\{a_0\gle \gu\})\cup \{\gu\gle \gu\})$,
applying Rule (\cref{ceq4hr2}) to $(((S_2-\{\gu\gleq a_0\})\cup \{a_0\geql \gu\})-\{a_0\gle \gu\})\cup \{\gu\gle \gu\}$ and $\gu\gle \gu$, we derive
\begin{equation} \notag
\dfrac{(((S_2-\{\gu\gleq a_0\})\cup \{a_0\geql \gu\})-\{a_0\gle \gu\})\cup \{\gu\gle \gu\}}
      {(((((S_2-\{\gu\gleq a_0\})\cup \{a_0\geql \gu\})-\{a_0\gle \gu\})\cup \{\gu\gle \gu\})-\{\gu\gle \gu\})\cup \{\square\}}.
\end{equation}
We put
\begin{equation} \notag
\mi{Tree}_2=\begin{array}[c]{c}
            S_2 \\[0.4mm]
            \hline \\[-3.8mm]
            (S_2-\{\gu\gleq a_0\})\cup \{a_0\geql \gu\} \\[0.4mm]
            \hline \\[-3.8mm]
            (((S_2-\{\gu\gleq a_0\})\cup \{a_0\geql \gu\})-\{a_0\gle \gu\})\cup \{\gu\gle \gu\} \\[0.4mm]
            \hline \\[-3.8mm]
            S_2'=(((((S_2-\{\gu\gleq a_0\})\cup \{a_0\geql \gu\})-\{a_0\gle \gu\})\cup \{\gu\gle \gu\})- \\
            \hfill \{\gu\gle \gu\})\cup \{\square\}.
            \end{array} 
\end{equation}
Hence, $\mi{Tree}_2$ is a finite {\it DPLL}-tree with the root $S_2$ constructed using Rules (\cref{ceq4hr1x}), (\cref{ceq4hr1111})--(\cref{ceq4hr66}), (\cref{ceq4hr8}) such that
for its only leaf $S_2'$, $\square\in S_2'$.

Case 2.2.2.1.2.2.1.2.1.2:
$a^*\neq a_0$.
Then $\mi{guards}(a^*)\cap \{a_0\gle \gu\}=\mi{guards}(a^*)\cap \mi{guards}(a_0)=\emptyset$,
$\mi{guards}(S_2,a^*)=\mi{guards}(S,a^*)\cup \{C^*-\{l^*\}\}\cup (\{a_0\gle \gu\}\cap \mi{guards}(a^*))=\mi{guards}(S,a^*)\cup \{C^*-\{l^*\}\}$,
$a^*\in \mi{atoms}(S)$, either $\mi{guards}(S,a^*)=\{\gz\gle a^*\}$ or $\mi{guards}(S,a^*)=\{\gz\gle a^*,a^*\gle \gu\}$,
$a^*\geql \gu, \gu\gleq a^*, C^*-\{l^*\}\in \mi{guards}(a^*)$, $a^*\geql \gu, \gu\gleq a^*\not\in \mi{guards}(S,a^*)=S\cap \mi{guards}(a^*)$, $a^*\geql \gu, \gu\gleq a^*\not\in S$,
$C^*-\{l^*\}\in \mi{OrdPropCl}^\gu$.
We get three cases for $C^*-\{l^*\}$.

Case 2.2.2.1.2.2.1.2.1.2.1:
$C^*-\{l^*\}=a^*\gle \gu$.
Then $S_2=(S-\{C^*\})\cup \{C^*-\{l^*\}\}\cup \{a_0\gle \gu\}=(S-\{C^*\})\cup \{a^*\gle \gu\}\cup \{a_0\gle \gu\}$,
either $\mi{guards}(S,a^*)=\{\gz\gle a^*\}$ or $\mi{guards}(S,a^*)=\{\gz\gle a^*,a^*\gle \gu\}$,
$\mi{guards}(S_2,a^*)=\mi{guards}(S,a^*)\cup \{C^*-\{l^*\}\}=\{\gz\gle a^*,a^*\gle \gu\}$;
$a^*$ is positively guarded in $S_2$;
$a_0\neq a^*$, 
$\mi{guards}(a_0)\cap \{a^*\gle \gu\}=\mi{guards}(a_0)\cap \mi{guards}(a^*)=\emptyset$,
$C^*\in S$, $C^*\not\in \mi{guards}(S)\supseteq \mi{guards}(S,a_0)=S\cap \mi{guards}(a_0)$, $C^*\not\in \mi{guards}(a_0)$,
$\mi{guards}(S_2,a_0)=((S-\{C^*\})\cup \{a^*\gle \gu\}\cup \{a_0\gle \gu\})\cap \mi{guards}(a_0)=
                      ((S-\{C^*\})\cap \mi{guards}(a_0))\cup (\{a^*\gle \gu\}\cap \mi{guards}(a_0))\cup (\{a_0\gle \gu\}\cap \mi{guards}(a_0))=
                      ((S\cap \mi{guards}(a_0))-\{C^*\})\cup \{a_0\gle \gu\}=(S\cap \mi{guards}(a_0))\cup \{a_0\gle \gu\}=\mi{guards}(S,a_0)\cup \{a_0\gle \gu\}=\{\gz\gle a_0,a_0\gle \gu\}$;
$a_0$ is positively guarded in $S_2$;
for all $a\in \mi{atoms}(S_2)-\{a^*,a_0\}=\mi{atoms}(S)-\{a^*,a_0\}$,
$a$ is positively guarded in $S$;
$a\neq a^*$, 
$\mi{guards}(a)\cap \{a^*\gle \gu\}=\mi{guards}(a)\cap \mi{guards}(a^*)=\emptyset$,
$a\neq a_0$,
$\mi{guards}(a)\cap \{a_0\gle \gu\}=\mi{guards}(a)\cap \mi{guards}(a_0)=\emptyset$,
$C^*\in S$, $C^*\not\in \mi{guards}(S)\supseteq \mi{guards}(S,a)=S\cap \mi{guards}(a)$, $C^*\not\in \mi{guards}(a)$,
$\mi{guards}(S_2,a)=((S-\{C^*\})\cup \{a^*\gle \gu\}\cup \{a_0\gle \gu\})\cap \mi{guards}(a)=
                    ((S-\{C^*\})\cap \mi{guards}(a))\cup (\{a^*\gle \gu\}\cap \mi{guards}(a))\cup (\{a_0\gle \gu\}\cap \mi{guards}(a))=(S\cap \mi{guards}(a))-\{C^*\}=S\cap \mi{guards}(a)=
                    \mi{guards}(S,a)$;
$a$ is positively guarded in $S_2$;
$S_2$ is positively guarded;
$\mi{guards}(S_2,a^*)=\{\gz\gle a^*,a^*\gle \gu\}\neq \{\gz\gle a^*\}$, 
$\mi{guards}(S_2,a_0)=\{\gz\gle a_0,a_0\gle \gu\}\neq \{\gz\gle a_0\}$, 
$a_0\in \mi{atoms}(S)$, $\mi{guards}(S,a_0)=\{\gz\gle a_0\}$, $a_0\in \mi{unsaturated}(S)=\{a \,|\, a\in \mi{atoms}(S), \mi{guards}(S,a)=\{\gz\gle a\}\}$,
$a^*, a_0\not\in \mi{unsaturated}(S_2)=\{a \,|\, a\in \mi{atoms}(S_2), \mi{guards}(S_2,a)=\{\gz\gle a\}\}=\{a \,|\, a\in \mi{atoms}(S_2), \mi{guards}(S_2,a)=\{\gz\gle a\}\}-\{a^*,a_0\}=
                                       \{a \,|\, a\in \mi{atoms}(S_2)-\{a^*,a_0\}, \mi{guards}(S_2,a)=\{\gz\gle a\}\}=\{a \,|\, a\in \mi{atoms}(S)-\{a^*,a_0\}, \mi{guards}(S,a)=\{\gz\gle a\}\}=
                                       \mi{unsaturated}(S)-\{a^*,a_0\}=\{a \,|\, a\in \mi{atoms}(S), \mi{guards}(S,a)=\{\gz\gle a\}\}-\{a^*,a_0\}\subset \mi{unsaturated}(S)$, 
$\mi{measure}(S_2)=(\mi{atoms}(S_2),\mi{unsaturated}(S_2),\mi{invalid}(S_2),\mi{count}(S_2))=(\mi{atoms}(S),                                                                               \linebreak[4]
                                                                                                            \mi{unsaturated}(S_2),\mi{invalid}(S_2),\mi{count}(S_2))\prec 
                   \mi{measure}(S)=(\mi{atoms}(S),                                                                                                                                         \linebreak[4]
                                                  \mi{unsaturated}(S),\mi{invalid}(S),\mi{count}(S))$;
by the induction hypothesis for $S_2$, there exists a finite {\it DPLL}-tree $\mi{Tree}_2$ with the root $S_2$ constructed using Rules (\cref{ceq4hr1x}), (\cref{ceq4hr1111})--(\cref{ceq4hr66}), (\cref{ceq4hr8}) satisfying
for every leaf $S_2'$ that either $\square\in S_2'$ or $S_2'\subseteq_{\mc F} \mi{OrdPropCl}$ is positive.

Case 2.2.2.1.2.2.1.2.1.2.2:
$C^*-\{l^*\}=a^*\geql \gu$.
Then $S_2=(S-\{C^*\})\cup \{C^*-\{l^*\}\}\cup \{a_0\gle \gu\}=(S-\{C^*\})\cup \{a^*\geql \gu\}\cup \{a_0\gle \gu\}$,
$\mi{guards}(S_2,a^*)=\mi{guards}(S,a^*)\cup \{C^*-\{l^*\}\}=\mi{guards}(S,a^*)\cup \{a^*\geql \gu\}$,
either $\mi{guards}(S,a^*)=\{\gz\gle a^*\}$ or $\mi{guards}(S,a^*)=\{\gz\gle a^*,a^*\gle \gu\}$.
We get two cases for $\mi{guards}(S,a^*)$.

Case 2.2.2.1.2.2.1.2.1.2.2.1:
$\mi{guards}(S,a^*)=\{\gz\gle a^*\}$.
Then $S\supseteq \mi{guards}(S,a^*)=\{\gz\gle a^*\}$;
$S_2\supseteq \mi{guards}(S_2)\supseteq \mi{guards}(S_2,a^*)=\mi{guards}(S,a^*)\cup \{a^*\geql \gu\}=\{\gz\gle a^*,a^*\geql \gu\}$,
$a^*\geql \gu\in \mi{guards}(S_2)$, $\gz\gle a^*\in S_2$,
$a^*\in \mi{atoms}(\gz\gle a^*)$, $a^*\geql \gu\neq \gz\gle a^*$, $\mi{simplify}(\gz\gle a^*,a^*,\gu)=\gz\gle \gu$,
applying Rule (\cref{ceq4hr4}) to $S_2$, $a^*\geql \gu$, and $\gz\gle a^*$, we derive
\begin{equation} \notag
\dfrac{S_2}
      {(S_2-\{\gz\gle a^*\})\cup \{\gz\gle \gu\}};
\end{equation}
$\gz\gle \gu\in (S_2-\{\gz\gle a^*\})\cup \{\gz\gle \gu\}$;
$\gz\gle \gu\in \mi{OrdPropLit}$ is a tautology;
$\gz\gle \gu$ is not a guard;
$\gz\gle \gu\not\in \mi{guards}((S_2-\{\gz\gle a^*\})\cup \{\gz\gle \gu\})$,
$\gz\gle \gu\in ((S_2-\{\gz\gle a^*\})\cup \{\gz\gle \gu\})-\mi{guards}((S_2-\{\gz\gle a^*\})\cup \{\gz\gle \gu\})$,
applying Rule (\cref{ceq4hr22}) to $(S_2-\{\gz\gle a^*\})\cup \{\gz\gle \gu\}$ and $\gz\gle \gu$, we derive
\begin{equation} \notag
\dfrac{(S_2-\{\gz\gle a^*\})\cup \{\gz\gle \gu\}}
      {((S_2-\{\gz\gle a^*\})\cup \{\gz\gle \gu\})-\{\gz\gle \gu\}}.
\end{equation}
We have that $S_2$ is simplified.
Hence, $\gz\gle \gu\not\in S_2$, $C^*, \gz\gle a^*\in S$, $C^*\not\in \mi{guards}(a^*)$, $\gz\gle a^*\in \mi{guards}(a^*)$, $C^*\neq \gz\gle a^*$, $a^*\geql \gu, a_0\gle \gu\not\in S$,
$a^*\geql \gu\neq a_0\gle \gu$,
$((S_2-\{\gz\gle a^*\})\cup \{\gz\gle \gu\})-\{\gz\gle \gu\}=((S_2-\{\gz\gle a^*\})-\{\gz\gle \gu\})\cup (\{\gz\gle \gu\}-\{\gz\gle \gu\})=S_2-\{\gz\gle a^*\}=
 ((S-\{C^*\})\cup \{a^*\geql \gu\}\cup \{a_0\gle \gu\})-\{\gz\gle a^*\}=((S-\{C^*\})-\{\gz\gle a^*\})\cup (\{a^*\geql \gu\}-\{\gz\gle a^*\})\cup (\{a_0\gle \gu\}-\{\gz\gle a^*\})=
 ((S-\{C^*\})-\{\gz\gle a^*\})\cup \{a^*\geql \gu\}\cup \{a_0\gle \gu\}$.
We put $S_2'=((S-\{C^*\})-\{\gz\gle a^*\})\cup \{a^*\geql \gu\}\cup \{a_0\gle \gu\}\subseteq_{\mc F} \mi{OrdPropCl}$.
We get that
$\mi{atoms}(\gz\gle a^*)=\{a^*\}$,
$a^*,a_0\in \mi{atoms}(S_2')=\mi{atoms}(((S-\{C^*\})-\{\gz\gle a^*\})\cup \{a^*\geql \gu\}\cup \{a_0\gle \gu\})=
                             \mi{atoms}((S-\{C^*\})-\{\gz\gle a^*\})\cup \mi{atoms}(a^*\geql \gu)\cup \mi{atoms}(a_0\gle \gu)=
                             \mi{atoms}(S-\{C^*\})\cup \mi{atoms}(a^*\geql \gu)\cup \mi{atoms}(a_0\gle \gu)=
                             \mi{atoms}(S-\{C^*\})\cup \mi{atoms}(C^*-\{l^*\})\cup \mi{atoms}(\Cn^*\gle \gu)=\mi{atoms}(S)$;
$S$ is simplified;
$(S-\{C^*\})-\{\gz\gle a^*\}\subseteq S$ is simplified;
$a^*\geql \gu, a_0\gle \gu\neq \square$ do not contain contradictions and tautologies;
$S_2'=((S-\{C^*\})-\{\gz\gle a^*\})\cup \{a^*\geql \gu\}\cup \{a_0\gle \gu\}$ is simplified;
$a^*\neq a_0$, 
$\mi{guards}(a^*)\cap \{a_0\gle \gu\}=\mi{guards}(a^*)\cap \mi{guards}(a_0)=\emptyset$,
$C^*\not\in \mi{guards}(a^*)$,
$\mi{guards}(S_2',a^*)=(((S-\{C^*\})-\{\gz\gle a^*\})\cup \{a^*\geql \gu\}\cup \{a_0\gle \gu\})\cap \mi{guards}(a^*)=
                       (((S-\{C^*\})-\{\gz\gle a^*\})\cap \mi{guards}(a^*))\cup (\{a^*\geql \gu\}\cap \mi{guards}(a^*))\cup (\{a_0\gle \gu\}\cap \mi{guards}(a^*))=
                       (((S\cap \mi{guards}(a^*))-\{\gz\gle a^*\})-\{C^*\})\cup \{a^*\geql \gu\}=((S\cap \mi{guards}(a^*))-\{\gz\gle a^*\})\cup \{a^*\geql \gu\}=
                       (\mi{guards}(S,a^*)-\{\gz\gle a^*\})\cup \{a^*\geql \gu\}=(\{\gz\gle a^*\}-\{\gz\gle a^*\})\cup \{a^*\geql \gu\}=\{a^*\geql \gu\}$;
$a^*$ is semi-positively guarded in $S_2'$;
$\mi{guards}(a_0)\cap \{\gz\gle a^*,a^*\geql \gu\}=\mi{guards}(a_0)\cap \mi{guards}(a^*)=\emptyset$,
$C^*\in S$, $C^*\not\in \mi{guards}(S)\supseteq \mi{guards}(S,a_0)=S\cap \mi{guards}(a_0)$, $C^*\not\in \mi{guards}(a_0)$,
$\mi{guards}(S_2',a_0)=(((S-\{C^*\})-\{\gz\gle a^*\})\cup \{a^*\geql \gu\}\cup \{a_0\gle \gu\})\cap \mi{guards}(a_0)=
                       (((S-\{C^*\})-\{\gz\gle a^*\})\cap \mi{guards}(a_0))\cup (\{a^*\geql \gu\}\cap \mi{guards}(a_0))\cup (\{a_0\gle \gu\}\cap \mi{guards}(a_0))=
                       (((S\cap \mi{guards}(a_0))-\{C^*\})-\{\gz\gle a^*\})\cup \{a_0\gle \gu\}=(S\cap \mi{guards}(a_0))\cup \{a_0\gle \gu\}=\mi{guards}(S,a_0)\cup \{a_0\gle \gu\}=
                       \{\gz\gle a_0,a_0\gle \gu\}$;
$a_0$ is positively guarded in $S_2$;
for all $a\in \mi{atoms}(S_2')-\{a^*,a_0\}=\mi{atoms}(S)-\{a^*,a_0\}$,
$a$ is positively guarded in $S$;
$a\neq a^*$, 
$\mi{guards}(a)\cap \{\gz\gle a^*,a^*\geql \gu\}=\mi{guards}(a)\cap \mi{guards}(a^*)=\emptyset$,
$a\neq a_0$,
$\mi{guards}(a)\cap \{a_0\gle \gu\}=\mi{guards}(a)\cap \mi{guards}(a_0)=\emptyset$,
$C^*\in S$, $C^*\not\in \mi{guards}(S)\supseteq \mi{guards}(S,a)=S\cap \mi{guards}(a)$, $C^*\not\in \mi{guards}(a)$,
$\mi{guards}(S_2',a)=(((S-\{C^*\})-\{\gz\gle a^*\})\cup \{a^*\geql \gu\}\cup \{a_0\gle \gu\})\cap \mi{guards}(a)=
                     (((S-\{C^*\})-\{\gz\gle a^*\})\cap \mi{guards}(a))\cup (\{a^*\geql \gu\}\cap \mi{guards}(a))\cup (\{a_0\gle \gu\}\cap \mi{guards}(a))=
                     (((S\cap \mi{guards}(a))-\{C^*\})-\{\gz\gle a^*\})=S\cap \mi{guards}(a)=\mi{guards}(S,a)$;
$a$ is positively guarded in $S_2'$;
$S_2'$ is semi-positively guarded;
by Lemma \ref{le66} for $S_2'$ and $a^*$, there exists a finite linear {\it DPLL}-tree $\mi{Tree}_2'$ with the root $S_2'$ constructed using Rules (\cref{ceq4hr1111111})--(\cref{ceq4hr4}), (\cref{ceq4hr8}) satisfying 
for its only leaf $S_2''$ that either $\square\in S_2''$, or $S_2''\subseteq_{\mc F} \mi{OrdPropCl}$ is semi-positively guarded, $\mi{atoms}(S_2'')\subseteq \mi{atoms}(S_2')-\{a^*\}$.
We get two cases for $S_2''$.

Case 2.2.2.1.2.2.1.2.1.2.2.1.1:
$\square\in S_2''$.
We put 
\begin{equation} \notag
\mi{Tree}_2=\begin{array}[c]{c}
            S_2 \\[0.4mm]
            \hline \\[-3.8mm]
            (S_2-\{\gz\gle a^*\})\cup \{\gz\gle \gu\} \\[0.4mm]
            \hline \\[-3.8mm]
            \mi{Tree}_2'.
            \end{array}
\end{equation}
Hence, $\mi{Tree}_2$ is a finite {\it DPLL}-tree with the root $S_2$ constructed using Rules (\cref{ceq4hr1x}), (\cref{ceq4hr1111})--(\cref{ceq4hr66}), (\cref{ceq4hr8}) such that
for its only leaf $S_2''$, $\square\in S_2''$.

Case 2.2.2.1.2.2.1.2.1.2.2.1.2:
$S_2''\subseteq_{\mc F} \mi{OrdPropCl}$ is semi-positively guarded, $\mi{atoms}(S_2'')\subseteq \mi{atoms}(S_2')-\{a^*\}$.
Then $a^*\in \mi{atoms}(S_2')$, $\mi{atoms}(S_2'')\subseteq \mi{atoms}(S_2')-\{a^*\}\subset \mi{atoms}(S_2')\subseteq \mi{atoms}(S)$, 
$\mi{measure}(S_2'')=(\mi{atoms}(S_2''),\mi{unsaturated}(S_2''),                                                                                                                           \linebreak[4]
                                                                \mi{invalid}(S_2''),\mi{count}(S_2''))\prec \mi{measure}(S)=(\mi{atoms}(S),\mi{unsaturated}(S),\mi{invalid}(S),            \linebreak[4]
                                                                                                                                                                               \mi{count}(S))$;
by the induction hypothesis for $S_2''$, there exists a finite {\it DPLL}-tree $\mi{Tree}_2''$ with the root $S_2''$ constructed using Rules (\cref{ceq4hr1x}), (\cref{ceq4hr1111})--(\cref{ceq4hr66}), (\cref{ceq4hr8}) satisfying
for every leaf $S_2'''$ that either $\square\in S_2'''$ or $S_2'''\subseteq_{\mc F} \mi{OrdPropCl}$ is positive.
We put
\begin{equation} \notag
\mi{Tree}_2=\begin{array}[c]{c}
            S_2 \\[0.4mm]
            \hline \\[-3.8mm]
            (S_2-\{\gz\gle a^*\})\cup \{\gz\gle \gu\} \\[0.4mm]
            \hline \\[-3.8mm]
            \mi{Tree}_2' \\[0.4mm]
            \hline \\[-3.8mm]
            \mi{Tree}_2''.
            \end{array}
\end{equation}
Hence, $\mi{Tree}_2$ is a finite {\it DPLL}-tree with the root $S_2$ constructed using Rules (\cref{ceq4hr1x}), (\cref{ceq4hr1111})--(\cref{ceq4hr66}), (\cref{ceq4hr8}) such that
for every leaf $S_2'''$, either $\square\in S_2'''$ or $S_2'''\subseteq_{\mc F} \mi{OrdPropCl}$ is positive.

Case 2.2.2.1.2.2.1.2.1.2.2.2:
$\mi{guards}(S,a^*)=\{\gz\gle a^*,a^*\gle \gu\}$.
Then $S_2\supseteq \mi{guards}(S_2)\supseteq \mi{guards}(S_2,a^*)=\mi{guards}(S,a^*)\cup \{a^*\geql \gu\}=\{\gz\gle a^*,a^*\gle \gu,a^*\geql \gu\}$,
$a^*\geql \gu\in \mi{guards}(S_2)$, $a^*\gle \gu\in S_2$,
$a^*\in \mi{atoms}(a^*\gle \gu)$, $a^*\geql \gu\neq a^*\gle \gu$, $\mi{simplify}(a^*\gle \gu,a^*,\gu)=\gu\gle \gu$,
applying Rule (\cref{ceq4hr4}) to $S_2$, $a^*\geql \gu$, and $a^*\gle \gu$, we derive
\begin{equation} \notag
\dfrac{S_2}
      {(S_2-\{a^*\gle \gu\})\cup \{\gu\gle \gu\}};
\end{equation}
$\gu\gle \gu\in (S_2-\{a^*\gle \gu\})\cup \{\gu\gle \gu\}$;
$\gu\gle \gu\in \mi{OrdPropLit}$ is a contradiction;
$\gu\gle \gu$ is not a guard;
$\gu\gle \gu\not\in \mi{guards}((S_2-\{a^*\gle \gu\})\cup \{\gu\gle \gu\})$,
$\gu\gle \gu\in ((S_2-\{a^*\gle \gu\})\cup \{\gu\gle \gu\})-\mi{guards}((S_2-\{a^*\gle \gu\})\cup \{\gu\gle \gu\})$,
applying Rule (\cref{ceq4hr2}) to $(S_2-\{a^*\gle \gu\})\cup \{\gu\gle \gu\}$ and $\gu\gle \gu$, we derive
\begin{equation} \notag
\dfrac{(S_2-\{a^*\gle \gu\})\cup \{\gu\gle \gu\}}
      {(((S_2-\{a^*\gle \gu\})\cup \{\gu\gle \gu\})-\{\gu\gle \gu\})\cup \{\square\}}.
\end{equation}
We put
\begin{equation} \notag
\mi{Tree}_2=\begin{array}[c]{c}
            S_2 \\[0.4mm]
            \hline \\[-3.8mm]
            (S_2-\{a^*\gle \gu\})\cup \{\gu\gle \gu\} \\[0.4mm]
            \hline \\[-3.8mm]
            S_2'=(((S_2-\{a^*\gle \gu\})\cup \{\gu\gle \gu\})-\{\gu\gle \gu\})\cup \{\square\}.
            \end{array} 
\end{equation}
Hence, $\mi{Tree}_2$ is a finite {\it DPLL}-tree with the root $S_2$ constructed using Rules (\cref{ceq4hr1x}), (\cref{ceq4hr1111})--(\cref{ceq4hr66}), (\cref{ceq4hr8}) such that
for its only leaf $S_2'$, $\square\in S_2'$.

Case 2.2.2.1.2.2.1.2.1.2.3:
$C^*-\{l^*\}=\gu\gleq a^*$.
Then $S_2=(S-\{C^*\})\cup \{C^*-\{l^*\}\}\cup \{a_0\gle \gu\}=(S-\{C^*\})\cup \{\gu\gleq a^*\}\cup \{a_0\gle \gu\}$,
$\mi{guards}(S_2,a^*)=\mi{guards}(S,a^*)\cup \{C^*-\{l^*\}\}=\mi{guards}(S,a^*)\cup \{\gu\gleq a^*\}$,
either $\mi{guards}(S,a^*)=\{\gz\gle a^*\}$ or $\mi{guards}(S,a^*)=\{\gz\gle a^*,a^*\gle \gu\}$.
We get two cases for $\mi{guards}(S,a^*)$.

Case 2.2.2.1.2.2.1.2.1.2.3.1:
$\mi{guards}(S,a^*)=\{\gz\gle a^*\}$.
Then $S\supseteq \mi{guards}(S,a^*)=\{\gz\gle a^*\}$;
$\mi{guards}(S_2)\supseteq \mi{guards}(S_2,a^*)=\mi{guards}(S,a^*)\cup \{\gu\gleq a^*\}=\{\gz\gle a^*,\gu\gleq a^*\}$,
$\gu\gleq a^*\in \mi{guards}(S_2)$, applying Rule (\cref{ceq4hr11111111}) to $S_2$ and $\gu\gleq a^*$, we derive
\begin{equation} \notag
\dfrac{S_2}
      {(S_2-\{\gu\gleq a^*\})\cup \{a^*\geql \gu\}};
\end{equation}
$a^*\in \mi{atoms}((S_2-\{\gu\gleq a^*\})\cup \{a^*\geql \gu\})$,
$(S_2-\{\gu\gleq a^*\})\cup \{a^*\geql \gu\}\supseteq \mi{guards}((S_2-\{\gu\gleq a^*\})\cup \{a^*\geql \gu\})\supseteq 
 \mi{guards}((S_2-\{\gu\gleq a^*\})\cup \{a^*\geql \gu\},a^*)=((S_2-\{\gu\gleq a^*\})\cup \{a^*\geql \gu\})\cap \mi{guards}(a^*)=
 ((S_2-\{\gu\gleq a^*\})\cap \mi{guards}(a^*))\cup (\{a^*\geql \gu\}\cap \mi{guards}(a^*))=((S_2\cap \mi{guards}(a^*))-\{\gu\gleq a^*\})\cup \{a^*\geql \gu\}=
 (\mi{guards}(S_2,a^*)-\{\gu\gleq a^*\})\cup \{a^*\geql \gu\}=(\{\gz\gle a^*,\gu\gleq a^*\}-\{\gu\gleq a^*\})\cup \{a^*\geql \gu\}=\{\gz\gle a^*,a^*\geql \gu\}$,
$a^*\geql \gu\in \mi{guards}((S_2-\{\gu\gleq a^*\})\cup \{a^*\geql \gu\})$, $\gz\gle a^*\in (S_2-\{\gu\gleq a^*\})\cup \{a^*\geql \gu\}$,
$a^*\in \mi{atoms}(\gz\gle a^*)$, $a^*\geql \gu\neq \gz\gle a^*$, $\mi{simplify}(\gz\gle a^*,a^*,\gu)=\gz\gle \gu$,
applying Rule (\cref{ceq4hr4}) to $(S_2-\{\gu\gleq a^*\})\cup \{a^*\geql \gu\}$, $a^*\geql \gu$, and $\gz\gle a^*$, we derive
\begin{equation} \notag
\dfrac{(S_2-\{\gu\gleq a^*\})\cup \{a^*\geql \gu\}}
      {(((S_2-\{\gu\gleq a^*\})\cup \{a^*\geql \gu\})-\{\gz\gle a^*\})\cup \{\gz\gle \gu\}};
\end{equation}
$\gz\gle \gu\in (((S_2-\{\gu\gleq a^*\})\cup \{a^*\geql \gu\})-\{\gz\gle a^*\})\cup \{\gz\gle \gu\}$;
$\gz\gle \gu\in \mi{OrdPropLit}$ is a tautology;
$\gz\gle \gu$ is not a guard;
$\gz\gle \gu\not\in \mi{guards}((((S_2-\{\gu\gleq a^*\})\cup \{a^*\geql \gu\})-\{\gz\gle a^*\})\cup \{\gz\gle \gu\})$,
$\gz\gle \gu\in ((((S_2-\{\gu\gleq a^*\})\cup \{a^*\geql \gu\})-\{\gz\gle a^*\})\cup \{\gz\gle \gu\})-\mi{guards}((((S_2-\{\gu\gleq a^*\})\cup \{a^*\geql \gu\})-\{\gz\gle a^*\})\cup \{\gz\gle \gu\})$,
applying Rule (\cref{ceq4hr22}) to $(((S_2-\{\gu\gleq a^*\})\cup \{a^*\geql \gu\})-\{\gz\gle a^*\})\cup \{\gz\gle \gu\}$ and $\gz\gle \gu$, we derive
\begin{equation} \notag
\dfrac{(((S_2-\{\gu\gleq a^*\})\cup \{a^*\geql \gu\})-\{\gz\gle a^*\})\cup \{\gz\gle \gu\}}
      {((((S_2-\{\gu\gleq a^*\})\cup \{a^*\geql \gu\})-\{\gz\gle a^*\})\cup \{\gz\gle \gu\})-\{\gz\gle \gu\}}.
\end{equation}
We have that $S_2$ is simplified.
Hence, $\gz\gle \gu\not\in S_2$, $\gu\gleq a^*\not\in S$, $C^*, \gz\gle a^*\in S$, $C^*\not\in \mi{guards}(a^*)$, $\gz\gle a^*\in \mi{guards}(a^*)$, $C^*\neq \gz\gle a^*$, 
$a^*\geql \gu, a_0\gle \gu\not\in S$, $a^*\geql \gu\neq a_0\gle \gu$,
$((((S_2-\{\gu\gleq a^*\})\cup \{a^*\geql \gu\})-\{\gz\gle a^*\})\cup \{\gz\gle \gu\})-\{\gz\gle \gu\}=
 ((((S_2-\{\gu\gleq a^*\})\cup \{a^*\geql \gu\})-\{\gz\gle a^*\})-\{\gz\gle \gu\})\cup (\{\gz\gle \gu\}-\{\gz\gle \gu\})=
 (((S_2-\{\gu\gleq a^*\})\cup \{a^*\geql \gu\})-\{\gz\gle \gu\})-\{\gz\gle a^*\}=
 (((S_2-\{\gu\gleq a^*\})-\{\gz\gle \gu\})\cup (\{a^*\geql \gu\}-\{\gz\gle \gu\}))-\{\gz\gle a^*\}=
 ((S_2-\{\gu\gleq a^*\})\cup \{a^*\geql \gu\})-\{\gz\gle a^*\}=
 ((S_2-\{\gu\gleq a^*\})-\{\gz\gle a^*\})\cup (\{a^*\geql \gu\}-\{\gz\gle a^*\})=
 ((S_2-\{\gu\gleq a^*\})-\{\gz\gle a^*\})\cup \{a^*\geql \gu\}=
 ((((S-\{C^*\})\cup \{\gu\gleq a^*\}\cup \{a_0\gle \gu\})-\{\gu\gleq a^*\})-\{\gz\gle a^*\})\cup \{a^*\geql \gu\}=
 ((((S-\{C^*\})-\{\gu\gleq a^*\})\cup (\{\gu\gleq a^*\}-\{\gu\gleq a^*\})\cup (\{a_0\gle \gu\}-\{\gu\gleq a^*\}))-\{\gz\gle a^*\})\cup \{a^*\geql \gu\}=
 (((S-\{C^*\})\cup \{a_0\gle \gu\})-\{\gz\gle a^*\})\cup \{a^*\geql \gu\}=
 ((S-\{C^*\})-\{\gz\gle a^*\})\cup (\{a_0\gle \gu\}-\{\gz\gle a^*\})\cup \{a^*\geql \gu\}=
 ((S-\{C^*\})-\{\gz\gle a^*\})\cup \{a^*\geql \gu\}\cup \{a_0\gle \gu\}$.
We put $S_2'=((S-\{C^*\})-\{\gz\gle a^*\})\cup \{a^*\geql \gu\}\cup \{a_0\gle \gu\}\subseteq_{\mc F} \mi{OrdPropCl}$.
We get from Case 2.2.2.1.2.2.1.2.1.2.2.1 that there exists a finite linear {\it DPLL}-tree $\mi{Tree}_2'$ with the root $S_2'$ constructed using Rules (\cref{ceq4hr1111111})--(\cref{ceq4hr4}), (\cref{ceq4hr8}) satisfying 
for its only leaf $S_2''$ that either $\square\in S_2''$, or $S_2''\subseteq_{\mc F} \mi{OrdPropCl}$ is semi-positively guarded, $\mi{atoms}(S_2'')\subseteq \mi{atoms}(S_2')-\{a^*\}$.
We get two cases for $S_2''$.

Case 2.2.2.1.2.2.1.2.1.2.3.1.1:
$\square\in S_2''$.
We put 
\begin{equation} \notag
\mi{Tree}_2=\begin{array}[c]{c}
            S_2 \\[0.4mm]
            \hline \\[-3.8mm]
            (S_2-\{\gu\gleq a^*\})\cup \{a^*\geql \gu\} \\[0.4mm]
            \hline \\[-3.8mm]
            (((S_2-\{\gu\gleq a^*\})\cup \{a^*\geql \gu\})-\{\gz\gle a^*\})\cup \{\gz\gle \gu\} \\[0.4mm]
            \hline \\[-3.8mm]
            \mi{Tree}_2'.
            \end{array}
\end{equation}
Hence, $\mi{Tree}_2$ is a finite {\it DPLL}-tree with the root $S_2$ constructed using Rules (\cref{ceq4hr1x}), (\cref{ceq4hr1111})--(\cref{ceq4hr66}), (\cref{ceq4hr8}) such that
for its only leaf $S_2''$, $\square\in S_2''$.

Case 2.2.2.1.2.2.1.2.1.2.3.1.2:
$S_2''\subseteq_{\mc F} \mi{OrdPropCl}$ is semi-positively guarded, $\mi{atoms}(S_2'')\subseteq \mi{atoms}(S_2')-\{a^*\}$.
Then $a^*\in \mi{atoms}(S_2')$, $\mi{atoms}(S_2'')\subseteq \mi{atoms}(S_2')-\{a^*\}\subset \mi{atoms}(S_2')\subseteq \mi{atoms}(S)$, 
$\mi{measure}(S_2'')=(\mi{atoms}(S_2''),\mi{unsaturated}(S_2''),                                                                                                                           \linebreak[4]
                                                                \mi{invalid}(S_2''),\mi{count}(S_2''))\prec \mi{measure}(S)=(\mi{atoms}(S),\mi{unsaturated}(S),\mi{invalid}(S),            \linebreak[4]
                                                                                                                                                                               \mi{count}(S))$;
by the induction hypothesis for $S_2''$, there exists a finite {\it DPLL}-tree $\mi{Tree}_2''$ with the root $S_2''$ constructed using Rules (\cref{ceq4hr1x}), (\cref{ceq4hr1111})--(\cref{ceq4hr66}), (\cref{ceq4hr8}) satisfying
for every leaf $S_2'''$ that either $\square\in S_2'''$ or $S_2'''\subseteq_{\mc F} \mi{OrdPropCl}$ is positive.
We put
\begin{equation} \notag
\mi{Tree}_2=\begin{array}[c]{c}
            S_2 \\[0.4mm]
            \hline \\[-3.8mm]
            (S_2-\{\gu\gleq a^*\})\cup \{a^*\geql \gu\} \\[0.4mm]
            \hline \\[-3.8mm]
            (((S_2-\{\gu\gleq a^*\})\cup \{a^*\geql \gu\})-\{\gz\gle a^*\})\cup \{\gz\gle \gu\} \\[0.4mm]
            \hline \\[-3.8mm]
            \mi{Tree}_2' \\[0.4mm]
            \hline \\[-3.8mm]
            \mi{Tree}_2''.
            \end{array}
\end{equation}
Hence, $\mi{Tree}_2$ is a finite {\it DPLL}-tree with the root $S_2$ constructed using Rules (\cref{ceq4hr1x}), (\cref{ceq4hr1111})--(\cref{ceq4hr66}), (\cref{ceq4hr8}) such that
for every leaf $S_2'''$, either $\square\in S_2'''$ or $S_2'''\subseteq_{\mc F} \mi{OrdPropCl}$ is positive.

Case 2.2.2.1.2.2.1.2.1.2.3.2:
$\mi{guards}(S,a^*)=\{\gz\gle a^*,a^*\gle \gu\}$.
Then $\mi{guards}(S_2)\supseteq \mi{guards}(S_2,a^*)=\mi{guards}(S,a^*)\cup \{\gu\gleq a^*\}=\{\gz\gle a^*,a^*\gle \gu,\gu\gleq a^*\}$,
$\gu\gleq a^*\in \mi{guards}(S_2)$, applying Rule (\cref{ceq4hr11111111}) to $S_2$ and $\gu\gleq a^*$, we derive
\begin{equation} \notag
\dfrac{S_2}
      {(S_2-\{\gu\gleq a^*\})\cup \{a^*\geql \gu\}};
\end{equation}
$a^*\in \mi{atoms}((S_2-\{\gu\gleq a^*\})\cup \{a^*\geql \gu\})$,
$(S_2-\{\gu\gleq a^*\})\cup \{a^*\geql \gu\}\supseteq \mi{guards}((S_2-\{\gu\gleq a^*\})\cup \{a^*\geql \gu\})\supseteq 
 \mi{guards}((S_2-\{\gu\gleq a^*\})\cup \{a^*\geql \gu\},a^*)=((S_2-\{\gu\gleq a^*\})\cup \{a^*\geql \gu\})\cap \mi{guards}(a^*)=
 ((S_2-\{\gu\gleq a^*\})\cap \mi{guards}(a^*))\cup (\{a^*\geql \gu\}\cap \mi{guards}(a^*))=((S_2\cap \mi{guards}(a^*))-\{\gu\gleq a^*\})\cup \{a^*\geql \gu\}=
 (\mi{guards}(S_2,a^*)-\{\gu\gleq a^*\})\cup \{a^*\geql \gu\}=(\{\gz\gle a^*,a^*\gle \gu,\gu\gleq a^*\}-\{\gu\gleq a^*\})\cup \{a^*\geql \gu\}=\{\gz\gle a^*,a^*\gle \gu,a^*\geql \gu\}$,
$a^*\geql \gu\in \mi{guards}((S_2-\{\gu\gleq a^*\})\cup \{a^*\geql \gu\})$, $a^*\gle \gu\in (S_2-\{\gu\gleq a^*\})\cup \{a^*\geql \gu\}$,
$a^*\in \mi{atoms}(a^*\gle \gu)$, $a^*\geql \gu\neq a^*\gle \gu$, $\mi{simplify}(a^*\gle \gu,a^*,\gu)=\gu\gle \gu$,
applying Rule (\cref{ceq4hr4}) to $(S_2-\{\gu\gleq a^*\})\cup \{a^*\geql \gu\}$, $a^*\geql \gu$, and $a^*\gle \gu$, we derive
\begin{equation} \notag
\dfrac{(S_2-\{\gu\gleq a^*\})\cup \{a^*\geql \gu\}}
      {(((S_2-\{\gu\gleq a^*\})\cup \{a^*\geql \gu\})-\{a^*\gle \gu\})\cup \{\gu\gle \gu\}};
\end{equation}
$\gu\gle \gu\in (((S_2-\{\gu\gleq a^*\})\cup \{a^*\geql \gu\})-\{a^*\gle \gu\})\cup \{\gu\gle \gu\}$;
$\gu\gle \gu\in \mi{OrdPropLit}$ is a contradiction;
$\gu\gle \gu$ is not a guard;
$\gu\gle \gu\not\in \mi{guards}((((S_2-\{\gu\gleq a^*\})\cup \{a^*\geql \gu\})-\{a^*\gle \gu\})\cup \{\gu\gle \gu\})$,
$\gu\gle \gu\in ((((S_2-\{\gu\gleq a^*\})\cup \{a^*\geql \gu\})-\{a^*\gle \gu\})\cup \{\gu\gle \gu\})-\mi{guards}((((S_2-\{\gu\gleq a^*\})\cup \{a^*\geql \gu\})-\{a^*\gle \gu\})\cup \{\gu\gle \gu\})$,
applying Rule (\cref{ceq4hr2}) to $(((S_2-\{\gu\gleq a^*\})\cup \{a^*\geql \gu\})-\{a^*\gle \gu\})\cup \{\gu\gle \gu\}$ and $\gu\gle \gu$, we derive
\begin{equation} \notag
\dfrac{(((S_2-\{\gu\gleq a^*\})\cup \{a^*\geql \gu\})-\{a^*\gle \gu\})\cup \{\gu\gle \gu\}}
      {(((((S_2-\{\gu\gleq a^*\})\cup \{a^*\geql \gu\})-\{a^*\gle \gu\})\cup \{\gu\gle \gu\})-\{\gu\gle \gu\})\cup \{\square\}}.
\end{equation}
We put
\begin{equation} \notag
\mi{Tree}_2=\begin{array}[c]{c}
            S_2 \\[0.4mm]
            \hline \\[-3.8mm]
            (S_2-\{\gu\gleq a^*\})\cup \{a^*\geql \gu\} \\[0.4mm]
            \hline \\[-3.8mm]
            (((S_2-\{\gu\gleq a^*\})\cup \{a^*\geql \gu\})-\{a^*\gle \gu\})\cup \{\gu\gle \gu\} \\[0.4mm]
            \hline \\[-3.8mm]
            S_2'=(((((S_2-\{\gu\gleq a^*\})\cup \{a^*\geql \gu\})-\{a^*\gle \gu\})\cup \{\gu\gle \gu\})- \\
            \hfill \{\gu\gle \gu\})\cup \{\square\}.
            \end{array} 
\end{equation}
Hence, $\mi{Tree}_2$ is a finite {\it DPLL}-tree with the root $S_2$ constructed using Rules (\cref{ceq4hr1x}), (\cref{ceq4hr1111})--(\cref{ceq4hr66}), (\cref{ceq4hr8}) such that
for its only leaf $S_2'$, $\square\in S_2'$.

Case 2.2.2.1.2.2.1.2.2:
For all $a\in \mi{atoms}(S_2)=\mi{atoms}(S)$, $C^*-\{l^*\}\not\in \mi{guards}(a)$.
Then $a_0\in \mi{atoms}(S)$, $C^*, C^*-\{l^*\}\not\in \mi{guards}(a_0)$, 
$\mi{guards}(S_2,a_0)=((S-\{C^*\})\cup \{C^*-\{l^*\}\}\cup \{a_0\gle \gu\})\cap \mi{guards}(a_0)=
                      ((S-\{C^*\})\cap \mi{guards}(a_0))\cup (\{C^*-\{l^*\}\}\cap \mi{guards}(a_0))\cup (\{a_0\gle \gu\}\cap \mi{guards}(a_0))=((S\cap \mi{guards}(a_0))-\{C^*\})\cup \{a_0\gle \gu\}=
                      \mi{guards}(S,a_0)\cup \{a_0\gle \gu\}=\{\gz\gle a_0,a_0\gle \gu\}$; 
$a_0$ is positively guarded in $S_2$;
for all $a\in \mi{atoms}(S_2)-\{a_0\}=\mi{atoms}(S)-\{a_0\}$,
$a$ is positively guarded in $S$;
$a\neq a_0$,
$\mi{guards}(a)\cap \{a_0\gle \gu\}=\mi{guards}(a)\cap \mi{guards}(a_0)=\emptyset$,
$C^*\in S$, $C^*\not\in \mi{guards}(S)\supseteq \mi{guards}(S,a)=S\cap \mi{guards}(a)$, $C^*, C^*-\{l^*\}\not\in \mi{guards}(a)$,
$\mi{guards}(S_2,a)=((S-\{C^*\})\cup \{C^*-\{l^*\}\}\cup \{a_0\gle \gu\})\cap \mi{guards}(a)=
                    ((S-\{C^*\})\cap \mi{guards}(a))\cup (\{C^*-\{l^*\}\}\cap \mi{guards}(a))\cup (\{a_0\gle \gu\}\cap \mi{guards}(a))=(S\cap \mi{guards}(a))-\{C^*\}=\mi{guards}(S,a)$;
$a$ is positively guarded in $S_2$;
$S_2$ is positively guarded;
$\mi{guards}(S_2,a_0)=\{\gz\gle a_0,a_0\gle \gu\}\neq \{\gz\gle a_0\}$, 
$\mi{guards}(S,a_0)=\{\gz\gle a_0\}$, $a_0\in \mi{unsaturated}(S)=\{a \,|\, a\in \mi{atoms}(S), \mi{guards}(S,a)=\{\gz\gle a\}\}$,
$a_0\not\in \mi{unsaturated}(S_2)=\{a \,|\, a\in \mi{atoms}(S_2), \mi{guards}(S_2,a)=\{\gz\gle a\}\}=\{a \,|\, a\in \mi{atoms}(S_2), \mi{guards}(S_2,a)=\{\gz\gle a\}\}-\{a_0\}=
                                  \{a \,|\, a\in \mi{atoms}(S_2)-\{a_0\}, \mi{guards}(S_2,a)=\{\gz\gle a\}\}=\{a \,|\, a\in \mi{atoms}(S)-\{a_0\}, \mi{guards}(S,a)=\{\gz\gle a\}\}=
                                  \mi{unsaturated}(S)-\{a_0\}=\{a \,|\, a\in \mi{atoms}(S), \mi{guards}(S,a)=\{\gz\gle a\}\}-\{a_0\}\subset \mi{unsaturated}(S)$, 
$\mi{measure}(S_2)=(\mi{atoms}(S_2),\mi{unsaturated}(S_2),\mi{invalid}(S_2),\mi{count}(S_2))=(\mi{atoms}(S),                                                                               \linebreak[4]
                                                                                                            \mi{unsaturated}(S_2),\mi{invalid}(S_2),\mi{count}(S_2))\prec 
                   \mi{measure}(S)=(\mi{atoms}(S),                                                                                                                                         \linebreak[4]
                                                  \mi{unsaturated}(S),\mi{invalid}(S),\mi{count}(S))$;
by the induction hypothesis for $S_2$, there exists a finite {\it DPLL}-tree $\mi{Tree}_2$ with the root $S_2$ constructed using Rules (\cref{ceq4hr1x}), (\cref{ceq4hr1111})--(\cref{ceq4hr66}), (\cref{ceq4hr8}) satisfying
for every leaf $S_2'$ that either $\square\in S_2'$ or $S_2'\subseteq_{\mc F} \mi{OrdPropCl}$ is positive.

Case 2.2.2.1.2.2.2:
$n\geq 1$.
Then $\Cn^*\geql \gu\in S_1$, $\Cn^*\gle \gu\in S_2$;
$\Cn^*\geql \gu=a_0\swedge\cdots\swedge a_n\geql \gu$ and $\Cn^*\gle \gu=a_0\swedge\cdots\swedge a_n\gle \gu$ are not guards;
$\Cn^*\geql \gu\not\in \mi{guards}(S_1)$, $\Cn^*\gle \gu\not\in \mi{guards}(S_2)$,
$\Cn^*\geql \gu\in S_1-\mi{guards}(S_1)$, $\Cn^*\gle \gu\in S_2-\mi{guards}(S_2)$;
for all $i\leq n$, $C^*\in S$, $C^*\not\in \mi{guards}(S)\supseteq \mi{guards}(S,a_i)=S\cap \mi{guards}(a_i)$, $C^*\not\in \mi{guards}(a_i)$.
We get two cases for $S_1$ and $S_2$.

Case 2.2.2.1.2.2.2.1:
$S_1$.
Then $\Cn^*\geql \gu\in S_1-\mi{guards}(S_1)$, $\Cn^*=a_0\swedge\cdots\swedge a_n$, $n\geq 1$,
applying Rule (\cref{ceq4hr111111}) to $S_1$ and $\Cn^*\geql \gu$, we derive
\begin{equation} \notag
\dfrac{S_1}
      {(S_1-\{\Cn^*\geql \gu\})\cup \{a_0\geql \gu,\dots,a_n\geql \gu\}}.
\end{equation}
Hence, $C^*\in S$, $l^*\in C^*$, $C^*=(C^*-\{l^*\})\cup \{l^*\}$, $l^*\not\in C^*-\{l^*\}\neq \square$;
$C^*$ is not unit;
$\Cn^*\geql \gu$ is unit;
$C^*\neq \Cn^*\geql \gu$;
for all $i\leq n$, $a_i\geql \gu\in \mi{guards}(a_i)$, $a_i\geql \gu\not\in \mi{guards}(S,a_i)=\{\gz\gle a_i\}=S\cap \mi{guards}(a_i)$, $\gz\gle a_i\in S$, $a_i\geql \gu\not\in S$;
$\{\gz\gle a_0,\dots,\gz\gle a_n\}\subseteq S$,
$\{a_0\geql \gu,\dots,a_n\geql \gu\}\cap S=\emptyset$,
$(S_1-\{\Cn^*\geql \gu\})\cup \{a_0\geql \gu,\dots,a_n\geql \gu\}=(((S-\{C^*\})\cup \{\Cn^*\geql \gu\})-\{\Cn^*\geql \gu\})\cup \{a_0\geql \gu,\dots,a_n\geql \gu\}=
 ((S-\{C^*\})-\{\Cn^*\geql \gu\})\cup (\{\Cn^*\geql \gu\}-\{\Cn^*\geql \gu\})\cup \{a_0\geql \gu,\dots,a_n\geql \gu\}=
 ((S-\{C^*\})-\{\Cn^*\geql \gu\})\cup \{a_0\geql \gu,\dots,a_n\geql \gu\}$.
We put $S_1'=((S-\{C^*\})-\{\Cn^*\geql \gu\})\cup \{a_0\geql \gu,\dots,a_n\geql \gu\}\subseteq_{\mc F} \mi{OrdPropCl}$.
We get that 
$S_1'=((S-\{C^*\})-\{\Cn^*\geql \gu\})\cup \{a_0\geql \gu,\dots,a_n\geql \gu\}=(S_1-\{\Cn^*\geql \gu\})\cup \{a_0\geql \gu,\dots,a_n\geql \gu\}$,
$\mi{atoms}(a_0\geql \gu,\dots,a_n\geql \gu)=\mi{atoms}(\Cn^*\geql \gu)=\mi{atoms}(\Cn^*)=\{a_0,\dots,a_n\}\subseteq \mi{atoms}(S_1')$, 
$\Cn^*\geql \gu\in S_1$,
$\{a_0,\dots,a_n\}\subseteq \mi{atoms}(S_1')=\mi{atoms}((S_1-\{\Cn^*\geql \gu\})\cup \{a_0\geql \gu,\dots,a_n\geql \gu\})=
                                             \mi{atoms}(S_1-\{\Cn^*\geql \gu\})\cup \mi{atoms}(a_0\geql \gu,\dots,a_n\geql \gu)=
                                             \mi{atoms}(S_1-\{\Cn^*\geql \gu\})\cup \mi{atoms}(\Cn^*\geql \gu)=\mi{atoms}((S_1-\{\Cn^*\geql \gu\})\cup \{\Cn^*\geql \gu\})=\mi{atoms}(S_1)\subseteq 
                                             \mi{atoms}(S)$;
$S_1$ is simplified; 
$S_1-\{\Cn^*\geql \gu\}\subseteq S_1$ is simplified;
for all $i\leq n$, $a_i\geql \gu\neq \square$ does not contain contradictions and tautologies;
$\{a_0\geql \gu,\dots,a_n\geql \gu\}$ is simplified;
$S_1'=(S_1-\{\Cn^*\geql \gu\})\cup \{a_0\geql \gu,\dots,a_n\geql \gu\}$ is simplified;
for all $i\leq n$, 
for all $j\leq n$ and $j\neq i$, $a_i\neq a_j$, $\mi{guards}(a_i)\cap \{\gz\gle a_j,a_j\geql \gu\}=\mi{guards}(a_i)\cap \mi{guards}(a_j)=\emptyset$;
$C^*\not\in \mi{guards}(a_i)$;
we have that $\Cn^*\geql \gu$ is not a guard;
$\Cn^*\geql \gu\not\in \mi{guards}(a_i)$,
$\mi{guards}(S_1',a_i)=(((S-\{C^*\})-\{\Cn^*\geql \gu\})\cup \{a_0\geql \gu,\dots,a_n\geql \gu\})\cap \mi{guards}(a_i)=
                       (((S-\{C^*\})-\{\Cn^*\geql \gu\})\cap \mi{guards}(a_i))\cup (\{a_0\geql \gu,\dots,a_n\geql \gu\}\cap \mi{guards}(a_i))=
                       (((S\cap \mi{guards}(a_i))-\{C^*\})-\{\Cn^*\geql \gu\})\cup \{a_i\geql \gu\}=(S\cap \mi{guards}(a_i))\cup \{a_i\geql \gu\}=\mi{guards}(S,a_i)\cup \{a_i\geql \gu\}=
                       \{\gz\gle a_i,a_i\geql \gu\}$;
by Lemma \ref{le88} for $S_1'$ and $\{a_0,\dots,a_n\}$, there exists a finite linear {\it DPLL}-tree $\mi{Tree}_1'$ with the root $S_1'$ constructed using Rules (\cref{ceq4hr22}) and (\cref{ceq4hr4}) satisfying
for its only leaf $S_1''$ that $S_1''=S_1'-\{\gz\gle a_0,\dots,\gz\gle a_n\}\subseteq_{\mc F} \mi{OrdPropCl}$ is simplified;  
$C^*\in S$, $\{\gz\gle a_0,\dots,\gz\gle a_n\}\subseteq S$, $C^*\neq \Cn^*\geql \gu$;
for all $i\leq n$, $C^*\not\in \mi{guards}(a_i)$, $\gz\gle a_i\in \mi{guards}(a_i)$, $C^*, \Cn^*\geql \gu\neq \gz\gle a_i$;
$\{C^*\}\cap \{\gz\gle a_0,\dots,\gz\gle a_n\}=\{\Cn^*\geql \gu\}\cap \{\gz\gle a_0,\dots,\gz\gle a_n\}=\emptyset$, 
$\{a_0\geql \gu,\dots,a_n\geql \gu\}\cap S=\emptyset$,
$S_1''=S_1'-\{\gz\gle a_0,\dots,\gz\gle a_n\}=(((S-\{C^*\})-\{\Cn^*\geql \gu\})\cup \{a_0\geql \gu,\dots,a_n\geql \gu\})-\{\gz\gle a_0,\dots,\gz\gle a_n\}=
       (((S-\{C^*\})-\{\Cn^*\geql \gu\})-\{\gz\gle a_0,\dots,\gz\gle a_n\})\cup (\{a_0\geql \gu,\dots,a_n\geql \gu\}-\{\gz\gle a_0,\dots,\gz\gle a_n\})=
       (((S-\{C^*\})-\{\Cn^*\geql \gu\})-\{\gz\gle a_0,\dots,\gz\gle a_n\})\cup \{a_0\geql \gu,\dots,a_n\geql \gu\}$,
$\mi{atoms}(\gz\gle a_0,\dots,\gz\gle a_n)=\{a_0,\dots,a_n\}$,
$\{a_0,\dots,a_n\}\subseteq \mi{atoms}(S_1'')=\mi{atoms}(S_1'-\{\gz\gle a_0,\dots,\gz\gle a_n\})=\mi{atoms}(S_1')=\mi{atoms}(S_1)\subseteq \mi{atoms}(S)$;
for all $i\leq n$,
$\mi{guards}(S_1'',a_i)=((((S-\{C^*\})-\{\Cn^*\geql \gu\})-\{\gz\gle a_0,\dots,\gz\gle a_n\})\cup \{a_0\geql \gu,\dots,a_n\geql \gu\})\cap \mi{guards}(a_i)=
                        ((((S-\{C^*\})-\{\Cn^*\geql \gu\})-\{\gz\gle a_0,\dots,\gz\gle a_n\})\cap \mi{guards}(a_i))\cup (\{a_0\geql \gu,\dots,a_n\geql \gu\}\cap \mi{guards}(a_i))=
                        ((((S\cap \mi{guards}(a_i))-\{\gz\gle a_0,\dots,\gz\gle a_n\})-\{C^*\})-\{\Cn^*\geql \gu\})\cup \{a_i\geql \gu\}=
                        ((S\cap \mi{guards}(a_i))-\{\gz\gle a_i\})\cup \{a_i\geql \gu\}=(\mi{guards}(S,a_i)-\{\gz\gle a_i\})\cup \{a_i\geql \gu\}=(\{\gz\gle a_i\}-\{\gz\gle a_i\})\cup \{a_i\geql \gu\}=
                        \{a_i\geql \gu\}$;
$a_i$ is semi-positively guarded in $S_1''$;
for all $a\in \mi{atoms}(S_1'')-\{a_0,\dots,a_n\}=\mi{atoms}(S_1)-\{a_0,\dots,a_n\}\subseteq \mi{atoms}(S)-\{a_0,\dots,a_n\}$,
$a$ is positively guarded in $S$;
for all $i\leq n$, $a\neq a_i$, $\mi{guards}(a)\cap \{\gz\gle a_i,a_i\geql \gu\}=\mi{guards}(a)\cap \mi{guards}(a_i)=\emptyset$;
$C^*\not\in \mi{guards}(S)\supseteq \mi{guards}(S,a)=S\cap \mi{guards}(a)$, $C^*\not\in \mi{guards}(a)$,
$\Cn^*\geql \gu\not\in \mi{guards}(a)$,
$\mi{guards}(S_1'',a)=((((S-\{C^*\})-\{\Cn^*\geql \gu\})-\{\gz\gle a_0,\dots,\gz\gle a_n\})\cup \{a_0\geql \gu,\dots,a_n\geql \gu\})\cap \mi{guards}(a)=
                      ((((S-\{C^*\})-\{\Cn^*\geql \gu\})-\{\gz\gle a_0,\dots,\gz\gle a_n\})\cap \mi{guards}(a))\cup (\{a_0\geql \gu,\dots,a_n\geql \gu\}\cap \mi{guards}(a))=
                      (((S\cap \mi{guards}(a))-\{C^*\})-\{\Cn^*\geql \gu\})-\{\gz\gle a_0,\dots,\gz\gle a_n\}=S\cap \mi{guards}(a)=\mi{guards}(S,a)$;
$a$ is positively guarded in $S_1''$;
$S_1''$ is semi-positively guarded; 
$a_0\in \mi{atoms}(S_1'')$, $\mi{guards}(S_1'',a_0)=\{a_0\geql \gu\}$;
by Lemma \ref{le66} for $S_1''$ and $a_0$, there exists a finite linear {\it DPLL}-tree $\mi{Tree}_1''$ with the root $S_1''$ constructed using Rules (\cref{ceq4hr1111111})--(\cref{ceq4hr4}), (\cref{ceq4hr8}) satisfying 
for its only leaf $S_1'''$ that either $\square\in S_1'''$, or $S_1'''\subseteq_{\mc F} \mi{OrdPropCl}$ is semi-positively guarded, $\mi{atoms}(S_1''')\subseteq \mi{atoms}(S_1'')-\{a_0\}$.
We get two cases for $S_1'''$.

Case 2.2.2.1.2.2.2.1.1:
$\square\in S_1'''$.
We put 
\begin{equation} \notag
\mi{Tree}_1=\begin{array}[c]{c}
            S_1 \\[0.4mm]
            \hline \\[-3.8mm]
            \mi{Tree}_1' \\[0.4mm]
            \hline \\[-3.8mm]
            \mi{Tree}_1''.
            \end{array}
\end{equation}
Hence, $\mi{Tree}_1$ is a finite {\it DPLL}-tree with the root $S_1$ constructed using Rules (\cref{ceq4hr1x}), (\cref{ceq4hr1111})--(\cref{ceq4hr66}), (\cref{ceq4hr8}) such that
for its only leaf $S_1'''$, $\square\in S_1'''$.

Case 2.2.2.1.2.2.2.1.2:
$S_1'''\subseteq_{\mc F} \mi{OrdPropCl}$ is semi-positively guarded,                                                                                                                       \linebreak[4]
                                                                     $\mi{atoms}(S_1''')\subseteq \mi{atoms}(S_1'')-\{a_0\}$.                                                              
Then $a_0\in \mi{atoms}(S_1'')$, $\mi{atoms}(S_1''')\subseteq \mi{atoms}(S_1'')-                                                                                                           \linebreak[4]
                                                                                \{a_0\}\subset \mi{atoms}(S_1'')\subseteq \mi{atoms}(S)$,                                                  
$\mi{measure}(S_1''')=(\mi{atoms}(S_1'''),\mi{unsaturated}(S_1'''),                                                                                                                        \linebreak[4]
                                                                   \mi{invalid}(S_1'''),\mi{count}(S_1'''))\prec \mi{measure}(S)=(\mi{atoms}(S),\mi{unsaturated}(S),\mi{invalid}(S),       \linebreak[4]
                                                                                                                                                                                    \mi{count}(S))$;
by the induction hypothesis for $S_1'''$, there exists a finite {\it DPLL}-tree $\mi{Tree}_1'''$ with the root $S_1'''$ constructed using Rules (\cref{ceq4hr1x}), (\cref{ceq4hr1111})--(\cref{ceq4hr66}), (\cref{ceq4hr8}) satisfying
for every leaf $S_1''''$ that either $\square\in S_1''''$ or $S_1''''\subseteq_{\mc F} \mi{OrdPropCl}$ is positive.
We put
\begin{equation} \notag
\mi{Tree}_1=\begin{array}[c]{c}
            S_1 \\[0.4mm]
            \hline \\[-3.8mm]
            \mi{Tree}_1' \\[0.4mm]
            \hline \\[-3.8mm]
            \mi{Tree}_1'' \\[0.4mm]
            \hline \\[-3.8mm]
            \mi{Tree}_1'''.
            \end{array}
\end{equation}
Hence, $\mi{Tree}_1$ is a finite {\it DPLL}-tree with the root $S_1$ constructed using Rules (\cref{ceq4hr1x}), (\cref{ceq4hr1111})--(\cref{ceq4hr66}), (\cref{ceq4hr8}) such that
for every leaf $S_1''''$, either $\square\in S_1''''$ or $S_1''''\subseteq_{\mc F} \mi{OrdPropCl}$ is positive.

Case 2.2.2.1.2.2.2.2:
$S_2$.
We get two cases for $C^*-\{l^*\}$.

Case 2.2.2.1.2.2.2.2.1:
There exists $a^*\in \mi{atoms}(S_2)=\mi{atoms}(S)$ such that $C^*-\{l^*\}\in \mi{guards}(a^*)$.  
Then $C^*\in S$, $C^*\not\in \mi{guards}(S)\supseteq \mi{guards}(S,a^*)=S\cap \mi{guards}(a^*)$, $C^*\not\in \mi{guards}(a^*)$;
we have that $\Cn^*\gle \gu$ is not a guard;
$\Cn^*\gle \gu\not\in \mi{guards}(a^*)$,
$\mi{guards}(S_2,a^*)=((S-\{C^*\})\cup \{C^*-\{l^*\}\}\cup \{\Cn^*\gle \gu\})\cap \mi{guards}(a^*)=
                      ((S-\{C^*\})\cap \mi{guards}(a^*))\cup (\{C^*-\{l^*\}\}\cap \mi{guards}(a^*))\cup (\{\Cn^*\gle \gu\}\cap \mi{guards}(a^*))=((S\cap \mi{guards}(a^*))-\{C^*\})\cup \{C^*-\{l^*\}\}=
                      (S\cap \mi{guards}(a^*))\cup \{C^*-\{l^*\}\}=\mi{guards}(S,a^*)\cup \{C^*-\{l^*\}\}$,
either $\mi{guards}(S,a^*)=\{\gz\gle a^*\}$ or $\mi{guards}(S,a^*)=\{\gz\gle a^*,a^*\gle \gu\}$,
$a^*\geql \gu, \gu\gleq a^*\in \mi{guards}(a^*)$, $a^*\geql \gu, \gu\gleq a^*\not\in \mi{guards}(S,a^*)=S\cap \mi{guards}(a^*)$, $a^*\geql \gu, \gu\gleq a^*\not\in S$,
$C^*-\{l^*\}\in \mi{OrdPropCl}^\gu$.
We get three cases for $C^*-\{l^*\}$.

Case 2.2.2.1.2.2.2.2.1.1:
$C^*-\{l^*\}=a^*\gle \gu$.
Then $S_2=(S-\{C^*\})\cup \{C^*-\{l^*\}\}\cup \{\Cn^*\gle \gu\}=(S-\{C^*\})\cup \{a^*\gle \gu\}\cup \{\Cn^*\gle \gu\}$,
either $\mi{guards}(S,a^*)=\{\gz\gle a^*\}$ or $\mi{guards}(S,a^*)=\{\gz\gle a^*,a^*\gle \gu\}$,
$\mi{guards}(S_2,a^*)=\mi{guards}(S,a^*)\cup \{C^*-\{l^*\}\}=\mi{guards}(S,a^*)\cup \{a^*\gle \gu\}=\{\gz\gle a^*,a^*\gle \gu\}$;
$a^*$ is positively guarded in $S_2$;
for all $a\in \mi{atoms}(S_2)-\{a^*\}=\mi{atoms}(S)-\{a^*\}$,
$a$ is positively guarded in $S$;
$a\neq a^*$, 
$\mi{guards}(a)\cap \{a^*\gle \gu\}=\mi{guards}(a)\cap \mi{guards}(a^*)=\emptyset$,
$C^*\in S$, $C^*\not\in \mi{guards}(S)\supseteq \mi{guards}(S,a)=S\cap \mi{guards}(a)$, $C^*\not\in \mi{guards}(a)$;
we have that $\Cn^*\gle \gu$ is not a guard;
$\Cn^*\gle \gu\not\in \mi{guards}(a)$,
$\mi{guards}(S_2,a)=((S-\{C^*\})\cup \{a^*\gle \gu\}\cup \{\Cn^*\gle \gu\})\cap \mi{guards}(a)=
                    ((S-\{C^*\})\cap \mi{guards}(a))\cup (\{a^*\gle \gu\}\cap \mi{guards}(a))\cup (\{\Cn^*\gle \gu\}\cap \mi{guards}(a))=(S\cap \mi{guards}(a))-\{C^*\}=S\cap \mi{guards}(a)=
                    \mi{guards}(S,a)$;
$a$ is positively guarded in $S_2$;
$S_2$ is positively guarded.
We get two cases for $\mi{guards}(S,a^*)$.

Case 2.2.2.1.2.2.2.2.1.1.1:
$\mi{guards}(S,a^*)=\{\gz\gle a^*\}$.
Then $\mi{guards}(S_2,a^*)=\{\gz\gle a^*,a^*\gle \gu\}\neq \{\gz\gle a^*\}$,
$a^*\in \mi{atoms}(S)$, $a^*\in \mi{unsaturated}(S)=\{a \,|\, a\in \mi{atoms}(S), \mi{guards}(S,a)=\{\gz\gle a\}\}$,
$a^*\not\in \mi{unsaturated}(S_2)=\{a \,|\, a\in \mi{atoms}(S_2), \mi{guards}(S_2,a)=\{\gz\gle a\}\}=\{a \,|\, a\in \mi{atoms}(S_2), \mi{guards}(S_2,a)=\{\gz\gle a\}\}-\{a^*\}=
                                  \{a \,|\, a\in \mi{atoms}(S_2)-\{a^*\}, \mi{guards}(S_2,a)=\{\gz\gle a\}\}=\{a \,|\, a\in \mi{atoms}(S)-\{a^*\}, \mi{guards}(S,a)=\{\gz\gle a\}\}=
                                  \mi{unsaturated}(S)-\{a^*\}=\{a \,|\, a\in \mi{atoms}(S), \mi{guards}(S,a)=\{\gz\gle a\}\}-\{a^*\}\subset \mi{unsaturated}(S)$, 
$\mi{measure}(S_2)=(\mi{atoms}(S_2),\mi{unsaturated}(S_2),\mi{invalid}(S_2),\mi{count}(S_2))=(\mi{atoms}(S),                                                                               \linebreak[4]
                                                                                                            \mi{unsaturated}(S_2),\mi{invalid}(S_2),\mi{count}(S_2))\prec 
                   \mi{measure}(S)=(\mi{atoms}(S),                                                                                                                                         \linebreak[4]
                                                  \mi{unsaturated}(S),\mi{invalid}(S),\mi{count}(S))$;
by the induction hypothesis for $S_2$, there exists a finite {\it DPLL}-tree $\mi{Tree}_2$ with the root $S_2$ constructed using Rules (\cref{ceq4hr1x}), (\cref{ceq4hr1111})--(\cref{ceq4hr66}), (\cref{ceq4hr8}) satisfying
for every leaf $S_2'$ that either $\square\in S_2'$ or $S_2'\subseteq_{\mc F} \mi{OrdPropCl}$ is positive.

Case 2.2.2.1.2.2.2.2.1.1.2:
$\mi{guards}(S,a^*)=\{\gz\gle a^*,a^*\gle \gu\}$.
Then $\mi{guards}(S_2,a^*)=\{\gz\gle a^*,a^*\gle \gu\}\neq \{\gz\gle a^*\}$, 
$\mi{guards}(S,a^*)=\{\gz\gle a^*,a^*\gle \gu\}\neq \{\gz\gle a^*\}$,
$a^*\not\in \mi{unsaturated}(S)=\{a \,|\, a\in \mi{atoms}(S), \mi{guards}(S,a)=\{\gz\gle a\}\}$,
$a^*\not\in \mi{unsaturated}(S_2)=\{a \,|\, a\in \mi{atoms}(S_2), \mi{guards}(S_2,a)=\{\gz\gle a\}\}=\{a \,|\, a\in \mi{atoms}(S_2), \mi{guards}(S_2,a)=\{\gz\gle a\}\}-\{a^*\}=
                                  \{a \,|\, a\in \mi{atoms}(S_2)-\{a^*\}, \mi{guards}(S_2,a)=\{\gz\gle a\}\}=\{a \,|\, a\in \mi{atoms}(S)-\{a^*\}, \mi{guards}(S,a)=\{\gz\gle a\}\}=
                                  \mi{unsaturated}(S)-\{a^*\}=\{a \,|\, a\in \mi{atoms}(S), \mi{guards}(S,a)=\{\gz\gle a\}\}-\{a^*\}=\mi{unsaturated}(S)$;
$\mi{guards}(S_2,a^*)=\mi{guards}(S,a^*)=\{\gz\gle a^*,a^*\gle \gu\}$;
for all $a\in \mi{atoms}(S_2)-\{a^*\}=\mi{atoms}(S)-\{a^*\}$, $\mi{guards}(S_2,a)=\mi{guards}(S,a)$;
$\mi{guards}(S_2)=\mi{guards}(S)$;
$C^*-\{l^*\}\in S_2$;
$C^*-\{l^*\}=a^*\gle \gu$ is a guard;
$C^*-\{l^*\}\in \mi{guards}(S_2)$,
$\Cn^*\gle \gu\not\in \mi{guards}(S_2)$,
$S_2-\mi{guards}(S_2)=((S-\{C^*\})\cup \{C^*-\{l^*\}\}\cup \{\Cn^*\gle \gu\})-\mi{guards}(S_2)=
                      ((S-\{C^*\})-\mi{guards}(S_2))\cup (\{C^*-\{l^*\}\}-\mi{guards}(S_2))\cup (\{\Cn^*\gle \gu\}-\mi{guards}(S_2))=
                      ((S-\{C^*\})-\mi{guards}(S))\cup \{\Cn^*\gle \gu\}\supseteq (S-\{C^*\})-\mi{guards}(S)$; 
$\Cn^*\gle \gu\neq \square$;
$\Cn^*\gle \gu=a_0\swedge\cdots\swedge a_n\gle \gu\in \mi{OrdPropLit}^\gu$;
for all $i\leq n$, $a_i\in \mi{atoms}(S_2)=\mi{atoms}(S)$, $\mi{guards}(S_2,a_i)=\mi{guards}(S,a_i)=\{\gz\gle a_i\}$;
$\mi{valid}(a_0\swedge\cdots\swedge a_n\gle \gu,S_2)$;
$\mi{valid}(\Cn^*\gle \gu,S_2)$;
for all $C\in (S-\{C^*\})-\mi{guards}(S)\subseteq S_2-\mi{guards}(S_2)$,
for all $l\in C$, 
$\mi{valid}(l,S)$ if and only if
$l\in \mi{OrdPropLit}^\gu$;
if either $l=\Cn\diamond \gu$ or $l=\gu\gleq \Cn$, $\Cn\in \mi{PropConj}$, and $\diamond\in \{\geql,\gle\}$, 
$\Cn=b_0\swedge\cdots\swedge b_k$, $b_j\in \mi{PropAtom}$;
for all $j\leq k$, 
$b_j\in \mi{atoms}(\Cn)=\mi{atoms}(l)\subseteq \mi{atoms}(C)\subseteq \mi{atoms}(S)=\mi{atoms}(S_2)$,
$\mi{guards}(S,b_j)=\mi{guards}(S_2,b_j)=\{\gz\gle b_j\}$ if and only if
$\mi{valid}(l,S_2)$;
$\mi{valid}(C,S)$ if and only if
$C\neq \square$; 
for all $l\in C$, $\mi{valid}(l,S)$;
if $C=\Cn\gle \gu\vee C^\natural$, $\Cn\in \mi{PropConj}$, and $\square\neq C^\natural\in \mi{OrdPropCl}$,
$C^\natural$ does not contain an order literal of the form $\Cn'\gle \gu$, $\Cn'\in \mi{PropConj}$, if and only if
$C\neq \square$; 
for all $l\in C$, $\mi{valid}(l,S_2)$;
if $C=\Cn\gle \gu\vee C^\natural$, $\Cn\in \mi{PropConj}$, and $\square\neq C^\natural\in \mi{OrdPropCl}$,
$C^\natural$ does not contain an order literal of the form $\Cn'\gle \gu$, $\Cn'\in \mi{PropConj}$, if and only if
$\mi{valid}(C,S_2)$;
$\mi{invalid}(S_2)=\{C \,|\, C\in S_2-\mi{guards}(S_2),\ \text{\it not}\ \mi{valid}(C,S_2)\}=\{C \,|\, C\in ((S-\{C^*\})-\mi{guards}(S))\cup \{\Cn^*\gle \gu\},\ \text{\it not}\ \mi{valid}(C,S_2)\}=
                   \{C \,|\, C\in (S-\{C^*\})-\mi{guards}(S),\ \text{\it not}\ \mi{valid}(C,S_2)\}=\{C \,|\, C\in (S-\{C^*\})-\mi{guards}(S),\ \text{\it not}\ \mi{valid}(C,S)\}=
                   \mi{invalid}(S)=\{C \,|\, C\in S-\mi{guards}(S),\ \text{\it not}\ \mi{valid}(C,S)\}=\emptyset$;
$\mi{count}(C^*)\geq 1$, 
$C^*\in S-\mi{guards}(S)$, $((S-\{C^*\})-\mi{guards}(S))\cup \{C^*\}=((S-\{C^*\})-\mi{guards}(S))\cup (\{C^*\}-\mi{guards}(S))=((S-\{C^*\})\cup \{C^*\})-\mi{guards}(S)=S-\mi{guards}(S)$,
$\mi{count}(S_2)=\sum_{C\in S_2-\mi{guards}(S_2)} \mi{count}(C)=\sum_{C\in ((S-\{C^*\})-\mi{guards}(S))\cup \{\Cn^*\gle \gu\}} \mi{count}(C)=
                 (\sum_{C\in (S-\{C^*\})-\mi{guards}(S)} \mi{count}(C))+\mi{count}(\Cn^*\gle \gu)<(\sum_{C\in (S-\{C^*\})-\mi{guards}(S)} \mi{count}(C))+\mi{count}(C^*)=
                 \sum_{C\in ((S-\{C^*\})-\mi{guards}(S))\cup \{C^*\}} \mi{count}(C)=\mi{count}(S)=\sum_{C\in S-\mi{guards}(S)} \mi{count}(C)$;
$\mi{measure}(S_2)=(\mi{atoms}(S_2),\mi{unsaturated}(S_2),\mi{invalid}(S_2),\mi{count}(S_2))=(\mi{atoms}(S),                                                                               \linebreak[4] 
                                                                                                            \mi{unsaturated}(S),\mi{invalid}(S),\mi{count}(S_2))\prec 
                   \mi{measure}(S)=(\mi{atoms}(S),                                                                                                                                         \linebreak[4]
                                                  \mi{unsaturated}(S),\mi{invalid}(S),\mi{count}(S))$;
by the induction hypothesis for $S_2$, there exists a finite {\it DPLL}-tree $\mi{Tree}_2$ with the root $S_2$ constructed using Rules (\cref{ceq4hr1x}), (\cref{ceq4hr1111})--(\cref{ceq4hr66}), (\cref{ceq4hr8}) satisfying
for every leaf $S_2'$ that either $\square\in S_2'$ or $S_2'\subseteq_{\mc F} \mi{OrdPropCl}$ is positive.

Case 2.2.2.1.2.2.2.2.1.2:
$C^*-\{l^*\}=a^*\geql \gu$.
Then $S_2=(S-\{C^*\})\cup \{C^*-\{l^*\}\}\cup \{\Cn^*\gle \gu\}=(S-\{C^*\})\cup \{a^*\geql \gu\}\cup \{\Cn^*\gle \gu\}$,
$\mi{guards}(S_2,a^*)=\mi{guards}(S,a^*)\cup \{C^*-\{l^*\}\}=\mi{guards}(S,a^*)\cup \{a^*\geql \gu\}$,
either $\mi{guards}(S,a^*)=\{\gz\gle a^*\}$ or $\mi{guards}(S,a^*)=\{\gz\gle a^*,a^*\gle \gu\}$.
We get two cases for $\mi{guards}(S,a^*)$.

Case 2.2.2.1.2.2.2.2.1.2.1:
$\mi{guards}(S,a^*)=\{\gz\gle a^*\}$.
Then $S\supseteq \mi{guards}(S,a^*)=\{\gz\gle a^*\}$;
$S_2\supseteq \mi{guards}(S_2)\supseteq \mi{guards}(S_2,a^*)=\mi{guards}(S,a^*)\cup \{a^*\geql \gu\}=\{\gz\gle a^*,a^*\geql \gu\}$,
$a^*\geql \gu\in \mi{guards}(S_2)$, $\gz\gle a^*\in S_2$,
$a^*\in \mi{atoms}(\gz\gle a^*)$, $a^*\geql \gu\neq \gz\gle a^*$, $\mi{simplify}(\gz\gle a^*,a^*,\gu)=\gz\gle \gu$,
applying Rule (\cref{ceq4hr4}) to $S_2$, $a^*\geql \gu$, and $\gz\gle a^*$, we derive
\begin{equation} \notag
\dfrac{S_2}
      {(S_2-\{\gz\gle a^*\})\cup \{\gz\gle \gu\}};
\end{equation}
$\gz\gle \gu\in (S_2-\{\gz\gle a^*\})\cup \{\gz\gle \gu\}$;
$\gz\gle \gu\in \mi{OrdPropLit}$ is a tautology;
$\gz\gle \gu$ is not a guard;
$\gz\gle \gu\not\in \mi{guards}((S_2-\{\gz\gle a^*\})\cup \{\gz\gle \gu\})$,
$\gz\gle \gu\in ((S_2-\{\gz\gle a^*\})\cup \{\gz\gle \gu\})-\mi{guards}((S_2-\{\gz\gle a^*\})\cup \{\gz\gle \gu\})$,
applying Rule (\cref{ceq4hr22}) to $(S_2-\{\gz\gle a^*\})\cup \{\gz\gle \gu\}$ and $\gz\gle \gu$, we derive
\begin{equation} \notag
\dfrac{(S_2-\{\gz\gle a^*\})\cup \{\gz\gle \gu\}}
      {((S_2-\{\gz\gle a^*\})\cup \{\gz\gle \gu\})-\{\gz\gle \gu\}}.
\end{equation}
We have that $S_2$ is simplified.
Hence, $\gz\gle \gu\not\in S_2$, $C^*, \gz\gle a^*\in S$, $C^*\not\in \mi{guards}(a^*)$, $\gz\gle a^*\in \mi{guards}(a^*)$, $C^*\neq \gz\gle a^*$, $a^*\geql \gu\not\in S$, 
$a^*\geql \gu\neq \Cn^*\gle \gu$,
$((S_2-\{\gz\gle a^*\})\cup \{\gz\gle \gu\})-\{\gz\gle \gu\}=((S_2-\{\gz\gle a^*\})-\{\gz\gle \gu\})\cup (\{\gz\gle \gu\}-\{\gz\gle \gu\})=S_2-\{\gz\gle a^*\}=
 ((S-\{C^*\})\cup \{a^*\geql \gu\}\cup \{\Cn^*\gle \gu\})-\{\gz\gle a^*\}=((S-\{C^*\})-\{\gz\gle a^*\})\cup (\{a^*\geql \gu\}-\{\gz\gle a^*\})\cup (\{\Cn^*\gle \gu\}-\{\gz\gle a^*\})=
 ((S-\{C^*\})-\{\gz\gle a^*\})\cup \{a^*\geql \gu\}\cup \{\Cn^*\gle \gu\}$.
We put $S_2'=((S-\{C^*\})-\{\gz\gle a^*\})\cup \{a^*\geql \gu\}\cup \{\Cn^*\gle \gu\}\subseteq_{\mc F} \mi{OrdPropCl}$.
We get that
$\mi{atoms}(\gz\gle a^*)=\{a^*\}$,
$a^*\in \mi{atoms}(S_2')=\mi{atoms}(((S-\{C^*\})-\{\gz\gle a^*\})\cup \{a^*\geql \gu\}\cup \{\Cn^*\gle \gu\})=
                         \mi{atoms}((S-\{C^*\})-\{\gz\gle a^*\})\cup \mi{atoms}(a^*\geql \gu)\cup \mi{atoms}(\Cn^*\gle \gu)=
                         \mi{atoms}(S-\{C^*\})\cup \mi{atoms}(a^*\geql \gu)\cup \mi{atoms}(\Cn^*\gle \gu)=
                         \mi{atoms}(S-\{C^*\})\cup \mi{atoms}(C^*-\{l^*\})\cup \mi{atoms}(\Cn^*\gle \gu)=\mi{atoms}(S)$;
$S$ is simplified;
$(S-\{C^*\})-\{\gz\gle a^*\}\subseteq S$ is simplified;
$a^*\geql \gu, \Cn^*\gle \gu\neq \square$ do not contain contradictions and tautologies;
$S_2'=((S-\{C^*\})-\{\gz\gle a^*\})\cup \{a^*\geql \gu\}\cup \{\Cn^*\gle \gu\}$ is simplified;
$C^*, \Cn^*\gle \gu\not\in \mi{guards}(a^*)$,
$\mi{guards}(S_2',a^*)=(((S-\{C^*\})-\{\gz\gle a^*\})\cup \{a^*\geql \gu\}\cup \{\Cn^*\gle \gu\})\cap \mi{guards}(a^*)=
                       (((S-\{C^*\})-\{\gz\gle a^*\})\cap \mi{guards}(a^*))\cup (\{a^*\geql \gu\}\cap \mi{guards}(a^*))\cup (\{\Cn^*\gle \gu\}\cap \mi{guards}(a^*))=
                       (((S\cap \mi{guards}(a^*))-\{\gz\gle a^*\})-\{C^*\})\cup \{a^*\geql \gu\}=((S\cap \mi{guards}(a^*))-\{\gz\gle a^*\})\cup \{a^*\geql \gu\}=
                       (\mi{guards}(S,a^*)-\{\gz\gle a^*\})\cup \{a^*\geql \gu\}=(\{\gz\gle a^*\}-\{\gz\gle a^*\})\cup \{a^*\geql \gu\}=\{a^*\geql \gu\}$;
$a^*$ is semi-positively guarded in $S_2'$;
for all $a\in \mi{atoms}(S_2')-\{a^*\}=\mi{atoms}(S)-\{a^*\}$,
$a$ is positively guarded in $S$;
$a\neq a^*$, 
$\mi{guards}(a)\cap \{\gz\gle a^*,a^*\geql \gu\}=\mi{guards}(a)\cap \mi{guards}(a^*)=\emptyset$;
we have that $\Cn^*\gle \gu$ is not a guard;
$\Cn^*\gle \gu\not\in \mi{guards}(a)$,
$C^*\in S$, $C^*\not\in \mi{guards}(S)\supseteq \mi{guards}(S,a)=S\cap \mi{guards}(a)$, $C^*\not\in \mi{guards}(a)$,
$\mi{guards}(S_2',a)=(((S-\{C^*\})-\{\gz\gle a^*\})\cup \{a^*\geql \gu\}\cup \{\Cn^*\gle \gu\})\cap \mi{guards}(a)=
                     (((S-\{C^*\})-\{\gz\gle a^*\})\cap \mi{guards}(a))\cup (\{a^*\geql \gu\}\cap \mi{guards}(a))\cup (\{\Cn^*\gle \gu\}\cap \mi{guards}(a))=
                     (((S\cap \mi{guards}(a))-\{C^*\})-\{\gz\gle a^*\})=S\cap \mi{guards}(a)=\mi{guards}(S,a)$;
$a$ is positively guarded in $S_2'$;
$S_2'$ is semi-positively guarded;
by Lemma \ref{le66} for $S_2'$ and $a^*$, there exists a finite linear {\it DPLL}-tree $\mi{Tree}_2'$ with the root $S_2'$ constructed using Rules (\cref{ceq4hr1111111})--(\cref{ceq4hr4}), (\cref{ceq4hr8}) satisfying 
for its only leaf $S_2''$ that either $\square\in S_2''$, or $S_2''\subseteq_{\mc F} \mi{OrdPropCl}$ is semi-positively guarded, $\mi{atoms}(S_2'')\subseteq \mi{atoms}(S_2')-\{a^*\}$.
We get two cases for $S_2''$.

Case 2.2.2.1.2.2.2.2.1.2.1.1:
$\square\in S_2''$.
We put 
\begin{equation} \notag
\mi{Tree}_2=\begin{array}[c]{c}
            S_2 \\[0.4mm]
            \hline \\[-3.8mm]
            (S_2-\{\gz\gle a^*\})\cup \{\gz\gle \gu\} \\[0.4mm]
            \hline \\[-3.8mm]
            \mi{Tree}_2'.
            \end{array}
\end{equation}
Hence, $\mi{Tree}_2$ is a finite {\it DPLL}-tree with the root $S_2$ constructed using Rules (\cref{ceq4hr1x}), (\cref{ceq4hr1111})--(\cref{ceq4hr66}), (\cref{ceq4hr8}) such that
for its only leaf $S_2''$, $\square\in S_2''$.

Case 2.2.2.1.2.2.2.2.1.2.1.2:
$S_2''\subseteq_{\mc F} \mi{OrdPropCl}$ is semi-positively guarded, $\mi{atoms}(S_2'')\subseteq \mi{atoms}(S_2')-\{a^*\}$.
Then $a^*\in \mi{atoms}(S_2')$, $\mi{atoms}(S_2'')\subseteq \mi{atoms}(S_2')-\{a^*\}\subset \mi{atoms}(S_2')\subseteq \mi{atoms}(S)$, 
$\mi{measure}(S_2'')=(\mi{atoms}(S_2''),\mi{unsaturated}(S_2''),                                                                                                                           \linebreak[4]
                                                                \mi{invalid}(S_2''),\mi{count}(S_2''))\prec \mi{measure}(S)=(\mi{atoms}(S),\mi{unsaturated}(S),\mi{invalid}(S),            \linebreak[4]
                                                                                                                                                                               \mi{count}(S))$;
by the induction hypothesis for $S_2''$, there exists a finite {\it DPLL}-tree $\mi{Tree}_2''$ with the root $S_2''$ constructed using Rules (\cref{ceq4hr1x}), (\cref{ceq4hr1111})--(\cref{ceq4hr66}), (\cref{ceq4hr8}) satisfying
for every leaf $S_2'''$ that either $\square\in S_2'''$ or $S_2'''\subseteq_{\mc F} \mi{OrdPropCl}$ is positive.
We put
\begin{equation} \notag
\mi{Tree}_2=\begin{array}[c]{c}
            S_2 \\[0.4mm]
            \hline \\[-3.8mm]
            (S_2-\{\gz\gle a^*\})\cup \{\gz\gle \gu\} \\[0.4mm]
            \hline \\[-3.8mm]
            \mi{Tree}_2' \\[0.4mm]
            \hline \\[-3.8mm]
            \mi{Tree}_2''.
            \end{array}
\end{equation}
Hence, $\mi{Tree}_2$ is a finite {\it DPLL}-tree with the root $S_2$ constructed using Rules (\cref{ceq4hr1x}), (\cref{ceq4hr1111})--(\cref{ceq4hr66}), (\cref{ceq4hr8}) such that
for every leaf $S_2'''$, either $\square\in S_2'''$ or $S_2'''\subseteq_{\mc F} \mi{OrdPropCl}$ is positive.

Case 2.2.2.1.2.2.2.2.1.2.2:
$\mi{guards}(S,a^*)=\{\gz\gle a^*,a^*\gle \gu\}$.
Then $S_2\supseteq \mi{guards}(S_2)\supseteq \mi{guards}(S_2,a^*)=\mi{guards}(S,a^*)\cup \{a^*\geql \gu\}=\{\gz\gle a^*,a^*\gle \gu,a^*\geql \gu\}$,
$a^*\geql \gu\in \mi{guards}(S_2)$, $a^*\gle \gu\in S_2$,
$a^*\in \mi{atoms}(a^*\gle \gu)$, $a^*\geql \gu\neq a^*\gle \gu$, $\mi{simplify}(a^*\gle \gu,a^*,\gu)=\gu\gle \gu$,
applying Rule (\cref{ceq4hr4}) to $S_2$, $a^*\geql \gu$, and $a^*\gle \gu$, we derive
\begin{equation} \notag
\dfrac{S_2}
      {(S_2-\{a^*\gle \gu\})\cup \{\gu\gle \gu\}};
\end{equation}
$\gu\gle \gu\in (S_2-\{a^*\gle \gu\})\cup \{\gu\gle \gu\}$;
$\gu\gle \gu\in \mi{OrdPropLit}$ is a contradiction;
$\gu\gle \gu$ is not a guard;
$\gu\gle \gu\not\in \mi{guards}((S_2-\{a^*\gle \gu\})\cup \{\gu\gle \gu\})$,
$\gu\gle \gu\in ((S_2-\{a^*\gle \gu\})\cup \{\gu\gle \gu\})-\mi{guards}((S_2-\{a^*\gle \gu\})\cup \{\gu\gle \gu\})$,
applying Rule (\cref{ceq4hr2}) to $(S_2-\{a^*\gle \gu\})\cup \{\gu\gle \gu\}$ and $\gu\gle \gu$, we derive
\begin{equation} \notag
\dfrac{(S_2-\{a^*\gle \gu\})\cup \{\gu\gle \gu\}}
      {(((S_2-\{a^*\gle \gu\})\cup \{\gu\gle \gu\})-\{\gu\gle \gu\})\cup \{\square\}}.
\end{equation}
We put
\begin{equation} \notag
\mi{Tree}_2=\begin{array}[c]{c}
            S_2 \\[0.4mm]
            \hline \\[-3.8mm]
            (S_2-\{a^*\gle \gu\})\cup \{\gu\gle \gu\} \\[0.4mm]
            \hline \\[-3.8mm]
            S_2'=(((S_2-\{a^*\gle \gu\})\cup \{\gu\gle \gu\})-\{\gu\gle \gu\})\cup \{\square\}.
            \end{array} 
\end{equation}
Hence, $\mi{Tree}_2$ is a finite {\it DPLL}-tree with the root $S_2$ constructed using Rules (\cref{ceq4hr1x}), (\cref{ceq4hr1111})--(\cref{ceq4hr66}), (\cref{ceq4hr8}) such that
for its only leaf $S_2'$, $\square\in S_2'$.

Case 2.2.2.1.2.2.2.2.1.3:
$C^*-\{l^*\}=\gu\gleq a^*$.
Then $S_2=(S-\{C^*\})\cup \{C^*-\{l^*\}\}\cup \{\Cn^*\gle \gu\}=(S-\{C^*\})\cup \{\gu\gleq a^*\}\cup \{\Cn^*\gle \gu\}$,
$\mi{guards}(S_2,a^*)=\mi{guards}(S,a^*)\cup \{C^*-\{l^*\}\}=\mi{guards}(S,a^*)\cup \{\gu\gleq a^*\}$,
either $\mi{guards}(S,a^*)=\{\gz\gle a^*\}$ or $\mi{guards}(S,a^*)=\{\gz\gle a^*,a^*\gle \gu\}$.
We get two cases for $\mi{guards}(S,a^*)$.

Case 2.2.2.1.2.2.2.2.1.3.1:
$\mi{guards}(S,a^*)=\{\gz\gle a^*\}$.
Then $S\supseteq \mi{guards}(S,a^*)=\{\gz\gle a^*\}$;
$\mi{guards}(S_2)\supseteq \mi{guards}(S_2,a^*)=\mi{guards}(S,a^*)\cup \{\gu\gleq a^*\}=\{\gz\gle a^*,\gu\gleq a^*\}$,
$\gu\gleq a^*\in \mi{guards}(S_2)$, applying Rule (\cref{ceq4hr11111111}) to $S_2$ and $\gu\gleq a^*$, we derive
\begin{equation} \notag
\dfrac{S_2}
      {(S_2-\{\gu\gleq a^*\})\cup \{a^*\geql \gu\}};
\end{equation}
$a^*\in \mi{atoms}((S_2-\{\gu\gleq a^*\})\cup \{a^*\geql \gu\})$,
$(S_2-\{\gu\gleq a^*\})\cup \{a^*\geql \gu\}\supseteq \mi{guards}((S_2-\{\gu\gleq a^*\})\cup \{a^*\geql \gu\})\supseteq 
 \mi{guards}((S_2-\{\gu\gleq a^*\})\cup \{a^*\geql \gu\},a^*)=((S_2-\{\gu\gleq a^*\})\cup \{a^*\geql \gu\})\cap \mi{guards}(a^*)=
 ((S_2-\{\gu\gleq a^*\})\cap \mi{guards}(a^*))\cup (\{a^*\geql \gu\}\cap \mi{guards}(a^*))=((S_2\cap \mi{guards}(a^*))-\{\gu\gleq a^*\})\cup \{a^*\geql \gu\}=
 (\mi{guards}(S_2,a^*)-\{\gu\gleq a^*\})\cup \{a^*\geql \gu\}=(\{\gz\gle a^*,\gu\gleq a^*\}-\{\gu\gleq a^*\})\cup \{a^*\geql \gu\}=\{\gz\gle a^*,a^*\geql \gu\}$,
$a^*\geql \gu\in \mi{guards}((S_2-\{\gu\gleq a^*\})\cup \{a^*\geql \gu\})$, $\gz\gle a^*\in (S_2-\{\gu\gleq a^*\})\cup \{a^*\geql \gu\}$,
$a^*\in \mi{atoms}(\gz\gle a^*)$, $a^*\geql \gu\neq \gz\gle a^*$, $\mi{simplify}(\gz\gle a^*,a^*,\gu)=\gz\gle \gu$,
applying Rule (\cref{ceq4hr4}) to $(S_2-\{\gu\gleq a^*\})\cup \{a^*\geql \gu\}$, $a^*\geql \gu$, and $\gz\gle a^*$, we derive
\begin{equation} \notag
\dfrac{(S_2-\{\gu\gleq a^*\})\cup \{a^*\geql \gu\}}
      {(((S_2-\{\gu\gleq a^*\})\cup \{a^*\geql \gu\})-\{\gz\gle a^*\})\cup \{\gz\gle \gu\}};
\end{equation}
$\gz\gle \gu\in (((S_2-\{\gu\gleq a^*\})\cup \{a^*\geql \gu\})-\{\gz\gle a^*\})\cup \{\gz\gle \gu\}$;
$\gz\gle \gu\in \mi{OrdPropLit}$ is a tautology;
$\gz\gle \gu$ is not a guard;
$\gz\gle \gu\not\in \mi{guards}((((S_2-\{\gu\gleq a^*\})\cup \{a^*\geql \gu\})-\{\gz\gle a^*\})\cup \{\gz\gle \gu\})$,
$\gz\gle \gu\in ((((S_2-\{\gu\gleq a^*\})\cup \{a^*\geql \gu\})-\{\gz\gle a^*\})\cup \{\gz\gle \gu\})-\mi{guards}((((S_2-\{\gu\gleq a^*\})\cup \{a^*\geql \gu\})-\{\gz\gle a^*\})\cup \{\gz\gle \gu\})$,
applying Rule (\cref{ceq4hr22}) to $(((S_2-\{\gu\gleq a^*\})\cup \{a^*\geql \gu\})-\{\gz\gle a^*\})\cup \{\gz\gle \gu\}$ and $\gz\gle \gu$, we derive
\begin{equation} \notag
\dfrac{(((S_2-\{\gu\gleq a^*\})\cup \{a^*\geql \gu\})-\{\gz\gle a^*\})\cup \{\gz\gle \gu\}}
      {((((S_2-\{\gu\gleq a^*\})\cup \{a^*\geql \gu\})-\{\gz\gle a^*\})\cup \{\gz\gle \gu\})-\{\gz\gle \gu\}}.
\end{equation}
We have that $S_2$ is simplified.
Hence, $\gz\gle \gu\not\in S_2$, $\gu\gleq a^*\not\in S$, $C^*, \gz\gle a^*\in S$, $C^*\not\in \mi{guards}(a^*)$, $\gz\gle a^*\in \mi{guards}(a^*)$, $C^*\neq \gz\gle a^*$, $a^*\geql \gu\not\in S$, 
$a^*\geql \gu\neq \Cn^*\gle \gu$,
$((((S_2-\{\gu\gleq a^*\})\cup \{a^*\geql \gu\})-\{\gz\gle a^*\})\cup \{\gz\gle \gu\})-\{\gz\gle \gu\}=
 ((((S_2-\{\gu\gleq a^*\})\cup \{a^*\geql \gu\})-\{\gz\gle a^*\})-\{\gz\gle \gu\})\cup (\{\gz\gle \gu\}-\{\gz\gle \gu\})=
 (((S_2-\{\gu\gleq a^*\})\cup \{a^*\geql \gu\})-\{\gz\gle \gu\})-\{\gz\gle a^*\}=
 (((S_2-\{\gu\gleq a^*\})-\{\gz\gle \gu\})\cup (\{a^*\geql \gu\}-\{\gz\gle \gu\}))-\{\gz\gle a^*\}=
 ((S_2-\{\gu\gleq a^*\})\cup \{a^*\geql \gu\})-\{\gz\gle a^*\}=
 ((S_2-\{\gu\gleq a^*\})-\{\gz\gle a^*\})\cup (\{a^*\geql \gu\}-\{\gz\gle a^*\})=
 ((S_2-\{\gu\gleq a^*\})-\{\gz\gle a^*\})\cup \{a^*\geql \gu\}=
 ((((S-\{C^*\})\cup \{\gu\gleq a^*\}\cup \{\Cn^*\gle \gu\})-\{\gu\gleq a^*\})-\{\gz\gle a^*\})\cup \{a^*\geql \gu\}=
 ((((S-\{C^*\})-\{\gu\gleq a^*\})\cup (\{\gu\gleq a^*\}-\{\gu\gleq a^*\})\cup (\{\Cn^*\gle \gu\}-\{\gu\gleq a^*\}))-\{\gz\gle a^*\})\cup \{a^*\geql \gu\}=
 (((S-\{C^*\})\cup \{\Cn^*\gle \gu\})-\{\gz\gle a^*\})\cup \{a^*\geql \gu\}=
 ((S-\{C^*\})-\{\gz\gle a^*\})\cup (\{\Cn^*\gle \gu\}-\{\gz\gle a^*\})\cup \{a^*\geql \gu\}=
 ((S-\{C^*\})-\{\gz\gle a^*\})\cup \{a^*\geql \gu\}\cup \{\Cn^*\gle \gu\}$.
We put $S_2'=((S-\{C^*\})-\{\gz\gle a^*\})\cup \{a^*\geql \gu\}\cup \{\Cn^*\gle \gu\}\subseteq_{\mc F} \mi{OrdPropCl}$.
We get from Case 2.2.2.1.2.2.2.2.1.2.1 that there exists a finite linear {\it DPLL}-tree $\mi{Tree}_2'$ with the root $S_2'$ constructed using Rules (\cref{ceq4hr1111111})--(\cref{ceq4hr4}), (\cref{ceq4hr8}) satisfying 
for its only leaf $S_2''$ that either $\square\in S_2''$, or $S_2''\subseteq_{\mc F} \mi{OrdPropCl}$ is semi-positively guarded, $\mi{atoms}(S_2'')\subseteq \mi{atoms}(S_2')-\{a^*\}$.
We get two cases for $S_2''$.

Case 2.2.2.1.2.2.2.2.1.3.1.1:
$\square\in S_2''$.
We put 
\begin{equation} \notag
\mi{Tree}_2=\begin{array}[c]{c}
            S_2 \\[0.4mm]
            \hline \\[-3.8mm]
            (S_2-\{\gu\gleq a^*\})\cup \{a^*\geql \gu\} \\[0.4mm]
            \hline \\[-3.8mm]
            (((S_2-\{\gu\gleq a^*\})\cup \{a^*\geql \gu\})-\{\gz\gle a^*\})\cup \{\gz\gle \gu\} \\[0.4mm]
            \hline \\[-3.8mm]
            \mi{Tree}_2'.
            \end{array}
\end{equation}
Hence, $\mi{Tree}_2$ is a finite {\it DPLL}-tree with the root $S_2$ constructed using Rules (\cref{ceq4hr1x}), (\cref{ceq4hr1111})--(\cref{ceq4hr66}), (\cref{ceq4hr8}) such that
for its only leaf $S_2''$, $\square\in S_2''$.

Case 2.2.2.1.2.2.2.2.1.3.1.2:
$S_2''\subseteq_{\mc F} \mi{OrdPropCl}$ is semi-positively guarded, $\mi{atoms}(S_2'')\subseteq \mi{atoms}(S_2')-\{a^*\}$.
Then $a^*\in \mi{atoms}(S_2')$, $\mi{atoms}(S_2'')\subseteq \mi{atoms}(S_2')-\{a^*\}\subset \mi{atoms}(S_2')\subseteq \mi{atoms}(S)$, 
$\mi{measure}(S_2'')=(\mi{atoms}(S_2''),\mi{unsaturated}(S_2''),                                                                                                                           \linebreak[4]
                                                                \mi{invalid}(S_2''),\mi{count}(S_2''))\prec \mi{measure}(S)=(\mi{atoms}(S),\mi{unsaturated}(S),\mi{invalid}(S),            \linebreak[4]
                                                                                                                                                                               \mi{count}(S))$;
by the induction hypothesis for $S_2''$, there exists a finite {\it DPLL}-tree $\mi{Tree}_2''$ with the root $S_2''$ constructed using Rules (\cref{ceq4hr1x}), (\cref{ceq4hr1111})--(\cref{ceq4hr66}), (\cref{ceq4hr8}) satisfying
for every leaf $S_2'''$ that either $\square\in S_2'''$ or $S_2'''\subseteq_{\mc F} \mi{OrdPropCl}$ is positive.
We put
\begin{equation} \notag
\mi{Tree}_2=\begin{array}[c]{c}
            S_2 \\[0.4mm]
            \hline \\[-3.8mm]
            (S_2-\{\gu\gleq a^*\})\cup \{a^*\geql \gu\} \\[0.4mm]
            \hline \\[-3.8mm]
            (((S_2-\{\gu\gleq a^*\})\cup \{a^*\geql \gu\})-\{\gz\gle a^*\})\cup \{\gz\gle \gu\} \\[0.4mm]
            \hline \\[-3.8mm]
            \mi{Tree}_2' \\[0.4mm]
            \hline \\[-3.8mm]
            \mi{Tree}_2''.
            \end{array}
\end{equation}
Hence, $\mi{Tree}_2$ is a finite {\it DPLL}-tree with the root $S_2$ constructed using Rules (\cref{ceq4hr1x}), (\cref{ceq4hr1111})--(\cref{ceq4hr66}), (\cref{ceq4hr8}) such that
for every leaf $S_2'''$, either $\square\in S_2'''$ or $S_2'''\subseteq_{\mc F} \mi{OrdPropCl}$ is positive.

Case 2.2.2.1.2.2.2.2.1.3.2:
$\mi{guards}(S,a^*)=\{\gz\gle a^*,a^*\gle \gu\}$.
Then $\mi{guards}(S_2)\supseteq \mi{guards}(S_2,a^*)=\mi{guards}(S,a^*)\cup \{\gu\gleq a^*\}=\{\gz\gle a^*,a^*\gle \gu,\gu\gleq a^*\}$,
$\gu\gleq a^*\in \mi{guards}(S_2)$, applying Rule (\cref{ceq4hr11111111}) to $S_2$ and $\gu\gleq a^*$, we derive
\begin{equation} \notag
\dfrac{S_2}
      {(S_2-\{\gu\gleq a^*\})\cup \{a^*\geql \gu\}};
\end{equation}
$a^*\in \mi{atoms}((S_2-\{\gu\gleq a^*\})\cup \{a^*\geql \gu\})$,
$(S_2-\{\gu\gleq a^*\})\cup \{a^*\geql \gu\}\supseteq \mi{guards}((S_2-\{\gu\gleq a^*\})\cup \{a^*\geql \gu\})\supseteq 
 \mi{guards}((S_2-\{\gu\gleq a^*\})\cup \{a^*\geql \gu\},a^*)=((S_2-\{\gu\gleq a^*\})\cup \{a^*\geql \gu\})\cap \mi{guards}(a^*)=
 ((S_2-\{\gu\gleq a^*\})\cap \mi{guards}(a^*))\cup (\{a^*\geql \gu\}\cap \mi{guards}(a^*))=((S_2\cap \mi{guards}(a^*))-\{\gu\gleq a^*\})\cup \{a^*\geql \gu\}=
 (\mi{guards}(S_2,a^*)-\{\gu\gleq a^*\})\cup \{a^*\geql \gu\}=(\{\gz\gle a^*,a^*\gle \gu,\gu\gleq a^*\}-\{\gu\gleq a^*\})\cup \{a^*\geql \gu\}=\{\gz\gle a^*,a^*\gle \gu,a^*\geql \gu\}$,
$a^*\geql \gu\in \mi{guards}((S_2-\{\gu\gleq a^*\})\cup \{a^*\geql \gu\})$, $a^*\gle \gu\in (S_2-\{\gu\gleq a^*\})\cup \{a^*\geql \gu\}$,
$a^*\in \mi{atoms}(a^*\gle \gu)$, $a^*\geql \gu\neq a^*\gle \gu$, $\mi{simplify}(a^*\gle \gu,a^*,\gu)=\gu\gle \gu$,
applying Rule (\cref{ceq4hr4}) to $(S_2-\{\gu\gleq a^*\})\cup \{a^*\geql \gu\}$, $a^*\geql \gu$, and $a^*\gle \gu$, we derive
\begin{equation} \notag
\dfrac{(S_2-\{\gu\gleq a^*\})\cup \{a^*\geql \gu\}}
      {(((S_2-\{\gu\gleq a^*\})\cup \{a^*\geql \gu\})-\{a^*\gle \gu\})\cup \{\gu\gle \gu\}};
\end{equation}
$\gu\gle \gu\in (((S_2-\{\gu\gleq a^*\})\cup \{a^*\geql \gu\})-\{a^*\gle \gu\})\cup \{\gu\gle \gu\}$;
$\gu\gle \gu\in \mi{OrdPropLit}$ is a contradiction;
$\gu\gle \gu$ is not a guard;
$\gu\gle \gu\not\in \mi{guards}((((S_2-\{\gu\gleq a^*\})\cup \{a^*\geql \gu\})-\{a^*\gle \gu\})\cup \{\gu\gle \gu\})$,
$\gu\gle \gu\in ((((S_2-\{\gu\gleq a^*\})\cup \{a^*\geql \gu\})-\{a^*\gle \gu\})\cup \{\gu\gle \gu\})-\mi{guards}((((S_2-\{\gu\gleq a^*\})\cup \{a^*\geql \gu\})-\{a^*\gle \gu\})\cup \{\gu\gle \gu\})$,
applying Rule (\cref{ceq4hr2}) to $(((S_2-\{\gu\gleq a^*\})\cup \{a^*\geql \gu\})-\{a^*\gle \gu\})\cup \{\gu\gle \gu\}$ and $\gu\gle \gu$, we derive
\begin{equation} \notag
\dfrac{(((S_2-\{\gu\gleq a^*\})\cup \{a^*\geql \gu\})-\{a^*\gle \gu\})\cup \{\gu\gle \gu\}}
      {(((((S_2-\{\gu\gleq a^*\})\cup \{a^*\geql \gu\})-\{a^*\gle \gu\})\cup \{\gu\gle \gu\})-\{\gu\gle \gu\})\cup \{\square\}}.
\end{equation}
We put
\begin{equation} \notag
\mi{Tree}_2=\begin{array}[c]{c}
            S_2 \\[0.4mm]
            \hline \\[-3.8mm]
            (S_2-\{\gu\gleq a^*\})\cup \{a^*\geql \gu\} \\[0.4mm]
            \hline \\[-3.8mm]
            (((S_2-\{\gu\gleq a^*\})\cup \{a^*\geql \gu\})-\{a^*\gle \gu\})\cup \{\gu\gle \gu\} \\[0.4mm]
            \hline \\[-3.8mm]
            S_2'=(((((S_2-\{\gu\gleq a^*\})\cup \{a^*\geql \gu\})-\{a^*\gle \gu\})\cup \{\gu\gle \gu\})- \\
            \hfill \{\gu\gle \gu\})\cup \{\square\}.
            \end{array} 
\end{equation}
Hence, $\mi{Tree}_2$ is a finite {\it DPLL}-tree with the root $S_2$ constructed using Rules (\cref{ceq4hr1x}), (\cref{ceq4hr1111})--(\cref{ceq4hr66}), (\cref{ceq4hr8}) such that
for its only leaf $S_2'$, $\square\in S_2'$.

Case 2.2.2.1.2.2.2.2.2:
For all $a\in \mi{atoms}(S_2)=\mi{atoms}(S)$, $C^*-\{l^*\}\not\in \mi{guards}(a)$.
Then, for all $a\in \mi{atoms}(C^*-\{l^*\})\subseteq \mi{atoms}(S_2)$, $C^*-\{l^*\}\not\in \mi{guards}(a)$;
$C^*-\{l^*\}$ is not a guard;
we have that $\Cn^*\gle \gu$ is not a guard;
$C^*\not\in \mi{guards}(S)$,
$\mi{guards}(S_2)=\{C \,|\, C\in (S-\{C^*\})\cup \{C^*-\{l^*\}\}\cup \{\Cn^*\gle \gu\}\ \text{\it is a guard}\}=\{C \,|\, C\in S-\{C^*\}\ \text{\it is a guard}\}=
                  \mi{guards}(S)-\{C^*\}=\{C \,|\, C\in S\ \text{\it is a guard}\}-\{C^*\}=\mi{guards}(S)$;
$S_2$ is positively guarded;
$\mi{unsaturated}(S_2)=\{a \,|\, a\in \mi{atoms}(S_2), \mi{guards}(S_2,a)=\{\gz\gle a\}\}=\mi{unsaturated}(S)=\{a \,|\, a\in \mi{atoms}(S), \mi{guards}(S,a)=\{\gz\gle a\}\}$;
$C^*-\{l^*\}, \Cn^*\gle \gu\not\in \mi{guards}(S_2)$,
$S_2-\mi{guards}(S_2)=((S-\{C^*\})\cup \{C^*-\{l^*\}\}\cup \{\Cn^*\gle \gu\})-\mi{guards}(S_2)=
                      ((S-\{C^*\})-\mi{guards}(S_2))\cup (\{C^*-\{l^*\}\}-\mi{guards}(S_2))\cup (\{\Cn^*\gle \gu\}-\mi{guards}(S_2))=
                      ((S-\{C^*\})-\mi{guards}(S))\cup \{C^*-\{l^*\}\}\cup \{\Cn^*\gle \gu\}\supseteq (S-\{C^*\})-\mi{guards}(S)$; 
$C^*-\{l^*\}\neq \square$;
$\mi{valid}(C^*,S)$;
for all $l\in C^*-\{l^*\}\subseteq C^*$, 
$\mi{valid}(l,S)$, $l\in \mi{OrdPropLit}^\gu$;
if either $l=\Cn\diamond \gu$ or $l=\gu\gleq \Cn$, $\Cn\in \mi{PropConj}$, and $\diamond\in \{\geql,\gle\}$, 
$\Cn=b_0\swedge\cdots\swedge b_k$, $b_j\in \mi{PropAtom}$; 
for all $j\leq k$, 
$b_j\in \mi{atoms}(\Cn)=\mi{atoms}(l)\subseteq \mi{atoms}(C^*-\{l^*\})\subseteq \mi{atoms}(S_2)=\mi{atoms}(S)$,
$\mi{guards}(S_2,b_j)=\mi{guards}(S,b_j)=\{\gz\gle b_j\}$;
$\mi{valid}(l,S_2)$;
if $C^*-\{l^*\}=\Cn\gle \gu\vee C^\natural\subseteq C^*$, $\Cn\in \mi{PropConj}$, and $\square\neq C^\natural\in \mi{OrdPropCl}$,
$\Cn\gle \gu\in C^*$, $C^*=(C^*-\{\Cn\gle \gu\})\cup \{\Cn\gle \gu\}=\Cn\gle \gu\vee (C^*-\{\Cn\gle \gu\})$,
$C^\natural\subseteq C^*$, $\Cn\gle \gu\not\in C^\natural$, $\square\neq C^\natural=C^\natural-\{\Cn\gle \gu\}\subseteq C^*-\{\Cn\gle \gu\}$; 
$C^*-\{\Cn\gle \gu\}$ does not contain an order literal of the form $\Cn'\gle \gu$, $\Cn'\in \mi{PropConj}$;
$C^\natural\subseteq C^*-\{\Cn\gle \gu\}$ does not contain an order literal of the form $\Cn'\gle \gu$, $\Cn'\in \mi{PropConj}$;
$\mi{valid}(C^*-\{l^*\},S_2)$;
$\Cn^*\gle \gu\neq \square$;
$\Cn^*\gle \gu=a_0\swedge\cdots\swedge a_n\gle \gu\in \mi{OrdPropLit}^\gu$;
for all $i\leq n$, $a_i\in \mi{atoms}(S_2)=\mi{atoms}(S)$, $\mi{guards}(S_2,a_i)=\mi{guards}(S,a_i)=\{\gz\gle a_i\}$;
$\mi{valid}(a_0\swedge\cdots\swedge a_n\gle \gu,S_2)$;
$\mi{valid}(\Cn^*\gle \gu,S_2)$;
for all $C\in (S-\{C^*\})-\mi{guards}(S)\subseteq S_2-\mi{guards}(S_2)$,
for all $l\in C$, 
$\mi{valid}(l,S)$ if and only if
$l\in \mi{OrdPropLit}^\gu$;
if either $l=\Cn\diamond \gu$ or $l=\gu\gleq \Cn$, $\Cn\in \mi{PropConj}$, and $\diamond\in \{\geql,\gle\}$, 
$\Cn=b_0\swedge\cdots\swedge b_k$, $b_j\in \mi{PropAtom}$;
for all $j\leq k$, 
$b_j\in \mi{atoms}(\Cn)=\mi{atoms}(l)\subseteq \mi{atoms}(C)\subseteq \mi{atoms}(S)=\mi{atoms}(S_2)$,
$\mi{guards}(S,b_j)=\mi{guards}(S_2,b_j)=\{\gz\gle b_j\}$ if and only if
$\mi{valid}(l,S_2)$;
$\mi{valid}(C,S)$ if and only if
$C\neq \square$; 
for all $l\in C$, $\mi{valid}(l,S)$;
if $C=\Cn\gle \gu\vee C^\natural$, $\Cn\in \mi{PropConj}$, and $\square\neq C^\natural\in \mi{OrdPropCl}$,
$C^\natural$ does not contain an order literal of the form $\Cn'\gle \gu$, $\Cn'\in \mi{PropConj}$, if and only if
$C\neq \square$; 
for all $l\in C$, $\mi{valid}(l,S_2)$;
if $C=\Cn\gle \gu\vee C^\natural$, $\Cn\in \mi{PropConj}$, and $\square\neq C^\natural\in \mi{OrdPropCl}$,
$C^\natural$ does not contain an order literal of the form $\Cn'\gle \gu$, $\Cn'\in \mi{PropConj}$, if and only if
$\mi{valid}(C,S_2)$;
$\mi{invalid}(S_2)=\{C \,|\, C\in S_2-\mi{guards}(S_2),\ \text{\it not}\ \mi{valid}(C,S_2)\}=
                   \{C \,|\, C\in ((S-\{C^*\})-\mi{guards}(S))\cup \{C^*-\{l^*\}\}\cup \{\Cn^*\gle \gu\},\ \text{\it not}\ \mi{valid}(C,S_2)\}=
                   \{C \,|\, C\in (S-\{C^*\})-\mi{guards}(S),\ \text{\it not}\ \mi{valid}(C,S_2)\}=\{C \,|\, C\in (S-\{C^*\})-\mi{guards}(S),\ \text{\it not}\ \mi{valid}(C,S)\}=
                   \mi{invalid}(S)=\{C \,|\, C\in S-\mi{guards}(S),\ \text{\it not}\ \mi{valid}(C,S)\}=\emptyset$;
$l^*\in C^*$, $\mi{count}(C^*)=\sum_{l\in C^*} \mi{count}(l)=\sum_{l\in (C^*-\{l^*\})\cup \{l^*\}} \mi{count}(l)=\mi{count}(C^*-\{l^*\})+1=(\sum_{l\in C^*-\{l^*\}} \mi{count}(l))+\mi{count}(l^*)$,
$C^*\in S-\mi{guards}(S)$, $((S-\{C^*\})-\mi{guards}(S))\cup \{C^*\}=((S-\{C^*\})-\mi{guards}(S))\cup (\{C^*\}-\mi{guards}(S))=((S-\{C^*\})\cup \{C^*\})-\mi{guards}(S)=S-\mi{guards}(S)$,
\begin{alignat*}{1}
\mi{count}(S_2) &= \sum_{C\in S_2-\mi{guards}(S_2)} \mi{count}(C)= \\
                &\phantom{\mbox{}=\mbox{}}
                   \sum_{C\in ((S-\{C^*\})-\mi{guards}(S))\cup \{C^*-\{l^*\}\}\cup \{\Cn^*\gle \gu\}} \mi{count}(C)\leq \\
                &\phantom{\mbox{}=\mbox{}}
                   \left(\sum_{C\in (S-\{C^*\})-\mi{guards}(S)} \mi{count}(C)\right)+\mi{count}(C^*-\{l^*\})+ \\
                &\phantom{\mbox{}=\mbox{}} \quad
                   \mi{count}(\Cn^*\gle \gu)< \\
                &\phantom{\mbox{}=\mbox{}}
                   \left(\sum_{C\in (S-\{C^*\})-\mi{guards}(S)} \mi{count}(C)\right)+\mi{count}(C^*-\{l^*\})+1= \\
                &\phantom{\mbox{}=\mbox{}}
                   \left(\sum_{C\in (S-\{C^*\})-\mi{guards}(S)} \mi{count}(C)\right)+\mi{count}(C^*)= \\
                &\phantom{\mbox{}=\mbox{}}
                   \sum_{C\in ((S-\{C^*\})-\mi{guards}(S))\cup \{C^*\}} \mi{count}(C)= \\
                &\phantom{\mbox{}=\mbox{}}
                   \mi{count}(S)=\sum_{C\in S-\mi{guards}(S)} \mi{count}(C);
\end{alignat*}
$\mi{measure}(S_2)=(\mi{atoms}(S_2),\mi{unsaturated}(S_2),\mi{invalid}(S_2),\mi{count}(S_2))=(\mi{atoms}(S),                                                                               \linebreak[4]
                                                                                                            \mi{unsaturated}(S),\mi{invalid}(S),\mi{count}(S_2))\prec 
                   \mi{measure}(S)=(\mi{atoms}(S),                                                                                                                                         \linebreak[4]
                                                  \mi{unsaturated}(S),\mi{invalid}(S),\mi{count}(S))$;
by the induction hypothesis for $S_2$, there exists a finite {\it DPLL}-tree $\mi{Tree}_2$ with the root $S_2$ constructed using Rules (\cref{ceq4hr1x}), (\cref{ceq4hr1111})--(\cref{ceq4hr66}), (\cref{ceq4hr8}) satisfying
for every leaf $S_2'$ that either $\square\in S_2'$ or $S_2'\subseteq_{\mc F} \mi{OrdPropCl}$ is positive.

We get in both Cases 2.2.2.1.2.2.1.1.1 and 2.2.2.1.2.2.1.1.2 that
there exists a finite {\it DPLL}-tree $\mi{Tree}_1$ with the root $S_1$ constructed using Rules (\cref{ceq4hr1x}), (\cref{ceq4hr1111})--(\cref{ceq4hr66}), (\cref{ceq4hr8}) satisfying
for every leaf $S_1'$ that either $\square\in S_1'$ or $S_1'\subseteq_{\mc F} \mi{OrdPropCl}$ is positive, and 
in all Cases
\settowidth{\llll}{2.2.2.1.2.2.1.2.1.2.3.2, and}
\!\!\parbox[t]{\llll}{
2.2.2.1.2.2.1.2.1.1.1, 
2.2.2.1.2.2.1.2.1.1.2,
2.2.2.1.2.2.1.2.1.1.3,
2.2.2.1.2.2.1.2.1.2.1,
2.2.2.1.2.2.1.2.1.2.2.1.1,
2.2.2.1.2.2.1.2.1.2.2.1.2,
2.2.2.1.2.2.1.2.1.2.2.2,
2.2.2.1.2.2.1.2.1.2.3.1.1,
2.2.2.1.2.2.1.2.1.2.3.1.2,
2.2.2.1.2.2.1.2.1.2.3.2, and
2.2.2.1.2.2.1.2.2,} \newline
there exists a finite {\it DPLL}-tree $\mi{Tree}_2$ with the root $S_2$ constructed using Rules (\cref{ceq4hr1x}), (\cref{ceq4hr1111})--(\cref{ceq4hr66}), (\cref{ceq4hr8}) satisfying
for every leaf $S_2'$ that either $\square\in S_2'$ or $S_2'\subseteq_{\mc F} \mi{OrdPropCl}$ is positive.
We further get in both Cases 2.2.2.1.2.2.2.1.1 and 2.2.2.1.2.2.2.1.2 that
there exists a finite {\it DPLL}-tree $\mi{Tree}_1$ with the root $S_1$ constructed using Rules (\cref{ceq4hr1x}), (\cref{ceq4hr1111})--(\cref{ceq4hr66}), (\cref{ceq4hr8}) satisfying
for every leaf $S_1'$ that either $\square\in S_1'$ or $S_1'\subseteq_{\mc F} \mi{OrdPropCl}$ is positive, and 
in all Cases
\settowidth{\llll}{2.2.2.1.2.2.2.2.1.3.2, and}
\!\!\parbox[t]{\llll}{
2.2.2.1.2.2.2.2.1.1.1,
2.2.2.1.2.2.2.2.1.1.2,
2.2.2.1.2.2.2.2.1.2.1.1,
2.2.2.1.2.2.2.2.1.2.1.2,
2.2.2.1.2.2.2.2.1.2.2,
2.2.2.1.2.2.2.2.1.3.1.1,
2.2.2.1.2.2.2.2.1.3.1.2,
2.2.2.1.2.2.2.2.1.3.2, and
2.2.2.1.2.2.2.2.2,} \newline
there exists a finite {\it DPLL}-tree $\mi{Tree}_2$ with the root $S_2$ constructed using Rules (\cref{ceq4hr1x}), (\cref{ceq4hr1111})--(\cref{ceq4hr66}), (\cref{ceq4hr8}) satisfying
for every leaf $S_2'$ that either $\square\in S_2'$ or $S_2'\subseteq_{\mc F} \mi{OrdPropCl}$ is positive.
We put
\begin{equation} \notag
\mi{Tree}=\dfrac{S}
                {\mi{Tree}_1\ \big|\ \mi{Tree}_2}.
\end{equation}
Hence, $\mi{Tree}$ is a finite {\it DPLL}-tree with the root $S$ constructed using Rules (\cref{ceq4hr1x}), (\cref{ceq4hr1111})--(\cref{ceq4hr66}), (\cref{ceq4hr8}) such that
for every leaf $S'$, either $\square\in S'$ or $S'\subseteq_{\mc F} \mi{OrdPropCl}$ is positive.
So, in Case 2.2.2.1.2.2, the statement holds.

Case 2.2.2.2:
$\emptyset\neq \mi{invalid}(S)\subseteq_{\mc F} \mi{OrdPropCl}$.
Then there exists $C^*\in \mi{invalid}(S)\subseteq S-\mi{guards}(S)$ not satisfying $\mi{valid}(C^*,S)$;
by Lemma \ref{le8}, there exists a finite linear {\it DPLL}-tree $\mi{Tree}'$ with the root $S$ constructed using Rules (\cref{ceq4hr1111})--(\cref{ceq4hr66}) satisfying
for its only leaf $S'$ that either $\square\in S'$, or $S'\subseteq_{\mc F} \mi{OrdPropCl}$ is semi-positively guarded, and exactly one of the following points holds.
\begin{enumerate}[\rm (a)]
\item
$S'=S$, $\mi{valid}(C^*,S)$;
\item
$S'=S-\{C^*\}$, $\mi{guards}(S')=\mi{guards}(S)$;
\item
there exists $C^{**}\in \mi{OrdPropCl}^\gu$, $\mi{atoms}(C^{**})\subseteq \mi{atoms}(C^*)$, satisfying 
$S'=(S-\{C^*\})\cup \{C^{**}\}$, $\mi{guards}(S')=\mi{guards}(S)$, $C^{**}\not\in S$, $\mi{valid}(C^{**},S')$;
\item
there exists $a^*\gle \gu\in \mi{OrdPropCl}^\gu$, $a^*\in \mi{atoms}(C^*)$, satisfying 
$S'=(S-\{C^*\})\cup \{a^*\gle \gu\}$, $\mi{guards}(S')=\mi{guards}(S)\cup \{a^*\gle \gu\}$, $a^*\gle \gu\not\in S$;
\item
there exists $S^{**}=\{a_0\geql \gu,\dots,a_n\geql \gu\}\subseteq \mi{OrdPropCl}^\gu$, $\{a_0,\dots,a_n\}\subseteq \mi{atoms}(C^*)$, $\{\gz\gle a_0,\dots,\gz\gle a_n\}\subseteq \mi{guards}(S)$, satisfying 
$S'=(S-(\{C^*\}\cup \{\gz\gle a_0,\dots,\gz\gle a_n\}))\cup S^{**}$, $\mi{guards}(S')=(\mi{guards}(S)-\{\gz\gle a_0,\dots,\gz\gle a_n\})\cup S^{**}$, $S^{**}\cap S=\emptyset$. 
\end{enumerate}
We get five cases for $S'$.

Case 2.2.2.2.1:
(a) $S'=S$ and $\mi{valid}(C^*,S)$.
This case is a contradiction with not $\mi{valid}(C^*,S)$.

Case 2.2.2.2.2:
(b) $S'=S-\{C^*\}$ and $\mi{guards}(S')=\mi{guards}(S)$.
We have that $S'$ is semi-positively guarded, and $S$ is positively guarded.
We get from Case 2.1.2.2.2 that $\mi{atoms}(S')=\mi{atoms}(S)$, $S'-\mi{guards}(S')\subseteq S-\mi{guards}(S)$;
for all $C\in S'-\mi{guards}(S')\subseteq S-\mi{guards}(S)$, $\mi{valid}(C,S)$ if and only if $\mi{valid}(C,S')$.
Then $\mi{unsaturated}(S')=\{a \,|\, a\in \mi{atoms}(S'), \mi{guards}(S',a)=\{\gz\gle a\}\}=\mi{unsaturated}(S)=\{a \,|\, a\in \mi{atoms}(S), \mi{guards}(S,a)=\{\gz\gle a\}\}$;
$C^*\not\in S'=S-\{C^*\}\supseteq S'-\mi{guards}(S')$, $C^*\in \mi{invalid}(S)$,
$C^*\not\in \mi{invalid}(S')=\{C \,|\, C\in S'-\mi{guards}(S'),\ \text{\it not}\ \mi{valid}(C,S')\}=\{C \,|\, C\in S'-\mi{guards}(S'),\ \text{\it not}\ \mi{valid}(C,S')\}-\{C^*\}=
                             \{C \,|\, C\in S'-\mi{guards}(S'),\ \text{\it not}\ \mi{valid}(C,S)\}-\{C^*\}\subseteq 
                             \mi{invalid}(S)-\{C^*\}=\{C \,|\, C\in S-\mi{guards}(S),\ \text{\it not}\ \mi{valid}(C,S)\}-\{C^*\}\subset \mi{invalid}(S)$;
$\mi{measure}(S')=(\mi{atoms}(S'),\mi{unsaturated}(S'),\mi{invalid}(S'),\mi{count}(S'))=(\mi{atoms}(S),\mi{unsaturated}(S),\mi{invalid}(S'),\mi{count}(S'))\prec 
                  \mi{measure}(S)=(\mi{atoms}(S),                                                                                                                                          \linebreak[4]
                                                 \mi{unsaturated}(S),\mi{invalid}(S),\mi{count}(S))$;
by the induction hypothesis for $S'$, there exists a finite {\it DPLL}-tree $\mi{Tree}''$ with the root $S'$ constructed using Rules (\cref{ceq4hr1x}), (\cref{ceq4hr1111})--(\cref{ceq4hr66}), (\cref{ceq4hr8}) satisfying
for every leaf $S''$ that either $\square\in S''$ or $S''\subseteq_{\mc F} \mi{OrdPropCl}$ is positive.
We put
\begin{equation} \notag
\mi{Tree}=\dfrac{\mi{Tree}'} 
                {\mi{Tree}''}.
\end{equation}
Hence, $\mi{Tree}$ is a finite {\it DPLL}-tree with the root $S$ constructed using Rules (\cref{ceq4hr1x}), (\cref{ceq4hr1111})--(\cref{ceq4hr66}), (\cref{ceq4hr8}) such that
for every leaf $S''$, either $\square\in S''$ or $S''\subseteq_{\mc F} \mi{OrdPropCl}$ is positive.

Case 2.2.2.2.3:
(c) There exists $C^{**}\in \mi{OrdPropCl}^\gu$, $\mi{atoms}(C^{**})\subseteq \mi{atoms}(C^*)$, such that
$S'=(S-\{C^*\})\cup \{C^{**}\}$, $\mi{guards}(S')=\mi{guards}(S)$, $C^{**}\not\in S$, $\mi{valid}(C^{**},S')$.
We have that $S'$ is semi-positively guarded, and $S$ is positively guarded.
We get from Case 2.1.2.2.3 that $\mi{atoms}(S')=\mi{atoms}(S)$, $(S-\{C^*\})-\mi{guards}(S')\subseteq S-\mi{guards}(S)$;
for all $C\in (S-\{C^*\})-\mi{guards}(S')\subseteq S-\mi{guards}(S)$, $\mi{valid}(C,S)$ if and only if $\mi{valid}(C,S')$.
Then $\mi{unsaturated}(S')=\{a \,|\, a\in \mi{atoms}(S'), \mi{guards}(S',a)=\{\gz\gle a\}\}=\mi{unsaturated}(S)=\{a \,|\, a\in \mi{atoms}(S), \mi{guards}(S,a)=\{\gz\gle a\}\}$;
$C^*\not\in S-\{C^*\}$,
$C^*\in S$, $C^*\neq C^{**}$,
$C^*\not\in S'=(S-\{C^*\})\cup \{C^{**}\}\supseteq S'-\mi{guards}(S')$,
$C^{**}\not\in S\supseteq \mi{guards}(S)=\mi{guards}(S')$, 
$((S-\{C^*\})\cup \{C^{**}\})-\mi{guards}(S')=((S-\{C^*\})-\mi{guards}(S'))\cup (\{C^{**}\}-\mi{guards}(S'))=((S-\{C^*\})-\mi{guards}(S'))\cup \{C^{**}\}$, $C^*\in \mi{invalid}(S)$,
$C^*\not\in \mi{invalid}(S')=\{C \,|\, C\in S'-\mi{guards}(S'),\ \text{\it not}\ \mi{valid}(C,S')\}=\{C \,|\, C\in S'-\mi{guards}(S'),\ \text{\it not}\ \mi{valid}(C,S')\}-\{C^*\}=
                             \{C \,|\, C\in ((S-\{C^*\})\cup \{C^{**}\})-\mi{guards}(S'),\ \text{\it not}\ \mi{valid}(C,S')\}-\{C^*\}=
                             \{C \,|\, C\in ((S-\{C^*\})-\mi{guards}(S'))\cup \{C^{**}\},\ \text{\it not}\ \mi{valid}(C,S')\}-\{C^*\}=
                             \{C \,|\, C\in (S-\{C^*\})-\mi{guards}(S'),\ \text{\it not}\ \mi{valid}(C,S')\}-\{C^*\}=
                             \{C \,|\, C\in (S-\{C^*\})-\mi{guards}(S'),\ \text{\it not}\ \mi{valid}(C,S)\}-\{C^*\}\subseteq 
                             \mi{invalid}(S)-\{C^*\}=\{C \,|\, C\in S-\mi{guards}(S),                                                                                                      \linebreak[4]
                                                                                      \text{\it not}\ \mi{valid}(C,S)\}-\{C^*\}\subset \mi{invalid}(S)$;
$\mi{measure}(S')=(\mi{atoms}(S'),\mi{unsaturated}(S'),                                                                                                                                    \linebreak[4]
                                                       \mi{invalid}(S'),\mi{count}(S'))=(\mi{atoms}(S),\mi{unsaturated}(S),\mi{invalid}(S'),\mi{count}(S'))\prec 
                  \mi{measure}(S)=(\mi{atoms}(S),\mi{unsaturated}(S),\mi{invalid}(S),\mi{count}(S))$;
by the induction hypothesis for $S'$, there exists a finite {\it DPLL}-tree $\mi{Tree}''$ with the root $S'$ constructed using Rules (\cref{ceq4hr1x}), (\cref{ceq4hr1111})--(\cref{ceq4hr66}), (\cref{ceq4hr8}) satisfying
for every leaf $S''$ that either $\square\in S''$ or $S''\subseteq_{\mc F} \mi{OrdPropCl}$ is positive.
We put
\begin{equation} \notag
\mi{Tree}=\dfrac{\mi{Tree}'} 
                {\mi{Tree}''}.
\end{equation}
Hence, $\mi{Tree}$ is a finite {\it DPLL}-tree with the root $S$ constructed using Rules (\cref{ceq4hr1x}), (\cref{ceq4hr1111})--(\cref{ceq4hr66}), (\cref{ceq4hr8}) such that
for every leaf $S''$, either $\square\in S''$ or $S''\subseteq_{\mc F} \mi{OrdPropCl}$ is positive.

Case 2.2.2.2.4:
(d) There exists $a^*\gle \gu\in \mi{OrdPropCl}^\gu$, $a^*\in \mi{atoms}(C^*)$, such that
$S'=(S-\{C^*\})\cup \{a^*\gle \gu\}$, $\mi{guards}(S')=\mi{guards}(S)\cup \{a^*\gle \gu\}$, $a^*\gle \gu\not\in S$.
We have that $S'$ is semi-positively guarded, and $S$ is positively guarded.
Then $C^*\in S$, $a^*\in \mi{atoms}(C^*)\subseteq \mi{atoms}(S)$,
$\mi{atoms}(S')=\mi{atoms}(\mi{guards}(S'))=\mi{atoms}(\mi{guards}(S)\cup \{a^*\gle \gu\})=\mi{atoms}(\mi{guards}(S))\cup \mi{atoms}(a^*\gle \gu)=\mi{atoms}(S)\cup \{a^*\}=\mi{atoms}(S)$;
$a^*$ is positively guarded in $S$;
either $\mi{guards}(S,a^*)=\{\gz\gle a^*\}$ or $\mi{guards}(S,a^*)=\{\gz\gle a^*,a^*\gle \gu\}$, 
$a^*\gle \gu\not\in S\supseteq \mi{guards}(S,a^*)$, $a^*\gle \gu\in \{\gz\gle a^*,a^*\gle \gu\}$, $\mi{guards}(S,a^*)\neq \{\gz\gle a^*,a^*\gle \gu\}$,
$\mi{guards}(S,a^*)=\{\gz\gle a^*\}$, $a^*\in \mi{unsaturated}(S)=\{a \,|\, a\in \mi{atoms}(S), \mi{guards}(S,a)=\{\gz\gle a\}\}$,
$a^*\in \mi{atoms}(S')$, 
$C^*\not\in \mi{guards}(S)\supseteq \mi{guards}(S,a^*)=S\cap \mi{guards}(a^*)$, $C^*\not\in \mi{guards}(a^*)$,
$\mi{guards}(S',a^*)=((S-\{C^*\})\cup \{a^*\gle \gu\})\cap \mi{guards}(a^*)=((S-\{C^*\})\cap \mi{guards}(a^*))\cup (\{a^*\gle \gu\}\cap \mi{guards}(a^*))=
                     ((S\cap \mi{guards}(a^*))-\{C^*\})\cup \{a^*\gle \gu\}=S\cap \mi{guards}(a^*)\cup \{a^*\gle \gu\}=\mi{guards}(S,a^*)\cup \{a^*\gle \gu\}=\{\gz\gle a^*,a^*\gle \gu\}\neq 
                     \{\gz\gle a^*\}$;
for all $a\in \mi{atoms}(S')-\{a^*\}=\mi{atoms}(S)-\{a^*\}$,
$a$ is positively guarded in $S$;
$a\neq a^*$, 
$\mi{guards}(a)\cap \{a^*\gle \gu\}=\mi{guards}(a)\cap \mi{guards}(a^*)=\emptyset$,
$C^*\not\in \mi{guards}(S)\supseteq \mi{guards}(S,a)=S\cap \mi{guards}(a)$, $C^*\not\in \mi{guards}(a)$,
$\mi{guards}(S',a)=((S-\{C^*\})\cup \{a^*\gle \gu\})\cap \mi{guards}(a)=((S-\{C^*\})\cap \mi{guards}(a))\cup (\{a^*\gle \gu\}\cap \mi{guards}(a))=(S\cap \mi{guards}(a))-\{C^*\}=S\cap \mi{guards}(a)=
                   \mi{guards}(S,a)$;
$a^*\not\in \mi{unsaturated}(S')=\{a \,|\, a\in \mi{atoms}(S'), \mi{guards}(S',a)=\{\gz\gle a\}\}=\{a \,|\, a\in \mi{atoms}(S'), \mi{guards}(S',a)=\{\gz\gle a\}\}-\{a^*\}=
                                 \{a \,|\, a\in \mi{atoms}(S')-\{a^*\}, \mi{guards}(S',a)=\{\gz\gle a\}\}=\{a \,|\, a\in \mi{atoms}(S)-\{a^*\}, \mi{guards}(S,a)=\{\gz\gle a\}\}=
                                 \mi{unsaturated}(S)-\{a^*\}=\{a \,|\, a\in \mi{atoms}(S), \mi{guards}(S,a)=\{\gz\gle a\}\}-\{a^*\}\subset \mi{unsaturated}(S)$, 
$\mi{measure}(S')=(\mi{atoms}(S'),\mi{unsaturated}(S'),\mi{invalid}(S'),\mi{count}(S'))=(\mi{atoms}(S),\mi{unsaturated}(S'),                                                               \linebreak[4]
                                                                                                                            \mi{invalid}(S'),\mi{count}(S'))\prec 
                  \mi{measure}(S)=(\mi{atoms}(S),\mi{unsaturated}(S),\mi{invalid}(S),                                                                                                      \linebreak[4]
                                                                                     \mi{count}(S))$;
by the induction hypothesis for $S'$, there exists a finite {\it DPLL}-tree $\mi{Tree}''$ with the root $S'$ constructed using Rules (\cref{ceq4hr1x}), (\cref{ceq4hr1111})--(\cref{ceq4hr66}), (\cref{ceq4hr8}) satisfying
for every leaf $S''$ that either $\square\in S''$ or $S''\subseteq_{\mc F} \mi{OrdPropCl}$ is positive.
We put
\begin{equation} \notag
\mi{Tree}=\dfrac{\mi{Tree}'} 
                {\mi{Tree}''}.
\end{equation}
Hence, $\mi{Tree}$ is a finite {\it DPLL}-tree with the root $S$ constructed using Rules (\cref{ceq4hr1x}), (\cref{ceq4hr1111})--(\cref{ceq4hr66}), (\cref{ceq4hr8}) such that
for every leaf $S''$, either $\square\in S''$ or $S''\subseteq_{\mc F} \mi{OrdPropCl}$ is positive.

Case 2.2.2.2.5:
(e) There exists $S^{**}=\{a_0\geql \gu,\dots,a_n\geql \gu\}\subseteq \mi{OrdPropCl}^\gu$, $\{a_0,\dots,a_n\}\subseteq \mi{atoms}(C^*)$, $\{\gz\gle a_0,\dots,\gz\gle a_n\}\subseteq \mi{guards}(S)$, such that
$S'=(S-(\{C^*\}\cup \{\gz\gle a_0,\dots,\gz\gle a_n\}))\cup S^{**}$, $\mi{guards}(S')=(\mi{guards}(S)-\{\gz\gle a_0,\dots,\gz\gle a_n\})\cup S^{**}$, $S^{**}\cap S=\emptyset$.
This case is the same as Case 2.1.2.2.5.

So, in both Cases 1 and 2, the statement holds.
The induction is completed.
%
%
%
\end{proof}

\subsection{Full proof of Theorem \ref{T2}}
\label{S7.12}

\begin{proof}
Let $S^F\subseteq \mi{OrdPropCl}$, $\mi{Tree}^F$ be a finite {\it DPLL}-tree with the root $S^F$; $V^F$ be a subset of the finite set of its leaves; 
$S^v\subseteq \mi{OrdPropCl}$, $v\in V^F$, be the label order clausal theory for $v$; $\mi{Tree}^v$, $v\in V^F$, be a finite {\it DPLL}-tree with the root $S^v$. 
By $\mi{Tree}^F[v/\mi{Tree}^v,v\in V^F]$ we denote the labelled rooted (arborescence) tree with the root $S^F$ built up from $\mi{Tree}^F$ by replacing every $v\in V^F$ with $\mi{Tree}^v$.
Note that $\mi{Tree}^F[v/\mi{Tree}^v,v\in V^F]$ is a finite {\it DPLL}-tree with the root $S^F$.
We get that
by Lemma \ref{le55}, there exists a finite linear {\it DPLL}-tree $\mi{Tree}_1^\circ$ with the root $S$ constructed using Rules (\cref{ceq4hr2}) and (\cref{ceq4hr22}) satisfying 
for its only leaf $S_1$ that either $\square\in S_1$ or $S_1\subseteq_{\mc F} \mi{OrdPropCl}$ is simplified;
if $\square\not\in S_1$, 
$S_1\subseteq_{\mc F} \mi{OrdPropCl}$ is simplified;
by Lemma \ref{le5} for $S_1$, there exists a finite {\it DPLL}-tree $\mi{Tree}_2^\circ$ with the root $S_1$ constructed using Rules (\cref{ceq4hr1}), (\cref{ceq4hr1111111})--(\cref{ceq4hr4}) satisfying 
for every leaf $S_2$ that 
either $\square\in S_2$ or $S_2\subseteq_{\mc F} \mi{OrdPropCl}$ is $\gz$-guarded;
if $\square\not\in S_2$,
$S_2\subseteq_{\mc F} \mi{OrdPropCl}$ is $\gz$-guarded;
by Lemma \ref{le7} for $S_2$, there exists a finite linear {\it DPLL}-tree $\mi{Tree}_3^\circ$ with the root $S_2$ constructed using Rules (\cref{ceq4hr1111111})--(\cref{ceq4hr4}), (\cref{ceq4hr7}), (\cref{ceq4hr8}) satisfying
for its only leaf $S_3$ that either $\square\in S_3$ or $S_3\subseteq_{\mc F} \mi{OrdPropCl}$ is positively guarded; 
if $\square\not\in S_3$,
$S_3\subseteq_{\mc F} \mi{OrdPropCl}$ is positively guarded;
by Lemma \ref{le9} for $S_3$, there exists a finite {\it DPLL}-tree $\mi{Tree}_4^\circ$ with the root $S_3$ constructed using Rules (\cref{ceq4hr1x}), (\cref{ceq4hr1111})--(\cref{ceq4hr66}), (\cref{ceq4hr8}) satisfying
for every leaf $S_4$ that 
either $\square\in S_4$ or $S_4\subseteq_{\mc F} \mi{OrdPropCl}$ is positive;
if $\square\not\in S_4$,
$S_4\subseteq_{\mc F} \mi{OrdPropCl}$ is positive; 
by Lemma \ref{le4} for $S_4$, there exists a finite {\it DPLL}-tree $\mi{Tree}_5^\circ$ with the root $S_4$ constructed using Rules (\cref{ceq4hr0}), (\cref{ceq4hr11}), (\cref{ceq4hr111}), (\cref{ceq4hr66}) 
with the following properties:
\begin{alignat}{1}
\label{eq555a}
& \text{if}\ S_4\ \text{is unsatisfiable, then}\ \mi{Tree}_5^\circ\ \text{is closed}; \\[1mm]
\label{eq555b}
& \text{if}\ S_4\ \text{is satisfiable, then}\ \mi{Tree}_5^\circ\ \text{is open, and} \\
\notag
& \phantom{\text{if}\ S_4\ \text{is satisfiable, then}\ \mbox{}}
                                               \text{there exists a model}\ {\mf A}_4\ \text{of}\ S_4\ \text{related to}\ \mi{Tree}_5^\circ.
\end{alignat}
We put,
\begin{IEEEeqnarray*}{RL}
\IEEEeqnarraymulticol{2}{l}{
\text{\it if}\ S_1\ \text{\it is simplified},} \\
\IEEEeqnarraymulticol{2}{l}{
\text{\it for every leaf}\ S_2\ \text{\it of}\ \mi{Tree}_2^\circ\ \text{\it which is $\gz$-guarded},} \\
\IEEEeqnarraymulticol{2}{l}{
\phantom{\text{\it for every\ \mbox{}}}
                    \mi{Tree}_3^\circ\ \text{\it with the root}\ S_2\ \text{\it and its only leaf}\ S_3,\, \text{\it and if}\ S_3\ \text{\it is positively guarded},} \\
\IEEEeqnarraymulticol{2}{l}{
\phantom{\text{\it for every\ \mbox{}}}
                    \mi{Tree}_4^\circ\ \text{\it with the root}\ S_3,} \\[1mm]
\mi{Tree}_4^\bullet &= \mi{Tree}_4^\circ[S_4/\mi{Tree}_5^\circ,\ S_4\ \text{\it is a leaf of}\ \mi{Tree}_4^\circ\ \text{\it which is positive}]; \\[1mm] 
\mi{Tree}_3^\bullet &= \left\{\begin{array}{ll}
                              \mi{Tree}_3^\circ           &\ \ \text{\it if}\ \square\in S_3, \\[1mm]
                              \dfrac{\mi{Tree}_3^\circ}
                                    {\mi{Tree}_4^\bullet} &\ \ \text{\it if}\ S_3\ \text{\it is positively guarded}; 
                              \end{array} 
                       \right. \\[1mm]
\mi{Tree}_2^\bullet &= \mi{Tree}_2^\circ[S_2/\mi{Tree}_3^\bullet,\ S_2\ \text{\it is a leaf of}\ \mi{Tree}_2^\circ\ \text{\it which is $\gz$-guarded}]; \\[1mm]
\mi{Tree}           &= \left\{\begin{array}{ll}
                              \mi{Tree}_1^\circ           &\ \ \text{\it if}\ \square\in S_1, \\[1mm]
                              \dfrac{\mi{Tree}_1^\circ}
                                    {\mi{Tree}_2^\bullet} &\ \ \text{\it if}\ S_1\ \text{\it is simplified}. 
                              \end{array} 
                       \right. 
\end{IEEEeqnarray*}
Note that if $S_1$ is simplified,
for every leaf $S_2$ of $\mi{Tree}_2^\circ$ which is $\gz$-guarded, $\mi{Tree}_3^\circ$ with the root $S_2$ and its only leaf $S_3$, and if $S_3$ is positively guarded,
$\mi{Tree}_4^\circ$ with the root $S_3$,
$\mi{Tree}_4^\bullet$ is a finite {\it DPLL}-tree with the root $S_3$ constructed using Rules (\cref{ceq4hr0}), (\cref{ceq4hr1x})--(\cref{ceq4hr66}), (\cref{ceq4hr8});
$\mi{Tree}_3^\bullet$ is a finite {\it DPLL}-tree with the root $S_2$ constructed using Rules (\cref{ceq4hr0}), (\cref{ceq4hr1x})--(\cref{ceq4hr8});
$\mi{Tree}_2^\bullet$ is a finite {\it DPLL}-tree with the root $S_1$ constructed using Rules (\cref{ceq4hr0})--(\cref{ceq4hr8});
$\mi{Tree}$ is a finite {\it DPLL}-tree with the root $S$ constructed using Rules (\cref{ceq4hr0})--(\cref{ceq4hr8}).
We distinguish two cases for $S$.

Case 1:
$S$ is unsatisfiable.
We have that either $\square\in S_1$ or $S_1\subseteq_{\mc F} \mi{OrdPropCl}$ is simplified.
We get two cases for $S_1$.

Case 1.1:
$\square\in S_1$.
Then $\mi{Tree}=\mi{Tree}_1^\circ$ with its only leaf $S_1$ is closed;
(\ref{eq55a}) holds and (\ref{eq55b}) holds trivially.

Case 1.2:
$S_1\subseteq_{\mc F} \mi{OrdPropCl}$ is simplified.
Then $\mi{Tree}=\dfrac{\mi{Tree}_1^\circ}
                      {\mi{Tree}_2^\bullet}$;
by Lemma \ref{le33}(ii) for $\mi{Tree}_1^\circ$ and $S_1$, $S_1$ is unsatisfiable;
for every leaf $S_2$ of $\mi{Tree}_2^\circ$,
we have that either $\square\in S_2$ or $S_2\subseteq_{\mc F} \mi{OrdPropCl}$ is $\gz$-guarded;
if $S_2\subseteq_{\mc F} \mi{OrdPropCl}$ is $\gz$-guarded,
by Lemma \ref{le33}(ii) for $S_1$, $\mi{Tree}_2^\circ$, and $S_2$, $S_2$ is unsatisfiable;
for $\mi{Tree}_3^\circ$ with the root $S_2$ and its only leaf $S_3$, we have that either $\square\in S_3$ or $S_3\subseteq_{\mc F} \mi{OrdPropCl}$ is positively guarded.
We get two cases for $S_3$.

Case 1.2.1:
$\square\in S_3$.
Then $\mi{Tree}_3^\bullet=\mi{Tree}_3^\circ$ with the root $S_2$ and its only leaf $S_3$ is closed.

Case 1.2.2:
$S_3\subseteq_{\mc F} \mi{OrdPropCl}$ is positively guarded.
Then $\mi{Tree}_3^\bullet=\dfrac{\mi{Tree}_3^\circ}
                                {\mi{Tree}_4^\bullet}$ with the root $S_2$;
by Lemma \ref{le33}(ii) for $S_2$, $\mi{Tree}_3^\circ$, and $S_3$, $S_3$ is unsatisfiable;
for $\mi{Tree}_4^\circ$ with the root $S_3$ and its every leaf $S_4$,
we have that either $\square\in S_4$ or $S_4\subseteq_{\mc F} \mi{OrdPropCl}$ is positive;
if $S_4\subseteq_{\mc F} \mi{OrdPropCl}$ is positive,
by Lemma \ref{le33}(ii) for $S_3$, $\mi{Tree}_4^\circ$, and $S_4$, $S_4$ is unsatisfiable;
by (\ref{eq555a}), $\mi{Tree}_5^\circ$ with the root $S_4$ is closed;
$\mi{Tree}_4^\bullet=\mi{Tree}_4^\circ[S_4/\mi{Tree}_5^\circ,\ S_4\ \text{\it is a leaf of}\ \mi{Tree}_4^\circ\ \text{\it which is positive}]$ with the root $S_3$ is closed;
$\mi{Tree}_3^\bullet=\dfrac{\mi{Tree}_3^\circ}
                           {\mi{Tree}_4^\bullet}$ with the root $S_2$ is closed.

We get in both Cases 1.2.1 and 1.2.2 that 
for every leaf $S_2$ of $\mi{Tree}_2^\circ$,
we have that either $\square\in S_2$ or $S_2\subseteq_{\mc F} \mi{OrdPropCl}$ is $\gz$-guarded;
if $S_2\subseteq_{\mc F} \mi{OrdPropCl}$ is $\gz$-guarded,
$\mi{Tree}_3^\bullet$ with the root $S_2$ is closed;
$\mi{Tree}_2^\bullet=\mi{Tree}_2^\circ[S_2/\mi{Tree}_3^\bullet,\ S_2\ \text{\it is a leaf of}\ \mi{Tree}_2^\circ\ \text{\it which is $\gz$-guarded}]$ is closed;
$\mi{Tree}=\dfrac{\mi{Tree}_1^\circ}
                 {\mi{Tree}_2^\bullet}$ is closed;
(\ref{eq55a}) holds and (\ref{eq55b}) holds trivially.

Case 2:
$S$ is satisfiable.
We have that either $\square\in S_1$ or $S_1\subseteq_{\mc F} \mi{OrdPropCl}$ is simplified.
Then, by Lemma \ref{le33}(i) for $\mi{Tree}_1^\circ$ with its only leaf $S_1$ and $S_1$, $S_1$ is satisfiable;
$\square$ is not satisfiable;
$\square\not\in S_1$;
$S_1\subseteq_{\mc F} \mi{OrdPropCl}$ is simplified;
$\mi{Tree}=\dfrac{\mi{Tree}_1^\circ}
                 {\mi{Tree}_2^\bullet}$;
by Lemma \ref{le33}(i) for $S_1$ and $\mi{Tree}_2^\circ$, there exists a leaf $S_2^*$ of $\mi{Tree}_2^\circ$ satisfying that $S_2^*$ is satisfiable;
either $\square\in S_2^*$ or $S_2^*\subseteq_{\mc F} \mi{OrdPropCl}$ is $\gz$-guarded;
$\square\not\in S_2^*$;
$S_2^*\subseteq_{\mc F} \mi{OrdPropCl}$ is $\gz$-guarded;
there exists a finite linear {\it DPLL}-tree ${\mi{Tree}_3^\circ}^*$ with the root $S_2^*$ constructed using Rules (\cref{ceq4hr1111111})--(\cref{ceq4hr4}), (\cref{ceq4hr7}), (\cref{ceq4hr8}) satisfying
for its only leaf $S_3^*$ that either $\square\in S_3^*$ or $S_3^*\subseteq_{\mc F} \mi{OrdPropCl}$ is positively guarded;
by Lemma \ref{le33}(i) for $S_2^*$, ${\mi{Tree}_3^\circ}^*$ with its only leaf $S_3^*$, and $S_3^*$, $S_3^*$ is satisfiable;
$\square\not\in S_3^*$;
$S_3^*\subseteq_{\mc F} \mi{OrdPropCl}$ is positively guarded;
there exists a finite {\it DPLL}-tree ${\mi{Tree}_4^\circ}^*$ with the root $S_3^*$ constructed using Rules (\cref{ceq4hr1x}), (\cref{ceq4hr1111})--(\cref{ceq4hr66}), (\cref{ceq4hr8}) satisfying
for every leaf $S_4$ that either $\square\in S_4$ or $S_4\subseteq_{\mc F} \mi{OrdPropCl}$ is positive;
${\mi{Tree}_4^\bullet}^*={\mi{Tree}_4^\circ}^*[S_4/\mi{Tree}_5^\circ,\ S_4\ \text{\it is a leaf of}\ {\mi{Tree}_4^\circ}^*\ \text{\it which is positive}]$ with the root $S_3^*$,
${\mi{Tree}_3^\bullet}^*=\dfrac{{\mi{Tree}_3^\circ}^*}
                               {{\mi{Tree}_4^\bullet}^*}$ with the root $S_2^*$;
by Lemma \ref{le33}(i) for $S_3^*$ and ${\mi{Tree}_4^\circ}^*$, there exists a leaf $S_4^*$ of ${\mi{Tree}_4^\circ}^*$ satisfying that $S_4^*$ is satisfiable;
either $\square\in S_4^*$ or $S_4^*\subseteq_{\mc F} \mi{OrdPropCl}$ is positive;
$\square\not\in S_4^*$;
$S_4^*\subseteq_{\mc F} \mi{OrdPropCl}$ is positive;
there exists a finite {\it DPLL}-tree ${\mi{Tree}_5^\circ}^*$ with the root $S_4^*$ constructed using Rules (\cref{ceq4hr0}), (\cref{ceq4hr11}), (\cref{ceq4hr111}), (\cref{ceq4hr66}) satisfying
by (\ref{eq555b}) for $S_4^*$ and ${\mi{Tree}_5^\circ}^*$ that ${\mi{Tree}_5^\circ}^*$ is open, and there exists a model ${\mf A}_4^*$ of $S_4^*$ related to ${\mi{Tree}_5^\circ}^*$; 
${\mi{Tree}_4^\bullet}^*={\mi{Tree}_4^\circ}^*[S_4/\mi{Tree}_5^\circ,\ S_4\ \text{\it is a leaf of}\ {\mi{Tree}_4^\circ}^*\ \text{\it which is positive}]$ is open;
by Lemma \ref{le33}(ii) for $S_3^*$, ${\mi{Tree}_4^\circ}^*$, $S_4^*$, and ${\mf A}_4^*$, there exists a model ${\mf A}_3^*$ of $S_3^*$ 
related to ${\mi{Tree}_4^\circ}^*$ and ${\mi{Tree}_4^\bullet}^*={\mi{Tree}_4^\circ}^*[S_4/\mi{Tree}_5^\circ,\ S_4\ \text{\it is a leaf of}\ {\mi{Tree}_4^\circ}^*\ \text{\it which is positive}]$;
${\mi{Tree}_3^\bullet}^*=\dfrac{{\mi{Tree}_3^\circ}^*}
                               {{\mi{Tree}_4^\bullet}^*}$ is open;
by Lemma \ref{le33}(ii) for $S_2^*$, ${\mi{Tree}_3^\circ}^*$, $S_3^*$, and ${\mf A}_3^*$, there exists a model ${\mf A}_2^*$ of $S_2^*$ 
related to ${\mi{Tree}_3^\circ}^*$ and ${\mi{Tree}_3^\bullet}^*=\dfrac{{\mi{Tree}_3^\circ}^*}
                                                                      {{\mi{Tree}_4^\bullet}^*}$;
$\mi{Tree}_2^\bullet=\mi{Tree}_2^\circ[S_2/\mi{Tree}_3^\bullet,\ S_2\ \text{\it is a leaf of}\ \mi{Tree}_2^\circ\ \text{\it which is $\gz$-guarded}]$ is open;
by Lemma \ref{le33}(ii) for $S_1$, $\mi{Tree}_2^\circ$, $S_2^*$, and ${\mf A}_2^*$, there exists a model ${\mf A}_1$ of $S_1$ 
related to $\mi{Tree}_2^\circ$ and $\mi{Tree}_2^\bullet=\mi{Tree}_2^\circ[S_2/\mi{Tree}_3^\bullet,\ S_2\ \text{\it is a leaf of}\ \mi{Tree}_2^\circ\ \text{\it which is $\gz$-guarded}]$;
$\mi{Tree}=\dfrac{\mi{Tree}_1^\circ}
                 {\mi{Tree}_2^\bullet}$ is open;
by Lemma \ref{le33}(ii) for $\mi{Tree}_1^\circ$, $S_1$, and ${\mf A}_1$, there exists a model ${\mf A}$ of $S$ related to $\mi{Tree}_1^\circ$ and $\mi{Tree}=\dfrac{\mi{Tree}_1^\circ}
                                                                                                                                                                   {\mi{Tree}_2^\bullet}$;
(\ref{eq55a}) holds trivially and (\ref{eq55b}) holds.

So, in both Cases 1 and 2, (\ref{eq55a}) and (\ref{eq55b}) hold.
%
%
%
\end{proof}

\end{document}